\pdfoutput=1
\documentclass[12pt]{report}
\usepackage{amsmath} 
\usepackage{amssymb} 
\usepackage{fbe_tez}
\usepackage{lscape}
\usepackage{graphicx}
\usepackage[bookmarks=true,bookmarksnumbered=true]{hyperref}
\hypersetup{pdfborder = {0 0 1},urlbordercolor = {0 1 1},colorlinks = false}
\usepackage{color}

\usepackage{pstricks}

\newcommand{\pt}{$p_{T}$~} 
\newcommand{\ymax}{$|$y$|_{max}$~}
\newcommand{\capspace}{\vspace{5.2718mm}}
\newcommand{\pbinv}{pb$^{-1}$}
\newcommand{\proton}{proton-proton~}
\newcommand{\lumi}{cm$^{-2}$s$^{-1}$}
\newcommand{\s}{$\sqrt{s}$}
\newcommand{\dslash}{{\not} \partial} 

\newenvironment{dedication}
{
   \cleardoublepage
   \thispagestyle{empty}
   \vspace*{\stretch{1}}
   \hfill\begin{minipage}[t]{0.66\textwidth}
   \raggedright
}%
{
   \end{minipage}
   \vspace*{\stretch{3}}
   \clearpage
}
\title{MEASUREMENT OF THE DIFFERENTIAL DIJET PRODUCTION CROSS SECTION IN PROTON-PROTON COLLISIONS AT \s = 7 TeV }
\turkcebaslik{$\sqrt{s}$ = 7 TeV'DEK\.{I} PROTON-PROTON \c{C}ARPI\c{S}MALARINDA JET \c{C}\.{I}FT\.{I} OLU\c{S}UMUNUN D\.{I}FERANS\.{I}YEL TES\.{I}R KES\.{I}T\.{I} \"{O}L\c{C}\"{U}M\"{U} }
\degree{B.S., Physics, Bo\u{g}azi\c{c}i University, 2005 }
\author{Bora I\c{s}{\i}ldak}
\program{Physics}
\subyear{2011}
\supervisor{Prof. Erhan G\"{u}lmez}

\examineri{Prof. Metin Ar{\i}k}
\examinerii{Assoc. Prof. Kerem Canko\c{c}ak}
\examineriii{Prof. Serkant Ali \c{C}etin }
\examineriv{Prof. Osman Teoman Turgut}
\dateofapproval{7.12.2011}

\begin{document}
\pagenumbering{roman}
%
% Choose the relevant one
%\makemstitle        % For M.S. theses
\makephdtitle      % For Ph.D. theses
%\makeproposaltitle % For Proposals

\makeapprovalpage
\pagebreak
\begin{dedication}
\textit{To my dearest wife Ceren and beautiful daughter Birce,}
\end{dedication}
\pagebreak
\begin{acknowledgements}
I would like to express my deeply-felt gratitude to my thesis advisor Prof. Erhan G\"{u}lmez for his guidance and endless support throughout my studies. His well-known patience and prudence eased my graduate life. 

I would also like to thank Dr. Konstantinos Kousouris who has shared his profound knowledge with me. This thesis would be impossible without his help and wisdom. He taught me all aspects of a high energy experimental physics analysis. His coherent teaching and wide intelligence always pushed me forward in the past two years.

I am also grateful to Dr. Niki Saoulidou for her invaluable support during my thesis writing process. I am also happy to work with  Dr. Klaus Rabbertz who gave his time and consideration for my studies.

I want to express my thanks to both Prof. Ya\c{s}ar \"{O}nel and Prof. Mithat Kaya who played important roles in my
life at CERN. They were very encouraging and sincere. I specially thank Serhat I\c{s}t{\i}n, Cemile Ezer, \"{O}zlem
Kaya and Kaz{\i}m \c{C}aml{\i}bel for their friendships.

I specially thank my mother and father \c{C}i\u{g}dem, Mehmet I\c{s}{\i}ldak, my grandmother Melahat Elma, my sister
Banu and brother Bar{\i}\c{s} for their support throughout my life. They have always been behind me in every decision
I have made. I sincerely thank my family-in-law, G\"{u}l\"{u}mser, Vecihi and Yi\u{g}it Yavuz. Without their
encouragement, it would be hard to pursue my graduate studies abroad. They have always helped me to think positively
whenever I met with obstacles.

I gratefully acknowledge the Bogazici University Research Fund (No: 09B302P and 5883) for their kind support in this work.

\end{acknowledgements}

\begin{abstract}

A measurement of the double-differential inclusive dijet production cross section in proton-proton collisions at $\sqrt{s}=$ 7 TeV is presented as a function of the dijet invariant mass and jet rapidity. The data correspond to an integrated luminosity of 36 \pbinv, recorded with the CMS detector at the LHC in 2010. The measurement covers the dijet mass range 0.2 TeV to 3.5 TeV and jet rapidities up to $|y|=2.5$. It is found to be in good agreement with next-to-leading-order QCD predictions.
\end{abstract}
%
% The usage of the "foreword" and "preface" environments are similar
% the "abstract" and "acknowledgements". See FBE manual for the
% correct order of these pages in the thesis.
%
\begin{ozet}
7 TeV k\"{u}tle merkezi enerjisinde \c{c}arp{\i}\c{s}an protonlardan meydana gelen \c{c}ift jetlerin olu\c{s}um tesir kesiti, jet \c{c}iftininin de\u{g}i\c{s}mez k\"{u}tlesi ve jet rapiditesine g\"{o}re \"{o}l\c{c}\"{u}lm\"{u}\c{s}t\"{u}r. 36 \pbinv'l{\i}k toplam luminositeye denk gelen veri 2010 y{\i}l{\i}nda  CMS detekt\"{o}r\"{u} ile al{\i}nm{\i}\c{s}t{\i}r. \"{O}l\c{c}\"{u}m, jet \c{c}ifti de\u{g}i\c{s}mez k\"{u}tlesinde 0.2 TeV'den 3.5 TeV'e, jet rapiditelerinde ise $|y|$=2.5'a kadar olan de\u{g}erleri kapsamaktad{\i}r. Sonu\c{c} olarak \"{o}l\c{c}\"{u}m ve kuantum kromodinamik \"{o}ng\"{o}r\"{u}lerin tutarl{\i} oldu\u{g}u g\"{o}zlenmi\c{s}tir.
\end{ozet}

%    * \u{g} – ğ
%    * \u{G} – Ğ
%    * \c{c} – ç
%    * \c{C} – Ç
%    * \c{s} – ş
%    * \c{S} – Ş
%    * \”{u} – ü
%    * \”{U} – Ü
%    * \”{o} – ö
%    * \”{O} – Ö
%    * {\i} – ı
\tableofcontents

\listoffigures 

\listoftables 

\begin{symabbreviations}
\sym{APD}{Avalanche PhotoDiode}
\sym{ATLAS}{A Toroidal LHC ApparatuS}
\sym{CERN}{ (Conseil Europ\'{e}en pour la Recherche Nucl\'{e}aire)}
\sym{CMS}{Compact Muon Solenoid}
\sym{CP Violation}{Charge-Parity Violation}
\sym{CSC}{Cathode Strip Chambers }
\sym{CTEQ}{Co-ordinated Theoretical-Experimental project on QCD}
\sym{DAQ}{Data AcQusition}
\sym{DBS}{Dataset Bookkeeping System}
\sym{DIS}{Deep Inelastic Scattering}
\sym{DTC}{Drift Tube Chamber}
\sym{ECAL}{Electromagnetic CALorimeter}
\sym{HB}{Hadronic Barrel}
\sym{HCAL}{Hadronic CALorimeter}
\sym{HE}{Hadronic Endcap}
\sym{HF}{Hadronic Forward}
\sym{HLT}{High Level Trigger }
\sym{HO}{Hadronic Outer}
\sym{HSM}{Standard Model Higgs boson}
\sym{IC}{Iterative Cone}
\sym{IRC}{InfraRed and Collinear}
\sym{JES}{Jet Energy Scale}
\sym{LEP}{Large Electron Positron Collider}
\sym{LHC}{Large Hadron Collider}
\sym{LO}{Leading Order}
\sym{LSP}{Lightest Supersymmetric Particle}
\sym{MC}{Monte Carlo}
\sym{MET}{Missing Transverse Energy}
\sym{MPI}{Multiple Parton Interaction}
\sym{NLO}{Next-to-Leading Order}
\sym{NLO}{Next-to-Next-to-Leading Order}
\sym{PDF}{Parton Distribution Function}
\sym{pQCD}{perturbative Quantum Chromodynamics}
\sym{QCD}{Quantum ChromoDynamics}
\sym{QED}{Quantum ElectroDynamics}
\sym{RF}{Radio Frequency}
\sym{RPC}{Resistive Plate Chambers }
\sym{SM}{Standard Model}
\sym{SUSY}{SUperSYmmetry}
\sym{TEC}{Tracker Endcap Discs}
\sym{TIB}{Tracker Inner Barrel}
\sym{TID}{Tracker Inner Discs}
\sym{TOB}{Tracker Outer Barrel}
\sym{VPT}{Vacuum PhotoTriode}
\end{symabbreviations}

\chapter{INTRODUCTION}
Particle physics is the discipline which seeks the ultimate answers of these two joint questions: 
\begin{itemize}
\item What are the most fundamental constituents of matter?

\item How do these constituents interact with each other?
\end{itemize}

In ancient Greek, Democritus of Abdera stated that everything is composed of ``\textit{atomos}" which means ``uncuttable"
in old Greek. Nevertheless, after Democritus' bright idea, it had been almost 2000 years for humankind to reveal all matter is a composition of generic and fundamental constituents. In 1932, there were four known particles, three
of which constitute an atom. However, it did not take so much for this list to grow. Today, our knowledge
about fundamental particles and their interactions is far beyond 1930s'. Needless to say, this does not mean that
a complete and consistent theory which answers all questions is accomplished. There are many open questions to be
answered and many theories to be confirmed. The acknowledged method for giving satisfactory answers and for performing
reliable tests in high energy particle physics is to collide particles and observe the outcome.

Apart from the questions to be answered, the main goal of this thesis is to give a detailed description of
an analysis performed in Compact Muon Solenoid (CMS) which is one of the four experiments of the Large Hadron
Collider (LHC) physics program at CERN (\textit{Conseil Europ\'{e}en pour la Recherche Nucl\'{e}aire}). In this particular analysis, the predictions of the Quantum Chromodynamics (QCD); theory about the fundamental constituents of nuclei (partons and gluon), were tested. 

In QCD, outgoing scattered partons from the parton-parton scattering manifest themselves as
hadronic jets. Hence, events with two high transverse momentum jets (dijets) arise in proton-proton collisions.
The invariant mass $M_{JJ}^{2}$ of the two jets can be given in terms of proton momentum fractions $x_{1,2}$
carried by the scattering partons as follows;
\begin{equation}
M_{JJ}^{2}=x_1 \cdot x_2 \cdot s
\end{equation}
where $\sqrt{s}$ is the center-of-mass energy of the colliding protons. The dijet cross section as a function of
$M_{JJ}^{2}$ can be precisely calculated in perturbative QCD and it also allows sensitive searches for physics beyond
the Standard Model, such as dijet resonances or contact interactions. The measured cross-section challenges the QCD
predictions at a new collision energy and in an unexplored kinematic regime,  beyond the reach of previous collider
measurements \cite{D0_dijet,ATLAS_dijet}. So far, dedicated searches for dijet resonances and contact interactions with the CMS
detector have been reported in several articles \cite{dijet_resonance,dijet_angular_dist}. In this thesis, the
measurement of the double-differential inclusive dijet production cross section is described as a function of the dijet invariant
mass and jet rapidity at $\sqrt{s}$ = 7 TeV. The work explained in this dissertation was published \cite{DijetPaper} and presented at the several conferences \cite{LHC_poster_session, TROIA11}.

In the subsequent chapter, the standard model of the particle physics and the theory of the fundamental constituents
of nuclei, Quantum Chromodynamics, will be described. In the third chapter, the specifications of the
accelerator machine LHC and the detector CMS will be explained. The goal of the fourth chapter is to give a description
of both Monte Carlo programs used in this analysis, and the reconstruction algorithms of the jet objects which
are sprays of particles coming from a single parton will be discussed. The remaining chapters are dedicated to explain
the work carried out to perform analysis and the results obtained, respectively.

\pagenumbering{arabic}

\chapter{THEORY}

\section{Standard Model}
The ultimate objective of the particle physics is to give a prescription of all the phenomena related to the fundamental particles. The so called \textit{Standard Model of the Particle Physics} (SM) is the most extensive theory of particle properties and particle interactions.

\begin{figure}[ht]
  \centering
  \includegraphics[width=0.48\textwidth]{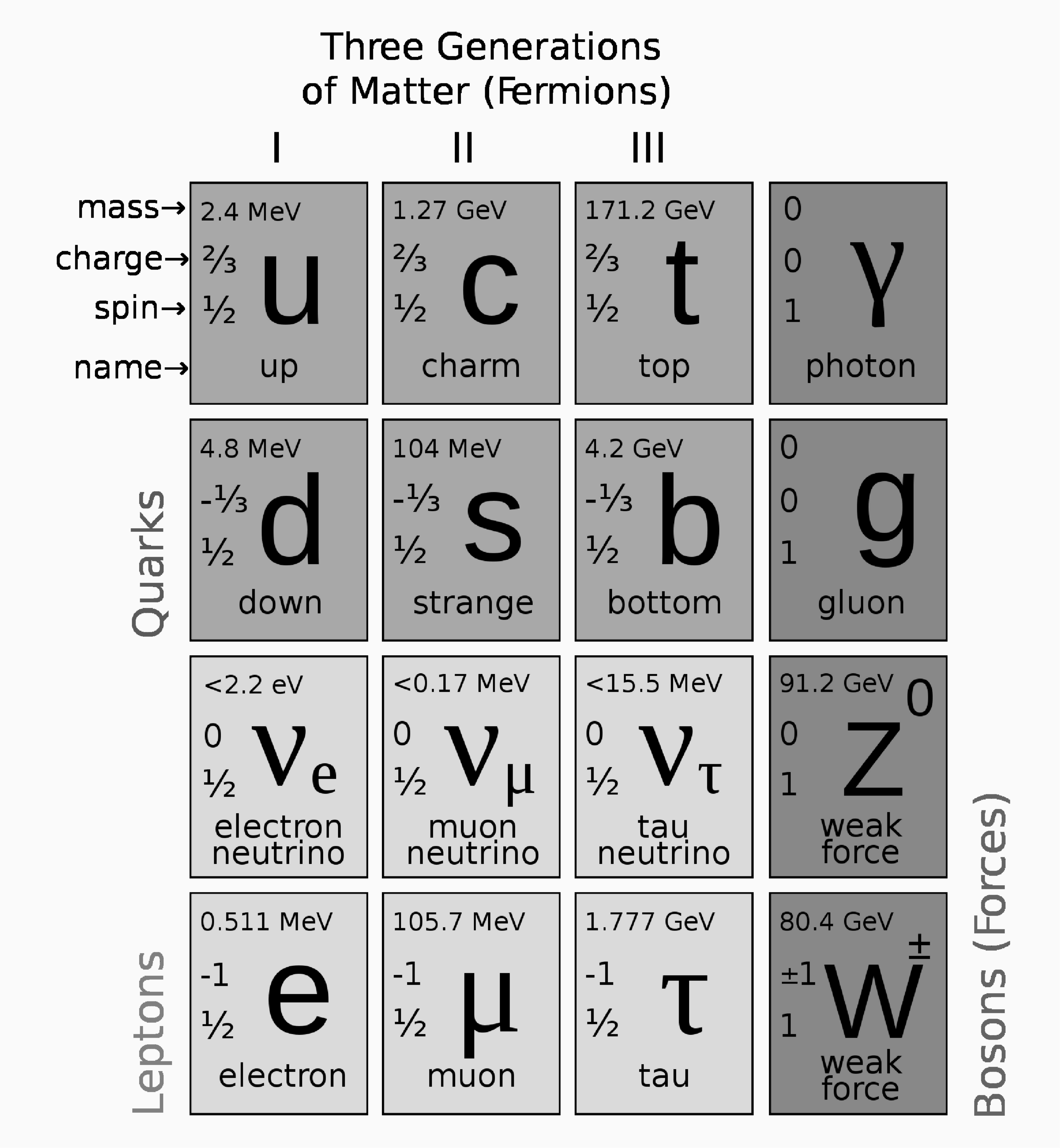}
  \includegraphics[width=0.48\textwidth]{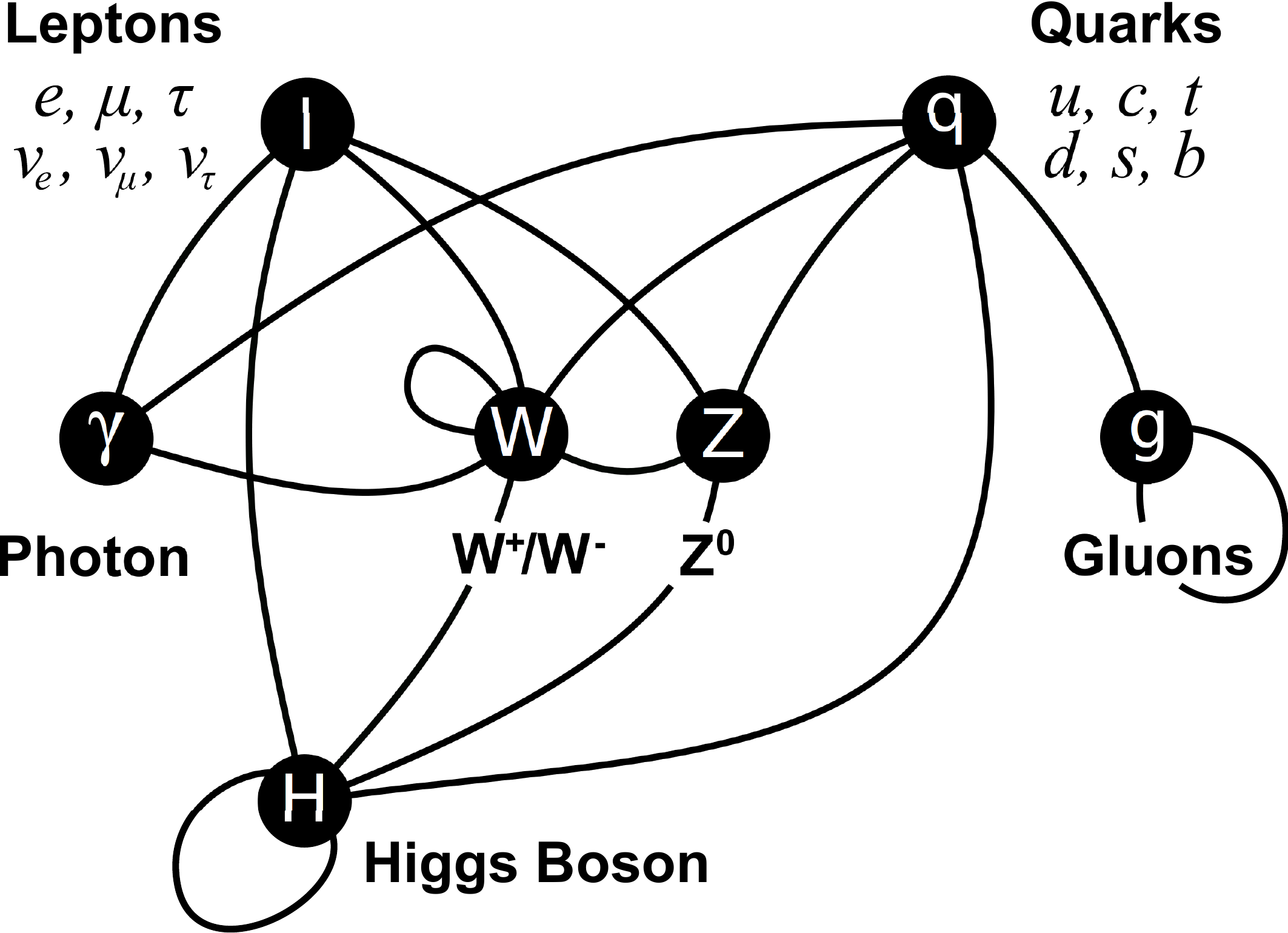}
  \capspace
  \caption{The Standard Model of elementary particles, with the gauge bosons in the rightmost column (left). Interactions between fundamental particles (right).}
  \label{fig:StdModelParticles}
\end{figure}

SM consists of quarks, leptons and force carrying bosons. For each quark and lepton, there is a corresponding antiparticle. According to the SM, all phenomena of particles can be explained by these fundamental particles along with the appropriate quantum field theory. There are three types of interactions in SM. In fact, two of them are two different aspects of a single type of interaction. The electromagnetic interaction between two charged particles is explained by photon exchange. However, the electromagnetic theory which uses electric and magnetic fields is just an effective theory of infinitely many photon exchanging charged particles. The weak force or weak interaction is propagated by W$^{\pm}$ and Z$^{0}$ bosons. In 1970's, a unified description of the electromagnetic and the weak interaction was given by Abdus Salam, Sheldon Glashow and Steven Weinberg. The other interaction which is known to be the force holding the nucleus together is the strong interaction mediated by the gluons. Hence the Standard Model consists of two parts; the electroweak and strong interactions. A visual summary of SM and the type of interactions are shown in Figure \ref{fig:StdModelParticles}.

\section{Quantum Chromodynamics}\label{Quantum Chromodynamics}
Quantum Chromodynamics (QCD) is the non-Abelian Yang-Mills theory of quarks and gluons. In quantum field theory, all the story begins with the necessity of the gauge invariance, especially the local gauge invariance.
It was first proposed in 1964 by Oscar W. Greenberg that quarks, if they exist and constitutes the proton, should have an additional quantum property to rescue the Pauli exclusion principle. This property, then, was named with colors (red, green, blue) referring to the three different quantum states of a given quark. It is like the electric charge where it is a single property which is represented by positive or negative values. The three colors are just names, and they have nothing to do with the colors we see in our daily lives. They are just labels to differentiate the three different quantum states of a given quark.

If quarks are fermionic particles with spin $1/2$, they must obey the Dirac equation. However, a quark field which can be described as a Dirac spinor comes with three color states. Thus, we can write the quark field $\psi$ as a three component column vector.
\begin{equation}
\vec{\psi}=\left( 
\begin{array}{c}
\psi_{1} \\ \psi_{2} \\ \psi_{3}
\end{array}\right) 
\end{equation}
where each element is a 4-component spinor.

Then, the Lagrangian density can be written as;
\begin{equation}
\boldsymbol{\mathcal{L}}=\overline{\psi} \left[i \hbar c \boldsymbol{\dslash_{\mu}}-\boldsymbol{M}c^2 \right] \psi
\end{equation}

where
\begin{equation}
\overline{\psi}= \left(\psi_1^{\dagger}\gamma^{0},~\psi_2^{\dagger}\gamma^{0},~\psi_3^{\dagger}\gamma^{0} \right) \\
\end{equation}
\begin{equation}
\boldsymbol{\dslash_{\mu}} =\left( \begin{array}{ccc}
\gamma^{\mu}\partial_{\mu} & 0 & 0 \\
0 & \gamma^{\mu}\partial_{\mu} & 0 \\
0 & 0 & \gamma^{\mu}\partial_{\mu}
\end{array} \right) 
\end{equation}
\begin{equation}
\boldsymbol{M} =\left( \begin{array}{ccc}
m & 0 & 0 \\
0 & m & 0 \\
0 & 0 & m
\end{array} \right) 
\end{equation}
assuming the three different color states of a given quark have the same mass.

Now, the argument is that the Lagrangian density for the quark field must be invariant under a unitary transformation.
\begin{equation}
\psi^{~\prime}= \boldsymbol{U}\psi
\end{equation}
where $\boldsymbol{U}$ is a 3$\times$3 matrix in color case. A unitary operator (a 3 by 3 matrix in this case) can be expressed as;
\begin{equation}
\boldsymbol{U}=e^{i\boldsymbol{H}}
\end{equation} 
where $\boldsymbol{H}$ is a 3$\times$3 Hermitian matrix. Moreover, any Hermitian matrix can be written as a linear combination of nine 3$\times$3 matrices;
\begin{equation}
\boldsymbol{H}=a_0\boldsymbol{I}+\vec{a}\cdot\boldsymbol{\vec{\lambda}}
\end{equation}
Here, $\vec{a}\cdot\boldsymbol{\vec{\lambda}}$ is a shorthand notation for $\sum^8_{i=1}a_i\lambda_i$, where $\lambda_i$ represents the so called Gell-Mann matrices which is a representation of the infinitesimal generators of the SU(3) group. They obey the following commutation relation;
\begin{equation}
\left[\lambda_i , \lambda_j \right]=2if^{ijk}\lambda_k
\end{equation}
where $f^{ijk}$ are the structure constants of the SU(3) group, and antisymmetric in indices with the following non-vanishing values;
\begin{equation}
f^{123}=1,~~f^{147}=f^{165}=f^{246}=f^{257}=f^{345}=f^{376}=\dfrac{1}{2},~~f^{458}=f^{678}=\dfrac{\sqrt{3}}{2}
\end{equation}

Thus, the unitary transformation of the field boils down to a phase transformation of the following form;
\begin{equation}
\label{eqn::SU(3)phase_tranformation}
\psi^{~\prime}= e^{ia_{0}} \cdot e^{i\vec{a}\cdot\boldsymbol{\vec{\lambda}}}\psi
\end{equation}
However, it can easily be argued that the phase transformation may change from one observer to another in the spirit of relativity. In fact, there is no reason not to make the components of the $\vec{a}$ space-time dependent. Hence Equation \ref{eqn::SU(3)phase_tranformation} can be written in a more general form;
\begin{equation}
\label{eqn::SU(3)phase_tranformation_more_general}
\psi^{~\prime}= e^{ia_{0}(x^{\mu})} \cdot e^{-iq~\vec{a}(x^{\mu})\cdot\boldsymbol{\vec{\lambda}}/\hbar c}\psi
\end{equation}
Where the factor $-q/\hbar c$ is introduced for future convenience. The first part is just the 3$\times$3 matrix representation of the U(1) group, and we know that all the electromagnetic theory for a Dirac particle can be generated from this symmetry by requiring a local gauge invariance. The second part of the transformation, $ e^{-iq~\vec{a}(x^{\mu})\cdot\boldsymbol{\vec{\lambda}}/\hbar c}$, is the non-Abelian part of the theory because of the non-commuting $\lambda$ matrices.
Again, it should be required that the SU(3) transformation, $\boldsymbol{S}=$ $e^{-iq~\vec{a}(x^{\mu})\cdot\boldsymbol{\vec{\lambda}}/\hbar c}$, must leave the Lagrangian density invariant.
\begin{eqnarray*}
&\psi^{~\prime}= \boldsymbol{S} \psi\\
&\overline{\psi}^{~\prime}=\overline{\psi} \boldsymbol{S}^{-1}
\end{eqnarray*}
Therefore, the transformed Lagrangian density becomes;
\begin{eqnarray}
\boldsymbol{\mathcal{L}}^{~\prime}=\overline{\psi}\boldsymbol{S}^{-1} \left[i \hbar c \gamma^{\mu}\boldsymbol{\partial_{\mu}}-\boldsymbol{M}c^2 \right]\boldsymbol{S} \psi \\
=\overline{\psi}\boldsymbol{S}^{-1} \left[i \hbar c \gamma^{\mu}(\boldsymbol{S}(\boldsymbol{\partial_{\mu}}\psi)+(\boldsymbol{\partial_{\mu}}\boldsymbol{S})\psi)-\boldsymbol{M}c^2\boldsymbol{S} \psi \right]
\end{eqnarray}
As it can easily be seen, $\partial_{\mu}(\boldsymbol{S}\psi)\neq\boldsymbol{S}(\partial_{\mu}\psi)$, hence, there is an extra term coming from the derivative of $\boldsymbol{S}$. By looking at the structure of the equation 2.14, the covariant derivative can be introduced as follows;
\begin{equation}
D_{\mu}=\partial_{\mu}+i\dfrac{q}{\hbar c}\vec{\lambda}\cdot A_{\mu}
\end{equation}
The $A_{\mu}$ part changes with the SU(3) transformation such that the covariant derivative satisfies the identity;
\begin{equation}
D_{\mu}^{~\prime}(\boldsymbol{S}\psi)=\boldsymbol{S}(D_{\mu}\psi)
\end{equation}
The prime symbol on $D_{\mu}$ denotes the transformation; $A_{\mu}\rightarrow A_{\mu}^{~\prime}$. However, the transformation of $A_{\mu}$ is not trivial, nevertheless, it can be deduced from the identity 2.16 as follows;
\begin{equation}
\vec{\lambda}\cdot A_{\mu}^{~\prime}=\boldsymbol{S}(\vec{\lambda}\cdot A_{\mu})\boldsymbol{S}^{-1}+i(\dfrac{q}{\hbar c})(\partial_{\mu}\boldsymbol{S})\boldsymbol{S}^{-1}
\end{equation}
In order to bring $\boldsymbol{S}$ and $\boldsymbol{S}^{-1}$ terms together, the commutator $[\boldsymbol{S},\vec{\lambda}\cdot A_{\mu}]$ should be evaluated. For the simplicity, an expansion of $\boldsymbol{S}$ for small values of $\vec{a}(x^{\mu})$ will suffice.
\begin{equation}
\boldsymbol{S}\cong 1-\dfrac{iq}{\hbar c} \vec{\lambda} \cdot \vec{a}(x^{\mu}) ,~~~~~~ \boldsymbol{S}^{-1}\cong 1+\dfrac{iq}{\hbar c}\vec{\lambda}\cdot \vec{a}(x^{\mu}),~~~~~~\partial_{\mu} \boldsymbol{S}\cong -\dfrac{iq}{\hbar c}\vec{\lambda}\cdot(\partial_{\mu} \vec{a}(x^{\mu}))
\end{equation}
With this approximation, Equation 2.17 becomes;
\begin{equation}
\vec{\lambda}\cdot A_{\mu}^{~\prime}\cong \vec{\lambda}\cdot A_{\mu}-\dfrac{iq}{\hbar c}[ \vec{\lambda} \cdot \vec{a}(x^{\mu}),\vec{\lambda}\cdot A_{\mu}]+\vec{\lambda}\cdot (\partial_{\mu}\vec{a}(x^{\mu}))
\end{equation}
By using the commutation relation of the $\lambda$ matrices (Equation 2.9), $[ \vec{\lambda} \cdot \vec{a}(x^{\mu}),\vec{\lambda}\cdot A_{\mu}]$ term can be found as;
\begin{equation}
[ \vec{\lambda} \cdot \vec{a}(x^{\mu}),\vec{\lambda}\cdot A_{\mu}]=2i\vec{\lambda}(\vec{a}(x^{\mu})\times A_{\mu})
\end{equation}
Therefore, the transformation of the vector field $A_{\mu}$ is given as;
\begin{equation}
A_{\mu}^{~\prime}\cong A_{\mu}+\dfrac{2q}{\hbar c}(\vec{a}(x^{\mu})\times A_{\mu})+ \partial_{\mu}\vec{a}(x^{\mu})
\end{equation}
where $(\vec{a}(x^{\mu})\times A_{\mu})_{i}=\sum_{j,k}^8 if_{ijk}\vec{a}_j(x^{\mu})(A_{\mu})_k $.

The new modified Lagrangian;
\begin{equation}
\boldsymbol{\mathcal{L}}= \left[i \hbar c \overline{\psi} \gamma^{\mu}\boldsymbol{\partial_{\mu}}\psi-\boldsymbol{M}c^2  \overline{\psi}\psi \right]-\left( q \overline{\psi} \gamma^{\mu}\lambda \psi \right)\cdot \boldsymbol{A_{\mu}}
\end{equation}
which is invariant under SU(3) gauge transformation. The cost of such a gauge invariance requirement is to introduce the gauge fields ($\boldsymbol{A_{\mu}}$) that correspond to the gluons in physics point of view. Now, the free gluon Lagrangian density must be added to the Lagrangian density for completeness. The free gluon Lagrangian is given as follows;
\begin{equation}
\boldsymbol{\mathcal{L}}_{gluons}=-\dfrac{1}{16\pi}\boldsymbol{F_{\mu \nu}}\boldsymbol{F^{\mu \nu}}
\end{equation}
where $\boldsymbol{F^{\mu \nu}}$ is given as;
\begin{equation}
\boldsymbol{F^{\mu \nu}}=\partial^{\mu}\boldsymbol{A^{\nu}}-\partial^{\nu}\boldsymbol{A^{\mu}}+\dfrac{2q}{\hbar c}(\boldsymbol{A^{\mu}}\times \boldsymbol{A^{\nu}})
\end{equation}
with the cross product defined as in Equation 2.21.
In order to define the gluon propagator, the choice of gauge must be fixed. This gauge fixing procedure pronounces itself as two terms in the Lagrangian density. One is the so called ``\textit{gauge fixing}" term, and the other is the ``\textit{Fadeev-Popov ghosts}" term. The gauge fixing term is given as a class of gauges named ``$\boldsymbol{R_{\xi}}$ \textit{gauges}" which is the generalization of the Lorenz gauge, and it is expressed as;
\begin{equation}
\boldsymbol{\mathcal{L}}_{gauge-fixing}=-\dfrac{1}{2\xi}(\partial_{\mu}\boldsymbol{A^{\mu}})^2
\end{equation}

In a non-abelian theory, there is also need for the ghost field Lagrangian density in the form of;
\begin{equation}
\boldsymbol{\mathcal{L}}_{ghost}=-\eta^{a \dagger} \partial_{\mu} \boldsymbol{D}_{\mu}^{ab}\eta^{b}
\end{equation}
where $\eta^{a}$ is a complex scalar field.\\

Altogether, the full QCD Lagrangian density is given as the sum of classical and the gauge-fixing part with the ghost field terms.
\begin{equation}
\boldsymbol{\mathcal{L}}_{QCD}=\boldsymbol{\mathcal{L}}_{classical}+\boldsymbol{\mathcal{L}}_{gauge-fixing}+\boldsymbol{\mathcal{L}}_{ghost}
\end{equation}
Then, the appropriate Feynman rules for QCD can be derived from this Lagrangian density \cite{Quantum Chromodynamics}.
\subsection{The Running Coupling Constant}
\label{Running Coupling}
For fundamental particles, the cross section of a specified process is given by
\begin{equation}
\sigma=\int \dfrac{1}{flux}~|\mathcal{M}^{2}|~d\Phi
\end{equation}
where, $\sigma$ is the cross section, \textit{flux} is the incoming particle flux, $\mathcal{M}$ is the matrix element for a given process calculated from the Feynman diagrams and $d\Phi$ is the phase space volume for the process. For example, in an electromagnetic interaction, a factor of $\alpha=1/137$ (coupling constant) is being introduced for each vertex in the relevant Feynman diagram while calculating the matrix element $\mathcal{M}$. This means that higher order diagrams with more vertices contribute less and less to the cross section. However, in QCD, the coupling constant, from the force between two protons, has been experimentally found out to be greater than one which has catastrophic consequences. If the coupling constant is greater than one, the more complicated Feynman diagrams with more vertices contribute more and more to the cross section which is the physical observable. 

In QED, a diagram which includes a loop diagram,
\begin{figure}[ht]
\centering
\includegraphics[width=0.50\textwidth]{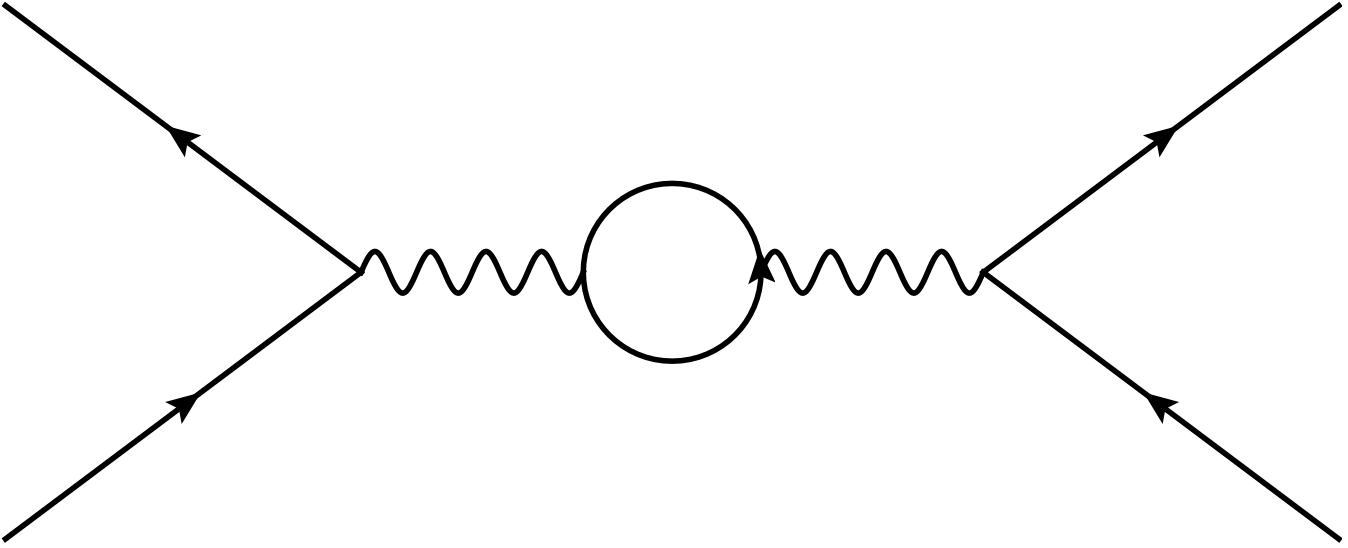}
\capspace
\caption{Electron-positron bubble diagram which creates a vacuum polarization in QED.}
\label{vacuum polarization diagram in QED}
\end{figure}

makes the effective charge of the electron ($e=\sqrt{4\pi\alpha}$);
\begin{equation}
\alpha(|q^{2}|)=\alpha(0)\lbrace 1+\dfrac{\alpha(0)}{3 \pi} \ln( |q^{2}|/(mc)^{2}) \rbrace
\end{equation}

Hence, higher order corrections coming from bubble-chain diagrams can be explicitly written as;
\begin{eqnarray}
\alpha(|q^{2}|)&=&\alpha(0)\left(1+\dfrac{\alpha(0)}{3 \pi} \ln( |q^{2}|/(mc)^{2}+\dfrac{\alpha^{2}(0)}{(3 \pi)^{2}} \ln^{2}( |q^{2}|/(mc)^{2}+\ldots \right) \\
\nonumber \\
&=&\dfrac{\alpha(0)}{1-(\alpha(0)/3\pi)\ln( |q^{2}|/(mc)^{2})} 
\end{eqnarray} 

In QCD case, there are also gluon-gluon couplings which lead to the gluon bubbles.
\begin{figure}[ht]
\centering
\includegraphics[width=0.40\textwidth]{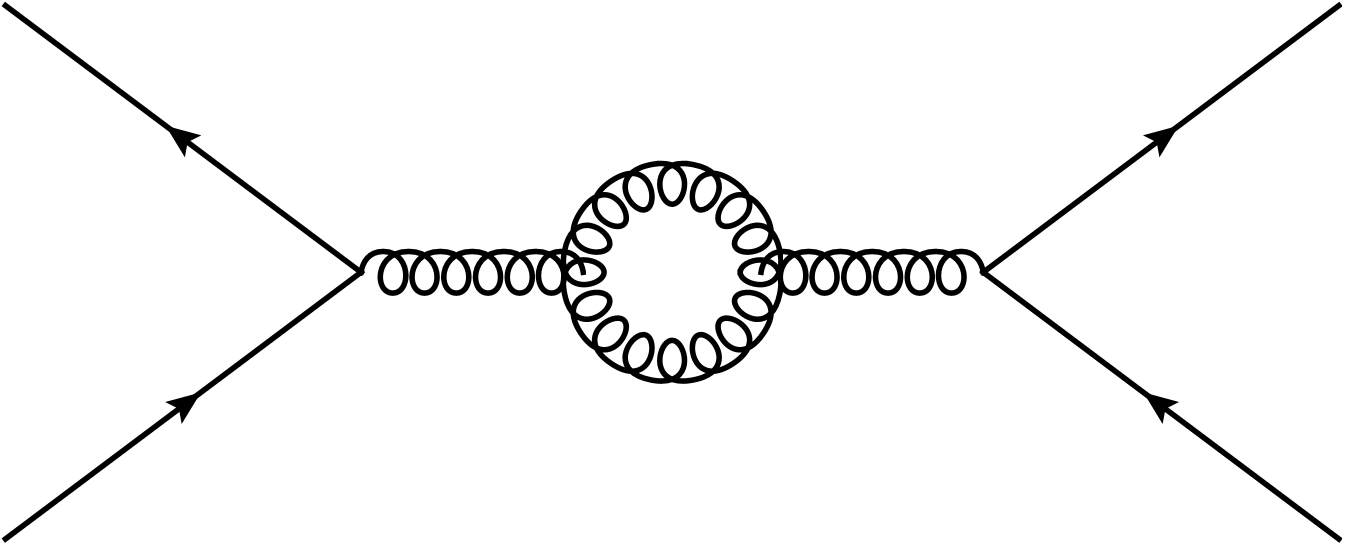}
\hspace{10mm}
\includegraphics[width=0.40\textwidth]{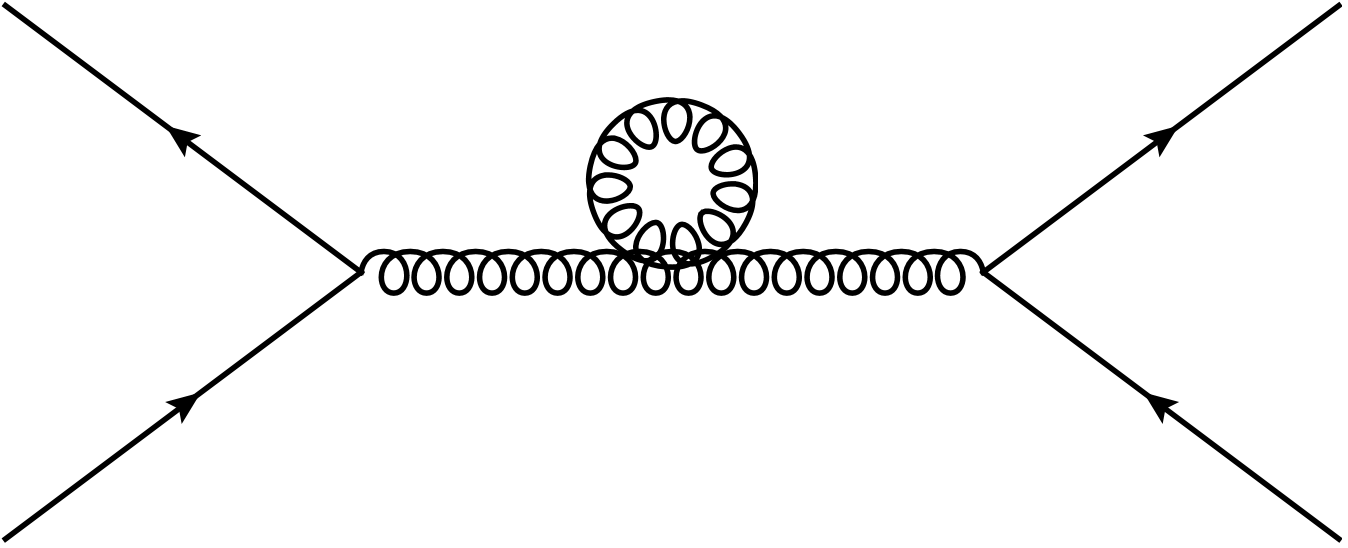}
\capspace
\caption{Gluon-gluon bubble diagram in QCD.}
\label{vacuum polarization diagram in QCD}
\end{figure}

If the renormalization equations are solved in the \textit{leading order} (LO), it can be found that the strong coupling depends on the momentum transfer as follows;
\begin{equation}
\alpha_{s}(|q^{2}|)=\dfrac{\alpha_{s}(\mu^{2})}{1+\alpha_{s}(\mu^{2})~b~\ln(|q^{2}|/\mu^{2})}
\label{running_alpha}
\end{equation}
where $b$ is $(11n_{c}-2n_{f})/12\pi$, $n_{c}$ is the number of colors, $n_{f}$ is the number of quark flavors.\\

As opposed to QED case, the gluon bubble contribution creates an anti-screening effect since the \textit{b} term in Equation \ref{running_alpha} is positive with $n_{c}=n_{f}=3$ in the Standard Model. Also, there is a new parameter $\mu$ in the equation. In QED, it is natural to use the coupling strength for the fully screened charge where $q^{2}=0$. This is the charge we have already known from Coulomb and Milikan. In QCD, it is not possible to start from $q^{2}=0$ since it is where $\alpha_{s}$ is large. Therefore, a point where $\alpha_{s}$ is small enough to perform perturbative expansion should be chosen. Given that $\mu^{2}$ is large enough to satisfy $\alpha(\mu^{2})<1$, it does not matter which $\mu$ value is used in Equation \ref{running_alpha} (See Appendix \ref{AppendixA}). Hence, it is appropriate to define a new parameter $\Lambda$ to write Equation \ref{running_alpha} in terms of a single parameter. \\

If the new parameter $\Lambda$ is defined as follows;
\begin{equation}
\ln\Lambda^{2}=\ln\mu^{2}-12\pi/\left[(11n_{c}-2n_{f})\alpha_{s}(\mu^{2})\right]
\end{equation}
Equation \ref{running_alpha} becomes
\begin{equation}
\alpha_{s}(|q^{2}|)=\dfrac{12 \pi}{(11n_{c}-2n_{f})\ln(|q^{2}|/\Lambda^{2})}
\label{running_alpha_modified}
\end{equation}
 Equation \ref{running_alpha_modified} shows the $q^{2}$ dependence of $\alpha_{s}$. For large values of $q^{2}$, $\alpha_{s}$ gets smaller which means that the ``strong" force becomes relatively weak at short distances. This is the essence of the \textit{asymptotic freedom}. If $q^{2}$ decreases, the $\alpha_{s}$ value increases. A small $q^{2}$ means that the interaction occurred over a relatively large distance. Thus, the force between two colored particles becomes much stronger. This effect is known as \textit{color confinement} and it is the reason for the colored particles not to be able to be observed individually above a distance of $1/\Lambda_{QCD}$. 
\subsection{Structure of the Proton and Parton Distribution Functions}
In the simplistic approach, protons are made of two up quarks and a down quark. Partons carry a certain fraction of the momentum of the proton, characterized by parton distribution functions (PDFs).  However, the first measurements of PDFs revealed that the total momentum of the three quarks (``valence quarks") is only about 35\% of the momentum of the proton. It was understood that the remaining fraction of the total momentum comes from gluons (~50\%), while  15\% of the momentum comes from the ``sea quarks". These are pairs of quarks and antiquarks, that can pop in and out of the vacuum because of the interactions between two gluons. Figure \ref{inner_struct_proton} shows a cartoon of the proton structure, as much as it is currently understood.

Each of these partons inside the proton carries a fraction of total momentum given by the PDF. PDF is defined as the  probability density for finding a particle within a longitudinal momentum fraction interval at a momentum transfer of $Q^{2}$. The function $f_i(x_i)$ is the probability for a parton to carry a fraction of the hadron between $x_i$ and $x_i+\delta x_i$.

\begin{figure}[ht]
\centering
\includegraphics[width=0.25\textwidth]{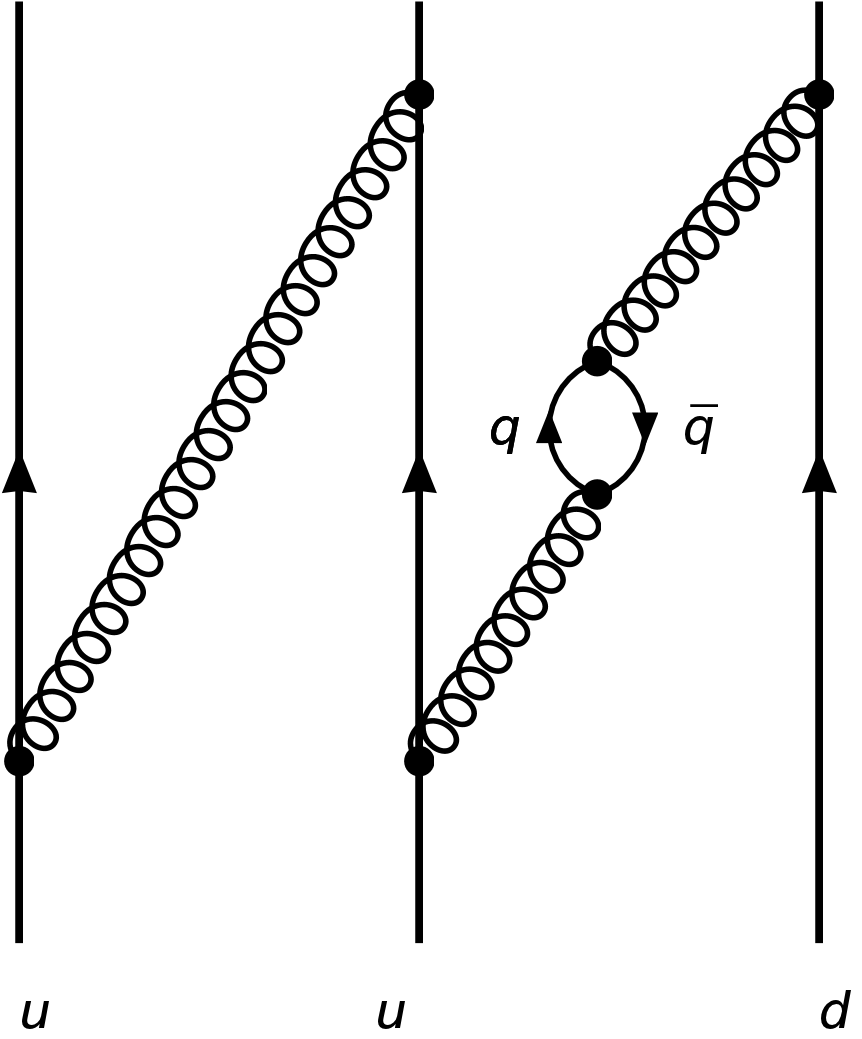}
\capspace
\caption{An illustration of the inner structure of the proton. At any given time, there might be one or several quark anti-quark pairs inside the proton.}
\label{inner_struct_proton}
\end{figure}

\begin{figure}[ht]
\centering
\includegraphics[width=0.50\textwidth]{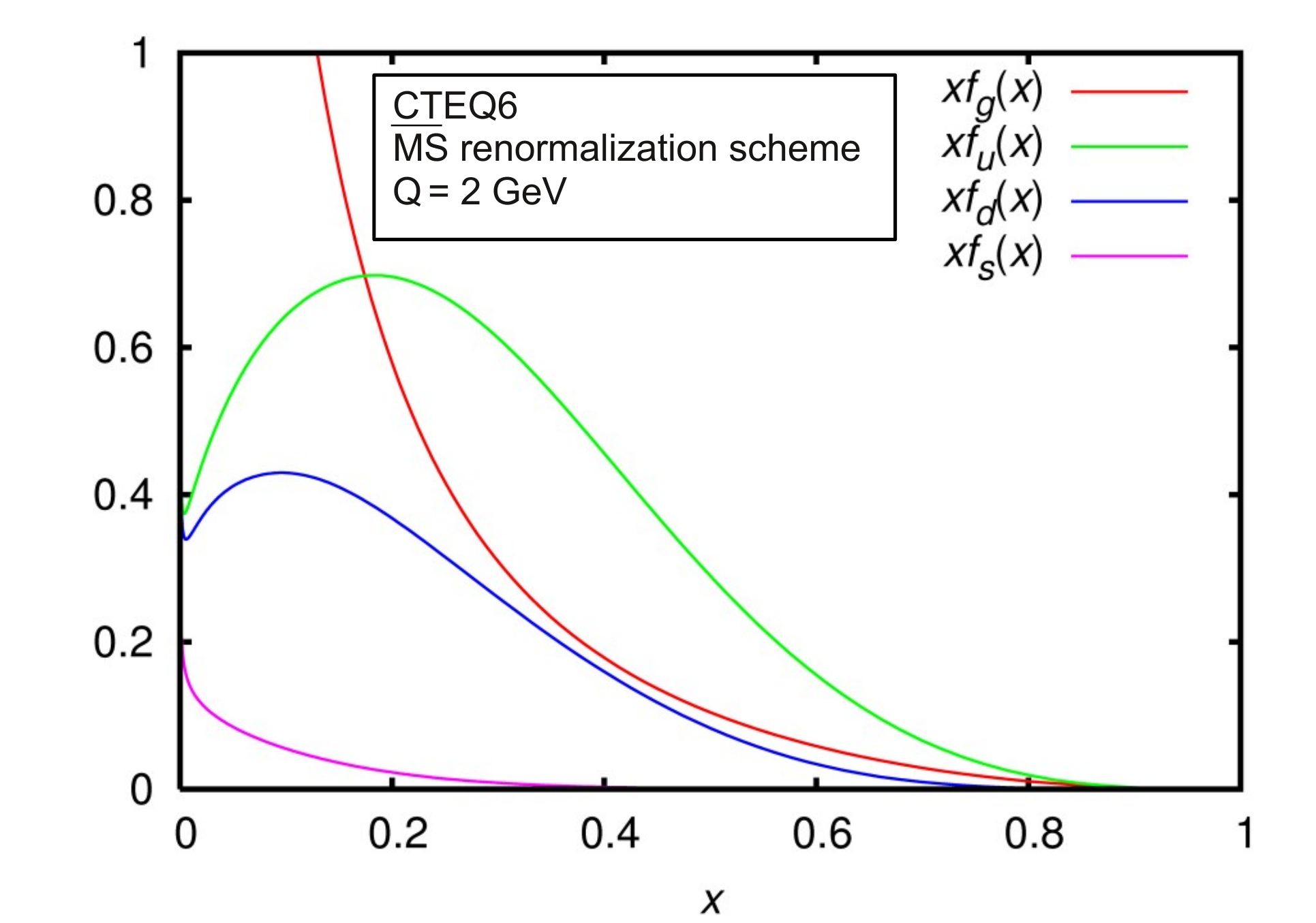}
\includegraphics[width=0.40\textwidth]{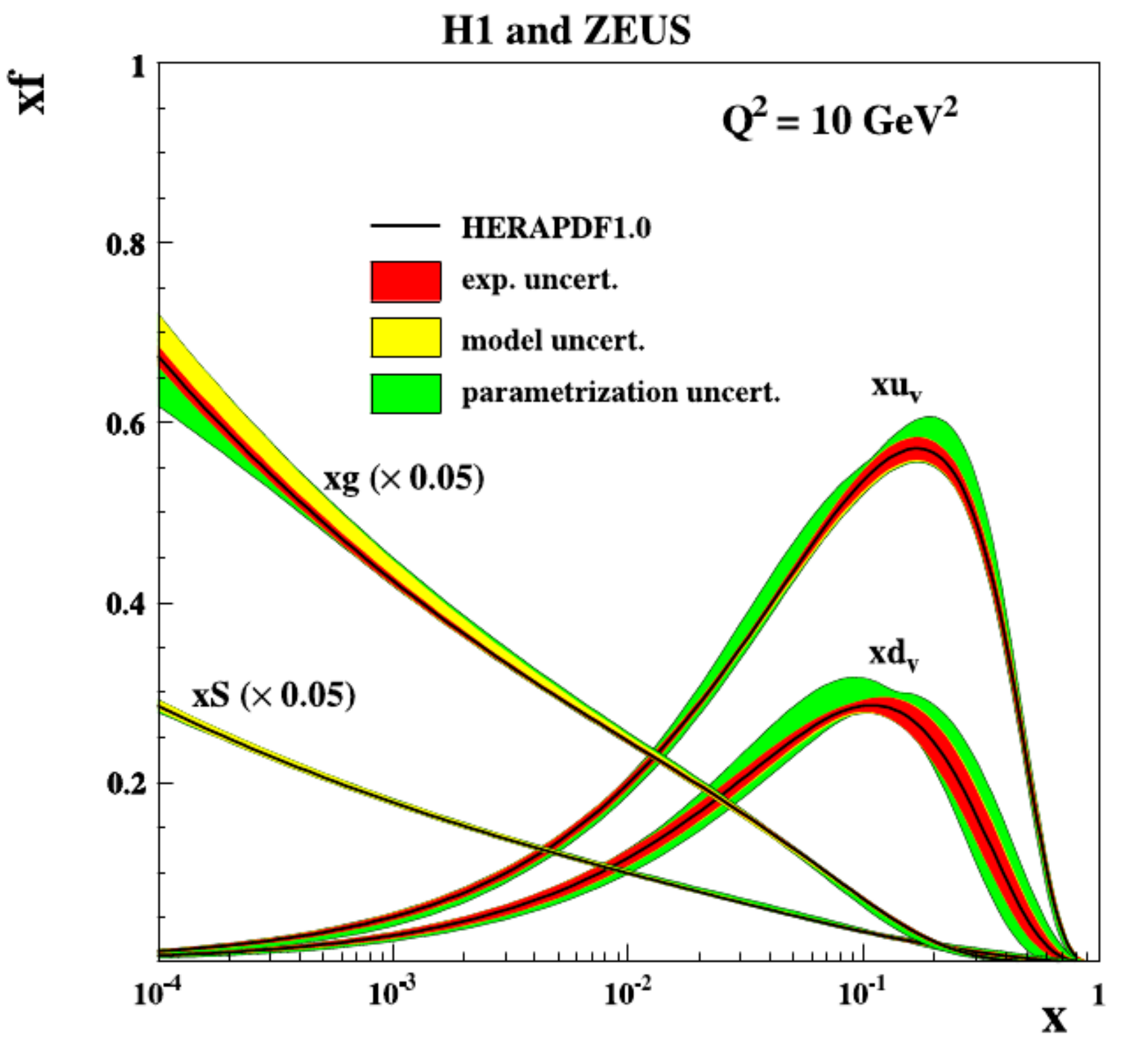}
\capspace
\caption{The CTEQ6 parton distribution functions in the $\overline{MS}$ renormalization scheme and Q = 2 GeV for gluons (red), up (green), down (blue), and strange (violet) quarks (left). The parton distribution functions from the HERAPDF1.0 at $Q^{2}$ = 10 GeV$^{2}$ (right).}
\label{inner_struct_proton}
\end{figure}

\clearpage
\subsection{Parton Parton Scattering and the Two-Jet Production in Hadron-Hadron Collisions}
\begin{figure}[ht]
\centering 
\includegraphics[width=0.60\textwidth]{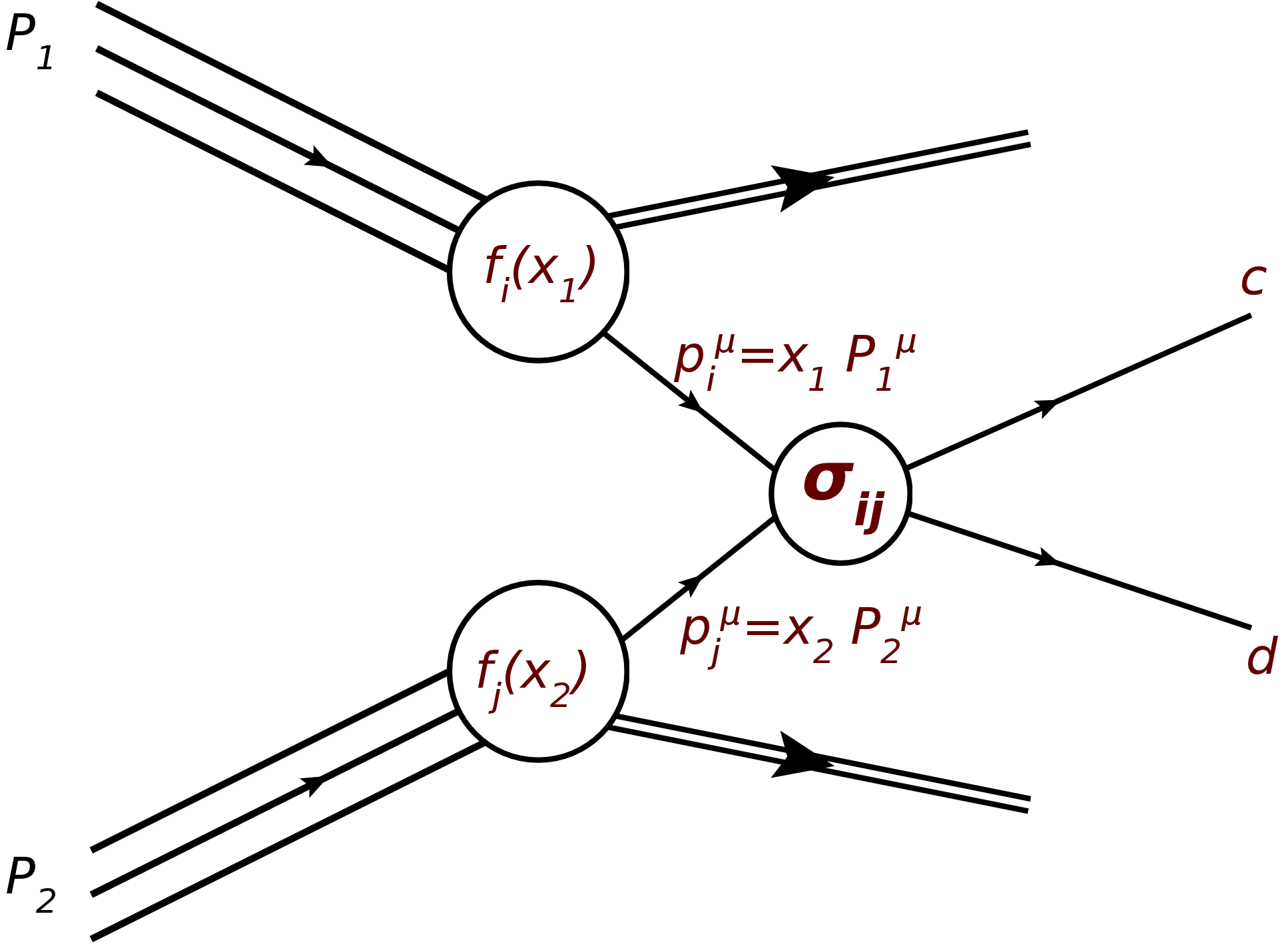}
\capspace
\caption{A schematic diagram of hadron-hadron collision which shows the hard subprocess of internal partons for the production of final states c and d.} 
\label{parton parton scattering} 
\end{figure}

\begin{figure}[ht]
  \centering
  \includegraphics[width=0.22\textwidth]{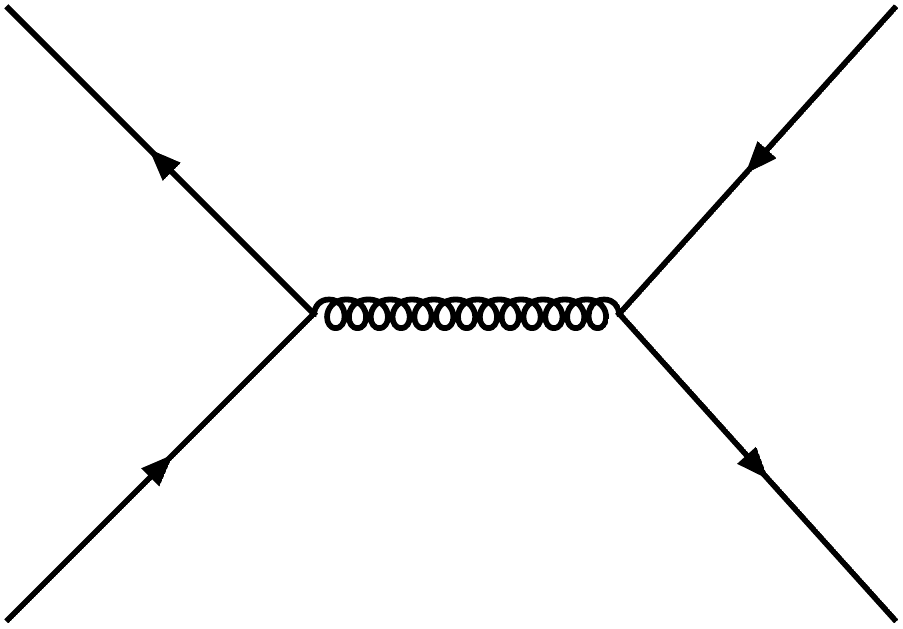}
  \hspace{6mm}
  \includegraphics[width=0.20\textwidth]{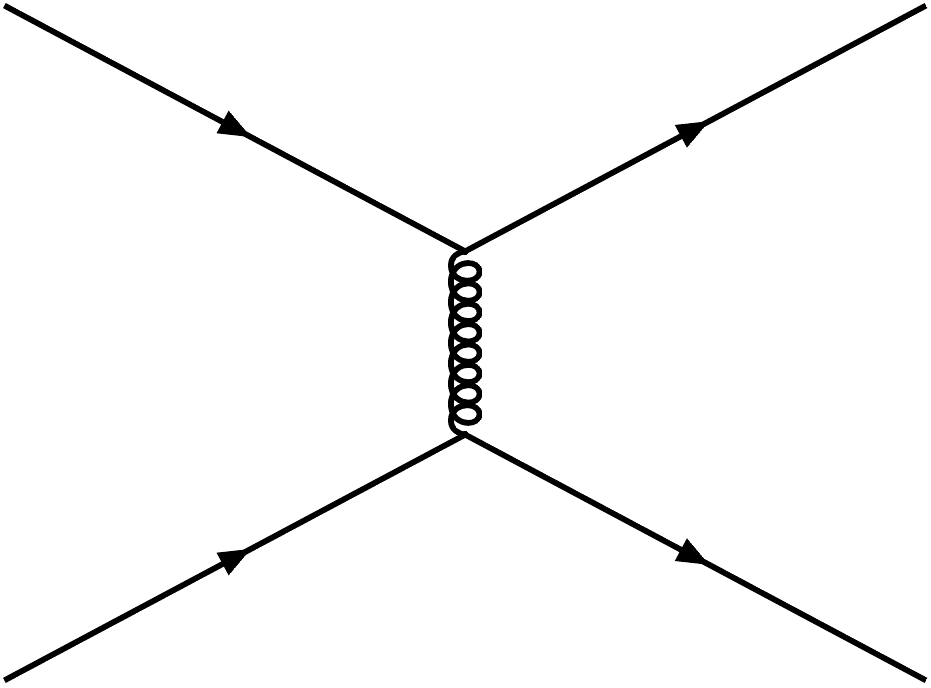}
  \hspace{6mm}
  \includegraphics[width=0.20\textwidth]{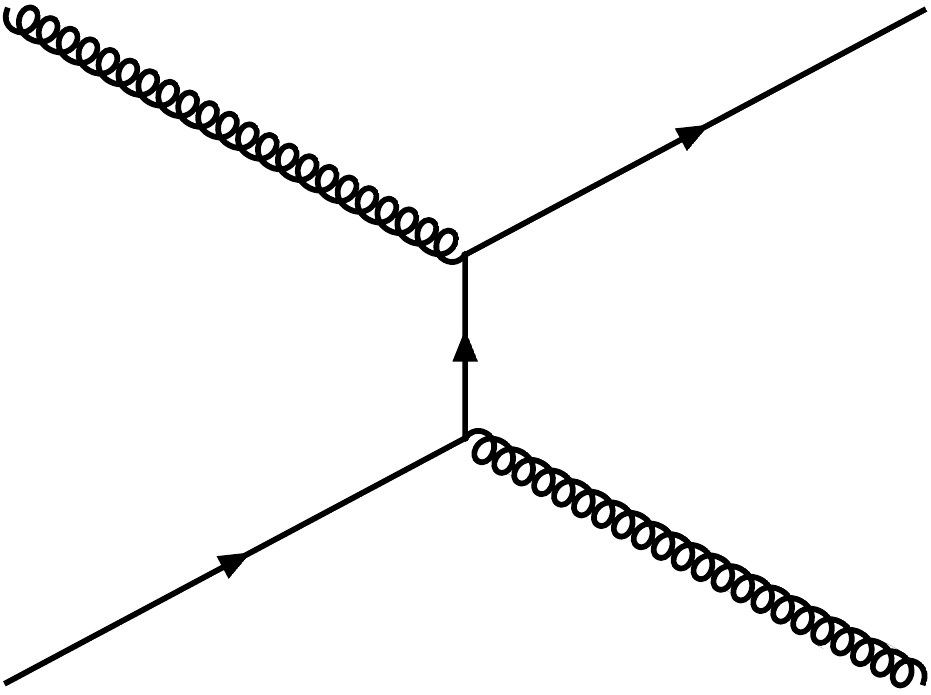}\\

  \vspace{15mm}  
  \includegraphics[width=0.20\textwidth]{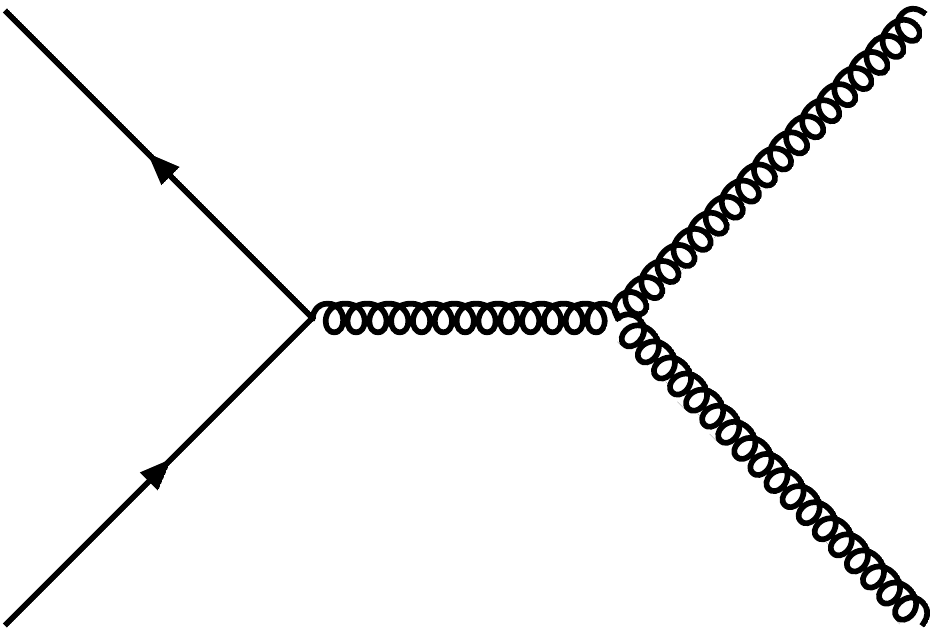}
  \hspace{6mm}
  \includegraphics[width=0.20\textwidth]{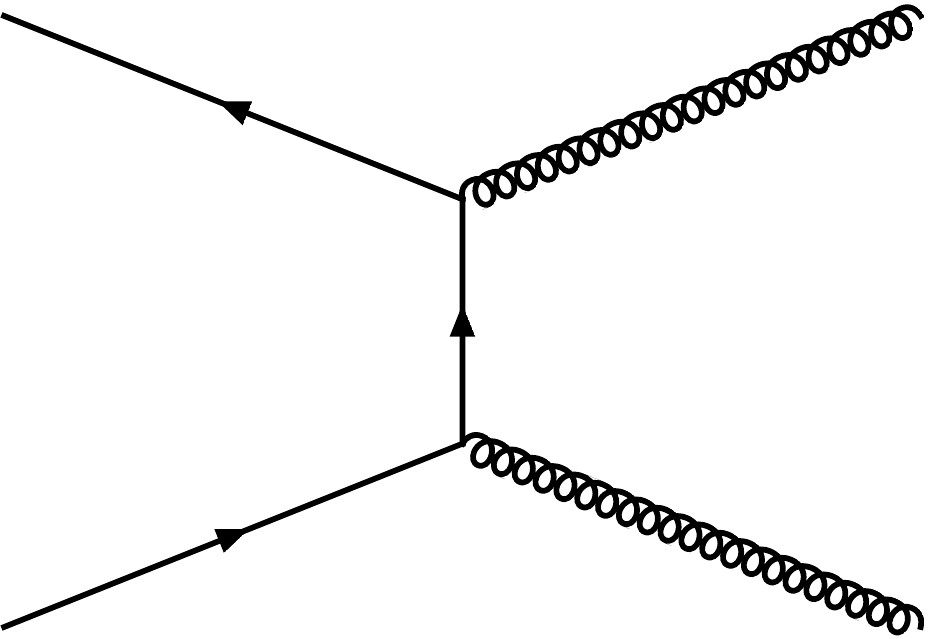}
  \hspace{6mm}
  \includegraphics[width=0.20\textwidth]{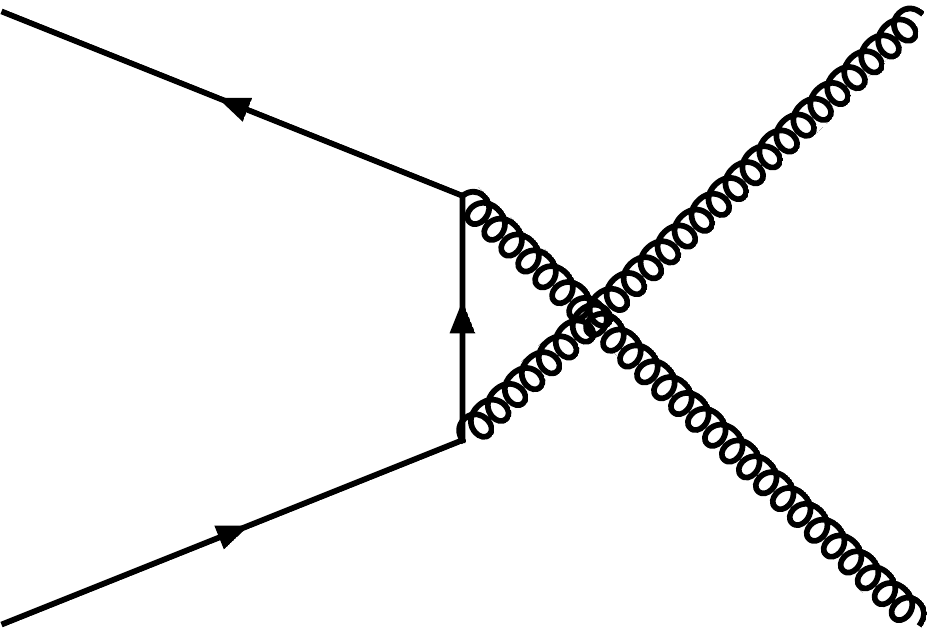}\\

  \vspace{15mm}  
  \includegraphics[width=0.20\textwidth]{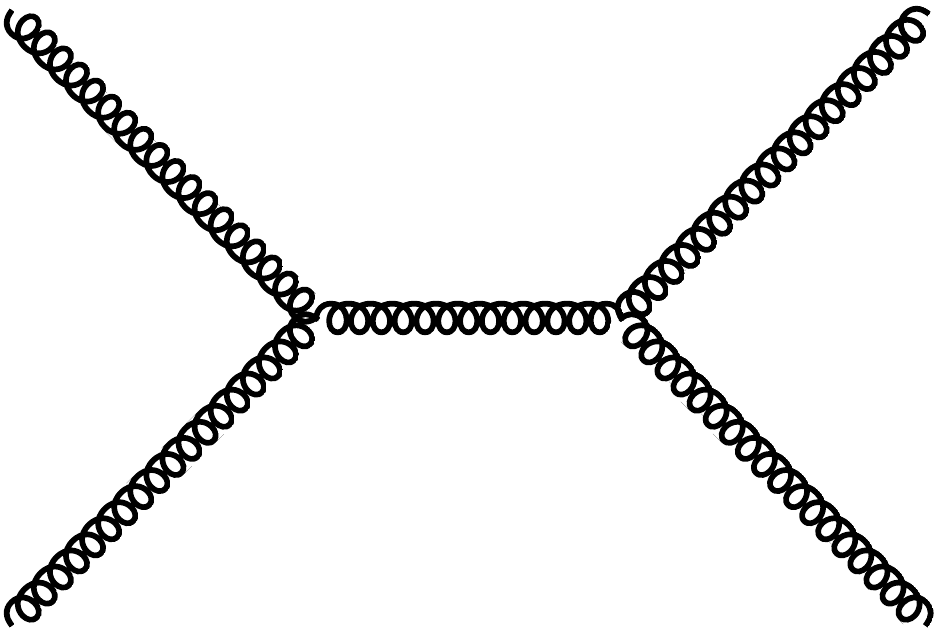}
  \hspace{6mm}
  \includegraphics[width=0.20\textwidth]{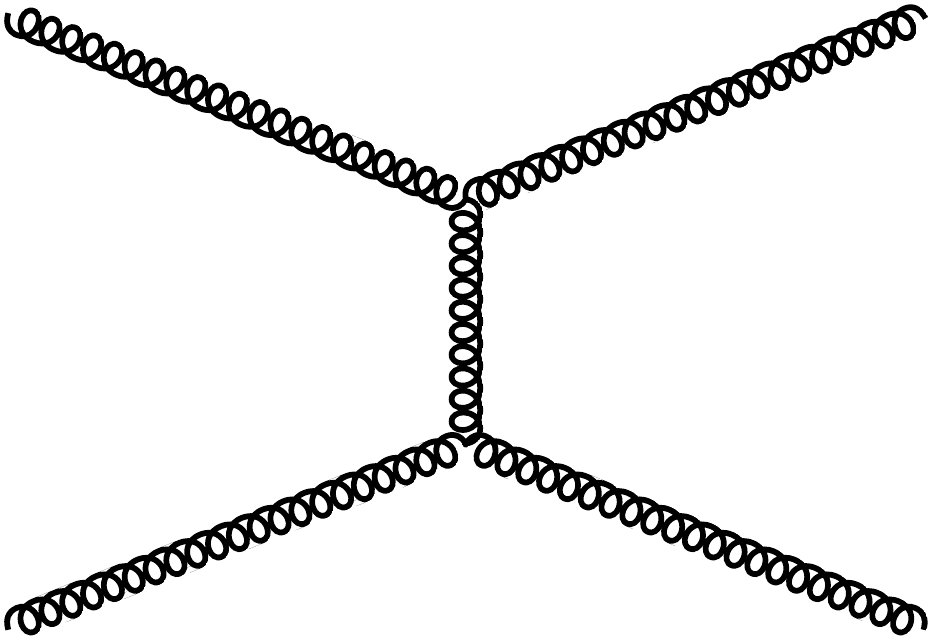}
  \hspace{6mm}
  \includegraphics[width=0.20\textwidth]{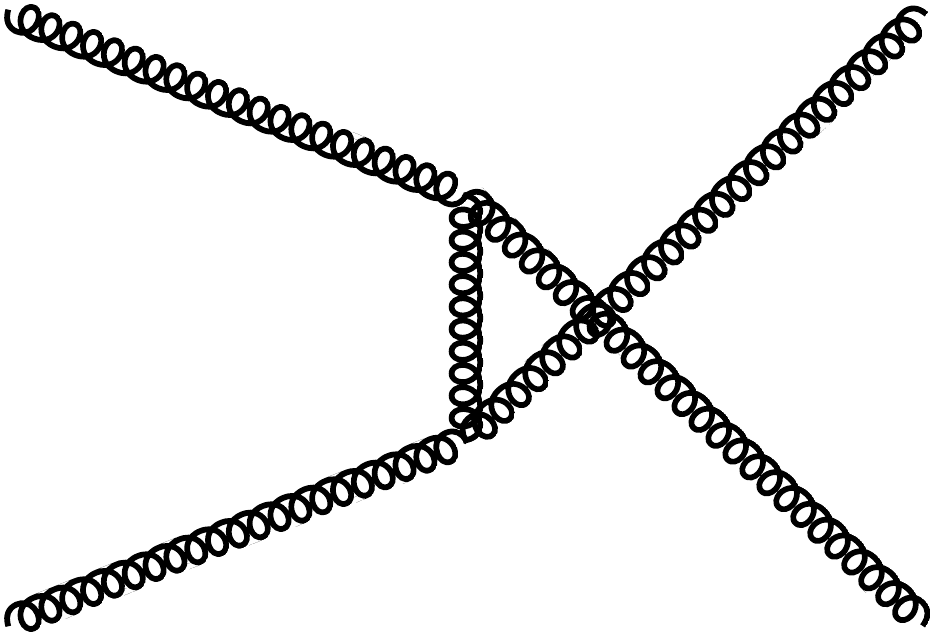}
  \capspace
  \caption{Feynman diagrams which contribute to parton-parton scattering.}
  \label{fig:QCD_diagrams}
\end{figure}

When two hadrons collide inelastically, the actual interaction occurs in terms of a momentum transfer between two constituents of the colliding hadrons. In LHC, two protons collide so harshly that the partons inside the protons go into a \textit{deep inelastic scattering} (DIS) (Figure \ref{parton parton scattering}). The differential cross section for the collision of two hadrons, $P_1$ and $P_2$, to give particles $c$ and $d$ is given by
\begin{equation}
d\sigma(P_{1}P_{2}\rightarrow cd)=\int_{0}^{1} dx_{1}dx_{2} \displaystyle \sum_{q_{i},q_{j}} f_{i}(x_{1},\mu^{2}_{F}) f_{j}(x_{2},\mu^{2}_{F}) d\hat{\sigma}_{(q_{i}q_{j}\rightarrow cd)}(Q^{2},\mu^{2}_{F})
\label{diff_cross_section} 
\end{equation} 
where the momenta of the partons which are strongly interacted are $p_{i}^{\mu}=x_{1}P_{1}$ and $p_{i}^{\mu}=x_{1}P_{1}$, $x_{1}$ and $x_{2}$ are fractions of hadron momentum carried by interacting partons. The function $f_{i}(x,\mu_{F}^{2})$ is the quark gluon parton PDFs are defined at a factorization scale of $\mu_{F}$ which is typically at  the order of $Q$ - a hard scale characteristic of the parton scattering. $d\hat{\sigma}(q_{i}q_{j}\rightarrow cd)$  is the short-distance (partonic) differential cross section for the scattering of partons of type \textit{i} and \textit{j}.
The outgoing partons, $c$ and $d$ are observed as jets since they eventually turn into a spray of color singlet particles, namely hadrons. From the conservation of momentum, the momenta of outgoing partons will be the same in magnitude but opposite in direction. For 2$\rightarrow$2 parton scattering (a($p_{1}^{\mu}$)+b($p_{2}^{\mu}$)$\rightarrow$ c($p_{3}^{\mu}$)+d($p_{4}^{\mu}$)), the differential cross section is given by \cite{QCD_Ellis}
\begin{equation}
\dfrac{E_3 E_4 d^{6}\hat{\sigma}}{d^{3}p_3 d^{3}p_4}=\dfrac{1}{2 \hat{s}} \dfrac{1}{16 \pi^{2}}\overline{\sum} |\mathcal{M}|^{2}~\delta^{4}(p_1 +p_2 -p_3 -p_4)
\label{two_jet_scattering_cross_sections} 
\end{equation}
where $\overline{\sum}$ denotes the averaged sum over the initial and the final state spins and colors. All the
leading-order parton-parton contributions can be derived from the diagrams shown in the Figure
\ref{fig:QCD_diagrams} by including all the crossing symmetries of the given diagrams. The expressions for the total
scattering amplitude $\overline{\sum} |\mathcal{M}|^{2}$ is given in Table \ref{scattering_amplitudes}, where
$\hat{s}=(p_{1}^{\mu}+p_{1}^{\mu})^{2}$, $\hat{u}=(p_{1}^{\mu}-p_{3}^{\mu})^{2}$ and
$\hat{t}=(p_{2}^{\mu}-p_{3}^{\mu})^{2}$.

\clearpage

Then, the two-jet inclusive cross section can be written as a convolution of the PDFs with the Equation \ref{diff_cross_section}
\begin{equation}
\dfrac{d^{3}\sigma}{dy_3 dy_4 dp_{T}^{2}} =\dfrac{1}{16 \pi s^{2}} \displaystyle \sum_{i,j,k,l=q,\bar{q},g} \dfrac{f_i(x_1,\mu^{2})}{x_1} \dfrac{f_i(x_2,\mu^{2})}{x_2} \times \overline{\sum} |\mathcal{M}(ij\rightarrow kl)|^{2} \dfrac{1}{1+\delta_{kl}}
\end{equation}
where $y_3$ and $y_4$ represents the rapidities of the outgoing partons in the laboratory frame. The momentum fractions $x_1$, $x_2$ are determined by using the conservation of momentum
\begin{equation}
x_1=\dfrac{1}{2} x_T(e^{y_3}+e^{y_4}), ~~~~\dfrac{1}{2} x_T(e^{-y_3}+e^{-y_4})
\end{equation}
where $x_T=2p_T/\sqrt{s}$.
\begin{table}[h]
 \caption{The differential cross sections for the various constituent quark-quark, quark-gluon, and gluon-gluon subprocesses \cite{diff_cross_sections_qq_gg}.}
\centering
\capspace
\resizebox{7cm}{!}{
\renewcommand{\arraystretch}{1.8}
\begin{tabular}{|c||c|}
\hline
Process            & $\overline{\sum} |\mathcal{M}|^{2}/g^{4}$\\
\hline
\hline
$qq\rightarrow qq$ &$\dfrac{4}{9} \dfrac{\hat{s}^{2}+\hat{u}^{2}}{\hat{t}^{2}}$  \\
\hline
$qq^{'} \rightarrow qq^{'}$ & $\dfrac{4}{9} \left(\dfrac{\hat{s}^{2}+\hat{u}^{2}}{\hat{t}^{2}}+\dfrac{\hat{s}^{2}+\hat{t}^{2}}{\hat{u}^{2}} \right) -\dfrac{8}{27} \dfrac{\hat{s}^{2}}{\hat{u} \hat{t}}$  \\
\hline
$q\bar{q^{'}}\rightarrow q\bar{q^{'}}$ &$\dfrac{4}{9} \dfrac{\hat{s}^{2}+\hat{u}^{2}}{\hat{t}^{2}}$  \\
\hline
$q\bar{q}\rightarrow q\bar{q}$ & $\dfrac{4}{9} \left(\dfrac{\hat{s}^{2}+\hat{u}^{2}}{\hat{t}^{2}}+\dfrac{\hat{u}^{2}+\hat{t}^{2}}{\hat{s}^{2}} \right) -\dfrac{8}{27} \dfrac{\hat{u}^{2}}{\hat{s} \hat{t}}$\\
\hline
$qq^{'}\rightarrow q^{'}\bar{q^{'}}$ &$\dfrac{4}{9} \dfrac{\hat{t}^{2}+\hat{u}^{2}}{\hat{s}^{2}}$  \\
\hline
$qg\rightarrow qg$ &$-\dfrac{4}{9} \dfrac{\hat{s}^{2}+\hat{u}^{2}}{\hat{s}\hat{u}}+\dfrac{\hat{s}^{2}+\hat{u}^{2}}{\hat{t}^{2}}$  \\
\hline
$gg\rightarrow gg$ &$\dfrac{2}{9} \left(3-\dfrac{\hat{t}\hat{u}}{\hat{s}^{2}} -\dfrac{\hat{s}\hat{u}}{\hat{t}^{2}}-\dfrac{\hat{s}\hat{t}}{\hat{u}^{2}}\right)$  \\
\hline
$q\bar{q}\rightarrow gg$ &$\dfrac{32}{27} \dfrac{\hat{t}^{2}+\hat{u}^{2}}{\hat{t}\hat{u}}-\dfrac{8}{3}\dfrac{\hat{t}^{2}+\hat{u}^{2}}{\hat{s}^{2}}$ \\
\hline
$gg\rightarrow q\bar{q}$ &$\dfrac{1}{6} \dfrac{\hat{t}^{2}+\hat{u}^{2}}{\hat{t}\hat{u}}-\dfrac{8}{3}\dfrac{\hat{t}^{2}+\hat{u}^{2}}{\hat{s}^{2}}$ \\
\hline
\end{tabular}
\label{scattering_amplitudes} 
}
 
\end{table}
	
\clearpage
\subsection{Relativistic Kinematics of The Dijet System}
Now, it is convenient to consider the kinematics of the two resultant jets in detail. Since the momentum fractions of the two incoming partons are not the same, the center of mass frame of the parton-parton system is boosted along the $z$-axis in laboratory frame. The rapidity, $y=\dfrac{1}{2}ln \left( \dfrac{E+p_z}{E-p_z} \right)$, transforms under boosts along $z$-axis as follows;
\begin{eqnarray}
E  & \rightarrow & \gamma(E+\beta p_z)\\
\nonumber \\
p_z& \rightarrow & \gamma(p_z+\beta E)\\
\nonumber \\
y  & \rightarrow & \dfrac{1}{2}ln \left( \dfrac{\gamma(E+\beta p_z)+\gamma(p_z+\beta E)}{\gamma(E+\beta p_z)-\gamma(p_z+\beta E)} \right)\\
\nonumber \\
& = & \dfrac{1}{2}ln \left( \dfrac{E(\beta +1)+p_z(\beta +1)}{E(1-\beta)+p_z(1-\beta)} \right)\\
\nonumber \\
& = & \dfrac{1}{2}ln \left( \dfrac{E+p_z}{E-p_z} \cdot \dfrac{1+\beta}{1-\beta}\right)\\
\nonumber \\
& = & \dfrac{1}{2}ln \left( \dfrac{E+p_z}{E-p_z} \right)+ \dfrac{1}{2}ln \left(\dfrac{1+\beta}{1-\beta}\right)\\
\nonumber \\
& = & y+y_{boost}
\end{eqnarray}
It means that the rapidity is additive under Lorentz transformations. Given that the rapidities of two jets are back-to-back in the center of mass frame of the two-jet system, they are given in the laboratory frame as follows
\begin{eqnarray}
y_3^{lab}&=&y_3^{CM}+y_{boost}\\
%\nonumber \\
y_4^{lab}&=&-y_3^{CM}+y_{boost}
\end{eqnarray}

\begin{figure}[ht]
  \centering
  \includegraphics[width=0.46\textwidth]{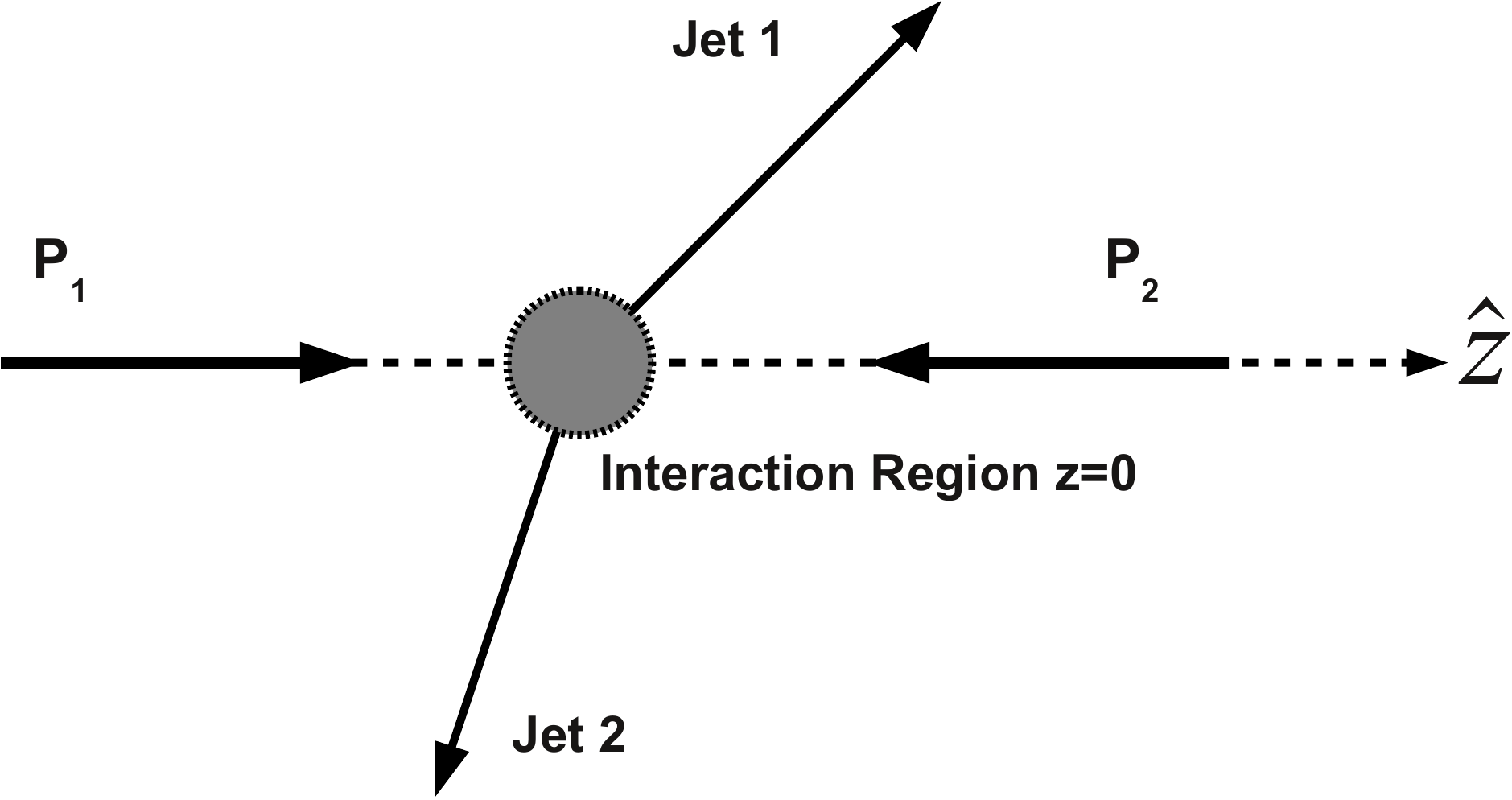}
  \hspace{6mm}
  \includegraphics[width=0.46\textwidth]{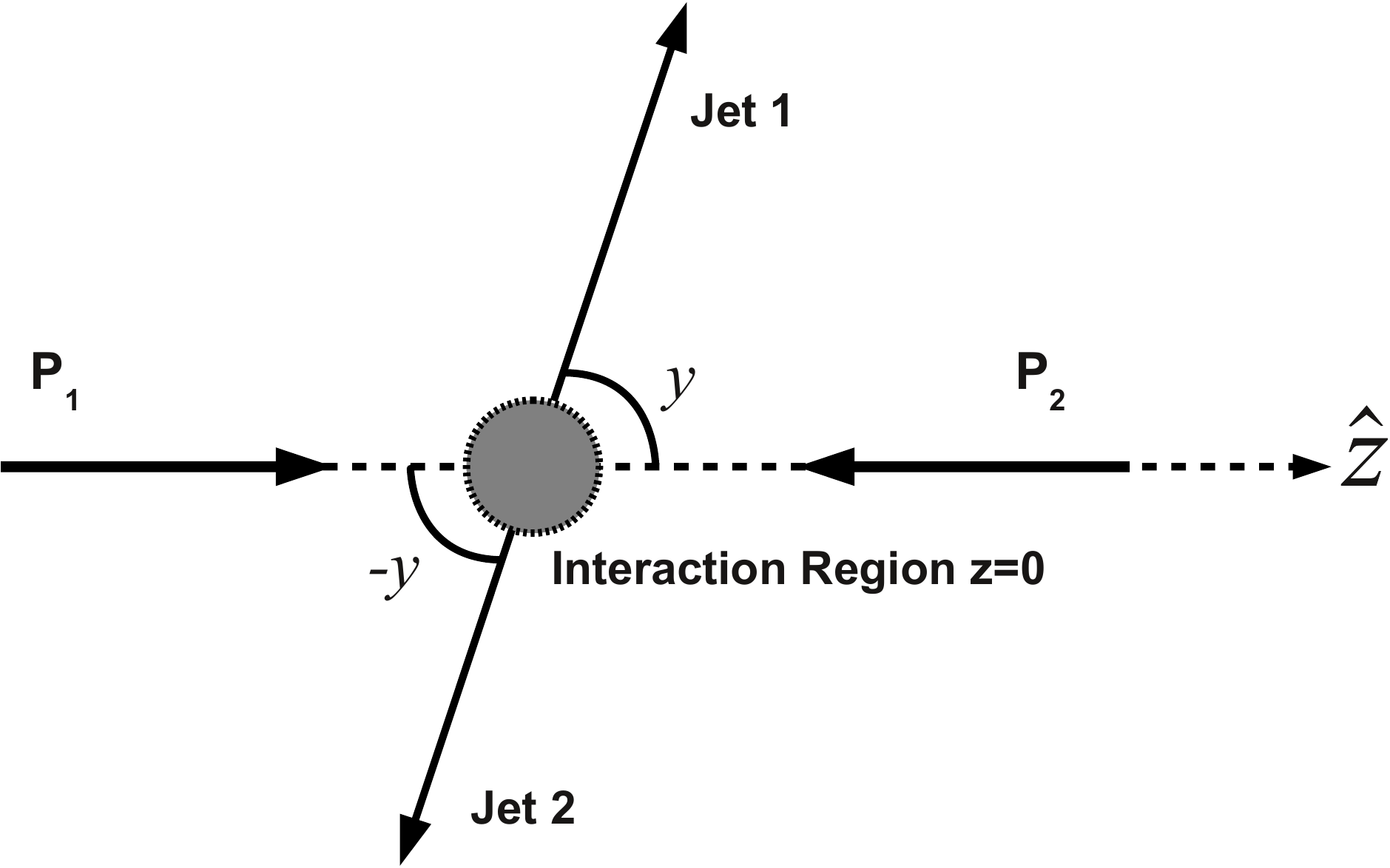}
  \capspace
  \caption{An illustration of the two jet $\hat{z}-y$ plane geometry in the laboratory frame (left) and in the parton-parton center of mass frame (right).}
  \label{jet_kinematics_lab}
\end{figure}

From Equations 2.45 and 2.46, the rapidities in the center of mass frame can be extracted by the rapidities in the laboratory frame.
\begin{eqnarray}
y_3^{CM}=\dfrac{y_3^{lab}-y_4^{lab}}{2}\\
y_4^{CM}=-y_3^{CM}=\dfrac{y_4^{lab}-y_3^{lab}}{2}\\
y_{boost}=\dfrac{y_3^{lab}+y_4^{lab}}{2}
\end{eqnarray}
For a massless parton ($\beta=1$) the rapidity in the center of mass frame is given by
\begin{equation}
y^{*}=\dfrac{1}{2}\ln \left(\dfrac{1+cos\theta^{*}}{1-cos\theta^{*}}\right)
\end{equation}
This leads to
\begin{equation}
\cos\theta^{*}=\tanh(y^{*})
\end{equation}
where $\theta^{*}$ is the scattering angle from the collision axis of two partons in the center of mass frame.
The invariant mass of the system is derived by starting from the basic relativistic kinematics
\begin{eqnarray}
M_{jj}^2&=&(p_1^{\mu}+p_2^{\mu})^{2}=(p_1^{0}+p_2^{0})^{2}-(\vec{\boldsymbol{p_1}}+\vec{\boldsymbol{p_2}})^{2}\\
&=&E_1^{2}+E_2^{2}+2E_{1}E_{2}-|\vec{\boldsymbol{p_1}}|^{2}-|\vec{\boldsymbol{p_1}}|^{2}-2 \vec{\boldsymbol{p_1}} \cdot \vec{\boldsymbol{p_2}}\\
\vec{\boldsymbol{p_1}} \cdot \vec{\boldsymbol{p_2}}&=&{p_{1}}_{x}{p_{2}}_{x}+{p_{1}}_{y}{p_{2}}_{y}+{p_{1}}_{z}{p_{2}}_{z}\\
{p_{i}}_{x}&=&{p_T}_i \cos \phi_1 \\
{p_{i}}_{y}&=&{p_T}_i \sin \phi_1\\
\dfrac{{p_{i}}_{z}}{E_i}&=&\beta_i \cos \theta_i=E \tanh y_i\\
{p_T}_i&=&{\beta_T}_i {E_T}_i\\
\text{where $i=1,2$} \nonumber \\
{E_T}_i^{2}&=&E_i^{2}-{p_{i}}_{z}^{2}=E_i^{2}-E_i^{2}{\tanh^{2} y_i}=E_i^{2}(1-{\tanh^{2} y_i})\\
\Rightarrow E_i&=&{E_T}_i \cosh y_i
\end{eqnarray}
so the scalar product of $\boldsymbol{\vec{p_1}\cdot\vec{p_2}}$ becomes
\begin{eqnarray}
\boldsymbol{\vec{p_1}\cdot\vec{p_2}}&=&(\dfrac{{\beta_T}_1 {\beta_T}_2 E_1 E_2 }{\cosh y_1 \cosh y_2}  )(\cos \phi_1 \cos \phi_2+\sin \phi_1 \sin \phi_2)\\&+&E_1 E_2 \tanh y_1 \tanh y_2 \nonumber \\
\nonumber \\
&=&\dfrac{{E_T}_1 {E_T}_2}{\cosh y_1 \cosh y_2}({\beta_T}_1 {\beta_T}_2 \cos\Delta \phi + \sinh y_1 \sinh y_2)
\end{eqnarray}
Thus
\begin{eqnarray}
M_{jj}^2&=&m_1^{2}+m_2^{2}+2{E_T}_1{E_T}_2(\cosh \Delta y-{\beta_T}_1 {\beta_T}_2\cos\Delta \phi)
\end{eqnarray}
in the limit where ${\beta_T}_1={\beta_T}_2\approx1$ it becomes
\begin{eqnarray}
M_{jj}^2&=&2{p_T}_1{p_T}_2(\cosh \Delta y - \cos\Delta \phi)
\end{eqnarray}
It is worth noting that even the outgoing jets are almost massless, the polar and azimuthal separation between them govern the invariant mass of the dijet system.
\section{Non-Perturbative Corrections} 
Since the NLO pQCD calculations provide predictions at the \textit{parton level}, whereas the experimental data are corrected to the \textit{particle level}, the non-perturbative effects must be taken into account when a comparison between the data and the theoretical predictions is desired. In this study, two Monte Carlo event generators, PYTHIA 6.4~\cite{Pythia} (both D6T and Z2 tunes) and Herwig++~\cite{Herwig++} which have different hadronization and multiple parton interactions (MPI) models are employed in order to extract the non-perturbative corrections to the NLO calculation.

With the very steep falling nature of the dijet mass spectrum, it is very important to have enough number of events to describe the spectrum correctly. Current event generator programs, by default, produce events due to their cross section weights which can be interpreted as the probability of producing an event from the generator point of view. Hence, it is practically impossible to populate the entire phase space since the cross section values for the dijet mass spectrum lies between $10^{10}$ pb to $10^{-11}$ pb in the order of magnitude. However, there is an appropriate technique of constructing these steeply falling spectra which is called as $\hat{p}_{T}$ slicing. In this technique, only a certain region of the phase space is taken into account during the calculation of the matrix element. Thus, generator program allows the transverse momentum exchange between two hardly interacted partons to be in a given range namely $\hat{p}_{T}$ bins.

\begin{table}[!h]
 \caption{Number of events used for each $\hat{p}_{T}$ slice and corresponding cross sections which is necessary to construct a correct dijet mass spectrum.}
\centering
\capspace
\resizebox{8cm}{!}{
\begin{tabular}{|l|c|c|}
\hline 
$\hat{p_{T}}$ bin (GeV) & Number of Events & Cross Sections (pb)\tabularnewline
\hline
\hline 
0-15       & 30M & 4.844e+10 \tabularnewline
\hline 
15-20      & 30M & 5.794e+8  \tabularnewline
\hline 
20-30      & 30M & 2.361e+8  \tabularnewline
\hline 
30-50     & 30M & 5.311e+7  \tabularnewline
\hline 
50-80      & 15M & 6.358e+6  \tabularnewline
\hline 
80-120     & 10M & 7.849e+5  \tabularnewline
\hline 
120-170    & 10M & 1.151e+5  \tabularnewline
\hline 
170-230    & 10M & 2.014e+4  \tabularnewline
\hline 
230-300    & 10M & 4.094e+3  \tabularnewline
\hline 
300-380    & 10M & 9.346e+2  \tabularnewline
\hline 
380-470    & 10M & 2.338e+2  \tabularnewline
\hline 
470-600    & 10M & 7.021e+1  \tabularnewline
\hline 
600-800    & 10M & 1.557e+1  \tabularnewline
\hline 
800-1000   & 10M & 1.843e+0  \tabularnewline
\hline 
1000-1400  & 10M & 3.318e-1  \tabularnewline
\hline 
1400-1800  & 10M & 1.086e-2  \tabularnewline
\hline 
1800-2200  & 10M & 3.499e-4  \tabularnewline
\hline 
2200-2600  & 10M & 7.549e-6  \tabularnewline
\hline 
2600-3000  & 10M & 6.465e-8  \tabularnewline
\hline 
3000-3500  & 10M & 6.295e-11 \tabularnewline
\hline
\end{tabular}
 \label{table:pthatslices}
 }
\end{table}

\begin{figure}[h]
   \begin{center}
     \includegraphics[width=0.45\textwidth]{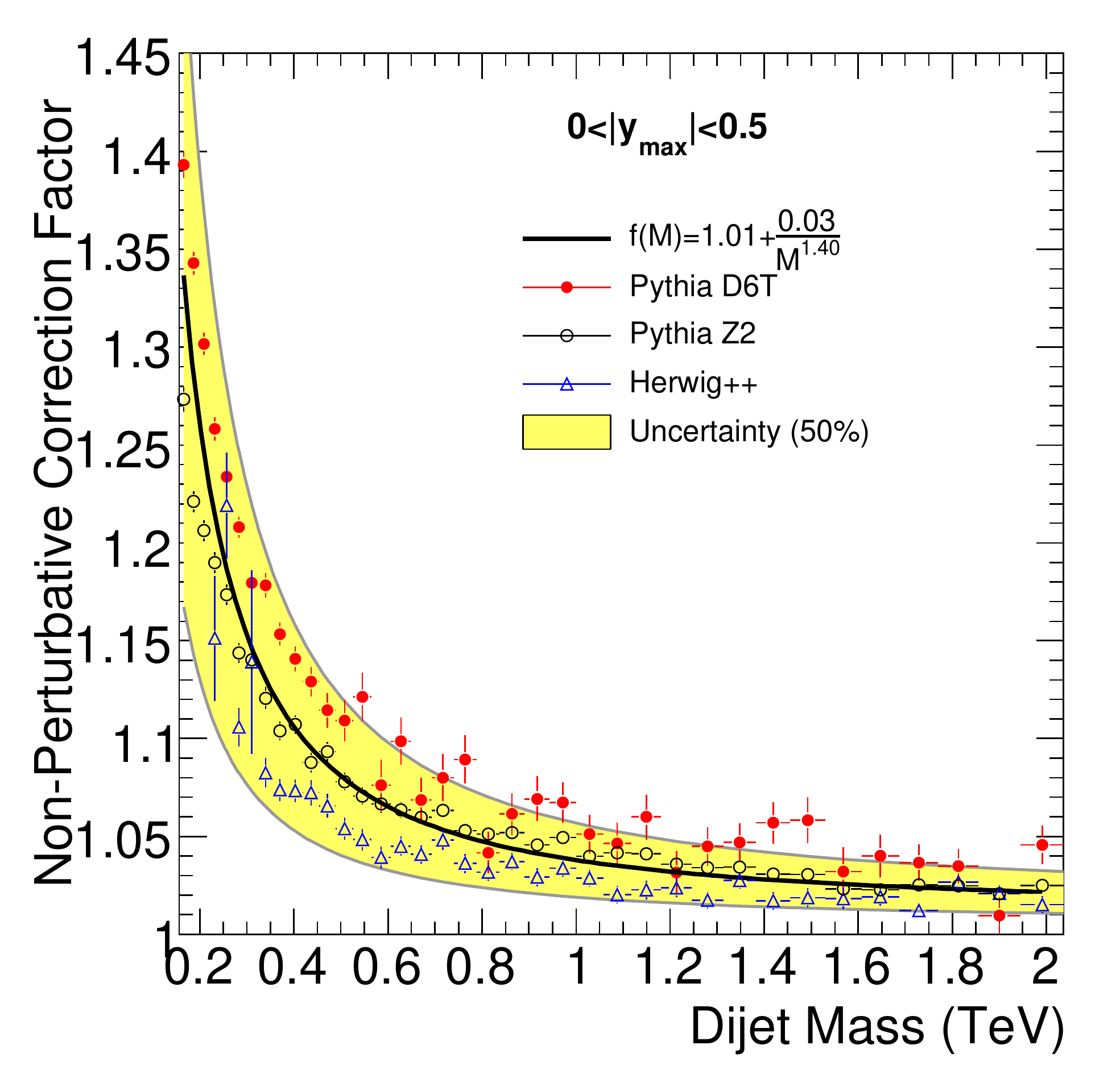}
     \includegraphics[width=0.45\textwidth]{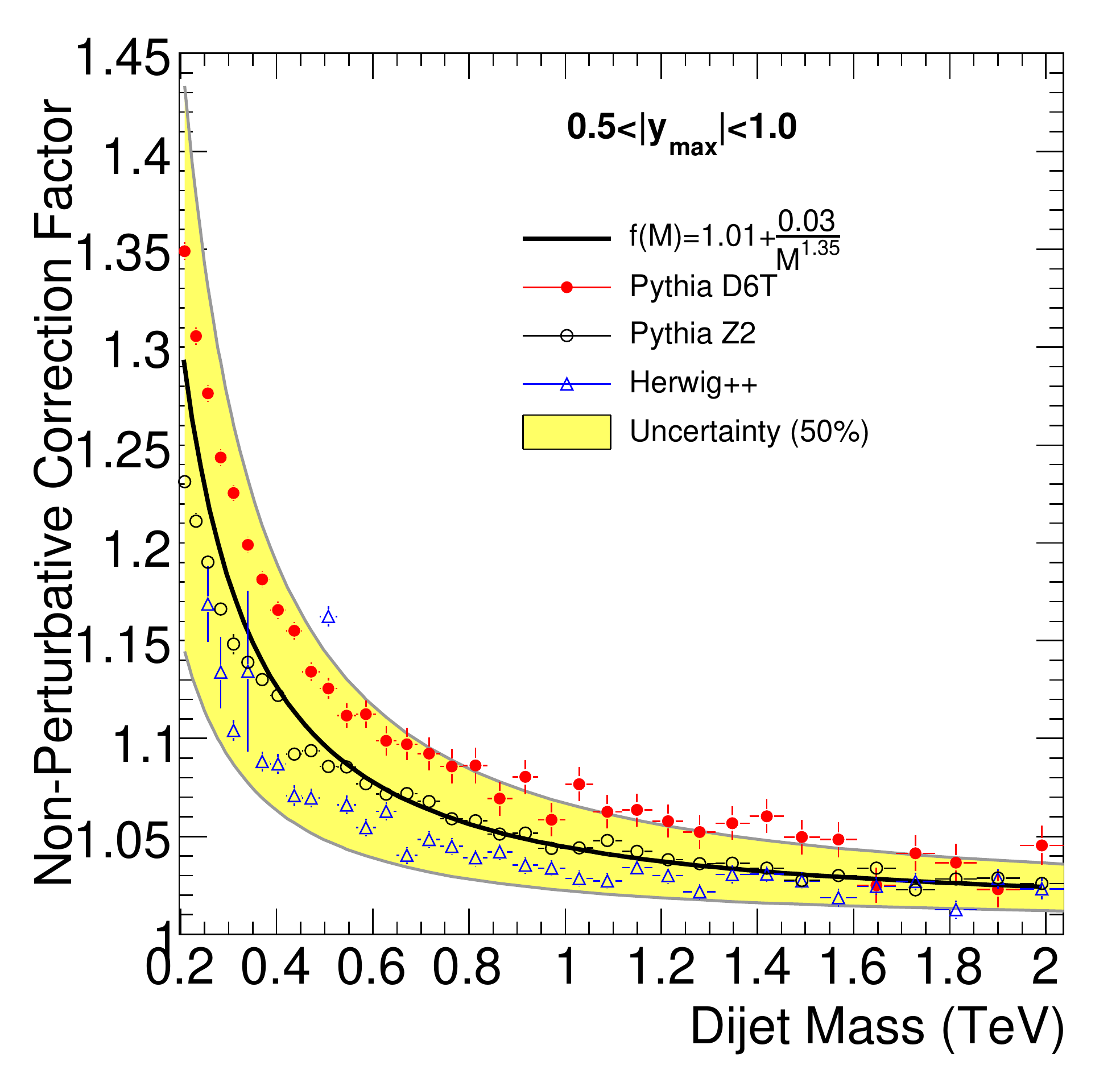}
     \includegraphics[width=0.45\textwidth]{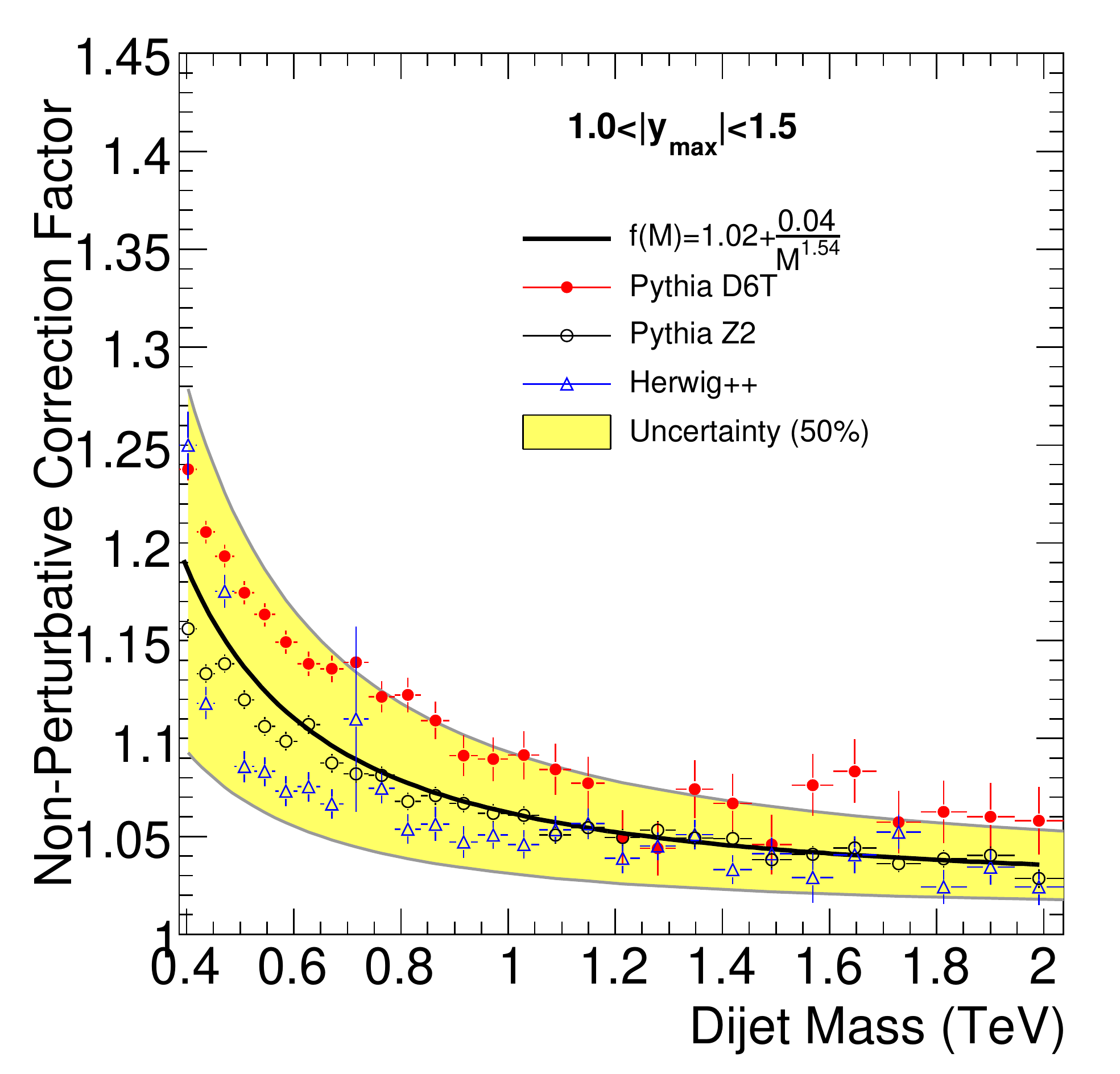}
     \includegraphics[width=0.45\textwidth]{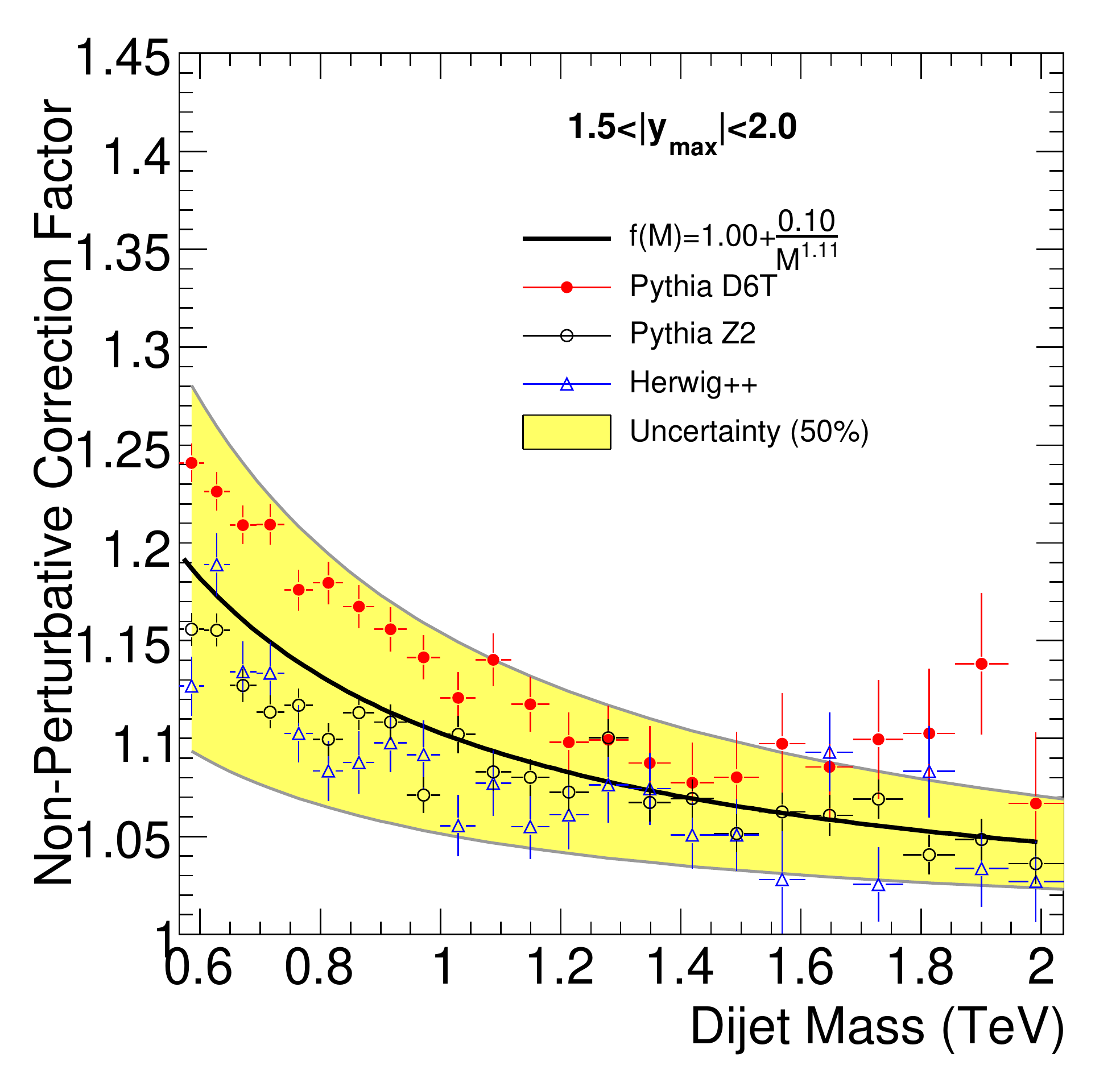}
     \includegraphics[width=0.45\textwidth]{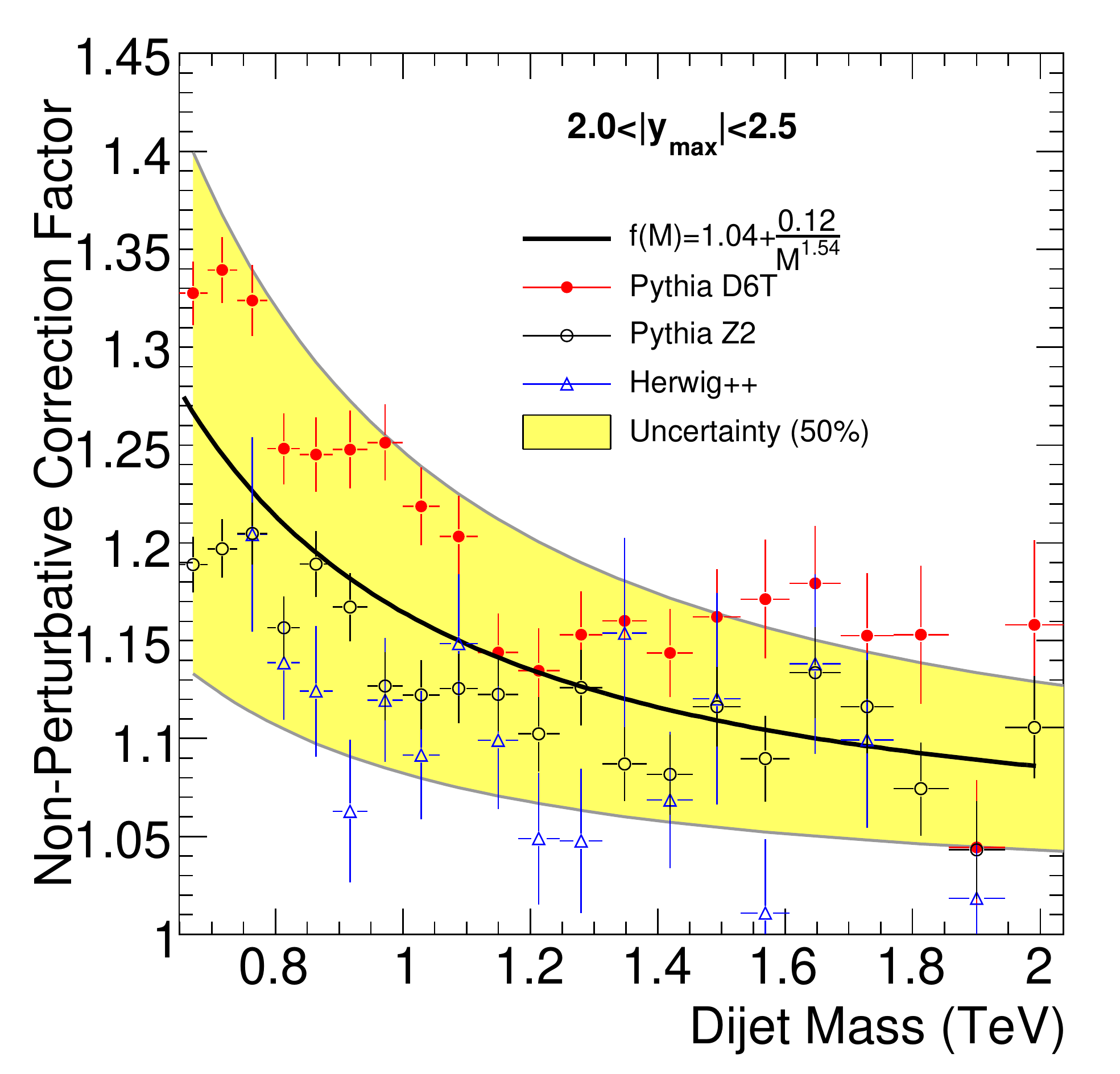}  
     \capspace 
     \caption{Non-Perturbative Corrections extracted from PYTHIA 6.4 (both for D6T and Z2 tunes) and HERWIG++ for each $|y|_{max}$ interval. The band represents the assigned systematic uncertainty.}
     \label{NP_Corrections}
   \end{center}
 \end{figure}

For the derivation of non-perturbative corrections, it is necessary to have two different datasets where the hadronic and the partonic final states are kept, respectively. This can be achieved by setting the parameters \textbf{\texttt{MSTP(81)=0}} and \textbf{\texttt{MSTJ(1)=0}}\footnote{for PYTHIA 6.4 Z2 tune the MSTP(81) parameter should be set to 20 instead of 0.} where the first one switches off MPI and the latter switches off hadronization. It should be noted that these effects cannot be disentangled since the multi-parton interactions may directly feed the hadronization process. In this study, 20 $\hat{p}_{T}$ bins are used to construct the dijet mass spectrum (Table~\ref{table:pthatslices}) for both cases. At the end the ratio of the two different mass spectra is fit to a parametrized function of the dijet mass which is considered as the non-perturbative correction to the NLO calculation. The ratio is formally expressed as below:
\begin{equation}
C_{i}=\dfrac{N_{i}^{MPI+Had}}{N_{i}^{No~MPI+No~Had}}
\label{Eqn:NP_Corrections}
\end{equation}
where $C_{i}$ is the non-perturbative correction factor for the bin \textit{i}, $N_{i}^{MPI+Had}$ is the number of events obtained for the bin \textit{i} at the hadron level and $N_{i}^{No~MPI+No~Had}$ is the number of events obtained for the same bin at the parton level. Finally, the weighted mean of the corrections extracted from PYTHIA 6.4 (both D6T and Z2 tunes) and Herwig++ is used as the overall correction factors, and the half of the difference from the unity is set as the systematic uncertainty. The overall correction factors are fit to a parametrized function of the dijet mass as follows:
\begin{equation}
f(m)=A+\dfrac{B}{m^{C}}
\label{Eqn:NP_Correction_Function}
\end{equation}
Table~\ref{table:NP_Corrections_parameters} summarizes the fit values for the non-perturbative correction for all $|y|_{max}$ bins, while Figure \ref{NP_Corrections} shows the NP corrections from each MC sample, the average fit and the assigned systematic uncertainty. 
\begin{table}[h]
\centering
\caption{Parameters of the overall correction for each $y_{Max}$ bin.}
\capspace
\normalsize
\begin{tabular}{|c|c|c|c|}
\hline $y_{Max}$ bin & A & B & C \\ 
\hline
\hline  $0  <y_{Max}<0.5$ & 1.01  & 0.03 & 1.40 \\ 
\hline  $0.5<y_{Max}<1.0$ & 1.01  & 0.03 & 1.35 \\ 
\hline  $1.0<y_{Max}<1.5$ & 1.02  & 0.04 & 1.54 \\ 
\hline  $1.5<y_{Max}<2.0$ & 1.00  & 0.10 & 1.11 \\ 
\hline  $2.0<y_{Max}<2.5$ & 1.04  & 0.12 & 1.54 \\ 
\hline 
\end{tabular} 
 \label{table:NP_Corrections_parameters}
\end{table}

\clearpage
\chapter{EXPERIMENTAL APPARATUS}

\section{CERN Large Hadron Collider }
The \textsl{Large Hadron Collider} LHC \cite{Brüning:782076} is a \proton collider ring which was built by the European Laboratory for Particle Physics (\textit{Organisation Européenne pour la Recherche Nucléaire}). The collider is 27 km long ring accelerator constructed inside the old Large Electron Positron Collider (LEP) tunnel located about 100 meter underground at the French-Swiss border near Geneva.
The design of LHC aims to collide protons with an energy of 7 TeV for each which corresponds to a center of mass energy of \s=14 TeV. The main goal of LHC programme is to investigate the electroweak symmetry breaking that is explained by the Higgs mechanism for the time being. Although the current explanation of electroweak symmetry breaking is under investigation at LHC, there are also studies on alternatives to Higgs mechanism associated with more symmetries, new forces, new particles or even new phenomena which probably have not been proposed yet. Moreover, it is hoped that puzzling questions of the standard model and the modern cosmology such as about the dark matter, charge-parity (CP) violation, matter-antimatter imbalance of the universe and possible existence of extra dimensions would be answered. The LHC is noted not only for \s=14 TeV center of mass energy of \proton collisions but also for its design luminosity of $\mathcal{L}$=10$^{34}$ \lumi. Before reaching its maximum capacity, it will be operated at a lower luminosity of about $\mathcal{L}$=2$\times$10$^{33}$ \lumi.

The collider consists of two rings where two counter-rotating proton bunches travel. Superconducting RF cavities are responsible to accelerate these rotating proton bunches and to keep the protons together in the bunch. There are eight RF cavities for each beam that give 2 MV at 400 MHz. In each turn a single proton gains 0.5 MeV of energy through these eight RF cavities. There are 2808 bunches in one beam each spaced 25 ns apart. In each bunch, there are approximately 1.15$\times$10$^{11}$ protons. Although 25 ns spacing means a frequency of 40 MHz, the large gaps in between beams lower the bunch crossing frequency down to $\sim$30 MHz (Figure \ref{fig:bunch}).

\begin{figure}[ht]
  \centering
  \includegraphics[width=0.70\textwidth]{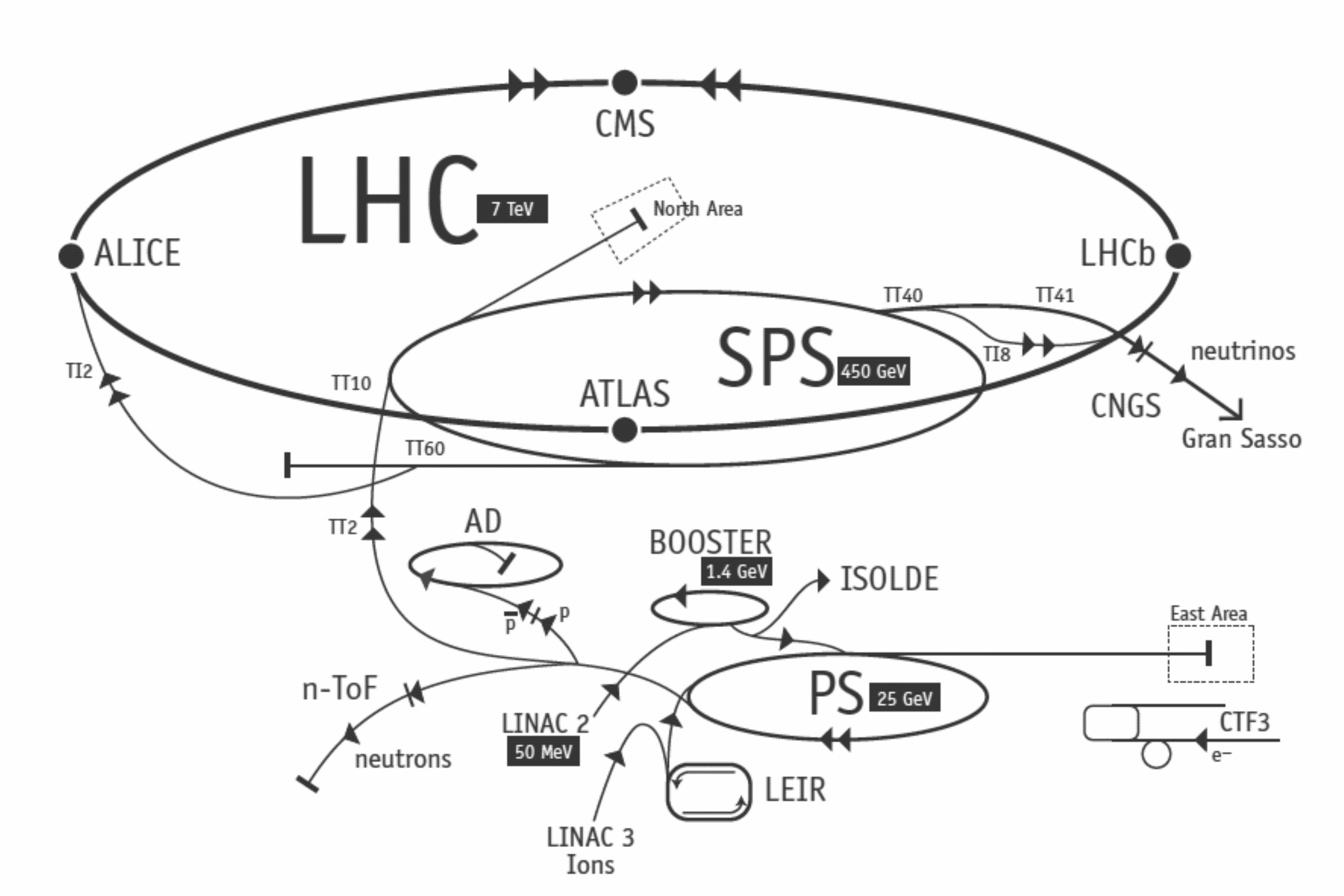}
  \capspace
  \caption{ General view of the Large Hadron Collider.}
  \label{fig:LHC}
\end{figure}
\vspace{20mm}
\begin{figure}[ht]
  \centering
  \includegraphics[width=0.65\textwidth]{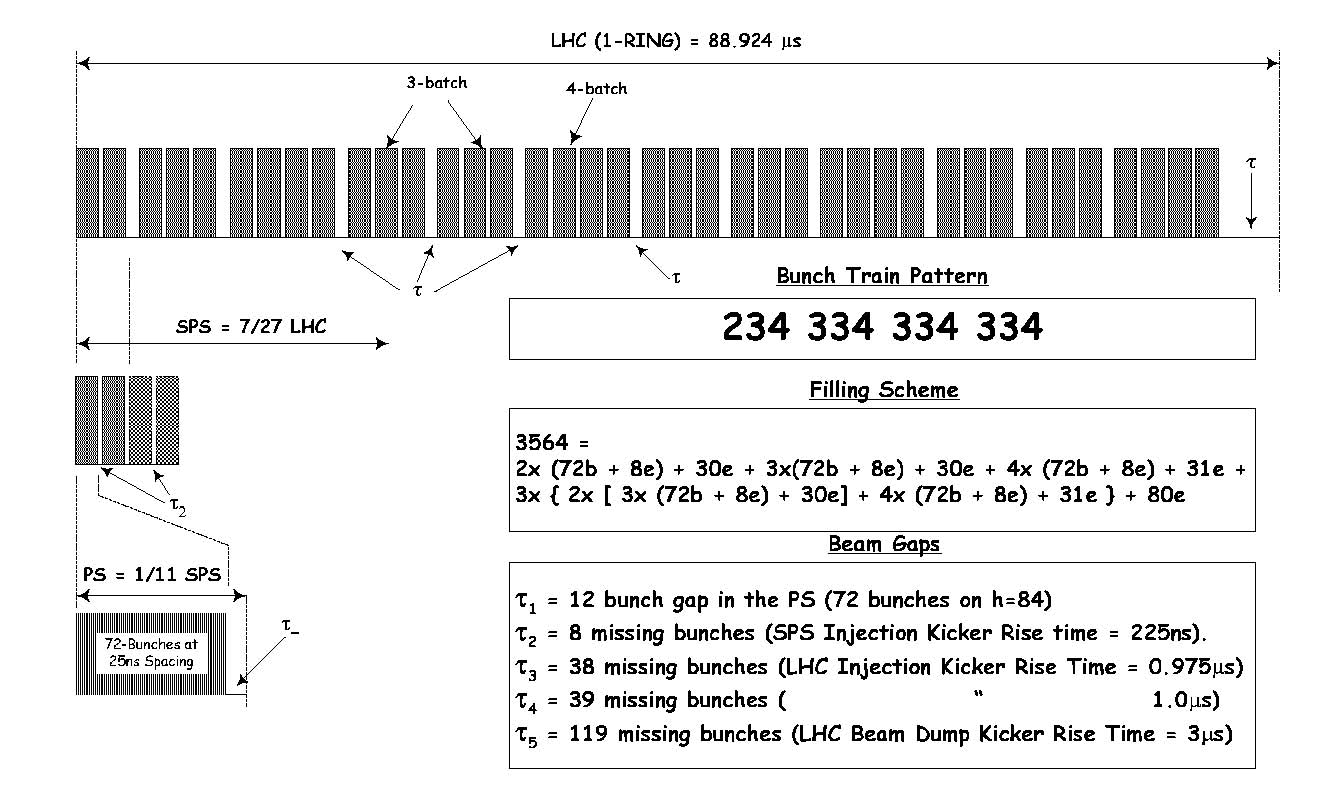}
  \capspace
  \caption{The bunch structure of the LHC beam. Notice the large gap at the end.}
  \label{fig:bunch}
\end{figure}
\clearpage
\section{Compact Muon Solenoid}
The Compact Muon Solenoid (CMS) \cite{CMS_TDR1,CMSExperiment} is one of the two general purpose detectors\footnote{The other is ATLAS.} in the Large Hadron Collider experiment which will study a variety of physics phenomena at the Tera electron Volt (TeV) energy scale. Its surface complex is located in Cessy, France and its experimental cavern is directly under that at the point 5 of the LHC (Figure \ref{fig:LHC}).
\subsection{Physics Goals of the CMS}
\begin{figure}[ht]
  \centering
  \includegraphics[width=0.48\textwidth]{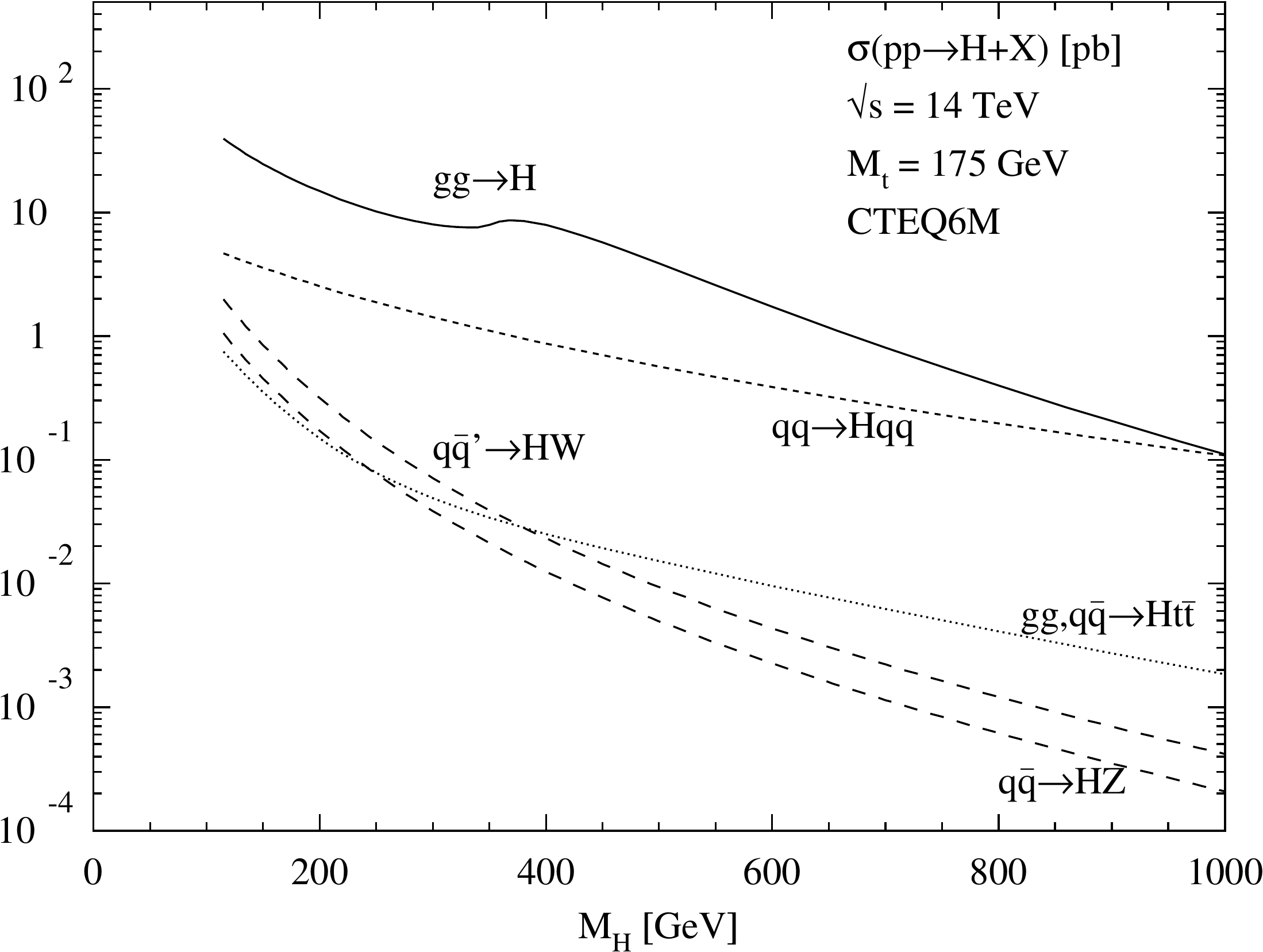}
  \includegraphics[width=0.48\textwidth]{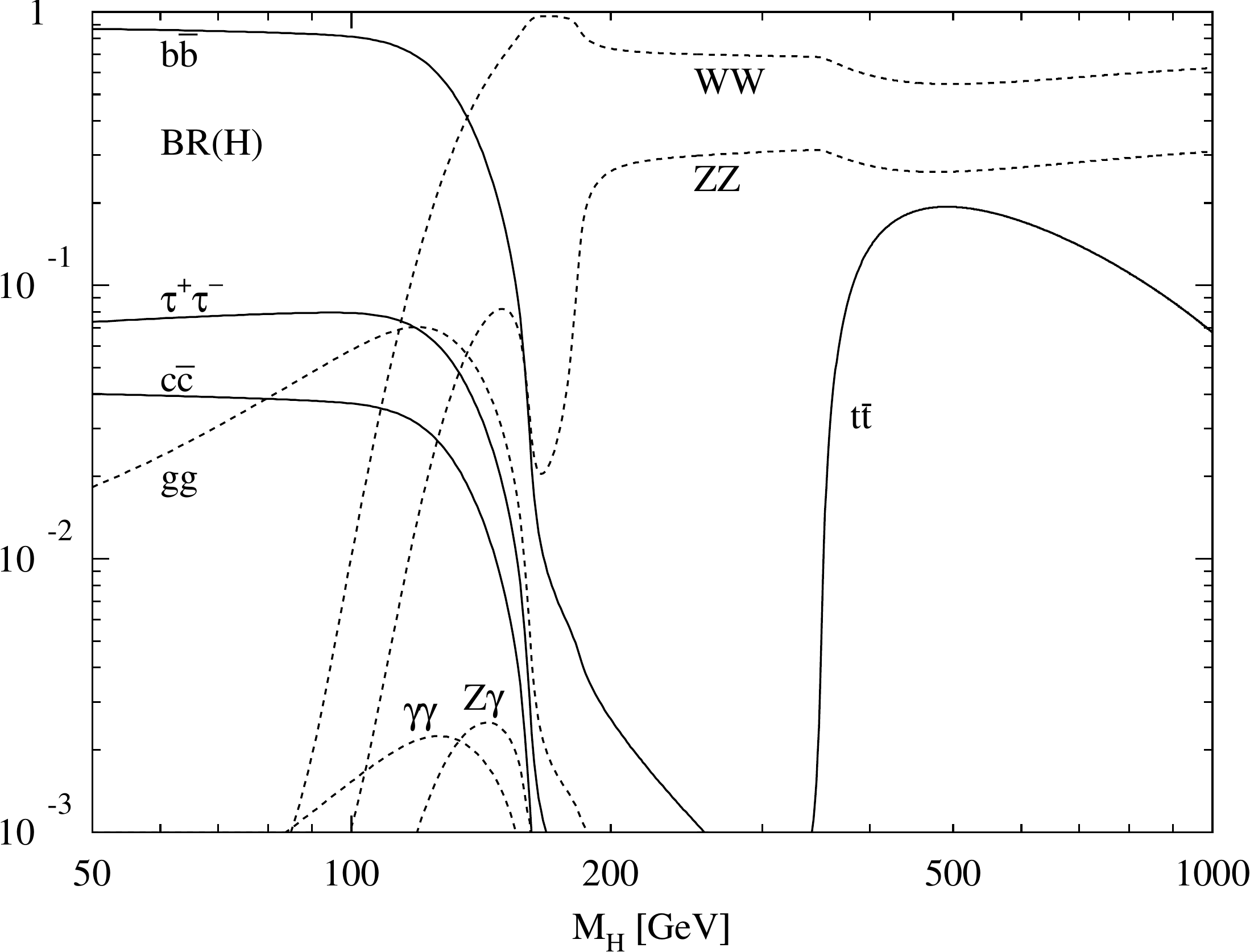}
  \capspace
  \caption{Higgs production cross sections at the LHC for the various production mechanisms as a function of the Higgs mass (left). Branching ratios of the dominant decay modes of the SM Higgs particle (right)}
  \label{fig:HiggsProdcutionAndDecay}
\end{figure}

The primary objective of the CMS is to search for the proposed signature of the Higgs boson which is predicted by the Standard Model of particle physics and is believed to be responsible for the electroweak symmetry breaking. After A. Salam and S. Weinberg revised the electroweak theory of S. Glashow in the light of local gauge invariance, it was predicted that there should be massive gauge bosons unlike photon. In early seventies, Gerard 't Hooft showed that the gauge theories are automatically renormalizable. The solution was found by adding an extra field which keeps the local gauge invariance of the Lagrangian of the system while giving mass to the gauge boson by a mechanism called ``Higgs Mechanism". It suggests that picking a particular ground state of the system may not conserve the symmetries of the Lagrangian. Then, if the Lagrangian is rewritten in terms of the picked ground states, there remains a massive scalar particle (Higgs particle) and a massive gauge field (Higgs field). Then the predicted intermediate vector bosons W$^(\pm)$ and Z$^0$ observed by UA1 and UA2 collaborations in 1983.
From Figure \ref{fig:HiggsProdcutionAndDecay} (right-side plot) it can be observed that the Higgs boson has many decay channels with branching ratios depending on the Higgs boson mass. It can be seen that the decays of the Standard Model Higgs boson into a pair of photons (HSM$\rightarrow\gamma\gamma$) is significant if the Higgs boson mass is below 150 GeV/c$^2$. Although the dominant decay mode of the Higgs is \textit{bb} in this mass range, the $\gamma \gamma$ decay mode can be well identified experimentally \cite{CMS_TDR2}.

Beyond the Standard Model, SUperSYmmetry (SUSY) \cite{SUSY} is considered to be a good candidate for solving the problem of the potentially diverging mass of the Higgs boson, the so-called `hierarchy problem'. Supersymmetry offers that, for each boson, there is a corresponding fermion with exactly the same quantum numbers. The useful point of such a fantastic claim is that the extra loop correction due to the supersymmetric partner coming in with a minus sign cancels the quadratic divergences. Another strong motivation of supersymmetry is that the difference between $M_{SM}$ and $M_{SUSY}$ should be in the order of the mass of the Higgs boson, and this is why SUSY particles are expected to be discovered at TeV colliders. Moreover, requiring the conservation of R-parity prevents the so-called Lightest Supersymmetric Particle (LSP) to decay further, offering a candidate for the dark matter problem in the universe.

Besides that, all predictions of Standard Model will be tested at the kinematic regime due to the high center of mass energy of \proton collisions at the LHC. One of its prediction about the  parton - parton scattering cross section as a function of the invariant mass of the scattered parton-parton system is the main discussion of this thesis.

\clearpage
\section{CMS Design and Construction}
\begin{figure}[ht]
  \centering
  \includegraphics[width=0.98\textwidth]{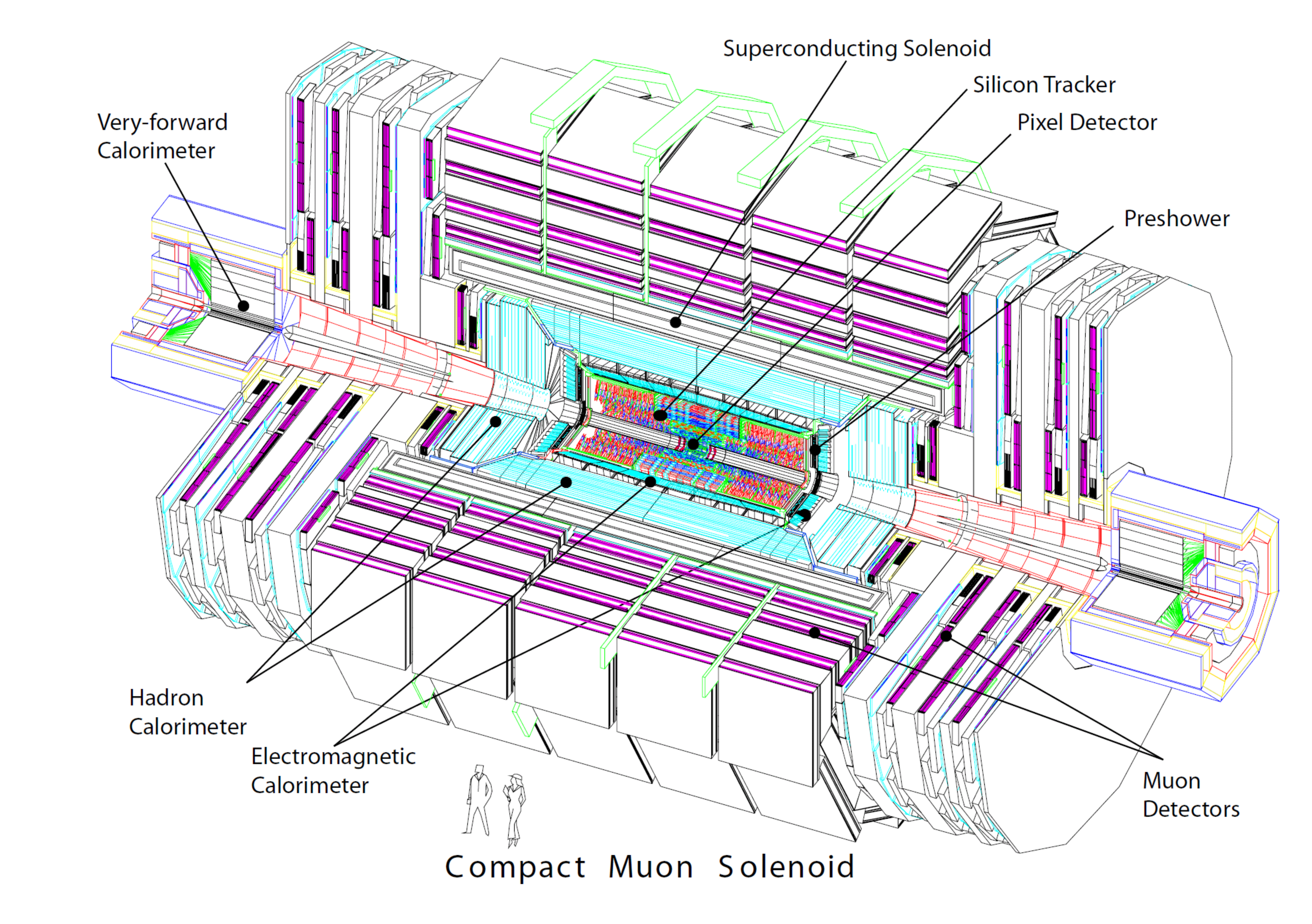}
  \capspace
  \caption{An exploded view of the CMS detector \cite{CMS_TDR1}.}
  \label{fig:CMS}
\end{figure}

CMS takes the name from its compactness\footnote{The solenoid of CMS encloses all the calorimetric systems.} with respect to its sister ATLAS and the capability of measuring the momentum of high energy muons accurately thanks to solenoid type magnet that can reach to a maximum field strength of 4 Tesla. CMS has a diameter of 15m and a length of 28.7m, small compared to ATLAS (25m diameter and 44m length). However, CMS weighs 14000 tonnes which is about two times heavier than ATLAS. The detector was designed as concentric layers of sub-detectors; the silicon tracker, electromagnetic calorimeter, hadronic calorimeter and the muon system at the outermost part (Figure \ref{fig:CMS}).

The coordinate convention of the CMS detector is as follows; the interaction point is accepted as the origin, with the x-axis pointing radially inward toward the center of the LHC and y-axis pointing vertically upwards. z-axis points along the
beam line in the direction of the Jura Mountains from LHC Point 5. The azimuthal angle $\phi$ is measured from the x-axis in the $x-y$ plane. The polar angle $\theta$ is measured from the z-axis. The pseudorapidity ($\eta$), which is commonly used in particle physics, is defined as;
\begin{equation}\label{ECAL Resolution}
\eta =-ln\left[tan\left(\dfrac{\theta}{2}\right)\right]
\end{equation}
\clearpage
	\subsection{Magnet System}
\begin{figure}[ht]
  \centering
  \includegraphics[width=0.60\textwidth]{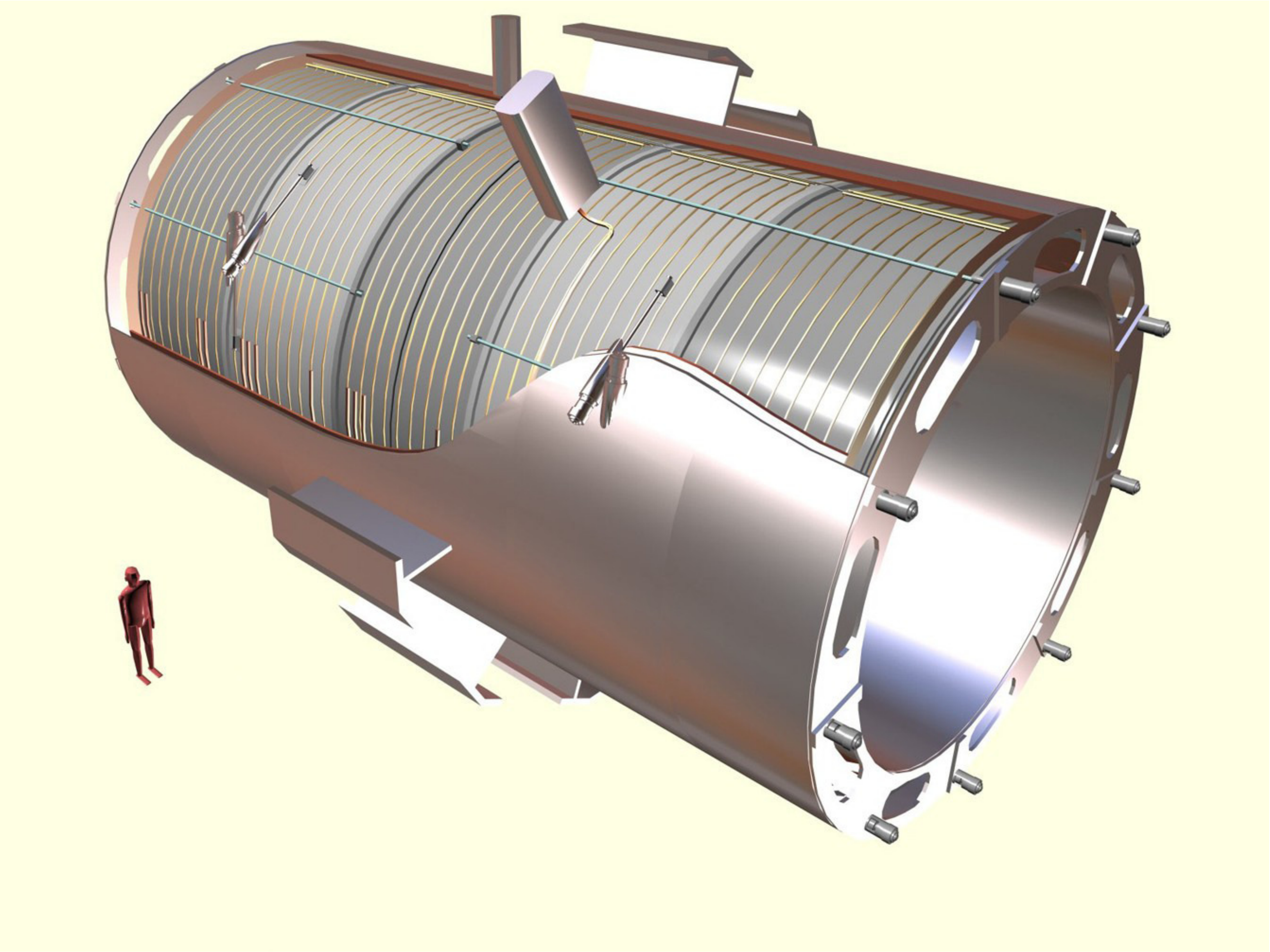}
  \capspace
  \caption{Artistic view of the 5 modules composing the cold mass inside the cryostat, with details of the supporting system (vertical, radial and longitudinal tie rods) \cite{CMSExperiment}.}
  \label{fig:Magnet}
\end{figure}

	The muon detection requirements drive the specifications and field configuration of the CMS magnet. High magnetic field is needed to make accurate momentum measurements of the TeV scale muons. CMS employs a large superconducting solenoid type magnet that can reach a magnetic field of about 4 T. The dimensions of magnet are 6 m in diameter and 12.5 m in length, and it weighs 12000 tonnes (Figure \ref{fig:Magnet}). The stored energy in the magnet is  2.6 GJ at full current. The flux is returned via a 10 000-t yoke integrated to the muon system. Three layers of iron return yokes are interlaced with the muon detectors to host and return the magnetic flux. Moreover, the return yokes are responsible for filtering. They absorb hadrons and let only muons and weakly interacting particles to pass through. 
	
	\subsection{Central Tracking System}
\begin{figure}[ht]
  \centering
  \includegraphics[width=0.90\textwidth]{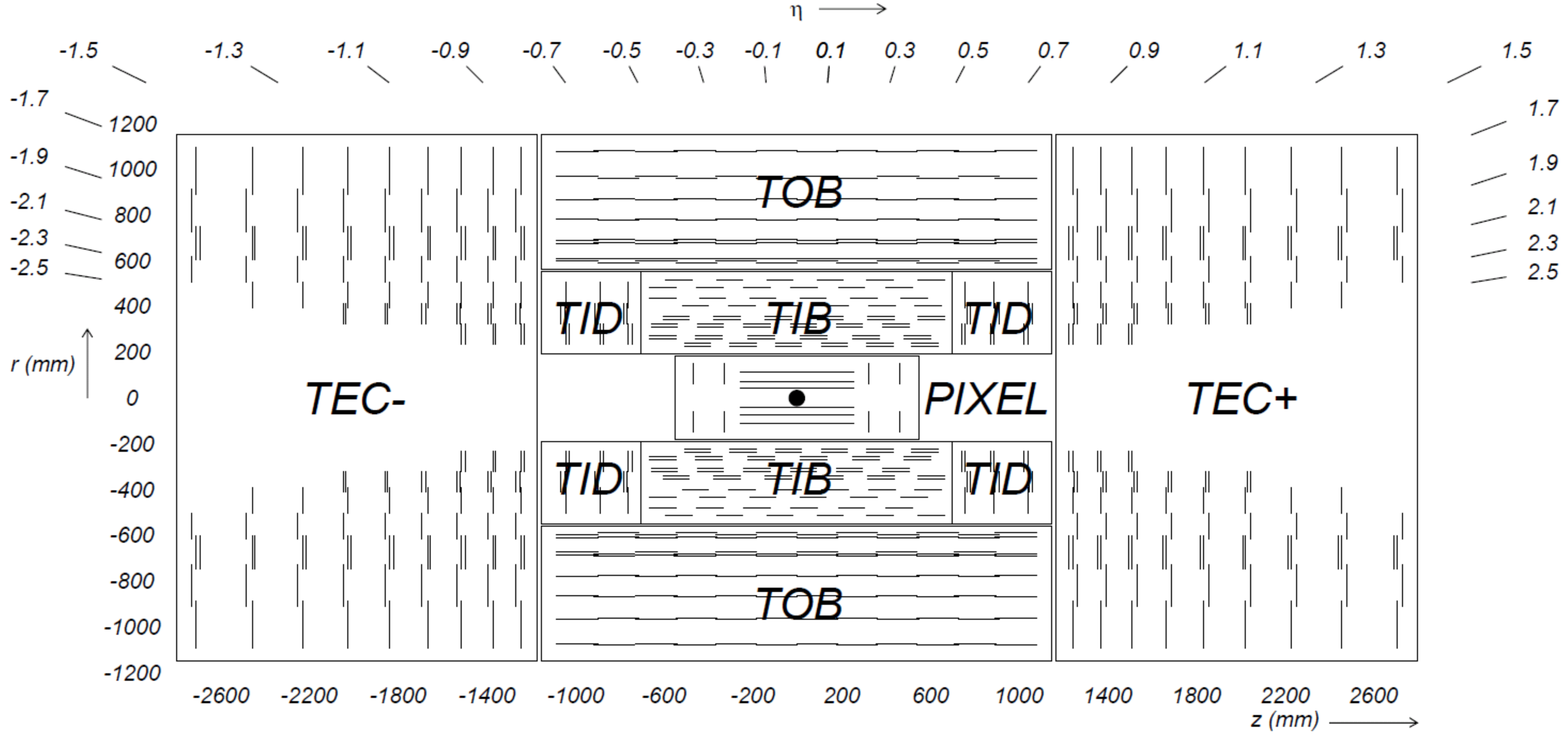}
  \capspace
  \caption{Schematic cross section through the CMS tracker. Each line represents a detector module. Double lines indicate back-to-back modules which deliver stereo hits \cite{CMSExperiment}.}
  \label{fig:Tracker}
\end{figure}

The silicon tracker is the most inner part of the CMS detector which is responsible for extracting the trajectories of charged particles and measuring their momenta. High particle flux expected from \proton collisions brings the necessity of constructing a fast and radiation hard tracker system. The system is based on a silicon technology which endures the violent radiation for several years and its design is totally driven by requirements of the LHC physics programme. The most critical specifications are: precise reconstruction of charged particle trajectories above      1 GeV/c and good secondary vertex resolution for heavy flavor identification. Mainly, the tracker consists of two parts; the inner tracker is a silicon pixel detector with a total surface of 1 m$^2$ and 66 million pixels, a silicon strip sensors in the outer region covering a total area of 200 m$^2$ surrounds the inner tracker (Figure \ref{fig:Tracker}). The dimensions of the tracker are 5.8 m in length and 2.5 m in diameter. Thus, the  acceptance is up to $|\eta|<$2.5. The homogeneous magnetic field of nominal 4 T is present for the whole volume of tracker. 

\begin{figure}[ht]
  \centering
  \includegraphics[width=0.45\textwidth]{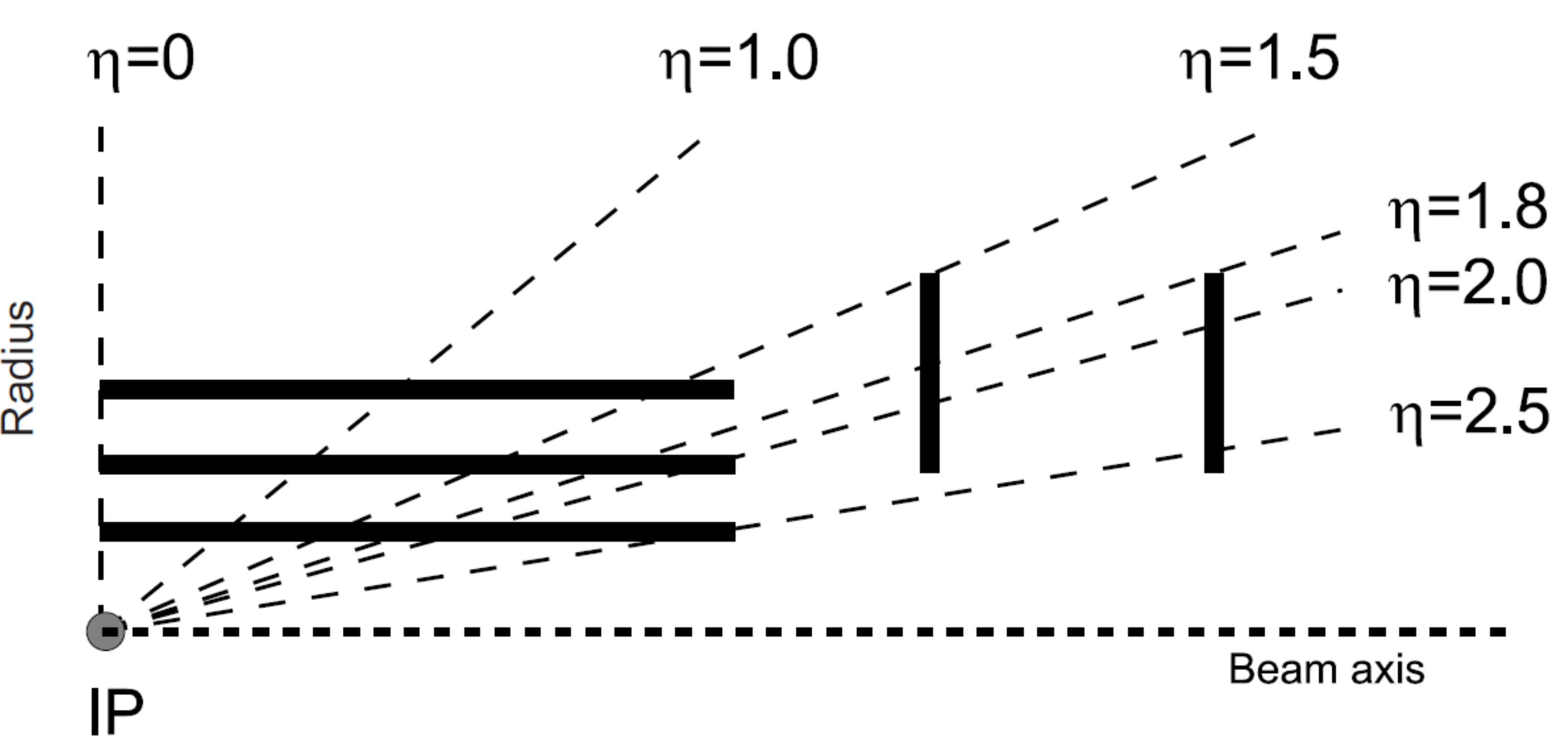}
  \hspace{10mm}
  \includegraphics[width=0.45\textwidth]{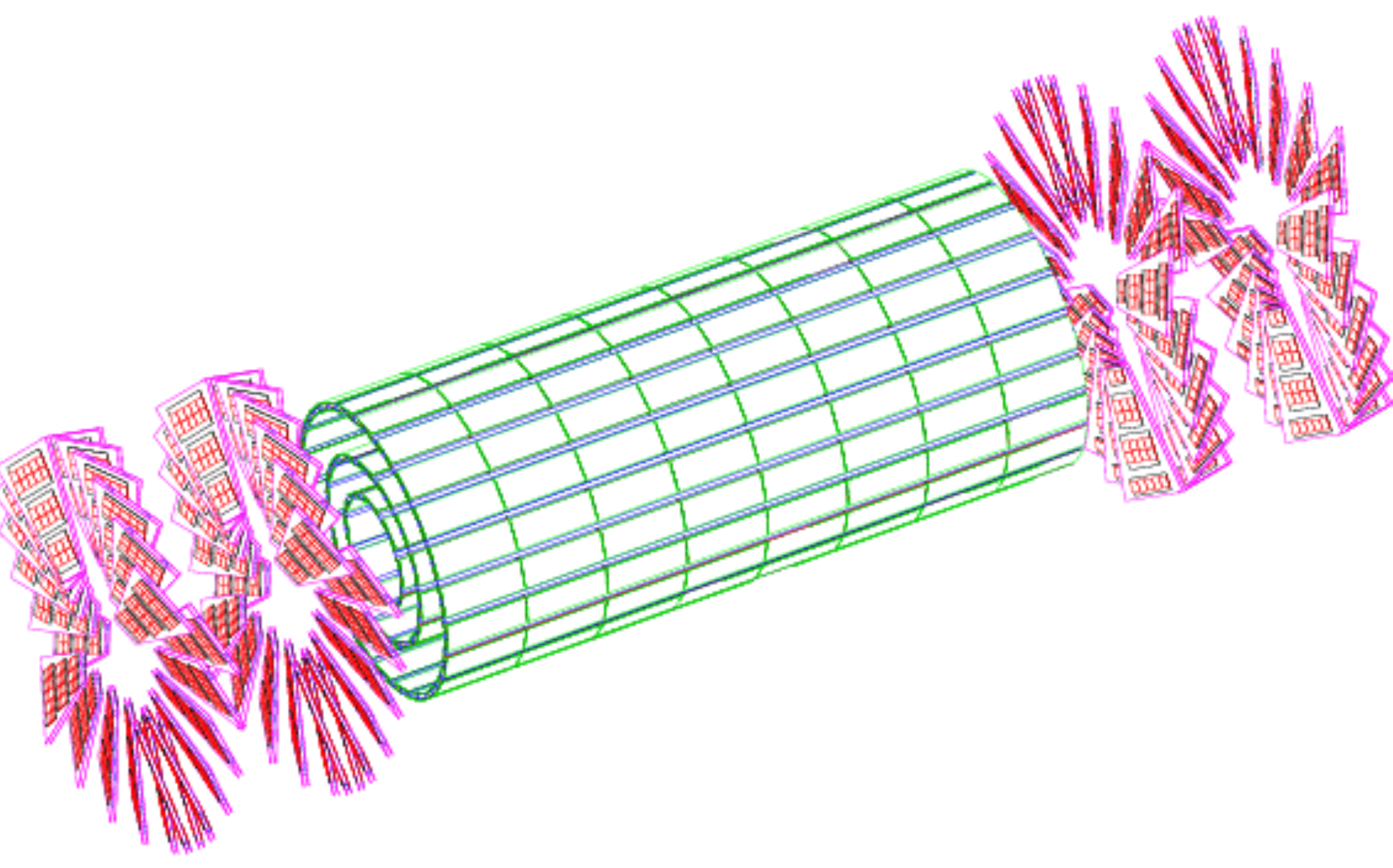}\\
  \vspace{20mm}
  \includegraphics[width=0.90\textwidth]{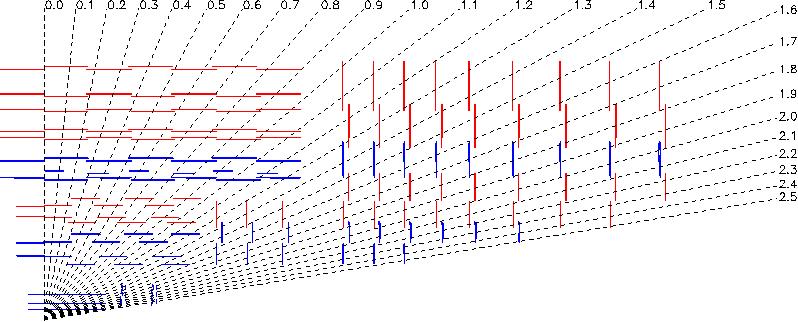}
  \capspace
  \caption{Geometrical layout of the pixel detector(top) and schematic layout of the silicon microstrip detector (bottom).}
  \label{fig:PixelAndSilicon}
\end{figure}

Three pixel layers are positioned at a radial distance of 4.4, 7.3 and 10.2 cm from the beam axis (Figure \ref{fig:PixelAndSilicon}, top). On each side of the barrel two discs complement the tracker at z = $\pm$32.5 and z = $\pm$46.5 cm. The size of each pixel is 100$\times$150 $\mu$m$^2$ and there is a total number of 65 million pixels. The silicon strip tracker encircles the pixel tracker (Figure \ref{fig:PixelAndSilicon}, bottom). The inner and outer part is different. The inner barrel (TIB) consists of four layers ranging from 20 to 55 cm and covering $|z|<$ 65 cm. Three tracker inner discs (TID) are positioned at each end in the region of 65 $<|z|<$ 110 cm. The inner strip tracker performs four spatial point measurements for each trajectory. The resolution for a single point measurement is 23 to 34 $\mu$m. The inner part is beset by the tracker outer barrel (TOB) comprising 6 layers, which extends from 55 to 116 cm in radius and $\pm$118 cm in z. The barrel part performs 6 measurements with a single point resolution in between 35 and 53 $\mu$m. The outermost part of the tracker is accomplished by 9 endcap discs (TEC) on each side ranging from 124 cm$<|z|<$ 282 cm and 22.5 cm $<$ r $<$ 113.5 cm. The endcaps can perform up to 9 measurements in total.

	\subsection{Electromagnetic Calorimeter}
	The electromagnetic calorimeter (ECAL) is the sub-detector of CMS which measures the energy of electrons and photons produced in the \proton collisions. The main design motivation is to perform a good reconstruction of di-photon decays of postulated Higgs boson if it is below 150 GeV. The CMS electromagnetic calorimeter is made of 61200 lead tungstate (PbWO$_{4}$) crystals in the barrel part and 7324 crystals in each of the two endcaps. PbWO$_{4}$ crystals have a density of 8.28 g/cm$^3$, radiation length of 0.89 cm and a Moli\`{e}re radius of 2.2 cm. These properties of PbWO$_{4}$ crystals allow a fine granularity and compactness. A preshower detector is placed in front of the endcap crystals. Avalanche photodiodes (APDs) are used as photodetectors in the barrel part and vacuum phototriodes (VPTs) in the endcaps. The barrel region covers a pseudorapidity interval of $|\eta|<$ 1.48 and it is segmented 360 - fold in $\phi$ and 2 $\times$ 85 - fold in $\eta$. In total, there are 61200 crystals installed in the ECAL barrel (EB). The front face of the crystals has a cross section of 26 $\times$ 26 mm$^2$ (0.0174 $\times$ 0.0174 in $\Delta \eta \times \Delta \phi$) and a length of 230 mm (25.8 $X_{0}$). Crystals front faces are located at a radial distance of 1.29 m from the beam axis. The single crystals in ECAL barrel are grouped into supermodules containing 17 crystals each. The ECAL endcaps (EE) cover the rapidity range from 1.48 to 3.00, and supermodules are formed out of 25 crystals. The face cross section here is 28.62 $\times$ 28.62 mm$^2$. Their length is 220 mm (24.7 $X_{0}$). The endcaps are installed with a preshower detector to identify neutral pions ($\pi^{0}$), which range from $\eta$=1.65 to $\eta$=2.60. The total thickness of the preshower detector is 20 cm (3$X_{0}$).
The expected energy resolution, for both EB and EE, is given according to formula;
\begin{equation}\label{ECAL Resolution}
\left( \dfrac{\sigma_{E}}{E}\right)^{2} = \left( \dfrac{S}{\sqrt{E}}\right)^{2}+\left( \dfrac{N}{E}\right)^{2}+C^{2}
\end{equation}
There are several reasons which affect the resolution. The stochastic term ($S$) depends on the event-to-event fluctuations, photo-statistics, and other fluctuations in the energy deposited in the preshower absorber. The constant term  ($C$) comes from the light collection non-uniformity, errors on the inter-calibration among the modules, and the energy leakage from the back of the crystal. The noise term  ($N$) accounts for the electronic, digitization, and pileup noise.

\begin{figure}[ht]
  \centering
  \includegraphics[width=0.75\textwidth]{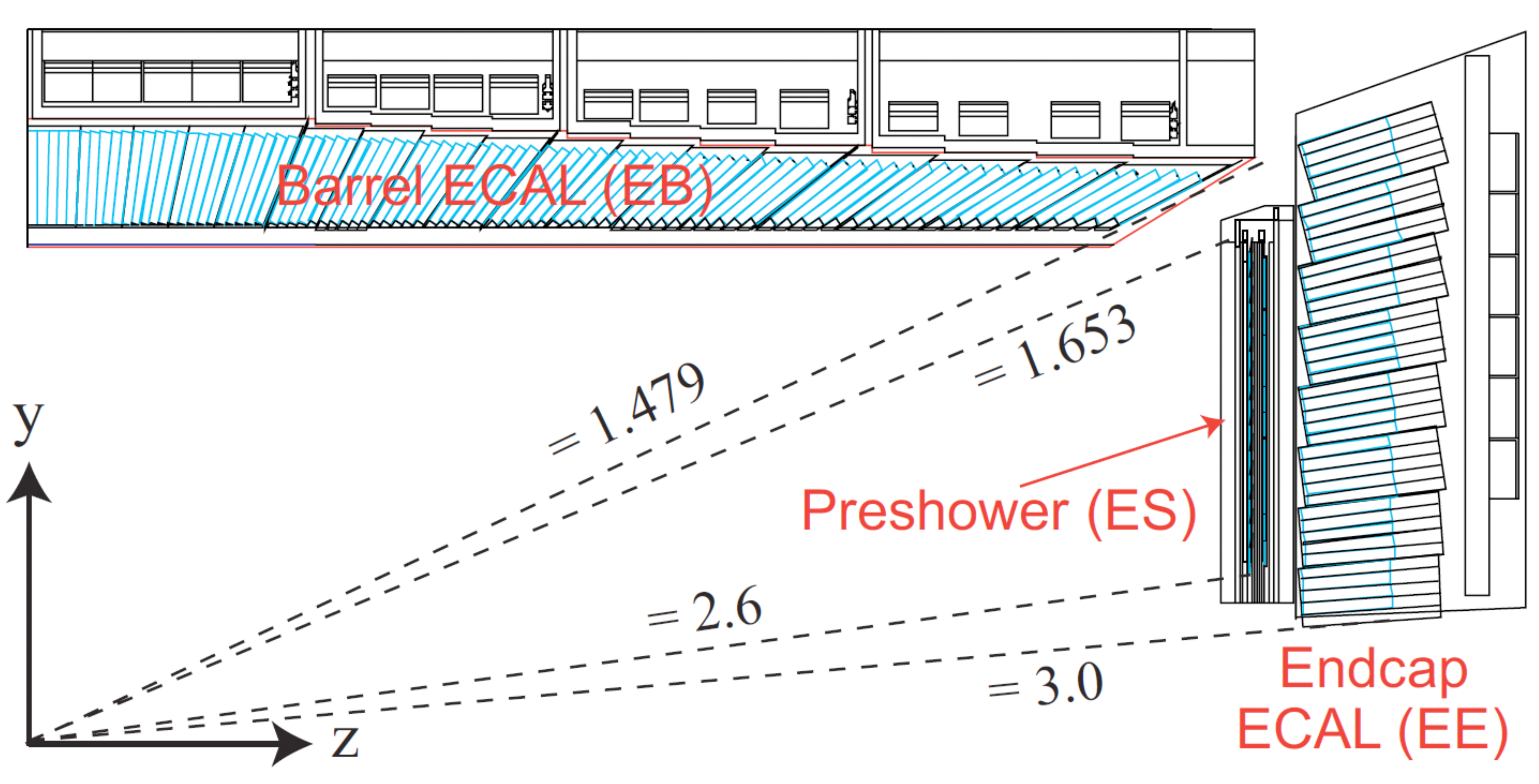}\\
  \includegraphics[width=0.75\textwidth]{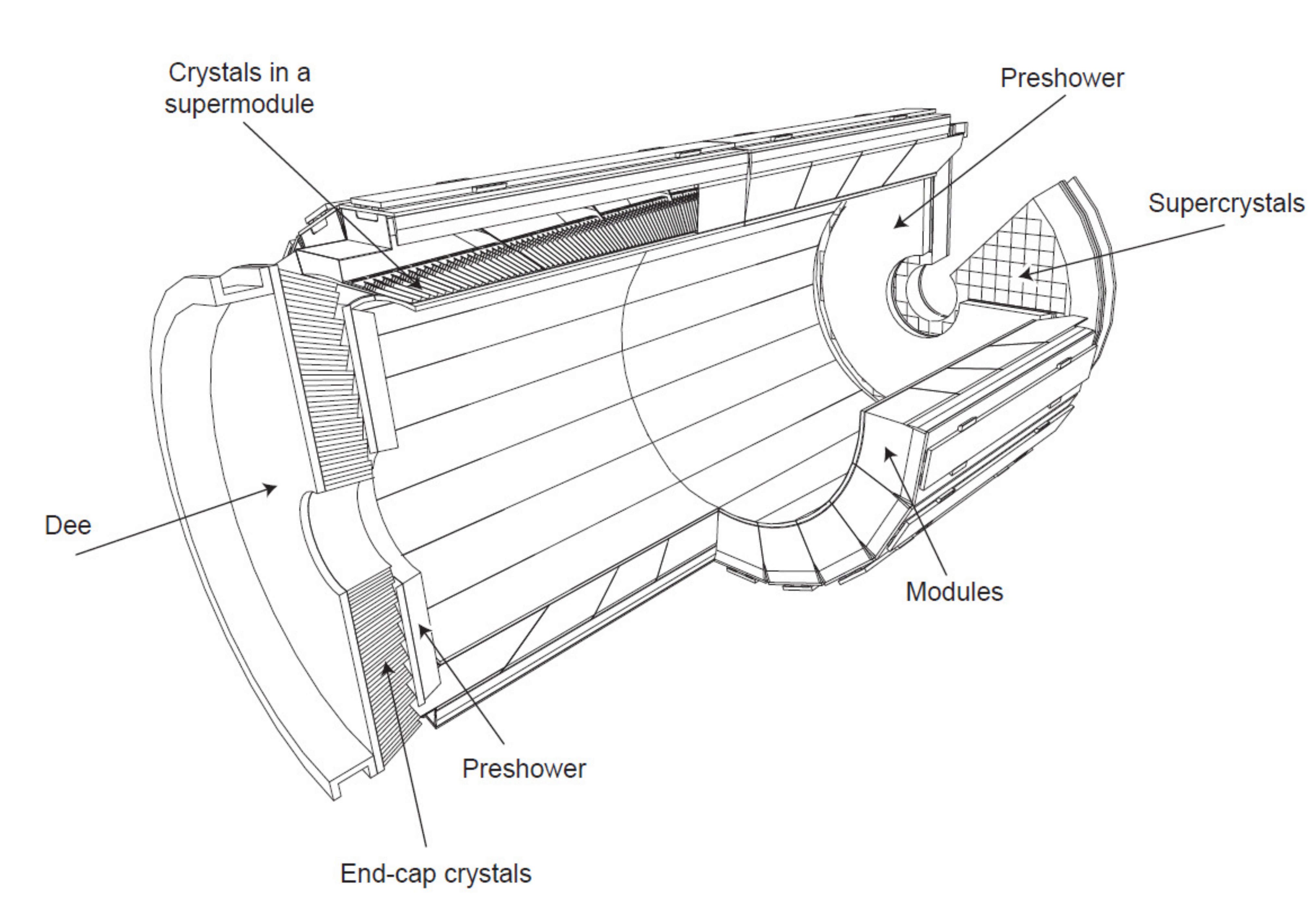}
  \capspace
  \caption{Geometric view of one quarter of the ECAL (top). Layout of the CMS electromagnetic calorimeter presenting the arrangement of crystal modules, supermodules, endcaps and the preshower in front (bottom) \cite{CMSExperiment}.}
  \label{fig:ECAL}
\end{figure}
\clearpage
	\subsection{Hadronic Calorimeter (HCAL)}
\begin{figure}[ht]
  \centering
  \includegraphics[width=0.90\textwidth]{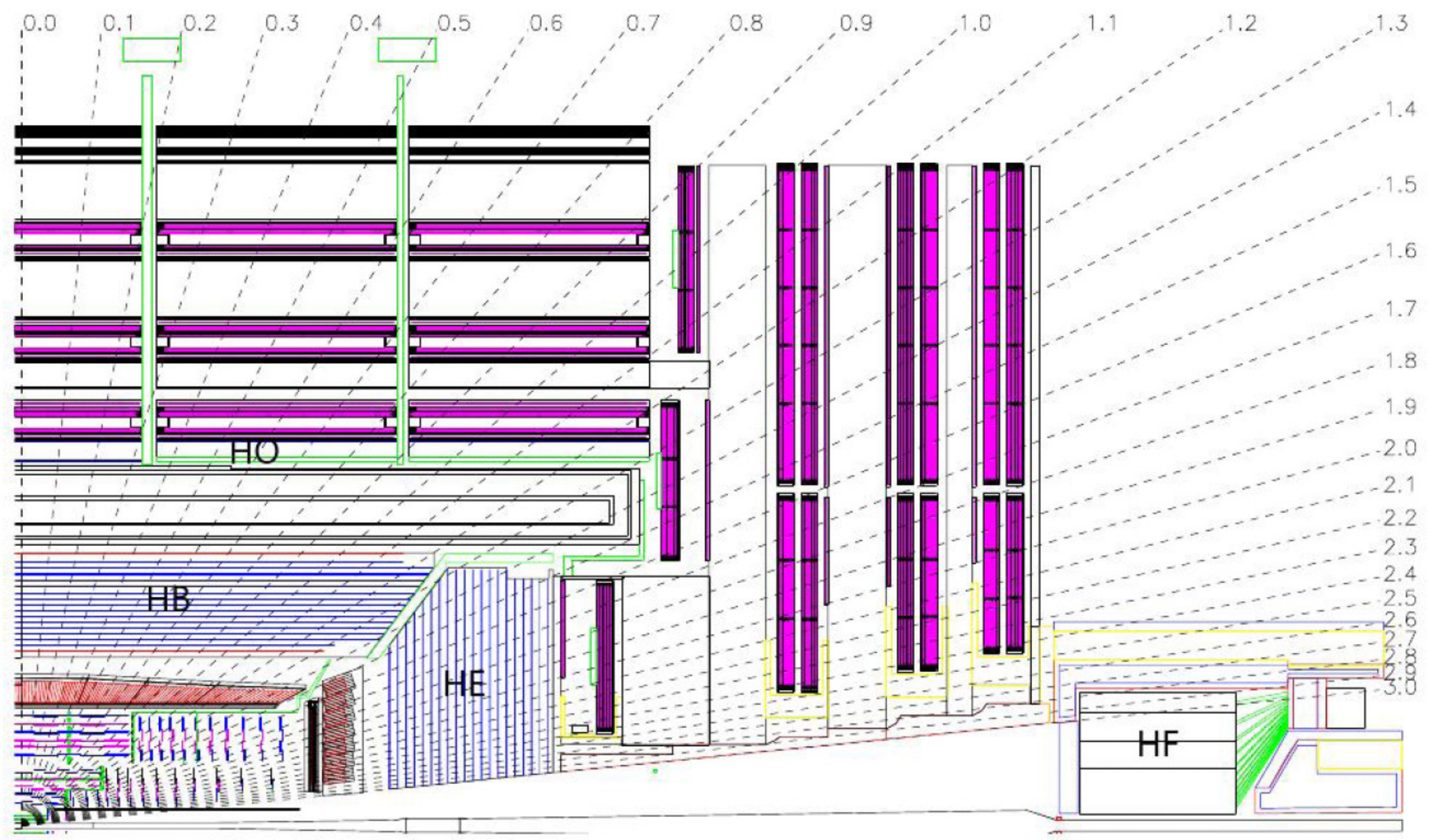}
  \capspace
  \caption{Longitudinal view of one quarter of the detector in the $r\eta$ - plane, showing the positions of the HCAL parts: hadron barrel (HB), hadron outer (HO), hadron endcap (HE) and hadron forward (HF)  \cite{CMSExperiment}.}
  \label{fig:HCAL}
\end{figure}

Since LHC is a hadron - hadron collider, the major fraction of the produced particles are hadrons. The hadronic calorimeter (HCAL) is the part of the CMS calorimeter system that is responsible for measuring the hadronic particle energies and the determination of missing transverse energy. The calorimeter ranges from 1.77 m to 2.95 m in radial dimension and is divided into four parts: barrel (HB), endcap (HE), outer barrel (HO), and hadronic forward (HF) calorimeter. Figure \ref{fig:HCAL} gives a schematic overview on the HCAL sub-detector. It reaches from the outer surface of the ECAL at 1.77 m to the inner surface of the magnet at 2.95 m radially. The hadron barrel part of the HCAL covers a region of $|\eta|<$ 1.4 and consists of 2304 towers, resulting in a segmentation of $\Delta\eta\times\Delta\phi$ = 0.087 $\times$ 0.087. Each tower is made up of alternating layers of non-magnetic brass absorber and plastic scintillator material. The reason for the absorber material to be non-magnetic is that it must not affect the magnetic field. The two hadron endcaps cover a region of 1.3 $<|\eta|<$ 3.0. They are positioned in the end parts of the CMS detector and thus are allowed to contain magnetic material. Here iron is used as the absorber material. The granularity begins from $\Delta\eta\times\Delta\phi$ = 0.087 $\times$ 0.087 at $|\eta|<$ 1.6 up to $\Delta\eta\times\Delta\phi$ = 0.17 $\times$ 0.17 at 1.6  $<|\eta|<$ 3.0.

An additional calorimeter, the outer barrel, is needed to be placed outside of the magnet to absorb escaping hadron showers from particles with transverse energies above 500 GeV. Without the outer barrel, these particles would cause a large missing transverse energy which is not convenient for many physics analysis purposes. The granularity and $\eta$ range of outer barrel is the same as the hadron barrel. The forward region of 3.0 $<|\eta|<$ 5.0 is covered by the steel/quartz fiber hadron forward calorimeter. The whole coverage of the HCAL is almost hermetic;~$0<\phi<2\pi$ in azimuth and $0<|\eta|<5.0$ in pseudorapidity.

In order to reconstruct jets from measured energies of hadrons, there exists different algorithms (See Section \ref{Jet Reconstruction}). One of these adds the $E_{T}$ of crystals close in $\eta-\phi$ to the crystal with the highest energy deposit and creates a proto-jet. This recombining procedure is repeated taking the proto-jet as the jet axis until the parameters of the proto-jet is stabilized.

In the head on collision of protons, the initial total transverse momentum and energy are zero. From the conservation of energy and momentum, the sum of these should stay as zero after the collision. If this sum differs after the collision, this is a clear indication that some particles have not been detected by the detector. Each cell in the calorimeter produces a four-vector, with an energy equal to the measured energy in the cell (massless), a direction pointing from the vertex to the center of the cell. The nonzero total transverse momentum is considered as the momentum imbalance arisen because of the non - detected particles. Thus, the vector of total transverse momentum, with the minus sign, is called as \textit{missing transverse energy} (MET).
\begin{equation}\label{MET}
\vec{E}_{T,miss}=-\sum_{i}\vec{E}^{i}_{T}
\end{equation}
where $\vec{E}_{T,miss}$ is the four vector of MET and $\vec{E}^{i}_{T}$ are the four vectors of the calorimeter cells.

The overall energy resolution (combined with ECAL) is given as;
\begin{equation}\label{HCAL Resolution}
\left( \dfrac{\sigma_{E}}{E}\right)^{2} = \left( \dfrac{120\%}{\sqrt{E}}\right)^{2}+(6.9\%)^{2}
\end{equation}
Both granularity and energy resolution of the HCAL are worse than ECAL. Therefore the overall precision of the calorimeter output is determined by the HCAL.

	\subsection{Muon Chamber}
\begin{figure}[ht]
  \centering
  \includegraphics[width=0.90\textwidth]{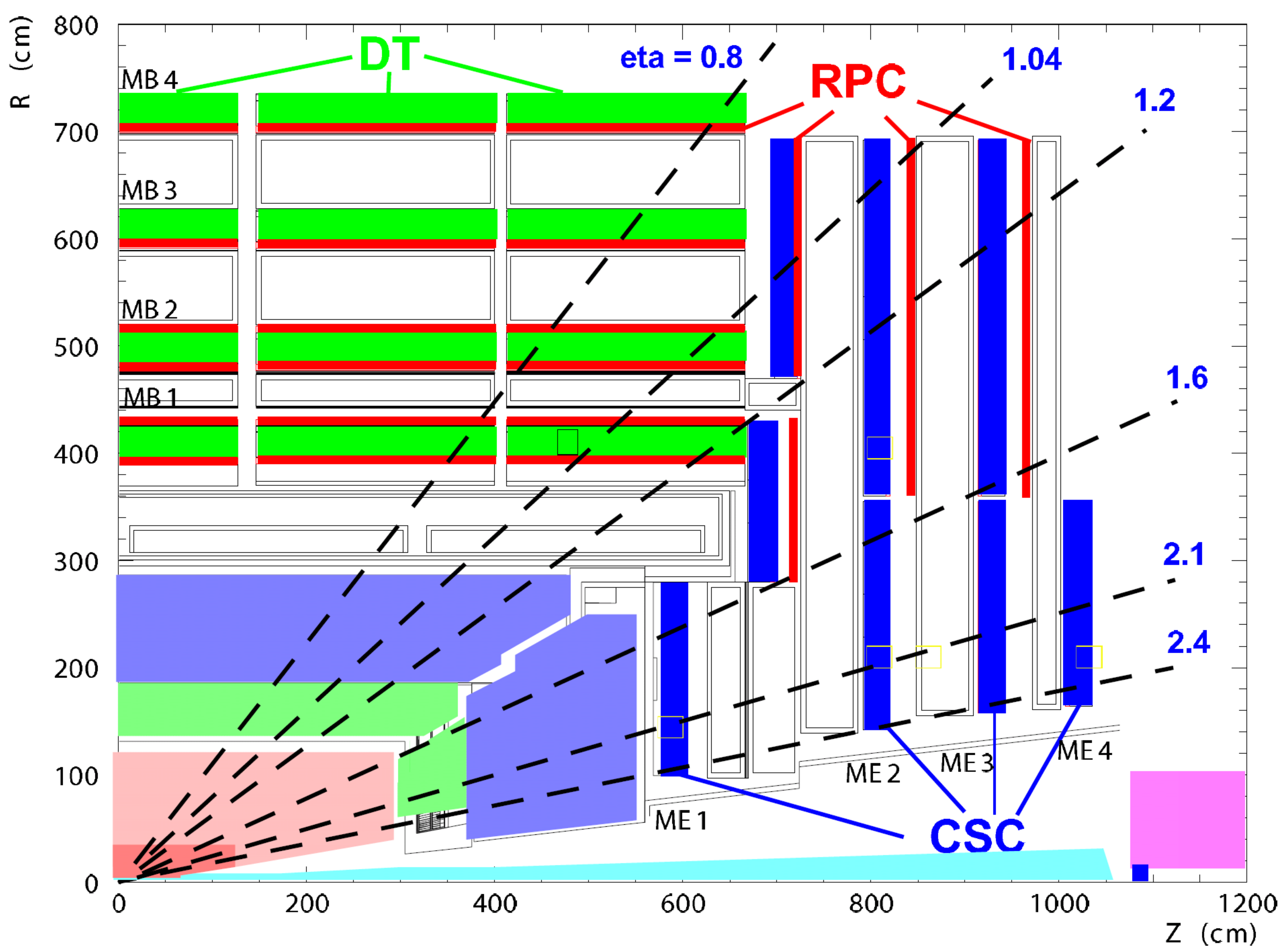}
  \capspace
  \caption{A longitudinal view of the muon system indicating the location of the three detector types contributing to the muon spectrometer \cite{CMSExperiment}.}
  \label{fig:Muon}
\end{figure}

If the Higgs boson mass is above 115 GeV/c$^2$, the $H\rightarrow ZZ \rightarrow 4\mu$ and $H\rightarrow WW \rightarrow 2\mu$ decay channels become dominant (Figure \ref{fig:HiggsProdcutionAndDecay}). Hence, a clear muon identification plays a crucial role in a prospective discovery of the Higgs boson. The muon system of the CMS detector is a tracking device in the outermost region. Only muons and non-interacting particles such as neutrinos can achieve pass through all  the calorimetric systems without leaving a large fraction of their energy. An important task of the muon system is to provide a fast recognition and an efficient reconstruction of the muons. Three different type of gaseous detectors are used to construct a complete muon system. In the barrel region ($|\eta|<$ 1.2) drift tube chambers (DTCs) are mounted. There are four layers of muon chambers positioned
in the return yoke, and cathode strip chambers (CSC) are located in the endcap (0.9 $<|\eta|<$ 2.4) (Figure \ref{fig:Muon}).
In order to improve muon trigger system and for a good measurement of the bunch crossing time, resistive plate chambers (RPC) are mounted in the barrel and endcap region ($|\eta|<$ 1.6). They provide a fast response, which is much shorter than the bunch
crossing time, but with coarser spatial resolution. In total, muons are measured up to four times. Once in the inner tracking system and three times in the muon system.

	\subsection{Trigger and Data Acquisition System}
At the design luminosity of 10$^{34}$ \lumi , each bunch, traveling oppositely inside the beam pipes, encounters the other 40 million times per second and they passes through each other. In other words, the bunch crossing frequency is 40 MHz. The average \proton collision per each bunch crossing is approximately 20. This corresponds to a data rate of 1.2 Tbyte/s, which is practically impossible to store in any tape or disk. Thus, the data are reduced first by the level-one trigger (L1 Trigger). L1 trigger system is implemented as a programmable hardware system that uses information from the muon chambers and calorimeters for selecting event signatures which are specified by physics interests. However, there exists a very limited time for L1 trigger to make a decision which means a rough form of the raw data from the calorimeters and muon system should be used. The  L1 trigger system is located 90 m outside the detector which causes a latency of 3.2 $\mu s$ between the bunch crossing and the L1 accept signal. The signal information accumulated in that time interval is buffered. The L1 trigger system reduces the event rate of 40 MHz down to 100 kHz to be passed to the High Level Trigger (HLT) system.
HLT system is a software system that uses the detector signals that pass the L1 trigger. It can run on high resolution data with complex reconstruction algorithms. The HLT decisions are based on the output signals from all sub-detectors. After the HLT decisions, the event rate decreases down to 100 Hz for mass storage which corresponds to a data rate of 150 Mbyte/s.
\begin{figure}[ht]
  \centering
  \includegraphics[width=0.90\textwidth]{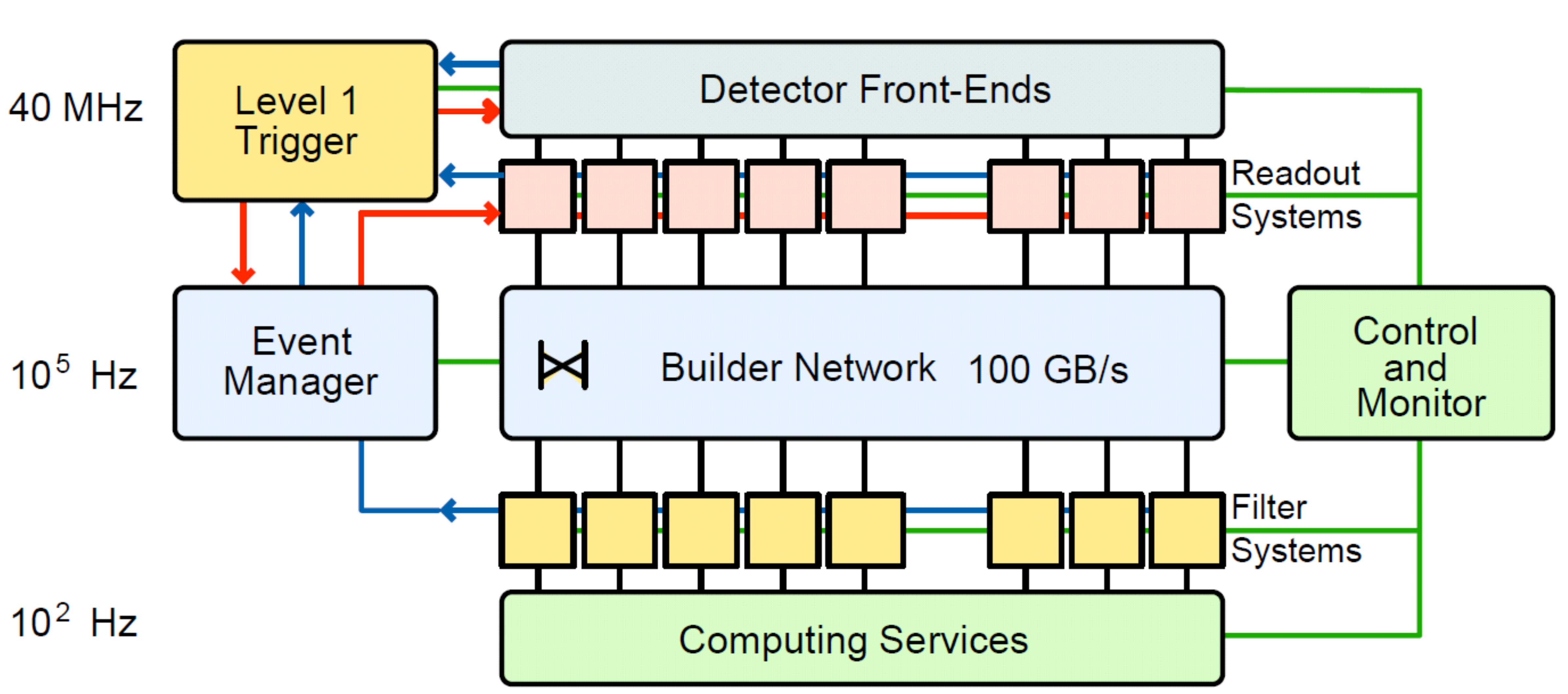}
  \capspace
  \caption{Architecture of the CMS Data Acqusition (DAQ) system.}
  \label{fig:TriDAQ}
\end{figure}

\clearpage
\chapter{EVENT GENERATION, DETECTOR SIMULATION AND JET RECONSTRUCTION}
Monte Carlo event generators are computer programs calculating particle physics processes numerically. There are two
main reasons to use Monte Carlo techniques in high energy particle physics. Firstly, cross section values must be
calculated for complex regions of the phase space in order to compare the experimental results with the theoretical
expectations, yet it is practically impossible to perform these calculations analytically. Secondly, it is always needed
to have realistic event simulations in order to have a foresight for both designing the experimental device and
analyzing its data. As discussed in the Section \ref{Quantum Chromodynamics}, QCD gives different descriptions for different scales of the
momentum transfer $Q$. First, a set of primary partons are produced by the hard subprocess that can be explained by a
fixed-order (LO or NLO or even NNLO) matrix element, then a parton shower evolves these primary partons while the scale
decreases down to a cut-off scale $Q_{0}\approx1 GeV$. Finally, the resultant partons of the shower arrange themselves
as color singlet (color neutral) hadrons. At this final step, the scale is too low, which means the strong coupling
constant $\alpha_{s}$ is large, and this hadronization process is non-perturbative. Hard matrix elements (ME) are
computed to a particular order in perturbation theory while the parton evolution process governed by the DGLAP (Dokshitzer-Gribov-Lipatov-Altarelli-Parisi) equations is done via the parton shower (PS) approach. Experimentally determined parton distribution functions are taken as input for the parton evolution. MPI and hadronization, the non-perturbative parts which are beyond our means of calculation, are evaluated by using dedicated phenomenological models. In an abstract sense, the whole process can be written as 
\begin{center}
(Hard subprocess $\oplus$ MPI) $\otimes$ Parton Shower $\otimes$ Hadronization
\end{center}
\clearpage
\begin{figure}[ht]
  \centering
  \includegraphics[width=0.80\textwidth]{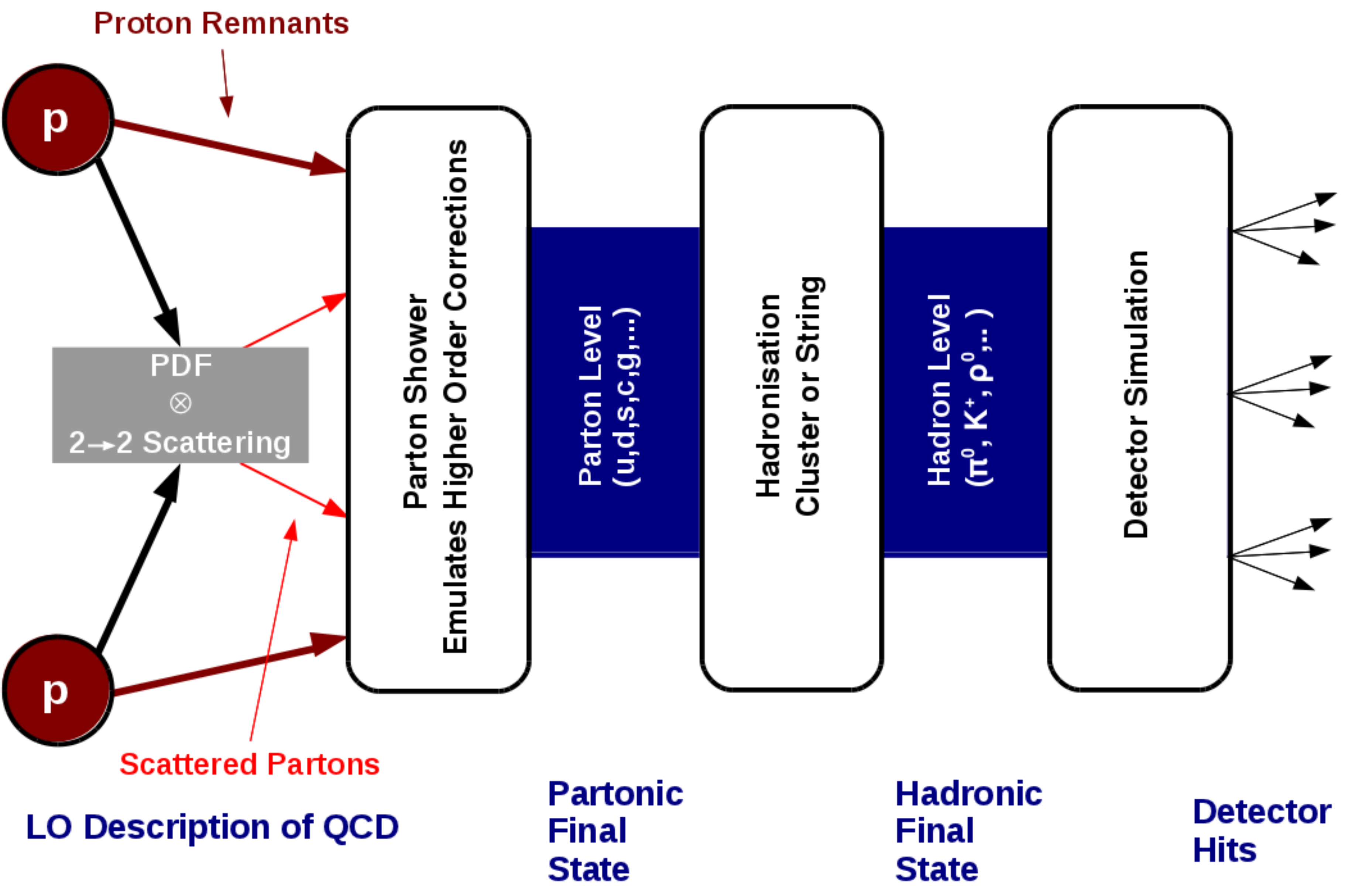}
  \capspace
  \caption{The basic steps in the generation of a simulated event.}
  \label{fig_MC_scheme}
\end{figure}

Monte Carlo generators compute single collisions (events), by this providing the opportunity not only to calculate inclusive quantities such as cross sections, but also to check different observables on an event-by-event basis.

\section{PYTHIA}
PYTHIA is a general-purpose leading-order MC event generator. It has been used extensively at LEP, HERA and the Tevatron for $e^{+}e^{-}$, $ep$ and $p\bar{p}$ physics. It contains an extensive subprocess library covering Standard Model physics with SUSY, Technicolor and many other exotic processes. The Lund string model \cite{LundStringModel} is used to describe the hadronization process. This model is based on a picture where quarks and antiquarks are linearly confined, located at the ends of a string, and gluons  are energy and momentum carrying kinks on this string. The production of a quark-antiquark pair in electron-positron annihilation can be given as a trivial example. At the annihilation point, the quark and antiquark move apart from each other, with half of the total energy for each in the center-of-mass frame. As the gluon string is stretched between them, its potential energy increases. At some critical point, when the potential energy reaches at the order of hadron masses, the strings are energetically more likely to be broken through the creation of a new quark-antiquark pair. The new antiquark at the end of the string segment which is connected to the original quark and the new quark  which is connected to the antiquark. Successive sequence of stretching and breaking continues until all the energy is converted into quark-antiquark pairs connected by short string segments, which can be considered as ``hadrons".

The generation of the underlying event is a complicated process. Different phenomenological models describing underlying event exist, with various degrees of complexities. Hence, there are different ``tunes" of Pythia. In CMS, two Pythia tunes are studued and used to generate MC samples: the Z2 \cite{Z2TUNE} tune which seems to be in a better agreement with collected data and the D6T \cite{D6T} tune is used as complementary for ``systematics"" studies.

\begin{figure}[ht]
  \centering
  \includegraphics[width=0.35\textwidth ,angle=90]{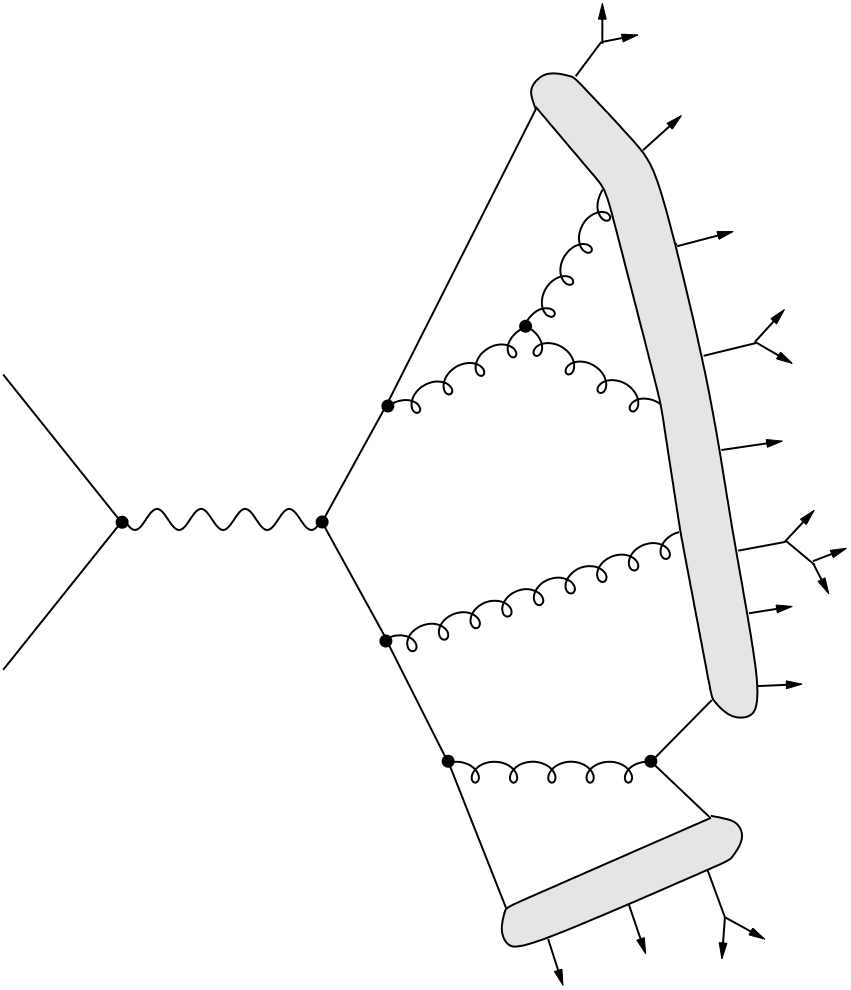}
  \hspace*{1cm}
  \includegraphics[width=0.40\textwidth]{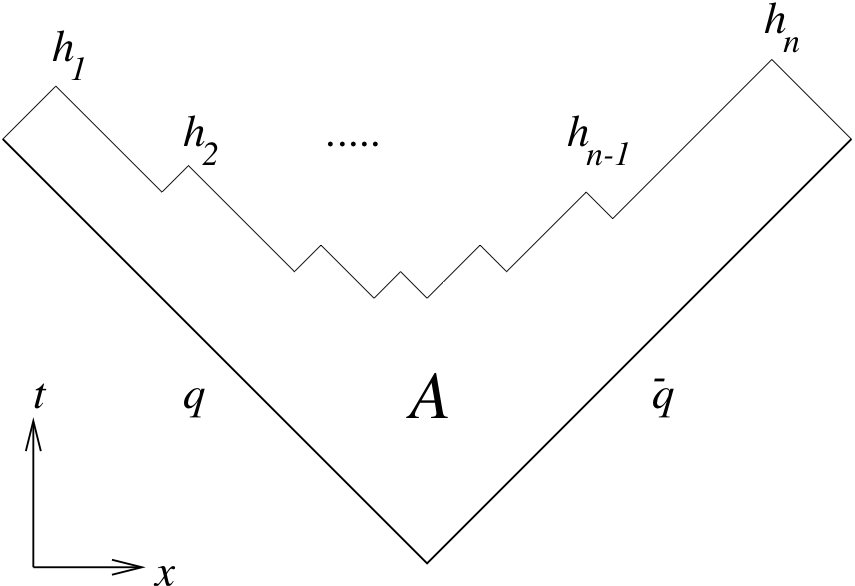}
  \capspace
  \caption{Schematic representation of the string model (left). Space-time picture of string hadronization (right).}
  \label{fig_Lund}
\end{figure}

\section{Herwig++}
Herwig++ is a C++ version of the Herwig (Hadron Emission Reactions With Interfering Gluons) \cite{Herwig} event generator, whose earlier versions were programmed in Fortran. It includes some improvements compared to Herwig6 such as the covariant formulation of the parton shower and mass-dependent splitting functions. Herwig also uses leading order calculations with different parton shower and hadronization models than Pythia uses. It has its own tunes and an implementation of a parton shower which uses a cluster hadronization model. That implementation of the different parton shower model gives the  possibility of extracting uncertainties by comparing Pythia and Herwig outputs. 
The cluster hadronization model is based on the pre-confinement property of QCD. It has been shown that at evolution scales, much less than the hard subprocess scale, $q\ll Q$, the partons in a shower form color singlet groups with an invariant mass distribution. Nevertheless, invariant mass distribution of the formed group is independent of the scale of the hard subprocess; it depends only on $q$ and $\Lambda_{QCD}$. Then, these clusters at the hadronization scale $Q_0$ can be identified as proto-hadrons which are candidates to decay into observed hadrons.
\begin{figure}[ht]
  \centering
  \includegraphics[width=0.35\textwidth ,angle=90]{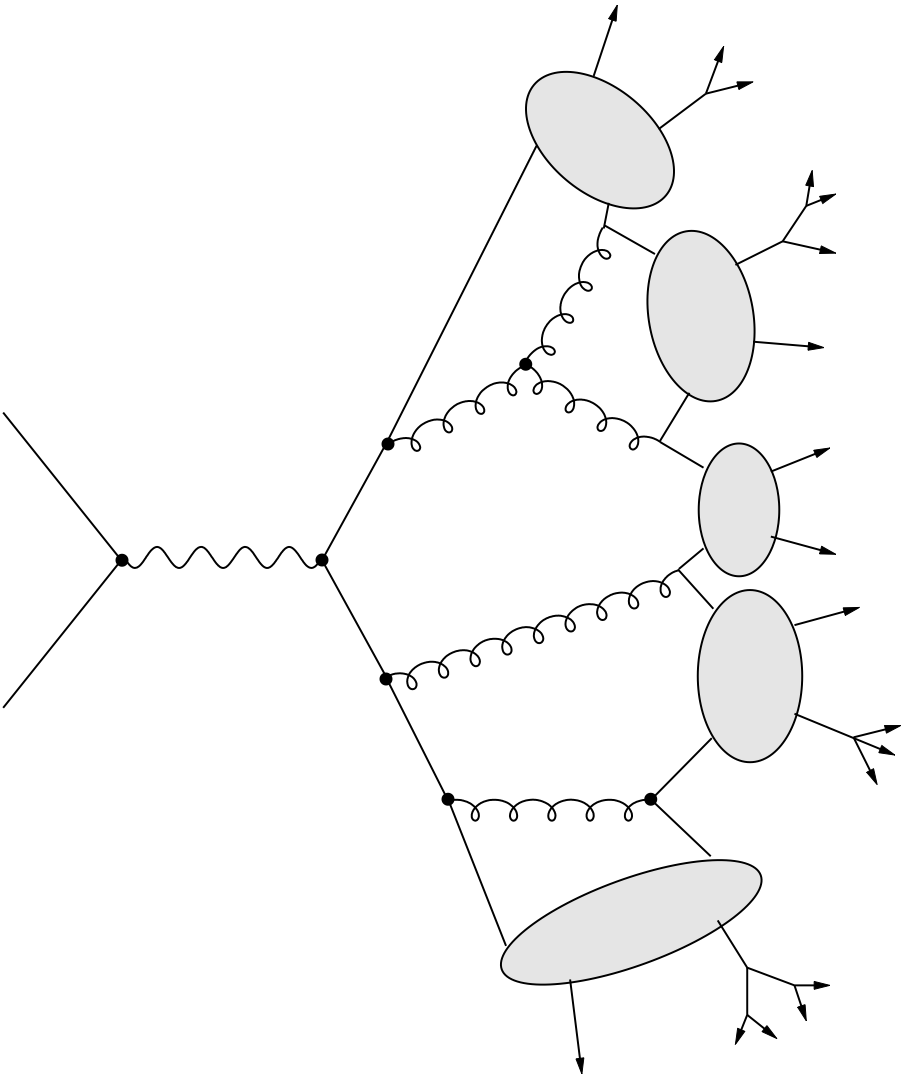}
  \capspace
  \caption{Cluster Hadronization Model.}
  \label{fig_Cluster}
\end{figure}

\section{Detector Simulation}
A realistic simulation of the CMS detector is based on the GEANT4 \cite{GEANT4} toolkit. It relies on a detailed description of the sub-detector volumes and materials, and the necessary information about the ``sensitive detector". It takes generated particles as input, passes them through the simulated geometry, and models physics processes that accompany particle passage through matter. Results of each particle's interactions with matter are recorded in the form of simulated hits. Energy loss by a given particle inside ``sensitive volume” of one of the sub-detectors, recorded with several other characteristics of the interaction is an example of a simulated hit. Generated particles are called as ``primary", and the particles originating from GEANT4-modeled interactions of a primary particle with matter are called as ``secondary". These simulated hits are then used as input to emulators which mimic the response of the detector readout and trigger electronics and digitize this information by also considering noise and other factors.

\section{Jet Reconstruction}
\label{Jet Reconstruction}
\subsection{Jet Reconstruction Algorithms}
As it was discussed in Chapter 2, scattered partons from the hard subprocess eventually turn into a spray of hadrons
due to color confinement. This spray of particles can be identified as jet objects in the detector through the application of  a set
of mathematical rules by taking detector entries as inputs. A jet is reconstructed by using energy depositions in calorimeter towers and track momentum information of charged particles, and by clustering this information in an appropriate way to assign a four vector to the jet object. Since a jet algorithm is a complicated combinatoric process and highly definition dependent, theorists and experimenters decide on the features of a good jet reconstruction algorithm that should satisfy the following \cite{Salam:2009jx} ;

\begin{itemize}

\item A jet algorithm should be simple to implement in an experimental analysis;
\item A jet algorithm should be simple to implement in the theoretical calculation;
\item A jet algorithm should be defined at any order of perturbation theory;
\item A jet algorithm should yield finite cross sections at any order of perturbation theory;
\item A jet algorithm should yield a cross section that is relatively insensitive to hadronization.
\end{itemize}

\begin{figure}[ht]
  \centering
  \includegraphics[width=0.70\textwidth]{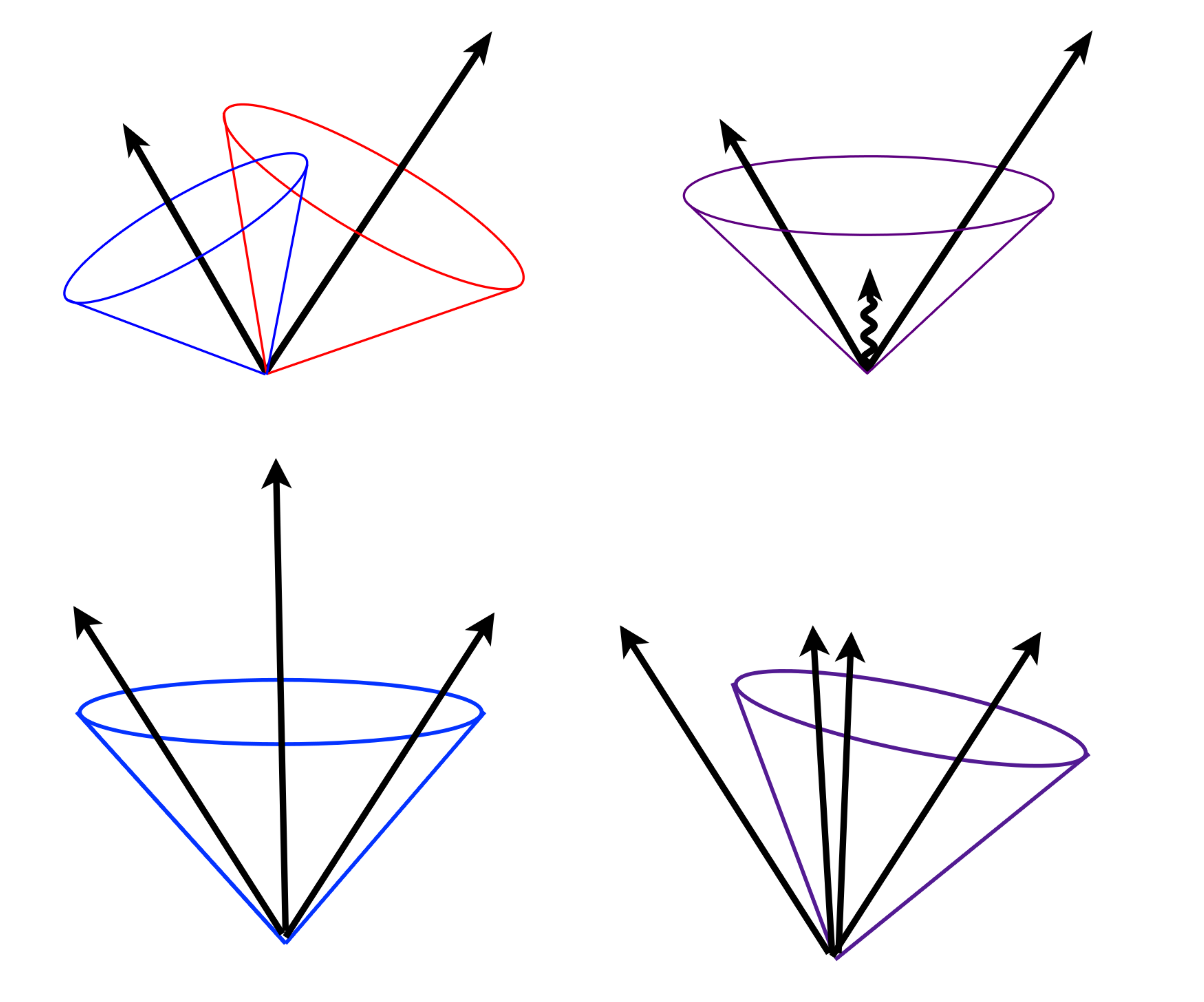}
  \capspace
  \caption{Illustration of the infrared sensitivity of a cursory designed jet algorithm (top). Illustration of the
product of a collinear unsafe jet algorithm. A collinear splitting changes the number of jets (bottom).}
  \label{fig_InfSafety}
\end{figure}

Infrared and collinear (IRC) safety is a fundamental requirement for jet algorithms. Infrared safety means that adding a
soft gluon should not change the results of the jet clustering. Collinear safety is splitting one parton into two
partons should not change the results of the jet clustering. The configurations of the infrared and collinear safety are shown separately in Figure \ref{fig_InfSafety}. Several algorithms have been developed during the past years, and three of them are officially chosen by the CMS collaboration: Iterative Cone, k$_{T}$ and anti-k$_{T}$.

\subsection{Iterative Cone Algorithm}
Although it lacks collinear and infrared safety, Iterative Cone (IC) algorithm is still present in the CMS
official reconstruction scheme for the practical purposes of high level trigger system. It is fast and it has a local
behavior, so it makes IC algorithm suitable to use in high level triggers. In order to reconstruct IC jets, an
iterative procedure is followed. The particle in the event with the biggest transverse energy is taken as seed, and a
cone with a radius $R=\sqrt{\delta\eta^{2}+\delta\phi^{2}}$ is built around it. All the objects contained in that cone
are merged into a proto-jet, whose direction and transverse energy of which are given as
\begin{equation}
E_{T}=\sum{i}E_{T}^{i};~~\eta=\frac{1}{E_{T}}\sum{i}E_{T}^{i}\cdot\eta_{i};~~\phi=\frac{1}{E_{T}}\sum{i}E_{T}^{i}
\cdot\phi_{i}
\end{equation}
After this determination, first proto-jet is used as the seed of the second iteration. This iterative procedure
continues until the desired minimum difference between the seed and the resultant proto-jet is achieved. Finally, the
last proto-jet is declared as ``the jet object".
\subsection{Generalised k$_{T}$ Algorithm}
\begin{figure}[ht]
  \centering
  \includegraphics[width=0.70\textwidth]{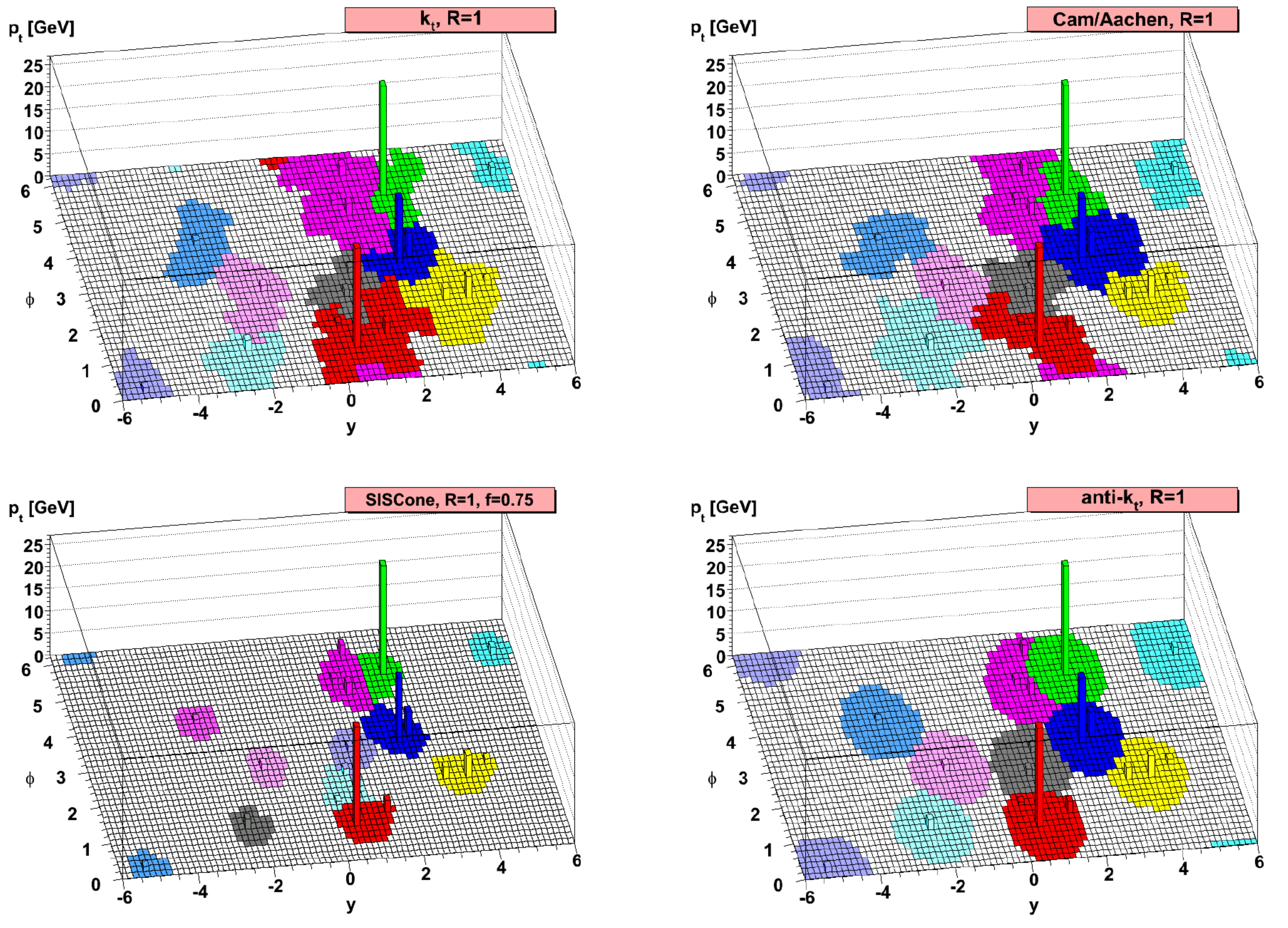}
  \capspace
  \caption{Illustration of  different jet algorithms in parton level \cite{anti_kT}.}
  \label{fig_diff_algos}
\end{figure}

The k$_{T}$, anti-k$_{T}$ and Cambridge-Aachen algorithms can be given in a single type, the generalized k$_{T}$
algorithm.The generalized k$_{T}$ algorithm represents a whole family of infrared- and collinear-safe algorithms
depending on a continuous parameter, denoted as $p$. The k$_{T}$ algorithm is based on a pair-wise recombination and it
combines two particles (or calorimeter towers) if their relative transverse momentum is less than a given threshold.
The distance $d_{ij}$ between the particle (or calorimeter tower) $i$ and $j$, and the distance $d_{iB}$ between the particle $i$ and beam (B) are defined as
\begin{equation}
d_{ij}=min(k_{ti}^{2p},k_{tj}^{2p})\Delta R_{ij}^{2}/R^{2};~~d_{iB}=k_{Ti}^{2};~~\Delta
R_{ij}^{2}=(y_i-y_j)^{2}+(\phi_i-\phi_j)^{2}
\end{equation}
where k$_{T}$ is the momentum of the particle and $p$ values of ${-1,0,1}$ represent anti-k$_{T}$, Cambridge-Aachen and
k$_{T}$ respectively. The k$_{T}$ algorithm with $p=1$ means that soft particles are clustered initially
\cite{kT_Algorithm}. If $p=-1$, anti-$_{T}$ algorithm is obtained and hard particles are clustered initially rather than
soft particles \cite{anti_kT}. If p =0, an energy dependent clustering algorithm which is called as Cambridge/Aachen
(CA) algorithm is obtained \cite{Cambridge_Aachen}. The behaviors of different jet algorithms are illustrated in
Figure \ref{fig_diff_algos}. As it can be seen in Figure 4.5, the anti-kT jet algorithm gives the best shape of
jets. In this thesis, jets are reconstructed using anti-k$_{T}$ algorithm with cone size parameter R=0.7.

\begin{figure}[ht]
  \centering
  \includegraphics[width=0.70\textwidth,angle=270]{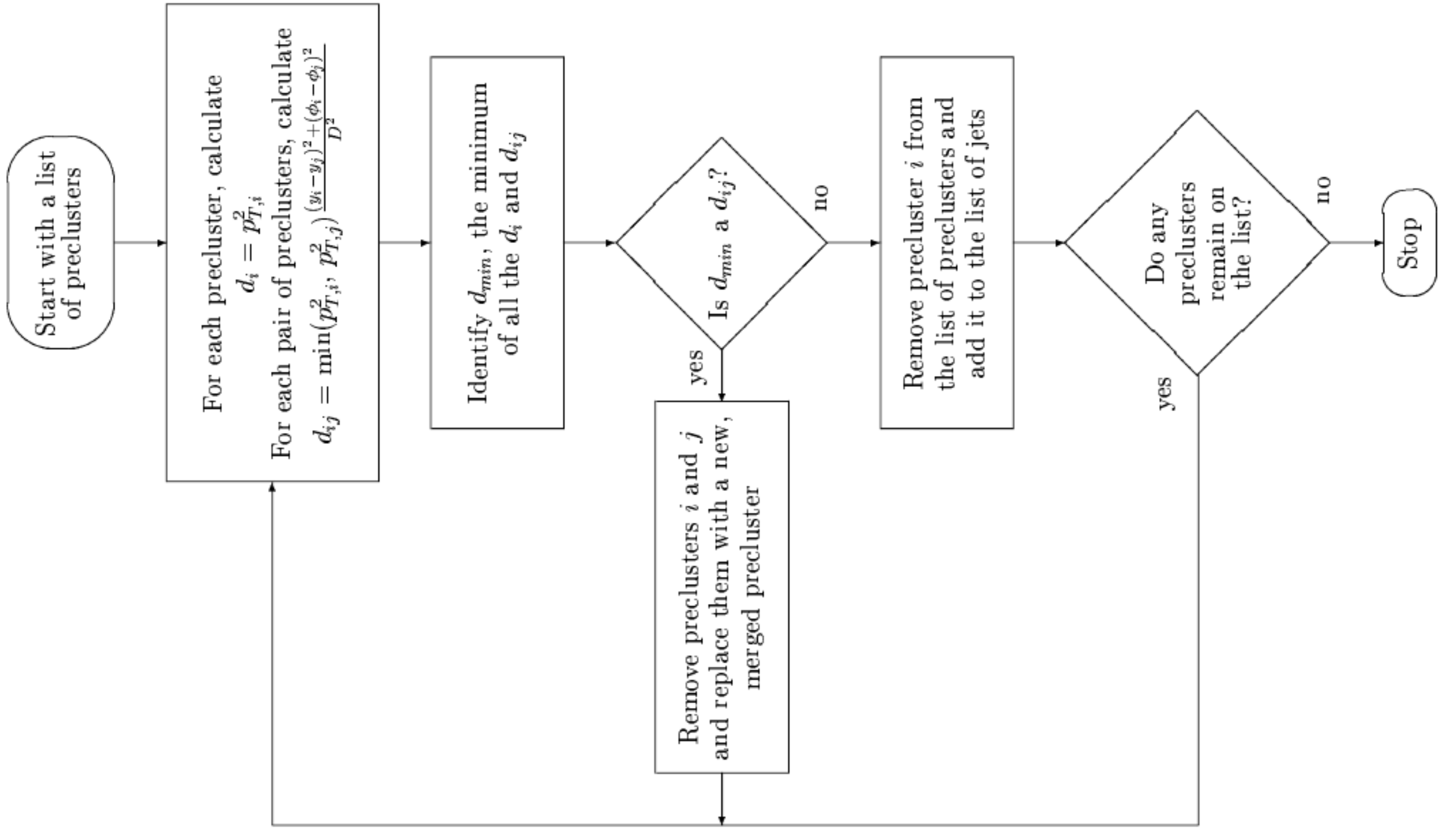}
 \capspace
  \caption{A flow chart for the k$_{T}$ algorithm.}
  \label{fig_kT_flow_chart}
\end{figure}
\clearpage
\subsection{NLOJet++ and fastNLO}
NLOJET++ \cite{NLOJet++} is a QCD event generator for hadron-hadron collisions, developed by Zolt\`{a}n Nagy, which can
calculate one-, two-, and three-jet observables at next-to-leading order. In case of the three-jet or inclusive jet
cross section, it extremely reduces the renormalization and factorization scale dependence with respect to a leading
order calculation. A slightly modified Catani-Seymour \cite{Catani_Seymour_dipole} dipole formalism is used in the
calculation to cancel infrared divergences which allows an extensive precision and flexibility during phase space
generation. Nevertheless, production of individual events which are suitable for detector simulation cannot be produced
by this program. 

Since precise computations in NLO are very time consuming or equivalently CPU consuming, a more efficient set-up in the
form of the fastNLO project \cite{fastNLO} has been setup. It allows the fast re-derivation of the considered cross
section for arbitrary input parton distribution functions and $\alpha_{s}$ values. This is done by separating the PDF
dependency from the hard matrix element calculation. The fastNLO package is attached to NLOJET++, which performs the
initial perturbative calculation in next-to-leading order.

\clearpage
\chapter{ANALYSIS}
This section is dedicated to describe the analysis in detail. Event and jet selections, trigger studies, spectrum construction and corrections for the smearing effect are discussed.\\
The differential cross section was calculated using equation \ref{master_formula} 
\begin{equation}\label{master_formula}
\dfrac{ \textnormal{d}^2 \sigma}{ \textnormal{d}M_{JJ} \textnormal{d}|y|_{max}}= \dfrac{\mathcal{C}}{\epsilon \cdot \mathcal{L}_{\textnormal{equiv}}}\cdot\dfrac{N}{\Delta M_{JJ}\Delta |y|_{max}}
\end{equation}
N is the number of dijet events, $\Delta M_{JJ}$  and $\Delta |y|_{max}$ are mass and rapidity bins respectively, $\mathcal{L}_{\textnormal{equiv}}$ is the equivalent luminosity for each dataset where the dijet event is coming from, $\epsilon$ is the efficiency factor for even selection (i.e., trigger and JetID), $\mathcal{C}$ is the correction factor for smearing effects due to the finite detector resolution. All these components will be discussed in the following sections.
\section{Data Set, Event Selection and Jet Selection}
\subsection{Data Set}
This analysis was performed with the 36 pb$^{-1}$ of data collected by the CMS in 2010 Run with High Level Jet triggers. The data are reconstructed with CMSSW (CMS Software) which is the official software framework of the CMS experiment. Good runs and good luminosity sections of them are declared by the data validation group at CMS was. The main concern of this official declaration is data taking condition of the detector with all its components. There are namely three primary data sets used in this analysis and their names and Dataset Bookkeeping System (DBS) identifications are listed in Table \ref{data_table0}.

\begin{table}[h]
  \centering
  \caption{Primary Data sets used in the analysis \label{data_table0}.}
  \capspace
  \begin{tabular}{ |c|c|c|}
    \hline
    ERA  & Primary Dataset  &  DBS Name  \\
    \hline
    \hline
    2010A   &  JetMETTau & /JetMETTau/Run2010A-Nov4ReReco\_v1/RECO  \\
    \hline
    2010A   &  JetMET    & /JetMET/Run2010A-Nov4ReReco\_v1/RECO  \\
    \hline
    2010B   &  Jet       & /Jet/Run2010B-Nov4ReReco\_v1/RECO  \\
    \hline
  \end{tabular}
\end{table}
\subsection{Event and Jet Selection}\label{Event and Jet Selection}
In order to calculate the invariant dijet mass and to construct the spectrum, each event in the data set must satisfy a certain set of selection criteria. First of all, an event should be triggered by one of these HLT Jet triggers; HLT\_Jet\_30U, HLT\_Jet\_50U, HLT\_Jet\_70U, HLT\_Jet\_140U. A brief definition of these triggers is given in Table \ref{table:HLTdescriptions}.\\
Each event should have at least two jets satisfying $0<$\ymax$<2.5$ condition where \ymax is defined as:
\begin{equation}\label{ymax_formula}
|y|_{max}=max(|y_{1}|,|y_{2}|)
\end{equation}
 \begin{figure}[h]
   \begin{center}
     \includegraphics[width=0.5\textwidth]{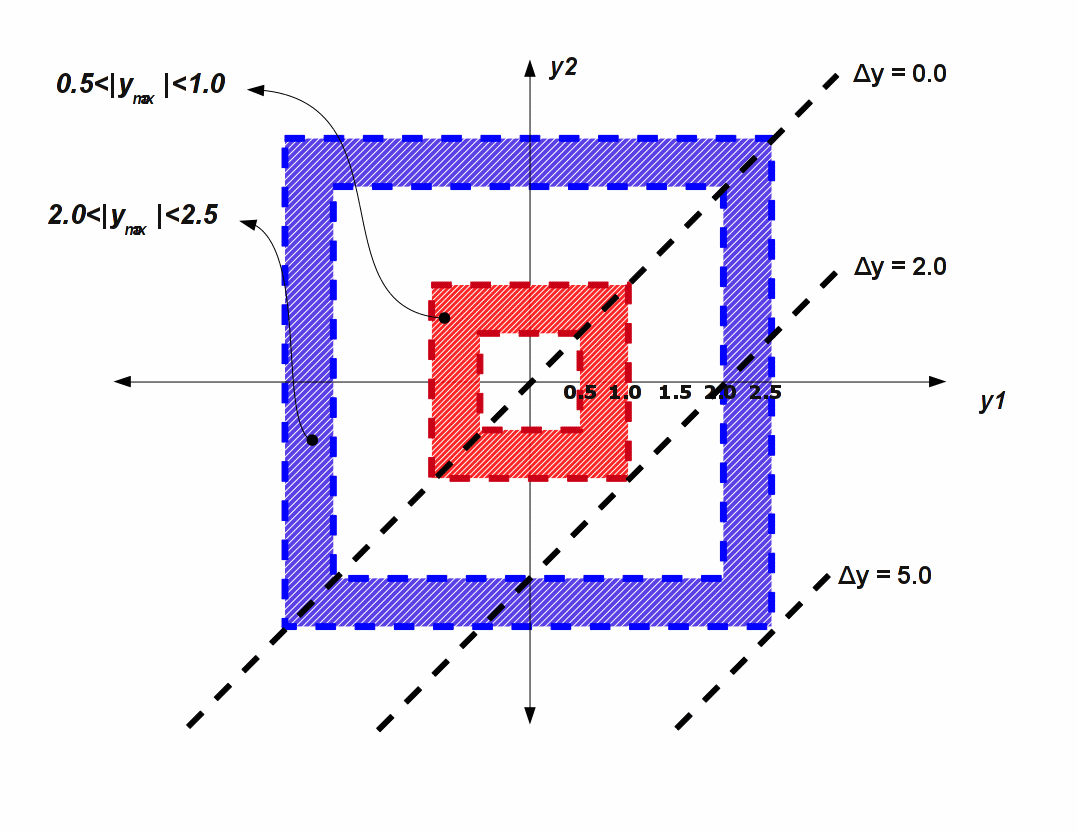}
     \capspace
     \caption{$|y|_{max}$ bins in $y_{1}$,$y_{2}$ phase space.}
     \label{fig.UnfodingForAll}
   \end{center}
 \end{figure}
\clearpage
Events are further required to have at least one well reconstructed primary vertex (PV) with $|$z(PV)$|<$24 cm and at least four tracks associated with the primary vertex fit (n$_{dof}\geq5$) where z$=$0 represents the central point of the detector. These PV selection criteria ensure that the event in interest is originated in the region of interaction. Additionally events with at least two reconstructed particle flow jets (PF Jets) with p$_{T}>$30 GeV (corrected) and both satisfying the loose PF JetID criteria are selected. PF JetID criteria are a set of criteria developed at CMS to reject most of the fake jets arising due to the calorimeter or readout noise or both \cite{PFJetID}. There are two types of PF JetID criteria called \emph{``loose PF JetID"} and \emph{``tight PF JetID"} and the loose one which requires at least two particles in a jet, one of which is a charged hadron, is used in this analysis. If any of the leading jet fails to satisfy loose PF JetID requirments, the event is not considered in the analysis.
\begin{table}[th]
  \centering
  \caption{L1 and High Level Jet Triggers.}
  \capspace
  \normalsize
  \small\addtolength{\tabcolsep}{-3pt}
  \begin{tabular}{|c|c|c|}
    \hline
    Trigger Path &  L1 seeds & Requirement \\
    \hline
    \hline
    HLT\_Jet30U     &  L1\_SingleJet20U  & requiring $\geq$ 1 jet at HLT with p$_T>$ 30 GeV\\ 
    \hline
    HLT\_Jet30U\_v3 &  L1\_SingleJet20U  & requiring $\geq$ 1 jet at HLT with p$_{T}>$ 30 GeV\\     
    \hline
    HLT\_Jet50U     &  L1\_SingleJet30U  & requiring $\geq$ 1 jet at HLT with p$_{T}>$ 50 GeV\\    
    \hline
    HLT\_Jet50U\_v3 &  L1\_SingleJet30U  & requiring $\geq$ 1 jet at HLT with p$_{T}>$ 50 GeV\\     
    \hline
    HLT\_Jet70U     &  L1\_SingleJet30U  & requiring $\geq$ 1 jet at HLT with p$_{T}>$ 70 GeV\\     
    \hline
    HLT\_Jet70U\_v2 &  L1\_SingleJet40U  & requiring $\geq$ 1 jet at HLT with p$_{T}>$ 70 GeV\\ 
    \hline
    HLT\_Jet70U\_v3 &  L1\_SingleJet40U  & requiring $\geq$ 1 jet at HLT with p$_{T}>$ 70 GeV\\
    \hline
    HLT\_Jet100U    &  L1\_SingleJet30U  & requiring $\geq$ 1 jet at HLT with p$_{T}>$ 100 GeV\\
    \hline
    HLT\_Jet100U\_v2 &  L1\_SingleJet60U & requiring $\geq$ 1 jet at HLT with p$_{T}>$ 100 GeV\\
    \hline
    HLT\_Jet100U\_v3 &  L1\_SingleJet60U & requiring $\geq$ 1 jet at HLT with p$_{T}>$ 100 GeV\\
    \hline
    HLT\_Jet140U\_v1 &  L1\_SingleJet60U & requiring $\geq$ 1 jet at HLT with p$_{T}>$ 140 GeV\\
    \hline
    HLT\_Jet140U\_v3 &  L1\_SingleJet60U & requiring $\geq$ 1 jet at HLT with p$_{T}>$ 140 GeV\\
    \hline

  \end{tabular}
  \label{table:HLTdescriptions}
\end{table}
\clearpage								
\section{Trigger Studies}
As it was discussed in  Section \ref{Event and Jet Selection}, events triggered with the HLT jet triggers are used in this analysis. Each of these triggers has different Level-1 seeds and different \pt thresholds. With the increasing instantaneous luminosity of proton proton collisions at LHC the number of hard scattering events has increased. As a result, the number of events that satisfy the requirements of HLT Jet triggers has increased. In order to keep the data writing rate in an acceptable value, the triggers with lower thresholds has been pre-scaled.  Furthermore, each of them has been introduced to DAQ system at different periods of the 2010 Run. Thus, each data sample with one of these triggers has different effective luminosity (Table \ref{data_table1}).
\begin{table}[h]
  \centering
  \caption{High Level Triggers used in the analysis accompanied by the effective luminosity, and the effective trigger prescales.
 \label{data_table1}}
  \capspace
  \normalsize
  \small\addtolength{\tabcolsep}{-3pt}
  \begin{tabular}{ |c|c|c|c|}
    \hline 
    Sample  & HLT Paths  & Eff. Luminosity & Eff. Prescale  \\ 
            & (OR) & \pbinv & \\
    \hline
    \hline
    Jet30U  & HLT\_Jet30U,  HLT\_Jet30U\_v3 & 0.4 & 111 \\ 
    \hline
    Jet50U  & HLT\_Jet50U,  HLT\_Jet50U\_v3 & 3.2 & 10.9 \\ 
    \hline
    Jet70U  & HLT\_Jet70U,  HLT\_Jet70U\_v2, HLT\_Jet70U\_v3 & 8.6 & 4.1 \\ 
    \hline
    Jet100U & HLT\_Jet100U, HLT\_Jet100U\_v2, HLT\_Jet100U\_v3 & 19.0 & 1.9 \\ 
    \hline
    Jet140U & HLT\_Jet100U, HLT\_Jet140U\_v1, HLT\_Jet140U\_v3 & 35.3 & 1 \\ 
    \hline 
  \end{tabular}
\end{table}
 It is an important task to determine the lower limits of parameter in interest for each of the data samples where the sample efficiency is 99\%. In this case the lower limits of the invariant dijet mass for all samples in each \ymax bin are determined. This is done by extracting the turn-on curve of every HLT jet trigger with respect to one step lower threshold trigger. The turn-on curves are constructed according the formula below;
\begin{equation}\label{trigger_eff_formula}
\epsilon_{A}= \dfrac{N_{Trigger A}}{N_{Trigger B}} \cdot \dfrac{\mathcal{L}_{A}}{\mathcal{L}_{B}}
\end{equation}
where $\epsilon_{A}$ is the efficiency of the higher threshold sample and $\mathcal{L}_{A,B}$ are the effective luminosities of two samples. After constructing the turn-on curve, it was parametrized by performing a fit to a sigmoid type of function in order to estimate the 99\% efficient mass point more precisely.

\begin{figure}[ht]
  \centering
  \includegraphics[width=0.40\textwidth]{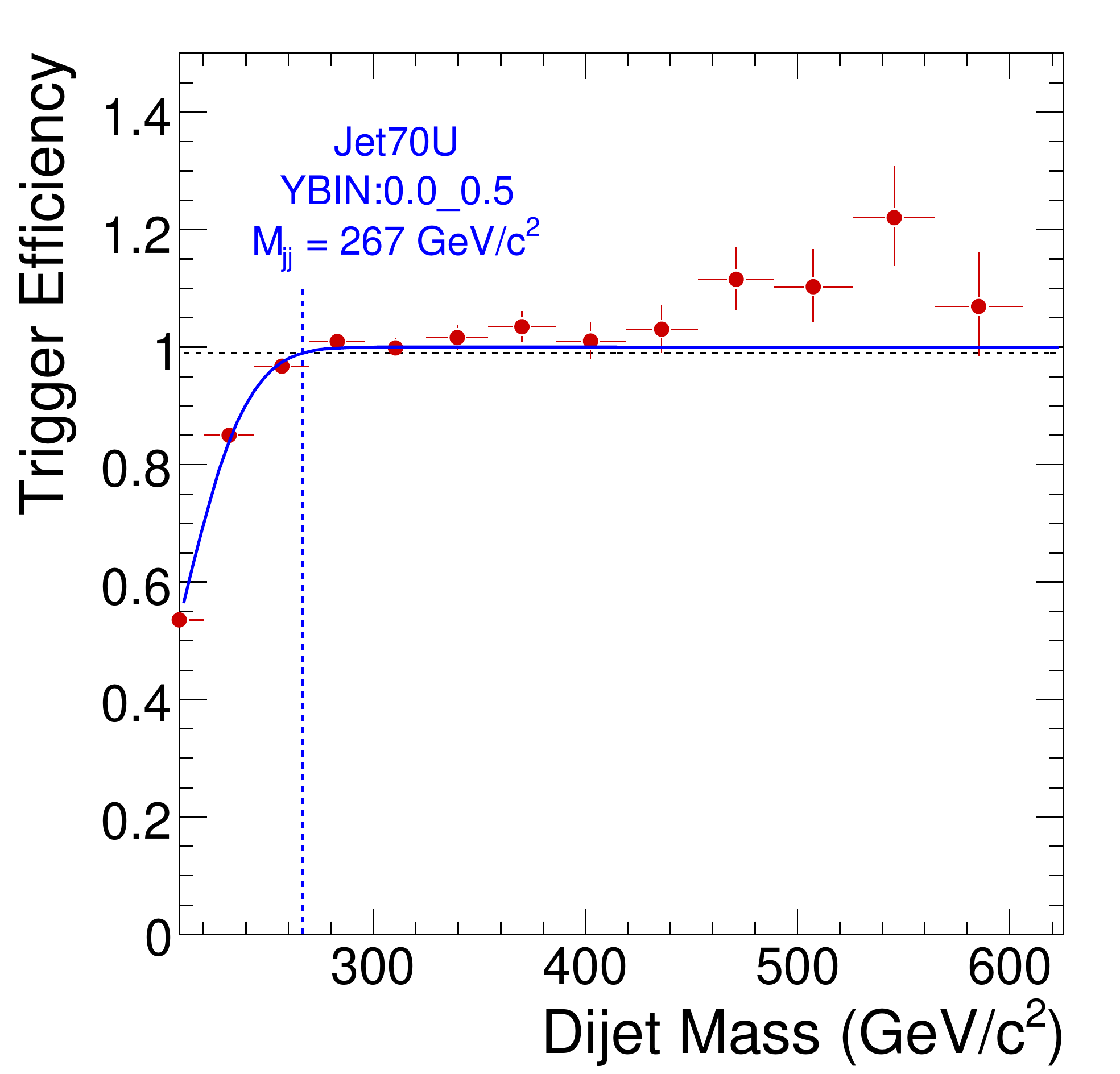}
  \includegraphics[width=0.40\textwidth]{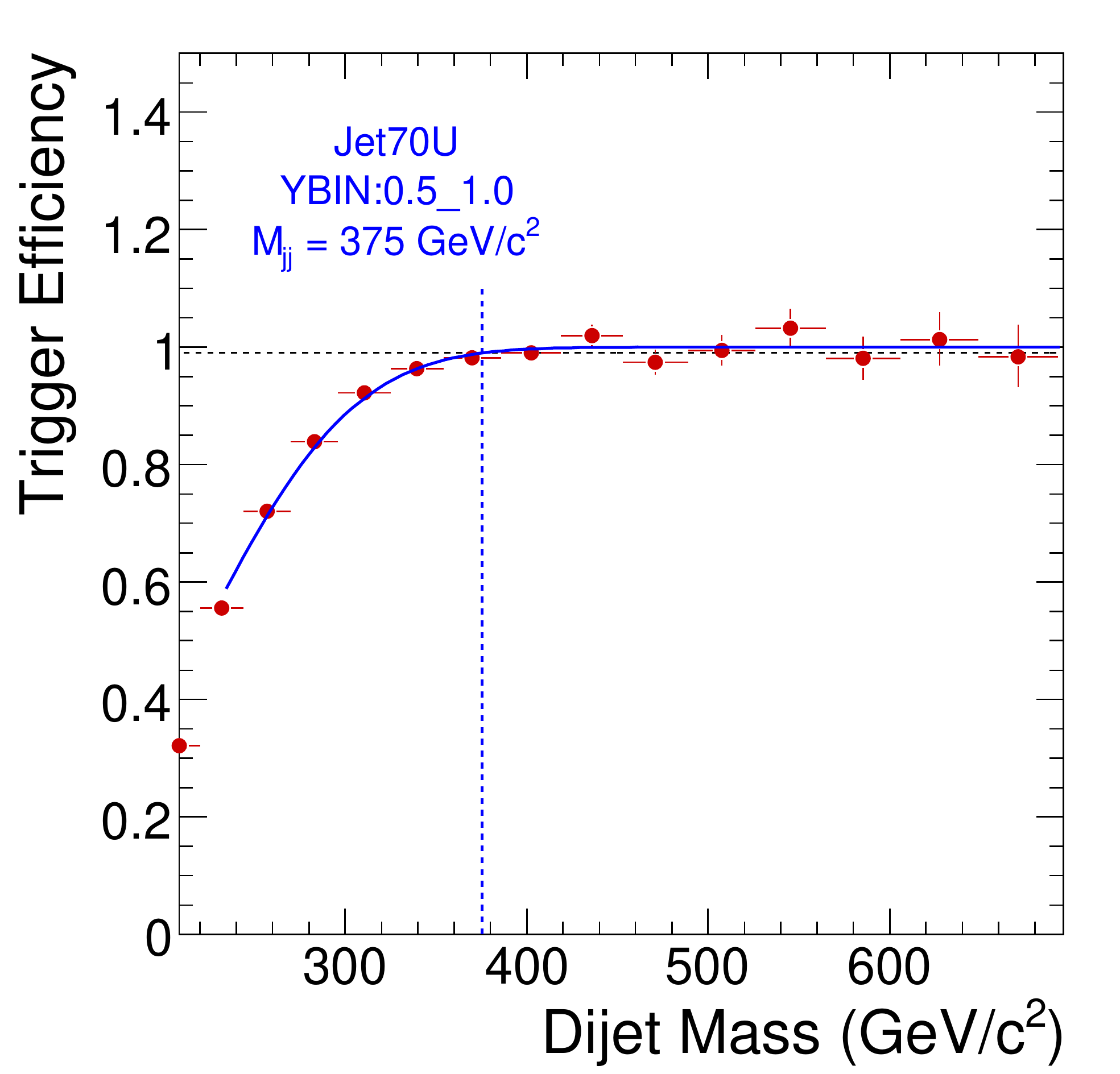} 
  \includegraphics[width=0.40\textwidth]{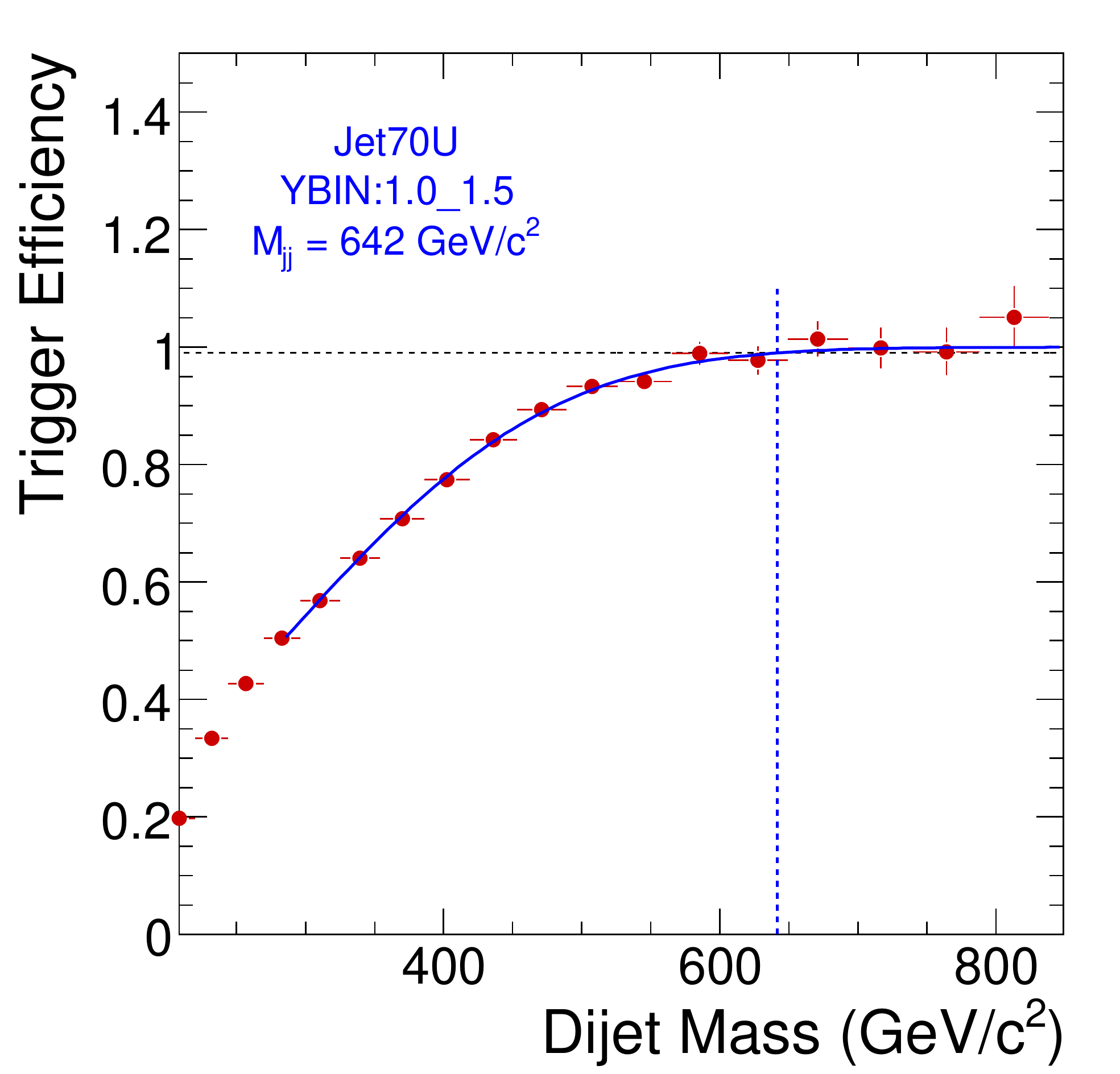} 
  \includegraphics[width=0.40\textwidth]{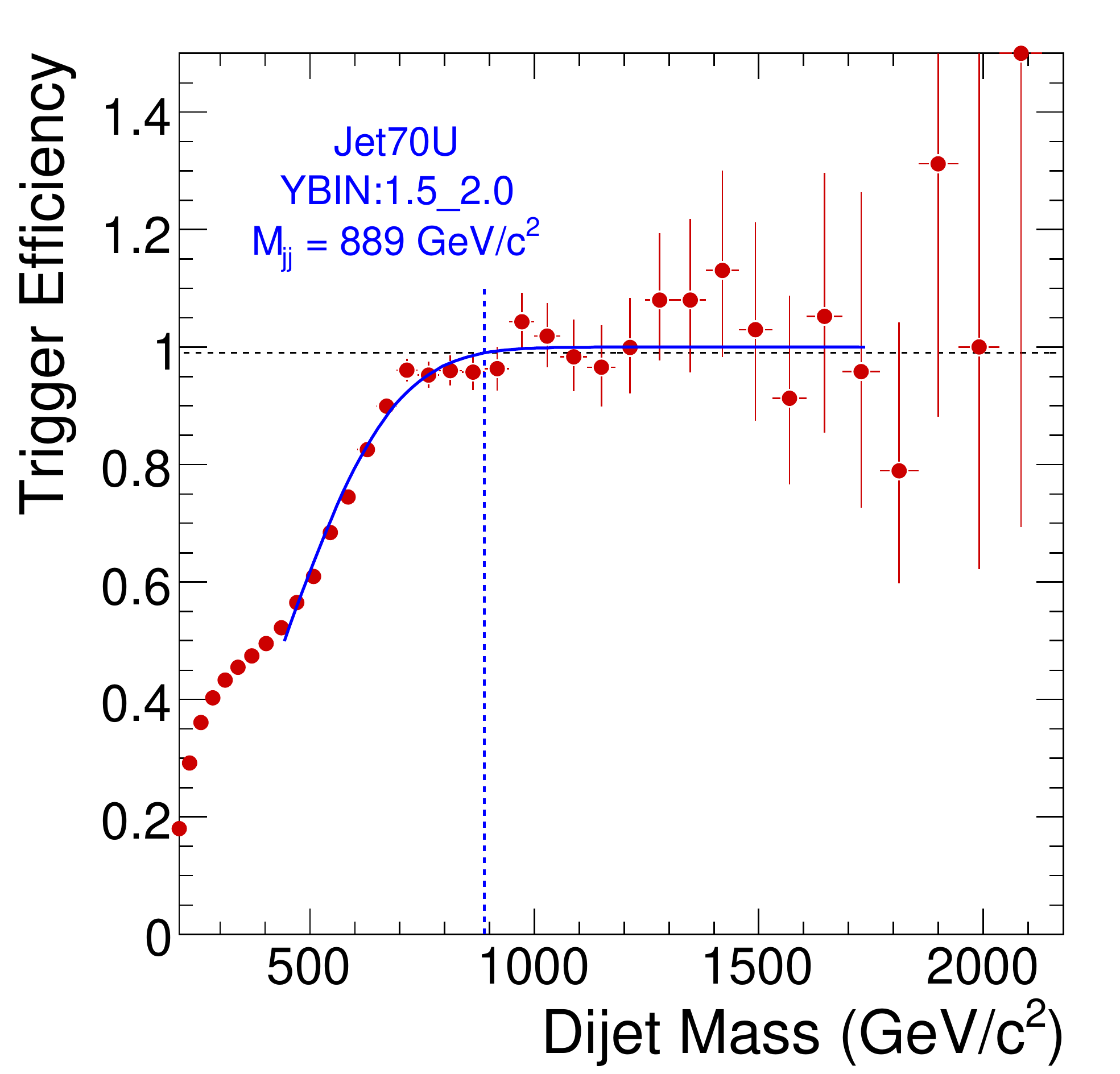}
  \includegraphics[width=0.40\textwidth]{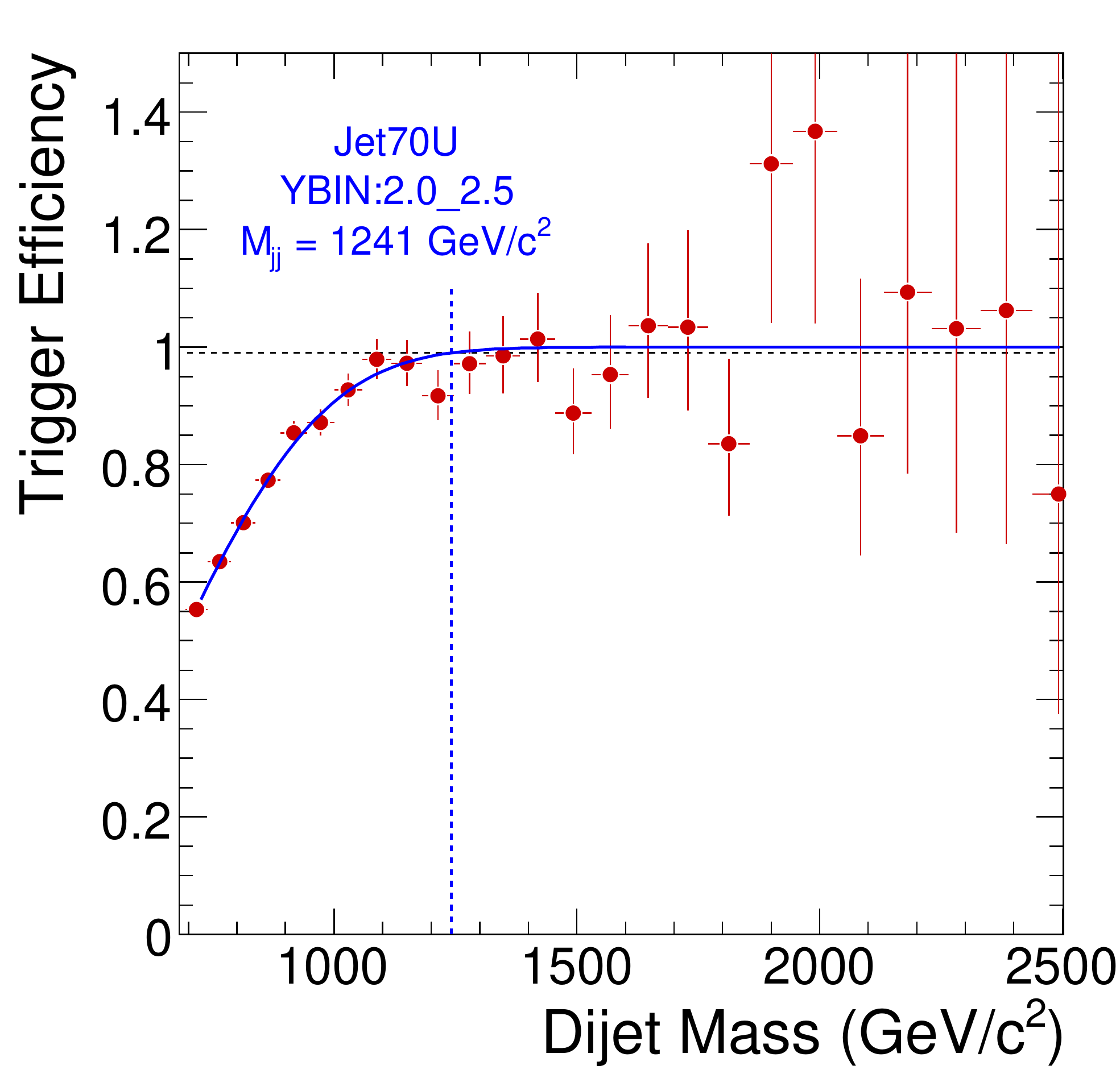}
  \capspace
  \caption{ Relative trigger efficiencies as a function of dijet mass for the five different \ymax bins and for the HLT\_Jet70U trigger. The  99\% efficiency point is determined by performing a fit with an error function.}
  \label{fig_data1}
\end{figure}

In this study, an adjusted form of the error function was used to perform this fit.
\begin{equation}\label{trigger_erf}
\epsilon_{A}(m_{jj})= \dfrac{1}{2} [ Erf( \alpha m_{jj}- \beta)+1] 
\end{equation}
Here $m_{jj}$ is the invariant mass of the dijet system, $\alpha$ and $\beta$ are the free parameters of the fit, and Erf is the well know error function. In Figure \ref{fig_data1} the turn-on curves for HLT\_Jet\_70U in all \ymax bin are given. All other plots of different samples can be found in Appendix \ref{AppendixA}. Also the 99\% efficiency mass points for all samples and rapidity bins are summarized on Table \ref{TriggerTurnOns}
\begin{table}[h]
  \centering
  \caption{Trigger efficiency turn-on masses for all jet samples and rapidity regions.}
  \capspace
  \normalsize
  \small\addtolength{\tabcolsep}{-3pt}
  \begin{tabular}{ |c|c|c|c|c|c|}
    \hline 
    Sample  & [0.0-0.5] & [0.5-1.0] & [1.0-1.5] & [1.5-2.0] & [2.0-2.5] \\ 
    \hline
    \hline
    Jet30U  & 156 GeV & 197 GeV & 386 GeV & 565 GeV & 649 GeV \\
    \hline
    Jet50U  & 220 GeV & 296 GeV & 489 GeV & 693 GeV & 890 GeV \\ 
    \hline
    Jet70U  & 270 GeV & 386 GeV & 649 GeV & 890 GeV & 1246 GeV \\ 
    \hline
    Jet100U & 386 GeV & 489 GeV & 838 GeV & 1058 GeV & 1687 GeV \\ 
    \hline
    Jet140U & 489 GeV & 649 GeV & 1058 GeV & 1607 GeV & 2231 GeV \\ 
    \hline 
  \end{tabular}.\label{TriggerTurnOns}
\end{table}
\clearpage

\begin{figure}[ht]
  \centering
  \includegraphics[width=0.70\textwidth]{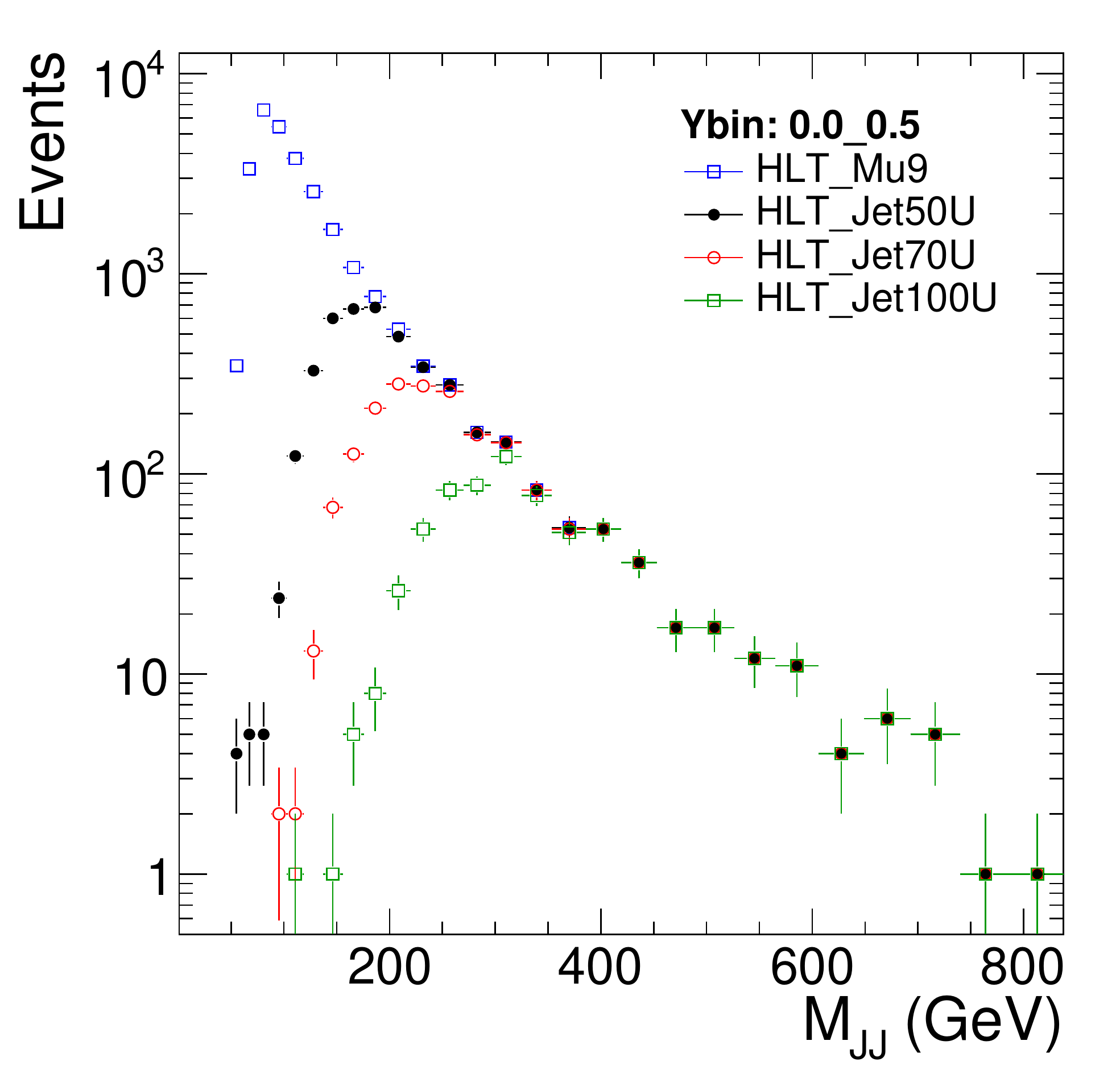}
  \includegraphics[width=0.70\textwidth]{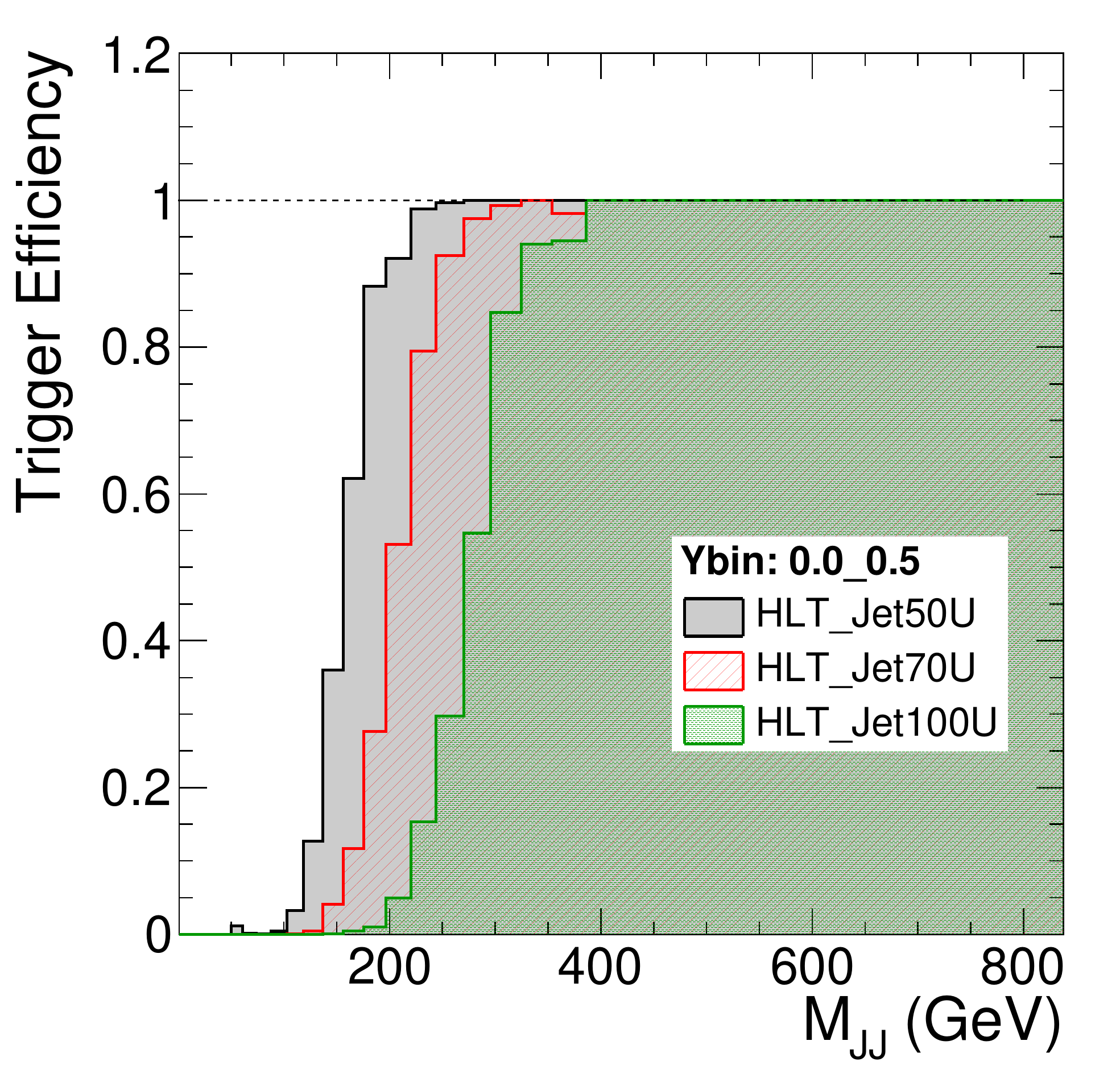} 
  \capspace
  \caption{ Trigger efficiencies measured with respect to the HLT\_Mu9 trigger.}
  \label{fig:TrigVsMuons}
\end{figure}

After determination of the 99\% efficiency point from the fit, the starting value of the next dijet mass bin is considered as the lower limit for the group obtained from that sample. It should be noted that all these turn-on curves are relative to the lower threshold trigger of the same kind, and they are adequate to construct a smooth and continuous spectrum. However, for an absolute efficiency, it should be checked with an orthogonal trigger. Such a measurement is shown in Figure \ref{fig:TrigVsMuons} and it is found to be 100\% efficient.

In Tables \ref{CutFlow_Jet30U} to \ref{CutFlow_Jet140U} number of events that have survived after each cut (event cut flow) is shown in all \ymax bins.
\begin{table}[htbH]
\footnotesize
  \centering
  \caption{Event cut flow for the Jet30U sample.
  \label{CutFlow_Jet30U}}
  \capspace
  \normalsize  
  \small\addtolength{\tabcolsep}{-3pt}
  \begin{tabular}{ |c|c|c|c|c|c|}
    \hline 
    Cut        & [0.0-0.5] & [0.5-1.0] & [1.0-1.5] & [1.5-2.0] & [2.0-2.5] \\ 
    \hline
    \ymax  & 107989 & 351063 & 506990 & 623448 & 769297 \\
    \hline
    \pt        & 107795 & 345634 & 485850 & 579603 & 687070 \\ 
    \hline
    Trigger efficiency cut       & 22290  & 38568  & 6728   & 3564   & 5903   \\ 
    \hline
    JetID      & 22282  & 38558  & 6722   & 3563   & 5901   \\ 
    \hline
  \end{tabular}
\end{table}
\begin{table}[htbH]
\footnotesize
  \centering
  \caption{Event cut flow for the Jet50U sample.
  \label{CutFlow_Jet50U}}
  \capspace
  \normalsize  
  \small\addtolength{\tabcolsep}{-3pt}
  \begin{tabular}{ |c|c|c|c|c|c|}
    \hline 
    Cut        & [0.0-0.5] & [0.5-1.0] & [1.0-1.5] & [1.5-2.0] & [2.0-2.5] \\ 
    \hline
    \ymax  & 119591 & 461123 & 766864 & 949419 & 1082527 \\
    \hline
    \pt        & 119591 & 461123 & 766570 & 946585 & 1073452 \\ 
    \hline
    Trigger efficiency cut       & 52729  & 66816  & 22539   & 12798   & 11319   \\ 
    \hline
    JetID      & 52704  & 66791  & 22505   & 12796   & 11315   \\ 
    \hline
  \end{tabular}
\end{table}
\begin{table}[htbH]
\footnotesize
  \centering
    \caption{Event cut flow for the Jet70U sample.
    \label{CutFlow_Jet70U}}
    \capspace
  \normalsize  
  \small\addtolength{\tabcolsep}{-3pt}
  \begin{tabular}{ |c|c|c|c|c|c|}
    \hline 
    Cut        & [0.0-0.5] & [0.5-1.0] & [1.0-1.5] & [1.5-2.0] & [2.0-2.5] \\ 
    \hline
    \ymax  & 177399 & 565583 & 775226 & 875831 & 900091 \\
    \hline
    \pt        & 177398 & 565576 & 775176 & 875581 & 899180 \\ 
    \hline
    Trigger efficiency cut       & 58564  & 53252  & 14799   & 9019   & 4326   \\ 
    \hline
    JetID      & 58528  & 53230  & 14752   & 9017   & 4325   \\ 
    \hline
  \end{tabular}
\end{table}
\begin{table}[hhtbH]
\footnotesize
  \centering
    \caption{Event cut flow for the Jet100U sample.
    \label{CutFlow_Jet100U}}
    \capspace
  \normalsize  
  \small\addtolength{\tabcolsep}{-3pt}
  \begin{tabular}{ |c|c|c|c|c|c|}
    \hline 
    Cut        & [0.0-0.5] & [0.5-1.0] & [1.0-1.5] & [1.5-2.0] & [2.0-2.5] \\ 
    \hline
    \ymax  & 79057 & 275526 & 402864 & 442696 & 399093 \\
    \hline
    \pt        & 79057 & 275525 & 402847 & 442676 & 399062 \\ 
    \hline
    Trigger efficiency cut       & 26823  & 39238  & 8799   & 7820   & 1263   \\ 
    \hline
    JetID      & 26804  & 39210  & 8743   & 7816   & 1261   \\ 
    \hline
  \end{tabular}
\end{table}
\begin{table}[htbH]
\footnotesize
  \centering
    \caption{Event cut flow for the Jet140U sample.
    \label{CutFlow_Jet140U}}
    \capspace
  \normalsize  
  \small\addtolength{\tabcolsep}{-3pt}
  \begin{tabular}{ |c|c|c|c|c|c|}
    \hline 
    Cut        & [0.0-0.5] & [0.5-1.0] & [1.0-1.5] & [1.5-2.0] & [2.0-2.5] \\ 
    \hline
    \ymax  & 66032 & 212202 & 297779 & 318654 & 265268 \\
    \hline
    \pt        & 66032 & 212199 & 297767 & 318631 & 265250 \\ 
    \hline
    Trigger efficiency cut       & 16781  & 18257  & 4437   & 1171   & 251   \\ 
    \hline
    JetID      & 16765  & 18239  & 4385   & 1170   & 251   \\ 
    \hline
  \end{tabular}
\end{table}
\clearpage
\clearpage

\section{Data Quality}
\subsection{Data and Monte Carlo Simulation Comparisons}
\begin{figure}[ht]
  \centering
  \includegraphics[width=0.40\textwidth]{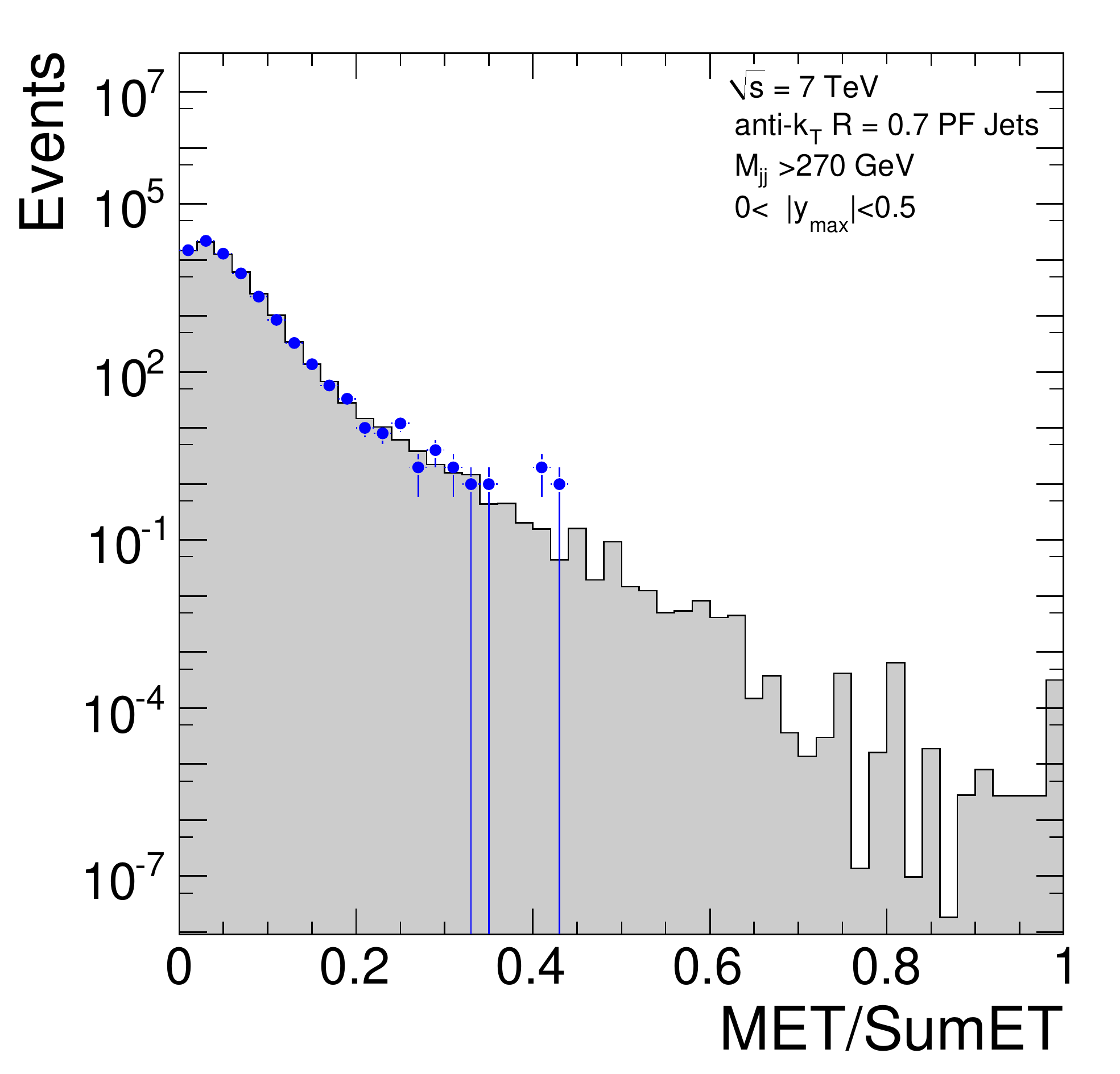}
  \includegraphics[width=0.40\textwidth]{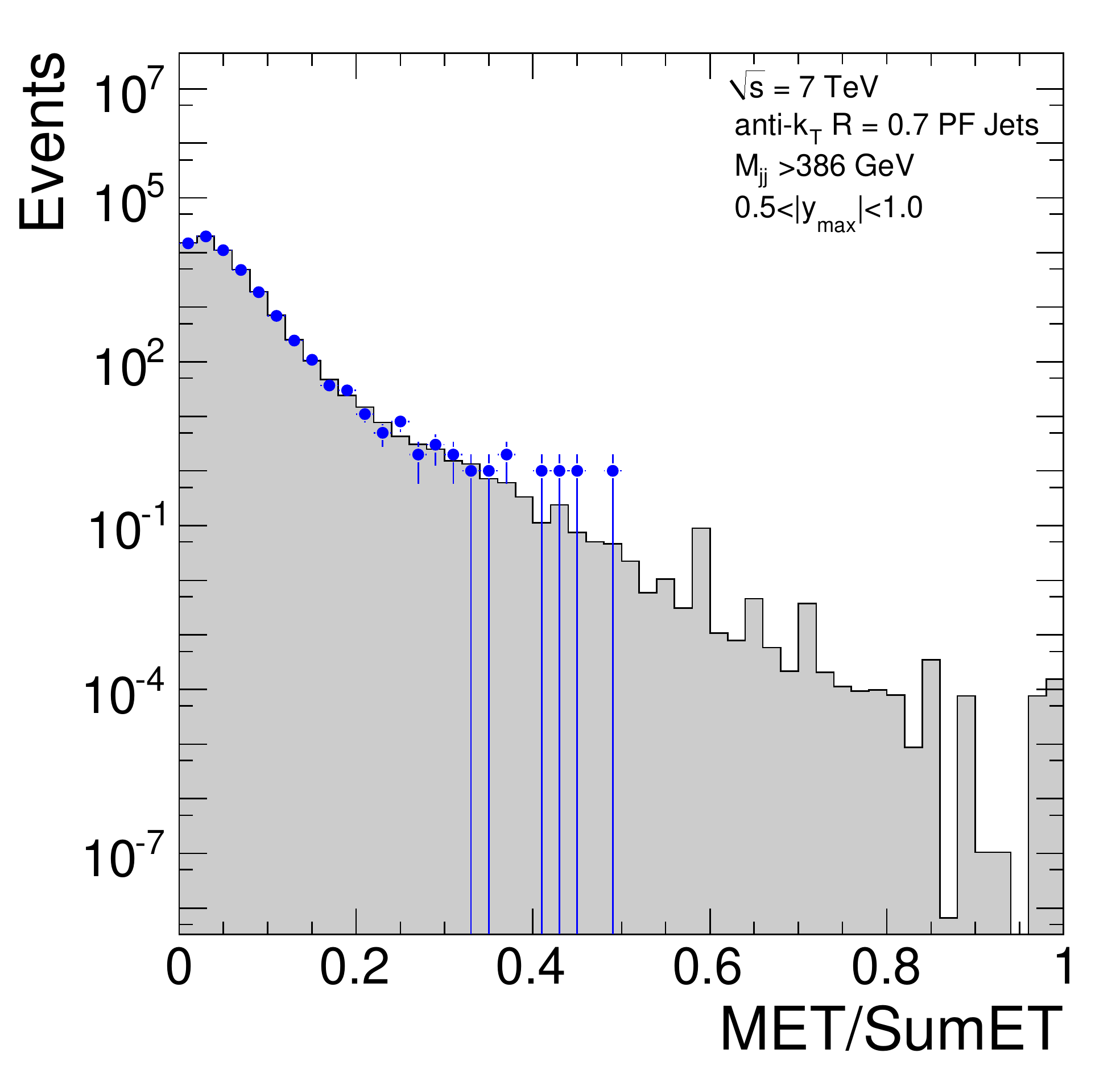} 
  \includegraphics[width=0.40\textwidth]{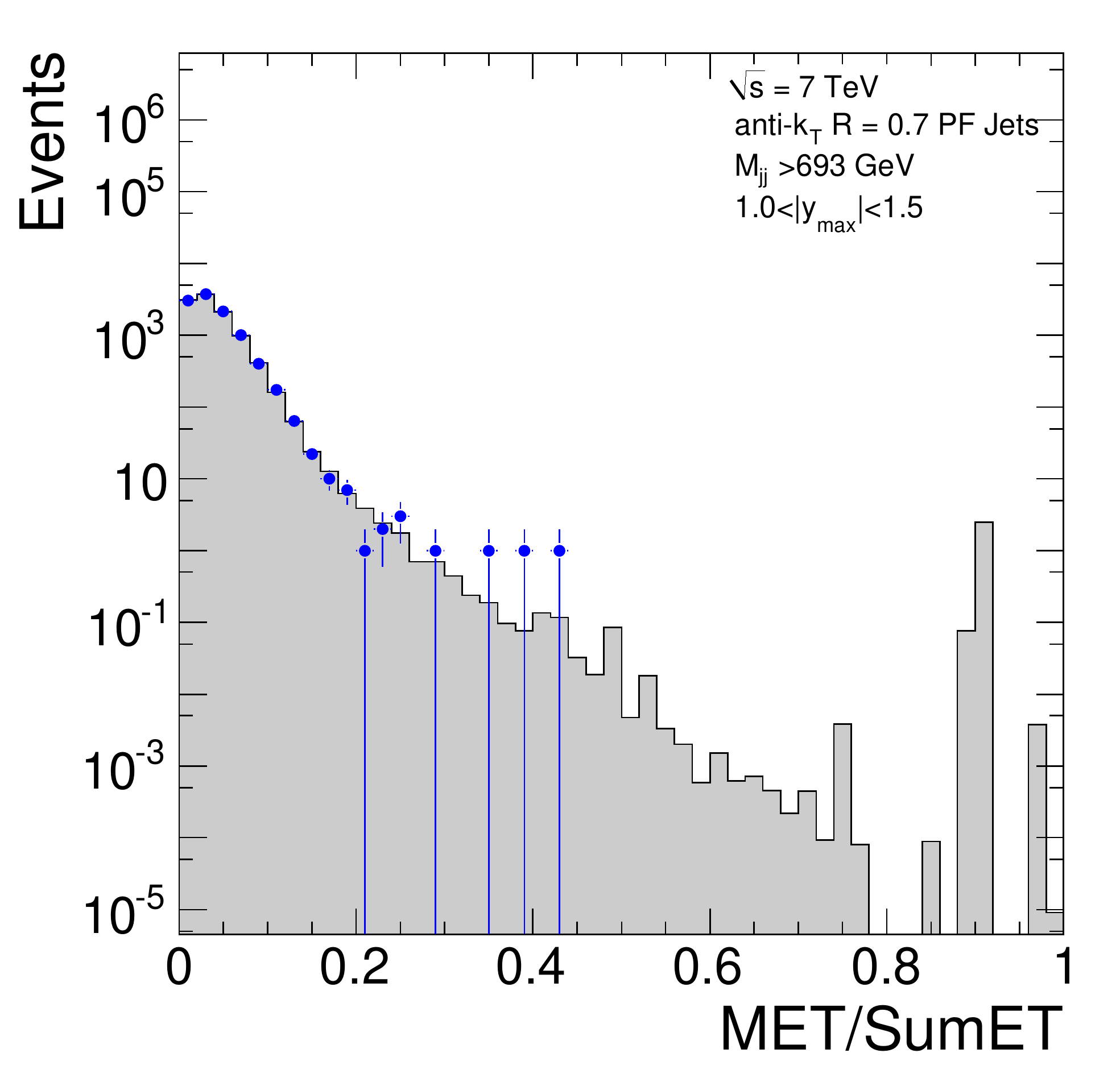} 
  \includegraphics[width=0.40\textwidth]{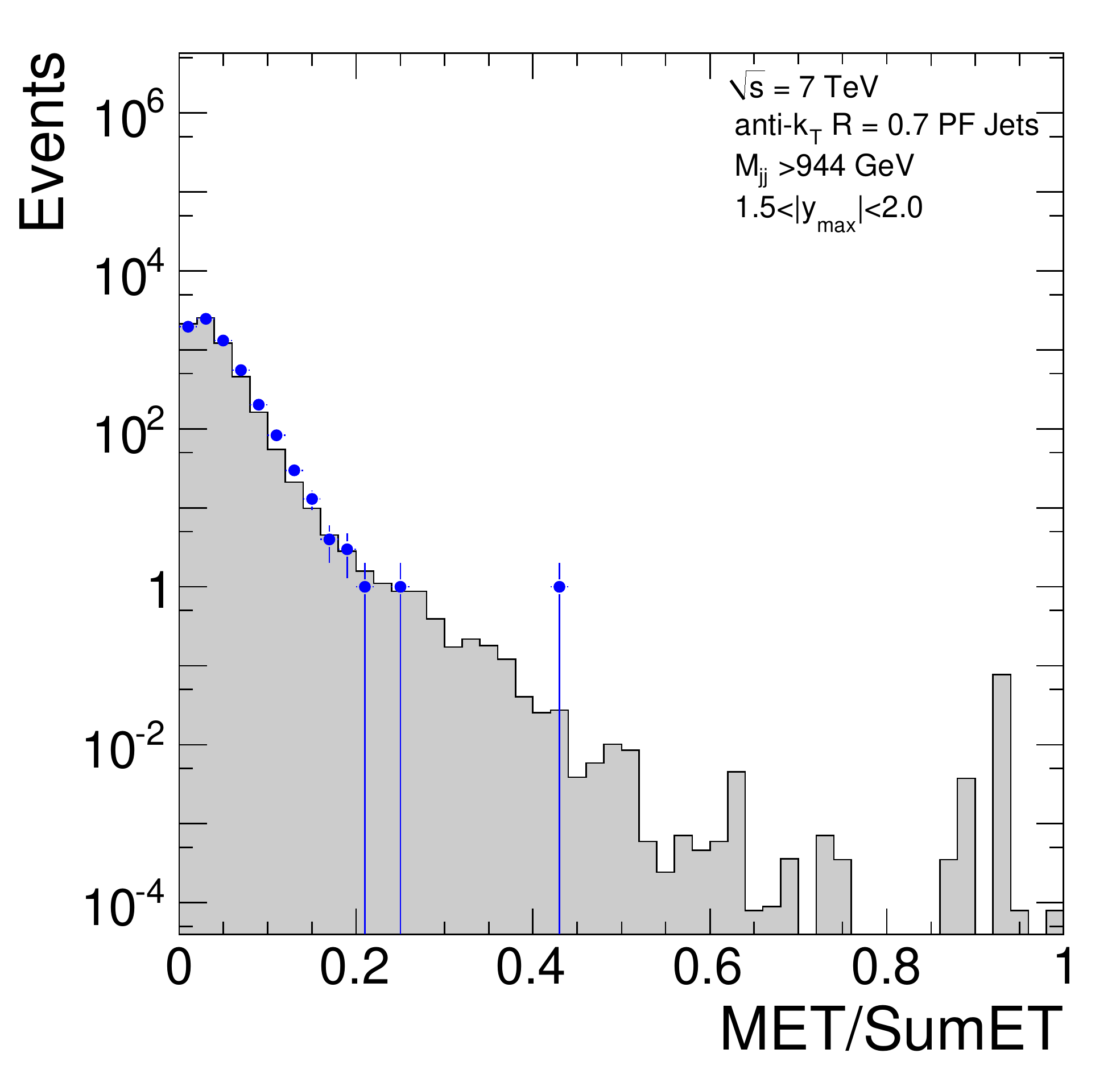}
  \includegraphics[width=0.40\textwidth]{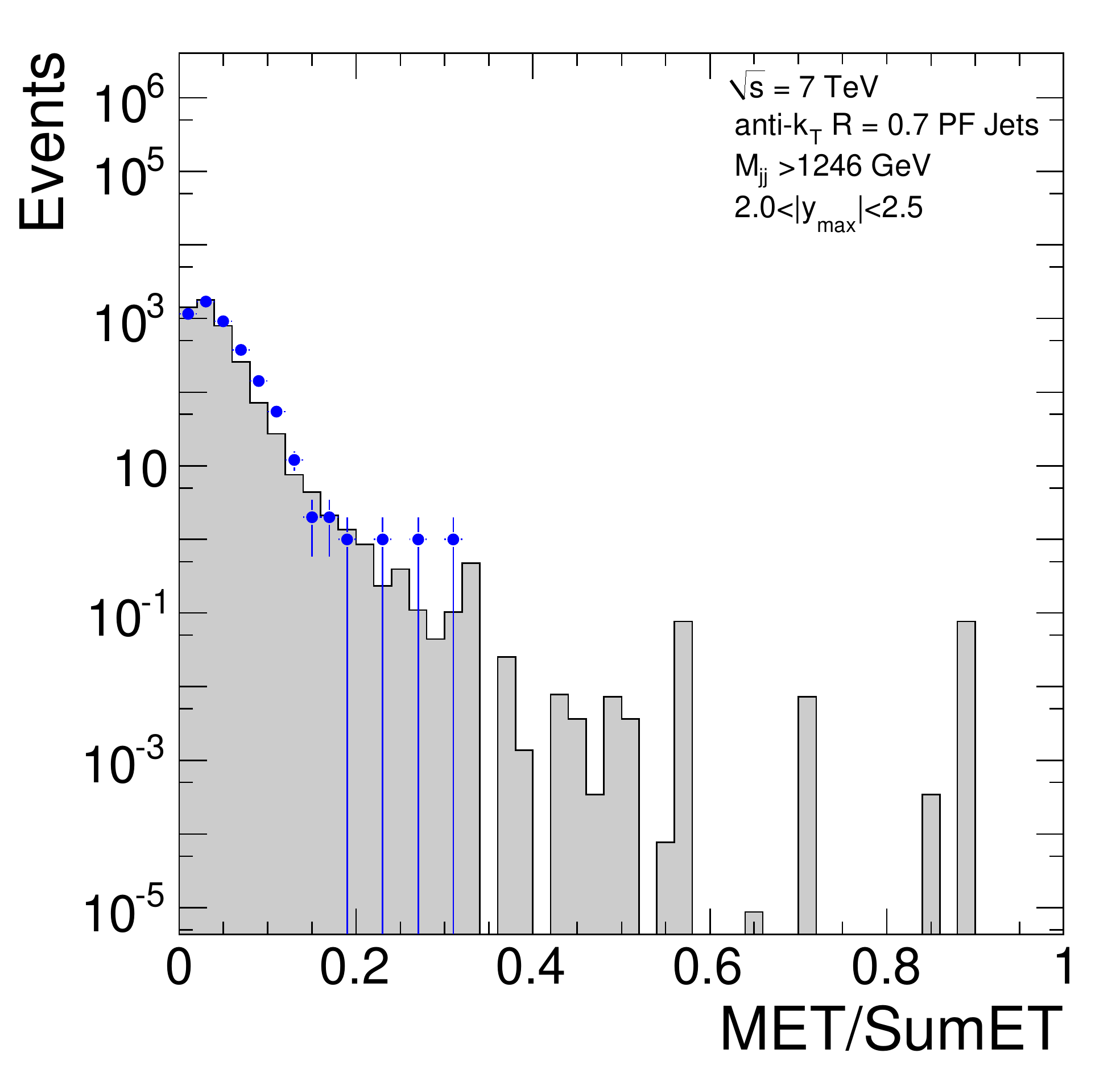}
  \capspace
  \caption{Ratio of the transverse missing energy to the total transverse energy of the event for the five different \ymax bins and for the Jet70U sample. The plots for data (points) and simulated (dashed histogram) events are compared.}
  \label{fig_data3}
\end{figure}

\begin{figure}[ht]
  \centering
  \includegraphics[width=0.40\textwidth]{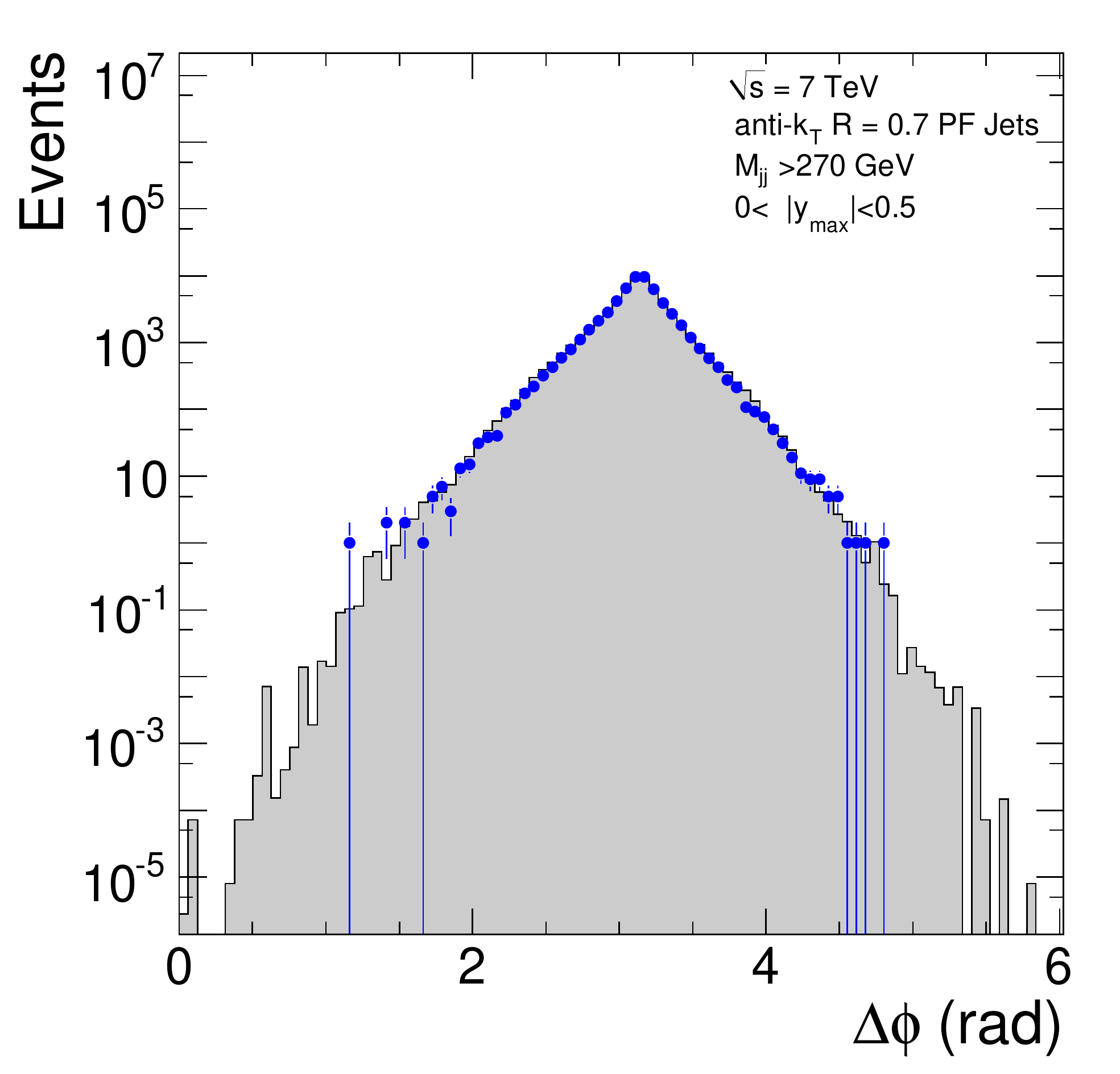}
  \includegraphics[width=0.40\textwidth]{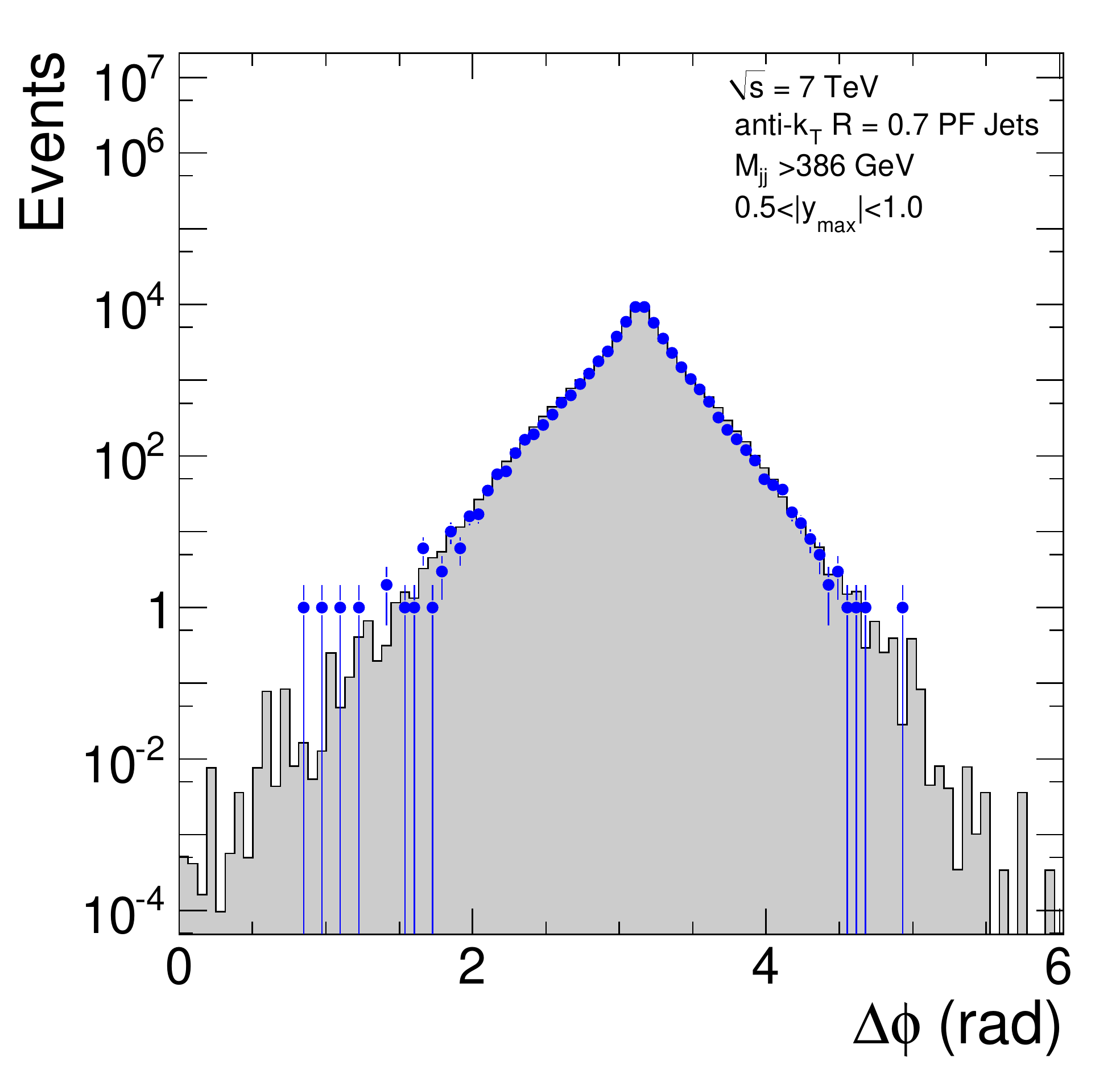} 
  \includegraphics[width=0.40\textwidth]{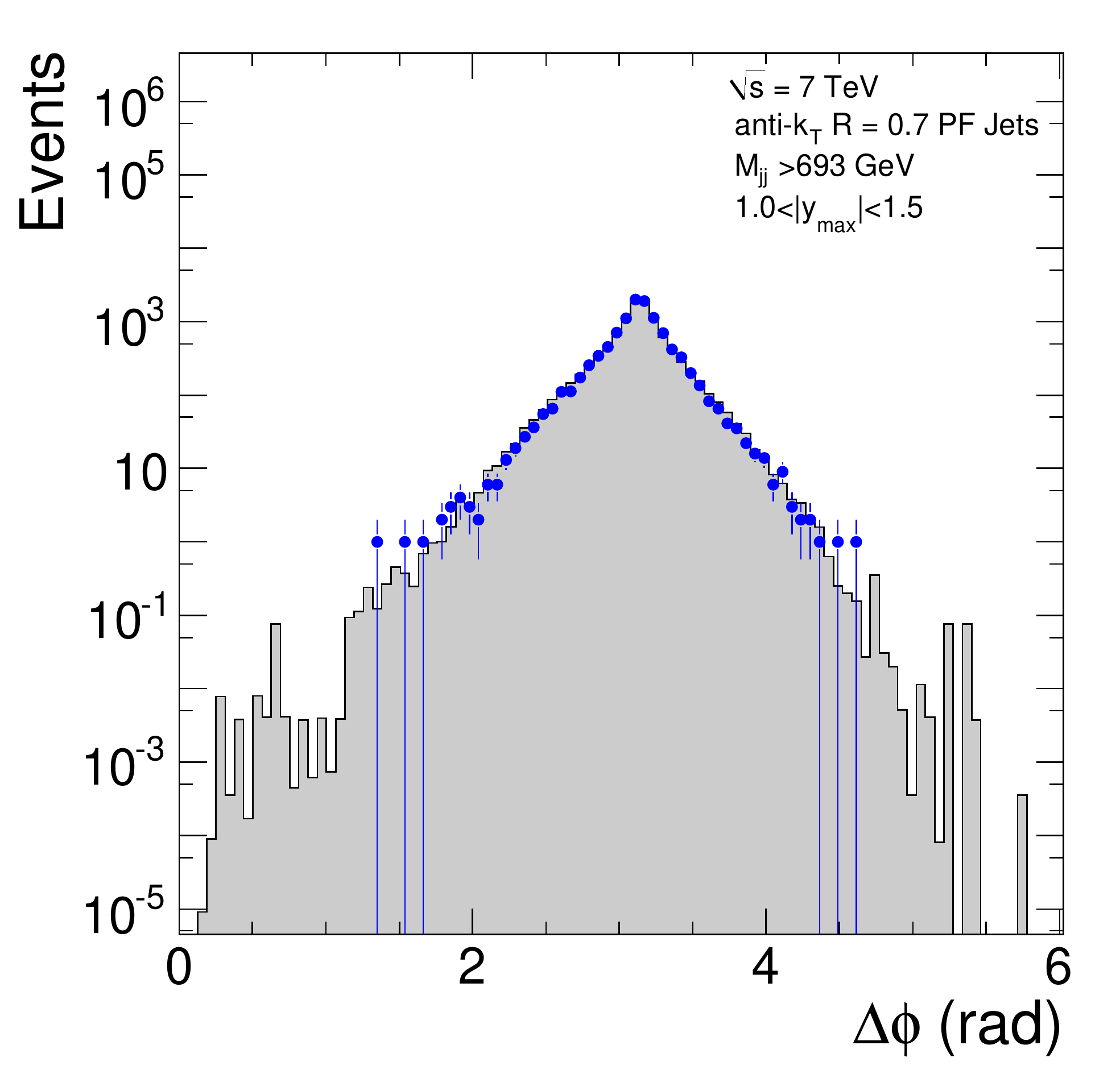} 
  \includegraphics[width=0.40\textwidth]{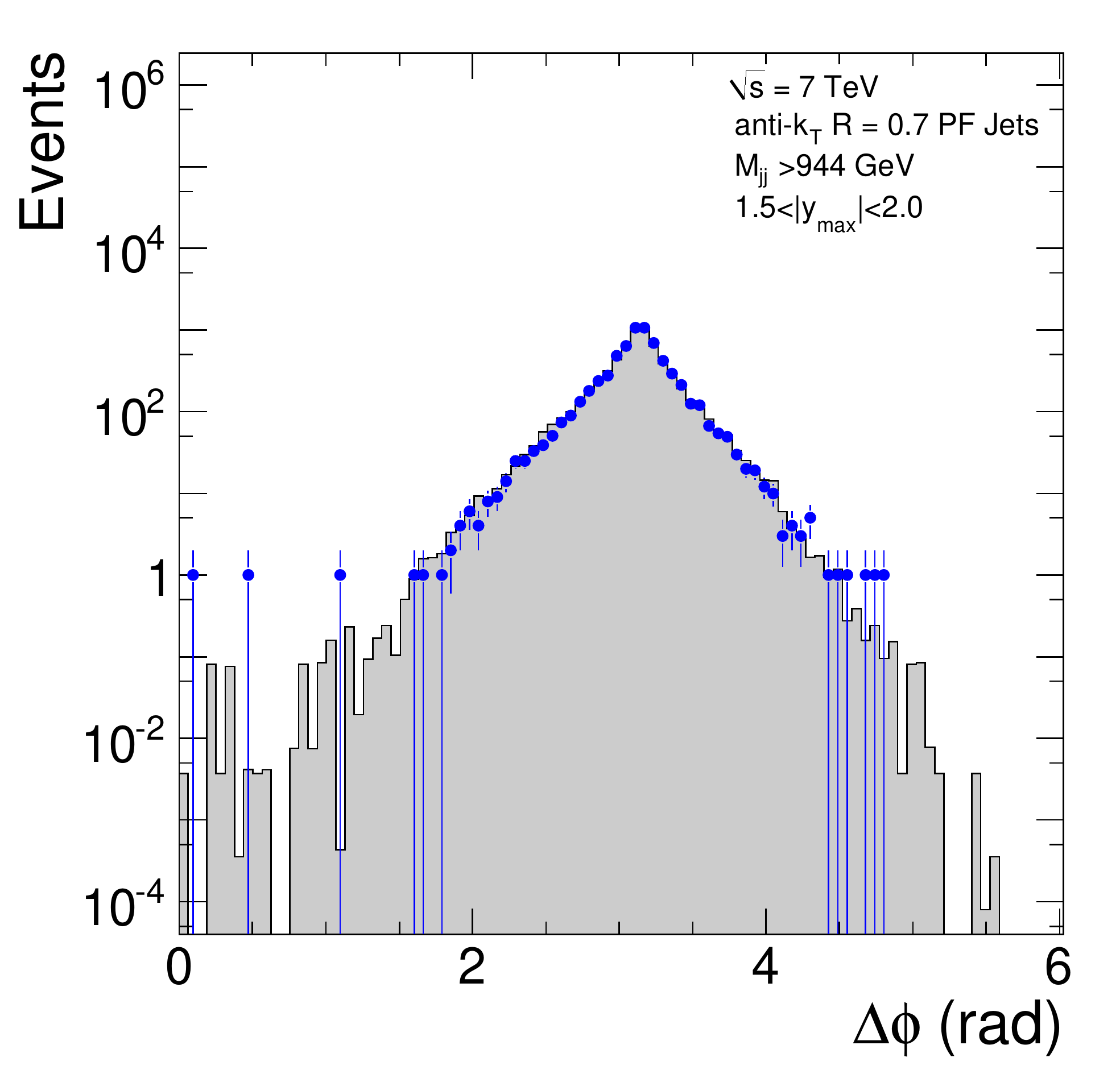}
  \includegraphics[width=0.40\textwidth]{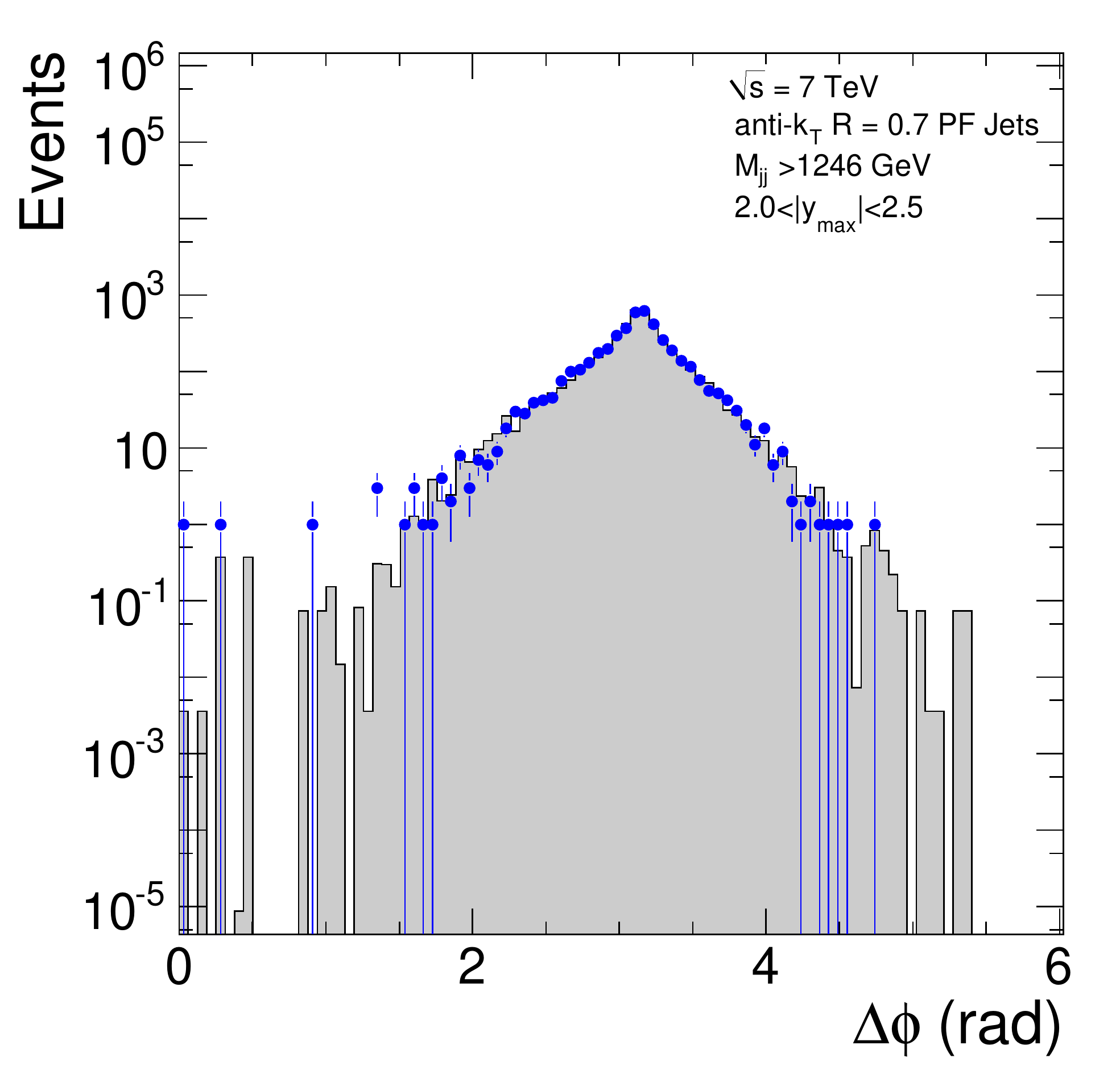}
  \capspace
  \caption{The angle between the two leading jets, $\Delta \phi$ for the five different \ymax bins and for the Jet70U sample. The plots for data (points) and simulated (dashed histogram) events are compared.}
  \label{fig_data4}
\end{figure}

\begin{figure}[ht]
  \centering
  \includegraphics[width=0.40\textwidth]{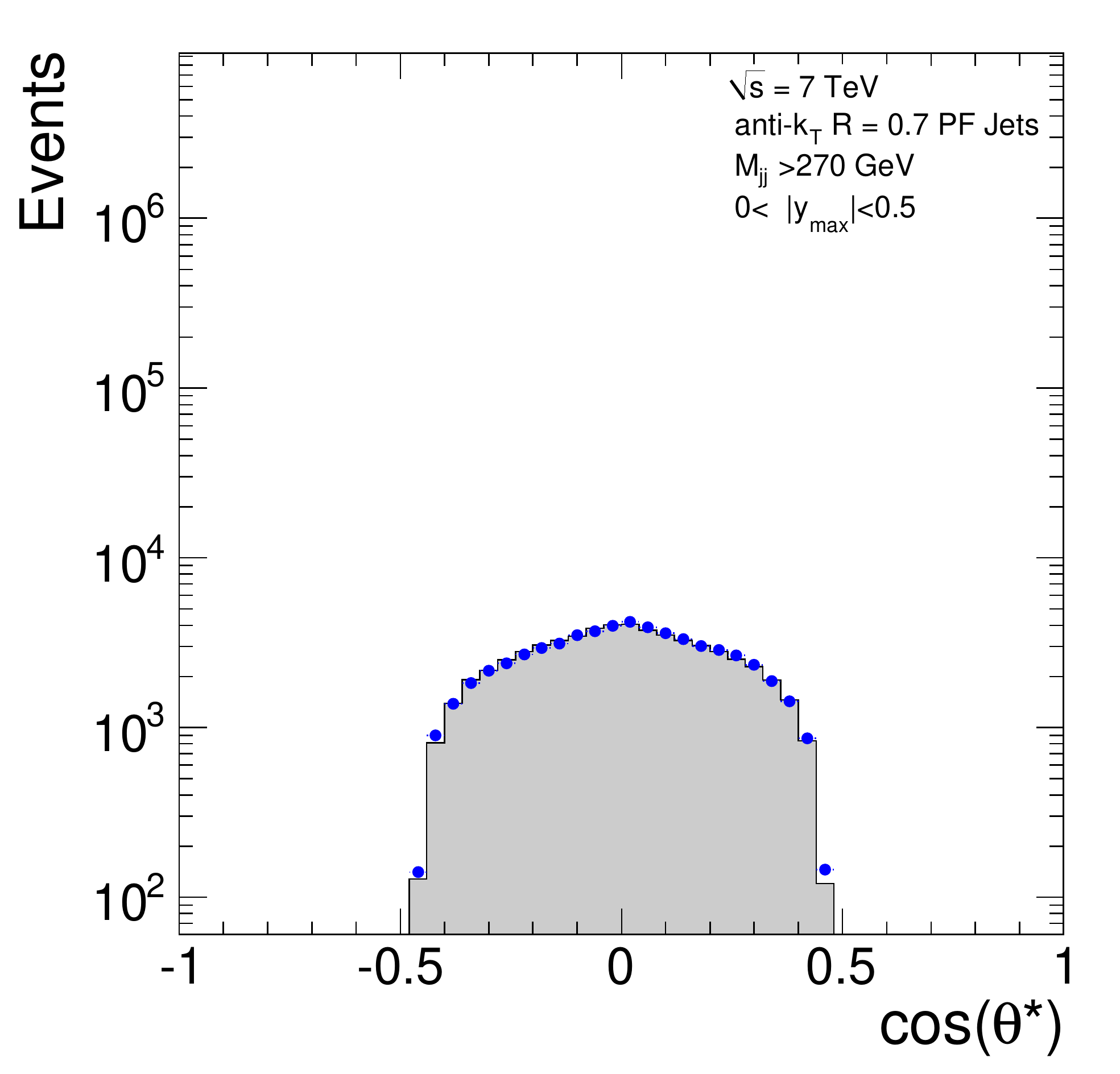} 
  \includegraphics[width=0.40\textwidth]{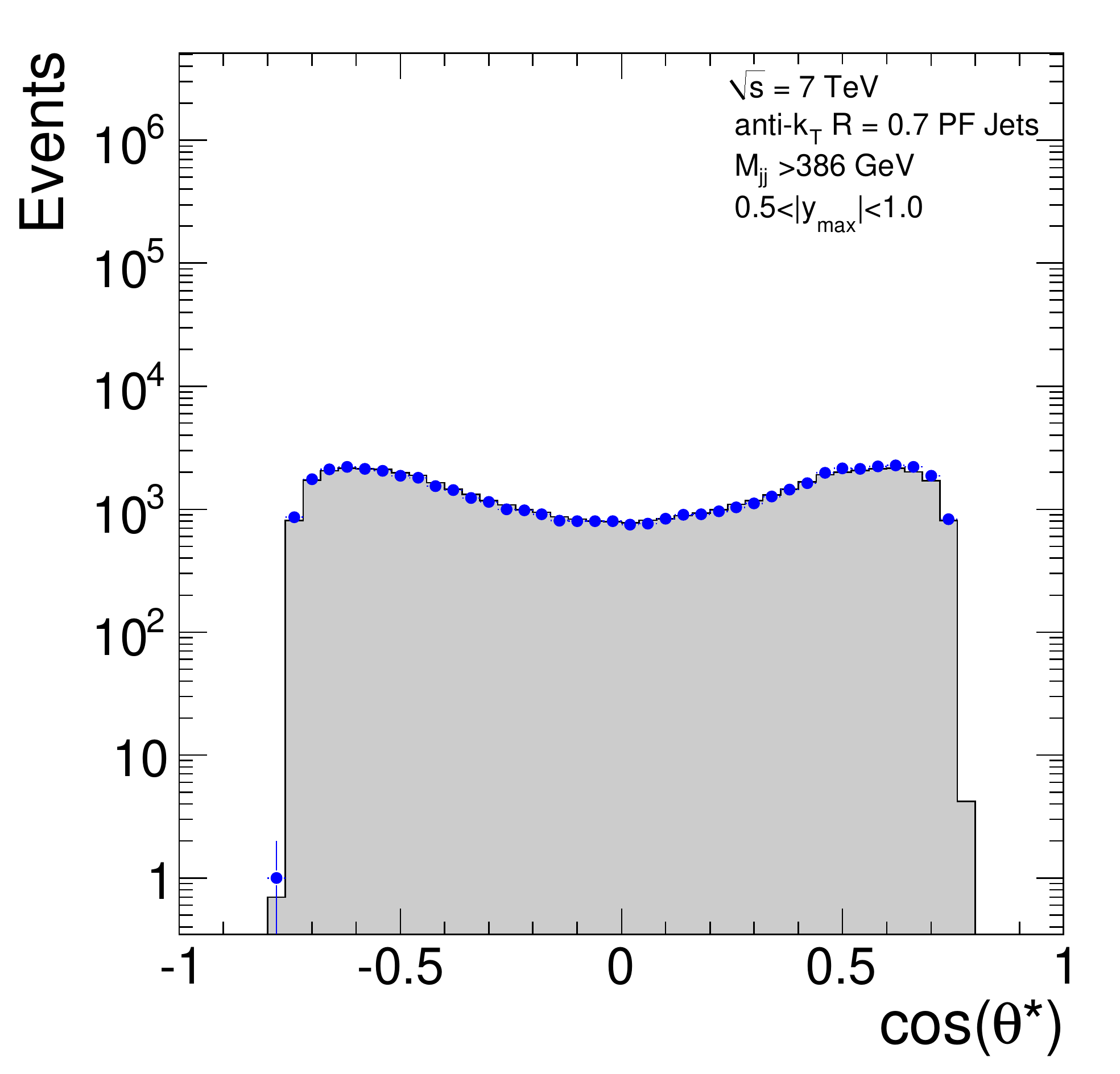}  
  \includegraphics[width=0.40\textwidth]{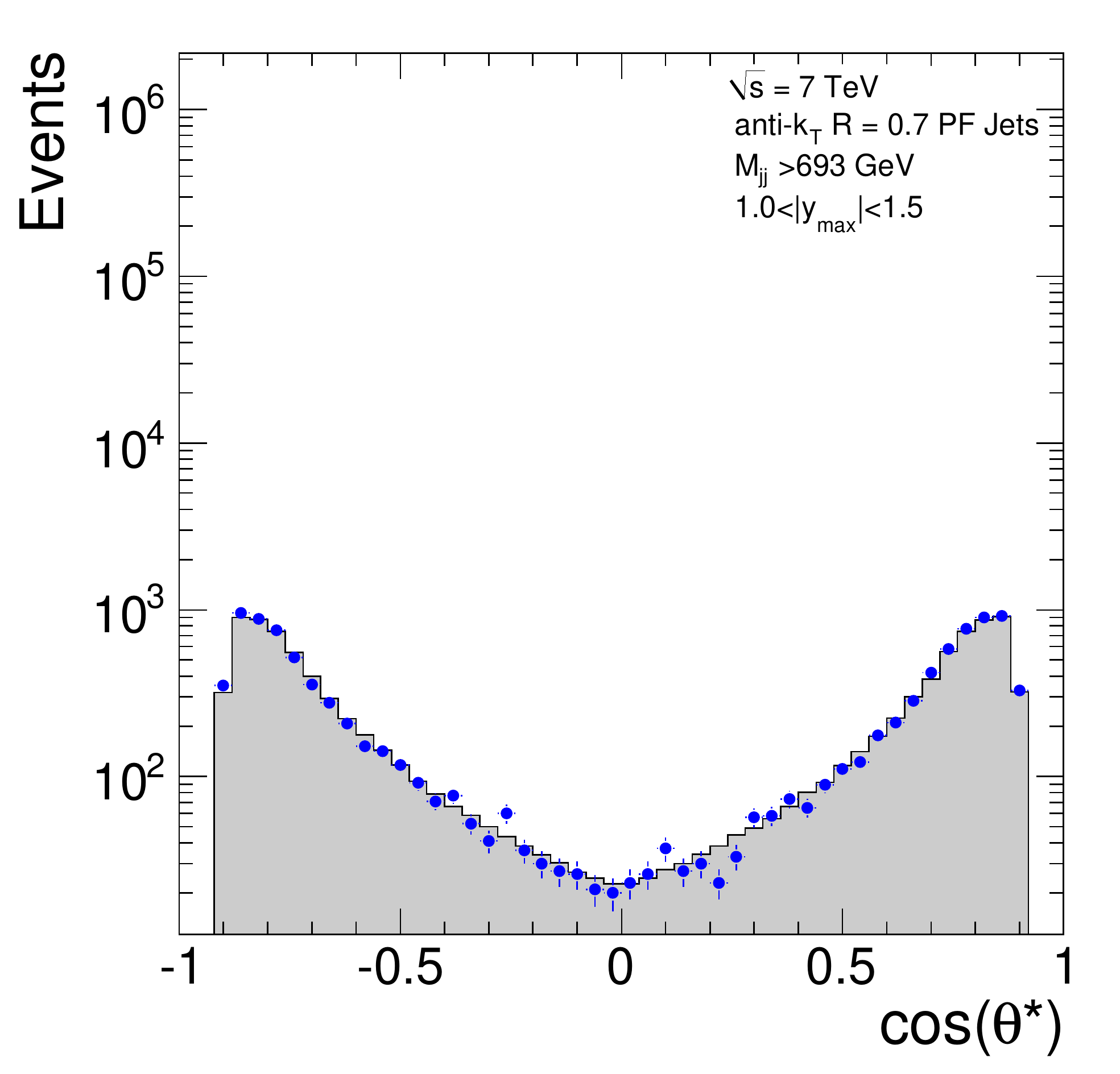}  
  \includegraphics[width=0.40\textwidth]{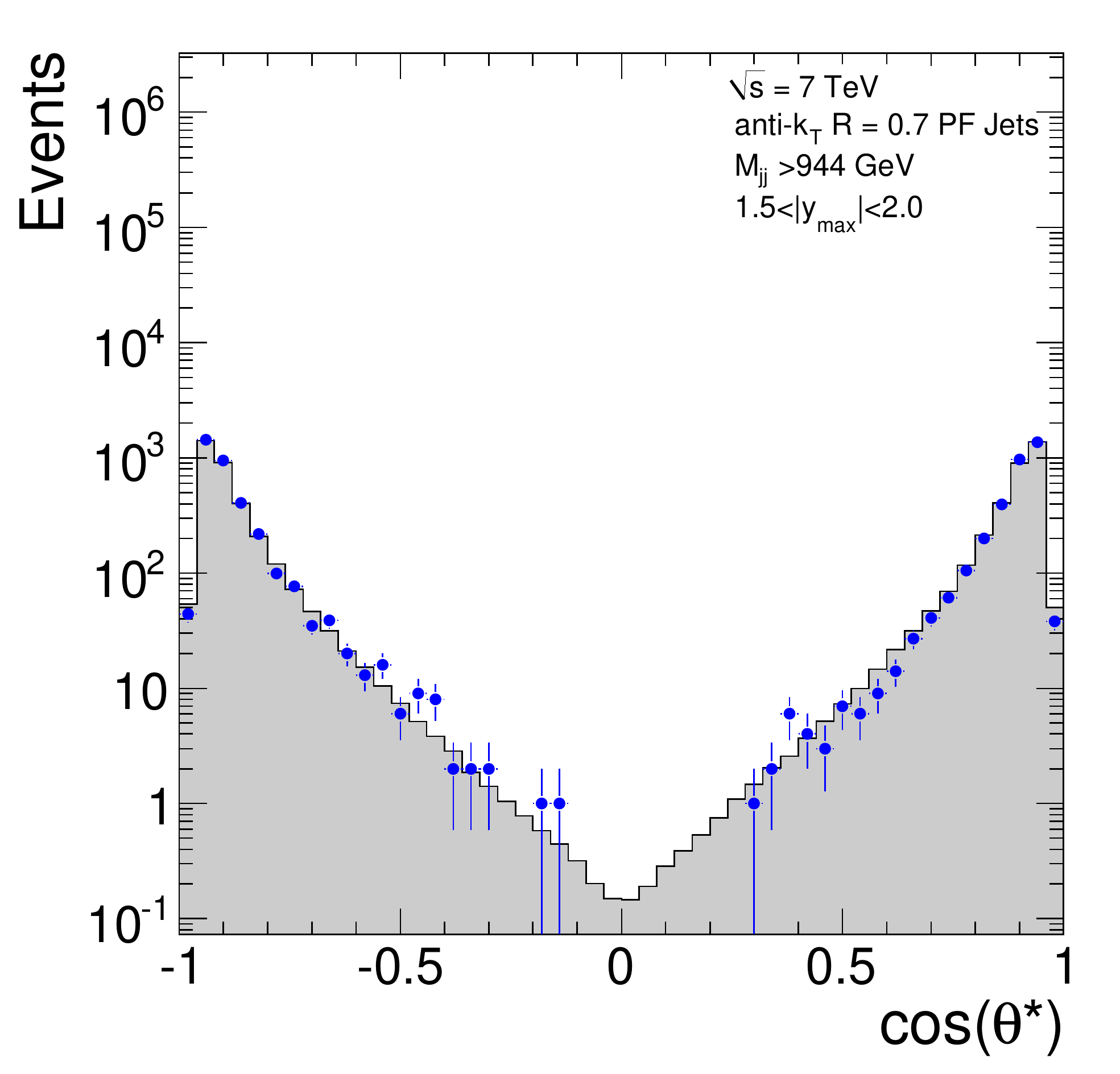} 
  \includegraphics[width=0.40\textwidth]{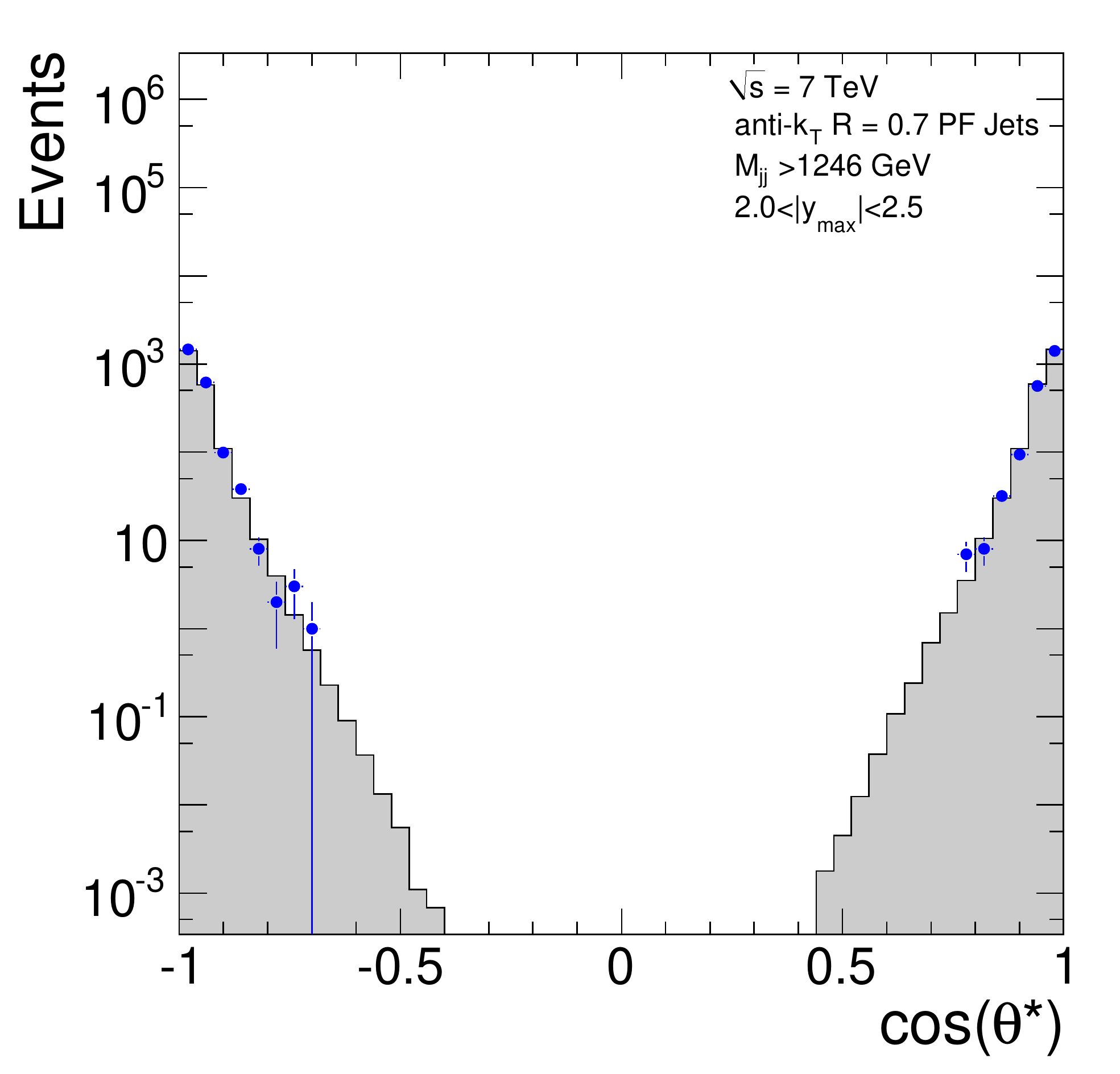}
  \capspace 
  \caption{The angle between beam axis and the dijet system at the center-of-mass frames, $cos(\theta^{*})$, for the five different \ymax bins and for the Jet70U sample. The plots for data (points) and simulated (dashed histogram) events are compared.}
\label{fig_data5}
\end{figure}

In order to examine and study the quality of our data and robustness of the event and jet selection, comparisons of event related and jet related variables with their Monte Carlo simulated predictions are performed. A lack of quality might be originated from the beam and detector related noise, detector pathologies, catastrophic reconstruction failures etc. As it was pointed before, two categories of distributions are examined; event related variables and jet related variables. The examined event related variables are;
 \begin{itemize}
 \item The ratio of the missing transverse energy in the event to the total transverse energy, $E_{T}^{miss}/\sum E_{T}$
 \item The azimuthal angle between the two leading jets, $\Delta \phi= \phi_1-\phi_2$
 \item The polar angle between the colliding partons and the scattered partons at the center-of-mass frame, $cos(\theta^{*})=tanh(y_1-y_2)$
 \end{itemize}
The first variable $E_{T}^{miss}/\sum E_{T}$ is sensitive to the detector originated noise which would end up with a significant energy imbalance in the event. Therefore, in the presence of noise, higher values of this variable is expected to be populated in the distribution. The second variable, $\Delta \phi= \phi_1-\phi_2$, is sensitive to both a general noise in the detector and a particular noise which could mimic a jet. In that case this value is expected to be away from the $\Delta \phi=\pi$ expectation. The third variable, $cos(\theta^{*})$, is also sensitive to a general noise in the detector and it shows the deviation from the expected value which might indicate a pathology in the data sample. In Figures \ref{fig_data3}-\ref{fig_data5} the comparisons between the data and simulated events for the Jet70U sample in five \ymax bins are shown. The rest of the plots can be found in Appendix B. All distributions for the data are in agreement with the simulated ones and no significant deviations from the expectations are observed.
\clearpage

\begin{figure}[ht]
  \centering
  \includegraphics[width=0.40\textwidth]{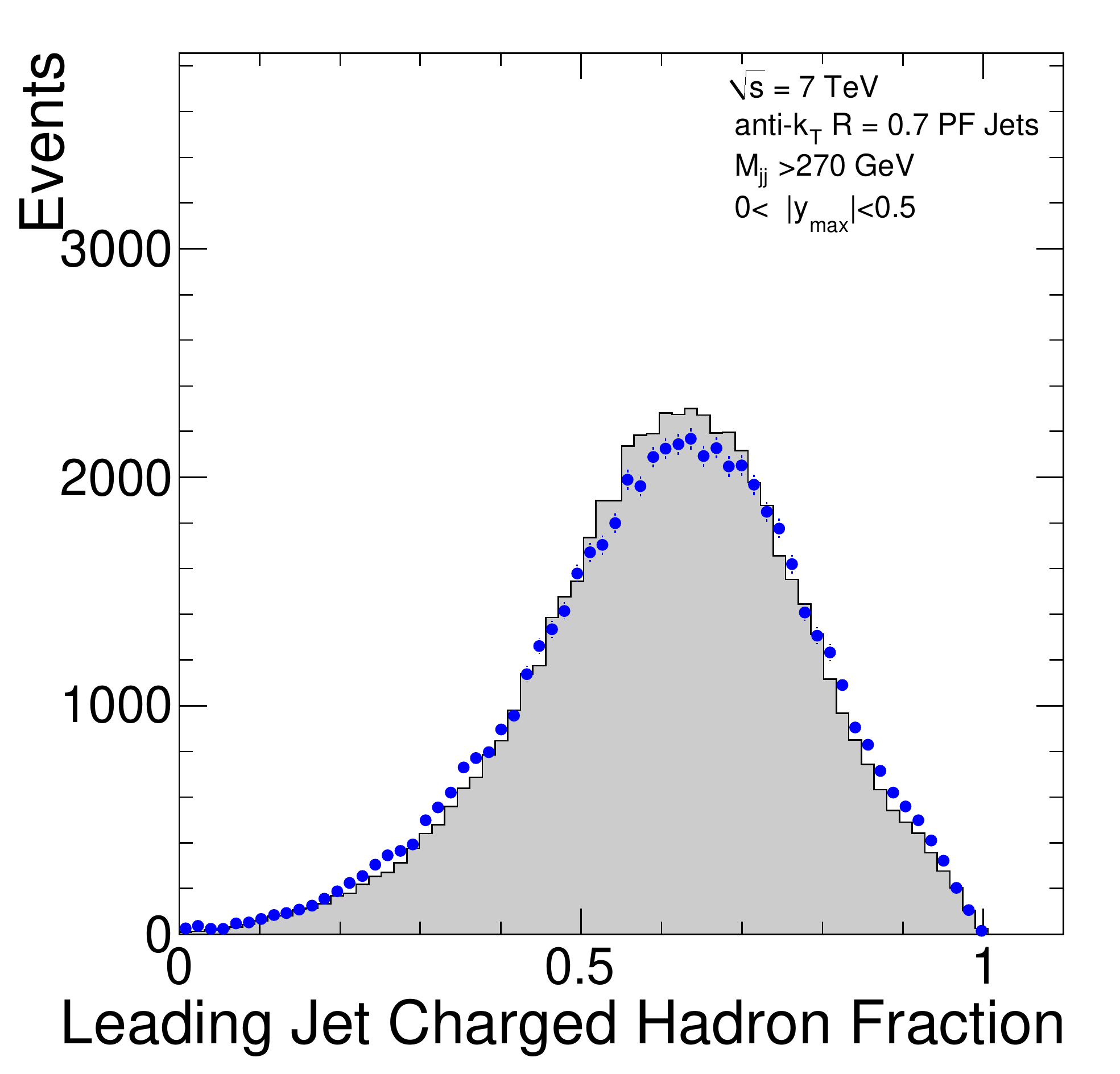}
  \includegraphics[width=0.40\textwidth]{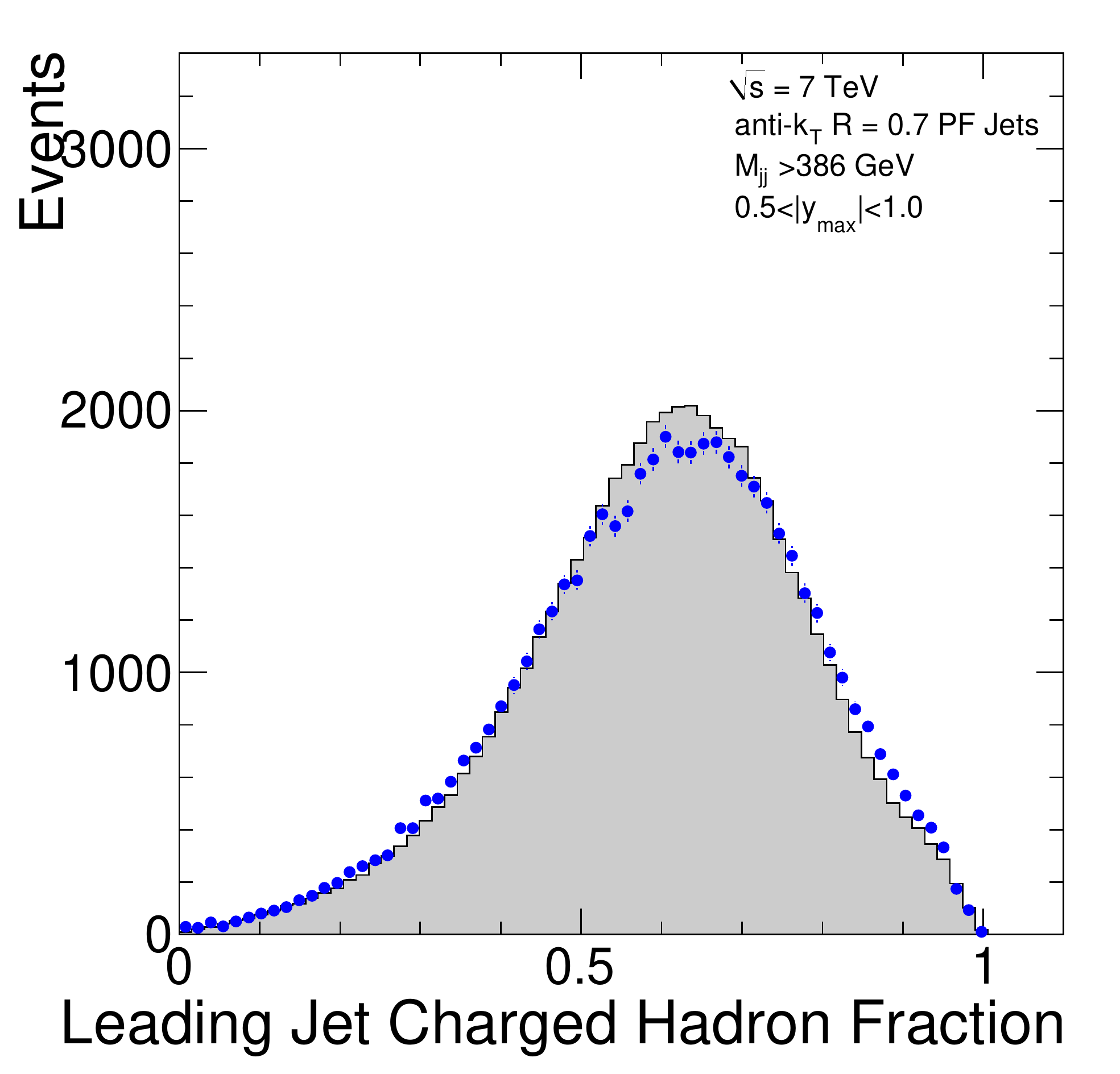}
  \includegraphics[width=0.40\textwidth]{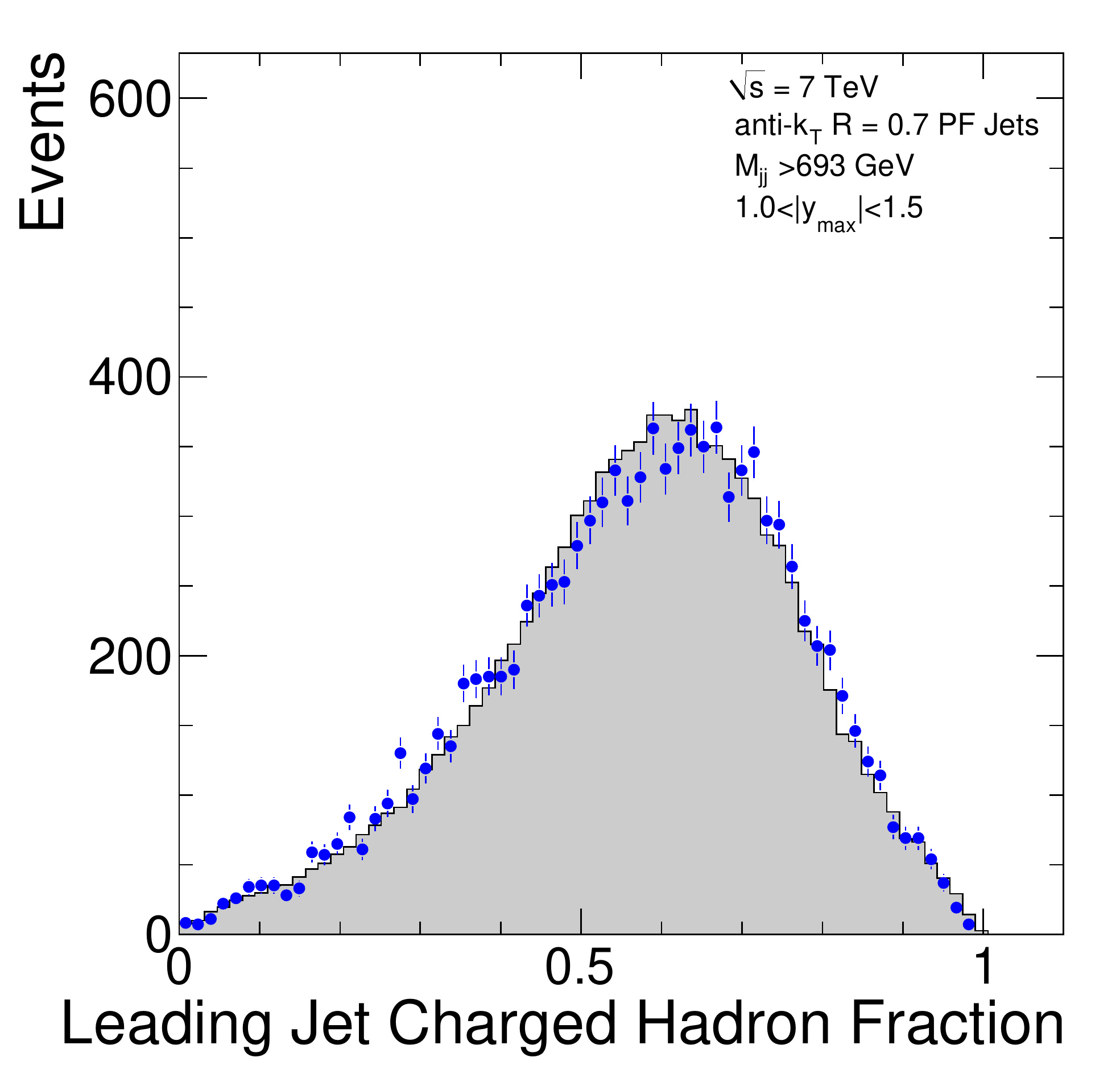} 
  \includegraphics[width=0.40\textwidth]{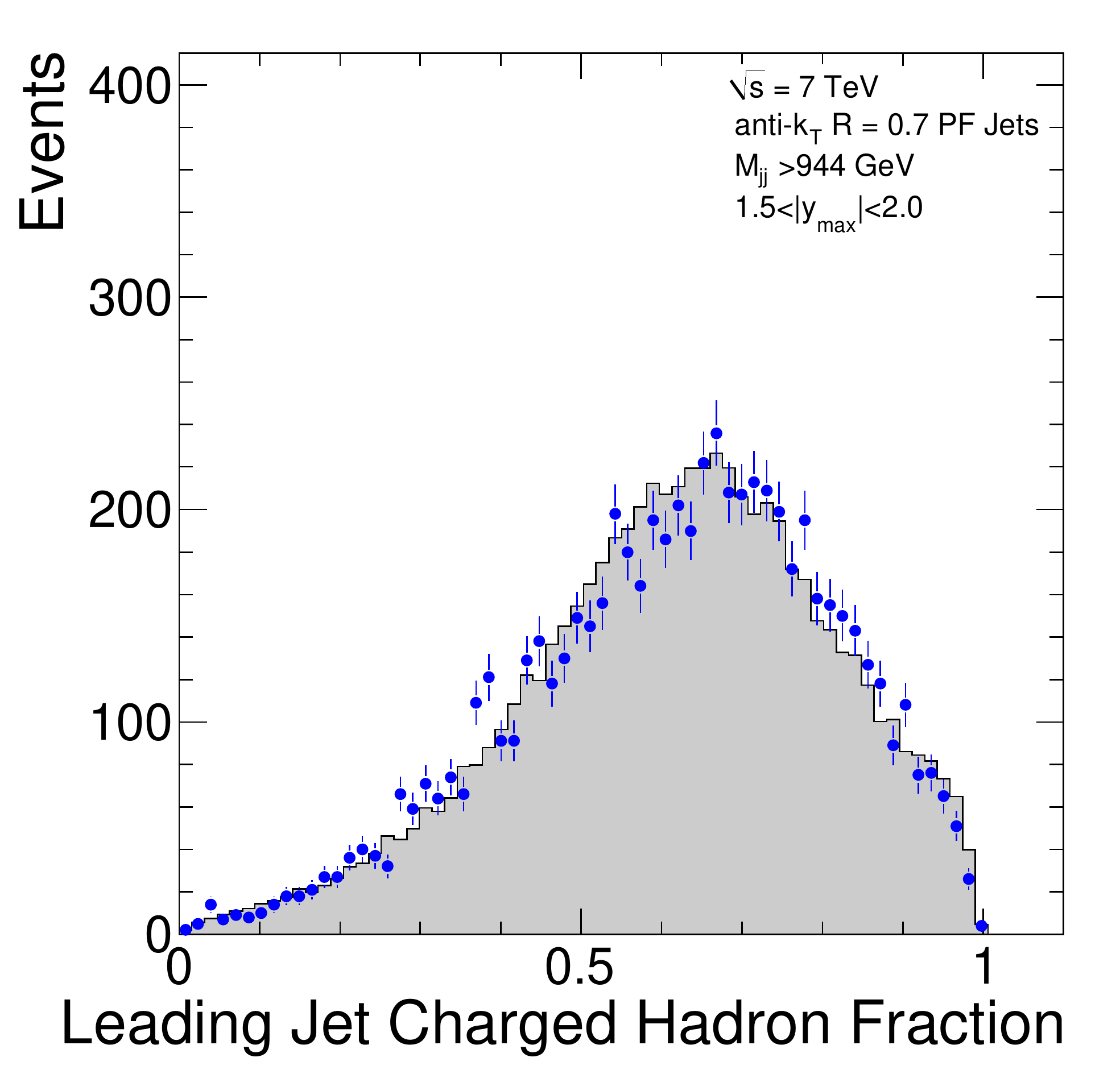}
  \includegraphics[width=0.40\textwidth]{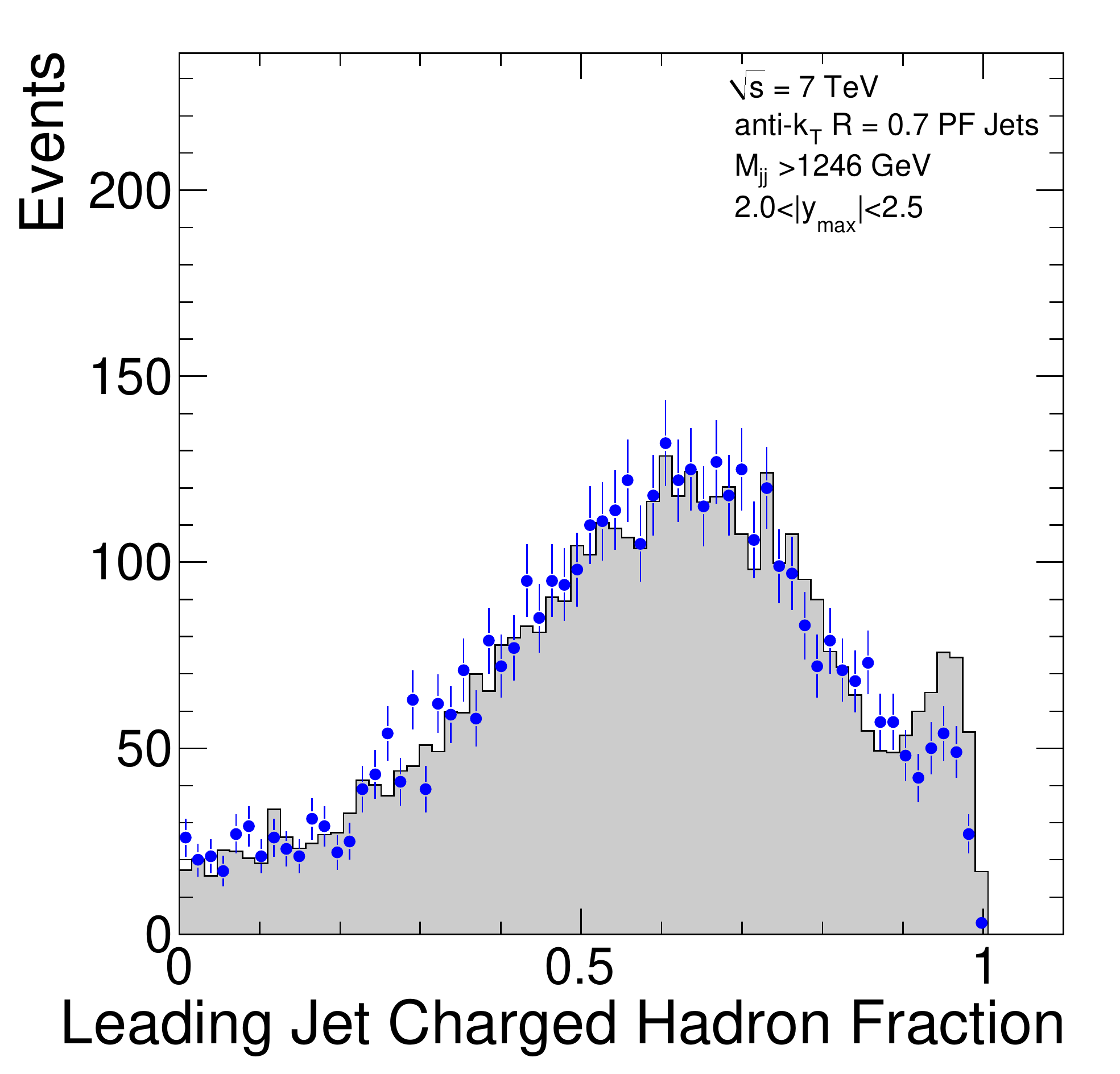}
  \capspace
  \caption{The charged hadron fraction of the leading jet for the five different \ymax bins and for the Jet70U sample. The plots for data (points) and simulated (dashed histogram) events are compared.}
  \label{fig_data6}
\end{figure}

\begin{figure}[ht]
  \centering
  \includegraphics[width=0.40\textwidth]{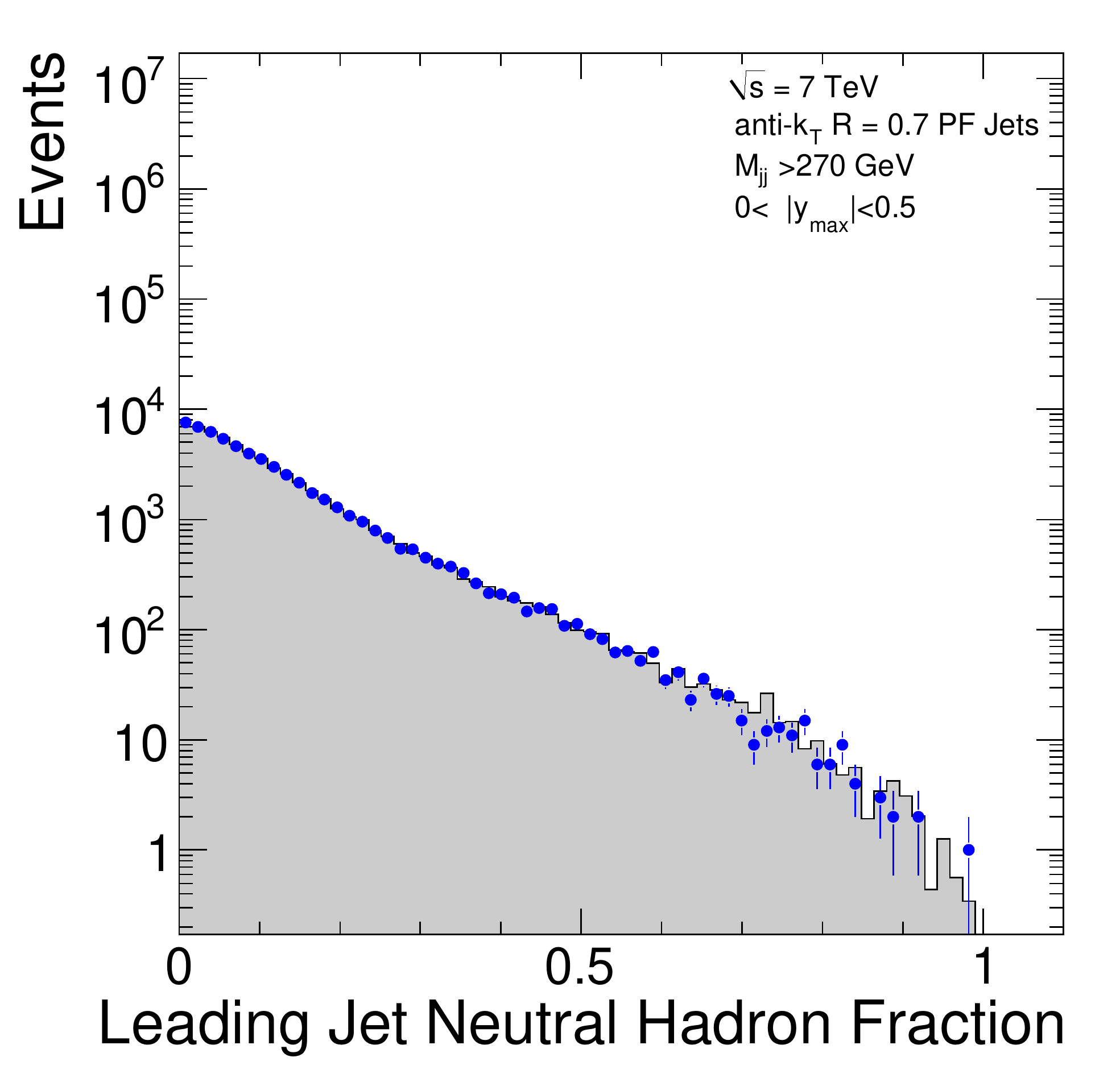}
  \includegraphics[width=0.40\textwidth]{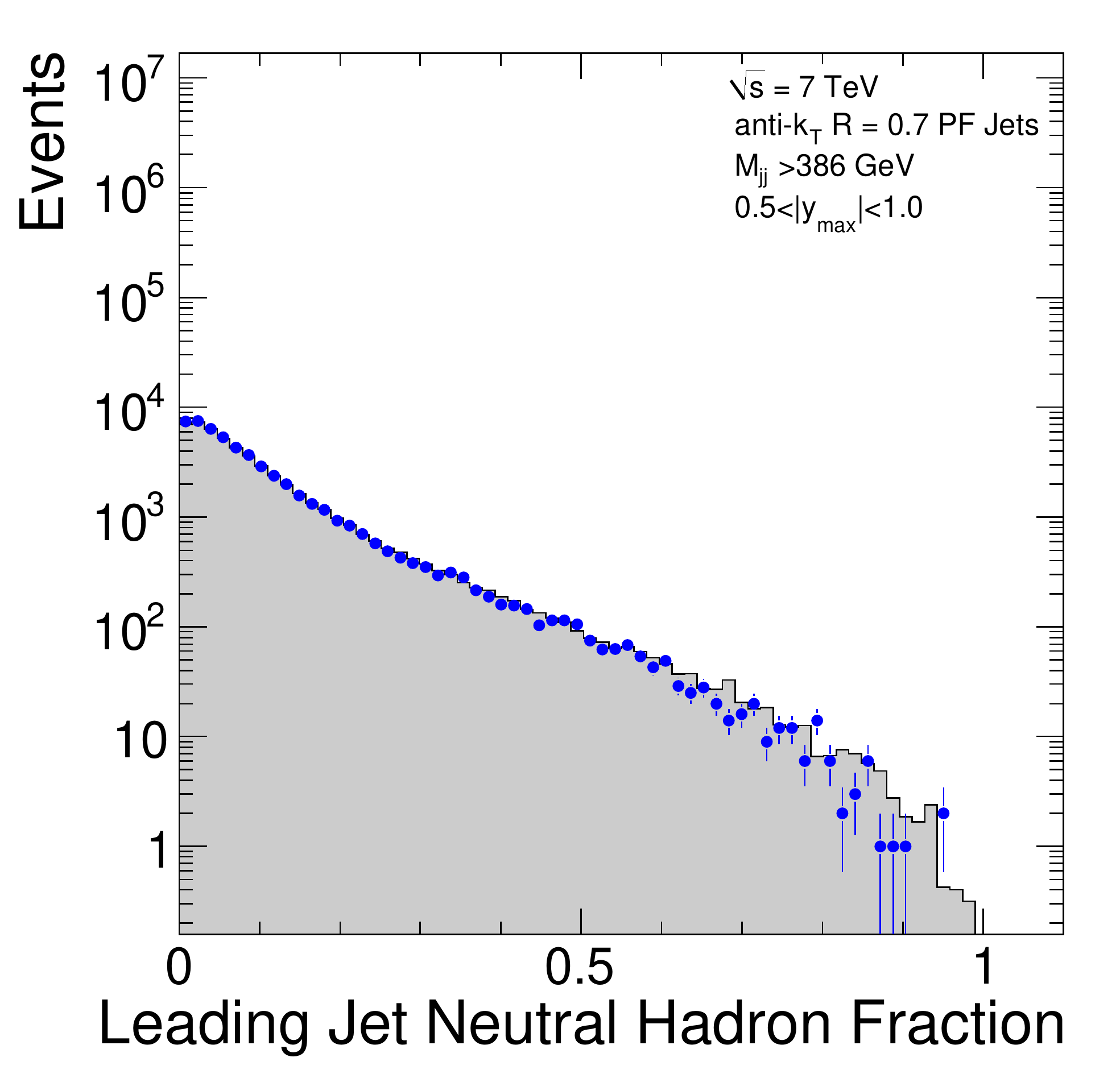} 
  \includegraphics[width=0.40\textwidth]{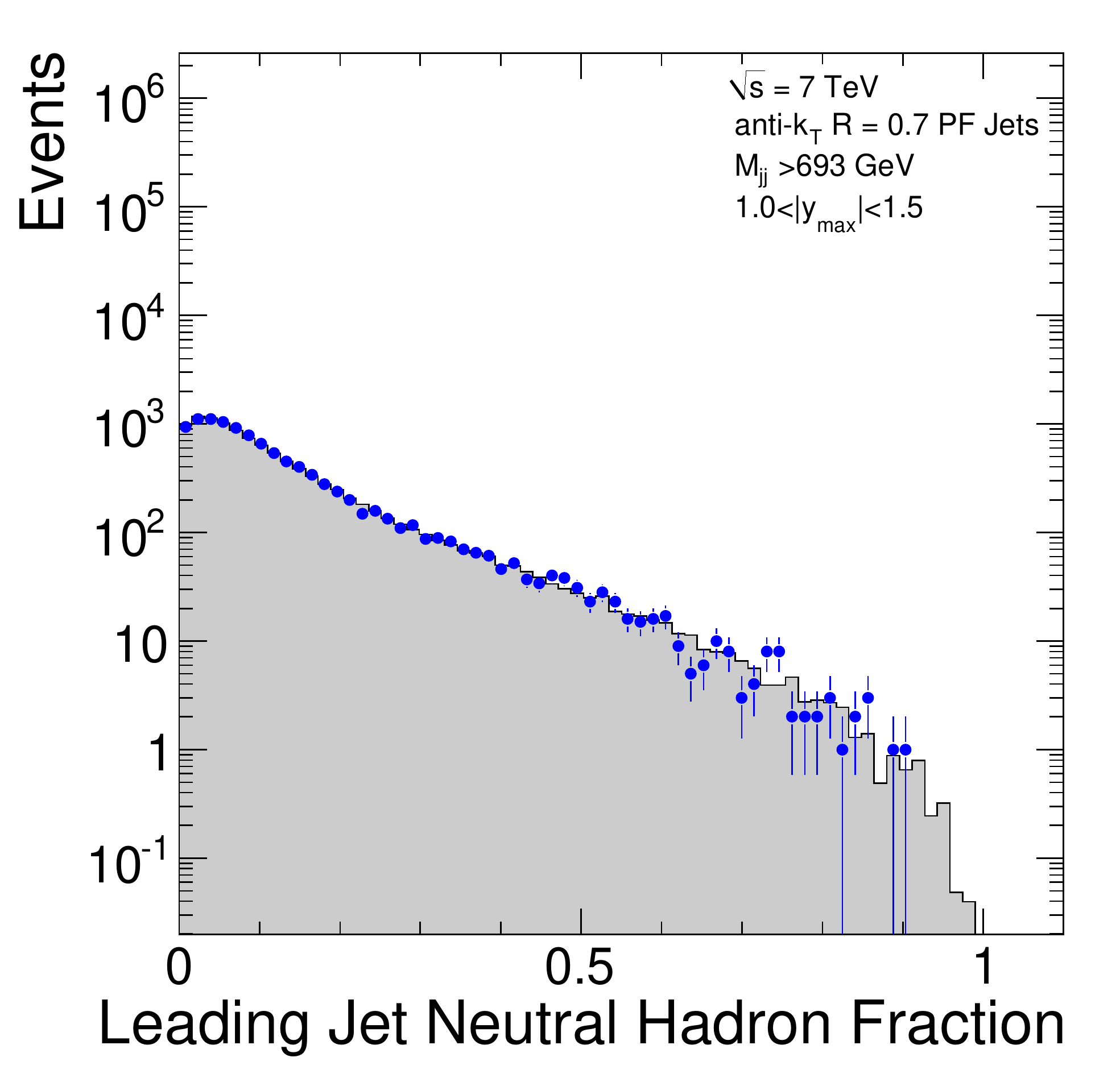} 
  \includegraphics[width=0.40\textwidth]{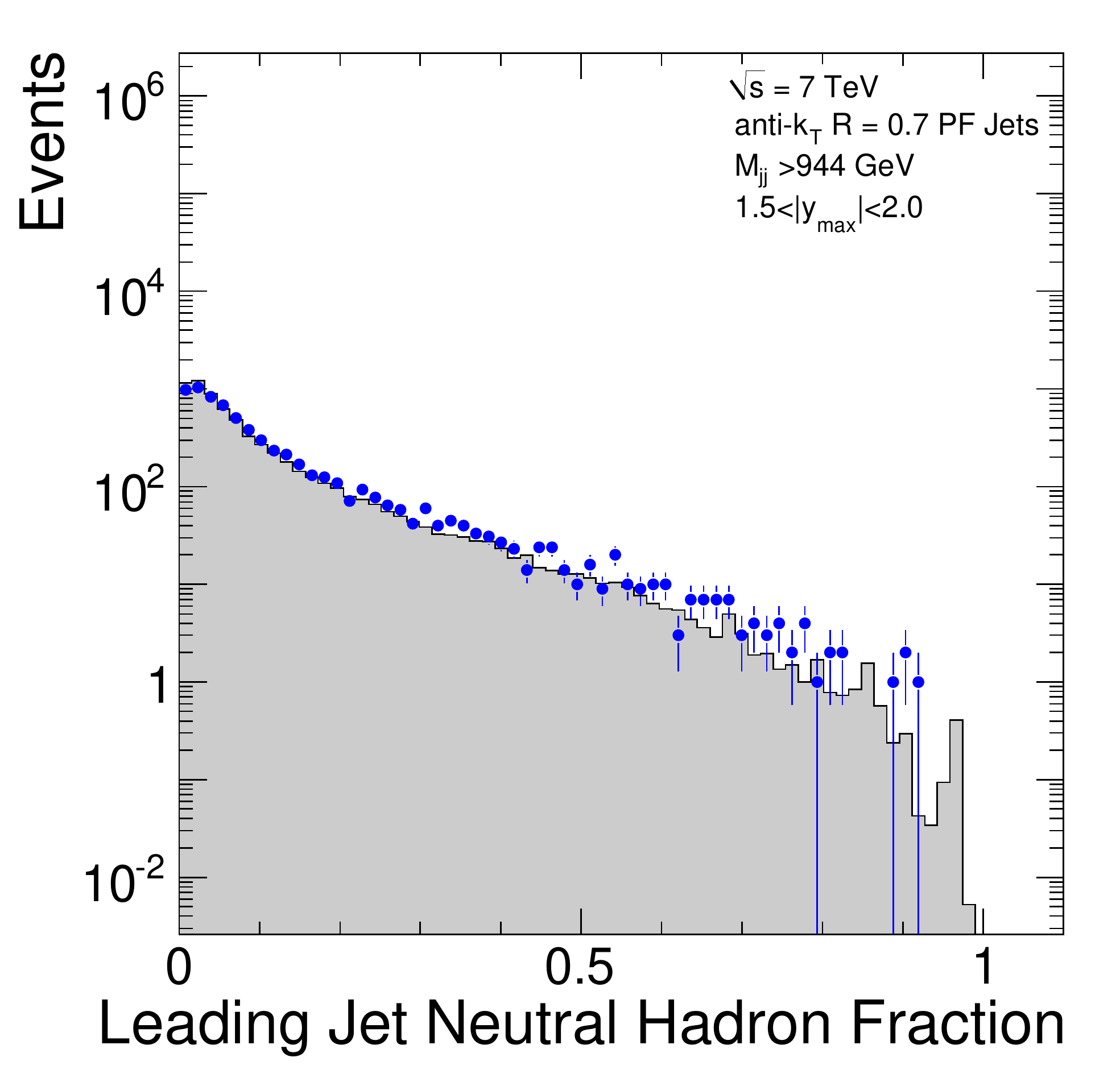}
  \includegraphics[width=0.40\textwidth]{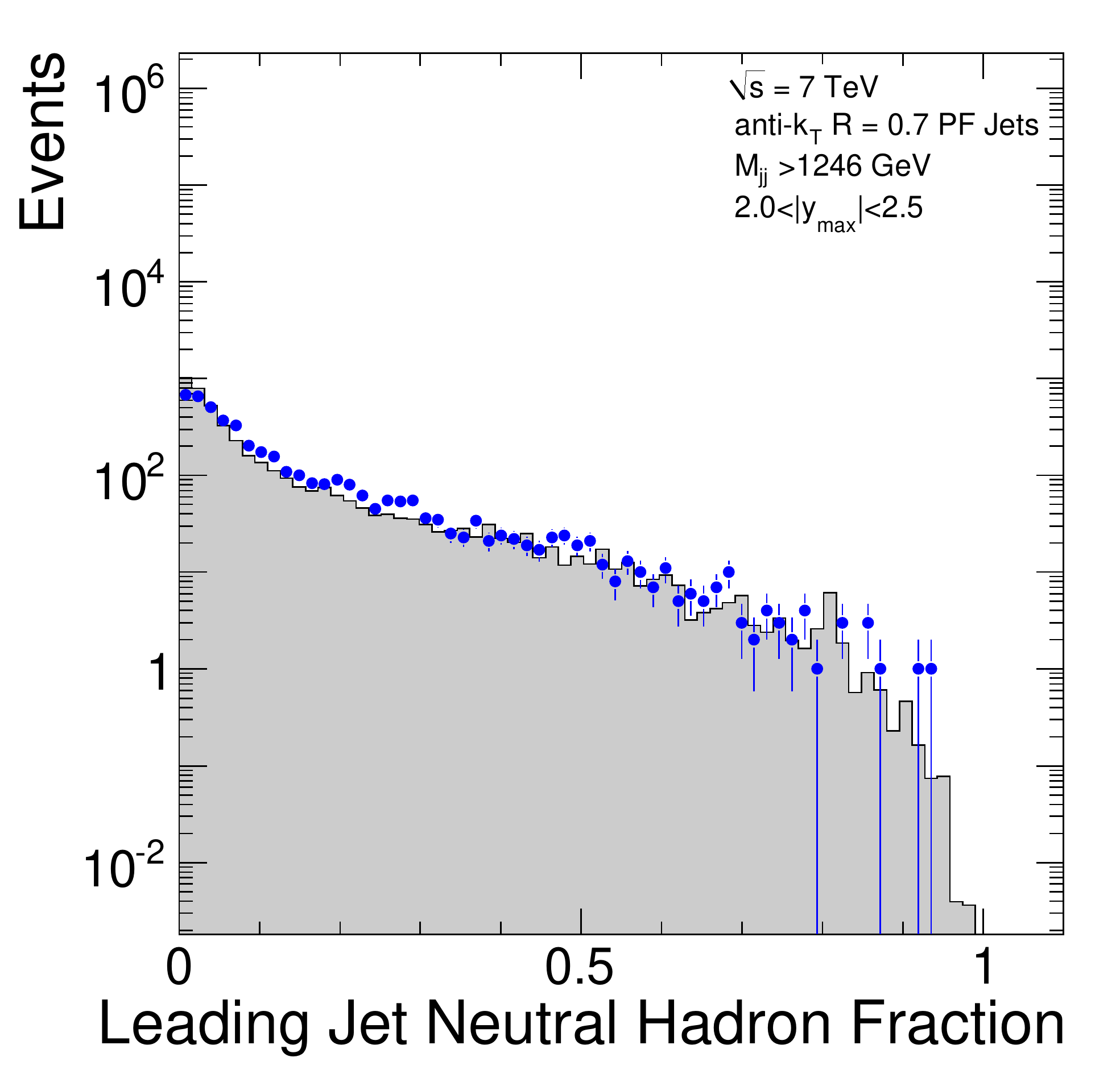}
  \capspace
  \caption{The neutral hadron fraction of the leading jet for the five different \ymax bins and for the Jet70U sample. The plots for data (points) and simulated (dashed histogram) events are compared.}
  \label{fig_data7}
\end{figure}

\begin{figure}[ht]
  \centering
  \includegraphics[width=0.40\textwidth]{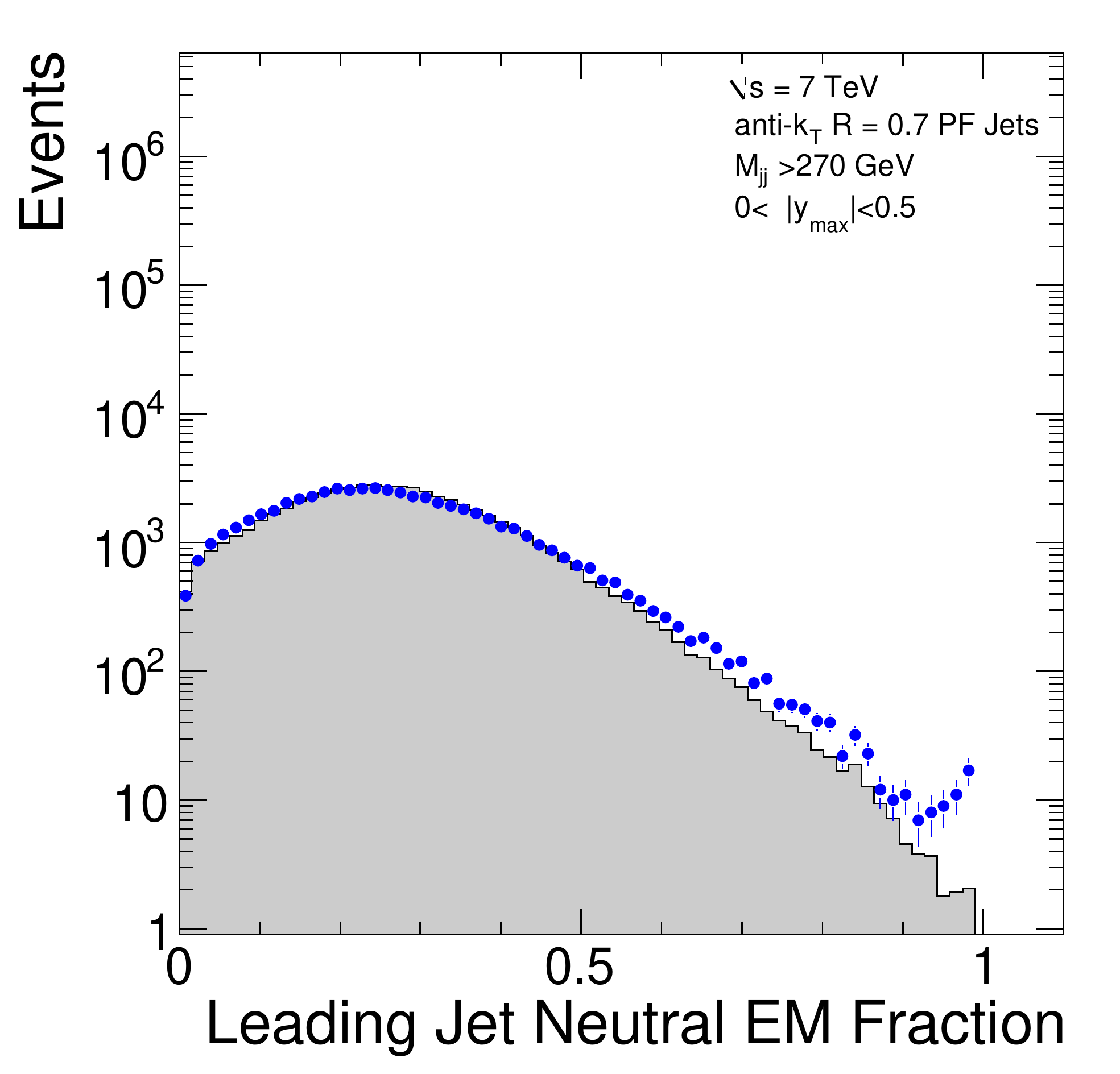}
  \includegraphics[width=0.40\textwidth]{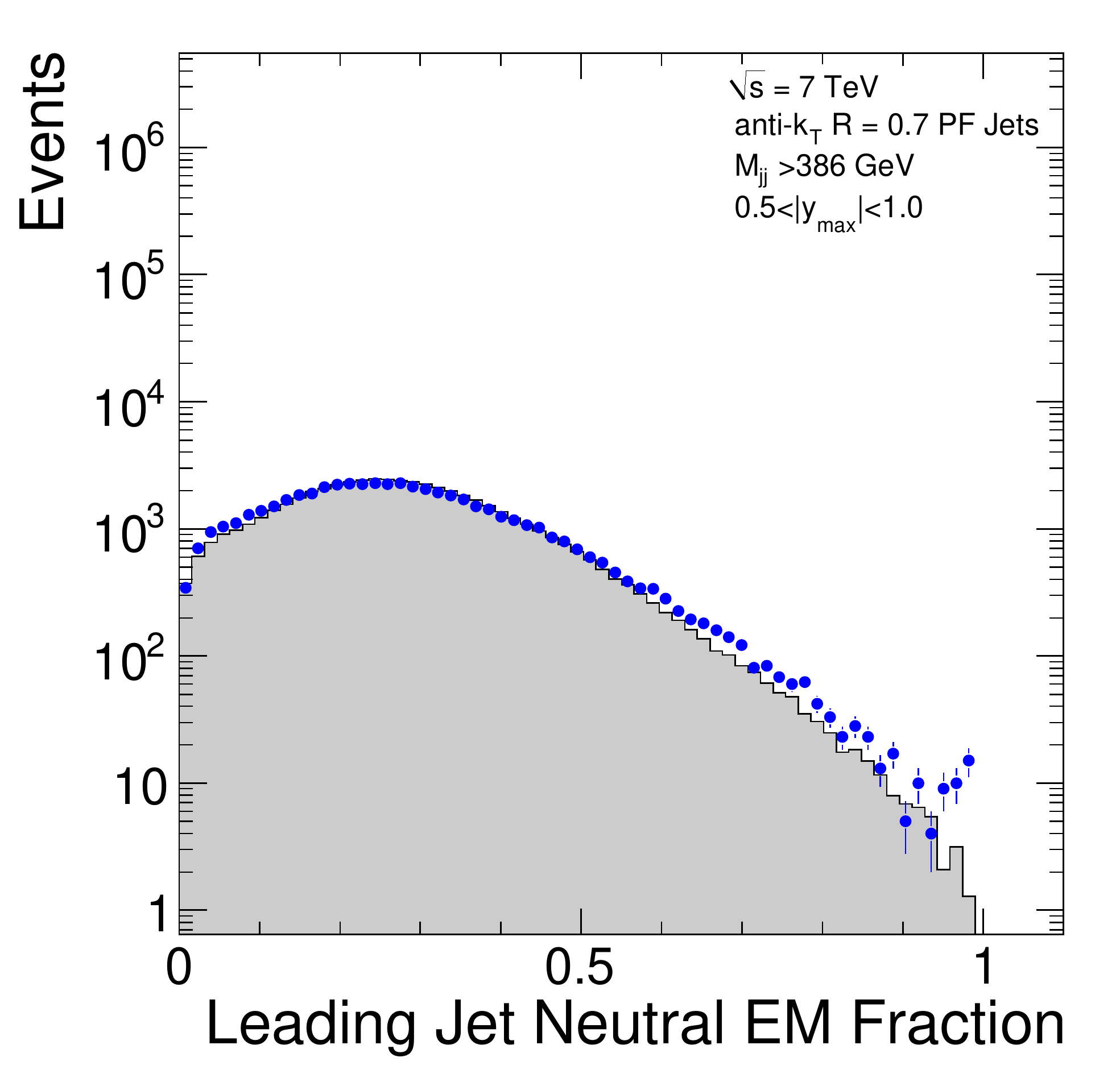} 
  \includegraphics[width=0.40\textwidth]{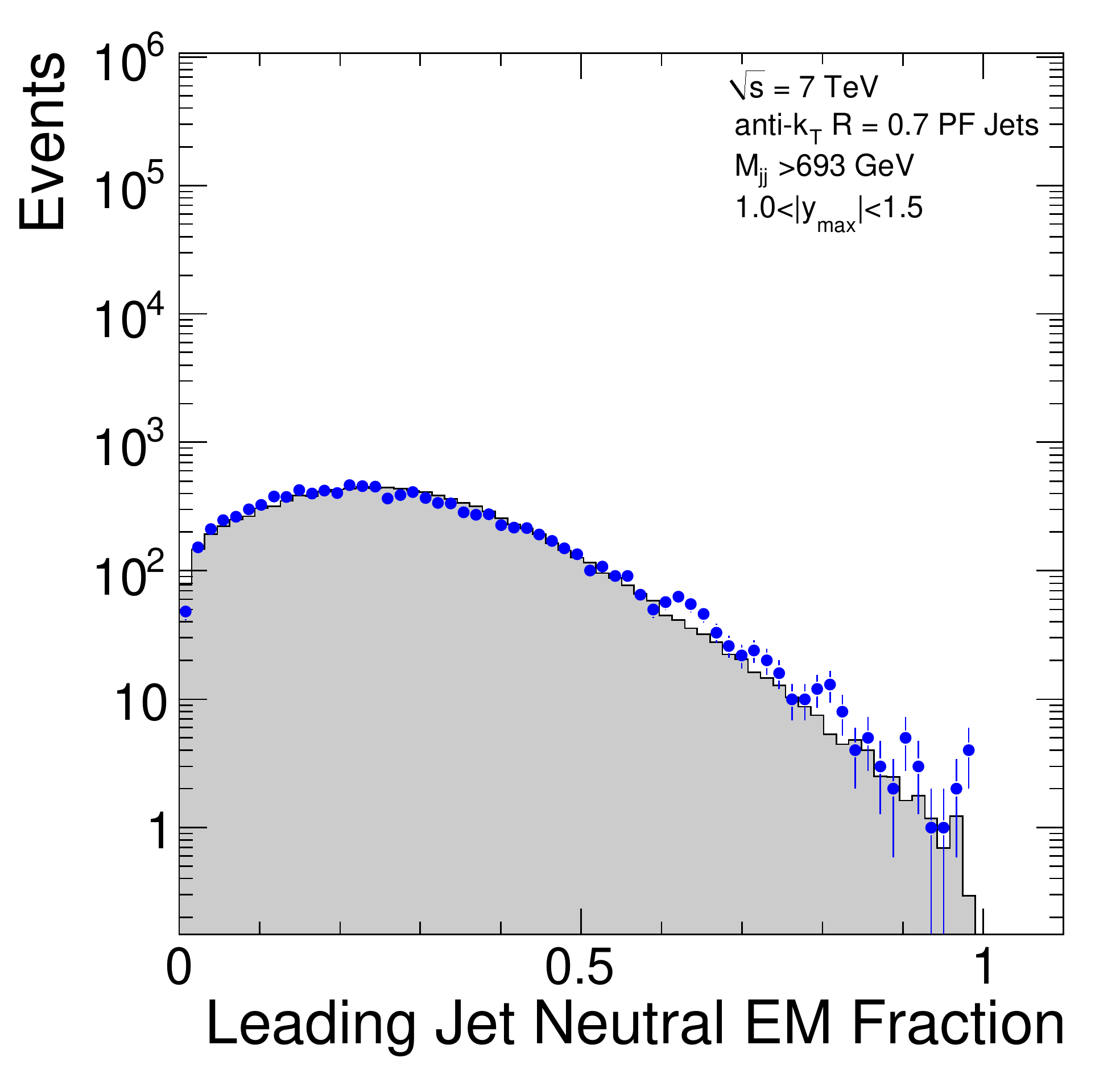} 
  \includegraphics[width=0.40\textwidth]{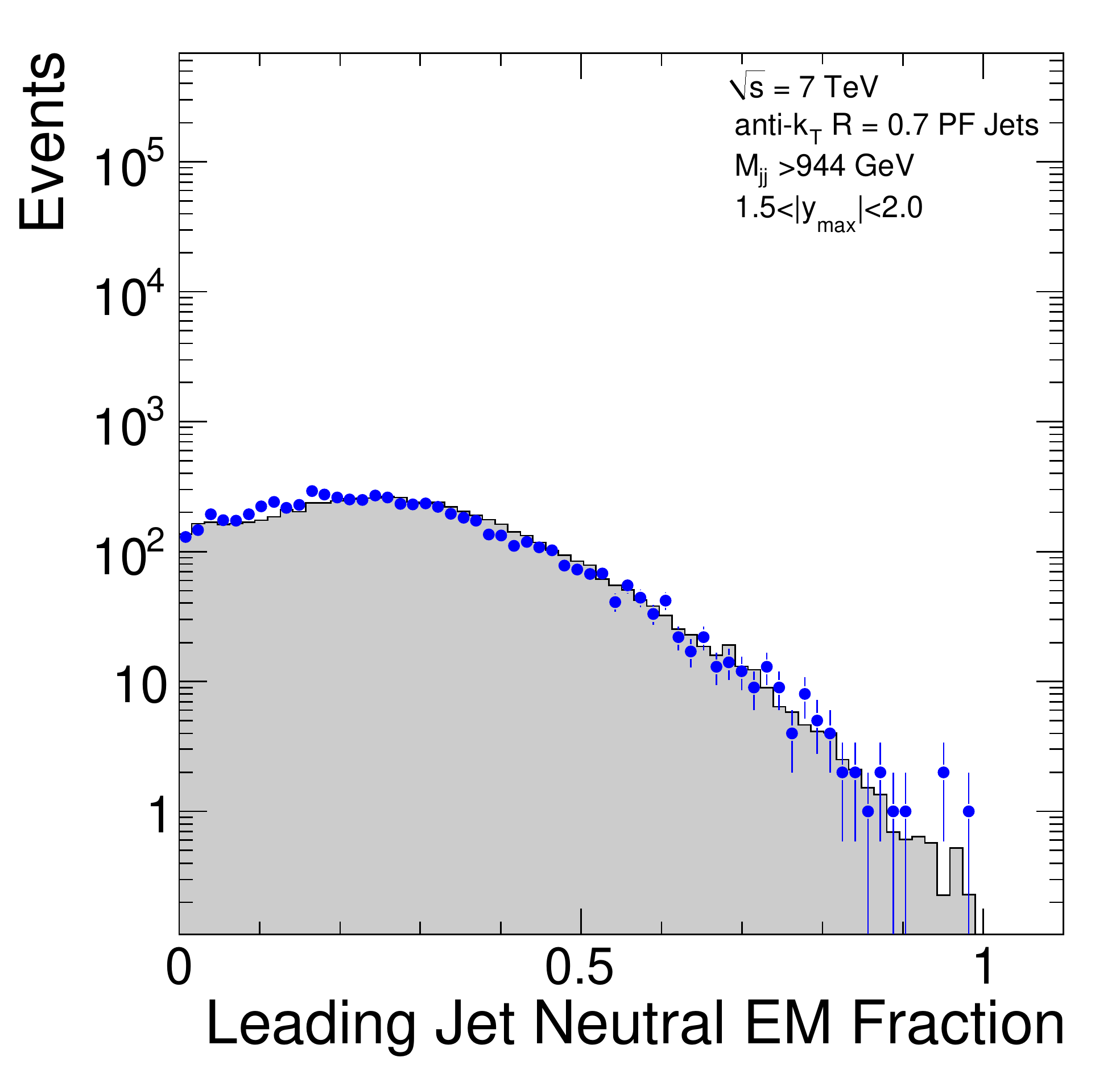}
  \includegraphics[width=0.40\textwidth]{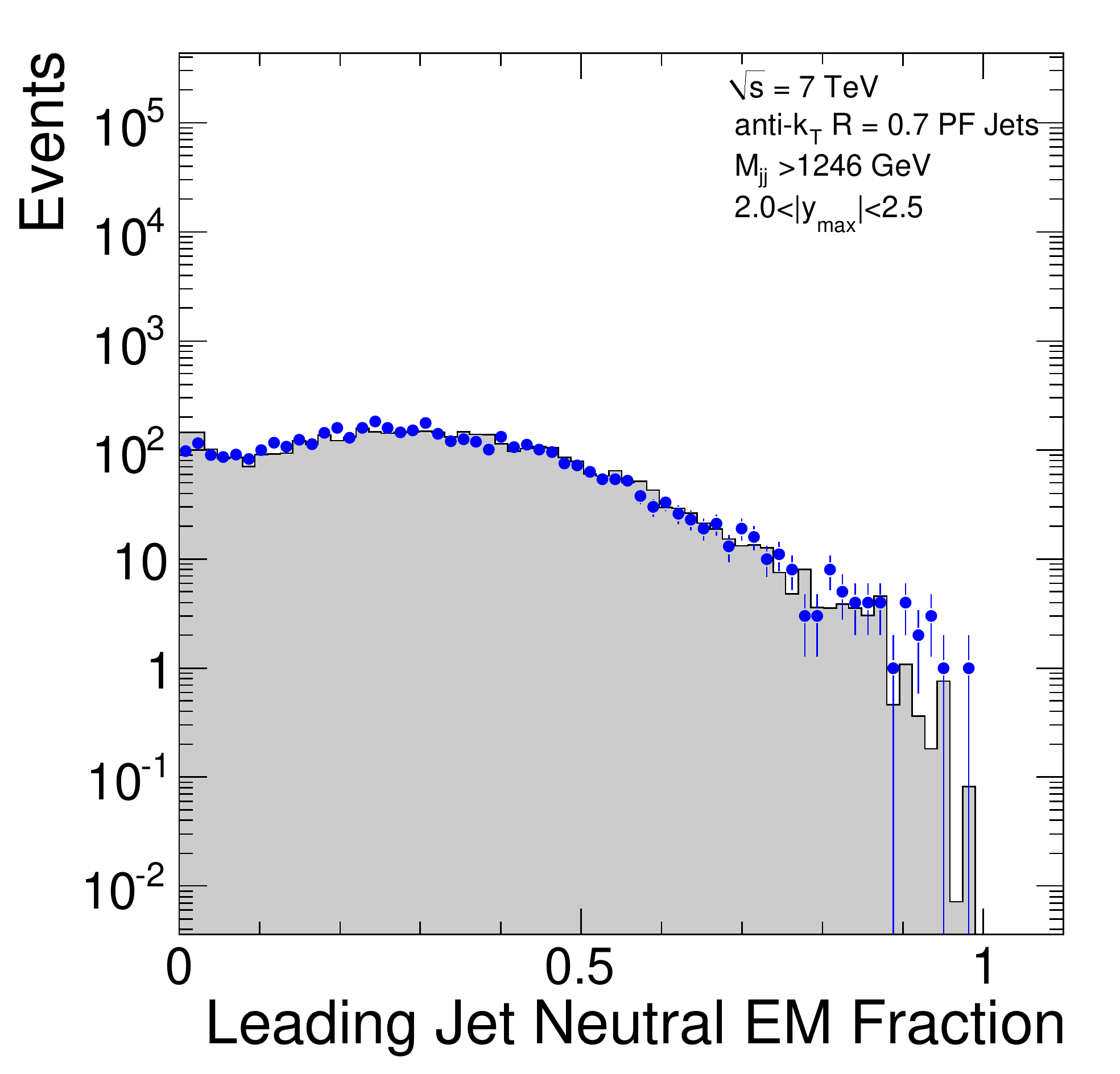}
  \capspace
  \caption{The neutral electromagnetic fraction of the leading jet for the five different \ymax bins and for the 
HLT\_Jet70U sample. The plots for data (points) and simulated (dashed histogram) events are compared.}
  \label{fig_data8}
\end{figure}

\begin{figure}[ht]
  \centering
  \includegraphics[width=0.40\textwidth]{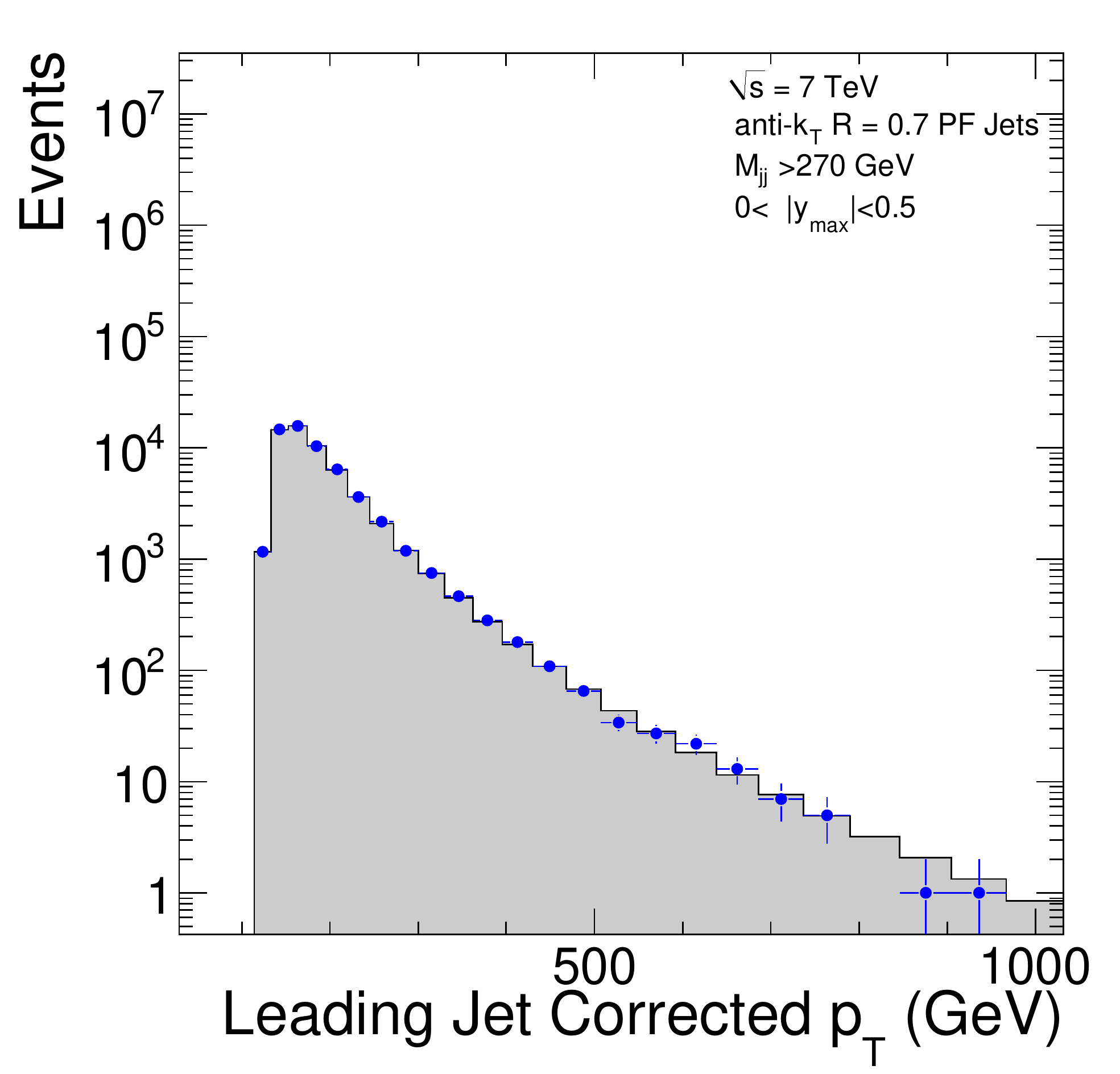}
  \includegraphics[width=0.40\textwidth]{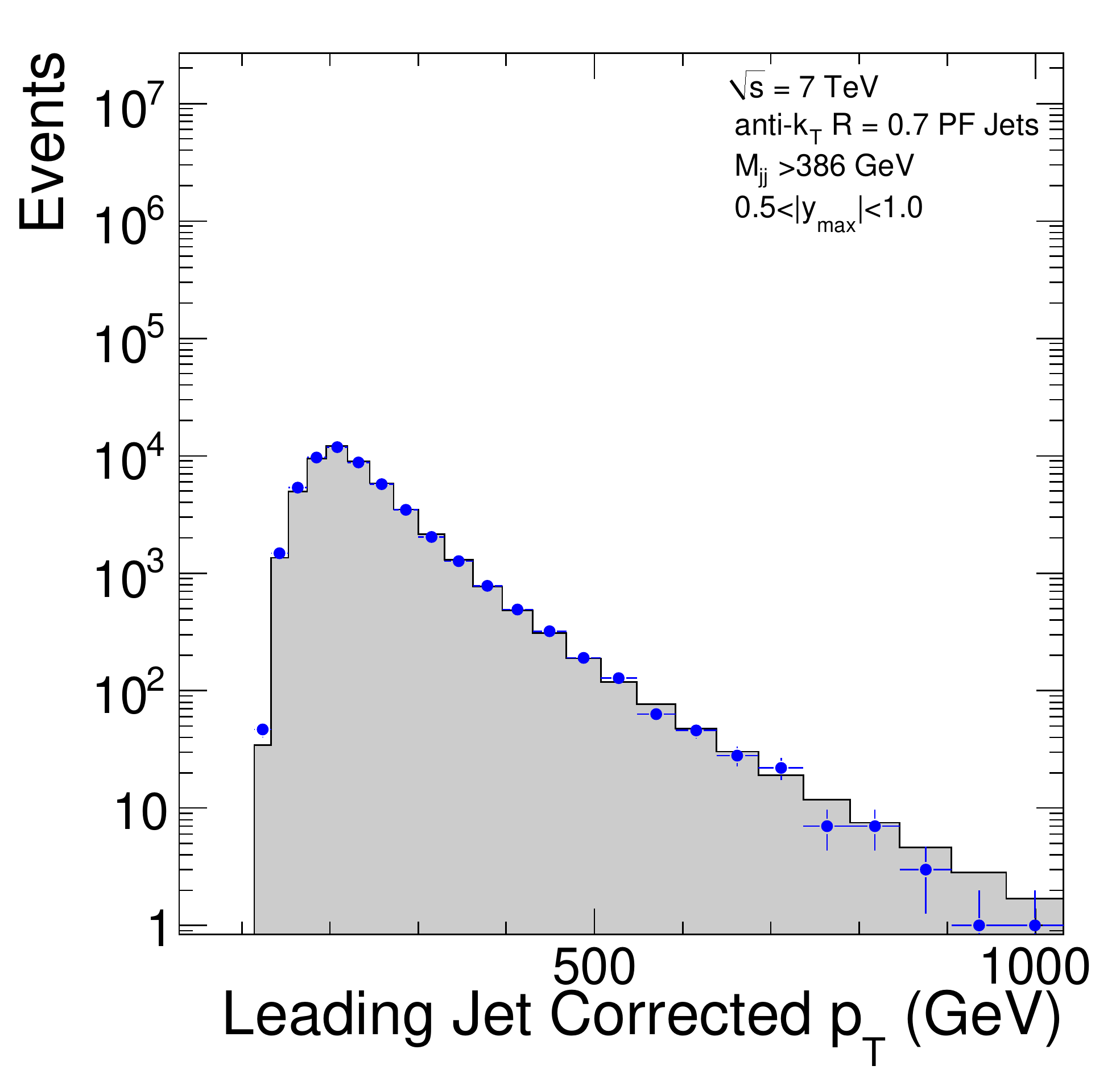} 
  \includegraphics[width=0.40\textwidth]{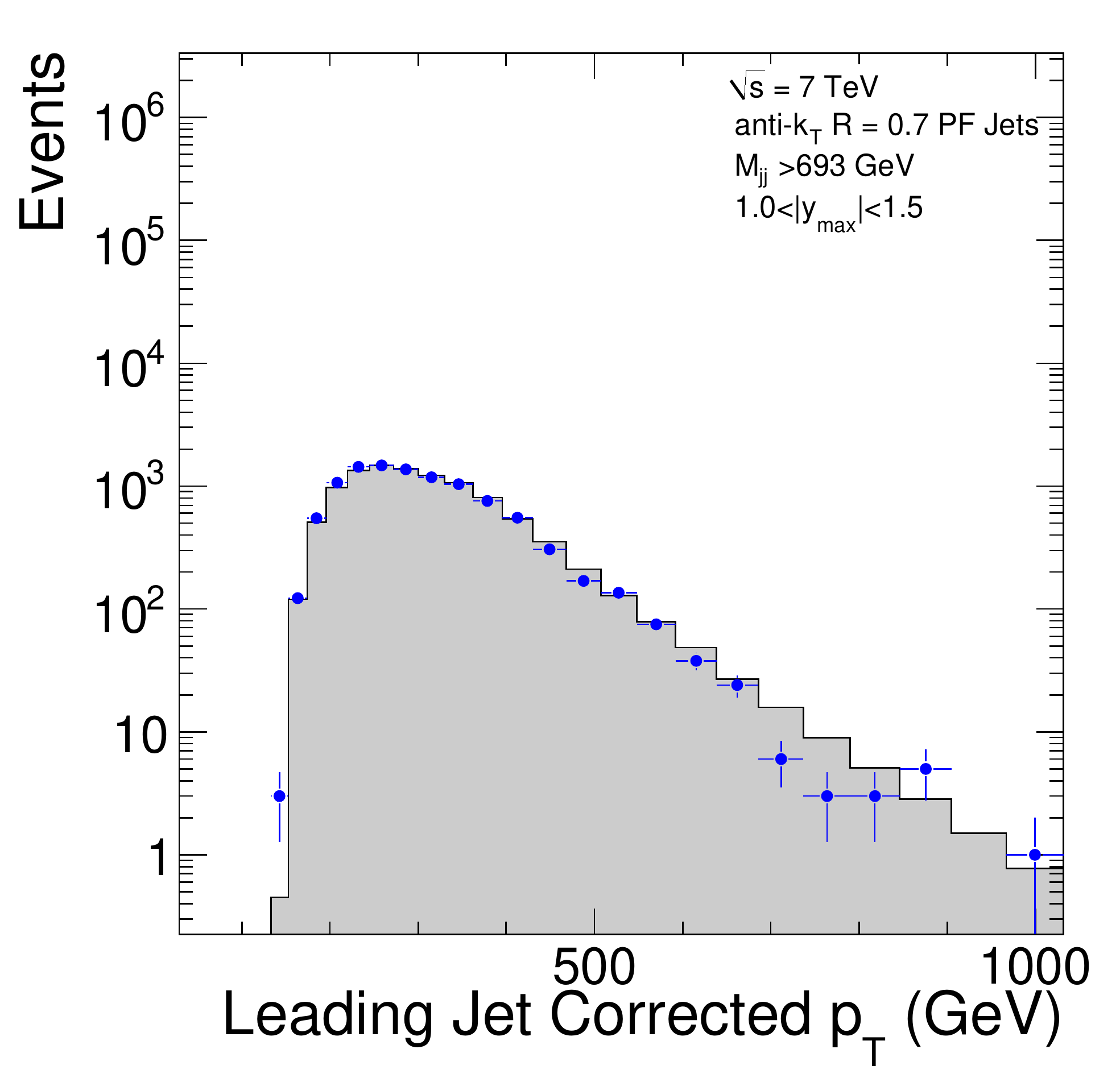} 
  \includegraphics[width=0.40\textwidth]{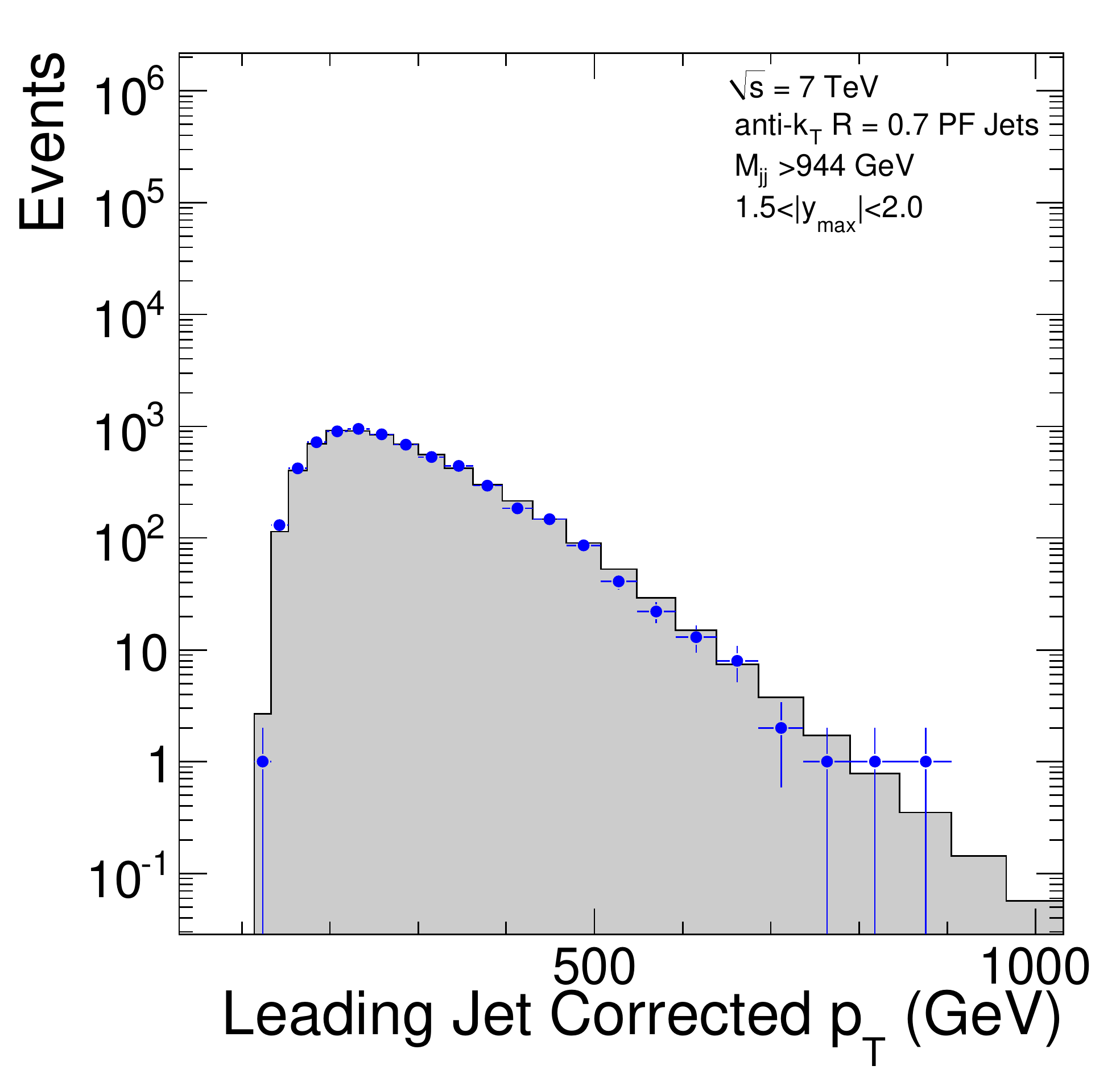}
  \includegraphics[width=0.40\textwidth]{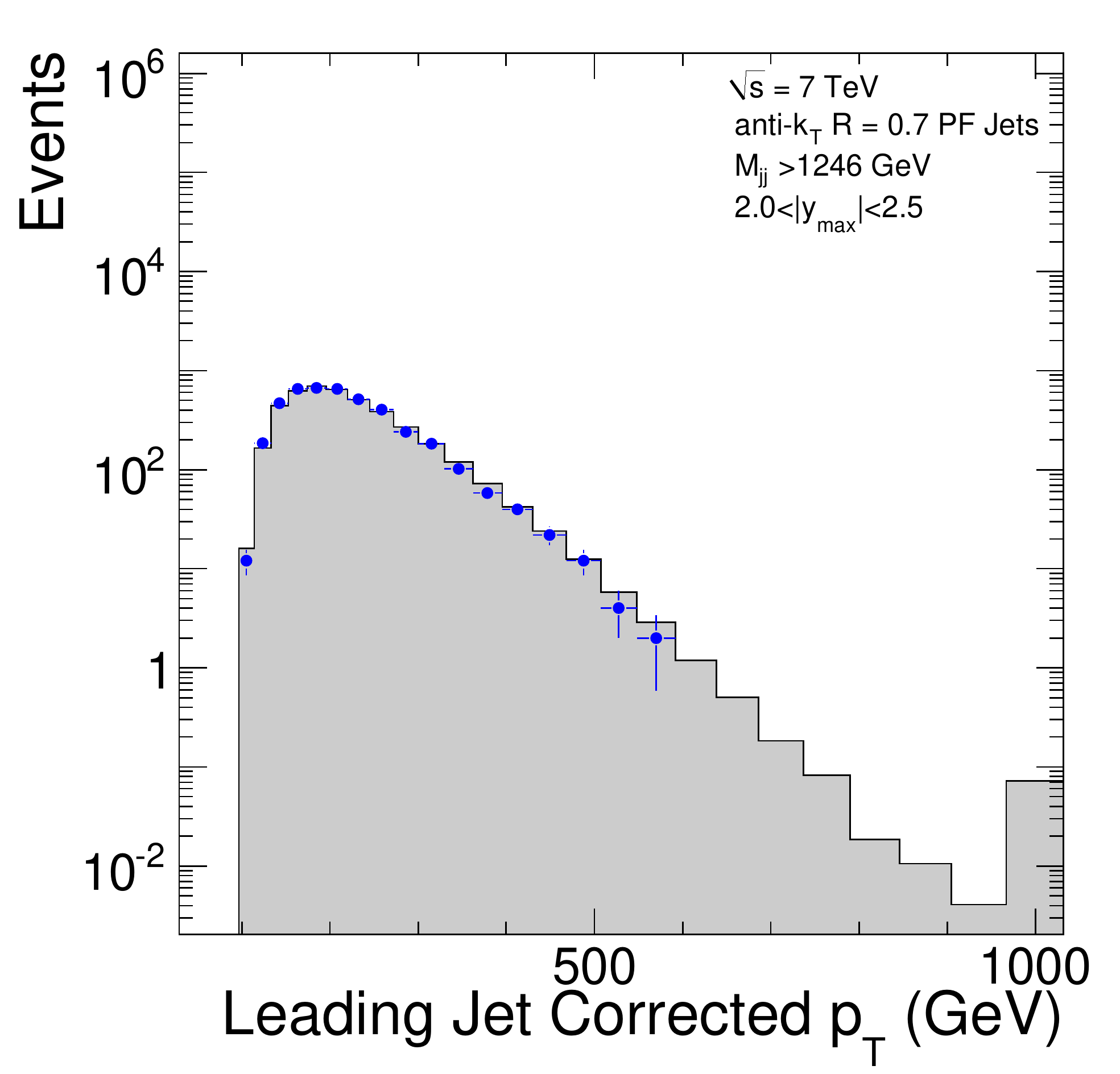}
  \capspace
  \caption{The \pt of the leading jet for the five different \ymax bins and for the Jet70U sample. The plots for data (points) and simulated (dashed histogram) events are compared.}
  \label{fig_data9}
\end{figure}

\begin{figure}[ht]
  \centering
  \includegraphics[width=0.40\textwidth]{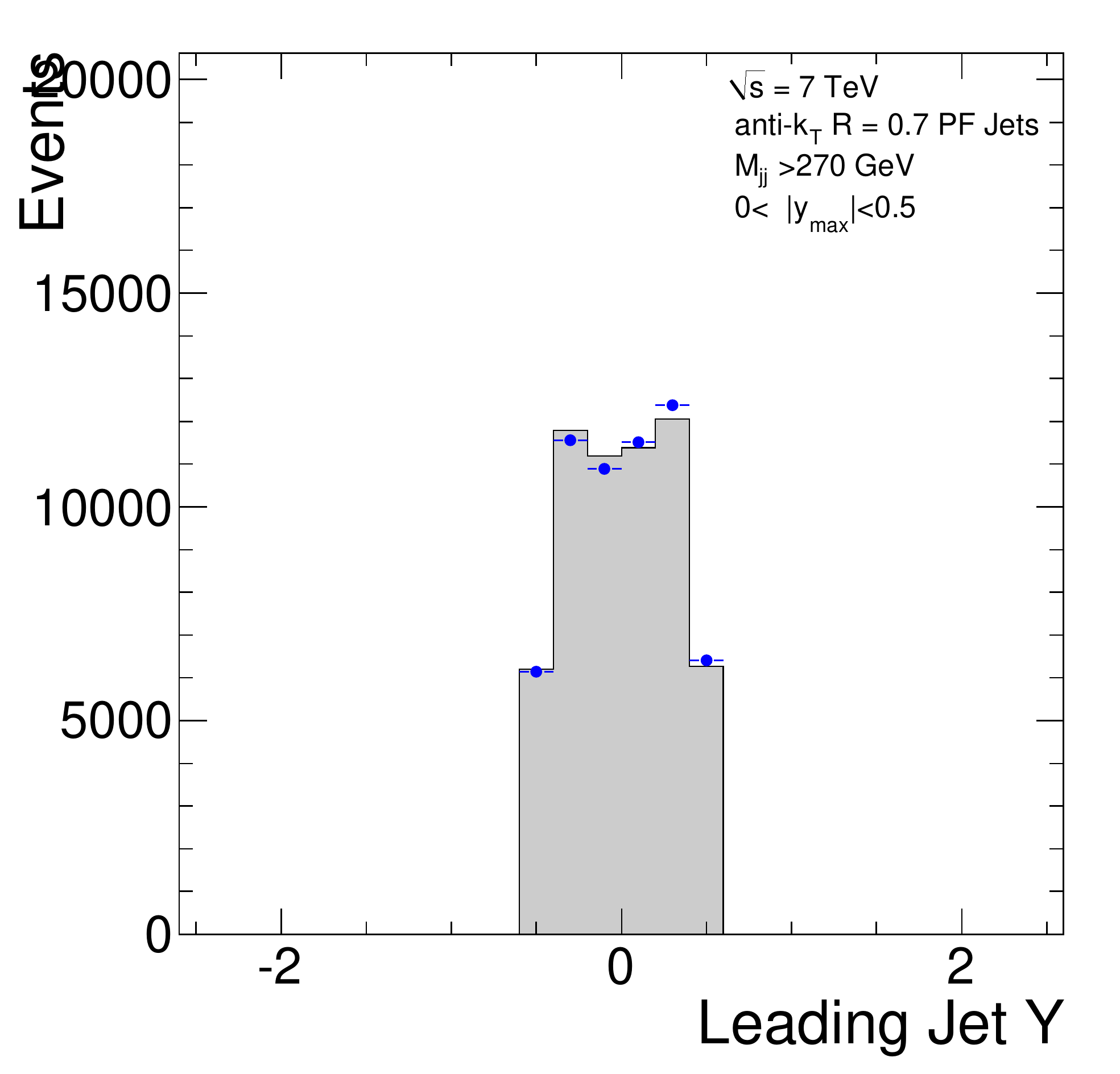}
  \includegraphics[width=0.40\textwidth]{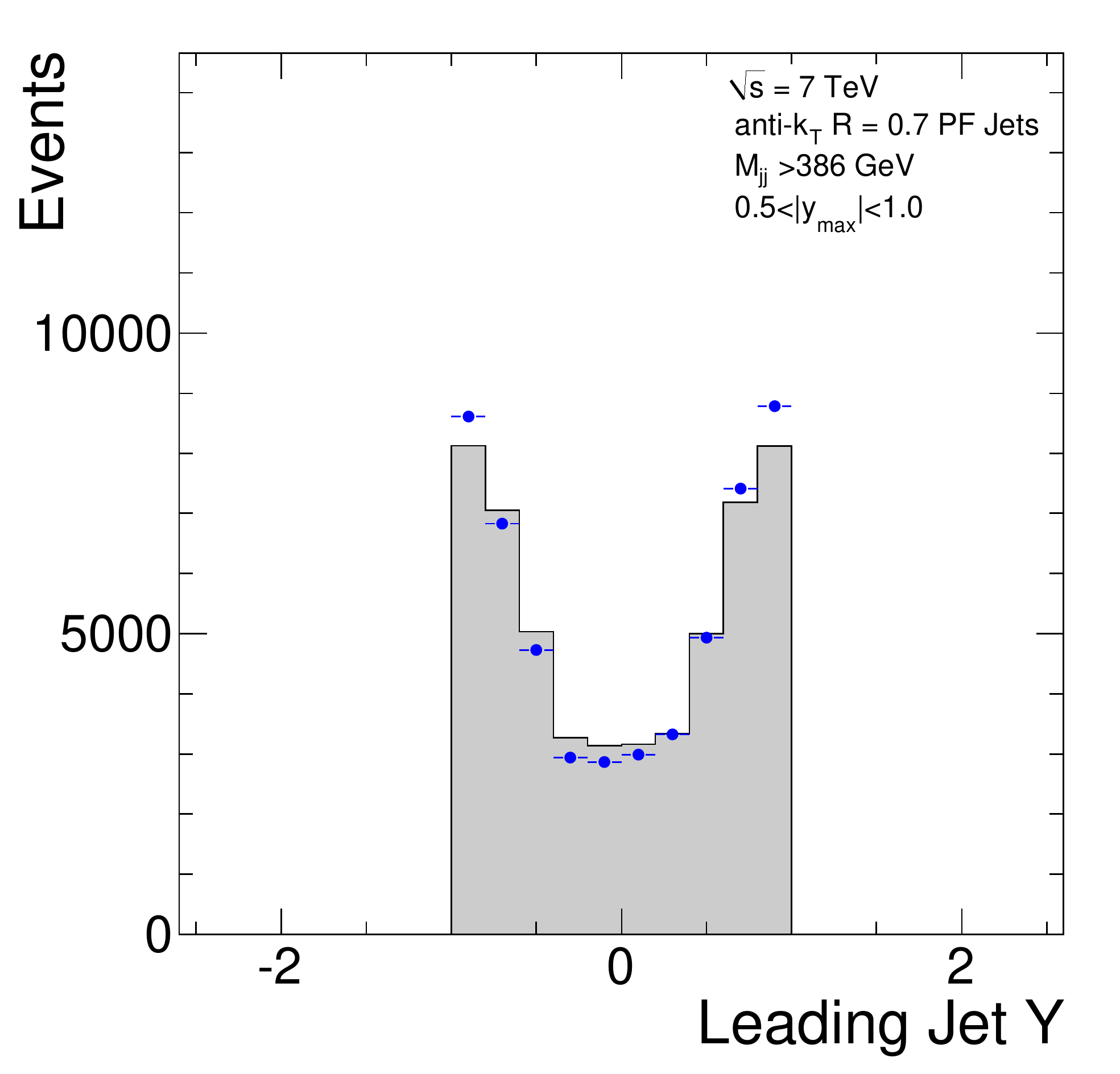} 
  \includegraphics[width=0.40\textwidth]{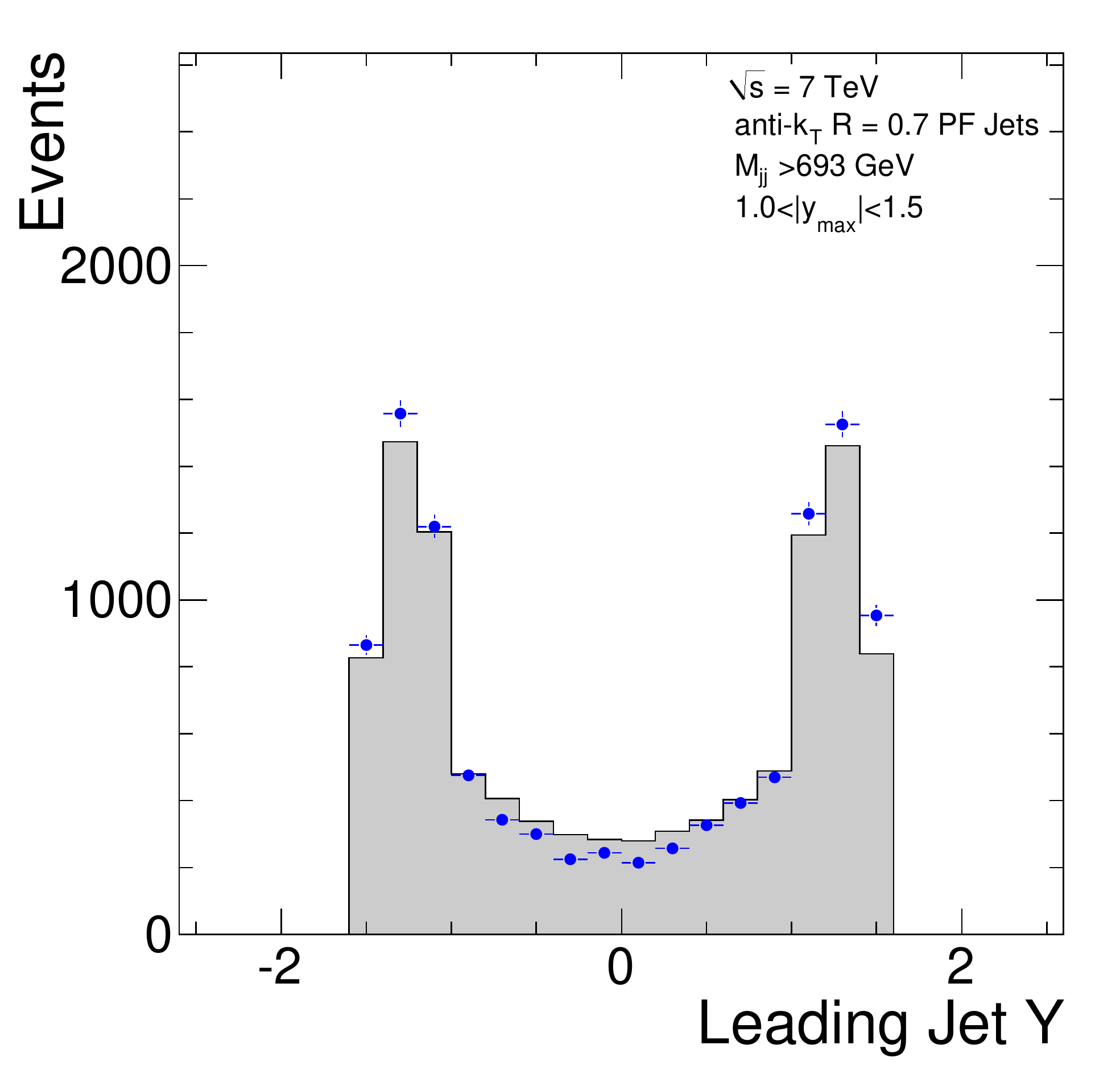} 
  \includegraphics[width=0.40\textwidth]{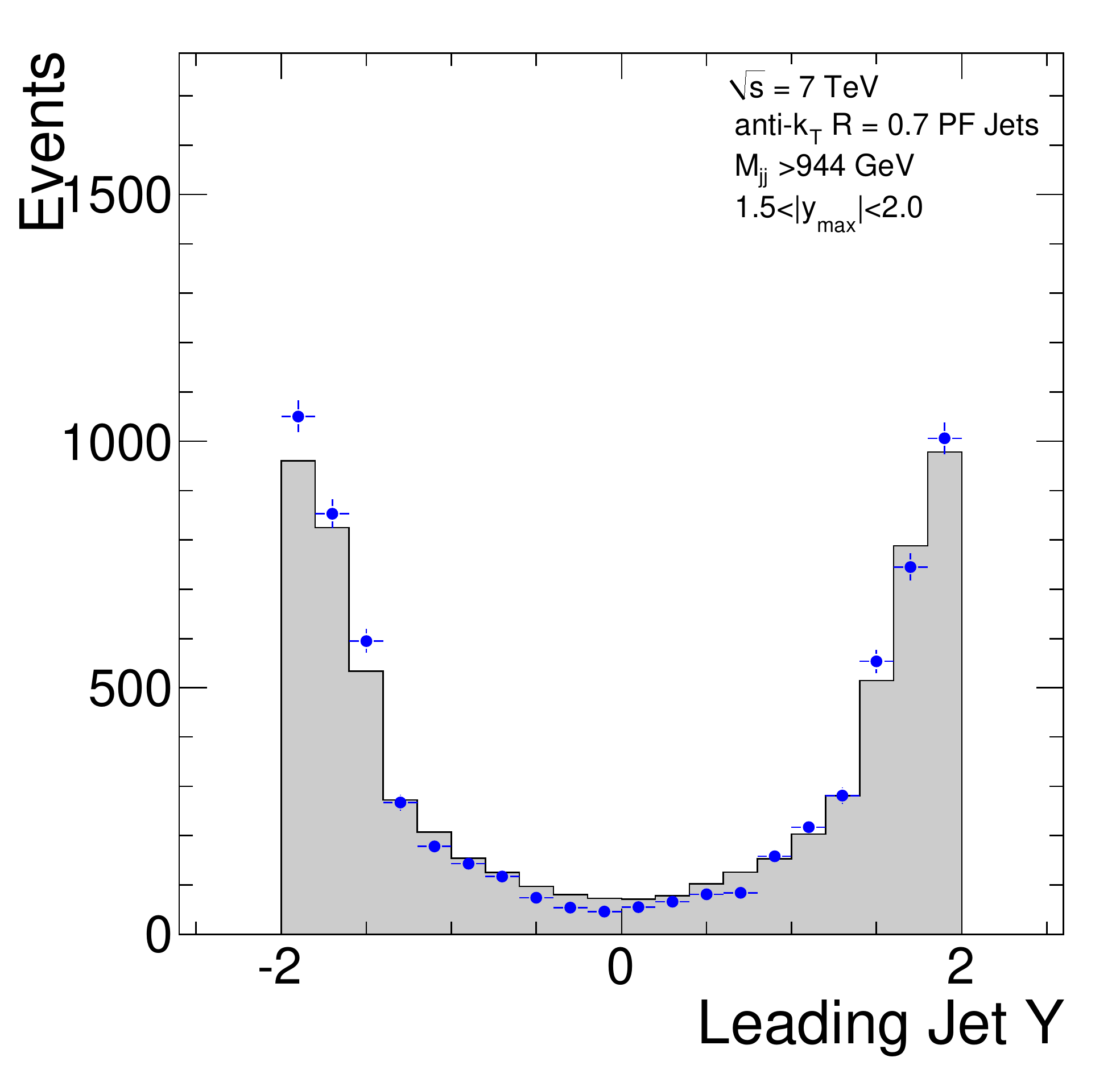}
  \includegraphics[width=0.40\textwidth]{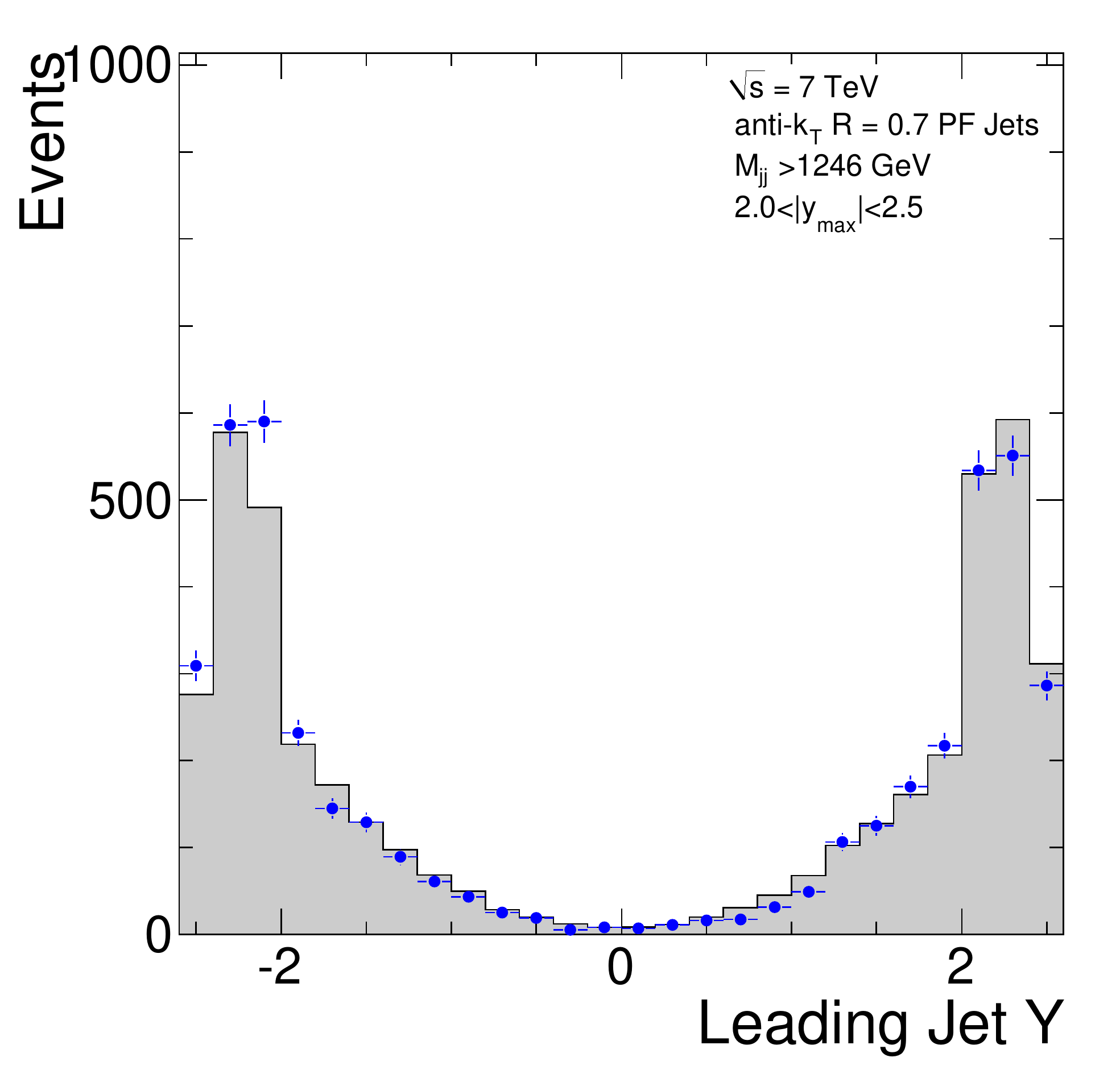}
  \capspace
  \caption{The $\eta$  of the leading jet  for the five different \ymax bins and for the Jet70U sample. The plots for data (points) and simulated (dashed histogram) events are compared.}
  \label{fig_data10}
\end{figure}

\begin{figure}[ht]
  \centering
  \includegraphics[width=0.40\textwidth]{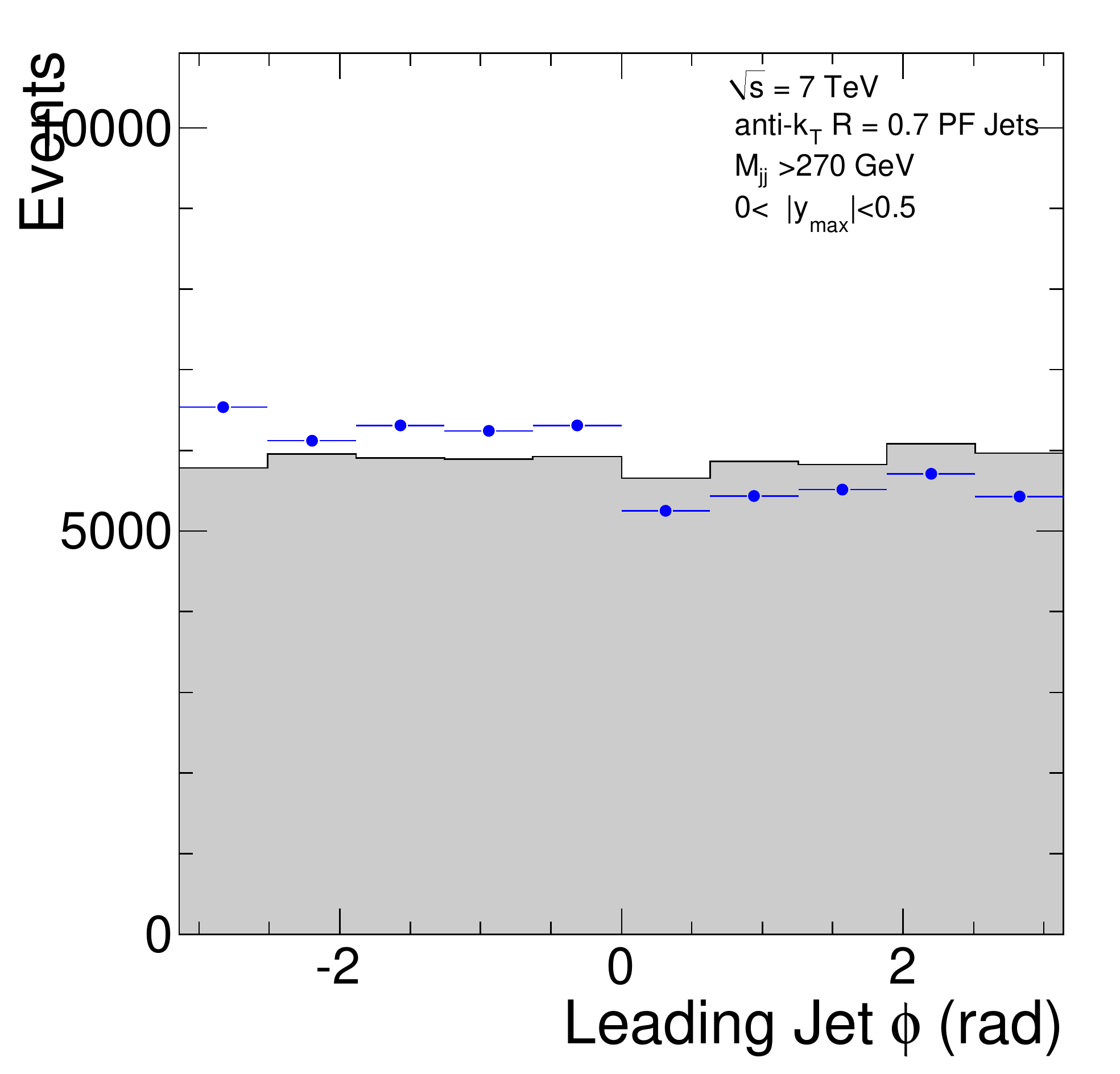}
  \includegraphics[width=0.40\textwidth]{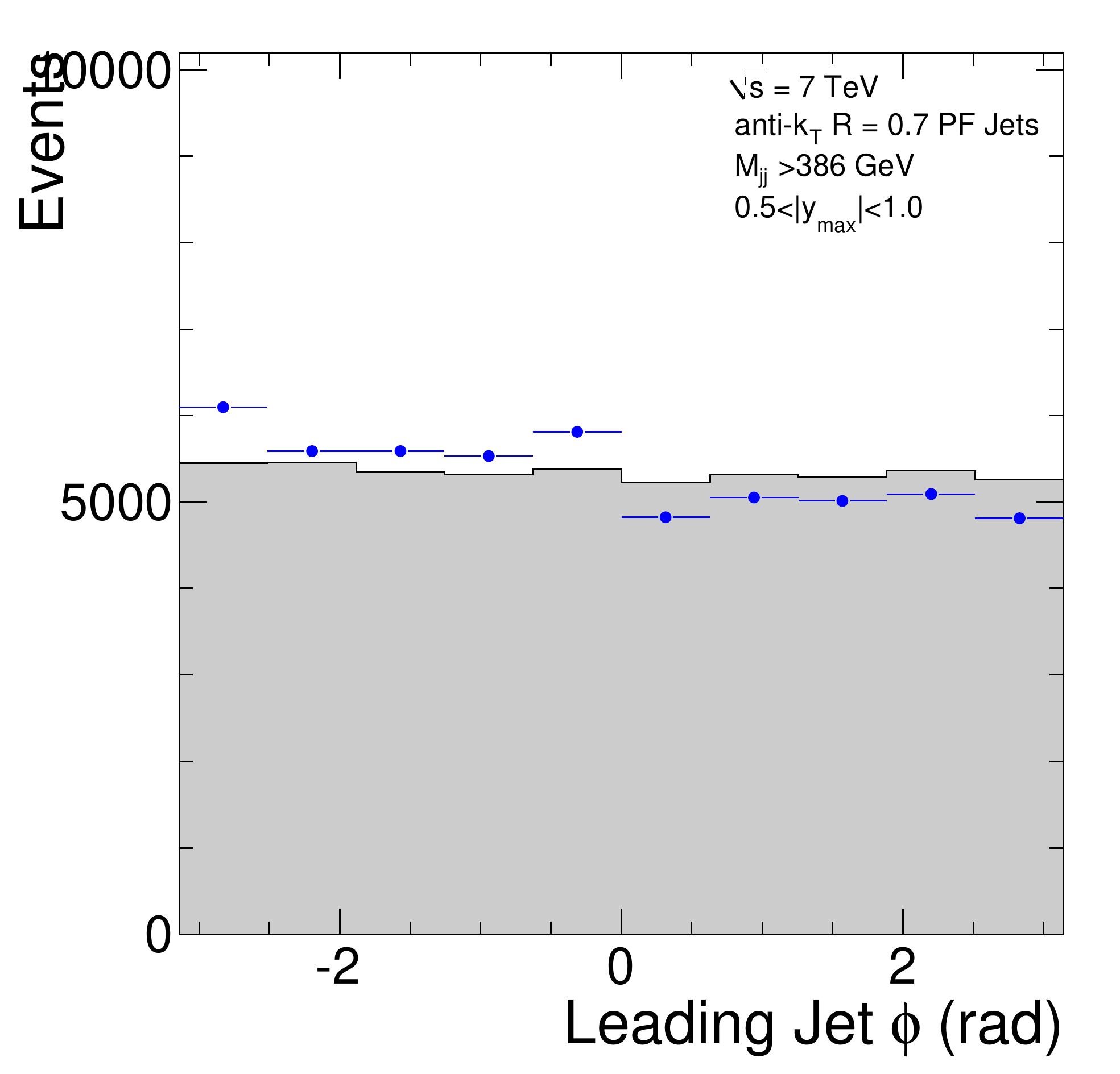} 
  \includegraphics[width=0.40\textwidth]{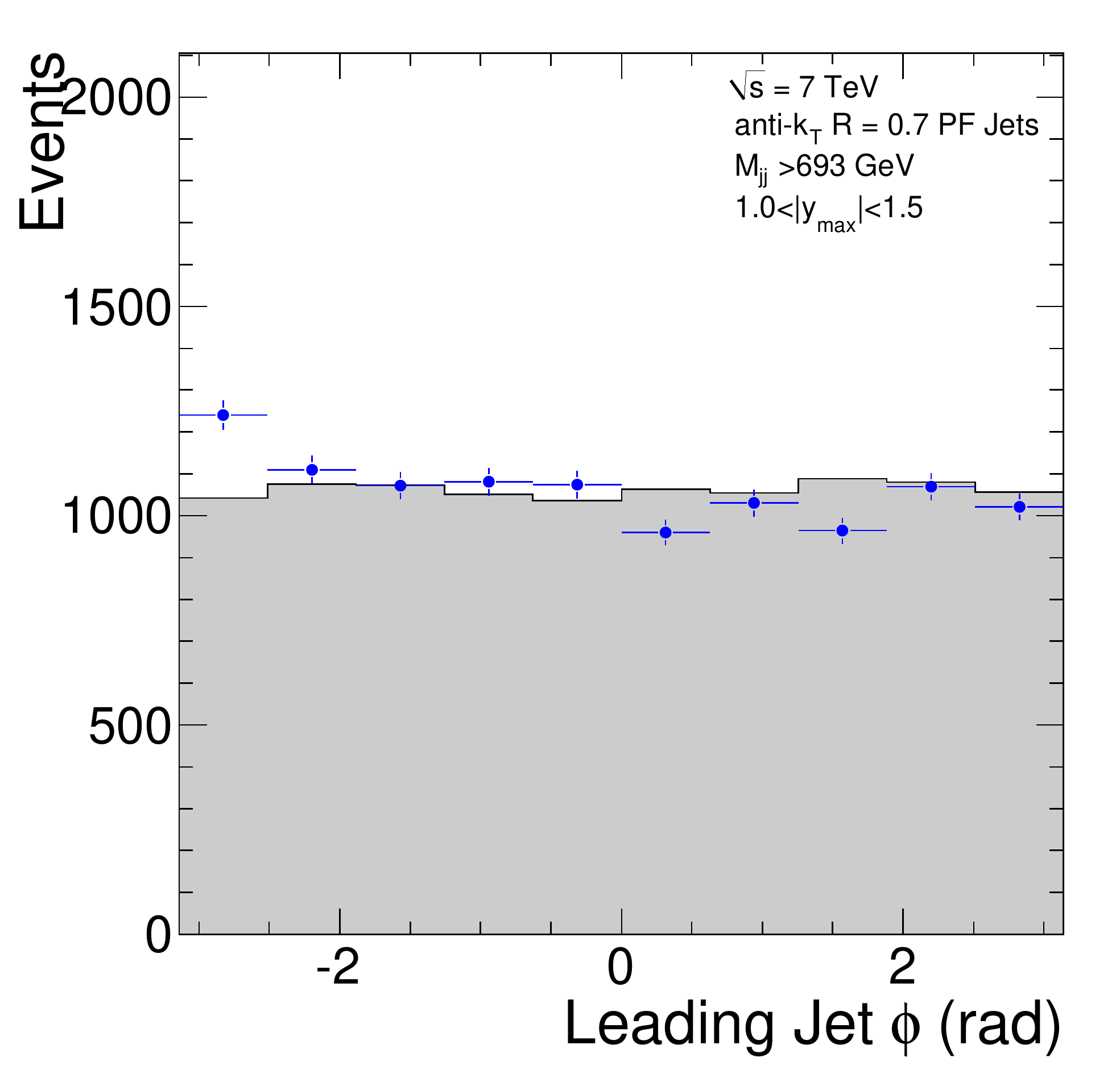} 
  \includegraphics[width=0.40\textwidth]{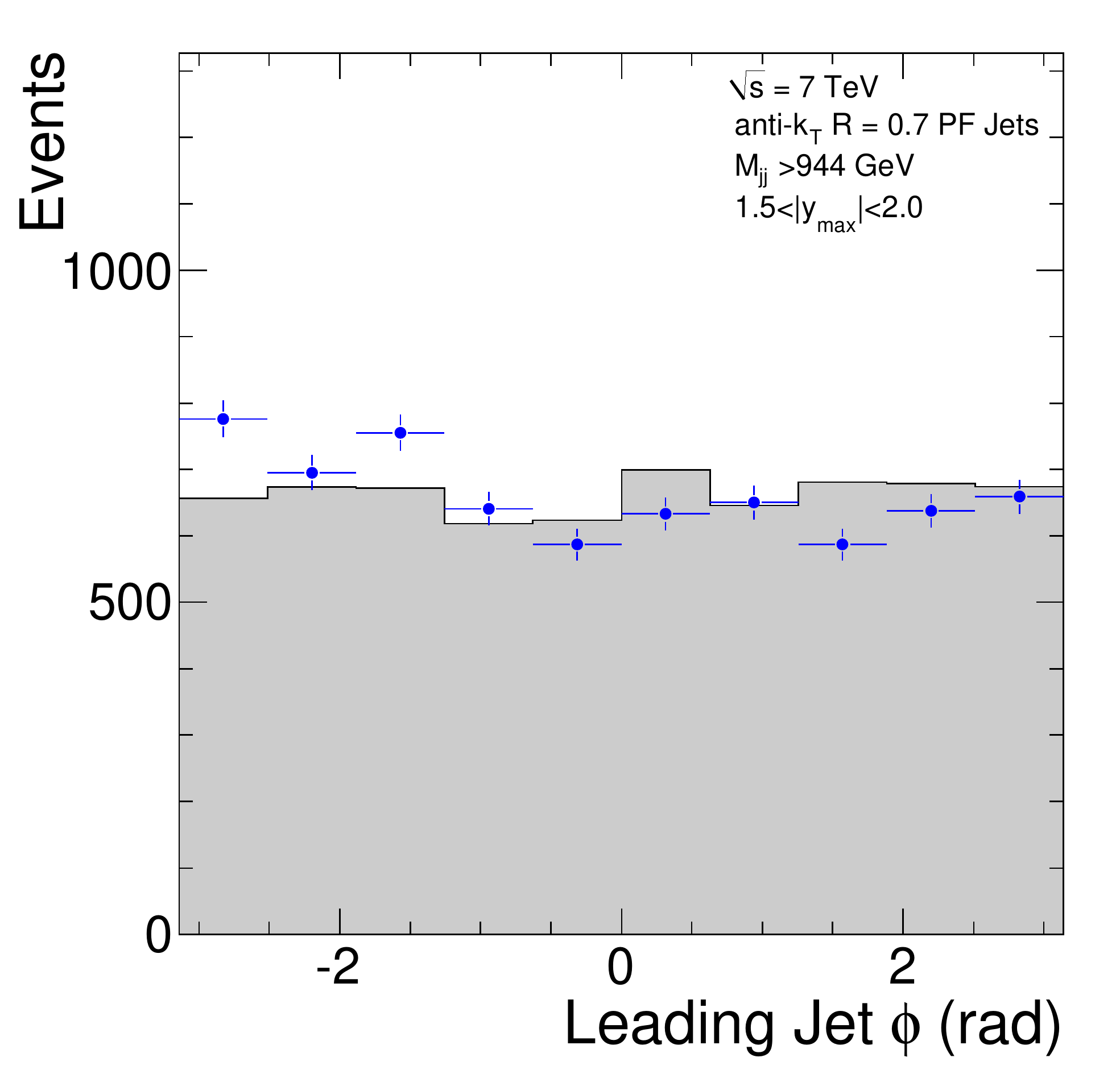}
  \includegraphics[width=0.40\textwidth]{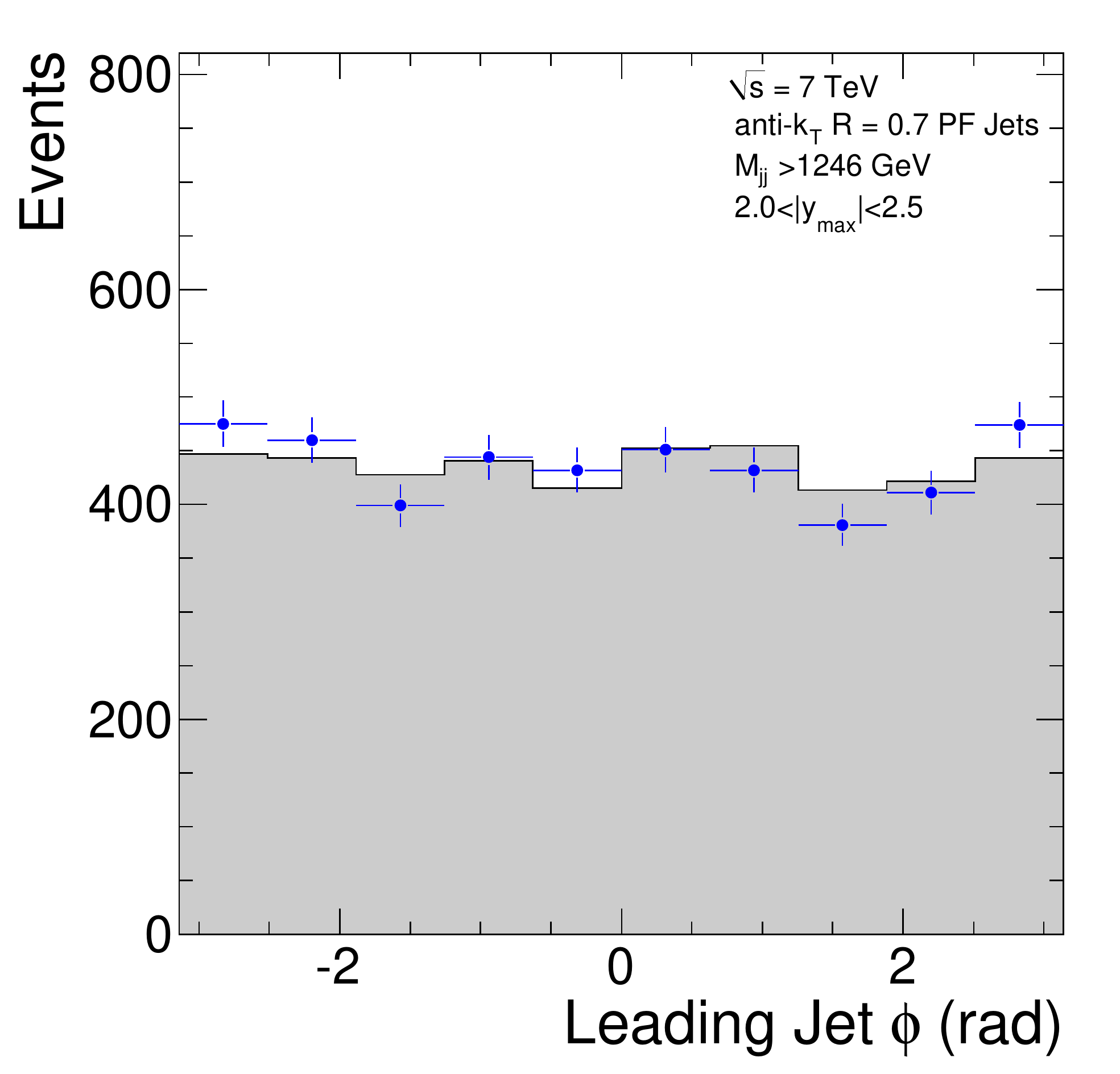}
  \capspace
  \caption{The $\phi$ of the leading jet for the five different \ymax bins and for the Jet70U sample. The plots for data (points) and simulated (dashed histogram) events are compared.}
  \label{fig_data11}
\end{figure}

Secondly, the following jet properties are examined;
 \begin{itemize}
 \item The charged hadron fraction, representing mostly the $\pi^{\pm}$ jet content
 \item The neutral hadron fraction, representing mostly the n jet content
 \item The neutral electromagnetic fraction, representing mostly the $\pi^{0}$ and the photons
 \end{itemize}
If there were noise in the hadronic calorimeter (HCAL), we would observe an excess of events in the neutral hadron fraction distribution of the data with respect to simulated events (MC). Similarly, if there were noise in the electromagnetic calorimeter, we would observe an excess of events in the neutral electromagnetic fraction distribution of the data with respect to simulated events (MC). In Figures \ref{fig_data6}-\ref{fig_data8} the distributions of these jet variables are shown for Jet70U sample and all the rest of the plots for different samples can be found in Appendix C. In general, a very good agreement between the data and the simulated events is observed. Moreover, comparisons of jet kinematic quantities (\pt , $\eta$, $\phi$) between the data and the MC events were studied, and they are shown in Figures \ref{fig_data9}-\ref{fig_data11} for Jet70U sample in all five \ymax bins. Again, a good agreement between the data and the simulated events is observed except for the $\phi$ distributions. This significant difference which pronounce itself as an asymmetry is due to the HCAL mis-calibration. A bias in the calibration scheme tends to over-calibrate jets in the $(-\pi,0)$ azimuthal range. Jets in this region are more likely to be the first leading one in \pt and whenever a jet, originally the second leading one, is promoted to the first leading one by the mis-calibration, the actual first leading one becomes the second one automatically on the other side. As a result, the asymmetry arise in the $\phi$ distributions of the jets. This property has an effect on jet resolution making it worse in the data than it is expected from the MC study and dealt with as a systematic uncertainty.
\clearpage

\clearpage

\clearpage

\clearpage

\clearpage
     
\clearpage

\clearpage

\clearpage

\clearpage

\clearpage
\subsection{Stability Over the Run Period}
All the checks presented in the previous section indicate that there are no significant pathologies in the data or an abnormal effect which is not modeled in the simulation. However, the time evolution of these basic data quantities should be checked in order to ensure that the quality of the recorded and selected data is stable during the run period. For that, all of the quantities introduced in the previous section have been examined as a function of the run number which is a time stamp in a sense.

\begin{figure}[ht]
  \centering
  \includegraphics[width=0.52\textwidth]{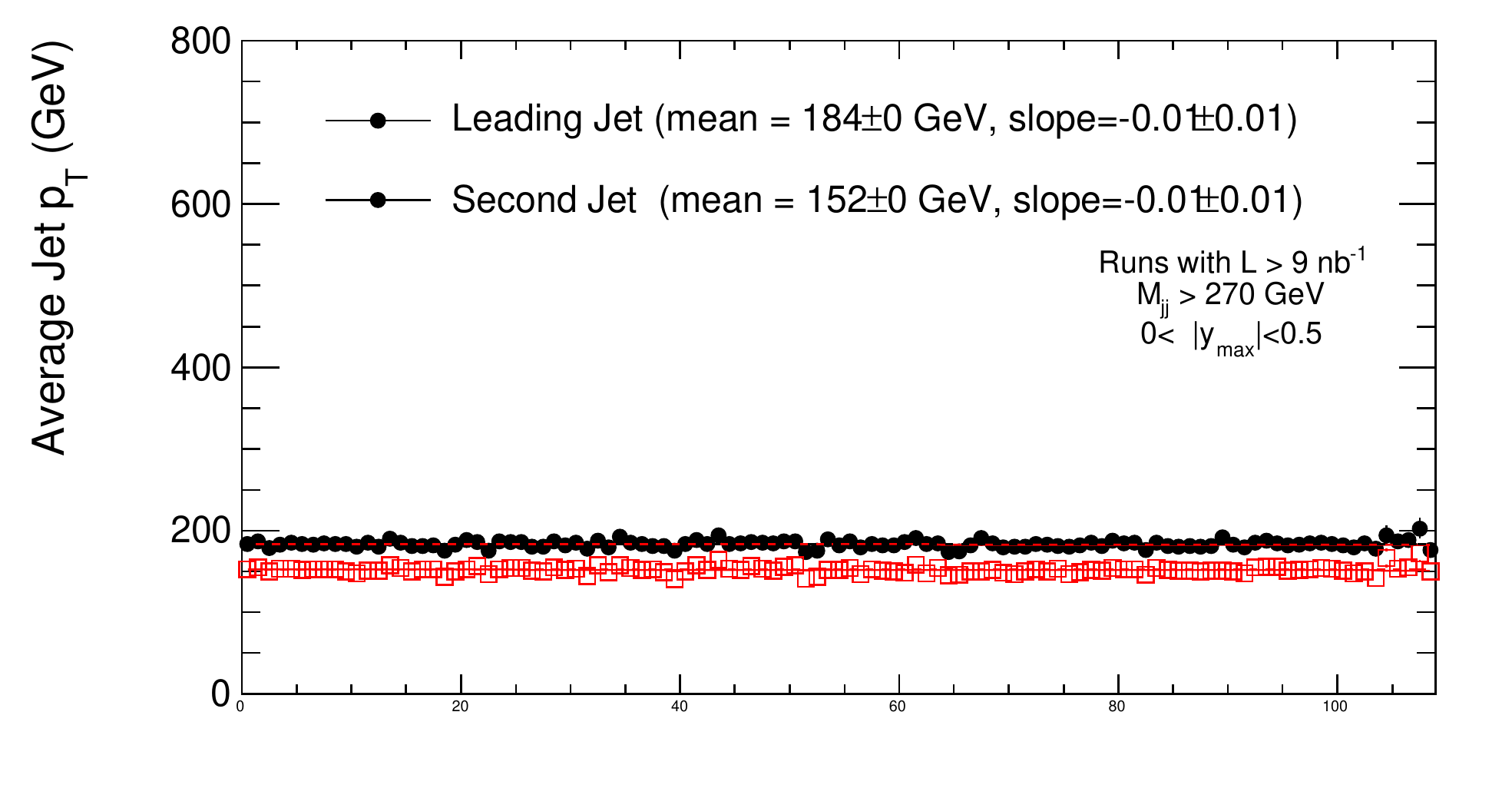}
  \includegraphics[width=0.52\textwidth]{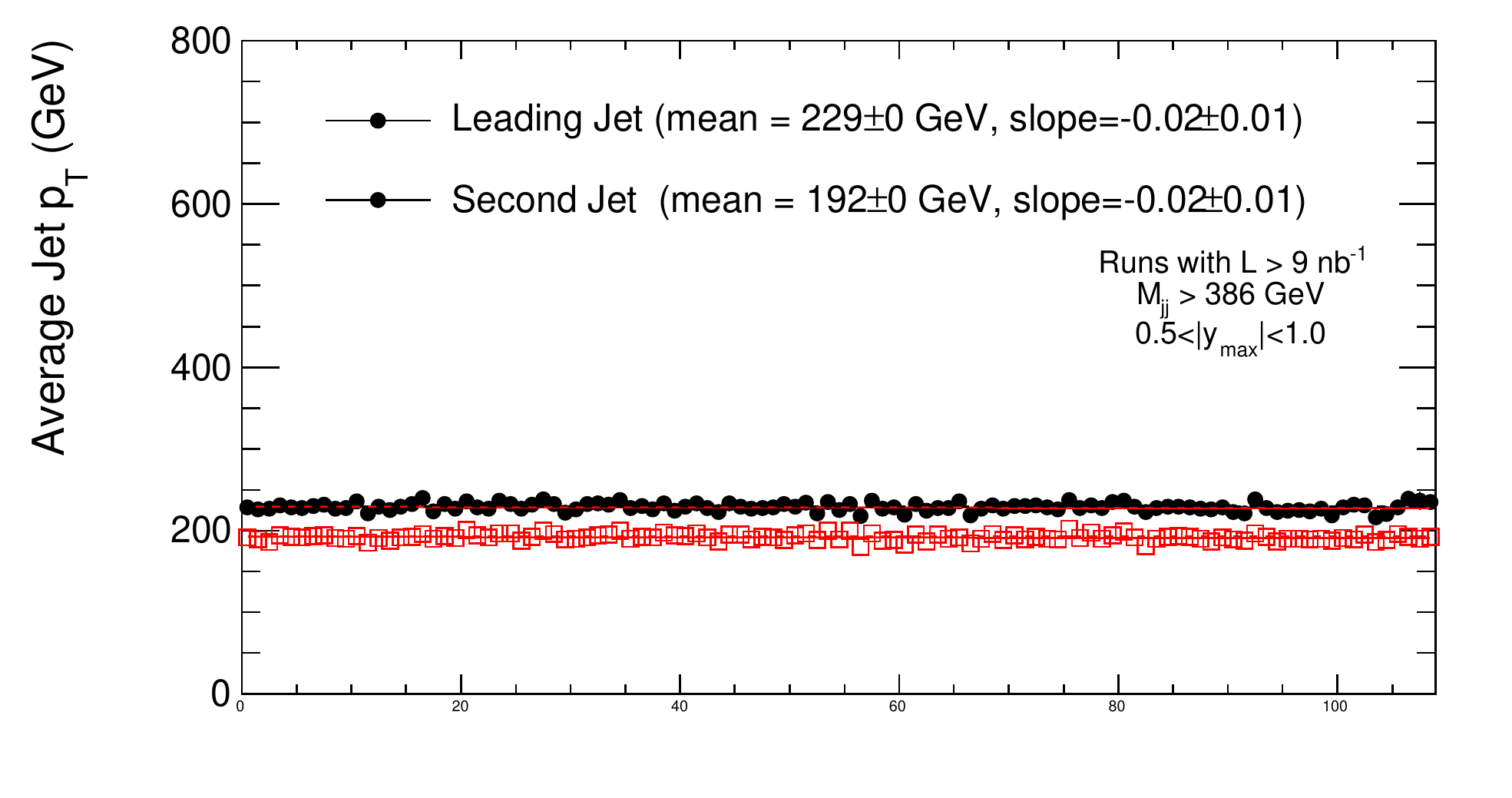}
  \includegraphics[width=0.52\textwidth]{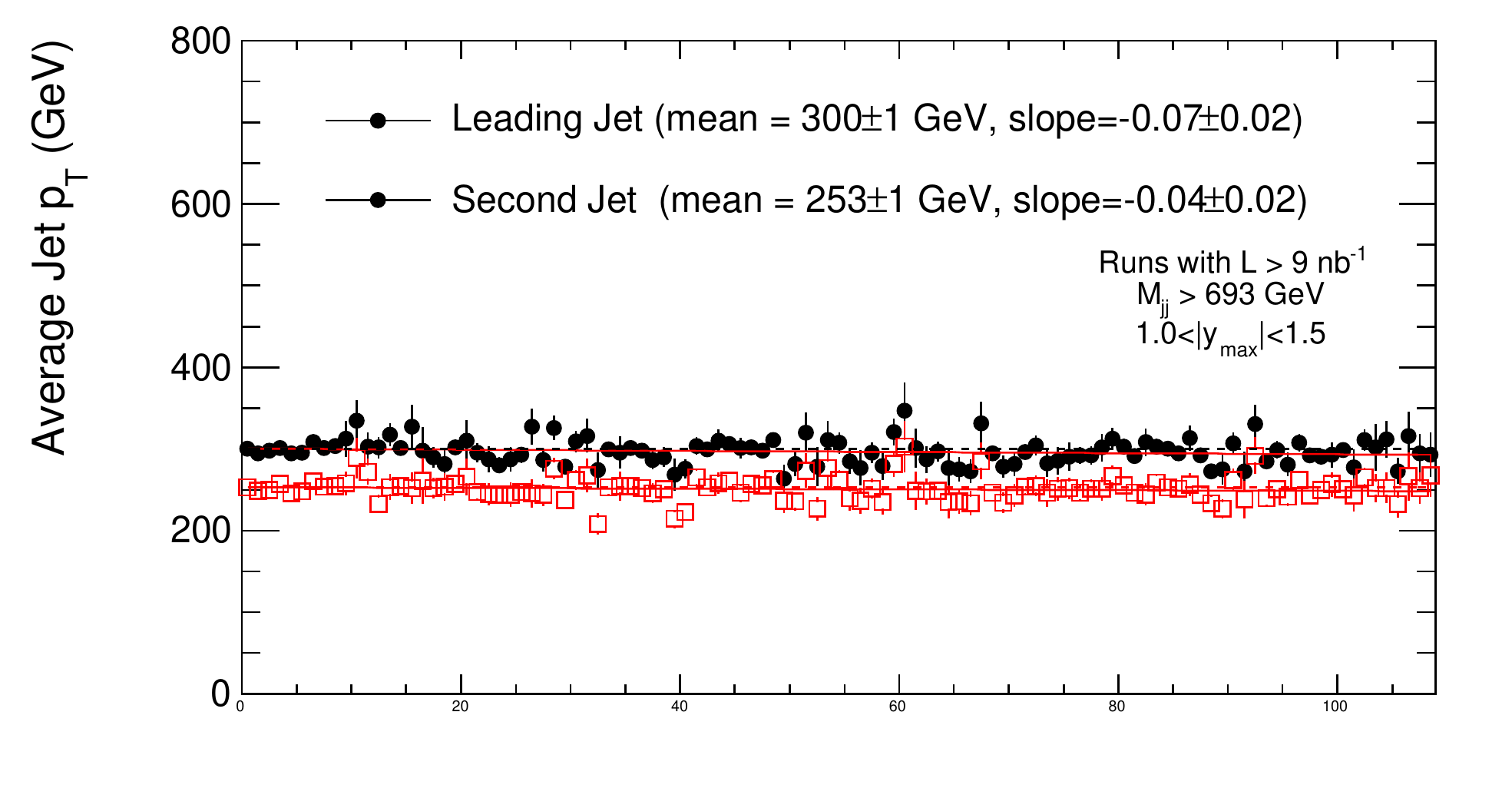}
  \includegraphics[width=0.52\textwidth]{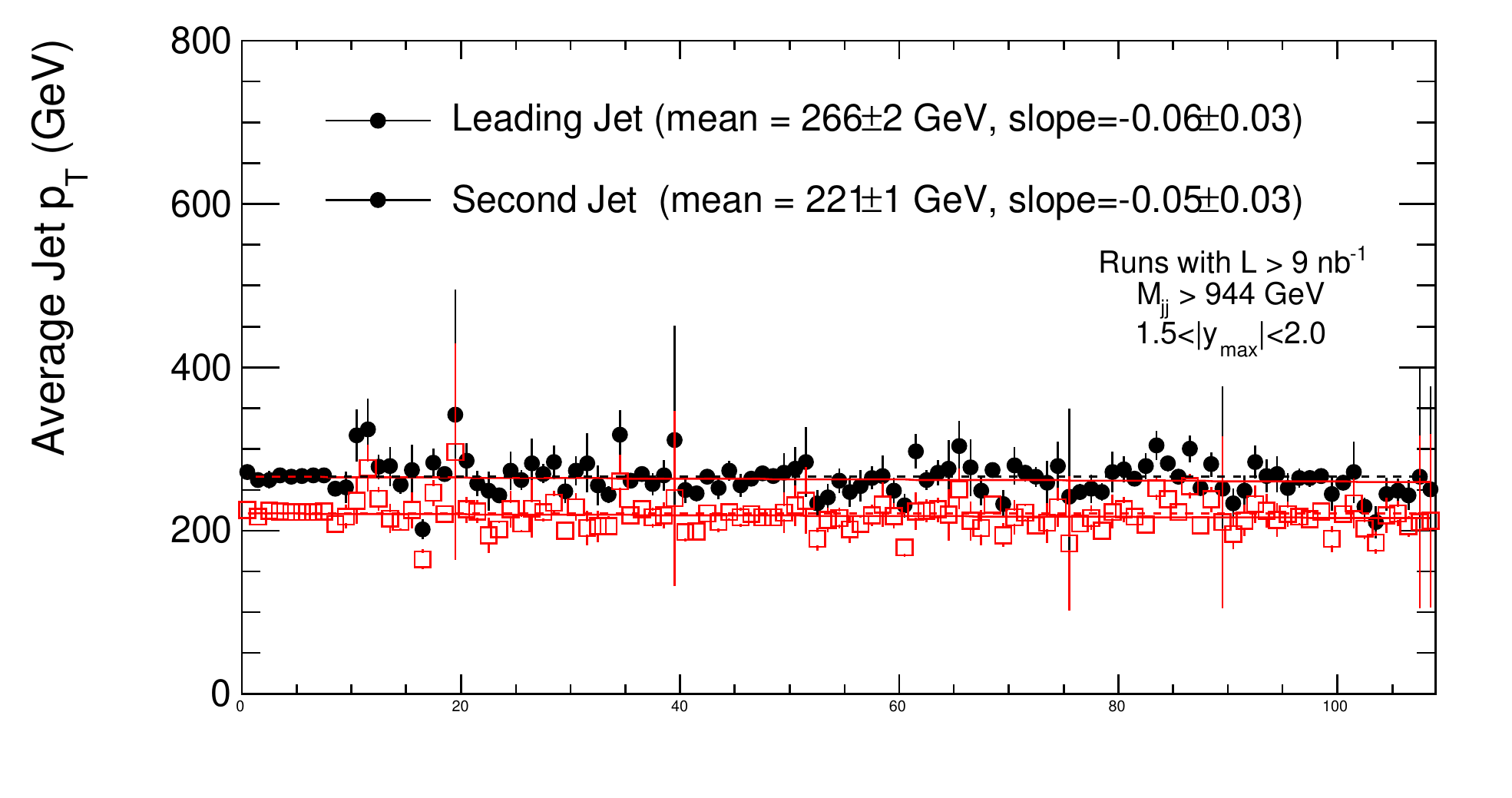}
  \includegraphics[width=0.53\textwidth]{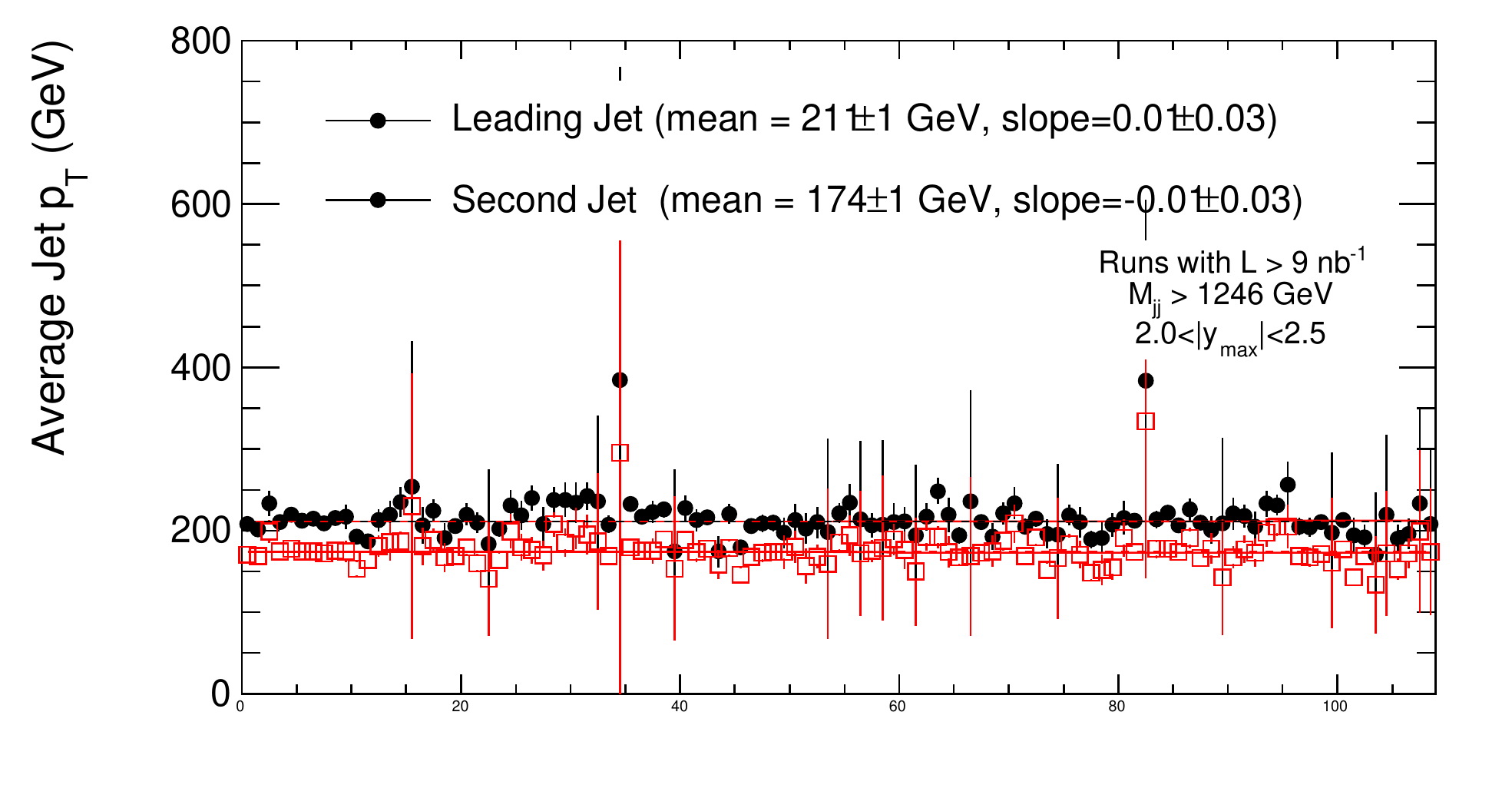}
  \capspace
  \caption{The \pt of the leading and second jet for the five different \ymax bins and for the Jet70U sample as a function of time (run number), fitted with a first degree polynomial.}
  \label{fig_data12}
\end{figure}

\begin{figure}[ht]
  \centering
  \includegraphics[width=0.52\textwidth]{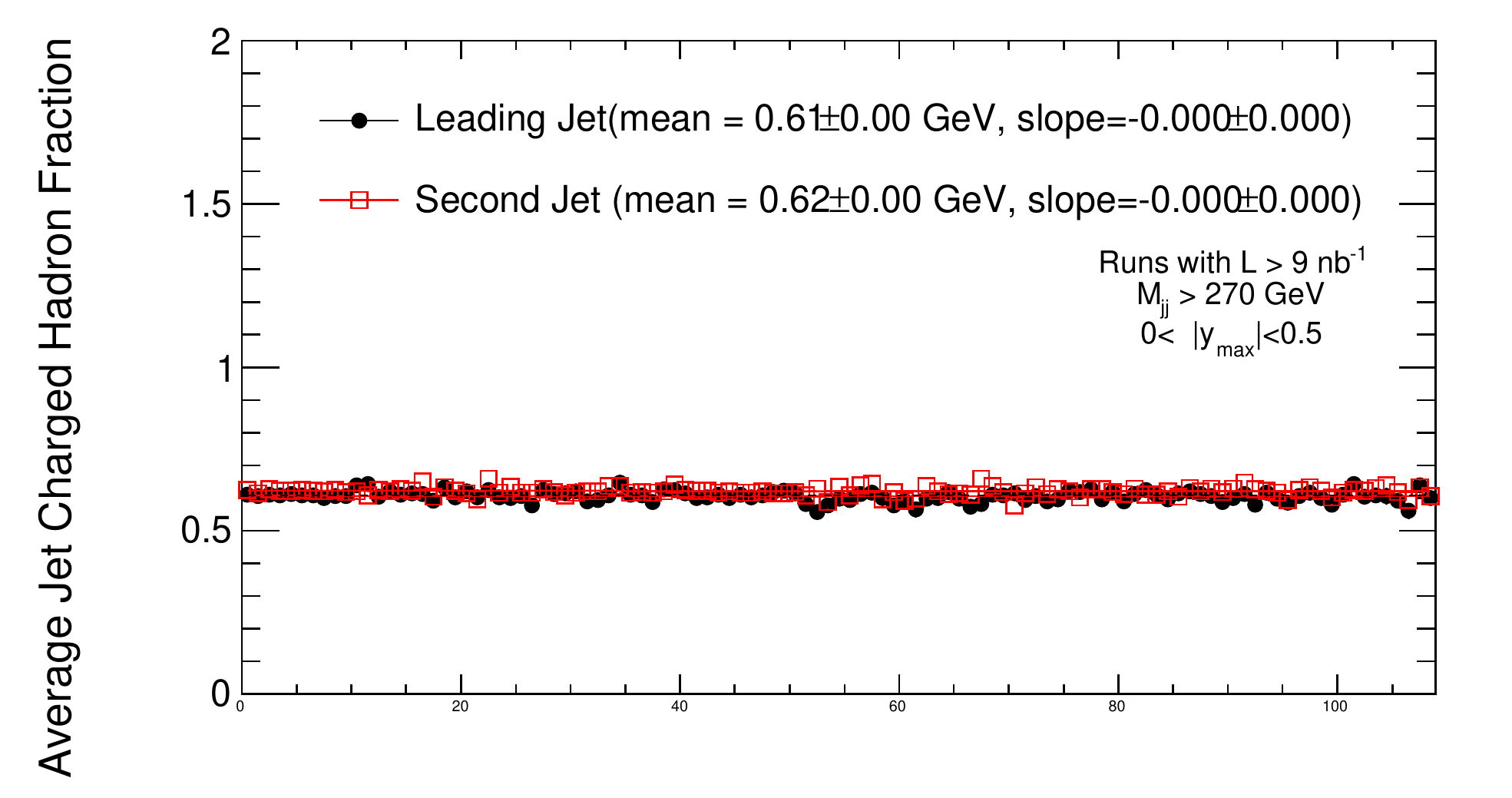}
  \includegraphics[width=0.52\textwidth]{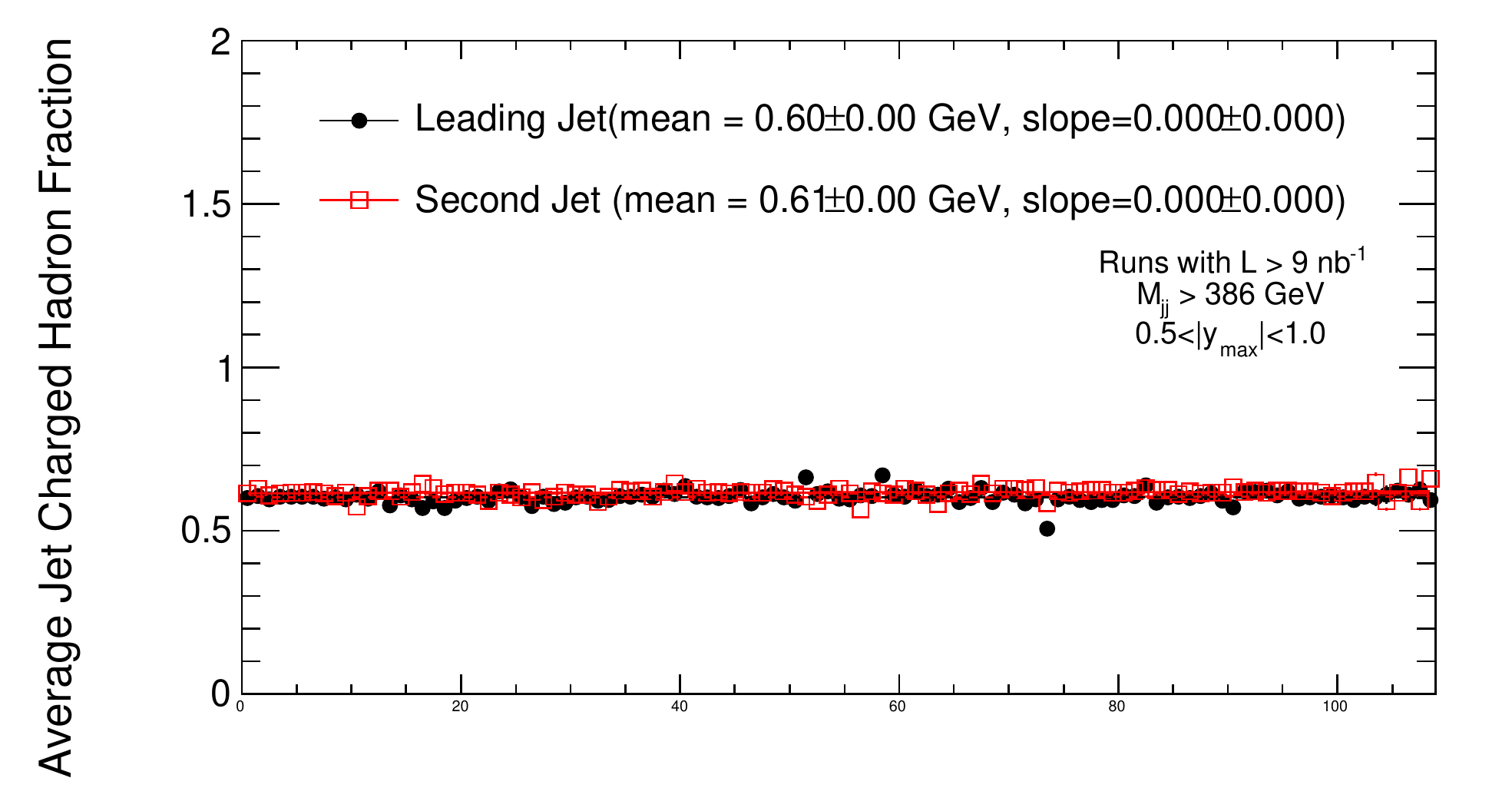} 
  \includegraphics[width=0.52\textwidth]{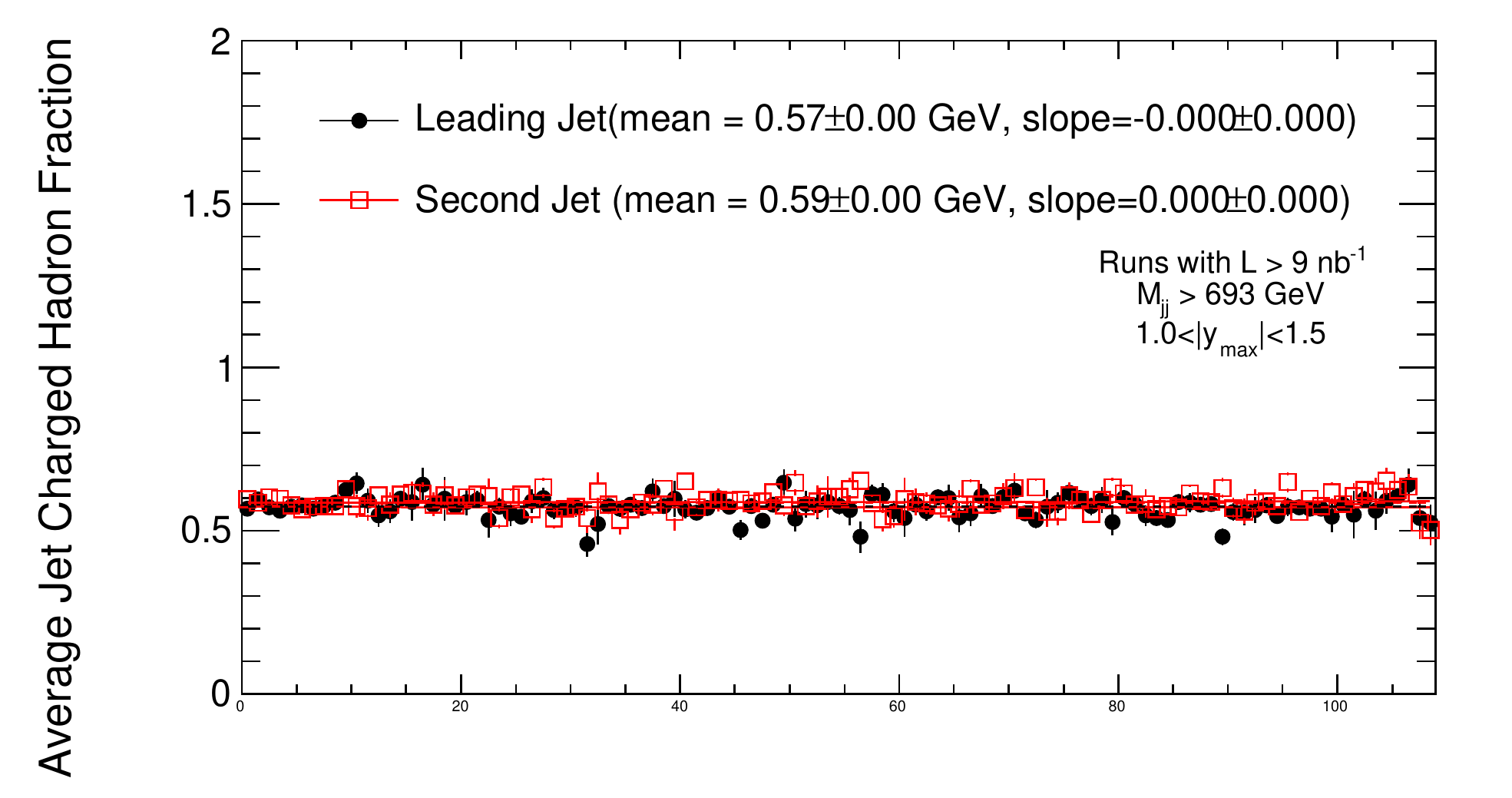} 
  \includegraphics[width=0.52\textwidth]{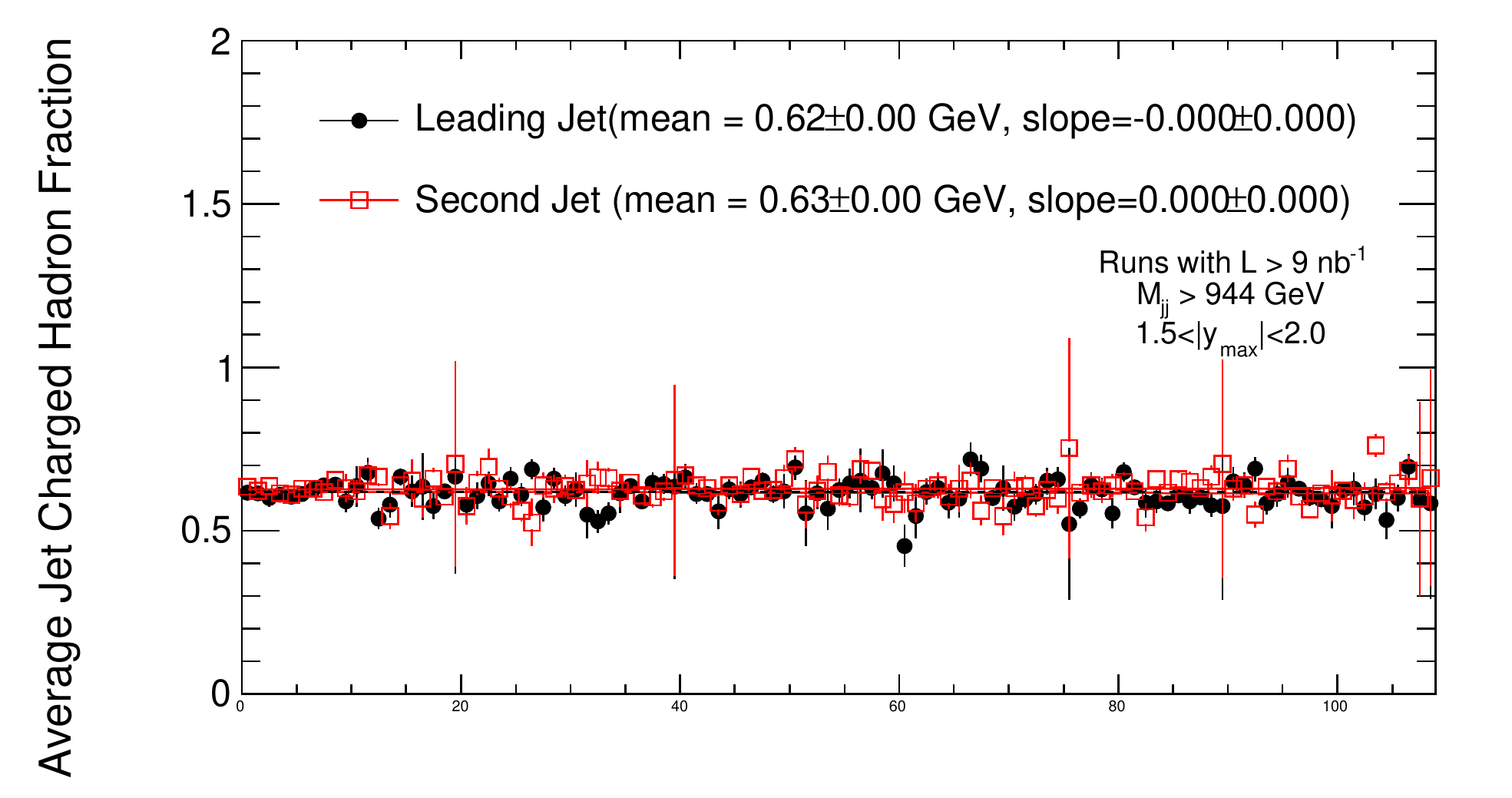}
  \includegraphics[width=0.53\textwidth]{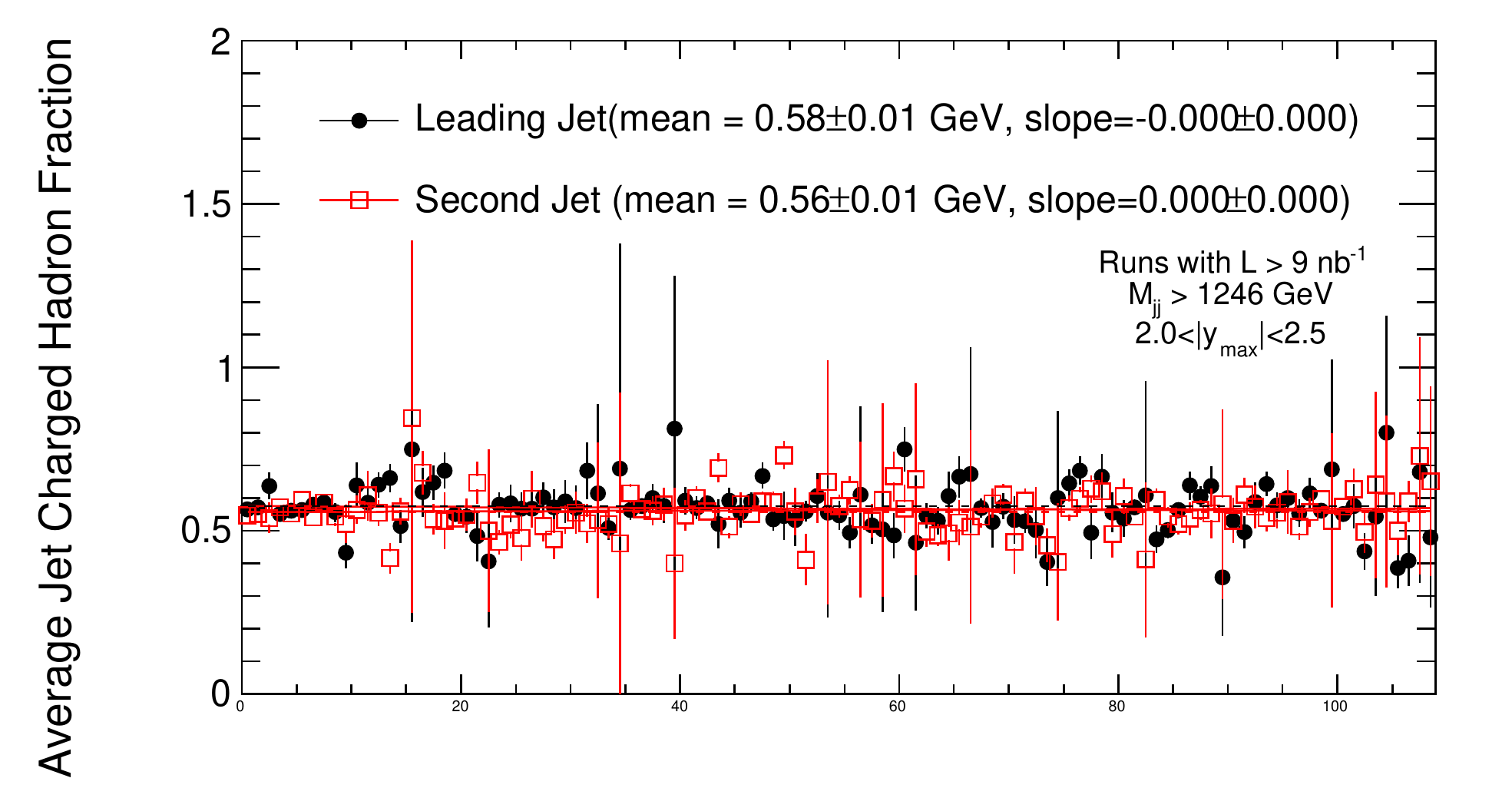}
  \capspace
  \caption{The charged hadron fraction of the leading and second jet for the five different $|y|_{max}$ bins and for the Jet70U sample as a function of time (run number), fitted with a first degree polynomial.}
  \label{fig_data13}
\end{figure}

\begin{figure}[ht]
  \centering
  \includegraphics[width=0.52\textwidth]{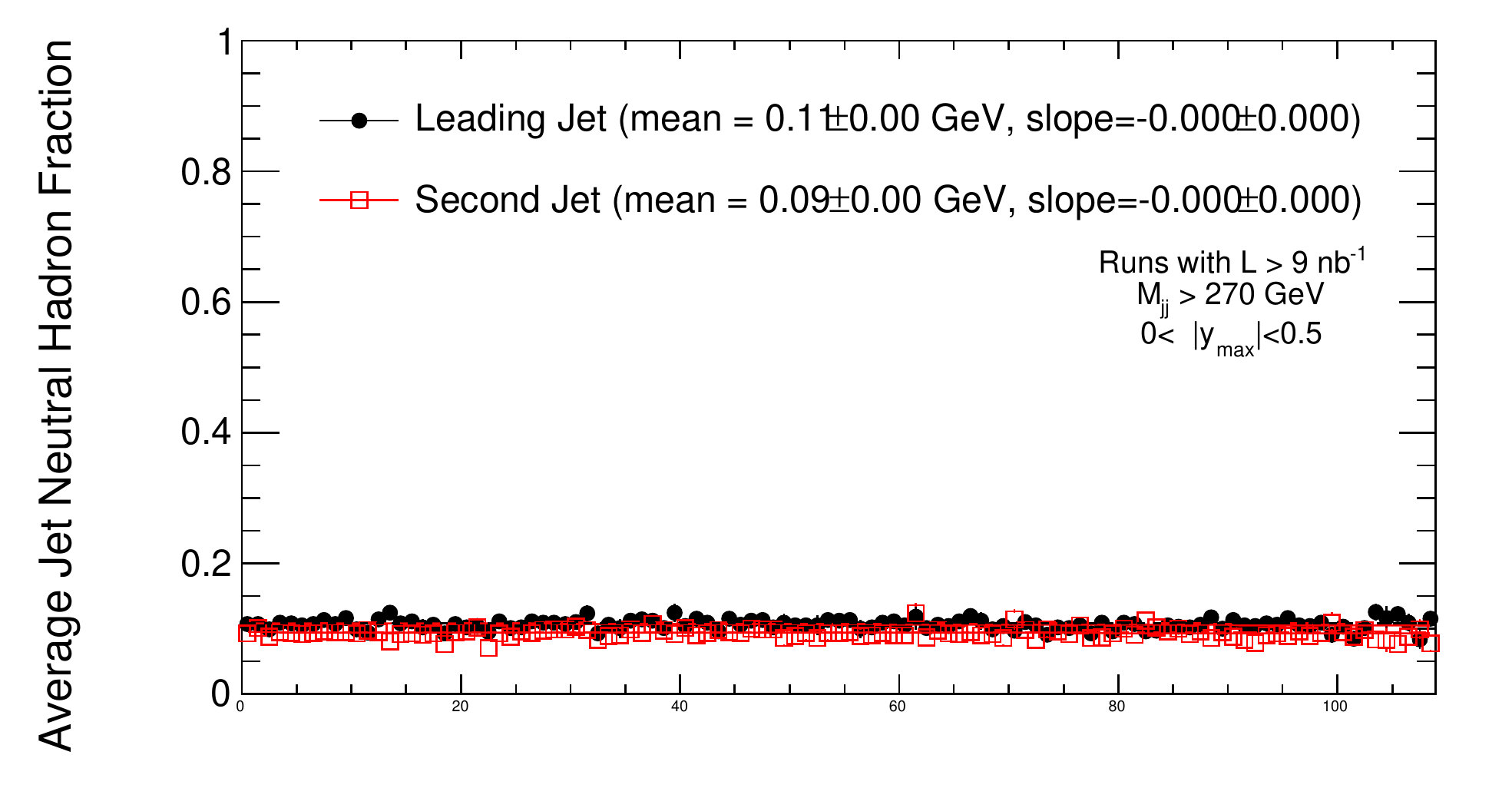}
  \includegraphics[width=0.52\textwidth]{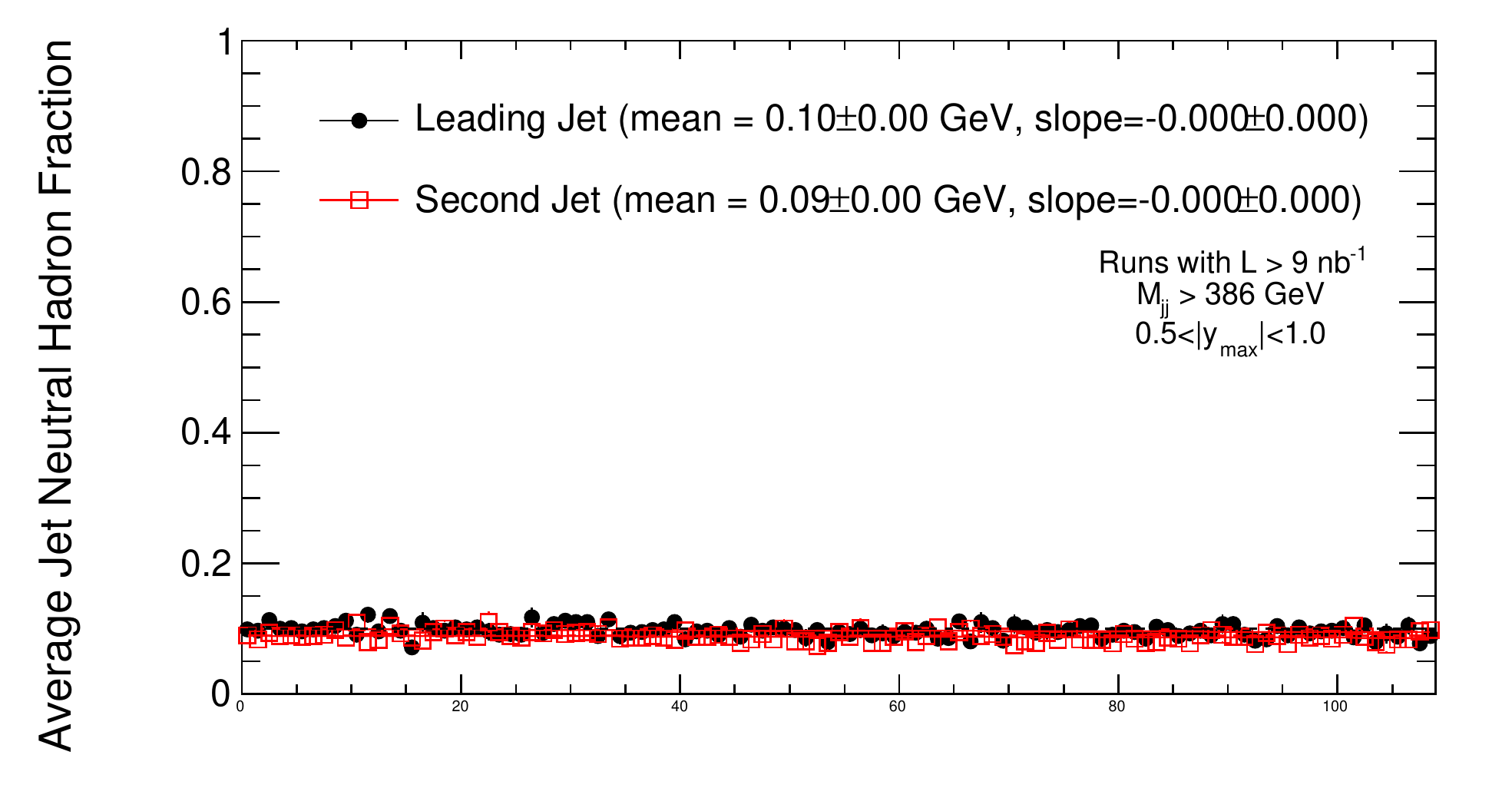} 
  \includegraphics[width=0.52\textwidth]{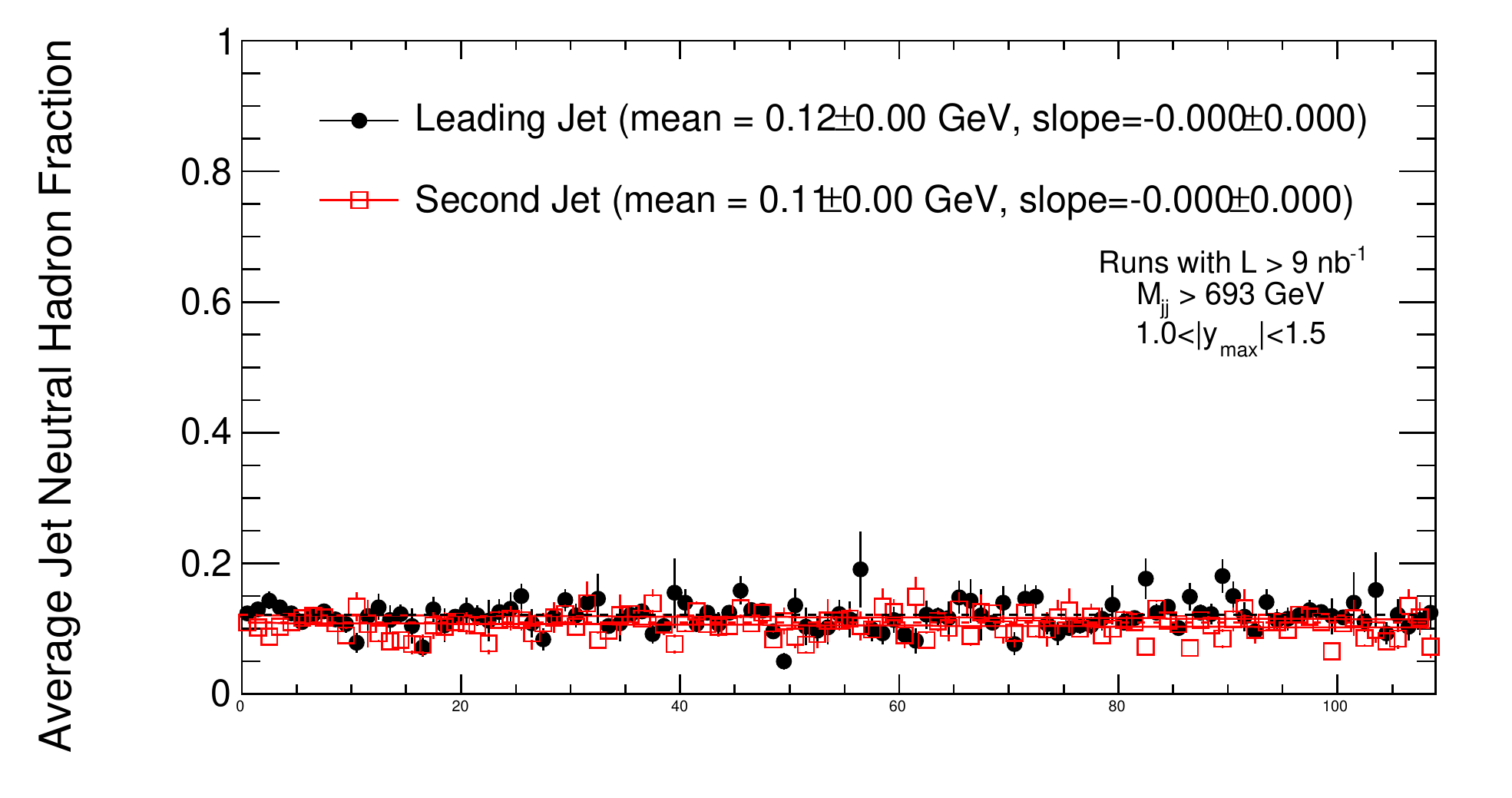} 
  \includegraphics[width=0.52\textwidth]{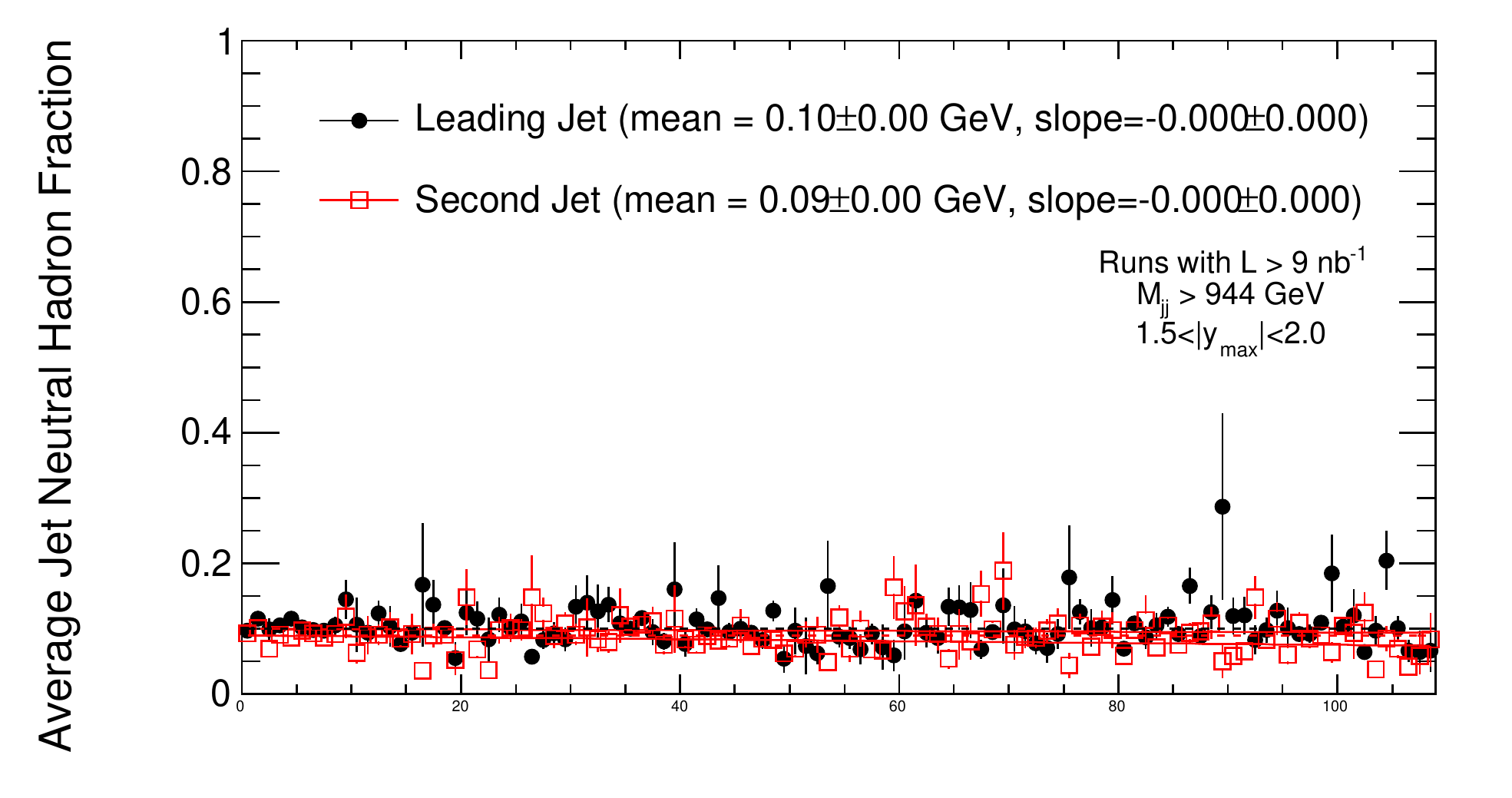}
  \includegraphics[width=0.53\textwidth]{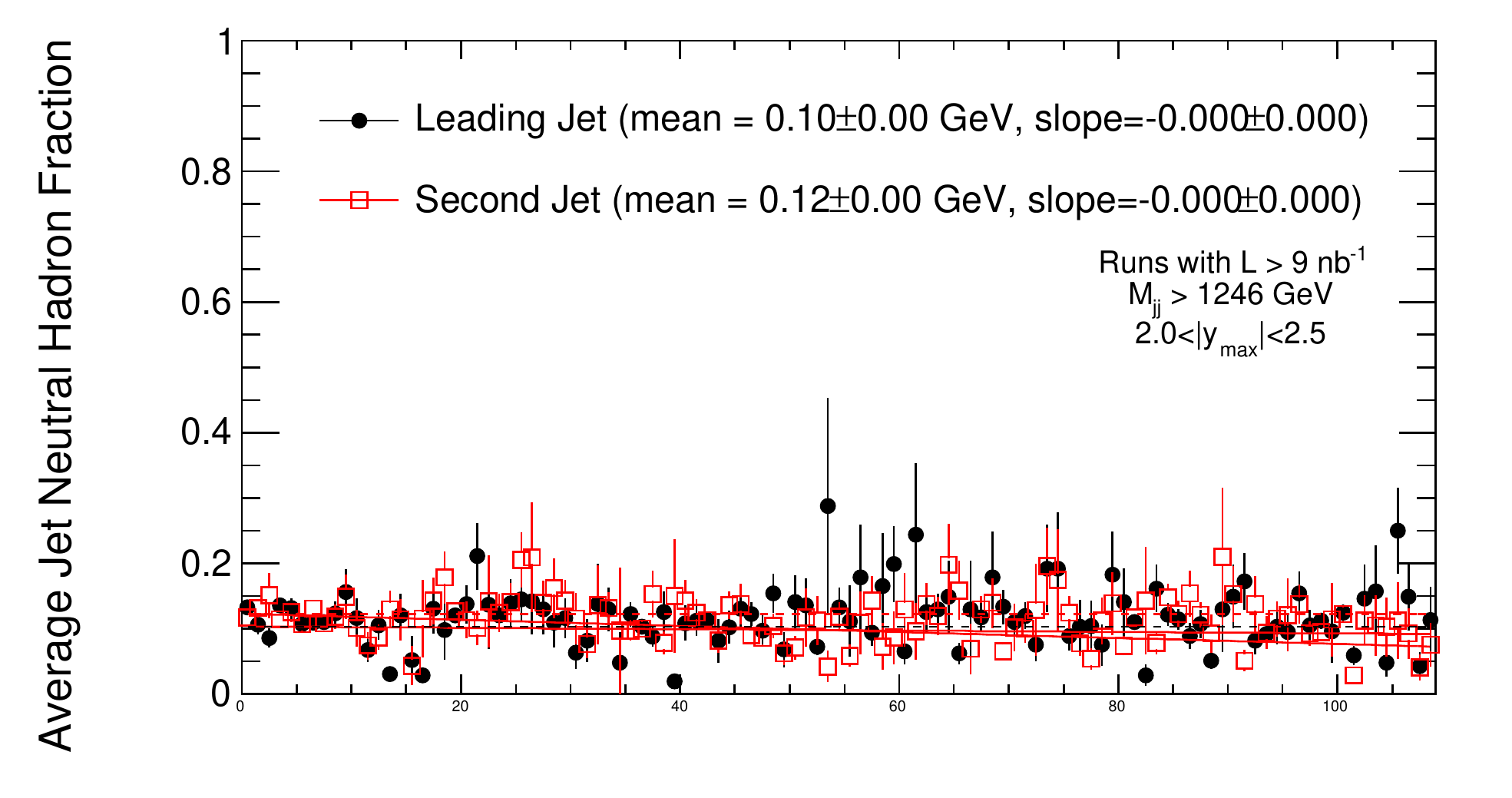}
  \capspace
  \caption{The neutral hadron fraction of the leading and second jet for the five different $|y|_{max}$ bins and for the Jet70U sample as a function of time (run number), fitted with a first degree polynomial.}
  \label{fig_data14}
\end{figure}

\begin{figure}[ht]
  \centering
  \includegraphics[width=0.52\textwidth]{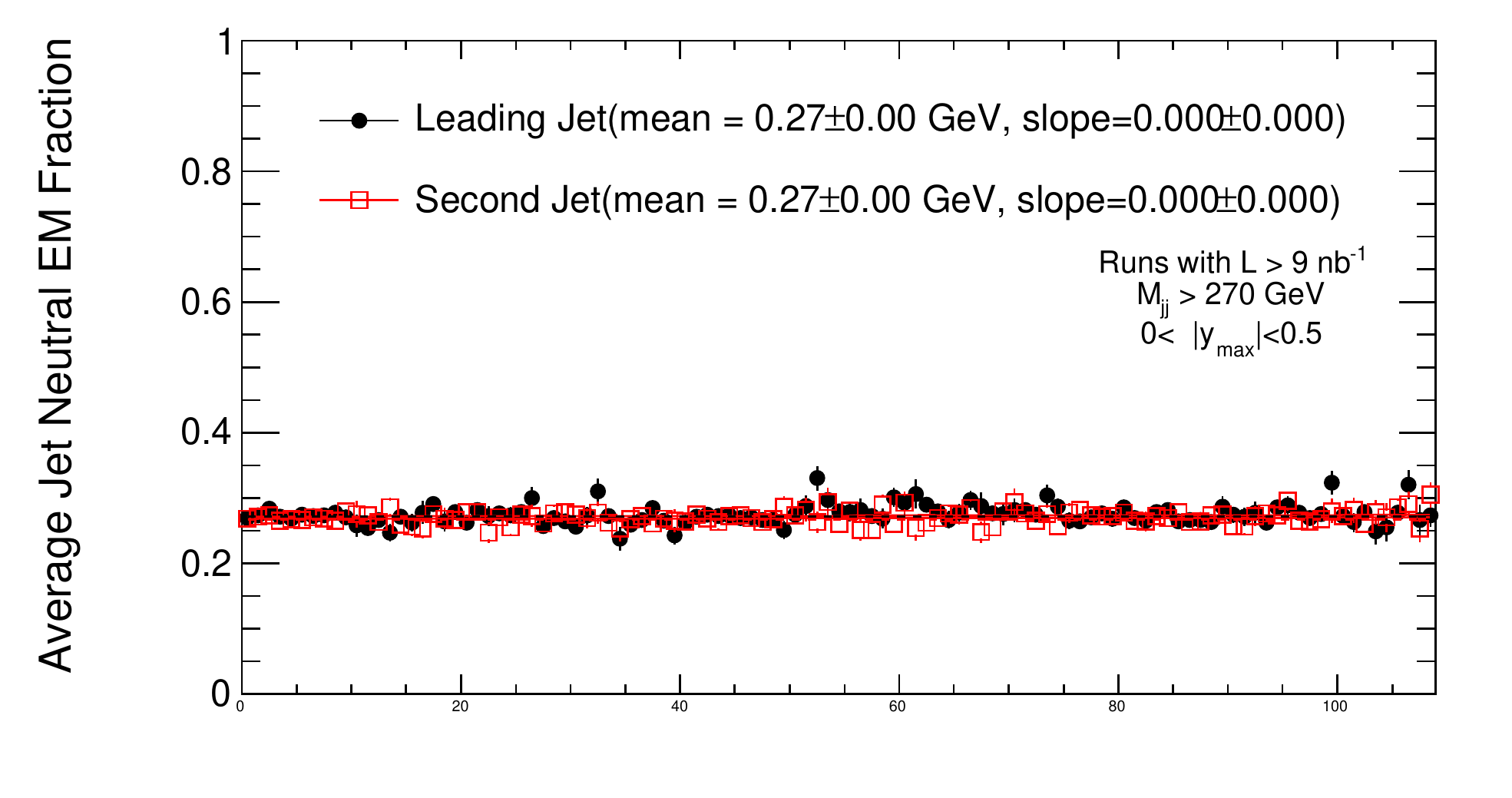}
  \includegraphics[width=0.52\textwidth]{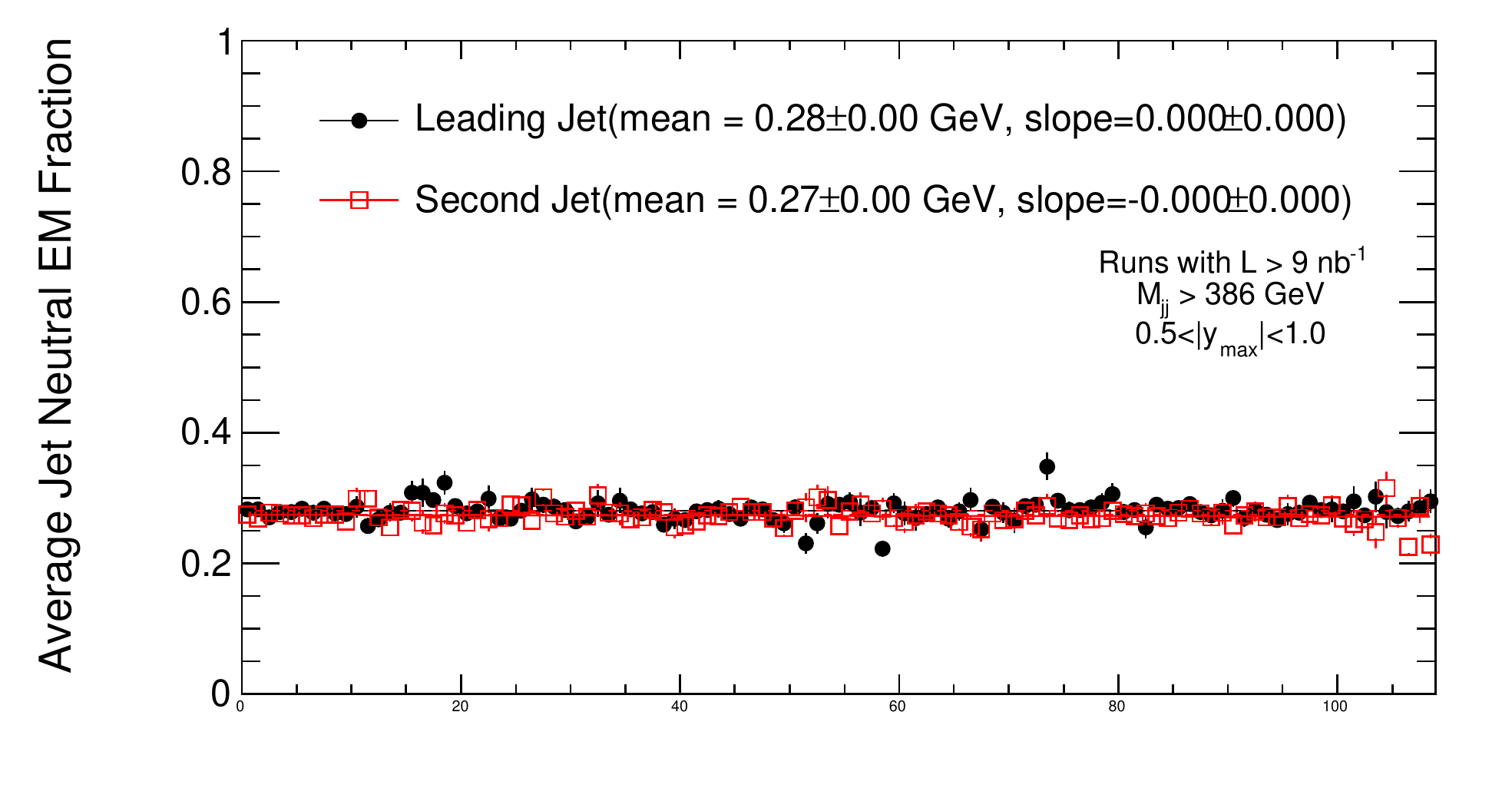} 
  \includegraphics[width=0.52\textwidth]{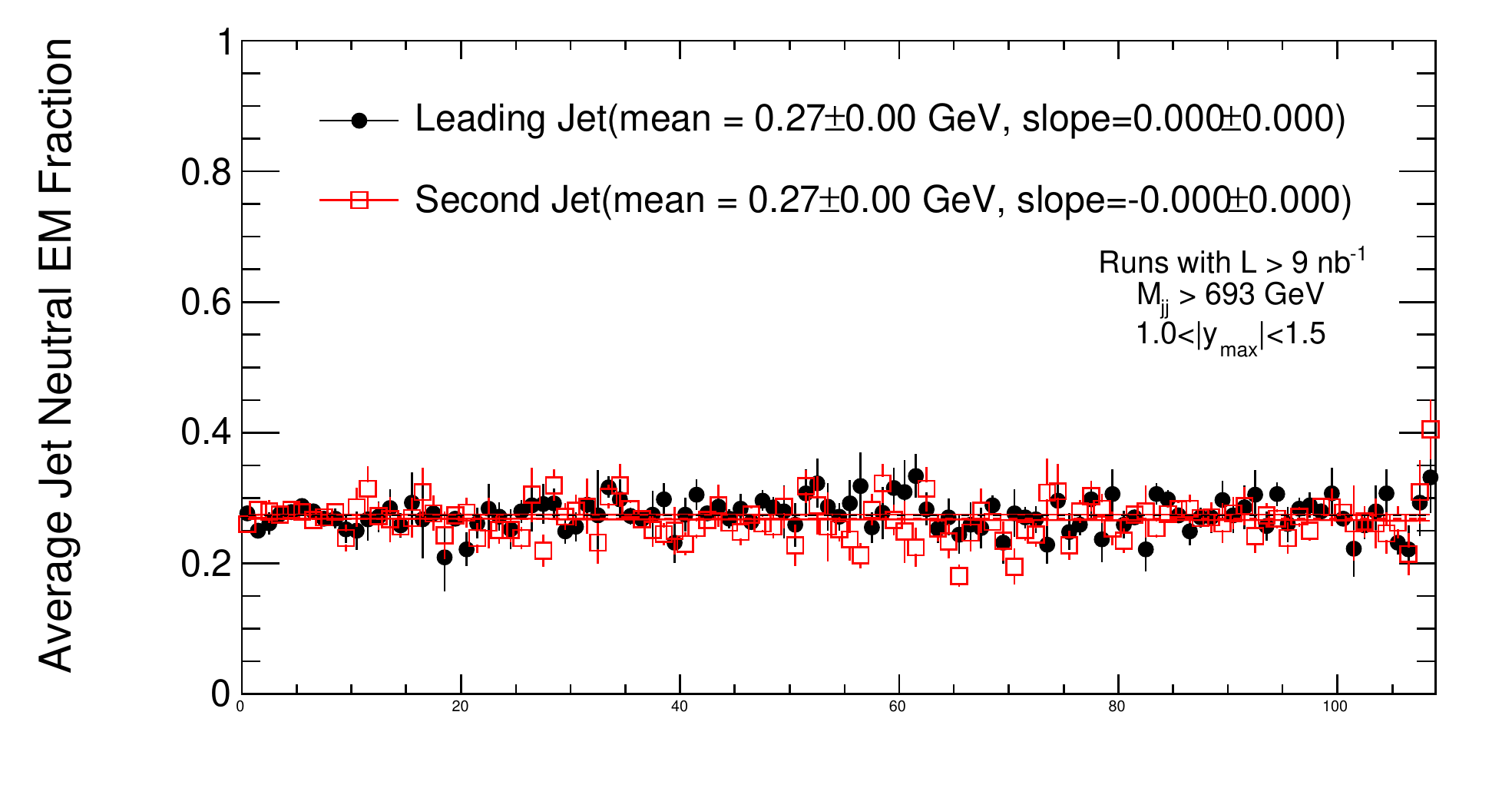} 
  \includegraphics[width=0.52\textwidth]{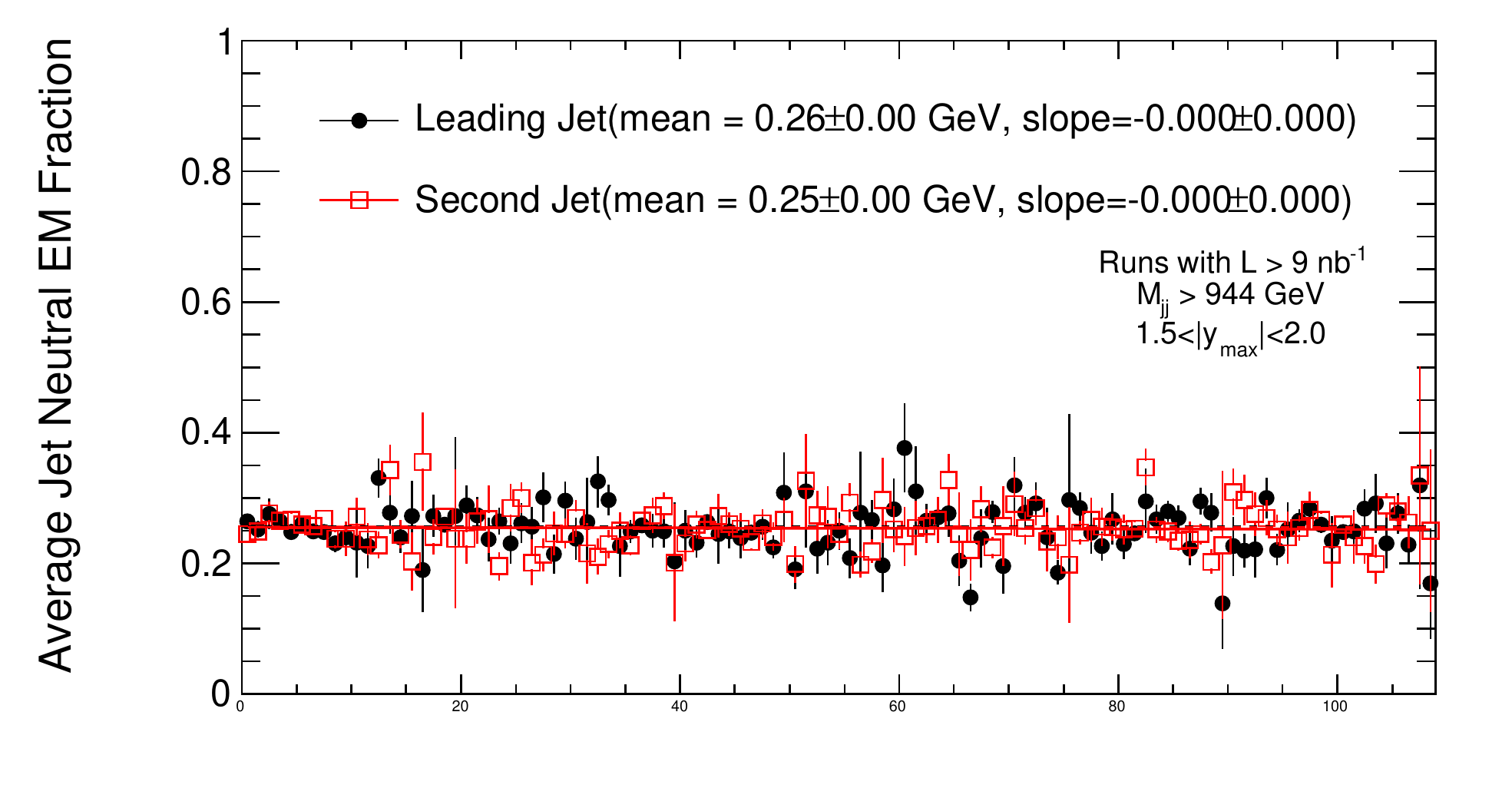}
  \includegraphics[width=0.53\textwidth]{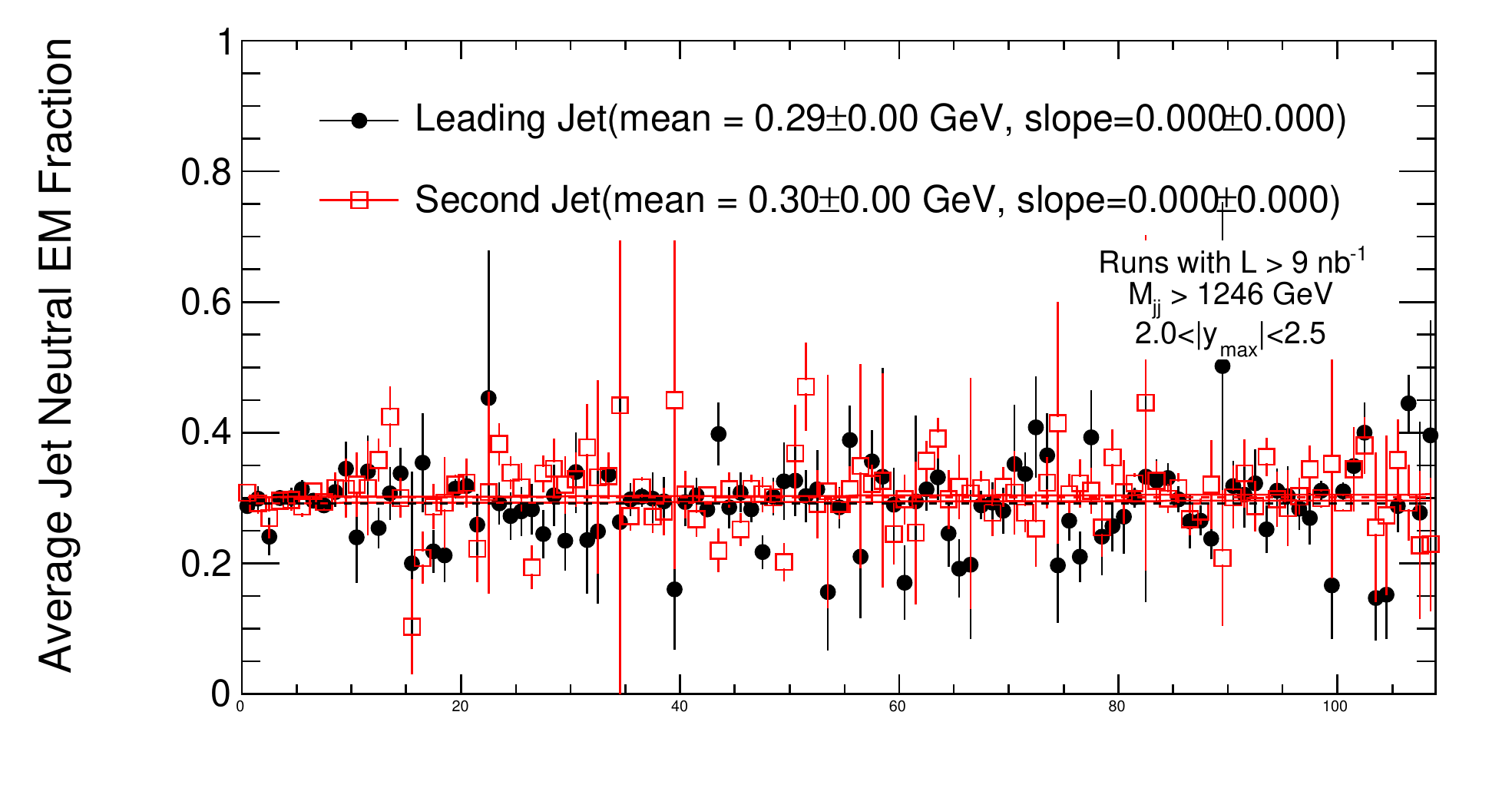}
  \capspace
  \caption{The neutral electromagnetic fraction of the leading and second jet for the five different $|y|_{max}$ bins and for the 
Jet70U sample as a function of time (run number), fitted with a first degree polynomial.}
  \label{fig_data15}
\end{figure}

In Figure \ref{fig_data12} the leading and second jet \pt for the Jet70U sample as a function of time (ordered run numbers and shown starting from 0 regardless of the actual run number for each plot) were shown for the runs with an integrated luminosity greater than 9 \pbinv, and then are fitted with a first degree polynomial The fit is consistent with the constant term and the slope is not statistically significant for all different \ymax bins which is an indication of a stable behavior. A stable behavior as a function of run number is again observed when examining the jet particle content plots, shown in Figures \ref{fig_data13}-\ref{fig_data15}. As it was observed for the jet $p_{T}$, the fit is consistent with the constant term and the slope is not statistically significant for all different \ymax bins which is an indication of a stable behavior. The plots for the rest of the samples are shown in Appendix D.
\clearpage

\clearpage

\clearpage

\clearpage

\clearpage
\section{Jet Energy Corrections}
A jet that is reconstructed and measured by the detector signals (reconstructed jet) has the energy usually different than that of the corresponding particle jet (generated jet). The generated jet is obtained from the Monte Carlo simulation by clustering - using the same jet reconstruction algorithm - the stable colorless particles arising at the end of the hadronization process which presumably occurs after the hard interaction. The reason for this energy to be different in the reconstructed jet than the generated jet is the non-uniform and non-linear response of the CMS calorimeters. Moreover, electronic noise and multiple \proton interactions in the same bunch crossing (pile-up events) can cause this extra amount of energy. The goal of the jet energy calibration is to find the relation between the energy measured in the detector jet and that of the corresponding particle jet. After finding the relation, a correction factor can be applied to each component of the jet momentum four-vector as a multiplicative factor $C(p_{T}^{raw},\eta)$ \cite{JES} (the index $\mu$ represents the components of four-vector).
\begin{equation}
P_{\mu}^{corrected}=C(p_{T}^{raw},\eta)\cdot P_{\mu}^{raw}
\end{equation}
In order to achieve this correction, CMS had adopted a successive factorized approach \cite{factorized_JES}. The order of the sequence is as follows ;
\begin{enumerate}
\item \textit{Offset}: Correction for pile-up, electronic noise, and jet energy lost by thresholds.
\item \textit{Relative($\eta$)}: Correction for variations in jet response with pseudo-rapidity relative to a control region.
\item \textit{Absolute (\pt)}: Correction to particle level versus jet \pt in the control region.
\item \textit{EMF}: Correction for variations in jet response with electromagnetic energy fraction.
\item \textit{Flavor}: Correction to particle level for different types of jet (light quark, c, b, gluon)
\item \textit{Underlying Event}: Luminosity independent underlying event energy in jet removed.
\item \textit{Parton}: Correction to parton level. 
\end{enumerate}
\begin{figure}[ht]
  \centering
  \includegraphics[width=0.98\textwidth]{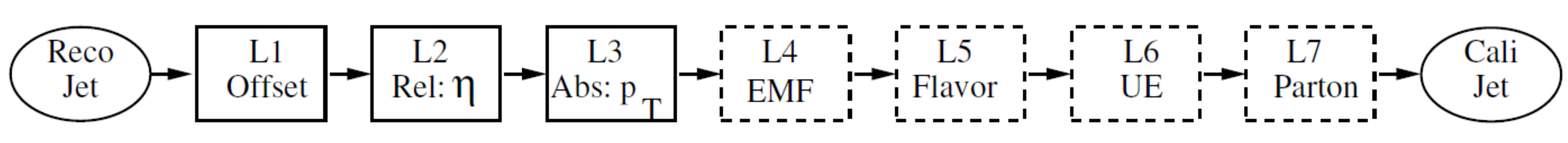}
  \capspace
  \caption{Schematic picture of a factorized multi-level jet correction, in which corrections to the reconstructed jet are applied in sequence to obtain the final calibrated jet. Required correction levels are shown in solid boxes and optional correction levels are shown in dashed boxes.}
  \label{fig:factorized_JES}
\end{figure}
\subsection{Offset Corrections}
The first step of factorized correction sequence starts with the offset correction which aims to estimate and subtract the extra amount of energy from the jet. This extra unwanted energy has presumably nothing to do with high-\pt scattering of partons. It includes contributions from the detector noise and from the multiple proton interactions (pile-up) in the same bunch crossing. For studying the noise, events are first collected with Zero Bias trigger (a random trigger with no conditions) and events with Minimum Bias trigger (a trigger requiring coincident hits in the Beam Scintillating Counters) are discarded. Since the Minimum Bias trigger is a sign of \proton interaction at a given bunch crossing, the remaining sample after the removal can be considered as a pure noise sample. In order to account for one additional event, Minimum Bias triggered events from the early runs are selected. In the early runs, the average number of \proton interaction per event is less than one, hence, a sample with Minimum Bias triggered events from the early runs can be taken as noise+one event sample. Figure \ref{fig:factorized_JES} shows the E$_{offset}$($\eta$) and p$_{T offset}$($\eta$) distributions for noise and noise+one pile-up samples. The contribution from the noise is less than 250 MeV and from the noise+one pile-up is less than 400 GeV in $p_{T}$. It increases up to 7 GeV in energy in the very forward region.
\clearpage
\begin{figure}[ht]
  \centering
  \includegraphics[width=0.40\textwidth ,angle=90]{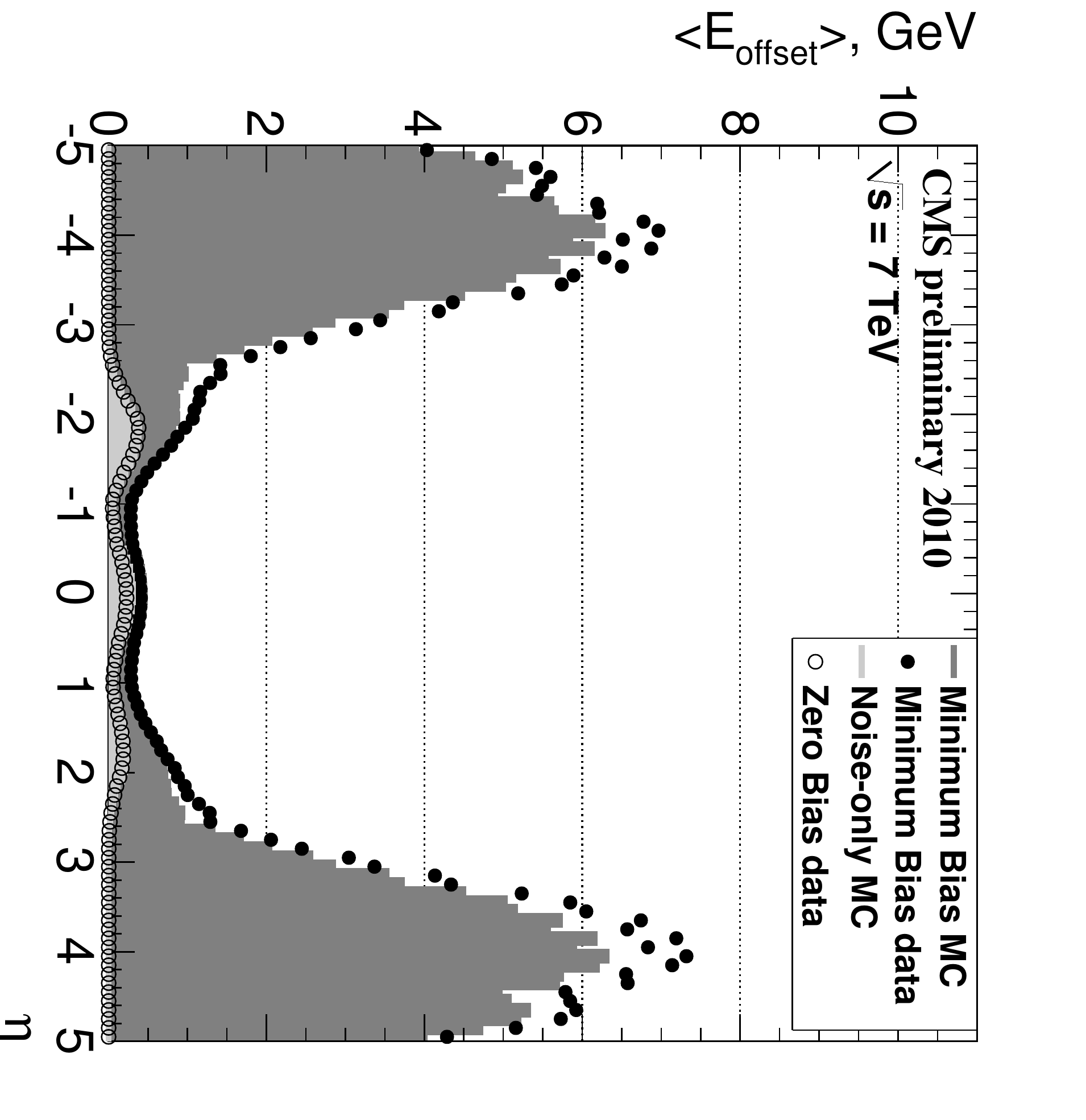}
  \includegraphics[width=0.40\textwidth ,angle=90]{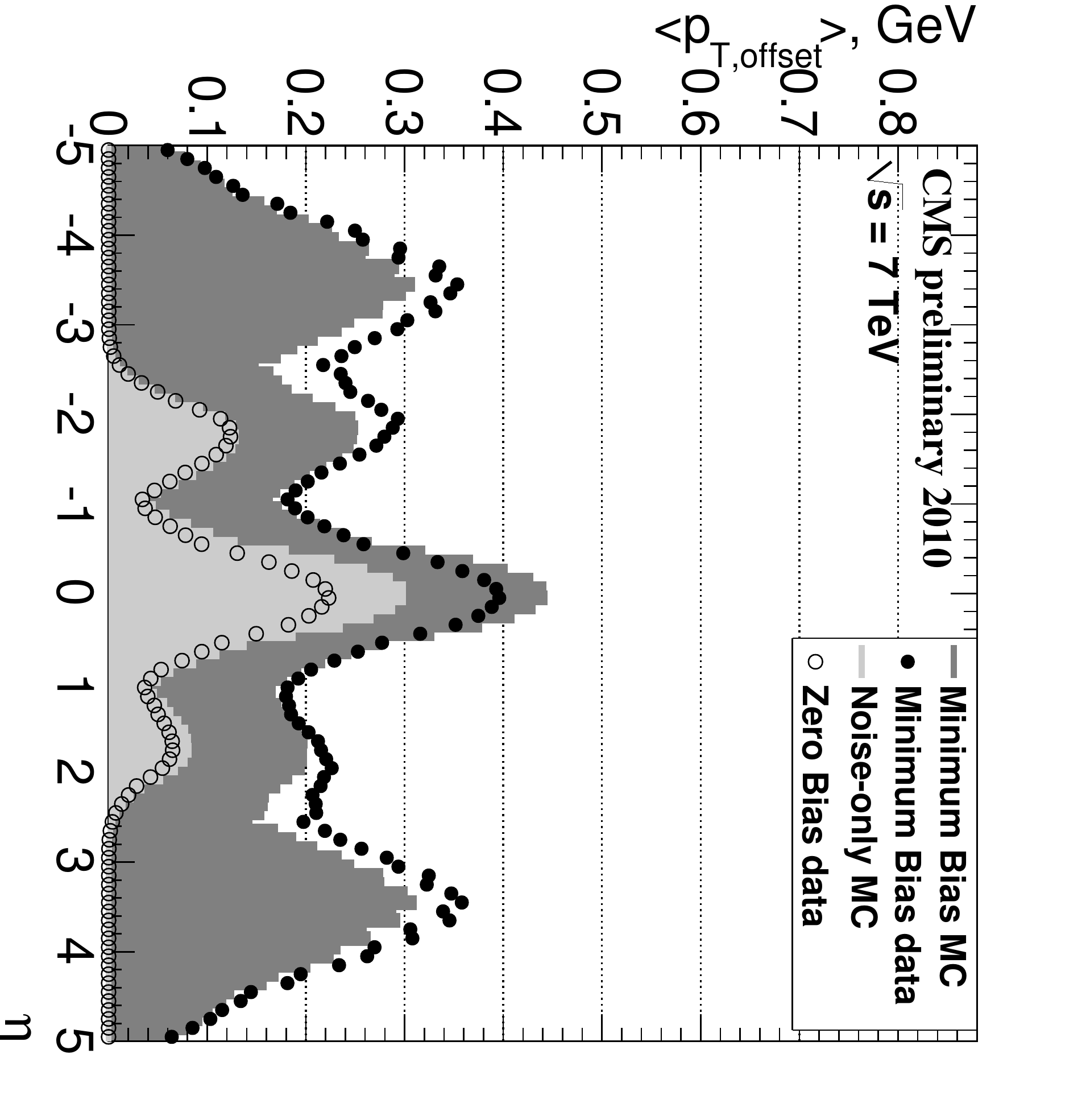}
  \capspace
  \caption{Offset contribution from the noise only and noise+one pile-up as function of $\eta$ in energy (left) and transverse momentum (right) \cite{JetPerformance}.}
  \label{fig:factorized_JES}
\end{figure}
\subsection{Relative Corrections: $\eta$ Dependence}
\begin{figure}[ht]
  \centering
  \includegraphics[width=0.4\textwidth]{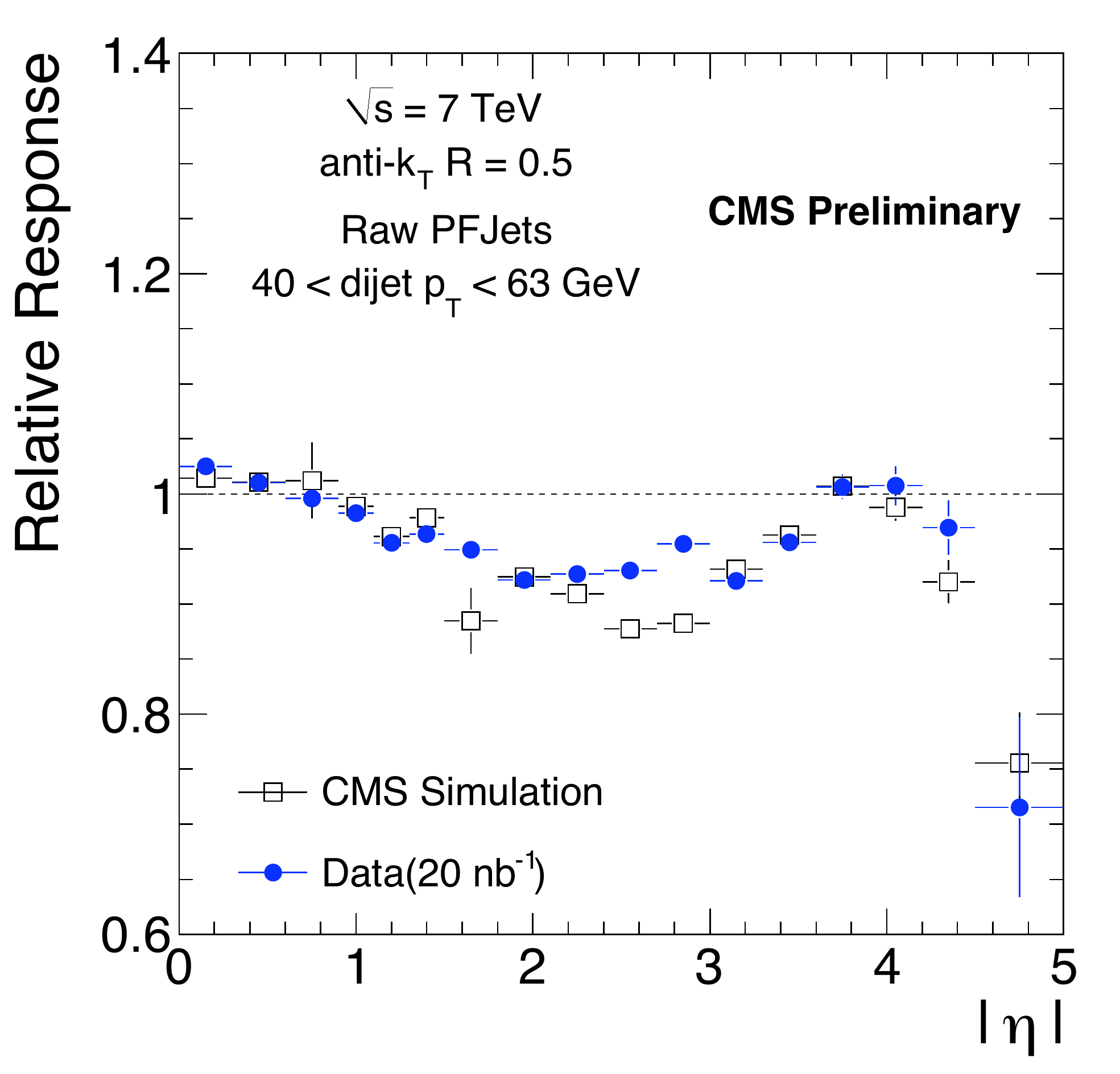}
  \includegraphics[width=0.4\textwidth]{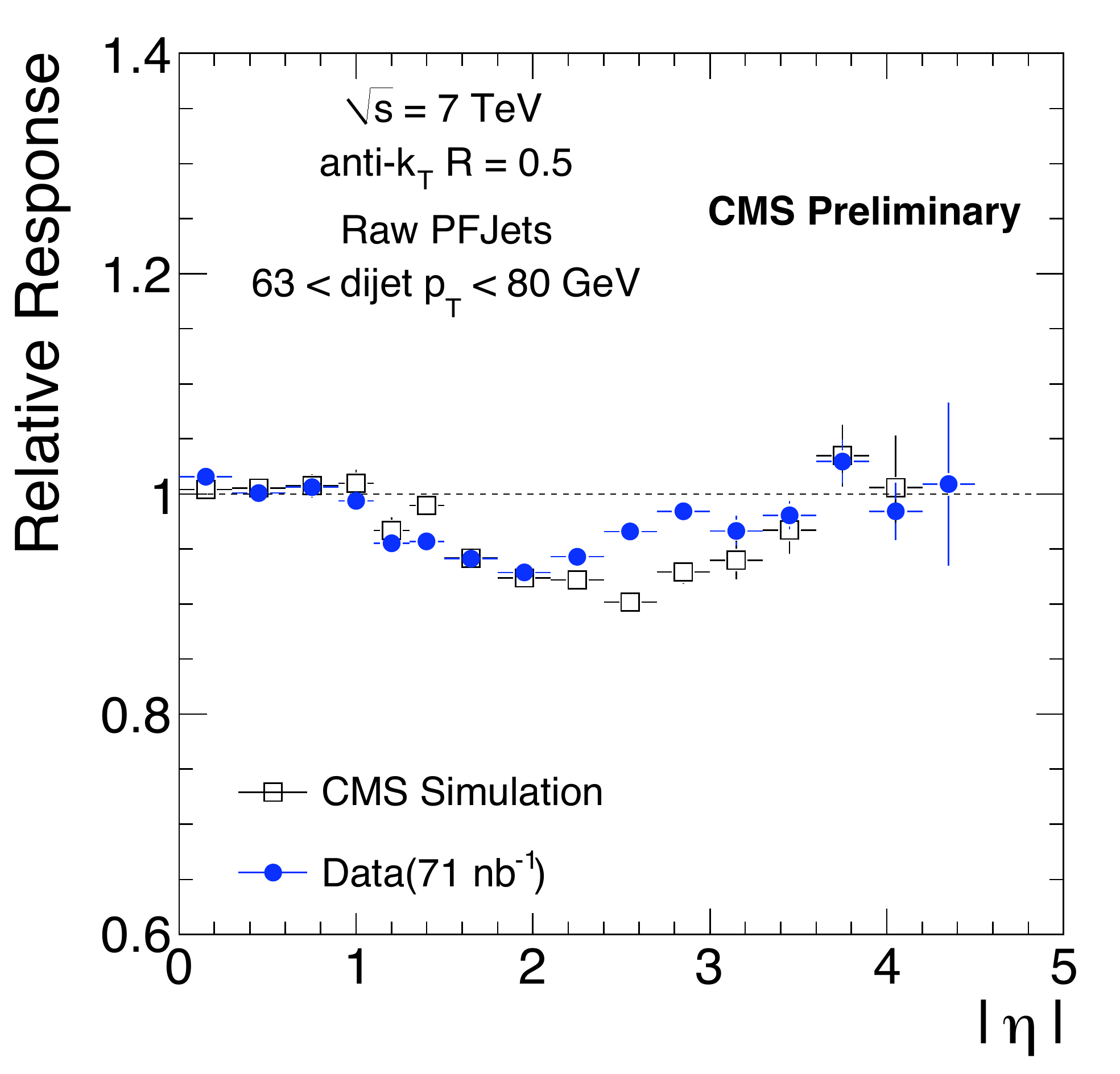}
  \includegraphics[width=0.4\textwidth]{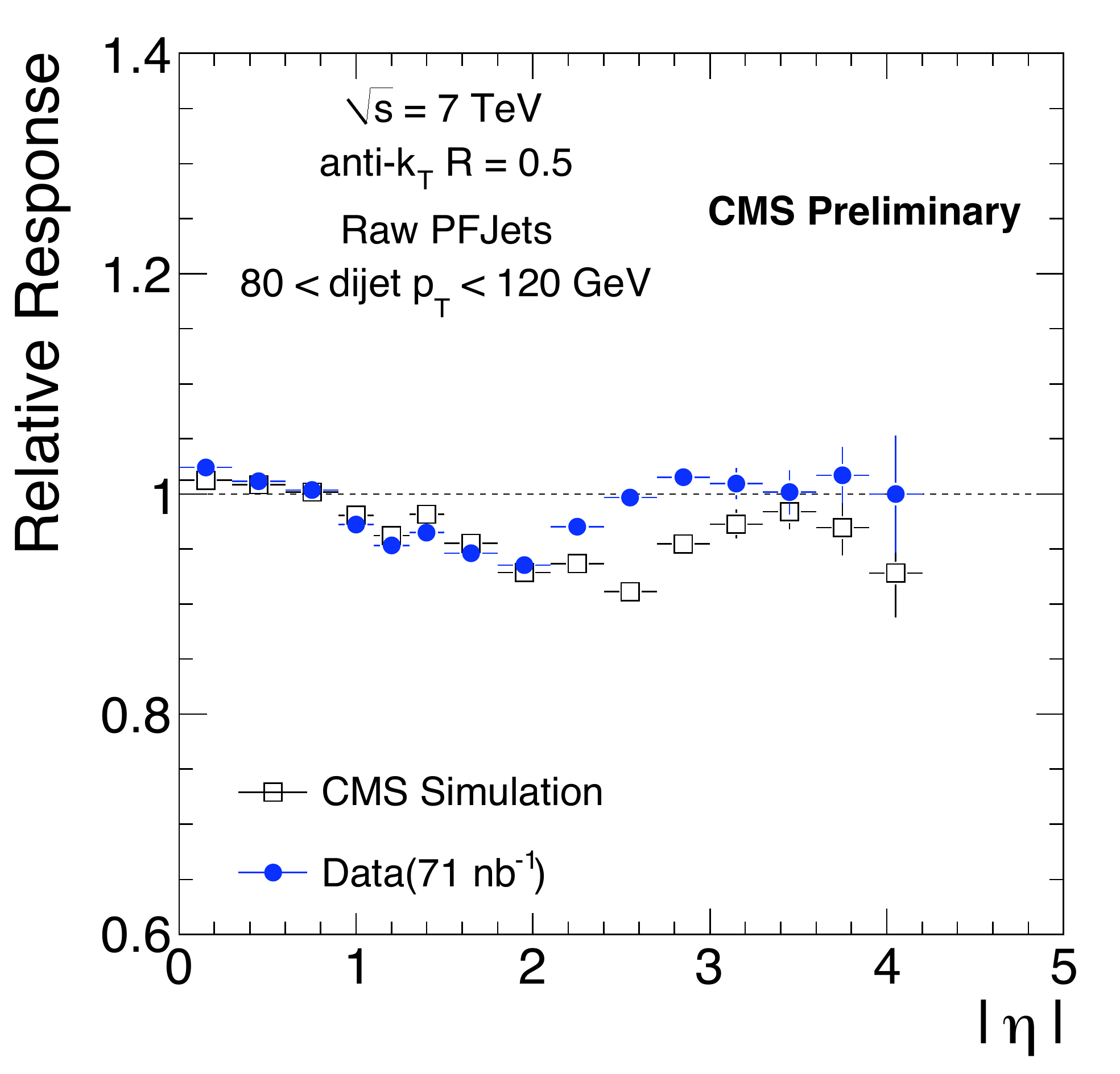}
  \includegraphics[width=0.4\textwidth]{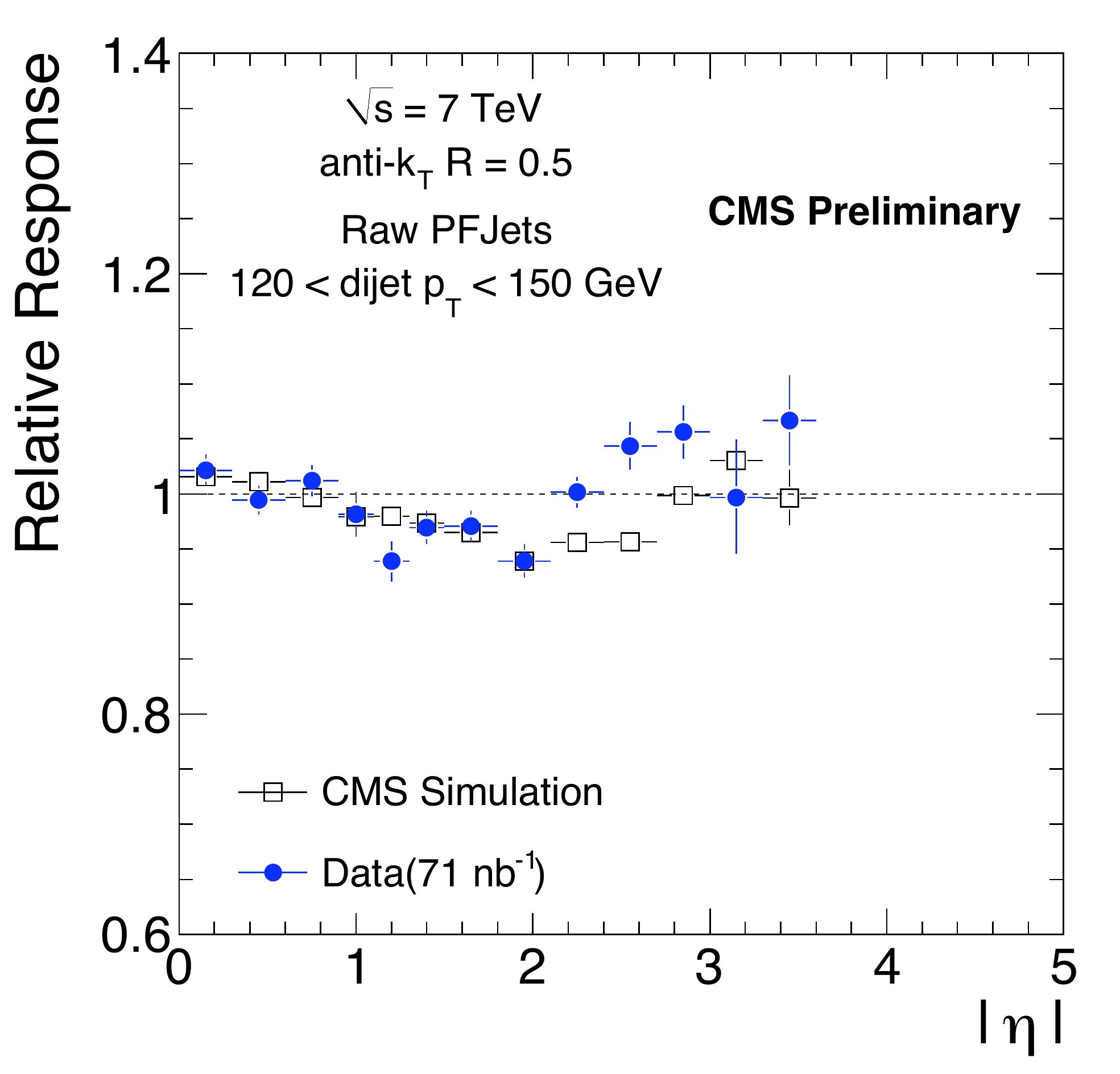}
  \capspace
  \caption{Relative jet response for PF jets as a function of $\eta$, for various p$_{T}^{dijet}$ bins  \cite{JetPerformance}.}
  \label{fig:RelativeResponse}
\end{figure}

For a given true \pt of the jet, the CMS detector's response changes as a function of $\eta$. The relative correction aims to make this dependence flat in $\eta$ and it should be applied after the offset correction. The derivation of the relative energy corrections employs the dijet \pt balance
technique which is  first used at SPPS \cite{UA2 Collaboration}, and later improved in the Tevatron experiments \cite{D0 Collaboration, CDF Collaboration}. The idea is to use \pt balance in back-to-back dijet events with one barrel jet in the central control region of the calorimeter, $|\eta|<$ 1.3, and the other probe jet at arbitrary $\eta$. The $|\eta|<$ 1.3 region is chosen as reference since the detector response to jets is uniform in this region \cite{CMS experiment}. The two leading jets must be separated by $\Delta\phi>$ 2.7 and no additional third jet in the event with $p_{T}^{3^{rd} jet}/p_{T}^{dijet}>$0.2 is allowed to increase the fraction of 2$\rightarrow$2 processes in the sample, where $p_{T}^{dijet}=(p_{T}^{probe}+p_{T}^{barrel})/2$ is an average uncorrected \pt of two leading jets. The quality of the dijet balance is given by:
\begin{equation}
B=\dfrac{p_{T}^{probe}-p_{T}^{barrel}}{p_{T}^{dijet}}
\end{equation}
The relative response in terms of the expectation value of B distribution, $<$B$>$, in a given $\eta^{probe}$ and $p_{T}^{dijet}$ bin is defined as below \cite{CMS Dijet Balance}.
\begin{equation}
R(\eta^{probe}, p_{T}^{dijet})=\dfrac{2+<B>}{2-<B>}
\end{equation}
The relative jet response as a function of $\eta$ obtained from the data and the Monte Carlo prediction are shown in Figure \ref{fig:RelativeResponse} for different $p_{T}^{dijet}$ ranges.
\subsection{Absolute Corrections: \pt Dependence}
Removal of the noise and pileup contributions and then making the response flat in $\eta$ allows us to apply an absolute calibration in \pt by using the $\gamma$+jet events. There are two methods for absolute calibration; \pt balancing method and the Missing E$_{T}$ Projection Fraction (MPF) method \cite{JetPerformance}. The latter is the main method in CMS. In MPF method, the primary assumption is that the $\gamma$+ jet events have no real missing E$_{T}$. Therefore, there is a perfect balance between the photon and the hadronic recoil in transverse
plane.
\begin{equation}\label{photonjetrecoil}
\vec{p}_{T}^{~\gamma}+\vec{p}_{T}^{~recoil}=0
\end{equation}
In fact, this is the ideal case for the balance. In real life, the detector response should be taken into account, and the balance is usually satisfied by introducing a quantity called missing transverse energy ($\vec{E}_{T}^{~missing}$). Thus, the Equation \ref{photonjetrecoil} can be rewritten for reconstructed events as;
\begin{equation}\label{photonjetrecoil_reco}
 R_{\gamma}\cdot\vec{p}_{T}^{~\gamma}+R_{recoil}\cdot\vec{p}_{T}^{~recoil}=-\vec{E}_{T}^{~missing}
\end{equation}
where $R_{\gamma}$ and $R_{recoil}$ are the detector responses to the photon and the hadronic recoil. The good calibration of photons enhances us to take $R_{\gamma}$=1, then solving above equations gives;
\begin{equation}\label{photonjetrecoil_reco}
 R_{MPF}\equiv\dfrac{R_{recoil}}{R_{\gamma}}=1+\dfrac{\vec{p}_{T}^{~\gamma}\cdot\vec{E}_{T}^{~missing}}{(p_{T}^{\gamma})^2}=R_{recoil}
\end{equation}

\begin{figure}[ht]
  \centering
  \includegraphics[width=0.49\textwidth]{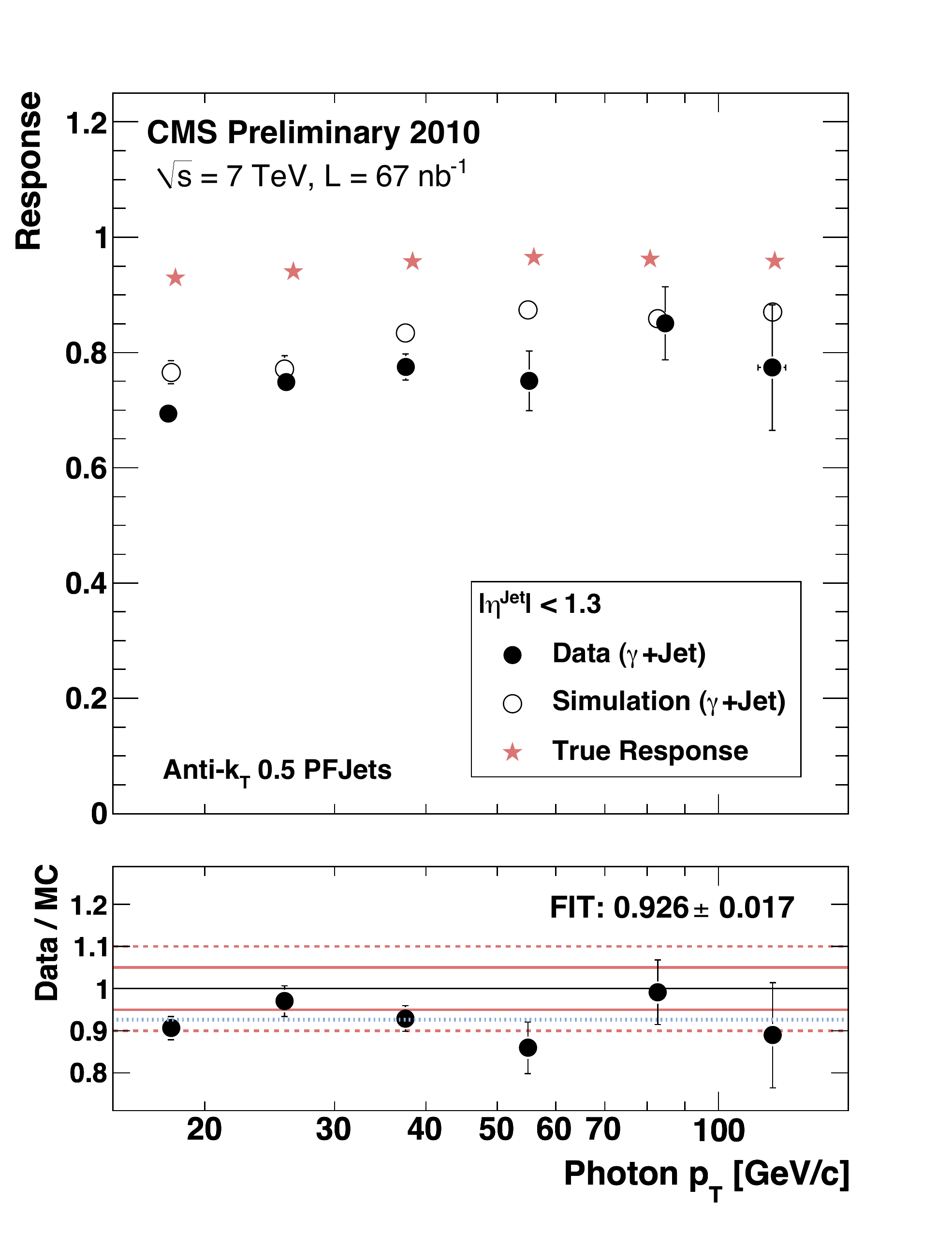}
  \includegraphics[width=0.49\textwidth]{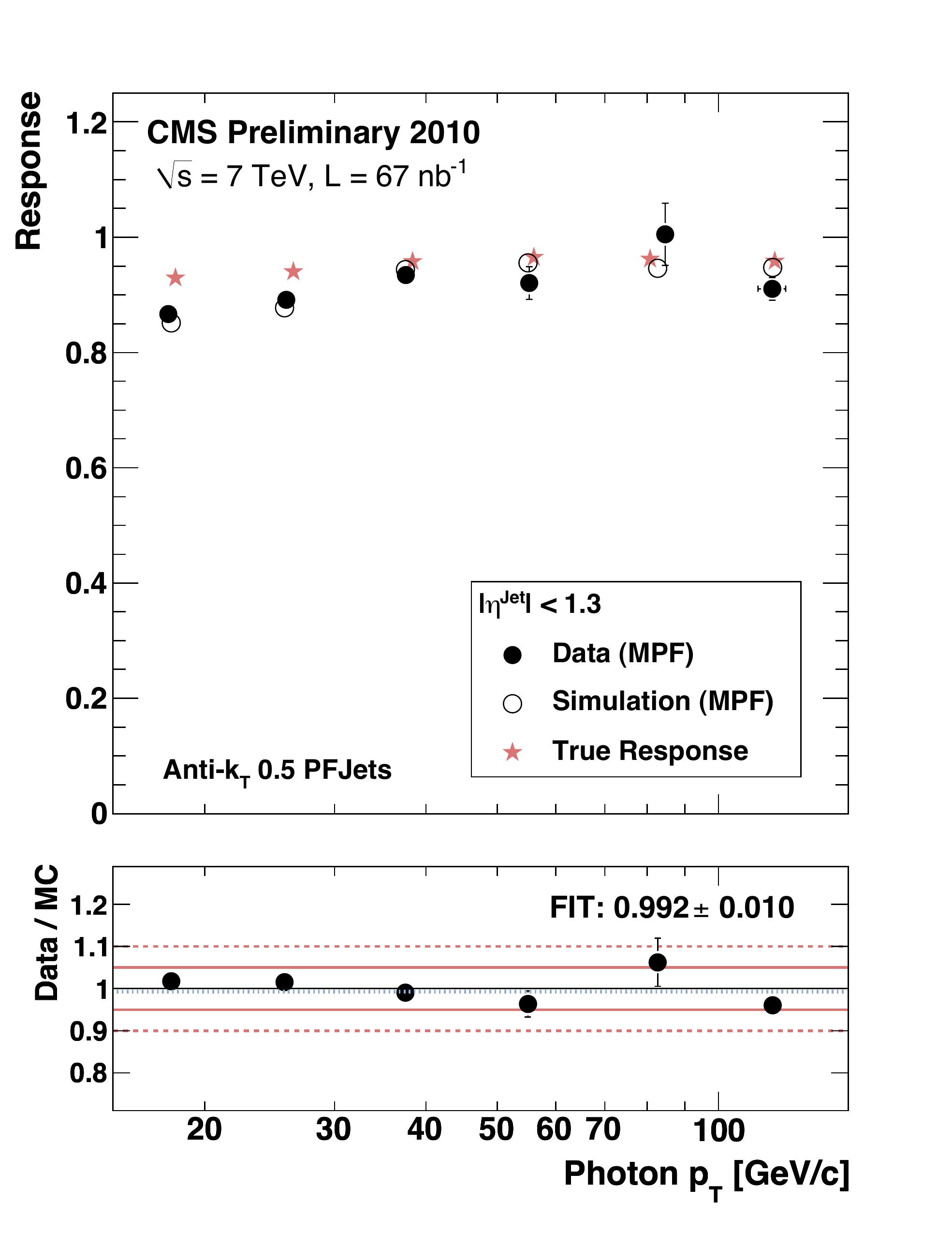}
  \capspace
  \caption{Response of $<p_{T}/p_{\gamma}>$ (left) and MPF response (right) as a function of photon \pt from data and simulation \cite{JetPerformance}.}
  \label{fig:AbsoluteCalib}
\end{figure}

Figure \ref{fig:AbsoluteCalib} shows the $<p_{T}/p_{\gamma}>$ response of and MPF response as a function of photon \pt .

\section{Corrections for the Smearing Effects}
\label{Unsmearing}
\begin{figure}[ht]
  \centering
  \includegraphics[width=0.40\textwidth]{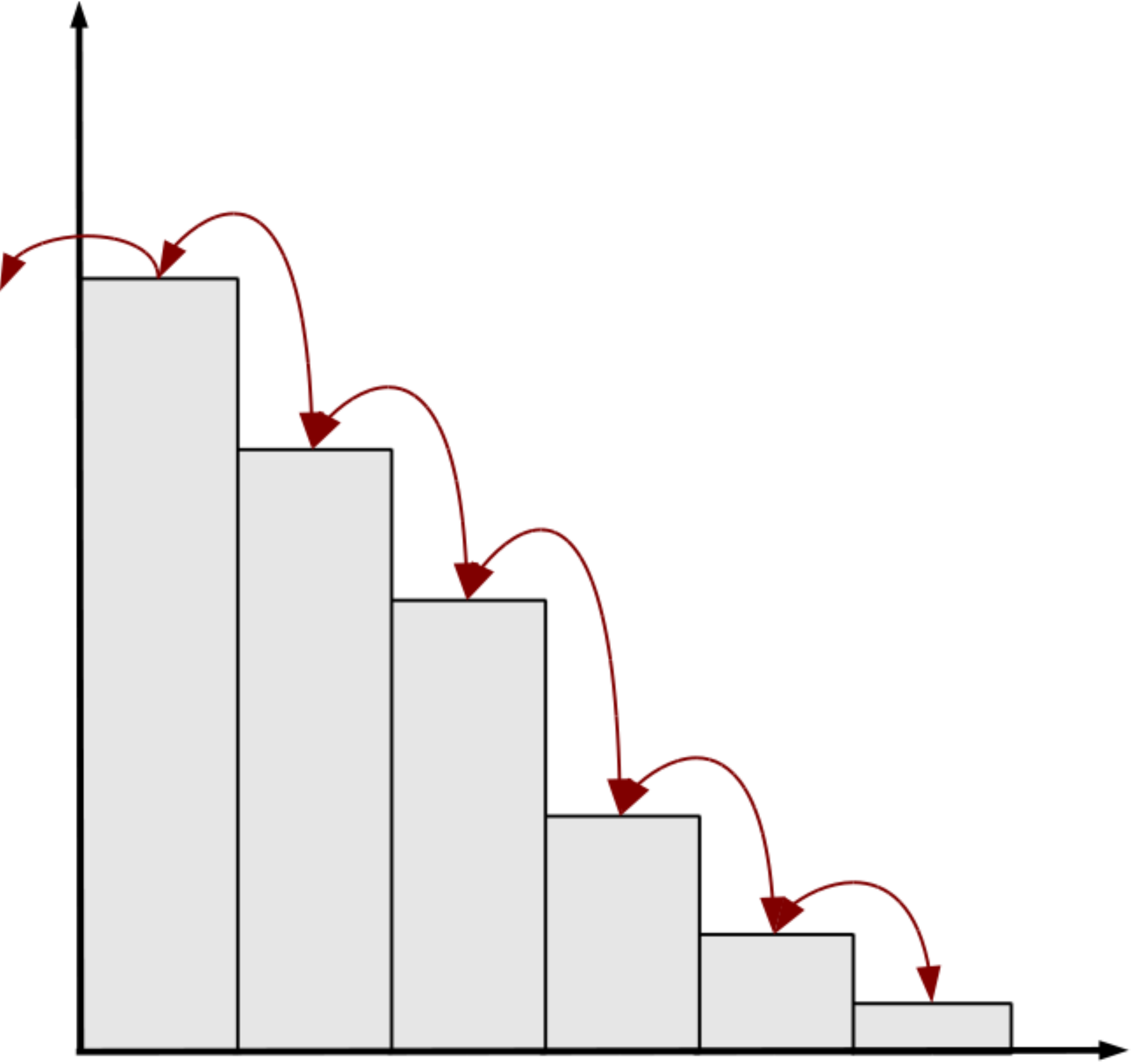}
  \includegraphics[width=0.45\textwidth]{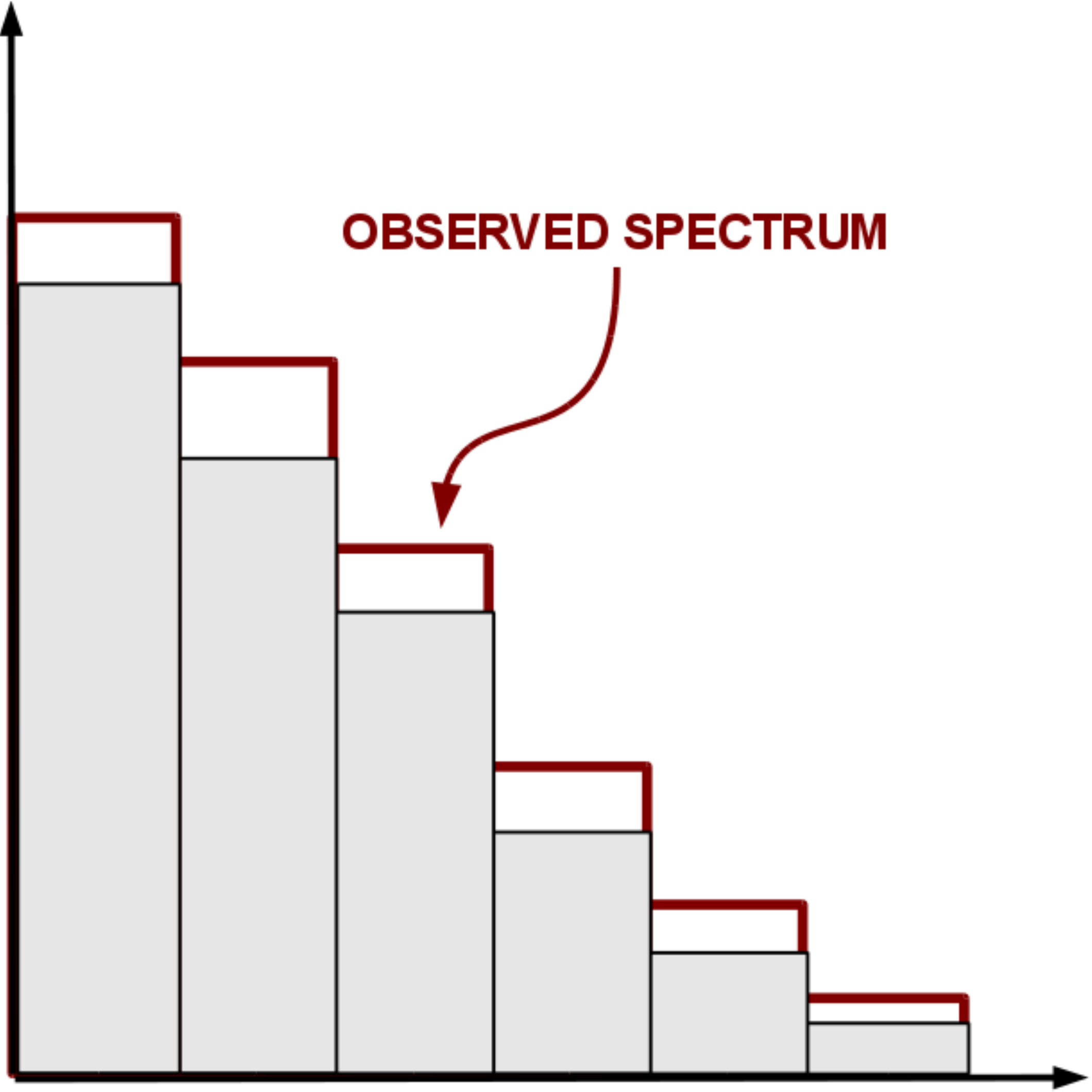}
  \capspace
  \caption{A cartoon illustration of smearing effects for a steeply falling distribution.}
  \label{fig_smearing}
\end{figure}

Due to the finite resolution of the detector, the measured spectrum, which is called \textsl{``at the detector level"}, is a smeared form of the actual distribution which is \textsl{``at the particle level"}. The smearing effect is considerable because of the very steep nature of the dijet mass spectrum. Each mass bin in the spectrum is contaminated by events that have migrated from neighboring bins and the original residents of the bin have a finite probability to migrate to the other bins depending on the detector resolution for the mass value at that bin. Since the event population of a bin with lower dijet mass value is far greater than the event population of a bin with higher dijet mass value, it is more likely for a bin with higher mass value to have a higher fraction of immigrant events. As a result of this fact, the measured spectrum is steeper than in the case that the resolution would be perfect. In Figure \ref{fig_smearing} a cartoon illustration of the smearing effect for a falling spectrum is shown.

The measured cross section can be modeled as the convolution of the particle level spectrum with the detector resolution: 
\begin{equation}
F(m^{reco})=\int^{\infty}_{0}f(m^{gen})R(m^{reco},m^{gen})dm^{gen}
\end{equation}
where $m^{reco}$ and $m^{gen}$ represent the measured mass value and the mass value at the particle level respectively.

In order to estimate the size of the smearing effect, a technique called \textsl{``forward smearing"} is adopted. In this technique, dijet mass values at the particle level are generated randomly according to the spectrum predicted by PYTHIA6 then smeared with the response function which effectively simulates the detector effects on the generated value. The dijet mass resolution is modeled by a Gaussian function centered at the generated mass:
\begin{equation}
R(m^{reco},m^{gen})=\dfrac{1}{\sqrt{2 \pi}\sigma(m^{gen})}\exp \left[ - \dfrac{1}{2} \dfrac{m_{gen}^{2}}{\sigma^{2}(m^{gen})} \right] 
\end{equation}
The $\sigma$ of the Gaussian is determined from the relative resolution parametrization. The details of the determination of resolution parameters will be discussed in the next section. Finally, the observed (``smeared") and the true (``generated") spectra are compared bin-to-bin in terms of the ratio of the bin contents, namely; $N_{true}/N_{observed}$. In Figure \ref{fig.UnfodingForAll}  the correction factors for unsmearing effect are shown for all five \ymax bins and it can be observed that the ratio is close to the unity in all rapidity bins. The shape of the curves is closely related to the mass dependence of the resolution and the spectrum slope. At lower dijet mass values the resolution is relatively worse than it is at higher mass values and the effect of smearing is larger. At higher mass values, even if the resolution improves, the spectrum becomes steeper which leads again to a larger smearing effect.
Since the effect of smearing is small it allows us to attempt an average unfolding using the ratio $N_{True}/N_{Observed}$ as a multiplicative correction factor (bin-by-bin correction). It is understood that the more advanced and statistically sound treatments of the unfolding problem yield the correct statistical uncertainties, which are underestimated in the simple approach described here.
However, the size of the effect is so small that the effect on the statistical uncertainty is negligible compared to the other systematic uncertainties of the measurement.

 \begin{figure}[h]
   \begin{center}
     \includegraphics[width=0.55\textwidth]{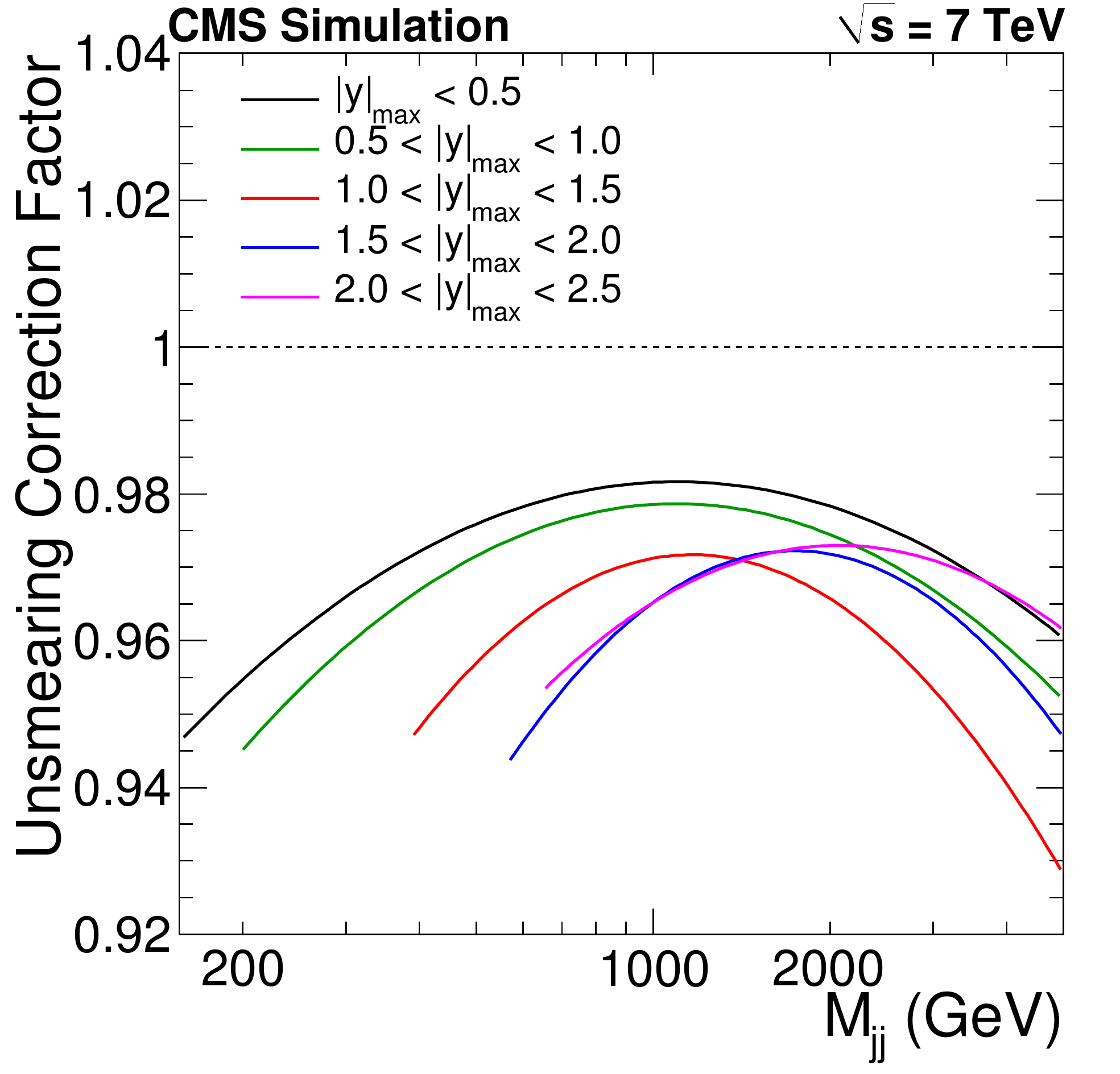}
     \capspace
     \caption{Unsmearing correction factors as a function of the dijet mass, in the various \ymax bins.}
     \label{fig.UnfodingForAll}
   \end{center}
 \end{figure}
\clearpage
\section{Dijet Mass Resolution}
\begin{figure}[h]
   \begin{center}
     \includegraphics[width=0.51\textwidth]{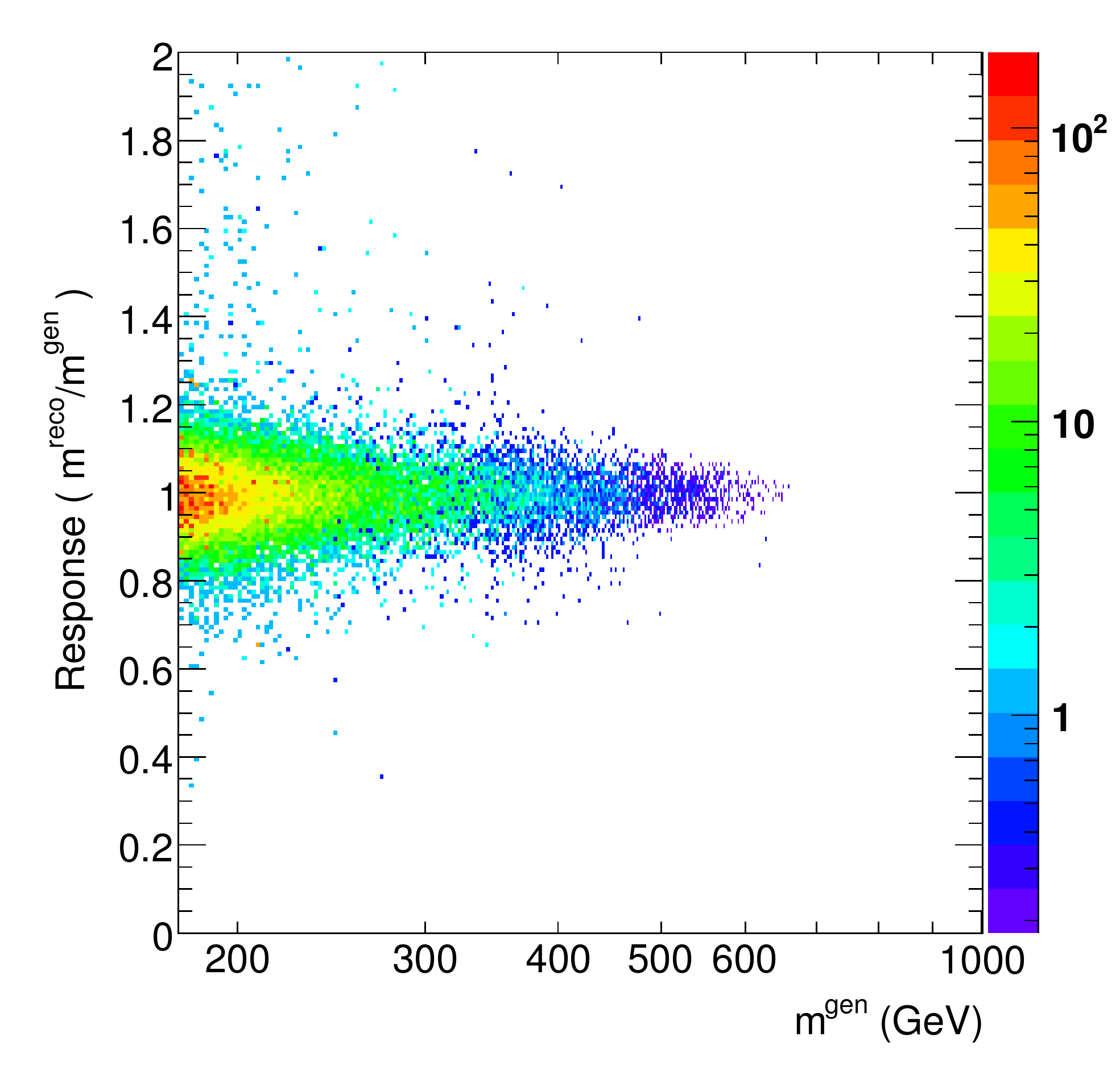}
     \includegraphics[width=0.47\textwidth]{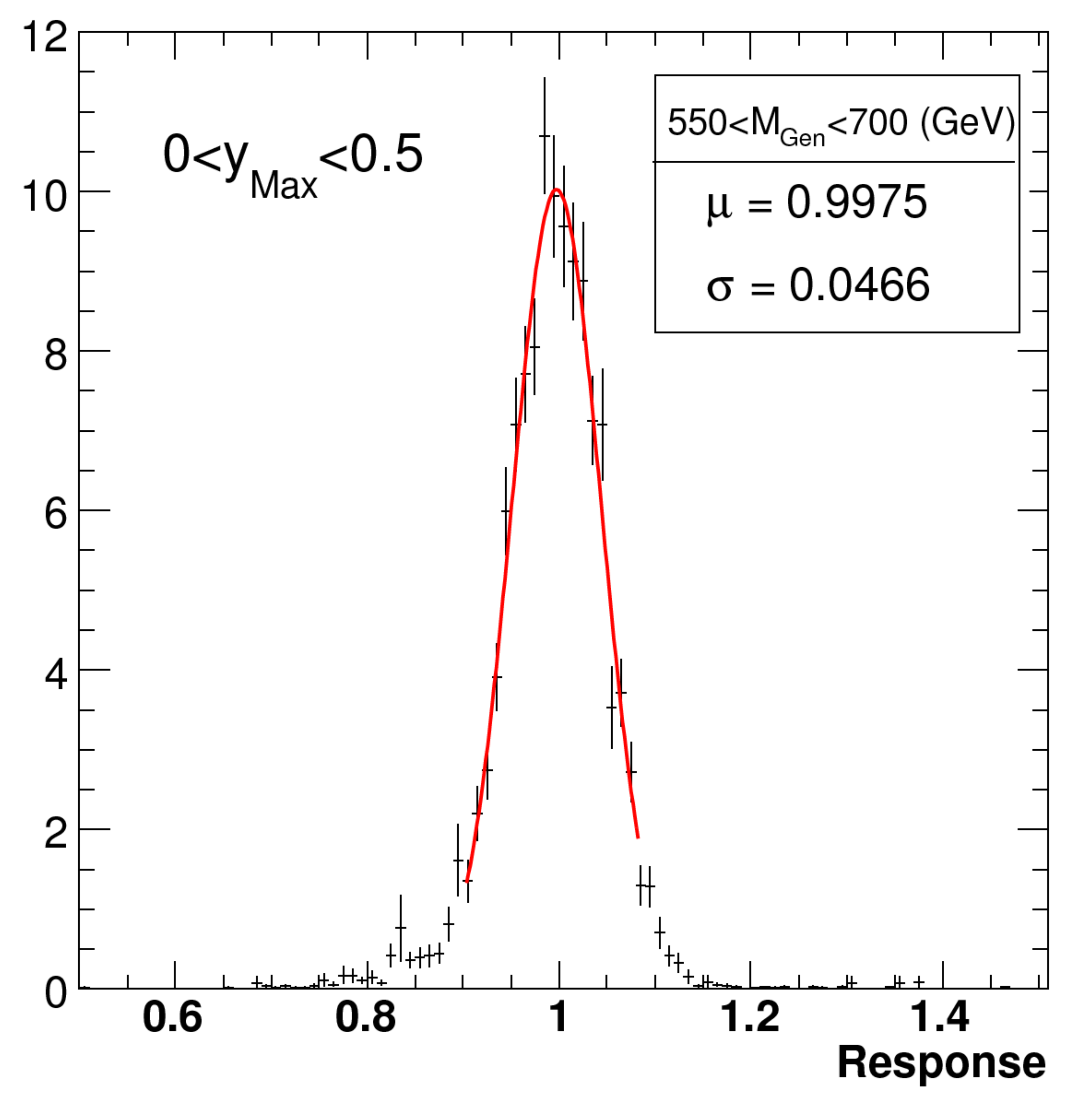}
     
     \caption{The response of the detector as a function of the generated mass in 0 $<$ \ymax $<$ 0.5 (left). The distribution of the ratio of the reconstructed dijet mass over the generated mass in 0 $<$ \ymax $<$ 0.5 bin, for 550 GeV$<M_{Gen}<$ 700 GeV (right).}
     \label{MassResponse}
   \end{center}
 \end{figure}

The dijet mass resolution is studied by using the full Monte Carlo simulation. The full Monte Carlo simulation means that the proton-proton collisions are simulated starting from the collision itself to the signals in the detector. These signals are reconstructed according to the reconstruction scheme. In simulated events, identical kinematic selection is applied to generated and reconstructed jets, and then the mass of  two generated jets $(m^{gen})$ are compared to the mass of two reconstructed jets $(m^{reco})$. The quantity $m^{reco}/m^{gen}$ (mass response) is recorded, in the bins of the generated dijet mass (Figure \ref{MassResponse}, left-side plot). The resulting distribution is projected onto the y-axis for a certain range of $m^{gen}$ value then each projection is fitted with a Gaussian in the range of  $\pm 1.5\cdot RMS $ around the mean value. The extracted $\sigma\left(\dfrac{m^{reco}}{m^{gen}}\right)$ represents the relative mass resolution. At the end, the relative mass resolution as a function of $m^{gen}$ is parametrized with a smooth continuous function of the following form:
\begin{equation}
\dfrac{\sigma(M^{gen})}{M^{gen}}=A+\dfrac{B}{\left(M^{gen}\right)^{C}}
\end{equation}

\begin{figure}[h]
   \begin{center}
     \includegraphics[width=0.40\textwidth]{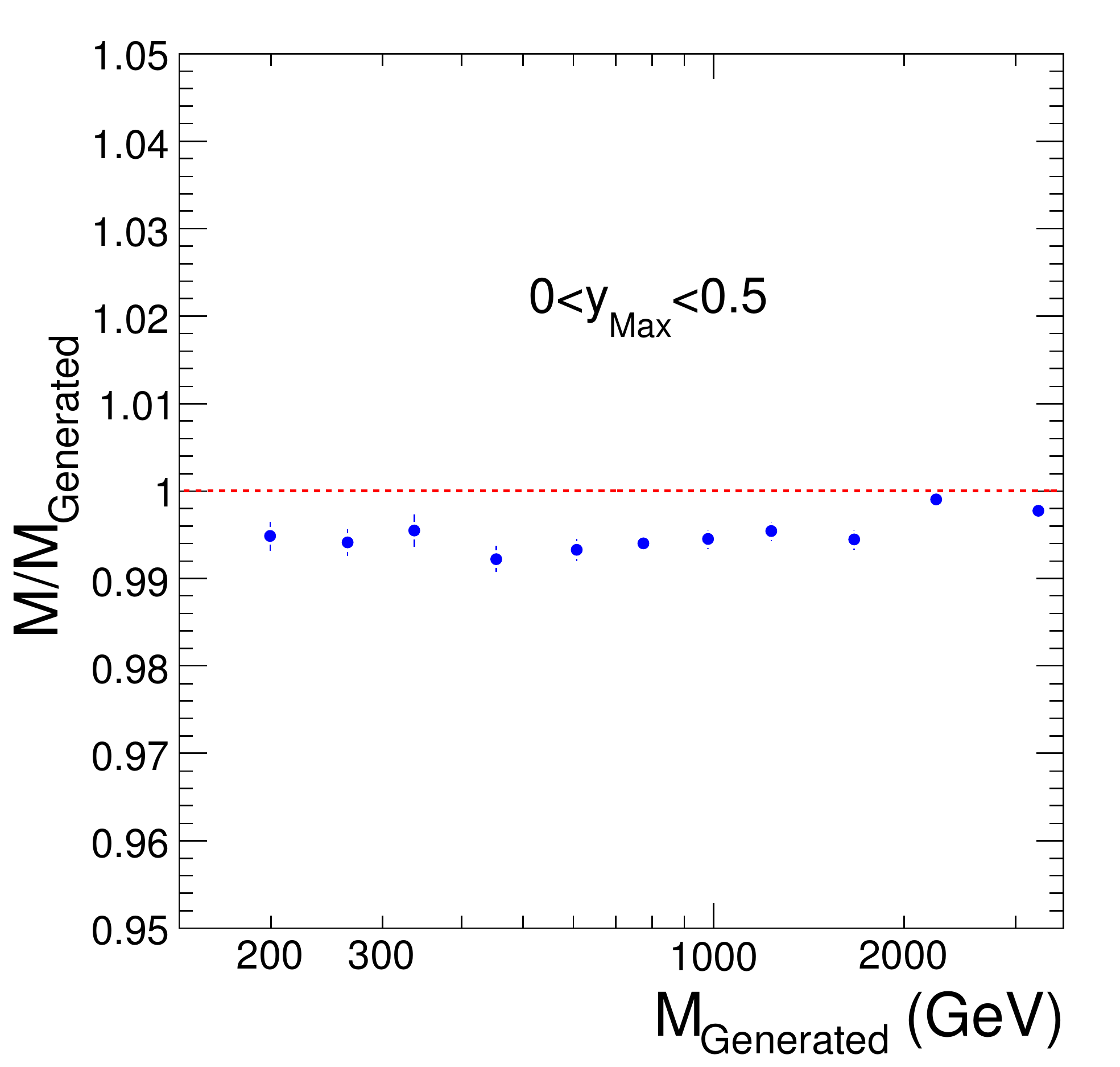}
     \includegraphics[width=0.40\textwidth]{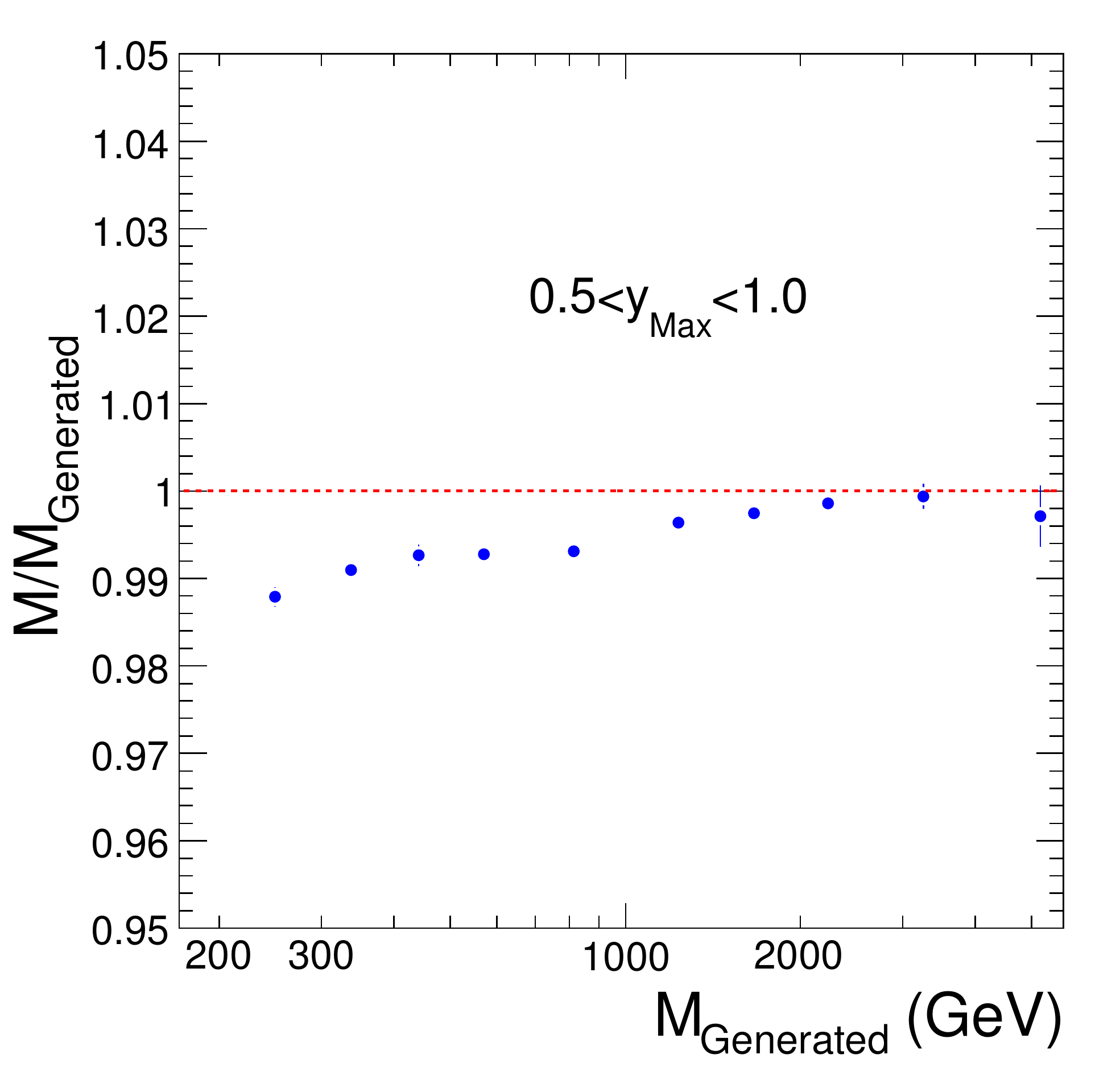}
     \includegraphics[width=0.40\textwidth]{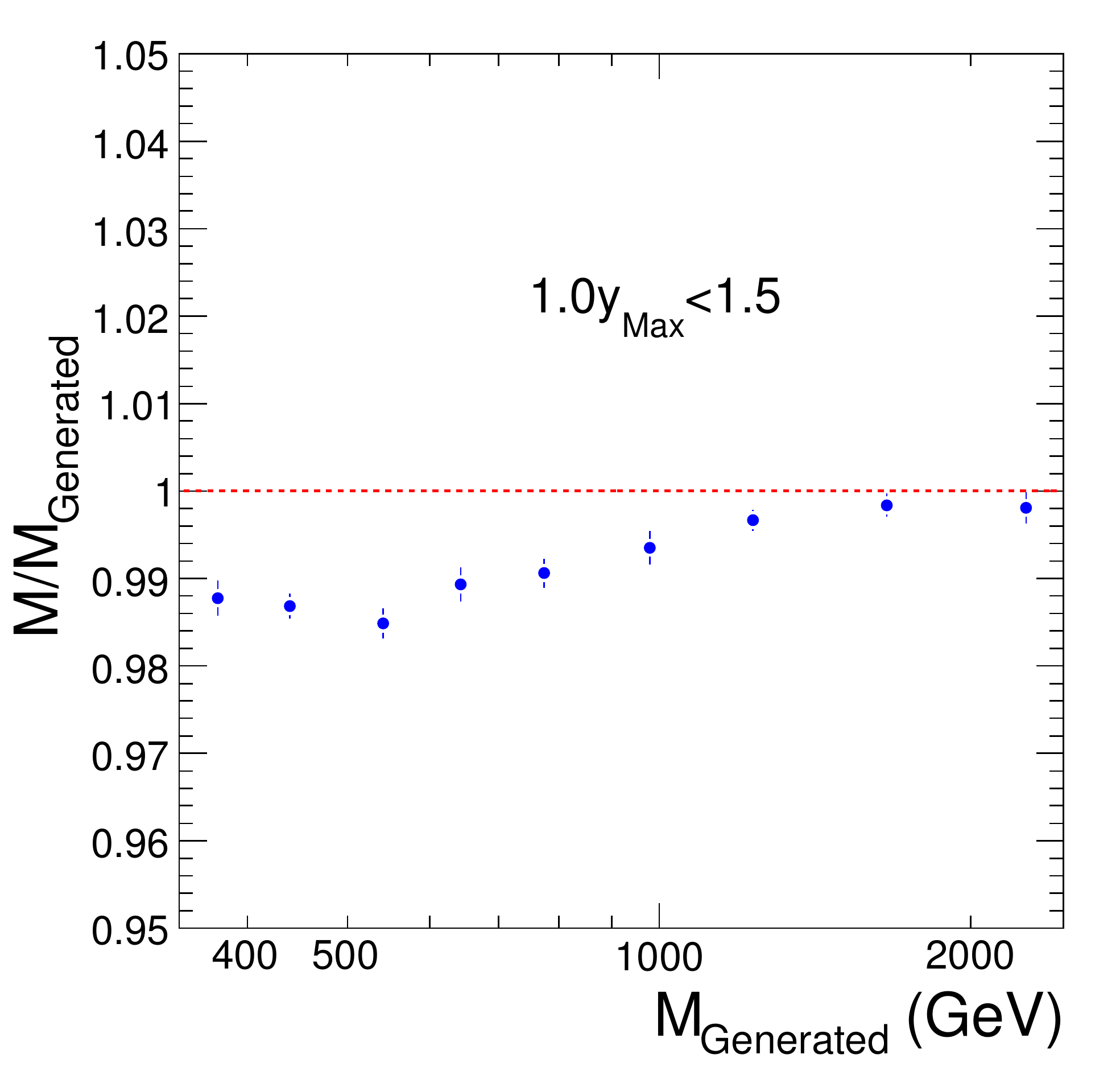}
     \includegraphics[width=0.40\textwidth]{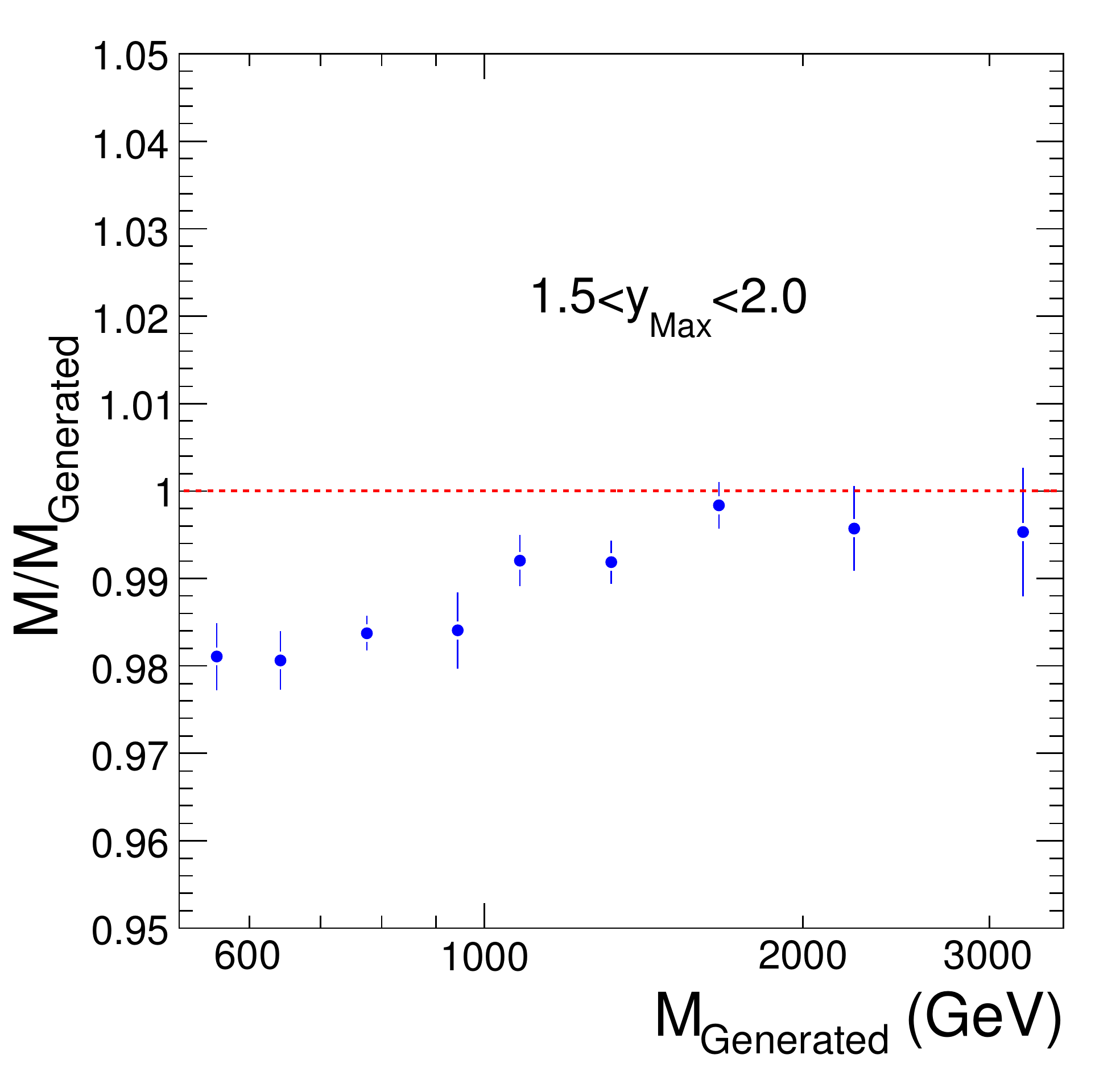}
     \includegraphics[width=0.40\textwidth]{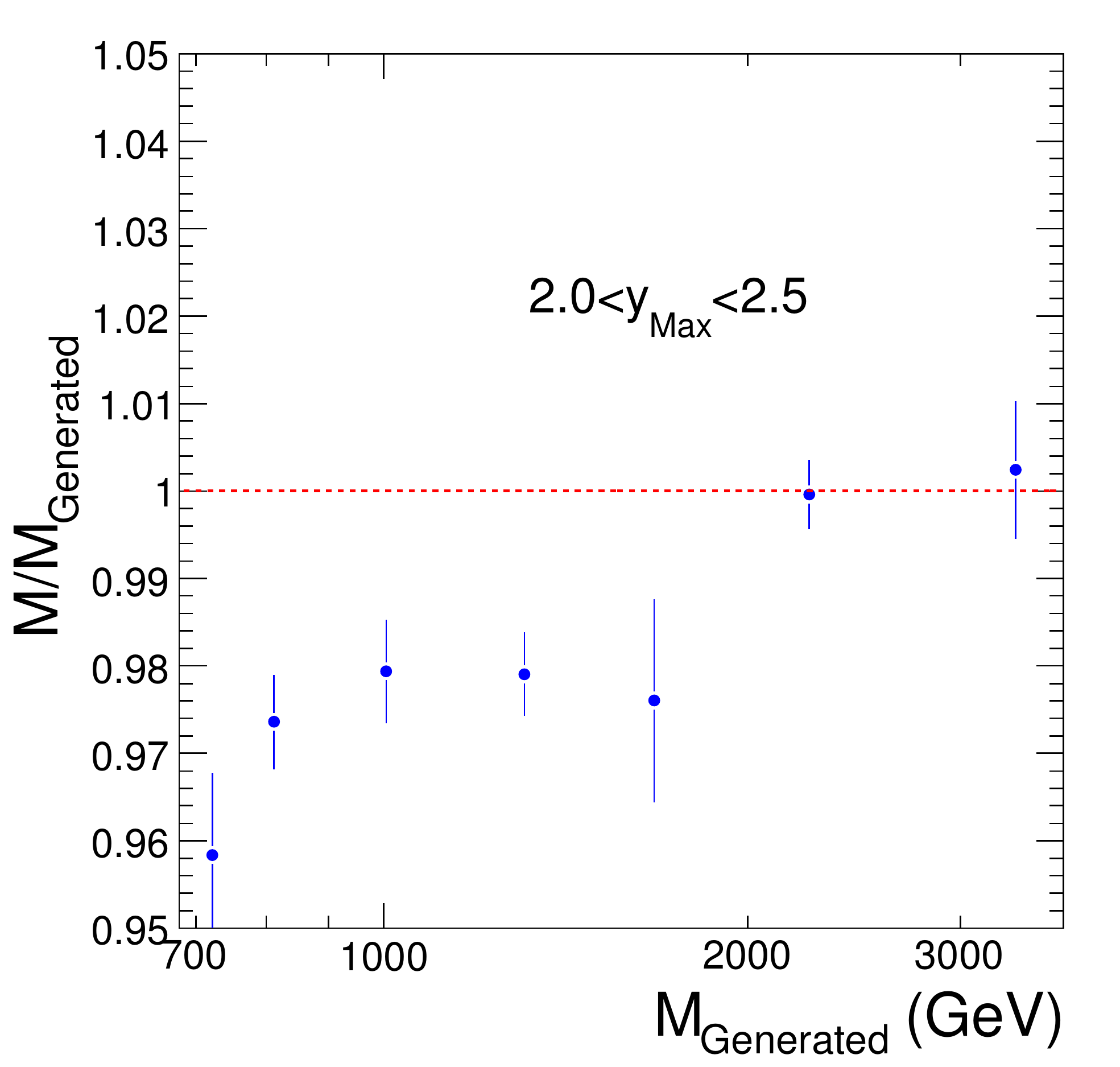} 
     \capspace
     \caption{Average ratio of the reconstructed dijet mass over the generated mass in all \ymax bins.}
     \label{MassClosure}
   \end{center}
 \end{figure}

\begin{figure}[h]
   \begin{center}
     \includegraphics[width=0.40\textwidth]{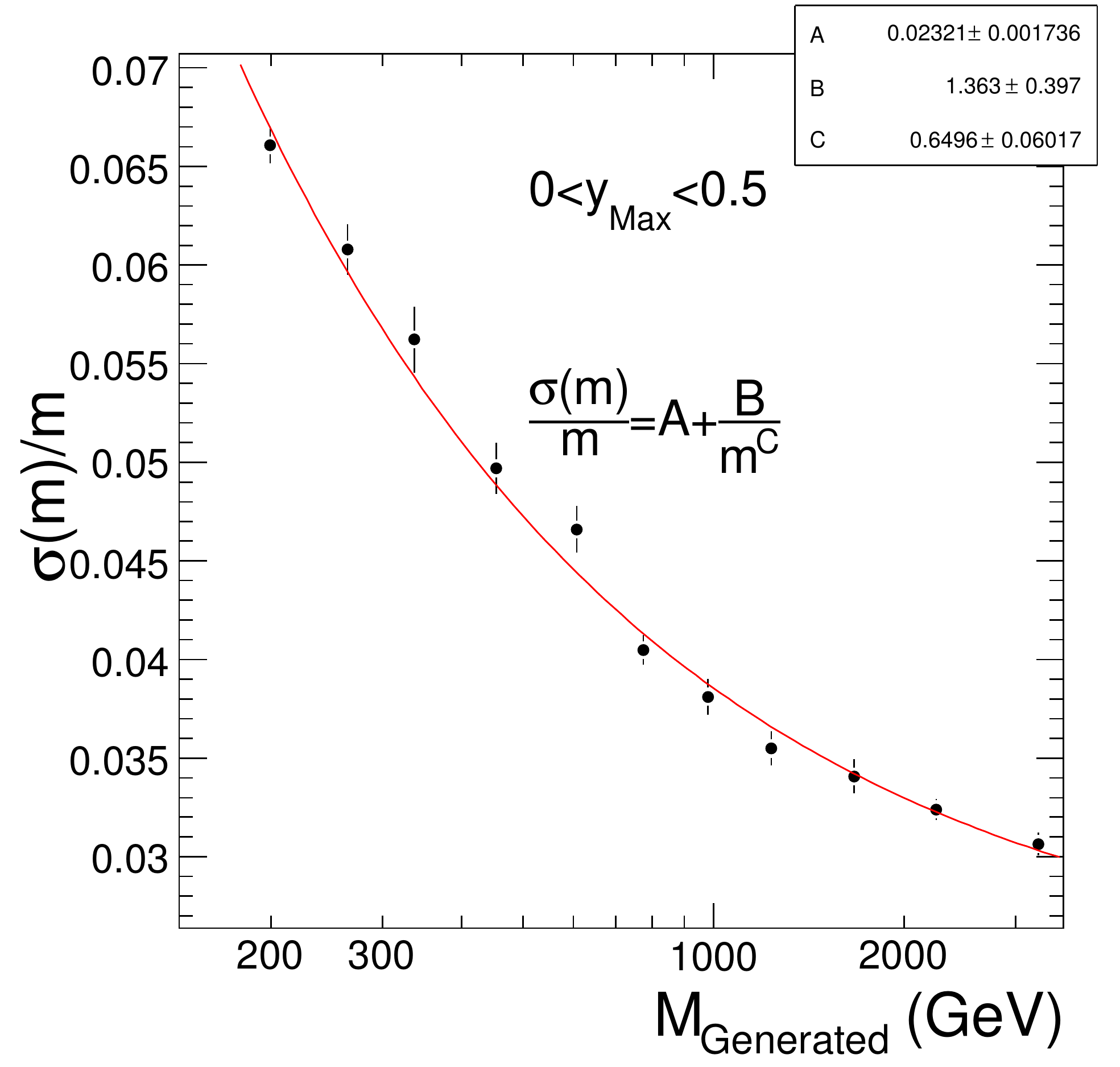}
     \includegraphics[width=0.40\textwidth]{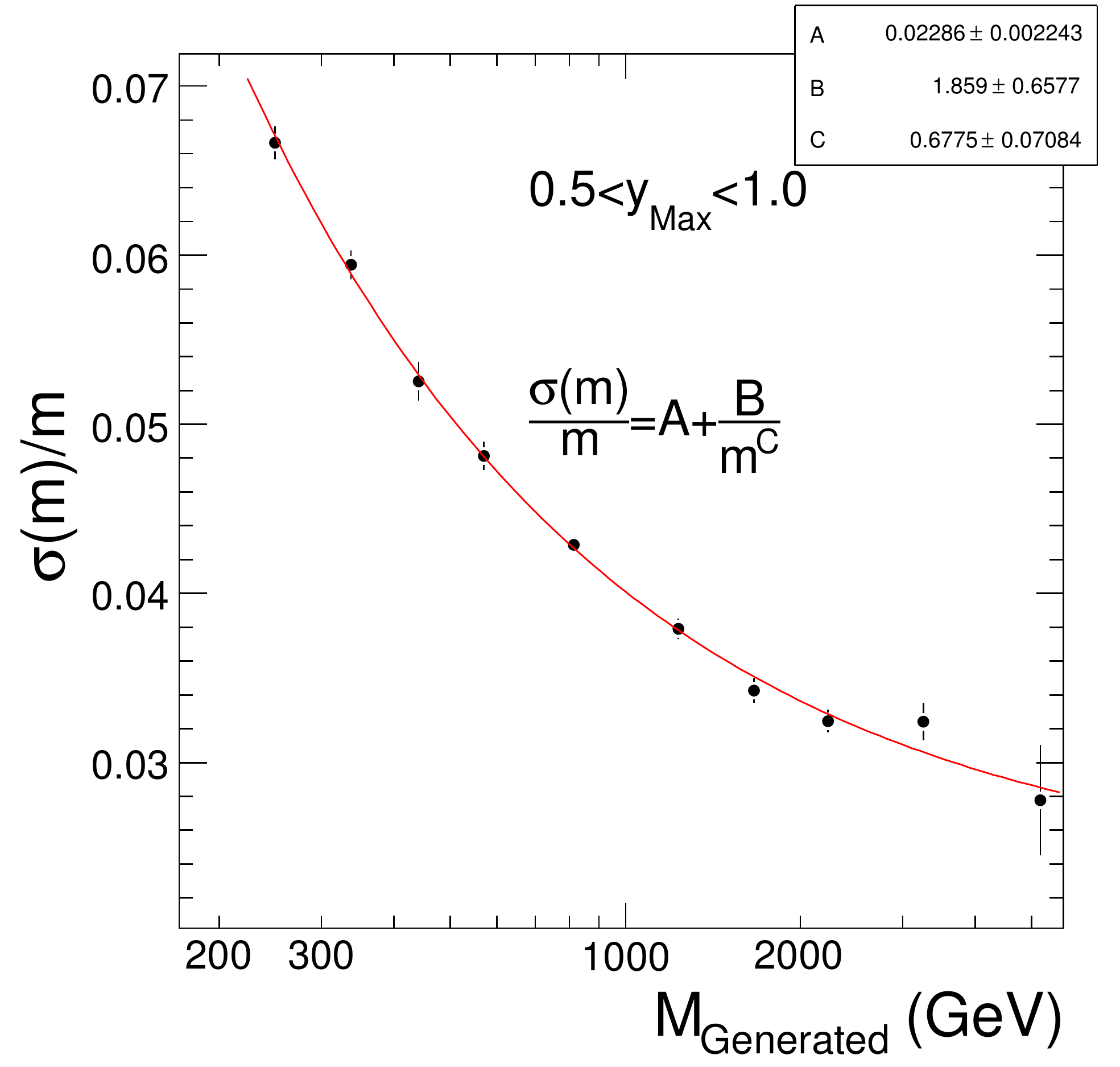}
     \includegraphics[width=0.40\textwidth]{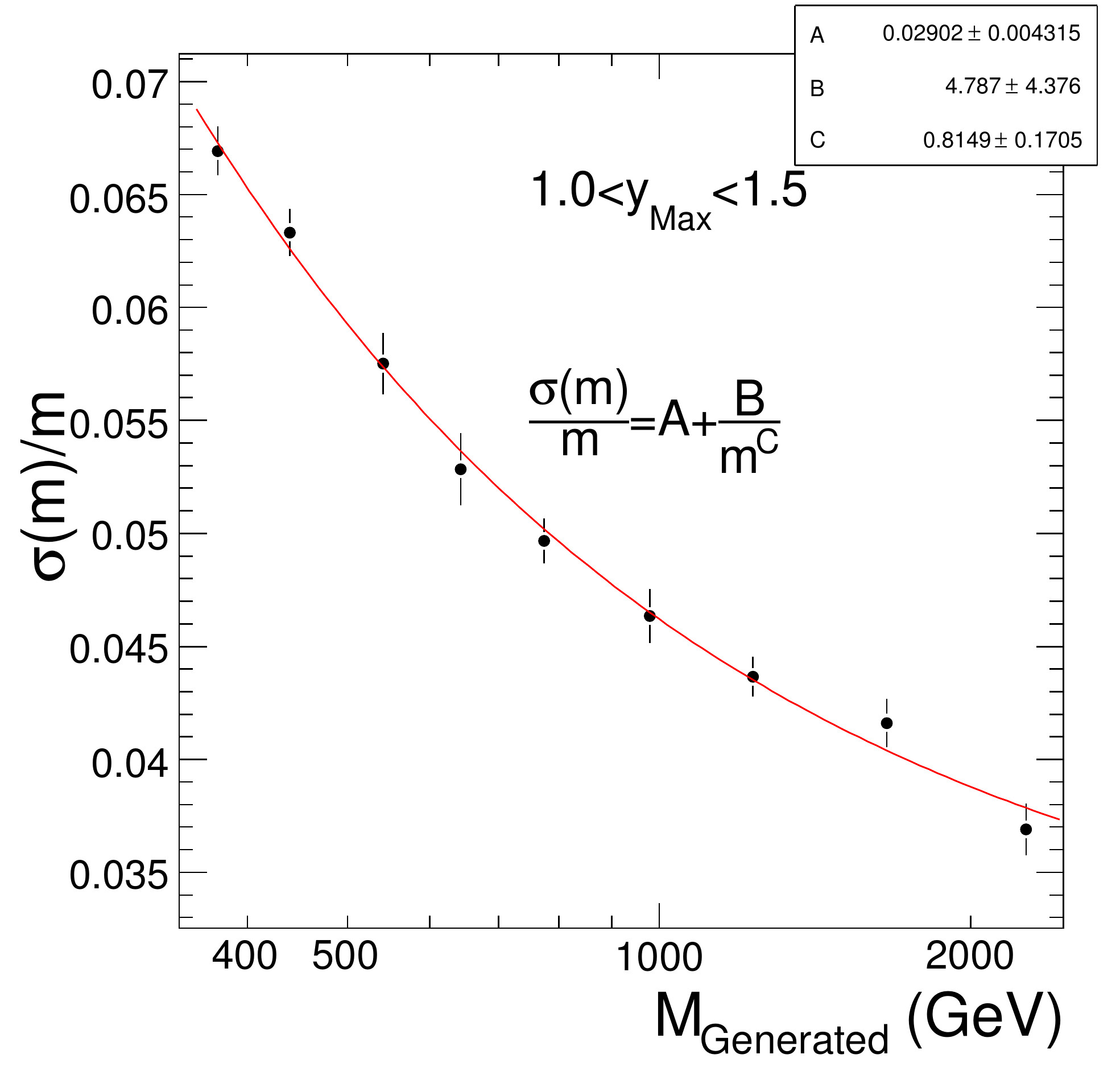}
     \includegraphics[width=0.40\textwidth]{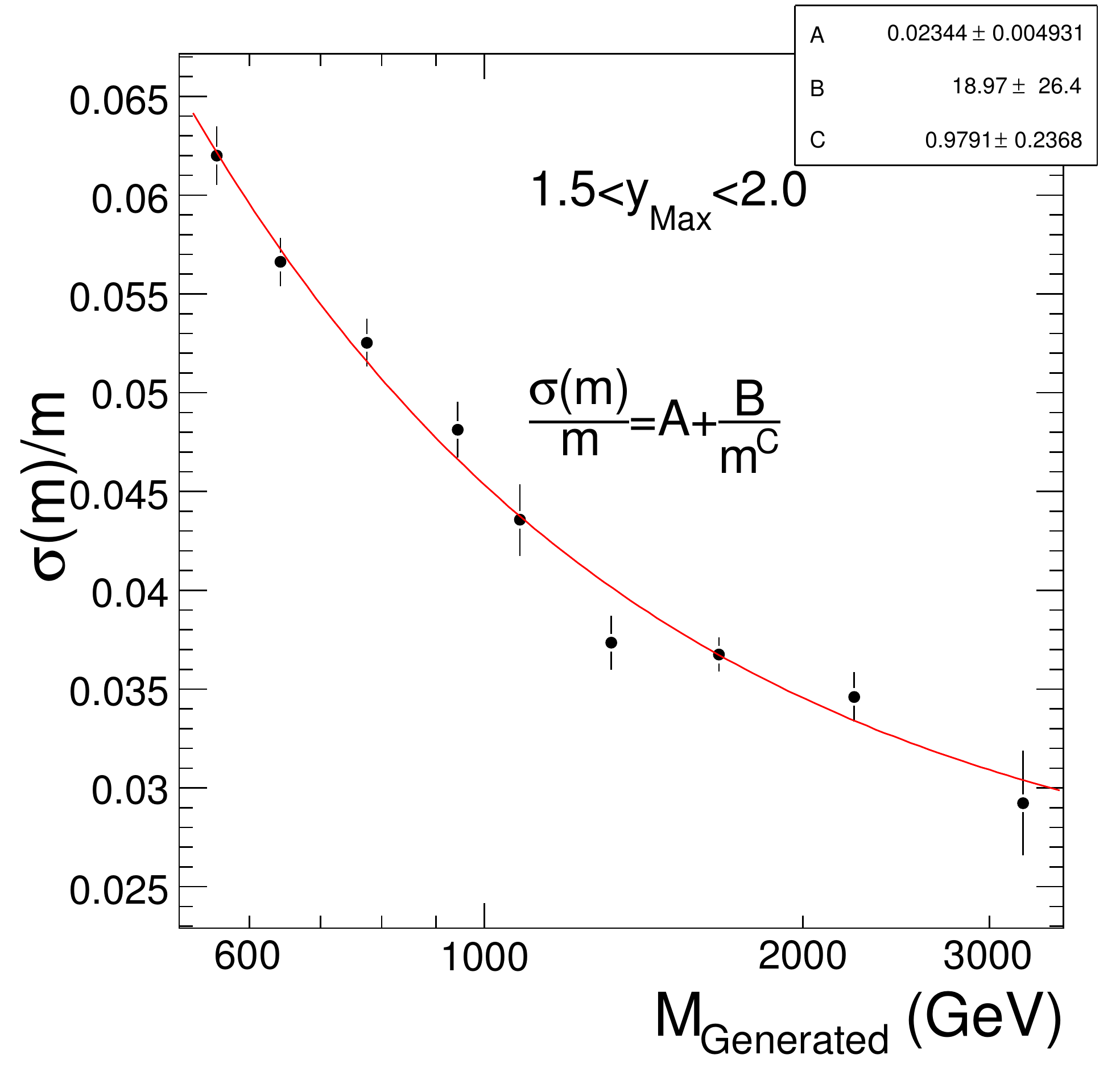}
     \includegraphics[width=0.40\textwidth]{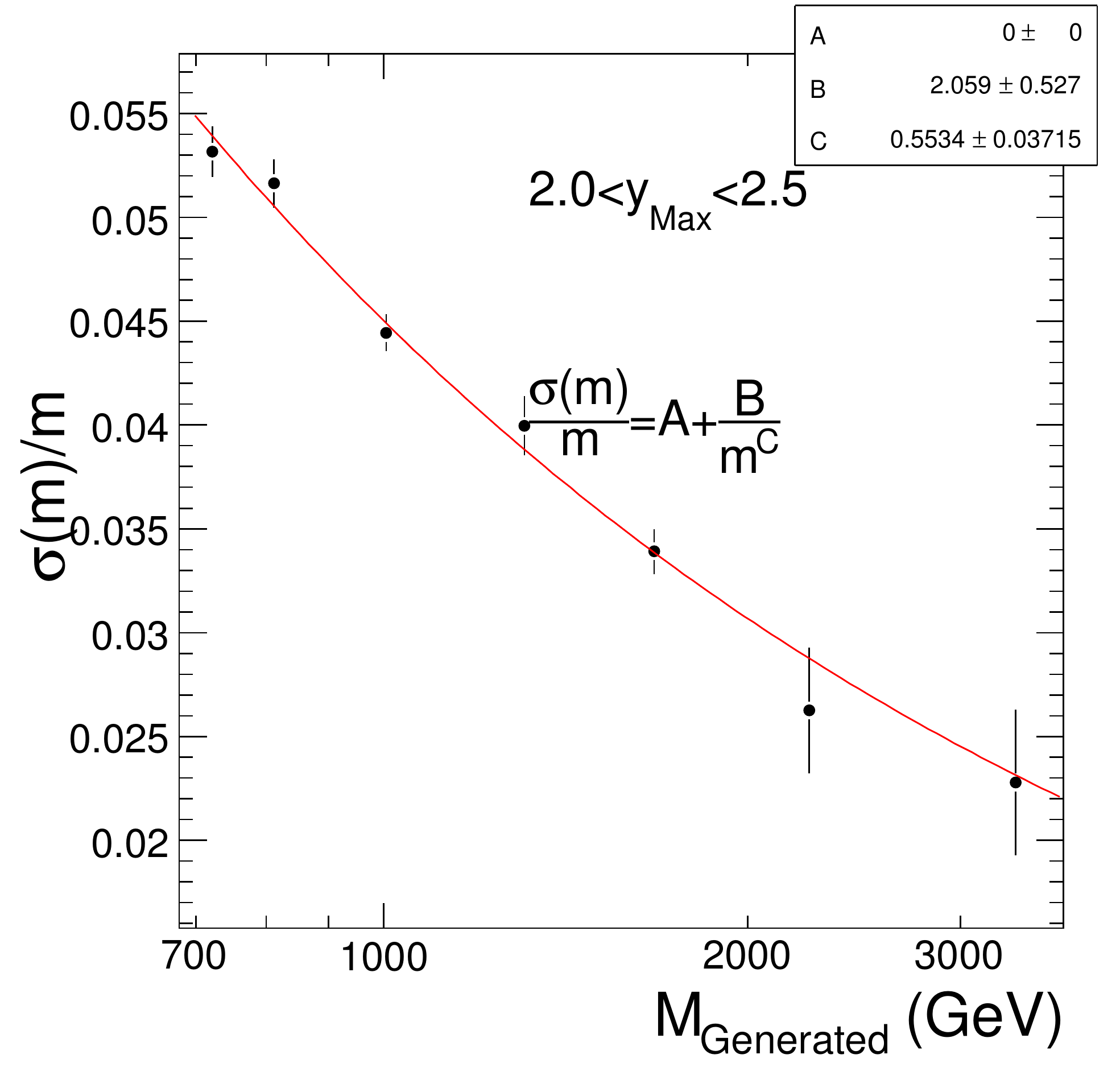}
     \capspace
     \caption{Relative dijet mass resolution, as a function of the generated mass in all \ymax bins.}
     \label{MassResolution}
   \end{center}
 \end{figure}

Figure \ref{MassClosure} shows the quantity $\langle M/M^{gen} \rangle$ as a function of $M^{gen}$ demonstrating that the reconstructed dijet mass agrees with the generated mass within 3\%. Figure \ref{MassResolution} shows the relative dijet mass resolution as a function of $M^{gen}$, fitted with the continuous function described above.
\clearpage
\section{Construction of the Dijet Mass Spectrum}
\begin{figure}[ht]
  \centering
  \includegraphics[width=0.32\textwidth]{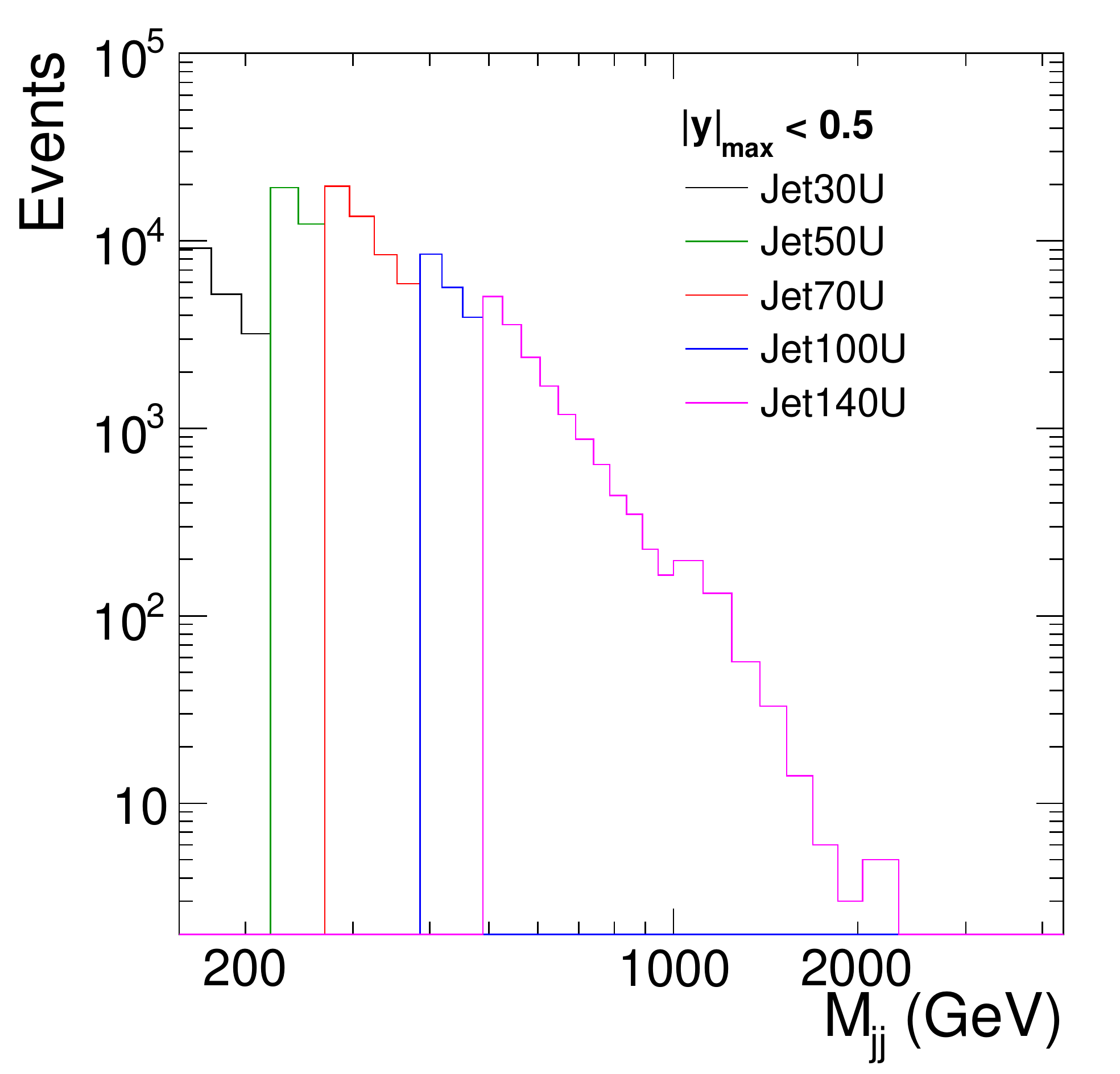}
  \includegraphics[width=0.32\textwidth]{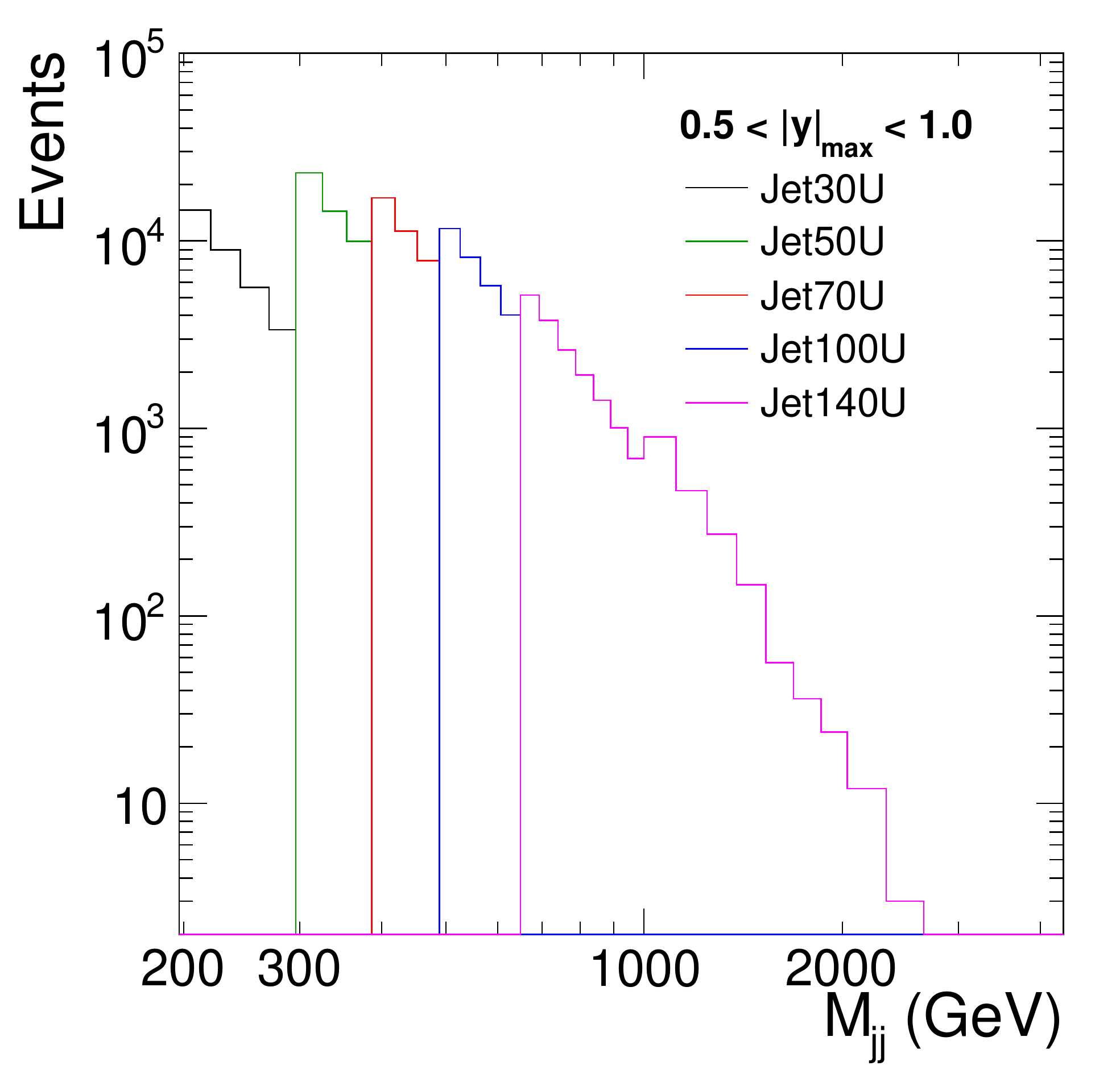}
  \includegraphics[width=0.32\textwidth]{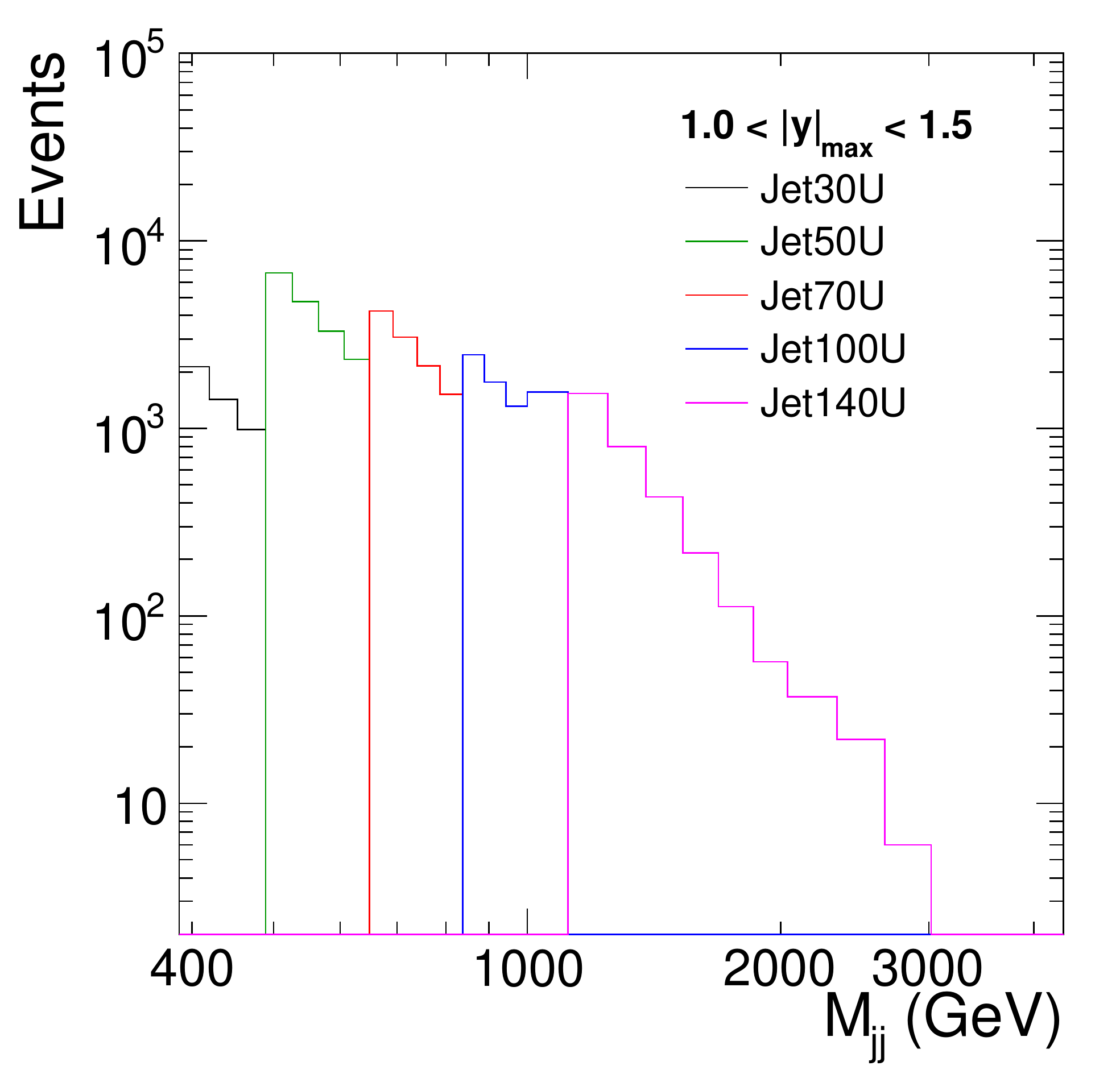}
  \includegraphics[width=0.32\textwidth]{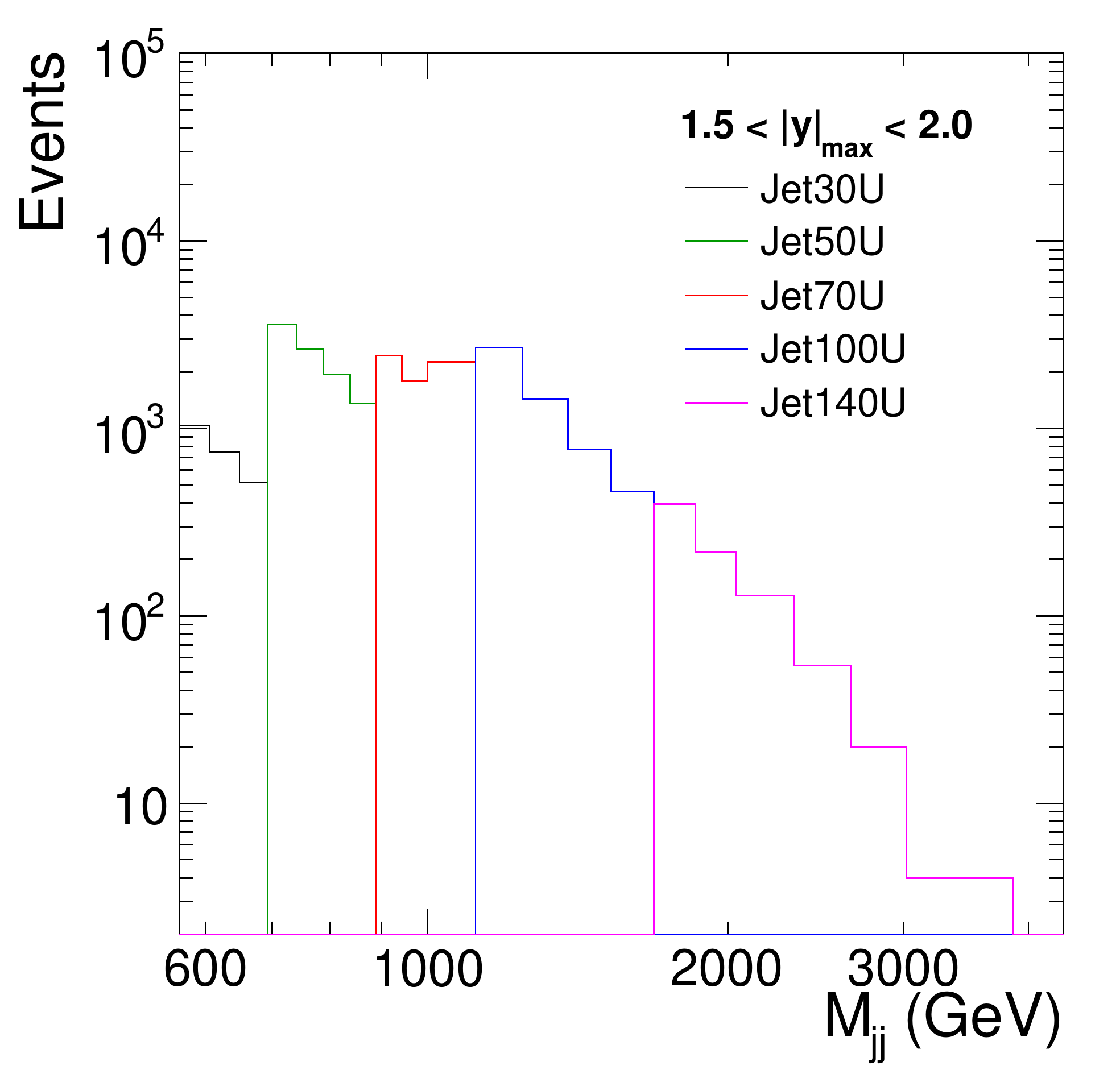}
  \includegraphics[width=0.32\textwidth]{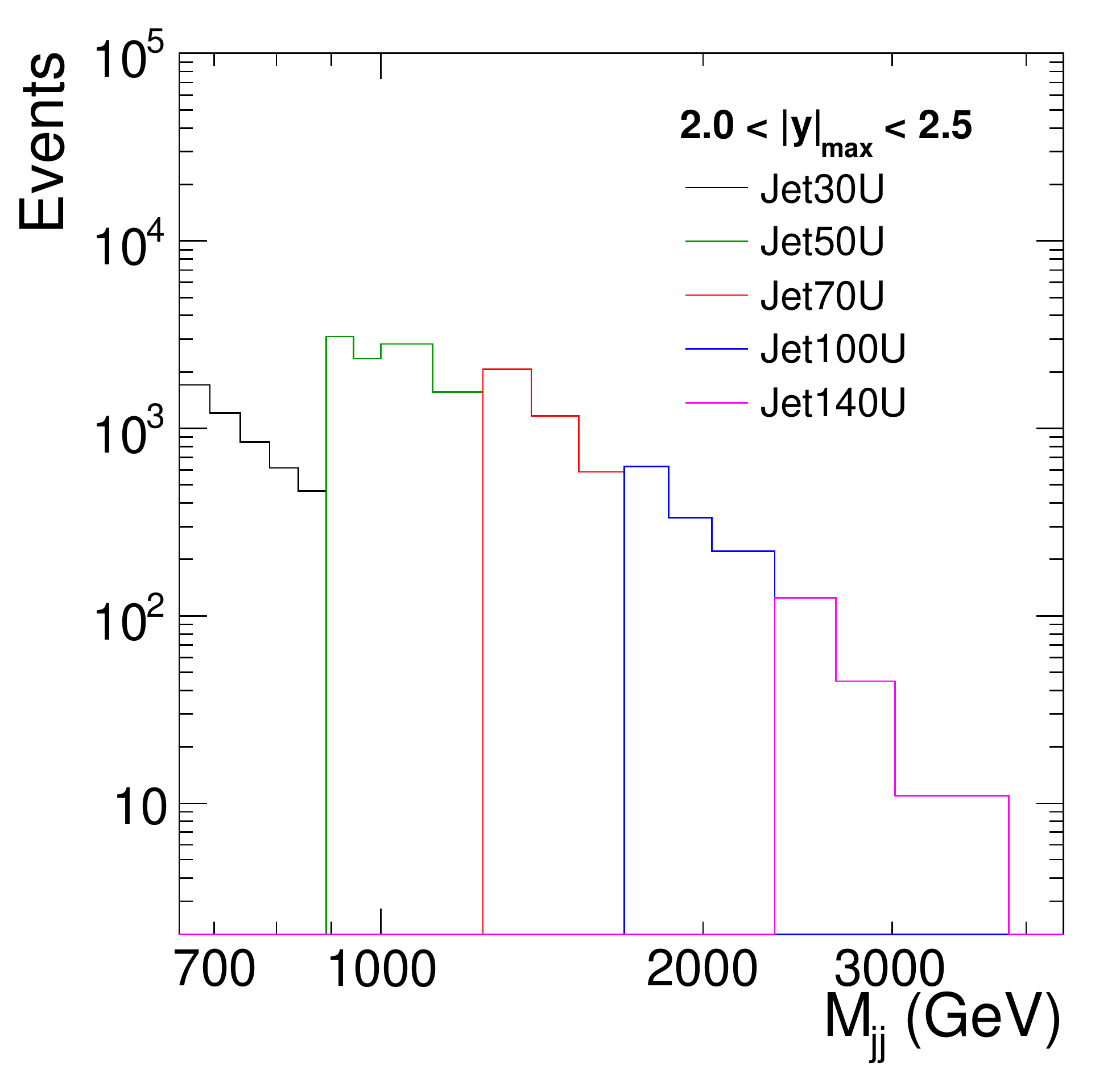}
  \capspace
  \caption{Event yield of the different samples. In each mass bin, only the contributing sample is shown.} 
  \label{fig:Yield}
\end{figure}

\begin{figure}[ht]
  \centering
  \includegraphics[width=0.8\textwidth]{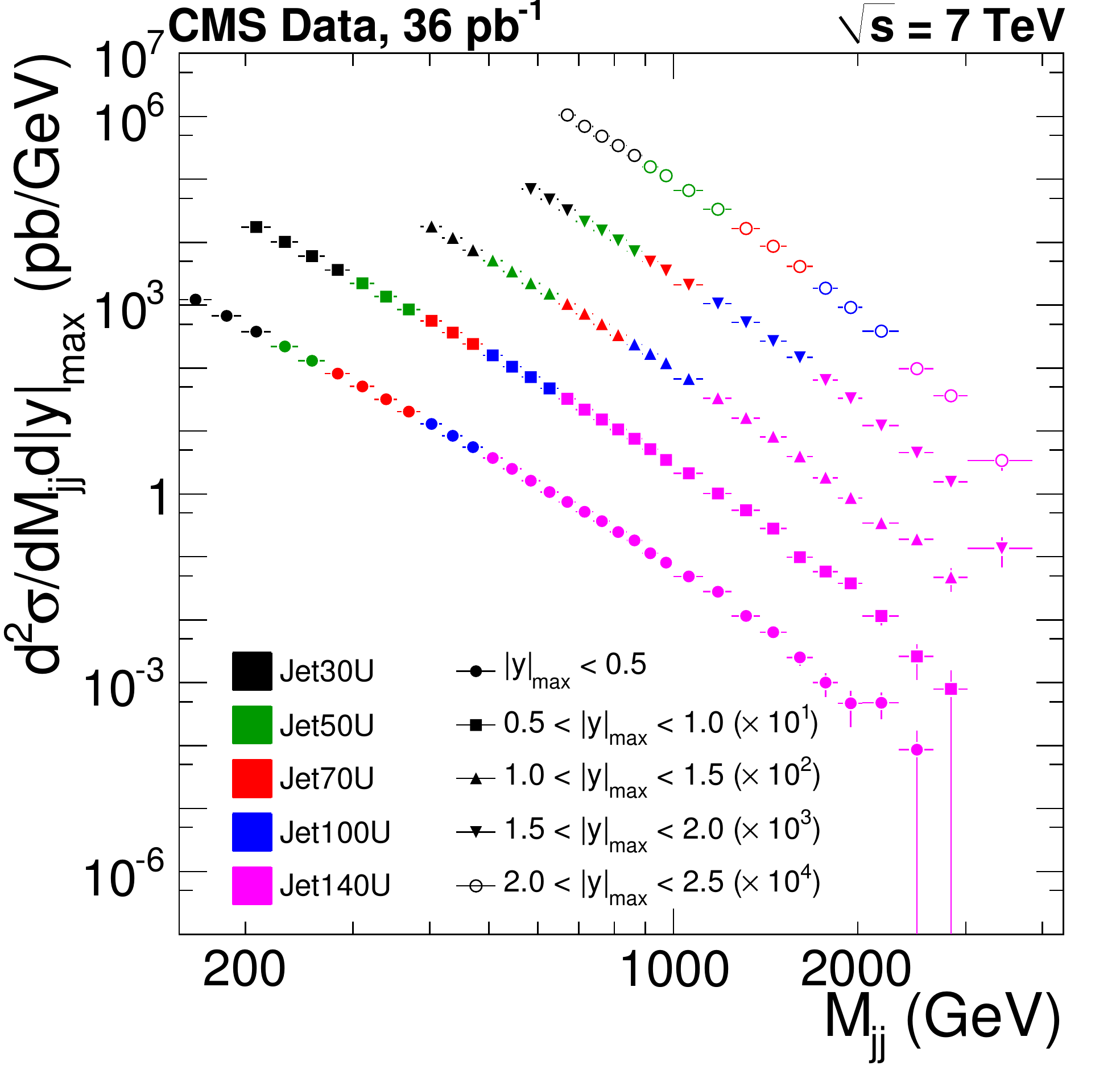}
  \capspace
  \caption{Dijet mass spectrum formed by the combination of the different samples. Each rapidity bin is further scaled by the number given in parentheses.} 
  \label{fig:Spectrum}
\end{figure}

Finally, the spectrum of the dijet invariant mass in each rapidity bin is constructed by all data samples by
combining them according to the trigger efficiencies. In order to use the maximum available number of events, each mass bin is populated by one and only one sample which is at least 99\% efficient and has the highest effective luminosity. The size of the dijet mass bins is approximately equal or larger than the width of the mass resolution at the bin center. Figure \ref{fig:Yield} shows how the data samples from different triggers are combined. At the end, the prescaled
samples are scaled up by a number so that they match the rate of the un-prescaled sample. The resulting dijet mass spectra can be shown in Figure \ref{fig:Spectrum}.
\clearpage

\clearpage
\chapter{SYSTEMATIC UNCERTAINTIES}

\section{Experimental Uncertainties}
\begin{figure}[ht]
  \centering
  \includegraphics[width=0.48\textwidth]{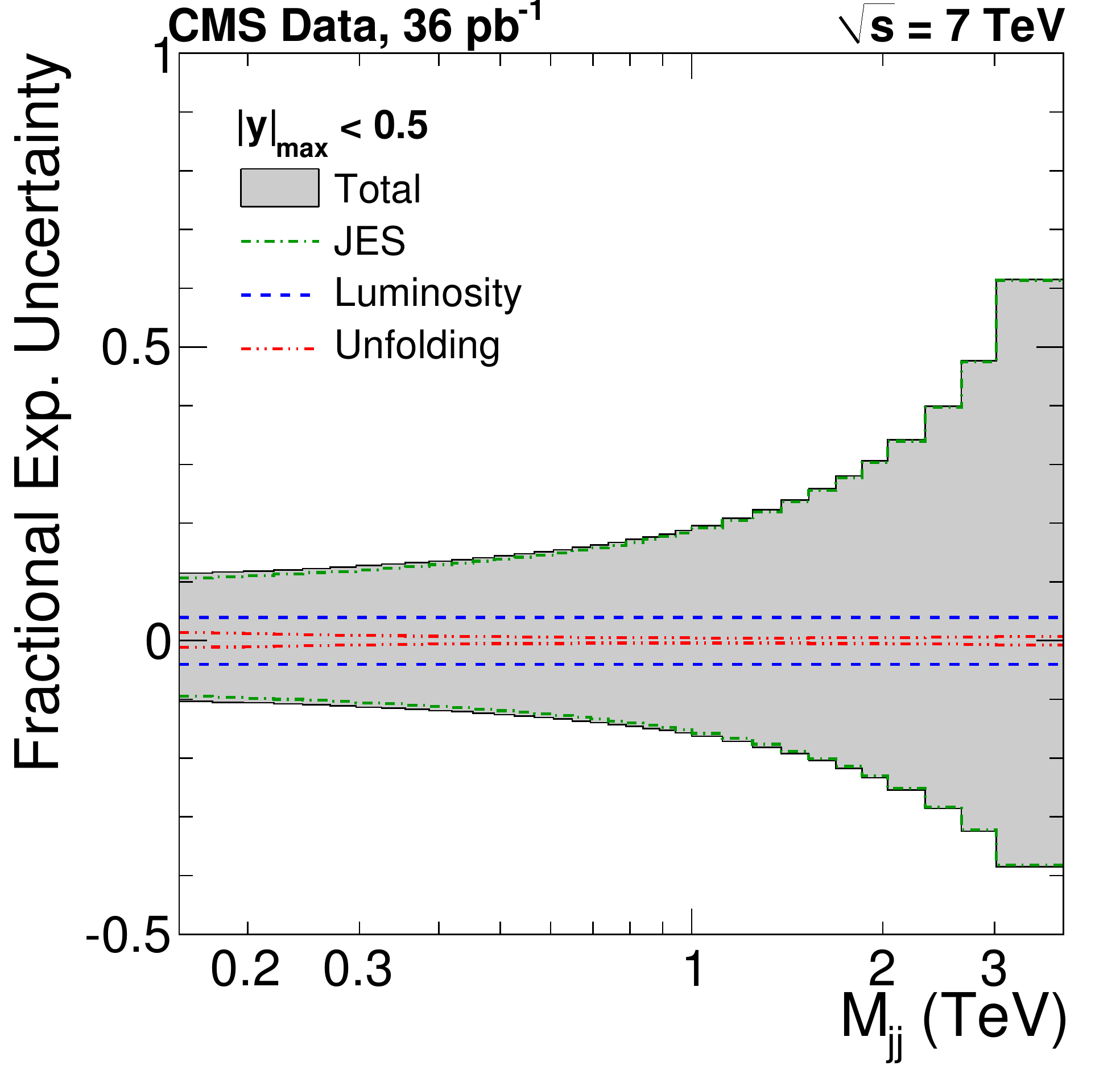}
  \includegraphics[width=0.48\textwidth]{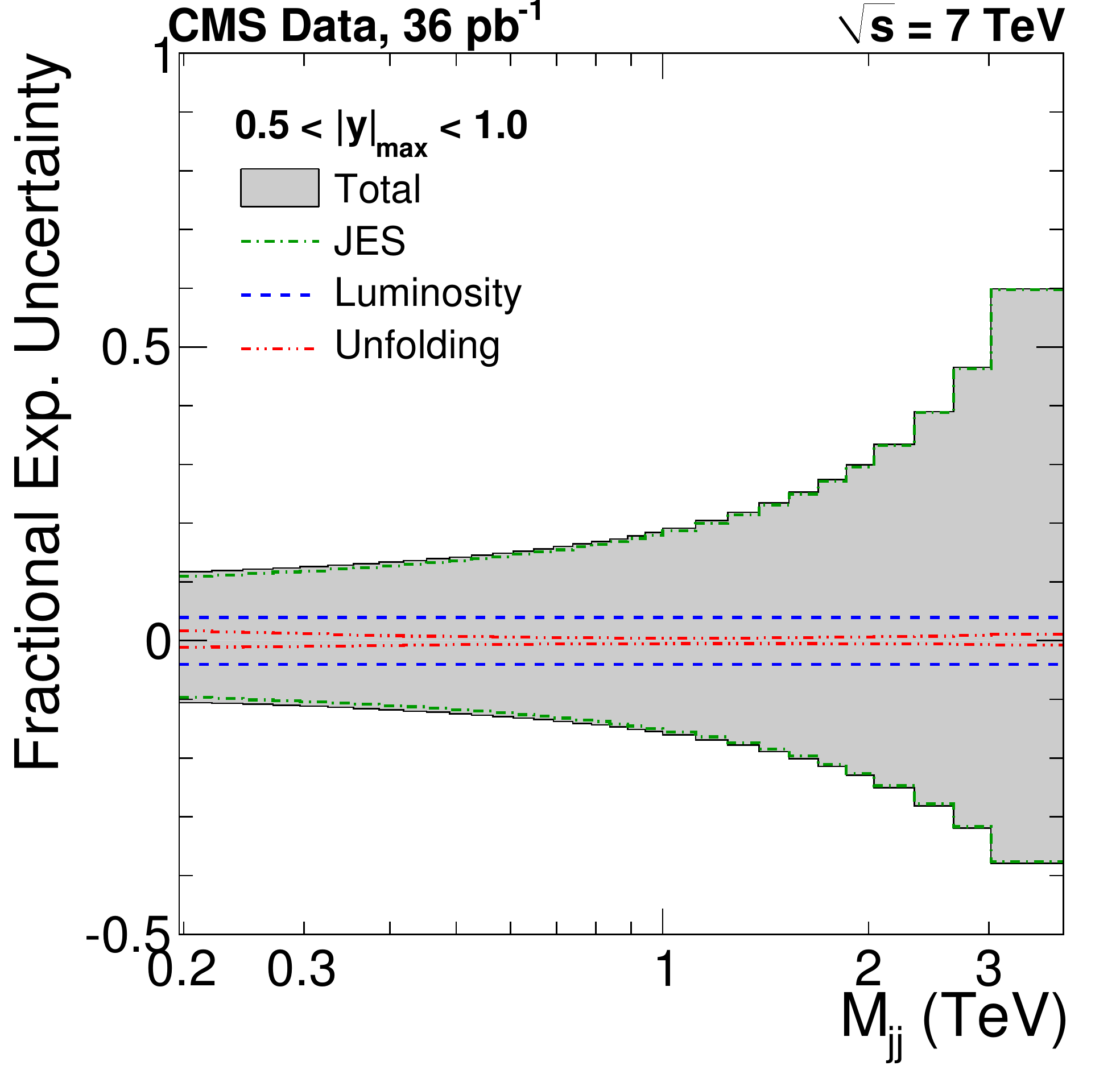}
  \includegraphics[width=0.48\textwidth]{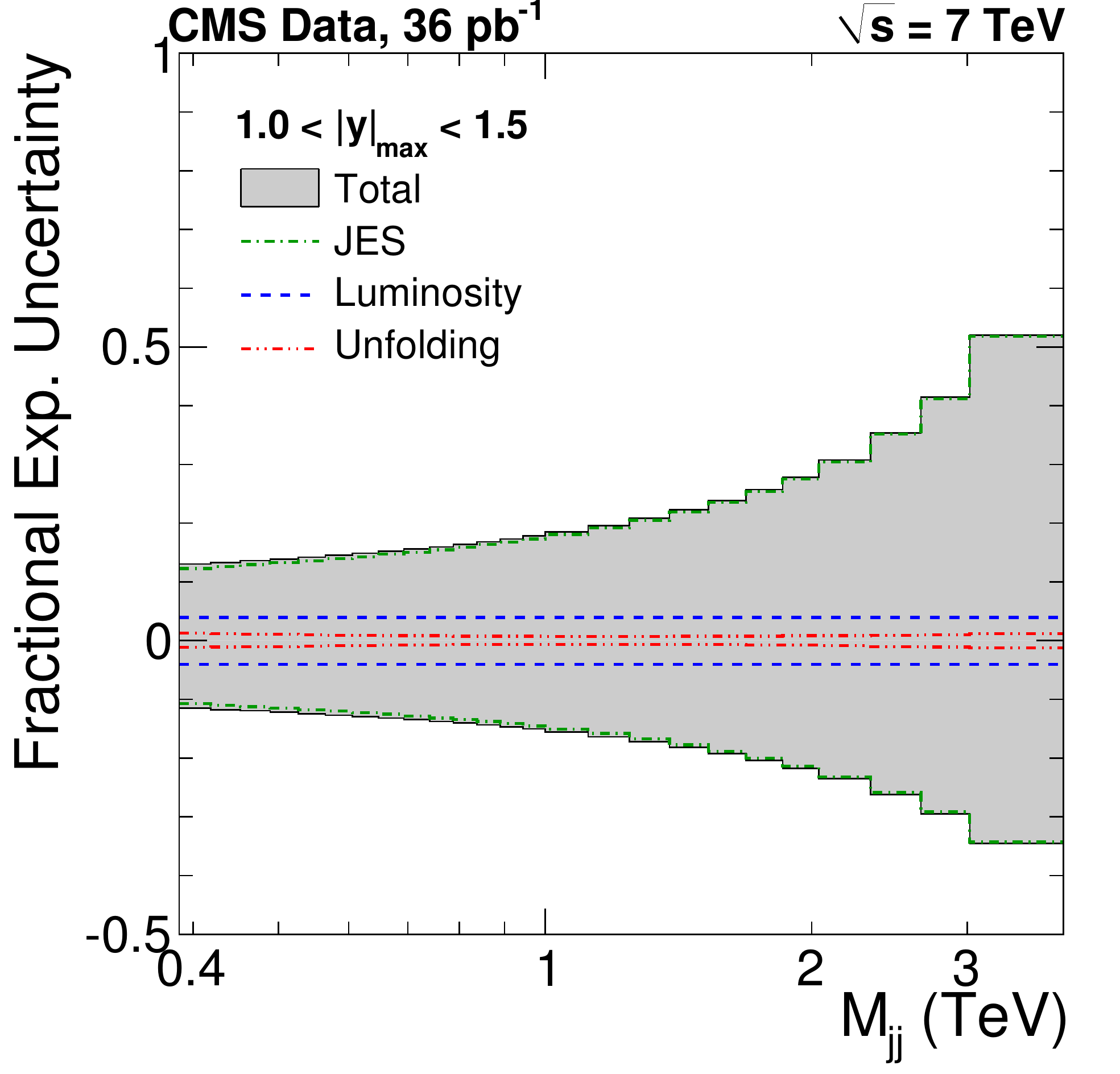}
  \includegraphics[width=0.48\textwidth]{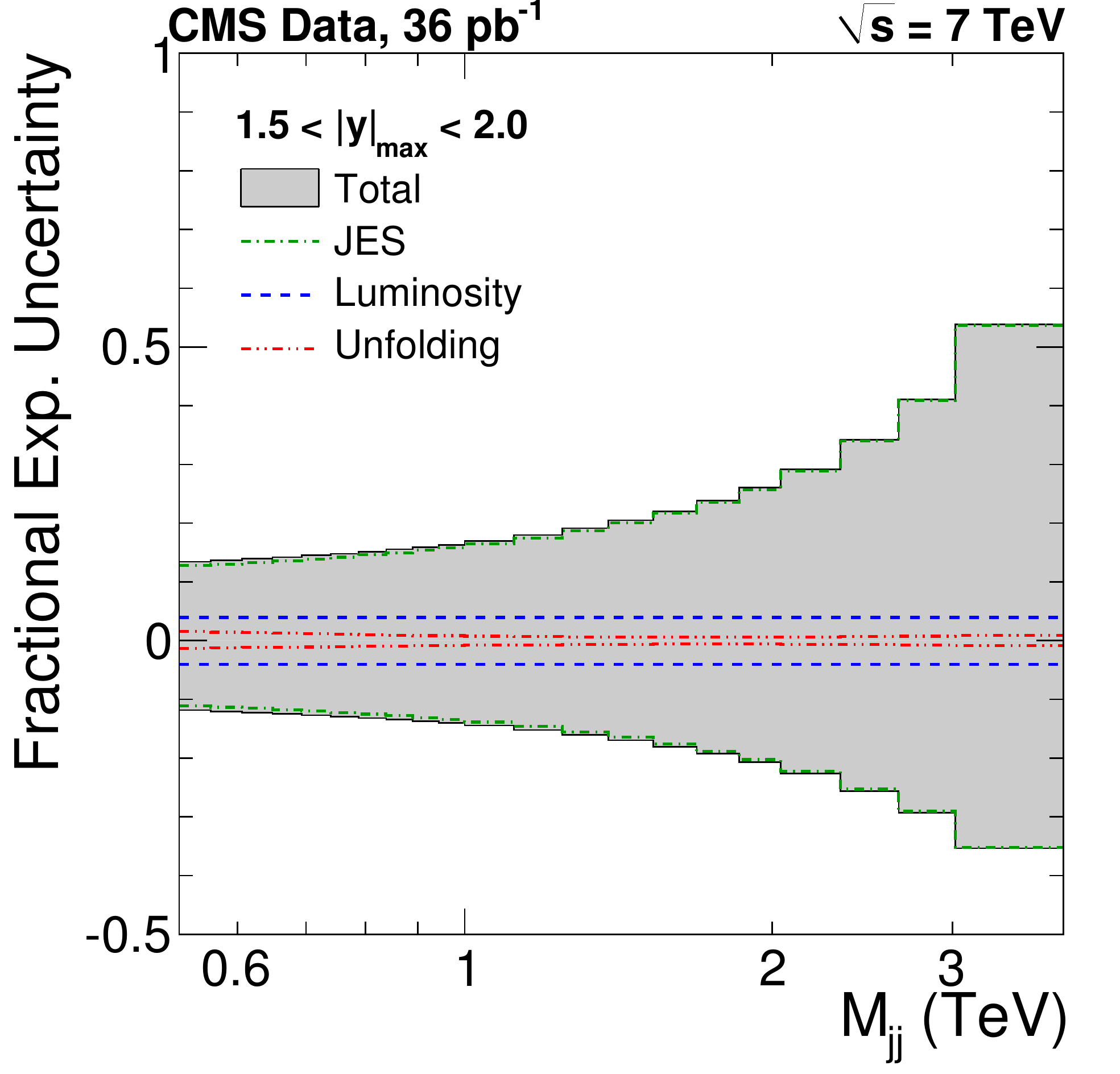}
  \includegraphics[width=0.48\textwidth]{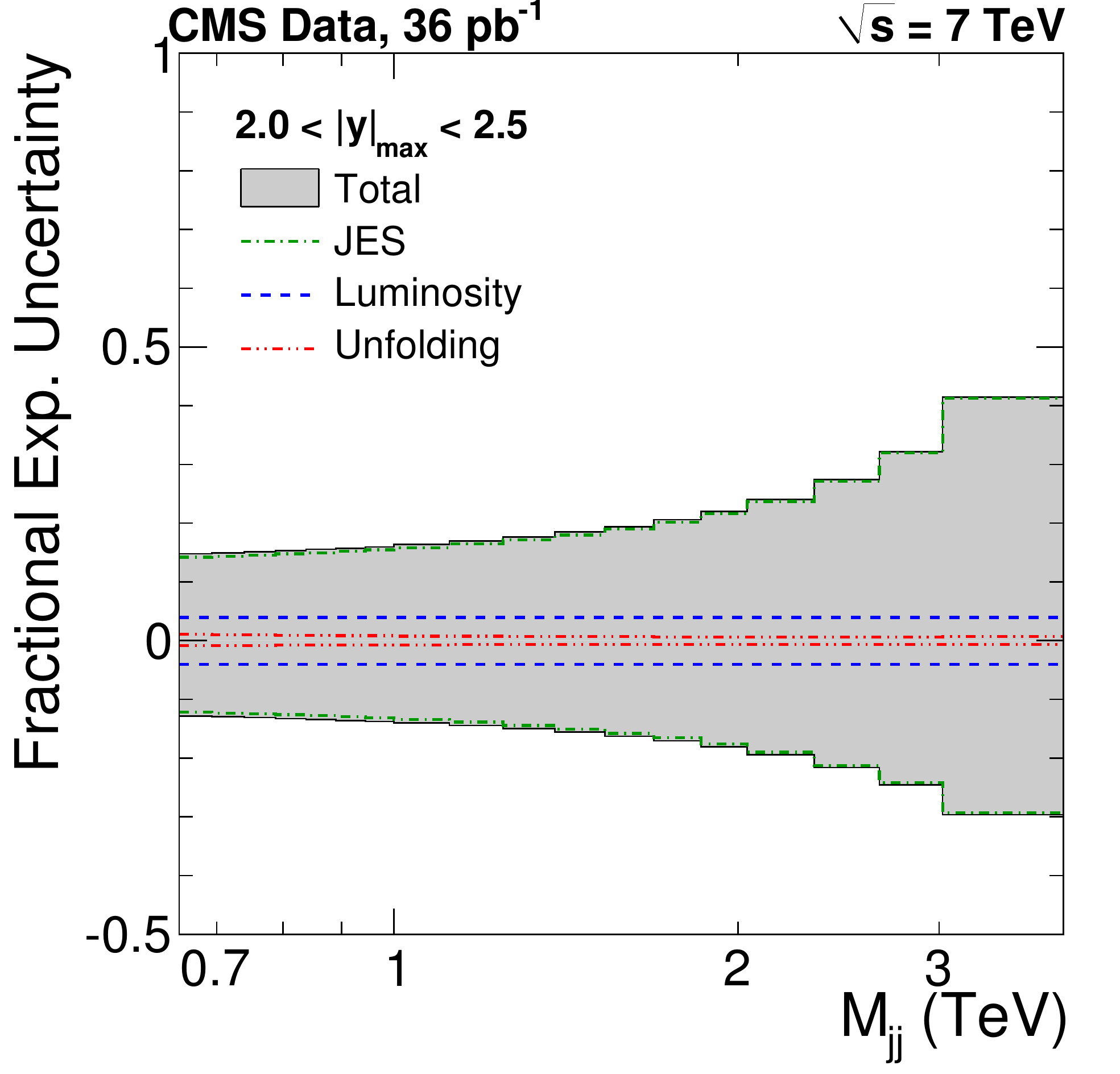}
  \capspace
  \caption{ Summary of the experimental systematic uncertainties: jet energy scale (green dashdotted line), luminosity (blue dashed line), unsmearing (red dash-double dotted line) and their sum in quadrature (filled).}
  \label{fig_ExpUnc}
\end{figure}

The experimental uncertainties are the uncertainties related with every step of the measurement process starting from the reconstruction to the spectrum construction. There are three main sources of experimental uncertainties; Jet Energy Scale uncertainty (JES Uncertainty), luminosity uncertainty and the uncertainty on the unsmearing corrections. The total experimental uncertainty is obtained by quadratic sum of these three independent uncertainties:
\begin{equation}\label{ExpUncQuad}
\sigma_{Experimental}^{2}=\sigma_{JES}^{2}+\sigma_{luminosity}^{2}+\sigma_{unsmearing}^{2}
\end{equation}
Figure \ref{fig_ExpUnc} shows all the independent experimental uncertainties and the total experimental uncertainty in all $|$y$|_{max}$ bins. The typical range is between $\sim$ 15\% at low mass values and $\sim$ 60\% at high mass values, approximately the same in all rapidity bins.
All individual components will be discussed below.
\subsection{Jet Energy Scale (JES) Uncertainty}
\begin{figure}[ht]
  \centering
  \includegraphics[width=0.48\textwidth]{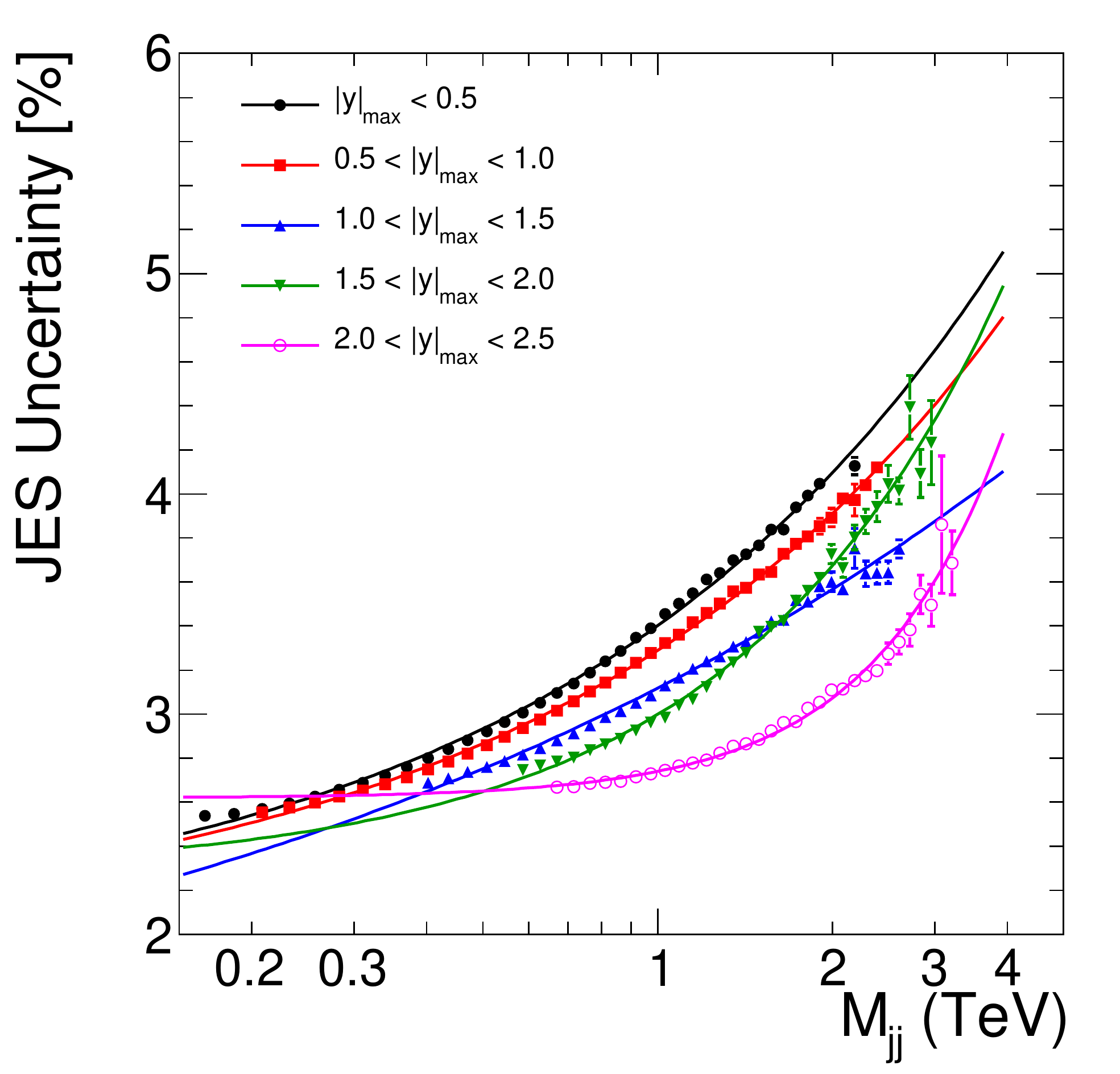}
  \includegraphics[width=0.48\textwidth]{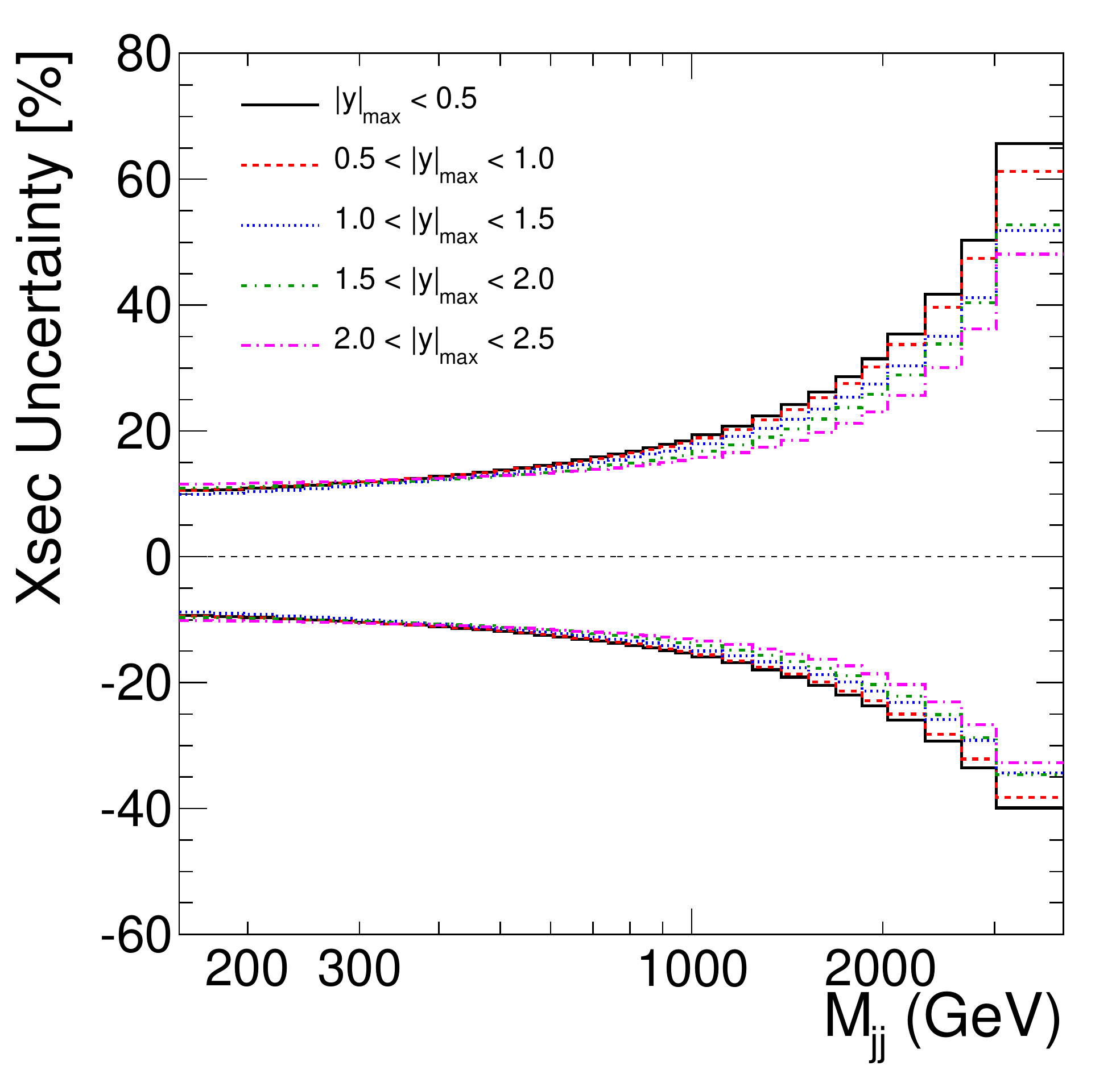}
  \caption{ Left: Average dijet mass scale uncertainty in all \ymax bins. Right: Cross section uncertainty due to the mass scale uncertainty.}
  \label{fig_AvgMassUnc}
\end{figure}

The JES uncertainty is the most dominant uncertainty source in all of three components of the experimental uncertainty. Due to the very steep fall of the dijet mass spectrum, a small uncertainty on the mass scale is translated into the cross section by a multiple of $\sim$5-7. The dijet mass spectrum can be described by a continuous function of the following form:
\begin{equation}\label{Eq.MassFit}
f(m)=A\cdot\left(m/\sqrt{s}\right)^{-a}\cdot\left(1-m/\sqrt{s}\right)^b
\end{equation}
In Equation \ref{Eq.MassFit}, a change of variable $m/\sqrt{s}=x$ yields the Equation \ref{Eq.MassFit_cv}.
\begin{equation}\label{Eq.MassFit_cv}
\tilde{f}(x)=A\cdot x^{-a}\cdot (1-x)^b
\end{equation}
Then the relative uncertainty can be calculated by a differentiation with respect to $x$
\begin{equation}\label{Eq.MassRelUnc}
\frac{\delta(\tilde{f}(x))}{\tilde{f}(x)}=\left[-a-b\cdot x \cdot (1-x)^{-1} \right] \frac{\delta x}{x} 
\end{equation}
Since the change of variable $m/\sqrt{s}=x$ is only a proportion, the relative uncertainty of mass scale $\delta m/m$ is equal to the $\delta x/x$. Hence, it can easily be seen that the relative uncertainty on the mass scale is pronounced in the cross section by a factor of $\pm \left[a+b\cdot x \cdot (1-x)^{-1} \right]$. However, this is a back of the envelope calculation, and a more rigorous study is needed to estimate the effect of JES uncertainty. On the other side, the JES uncertainty is given in terms of the \pt and the $\eta$ of a given jet. The uncertainty on the JES cannot be analytically mapped to the mass scale since the mass of the dijet system also depends on the polar and the azimuthal separation of jets. In other words, there might be many pairs of jets with different \pt and $\eta$ which give the same dijet mass value. Therefore, all jets in the selected events are systematically shifted by the respective uncertainty, and then a new value for the dijet mass is calculated. The average shift at each mass value is then fitted with a continuous function (Figure \ref{fig_AvgMassUnc}, right-side plot). It is important to note that for the increasing $|$y$|_{max}$, the uncertainty at a given mass value decreases. For a given dijet mass, two leading jets in a higher rapidity bin are more likely to have smaller \pt values which are accompanied by smaller uncertainties. Figure \ref{fig_AvgMassUnc} (right) shows the uncertainty induced on the cross section, due to the dijet mass scale uncertainty. This is calculated by the formula below:
\begin{equation}
\delta_\pm = \frac{\int_{m_1^\prime}^{m_2^\prime}{f(m)\,dm}}{\int_{m_1}^{m_2}{f(m)\,dm}}-1 
\end{equation}
where $\delta_\pm$ is the fractional change of the cross section, $m_{1,2}$ are the bin boundaries, $m_{1,2}^\prime=m_{1,2}\cdot[1\pm a_{JES}(m_{1,2})]$ are the shifted mass boundaries due to the relative mass scale change $a_{JES}(m)$ and $f(m)$ is a continuous fit on the measured spectrum. The resulting cross section uncertainty is asymmetric and almost fully correlated between the mass bins, ranging from 10\% at $M_{JJ}$=200 GeV/$c^2$ to 60\% at $M_{JJ}=$3 TeV/$c^2$.
\subsection{Luminosity Uncertainty}
The luminosity uncertainty is estimated to be 4\% \cite{CMS-DP-2011-002} and directly transferred to the cross section measurement. It is also correlated 100\%  across the mass bins.
\subsection{Unsmearing Uncertainty}
\begin{figure}[ht]
  \centering
  \includegraphics[width=0.48\textwidth]{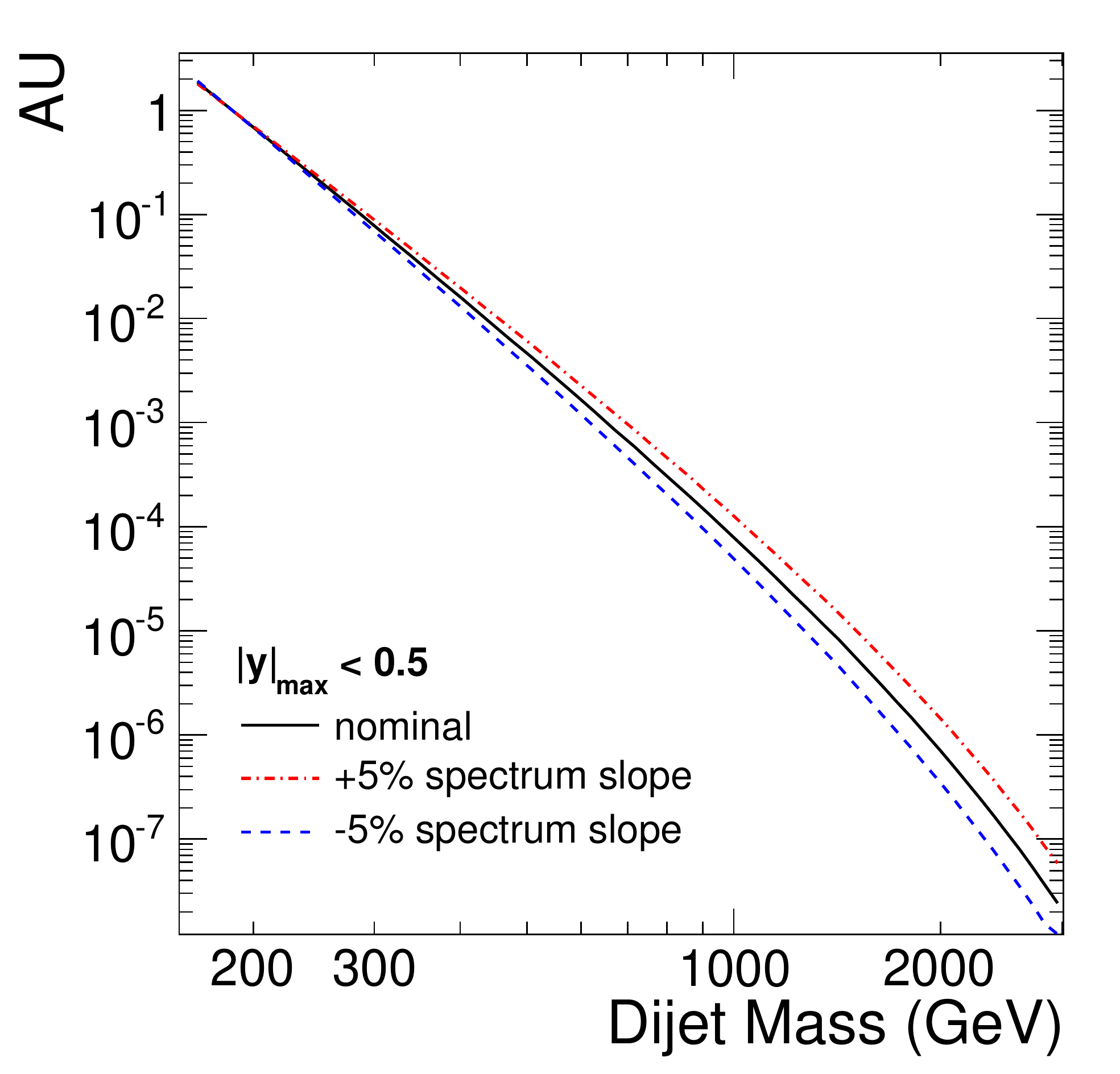}
  \caption{Simulated dijet mass spectrum (continuous line) in the rapidity bin $|y|_{max}<0.5$ which is used for the evaluation of the smearing effect. The dashed and the dashed-dotted lines correspond to softer and harder spectra respectively, systematically shifted by changing the slope by 5\%.} 
  \label{fig:UnfoldingSpectrum}
\end{figure}

\begin{figure}[ht]
  \centering
  \includegraphics[width=0.48\textwidth]{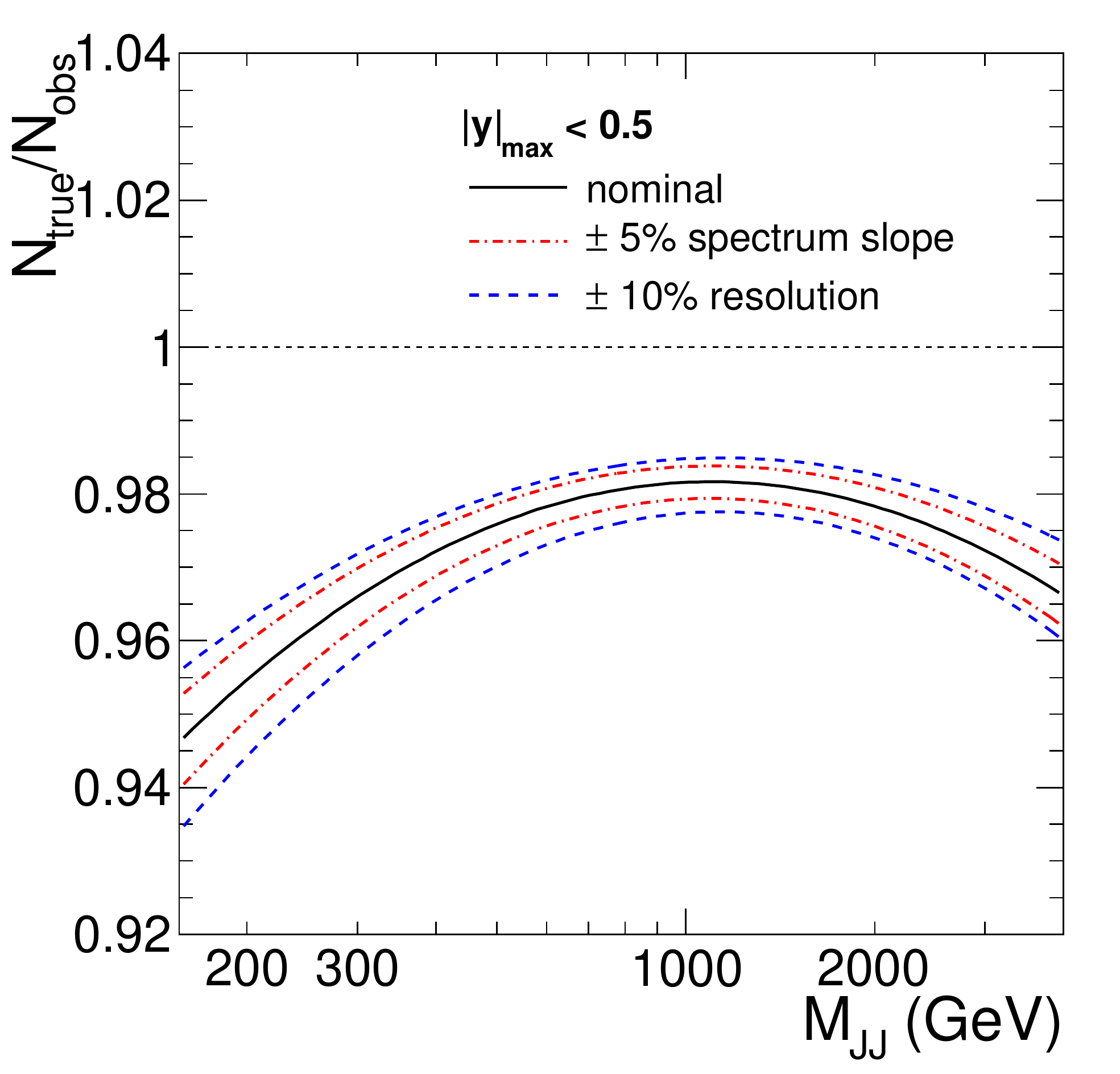}
  \includegraphics[width=0.48\textwidth]{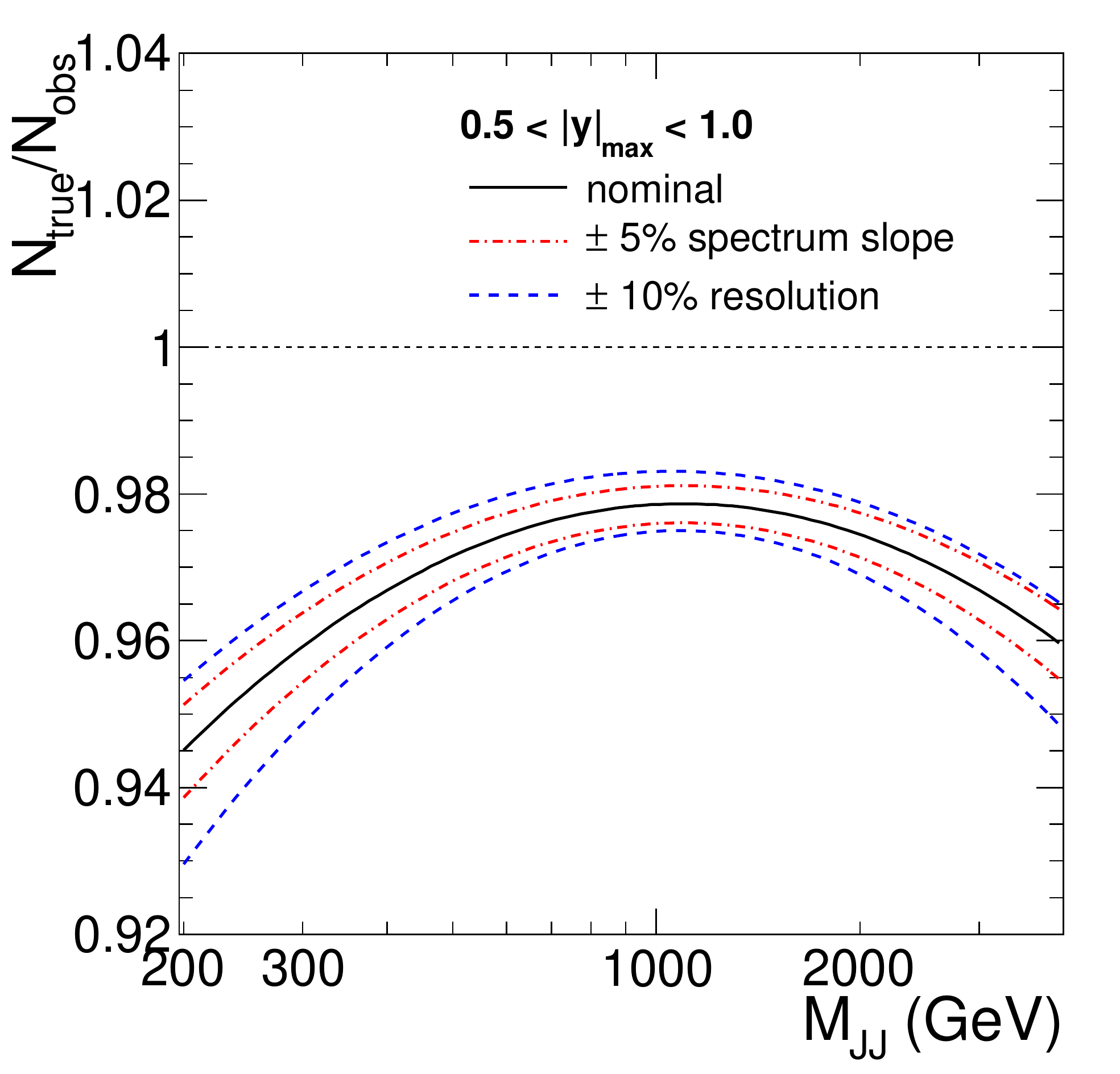}
  \includegraphics[width=0.48\textwidth]{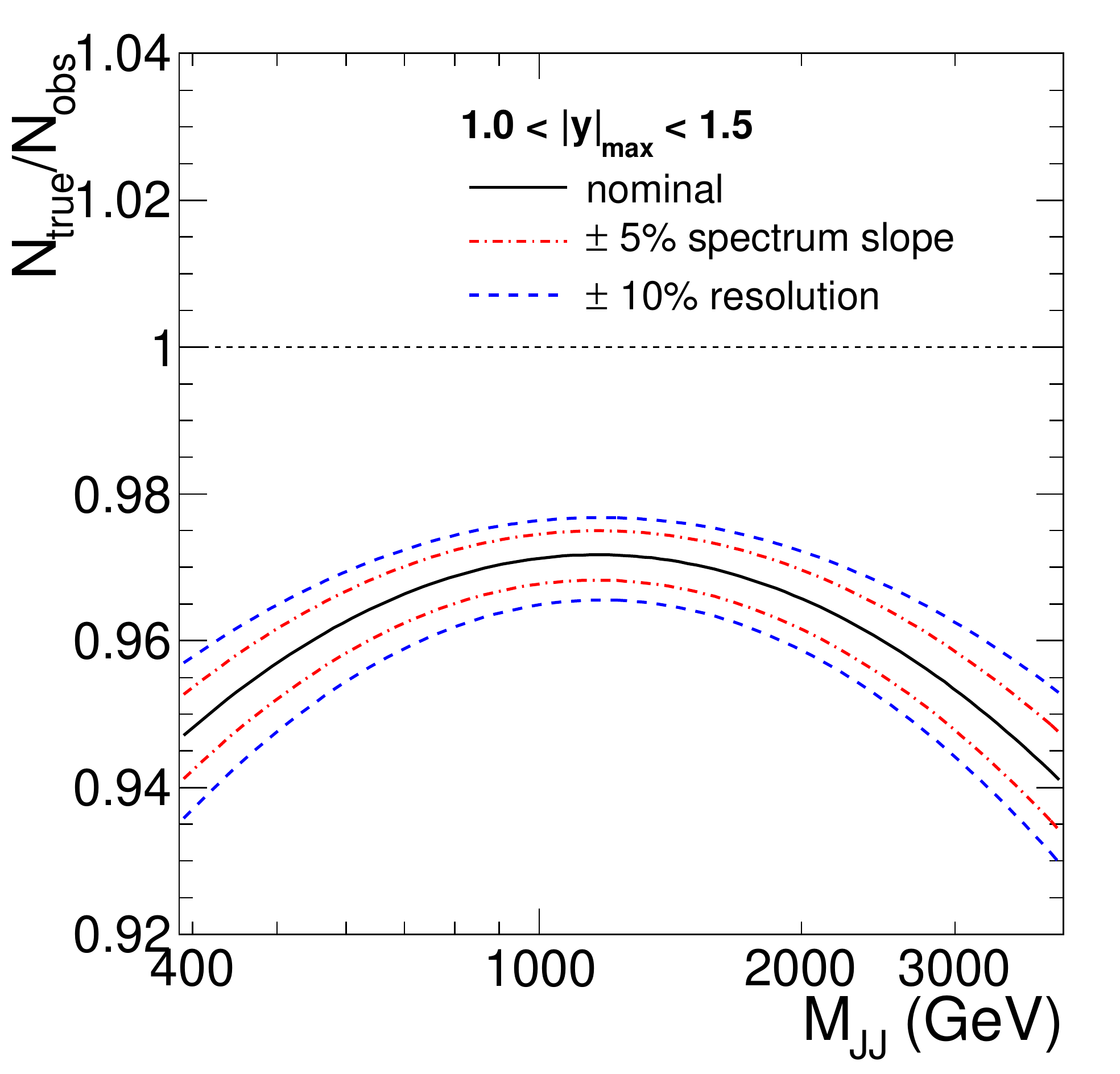}
  \includegraphics[width=0.48\textwidth]{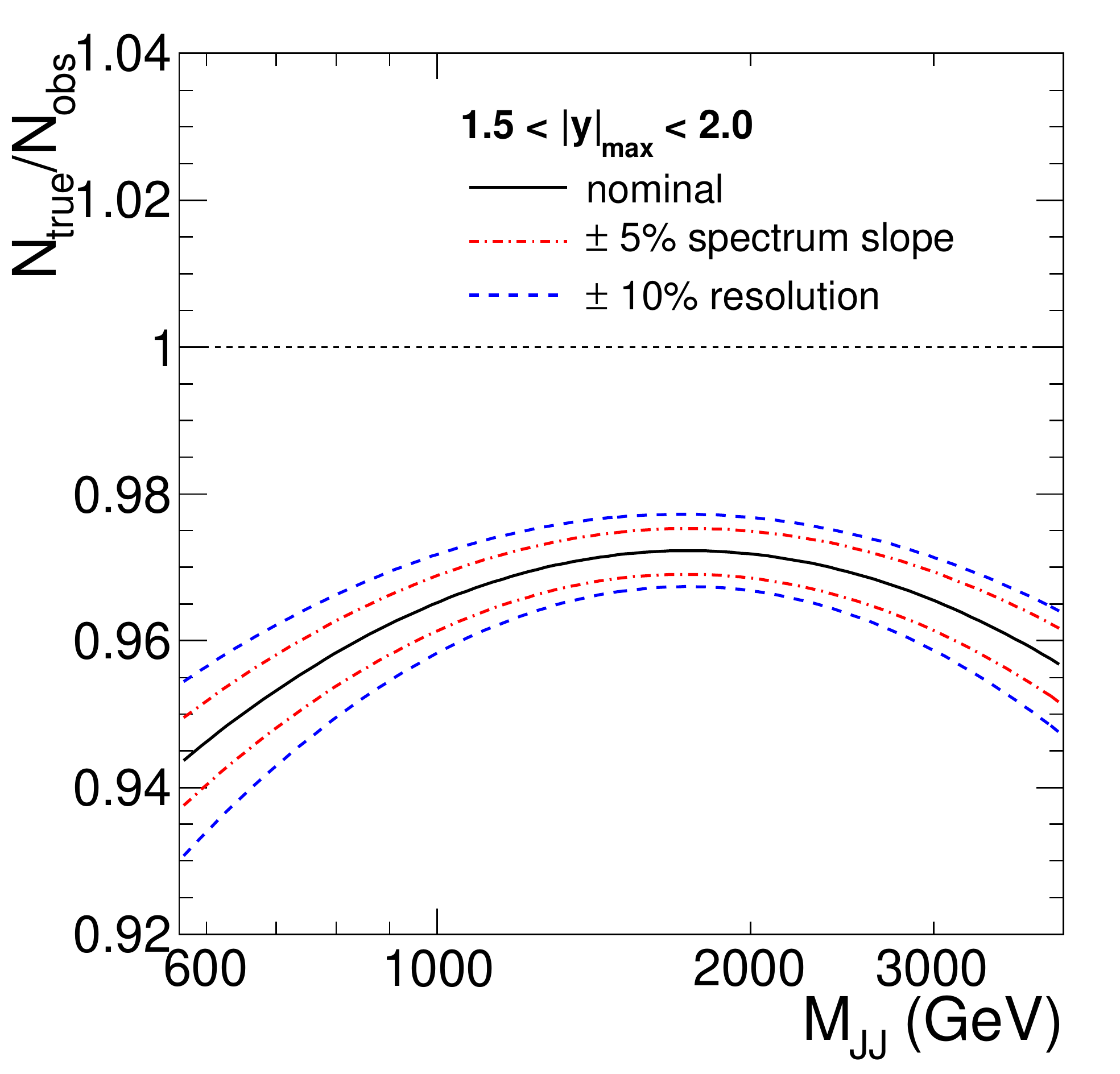}
  \includegraphics[width=0.48\textwidth]{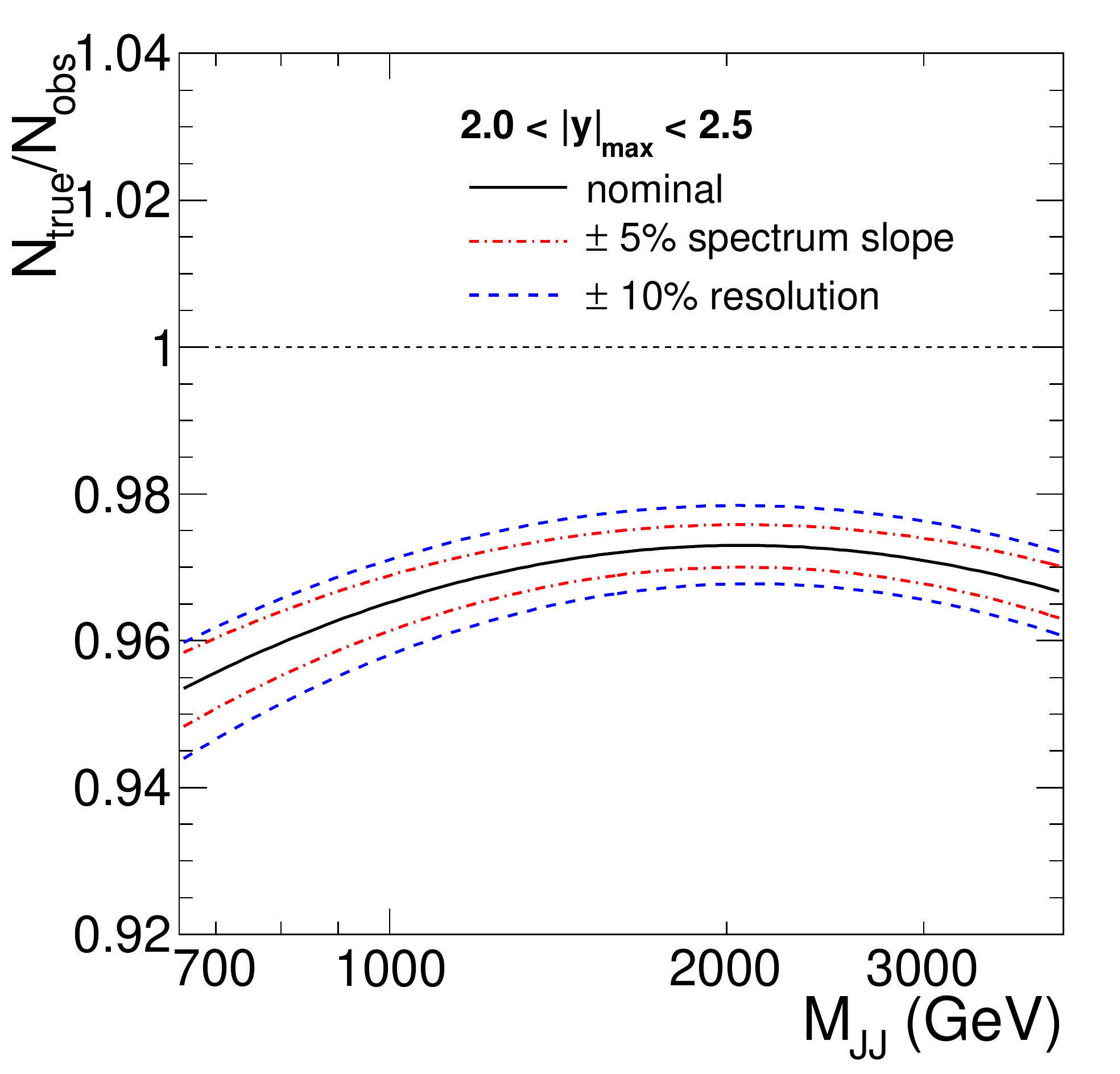}
  \caption{Unsmearing correction in the various rapidity bins. The uncertainty of the correction factor on the dijet mass resolution and the simulated spectrum are also shown.} 
  \label{fig:UnfoldingSystematics}
\end{figure}

There are two sources of the uncertainty on the correction factors for smearing effects. As discussed in Section \ref{Unsmearing}, the unsmearing corrections are derived by taking inputs from a toy Monte Carlo model of the dijet mass spectrum followed by a forward smearing due to the detector resolution for the dijet mass variable. In this technique, the slope of the spectrum and the resolution parameters depend on the pure Monte Carlo study. As a result, they may be slightly different than they are in reality. A reasonable approach to estimate the uncertainty on the unsmearing correction factors is by varying the spectrum slope and the resolution parameters. The spectrum slope is varied by 5\% and this is conservatively based on the comparison of the data and the theory. The analytical method to achieve the slope variation is just by varying the exponents in Equation \ref{Eq.MassFit} by exactly the same amount with the amount desired for the slope variation. The derivative of the Equation \ref{Eq.MassFit} is
\begin{equation}\label{Eq.MassFitDrvt}
\frac{\mathrm{d}f(m)}{\mathrm{d}m}=A\cdot\left(m/\sqrt{s}\right)^{-a}\cdot\left(1-m/\sqrt{s}\right)^b\cdot \left[-a(\frac{m}{\sqrt{s}})^{-1}-b(1-\frac{m}{\sqrt{s}})^{-1}\right]
\end{equation}
which is equivalent to the;
\begin{equation}\label{Eq.MassFitDrtvEq}
\frac{\mathrm{d}f(m)/\mathrm{d}m}{f(m)}=\left[-a(\frac{m}{\sqrt{s}})^{-1}-b(1-\frac{m}{\sqrt{s}})^{-1}\right]
\end{equation}
If both exponents $a$ and $b$ are scaled by 1.05 (5\%), then the Equation \ref{Eq.MassFitDrtvEq} becomes;
\begin{equation}\label{Eq.MassFitDrtvEqScaled}
\frac{\mathrm{d}\tilde{f}(m)/\mathrm{d}m}{\tilde{f}(m)}=-1.05\left[a(\frac{m}{\sqrt{s}})^{-1}+b(1-\frac{m}{\sqrt{s}})^{-1}\right]
\end{equation}
where $\tilde{f}(m)$ is the new function with scaled exponents. As it can easily be seen, the functional property of the Equation \ref{Eq.MassFit} allows us to vary the spectrum slope analytically (Figure \ref{fig:UnfoldingSpectrum}). 
The resolution is varied by 10\% where this number is motivated by the observed difference between data and simulation in the jet energy resolution \cite{JME-10-014}.
Figure \ref{fig:UnfoldingSystematics} shows the response of the unsmearing correction to the variations described above. Overall, the unsmearing uncertainty is in the order of 2-3\%, fully correlated across the mass bins.
\clearpage
\section{Theoretical Uncertainties}
\begin{figure}[ht]
  \centering
  \includegraphics[width=0.28\textwidth]{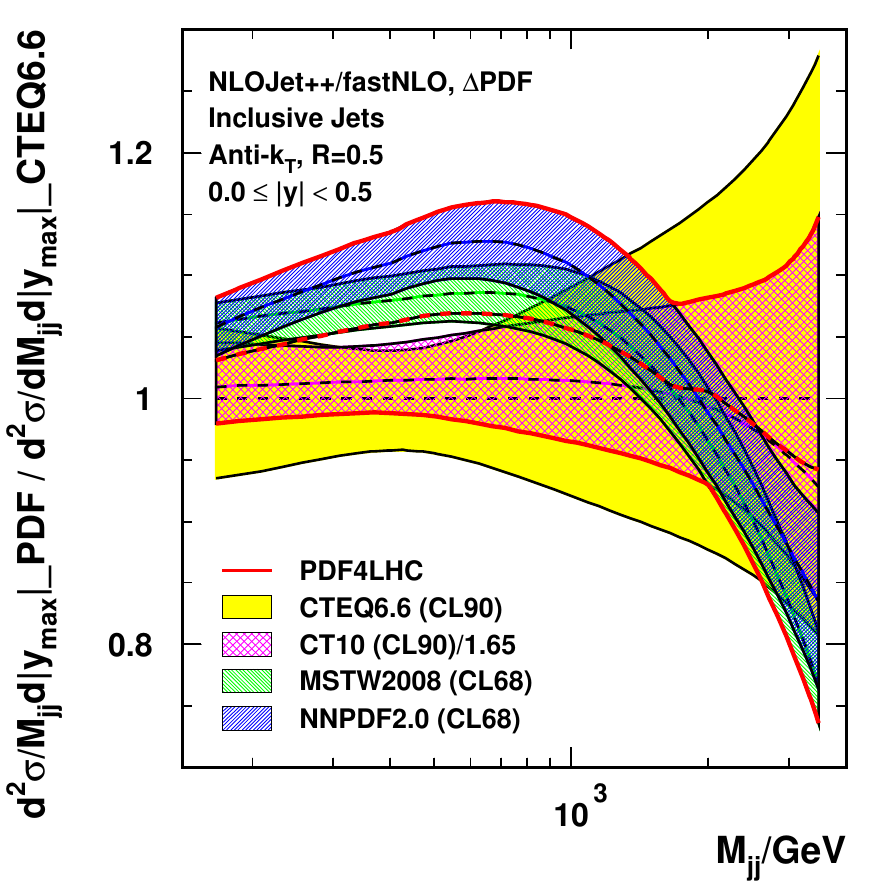}
  \includegraphics[width=0.28\textwidth]{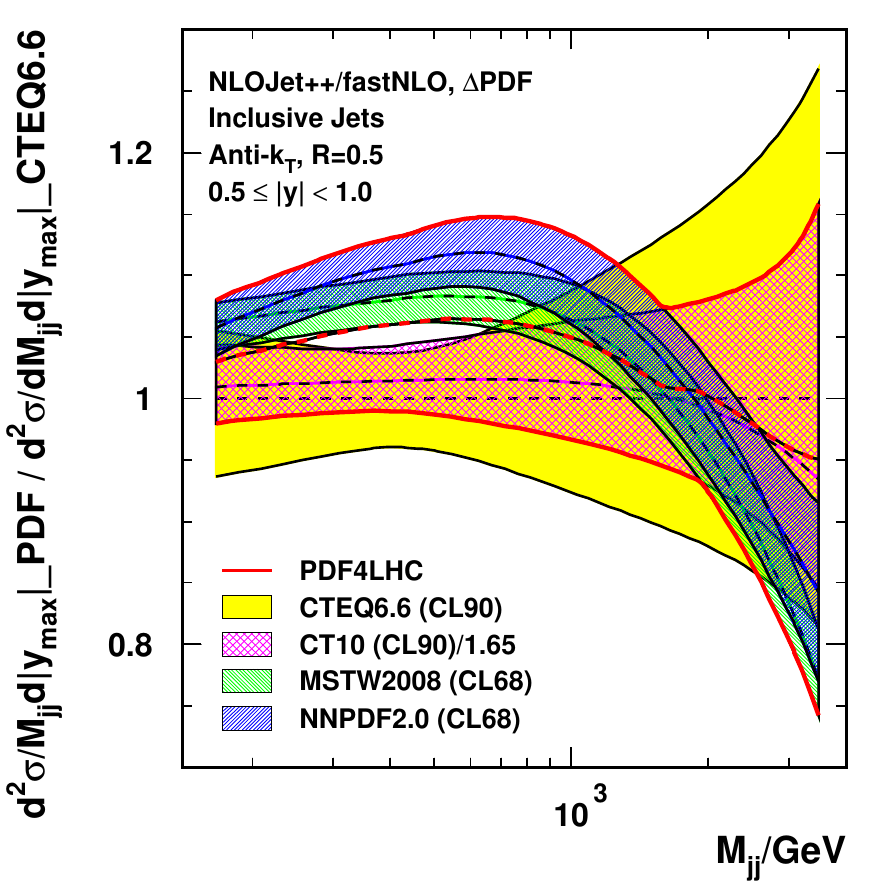}
  \includegraphics[width=0.28\textwidth]{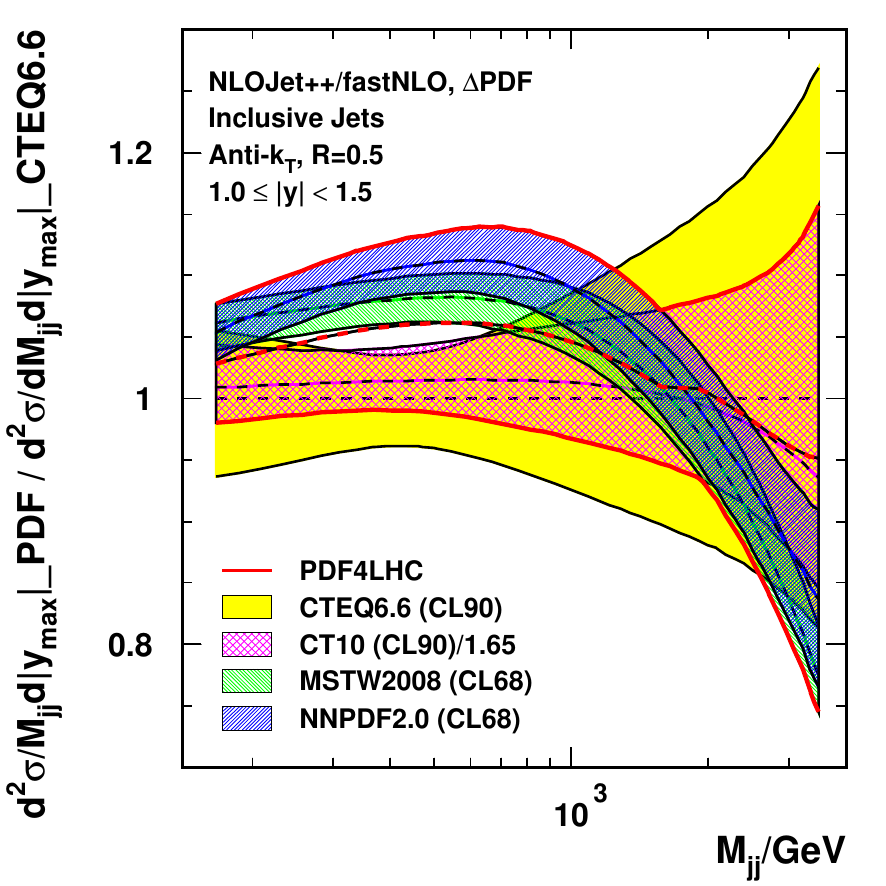}
  \includegraphics[width=0.28\textwidth]{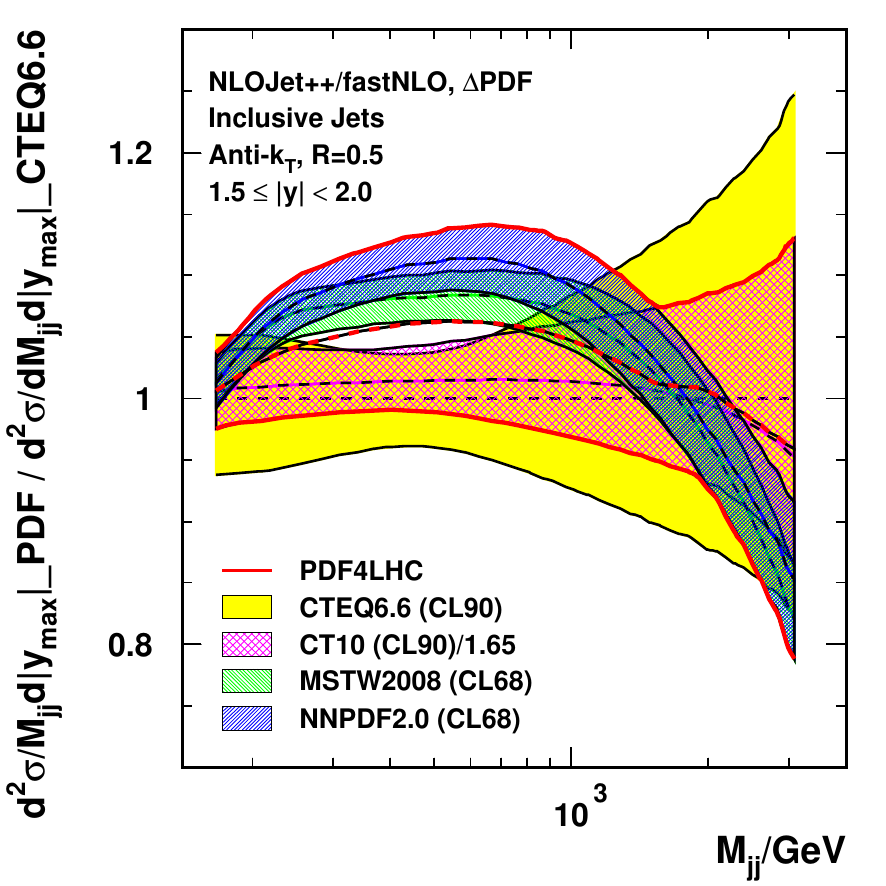}
  \includegraphics[width=0.28\textwidth]{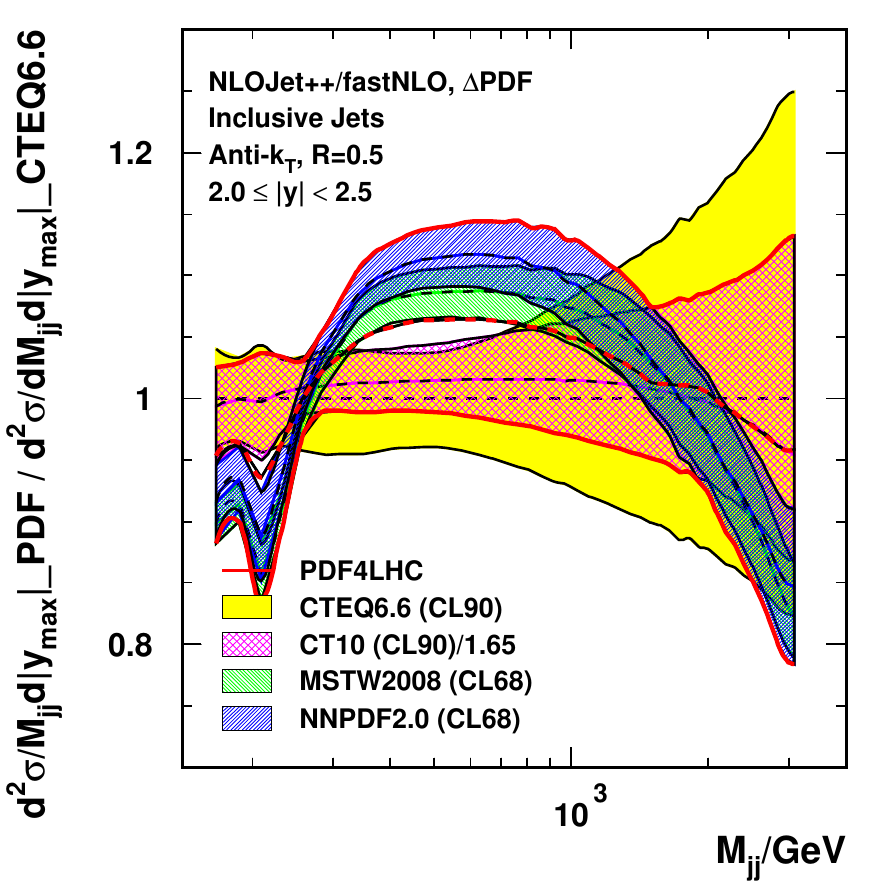}
  \caption{PDF Uncertainties according to PDF4LHC prescription. CT10, MSTW2008NLO, NNPDF2.0. are used to perform NLO calculations and the ratio between each of these three and CTEQ6.6 are used to set the PDF uncertainty.} 
  \label{fig:PDF4LHC}
\end{figure}

\begin{figure}[ht]
  \centering
  \includegraphics[width=0.48\textwidth]{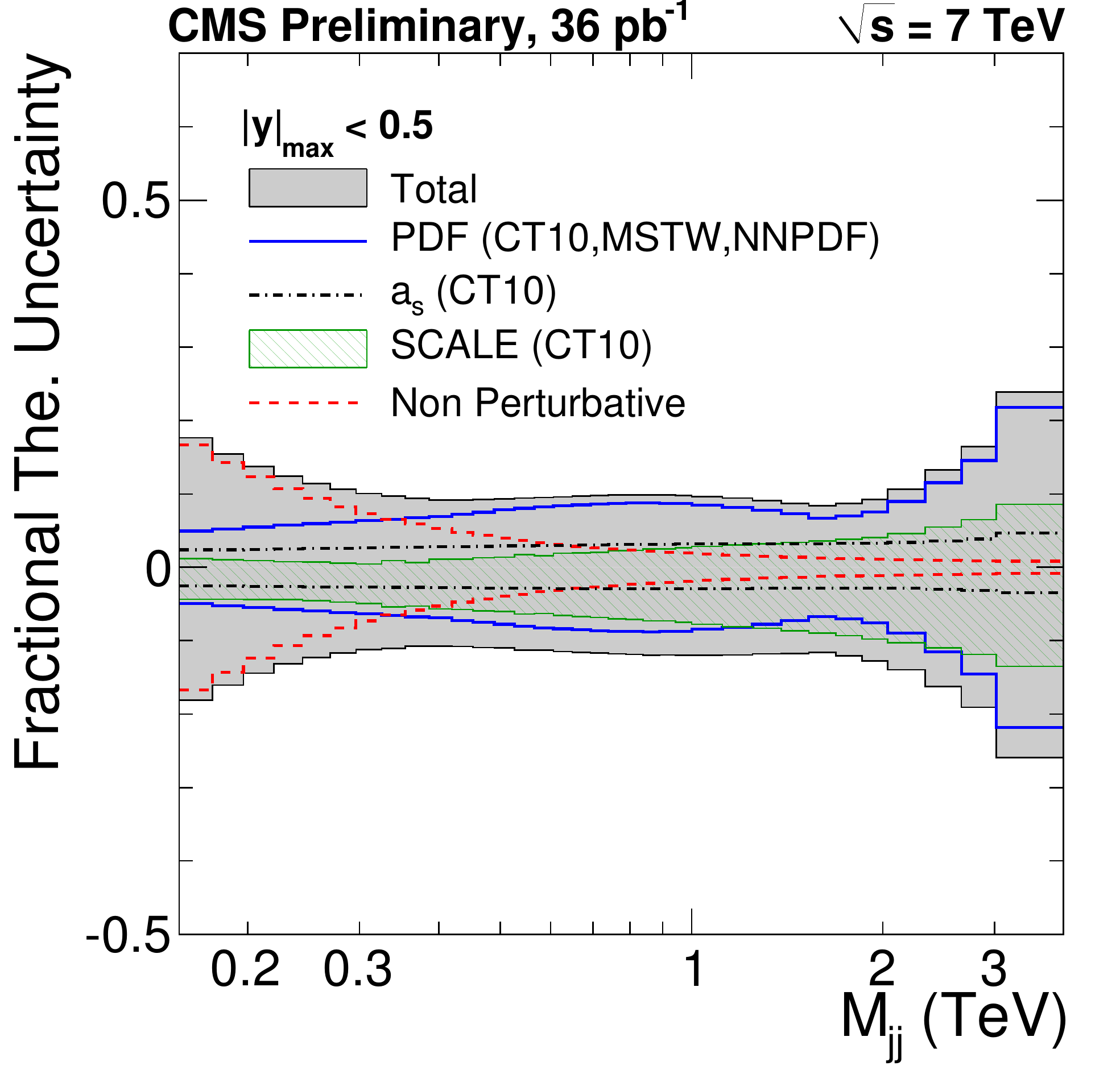}
  \includegraphics[width=0.48\textwidth]{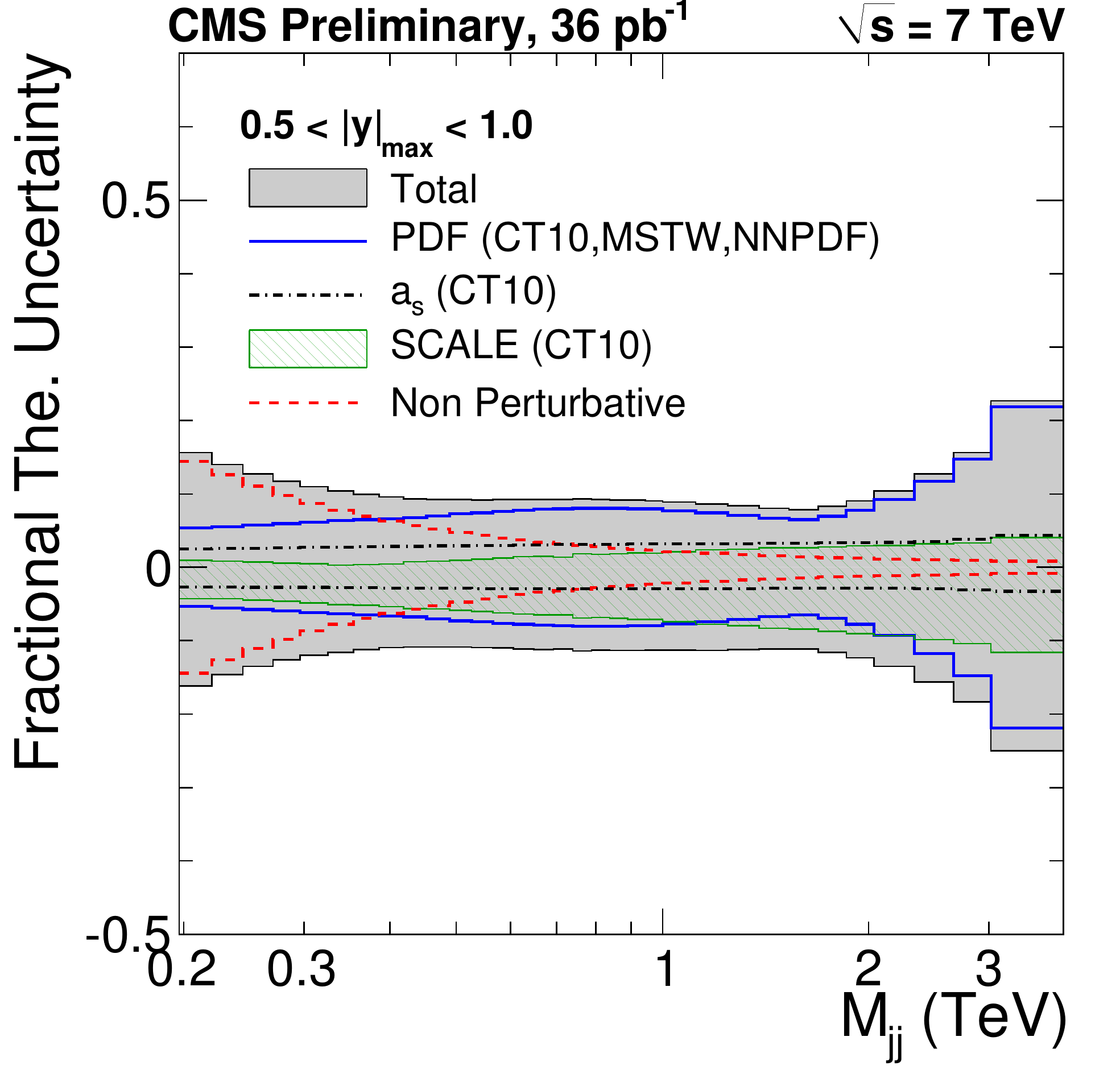}
  \includegraphics[width=0.48\textwidth]{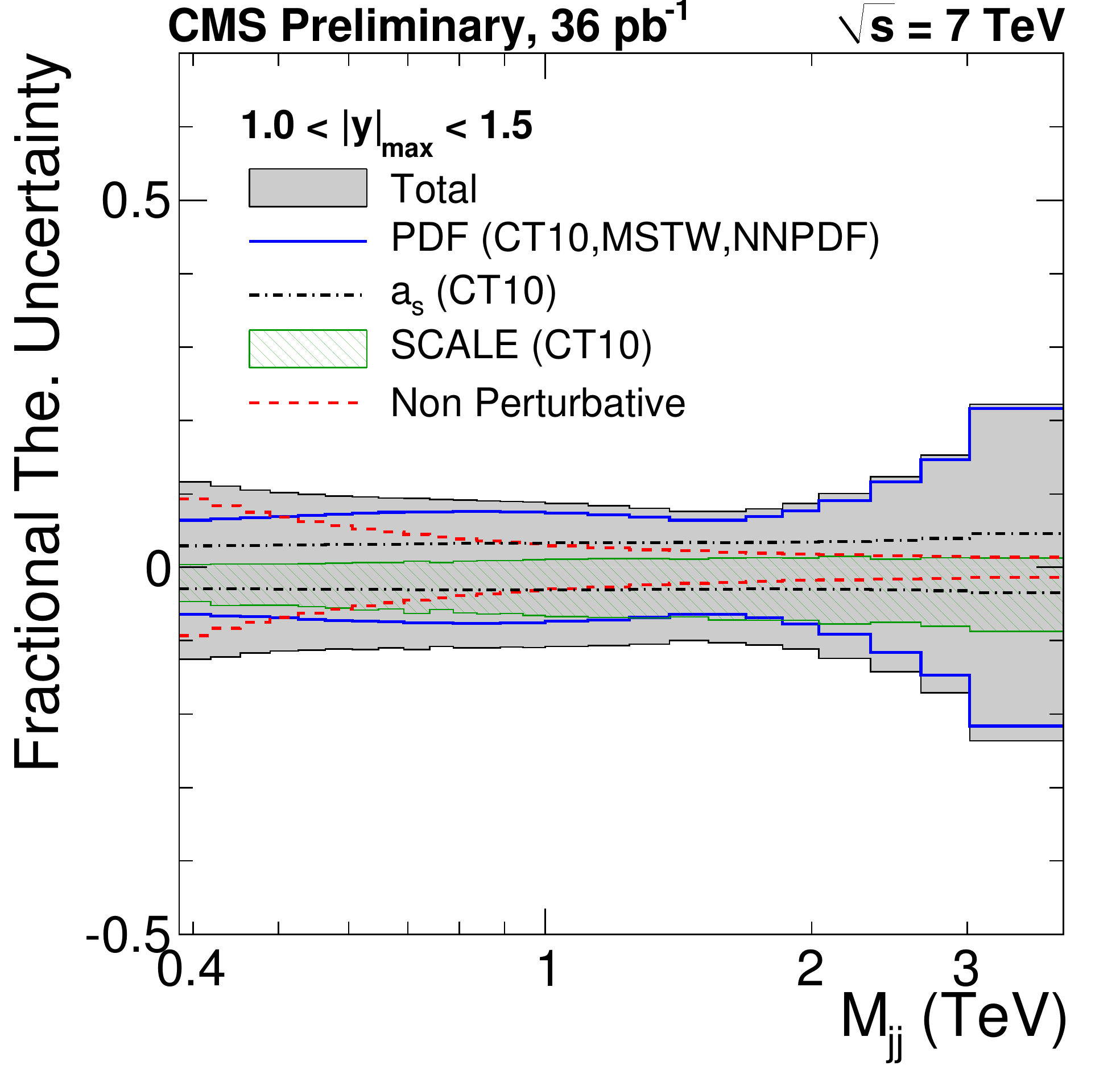}
  \includegraphics[width=0.48\textwidth]{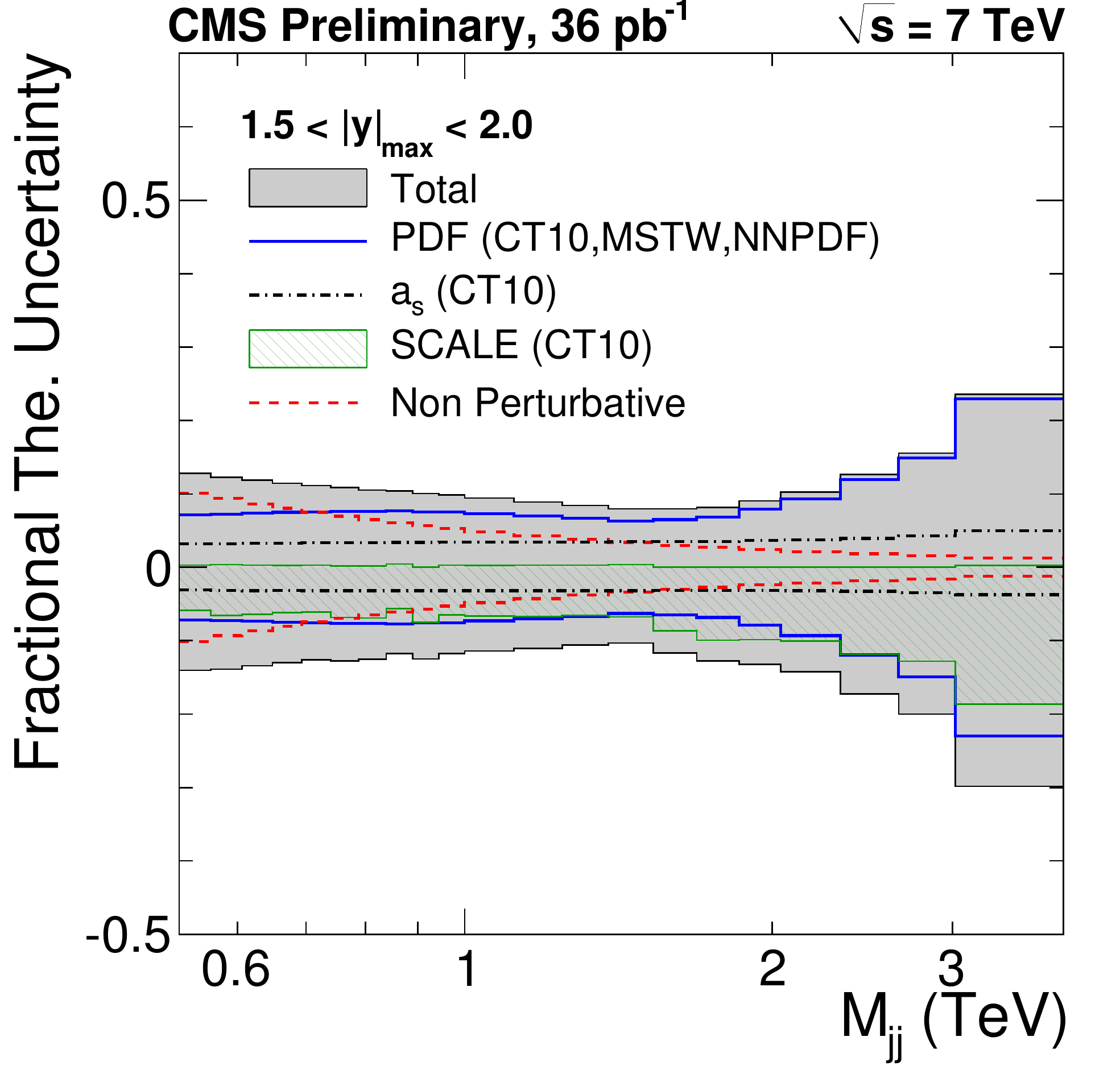}
  \includegraphics[width=0.48\textwidth]{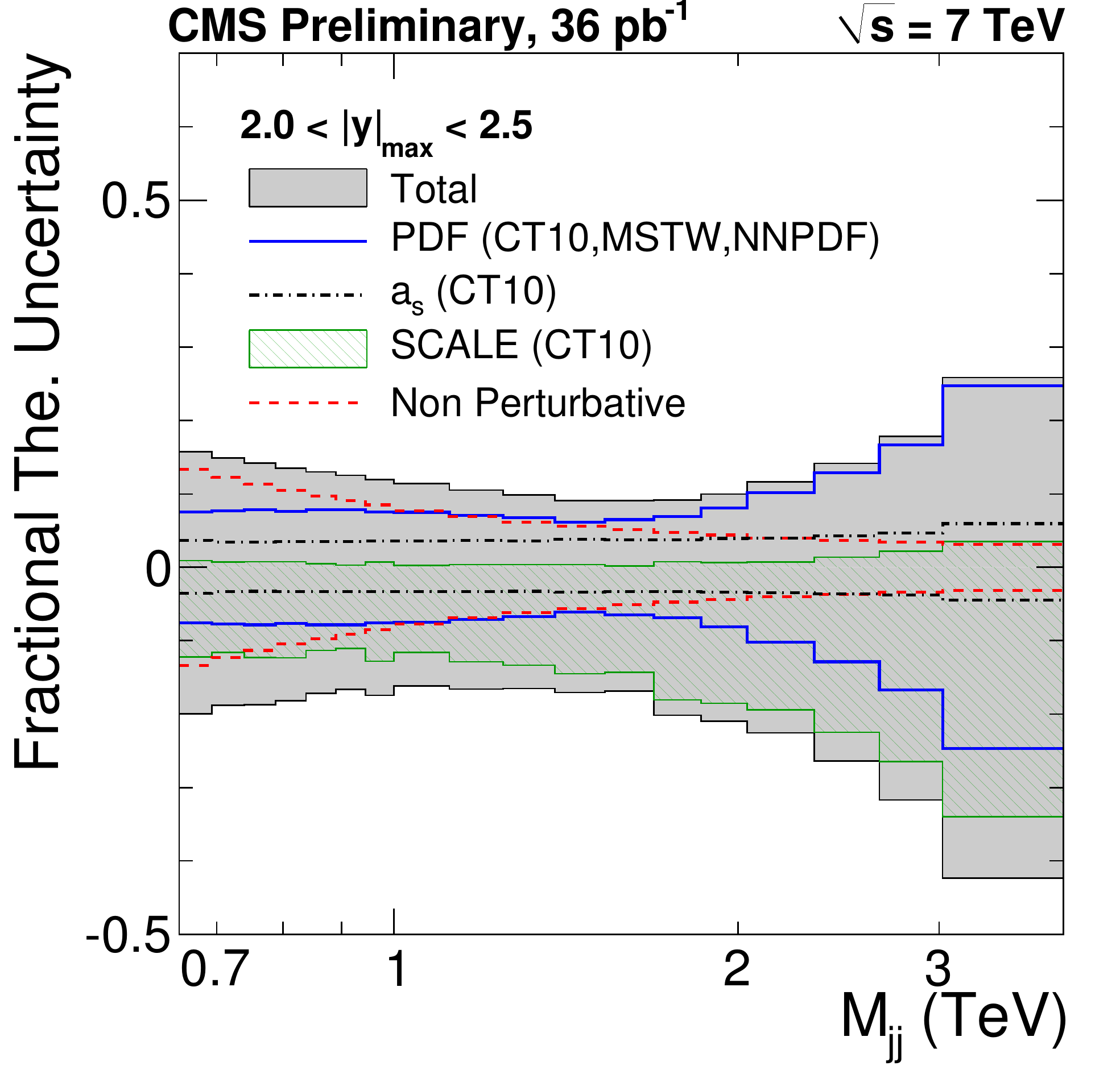}
  \caption{Summary of the theoretical systematic uncertainties: PDF (blue dashed-dotted line), scale variations (red dashed line), non-perturbative correction (green dashed-double dotted line) and the sum in quadrature (filled).}
  \label{fig:Theory_Uncertainties}
\end{figure}

The theoretical uncertainties are introduced due to the PDF dependence, the renormalization and factorization scale choice and by the non-perturbative corrections. The PDF uncertainty is estimated according to the PDF4LHC \cite{PDF4LHC} prescription through the variation of the CT10, MSTW2008NLO and NNPDF2.0 PDF sets (see Figure \ref{fig:PDF4LHC}). The renormalization and factorization scale uncertainty is estimated as the maximal deviation of the six point variation ($\mu_F/p_{T}^{ave}$, $\mu_R/p_{T}^{ave}$)=(1/2, 1/2), (2, 2), (1, 1/2), (1, 2), (1/2, 1), (2, 1). Finally, the non-perturbative correction uncertainty is estimated as half of the NP correction deviation from unity (see Figure \ref{NP_Corrections}). Overall, the PDF uncertainty dominates at high mass values, while the non-perturbative correction uncertainty is dominant at low masses. Figure \ref{fig:Theory_Uncertainties} shows the theoretical uncertainty decomposition in all rapidity bins.

\clearpage
\chapter{RESULTS AND CONCLUSIONS}

\section{Results}
\begin{figure}[ht]
  \centering
  \includegraphics[width=0.90\textwidth]{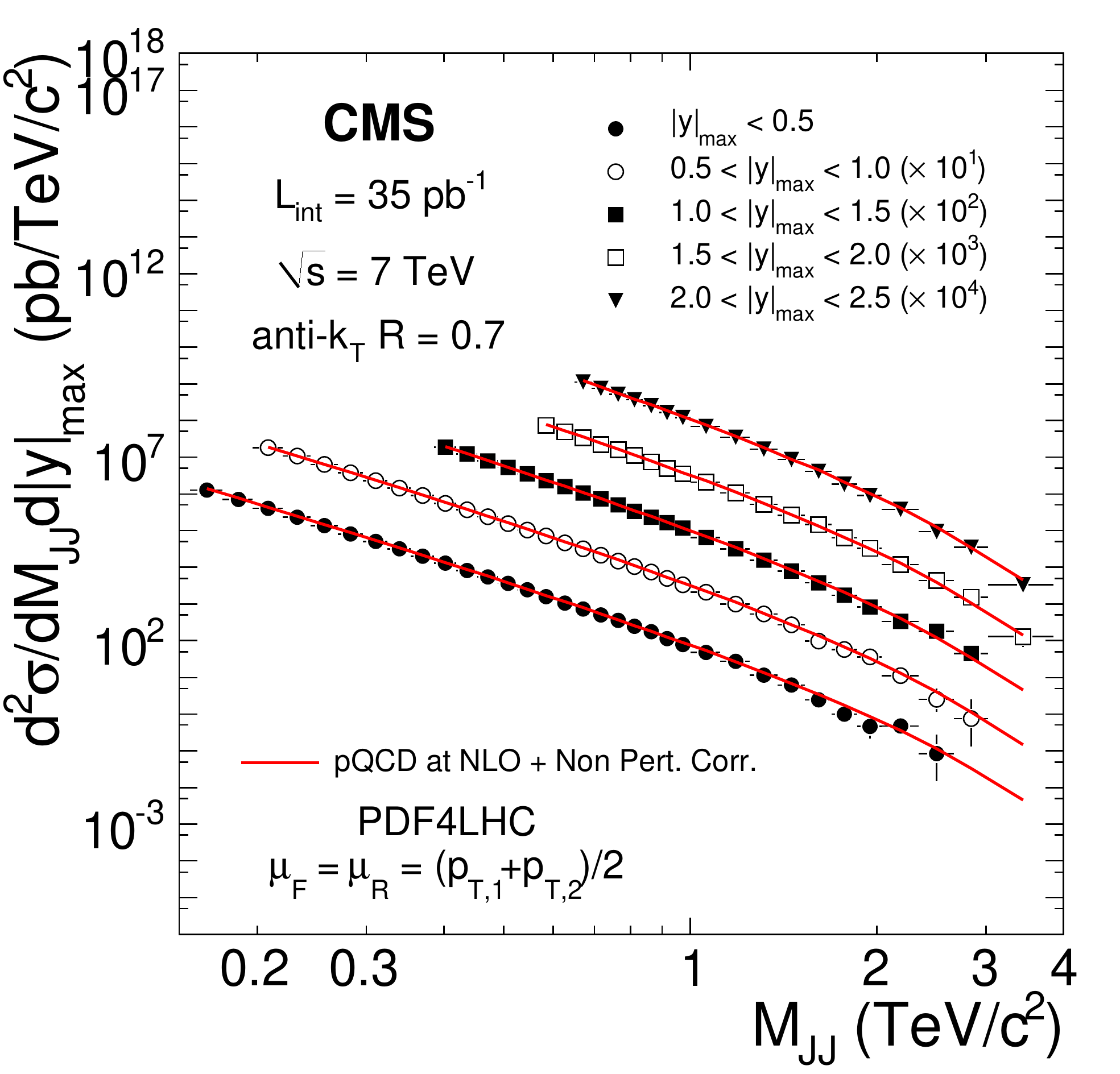}
  \capspace
  \caption{Measured double-differential dijet production cross sections (points scaled by the factors shown in the figure) as a function of the dijet invariant mass in bins of the variable $|$y$|_{max}$, compared to the theoretical predictions (curves). The horizontal error bars represent the bin widths, while the vertical error bars represent the statistical uncertainties of the data \cite{DijetPaper}.}
  \label{fig:Xsec}
\end{figure}

In this last chapter, the measured dijet mass cross-section is compared to the theory predictions at the particle level. The double differential cross-section in logarithmic scale is shown in Figure \ref{fig:Xsec} with statistical uncertainties only. The different \ymax bins are scaled for a better visualization. It is  observed that the dijet mass spectrum falls steeply and smoothly by many orders of magnitude, in agreement with the theory predictions, and the measurement covers the range from 0.2 TeV to 3.5 TeV. The exact mass ranges and the cross-section values are given in Tables \ref{tab:ybin0}-\ref{tab:ybin4}. The central values quoted in the tables for each bin are the mass value $m_0$ which satisfies the equation
\begin{equation}
f(m_0)(m_2-m_1)=\displaystyle \int_{m_1}^{_2} f(m) dm
\end{equation}
where $m_1$, $m_2$ are the bin boundaries and $f(m)$ is the continuous ﬁt of the cross section. The choice of the central value is adopted from the approach described in \cite{bin_centers}.

\begin{table*}[htbH]
\begin{center}
\caption{Double-differential dijet mass cross section in the rapidity range $|y|_\text{max}<0.5$. The reference mass is the point at which the cross section is drawn in Figures \ref{fig:Xsec} and \ref{fig:data_over_theory} and is calculated as described in the text. The experimental systematic uncertainties of the individual dijet-mass bins are almost 100\% correlated \cite{DijetPaper}.
\label{tab:ybin0}}
\capspace
%\footnotesize
\scriptsize
\begin{tabular}{|c|c|c|c|c|}
\hline
Mass Range & Reference Mass & Measured Cross Section & Statistical Uncertainty & Systematic Uncertainty \\
$(TeV)$ & $(TeV)$ & $(\text{pb}/TeV)$ & \% & \% \\
\hline
\hline
 [0.156, 0.176] &      0.165 &                     $1.32\times 10^{6}$ &   $-$1.1, $+$1.1 &     $-10$, $+$11 \\

 [0.176, 0.197] &      0.186 &                     $7.26\times 10^{5}$ &   $-$1.4, $+$1.4 &     $-10$, $+$12 \\

 [0.197, 0.220] &      0.208 &                     $4.12\times 10^{5}$ &   $-$1.8, $+$1.8 &     $-11$, $+$12 \\

 [0.220, 0.244] &      0.231 &                     $2.35\times 10^{5}$ &   $-$0.7, $+$0.7 &     $-11$, $+$12 \\

 [0.244, 0.270] &      0.256 &                     $1.39\times 10^{5}$ &   $-$0.9, $+$0.9 &     $-11$, $+$12 \\

 [0.270, 0.296] &      0.282 &                     $8.18\times 10^{4}$ &   $-$0.7, $+$0.7 &     $-11$, $+$13 \\

 [0.296, 0.325] &      0.310 &                     $5.08\times 10^{4}$ &   $-$0.9, $+$0.9 &     $-11$, $+$13 \\

 [0.325, 0.354] &      0.339 &                     $3.18\times 10^{4}$ &   $-$1.1, $+$1.1 &     $-11$, $+$13 \\

 [0.354, 0.386] &      0.369 &                     $2.04\times 10^{4}$ &   $-$1.3, $+$1.3 &     $-12$, $+$13 \\

 [0.386, 0.419] &      0.402 &                     $1.28\times 10^{4}$ &   $-$1.1, $+$1.1 &     $-12$, $+$14 \\

 [0.419, 0.453] &      0.435 &                     $8.22\times 10^{3}$ &   $-$1.4, $+$1.4 &     $-12$, $+$14 \\

 [0.453, 0.489] &      0.470 &                     $5.42\times 10^{3}$ &   $-$1.6, $+$1.6 &     $-12$, $+$14 \\

 [0.489, 0.526] &      0.507 &                     $3.65\times 10^{3}$ &   $-$1.4, $+$1.5 &     $-13$, $+$14 \\

 [0.526, 0.565] &      0.545 &                     $2.44\times 10^{3}$ &   $-$1.7, $+$1.7 &     $-13$, $+$15 \\

 [0.565, 0.606] &      0.585 &                     $1.58\times 10^{3}$ &   $-$2.1, $+$2.1 &     $-13$, $+$15 \\

 [0.606, 0.649] &      0.627 &                     $1.05\times 10^{3}$ &   $-$2.5, $+$2.5 &     $-13$, $+$16 \\

 [0.649, 0.693] &      0.670 &                     $7.35\times 10^{2}$ &   $-$2.9, $+$3.0 &     $-14$, $+$16 \\

 [0.693, 0.740] &      0.716 &                     $5.05\times 10^{2}$ &   $-$3.4, $+$3.5 &     $-14$, $+$16 \\

 [0.740, 0.788] &      0.763 &                     $3.62\times 10^{2}$ &   $-$4.0, $+$4.2 &     $-14$, $+$17 \\

 [0.788, 0.838] &      0.812 &                     $2.45\times 10^{2}$ &   $-$4.8, $+$5.0 &     $-15$, $+$17 \\

 [0.838, 0.890] &      0.863 &                     $1.77\times 10^{2}$ &   $-$5.5, $+$5.8 &     $-15$, $+$18 \\

 [0.890, 0.944] &      0.916 &                     $1.13\times 10^{2}$ &   $-$6.8, $+$7.2 &     $-15$, $+$18 \\

 [0.944, 1.000] &      0.971 &                     $7.94\times 10^{1}$ &   $-$7.9, $+$8.5 &     $-16$, $+$19 \\

 [1.000, 1.118] &      1.055 &                     $4.76\times 10^{1}$ &   $-$7.0, $+$7.5 &     $-16$, $+$20 \\

 [1.118, 1.246] &      1.178 &                     $2.72\times 10^{1}$ &   $-$8.9, $+$9.8 &     $-17$, $+$21 \\

 [1.246, 1.383] &      1.310 &                     $1.14\times 10^{1}$ &   $-$13, $+$15 &     $-18$, $+$22 \\

 [1.383, 1.530] &      1.452 &                     $6.24$ &   $-$17, $+$21 &     $-19$, $+$24 \\

 [1.530, 1.687] &      1.604 &                     $2.48$ &   $-$26, $+$35 &     $-20$, $+$26 \\

 [1.687, 1.856] &      1.766 &                     $9.85\times 10^{-1}$ &   $-$40, $+$60 &     $-22$, $+$28 \\

 [1.856, 2.037] &      1.941 &                     $4.59\times 10^{-1}$ &   $-$54, $+$97 &     $-23$, $+$31 \\

 [2.037, 2.332] &      2.170 &                     $4.69\times 10^{-1}$ &   $-$43, $+$68 &     $-25$, $+$34 \\

 [2.332, 2.659] &      2.479 &                     $8.45\times 10^{-2}$ &   $-$83, $+$230 &     $-29$, $+$40 \\
\hline
\end{tabular}
\end{center}
\end{table*}

\begin{table*}[htbH]
\begin{center}
\caption{Double-differential dijet mass cross section in the rapidity range $0.5<|y|_\text{max}<1.0$. The reference mass is the point at which the cross section is drawn in Figures \ref{fig:Xsec} and \ref{fig:data_over_theory} and is calculated as described in the text. The experimental systematic uncertainties of the individual dijet-mass bins are almost 100\% correlated \cite{DijetPaper}.
 \label{tab:ybin1}
}
\capspace
\scriptsize
%\footnotesize
\begin{tabular}{|c|c|c|c|c|}
\hline
Mass Range & Reference Mass & Measured Cross Section & Statistical Uncertainty & Systematic Uncertainty \\
$(TeV)$ & $(TeV)$ & $(\text{pb}/TeV)$ & \% & \% \\
\hline
\hline
 [0.197, 0.220] &      0.208 &                     $1.74\times 10^{6}$ &   $-$0.8, $+$0.9 &     $-11$, $+$12 \\

 [0.220, 0.244] &      0.231 &                     $1.02\times 10^{6}$ &   $-$1.1, $+$1.1 &     $-11$, $+$12 \\

 [0.244, 0.270] &      0.256 &                     $6.00\times 10^{5}$ &   $-$1.4, $+$1.4 &     $-11$, $+$12 \\

 [0.270, 0.296] &      0.282 &                     $3.64\times 10^{5}$ &   $-$1.7, $+$1.8 &     $-11$, $+$12 \\

 [0.296, 0.325] &      0.310 &                     $2.22\times 10^{5}$ &   $-$0.7, $+$0.7 &     $-11$, $+$13 \\

 [0.325, 0.354] &      0.339 &                     $1.38\times 10^{5}$ &   $-$0.8, $+$0.9 &     $-11$, $+$13 \\

 [0.354, 0.386] &      0.369 &                     $8.64\times 10^{4}$ &   $-$1.0, $+$1.0 &     $-12$, $+$13 \\

 [0.386, 0.419] &      0.402 &                     $5.42\times 10^{4}$ &   $-$0.8, $+$0.8 &     $-12$, $+$13 \\

 [0.419, 0.453] &      0.435 &                     $3.55\times 10^{4}$ &   $-$1.0, $+$1.0 &     $-12$, $+$14 \\

 [0.453, 0.489] &      0.470 &                     $2.34\times 10^{4}$ &   $-$1.1, $+$1.2 &     $-12$, $+$14 \\

 [0.489, 0.526] &      0.507 &                     $1.53\times 10^{4}$ &   $-$0.9, $+$0.9 &     $-12$, $+$14 \\

 [0.526, 0.565] &      0.545 &                     $1.01\times 10^{4}$ &   $-$1.1, $+$1.1 &     $-13$, $+$15 \\

 [0.565, 0.606] &      0.585 &                     $6.90\times 10^{3}$ &   $-$1.3, $+$1.3 &     $-13$, $+$15 \\

 [0.606, 0.649] &      0.627 &                     $4.60\times 10^{3}$ &   $-$1.6, $+$1.6 &     $-13$, $+$15 \\

 [0.649, 0.693] &      0.670 &                     $3.15\times 10^{3}$ &   $-$1.4, $+$1.4 &     $-13$, $+$16 \\

 [0.693, 0.740] &      0.716 &                     $2.14\times 10^{3}$ &   $-$1.7, $+$1.7 &     $-14$, $+$16 \\

 [0.740, 0.788] &      0.763 &                     $1.48\times 10^{3}$ &   $-$2.0, $+$2.0 &     $-14$, $+$16 \\

 [0.788, 0.838] &      0.812 &                     $1.04\times 10^{3}$ &   $-$2.3, $+$2.4 &     $-14$, $+$17 \\

 [0.838, 0.890] &      0.863 &                     $7.42\times 10^{2}$ &   $-$2.7, $+$2.8 &     $-15$, $+$17 \\

 [0.890, 0.944] &      0.916 &                     $5.01\times 10^{2}$ &   $-$3.2, $+$3.3 &     $-15$, $+$18 \\

 [0.944, 1.000] &      0.971 &                     $3.37\times 10^{2}$ &   $-$3.8, $+$4.0 &     $-15$, $+$18 \\

 [1.000, 1.118] &      1.055 &                     $2.08\times 10^{2}$ &   $-$3.4, $+$3.5 &     $-16$, $+$19 \\

 [1.118, 1.246] &      1.178 &                     $9.94\times 10^{1}$ &   $-$4.7, $+$4.9 &     $-17$, $+$20 \\

 [1.246, 1.383] &      1.310 &                     $5.38\times 10^{1}$ &   $-$6.1, $+$6.5 &     $-18$, $+$22 \\

 [1.383, 1.530] &      1.452 &                     $2.73\times 10^{1}$ &   $-$8.3, $+$9.0 &     $-19$, $+$23 \\

 [1.530, 1.687] &      1.604 &                     $9.70$ &   $-$13, $+$15 &     $-20$, $+$25 \\

 [1.687, 1.856] &      1.766 &                     $5.73$ &   $-$17, $+$20 &     $-21$, $+$27 \\

 [1.856, 2.037] &      1.941 &                     $3.66$ &   $-$20, $+$25 &     $-23$, $+$30 \\

 [2.037, 2.332] &      2.170 &                     $1.12$ &   $-$28, $+$38 &     $-25$, $+$33 \\

 [2.332, 2.659] &      2.479 &                     $2.52\times 10^{-1}$ &   $-$54, $+$97 &     $-28$, $+$39 \\

 [2.659, 3.019] &      2.819 &                     $7.62\times 10^{-2}$ &   $-$83, $+$230 &     $-32$, $+$47 \\
\hline

\end{tabular}
\end{center}
\end{table*}

\begin{table*}[htbH]
\begin{center}
\caption{Double-differential dijet mass cross section in the rapidity range $1.0<|y|_\text{max}<1.5$. The reference mass is the point at which the cross section is drawn in Figures \ref{fig:Xsec} and \ref{fig:data_over_theory} and is calculated as described in the text. The experimental systematic uncertainties of the individual dijet-mass bins are almost 100\% correlated \cite{DijetPaper}.
 \label{tab:ybin2}}
\capspace
\scriptsize
%\footnotesize
\begin{tabular}{|c|c|c|c|c|}
\hline
Mass Range & Reference Mass & Measured Cross Section & Statistical Uncertainty & Systematic Uncertainty \\
$(TeV)$ & $(TeV)$ & $(\text{pb}/TeV)$ & \% & \% \\
\hline
\hline
 [0.386, 0.419] &      0.402 &                     $1.84\times 10^{5}$ &   $-$2.2, $+$2.3 &     $-12$, $+$13 \\

 [0.419, 0.453] &      0.435 &                     $1.21\times 10^{5}$ &   $-$2.7, $+$2.8 &     $-12$, $+$13 \\

 [0.453, 0.489] &      0.470 &                     $7.77\times 10^{4}$ &   $-$3.3, $+$3.4 &     $-12$, $+$14 \\

 [0.489, 0.526] &      0.507 &                     $5.26\times 10^{4}$ &   $-$1.2, $+$1.2 &     $-12$, $+$14 \\

 [0.526, 0.565] &      0.545 &                     $3.56\times 10^{4}$ &   $-$1.5, $+$1.5 &     $-12$, $+$14 \\

 [0.565, 0.606] &      0.585 &                     $2.31\times 10^{4}$ &   $-$1.8, $+$1.8 &     $-13$, $+$15 \\

 [0.606, 0.649] &      0.627 &                     $1.60\times 10^{4}$ &   $-$2.1, $+$2.1 &     $-13$, $+$15 \\

 [0.649, 0.693] &      0.670 &                     $1.04\times 10^{4}$ &   $-$1.6, $+$1.6 &     $-13$, $+$15 \\

 [0.693, 0.740] &      0.716 &                     $7.20\times 10^{3}$ &   $-$1.8, $+$1.9 &     $-13$, $+$16 \\

 [0.740, 0.788] &      0.763 &                     $4.98\times 10^{3}$ &   $-$2.2, $+$2.2 &     $-14$, $+$16 \\

 [0.788, 0.838] &      0.812 &                     $3.35\times 10^{3}$ &   $-$2.6, $+$2.7 &     $-14$, $+$16 \\

 [0.838, 0.890] &      0.863 &                     $2.34\times 10^{3}$ &   $-$2.0, $+$2.1 &     $-14$, $+$17 \\

 [0.890, 0.944] &      0.916 &                     $1.65\times 10^{3}$ &   $-$2.4, $+$2.5 &     $-15$, $+$17 \\

 [0.944, 1.000] &      0.971 &                     $1.18\times 10^{3}$ &   $-$2.8, $+$2.9 &     $-15$, $+$18 \\

 [1.000, 1.118] &      1.055 &                     $6.61\times 10^{2}$ &   $-$2.6, $+$2.6 &     $-16$, $+$19 \\

 [1.118, 1.246] &      1.178 &                     $3.22\times 10^{2}$ &   $-$2.6, $+$2.7 &     $-16$, $+$20 \\

 [1.246, 1.383] &      1.310 &                     $1.57\times 10^{2}$ &   $-$3.6, $+$3.7 &     $-17$, $+$21 \\

 [1.383, 1.530] &      1.452 &                     $7.86\times 10^{1}$ &   $-$4.9, $+$5.1 &     $-18$, $+$22 \\

 [1.530, 1.687] &      1.603 &                     $3.80\times 10^{1}$ &   $-$6.8, $+$7.3 &     $-19$, $+$24 \\

 [1.687, 1.856] &      1.766 &                     $1.75\times 10^{1}$ &   $-$9.6, $+$11 &     $-20$, $+$26 \\

 [1.856, 2.037] &      1.941 &                     $8.32$ &   $-$13, $+$15 &     $-22$, $+$28 \\

 [2.037, 2.332] &      2.170 &                     $3.33$ &   $-$17, $+$20 &     $-23$, $+$31 \\

 [2.332, 2.659] &      2.478 &                     $1.83$ &   $-$21, $+$26 &     $-26$, $+$35 \\

 [2.659, 3.019] &      2.819 &                     $4.51\times 10^{-1}$ &   $-$40, $+$60 &     $-29$, $+$41 \\
\hline
\end{tabular}
\end{center}
\end{table*}

\begin{table*}[htbH]
\begin{center}
\caption{Double-differential dijet mass cross section in the rapidity range $1.5<|y|_\text{max}<2.0$. The reference mass is the point at which the cross section is drawn in Figures \ref{fig:Xsec} and \ref{fig:data_over_theory} and is calculated as described in the text. The experimental systematic uncertainties of the individual dijet-mass bins are almost 100\% correlated \cite{DijetPaper}.
 \label{tab:ybin3}}
\capspace
\scriptsize
%\footnotesize
\begin{tabular}{|c|c|c|c|c|}
\hline
Mass Range & Reference Mass & Measured Cross Section & Statistical Uncertainty & Systematic Uncertainty \\
$(TeV)$ & $(TeV)$ & $(\text{pb}/TeV)$ & \% & \% \\
\hline
\hline
 [0.565, 0.606] &      0.585 &                     $6.68\times 10^{4}$ &   $-$3.1, $+$3.2 &     $-12$, $+$14 \\

 [0.606, 0.649] &      0.627 &                     $4.52\times 10^{4}$ &   $-$3.7, $+$3.9 &     $-12$, $+$14 \\

 [0.649, 0.693] &      0.670 &                     $3.05\times 10^{4}$ &   $-$4.5, $+$4.7 &     $-12$, $+$14 \\

 [0.693, 0.740] &      0.716 &                     $2.02\times 10^{4}$ &   $-$1.7, $+$1.7 &     $-13$, $+$15 \\

 [0.740, 0.788] &      0.763 &                     $1.47\times 10^{4}$ &   $-$2.0, $+$2.0 &     $-13$, $+$15 \\

 [0.788, 0.838] &      0.812 &                     $1.04\times 10^{4}$ &   $-$2.3, $+$2.4 &     $-13$, $+$15 \\

 [0.838, 0.890] &      0.863 &                     $6.92\times 10^{3}$ &   $-$2.8, $+$2.8 &     $-13$, $+$16 \\

 [0.890, 0.944] &      0.916 &                     $4.77\times 10^{3}$ &   $-$2.1, $+$2.1 &     $-14$, $+$16 \\

 [0.944, 1.000] &      0.971 &                     $3.41\times 10^{3}$ &   $-$2.4, $+$2.4 &     $-14$, $+$16 \\

 [1.000, 1.118] &      1.055 &                     $2.04\times 10^{3}$ &   $-$2.1, $+$2.2 &     $-14$, $+$17 \\

 [1.118, 1.246] &      1.178 &                     $1.04\times 10^{3}$ &   $-$2.0, $+$2.0 &     $-15$, $+$18 \\

 [1.246, 1.383] &      1.310 &                     $5.20\times 10^{2}$ &   $-$2.7, $+$2.7 &     $-16$, $+$19 \\

 [1.383, 1.530] &      1.452 &                     $2.60\times 10^{2}$ &   $-$3.6, $+$3.8 &     $-17$, $+$21 \\

 [1.530, 1.687] &      1.604 &                     $1.45\times 10^{2}$ &   $-$4.7, $+$5.0 &     $-18$, $+$22 \\

 [1.687, 1.856] &      1.766 &                     $6.31\times 10^{1}$ &   $-$5.1, $+$5.3 &     $-19$, $+$24 \\

 [1.856, 2.037] &      1.941 &                     $3.24\times 10^{1}$ &   $-$6.8, $+$7.3 &     $-21$, $+$26 \\

 [2.037, 2.332] &      2.170 &                     $1.18\times 10^{1}$ &   $-$8.9, $+$9.7 &     $-23$, $+$29 \\

 [2.332, 2.659] &      2.479 &                     $4.37$ &   $-$14, $+$16 &     $-26$, $+$34 \\

 [2.659, 3.019] &      2.820 &                     $1.52$ &   $-$22, $+$28 &     $-29$, $+$41 \\

 [3.019, 3.854] &      3.344 &                     $1.31\times 10^{-1}$ &   $-$48, $+$79 &     $-35$, $+$54 \\
\hline
\end{tabular}
\end{center}
\end{table*}

\begin{table*}[htbH]
\begin{center}
\caption{Double-differential dijet mass cross section in the rapidity range $2.0<|y|_\text{max}<2.5$. The reference mass is the point at which the cross section is drawn in Figures \ref{fig:Xsec} and \ref{fig:data_over_theory} and is calculated as described in the text. The experimental systematic uncertainties of the individual dijet-mass bins are almost 100\% correlated \cite{DijetPaper}.
\label{tab:ybin4}}
\capspace
\scriptsize
%\footnotesize
\begin{tabular}{|c|c|c|c|c|}
\hline
Mass Range & Reference Mass & Measured Cross Section & Statistical Uncertainty & Systematic Uncertainty \\
$(TeV)$ & $(TeV)$ & $(\text{pb}/TeV)$ & \% & \% \\
\hline
\hline
 [0.649, 0.693] &      0.670 &                     $1.00\times 10^{5}$ &   $-$2.5, $+$2.5 &     $-13$, $+$15 \\

 [0.693, 0.740] &      0.716 &                     $6.70\times 10^{4}$ &   $-$2.9, $+$3.0 &     $-13$, $+$15 \\

 [0.740, 0.788] &      0.763 &                     $4.63\times 10^{4}$ &   $-$3.5, $+$3.6 &     $-13$, $+$15 \\

 [0.788, 0.838] &      0.812 &                     $3.29\times 10^{4}$ &   $-$4.1, $+$4.2 &     $-13$, $+$15 \\

 [0.838, 0.890] &      0.863 &                     $2.31\times 10^{4}$ &   $-$4.8, $+$5.0 &     $-13$, $+$16 \\

 [0.890, 0.944] &      0.916 &                     $1.52\times 10^{4}$ &   $-$1.8, $+$1.9 &     $-14$, $+$16 \\

 [0.944, 1.000] &      0.971 &                     $1.11\times 10^{4}$ &   $-$2.1, $+$2.2 &     $-14$, $+$16 \\

 [1.000, 1.118] &      1.055 &                     $6.41\times 10^{3}$ &   $-$1.9, $+$1.9 &     $-14$, $+$16 \\

 [1.118, 1.246] &      1.178 &                     $3.26\times 10^{3}$ &   $-$2.6, $+$2.6 &     $-14$, $+$17 \\

 [1.246, 1.383] &      1.310 &                     $1.59\times 10^{3}$ &   $-$2.2, $+$2.3 &     $-15$, $+$18 \\

 [1.383, 1.530] &      1.452 &                     $8.39\times 10^{2}$ &   $-$3.0, $+$3.1 &     $-16$, $+$18 \\

 [1.530, 1.687] &      1.604 &                     $4.01\times 10^{2}$ &   $-$4.2, $+$4.4 &     $-16$, $+$19 \\

 [1.687, 1.856] &      1.766 &                     $1.80\times 10^{2}$ &   $-$4.1, $+$4.2 &     $-17$, $+$21 \\

 [1.856, 2.037] &      1.941 &                     $8.96\times 10^{1}$ &   $-$5.6, $+$5.9 &     $-18$, $+$22 \\

 [2.037, 2.332] &      2.170 &                     $3.75\times 10^{1}$ &   $-$6.8, $+$7.2 &     $-19$, $+$24 \\

 [2.332, 2.659] &      2.479 &                     $9.44$ &   $-$9.4, $+$10 &     $-22$, $+$27 \\

 [2.659, 3.019] &      2.819 &                     $3.52$ &   $-$15, $+$17 &     $-25$, $+$32 \\

 [3.019, 3.854] &      3.338 &                     $3.29\times 10^{-1}$ &   $-$31, $+$43 &     $-30$, $+$41 \\
\hline
\end{tabular}
\end{center}
\end{table*}
\clearpage
\section{Data vs. Theory Comparison}
\begin{figure}[ht]
  \centering
  \includegraphics[width=0.70\textwidth]{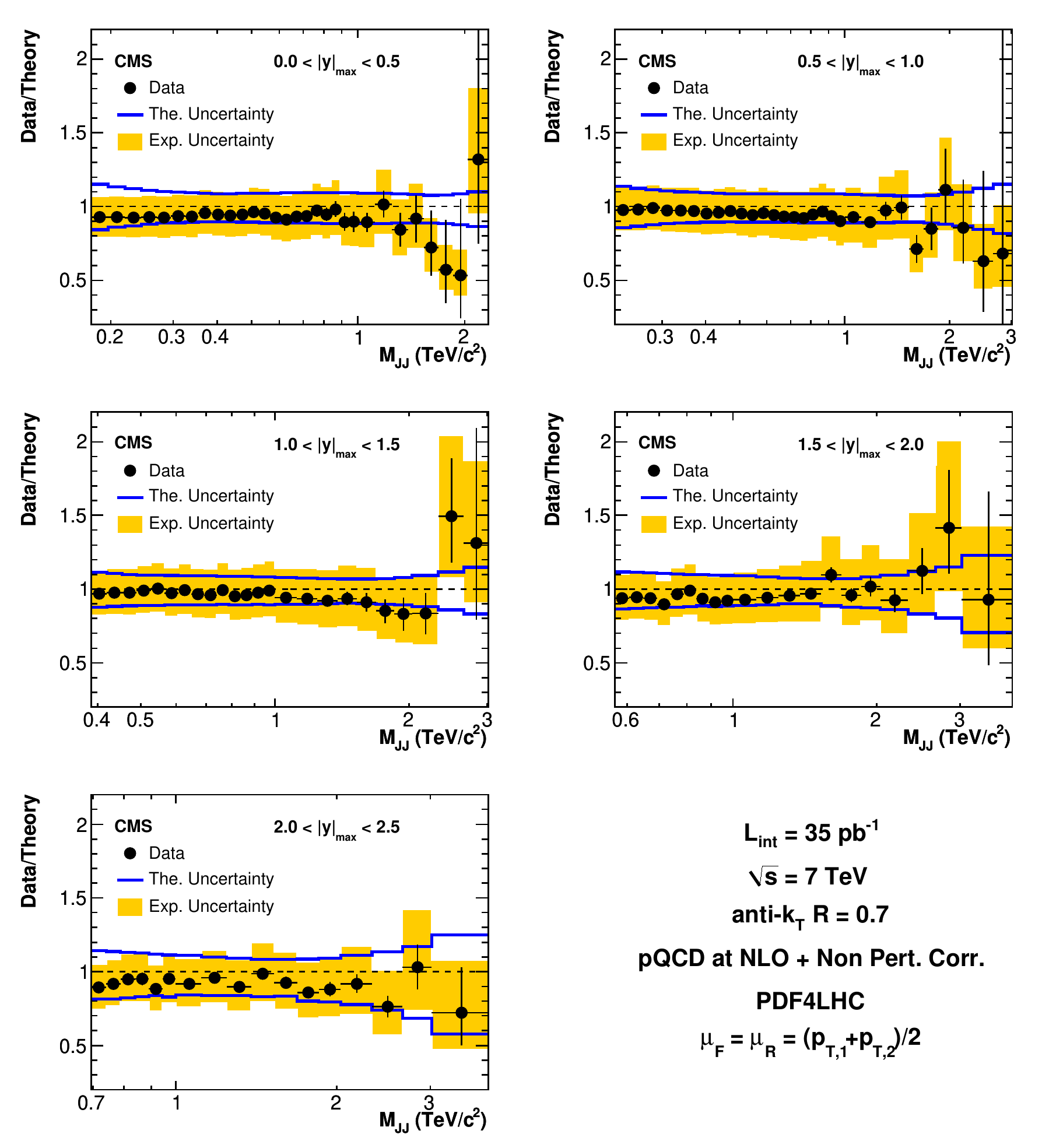}
  \capspace
  \caption{Ratio of the measured double differential dijet production cross-section over the theory prediction in the different rapidity bins. The solid band represents the experimental systematical uncertainty and is centered around the points. The error bars on the points represent the statistical uncertainty. The theoretical uncertainty is shown as lines centered around unity \cite{DijetPaper}.}
  \label{fig:data_over_theory}
\end{figure}

In Figure \ref{fig:data_over_theory} the ratio data/theory is shown superimposed with the experimental and theoretical uncertainties. Although the  experimental uncertainties are comparable to the theoretical uncertainties, they are not small enough to constrain the parameters in the theory. Nevertheless, an excellent agreement is observed, indicating that the QCD predictions describe the parton-parton scattering accurately in this kinematic regime.
\clearpage
\section{Conclusions}
The double differential dijet mass cross section was presented in this dissertation and is the most extensive measurement with the farthest range in both rapidity and energy to date. Good agreement between data and perturbative QCD was observed in all five rapidity regions which confirms the  Standard Model predictions. This measurement can then be used to reduce the uncertainties on the parton distribution functions. Such reduction will be useful for the future versions of Monte Carlo simulations and NLO calculation softwares. For most of the other analyses in particle physics, QCD events are often a major background, such as searches for the Higgs boson. Understanding the nature of the hard parton-parton scattering helps physicists to understand the background in their analyses.

\clearpage

\appendix      % Starts appendices
\chapter{Running Coupling Constant}
\label{AppendixA}
 In Section \ref{Running Coupling}, it was mentioned that the value of $\mu$ in Equation \ref{running_alpha} does not effect the result of the equation. It means that two different values of $\mu$ are simultaneously correct. In order to prove it let us start with the Equation \ref{running_alpha}.
 
\begin{equation}
\alpha_{s}(|q^{2}|)=\dfrac{\alpha_{s}(\mu^{2})}{1+\alpha_{s}(\mu^{2})~b~\ln(|q^{2}|/\mu^{2})}
\label{App:running_alpha}
\end{equation}

$\alpha_{s}(\mu^{2})$ term in the equation can be written as follows, referring to the original equation;

\begin{equation}
\alpha_{s}(\mu^{2})=\dfrac{\alpha_{s}(\mu_{0}^{2})}{1+\alpha_{s}(\mu_{0}^{2})~b~\ln(|q^{2}|/\mu_{0}^{2})}
\end{equation}

If we plug the equation above into Equation \ref{App:running_alpha}, we have

\begin{equation}
\alpha_{s}(|q^{2}|)=\dfrac{\dfrac{\alpha_{s}(\mu_{0}^{2})}{1+\alpha_{s}(\mu_{0}^{2})~b~\ln(|q^{2}|/\mu_{0}^{2})}}{1+\dfrac{\alpha_{s}(\mu_{0}^{2})}{1+\alpha_{s}(\mu_{0}^{2})~b~\ln(|q^{2}|/\mu_{0}^{2})}~b~\ln(|q^{2}|/\mu^{2})}
\label{App:running_alpha2}
\end{equation}

\begin{equation}
=\dfrac{\alpha_{s}(\mu_{0}^{2})}{1+\alpha_{s}(\mu_{0}^{2})~b~\ln(\mu_{0}/\mu^{2})+\alpha_{s}(\mu_{0}^{2})~b~\ln(|q^{2}|/\mu^{2})}=\dfrac{\alpha_{s}(\mu_{0}^{2})}{1+\alpha_{s}(\mu_{0}^{2})~b~\ln(|q^{2}|/\mu_{0}^{2})}
\end{equation}

\clearpage
\chapter{Trigger Efficiencies}
\begin{figure}[ht]
  \centering
  \includegraphics[width=0.355\textwidth]{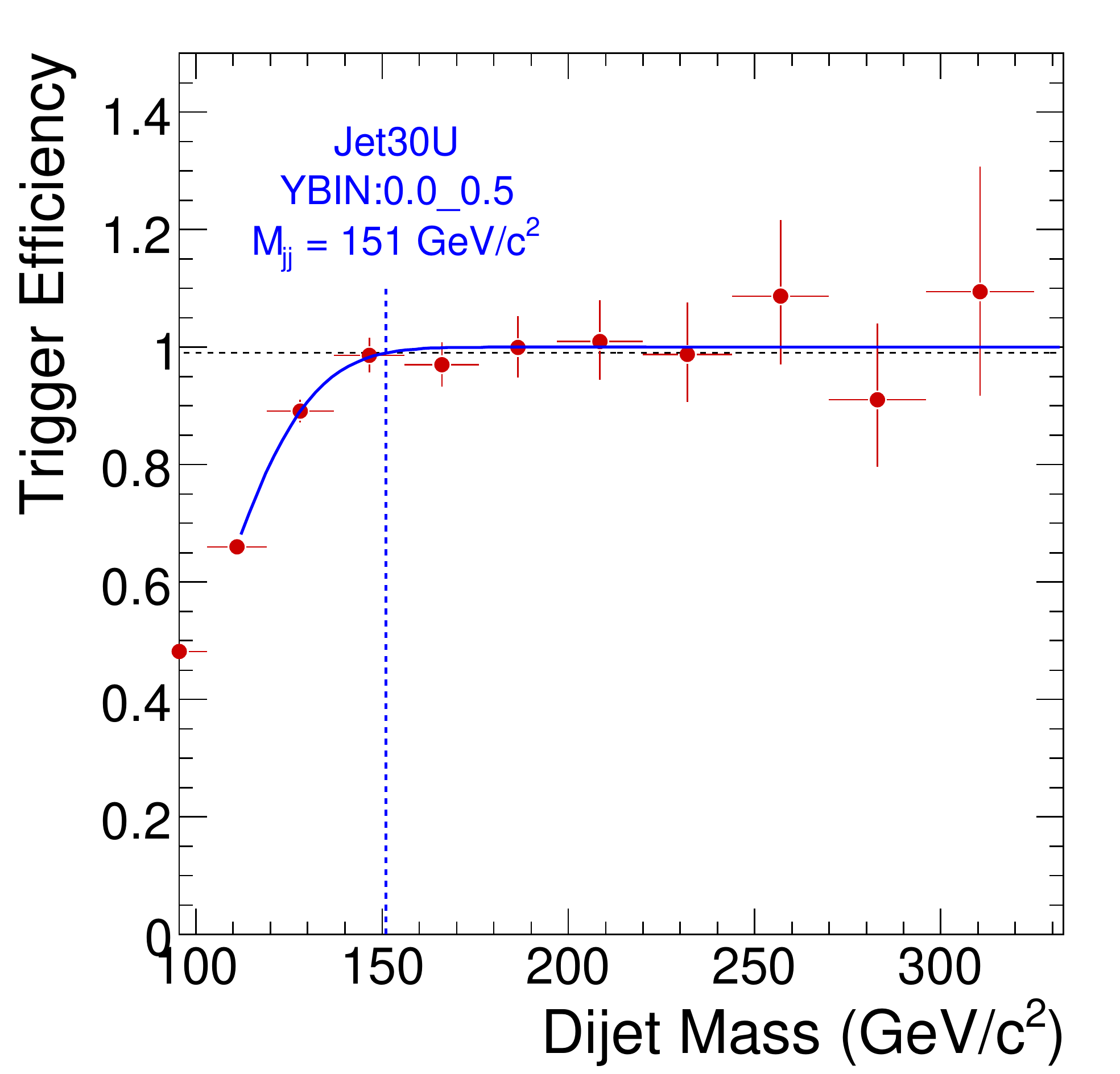}
  \includegraphics[width=0.355\textwidth]{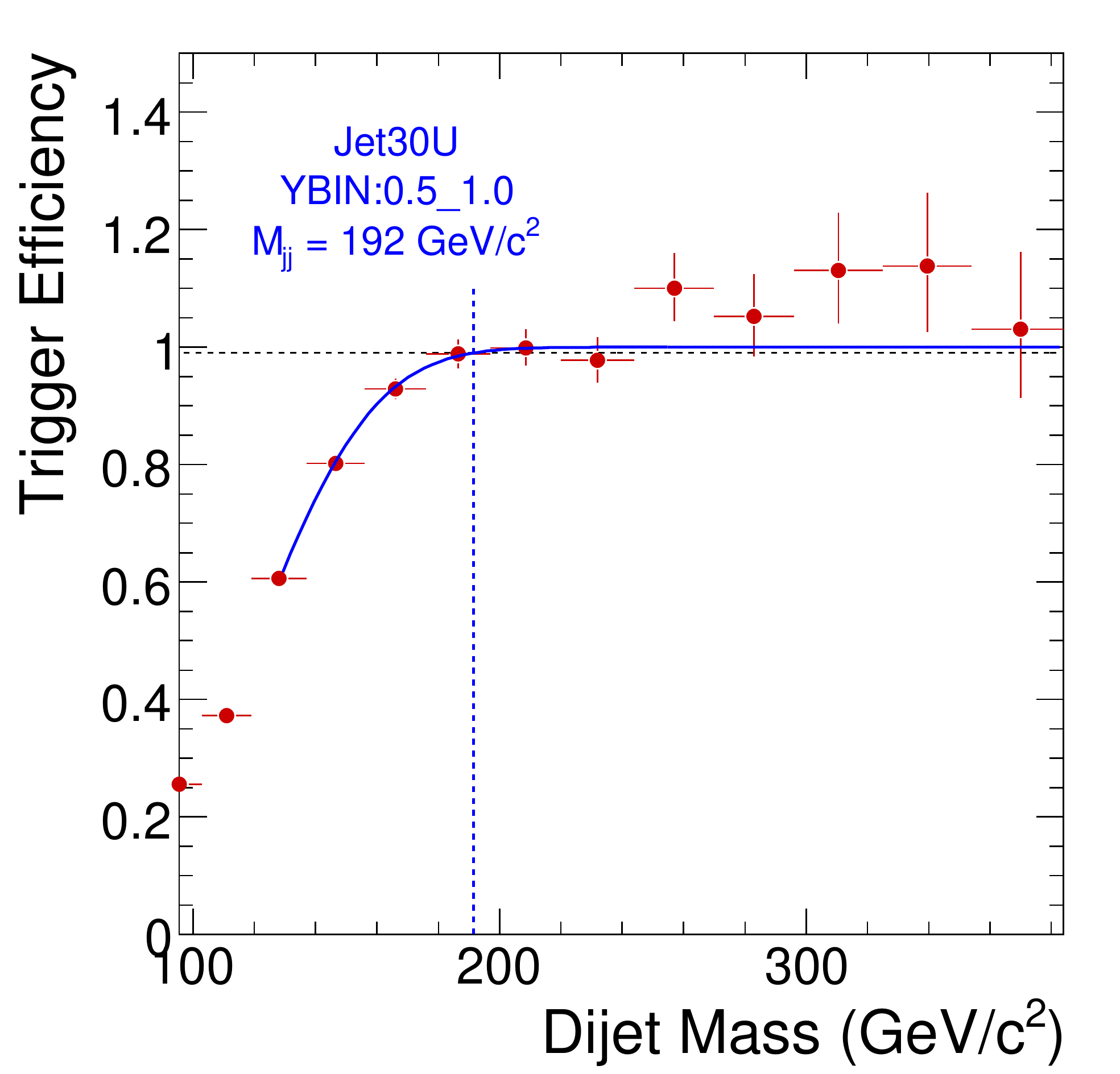} 
  \includegraphics[width=0.355\textwidth]{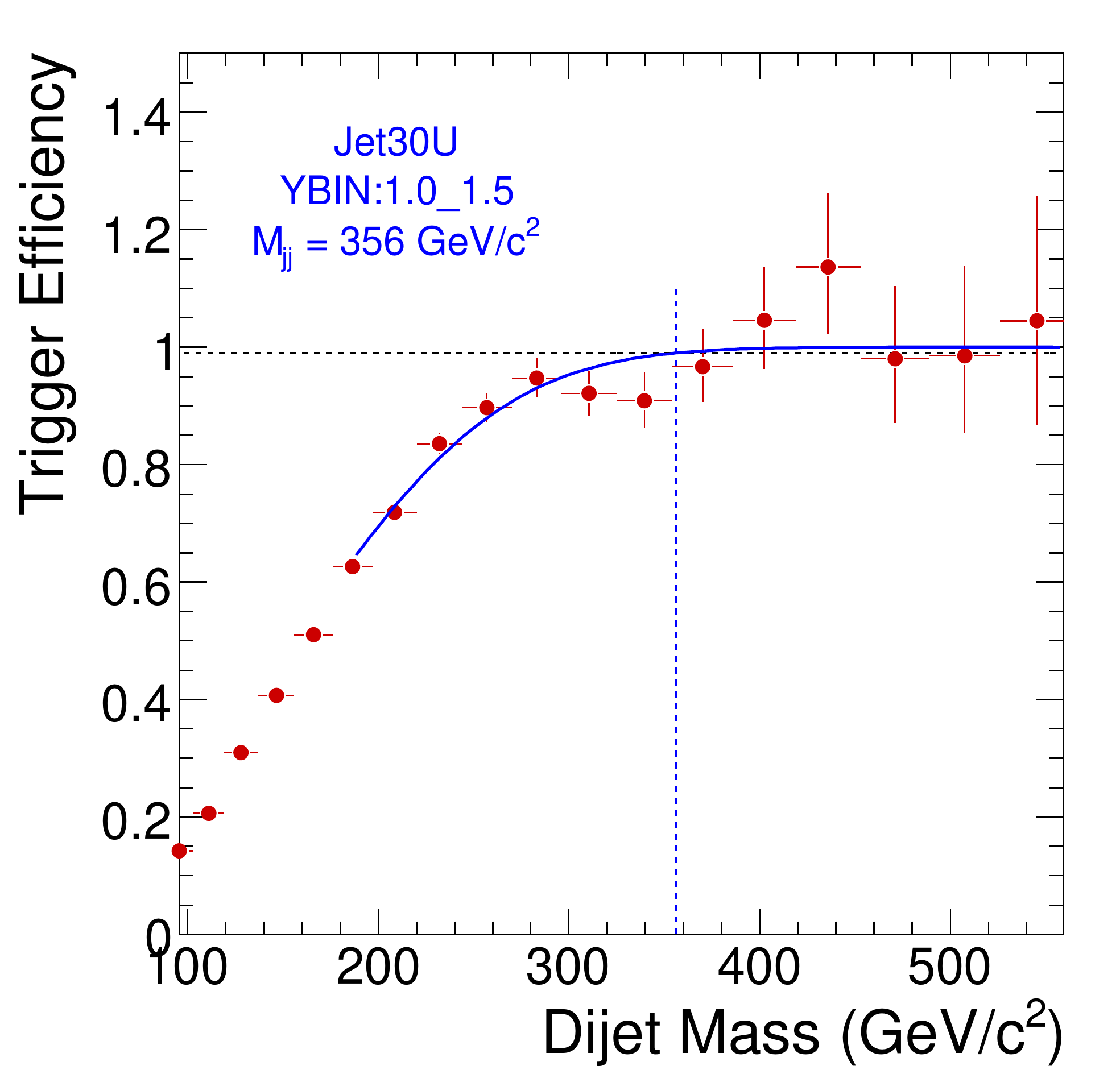} 
  \includegraphics[width=0.355\textwidth]{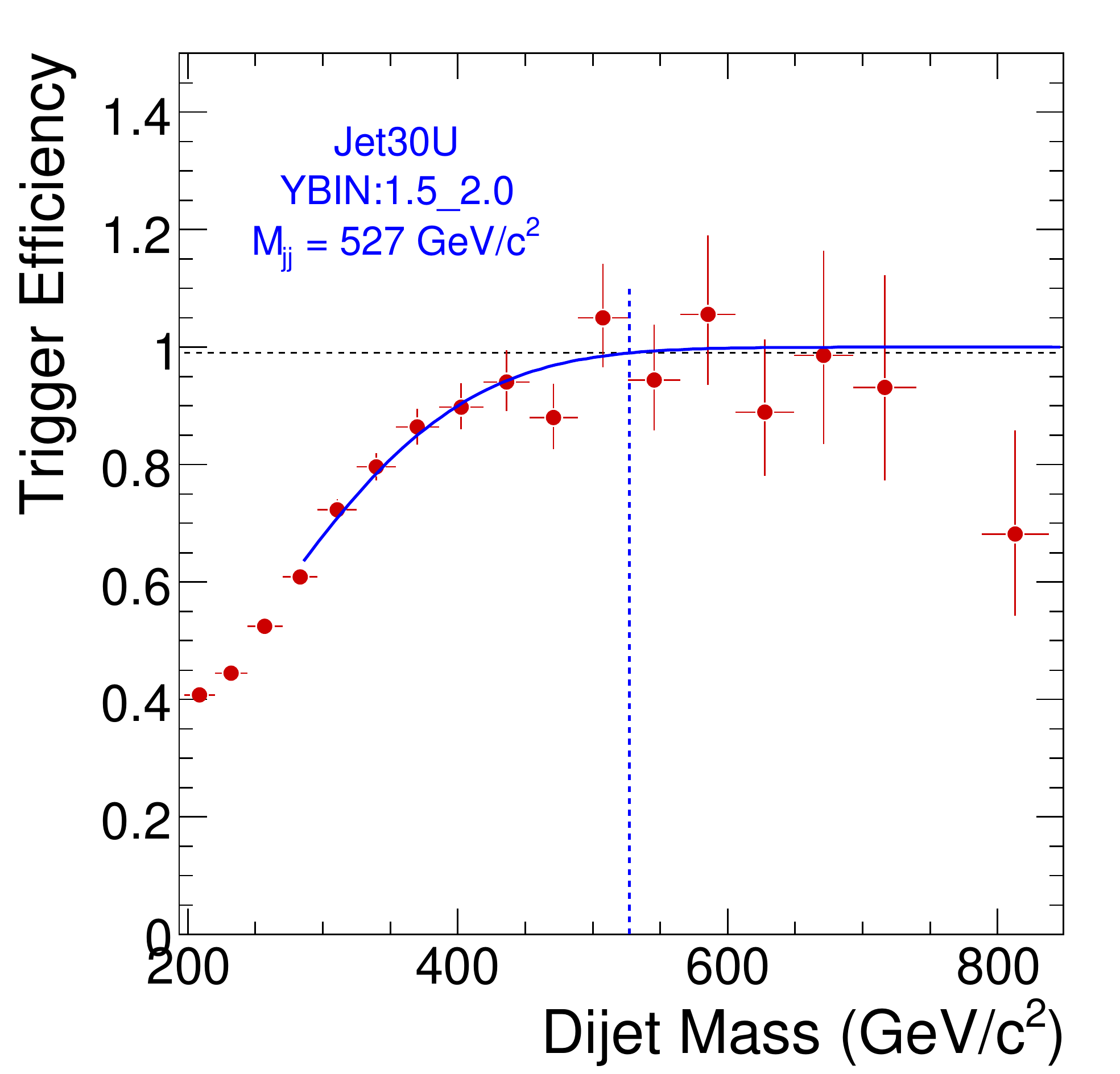}
  \includegraphics[width=0.355\textwidth]{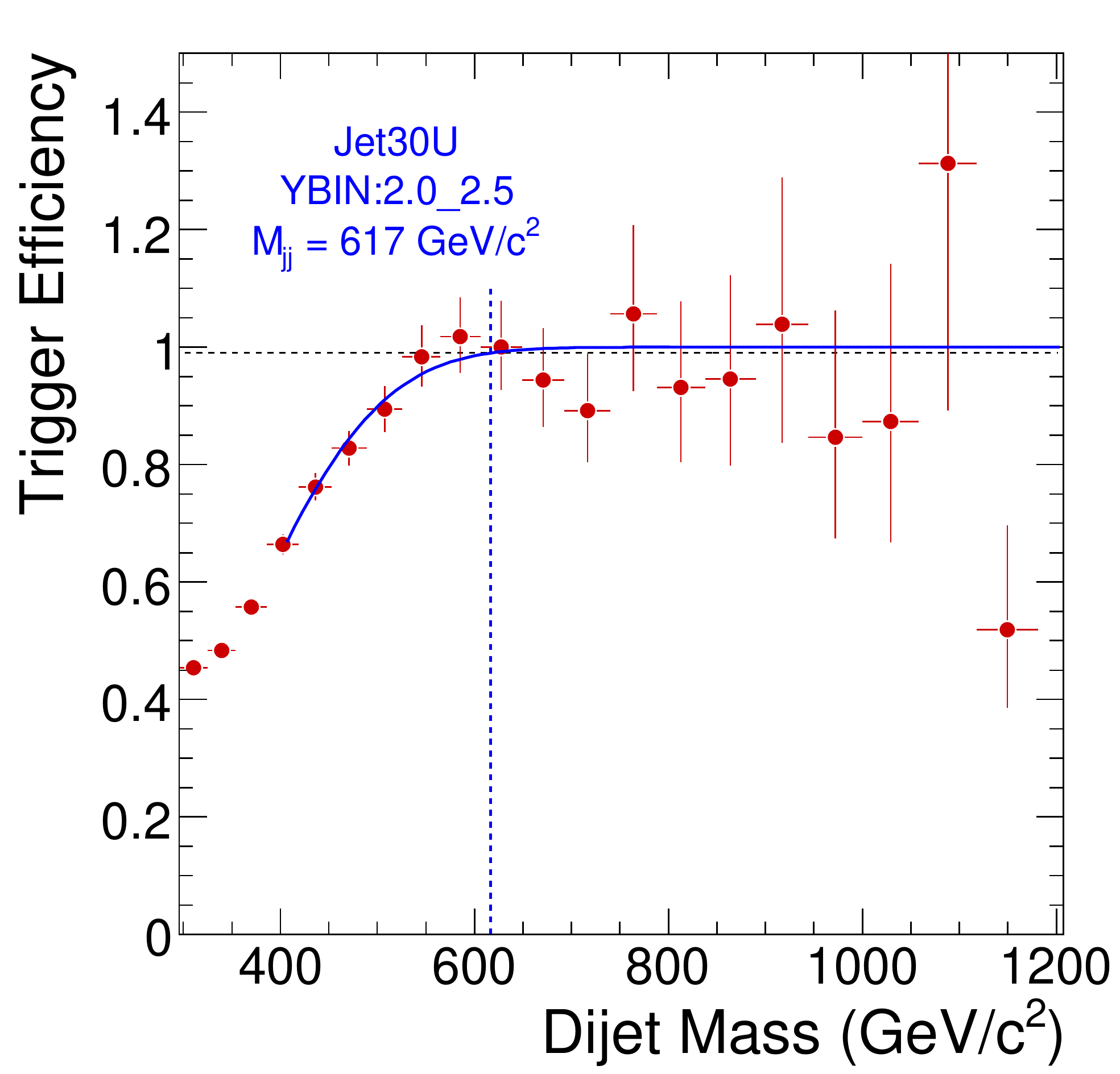}
  \capspace
  \caption{ Relative trigger efficiencies as a function of dijet mass for the five different $|y|_{max}$ bins and for the HLT\_Jet30U trigger. The  $100\%$ efficiency point is determined by performing a fit with an error function.}
  \label{fig_data_Appendix1}
\end{figure}

\begin{figure}[ht]
  \centering
  \includegraphics[width=0.48\textwidth]{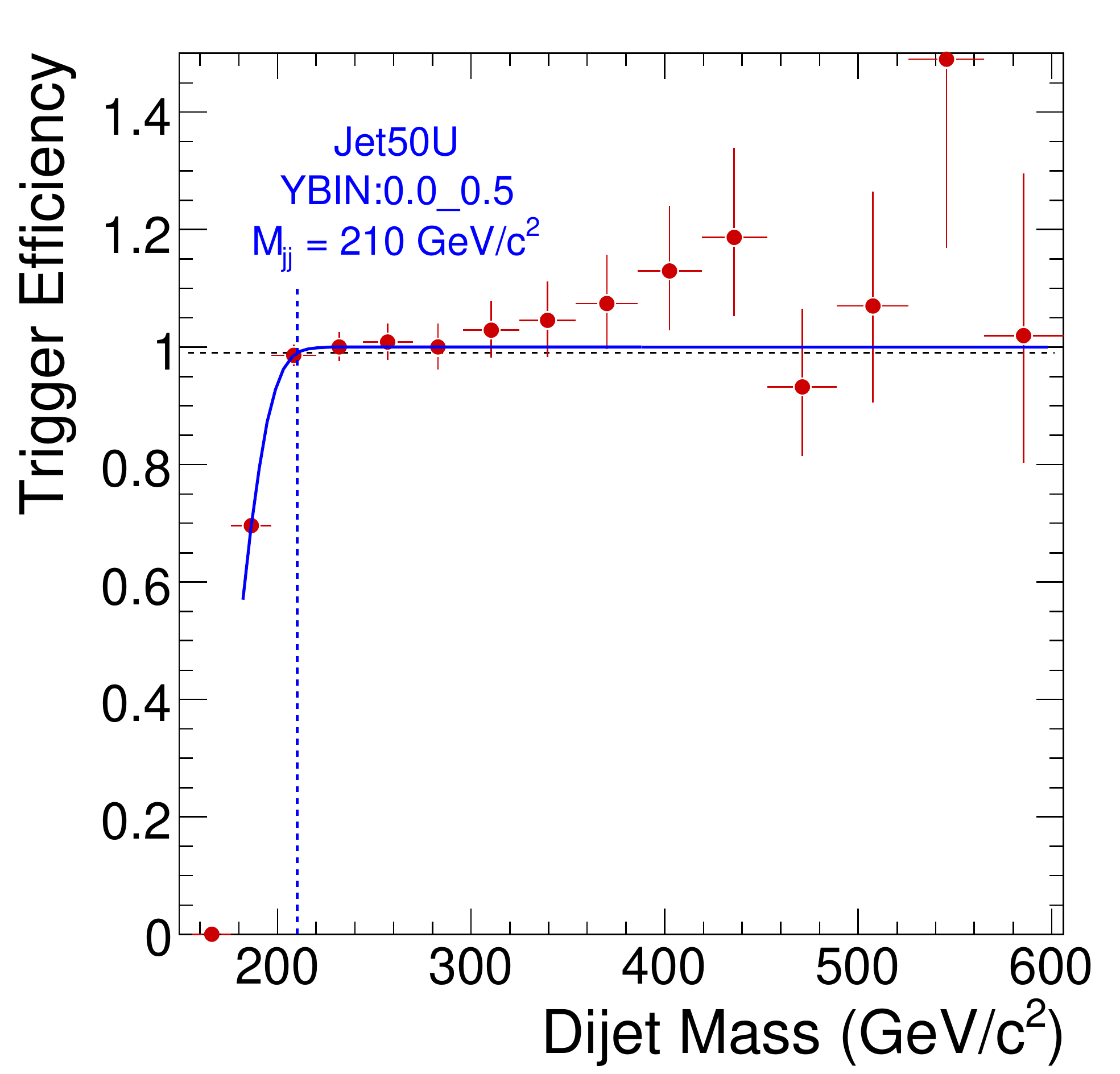}
  \includegraphics[width=0.48\textwidth]{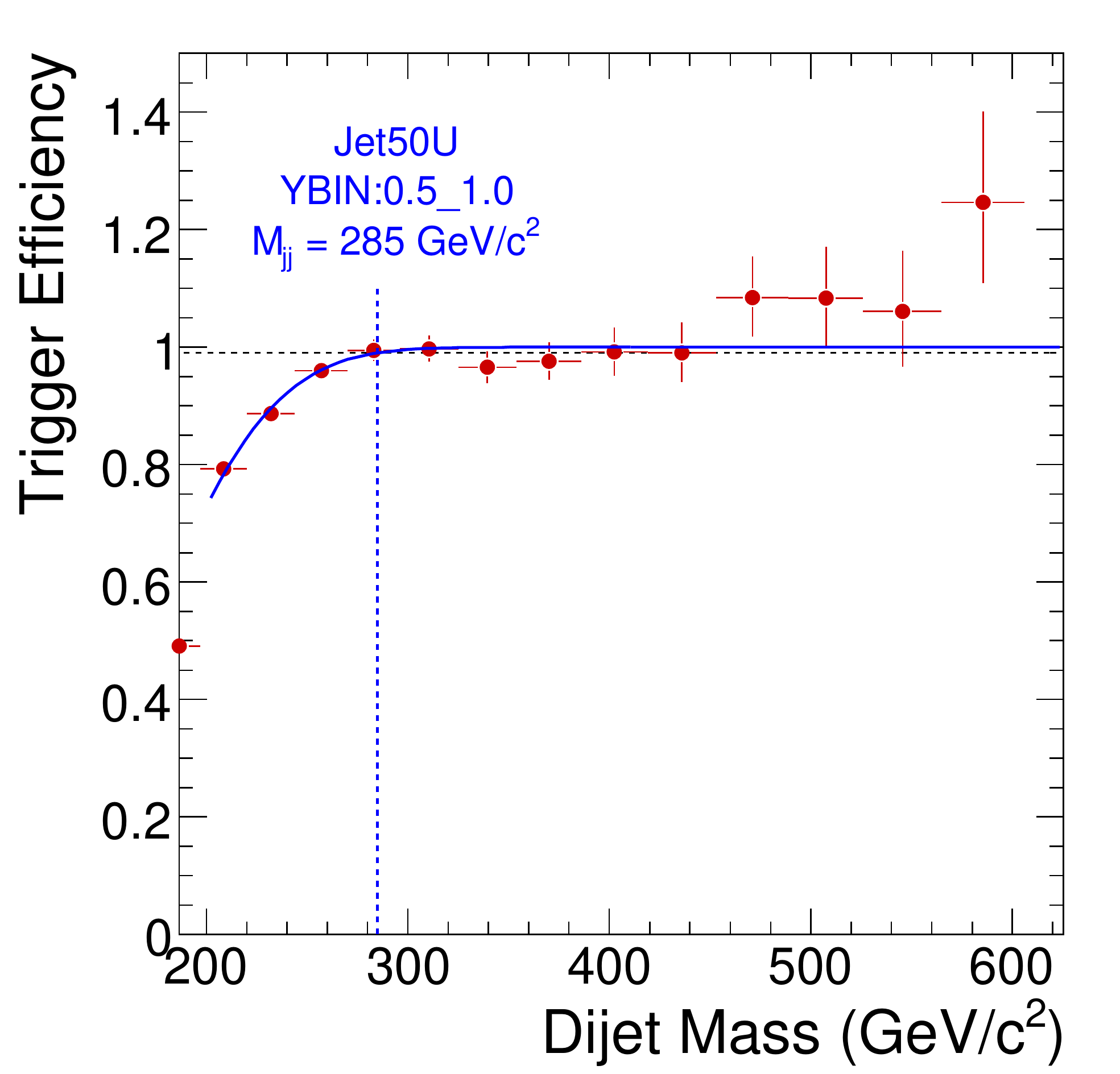} 
  \includegraphics[width=0.48\textwidth]{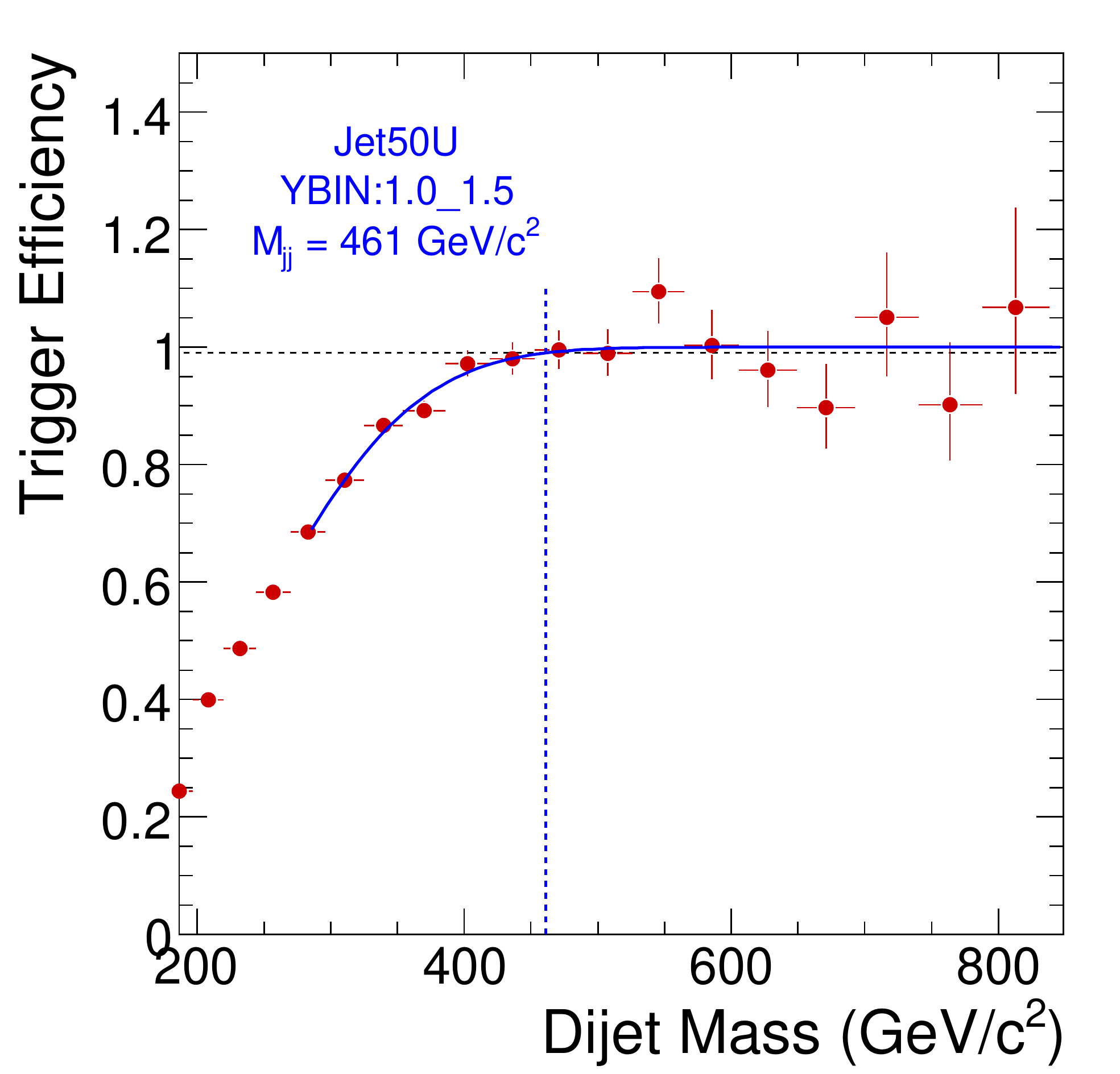} 
  \includegraphics[width=0.48\textwidth]{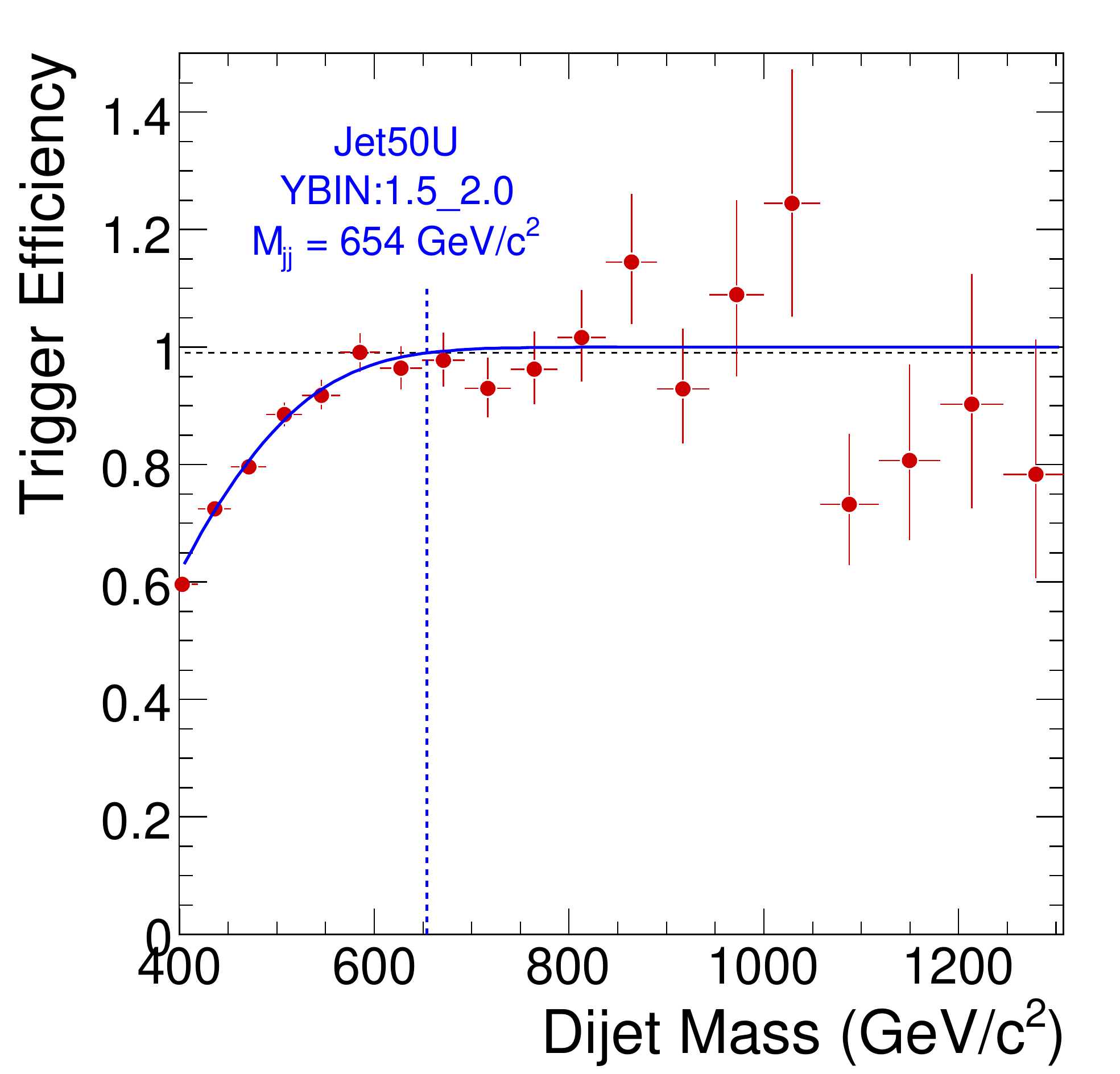}
  \includegraphics[width=0.48\textwidth]{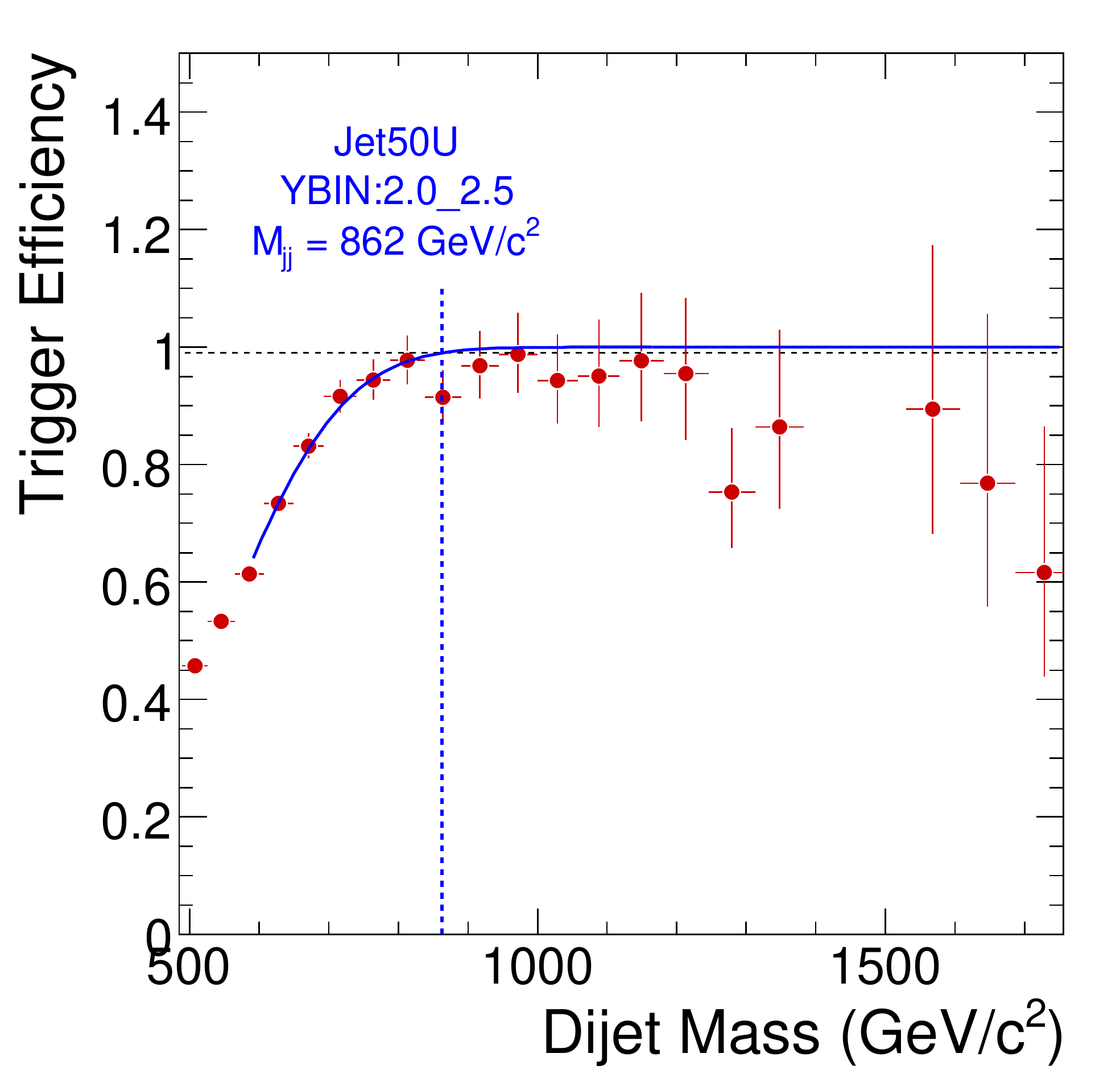}
  \capspace
  \caption{ Relative trigger efficiencies as a function of dijet mass for the five different $|y|_{max}$ bins and for the HLT\_Jet50U trigger. The  $100\%$ efficiency point is determined by performing a fit with an error function.}
  \label{fig_data_Appendix2}
\end{figure}

\begin{figure}[ht]
  \centering
  \includegraphics[width=0.48\textwidth]{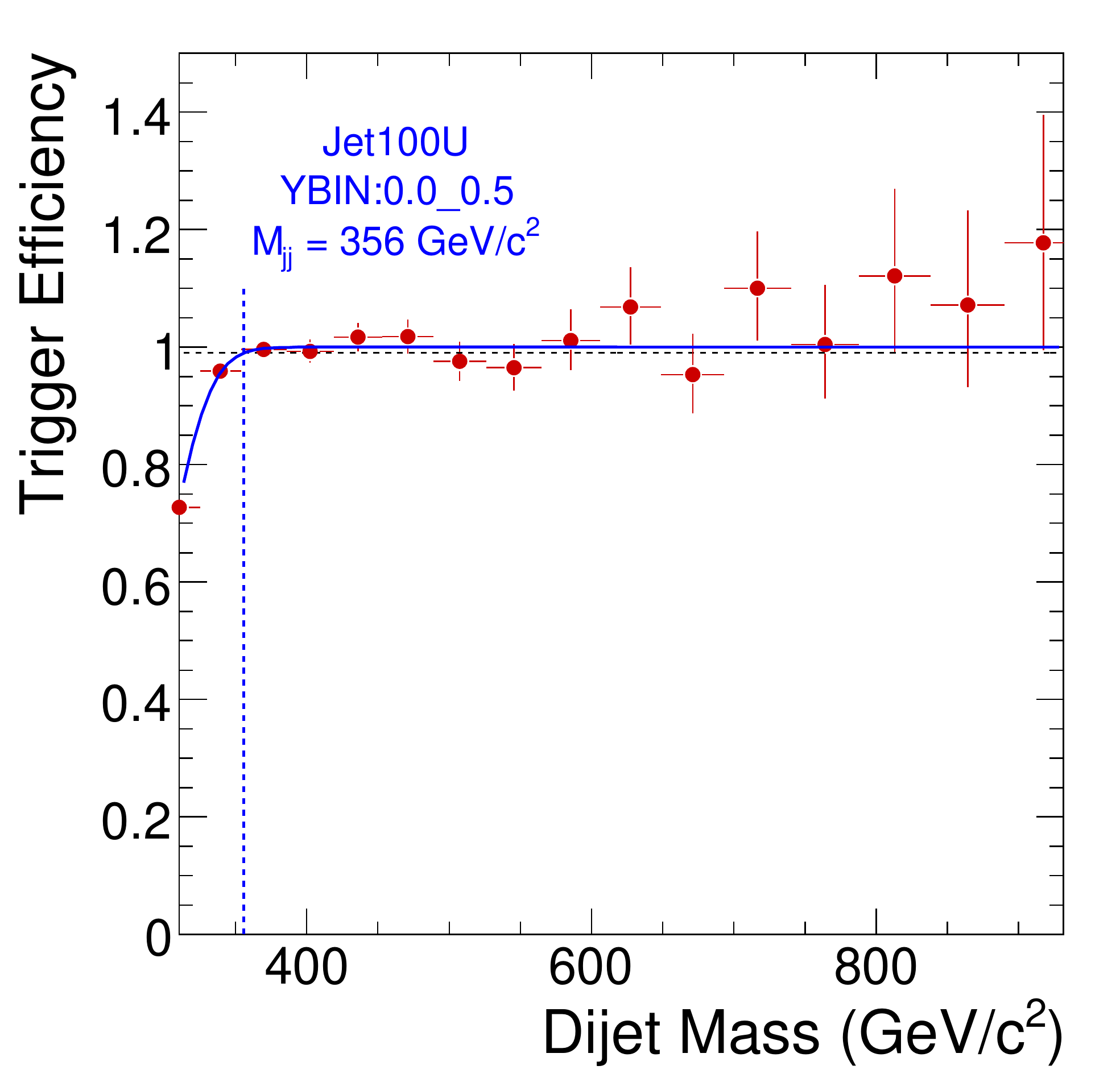}
  \includegraphics[width=0.48\textwidth]{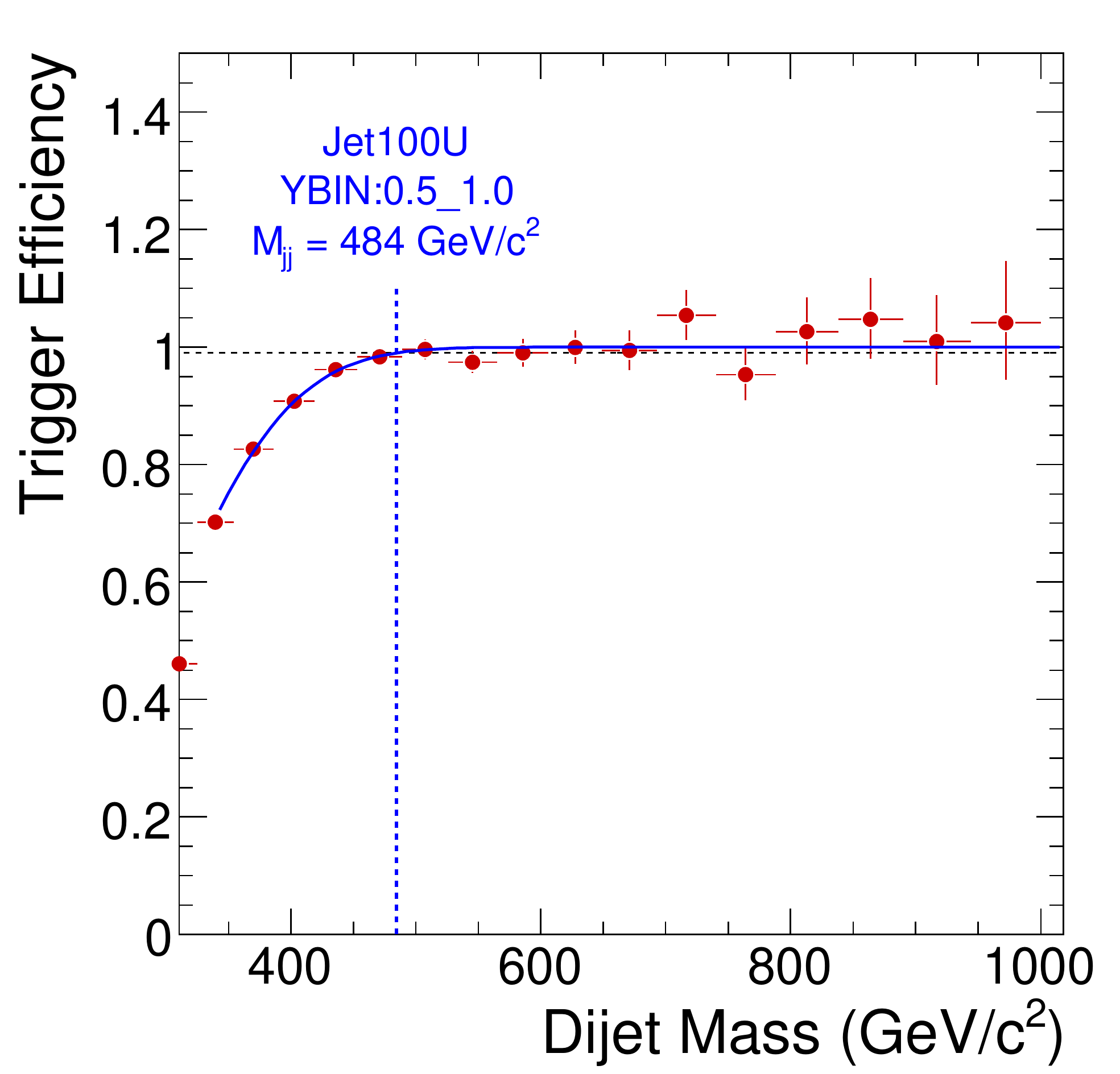} 
  \includegraphics[width=0.48\textwidth]{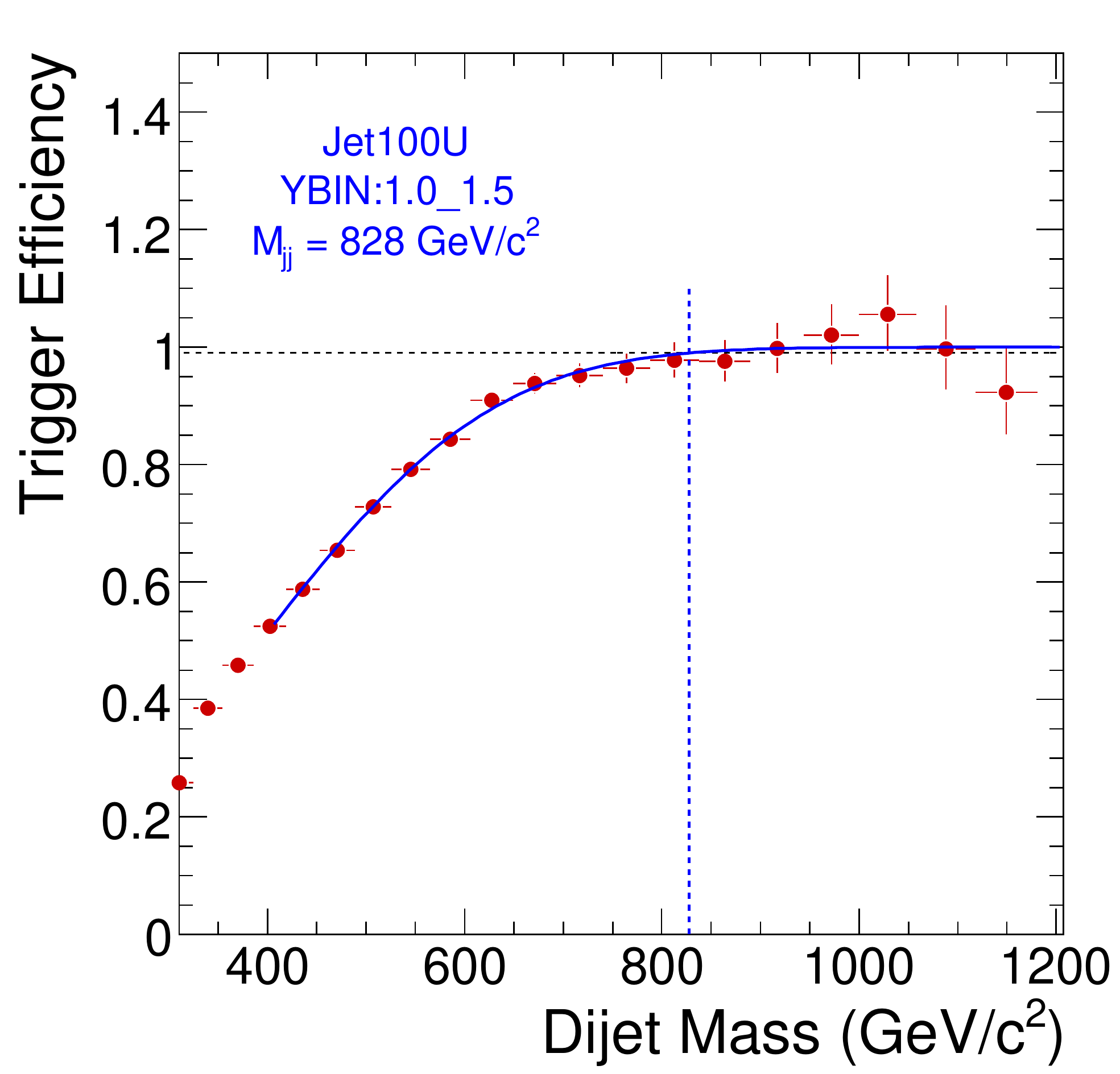} 
  \includegraphics[width=0.48\textwidth]{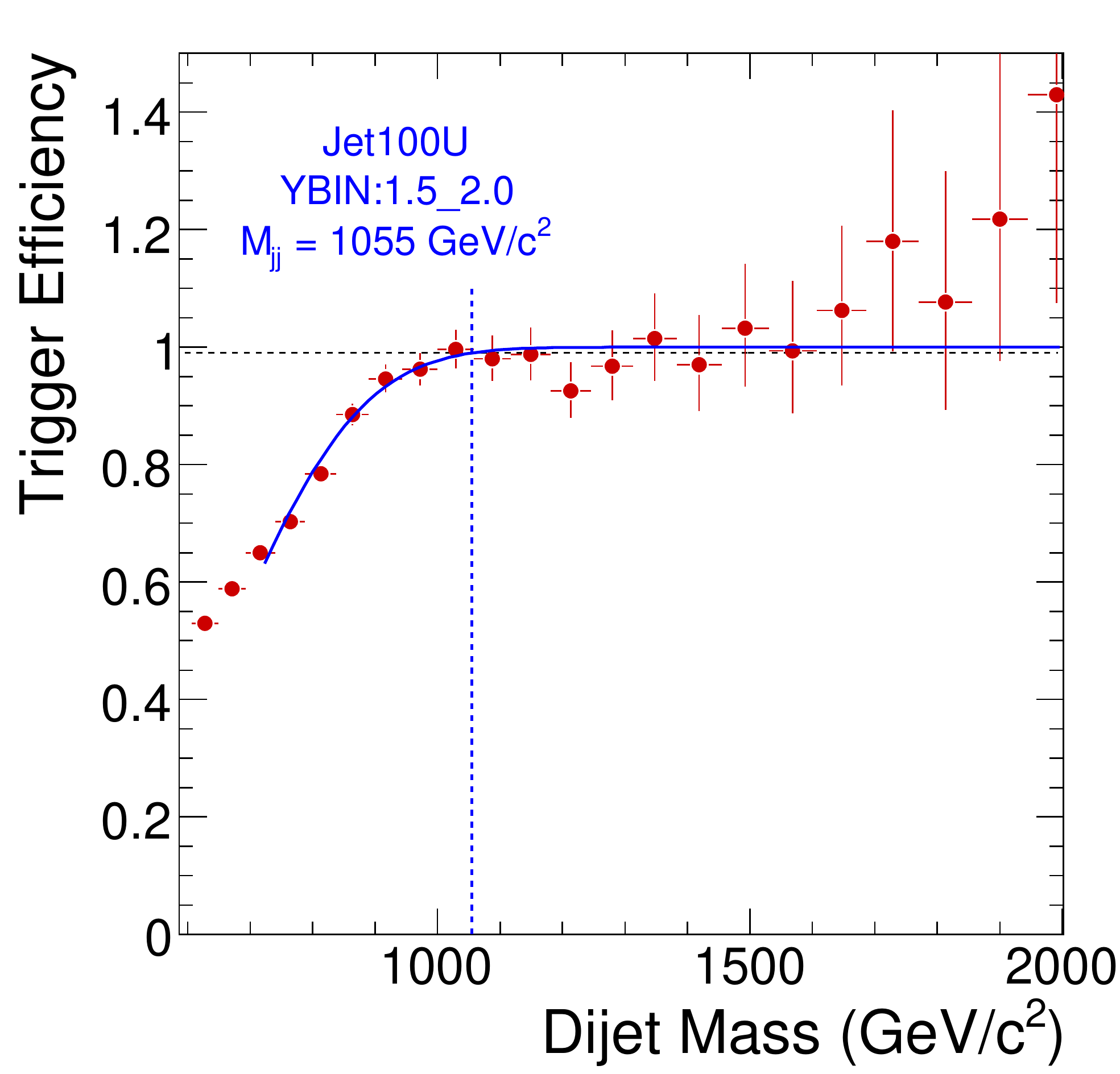}
  \includegraphics[width=0.48\textwidth]{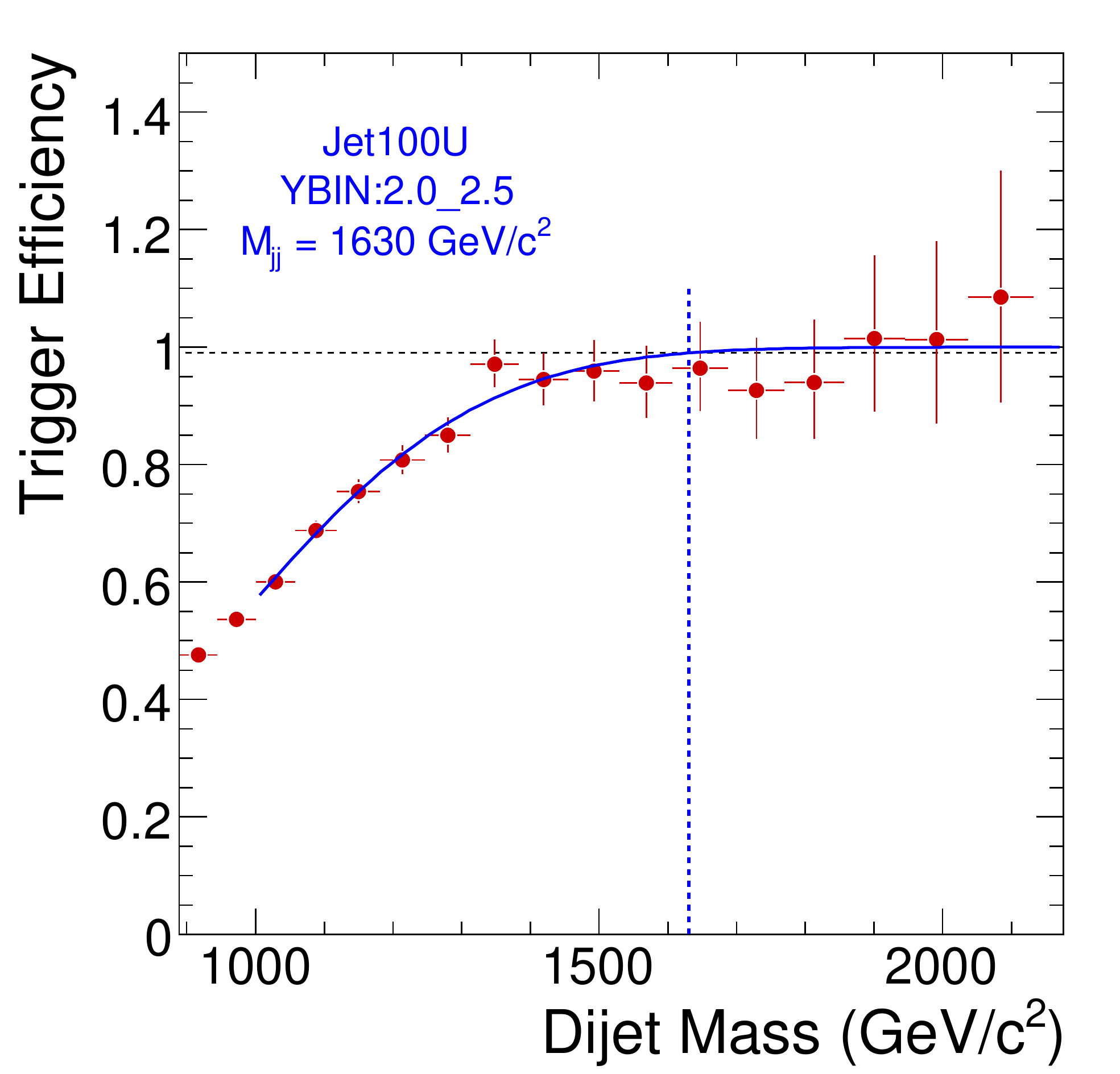}
  \capspace
  \caption{ Relative trigger efficiencies as a function of dijet mass for the five different $|y|_{max}$ bins and for the HLT\_Jet100U trigger. The  $100\%$ efficiency point is determined by performing a fit with an error function.}
  \label{fig_data_Appendix3}
\end{figure}

\begin{figure}[ht]
  \centering
  \includegraphics[width=0.48\textwidth]{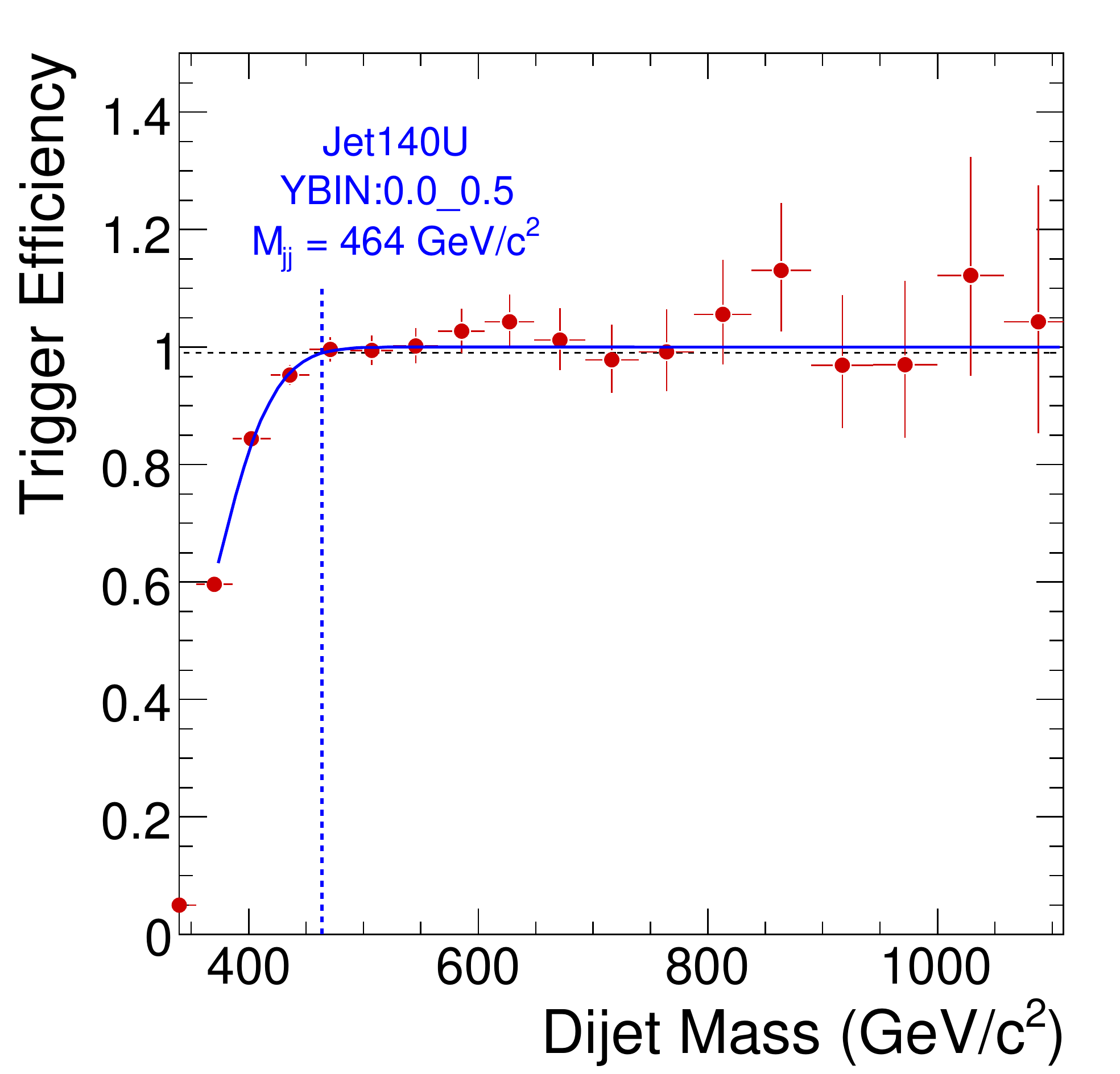}
  \includegraphics[width=0.48\textwidth]{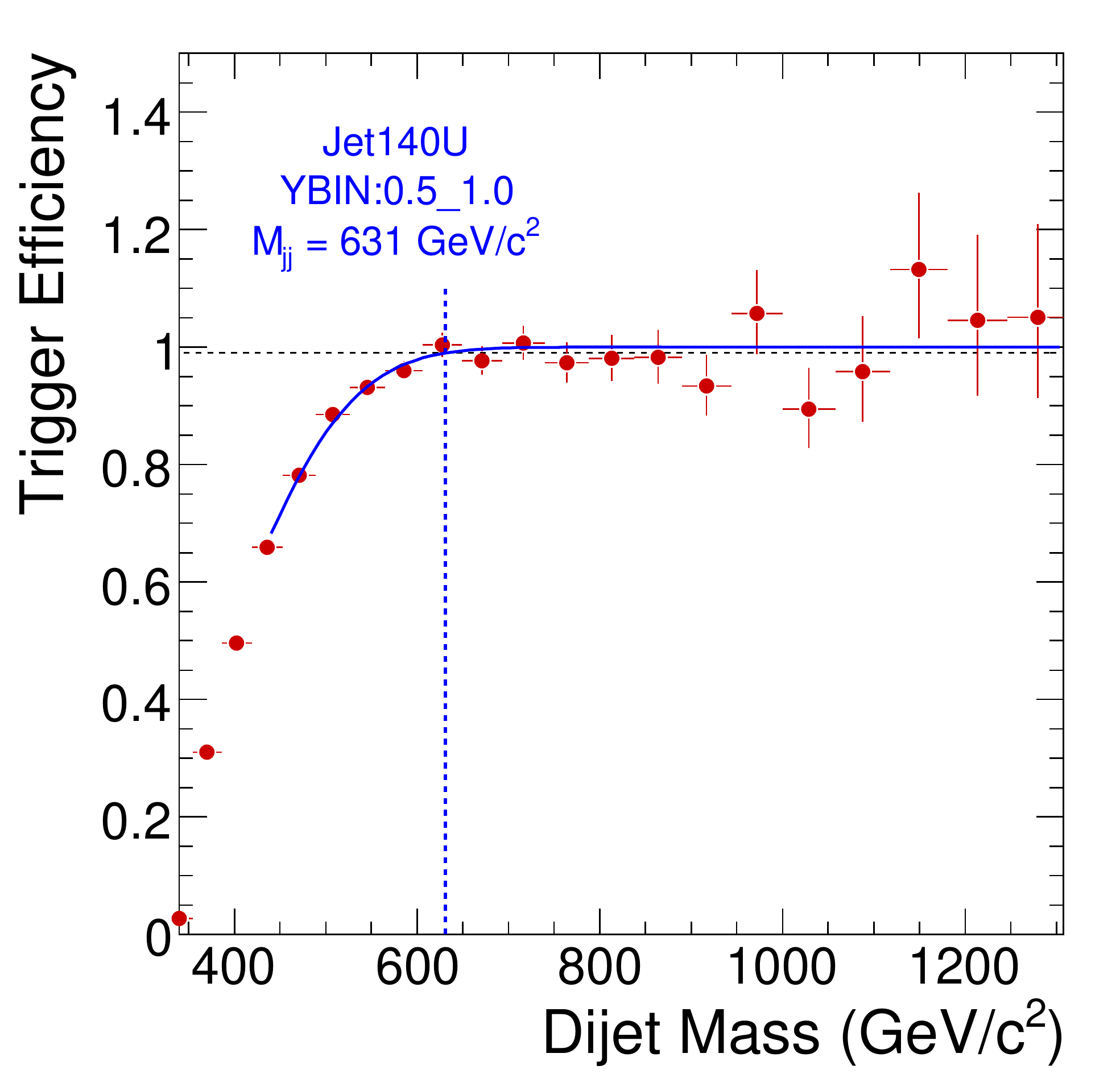} 
  \includegraphics[width=0.48\textwidth]{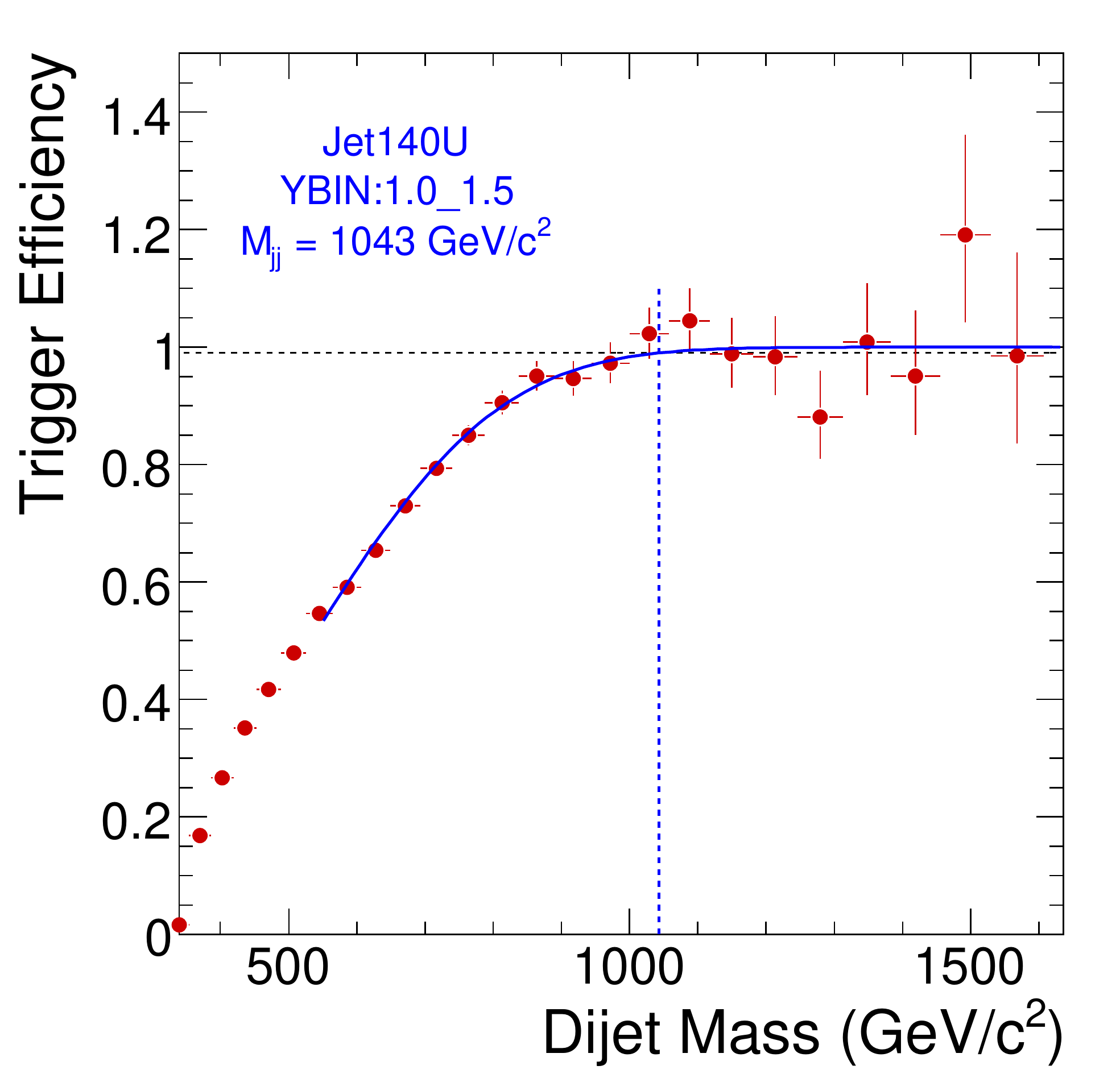} 
  \includegraphics[width=0.48\textwidth]{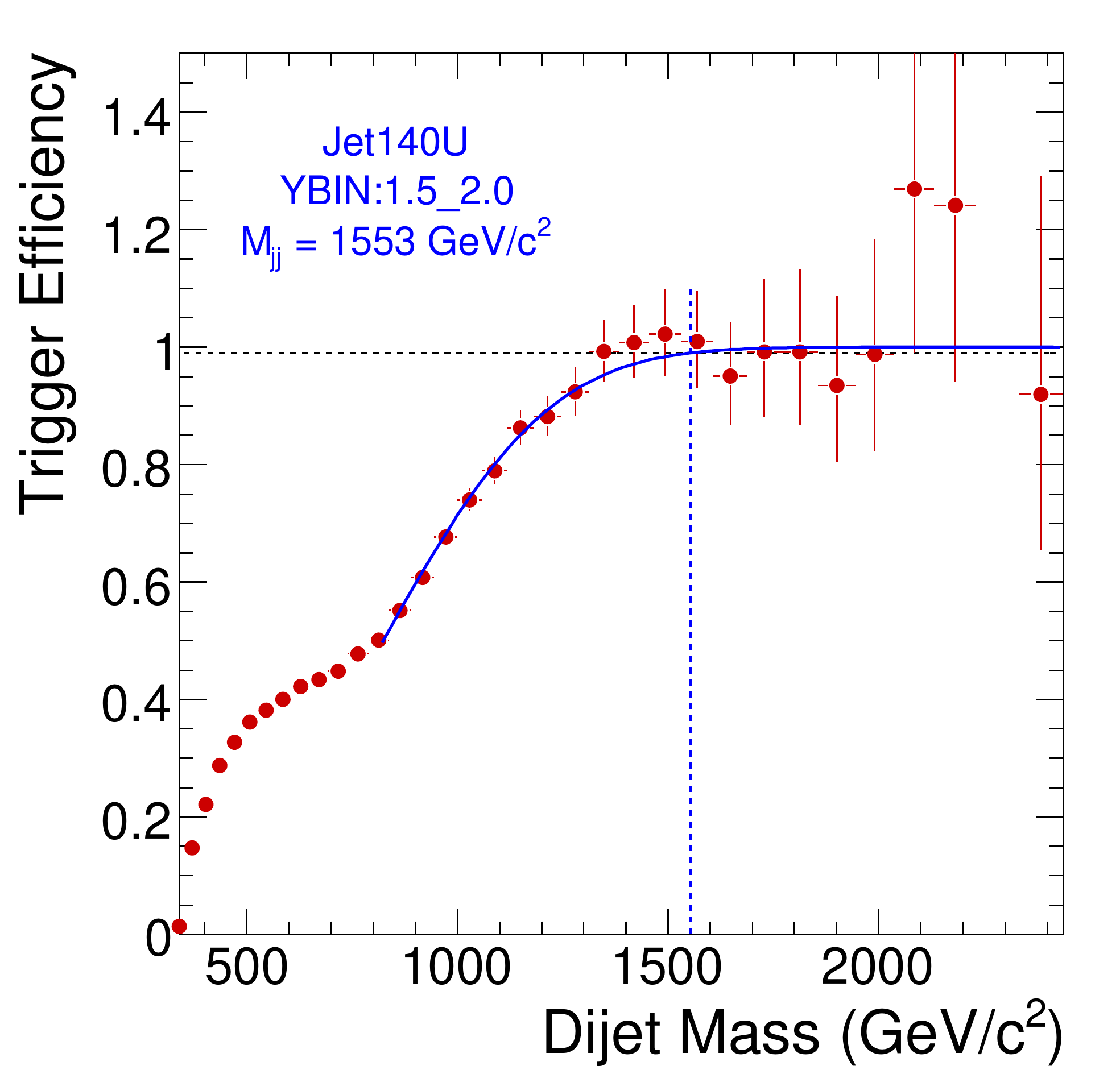}
  \includegraphics[width=0.48\textwidth]{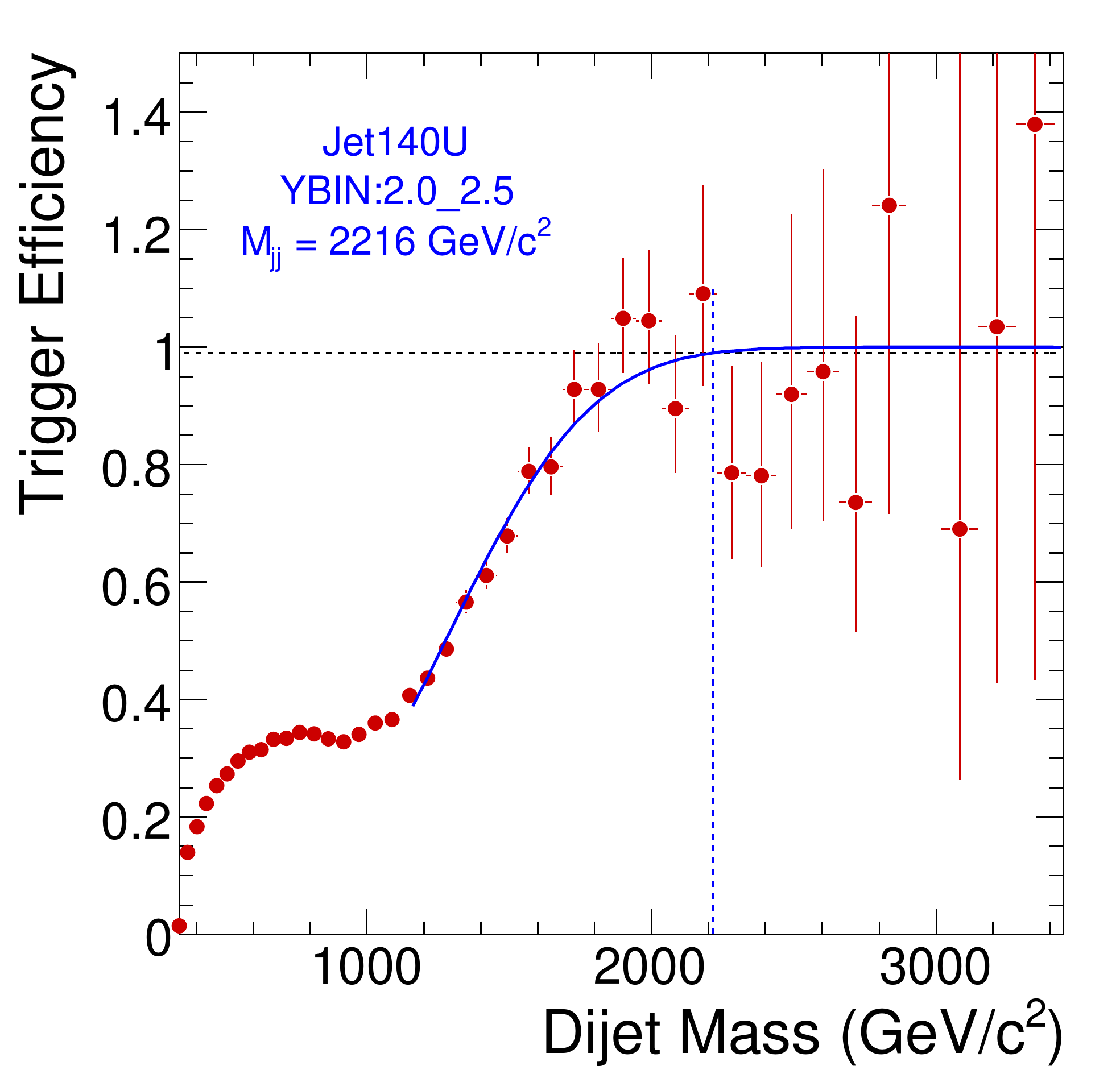}
  \capspace
  \caption{ Relative trigger efficiencies as a function of dijet mass for the five different $|y|_{max}$ bins and for the HLT\_Jet140U trigger. The  $100\%$ efficiency point is determined by performing a fit with an error function.}
  \label{fig_data_Appendix4}
\end{figure}
\clearpage
\section{Appendix C}

\begin{figure}[ht]
\centering

\includegraphics[width=0.35\textwidth]{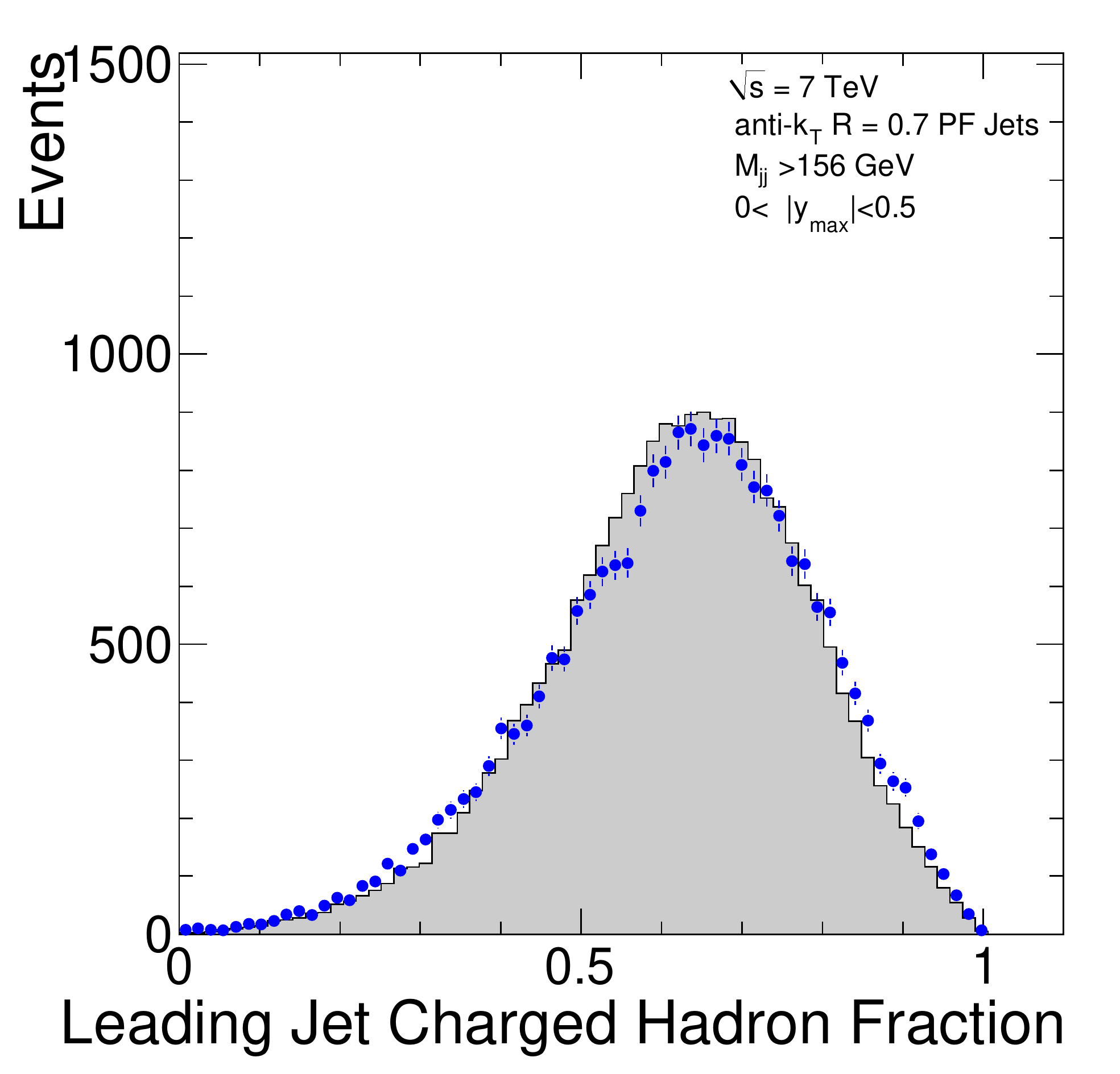} 
\includegraphics[width=0.35\textwidth]{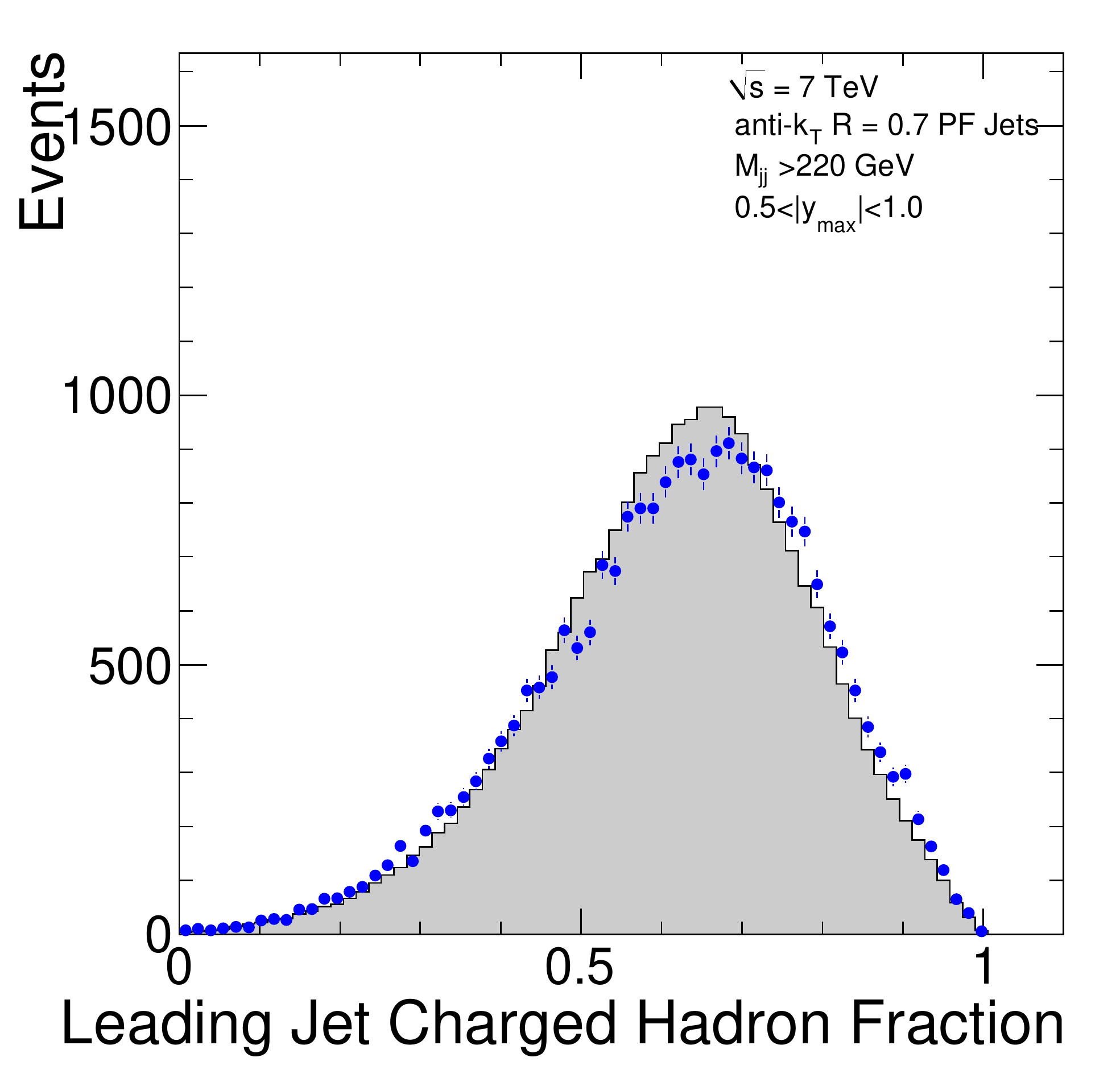}  
\includegraphics[width=0.35\textwidth]{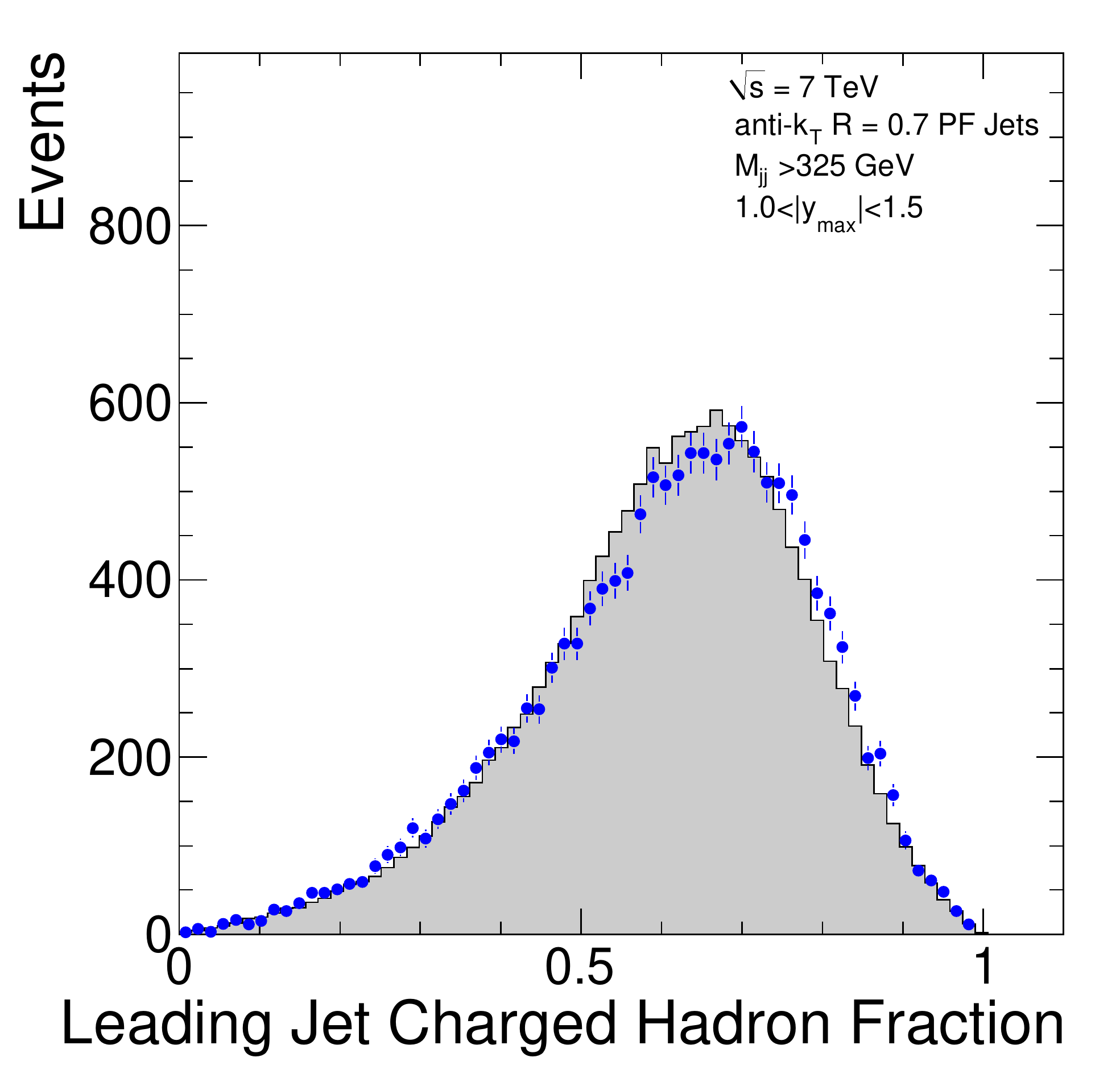}  
\includegraphics[width=0.35\textwidth]{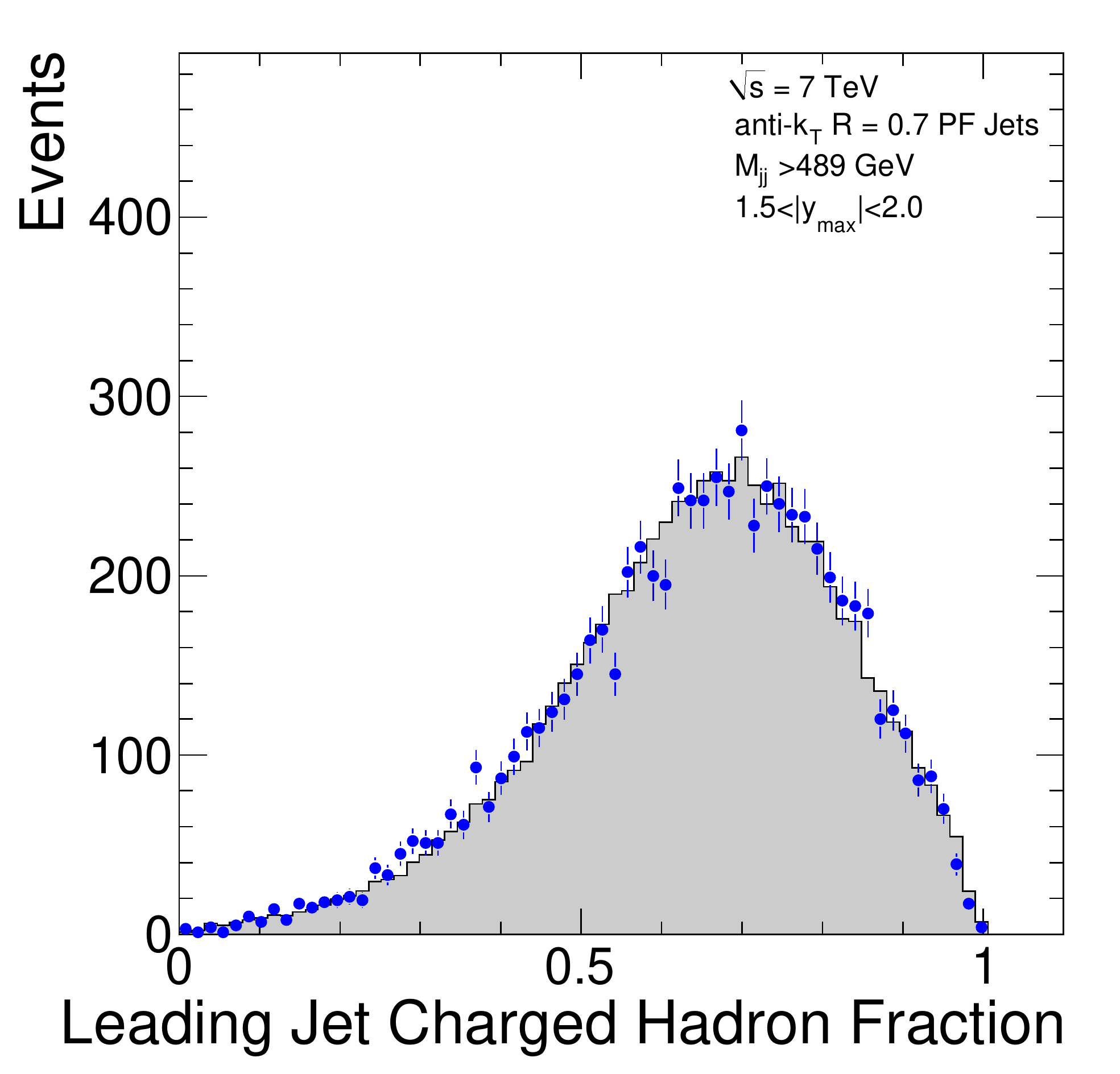} 
\includegraphics[width=0.35\textwidth]{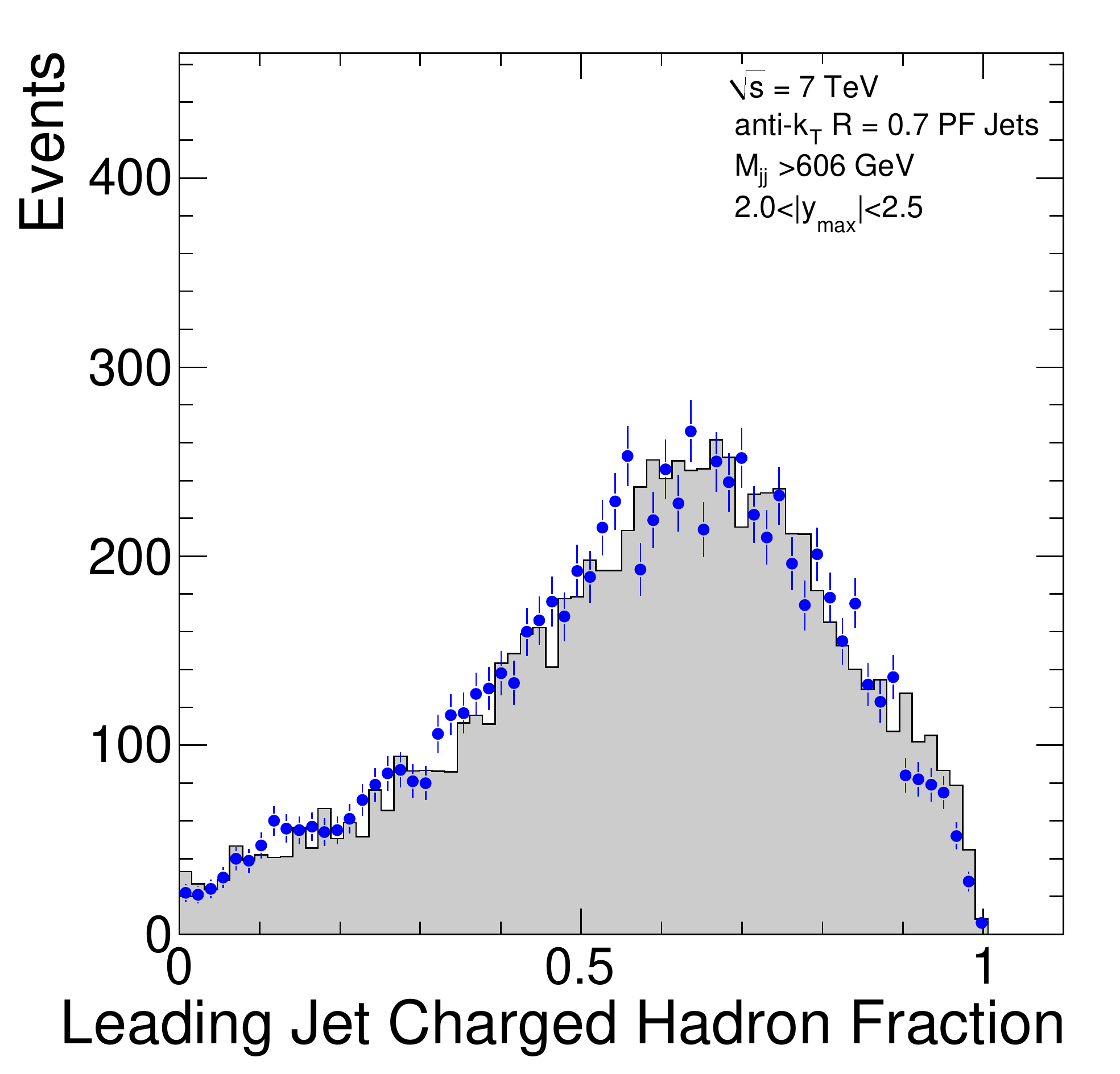}

\caption{ The charged hadron fraction of the leading jet  for the five different $y_{max}$ bins and for the 
HLT$_{-}$Jet30U trigger, for data (points) and simulated (dashed histogram) events.}
\label{fig_appc1}
\end{figure}

\begin{figure}[ht]
\centering

\includegraphics[width=0.48\textwidth]{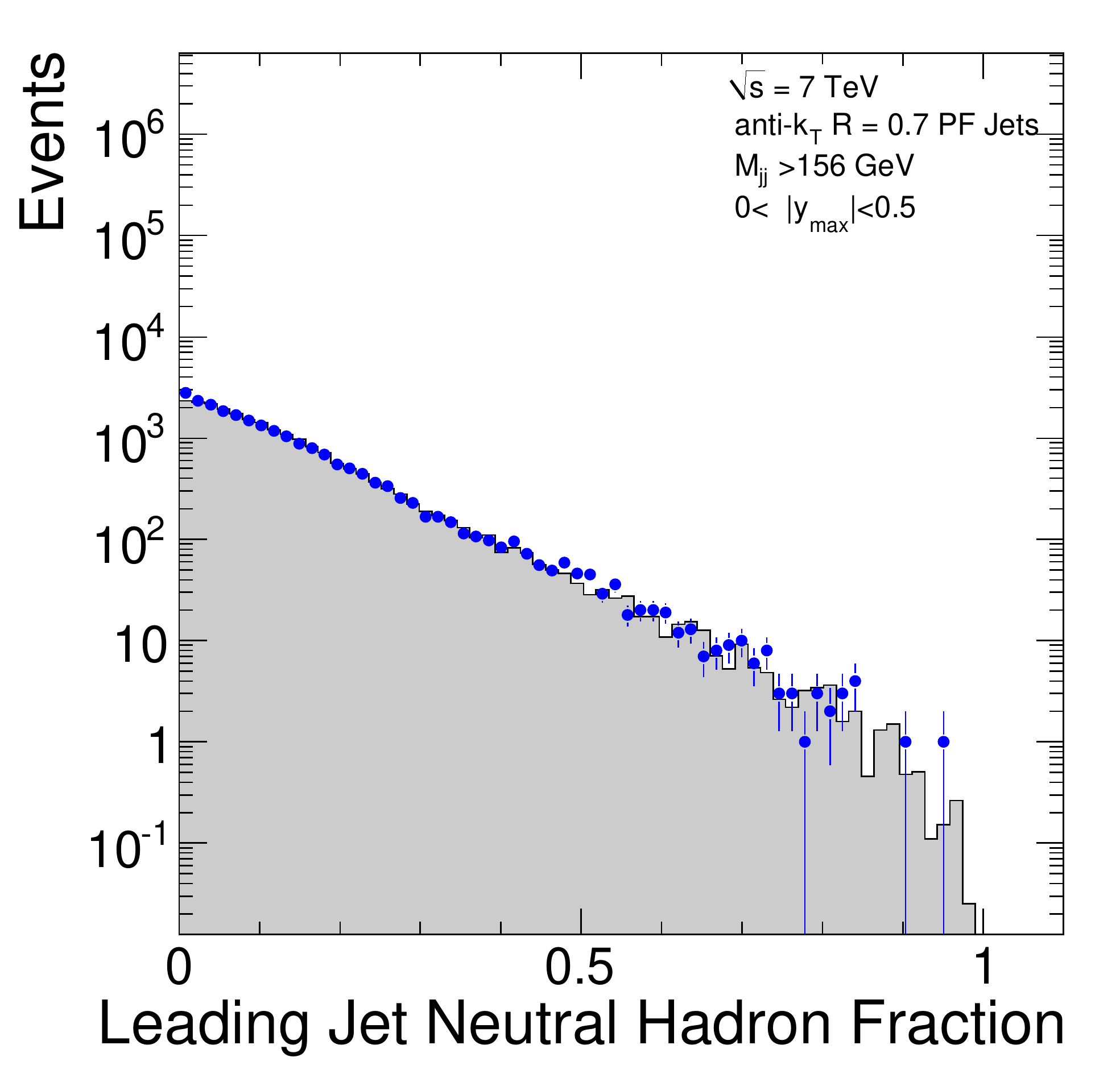} 
\includegraphics[width=0.48\textwidth]{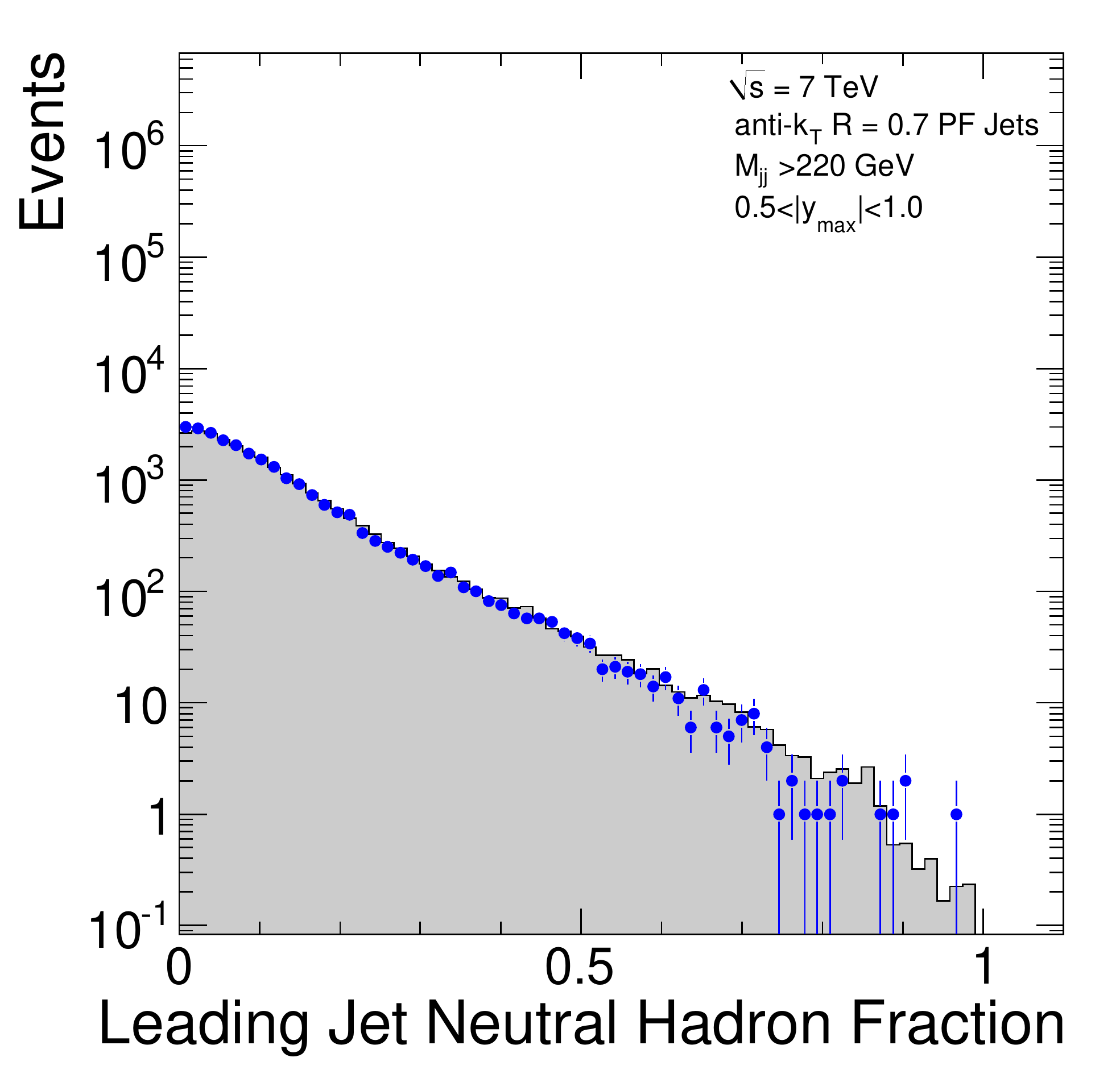}  
\includegraphics[width=0.48\textwidth]{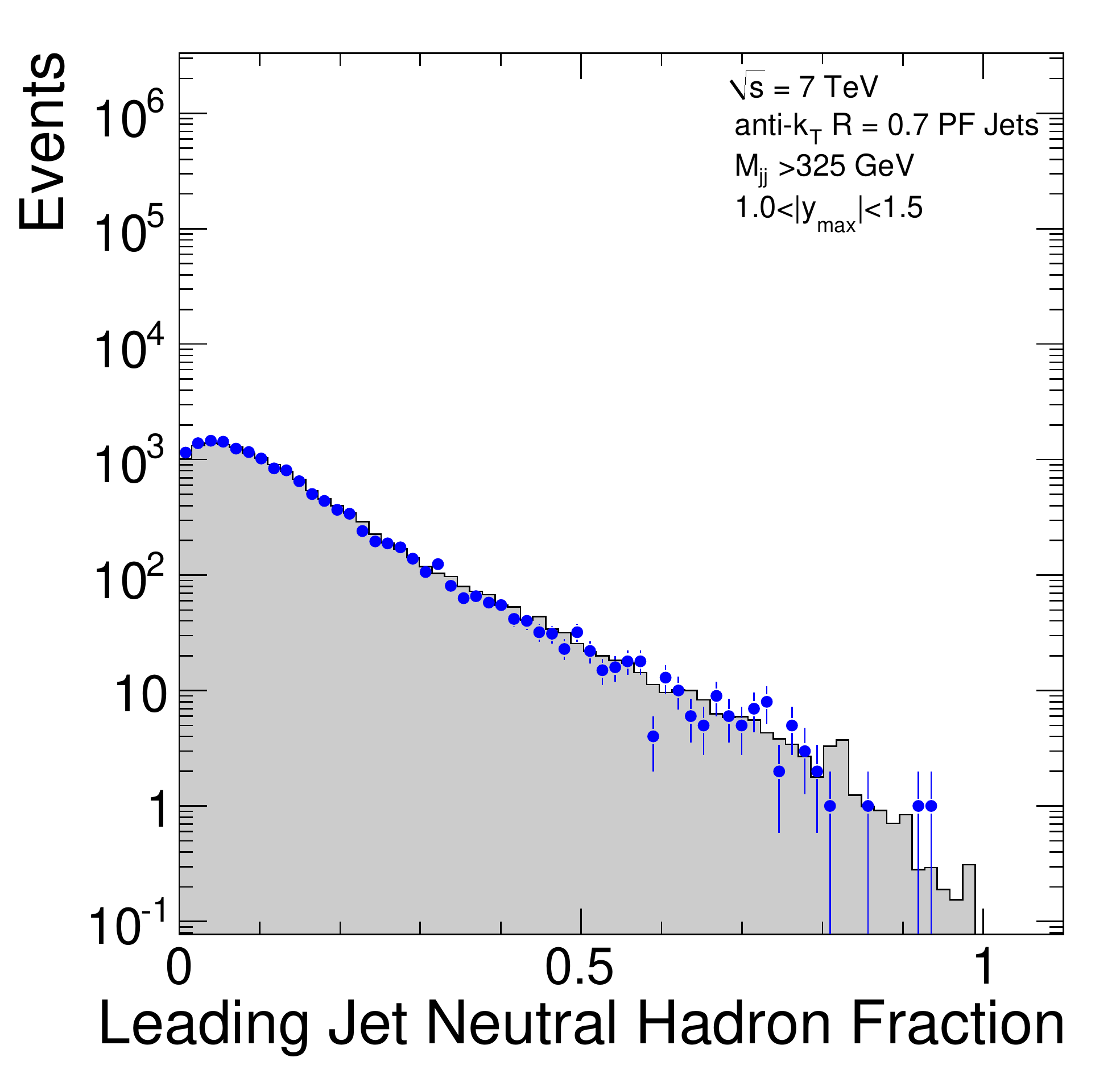}  
\includegraphics[width=0.48\textwidth]{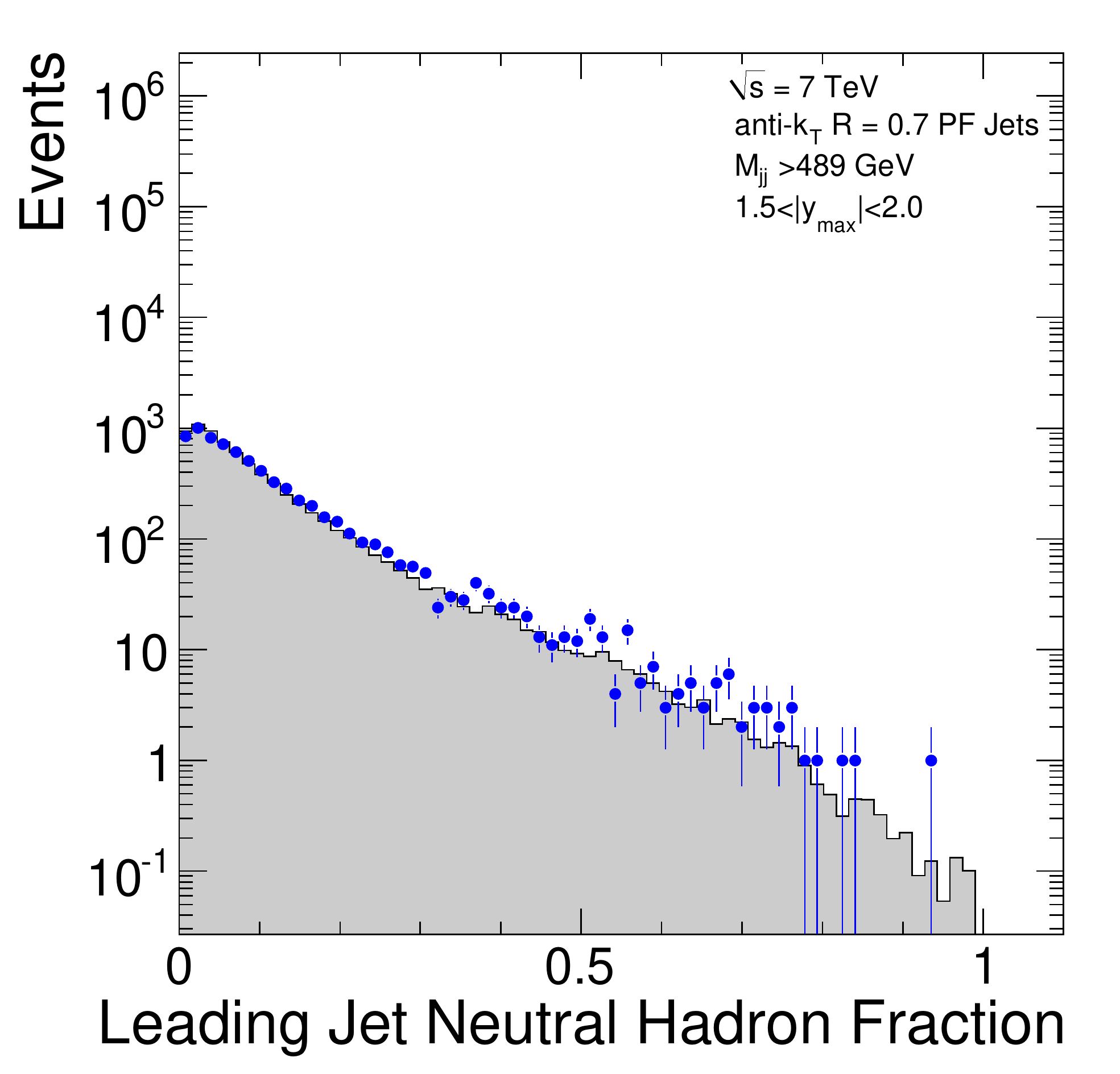} 
\includegraphics[width=0.48\textwidth]{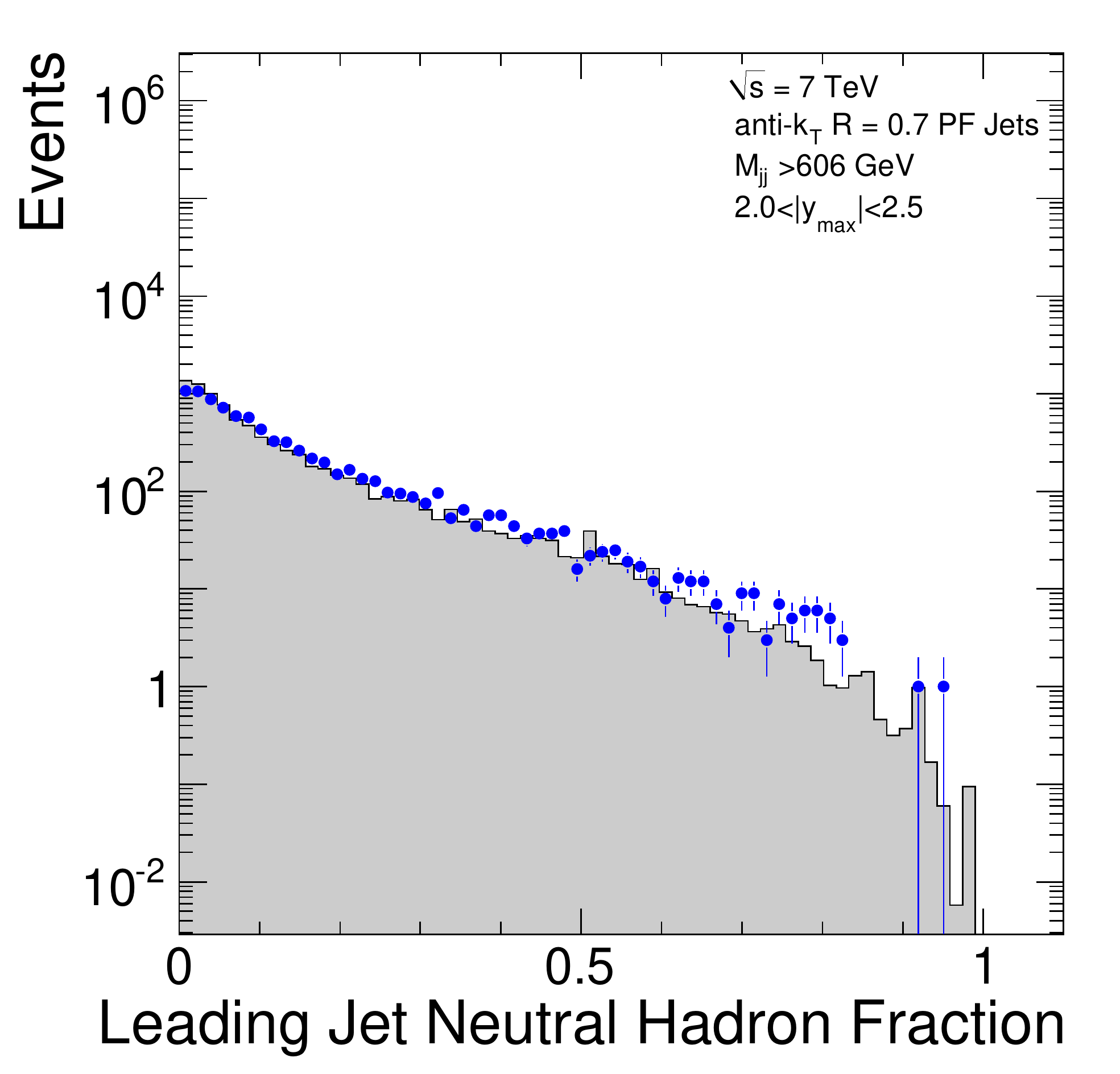}

\caption{ The neutral hadron fraction of the leading jet  for the five different $y_{max}$ bins and for the 
HLT$_{-}$Jet30U trigger, for data (points) and simulated (dashed histogram) events.}
\label{fig_appc2}
\end{figure}

\begin{figure}[ht]
\centering

\includegraphics[width=0.48\textwidth]{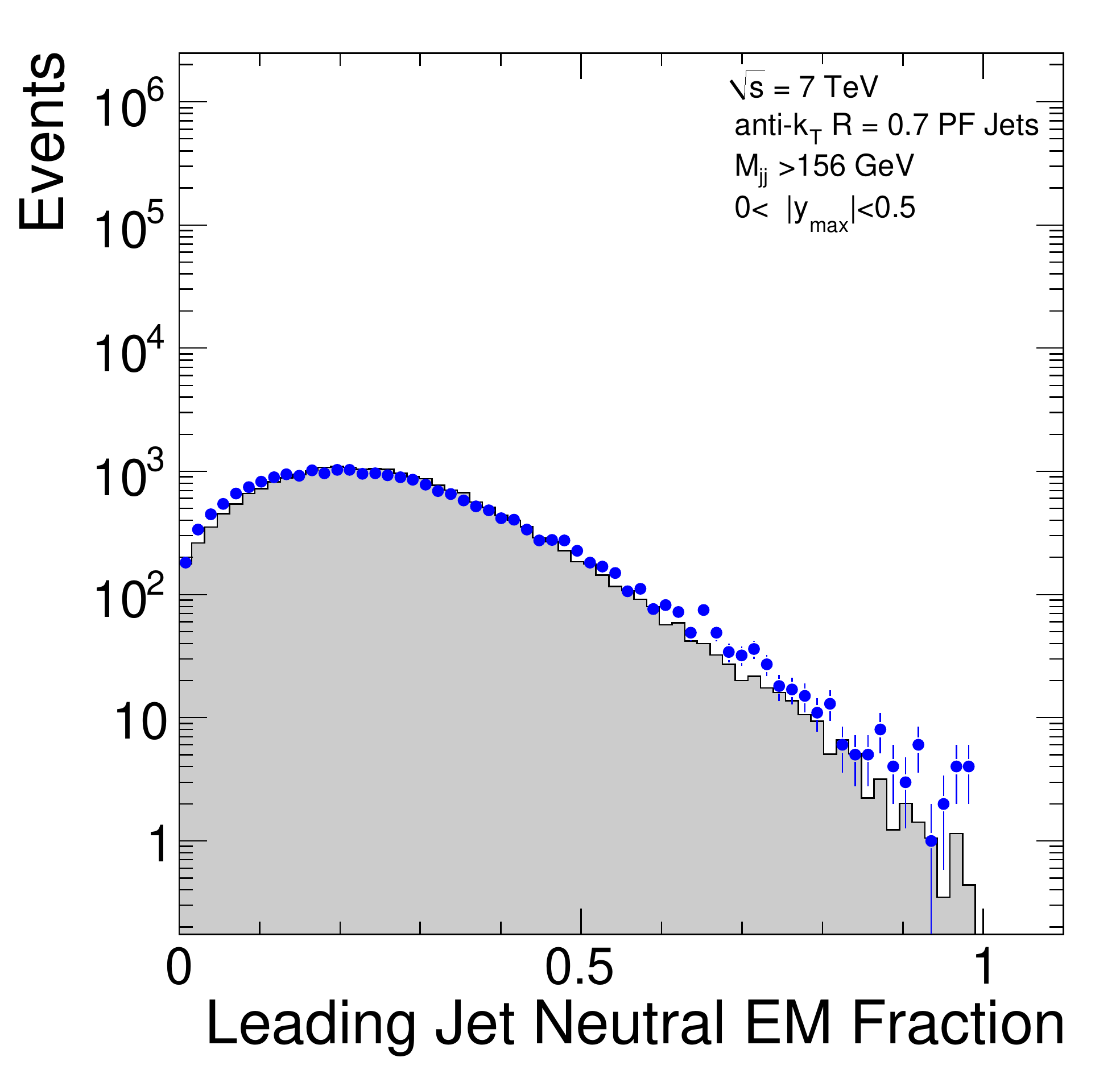} 
\includegraphics[width=0.48\textwidth]{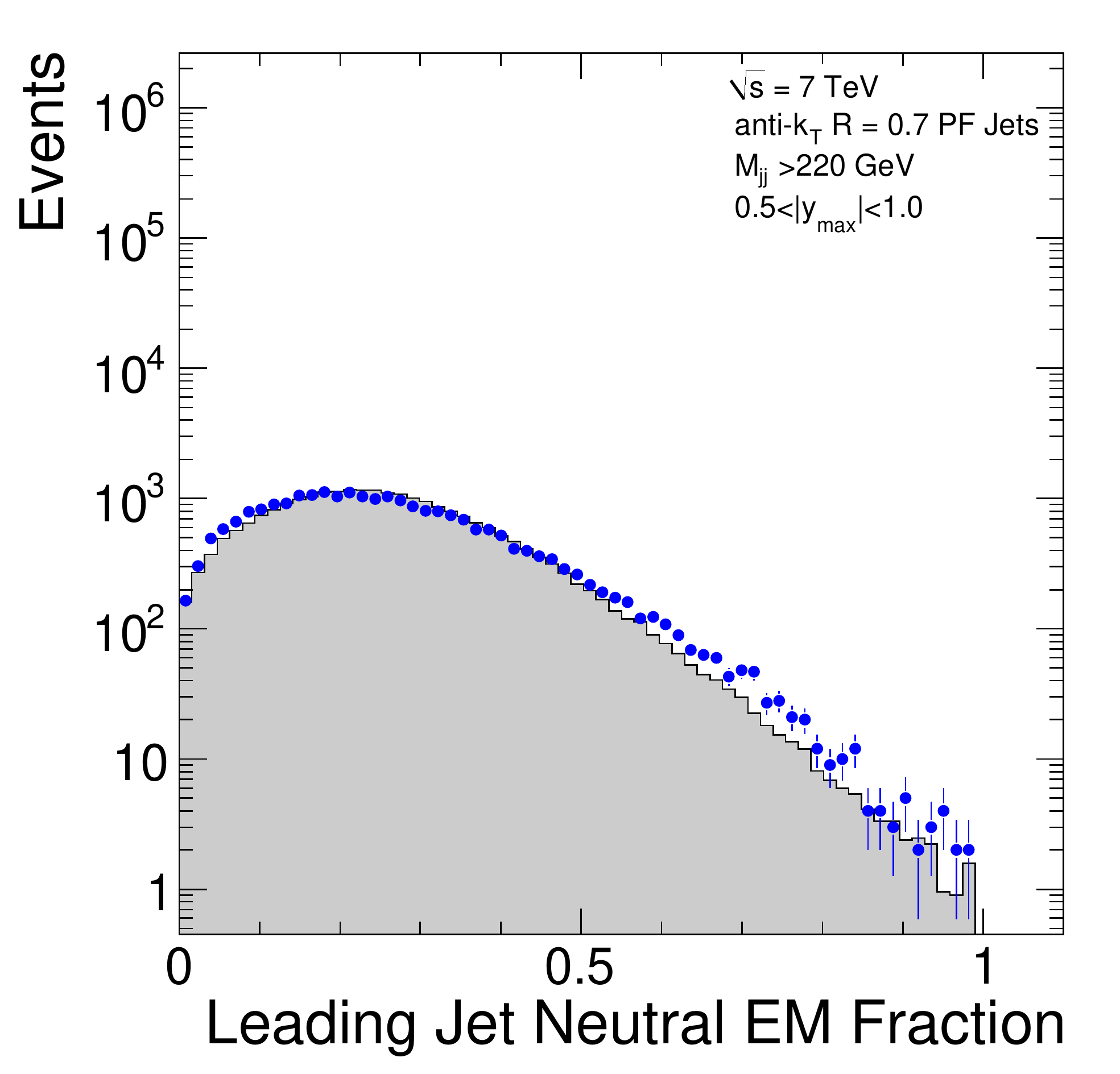}  
\includegraphics[width=0.48\textwidth]{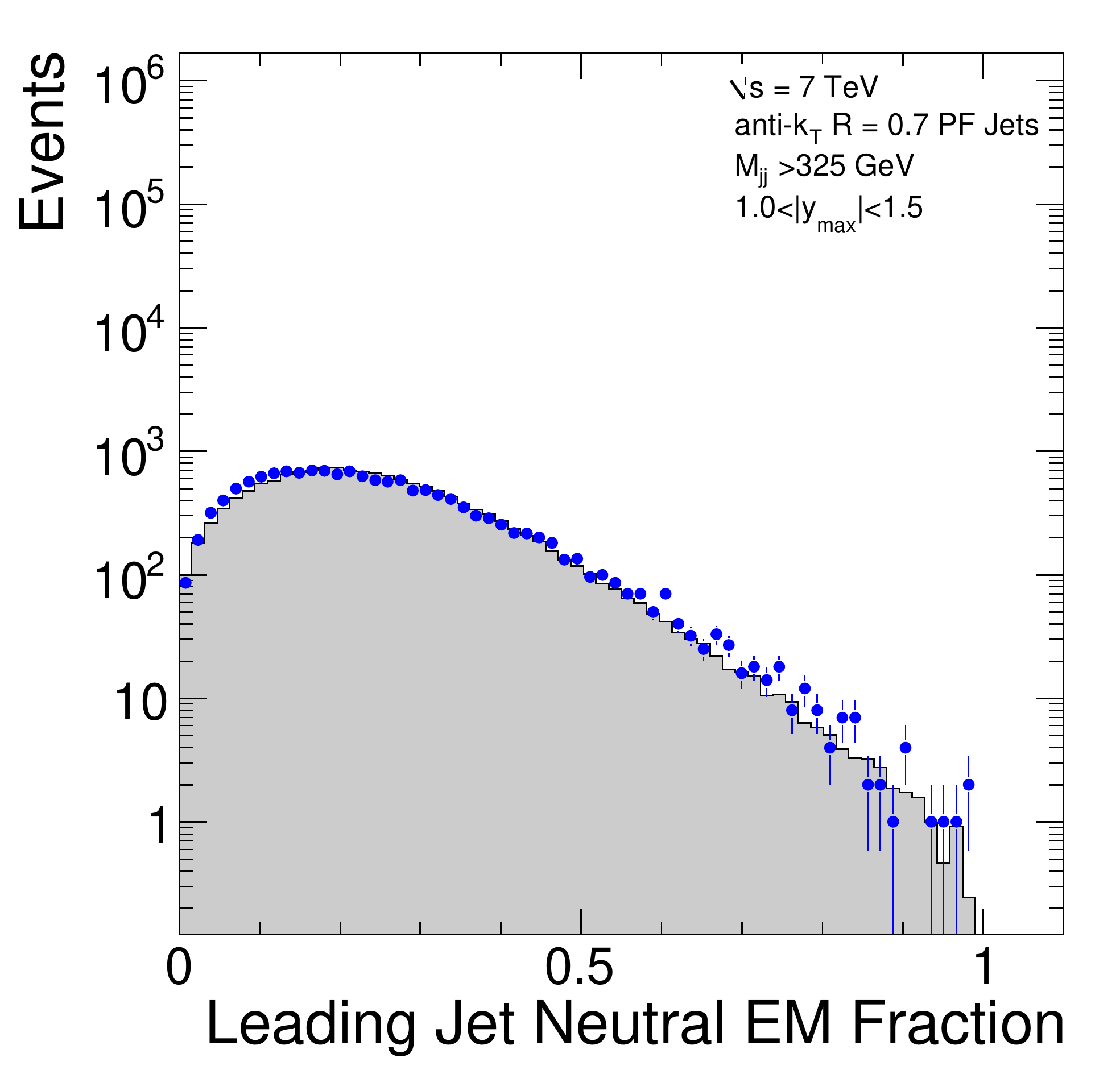}  
\includegraphics[width=0.48\textwidth]{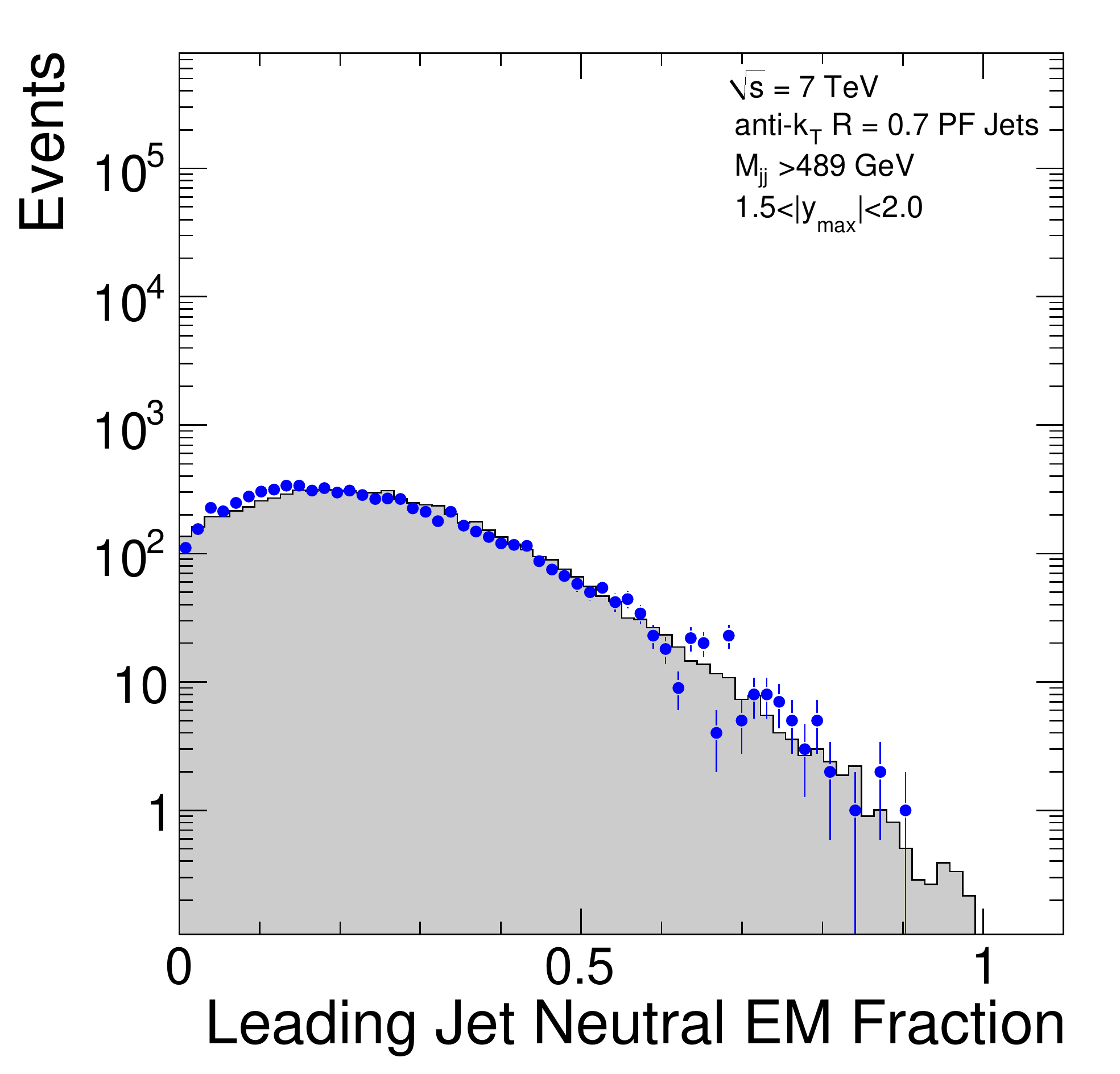} 
\includegraphics[width=0.48\textwidth]{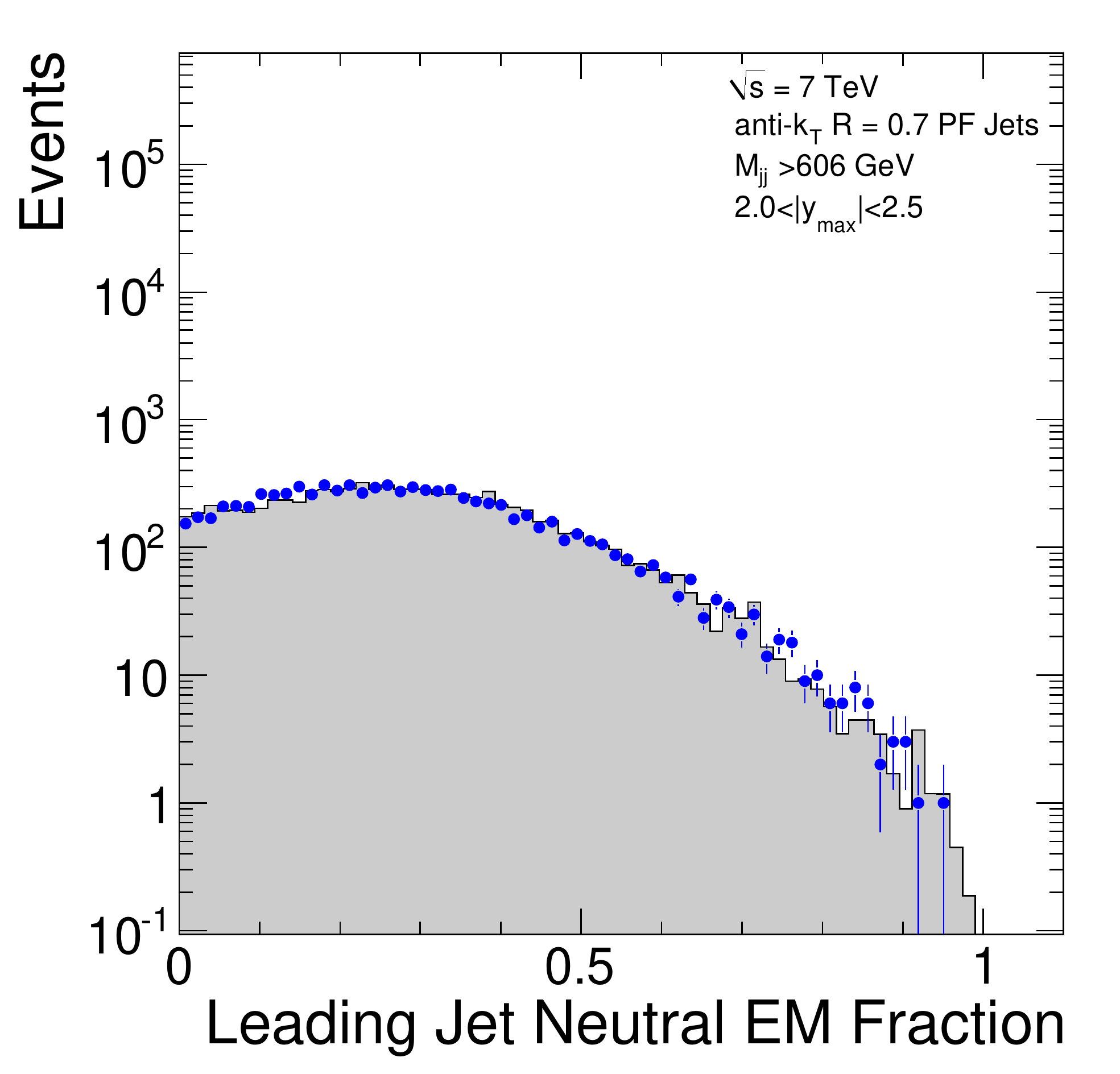}

\caption{ The neutral electromagnetic  fraction of the leading jet  for the five different $y_{max}$ bins and for the 
HLT$_{-}$Jet30U trigger, for data (points) and simulated (dashed histogram) events.}
\label{fig_appc3}
\end{figure}

 \begin{figure}[ht]
\centering

\includegraphics[width=0.48\textwidth]{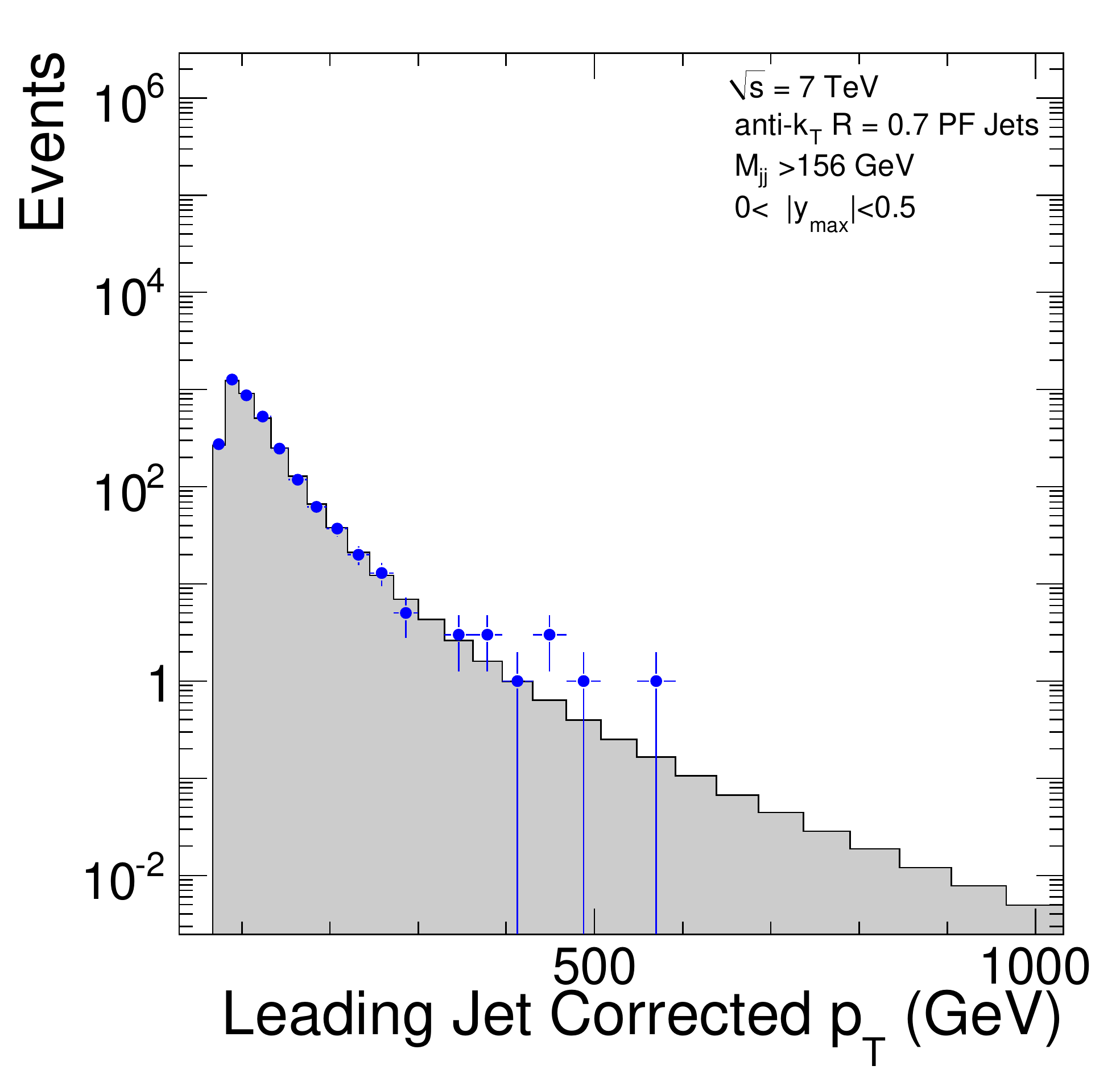} 
\includegraphics[width=0.48\textwidth]{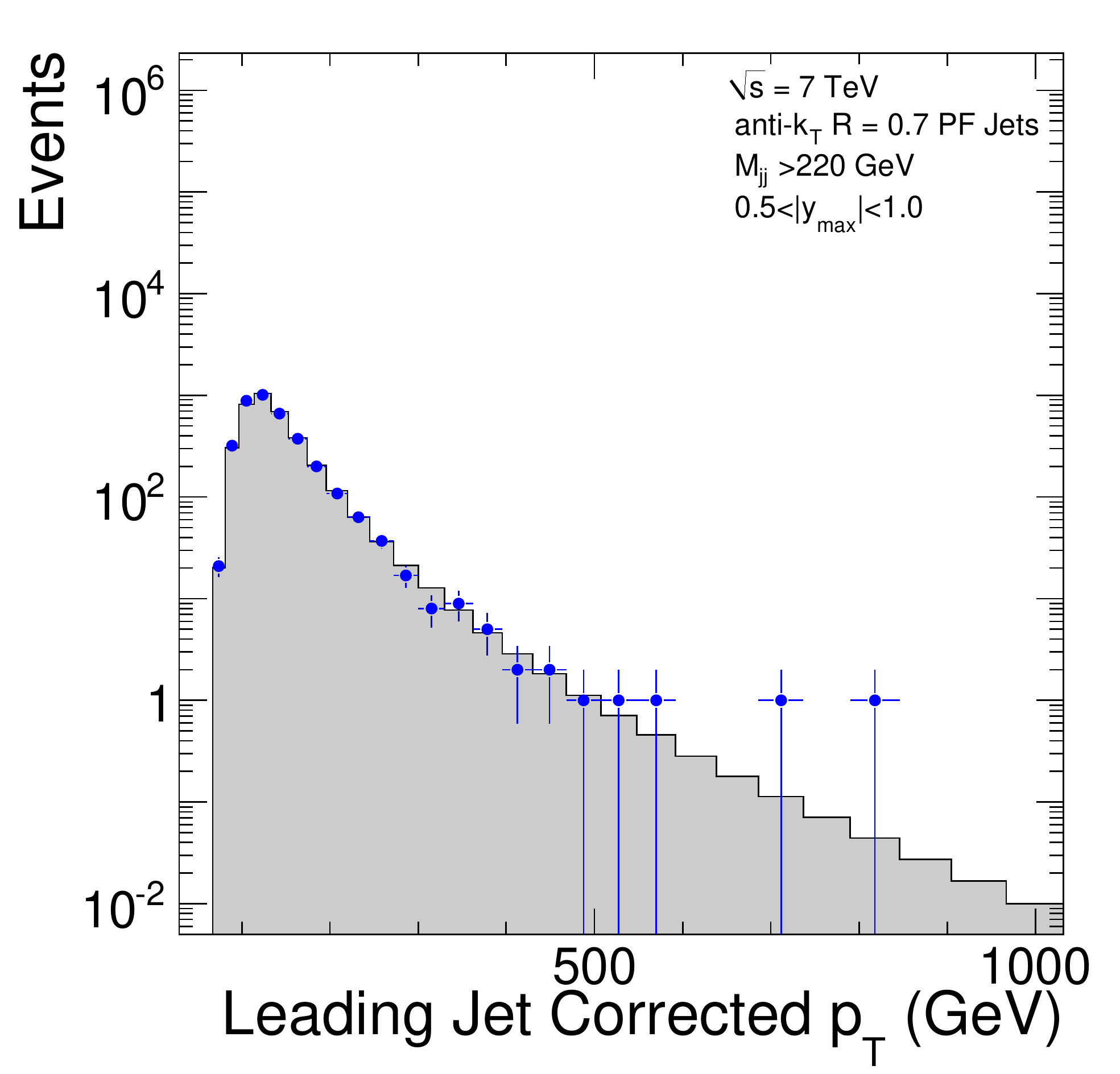}  
\includegraphics[width=0.48\textwidth]{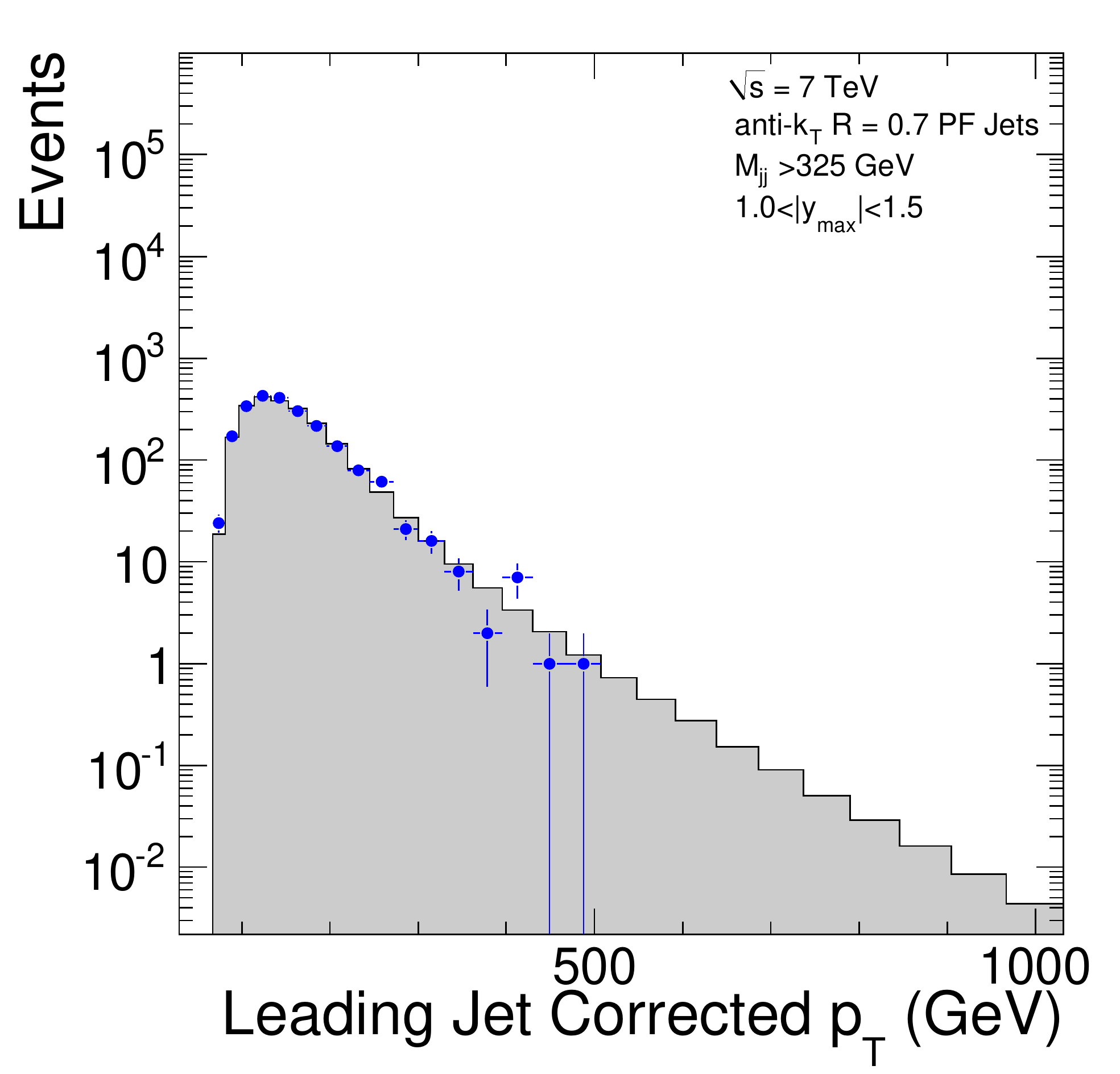}  
\includegraphics[width=0.48\textwidth]{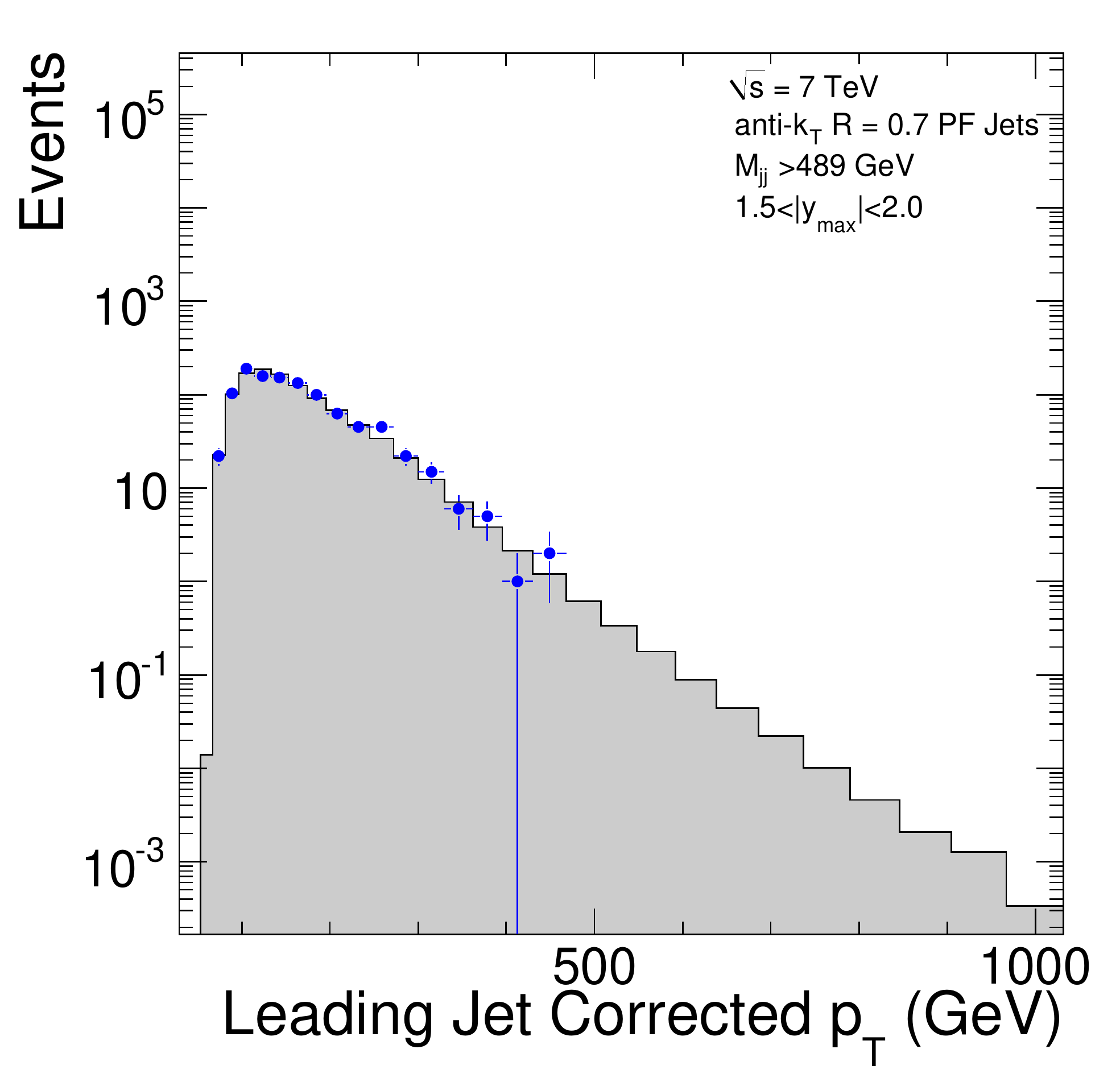} 
\includegraphics[width=0.48\textwidth]{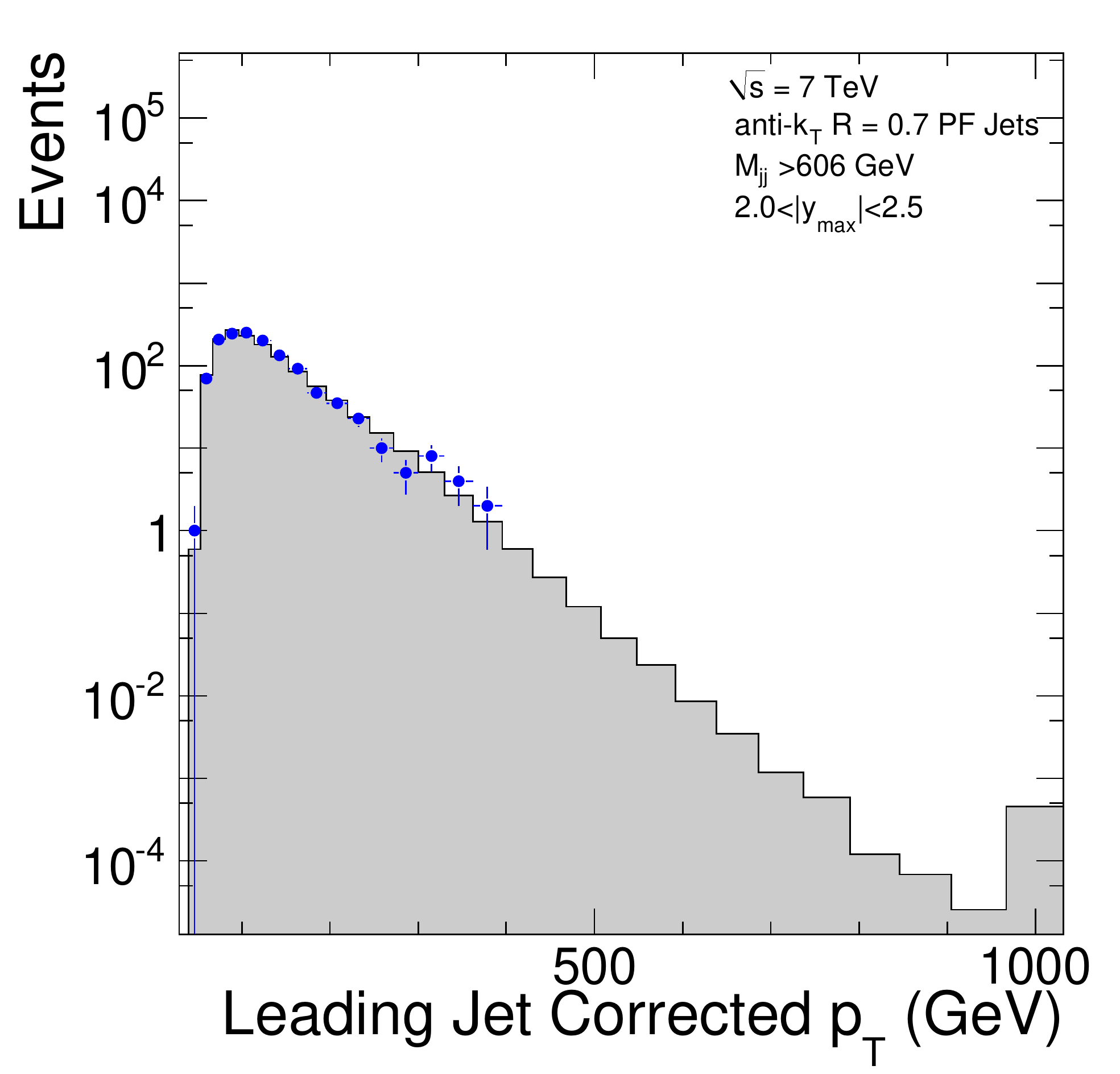}

\caption{ The $p_T$f of the leading jet  for the five different $y_{max}$ bins and for the 
HLT$_{-}$Jet30U trigger, for data (points) and simulated (dashed histogram) events.}
\label{fig_appc4}
\end{figure}

\begin{figure}[ht]
\centering

\includegraphics[width=0.48\textwidth]{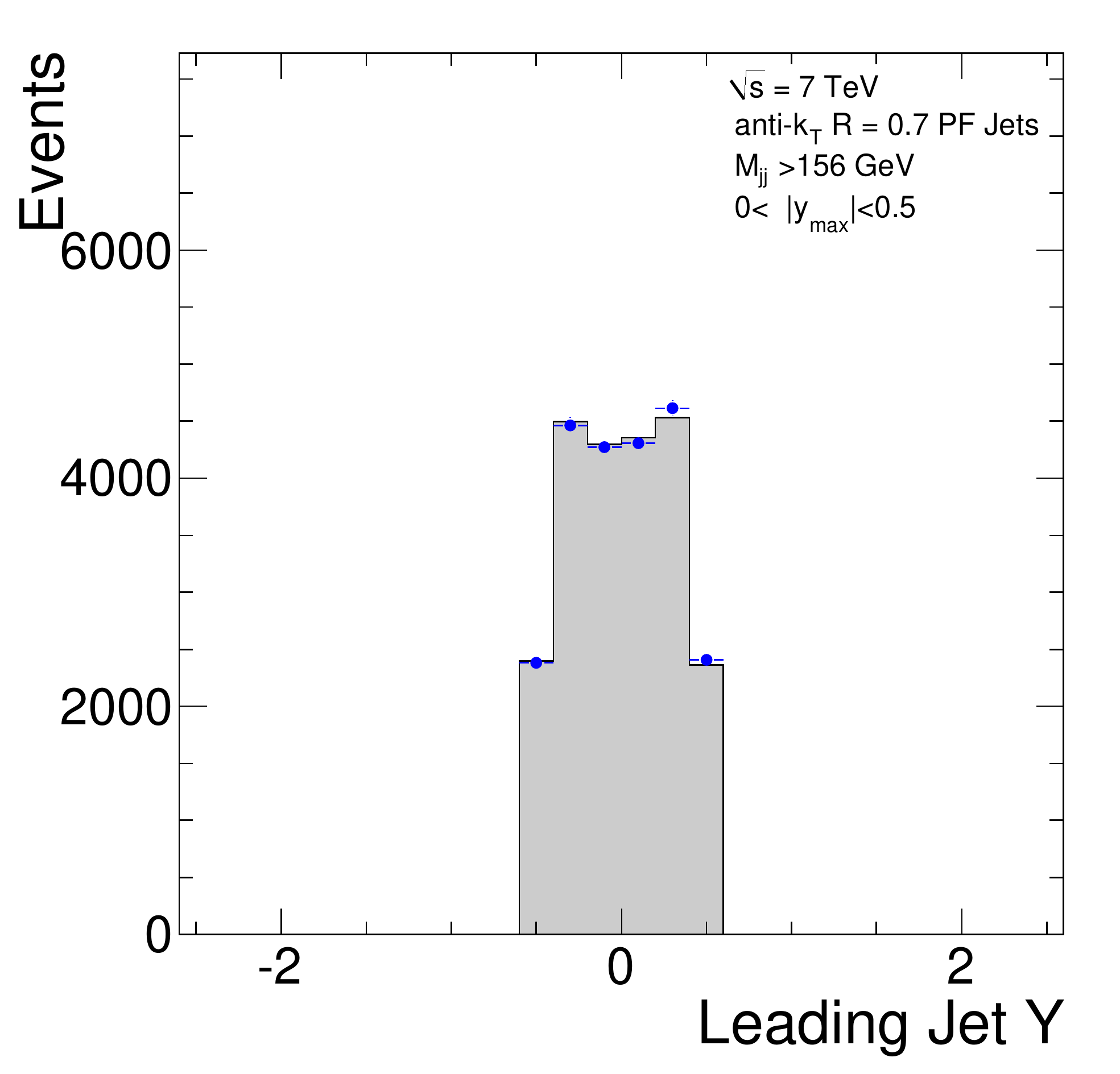} 
\includegraphics[width=0.48\textwidth]{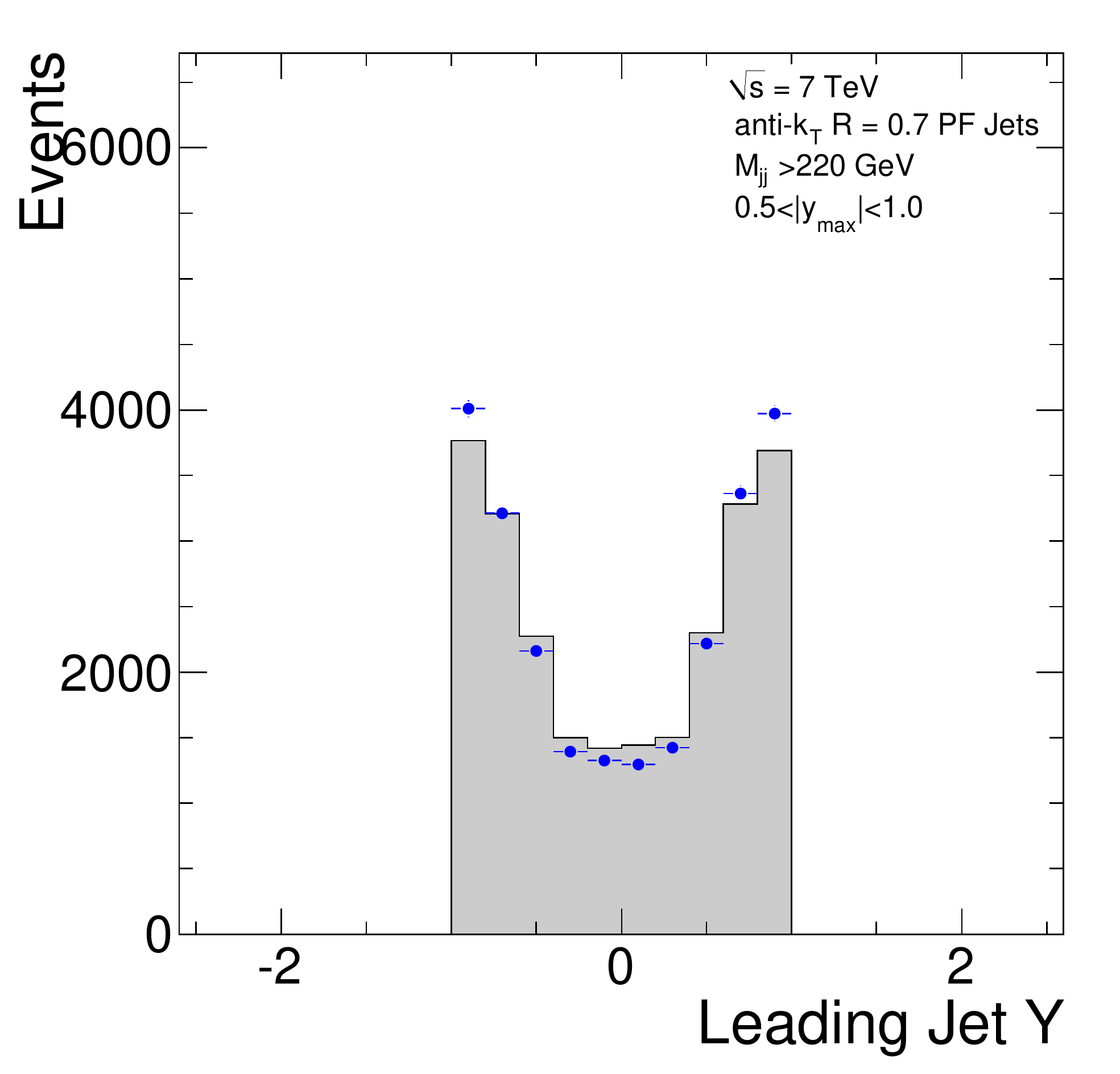}  
\includegraphics[width=0.48\textwidth]{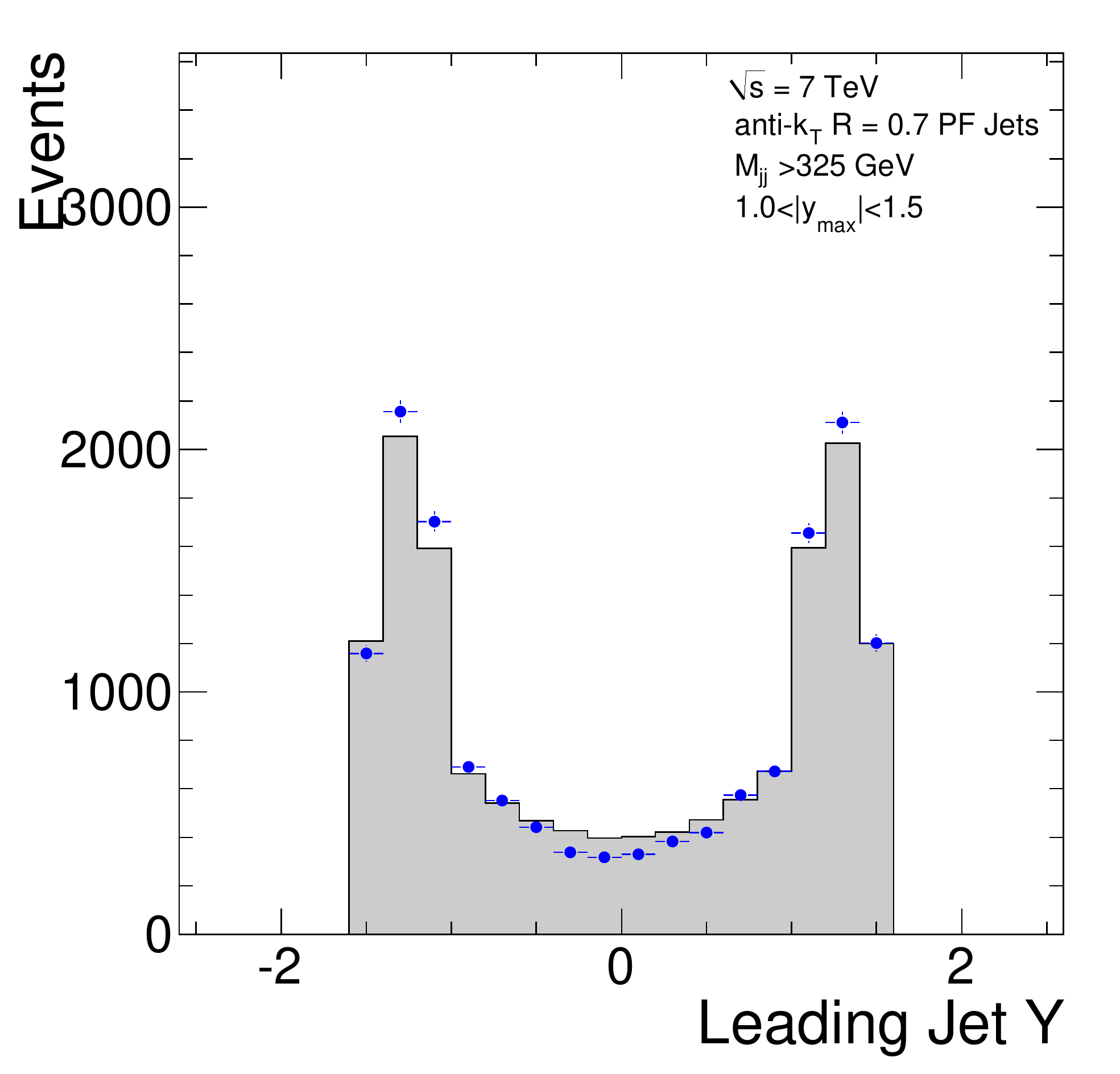}  
\includegraphics[width=0.48\textwidth]{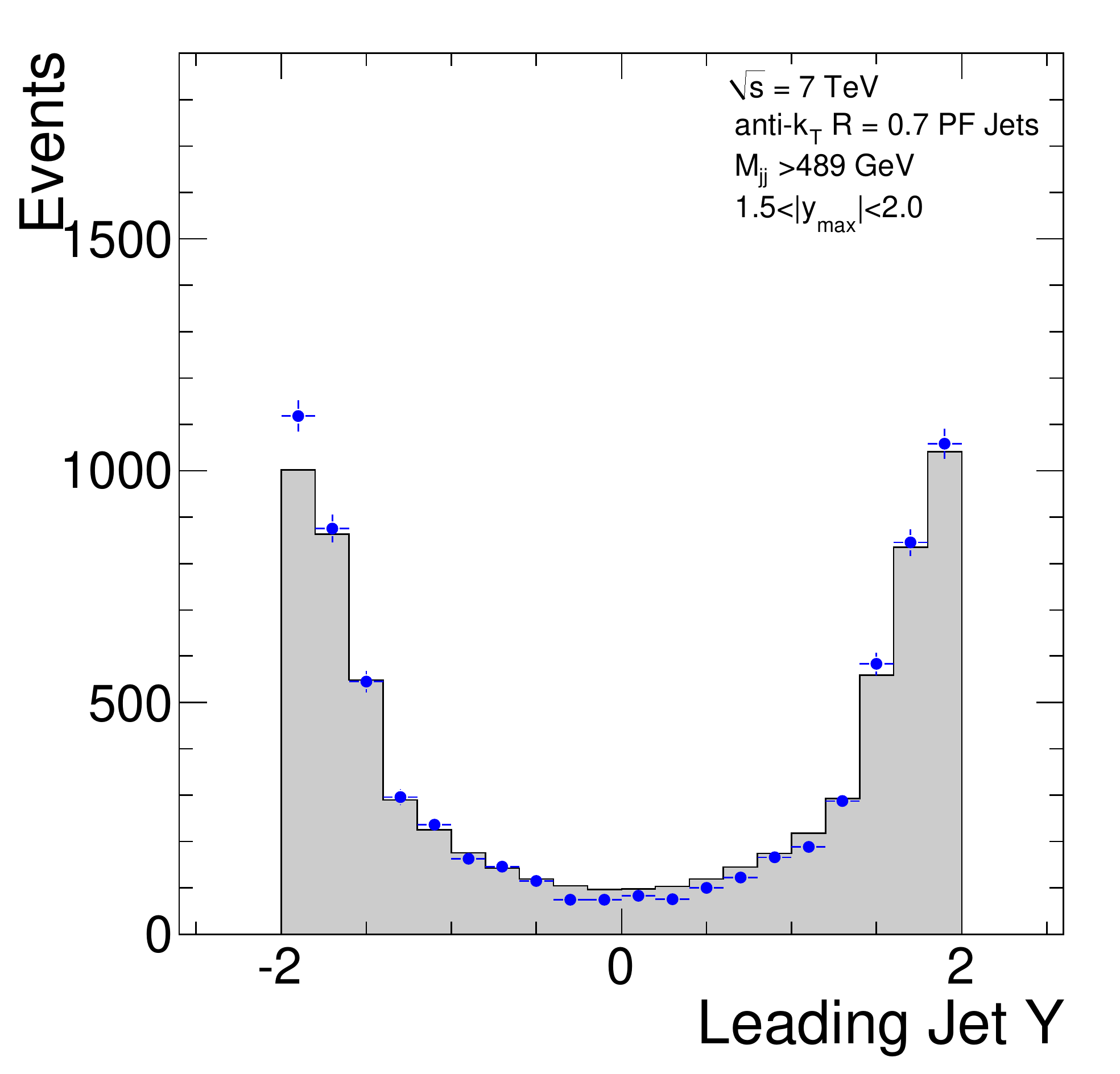} 
\includegraphics[width=0.48\textwidth]{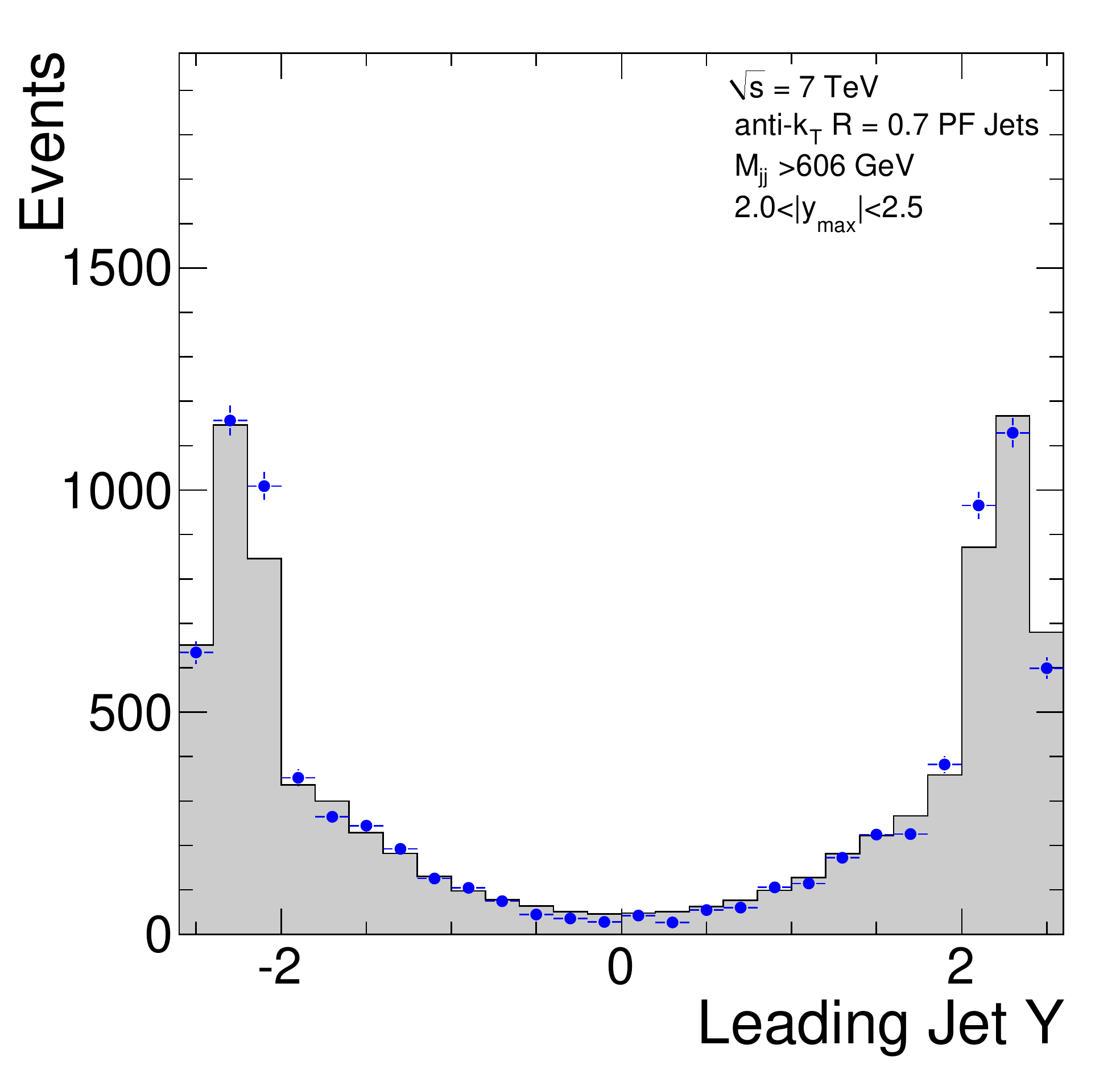}

\caption{ The $\eta$  of the leading jet  for the five different $y_{max}$ bins and for the 
HLT$_{-}$Jet30U trigger, for data (points) and simulated (dashed histogram) events.}
\label{fig_appc5}
\end{figure}

\begin{figure}[ht]
\centering

\includegraphics[width=0.48\textwidth]{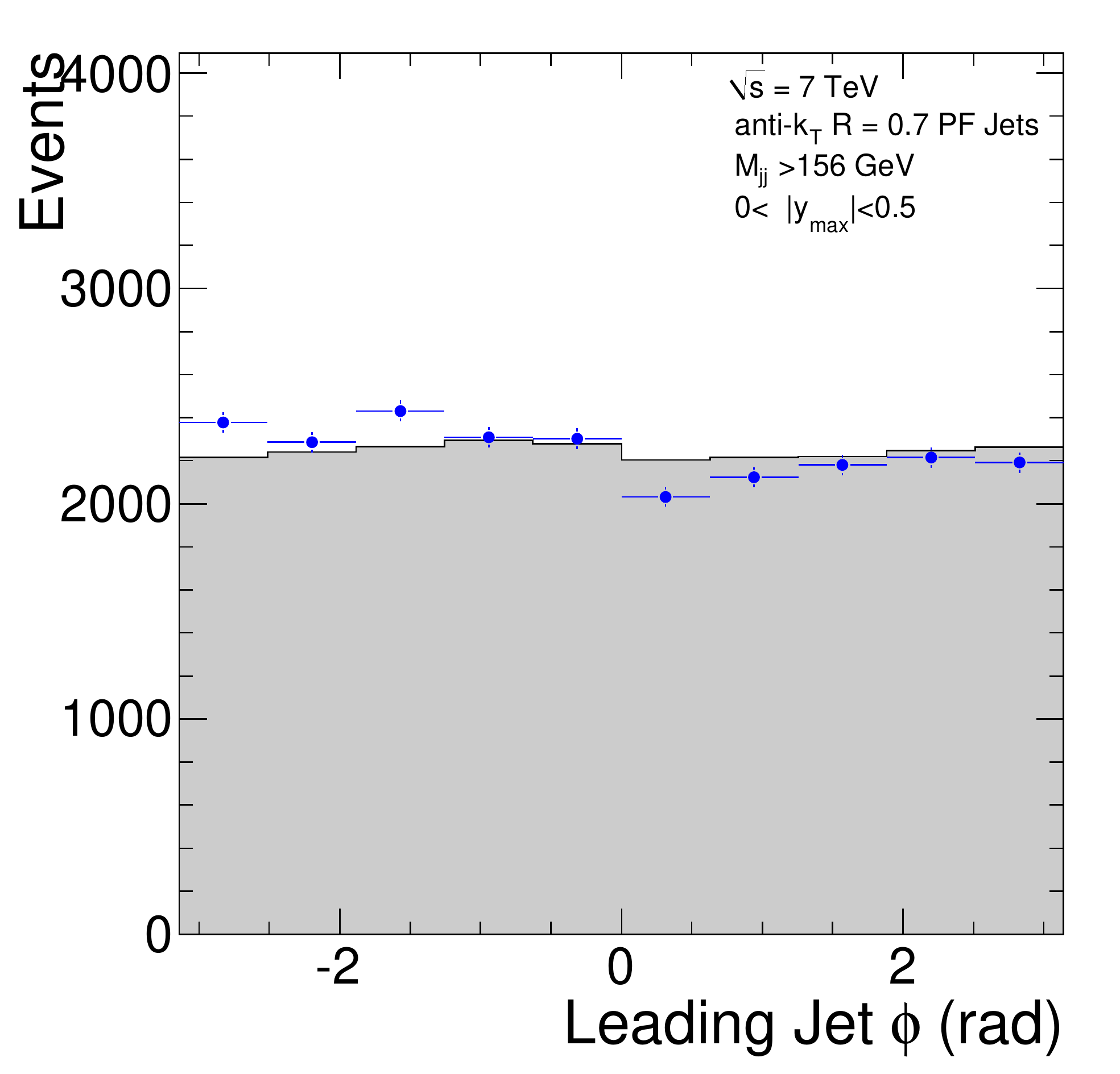} 
\includegraphics[width=0.48\textwidth]{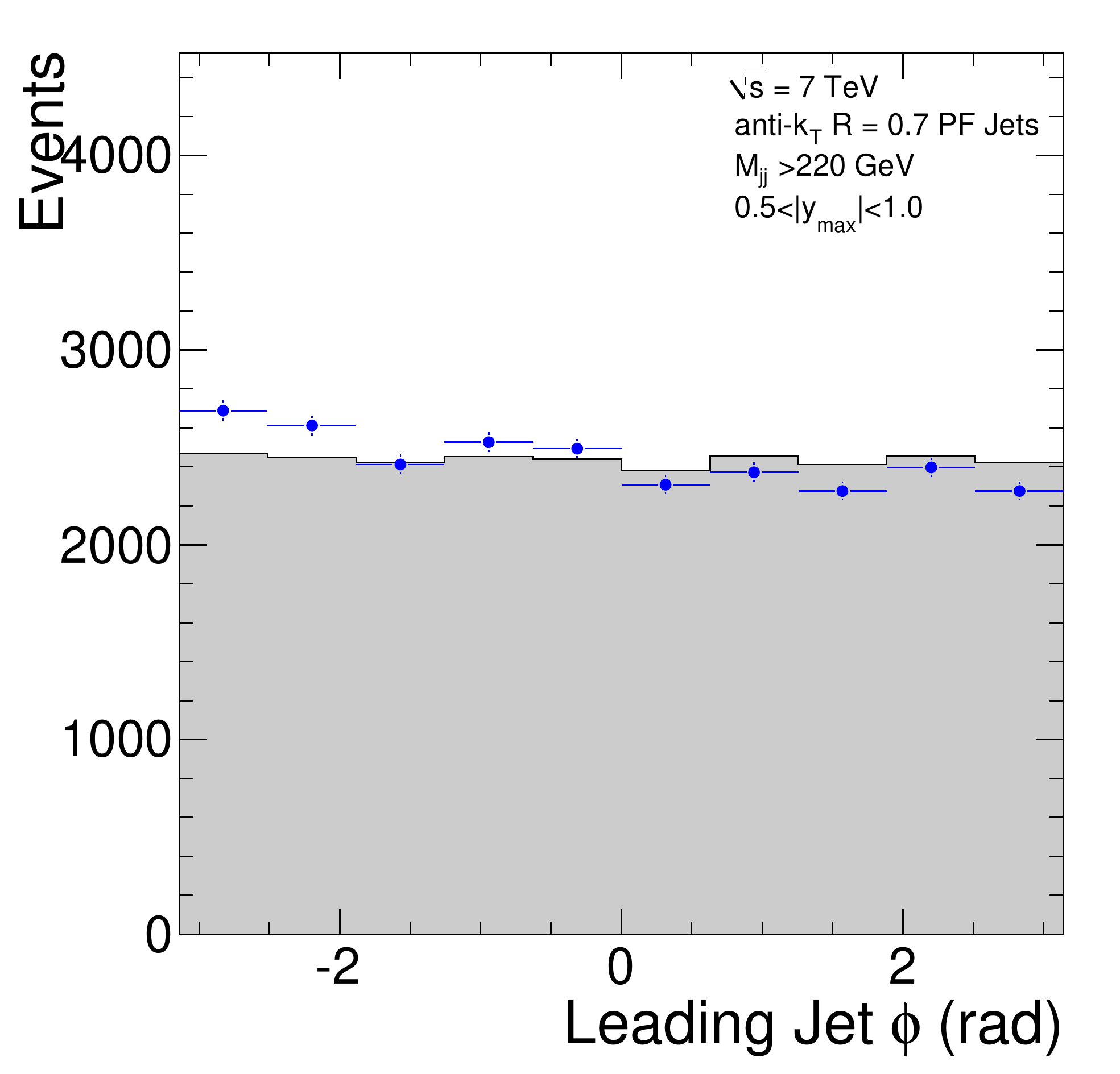}  
\includegraphics[width=0.48\textwidth]{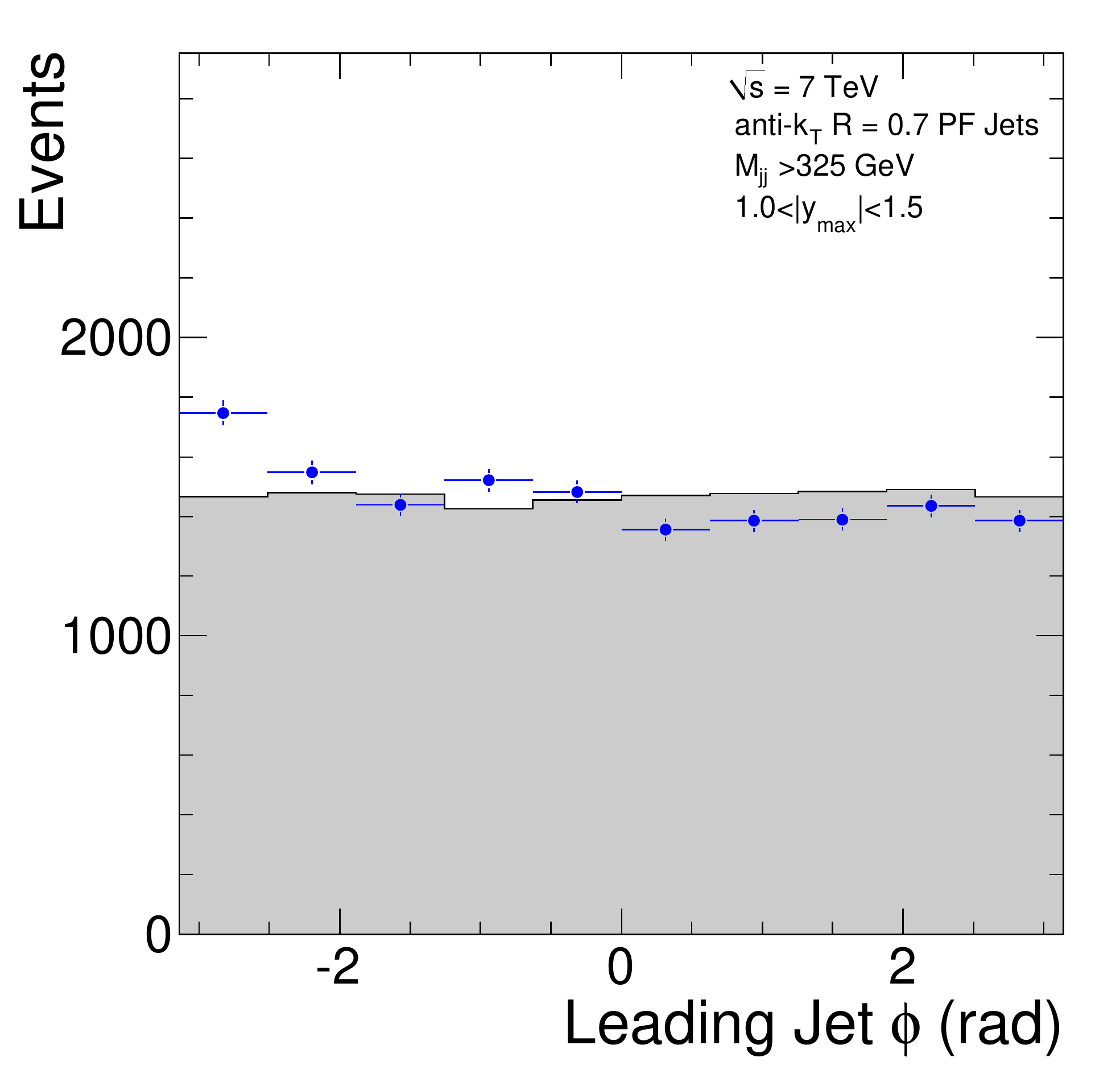}  
\includegraphics[width=0.48\textwidth]{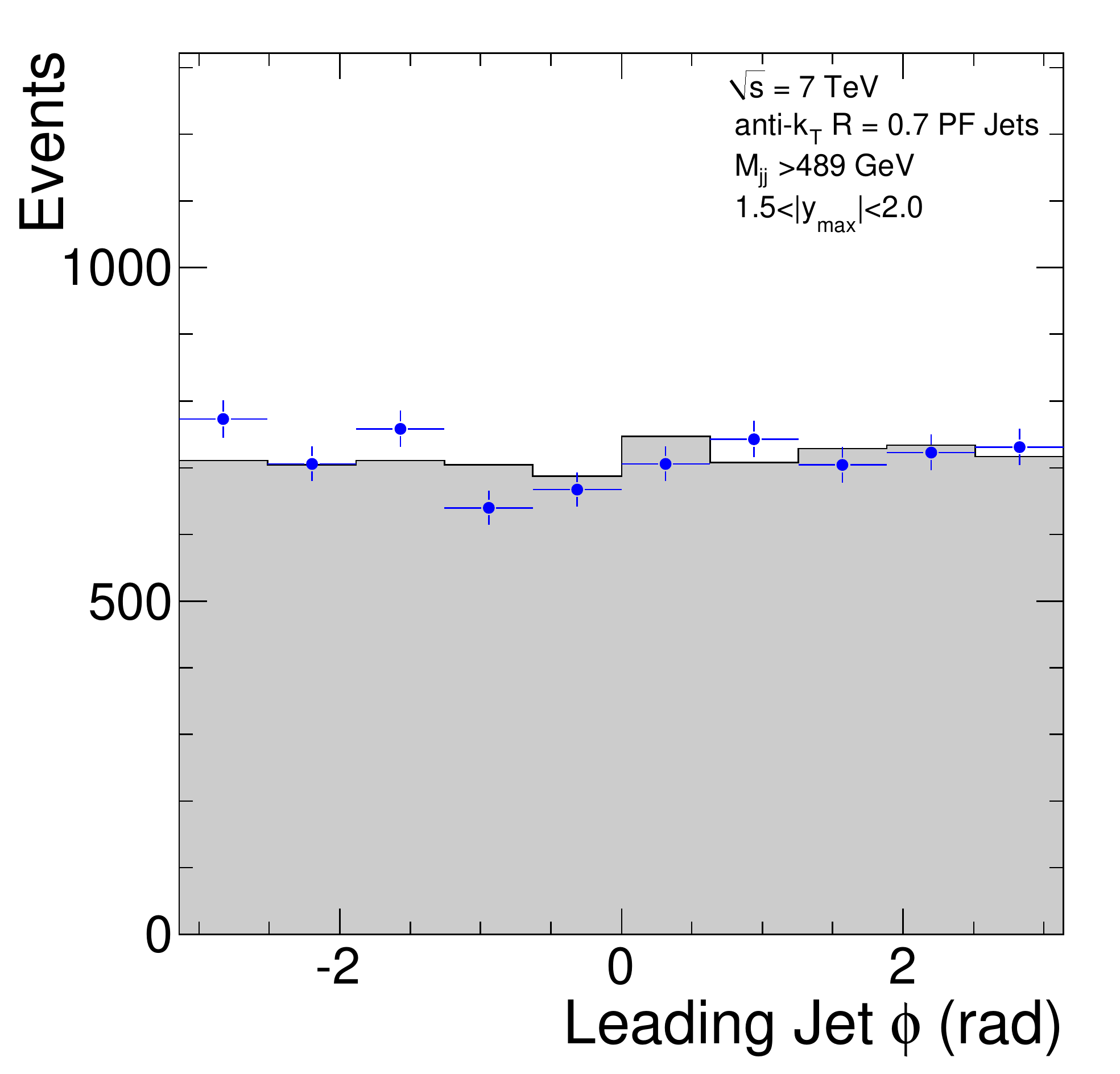} 
\includegraphics[width=0.48\textwidth]{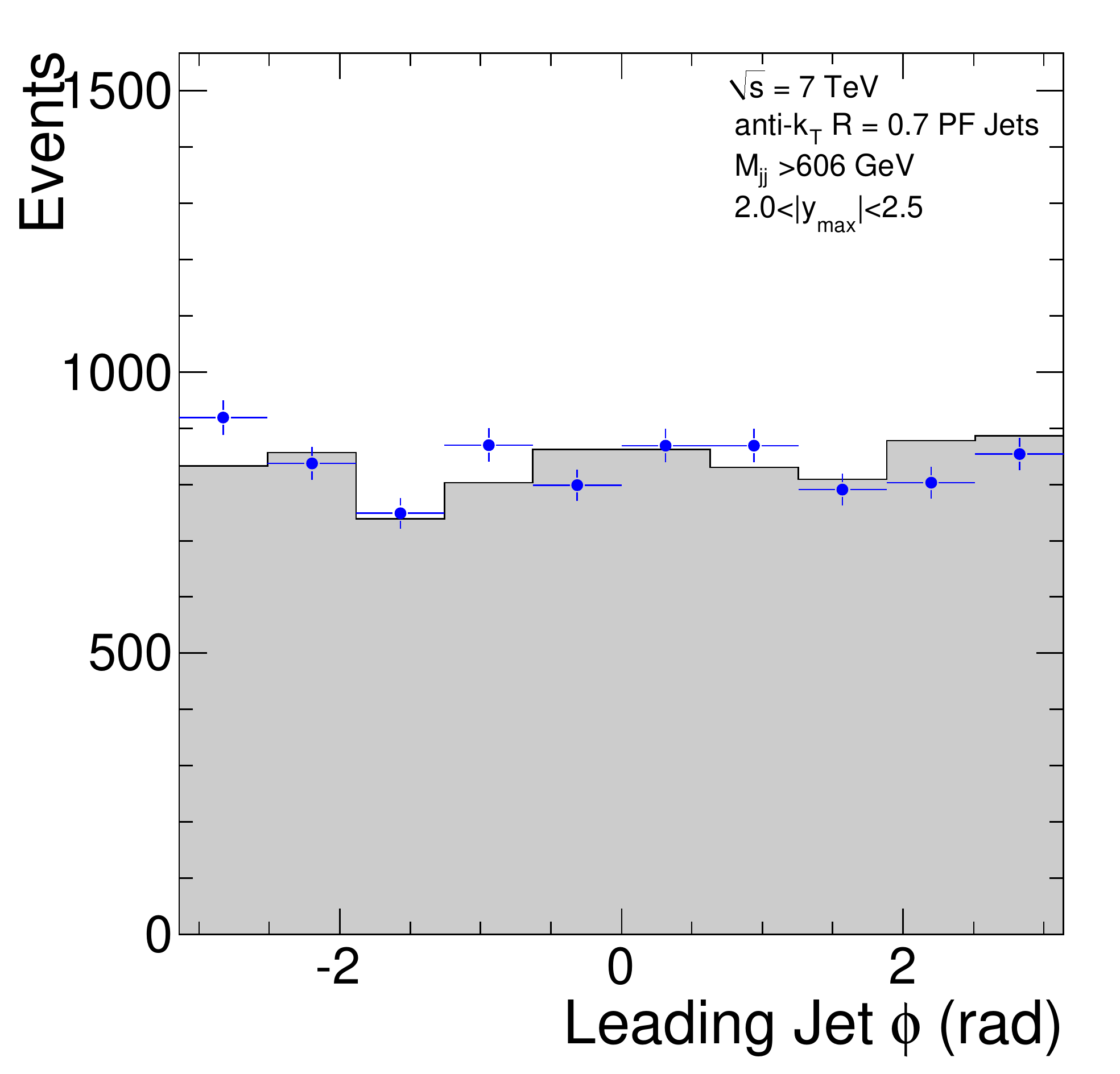}

\caption{ The $\phi$ of the leading jet  for the five different $y_{max}$ bins and for the 
HLT$_{-}$Jet30U trigger, for data (points) and simulated (dashed histogram) events.}
\label{fig_appc6}
\end{figure}

\clearpage
%%%%%%% JET 50

\begin{figure}[ht]
\centering

\includegraphics[width=0.48\textwidth]{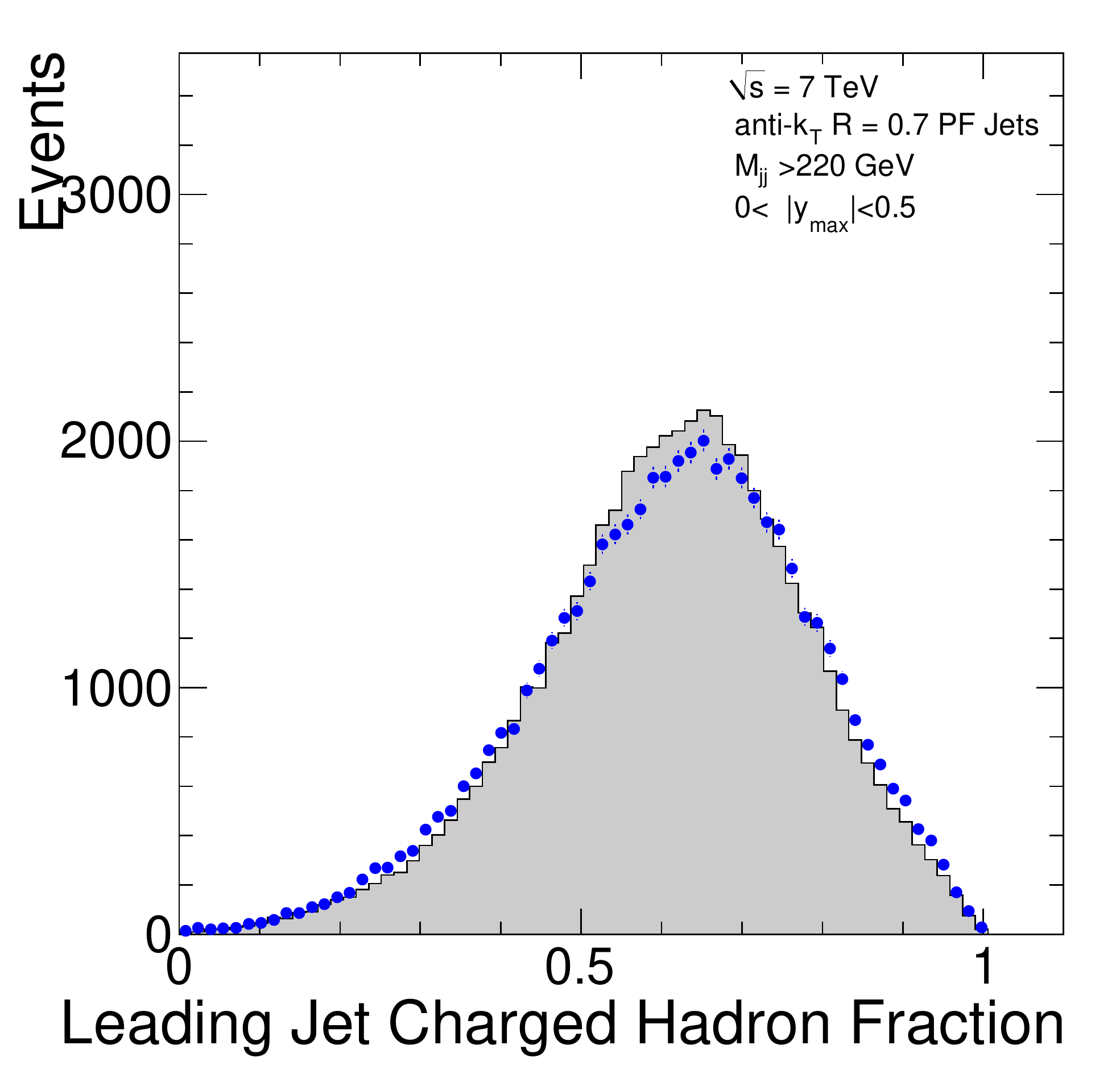} 
\includegraphics[width=0.48\textwidth]{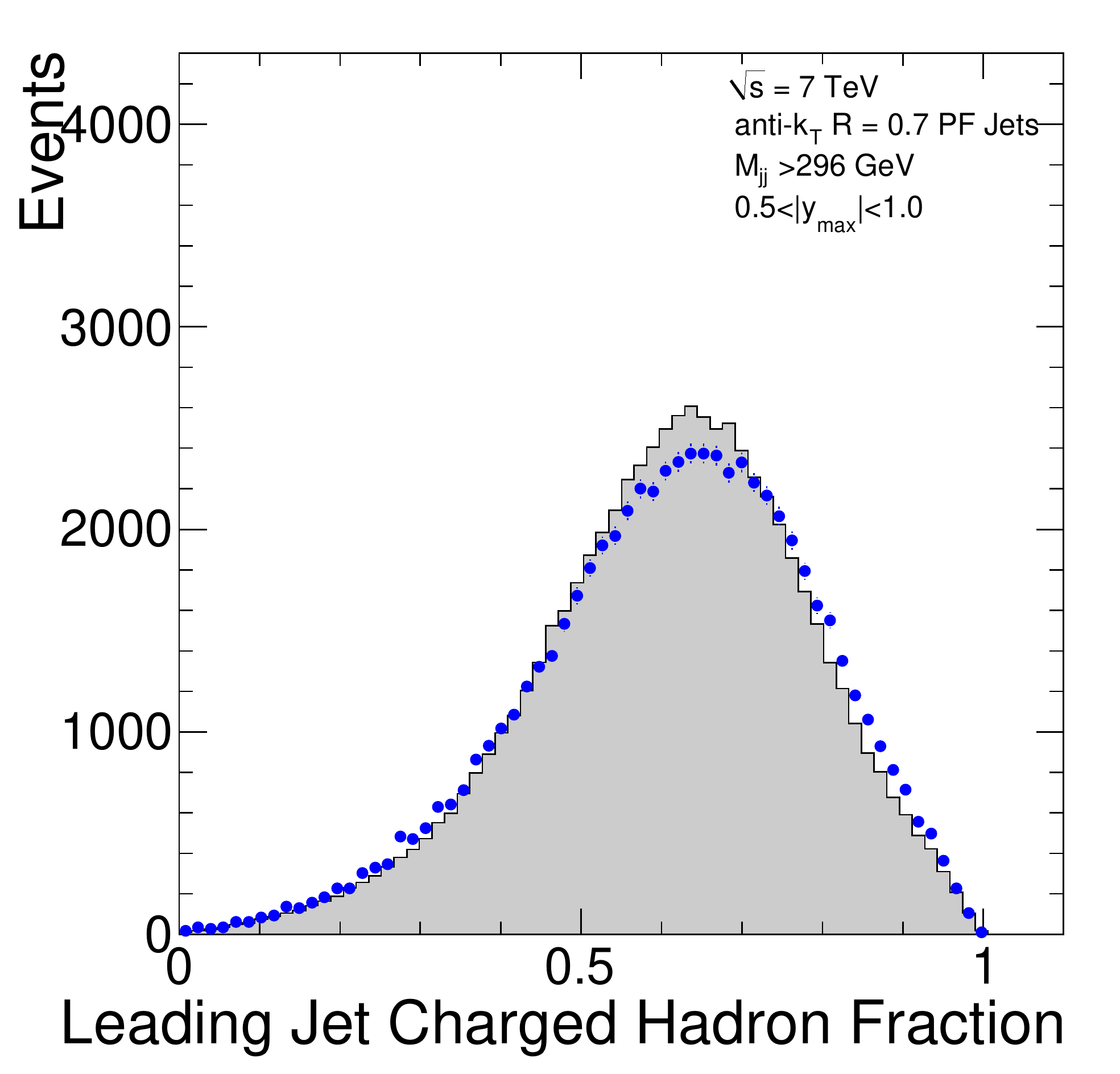} 
\includegraphics[width=0.48\textwidth]{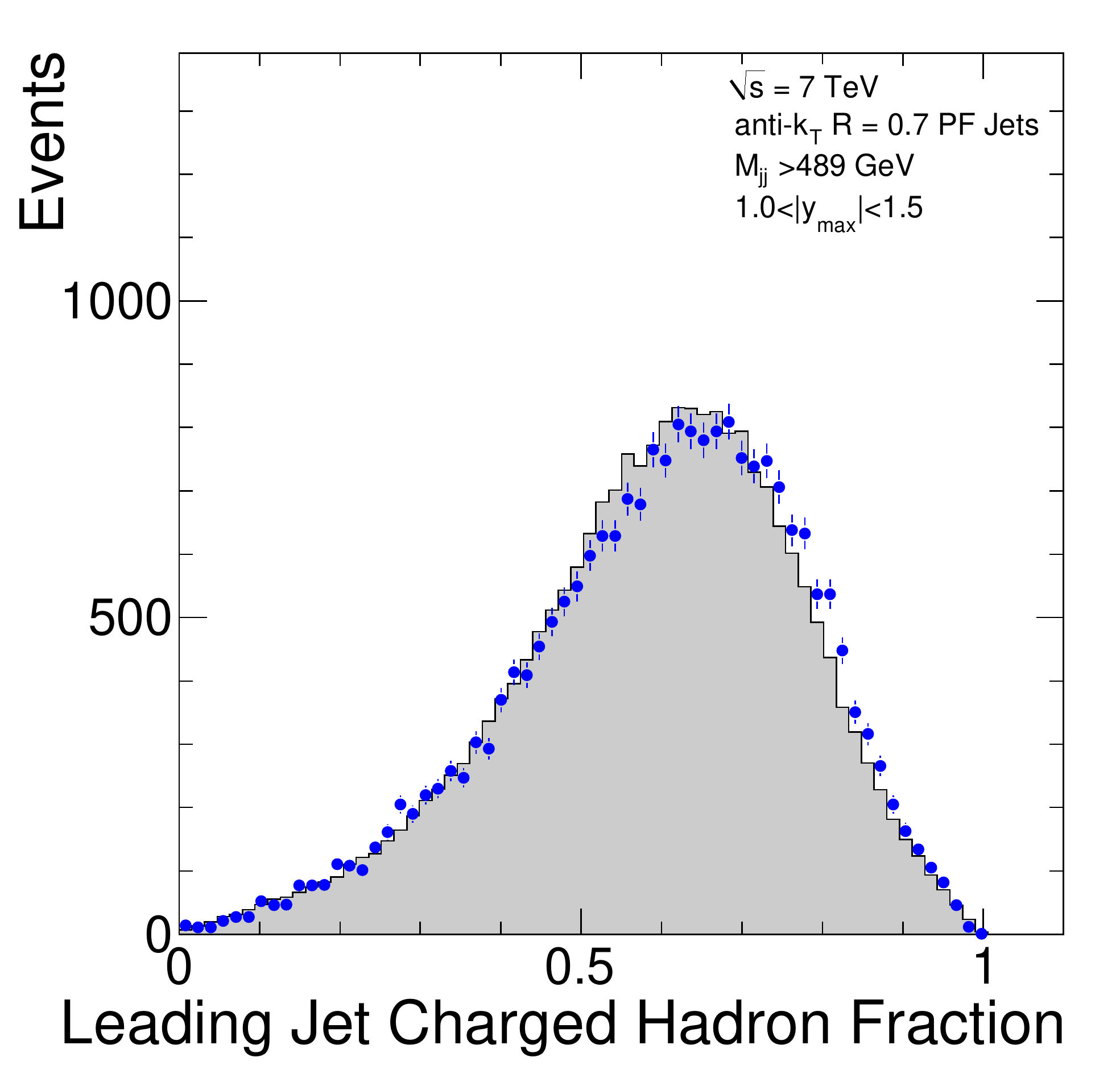} 
\includegraphics[width=0.48\textwidth]{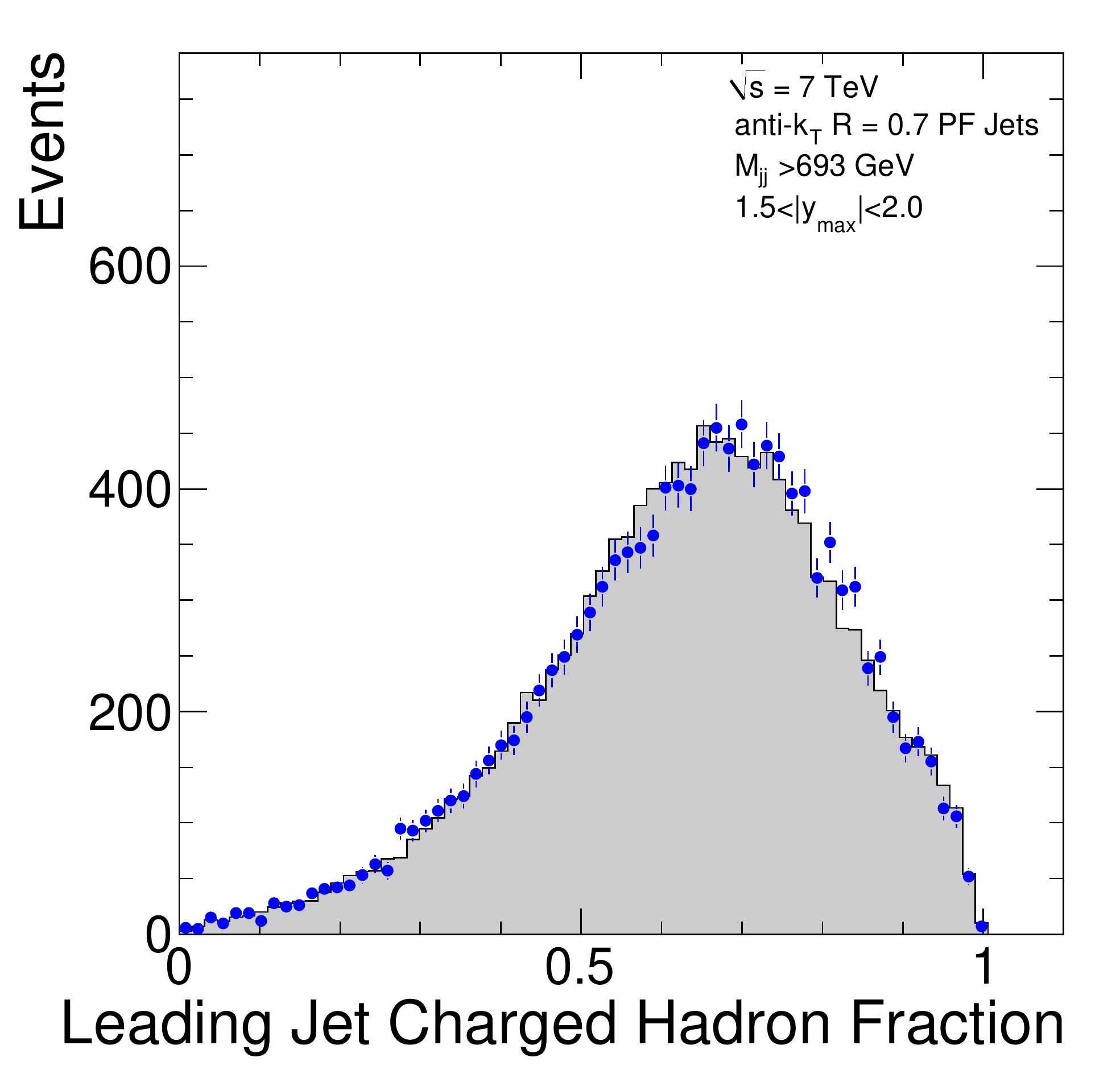} 
\includegraphics[width=0.48\textwidth]{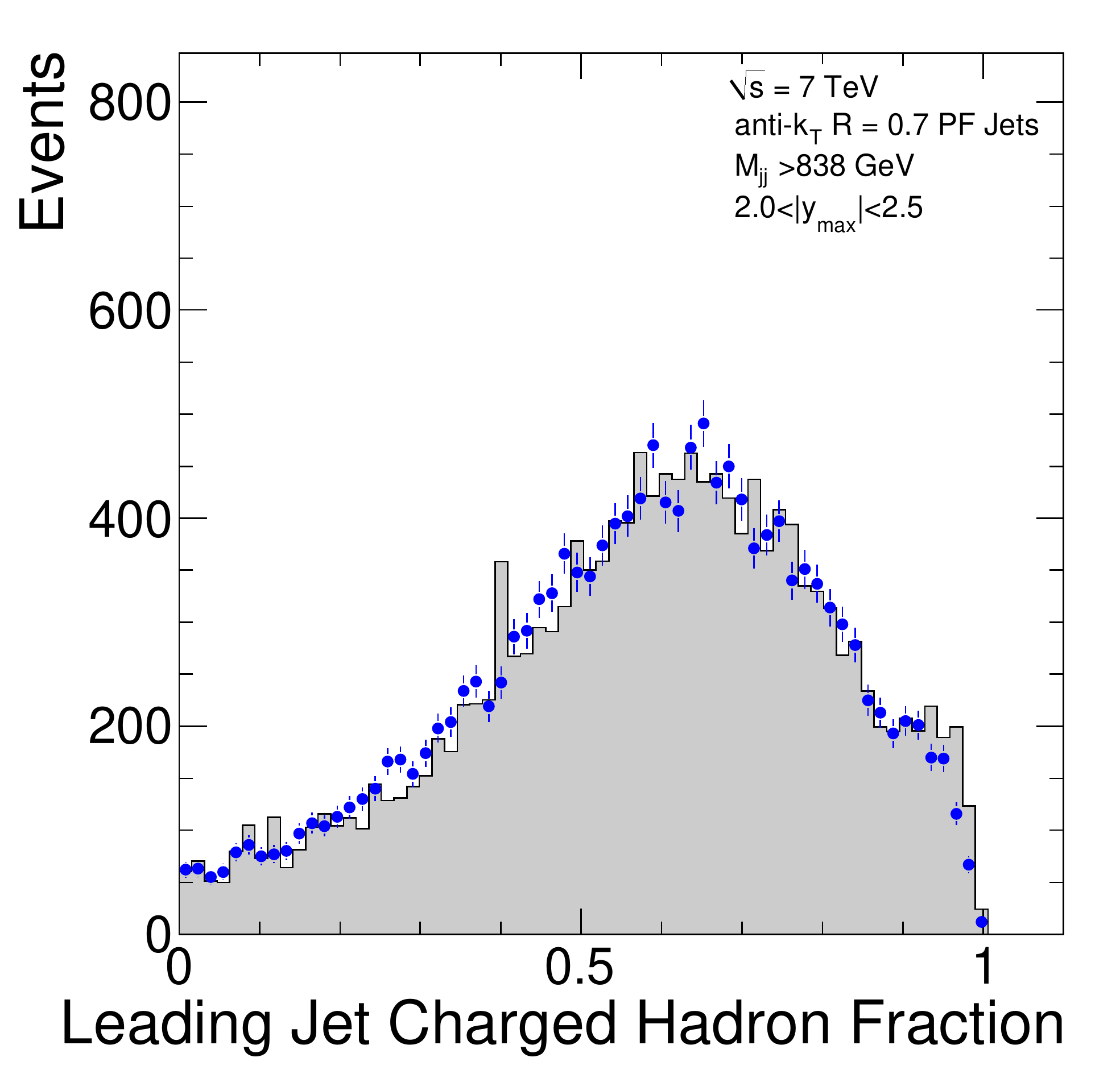}

\caption{ The charged hadron fraction of the leading jet  for the five different $y_{max}$ bins and for the
HLT$_{-}$Jet50U trigger, for data (points) and simulated (dashed histogram) events.}
\label{fig_appc7}
\end{figure}

\begin{figure}[ht]
\centering

\includegraphics[width=0.48\textwidth]{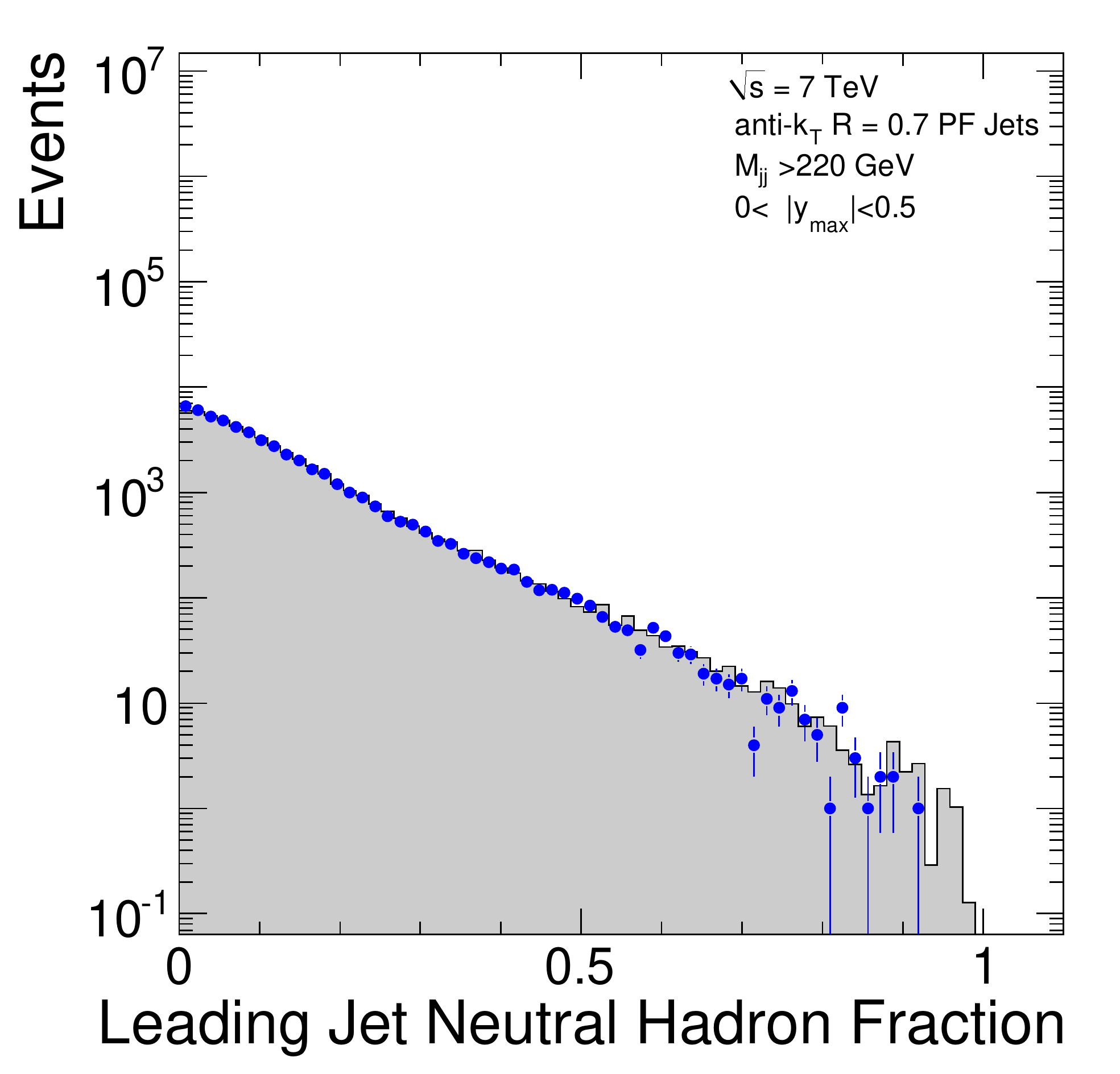} 
\includegraphics[width=0.48\textwidth]{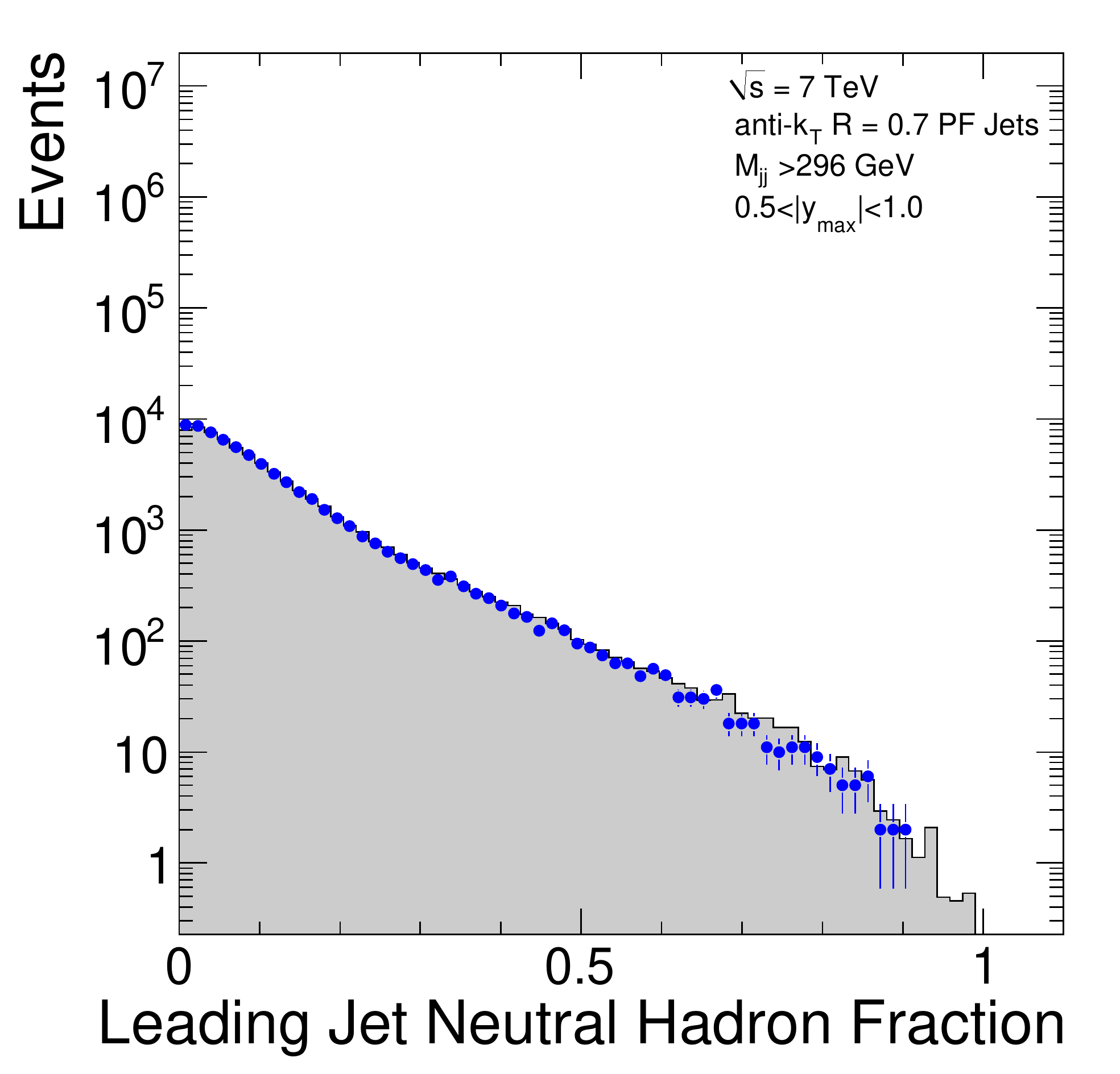} 
\includegraphics[width=0.48\textwidth]{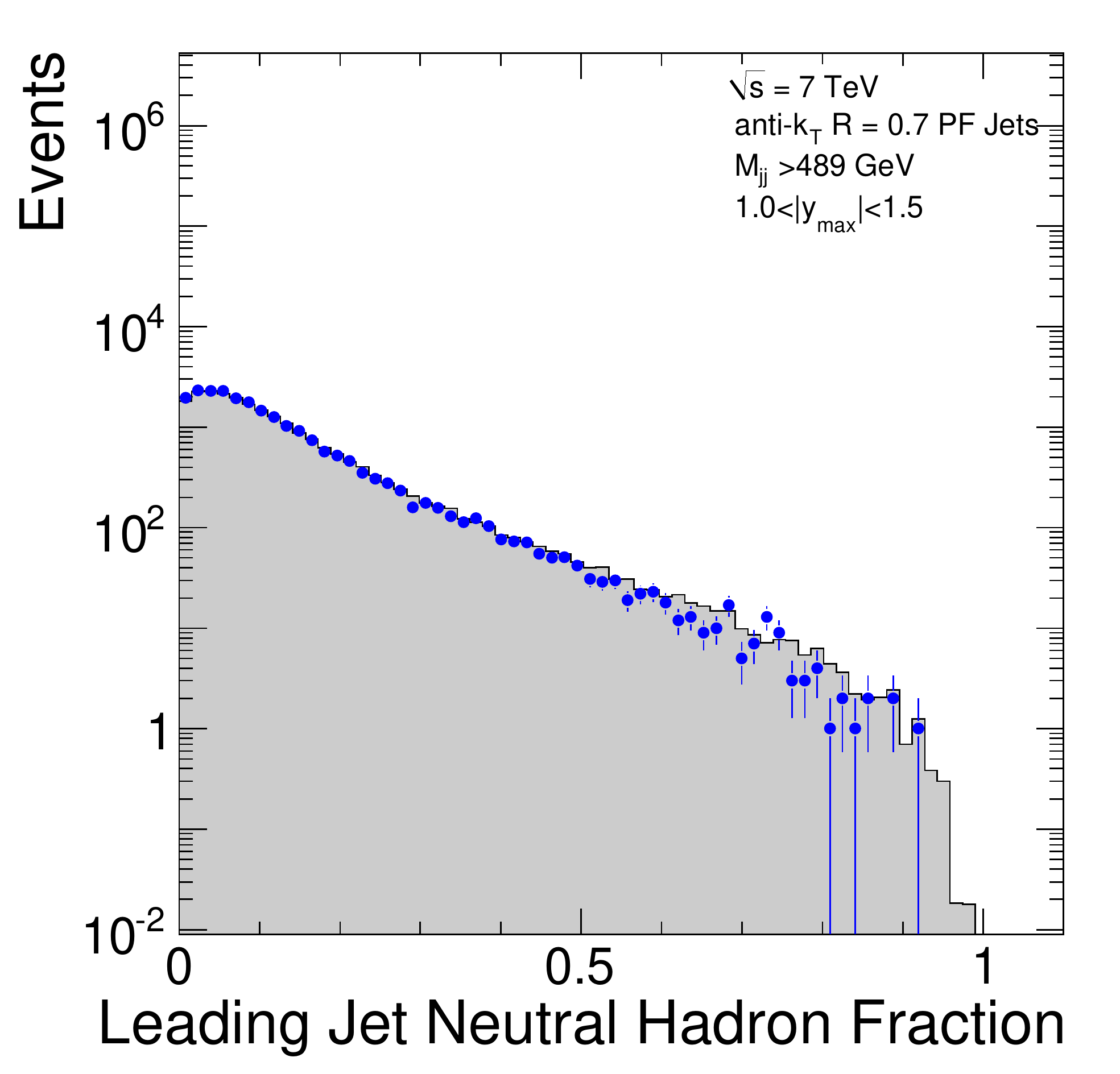} 
\includegraphics[width=0.48\textwidth]{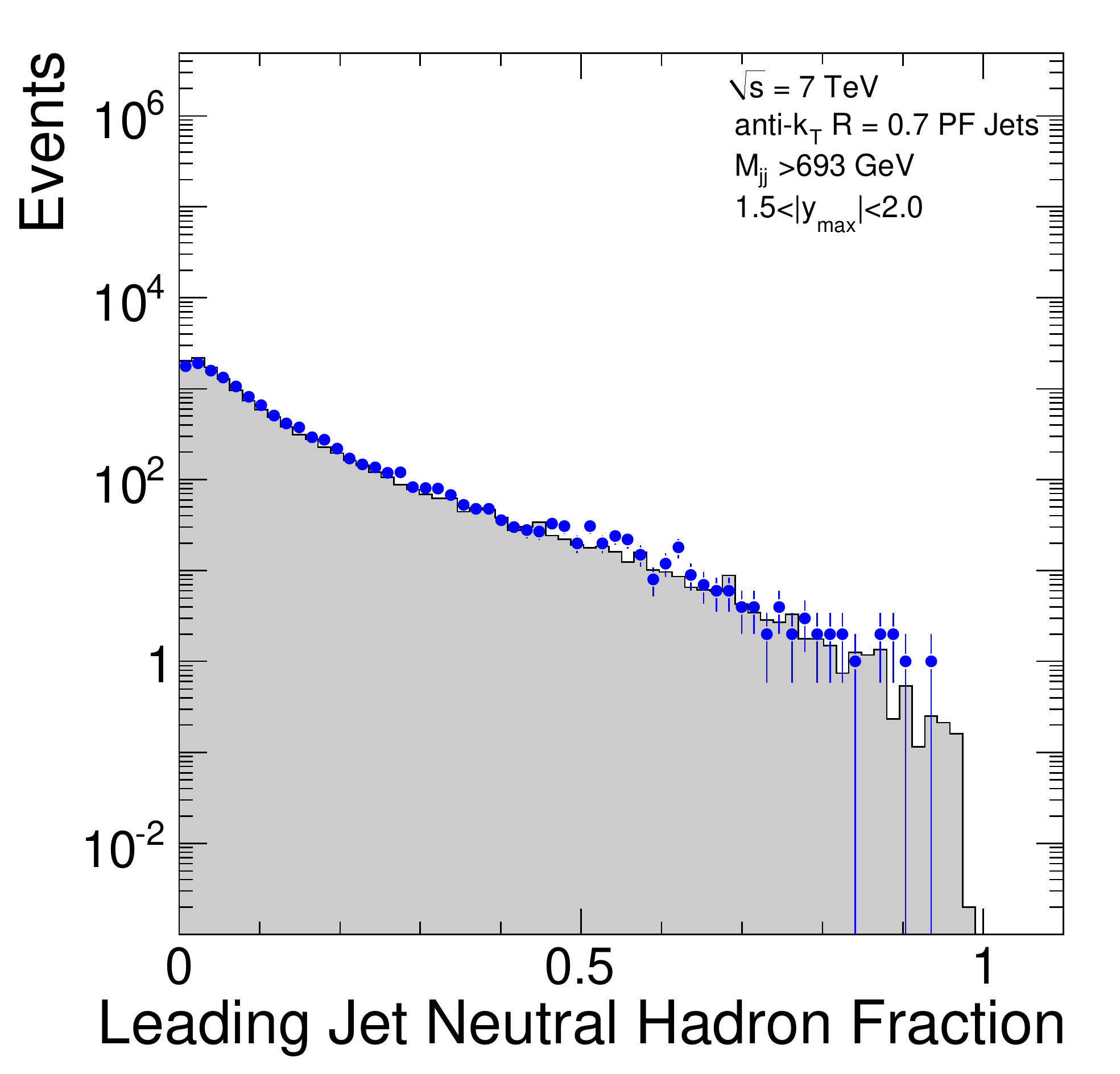} 
\includegraphics[width=0.48\textwidth]{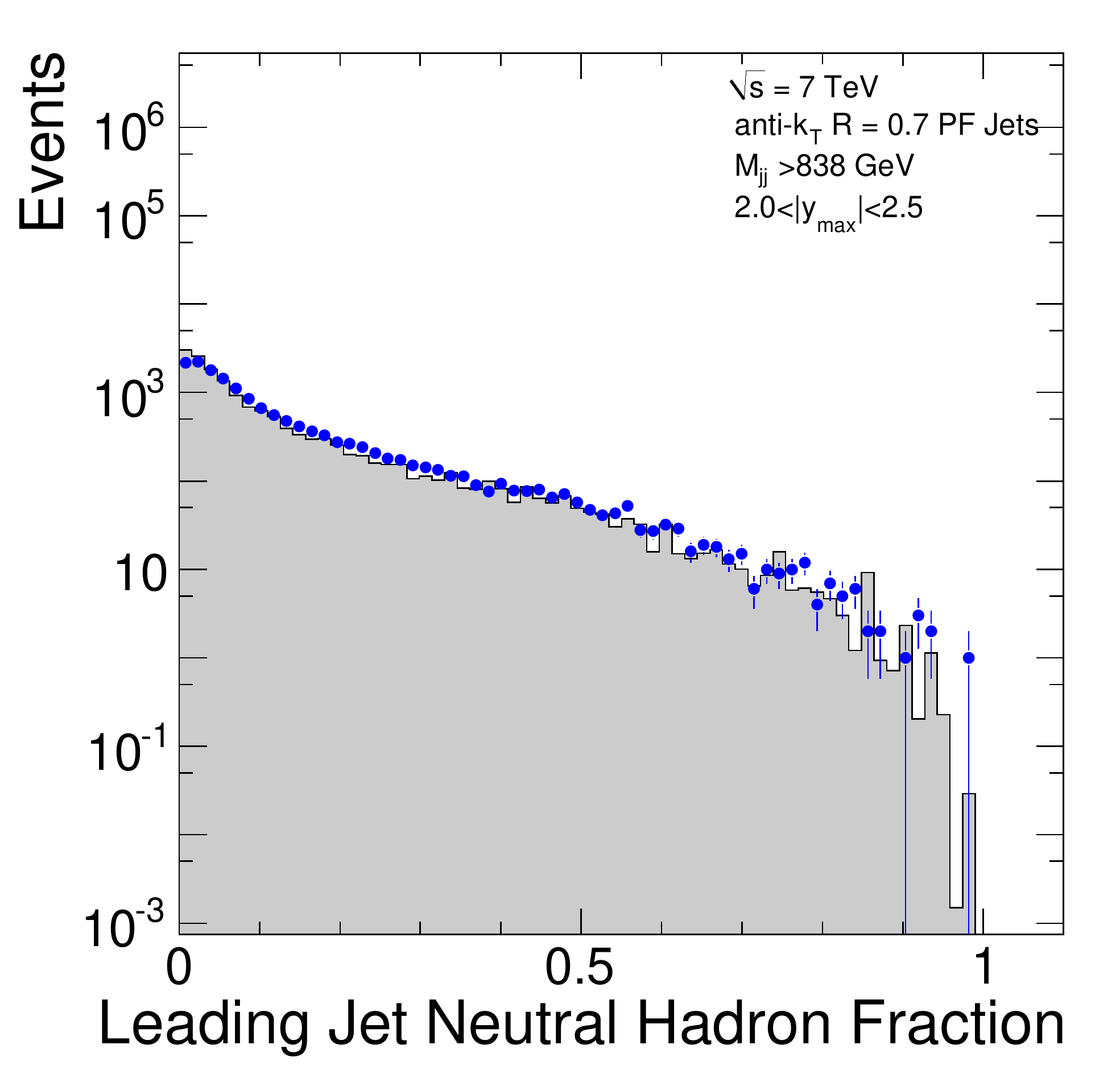}

\caption{ The neutral hadron fraction of the leading jet  for the five different $y_{max}$ bins and for the
HLT$_{-}$Jet50U trigger, for data (points) and simulated (dashed histogram) events.}
\label{fig_appc8}
\end{figure}

\begin{figure}[ht]
\centering

\includegraphics[width=0.48\textwidth]{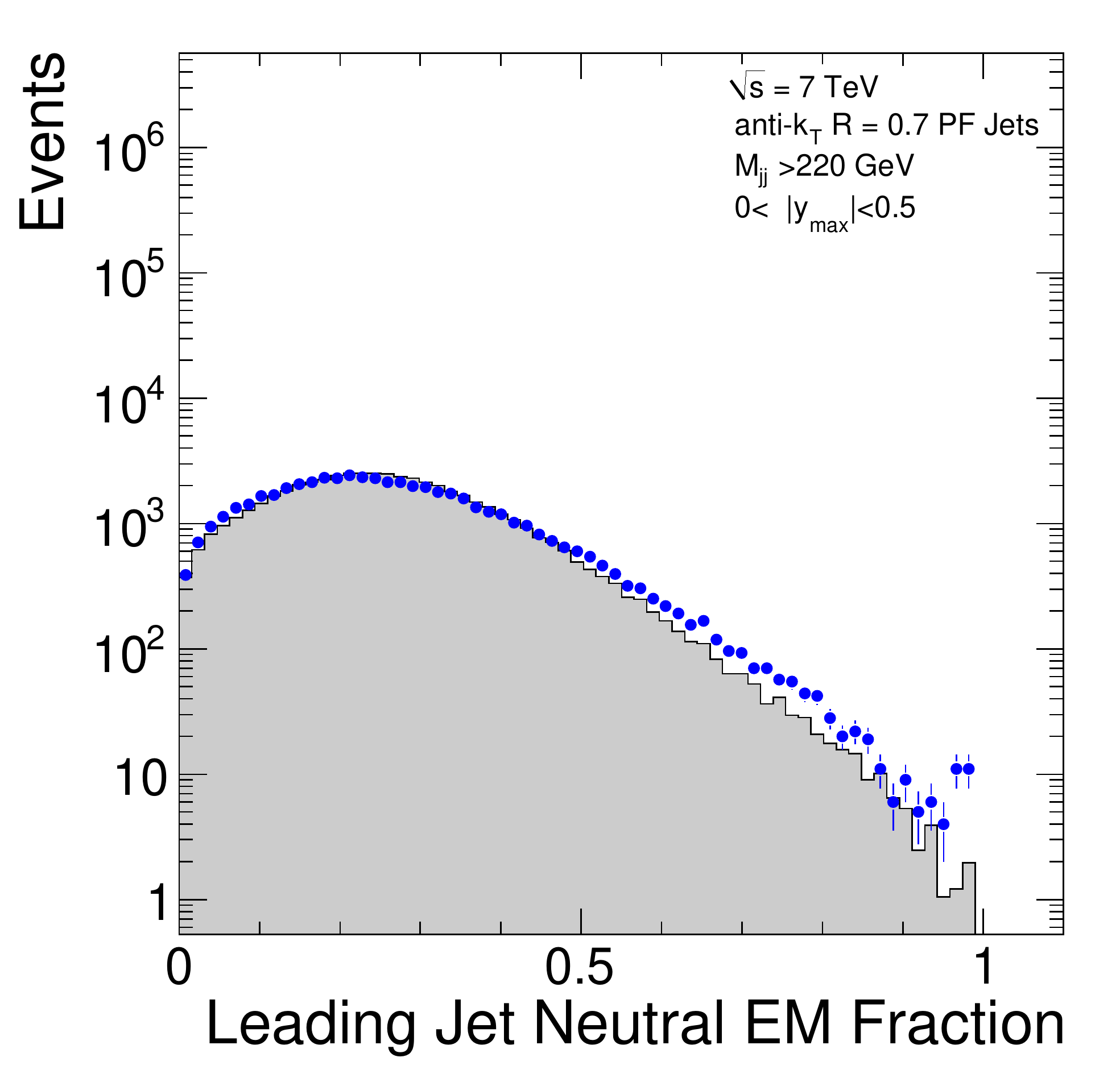} 
\includegraphics[width=0.48\textwidth]{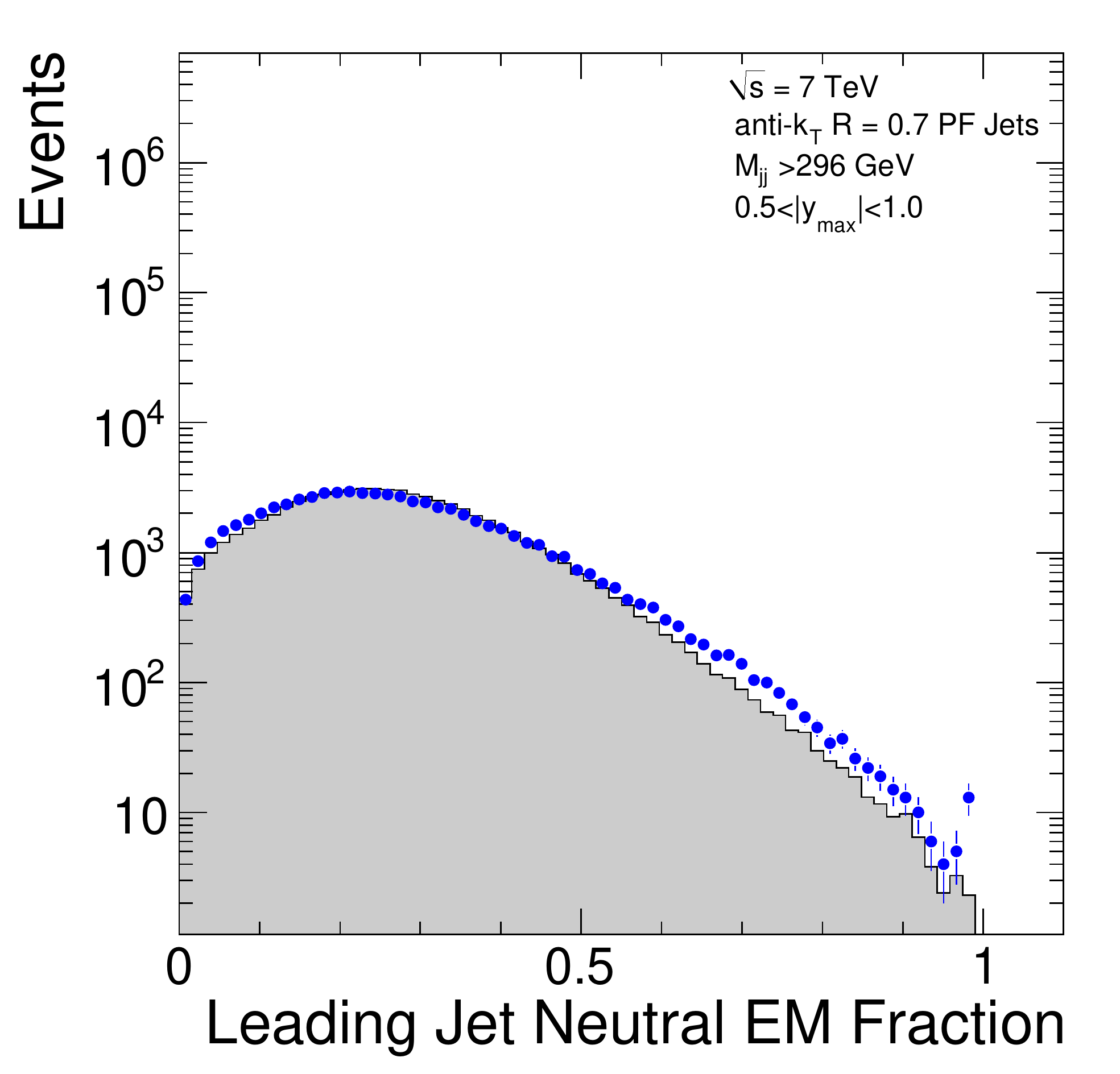} 
\includegraphics[width=0.48\textwidth]{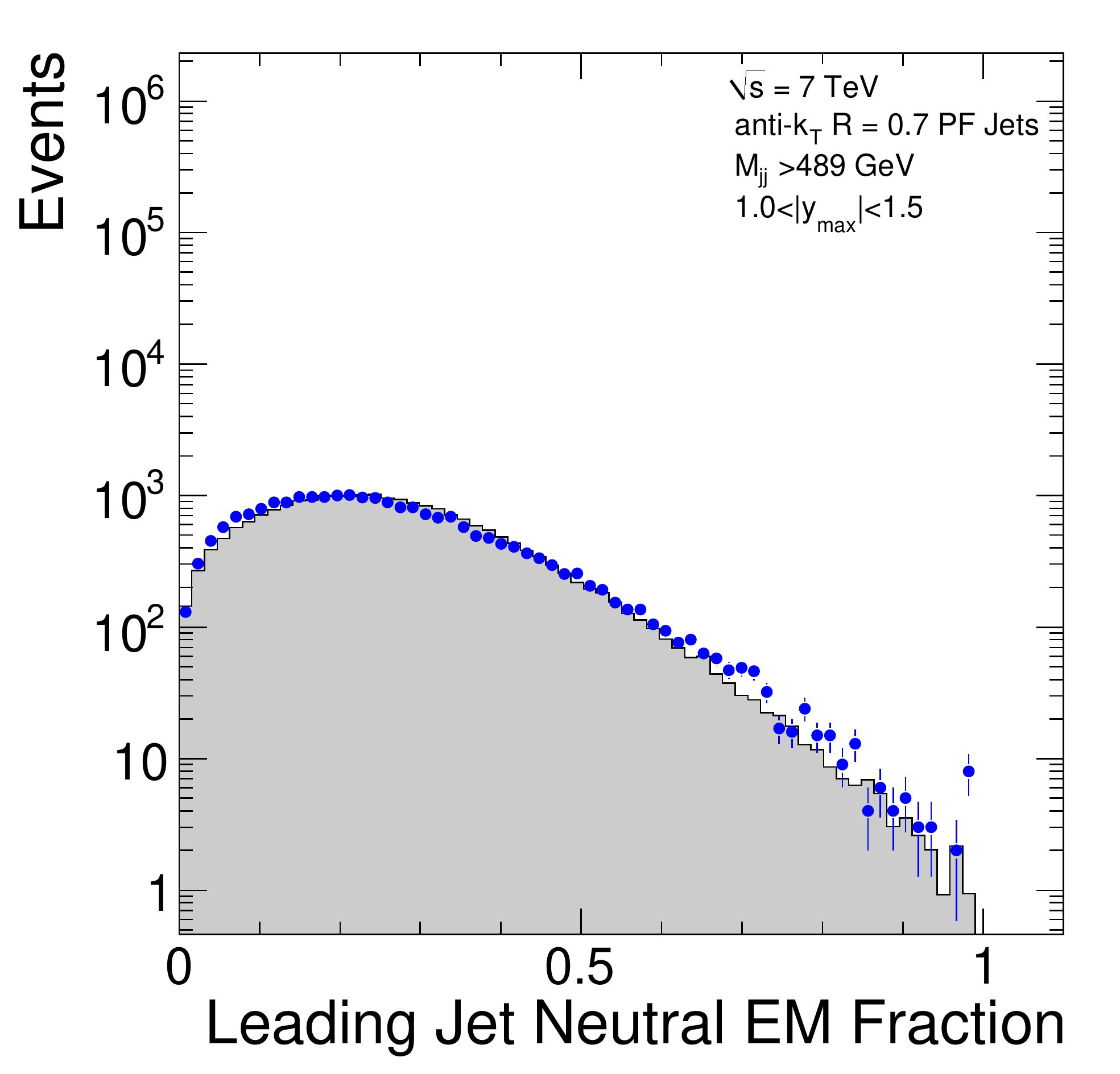} 
\includegraphics[width=0.48\textwidth]{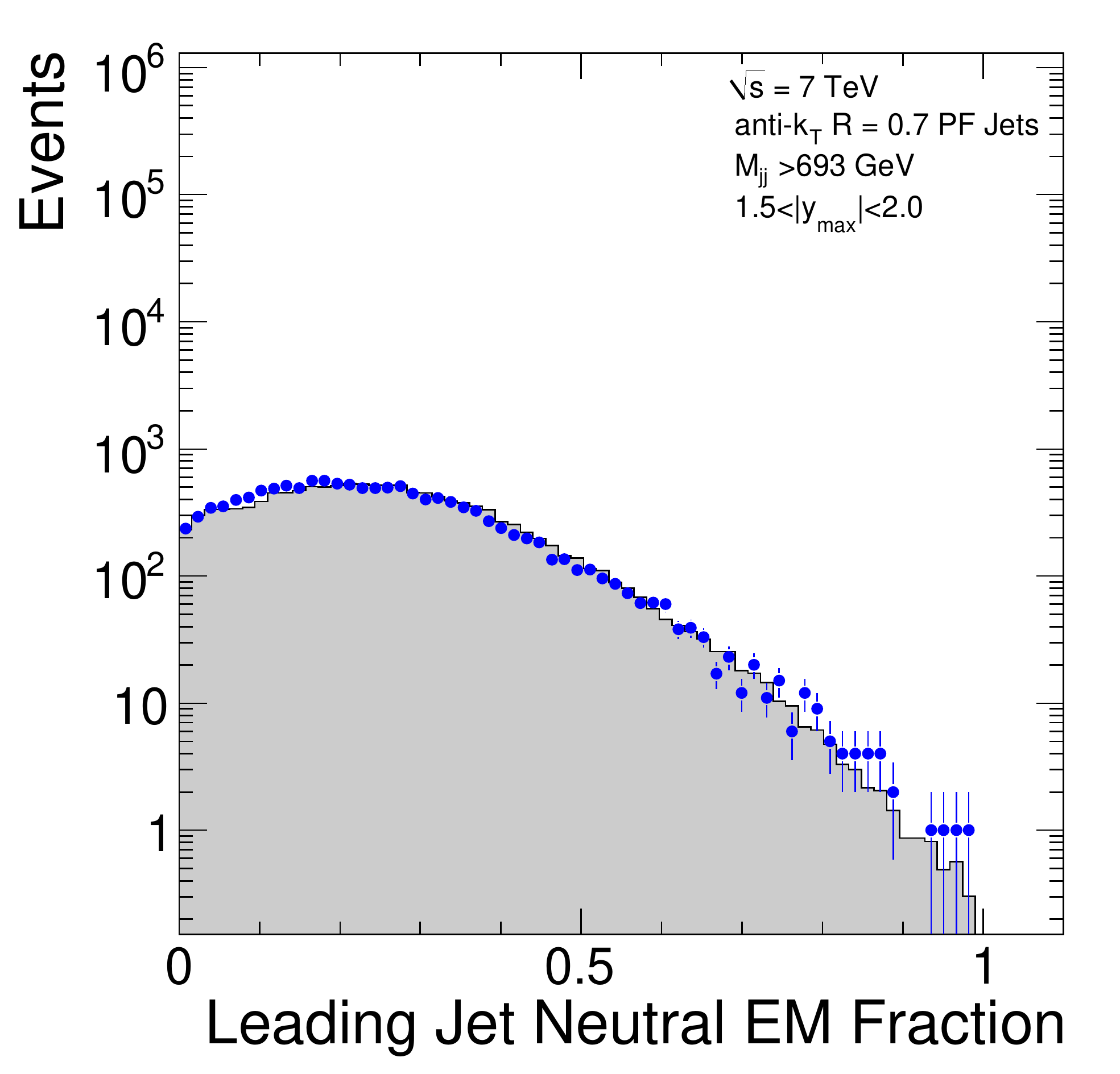} 
\includegraphics[width=0.48\textwidth]{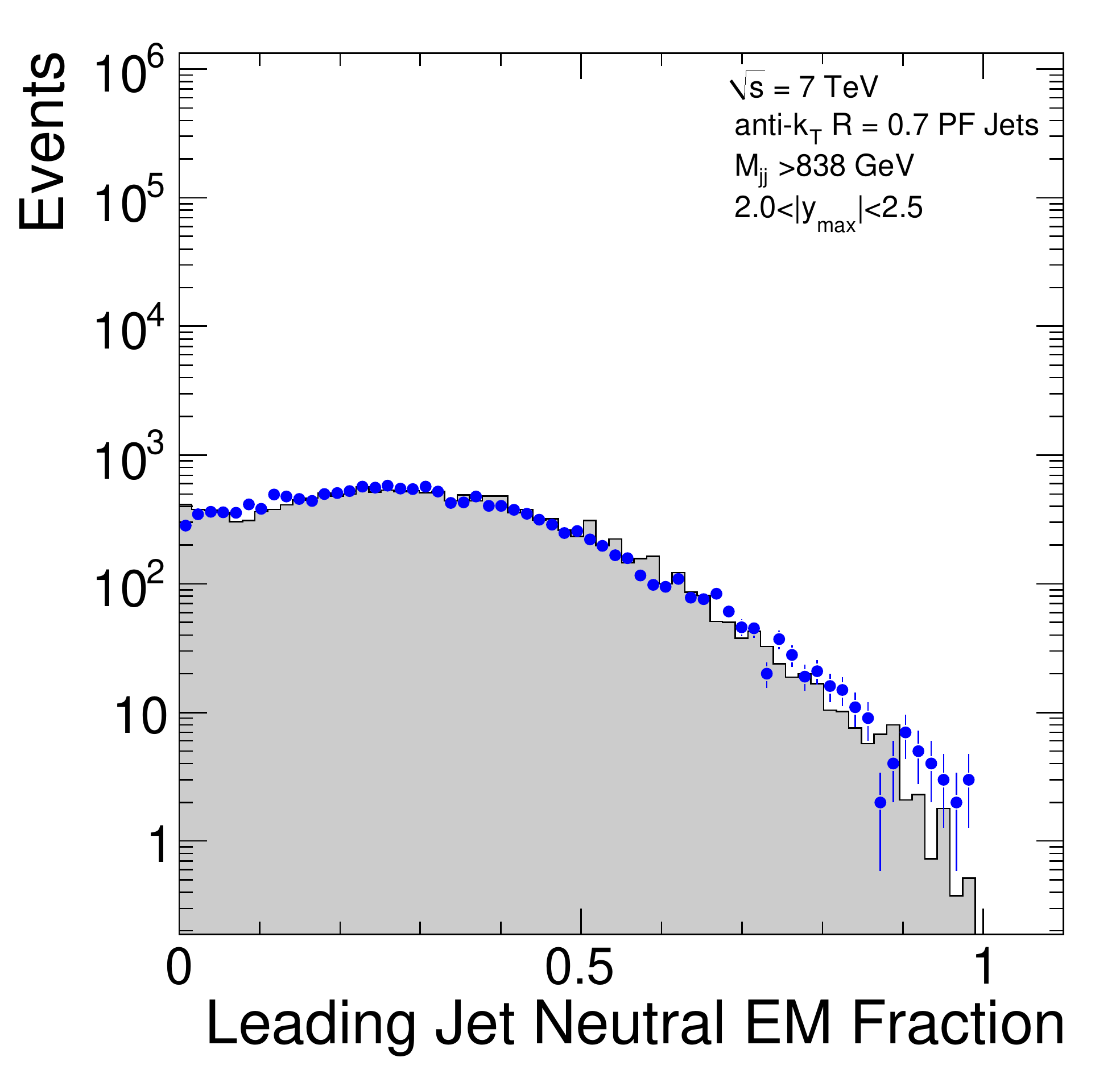}

\caption{ The neutral electromagnetic  fraction of the leading jet  for the five different $y_{max}$ bins and for the
HLT$_{-}$Jet50U trigger, for data (points) and simulated (dashed histogram) events.}
\label{fig_appc9}
\end{figure}

 \begin{figure}[ht]
\centering

\includegraphics[width=0.48\textwidth]{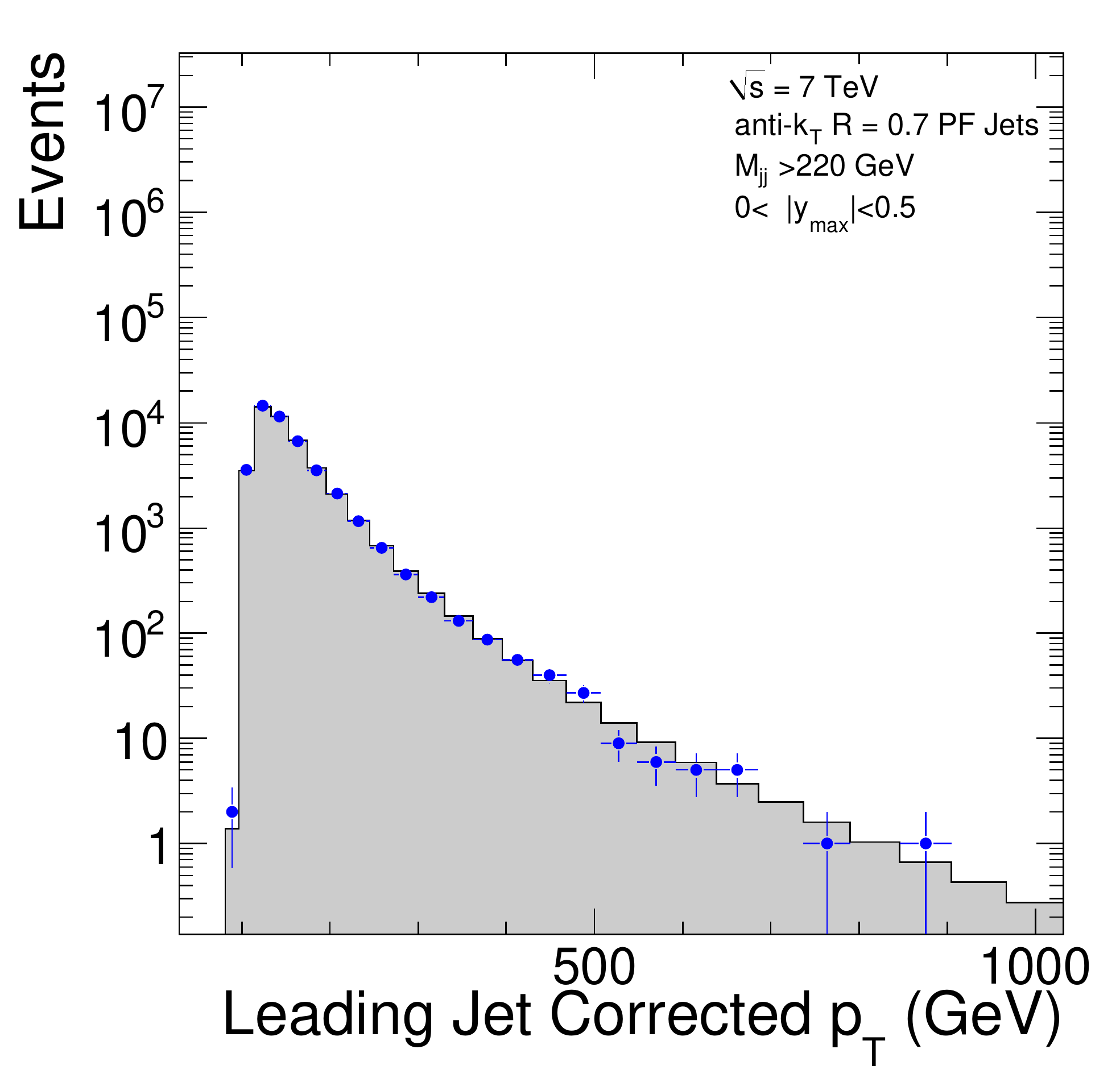} 
\includegraphics[width=0.48\textwidth]{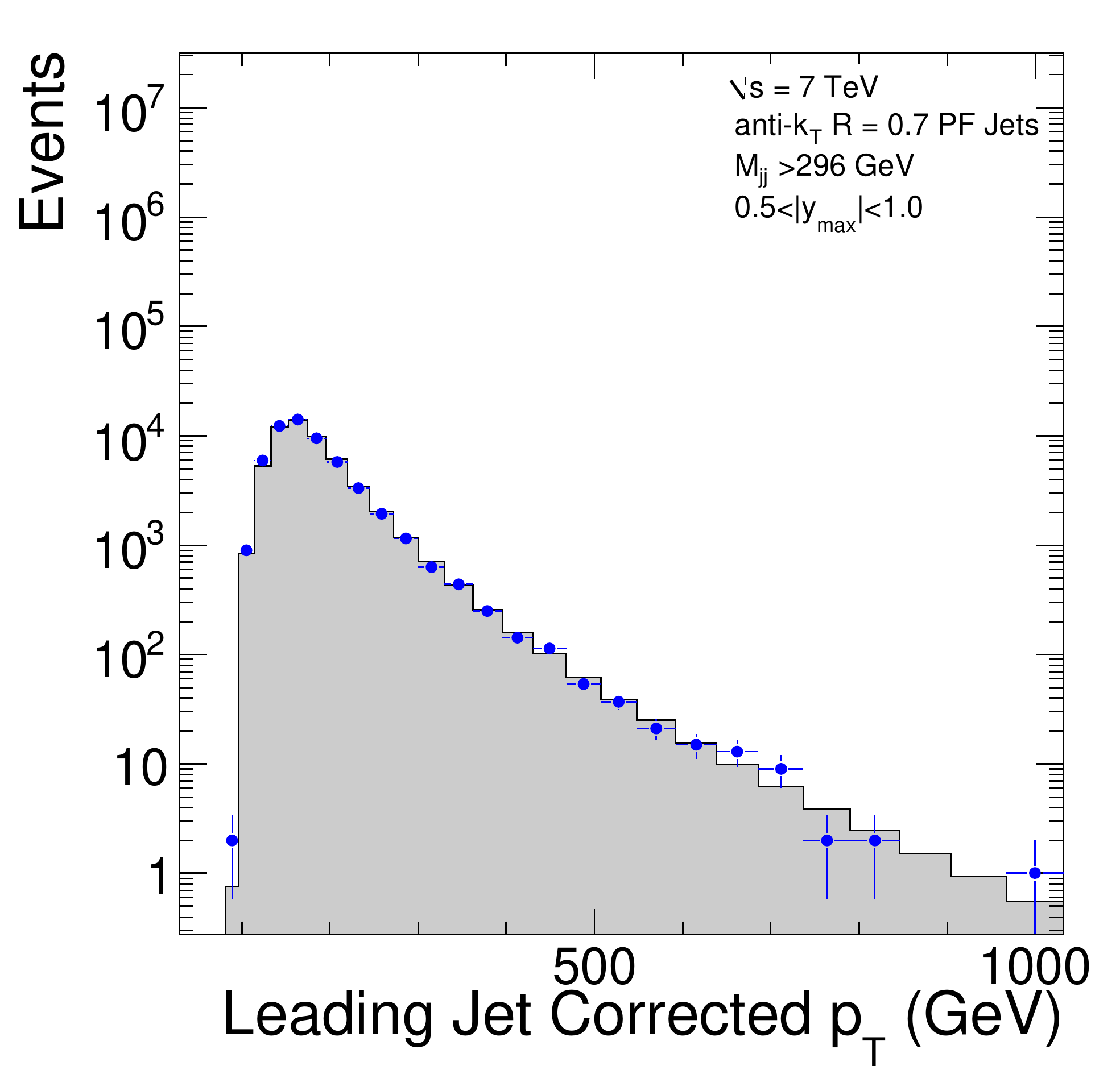} 
\includegraphics[width=0.48\textwidth]{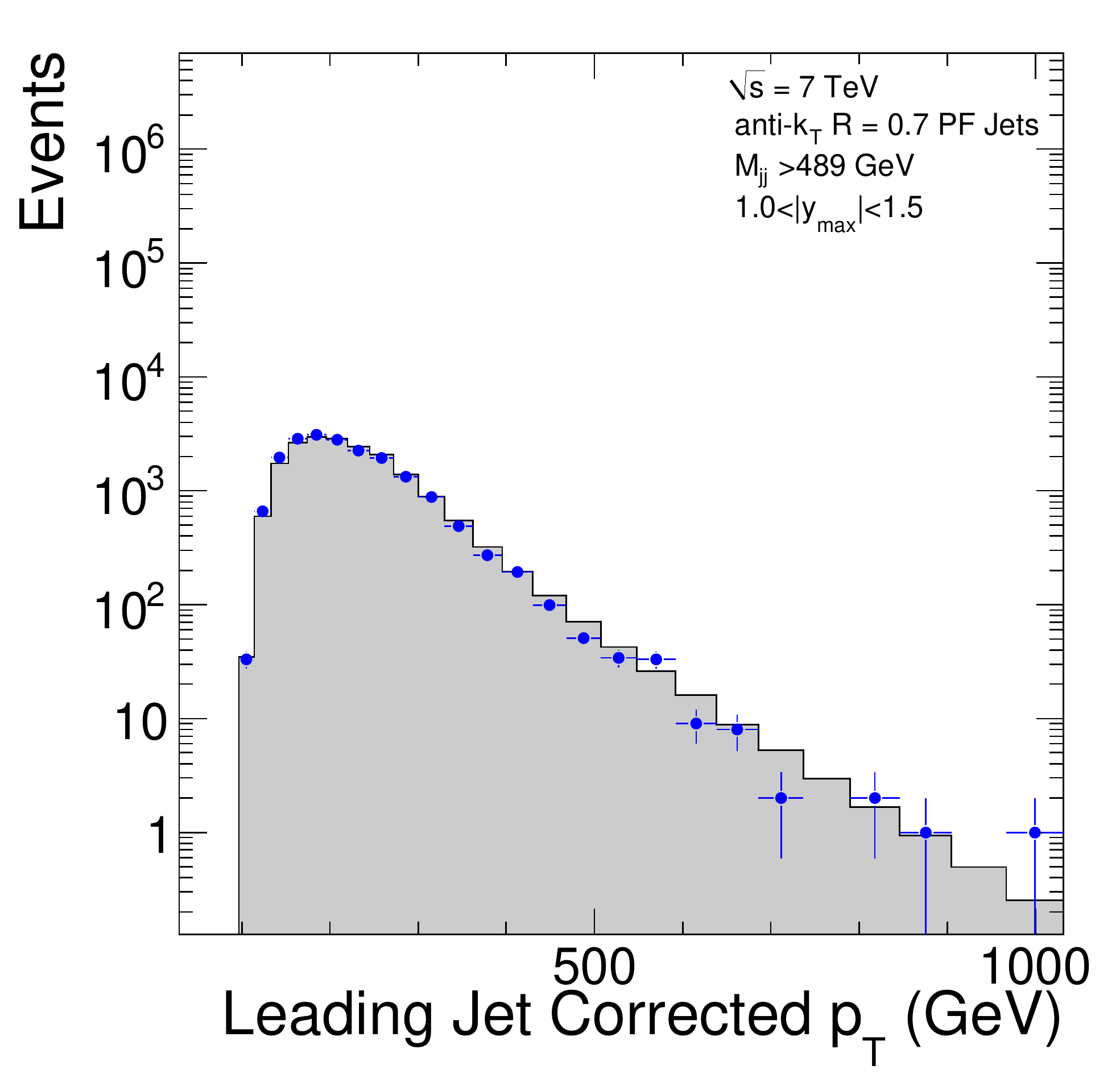} 
\includegraphics[width=0.48\textwidth]{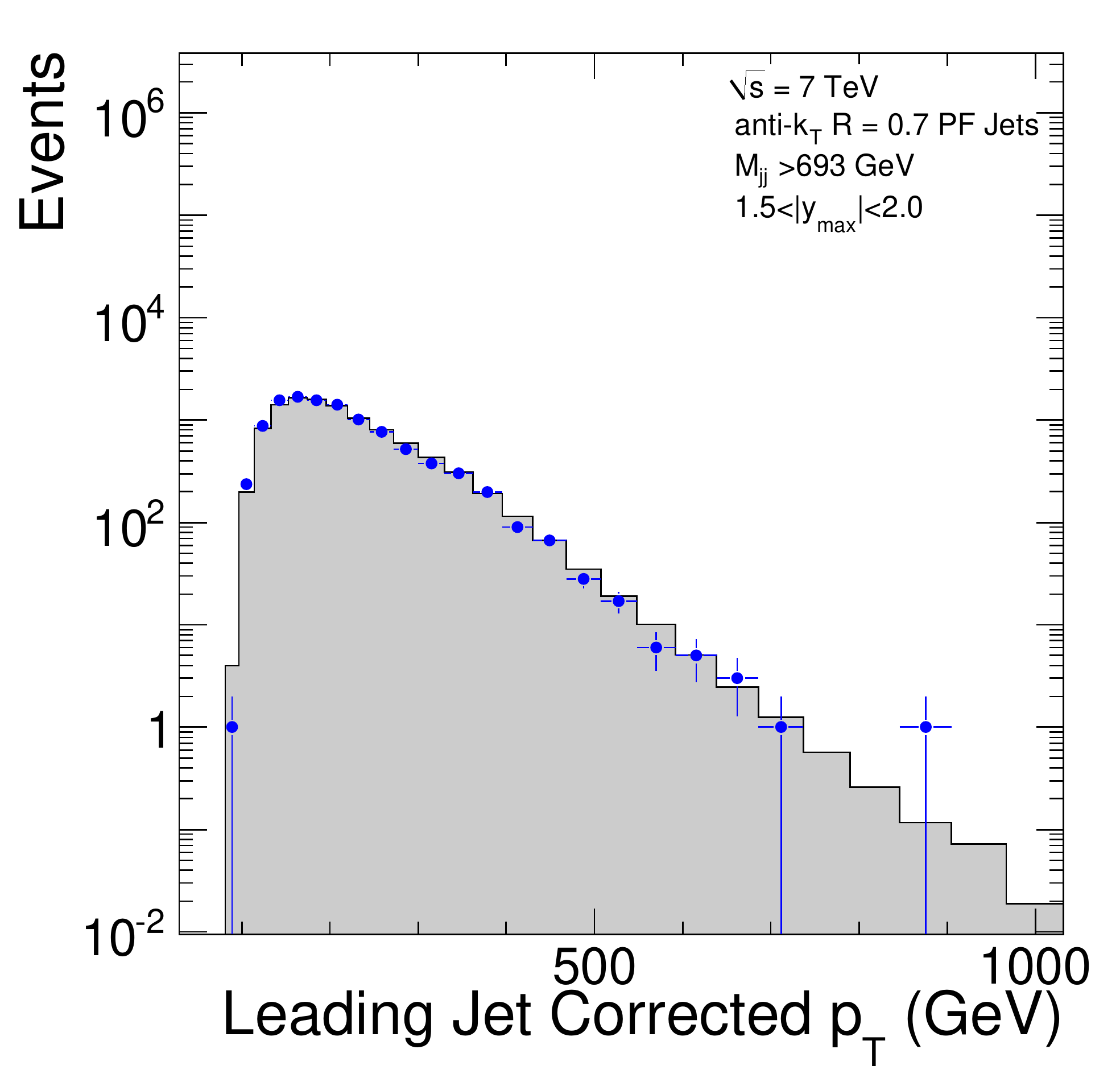} 
\includegraphics[width=0.48\textwidth]{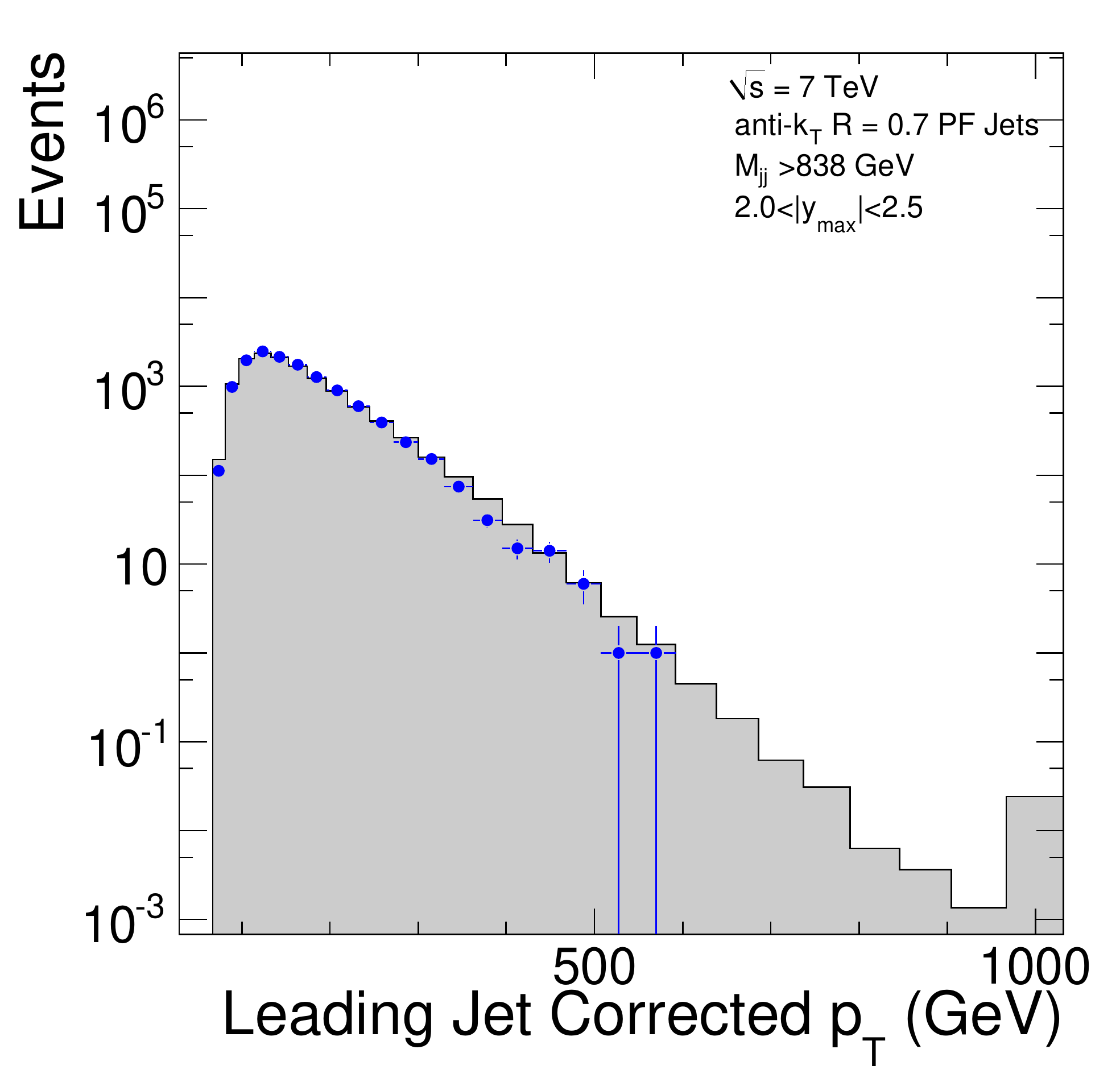}

\caption{ The $p_T$f of the leading jet  for the five different $y_{max}$ bins and for the
HLT$_{-}$Jet50U trigger, for data (points) and simulated (dashed histogram) events.}
\label{fig_appc10}
\end{figure}

\begin{figure}[ht]
\centering

\includegraphics[width=0.48\textwidth]{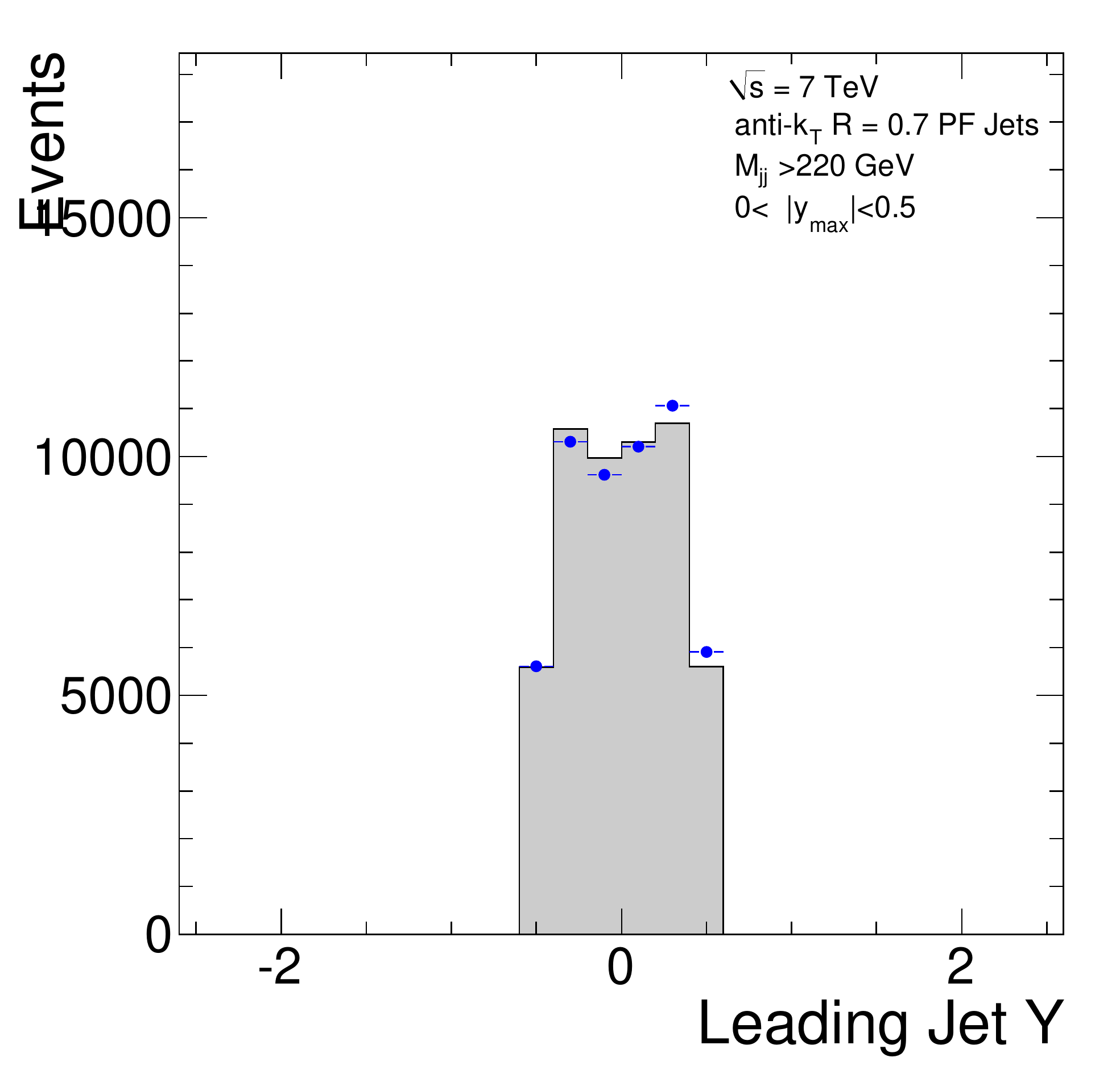} 
\includegraphics[width=0.48\textwidth]{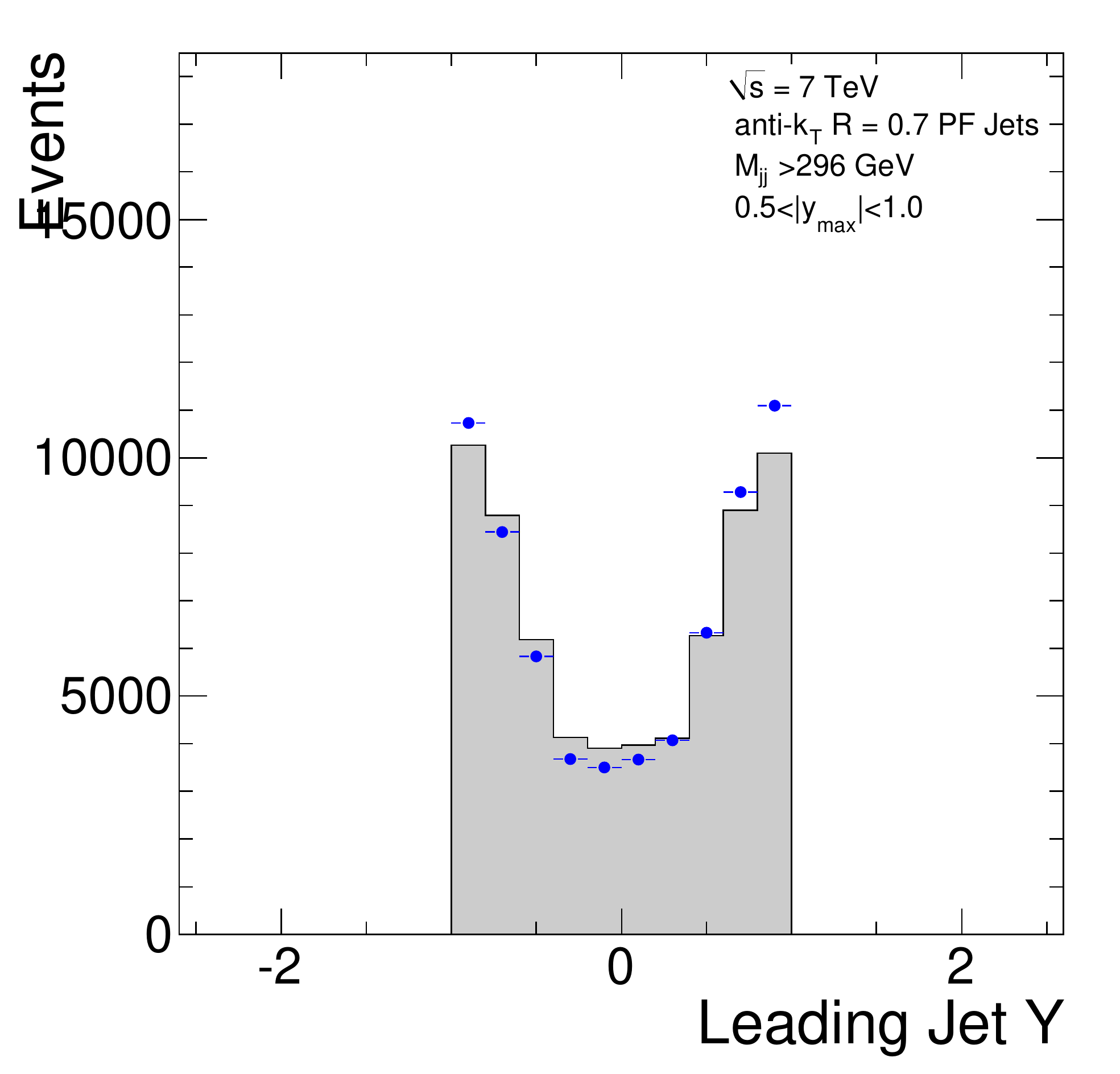} 
\includegraphics[width=0.48\textwidth]{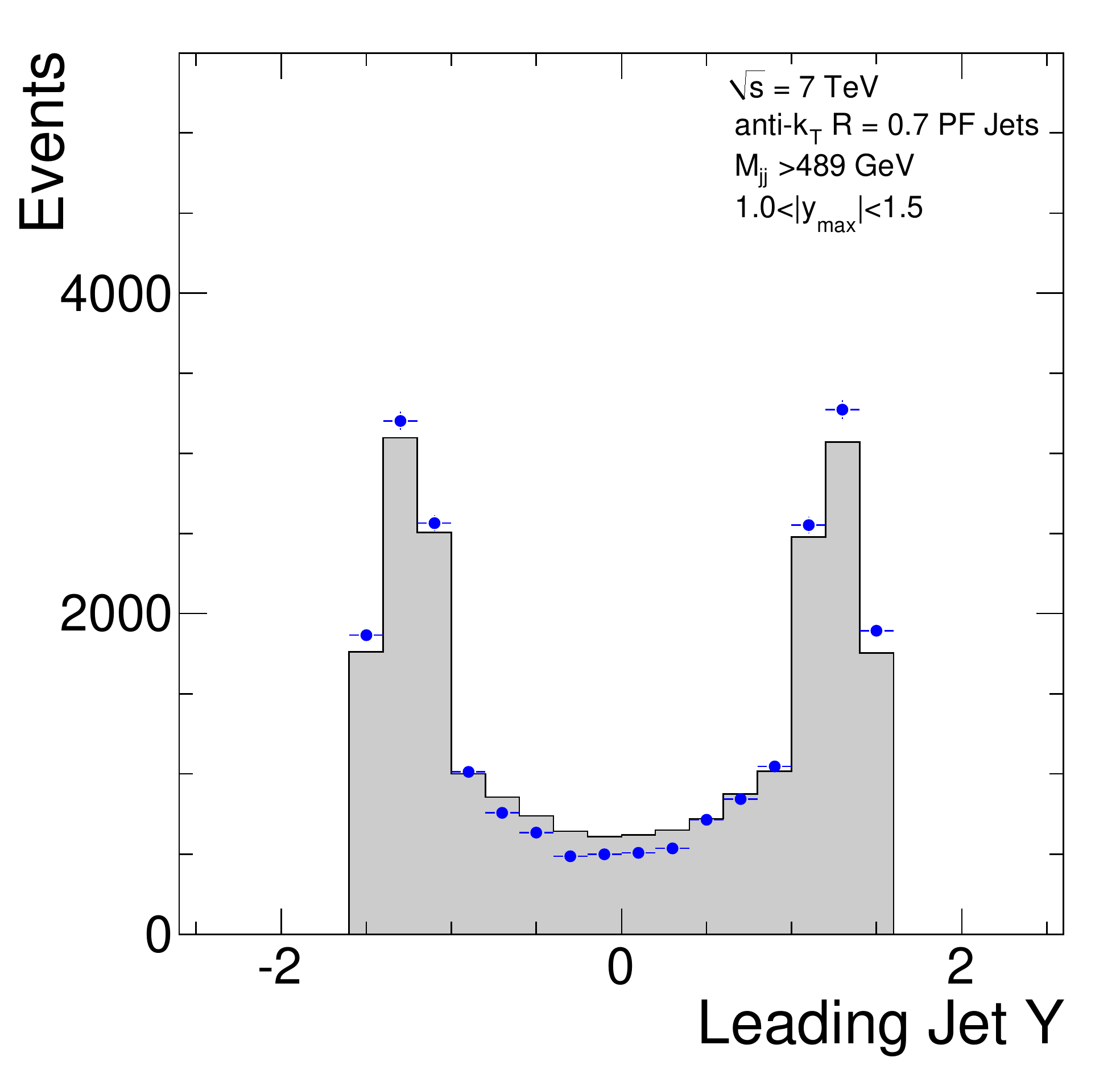} 
\includegraphics[width=0.48\textwidth]{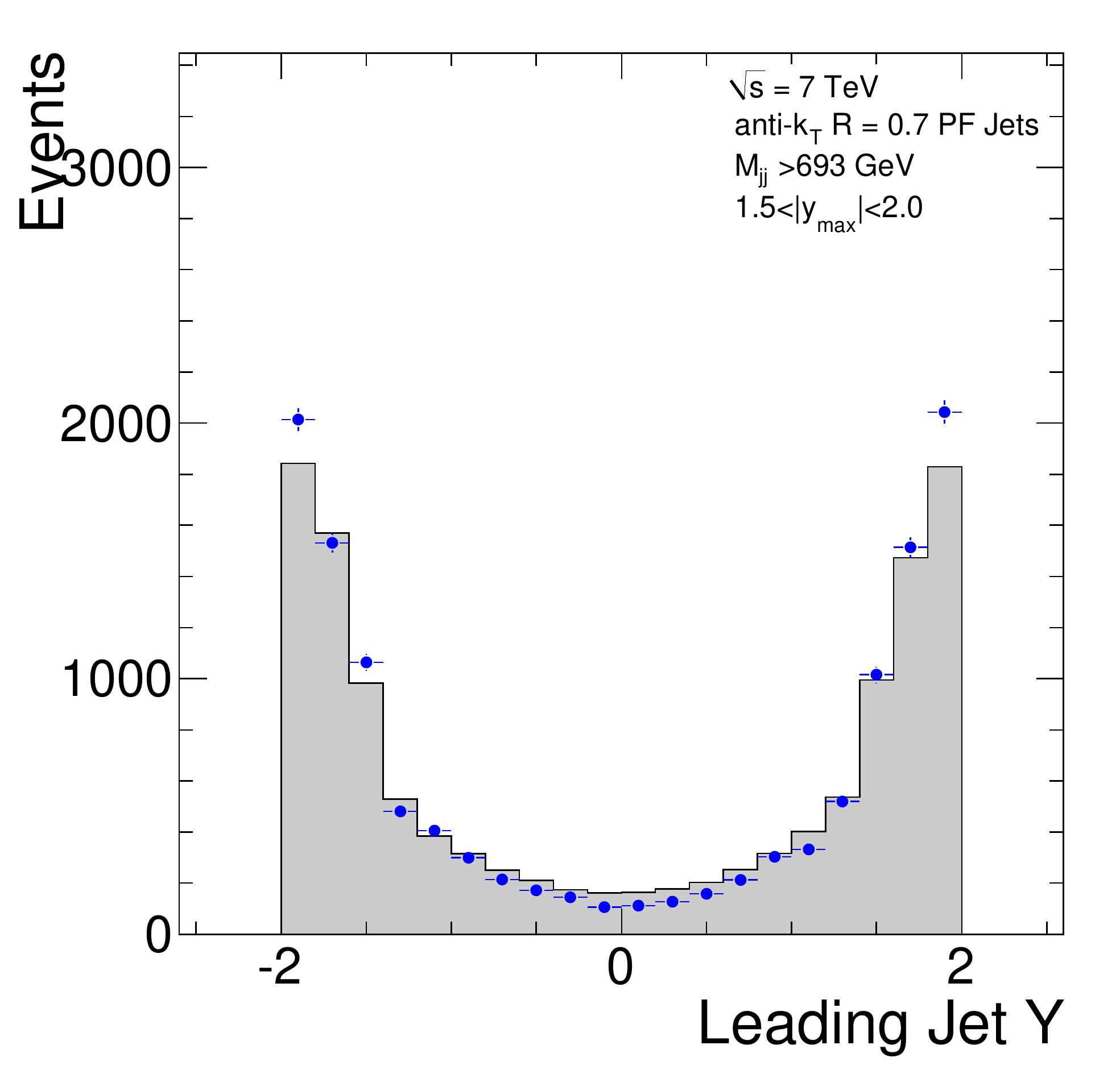} 
\includegraphics[width=0.48\textwidth]{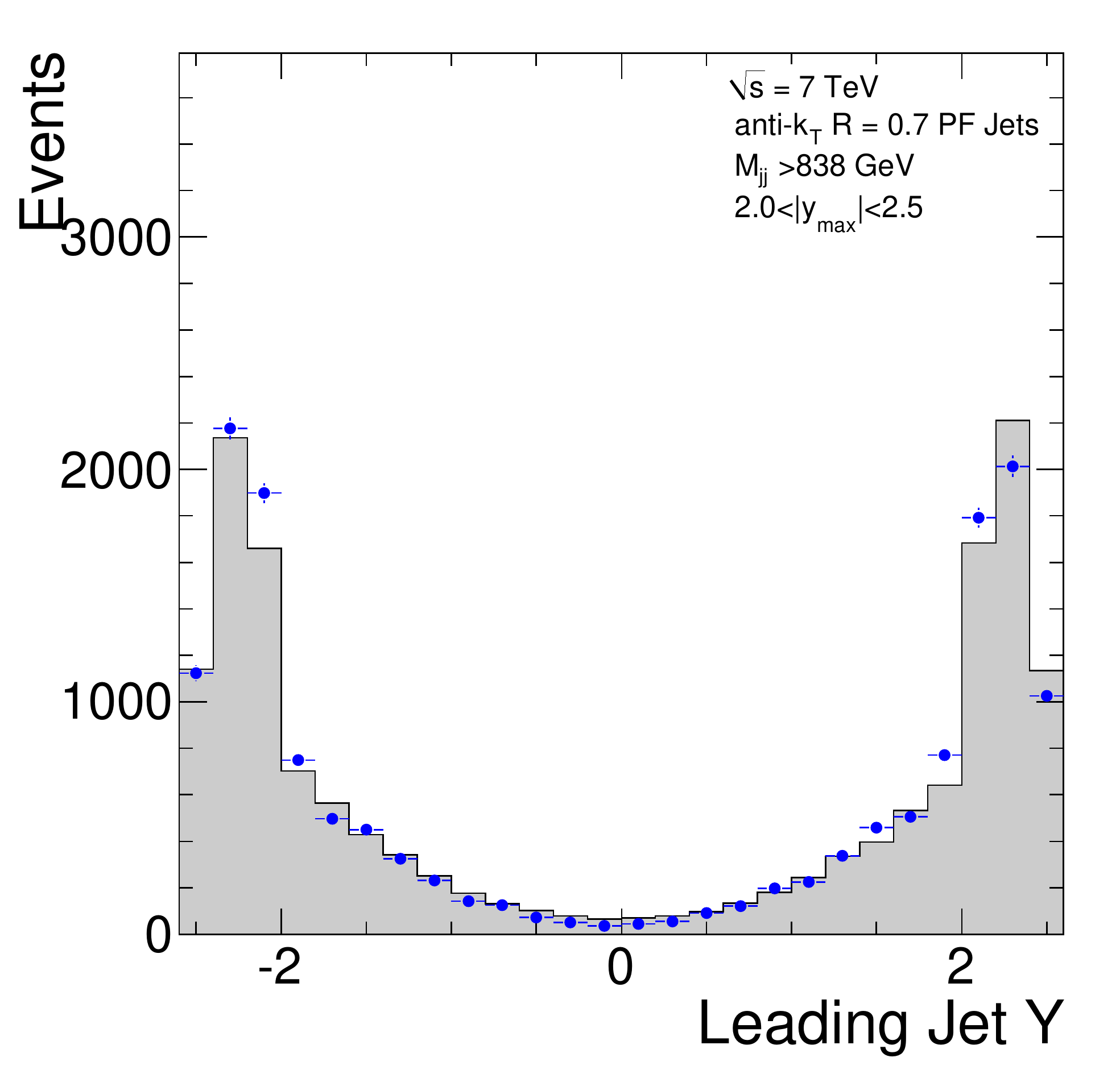}

\caption{ The $\eta$  of the leading jet  for the five different $y_{max}$ bins and for the
HLT$_{-}$Jet50U trigger, for data (points) and simulated (dashed histogram) events.}
\label{fig_appc11}
\end{figure}

\begin{figure}[ht]
\centering

\includegraphics[width=0.48\textwidth]{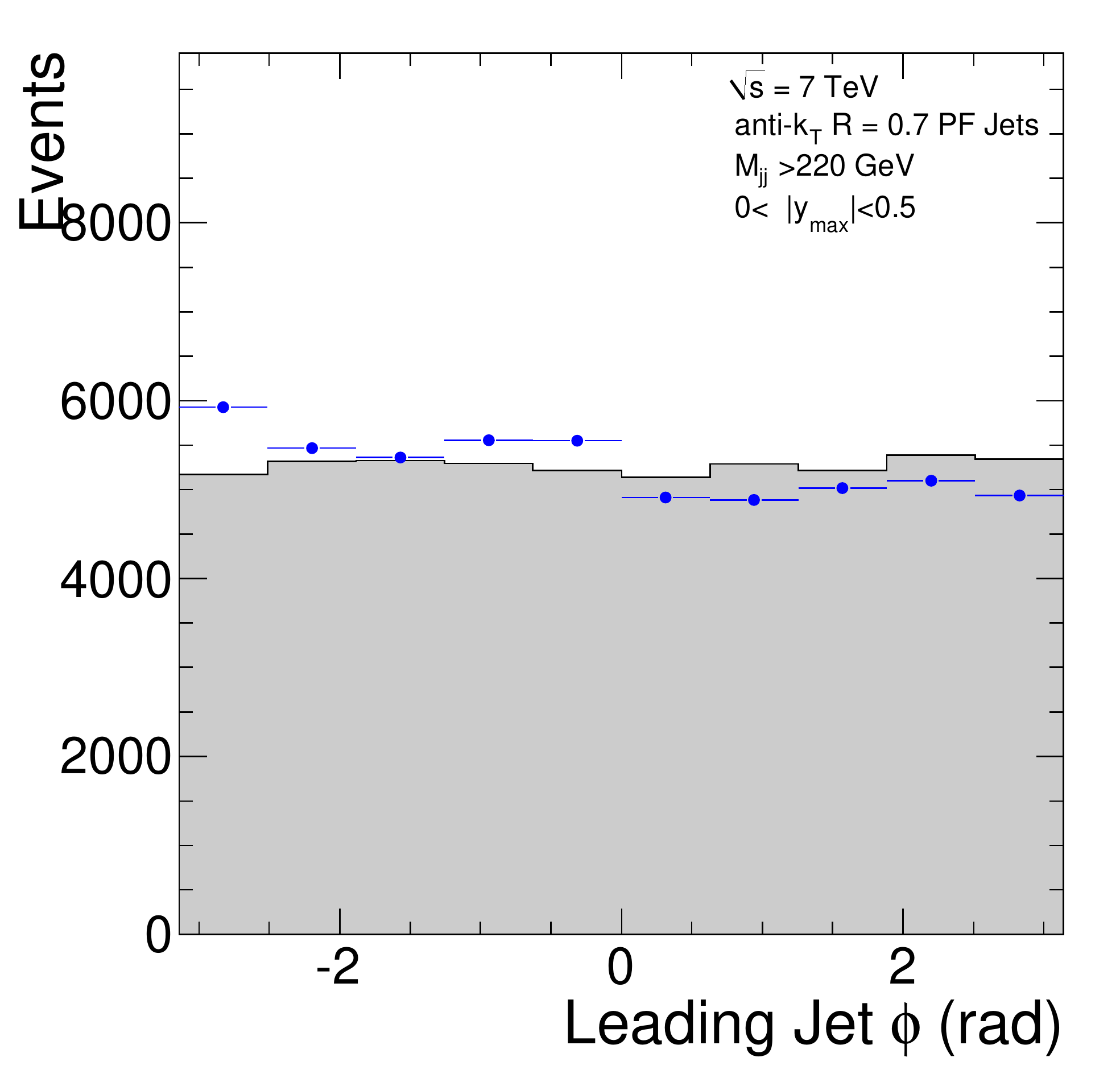} 
\includegraphics[width=0.48\textwidth]{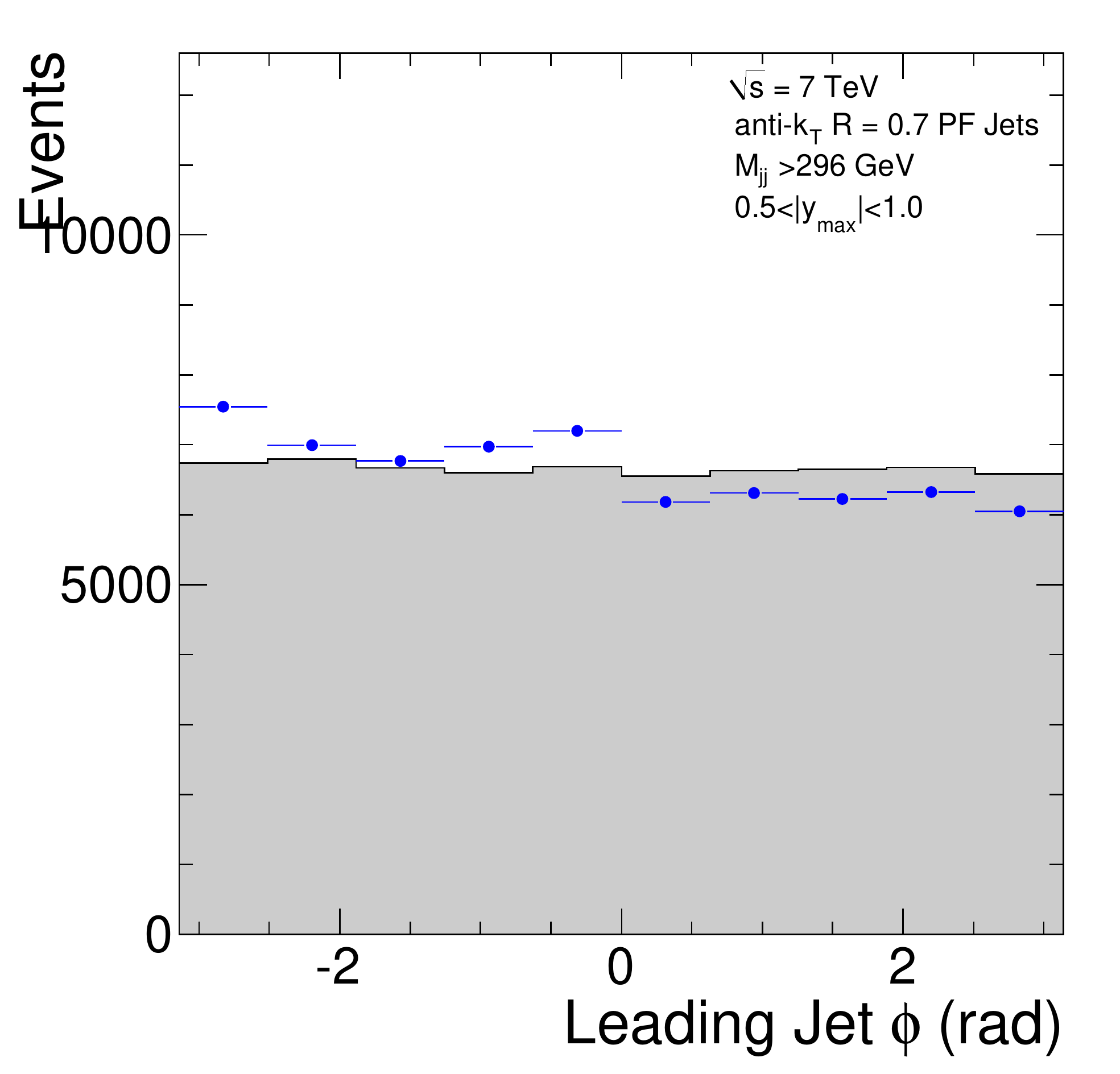} 
\includegraphics[width=0.48\textwidth]{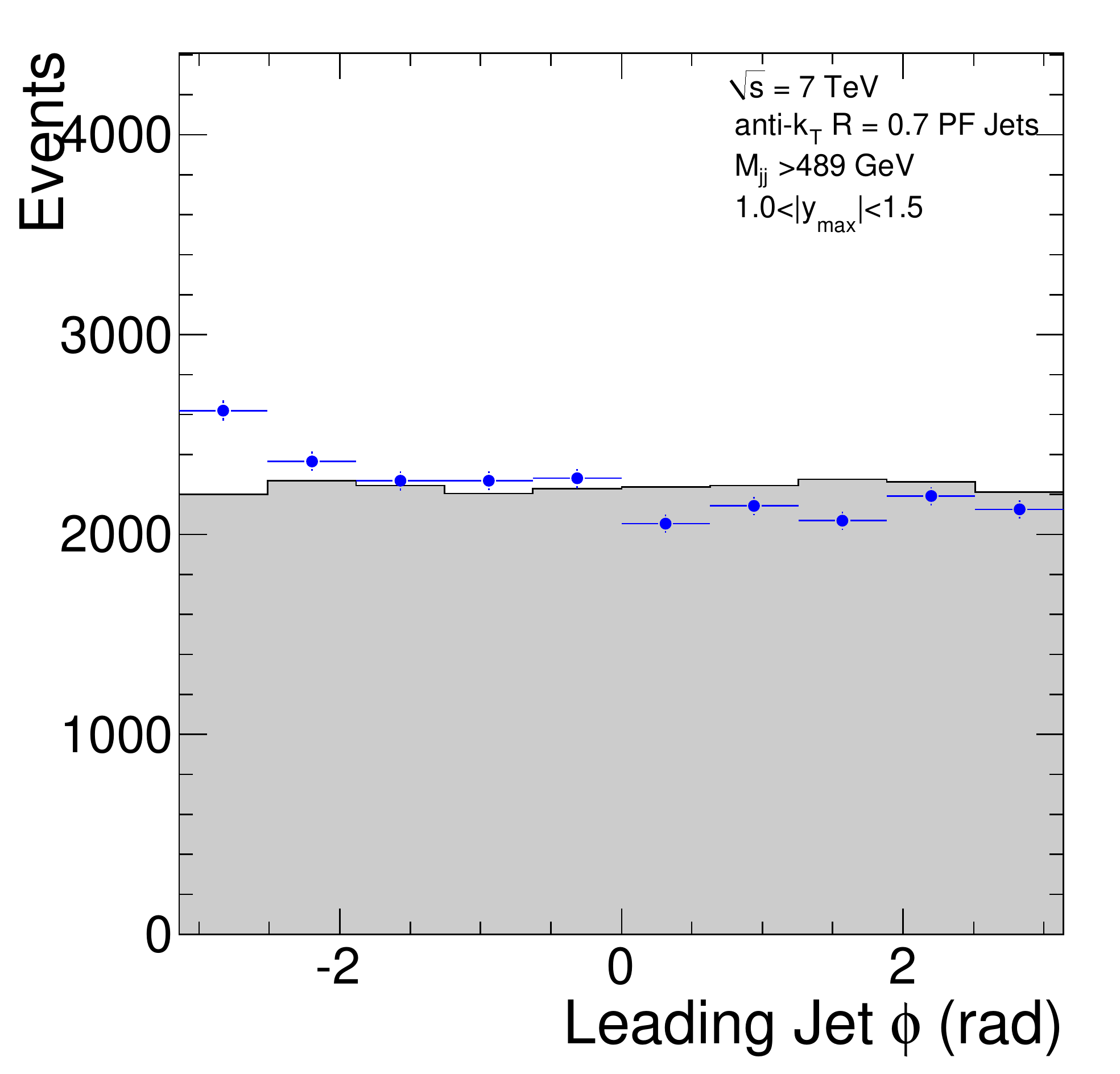} 
\includegraphics[width=0.48\textwidth]{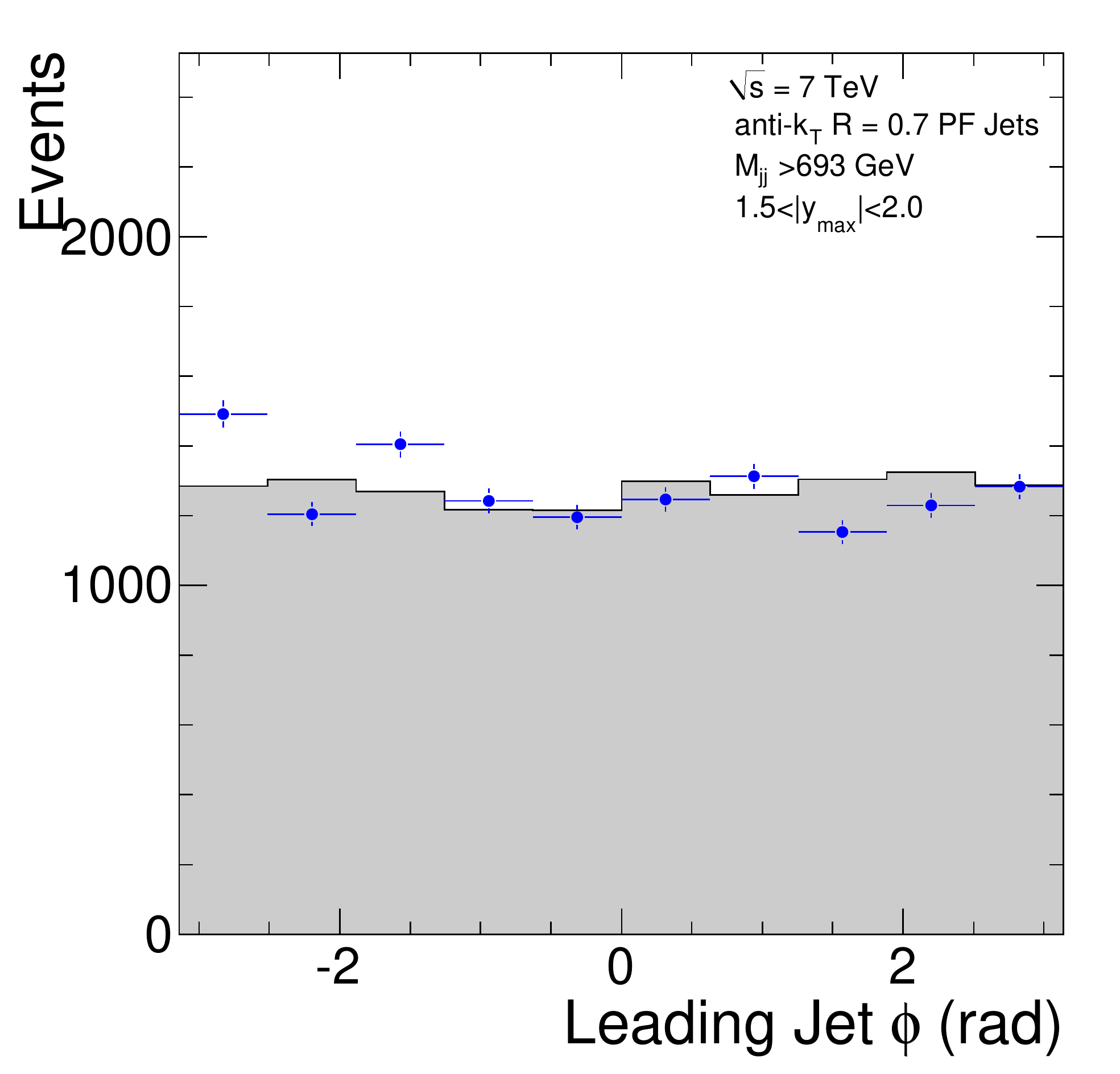} 
\includegraphics[width=0.48\textwidth]{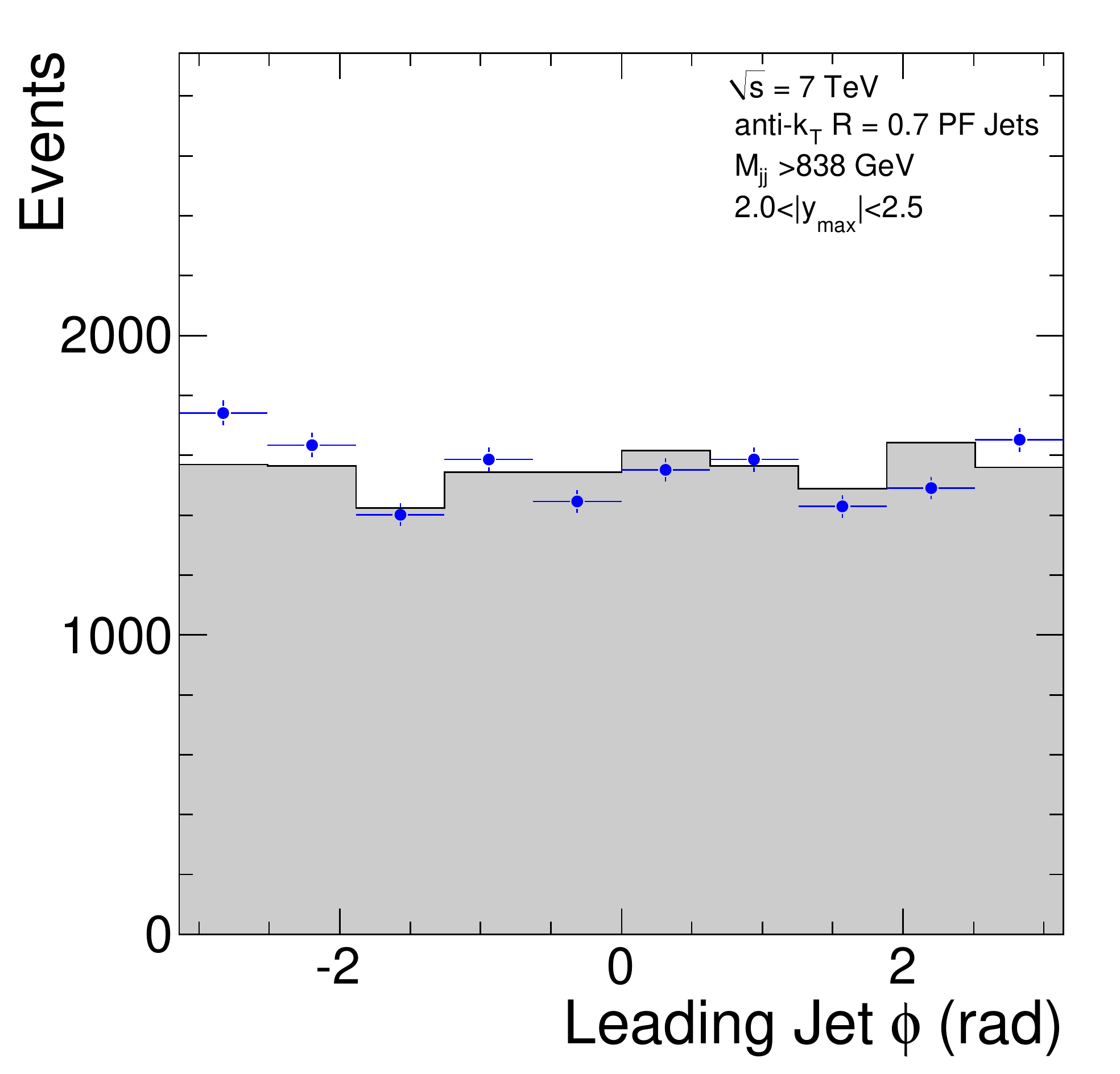}

\caption{ The $\phi$ of the leading jet  for the five different $y_{max}$ bins and for the
HLT$_{-}$Jet50U trigger, for data (points) and simulated (dashed histogram) events.}
\label{fig_appc12}
\end{figure}

\clearpage
%%%%%%%% 100U

\begin{figure}[ht]
\centering

\includegraphics[width=0.48\textwidth]{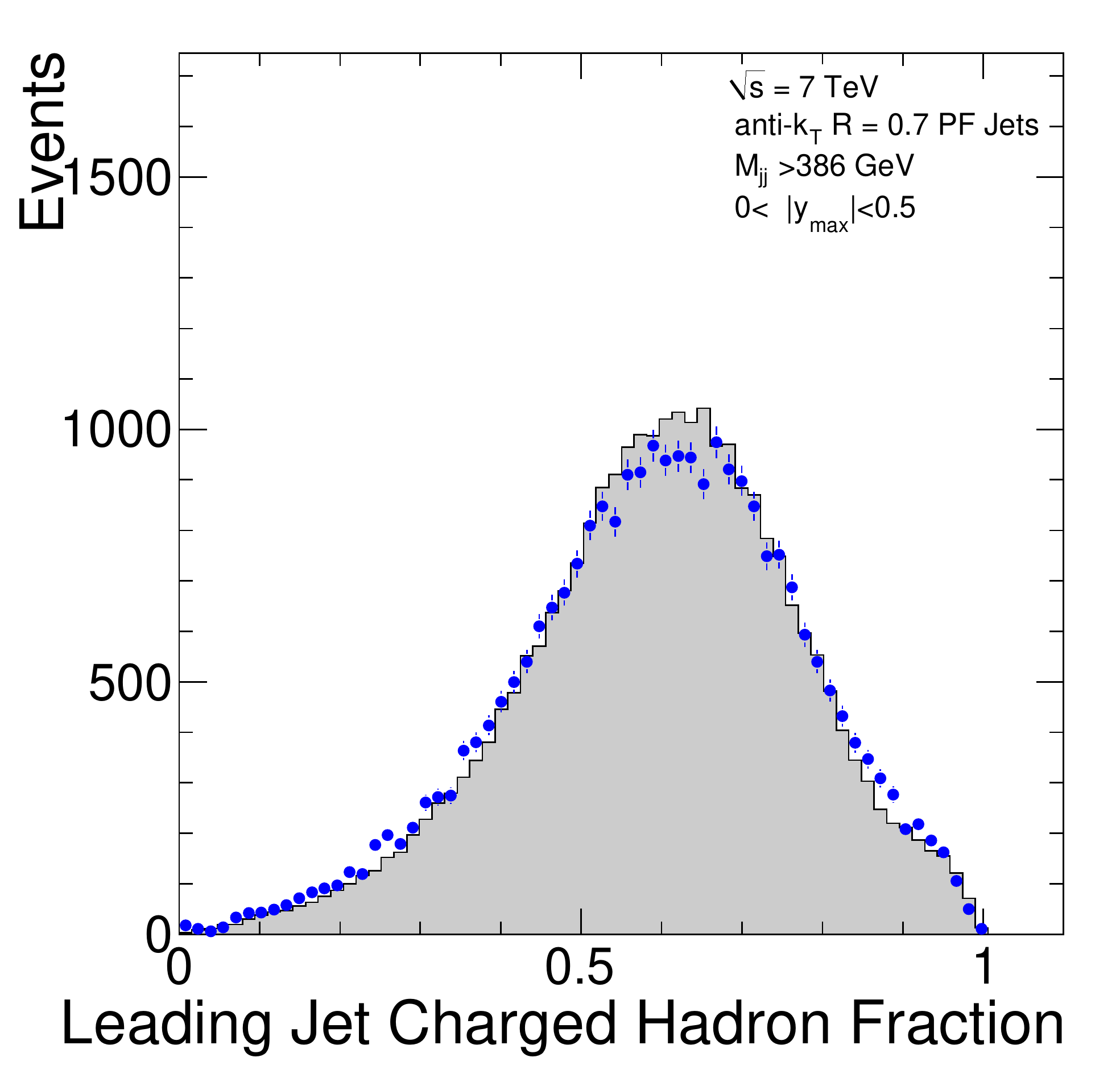} 
\includegraphics[width=0.48\textwidth]{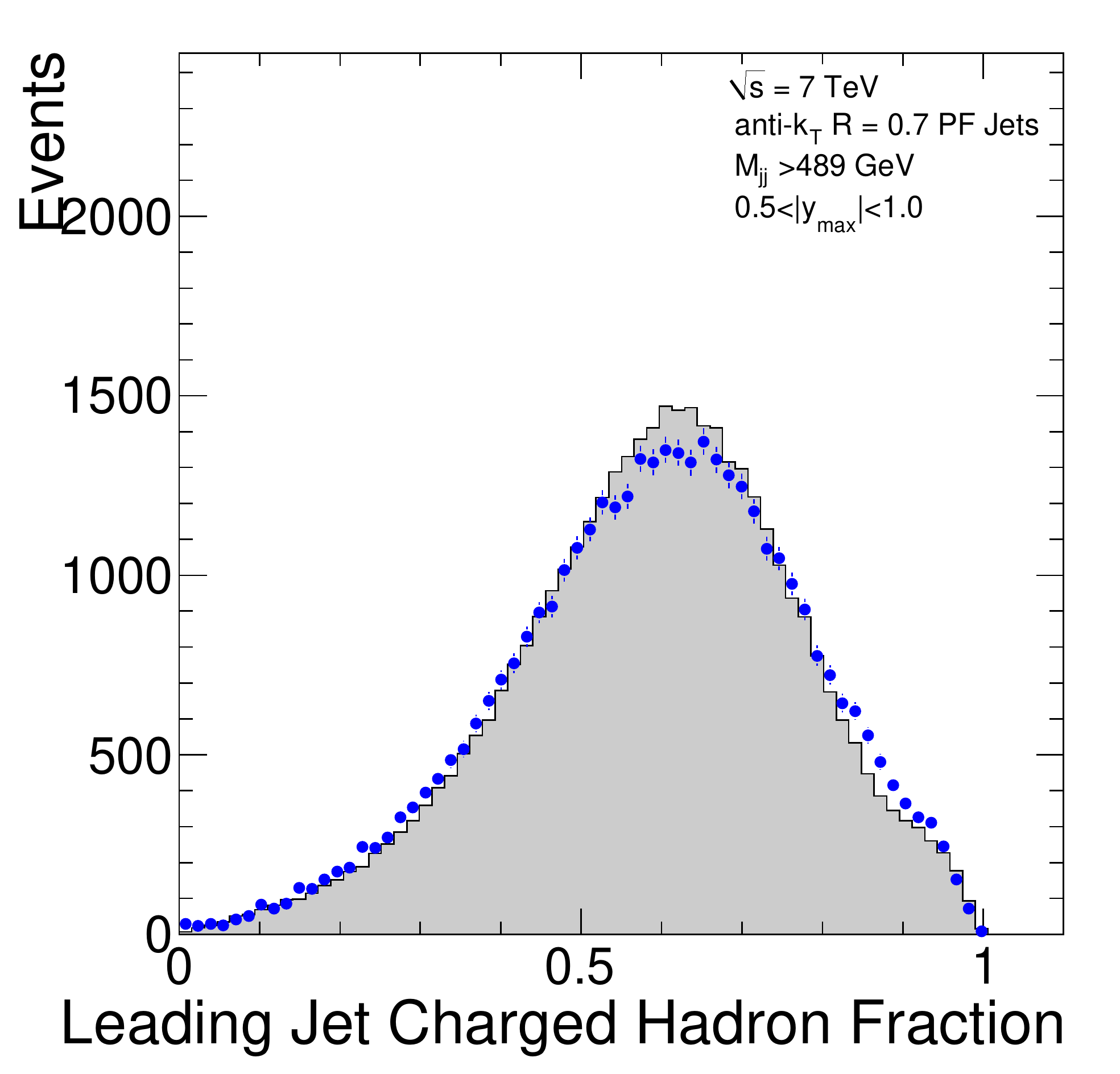} 
\includegraphics[width=0.48\textwidth]{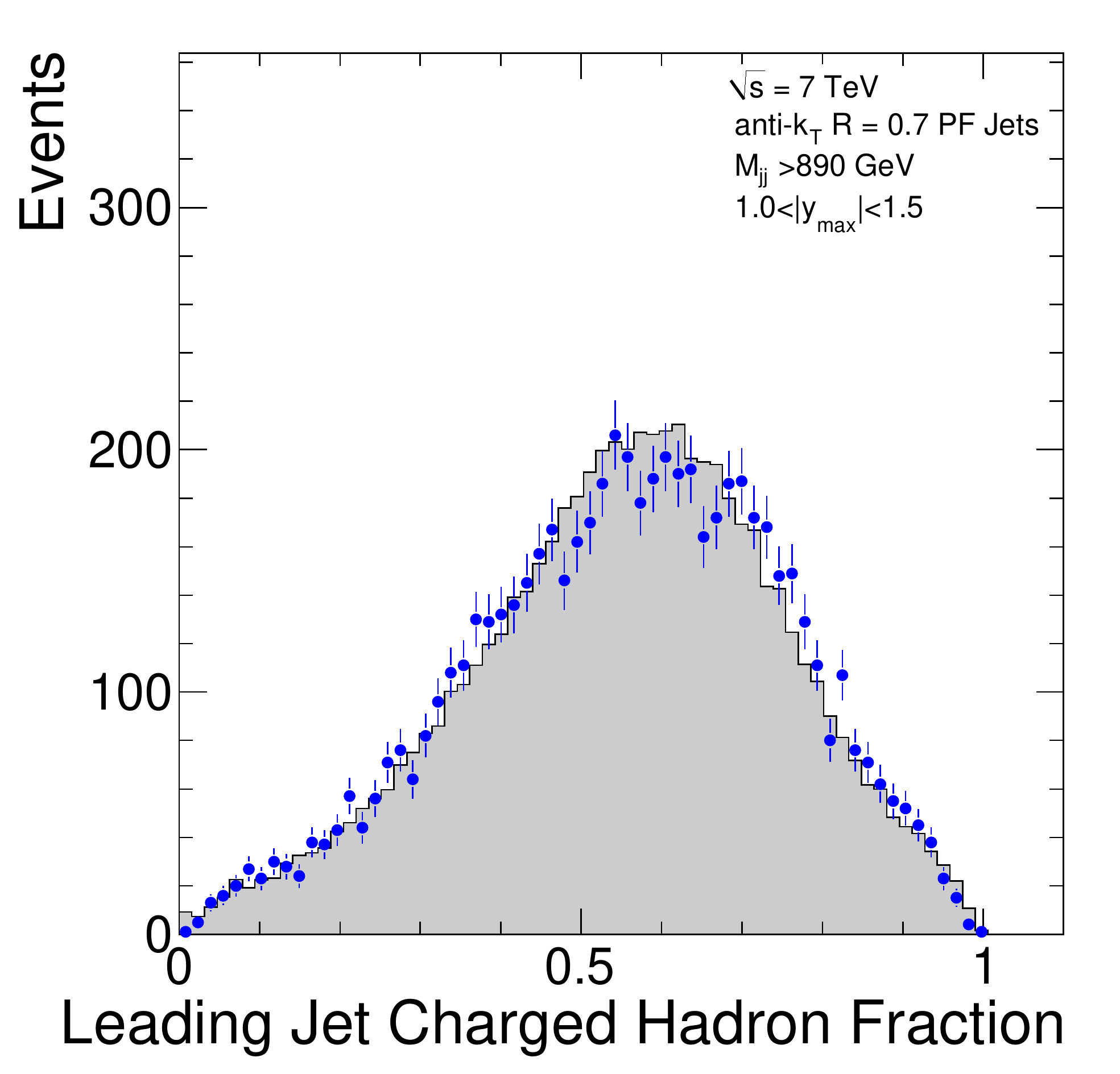} 
\includegraphics[width=0.48\textwidth]{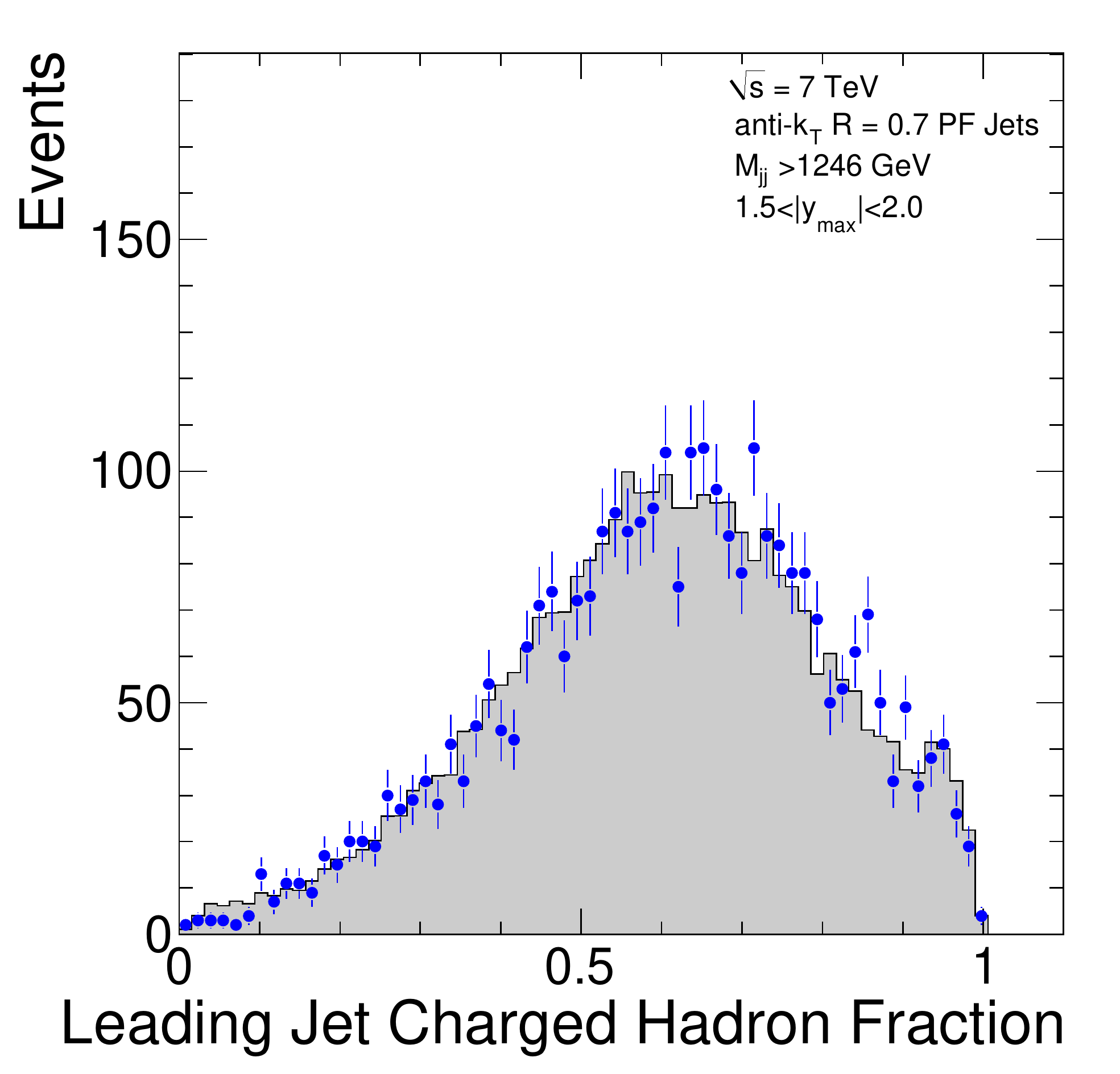} 
\includegraphics[width=0.48\textwidth]{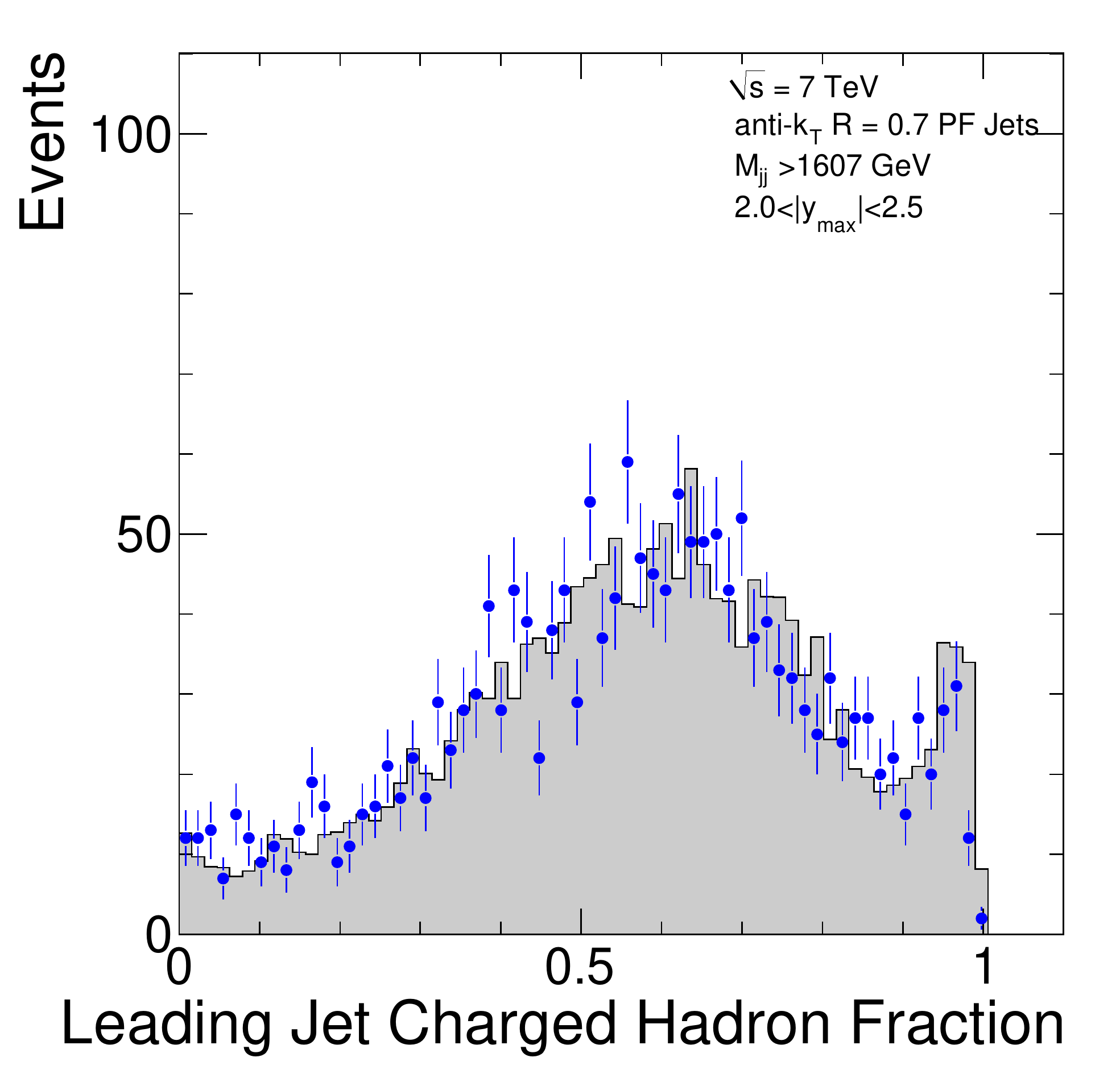}

\caption{ The charged hadron fraction of the leading jet  for the five different $y_{max}$ bins and for the
HLT$_{-}$Jet100U trigger, for data (points) and simulated (dashed histogram) events.}
\label{fig_appc13}
\end{figure}

\begin{figure}[ht]
\centering

\includegraphics[width=0.48\textwidth]{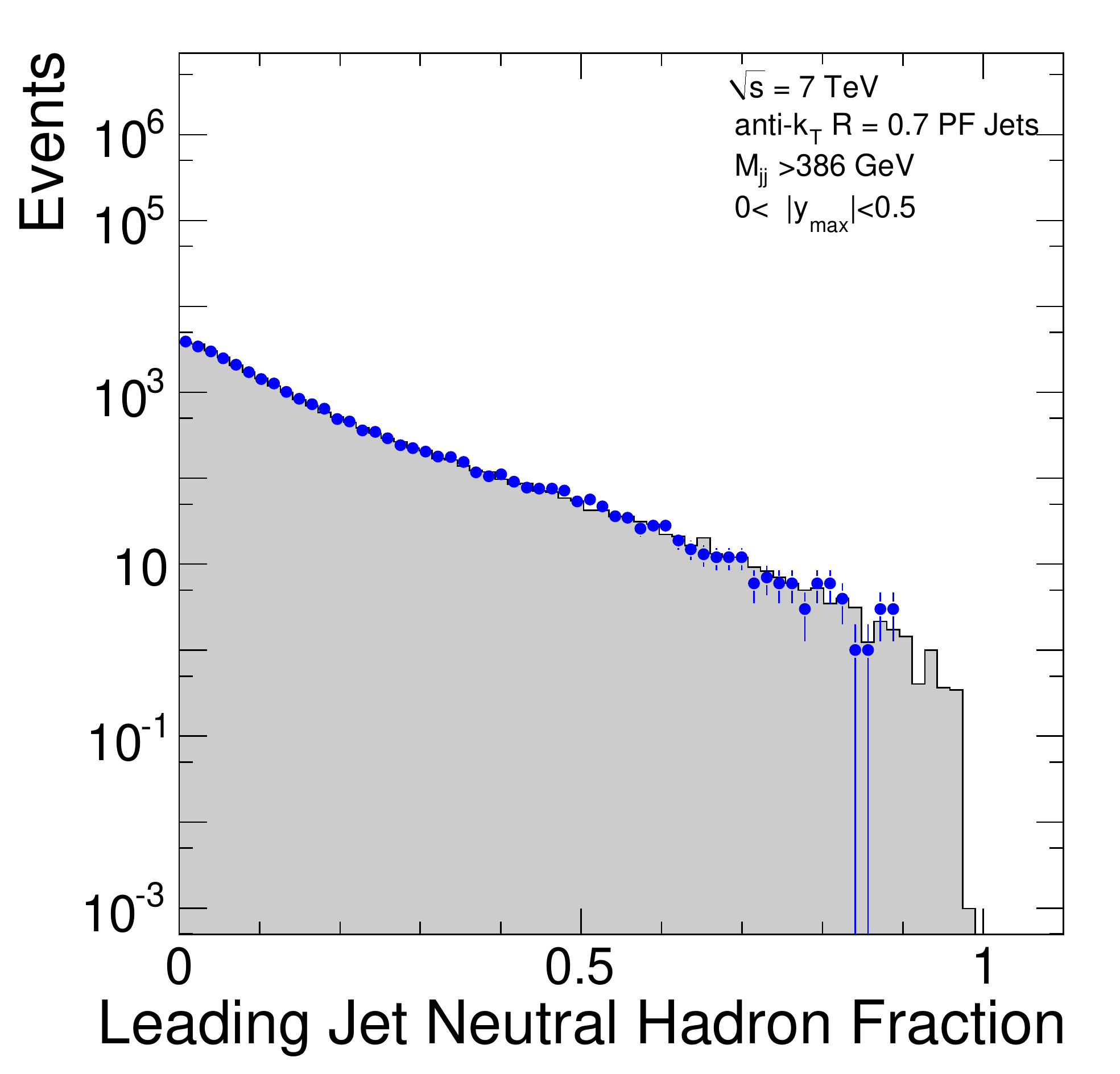} 
\includegraphics[width=0.48\textwidth]{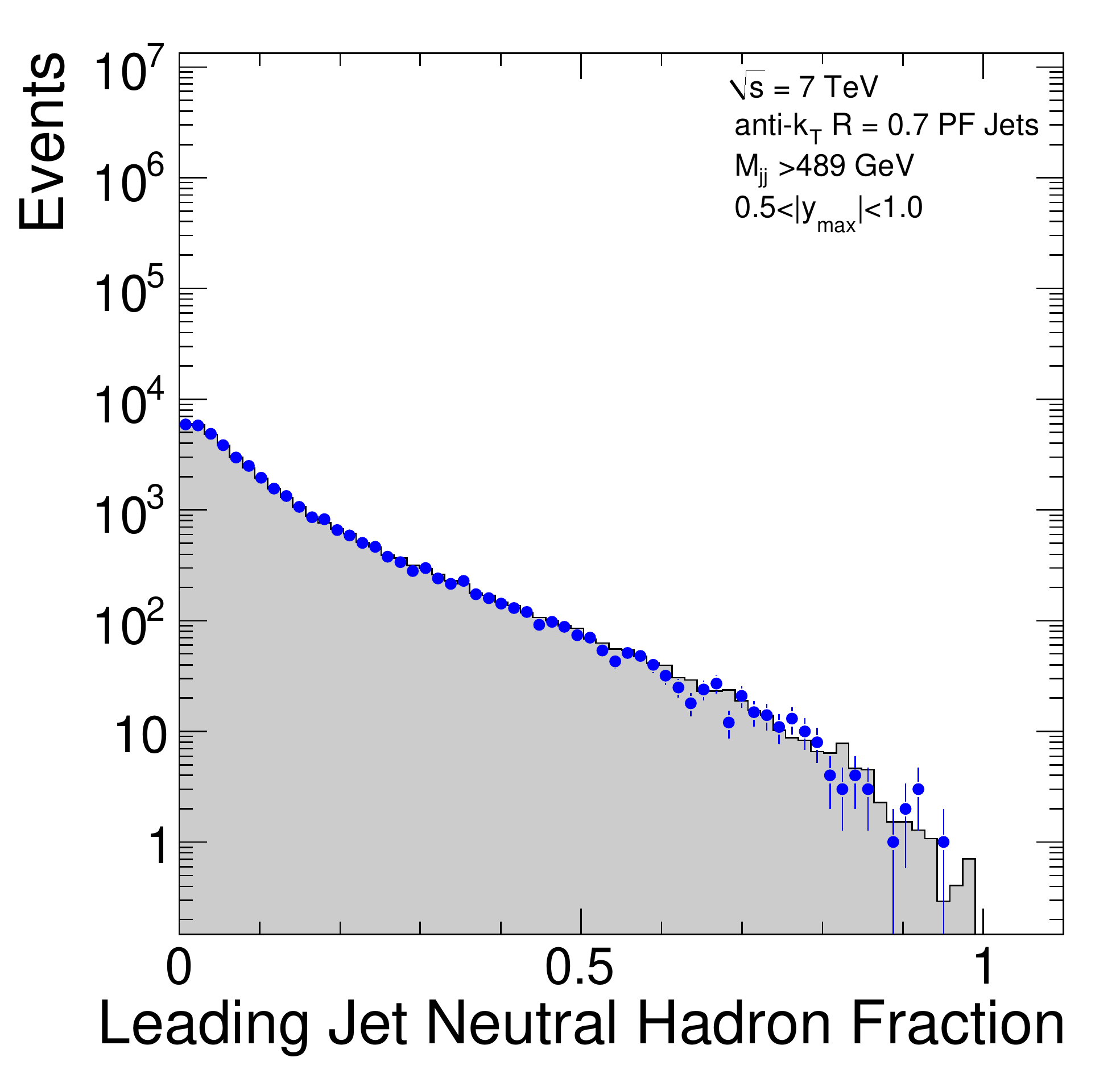} 
\includegraphics[width=0.48\textwidth]{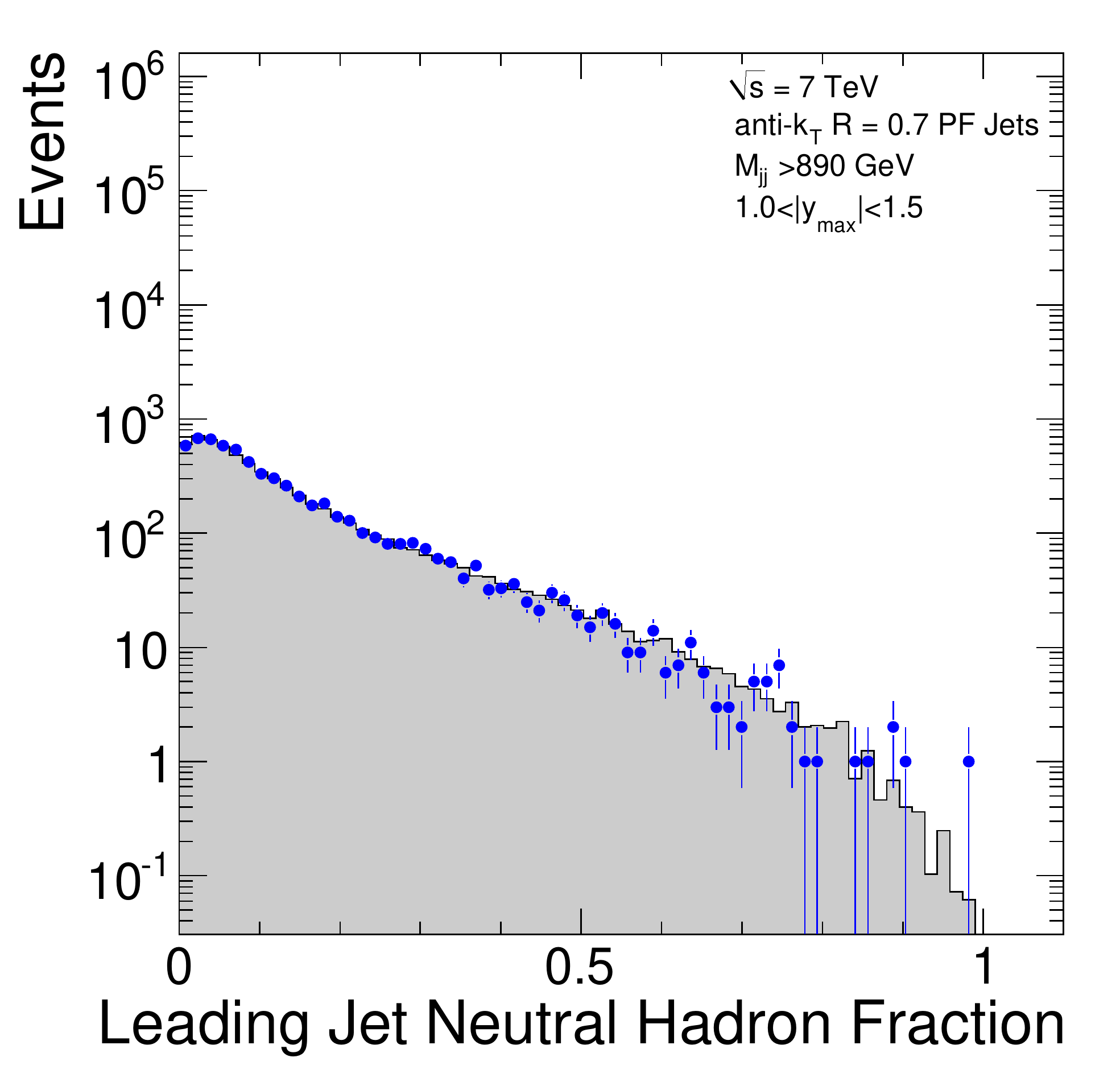} 
\includegraphics[width=0.48\textwidth]{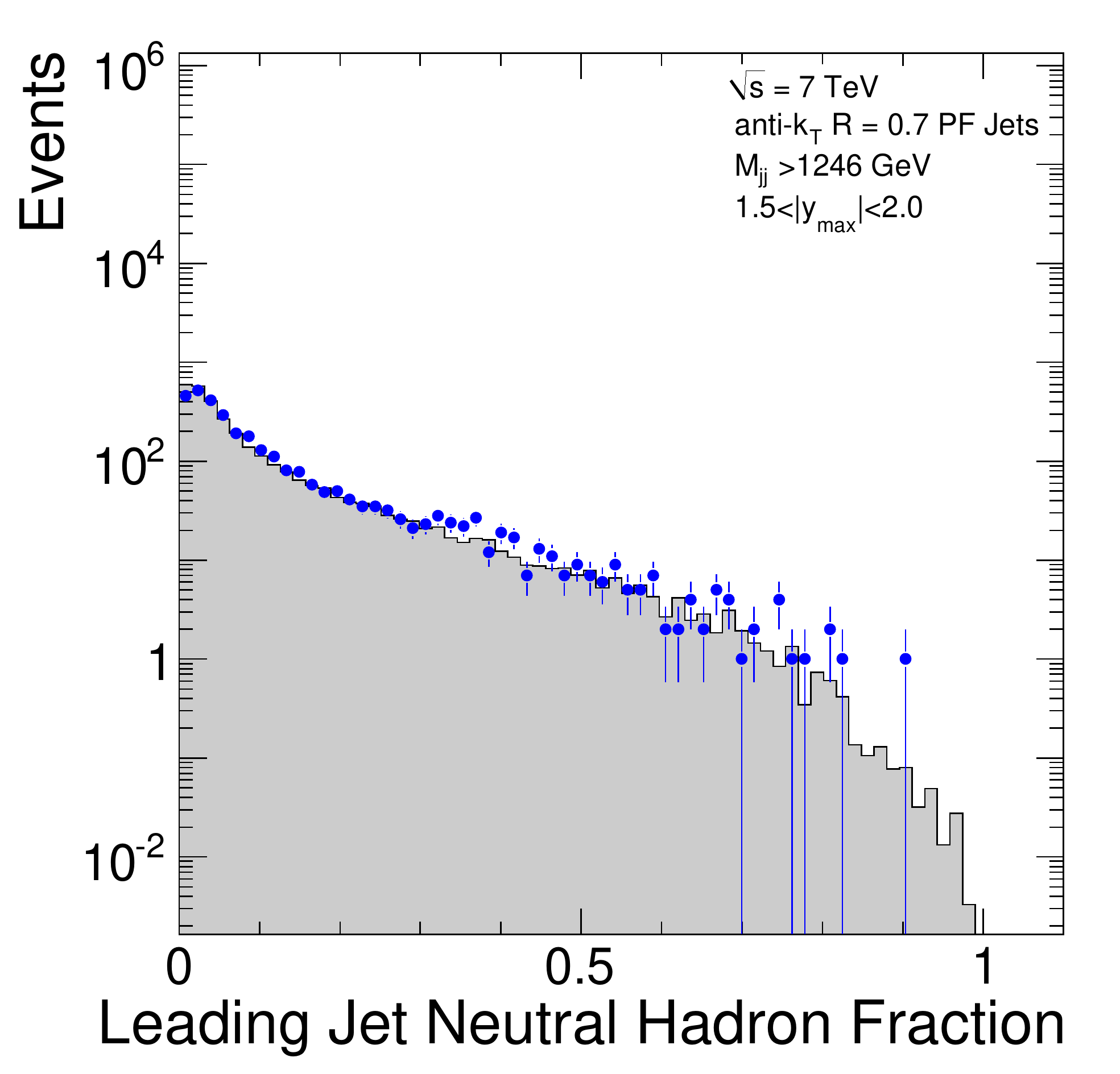} 
\includegraphics[width=0.48\textwidth]{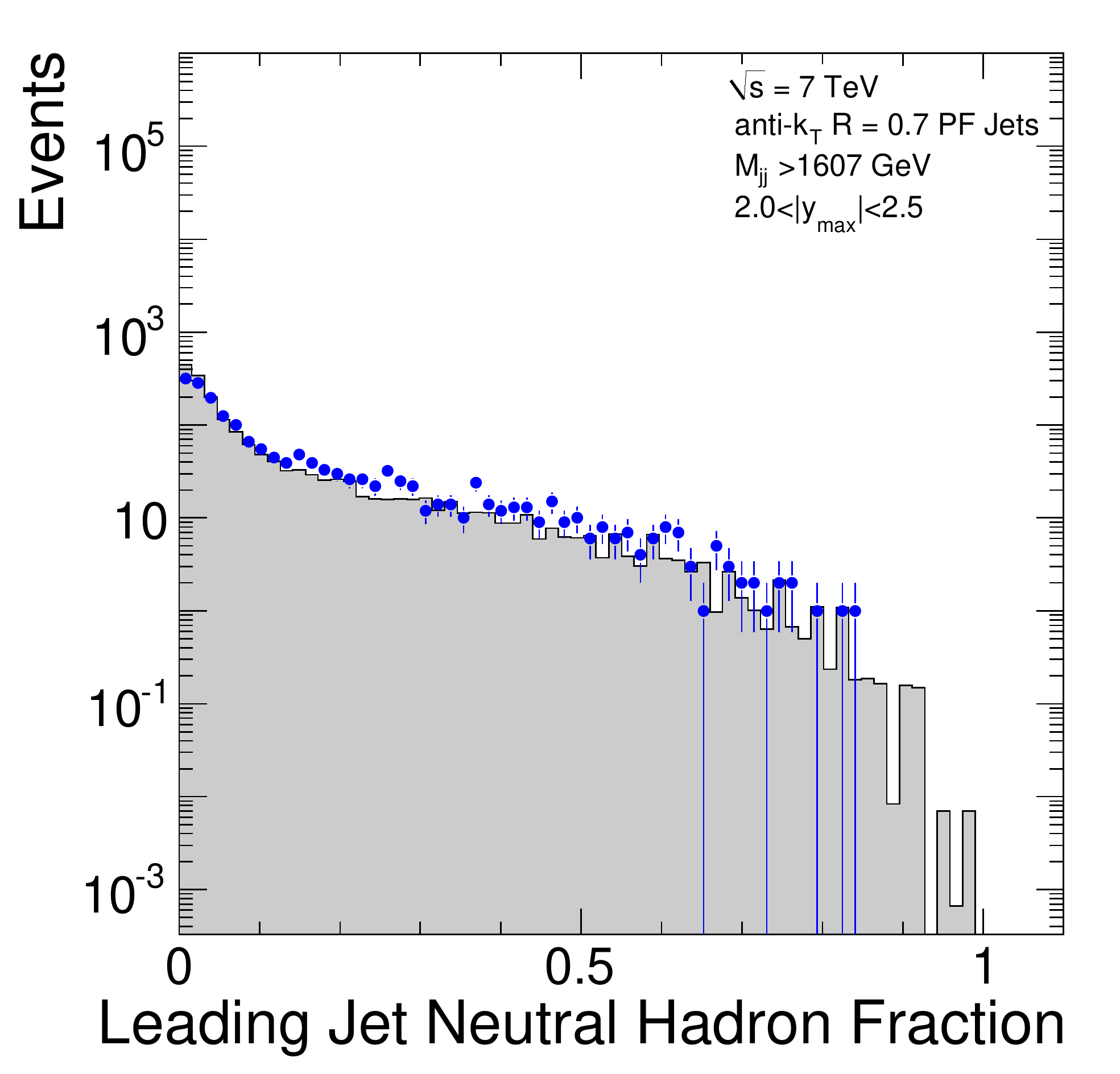}

\caption{ The neutral hadron fraction of the leading jet  for the five different $y_{max}$ bins and for the
HLT$_{-}$Jet100U trigger, for data (points) and simulated (dashed histogram) events.}
\label{fig_appc14}
\end{figure}

\begin{figure}[ht]
\centering

\includegraphics[width=0.48\textwidth]{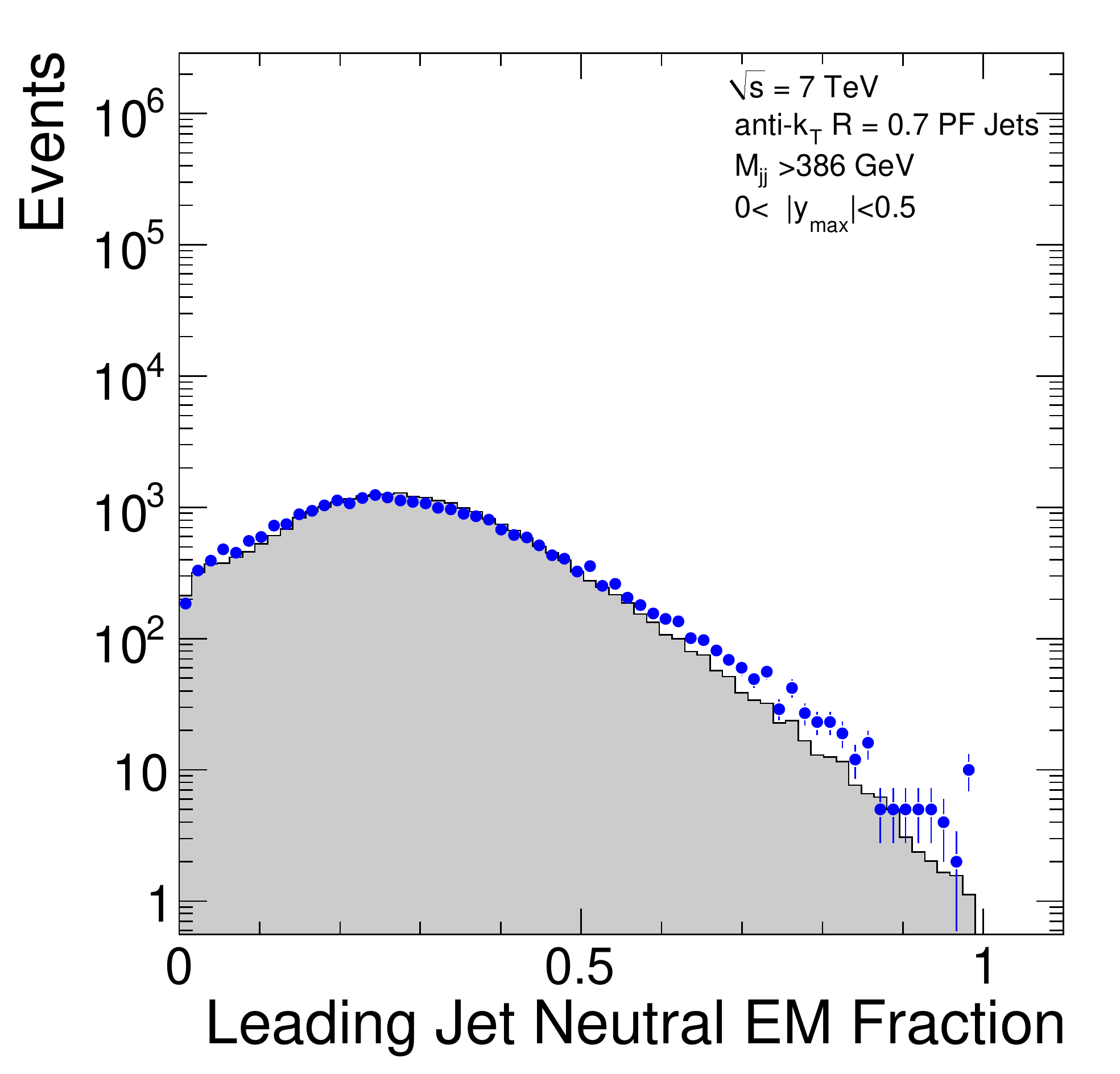} 
\includegraphics[width=0.48\textwidth]{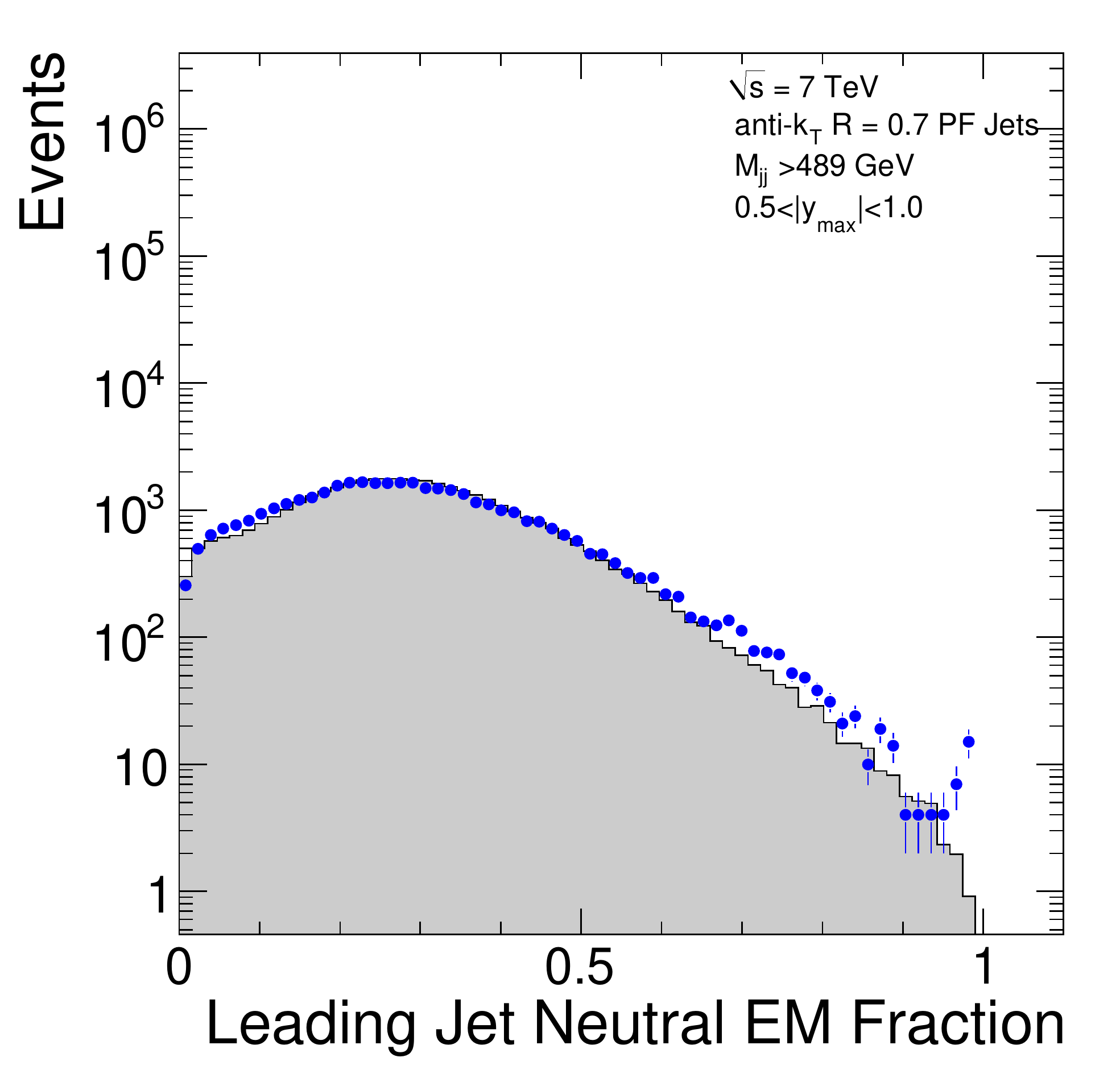} 
\includegraphics[width=0.48\textwidth]{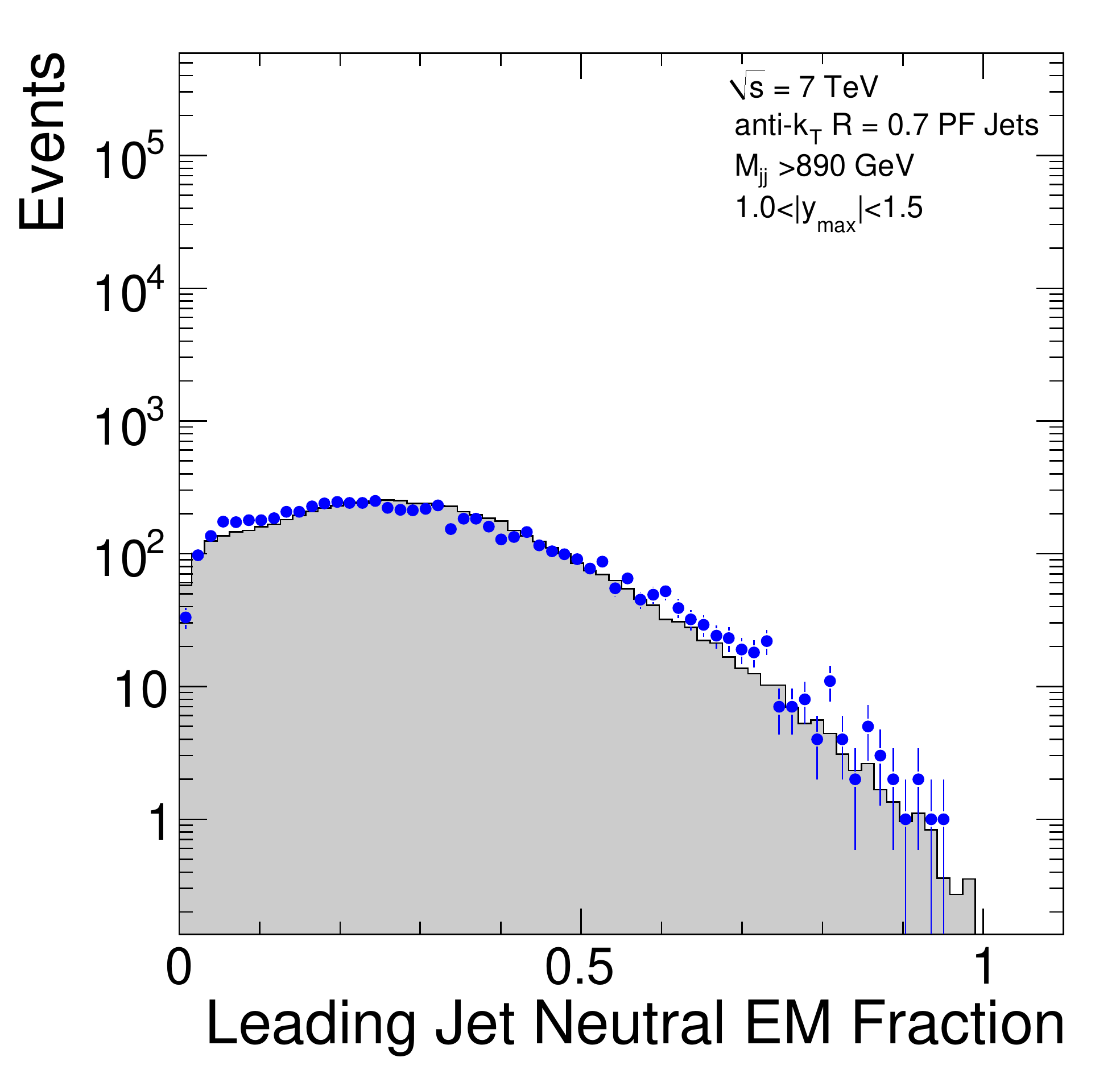} 
\includegraphics[width=0.48\textwidth]{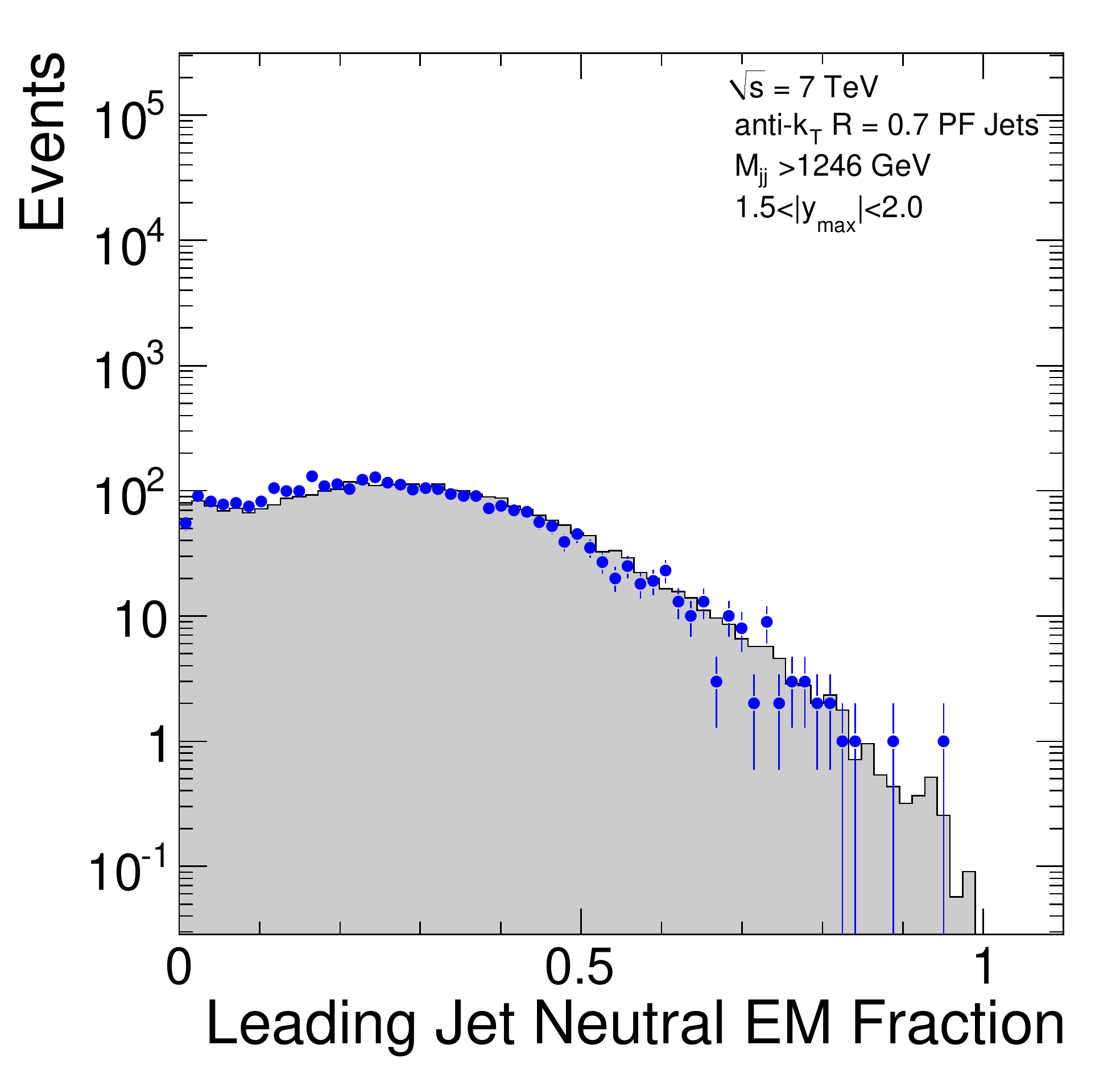} 
\includegraphics[width=0.48\textwidth]{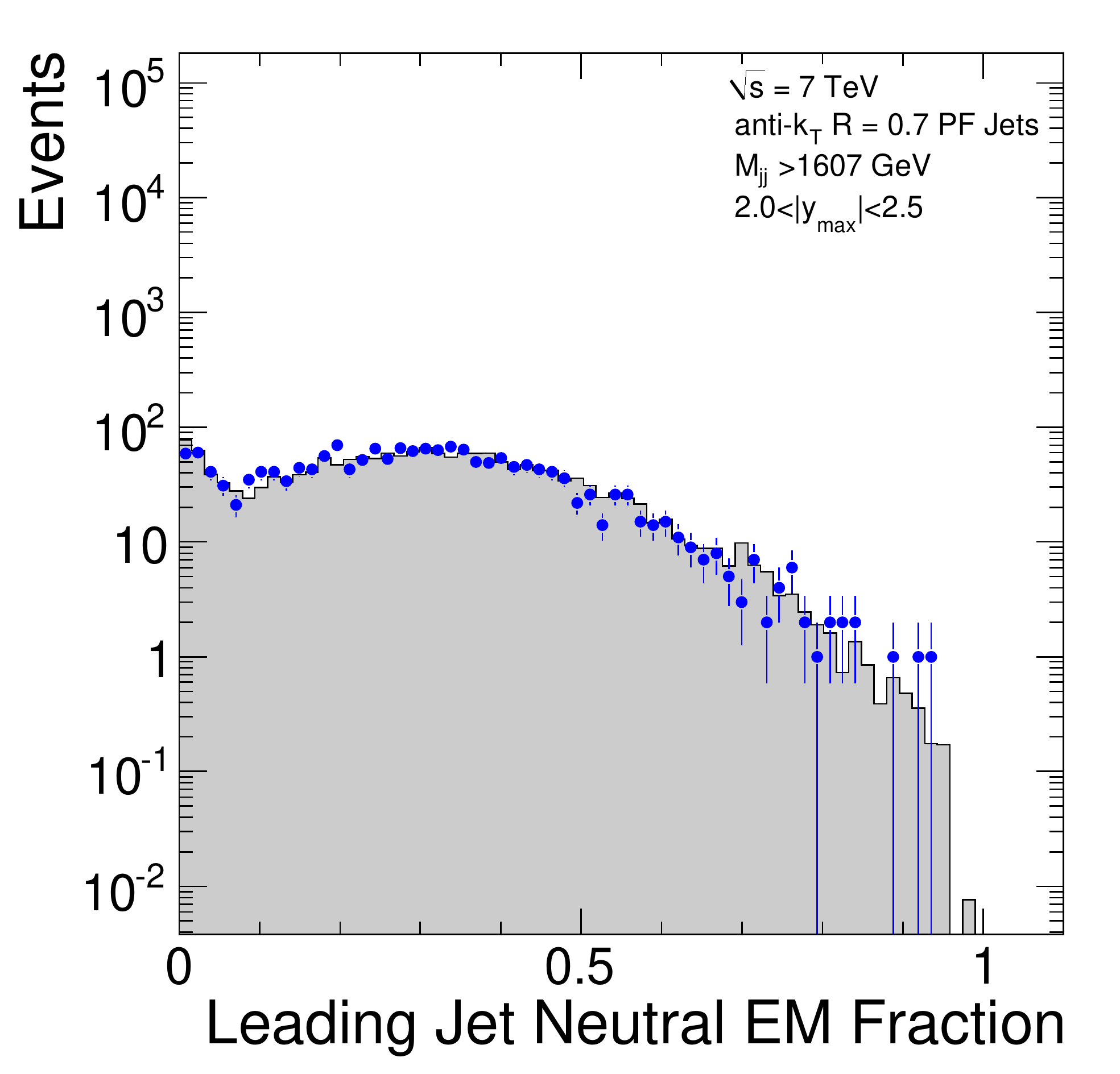}

\caption{ The neutral electromagnetic  fraction of the leading jet  for the five different $y_{max}$ bins and for the
HLT$_{-}$Jet100U trigger, for data (points) and simulated (dashed histogram) events.}
\label{fig_appc15}
\end{figure}

 \begin{figure}[ht]
\centering

\includegraphics[width=0.48\textwidth]{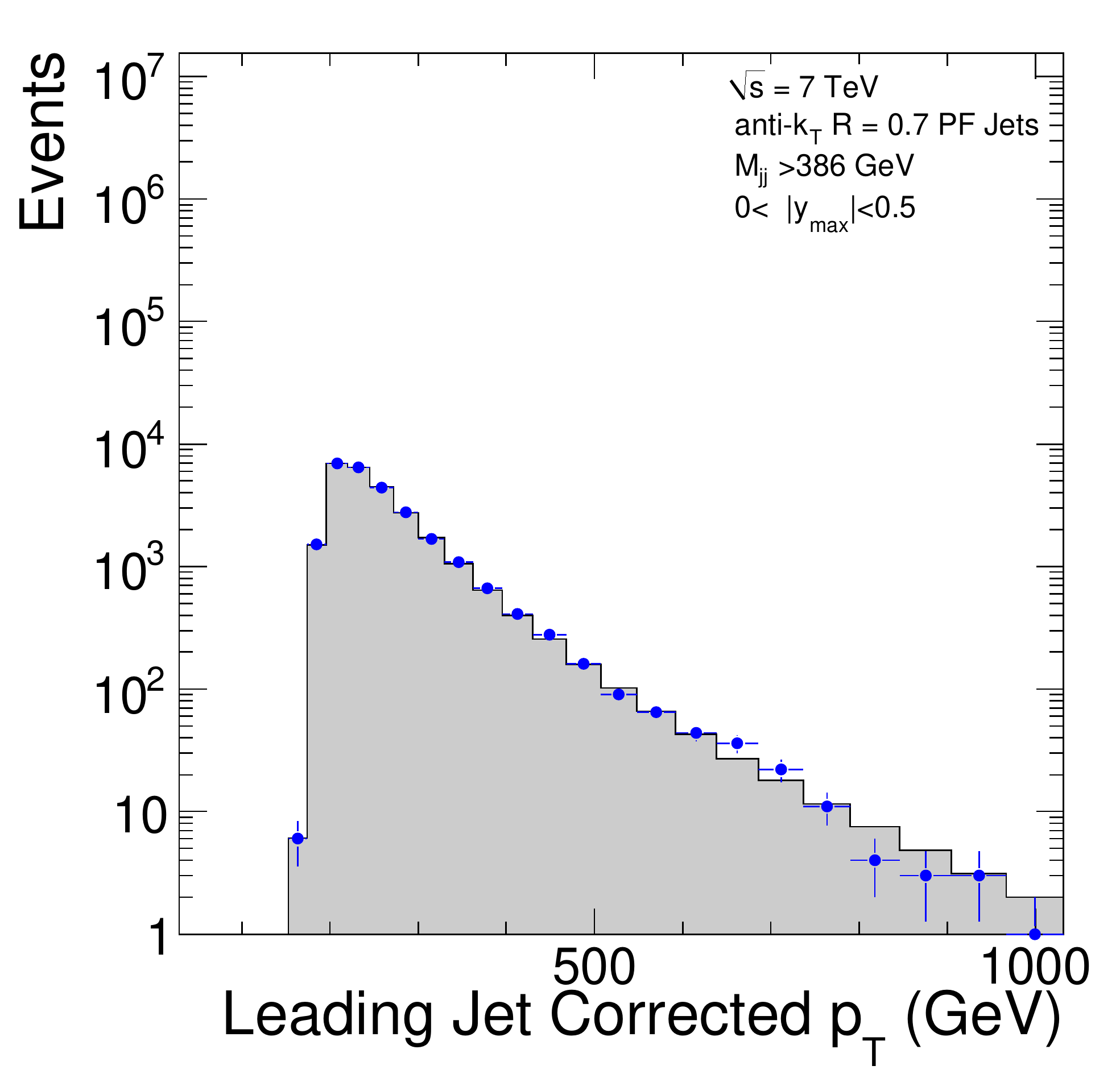} 
\includegraphics[width=0.48\textwidth]{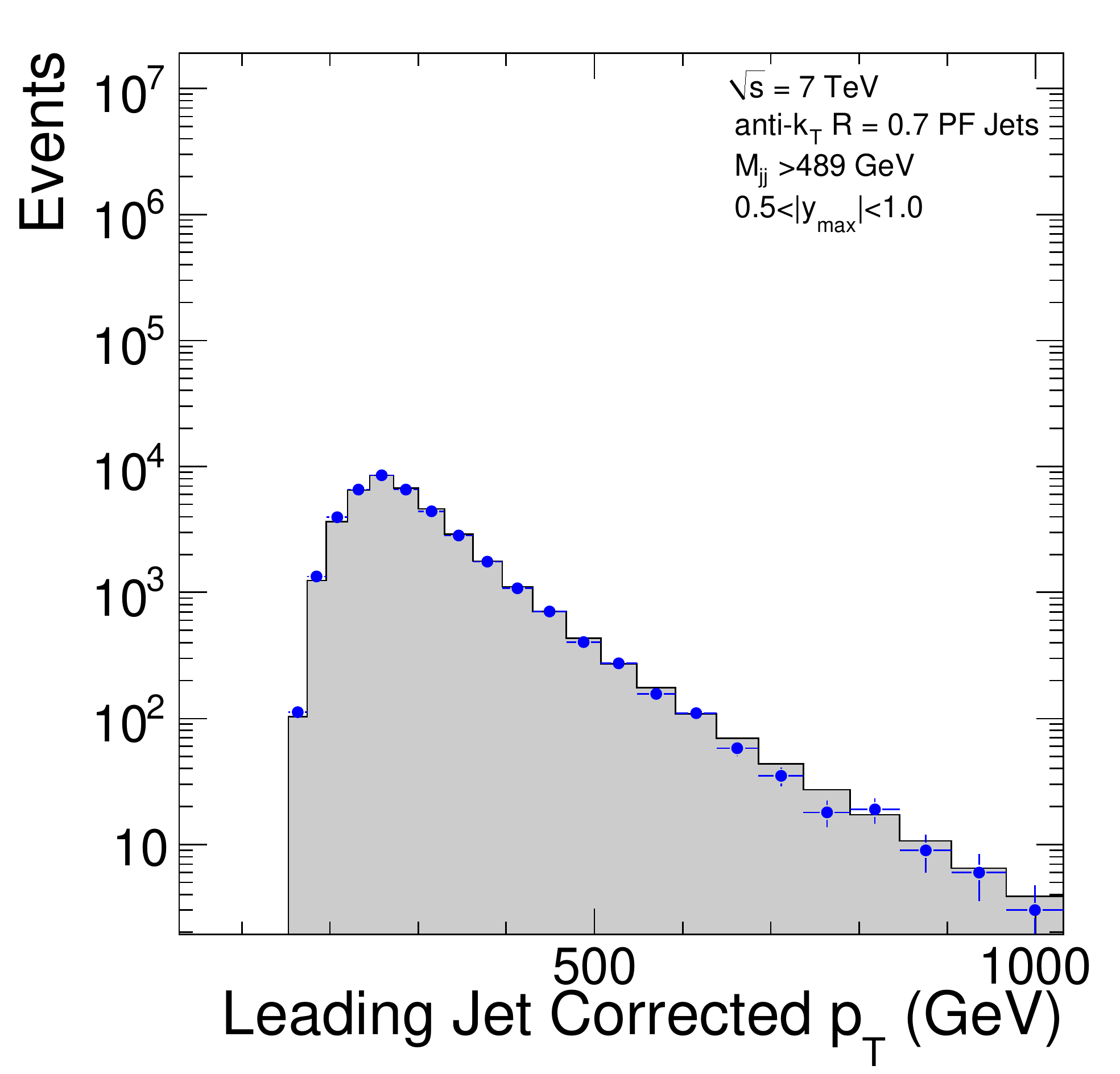} 
\includegraphics[width=0.48\textwidth]{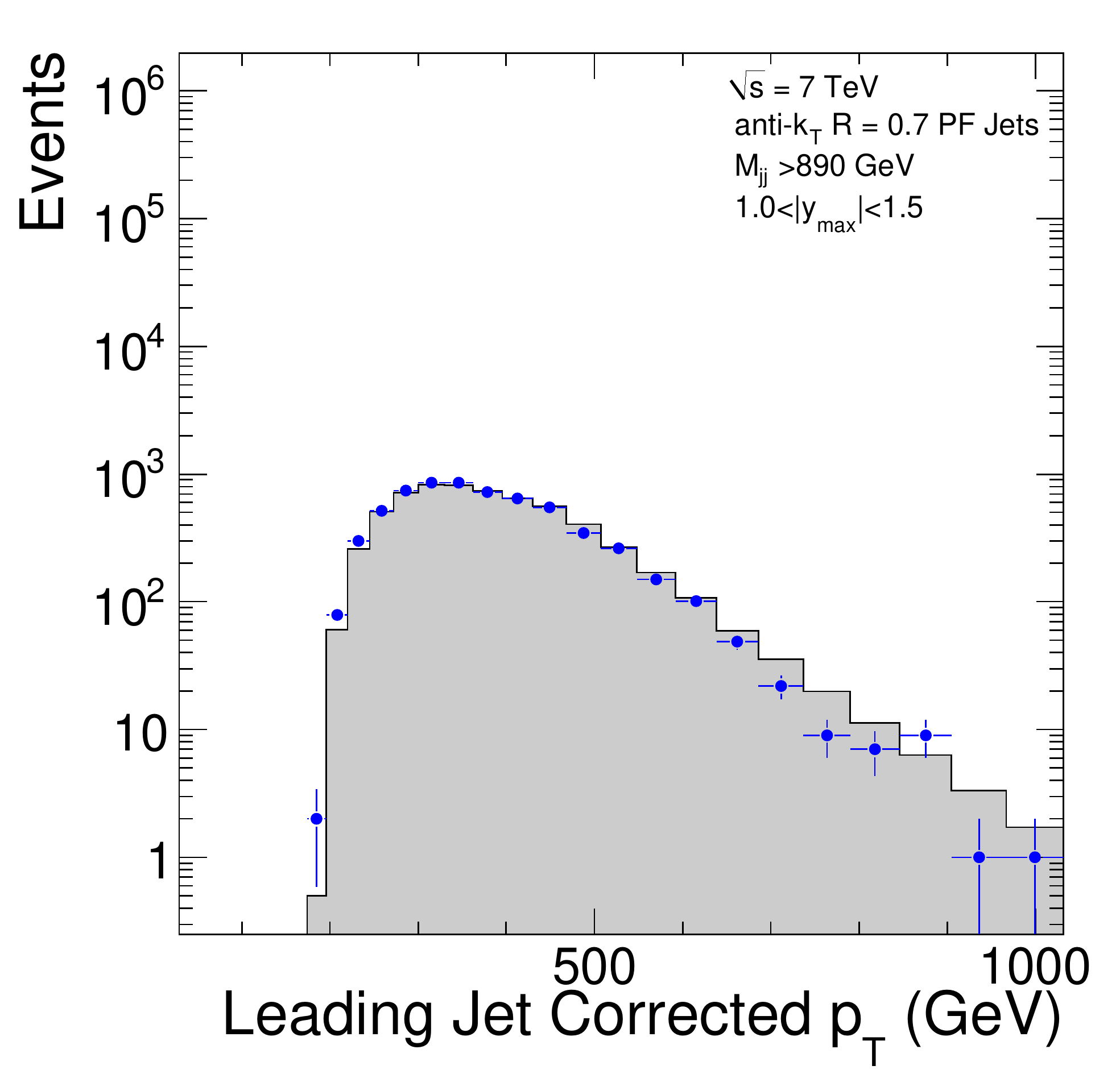} 
\includegraphics[width=0.48\textwidth]{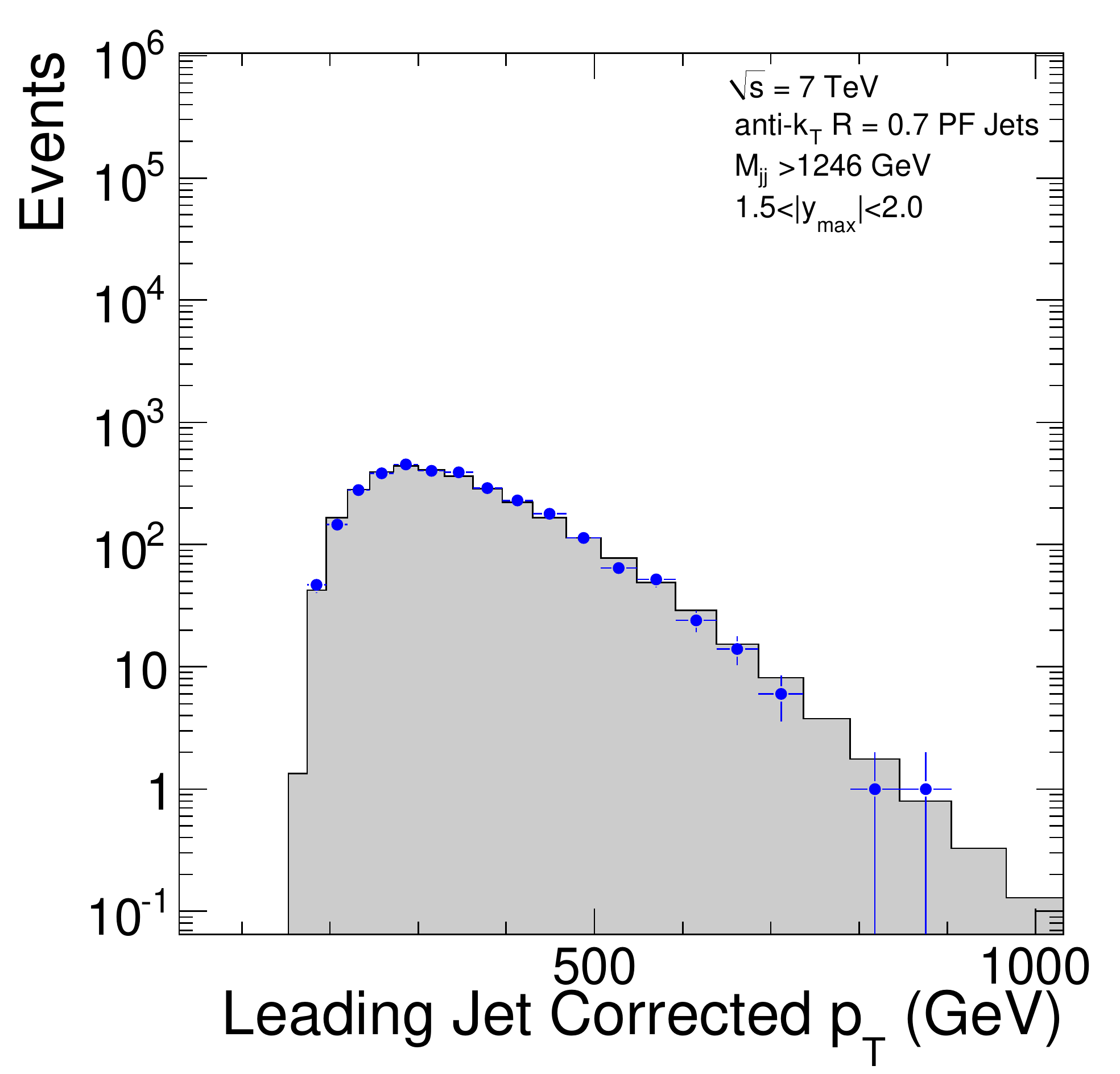} 
\includegraphics[width=0.48\textwidth]{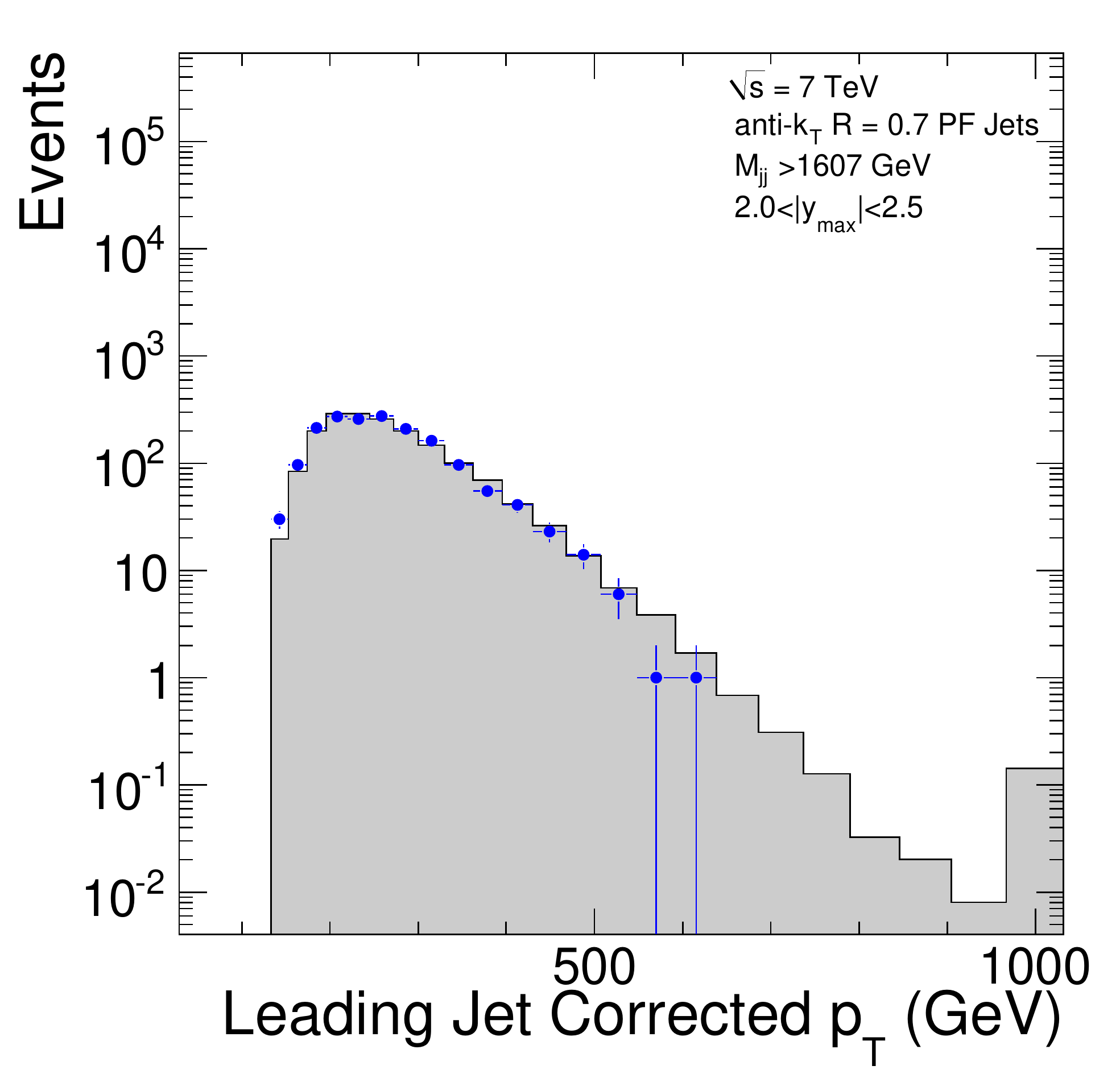}

\caption{ The $p_T$f of the leading jet  for the five different $y_{max}$ bins and for the
HLT$_{-}$Jet100U trigger, for data (points) and simulated (dashed histogram) events.}
\label{fig_appc16}
\end{figure}

\begin{figure}[ht]
\centering

\includegraphics[width=0.48\textwidth]{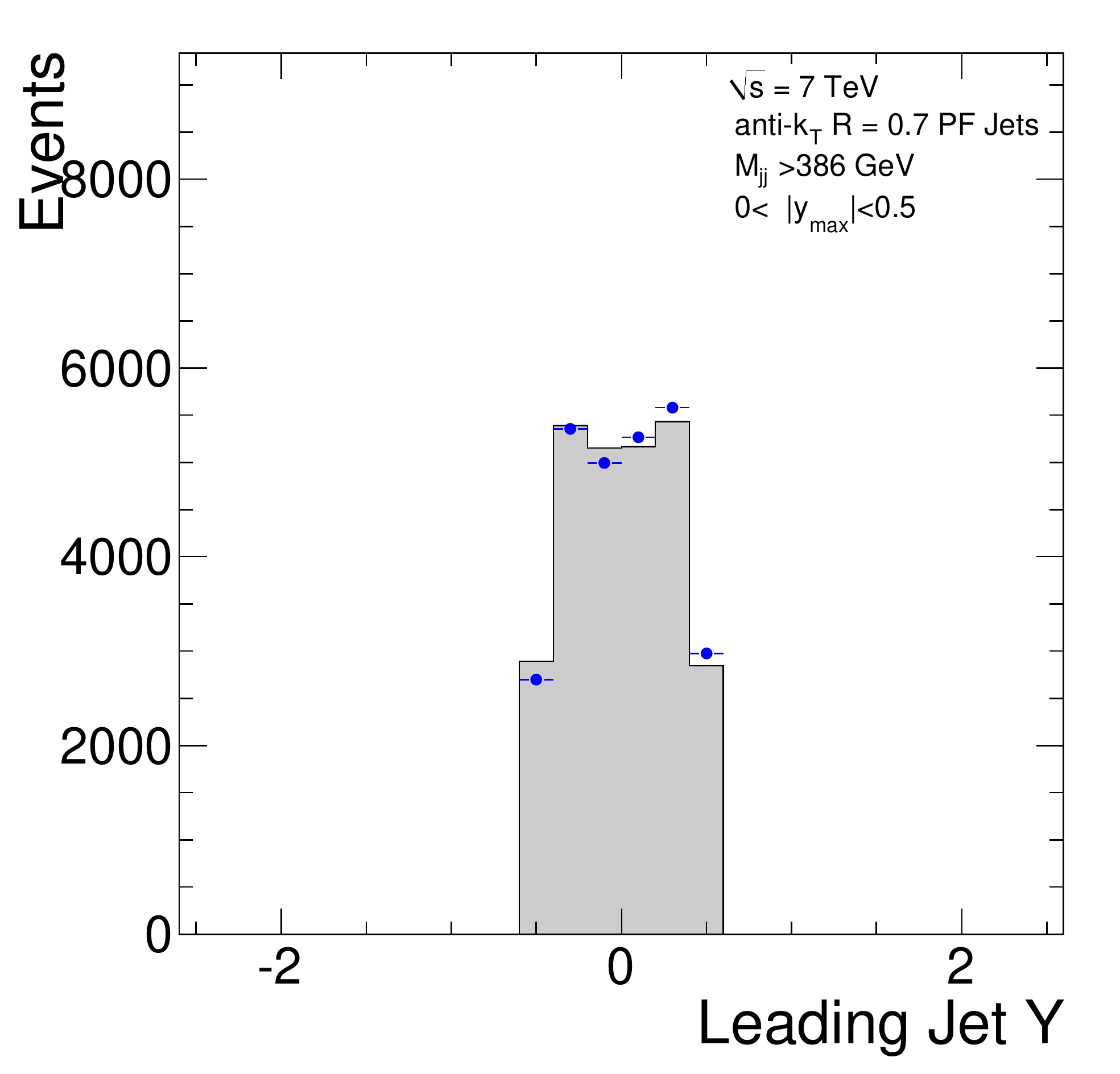} 
\includegraphics[width=0.48\textwidth]{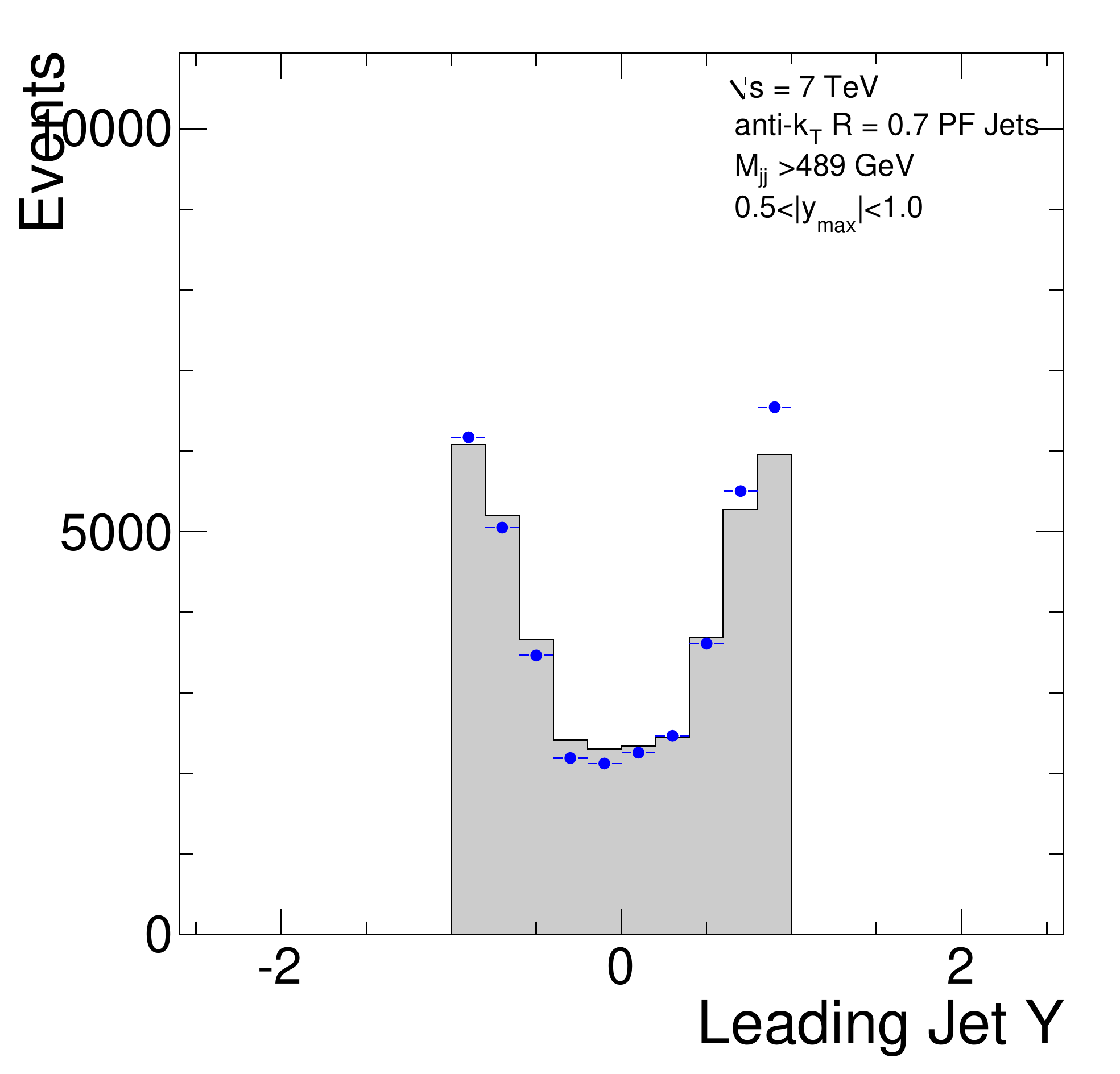} 
\includegraphics[width=0.48\textwidth]{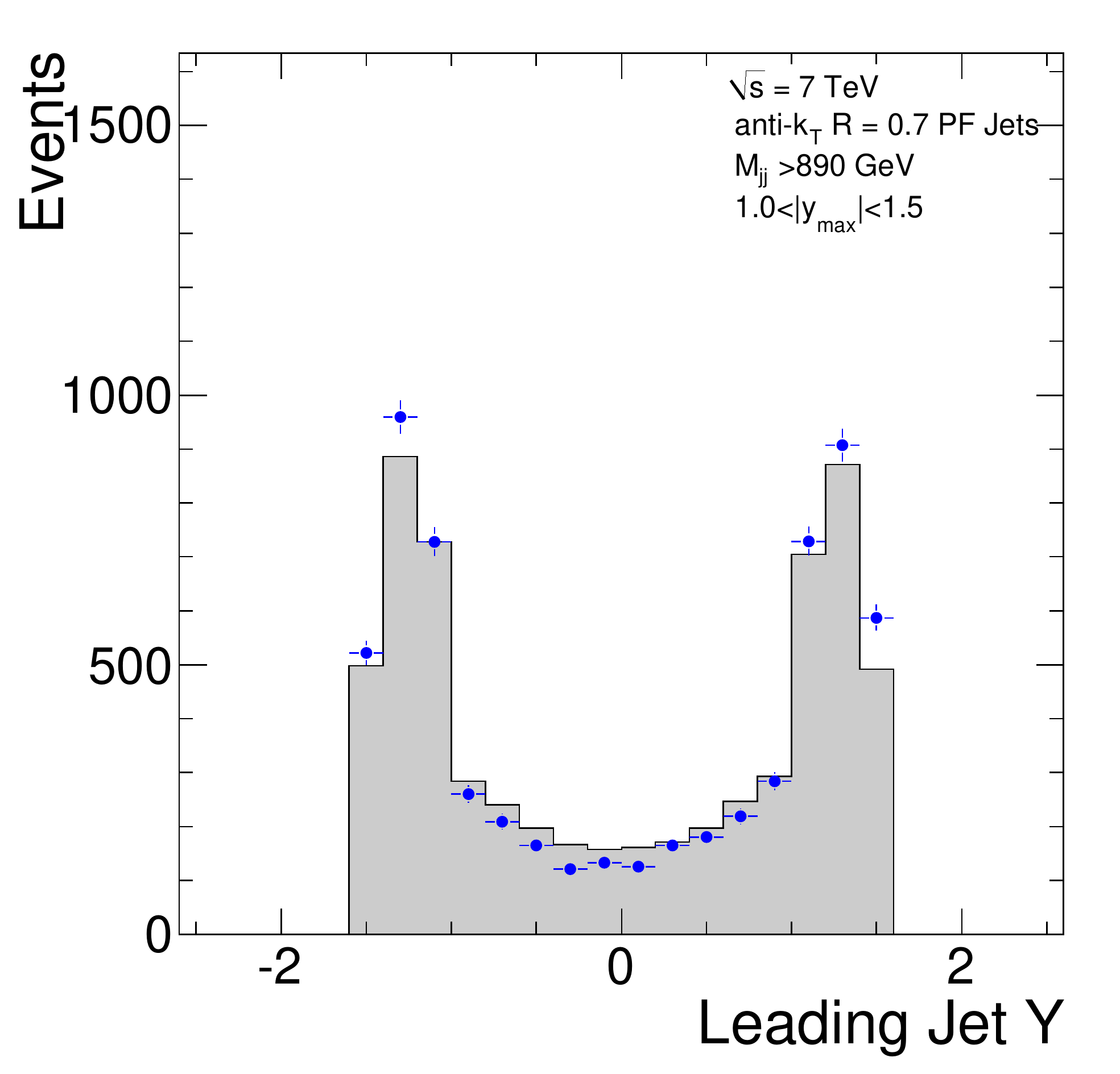} 
\includegraphics[width=0.48\textwidth]{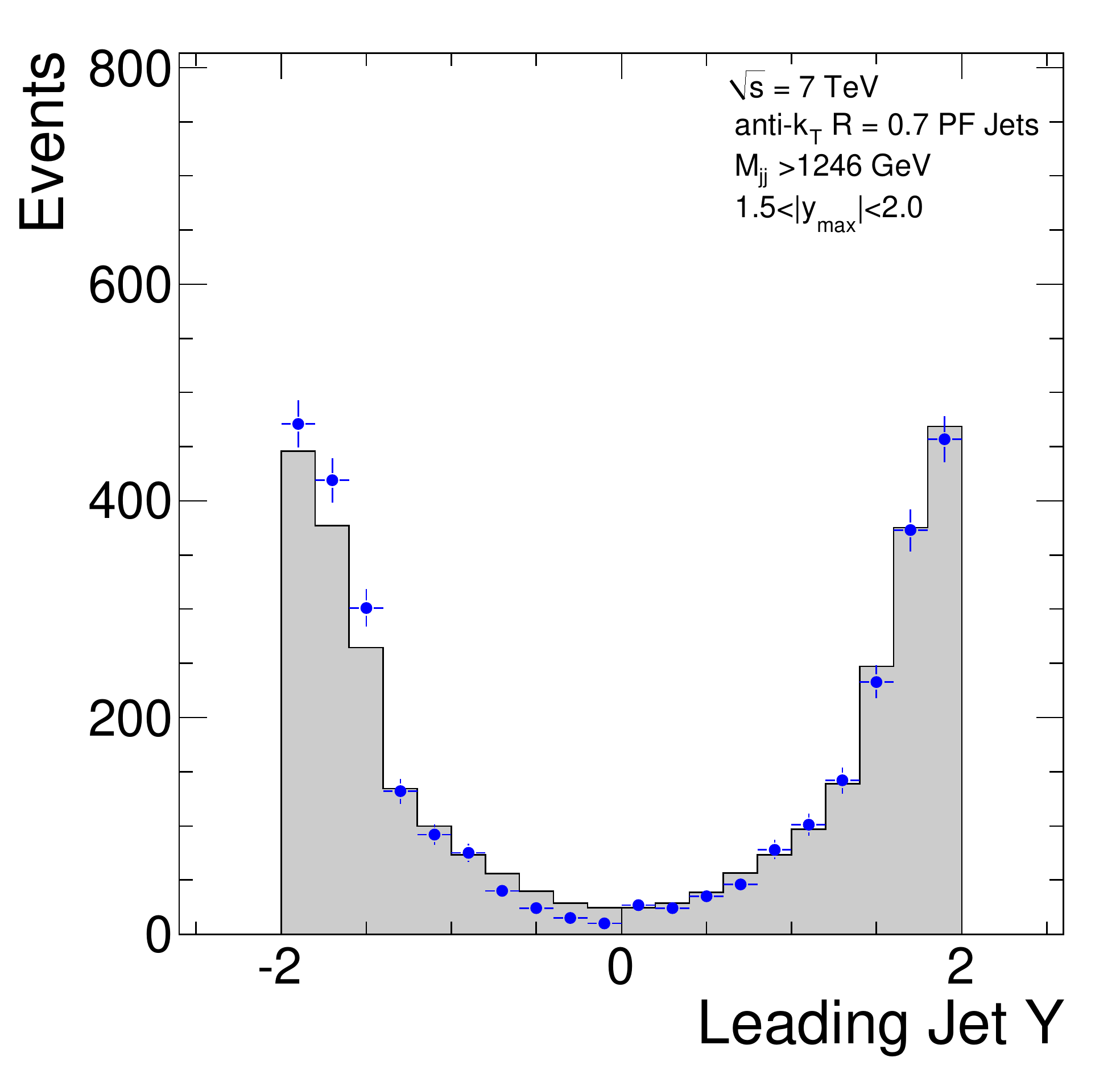} 
\includegraphics[width=0.48\textwidth]{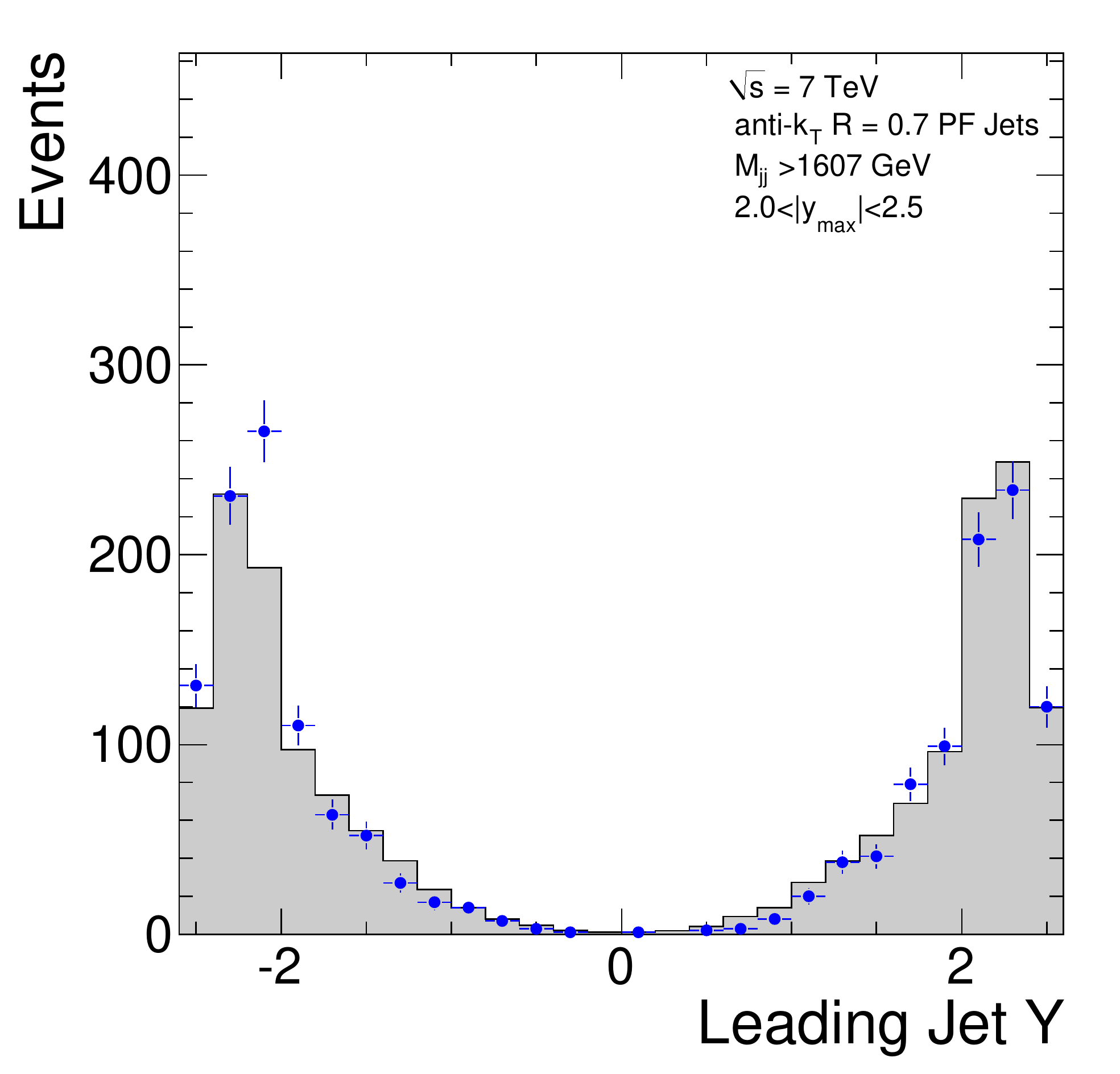}

\caption{ The $\eta$  of the leading jet  for the five different $y_{max}$ bins and for the
HLT$_{-}$Jet100U trigger, for data (points) and simulated (dashed histogram) events.}
\label{fig_appc17}
\end{figure}

\begin{figure}[ht]
\centering

\includegraphics[width=0.48\textwidth]{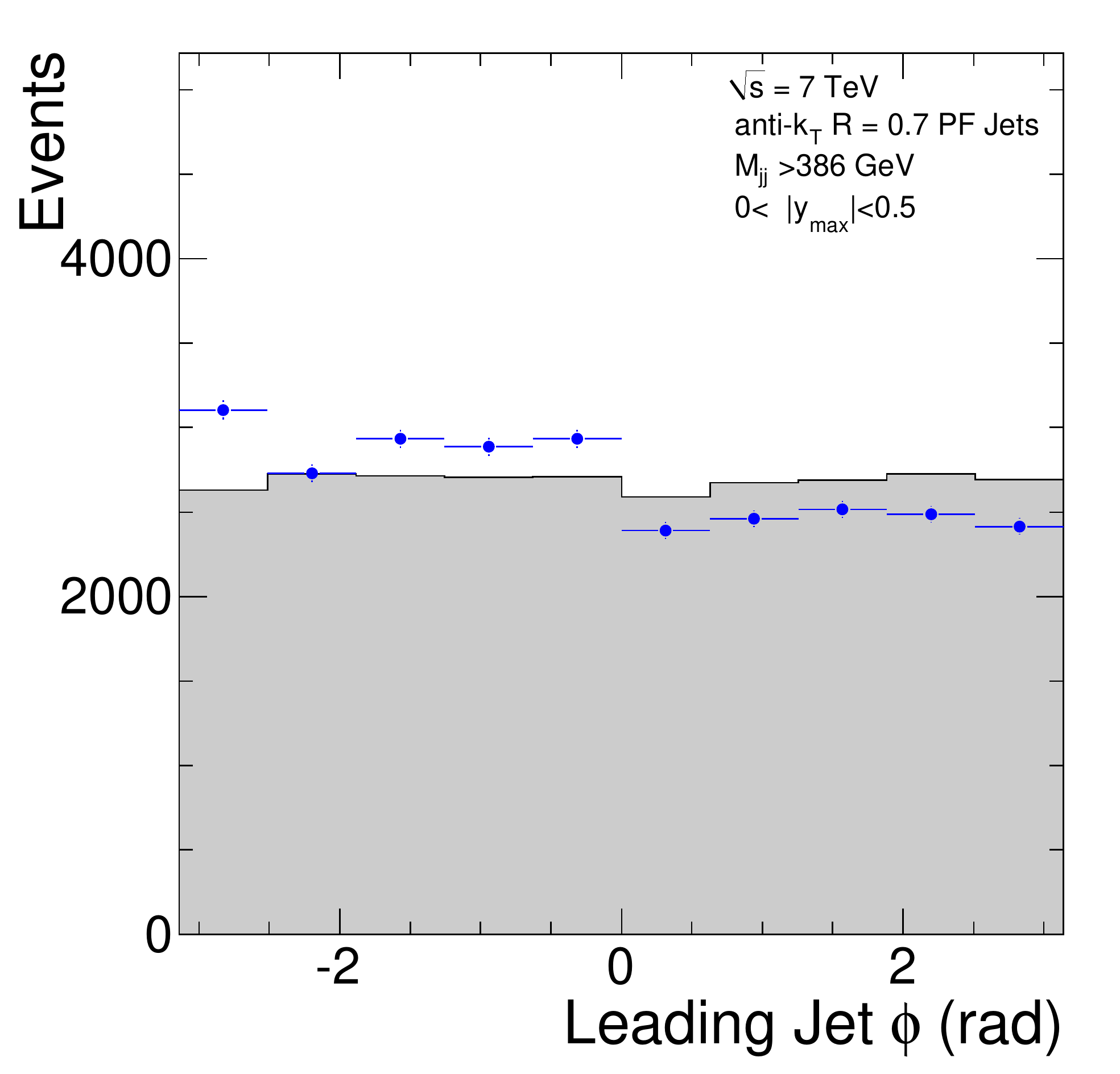} 
\includegraphics[width=0.48\textwidth]{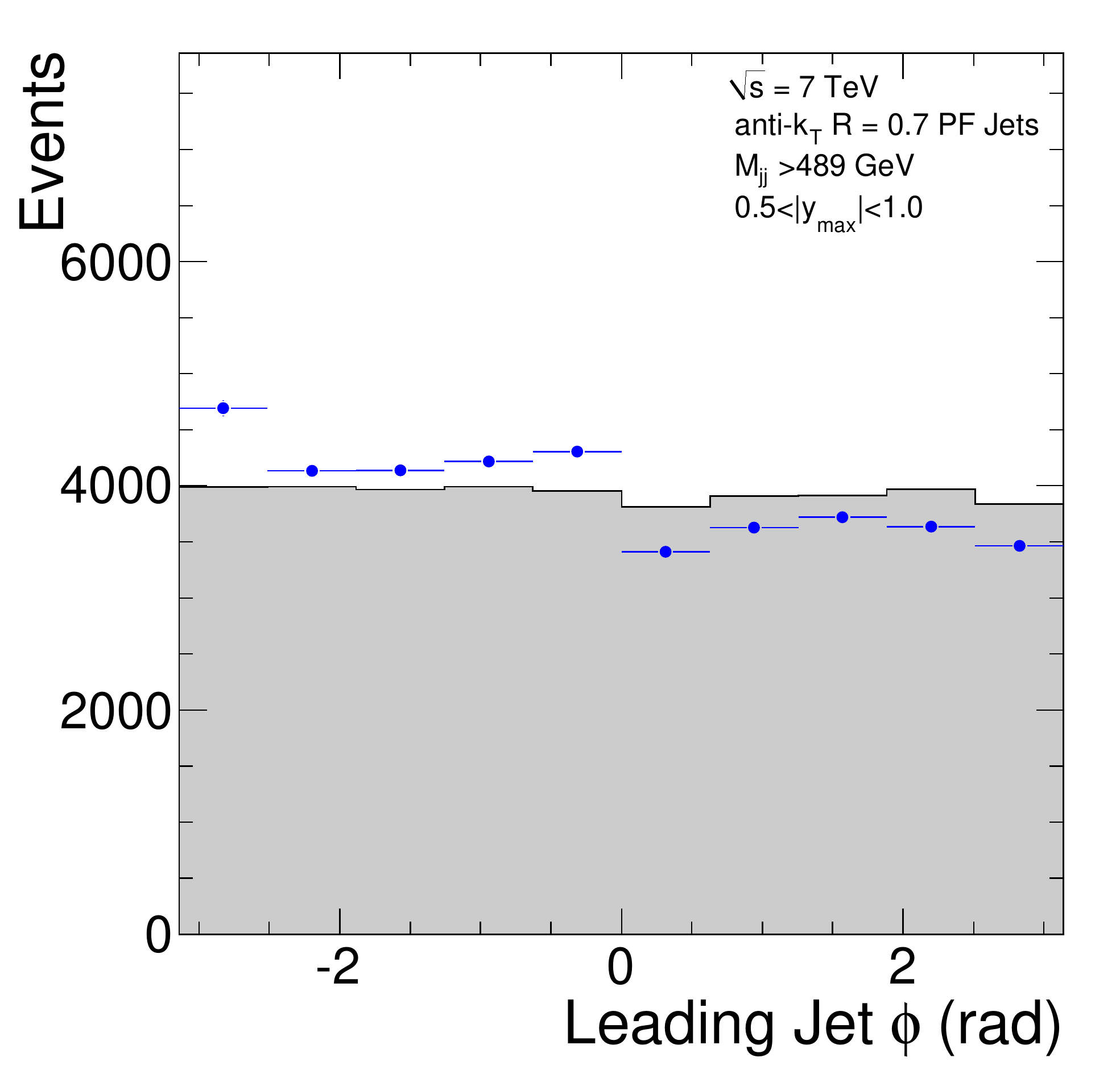} 
\includegraphics[width=0.48\textwidth]{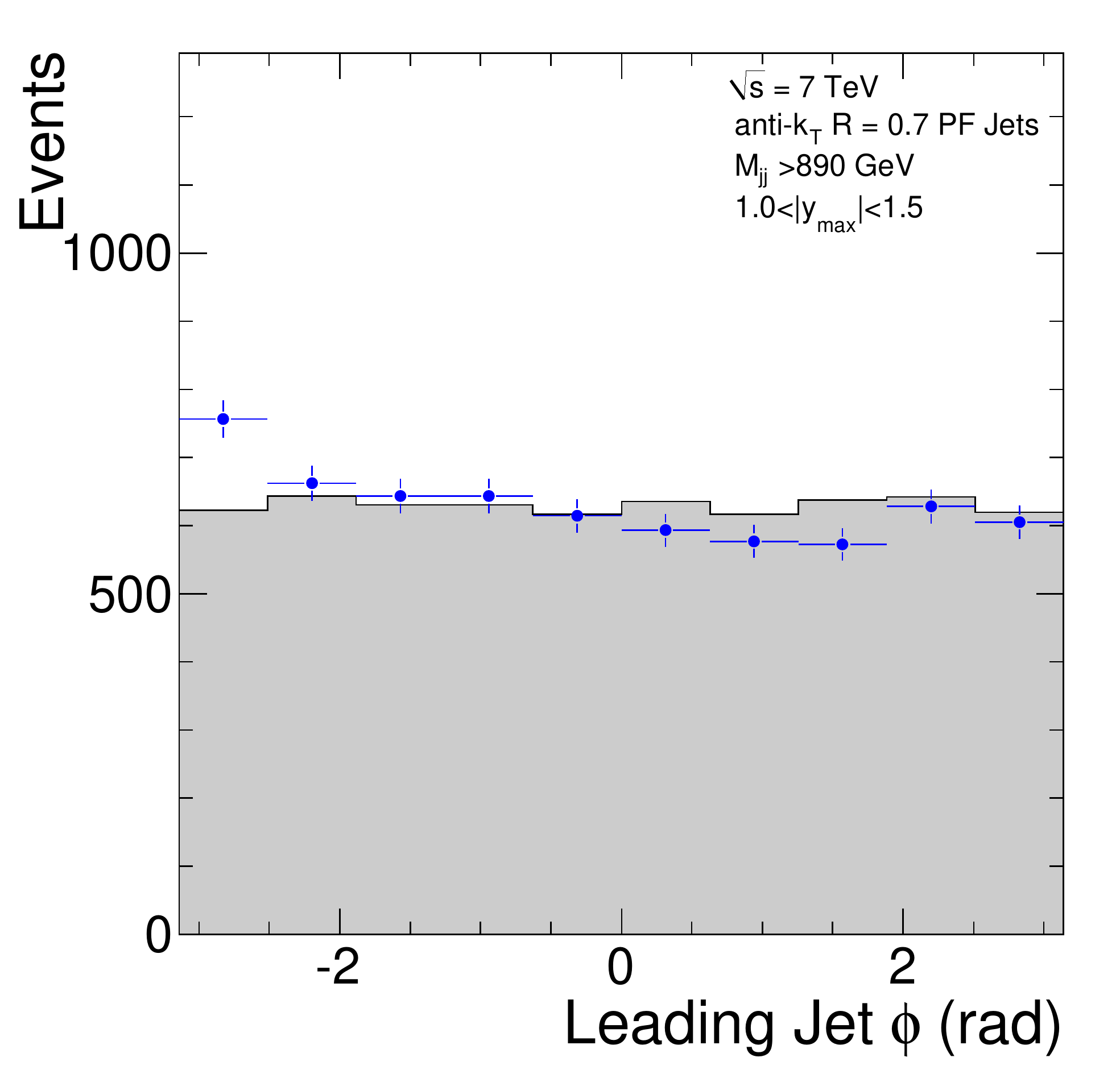} 
\includegraphics[width=0.48\textwidth]{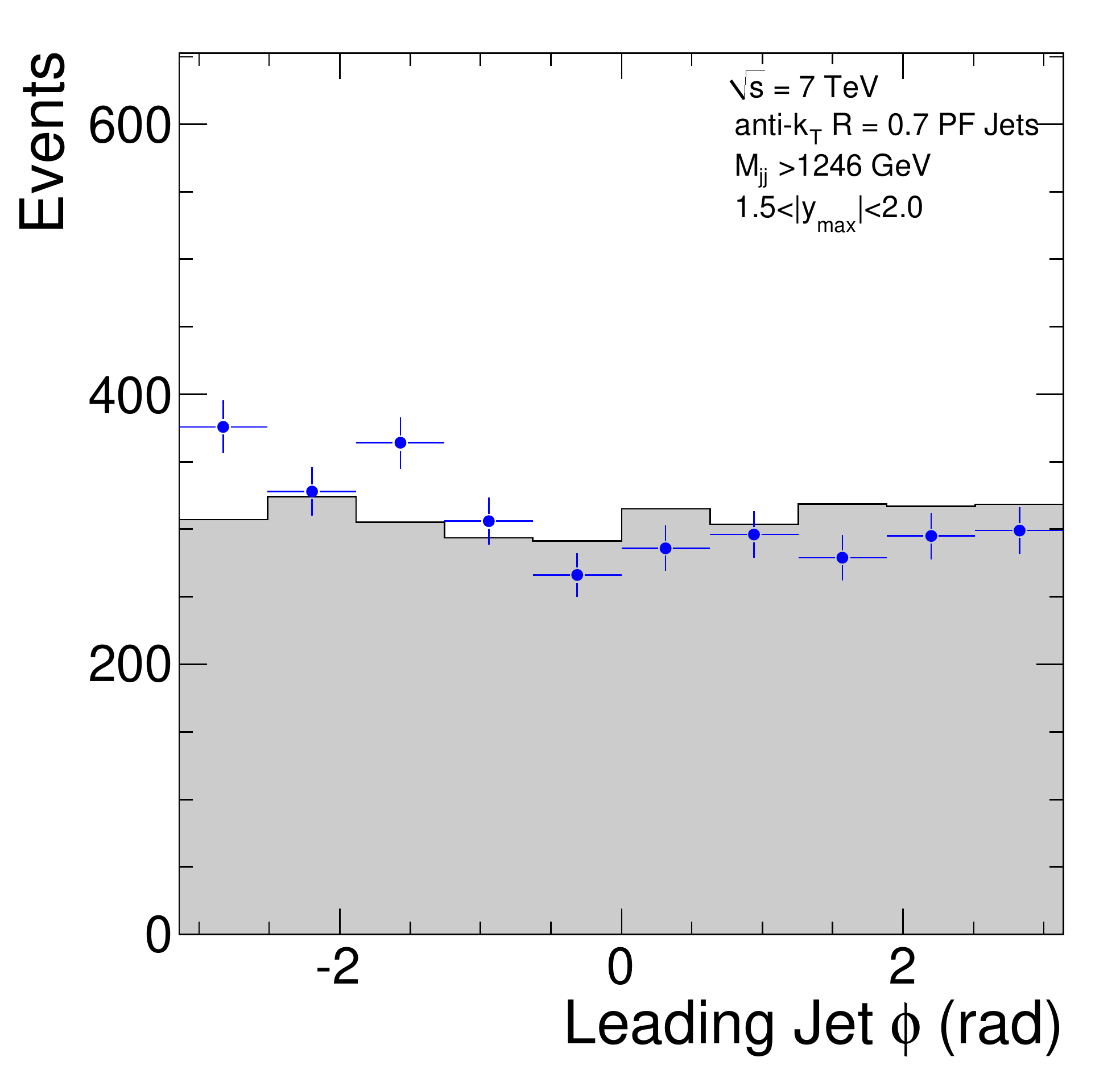} 
\includegraphics[width=0.48\textwidth]{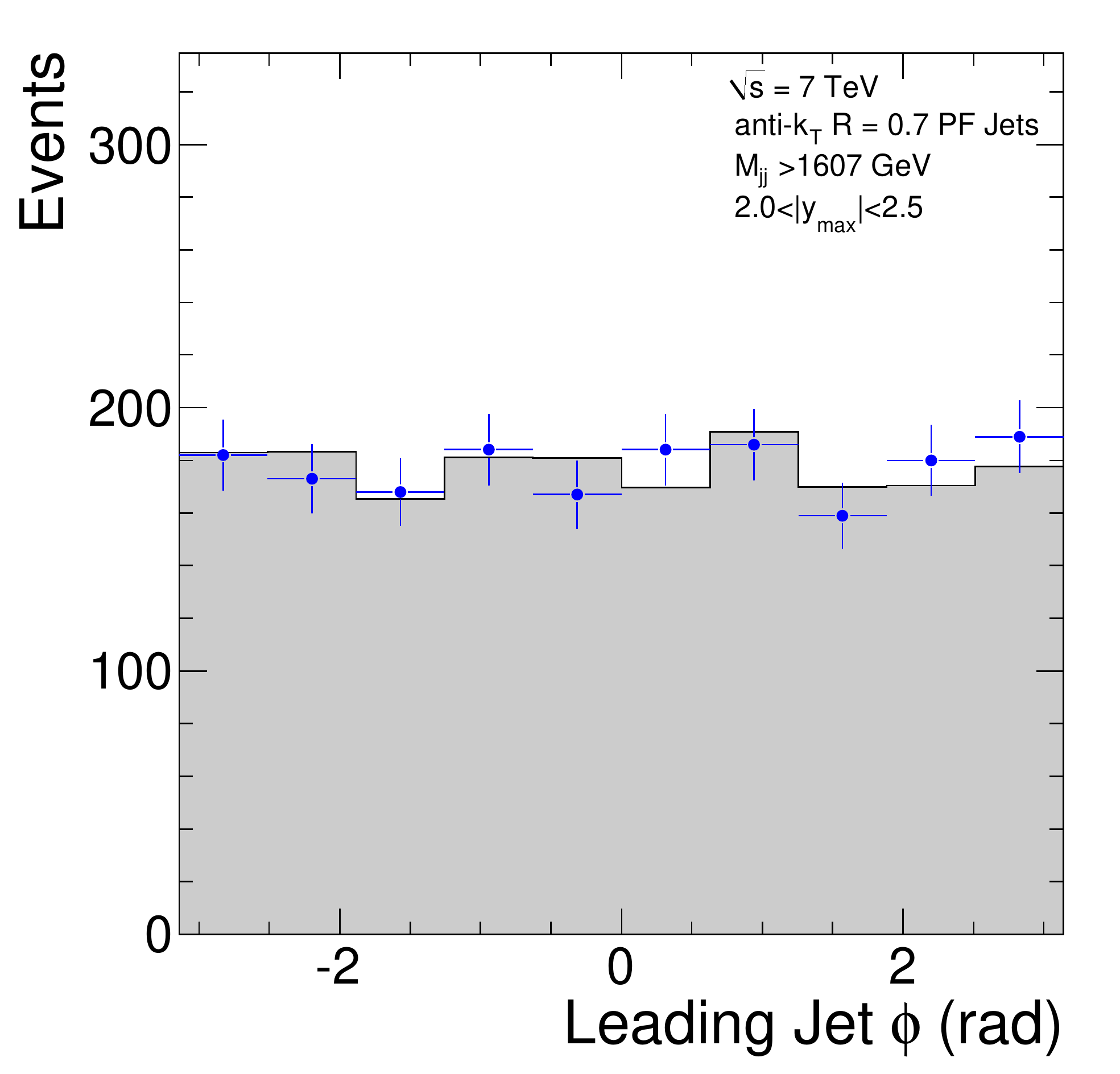}

\caption{ The $\phi$ of the leading jet  for the five different $y_{max}$ bins and for the
HLT$_{-}$Jet100U trigger, for data (points) and simulated (dashed histogram) events.}
\label{fig_appc18}
\end{figure}

\clearpage
%%%%%%%% 140U

\begin{figure}[ht]
\centering

\includegraphics[width=0.48\textwidth]{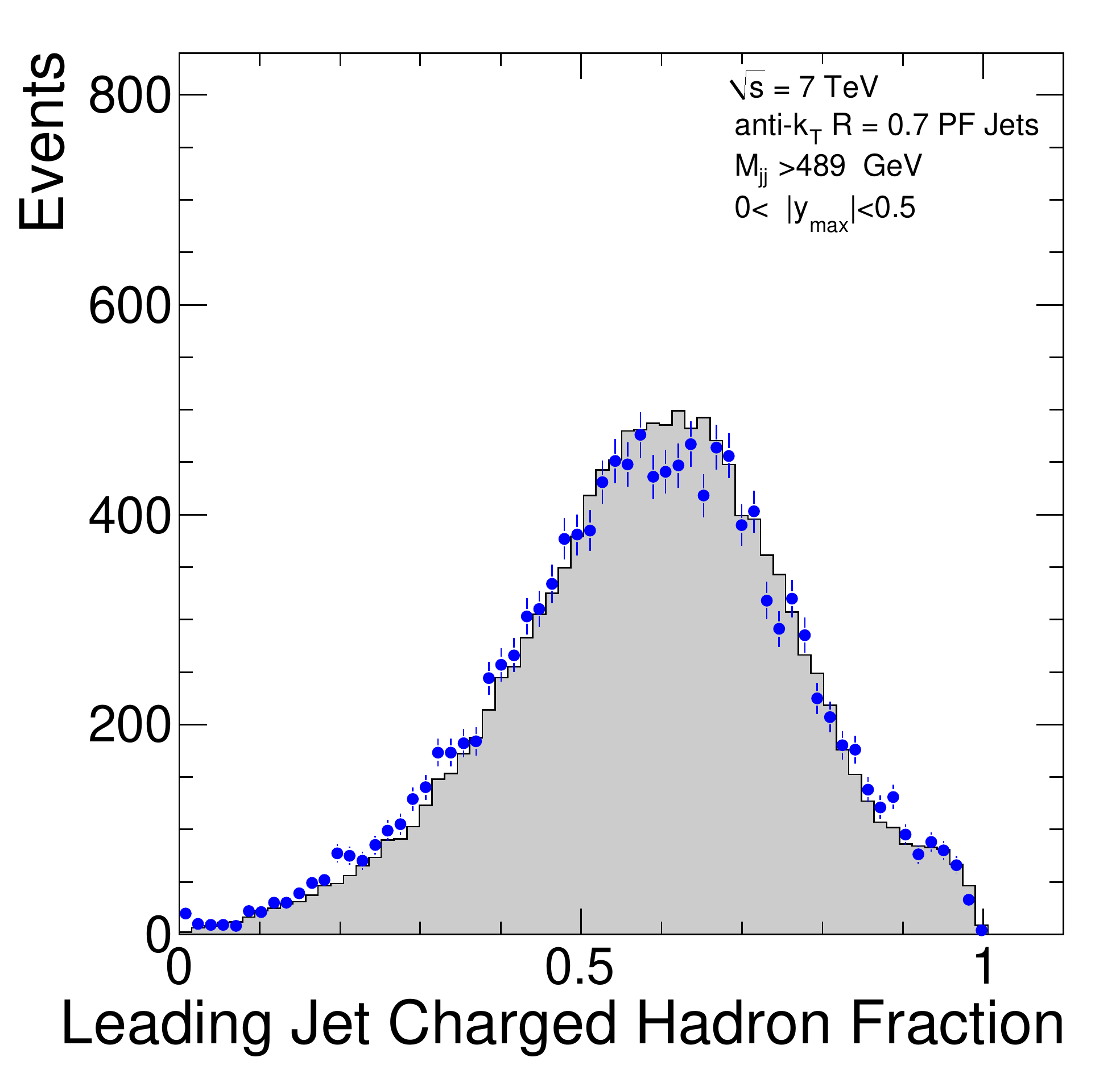} 
\includegraphics[width=0.48\textwidth]{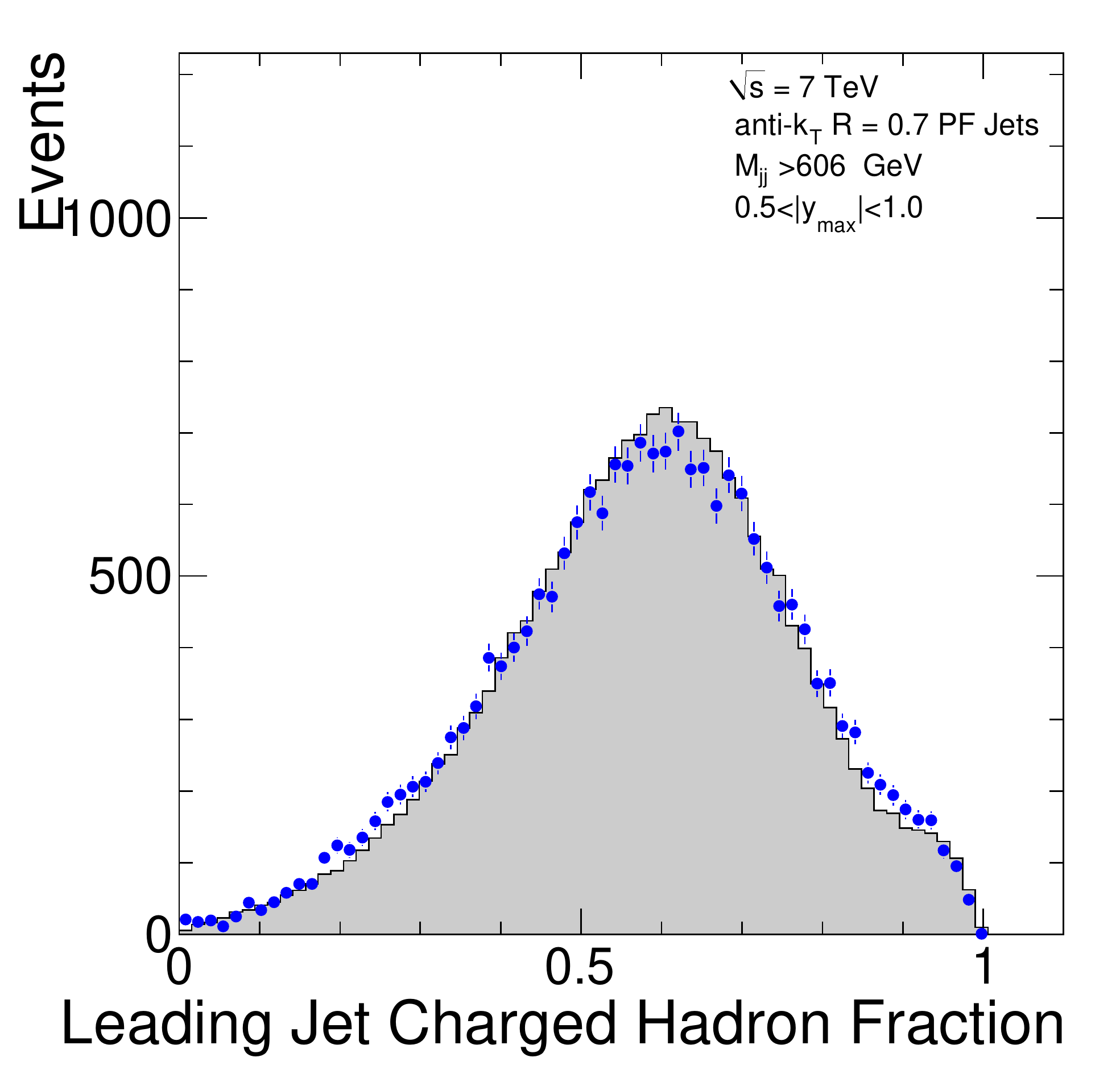} 
\includegraphics[width=0.48\textwidth]{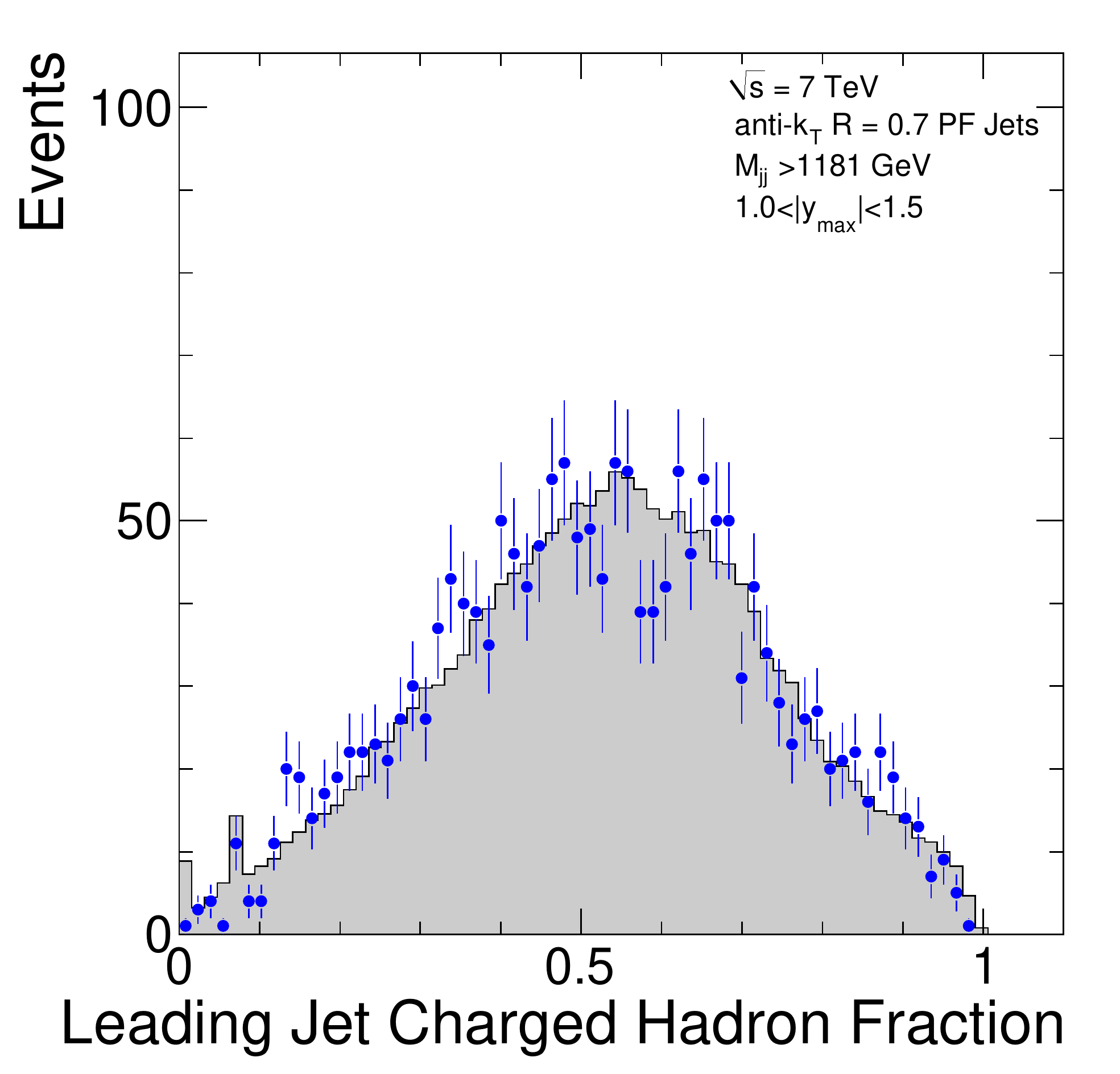} 
\includegraphics[width=0.48\textwidth]{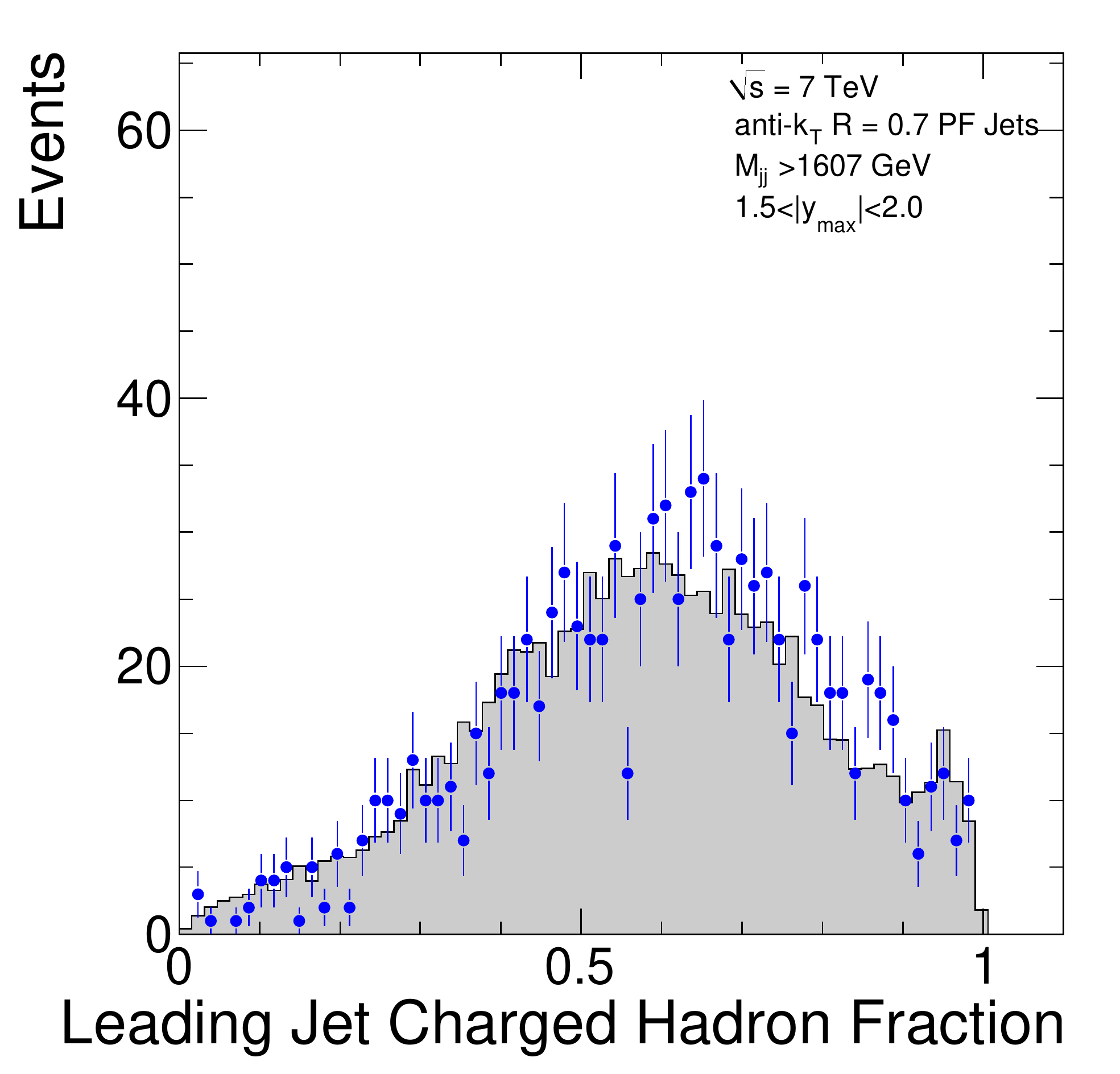} 
\includegraphics[width=0.48\textwidth]{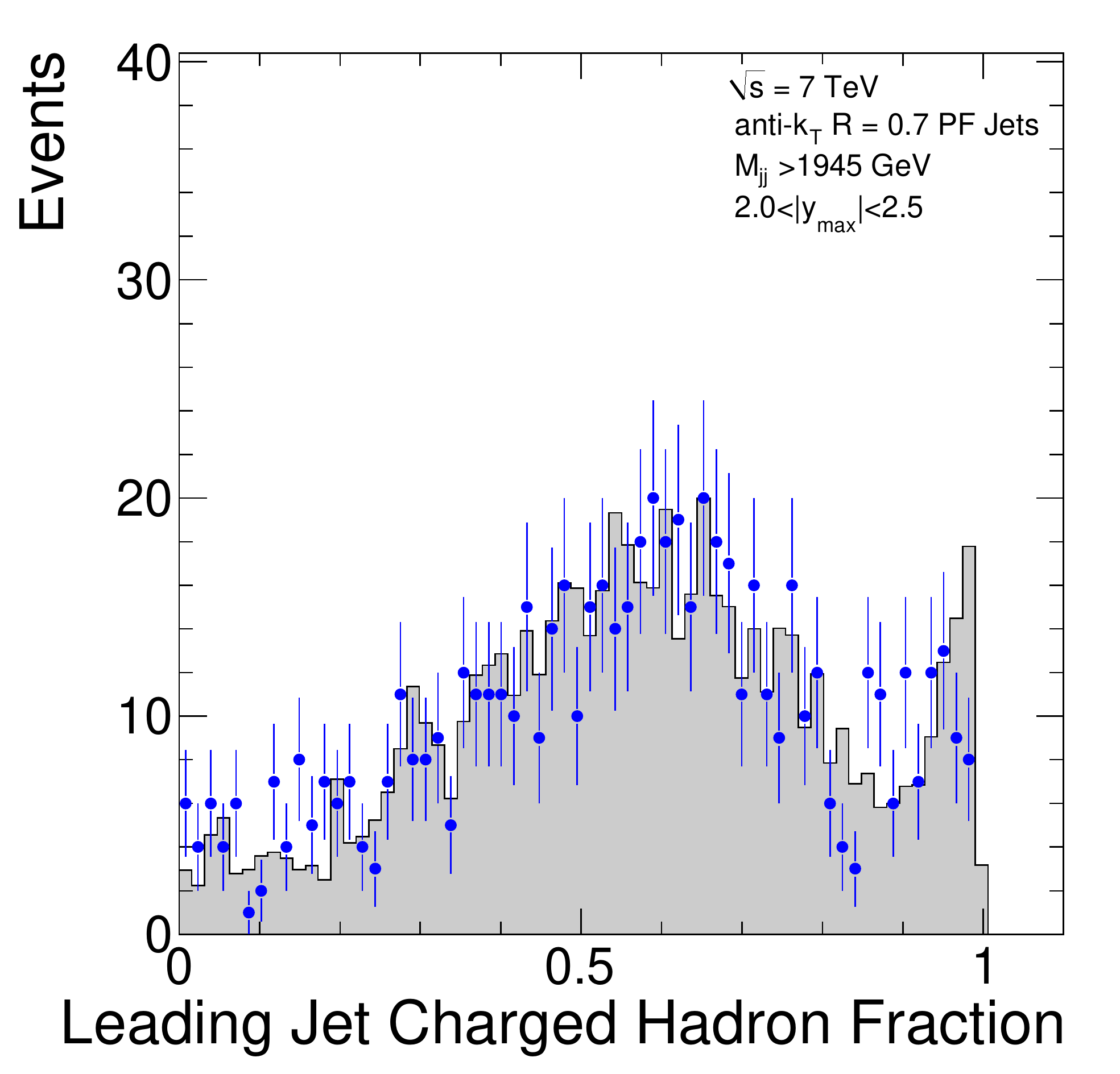}

\caption{ The charged hadron fraction of the leading jet  for the five different $y_{max}$ bins and for the
HLT$_{-}$Jet140U trigger, for data (points) and simulated (dashed histogram) events.}
\label{fig_appc19}
\end{figure}

\begin{figure}[ht]
\centering

\includegraphics[width=0.48\textwidth]{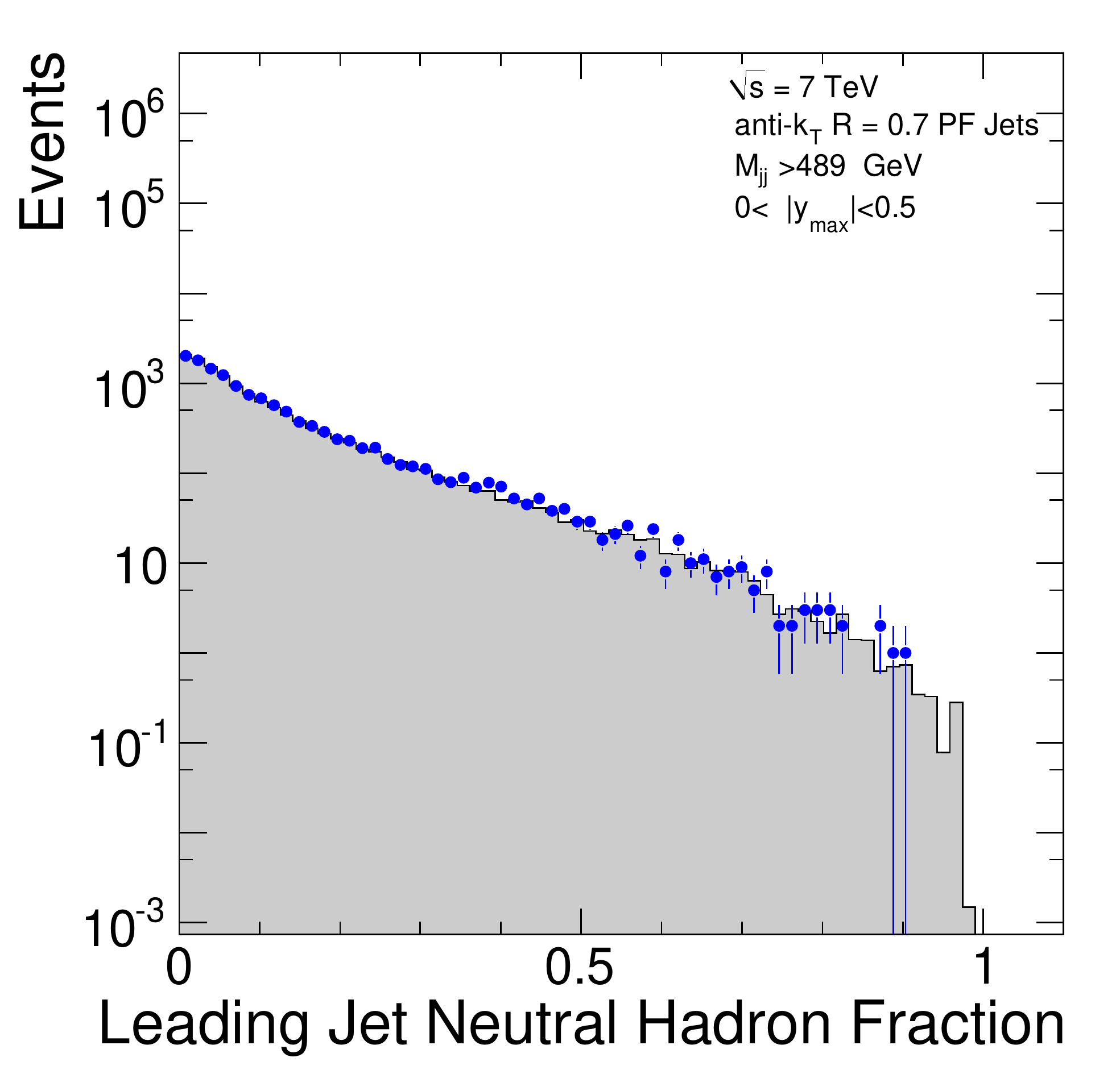} 
\includegraphics[width=0.48\textwidth]{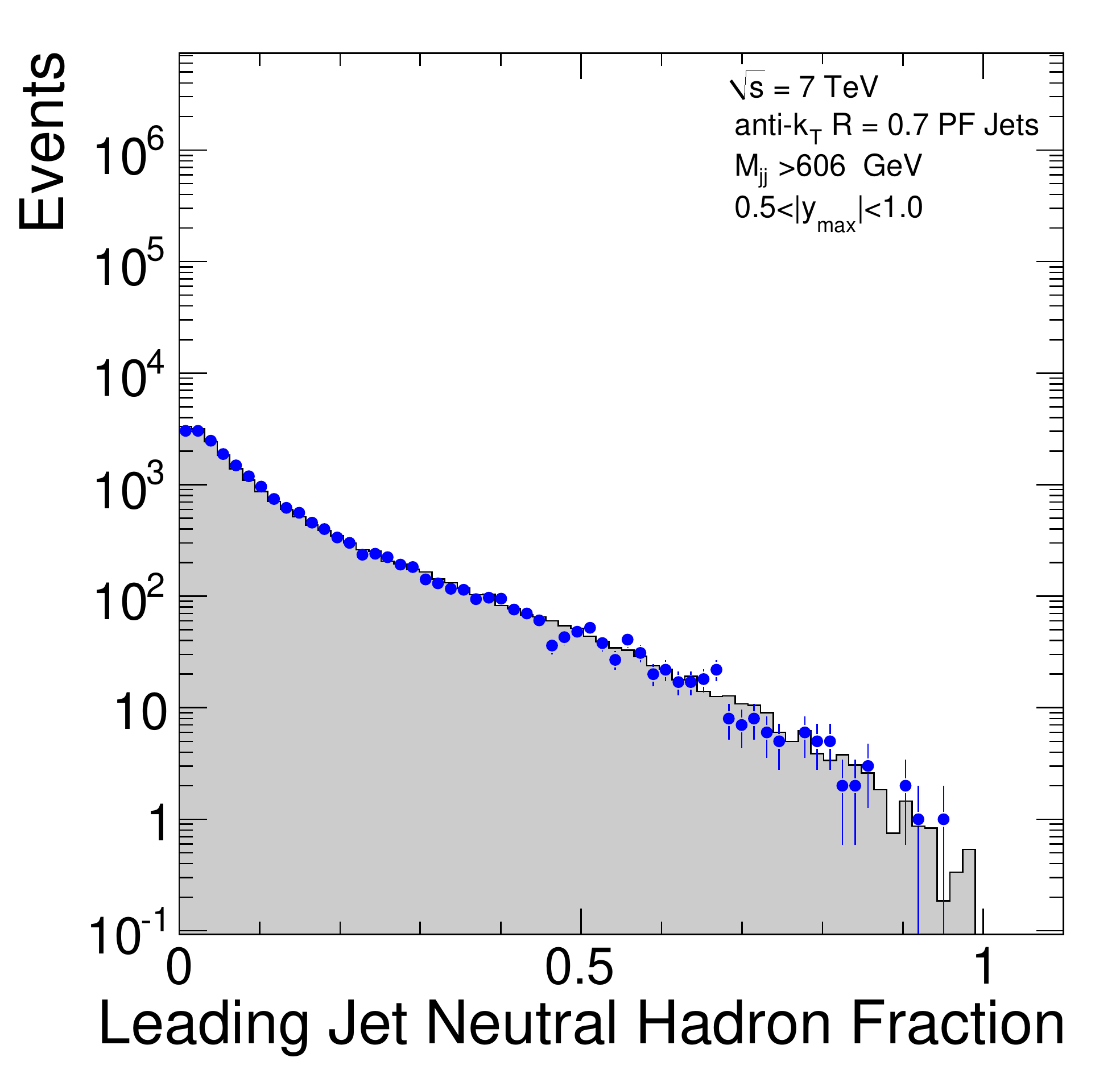} 
\includegraphics[width=0.48\textwidth]{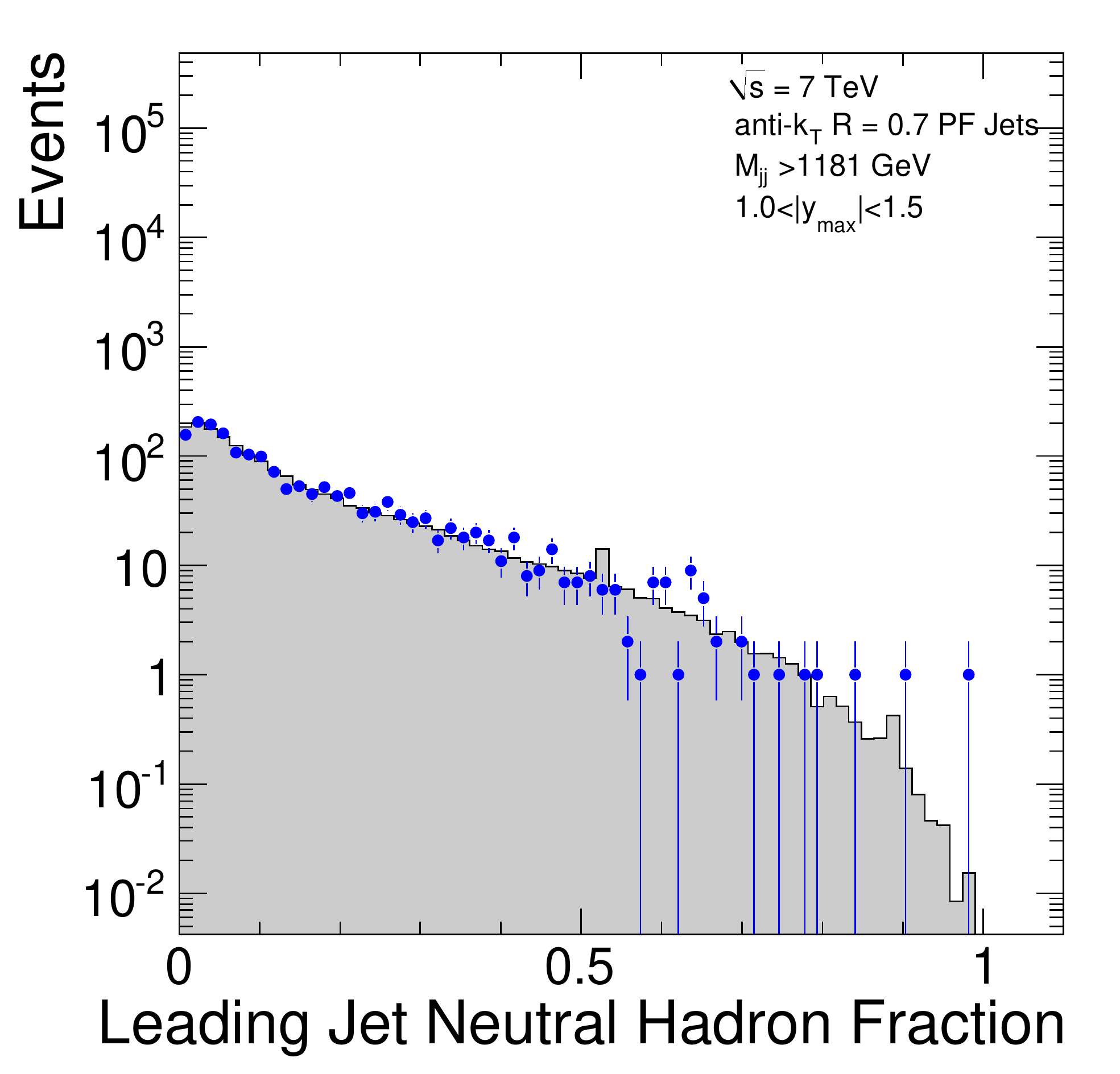} 
\includegraphics[width=0.48\textwidth]{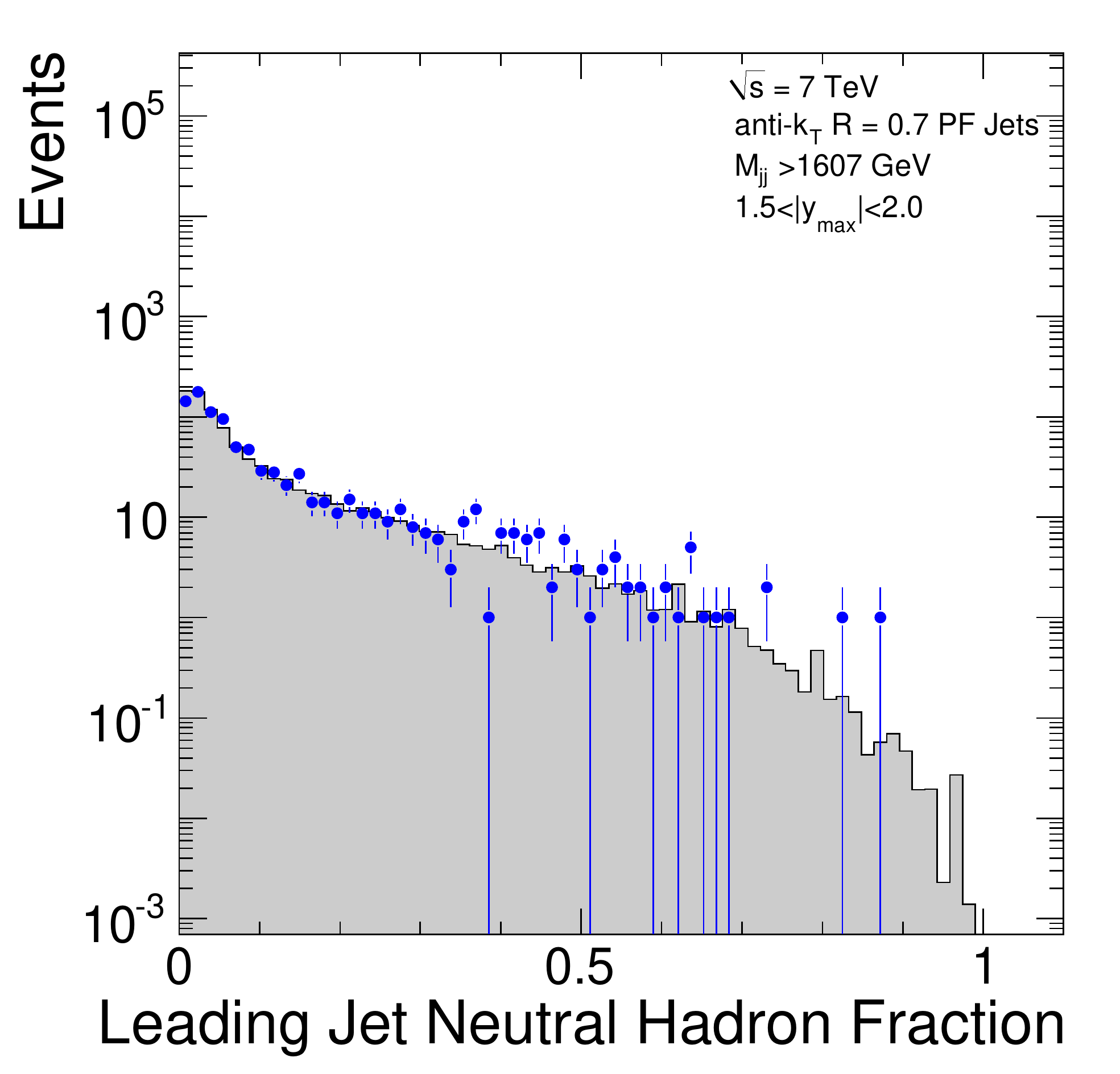} 
\includegraphics[width=0.48\textwidth]{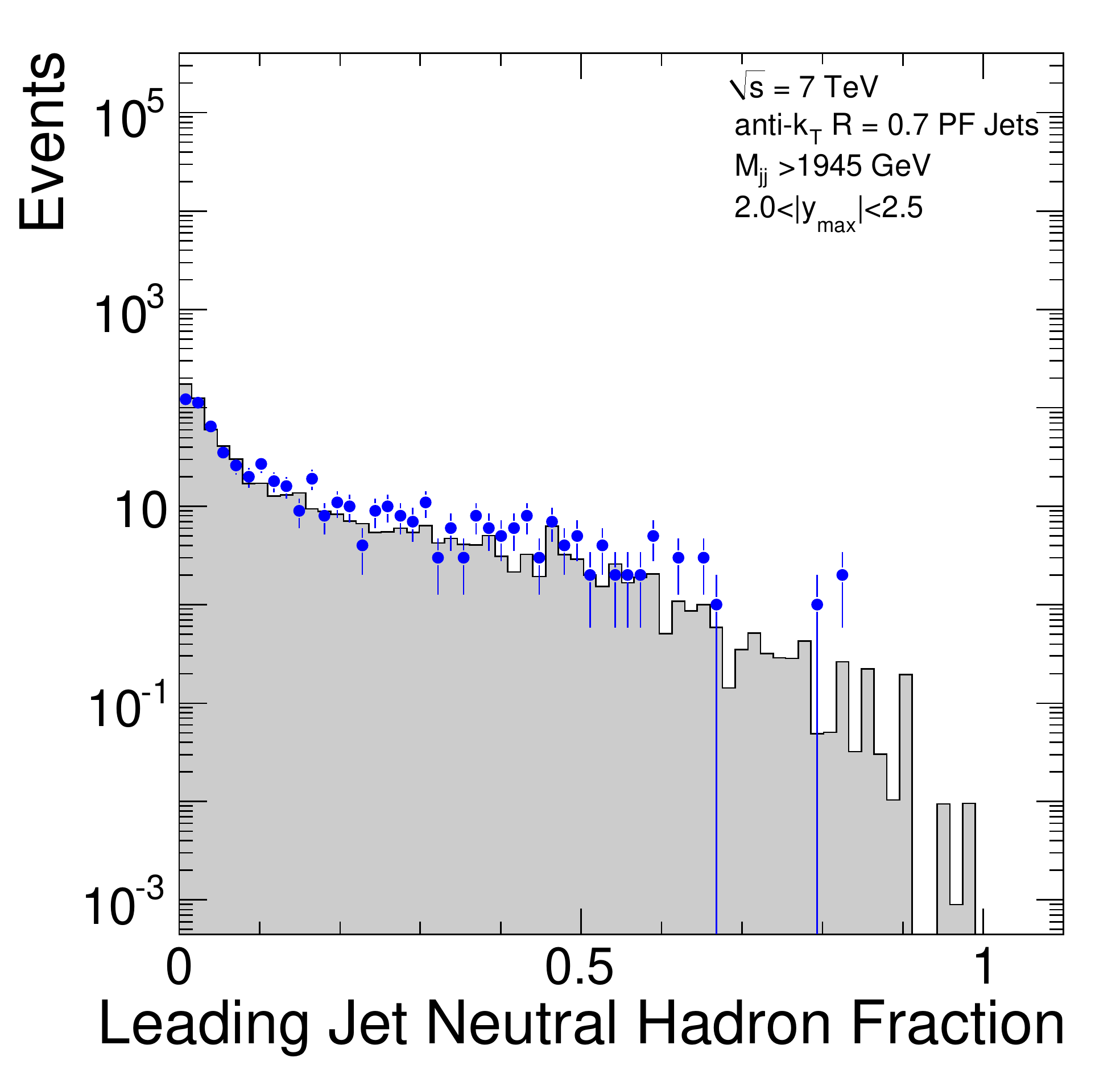}

\caption{ The neutral hadron fraction of the leading jet  for the five different $y_{max}$ bins and for the
HLT$_{-}$Jet140U trigger, for data (points) and simulated (dashed histogram) events.}
\label{fig_appc20}
\end{figure}

\begin{figure}[ht]
\centering

\includegraphics[width=0.48\textwidth]{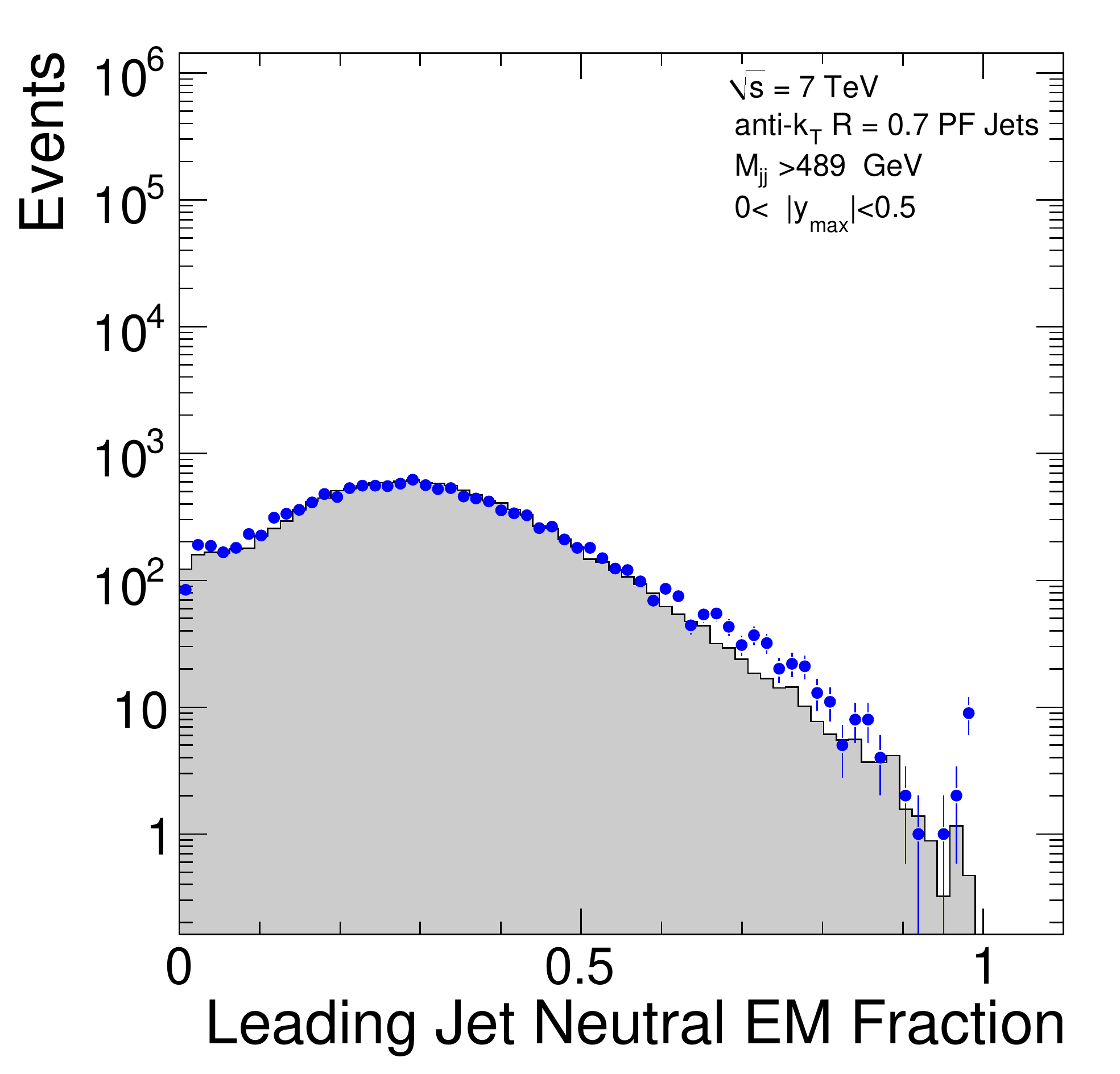} 
\includegraphics[width=0.48\textwidth]{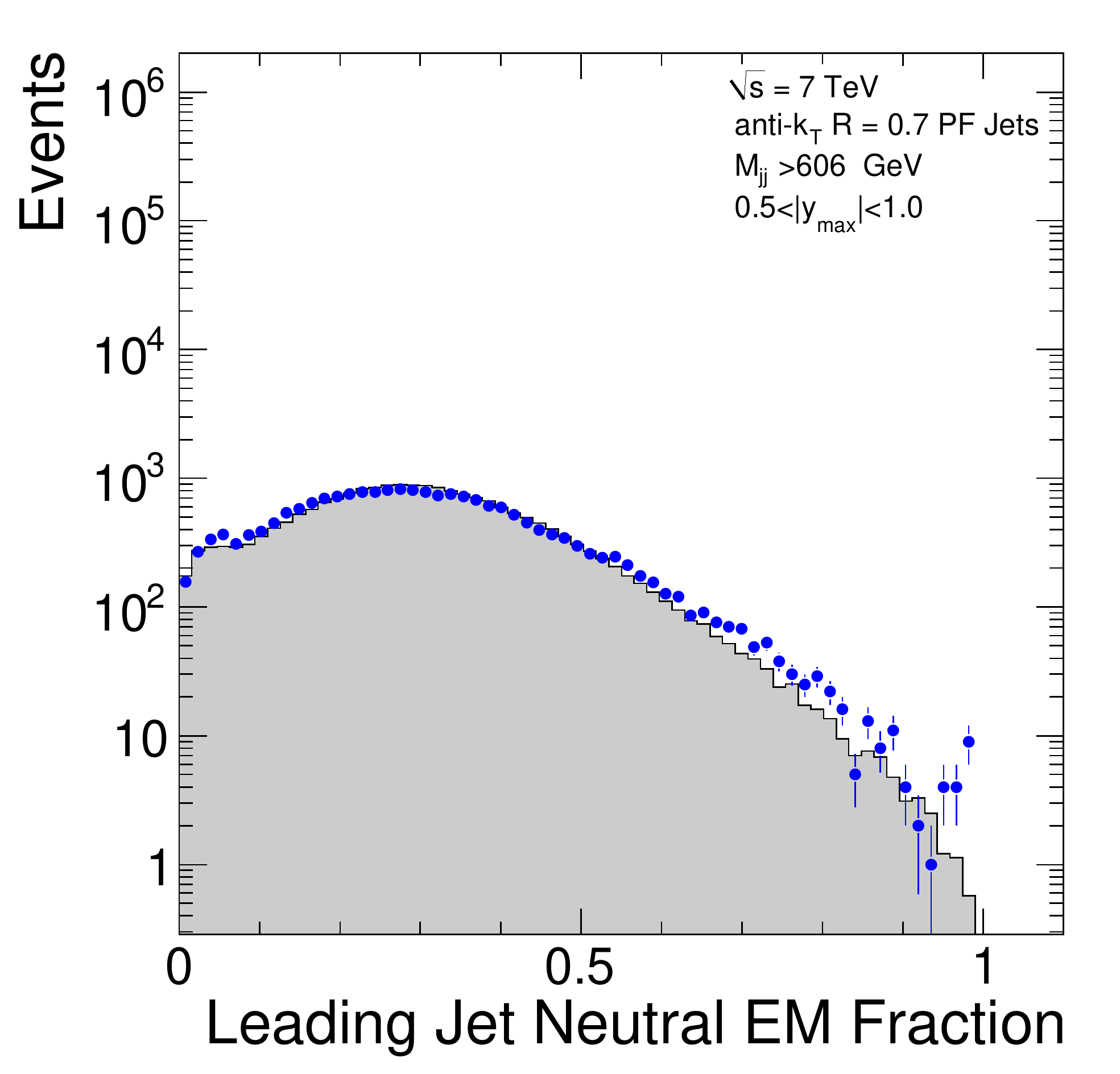} 
\includegraphics[width=0.48\textwidth]{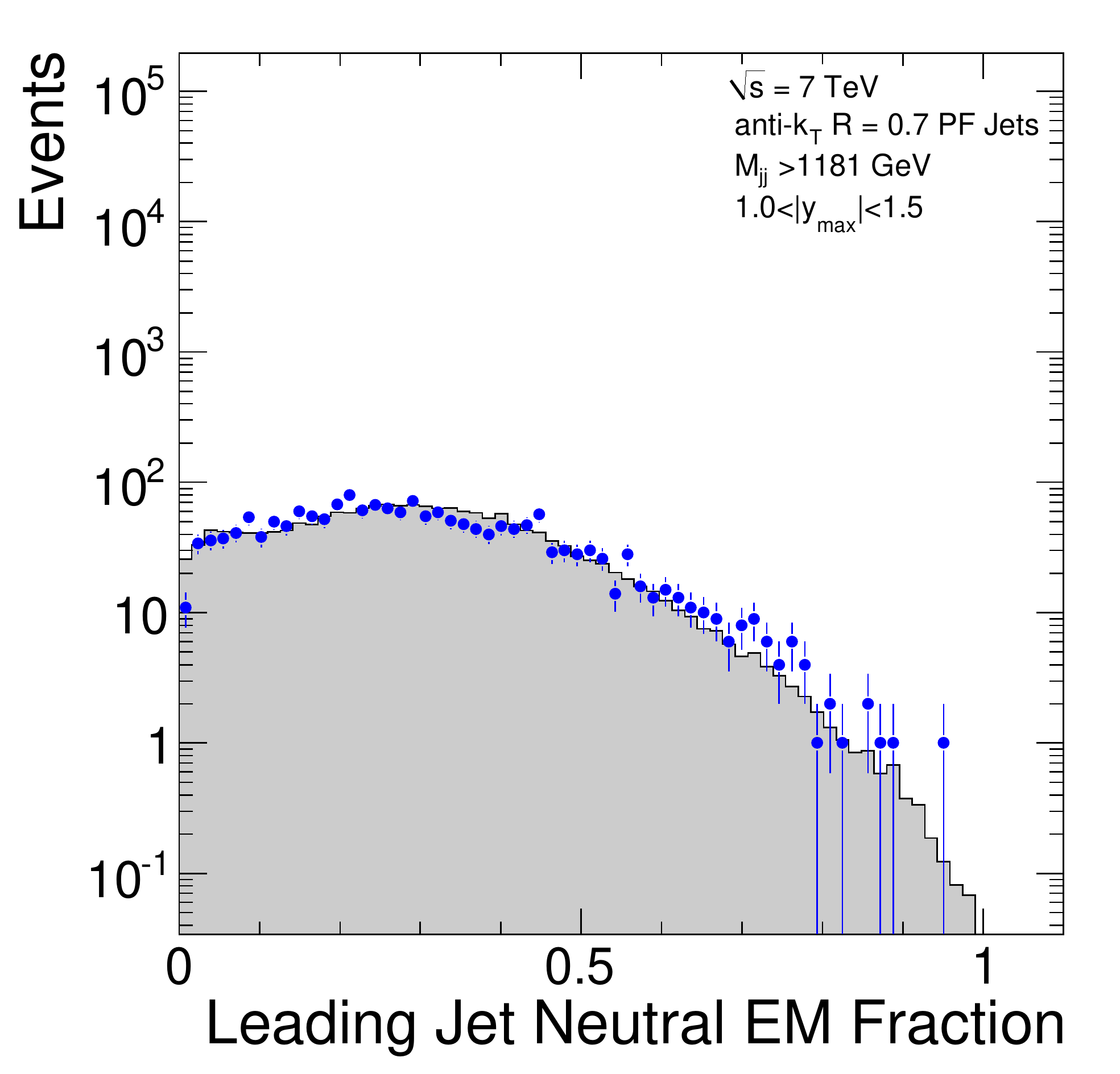} 
\includegraphics[width=0.48\textwidth]{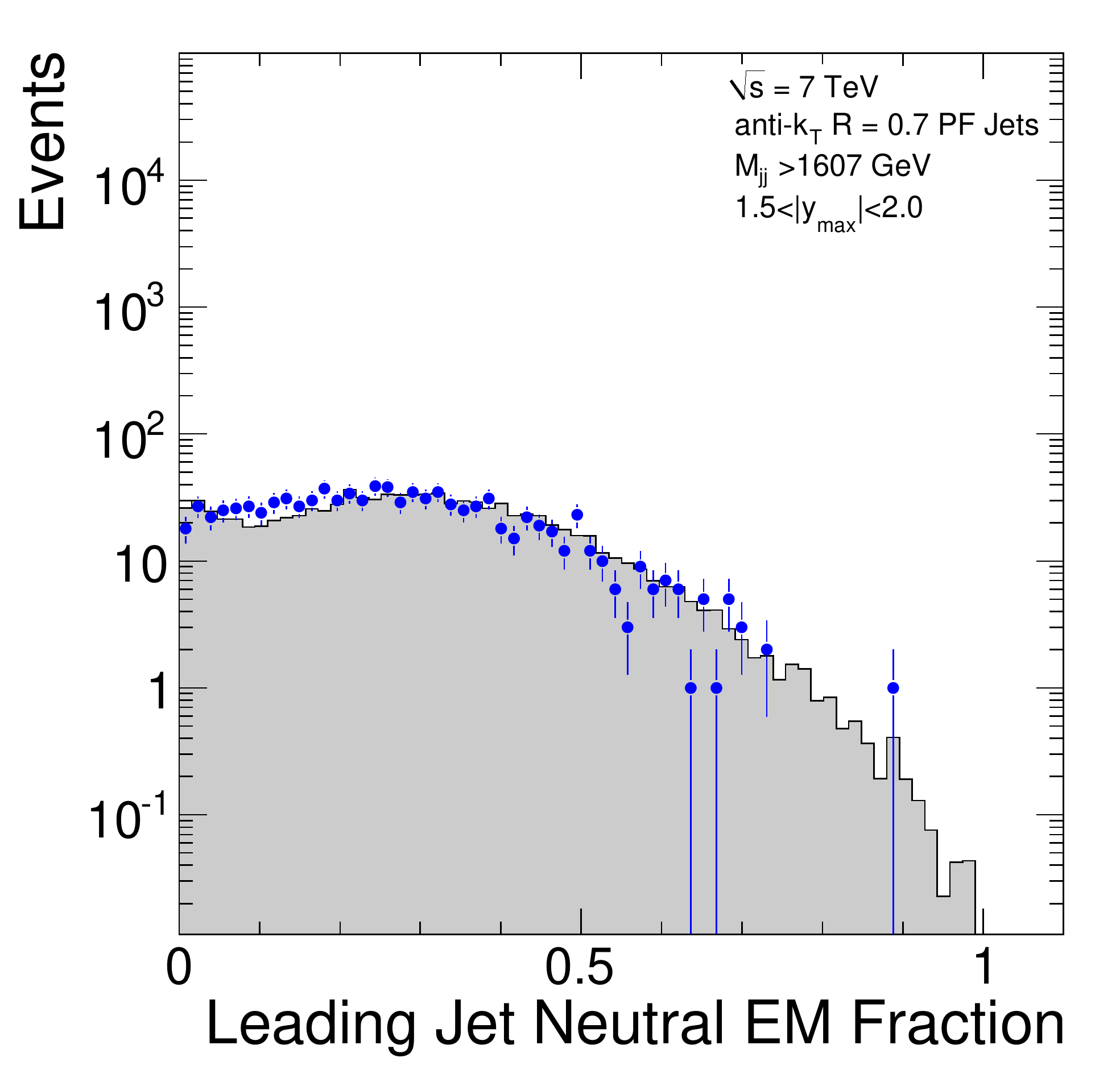} 
\includegraphics[width=0.48\textwidth]{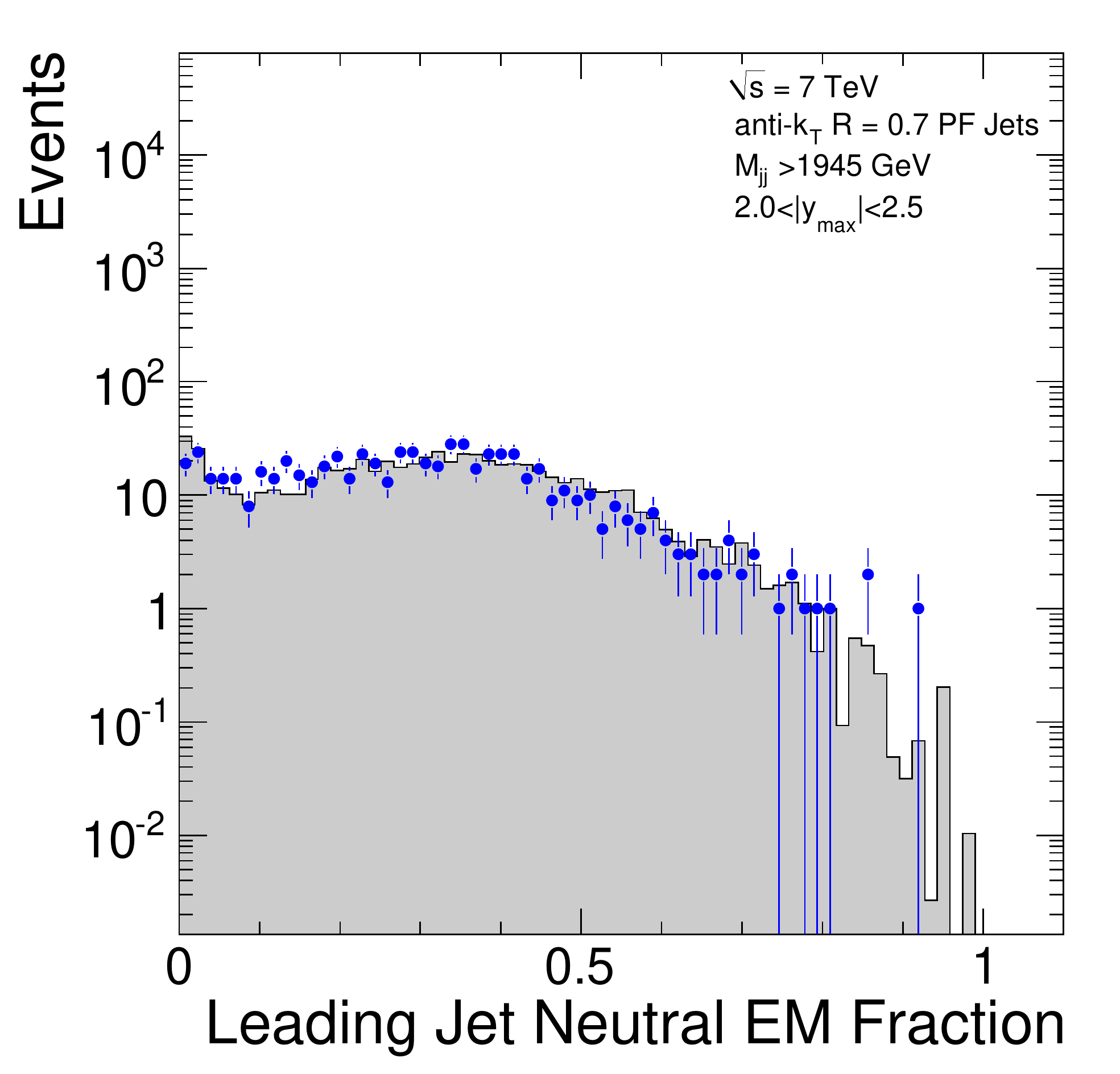}

\caption{ The neutral electromagnetic  fraction of the leading jet  for the five different $y_{max}$ bins and for the
HLT$_{-}$Jet140U trigger, for data (points) and simulated (dashed histogram) events.}
\label{fig_appc21}
\end{figure}

 \begin{figure}[ht]
\centering

\includegraphics[width=0.48\textwidth]{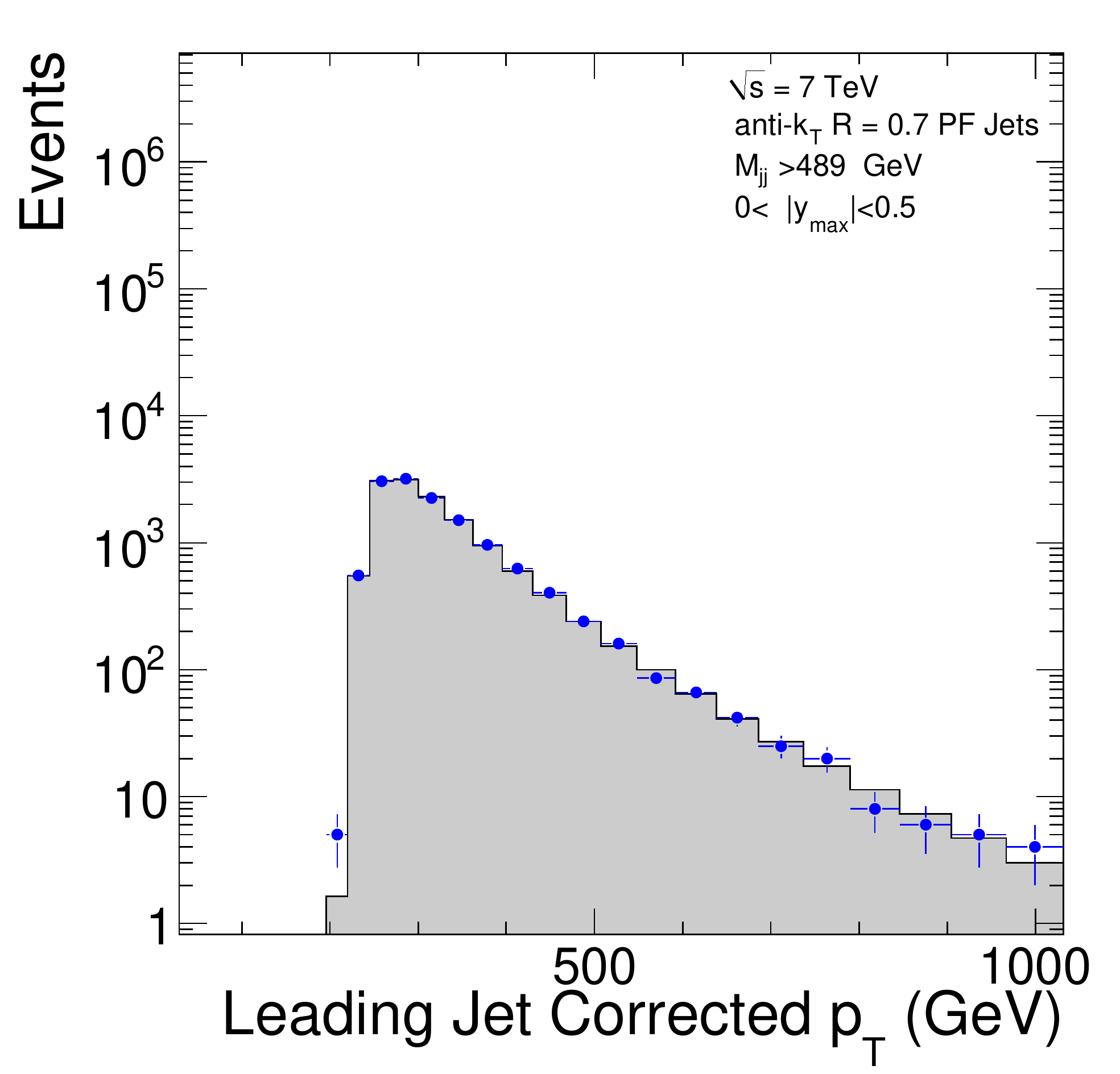} 
\includegraphics[width=0.48\textwidth]{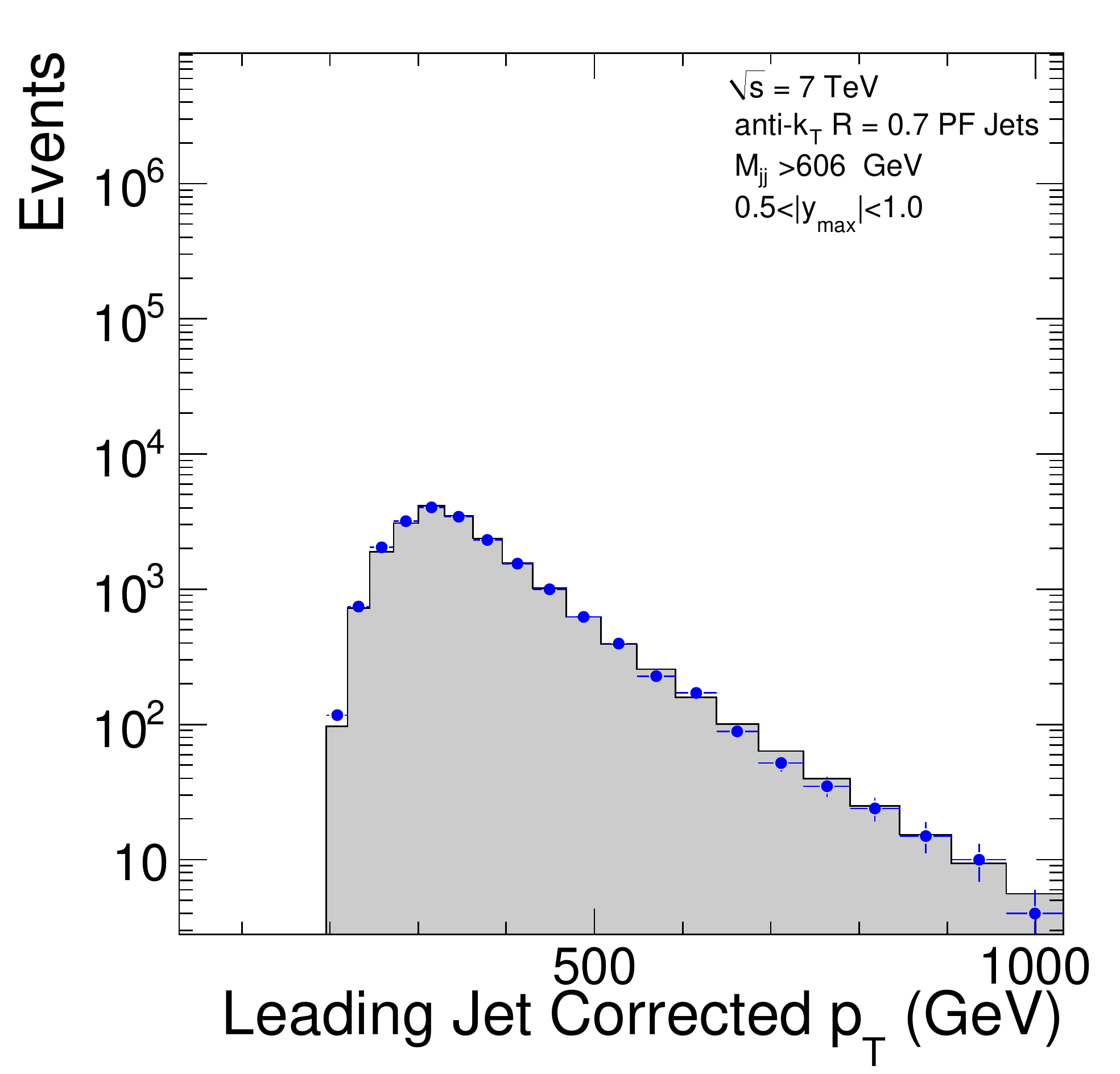} 
\includegraphics[width=0.48\textwidth]{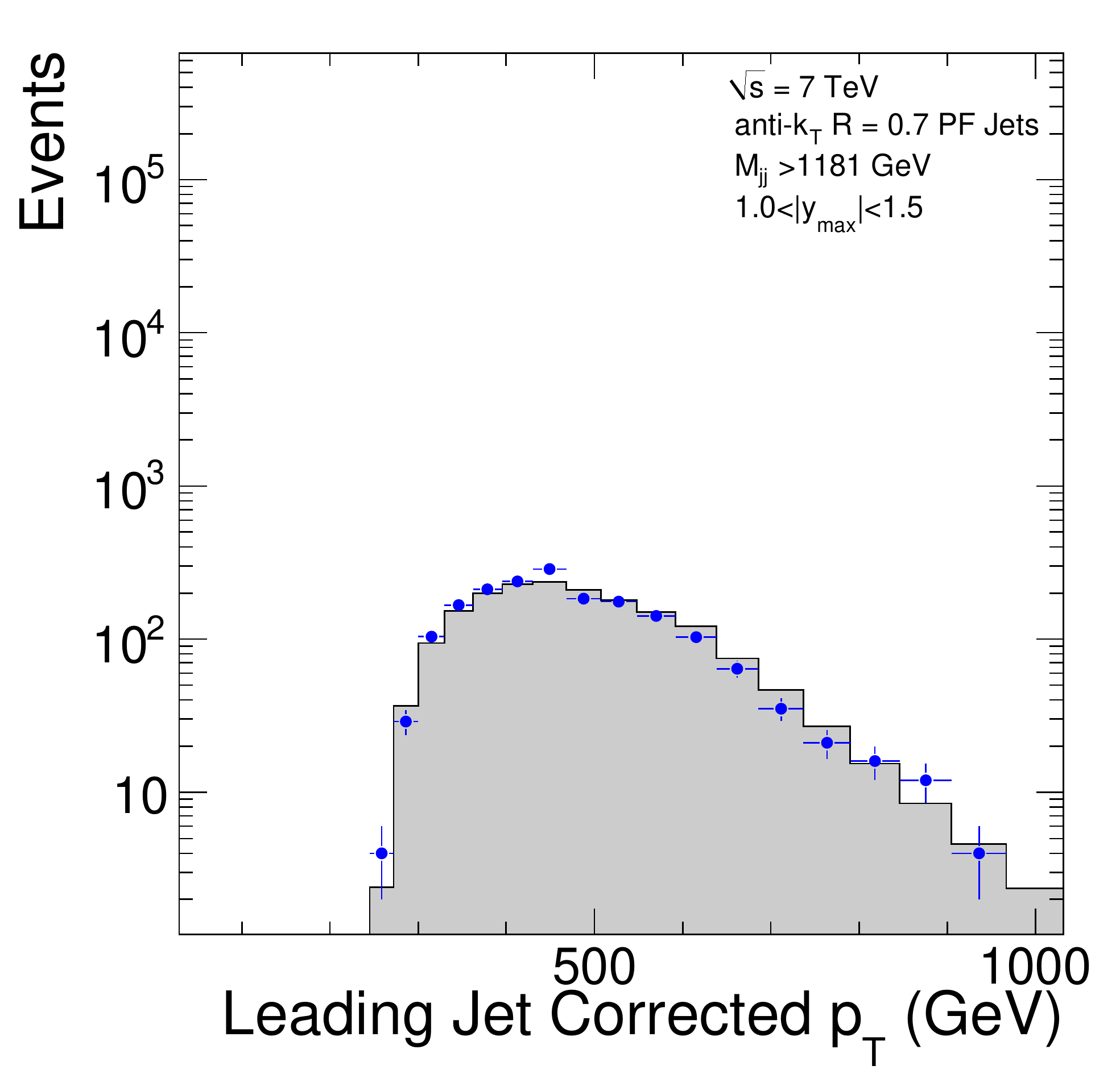} 
\includegraphics[width=0.48\textwidth]{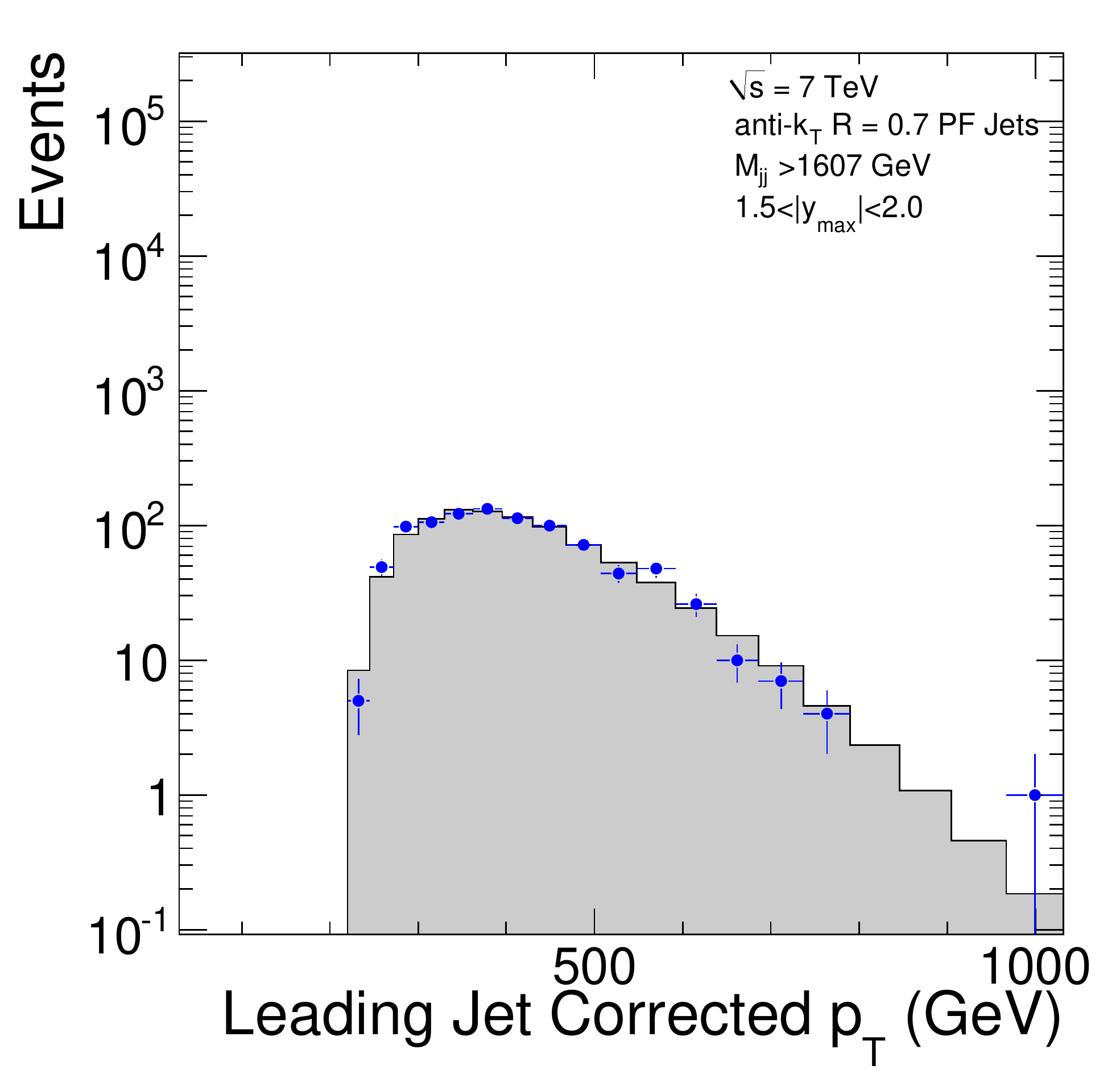} 
\includegraphics[width=0.48\textwidth]{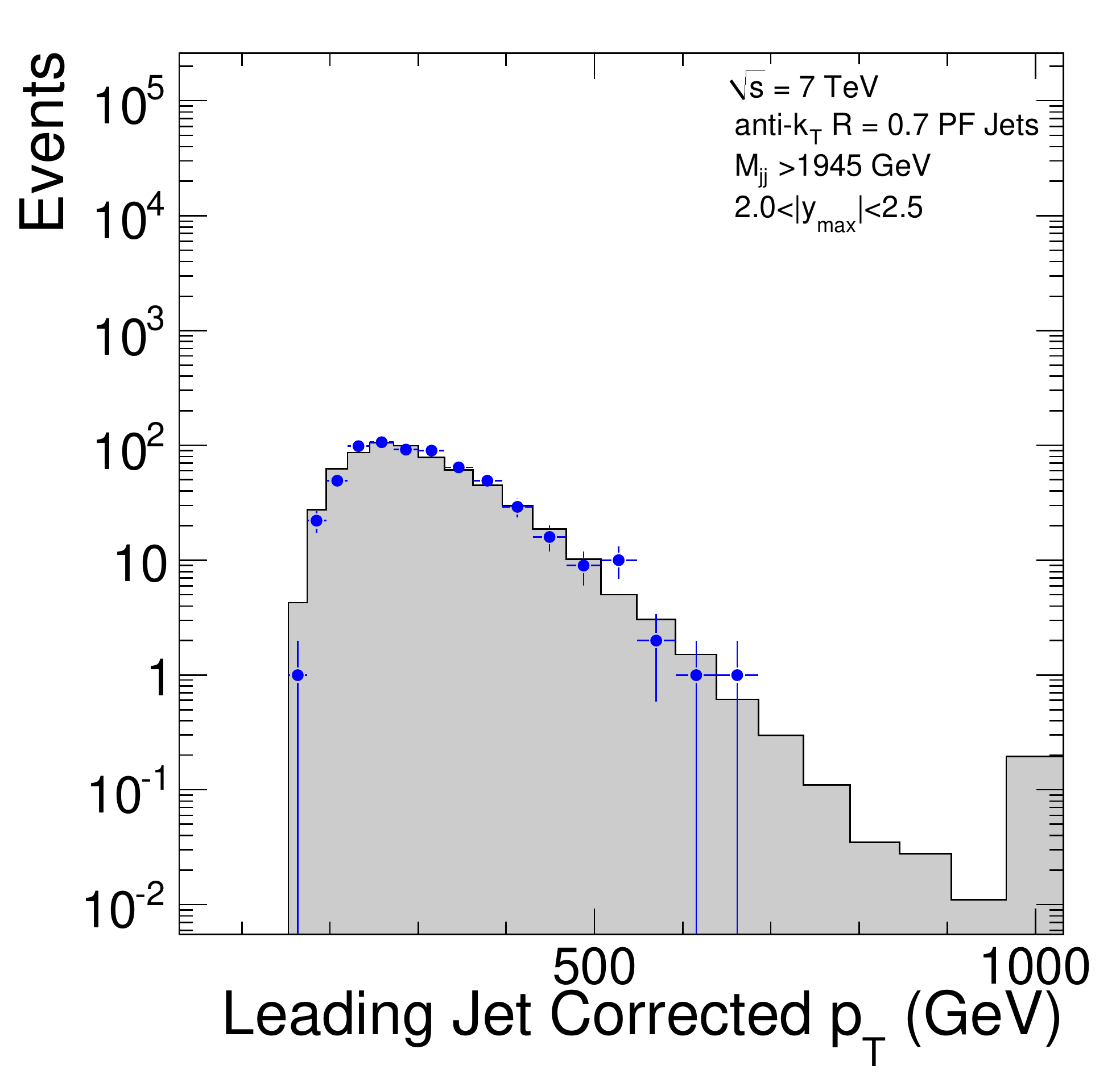}

\caption{ The $p_T$f of the leading jet  for the five different $y_{max}$ bins and for the
HLT$_{-}$Jet140U trigger, for data (points) and simulated (dashed histogram) events.}
\label{fig_appc22}
\end{figure}

\begin{figure}[ht]
\centering

\includegraphics[width=0.48\textwidth]{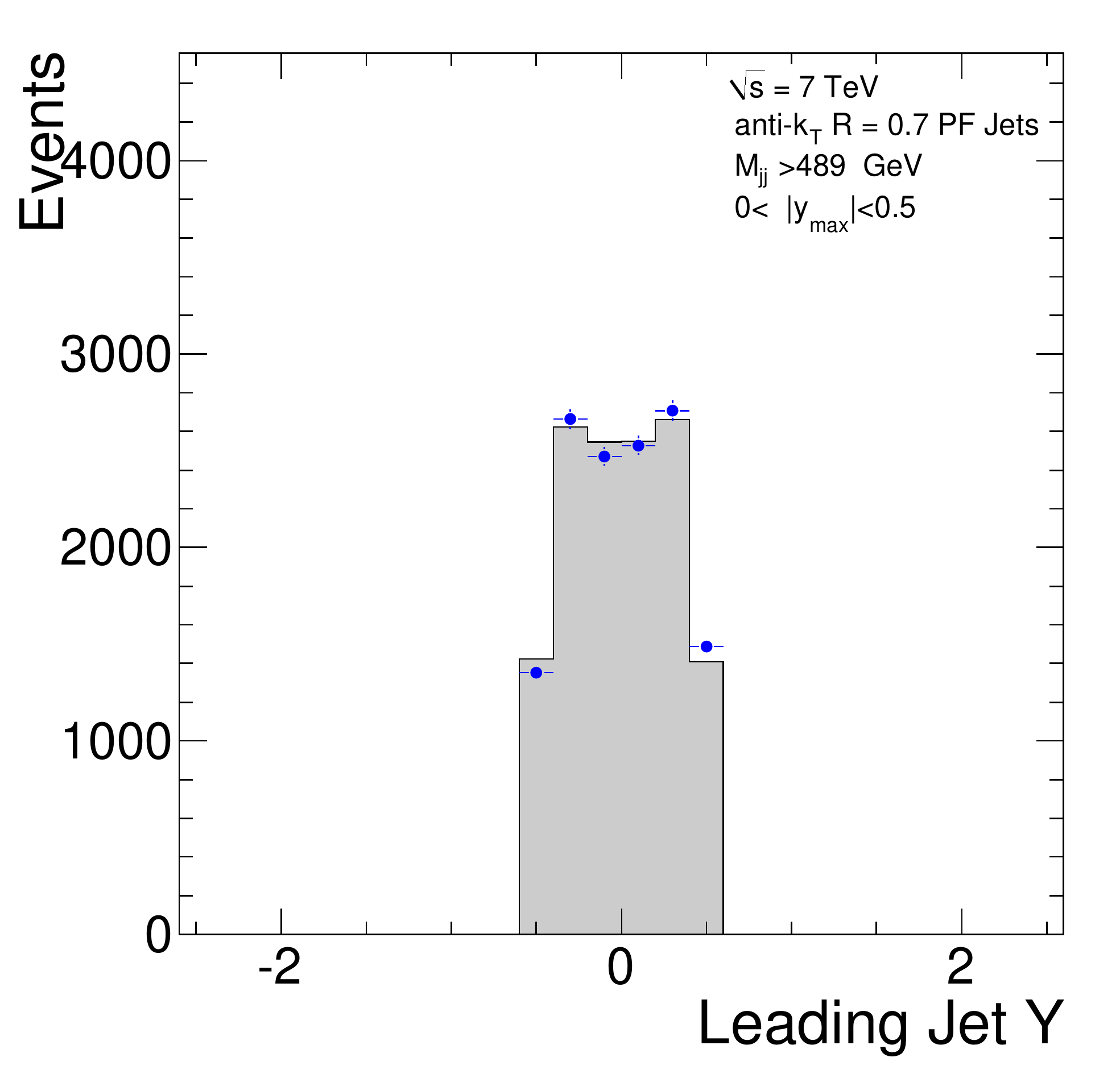} 
\includegraphics[width=0.48\textwidth]{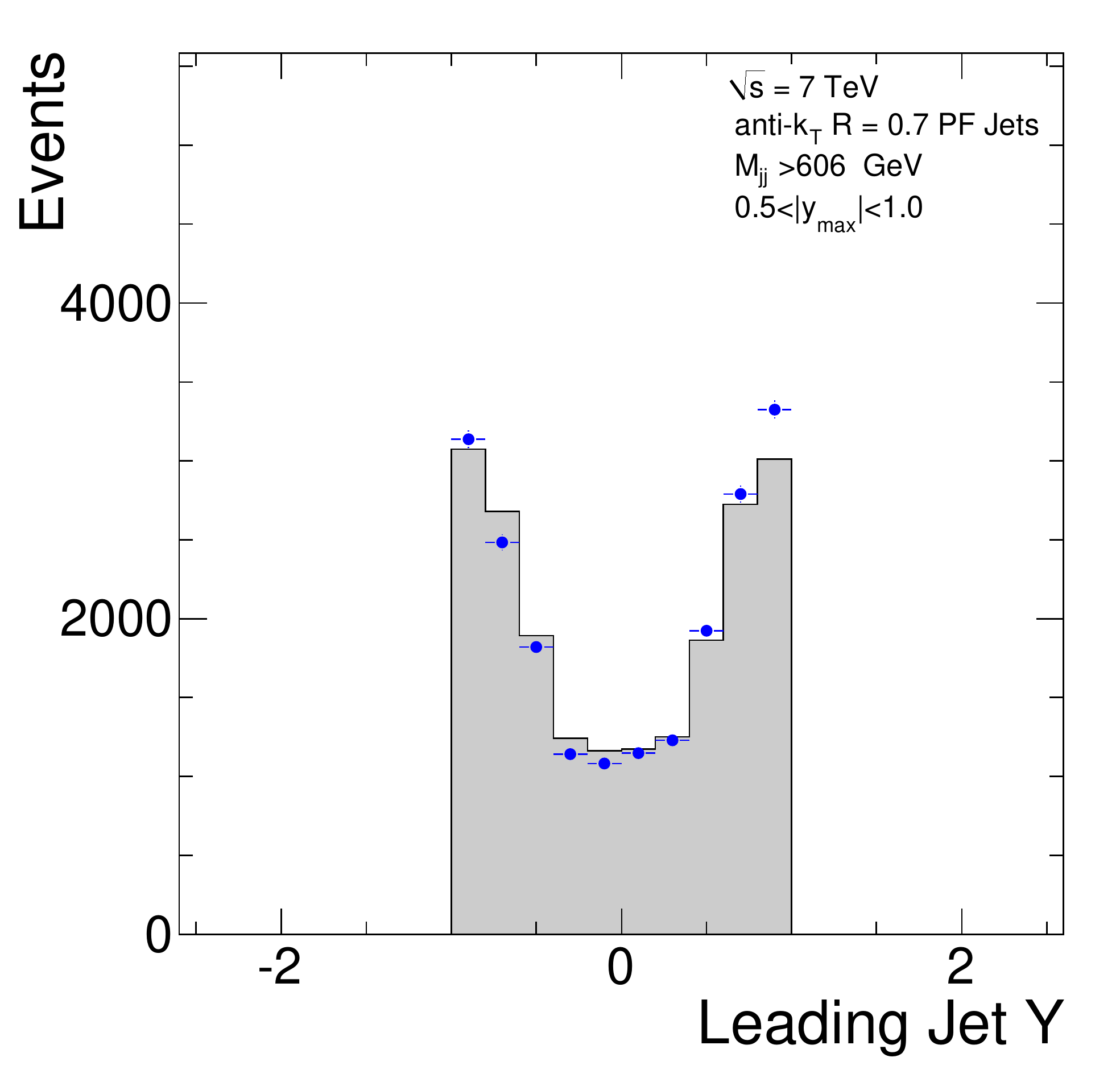} 
\includegraphics[width=0.48\textwidth]{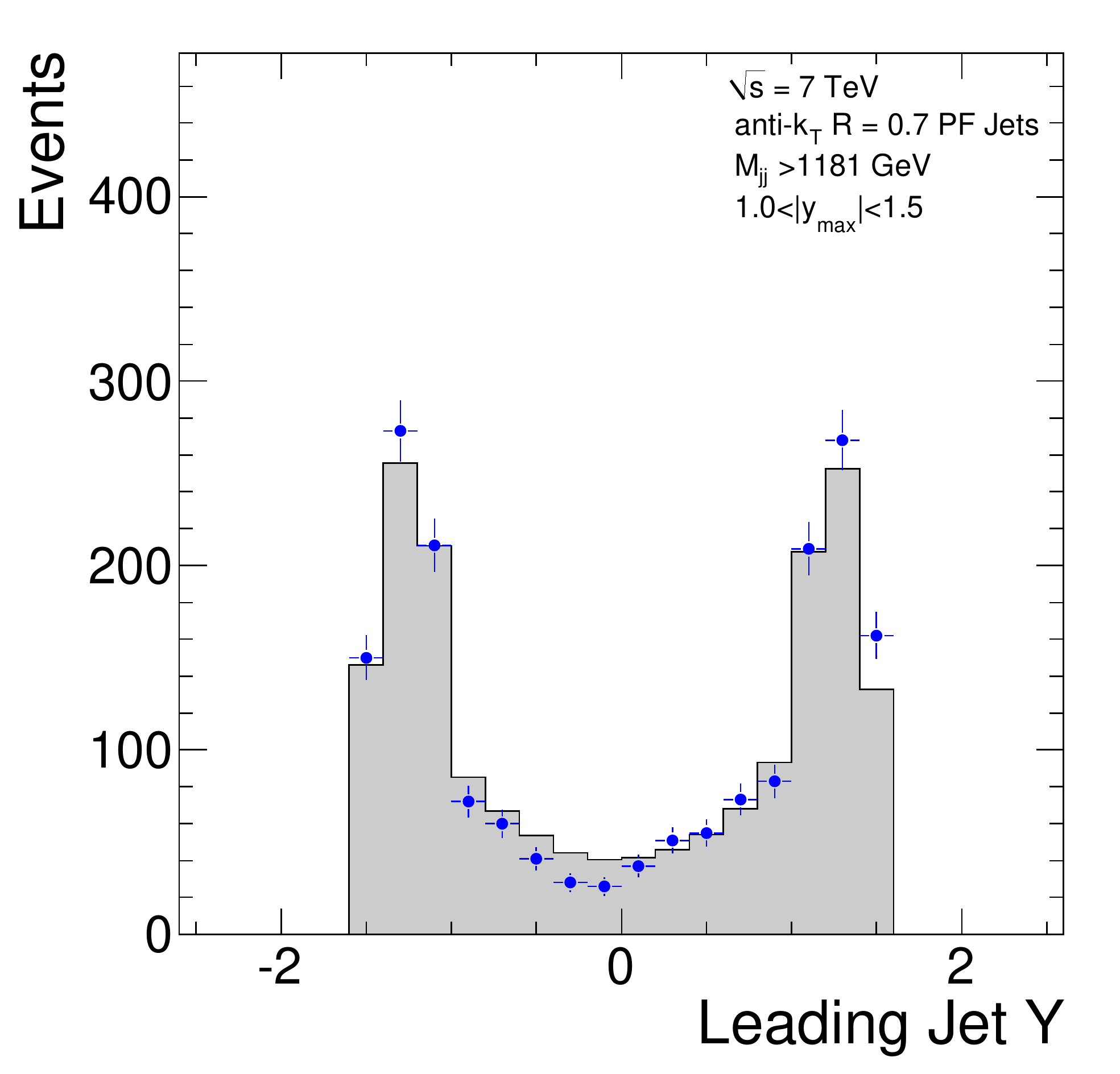} 
\includegraphics[width=0.48\textwidth]{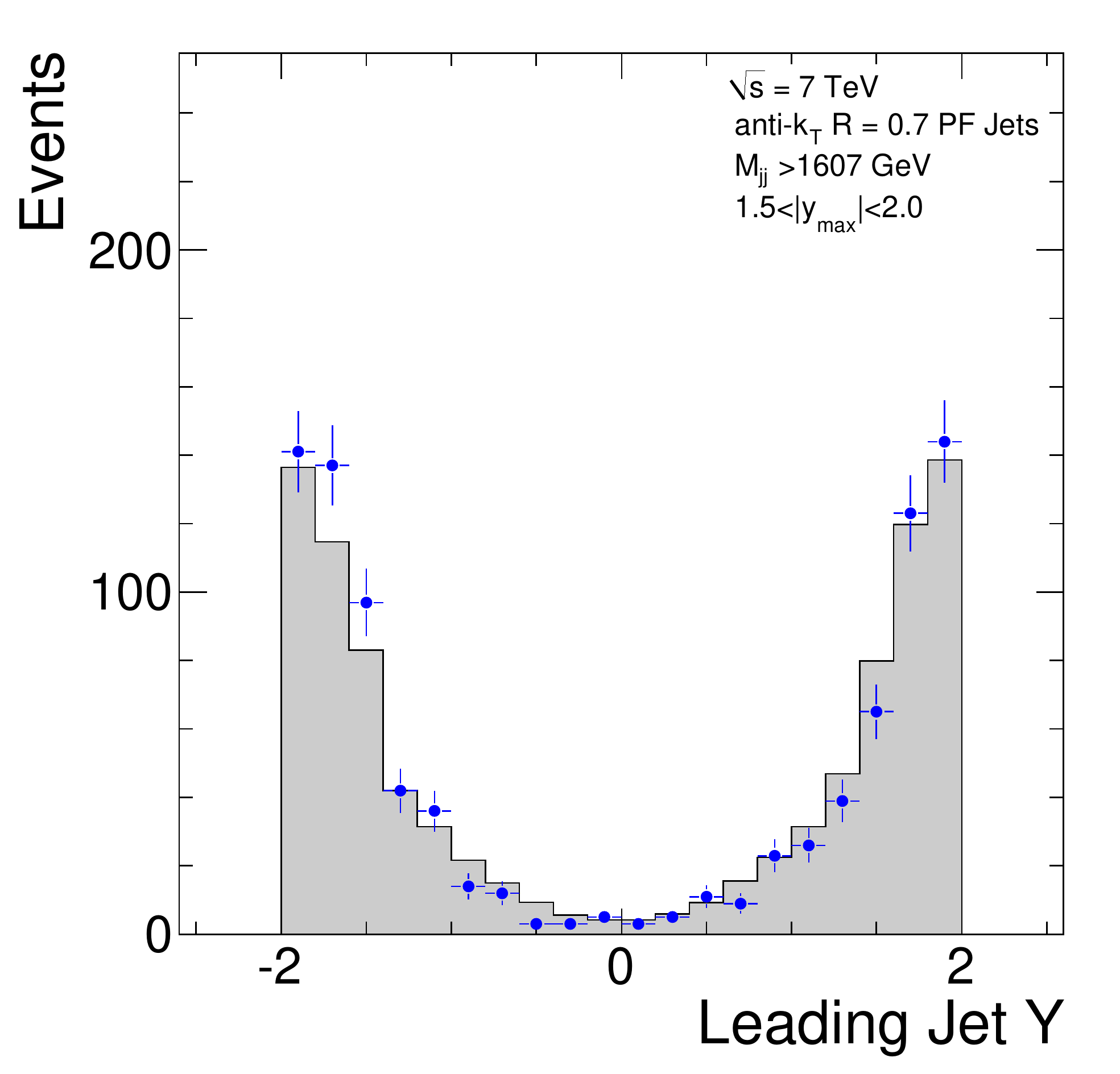} 
\includegraphics[width=0.48\textwidth]{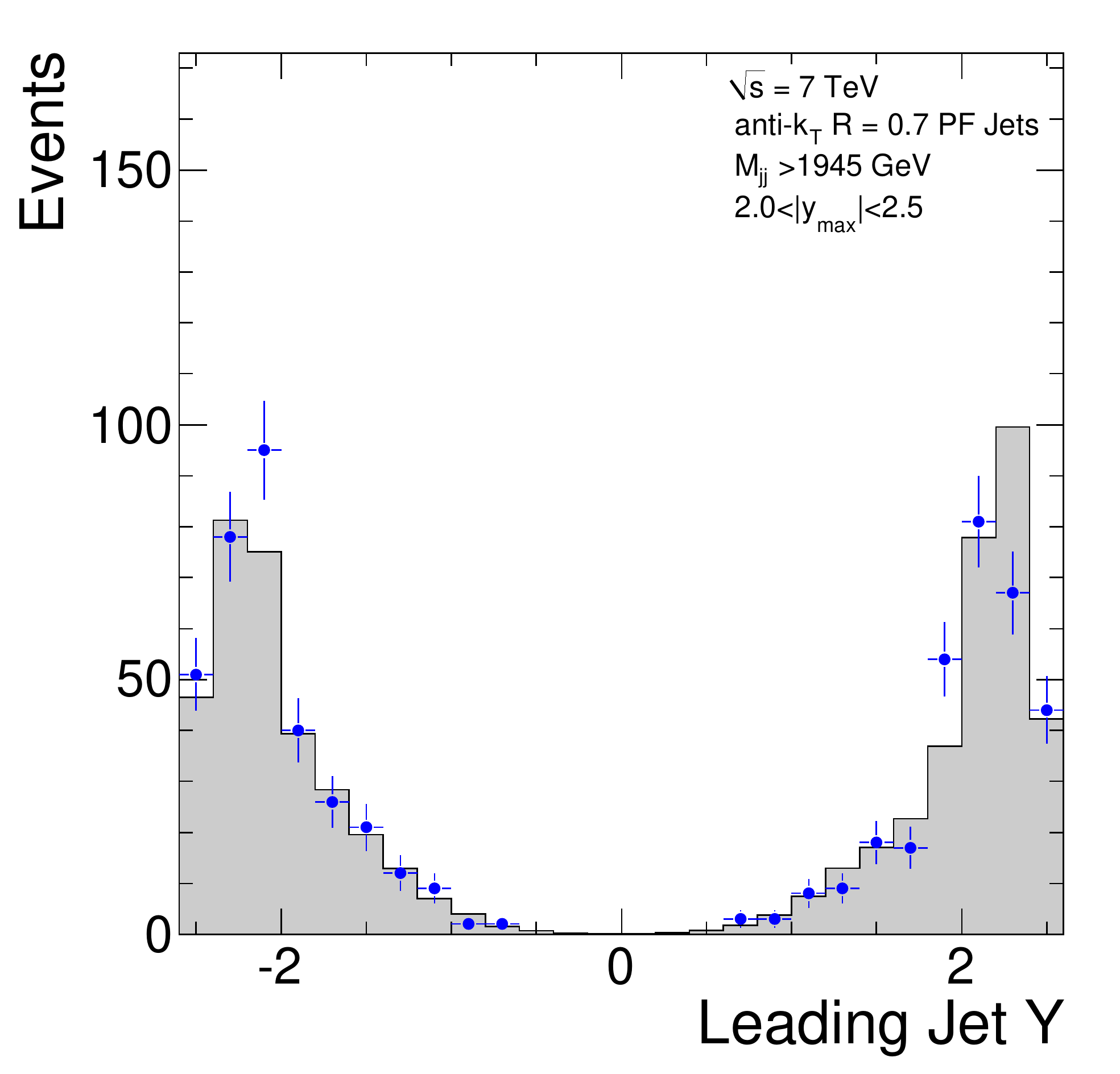}

\caption{ The $\eta$  of the leading jet  for the five different $y_{max}$ bins and for the
HLT$_{-}$Jet140U trigger, for data (points) and simulated (dashed histogram) events.}
\label{fig_appc23}
\end{figure}

\begin{figure}[ht]
\centering

\includegraphics[width=0.48\textwidth]{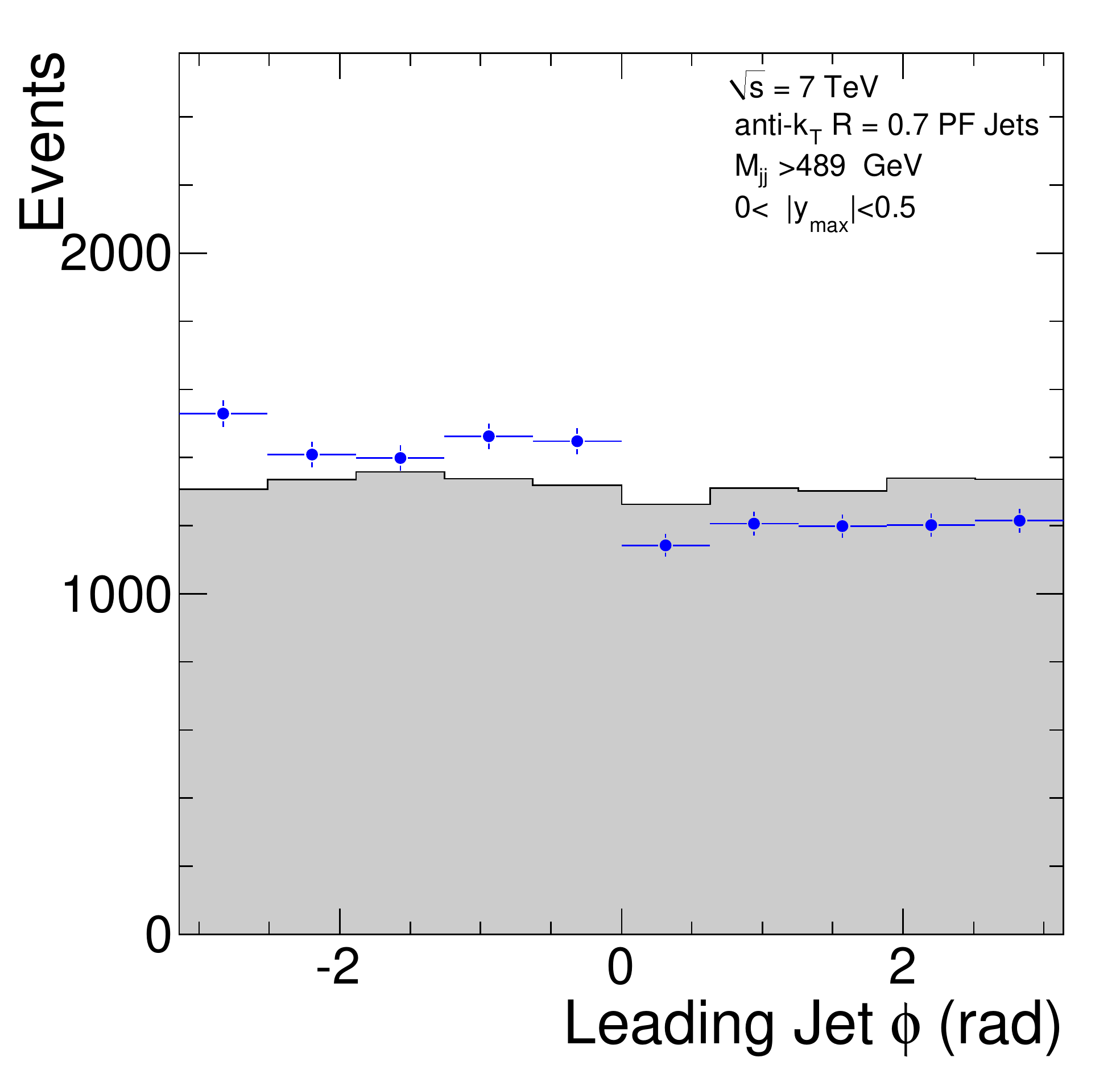} 
\includegraphics[width=0.48\textwidth]{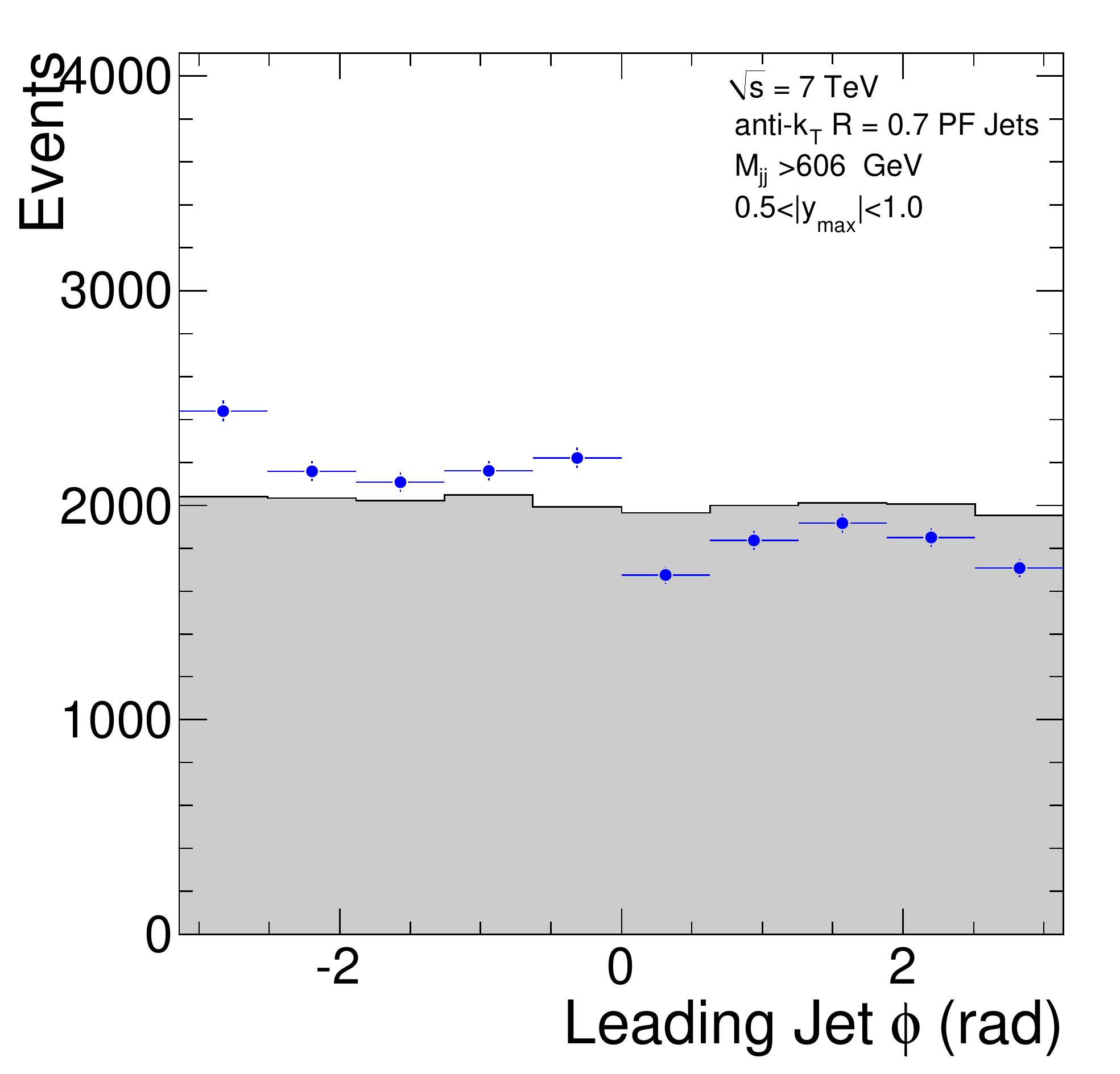} 
\includegraphics[width=0.48\textwidth]{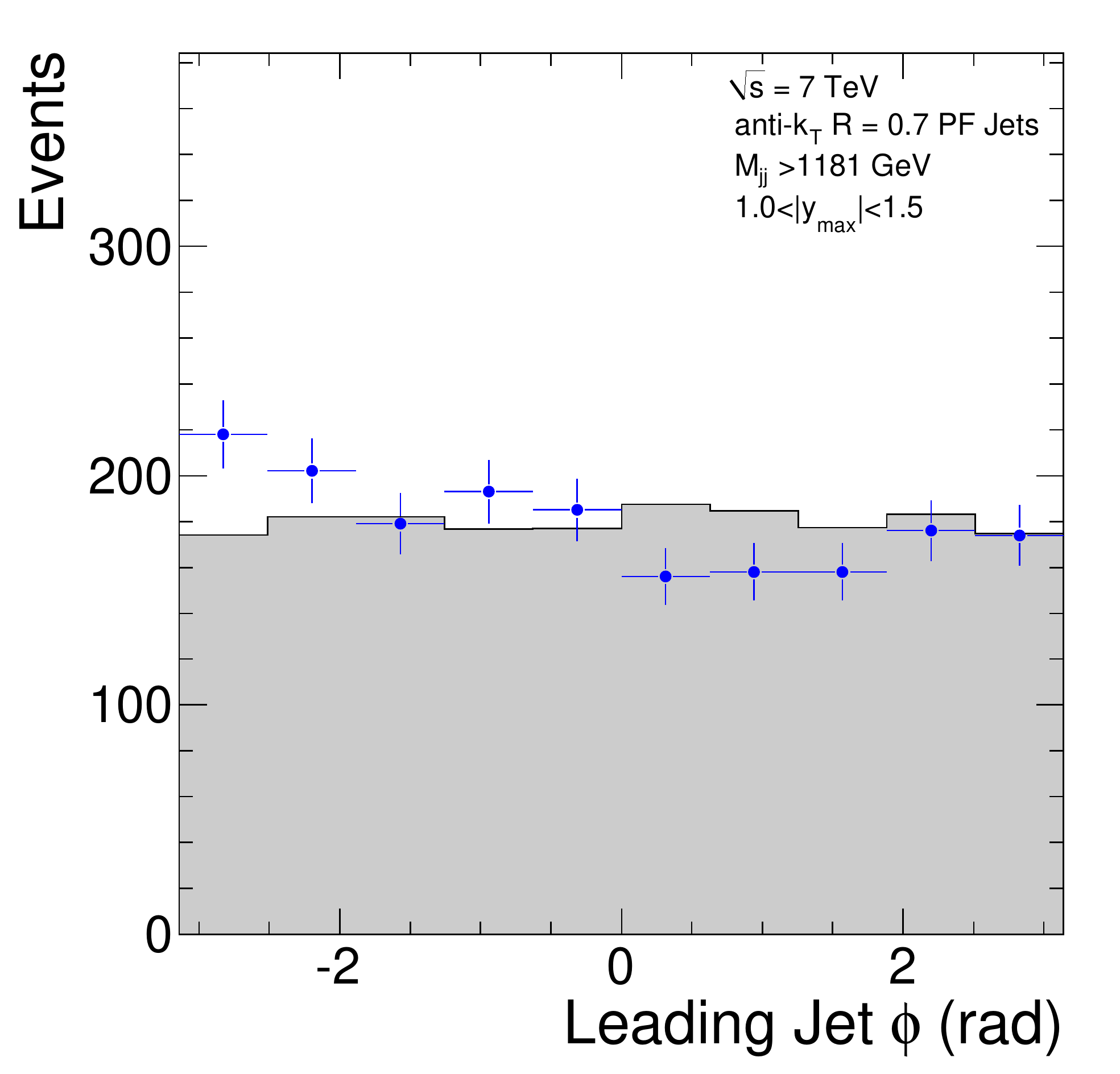} 
\includegraphics[width=0.48\textwidth]{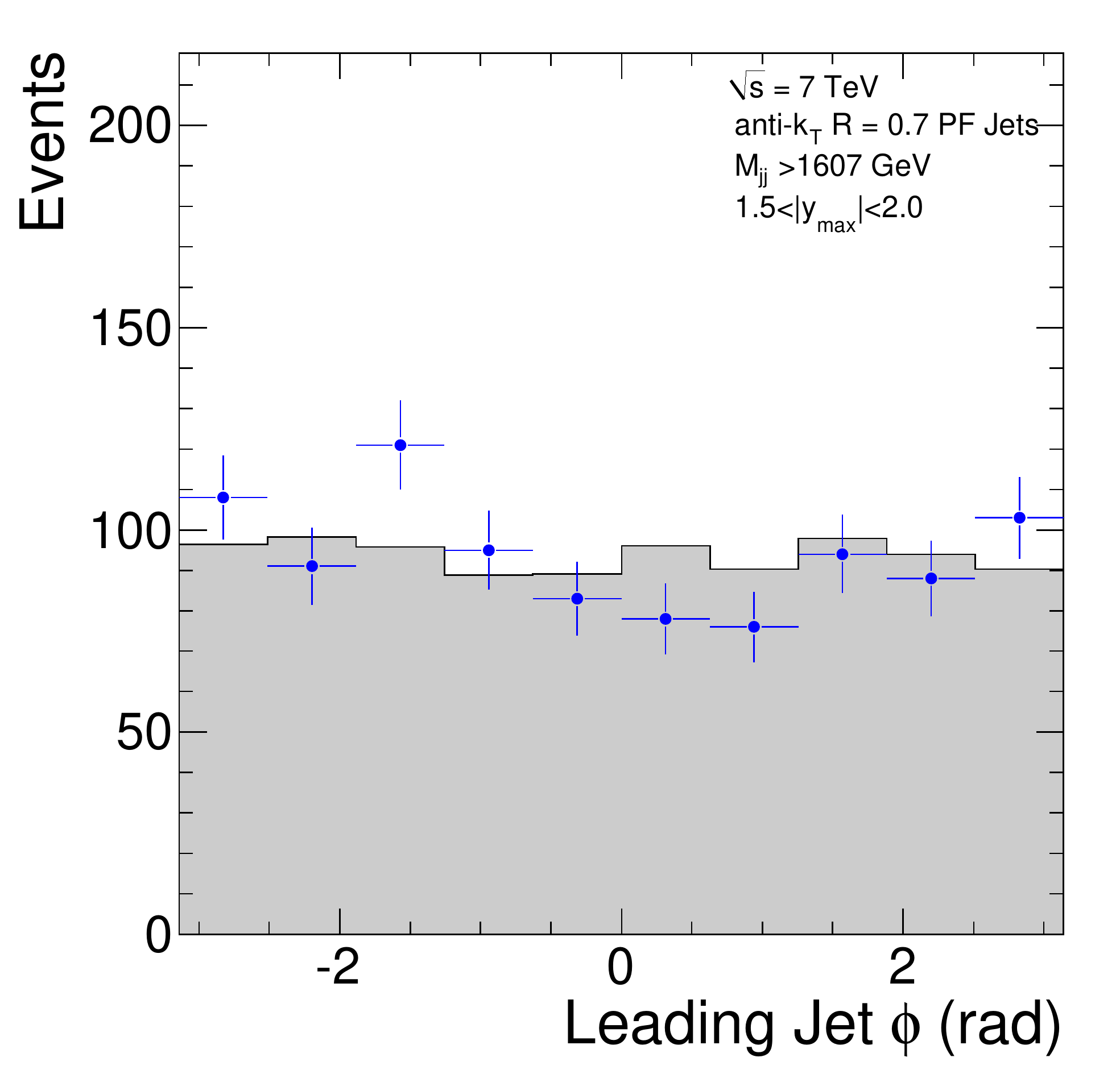} 
\includegraphics[width=0.48\textwidth]{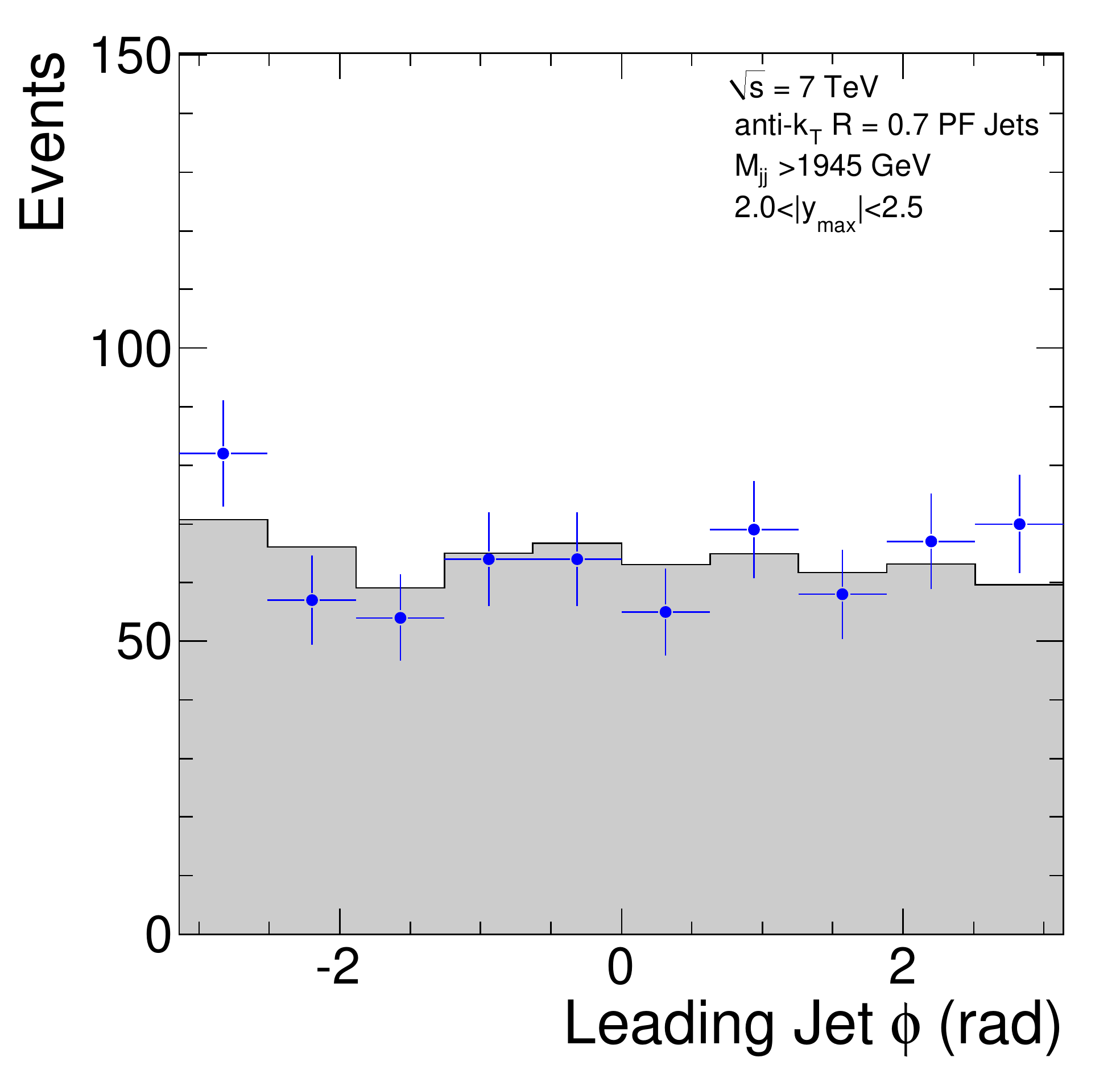}

\caption{ The $\phi$ of the leading jet  for the five different $y_{max}$ bins and for the
HLT$_{-}$Jet140U trigger, for data (points) and simulated (dashed histogram) events.}
\label{fig_appc24}
\end{figure}

\clearpage
\section{Appendix D}

\begin{figure}[h]
\centering

\includegraphics[width=0.49\textwidth]{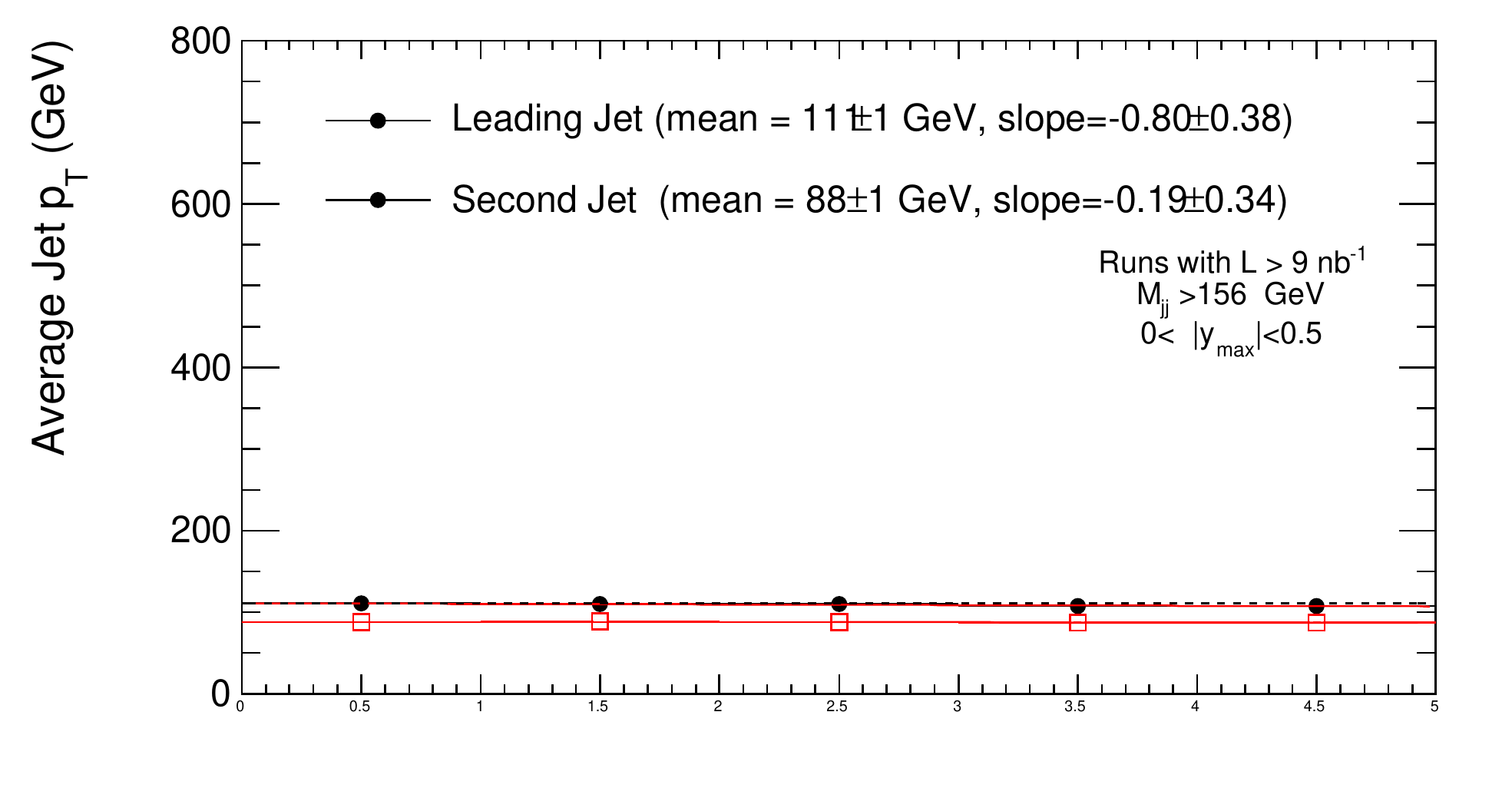} 
\includegraphics[width=0.49\textwidth]{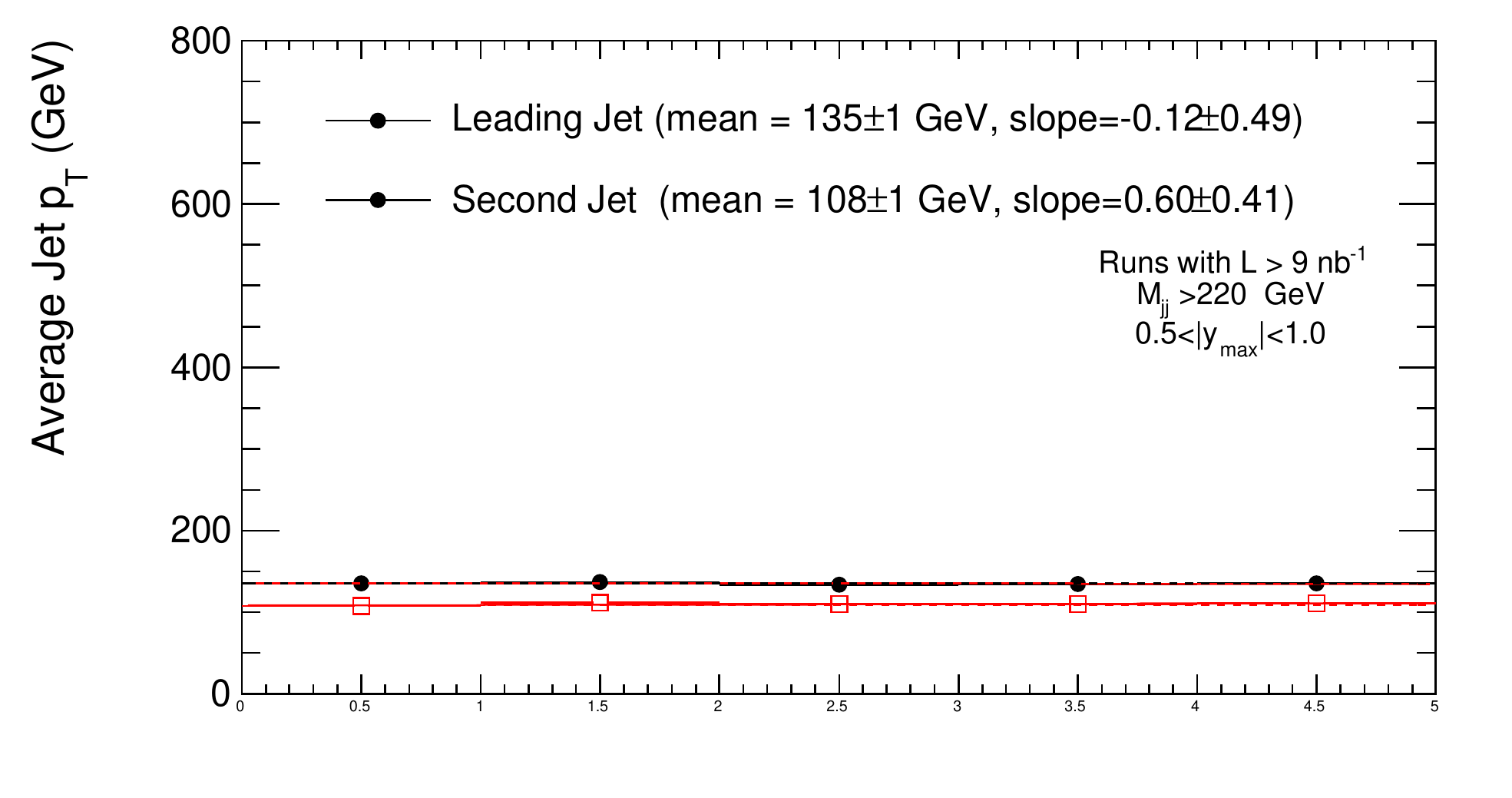}  
\includegraphics[width=0.49\textwidth]{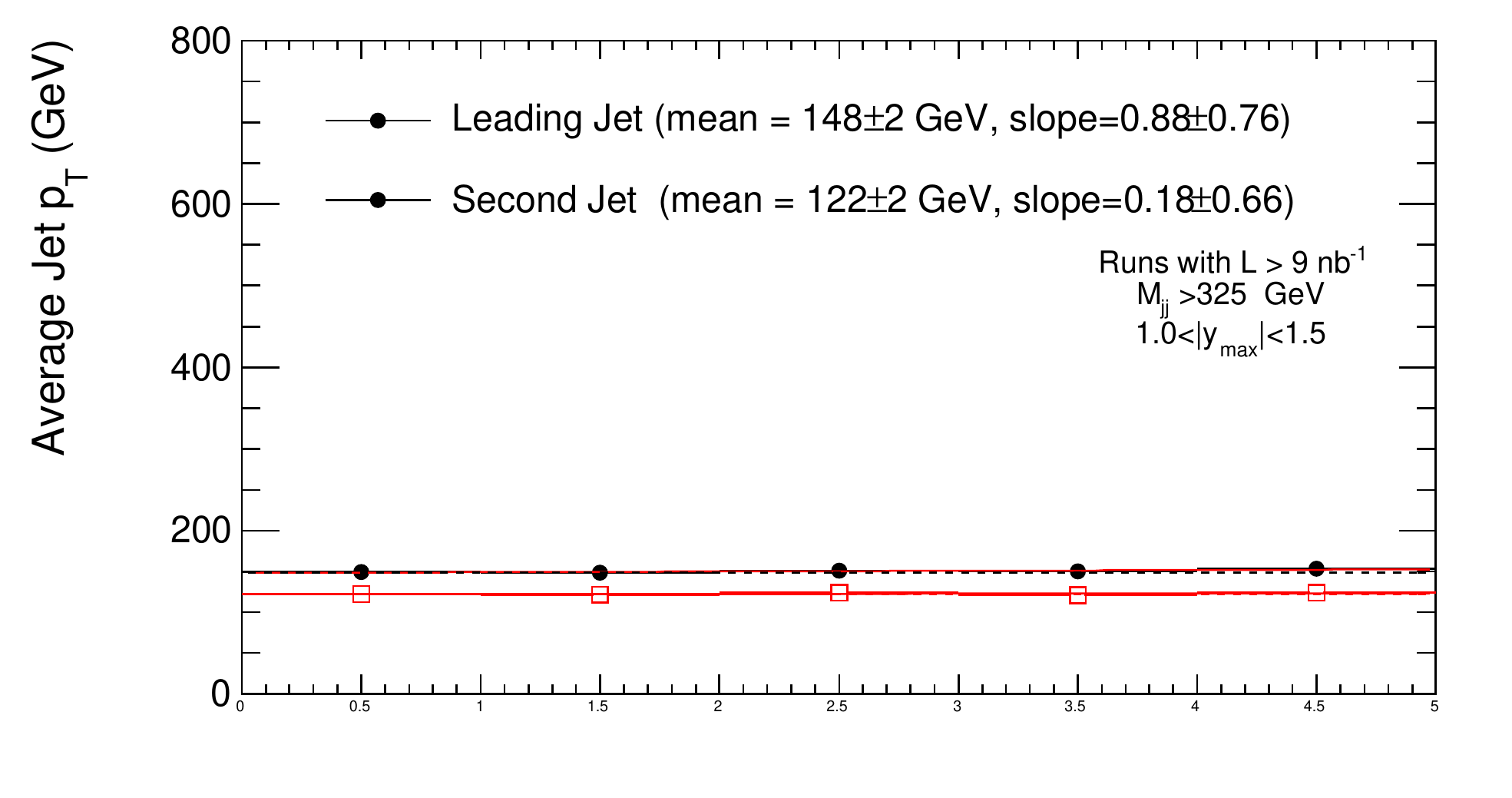}  
\includegraphics[width=0.49\textwidth]{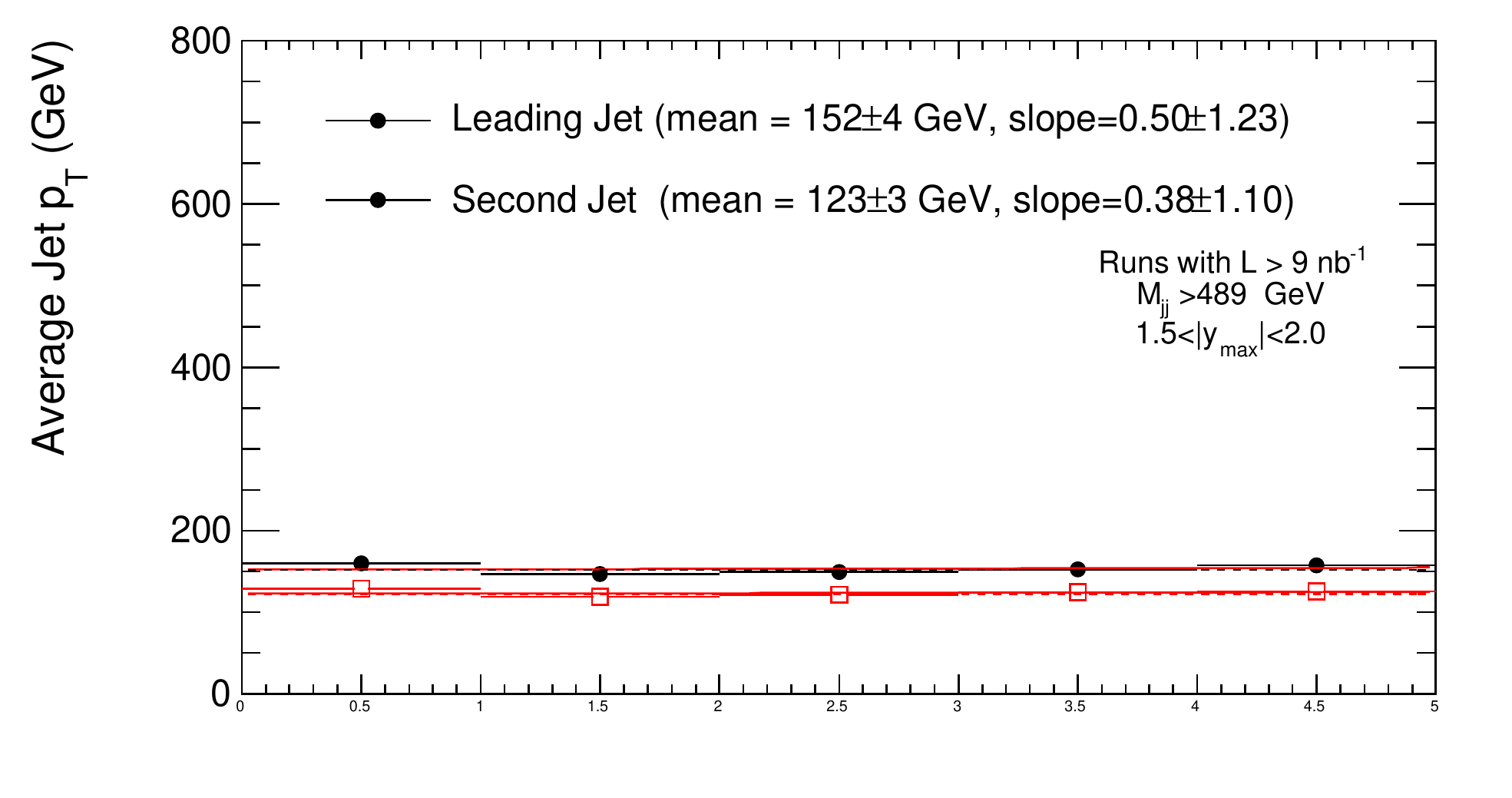} 
\includegraphics[width=0.49\textwidth]{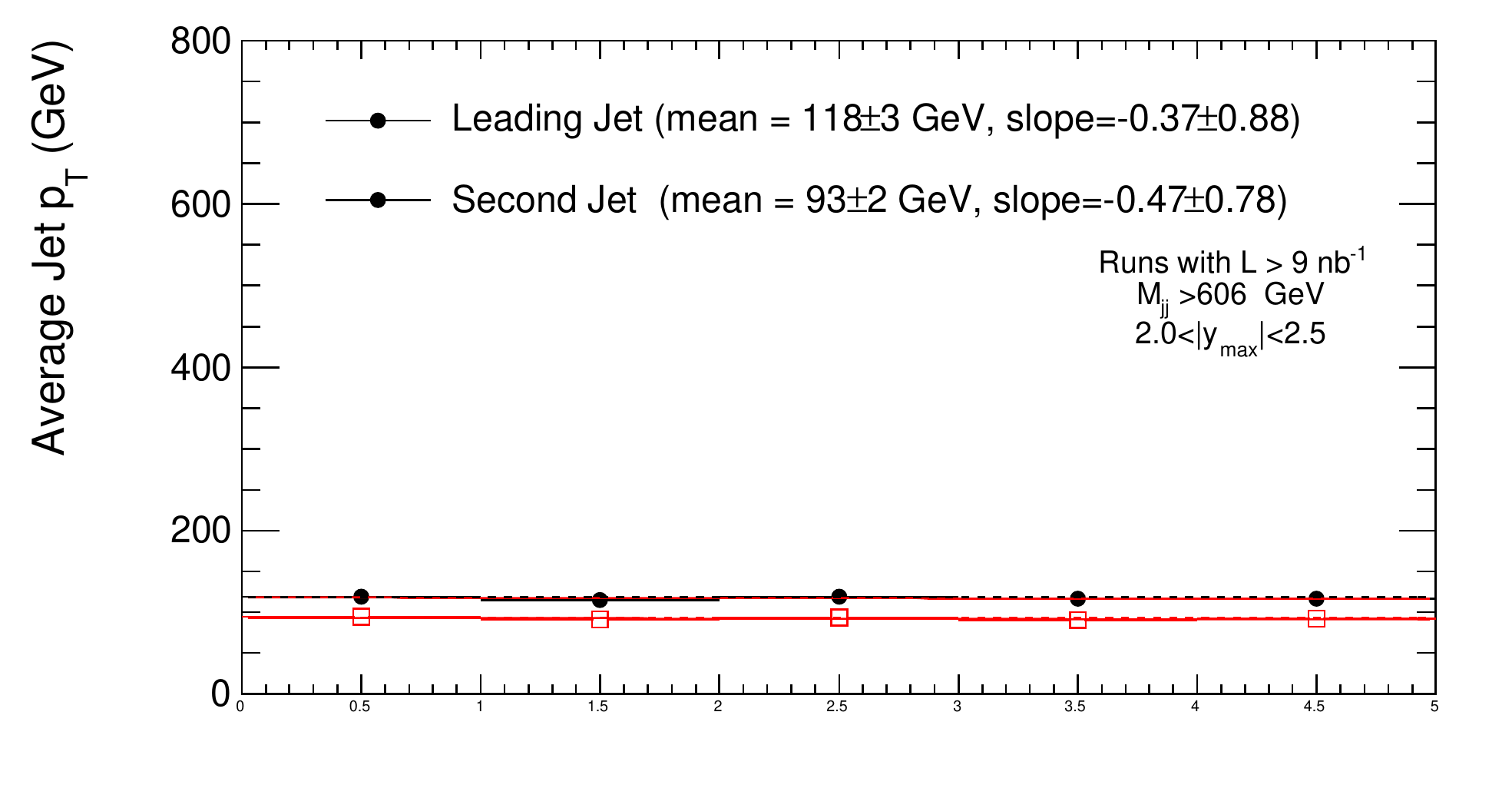} 
   
\capspace
\caption{ The $p_T$ of the leading and second jet  for the five different $y_{max}$ bins and for the 
HLT$_{-}$Jet30U trigger as a function of time (run number), fitted with a first degree polynomial. }
\label{fig_appd1}
\end{figure}

\begin{figure}[h]
\centering

\includegraphics[width=0.49\textwidth]{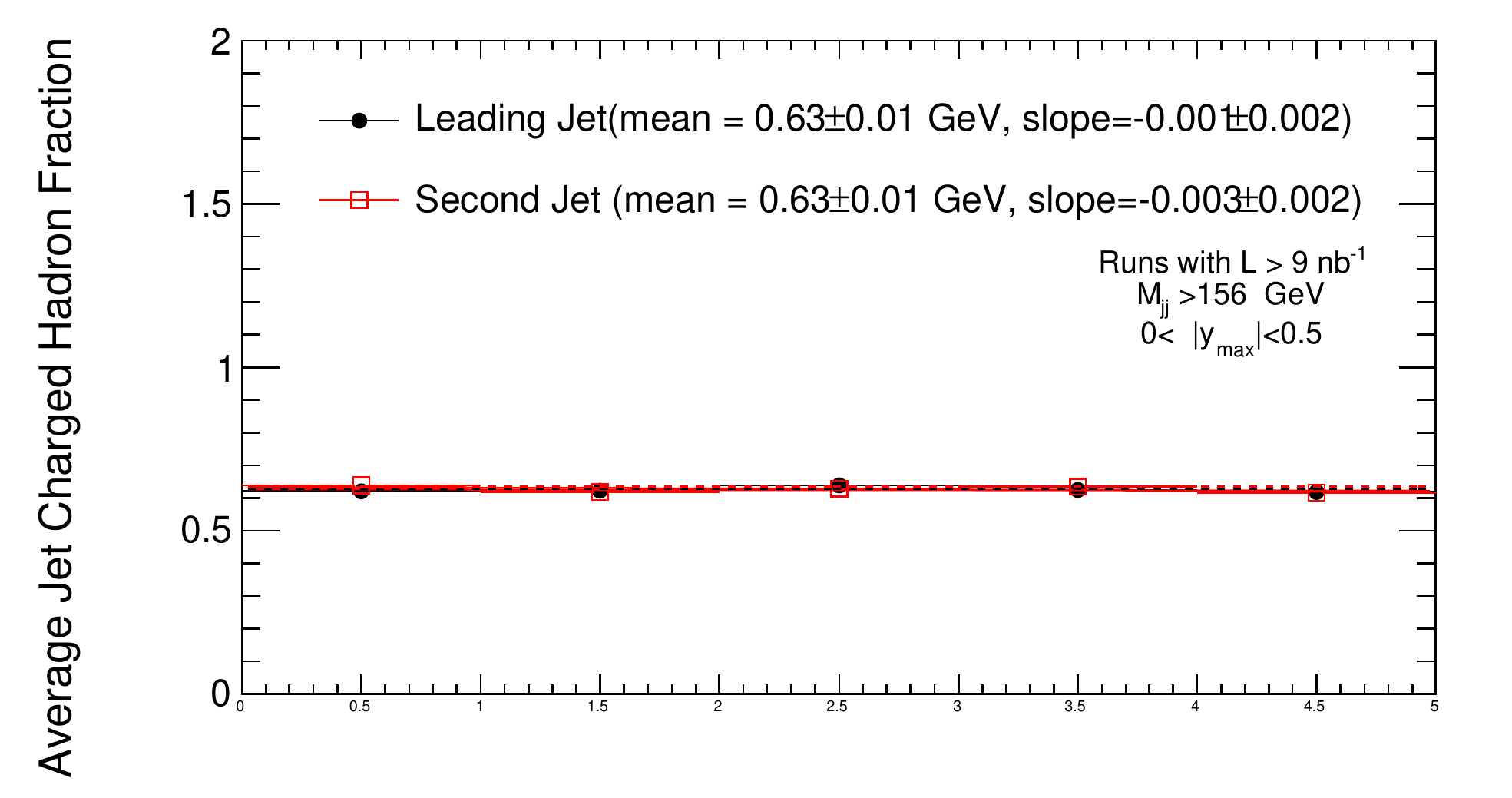} 
\includegraphics[width=0.49\textwidth]{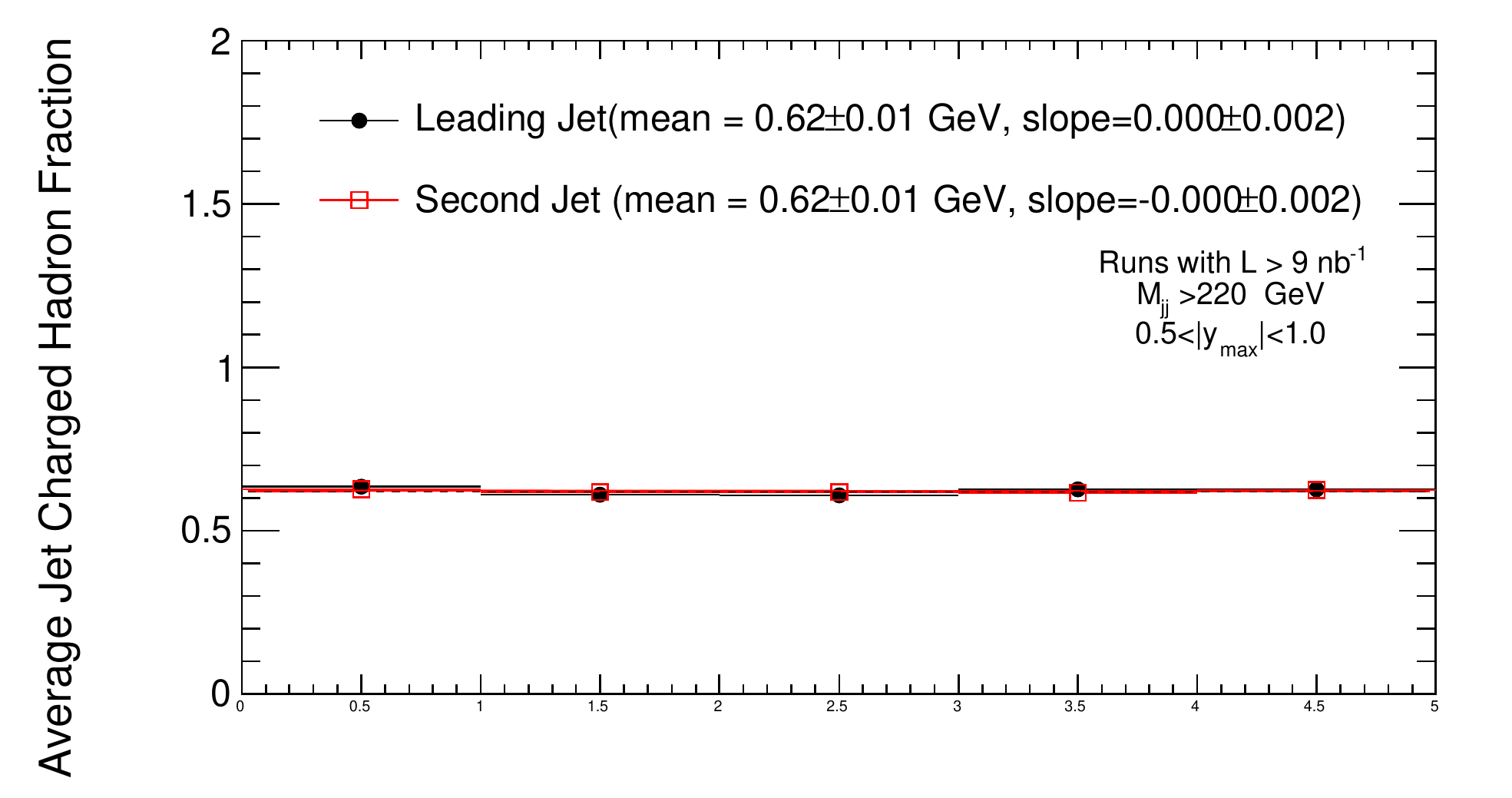}  
\includegraphics[width=0.49\textwidth]{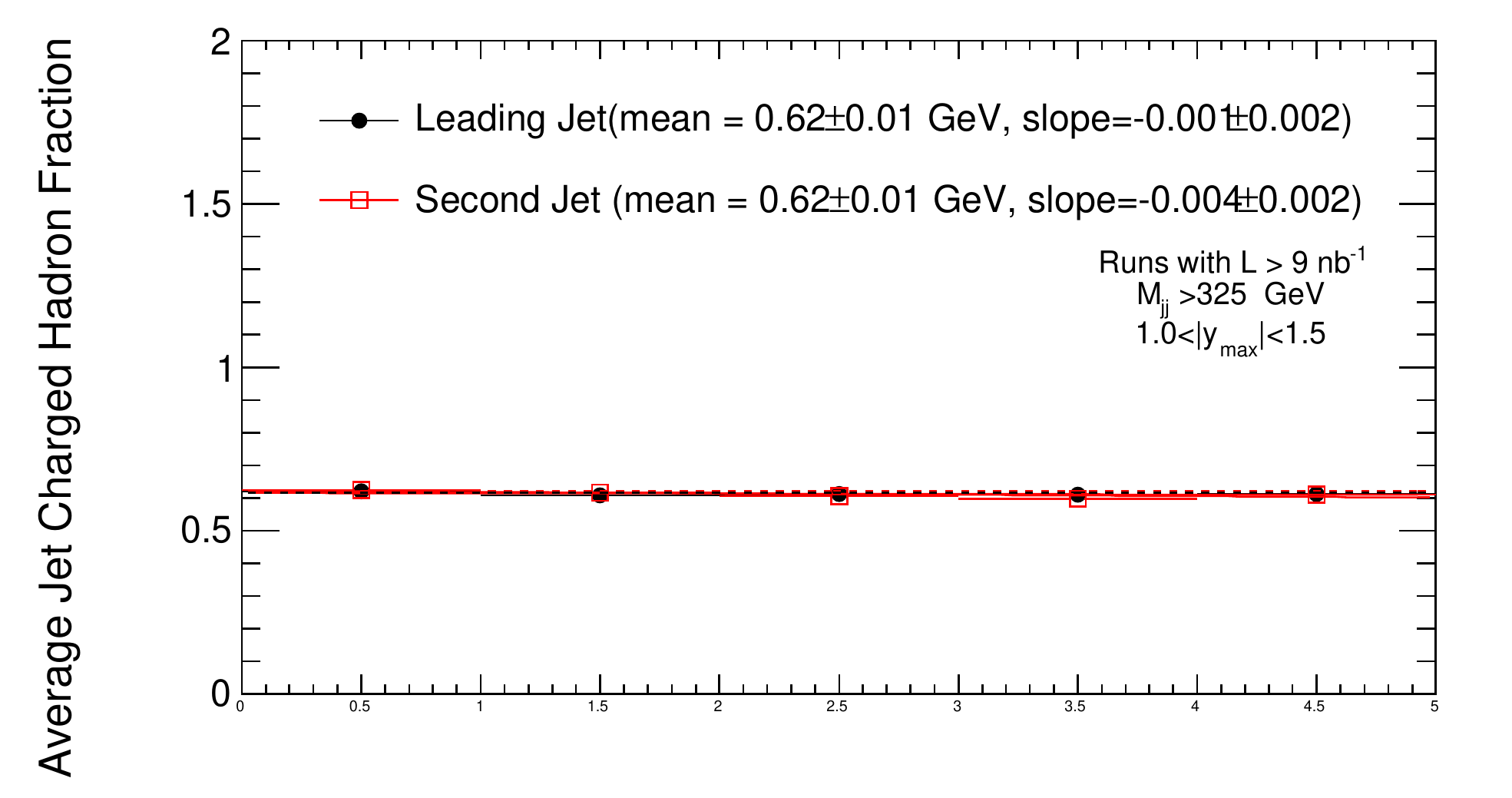}  
\includegraphics[width=0.49\textwidth]{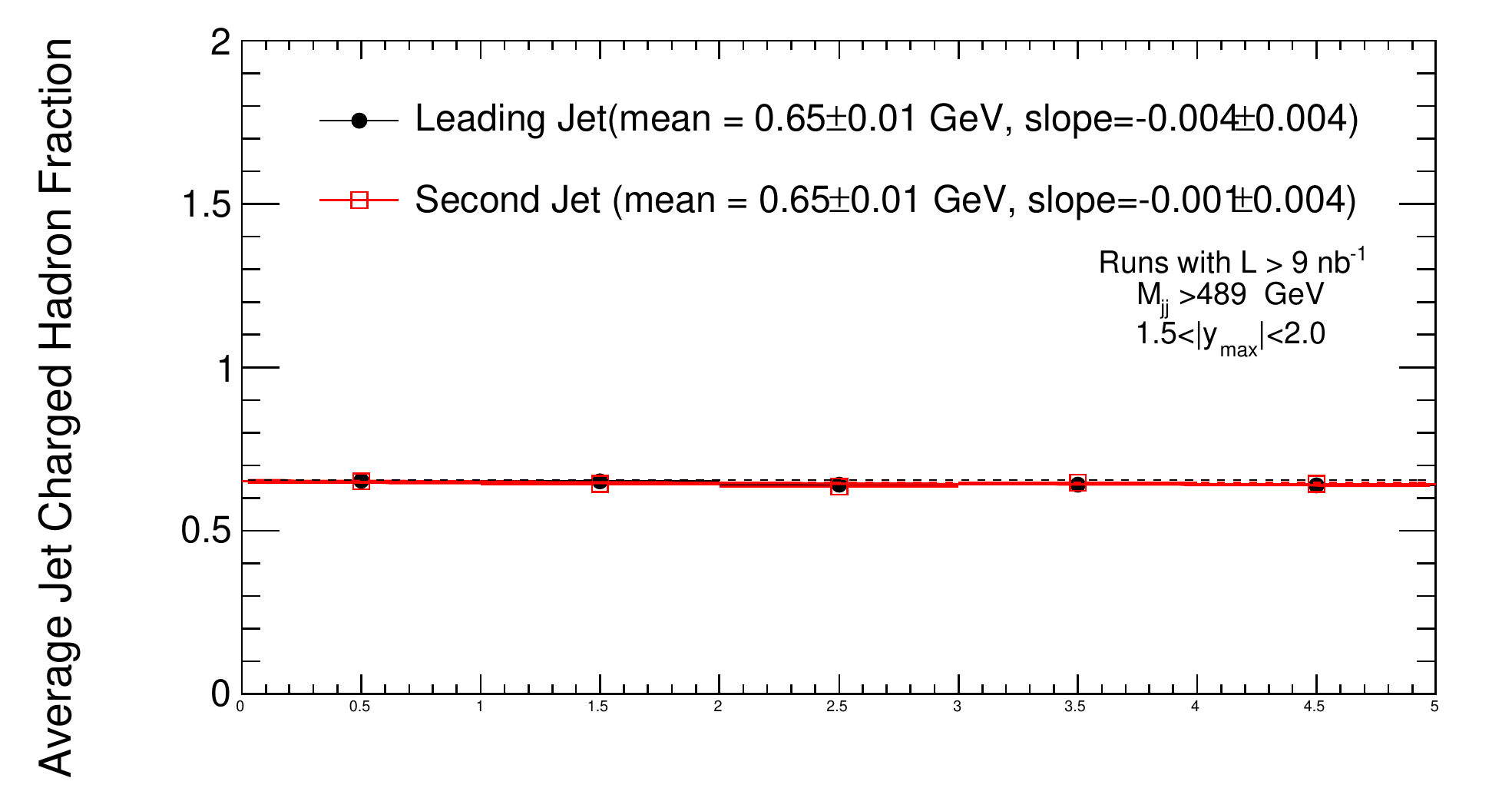} 
\includegraphics[width=0.49\textwidth]{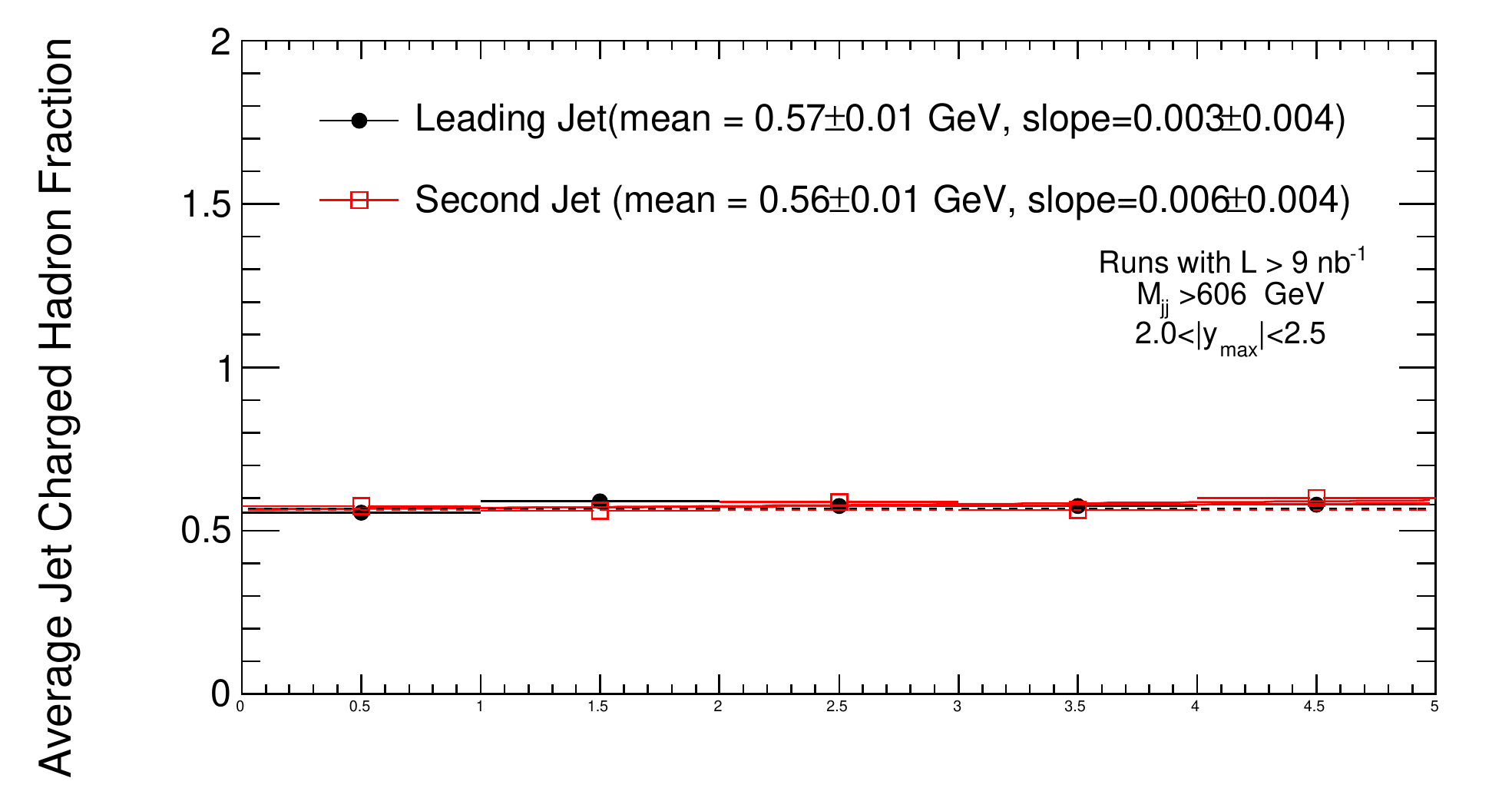} 
   
\capspace
\caption{ The charged hadron fraction  of the leading and second jet  for the five different $y_{max}$ bins and for the 
HLT$_{-}$Jet30U trigger as a function of time (run number), fitted with a first degree polynomial. }
\label{fig_appd2}
\end{figure}

\begin{figure}[h]
\centering

\includegraphics[width=0.49\textwidth]{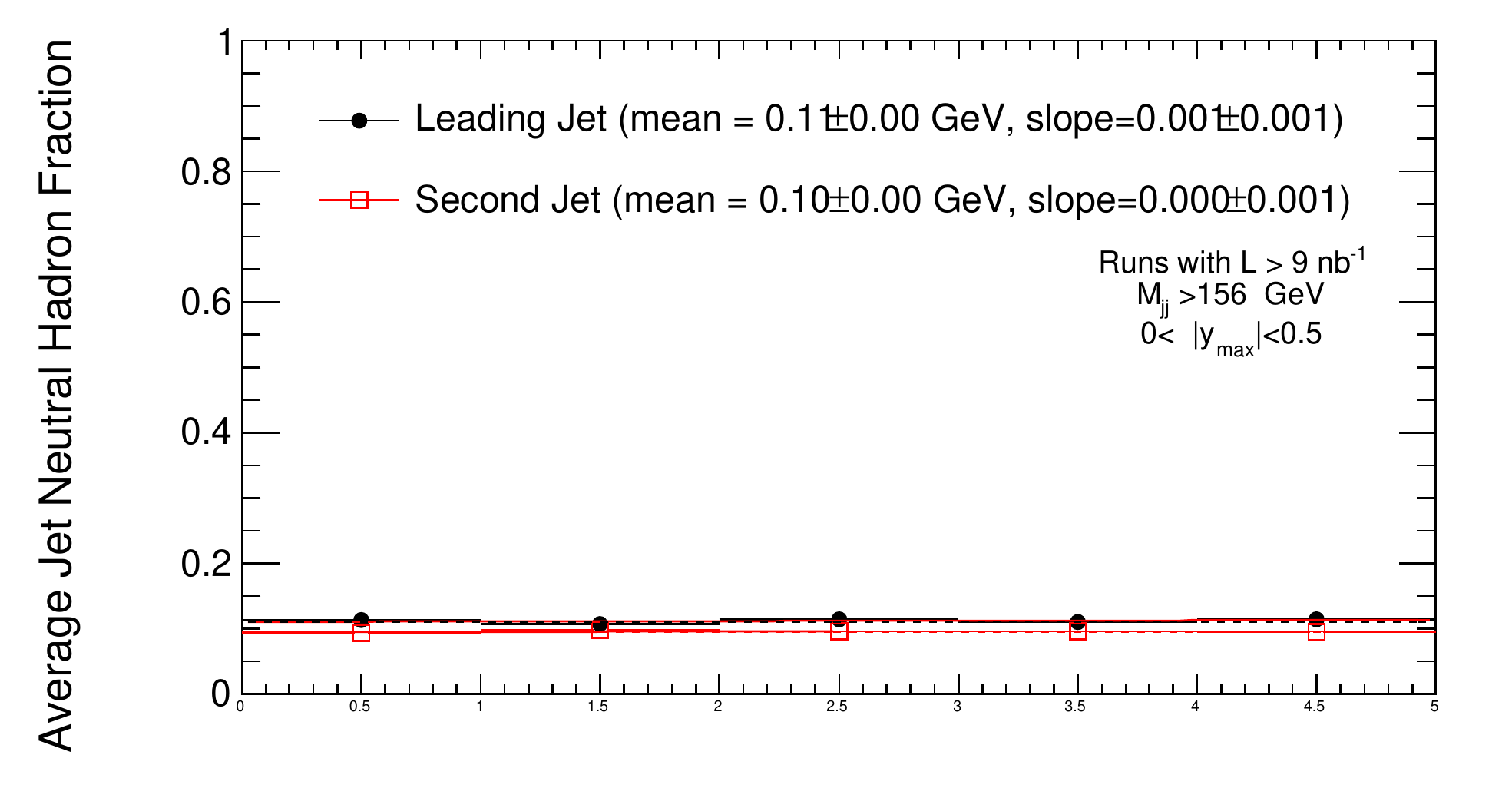} 
\includegraphics[width=0.49\textwidth]{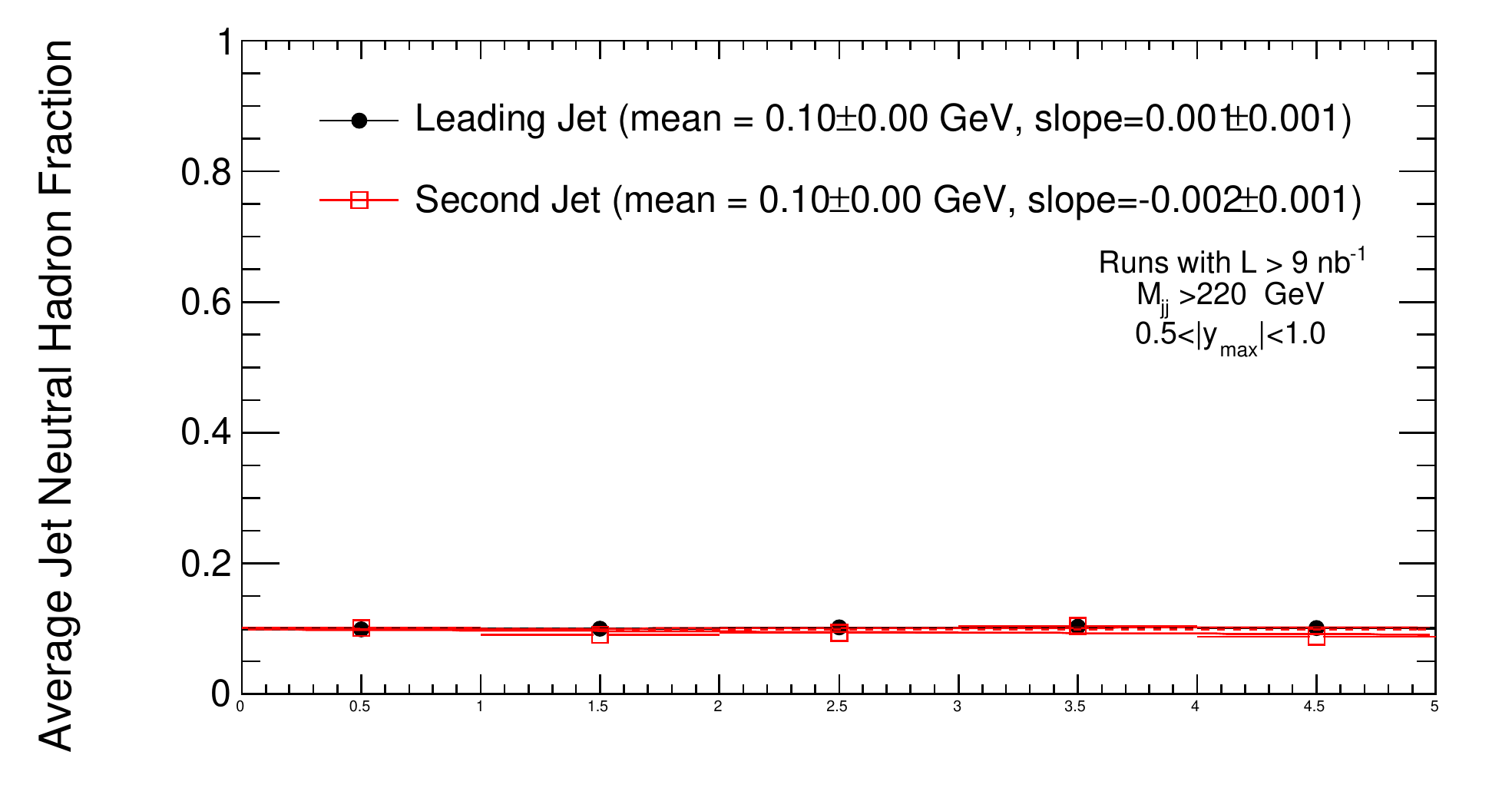}  
\includegraphics[width=0.49\textwidth]{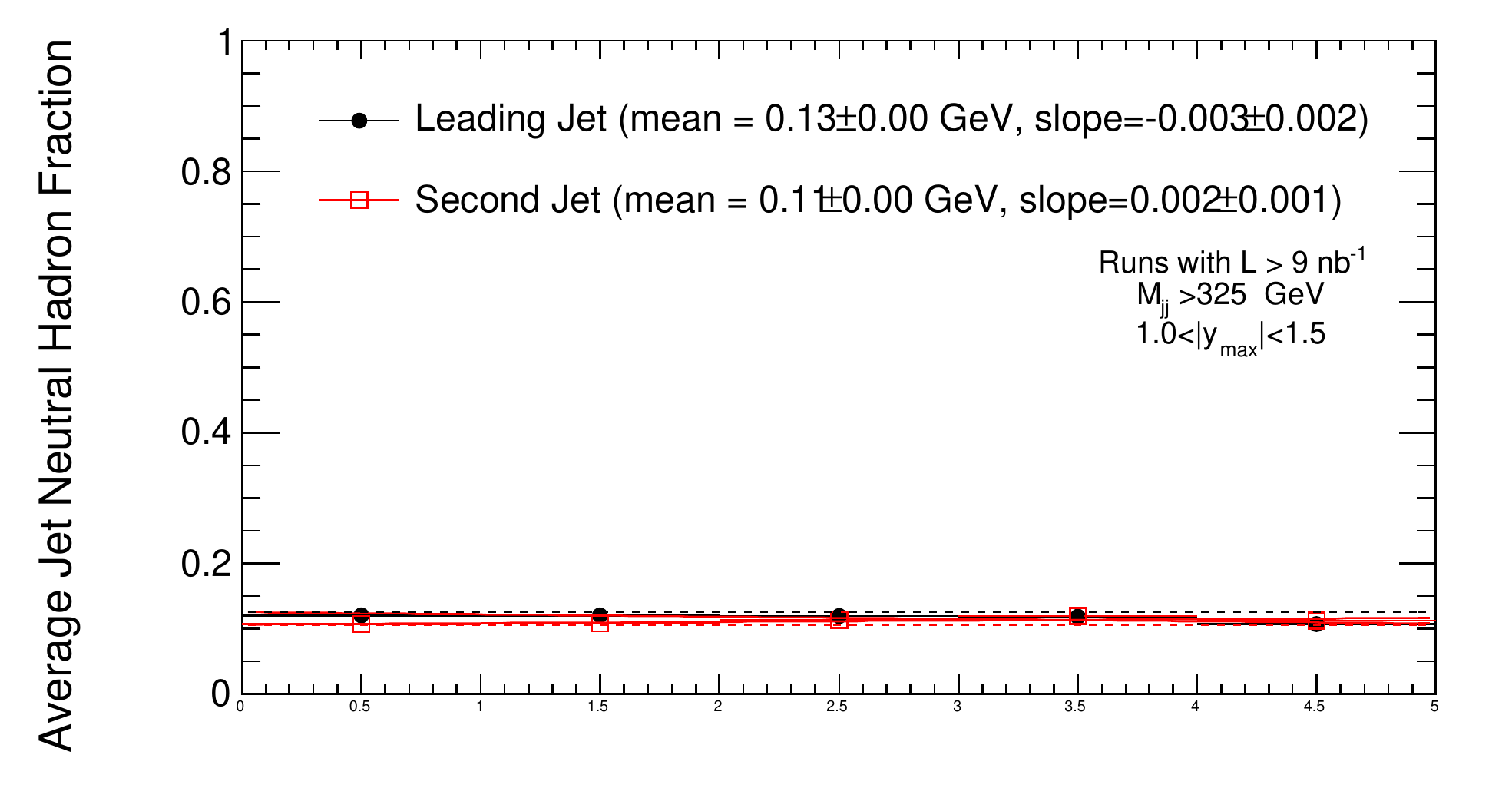}  
\includegraphics[width=0.49\textwidth]{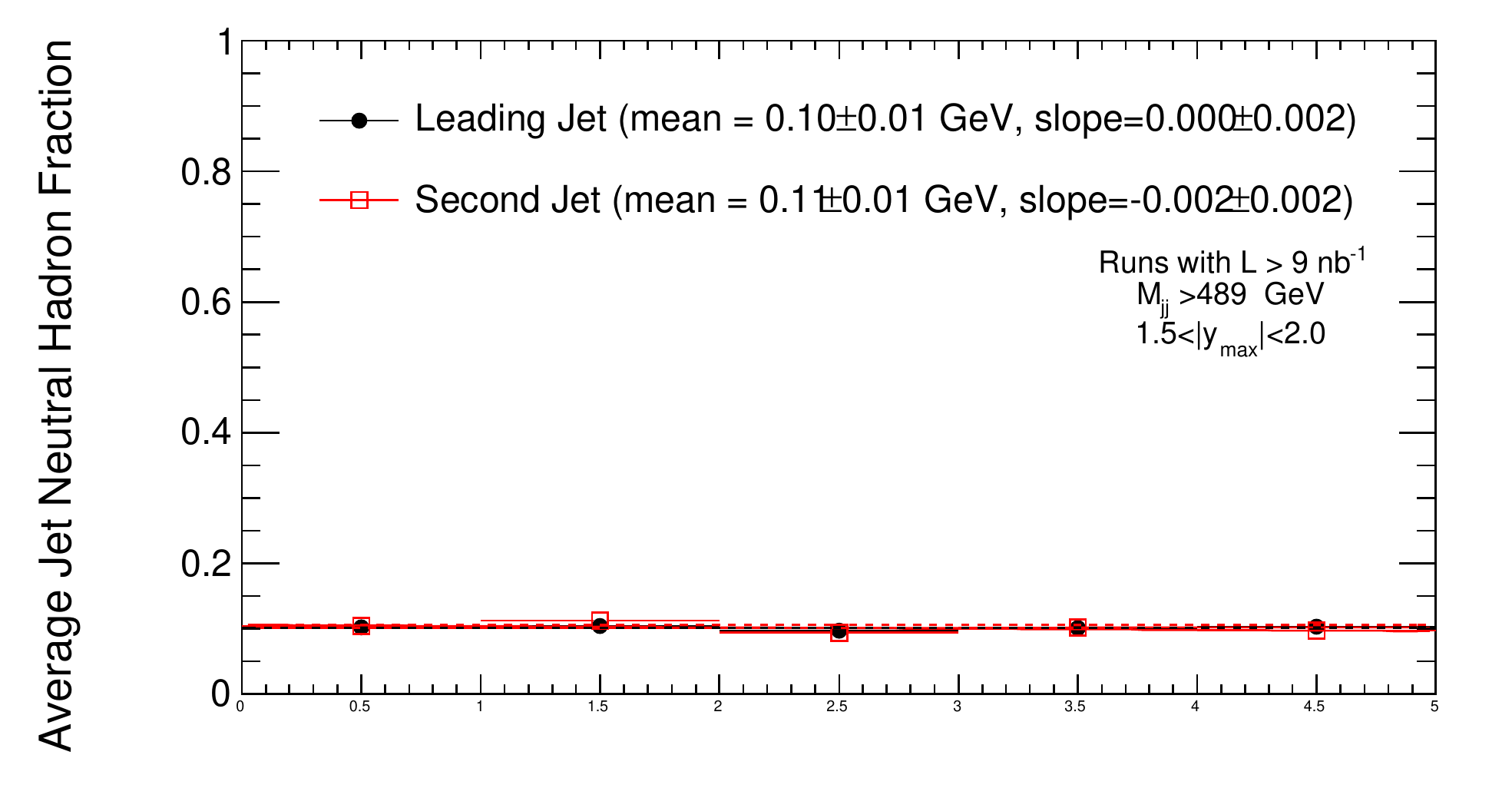} 
\includegraphics[width=0.49\textwidth]{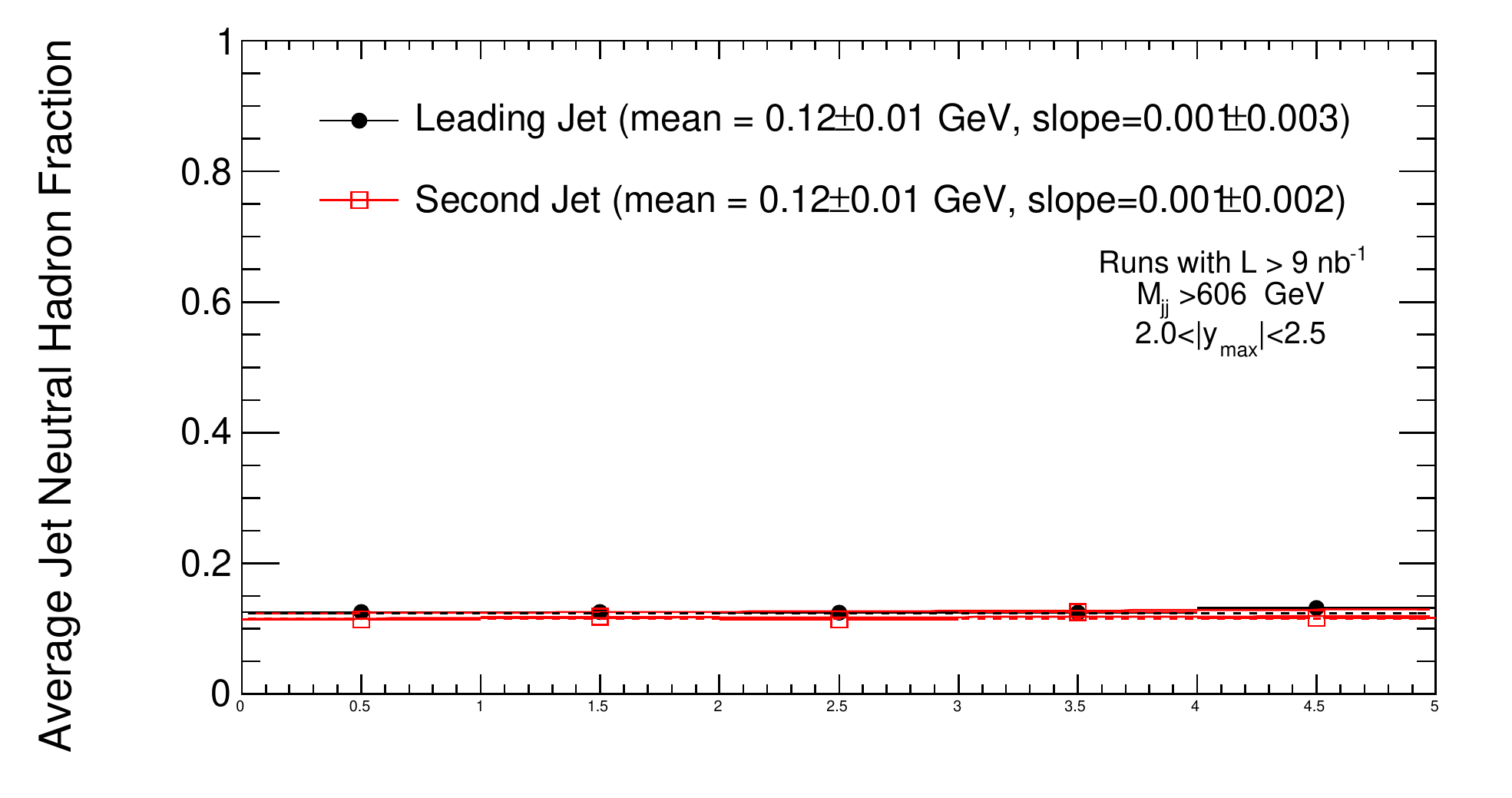} 
   
\capspace
\caption{ The neutral  hadron fraction  of the leading and second jet  for the five different $y_{max}$ bins and for the 
HLT$_{-}$Jet30U trigger as a function of time (run number), fitted with a first degree polynomial. }
\label{fig_appd3}
\end{figure}

 \clearpage

\begin{figure}[h]
\centering

\includegraphics[width=0.49\textwidth]{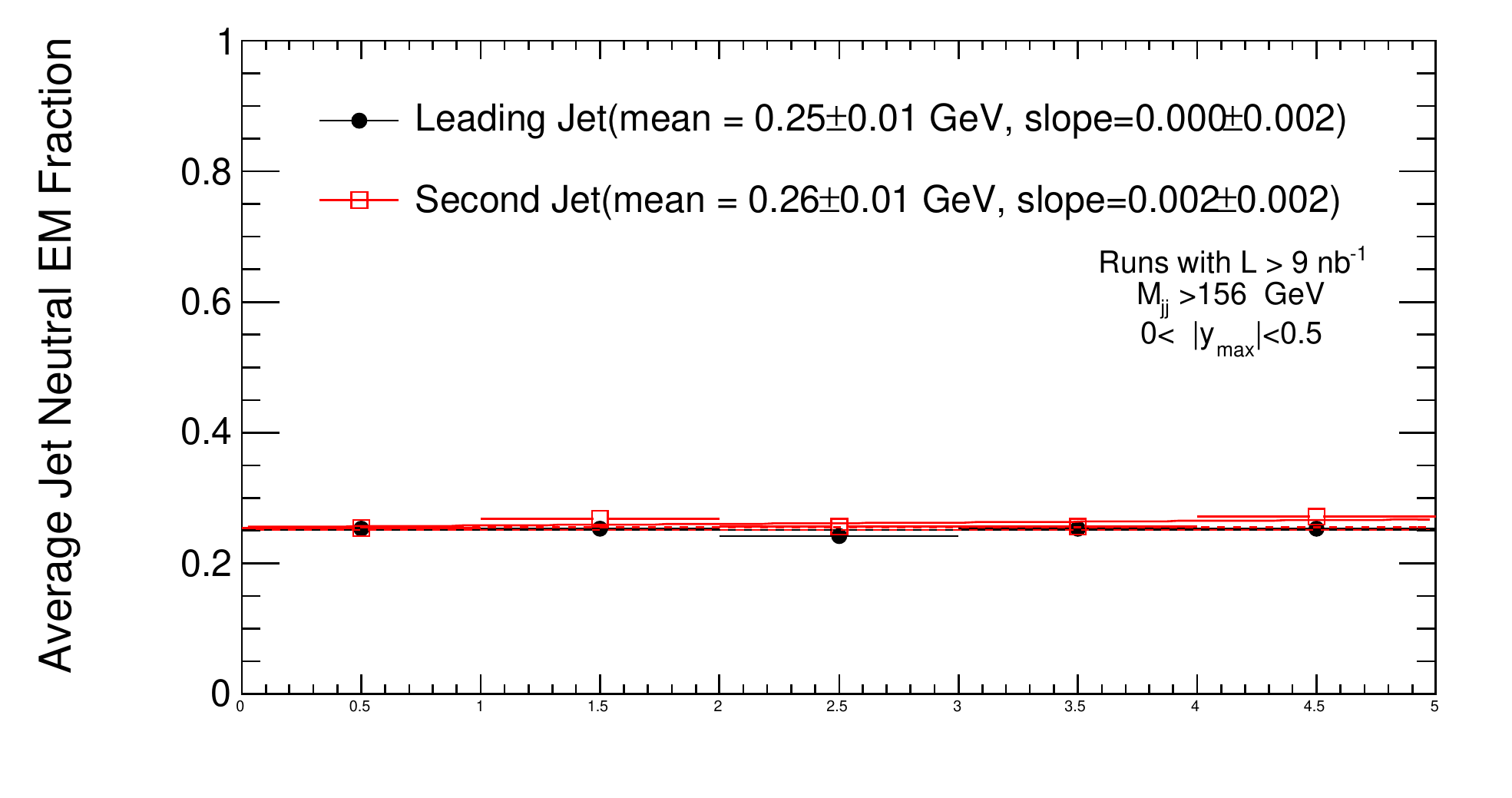} 
\includegraphics[width=0.49\textwidth]{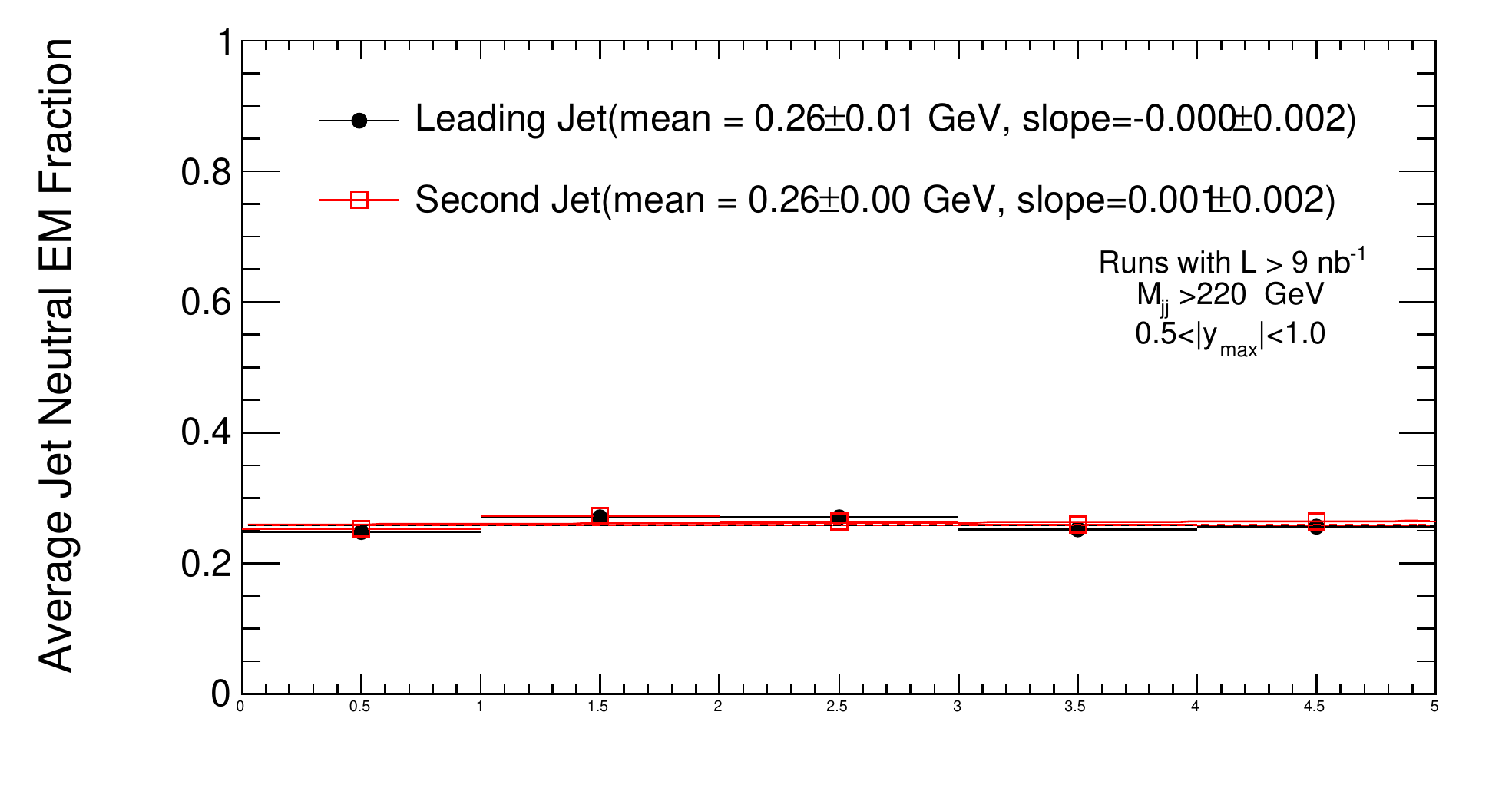}  
\includegraphics[width=0.49\textwidth]{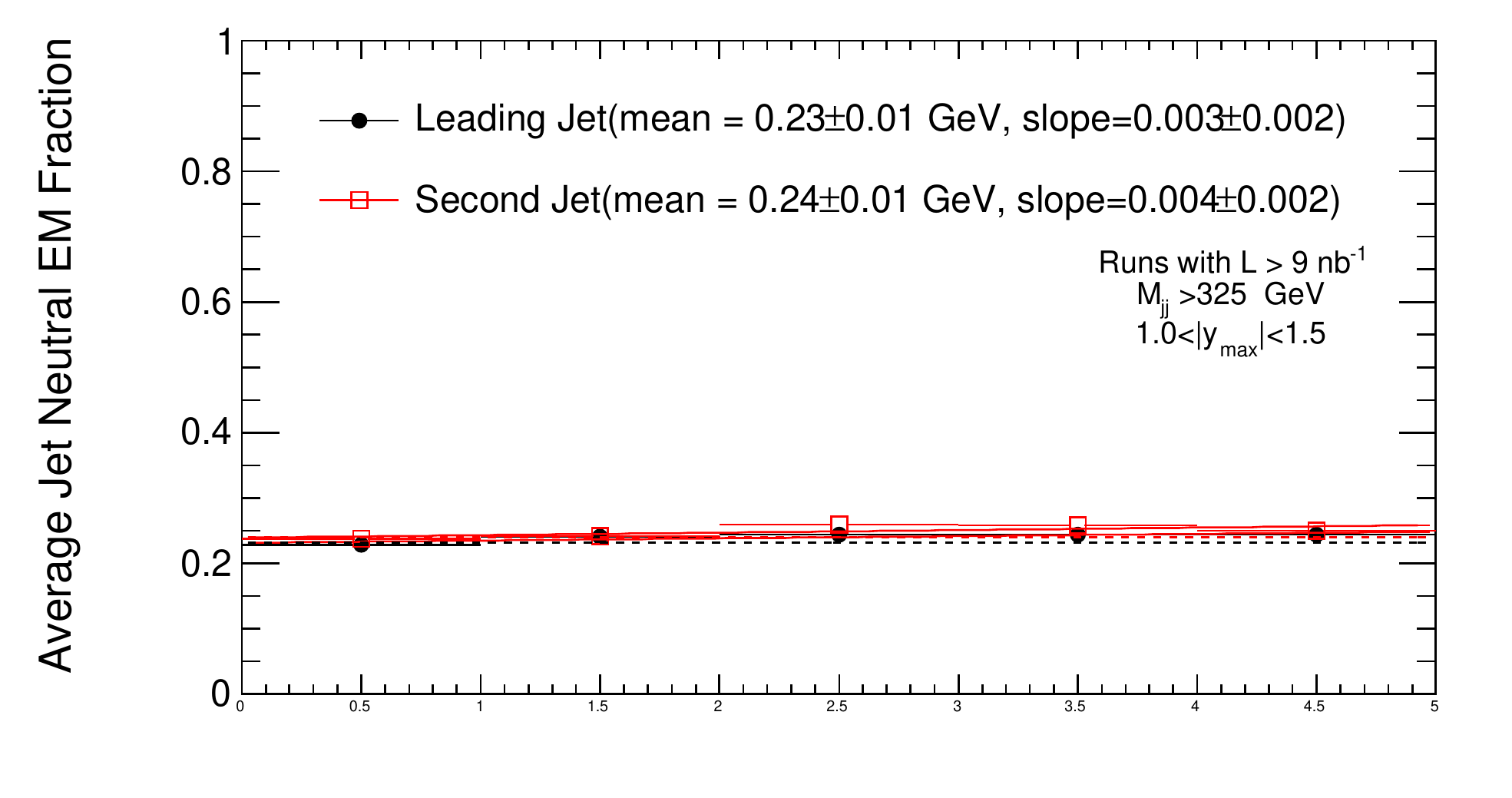}  
\includegraphics[width=0.49\textwidth]{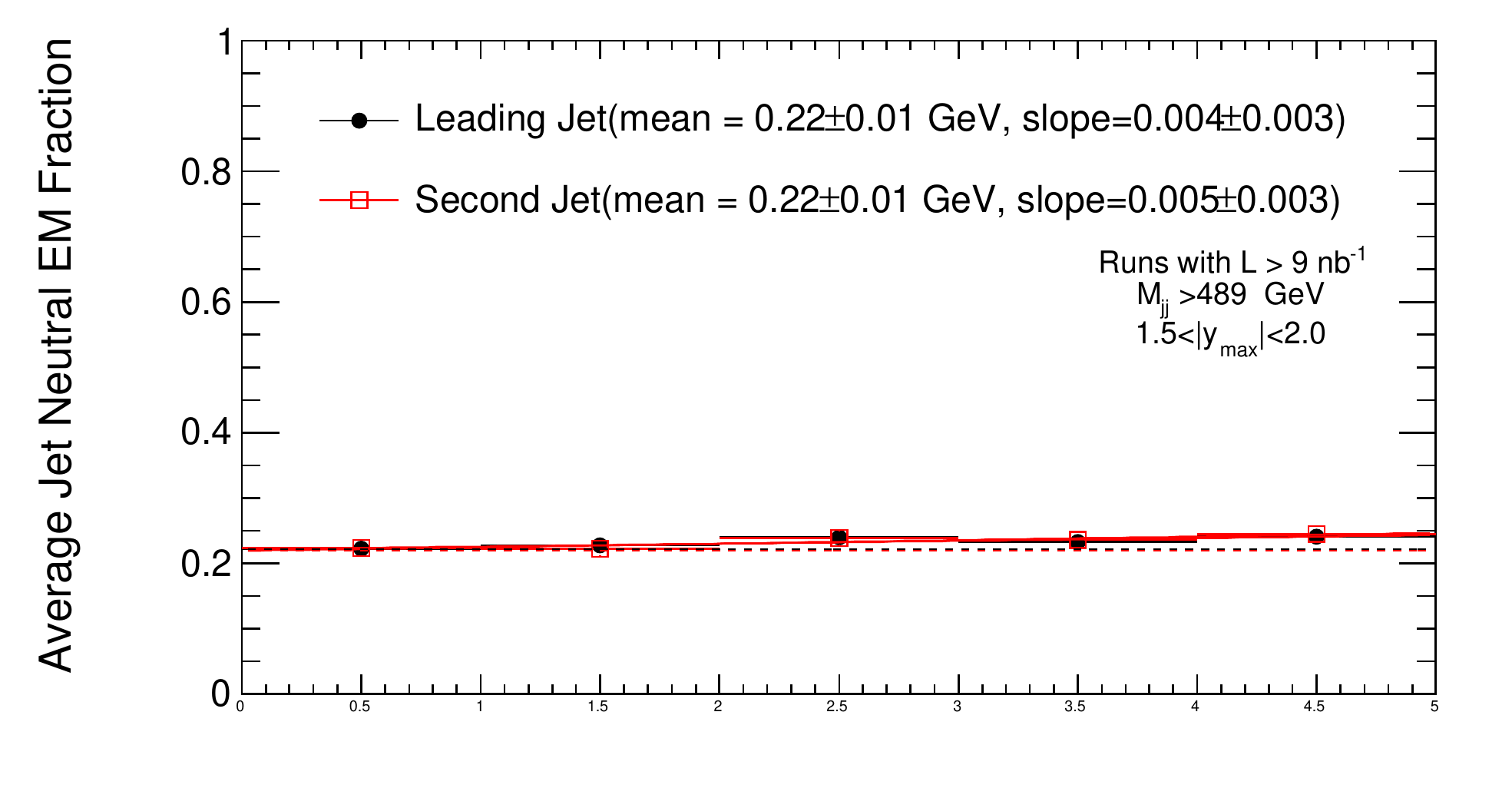} 
\includegraphics[width=0.49\textwidth]{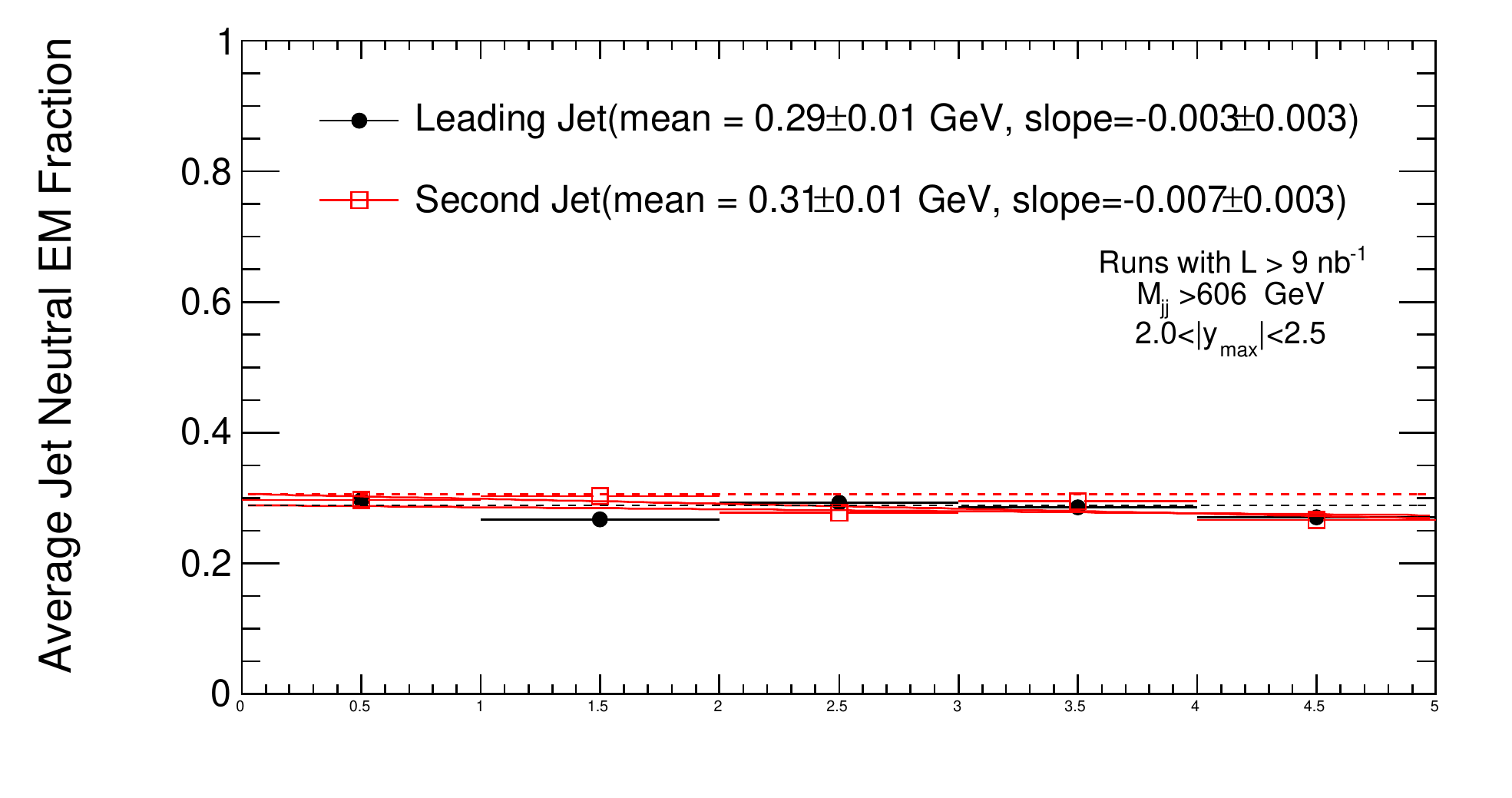} 
   
\capspace
\caption{ The neutral  electromagnetic fraction  of the leading and second jet  for the five different $y_{max}$ bins and for the 
HLT$_{-}$Jet30U trigger as a function of time (run number), fitted with a first degree polynomial. }
\label{fig_appd4}
\end{figure}

%%% jet  50

\begin{figure}[h]
\centering

\includegraphics[width=0.49\textwidth]{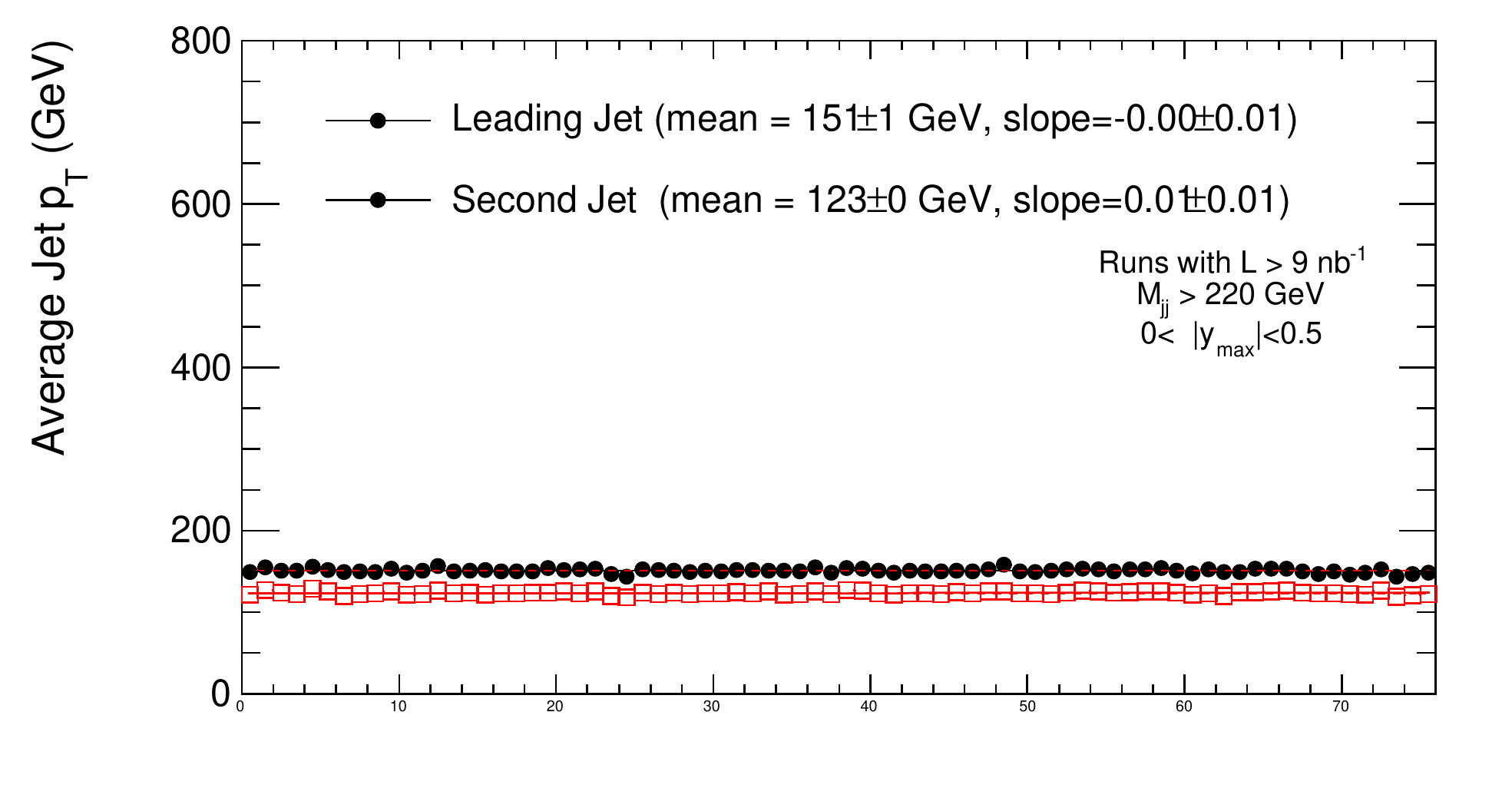} 
\includegraphics[width=0.49\textwidth]{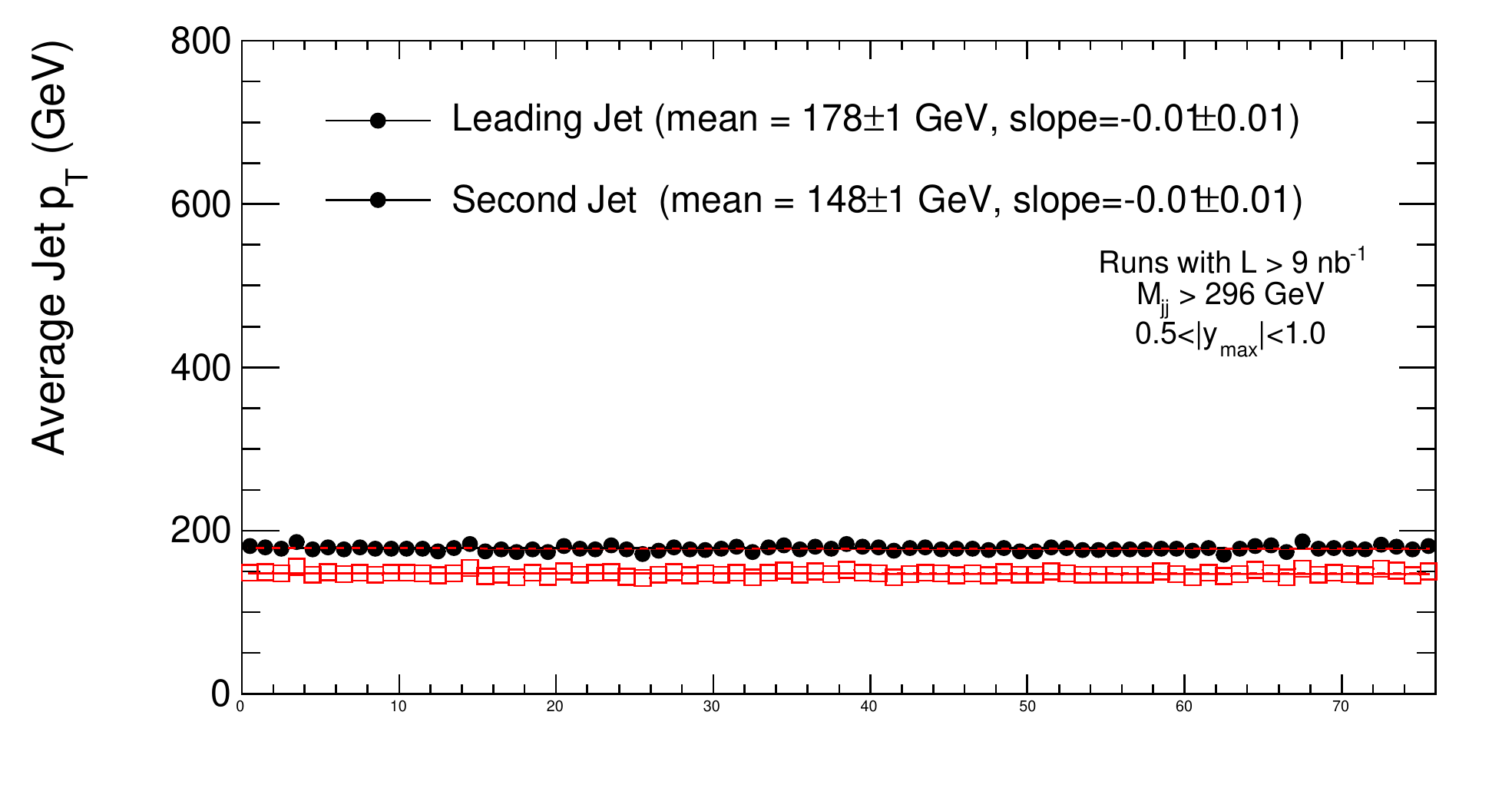} 
\includegraphics[width=0.49\textwidth]{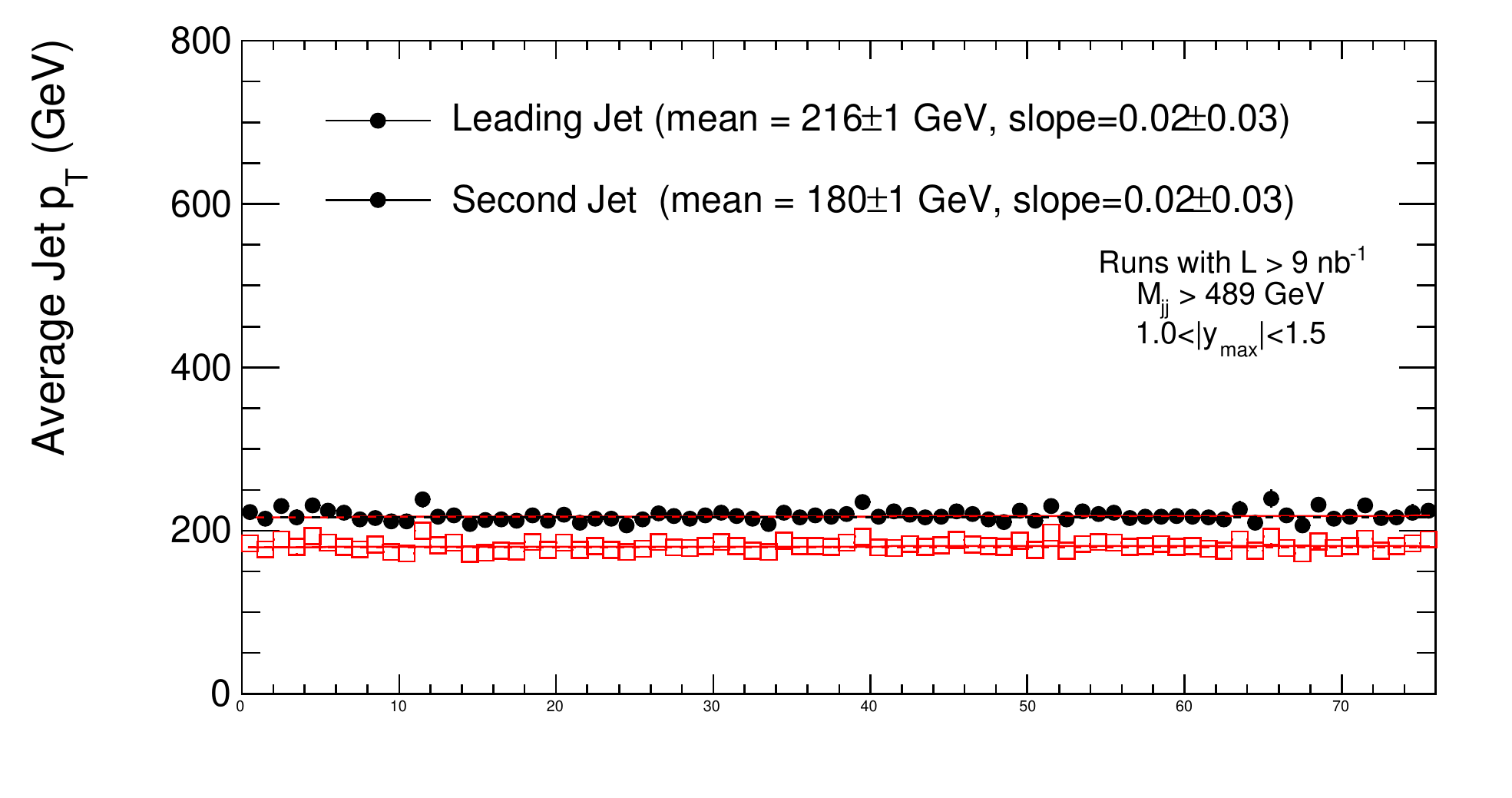} 
\includegraphics[width=0.49\textwidth]{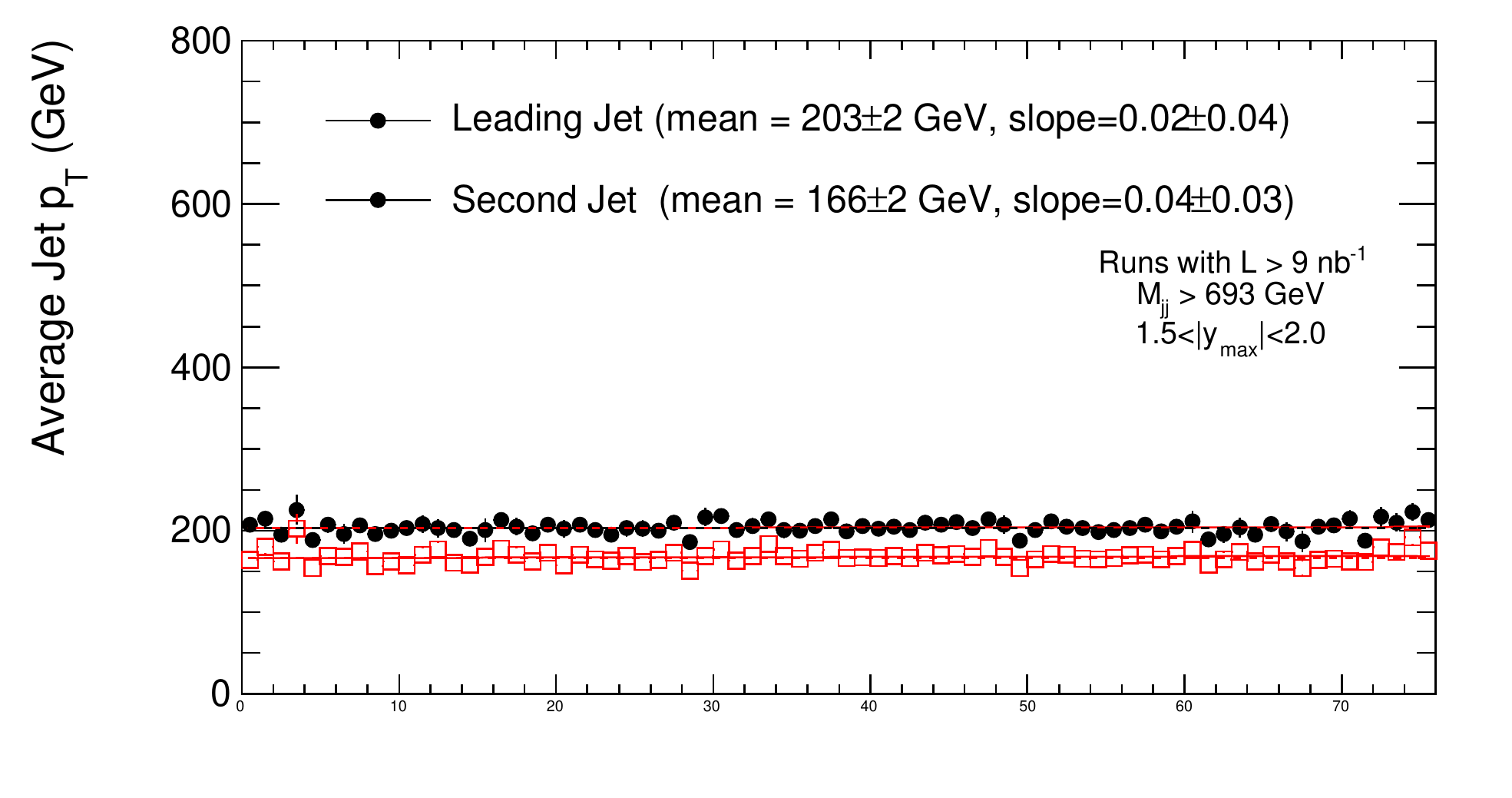} 
\includegraphics[width=0.49\textwidth]{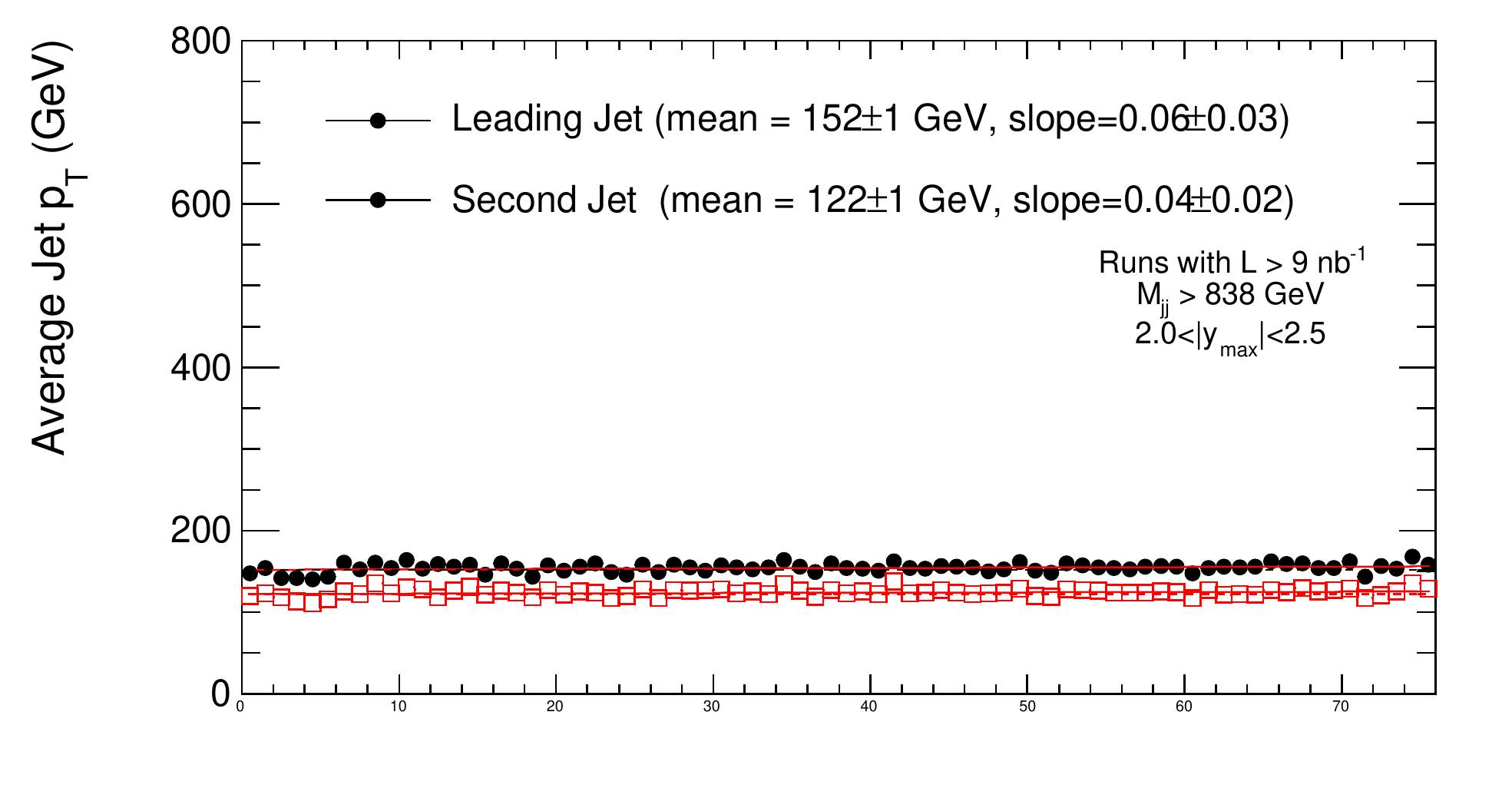} 
  
\capspace
\caption{ The $p_T$ of the leading and second jet  for the five different $y_{max}$ bins and for the
HLT$_{-}$Jet50U trigger as a function of time (run number), fitted with a first degree polynomial. }
\label{fig_appd5}
\end{figure}

\begin{figure}[h]
\centering

\includegraphics[width=0.49\textwidth]{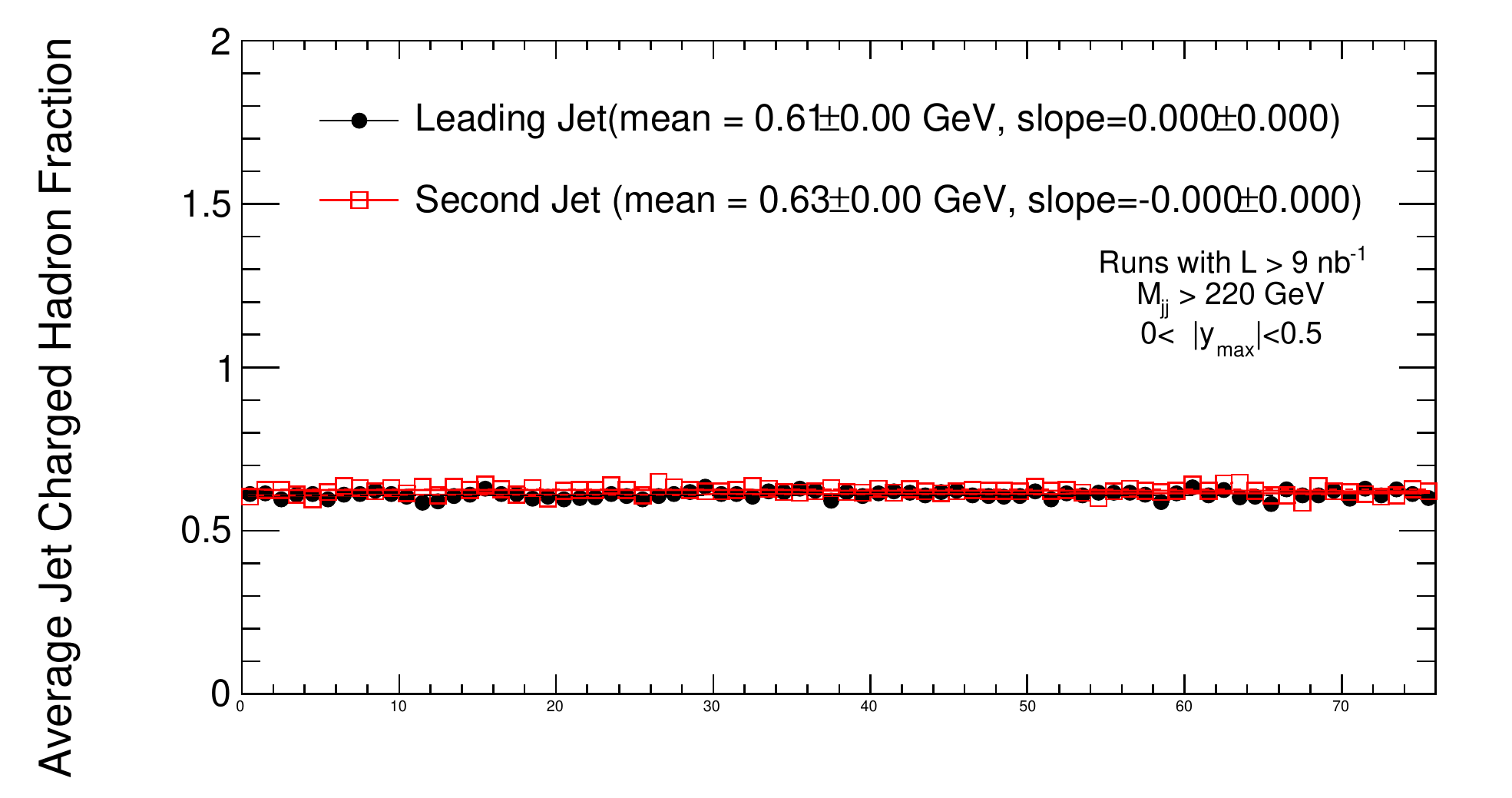} 
\includegraphics[width=0.49\textwidth]{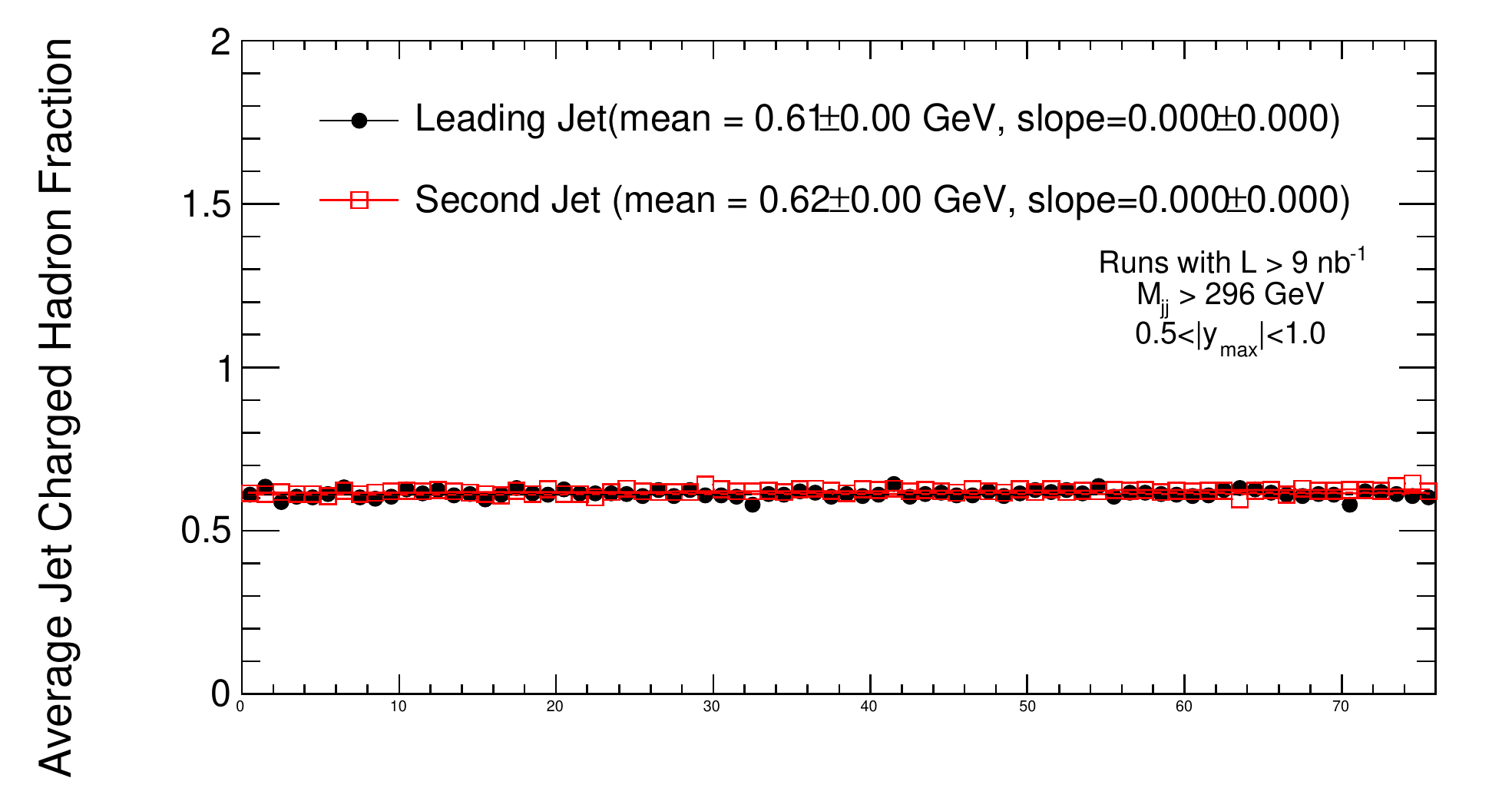} 
\includegraphics[width=0.49\textwidth]{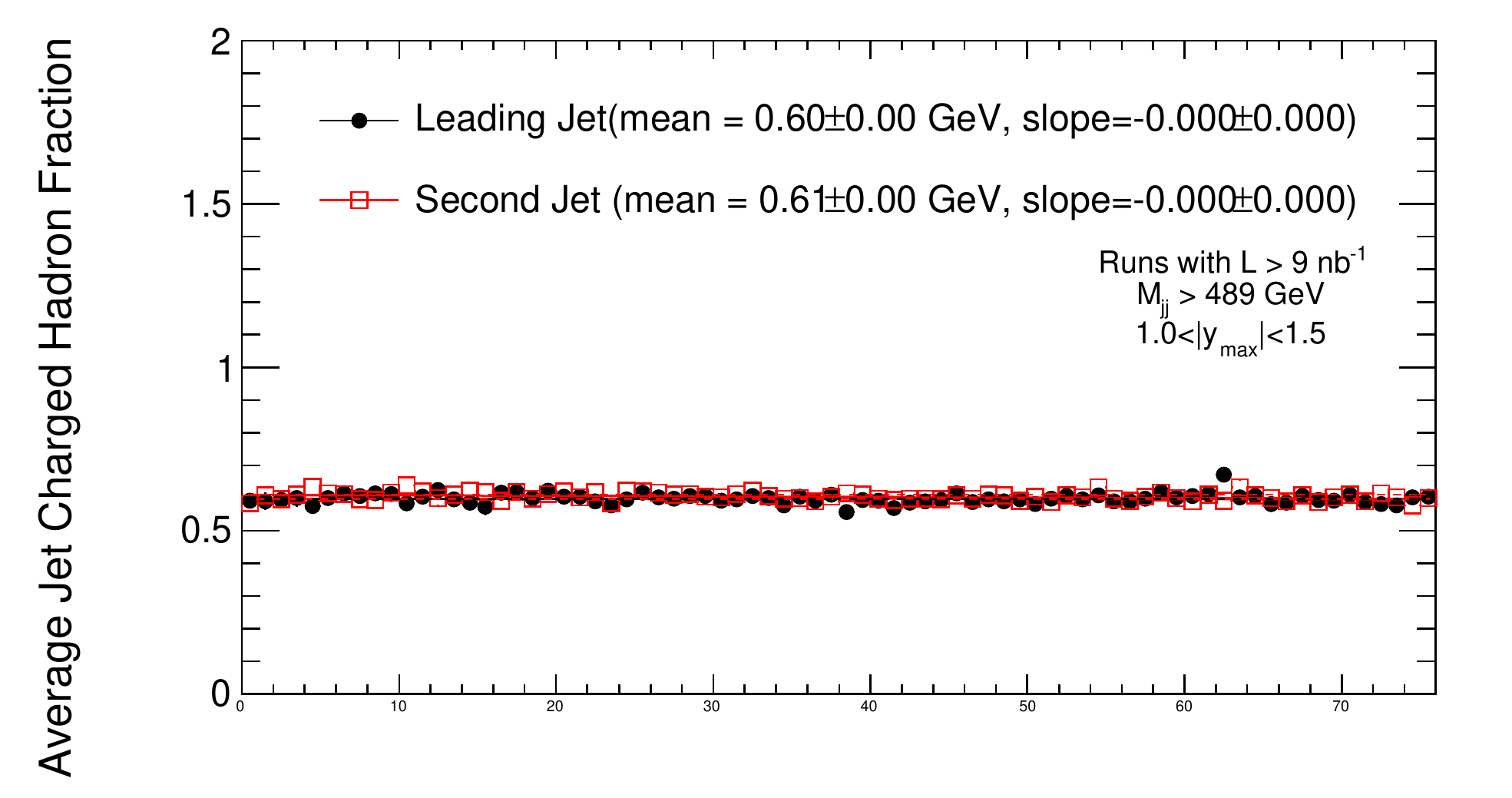} 
\includegraphics[width=0.49\textwidth]{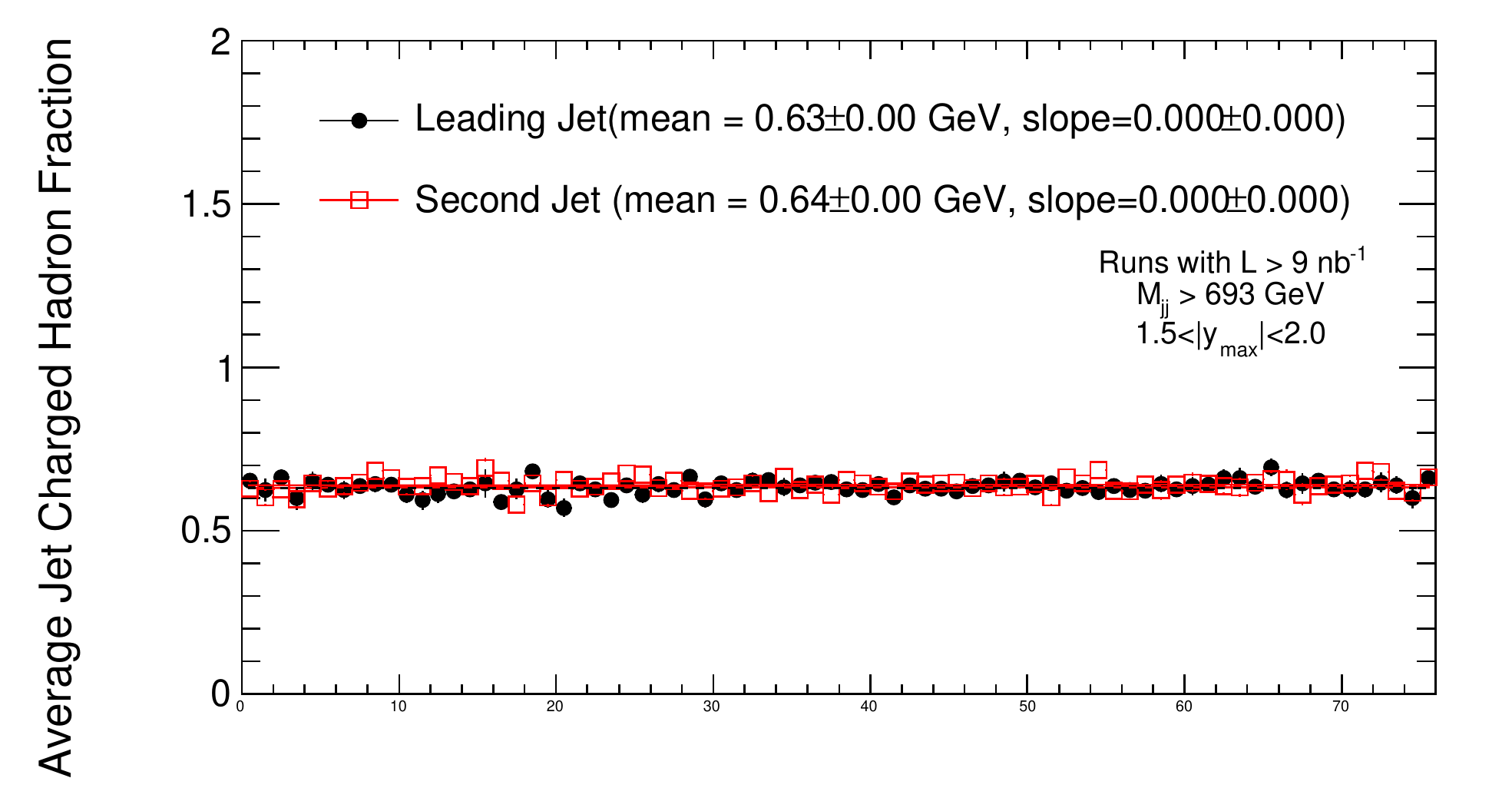} 
\includegraphics[width=0.49\textwidth]{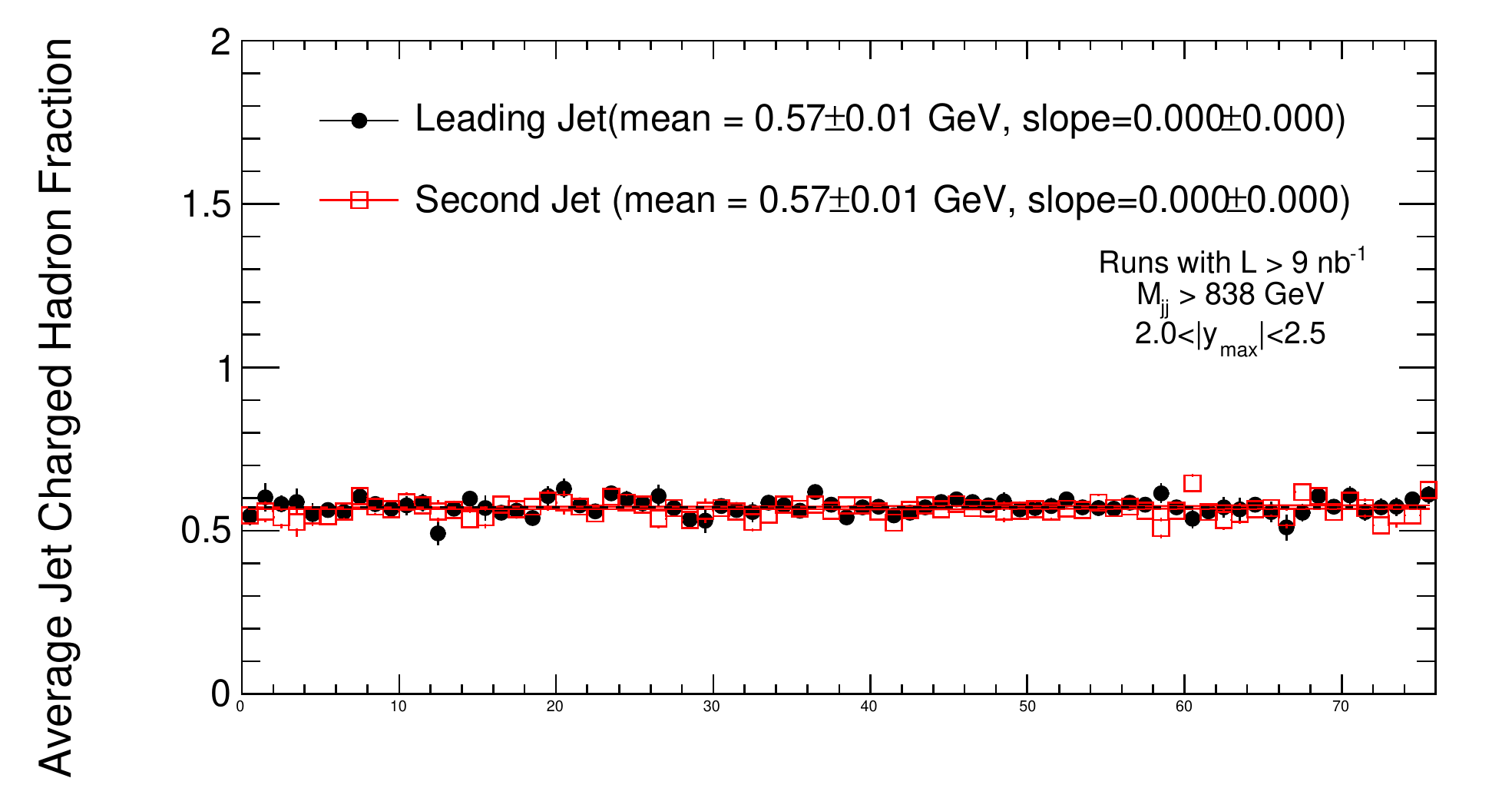} 
  
\capspace
\caption{ The charged hadron fraction  of the leading and second jet  for the five different $y_{max}$ bins and for the
HLT$_{-}$Jet50U trigger as a function of time (run number), fitted with a first degree polynomial. }
\label{fig_appd6}
\end{figure}

\clearpage

\begin{figure}[h]
\centering

\includegraphics[width=0.49\textwidth]{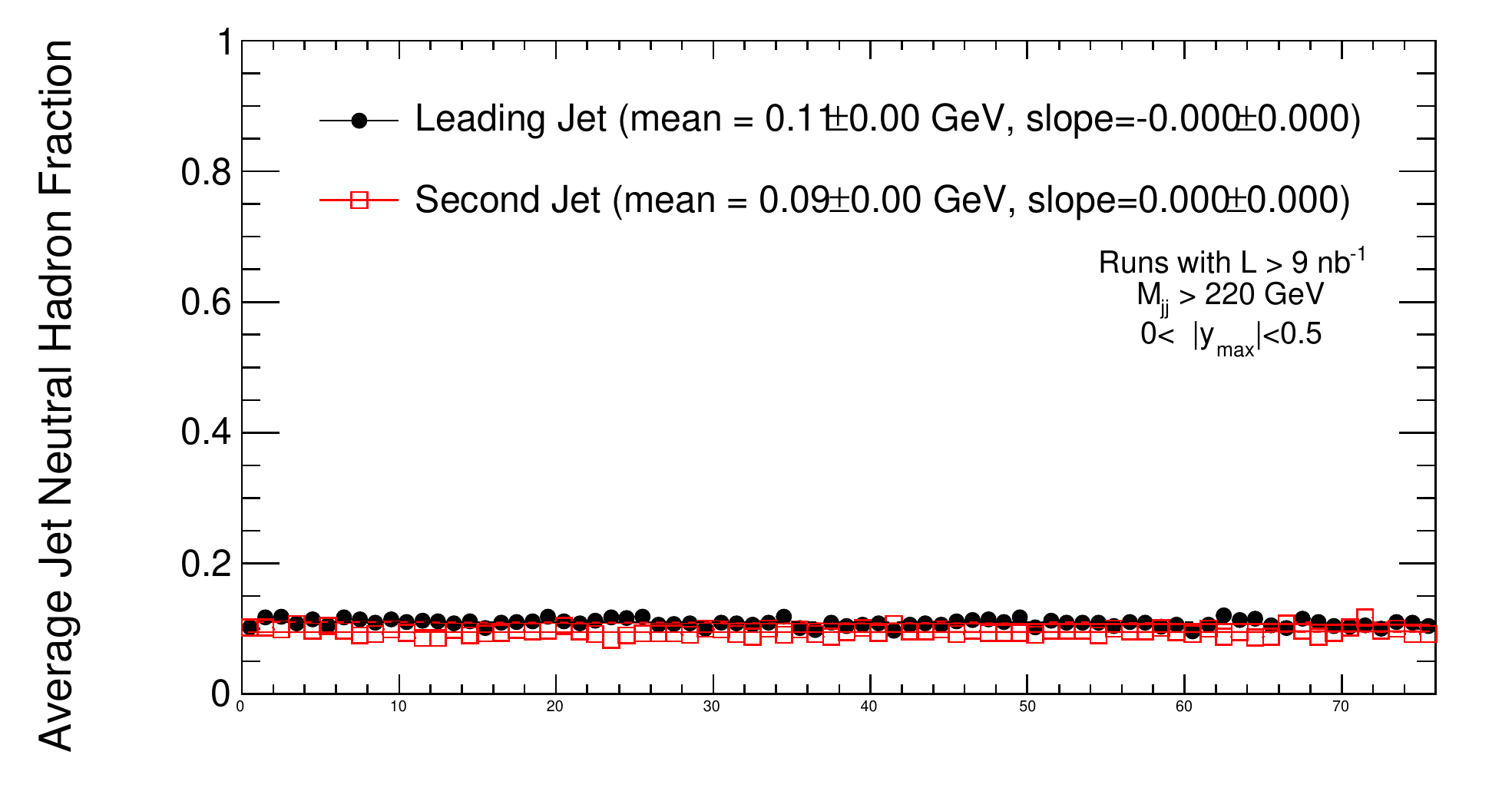} 
\includegraphics[width=0.49\textwidth]{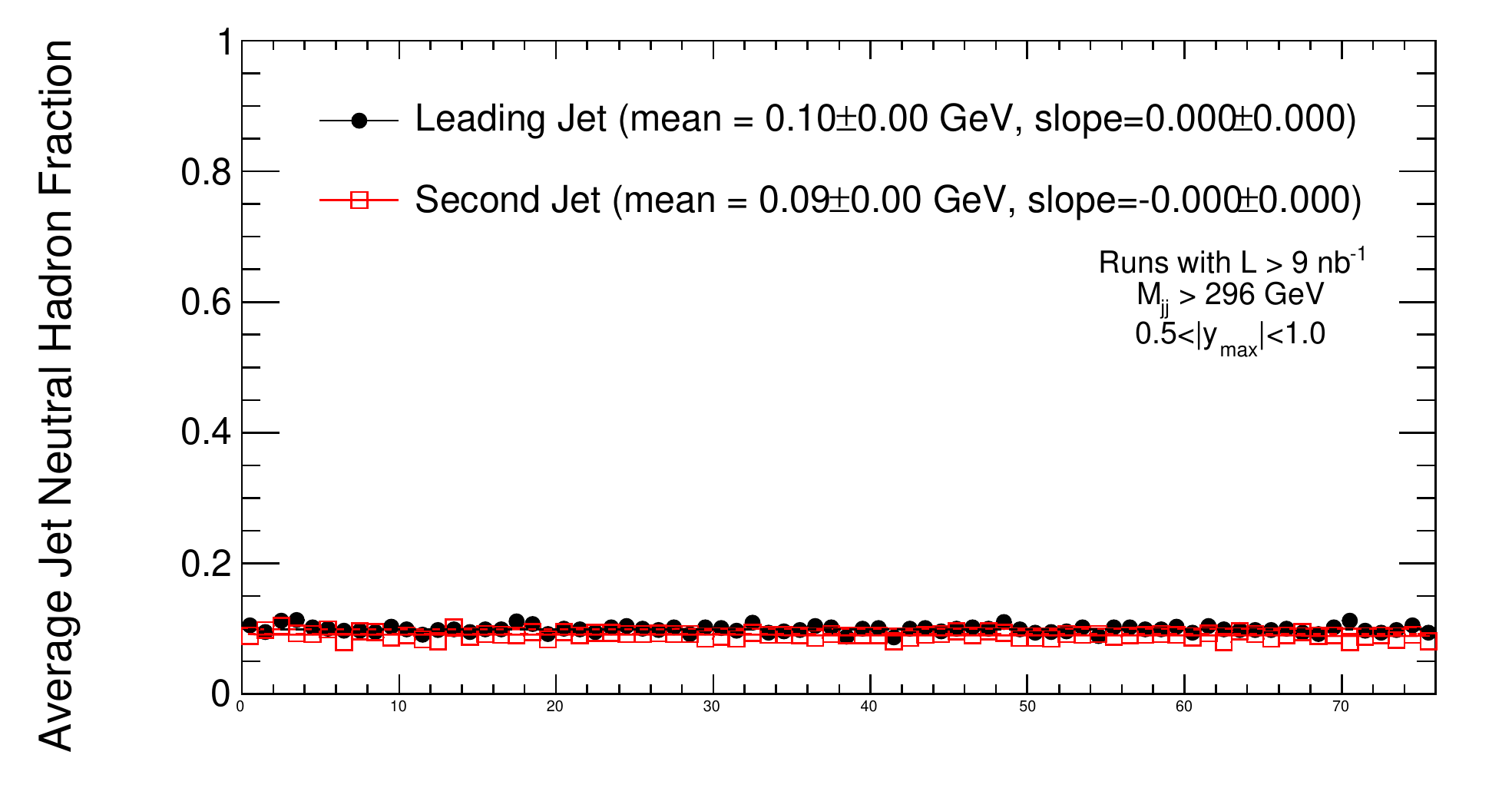} 
\includegraphics[width=0.49\textwidth]{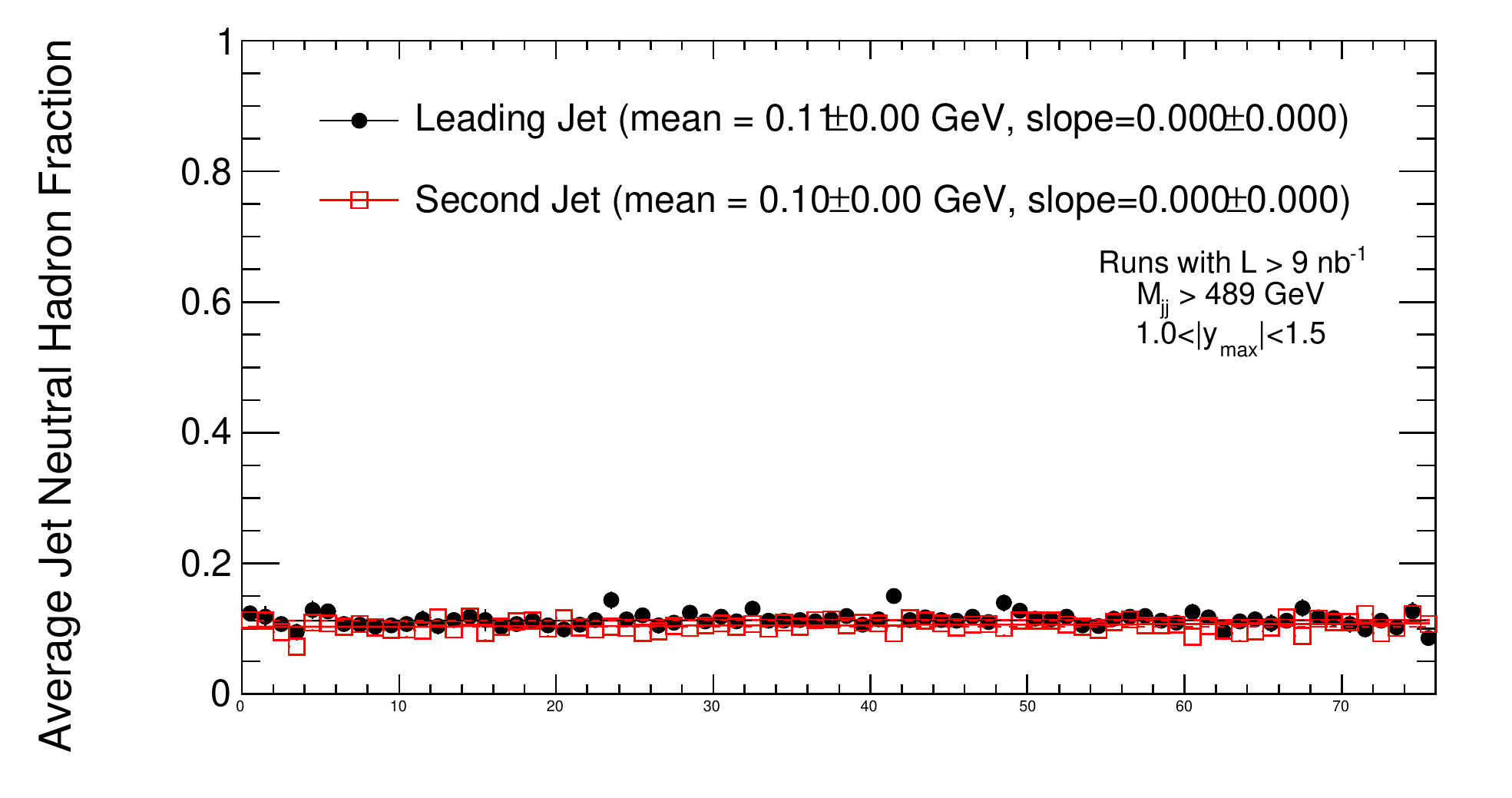} 
\includegraphics[width=0.49\textwidth]{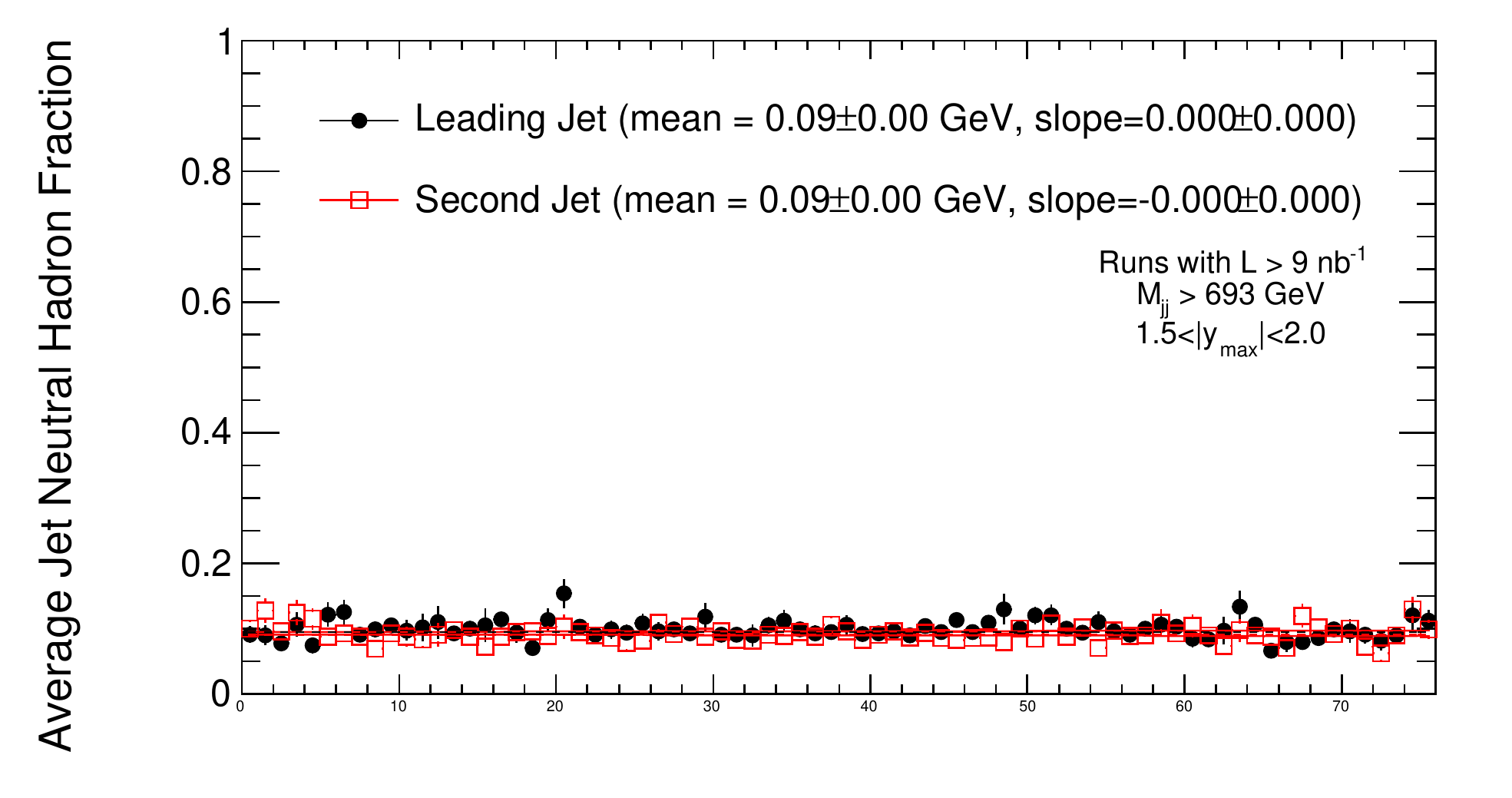} 
\includegraphics[width=0.49\textwidth]{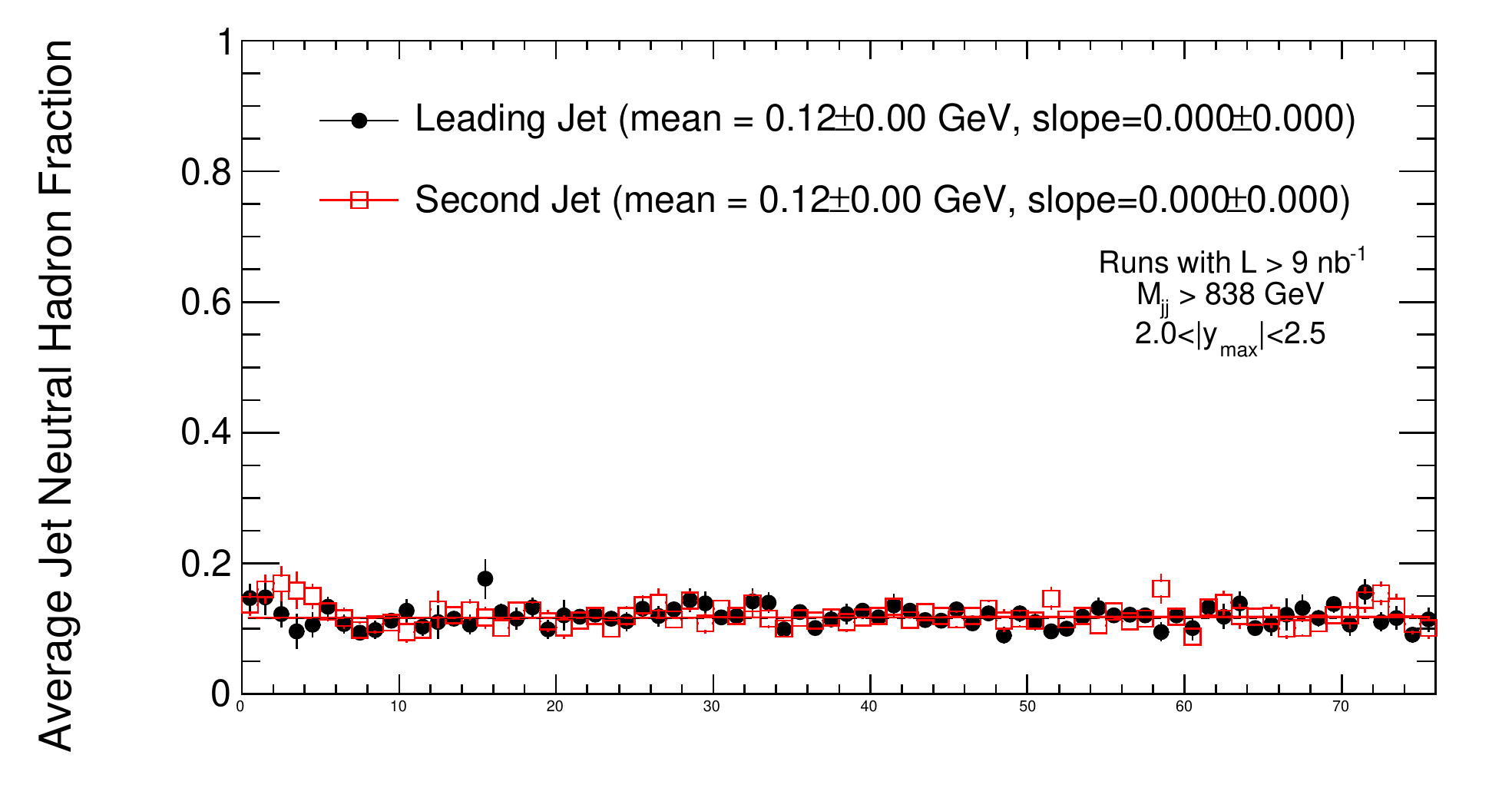} 
  
\capspace
\caption{ The neutral  hadron fraction  of the leading and second jet  for the five different $y_{max}$ bins and for the
HLT$_{-}$Jet50U trigger as a function of time (run number), fitted with a first degree polynomial. }
\label{fig_appd7}
\end{figure}

\begin{figure}[h]
\centering

\includegraphics[width=0.49\textwidth]{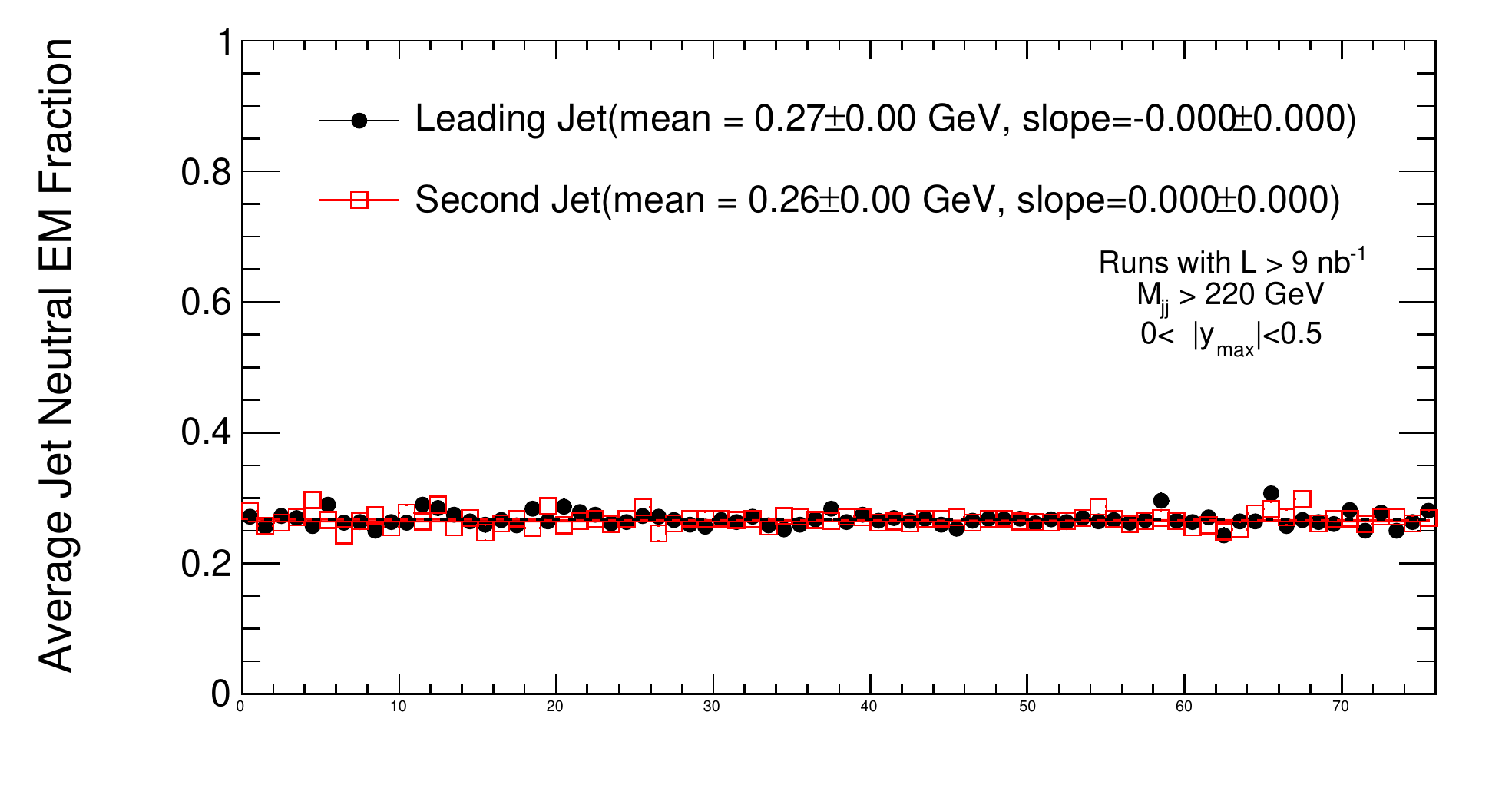} 
\includegraphics[width=0.49\textwidth]{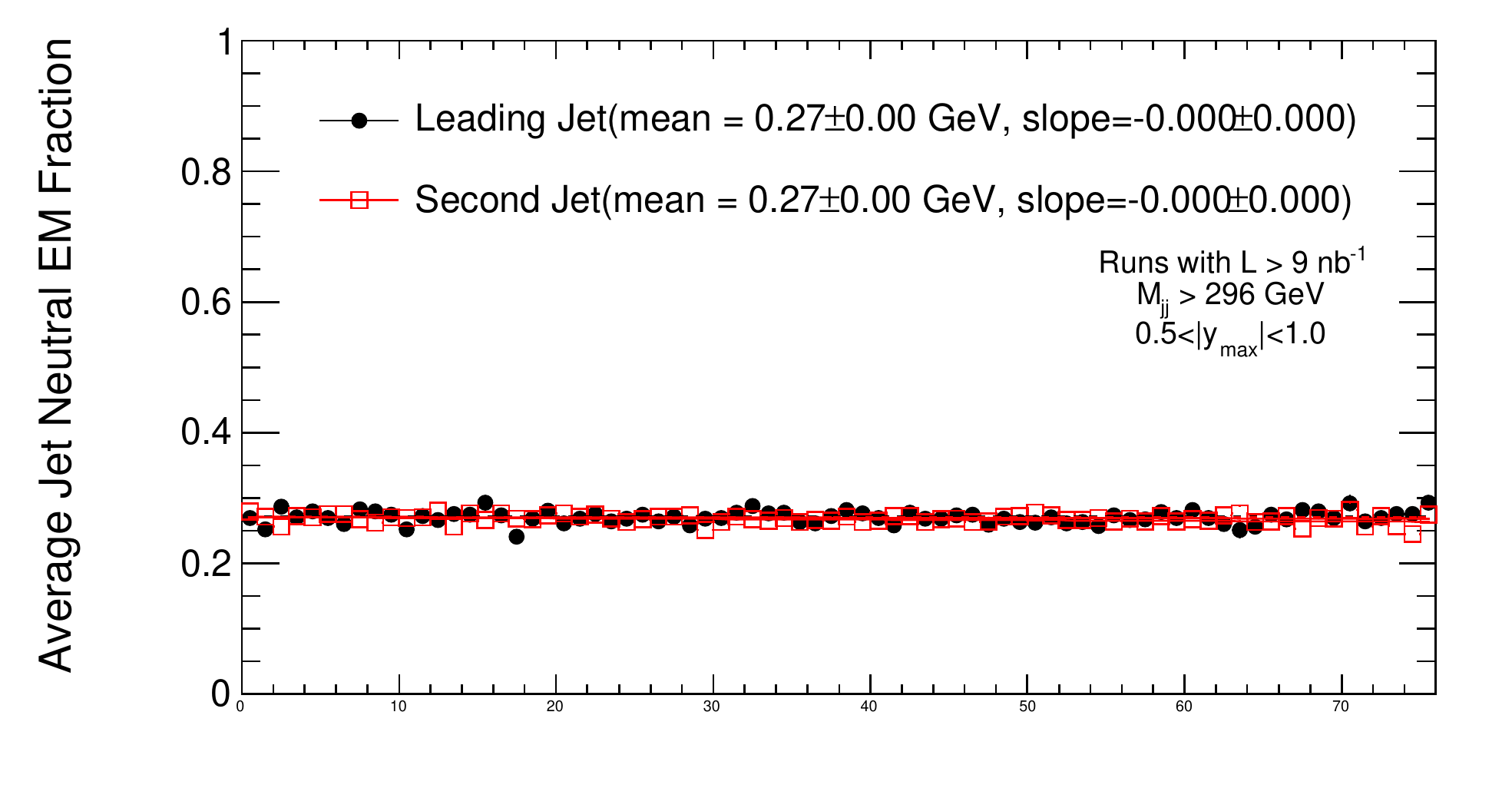} 
\includegraphics[width=0.49\textwidth]{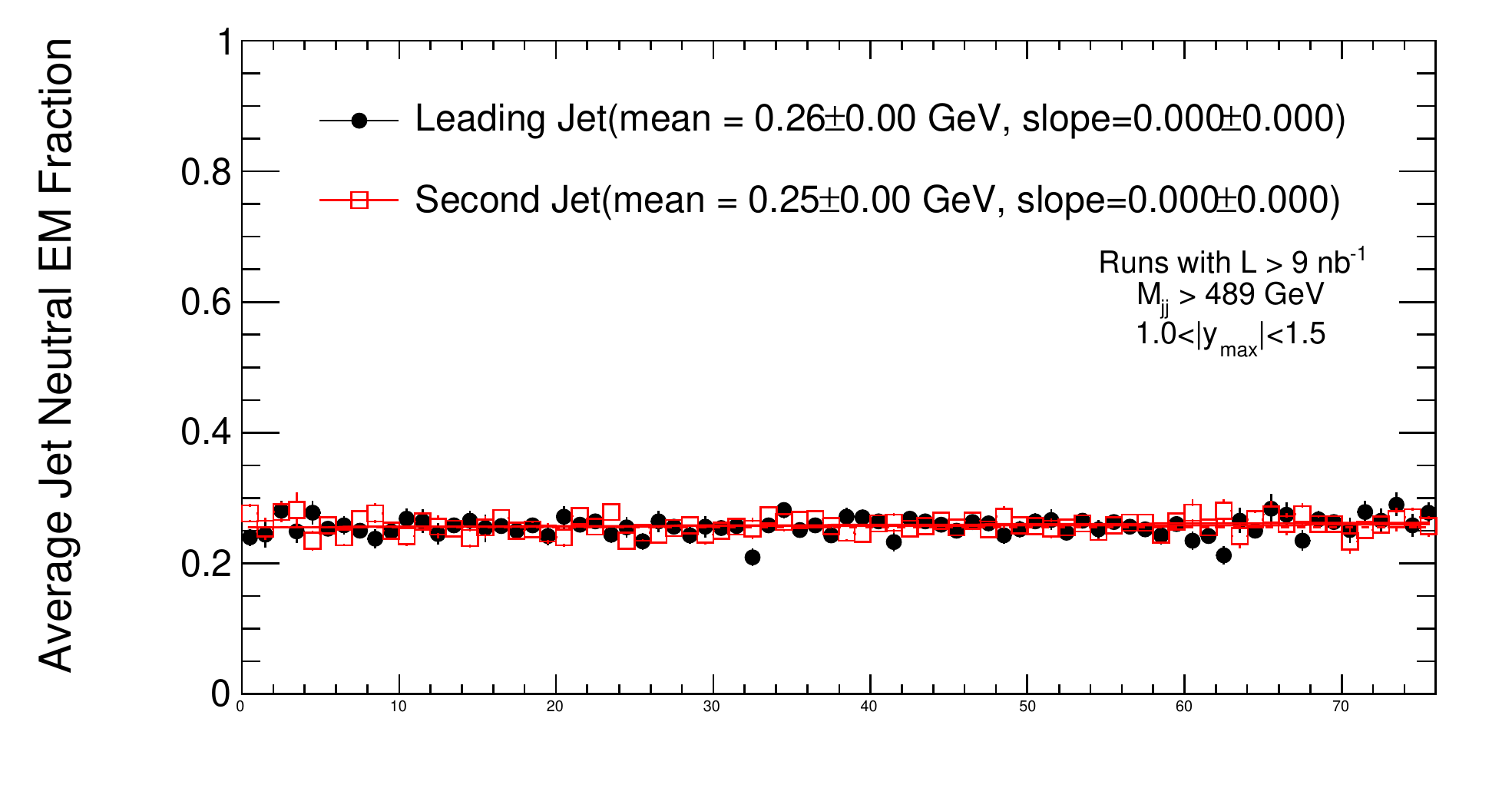} 
\includegraphics[width=0.49\textwidth]{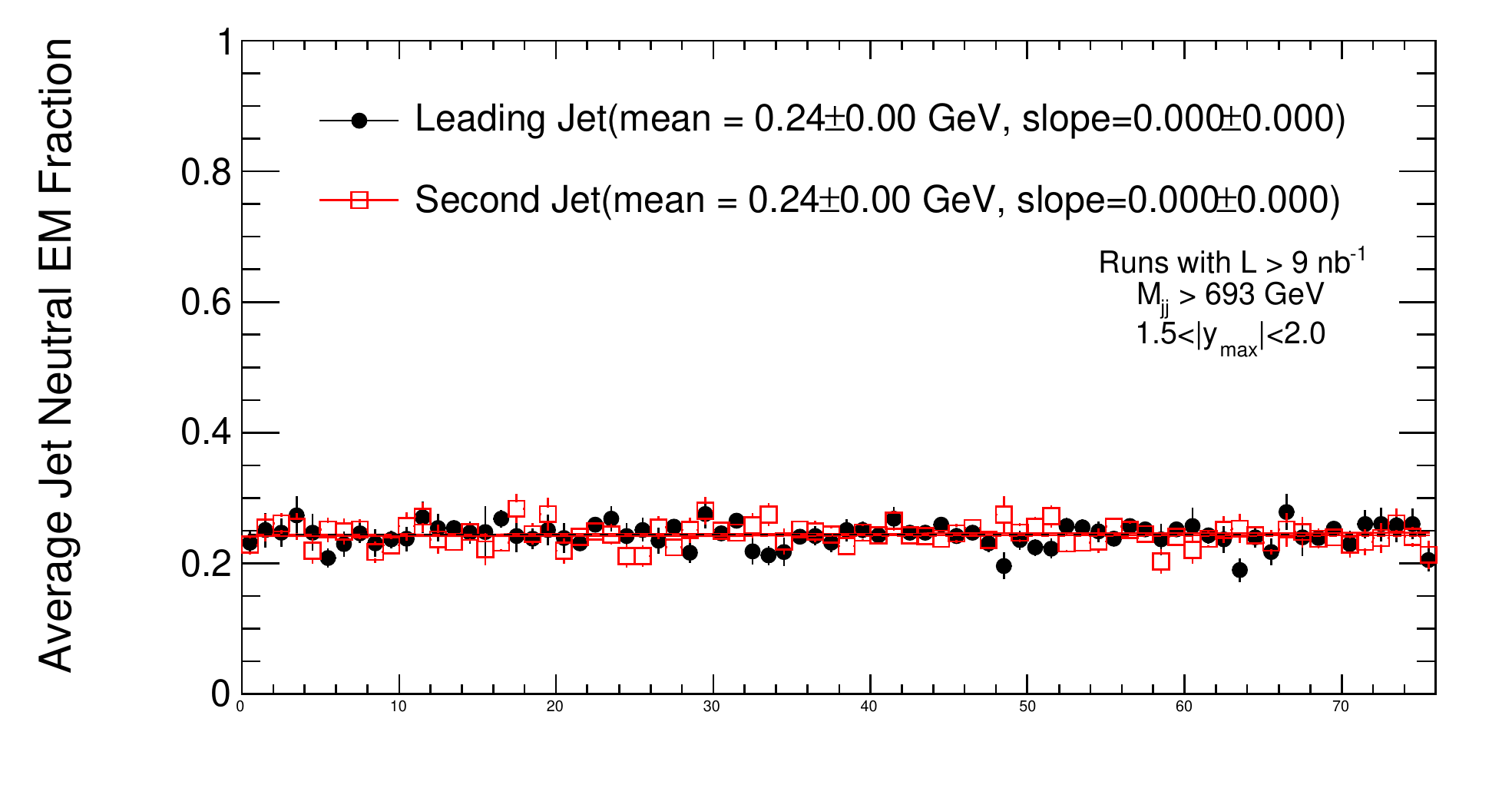} 
\includegraphics[width=0.49\textwidth]{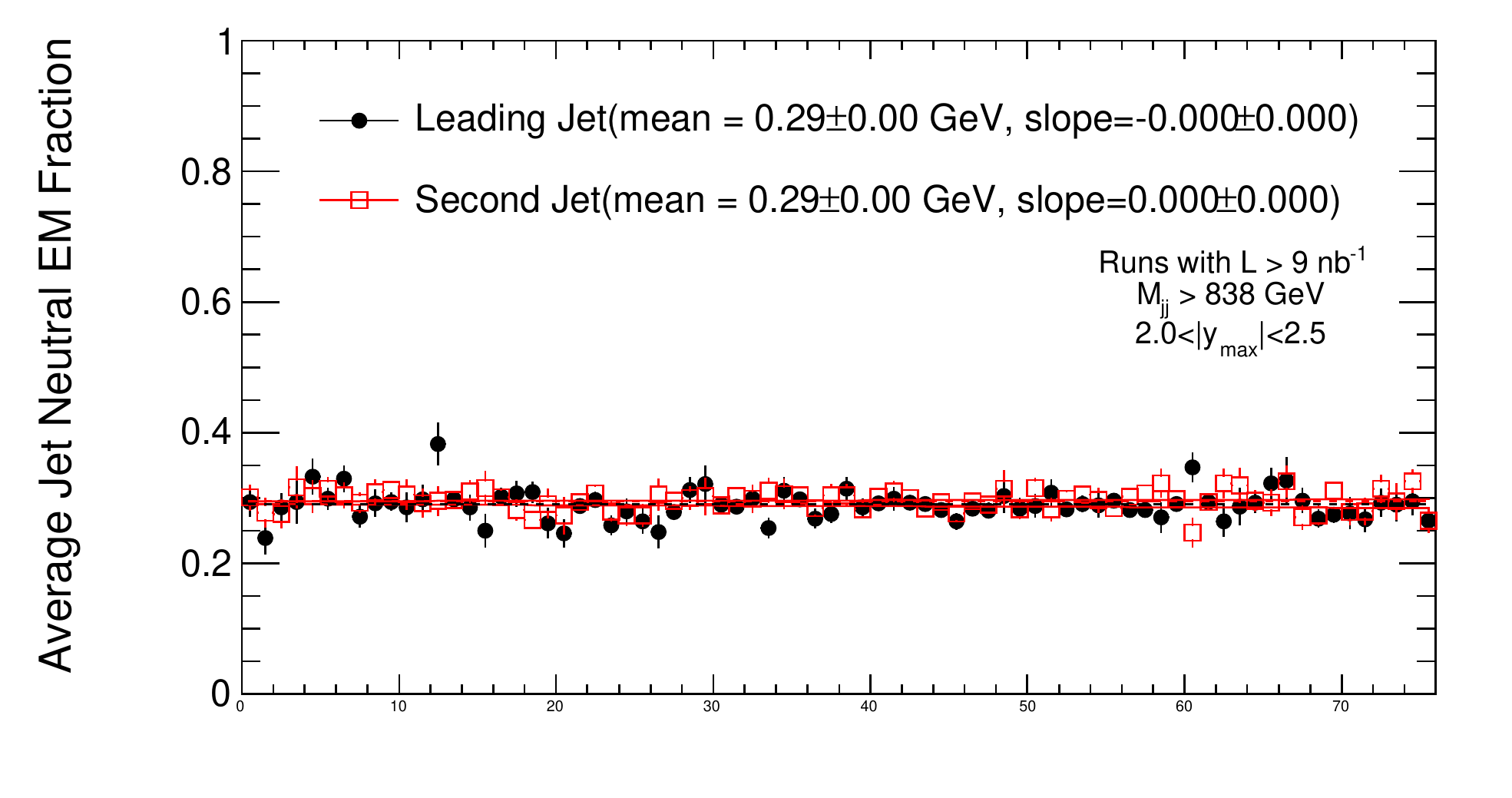} 
  
\capspace
\caption{ The neutral  electromagnetic fraction  of the leading and second jet  for the five different $y_{max}$ bins and for the
HLT$_{-}$Jet50U trigger as a function of time (run number), fitted with a first degree polynomial. }
\label{fig_appd8}
\end{figure}

%%% jet 100

\begin{figure}[h]
\centering

\includegraphics[width=0.49\textwidth]{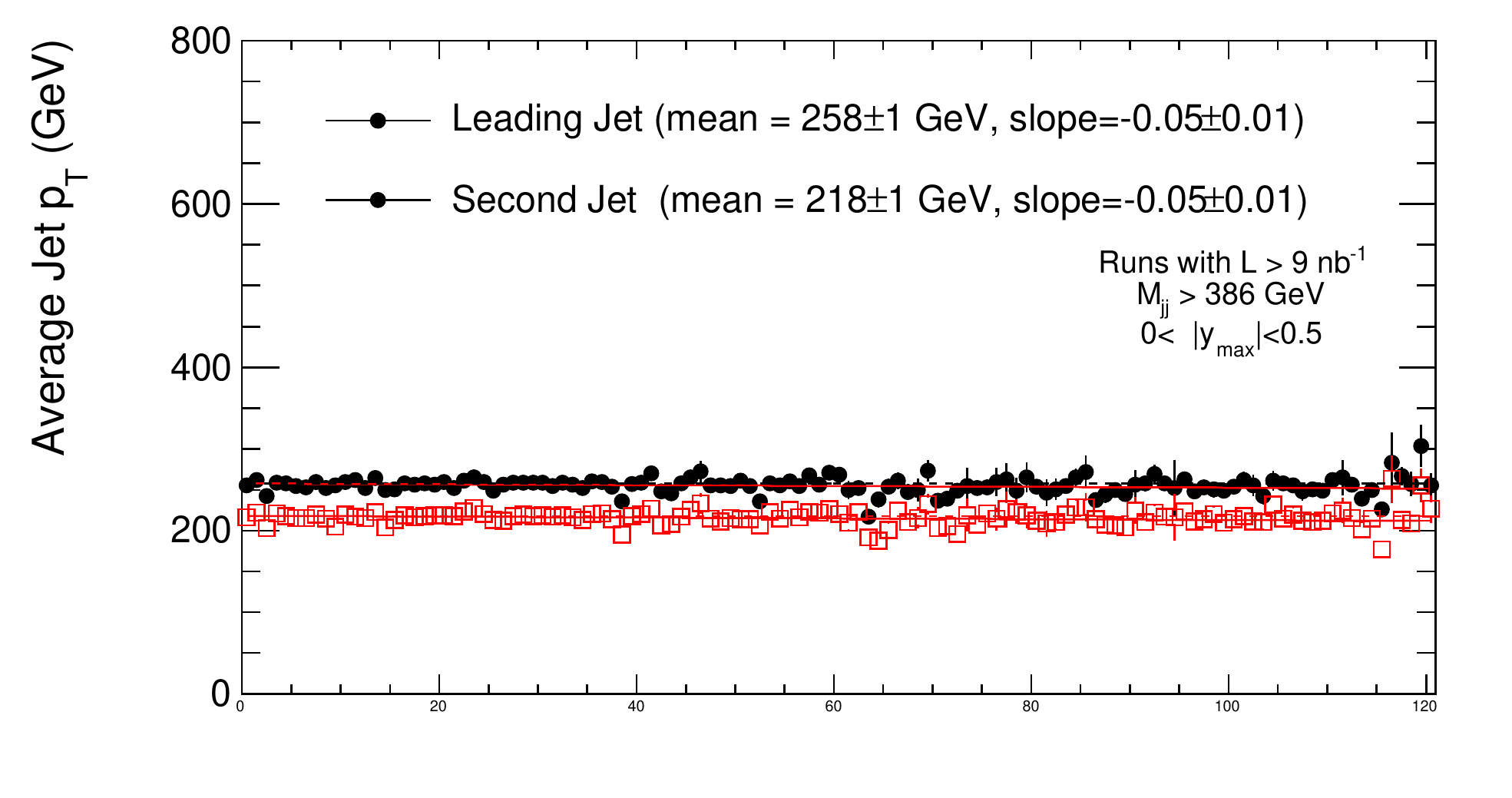} 
\includegraphics[width=0.49\textwidth]{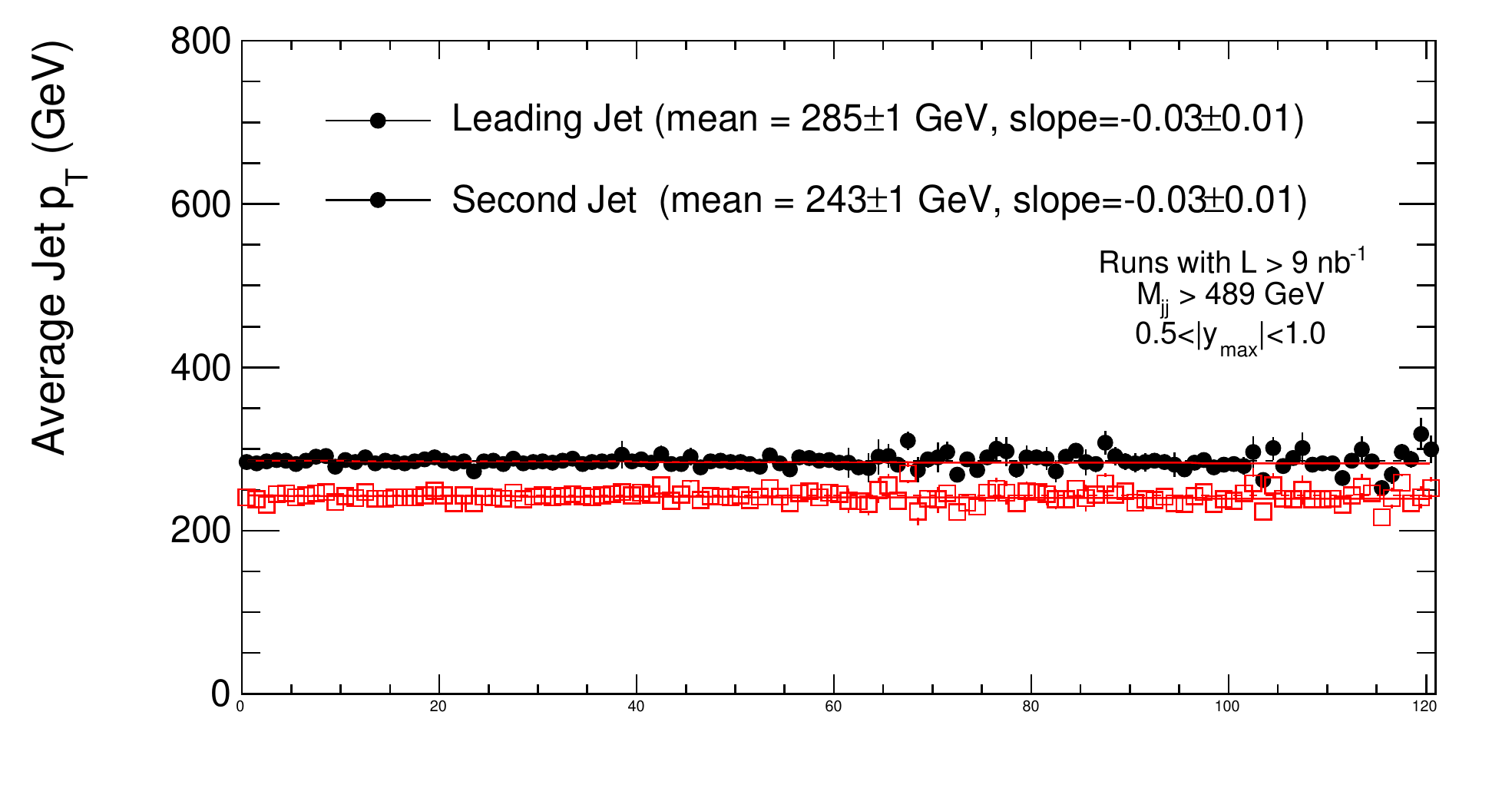} 
\includegraphics[width=0.49\textwidth]{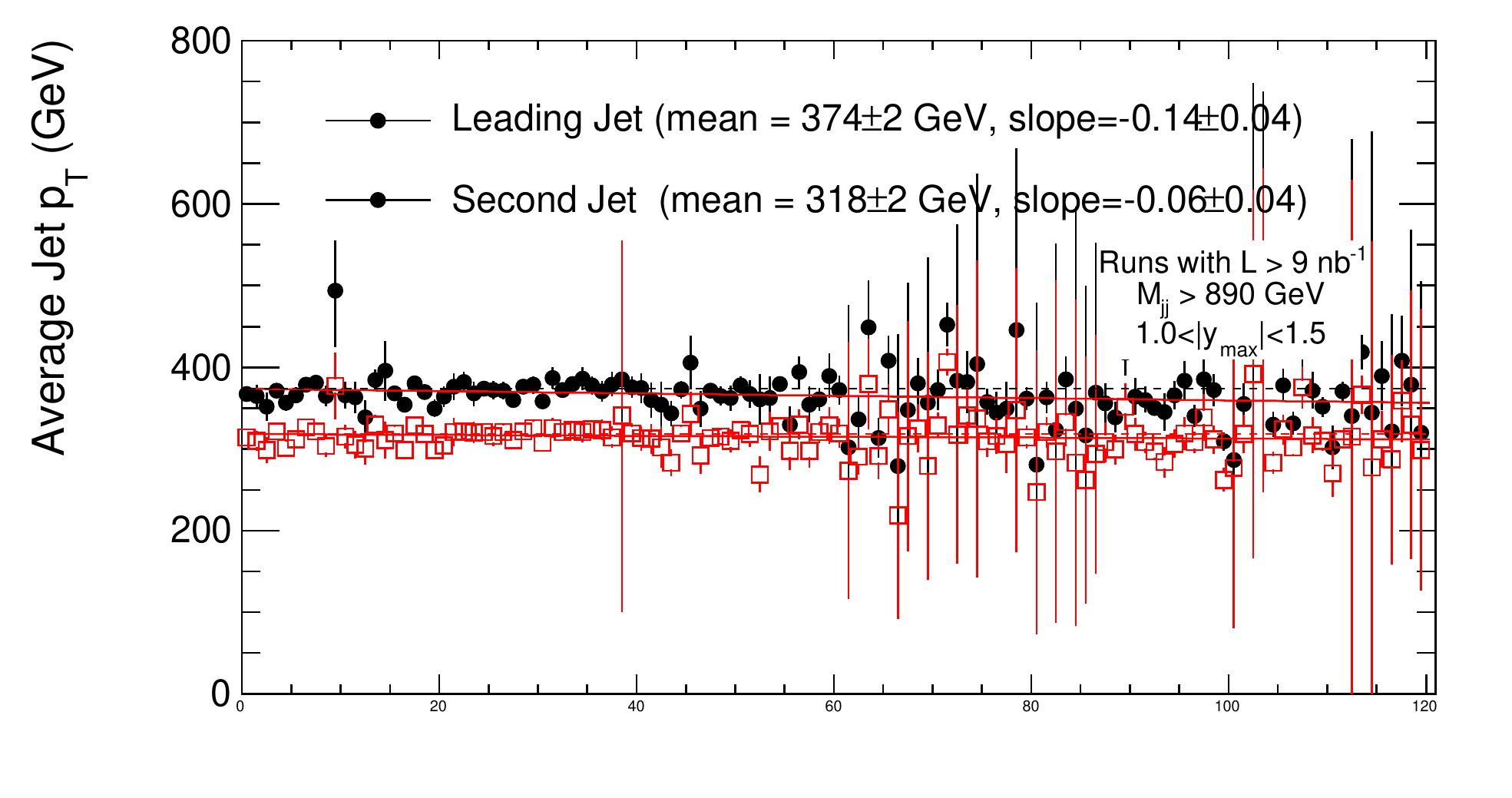} 
\includegraphics[width=0.49\textwidth]{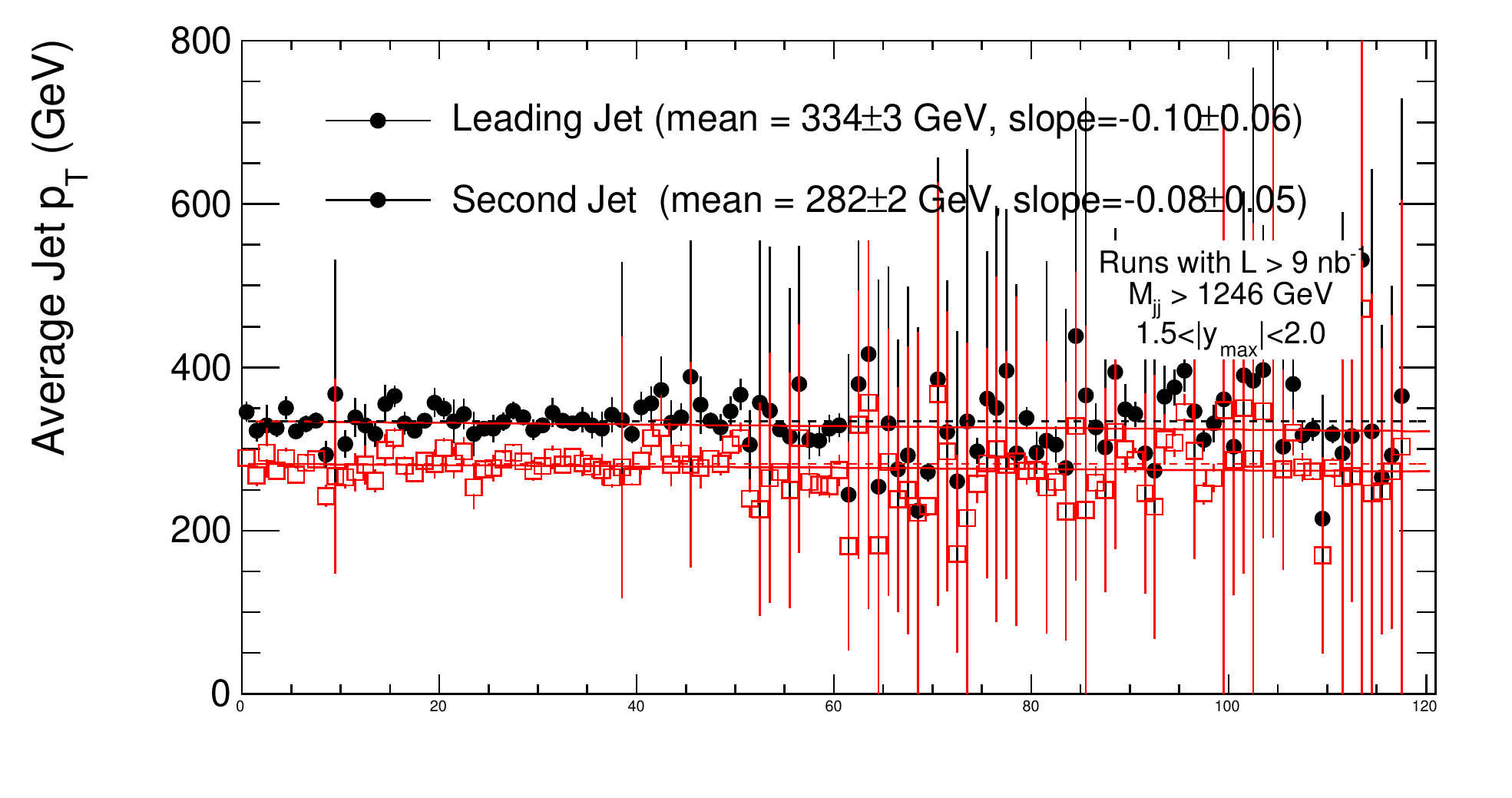} 
\includegraphics[width=0.49\textwidth]{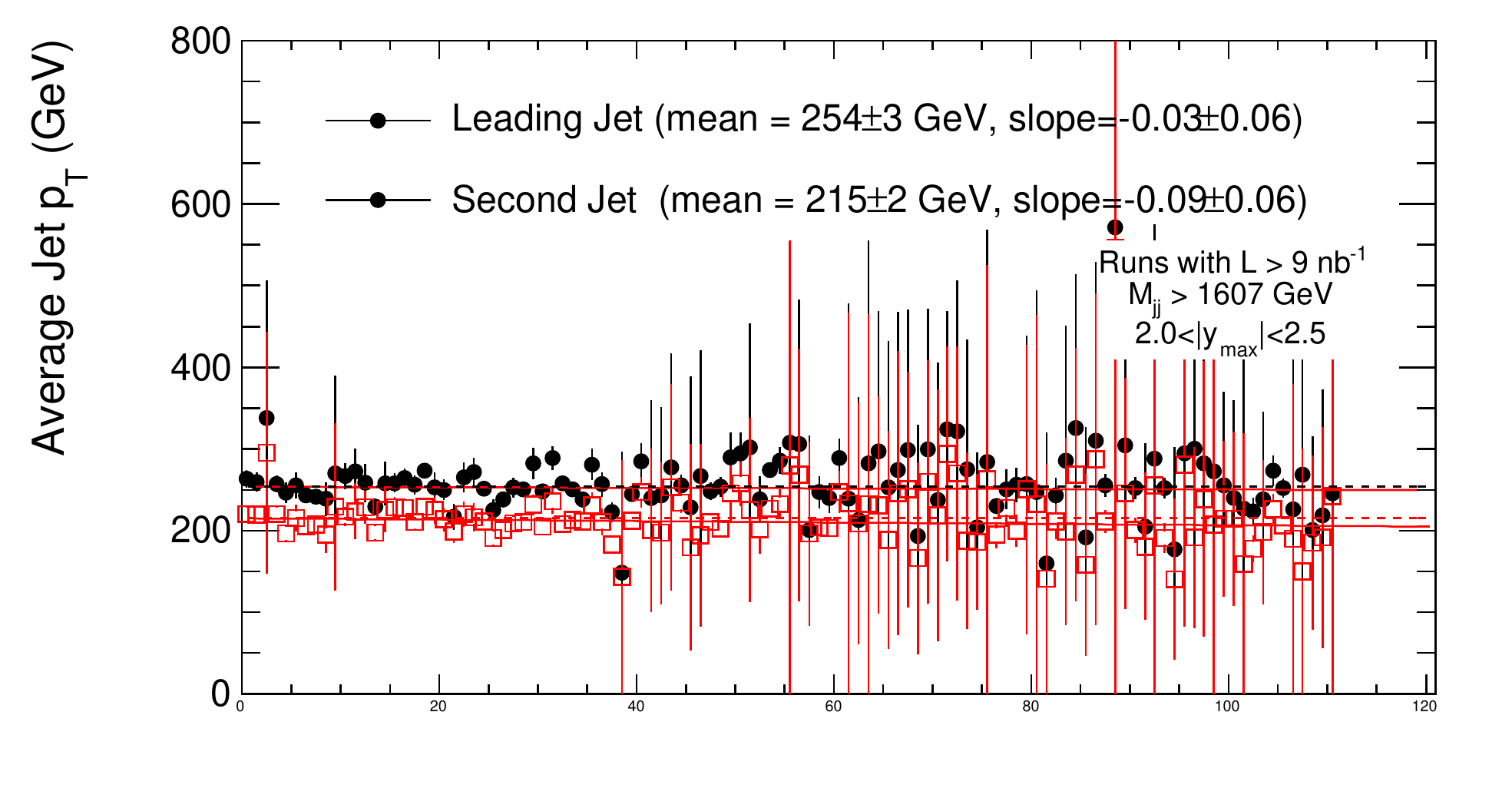} 
  
\capspace
\caption{ The $p_T$ of the leading and second jet  for the five different $y_{max}$ bins and for the
HLT$_{-}$Jet100U trigger as a function of time (run number), fitted with a first degree polynomial. }
\label{fig_appd9}
\end{figure}

\clearpage

\begin{figure}[h]
\centering

\includegraphics[width=0.49\textwidth]{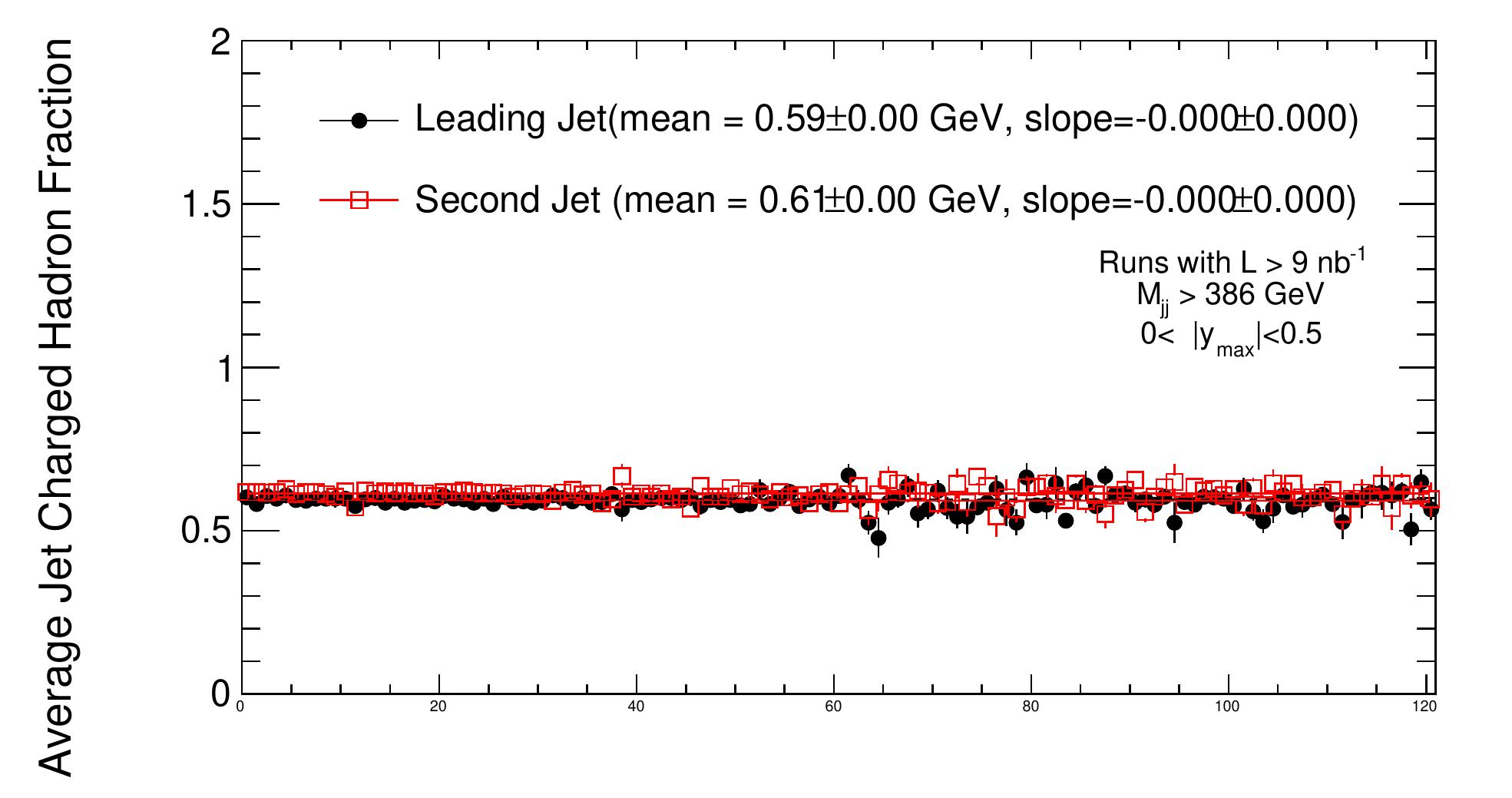} 
\includegraphics[width=0.49\textwidth]{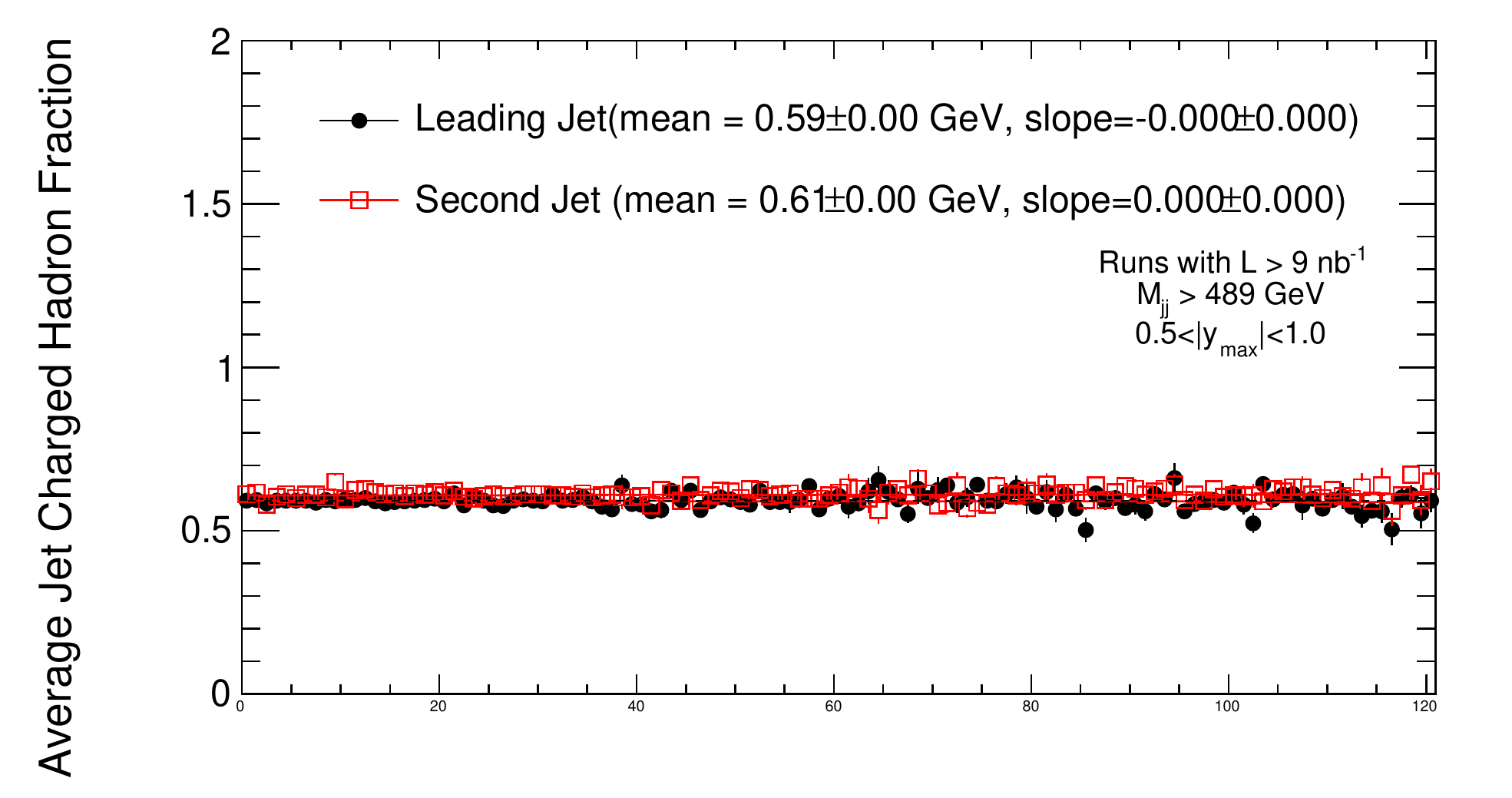} 
\includegraphics[width=0.49\textwidth]{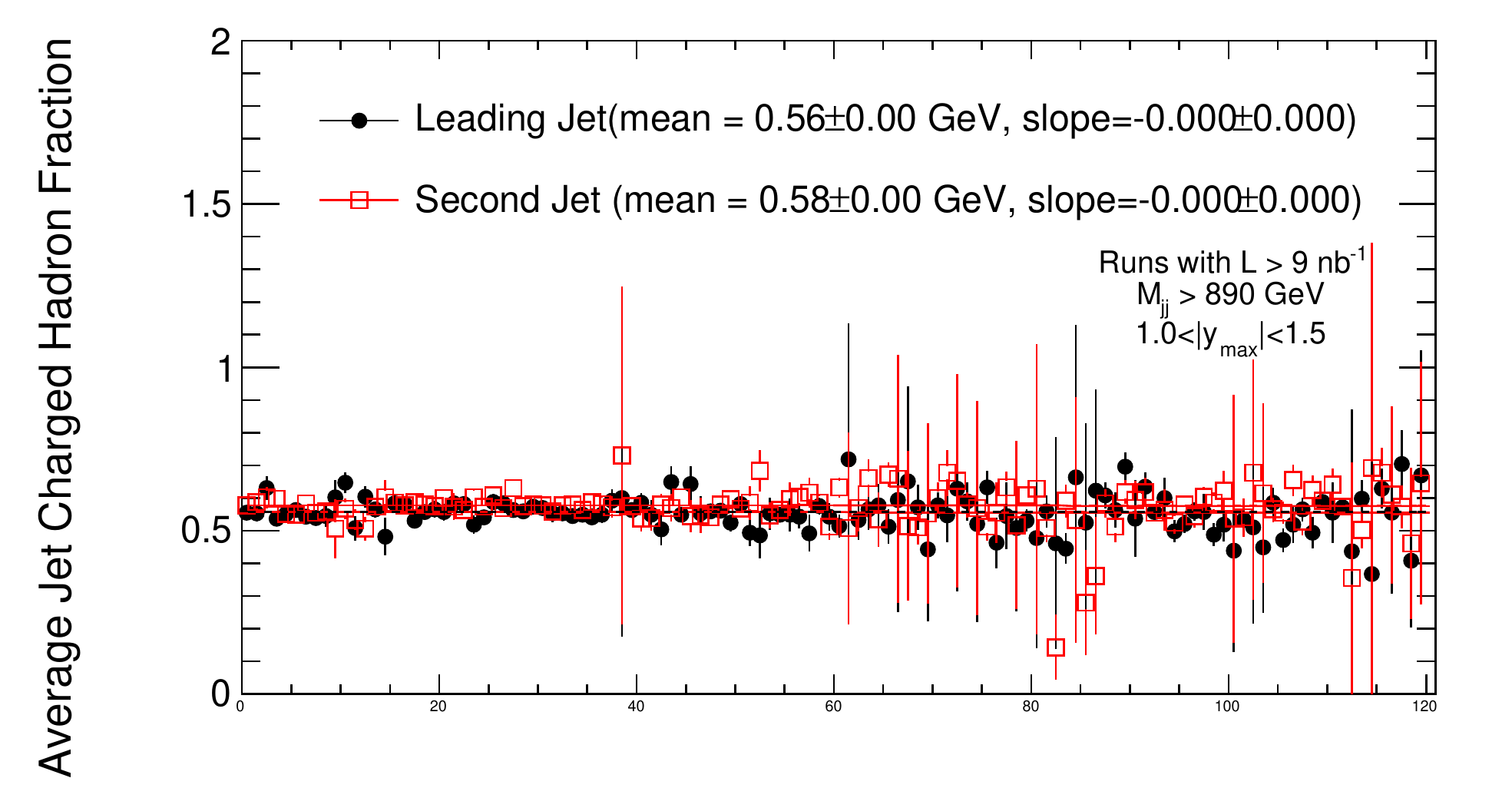} 
\includegraphics[width=0.49\textwidth]{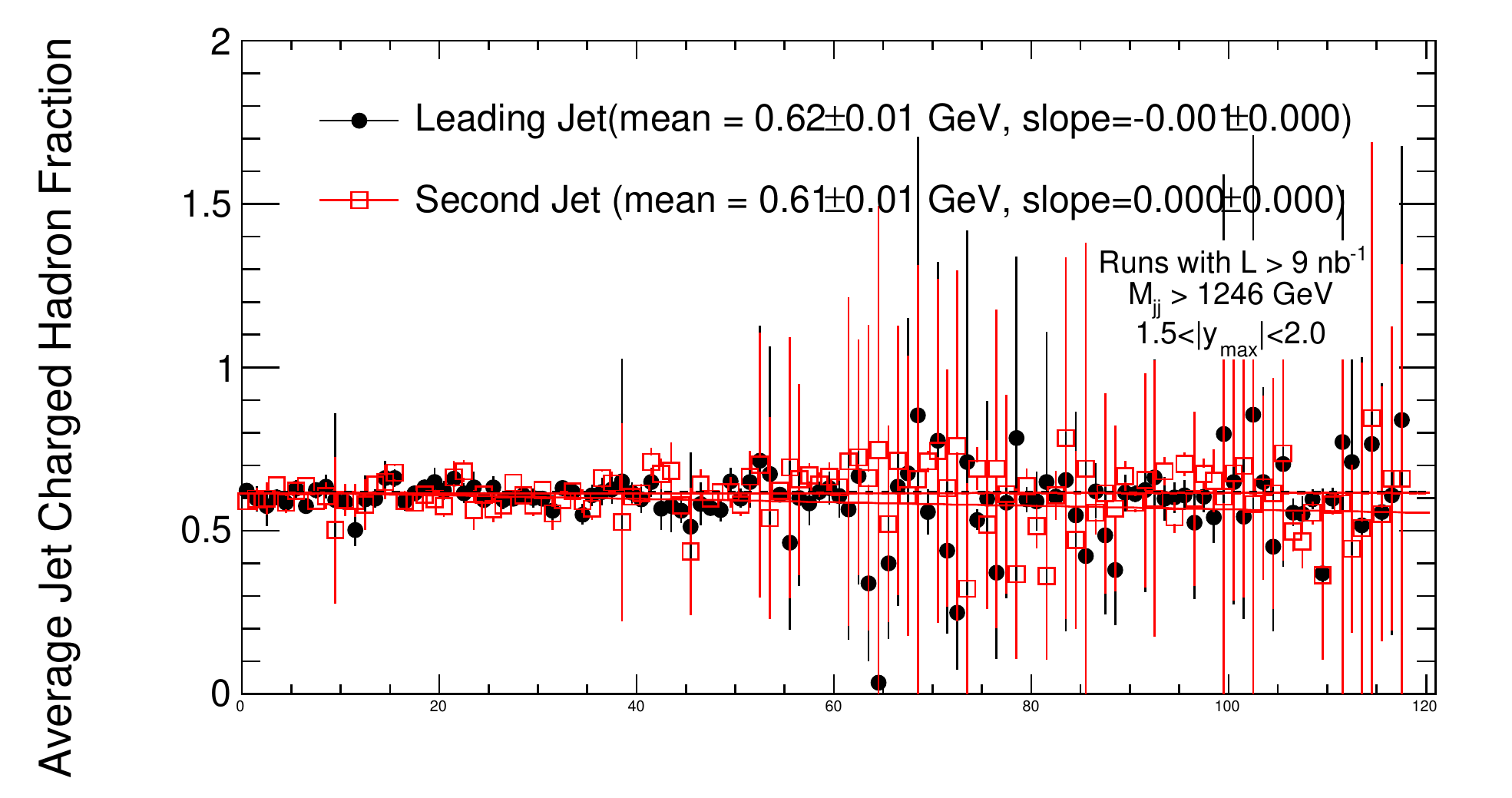} 
\includegraphics[width=0.49\textwidth]{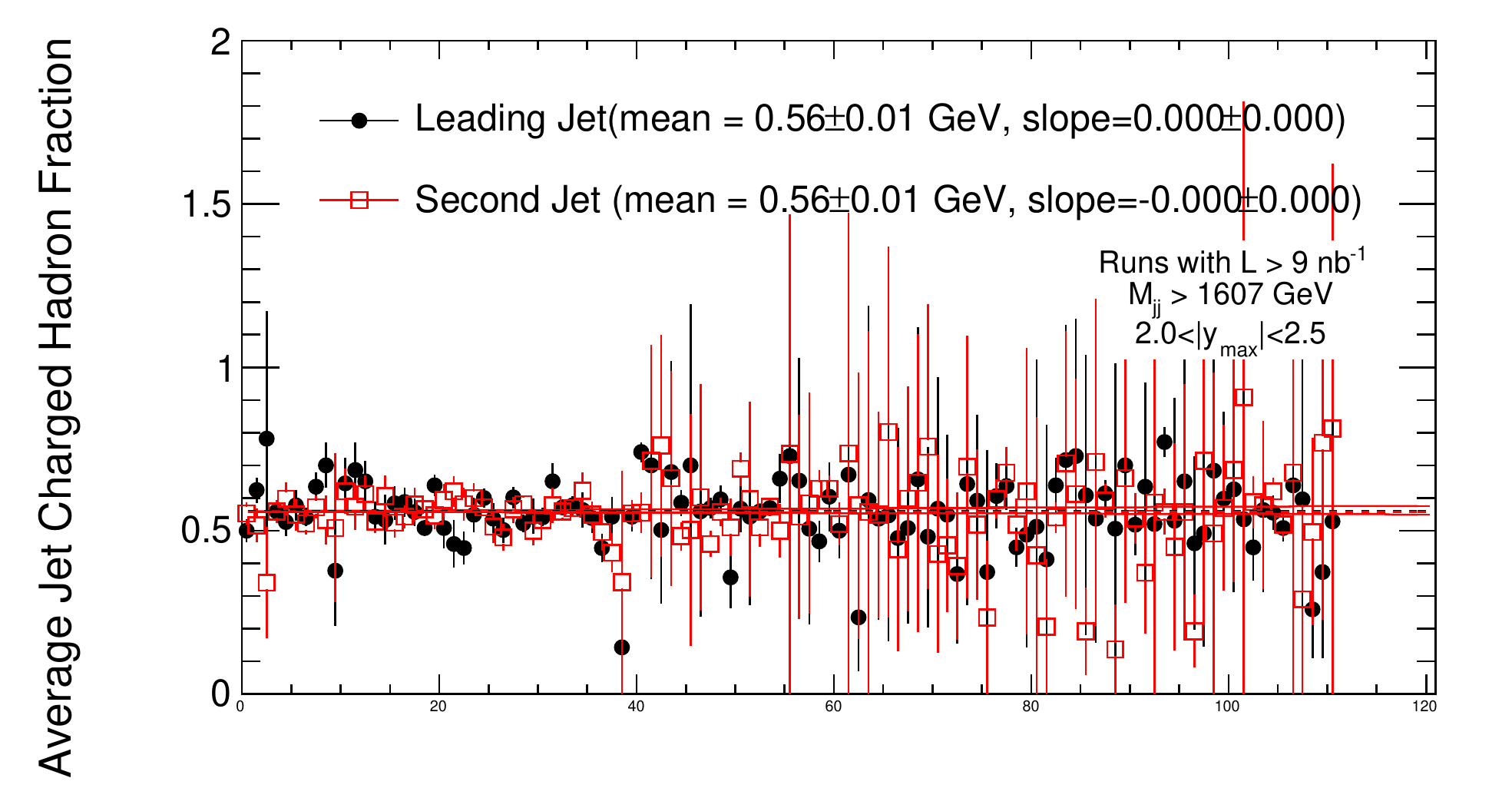} 
  
\capspace
\caption{ The charged hadron fraction  of the leading and second jet  for the five different $y_{max}$ bins and for the
HLT$_{-}$Jet100U trigger as a function of time (run number), fitted with a first degree polynomial. }
\label{fig_appd10}
\end{figure}

\begin{figure}[h]
\centering

\includegraphics[width=0.49\textwidth]{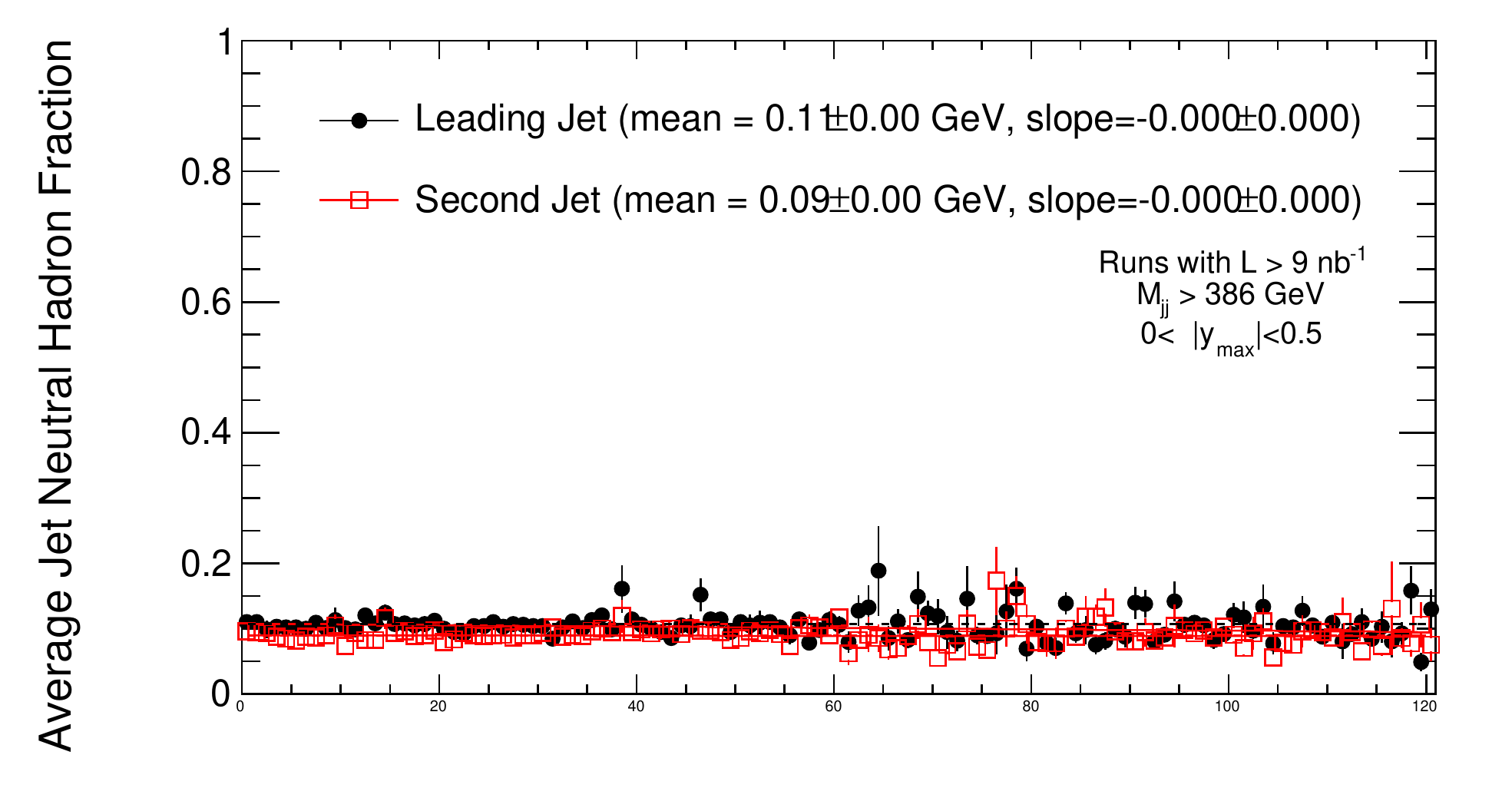} 
\includegraphics[width=0.49\textwidth]{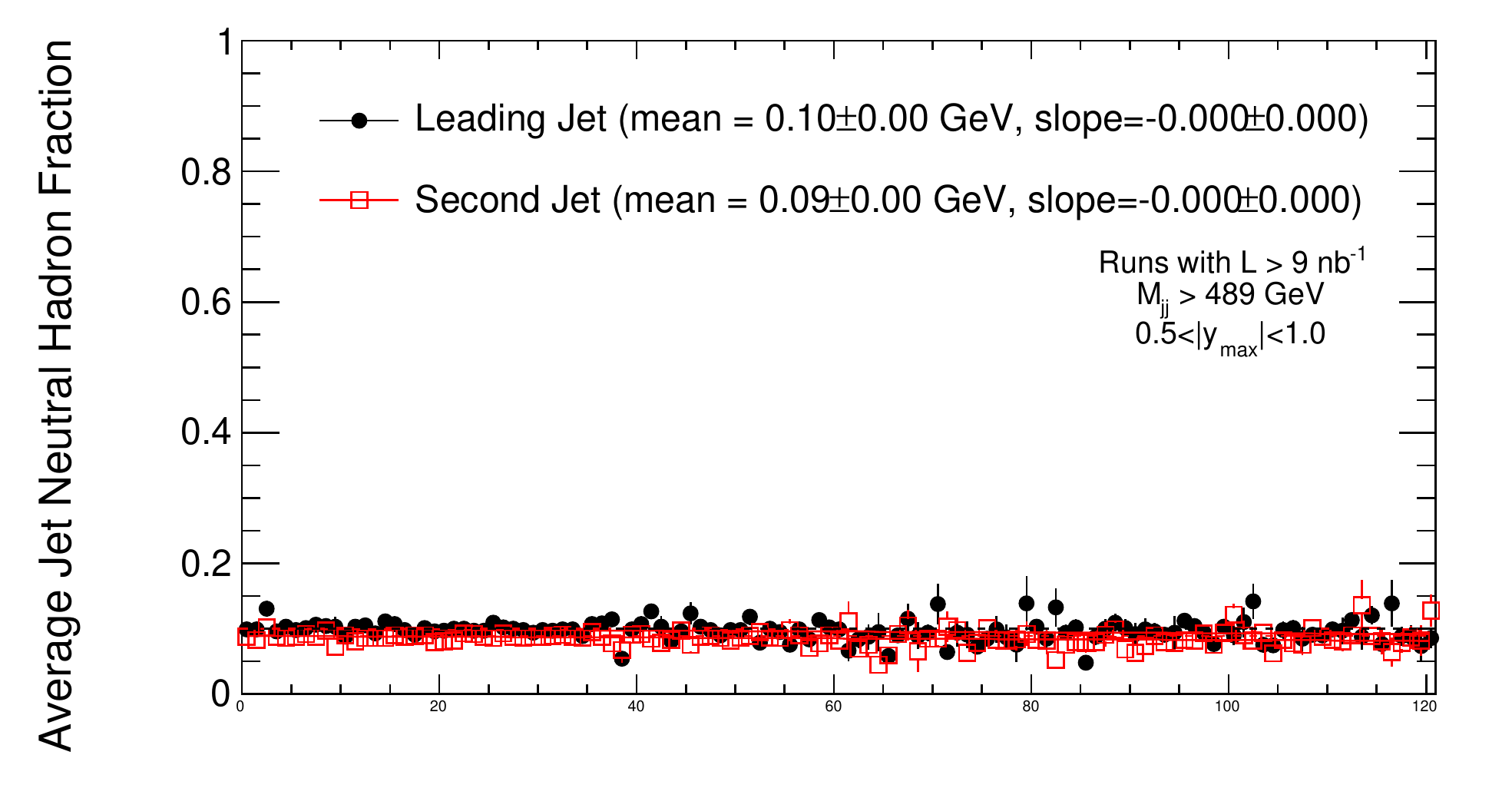} 
\includegraphics[width=0.49\textwidth]{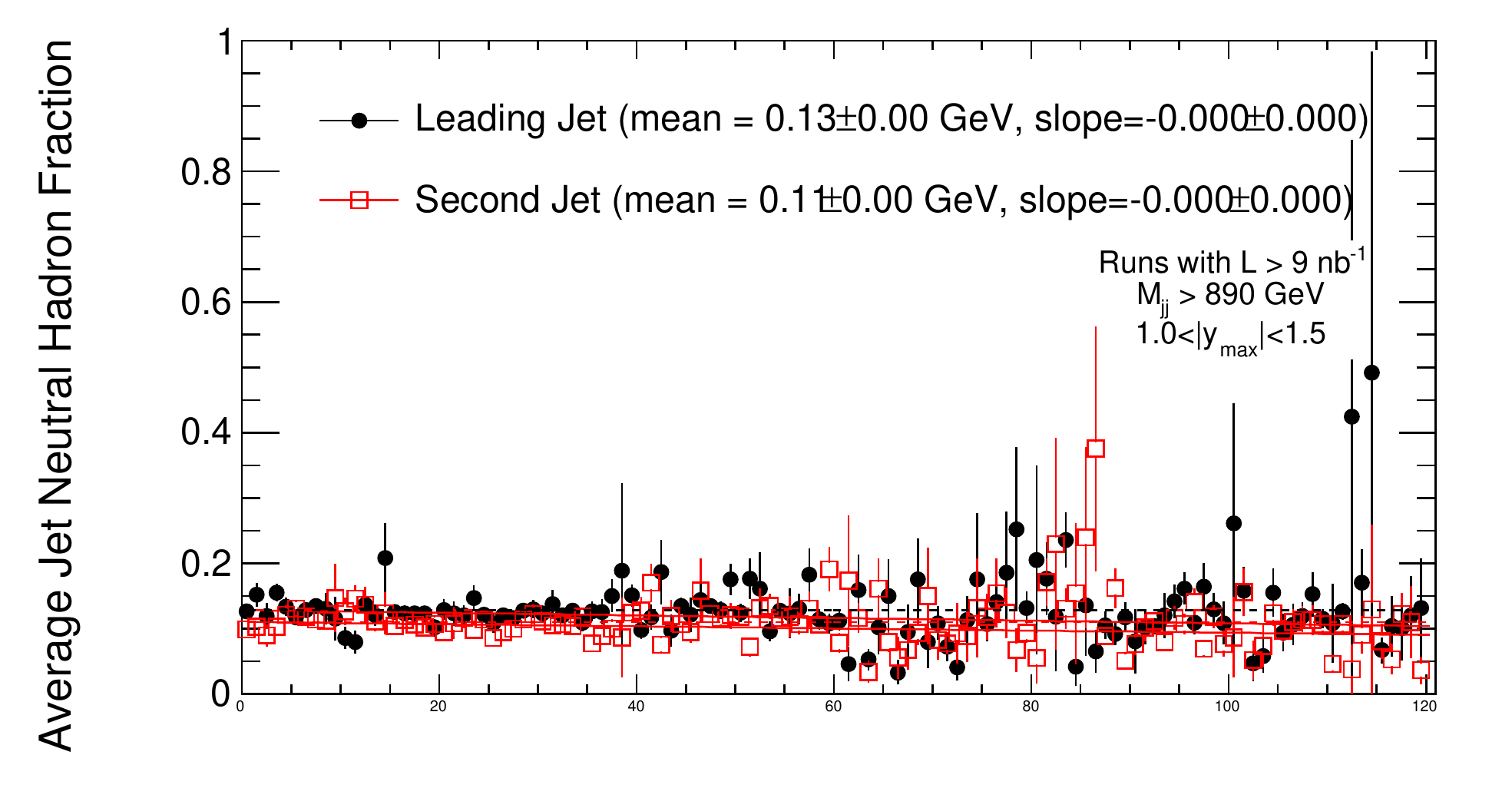} 
\includegraphics[width=0.49\textwidth]{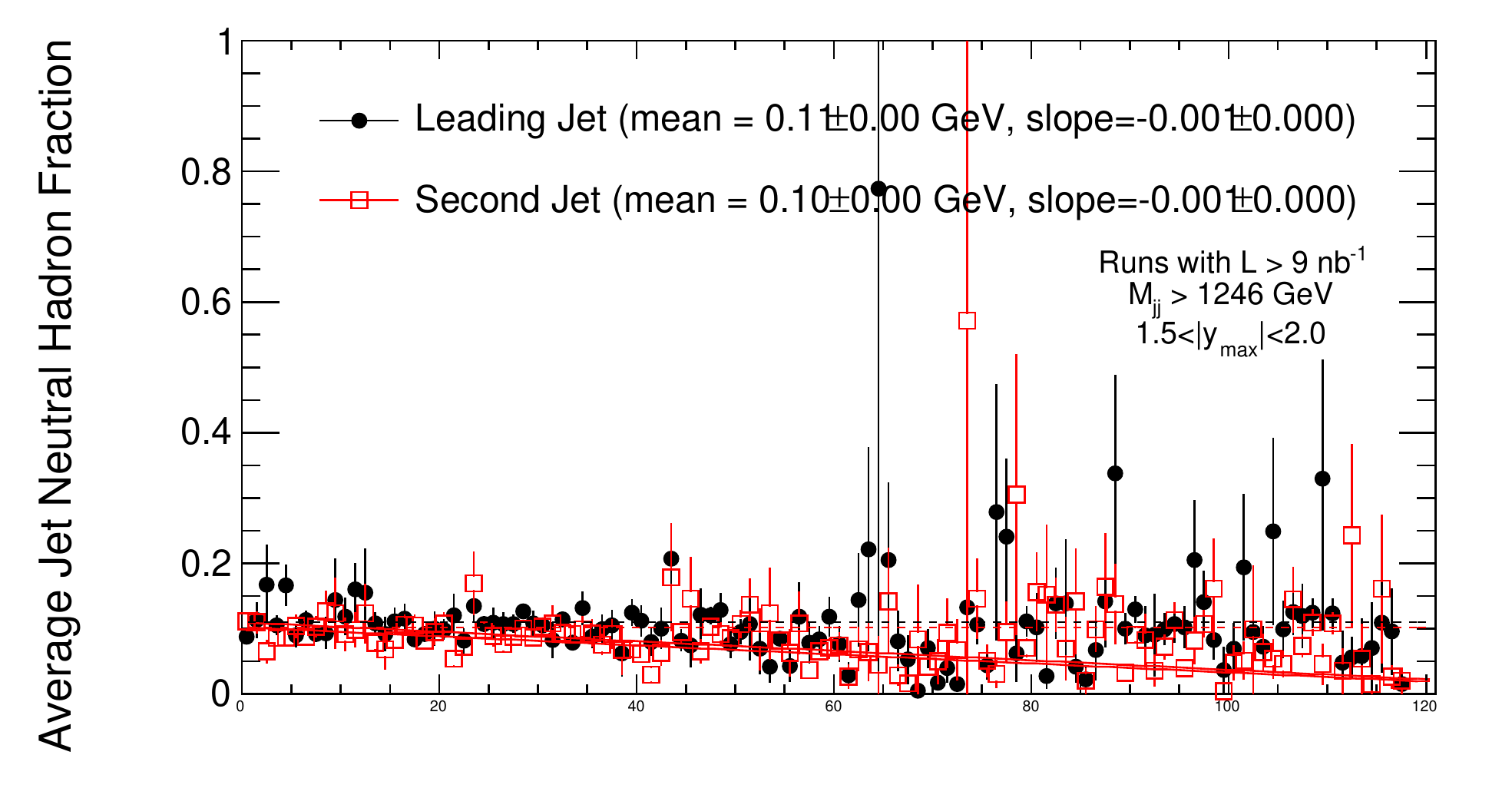} 
\includegraphics[width=0.49\textwidth]{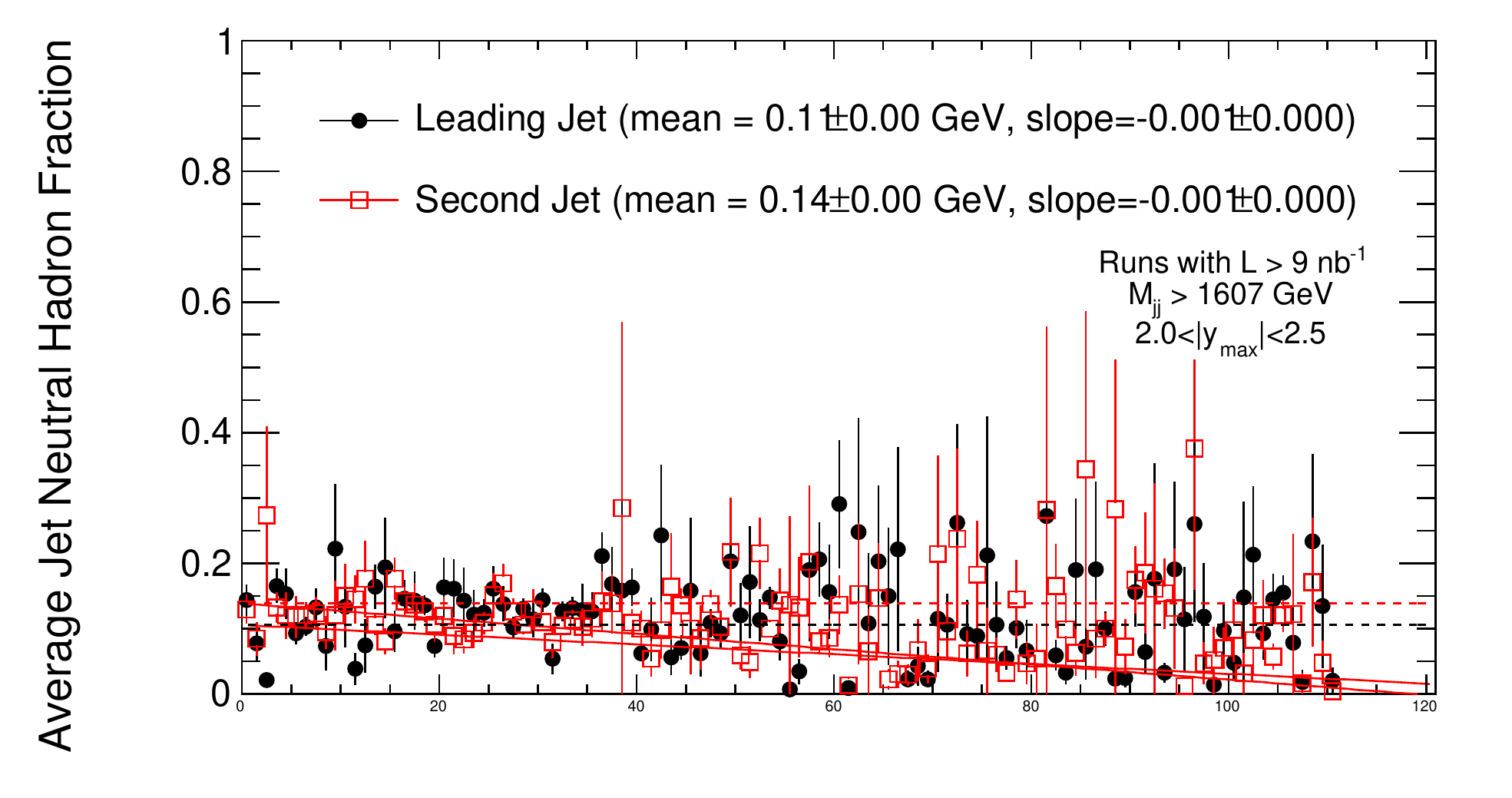} 
  
\capspace
\caption{ The neutral  hadron fraction  of the leading and second jet  for the five different $y_{max}$ bins and for the
HLT$_{-}$Jet100U trigger as a function of time (run number), fitted with a first degree polynomial. }
\label{fig_appd11}
\end{figure}

\begin{figure}[h]
\centering

\includegraphics[width=0.49\textwidth]{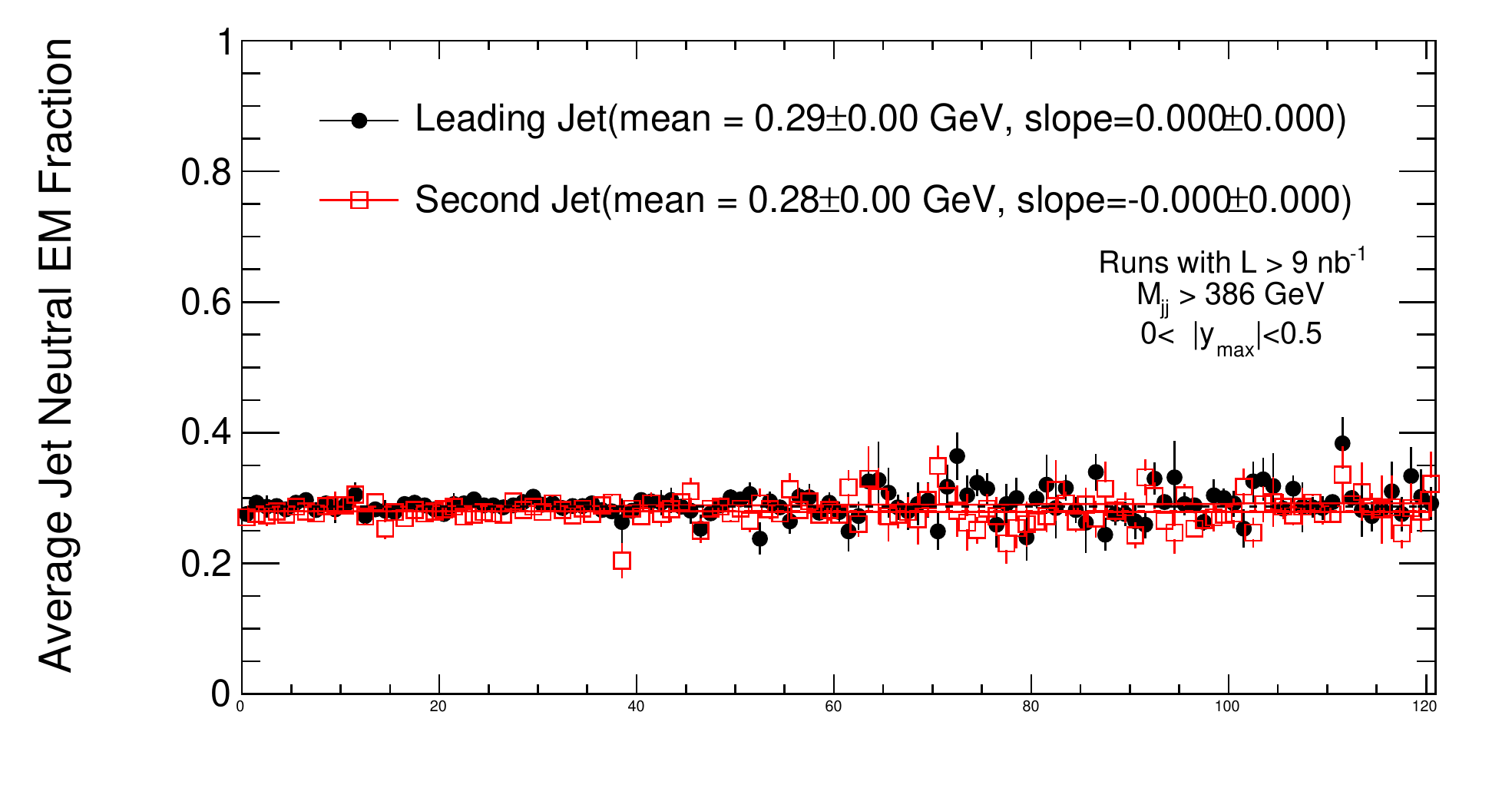} 
\includegraphics[width=0.49\textwidth]{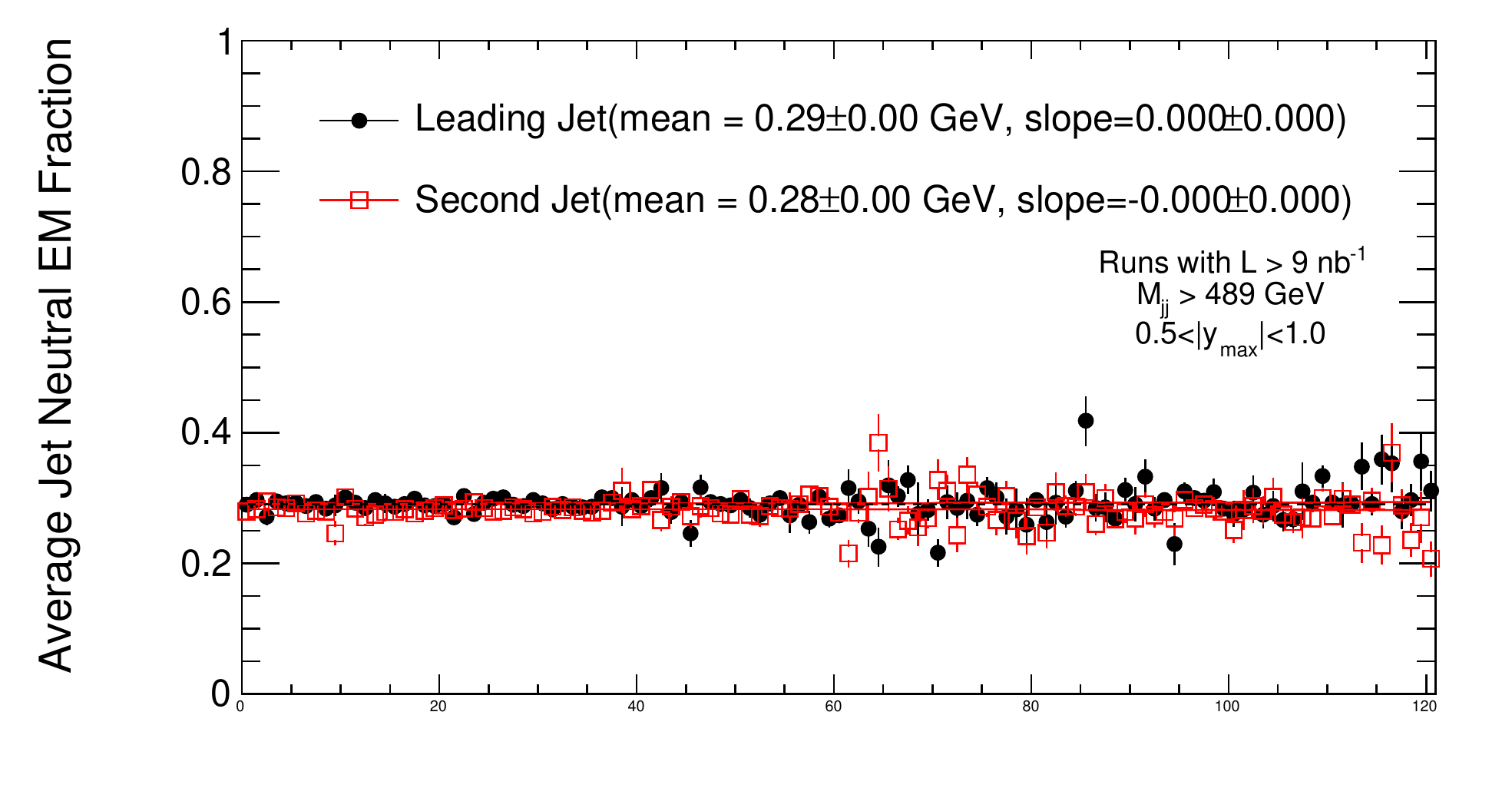} 
\includegraphics[width=0.49\textwidth]{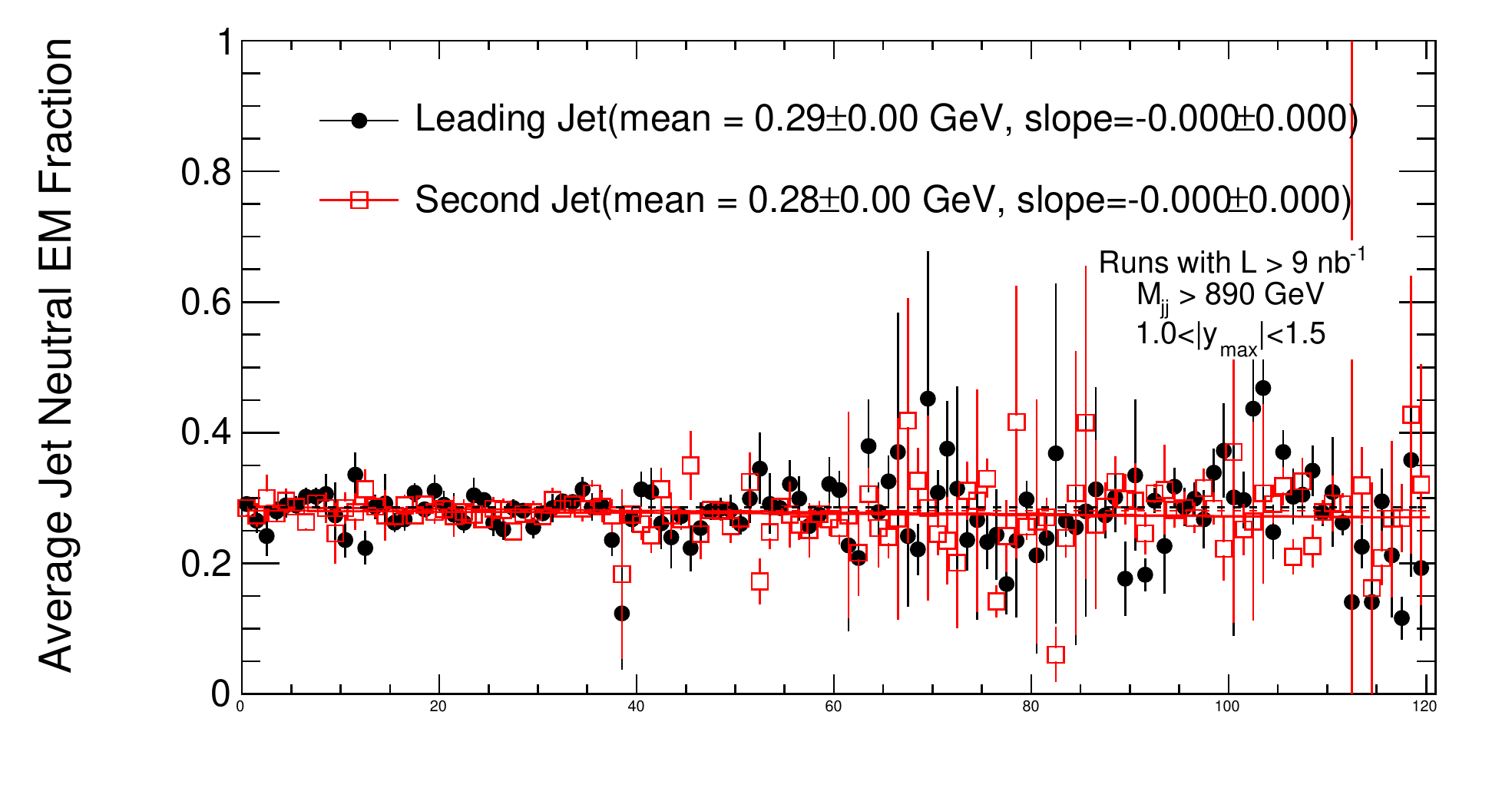} 
\includegraphics[width=0.49\textwidth]{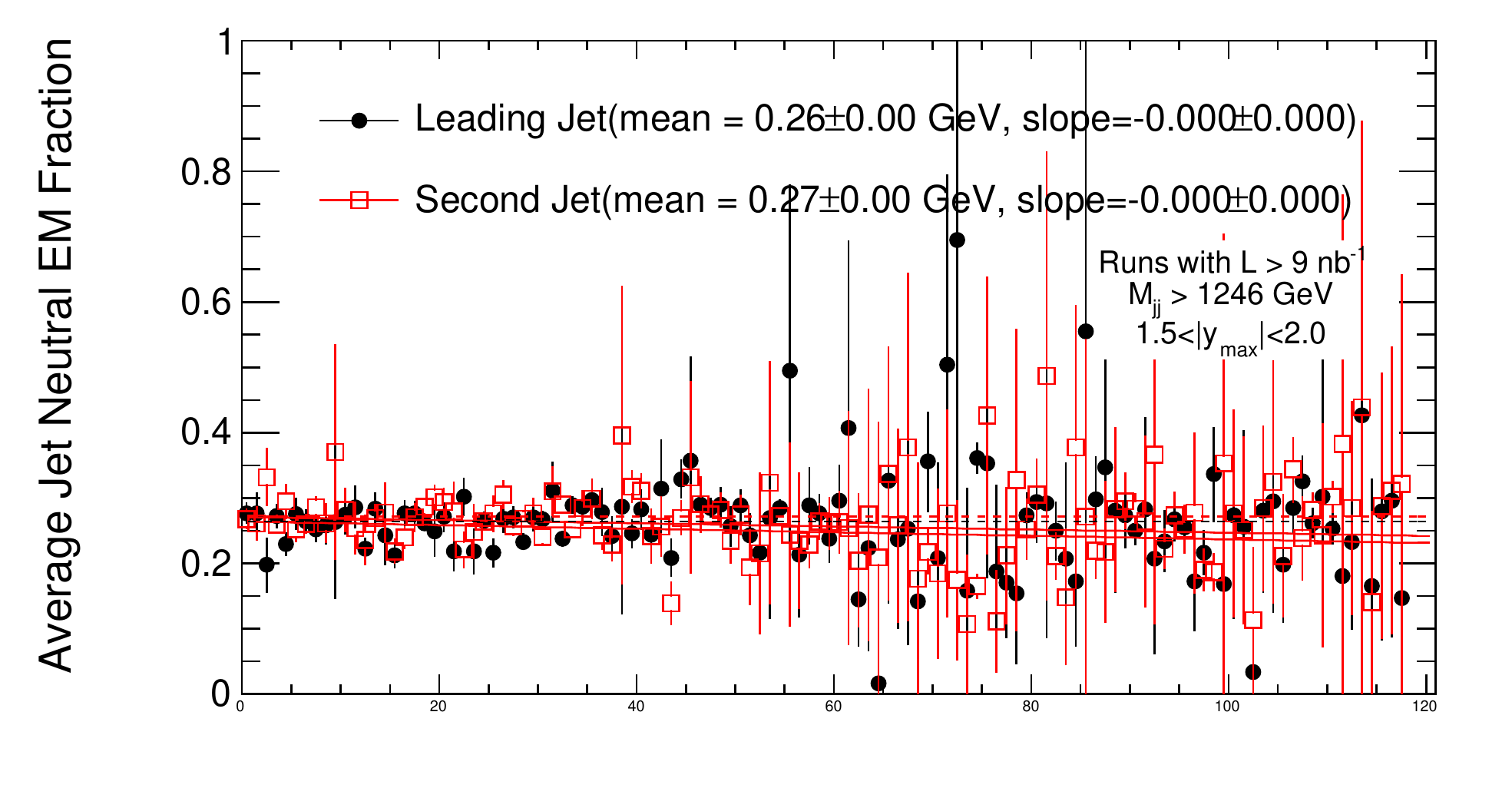} 
\includegraphics[width=0.49\textwidth]{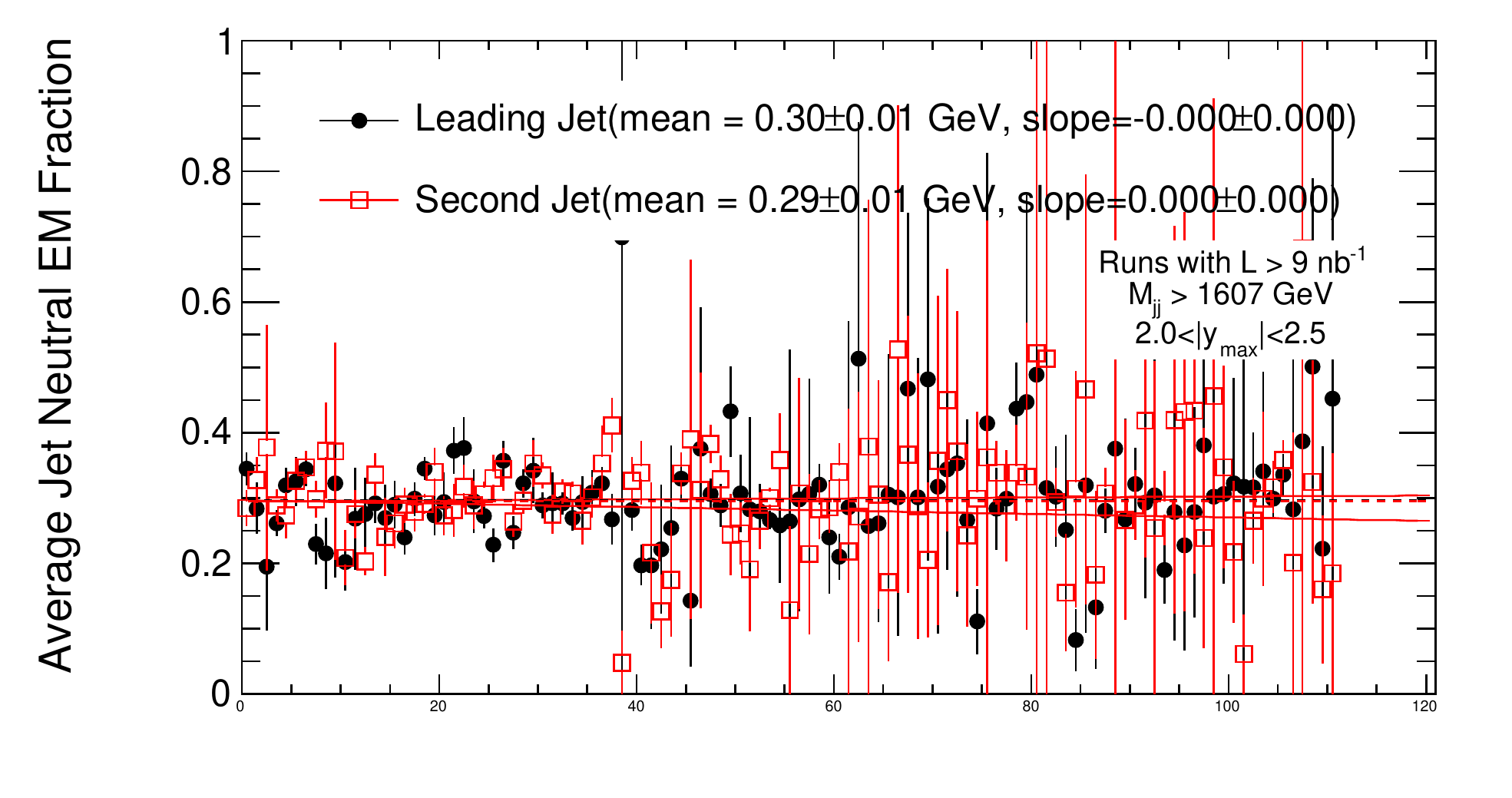} 
  
\capspace
\caption{ The neutral  electromagnetic fraction  of the leading and second jet  for the five different $y_{max}$ bins and for the
HLT$_{-}$Jet100U trigger as a function of time (run number), fitted with a first degree polynomial. }
\label{fig_appd12}
\end{figure}

\clearpage
%%% jet 140

\begin{figure}[h]
\centering

\includegraphics[width=0.49\textwidth]{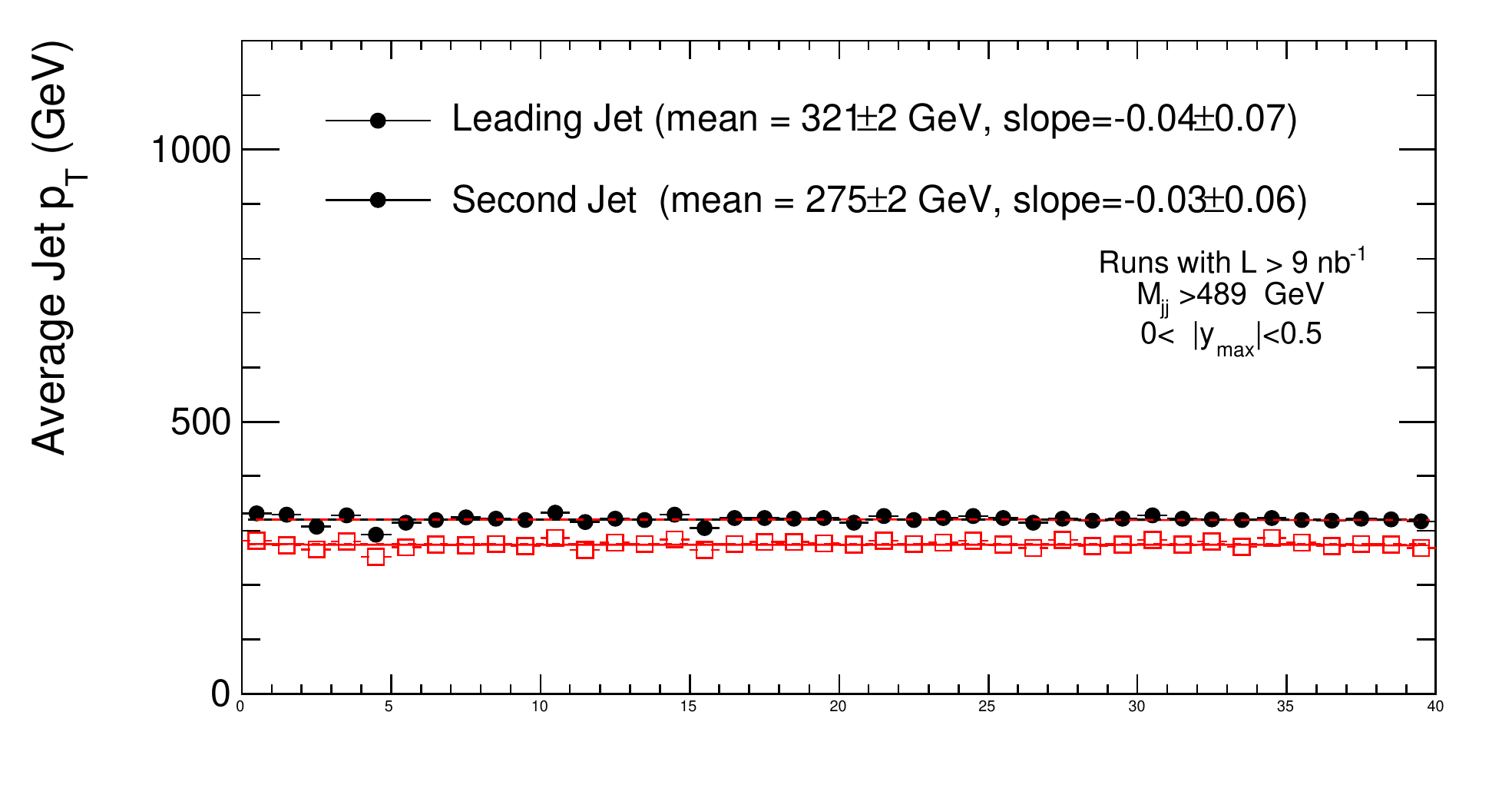} 
\includegraphics[width=0.49\textwidth]{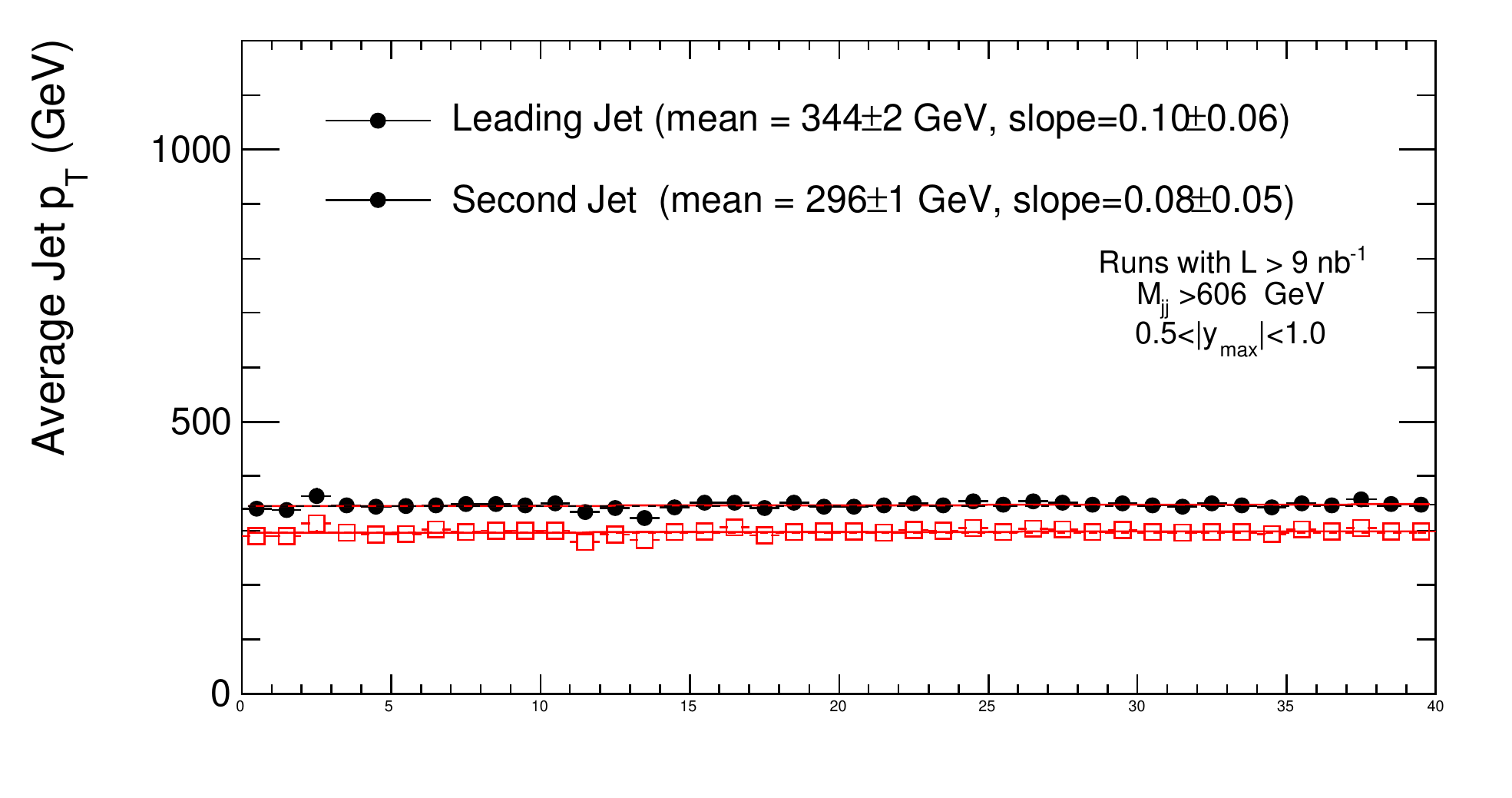} 
\includegraphics[width=0.49\textwidth]{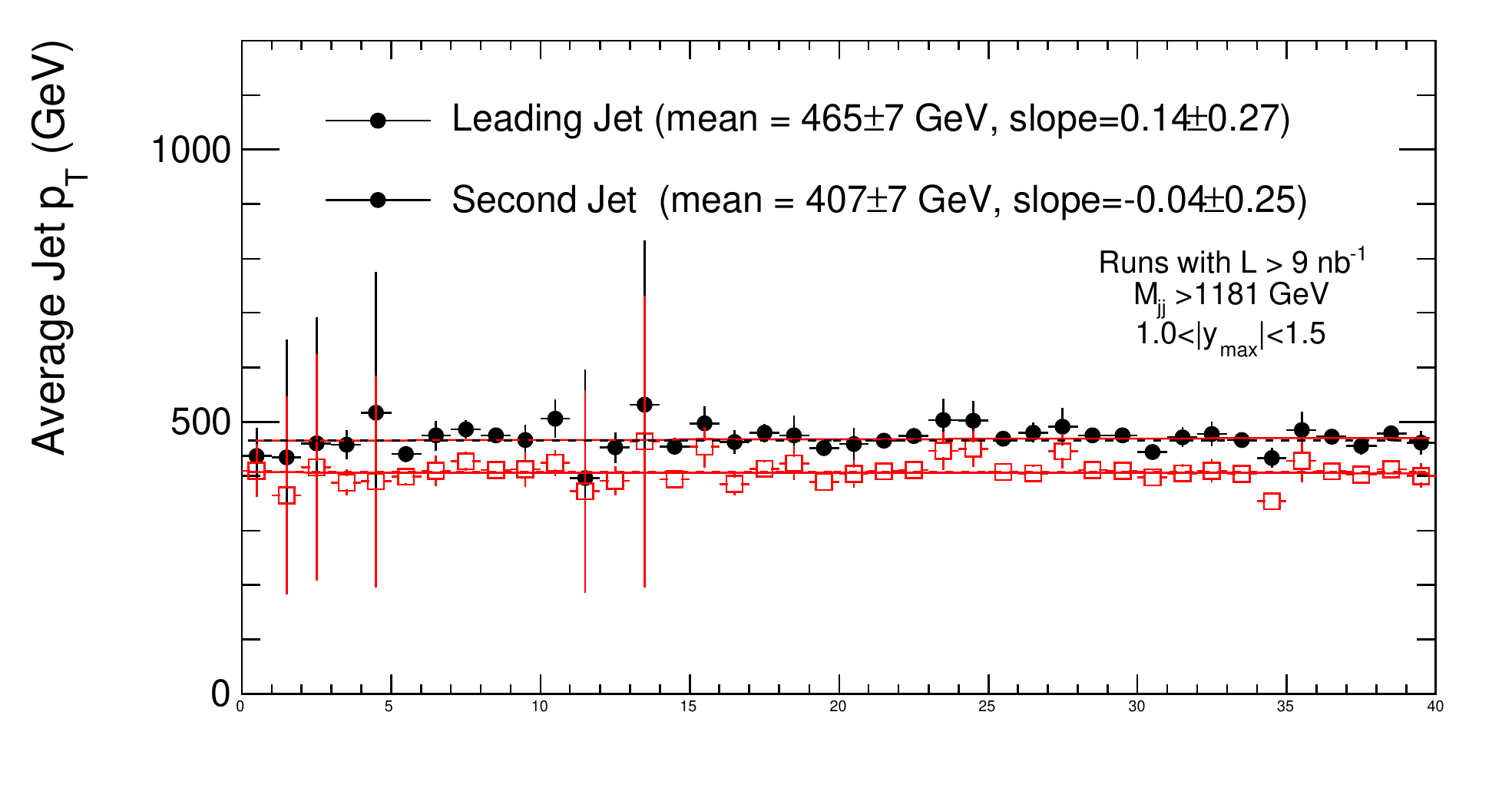} 
\includegraphics[width=0.49\textwidth]{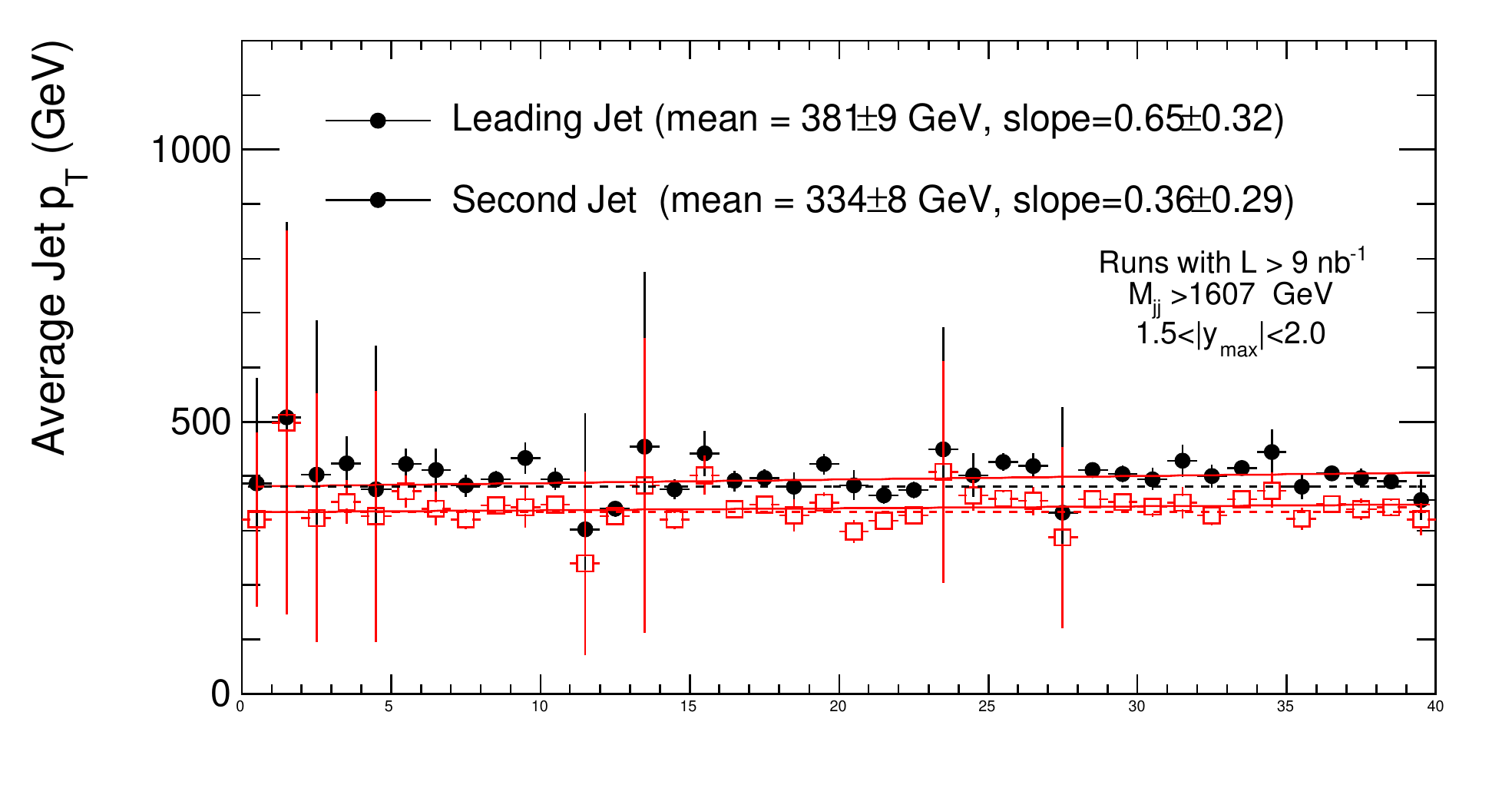} 
\includegraphics[width=0.49\textwidth]{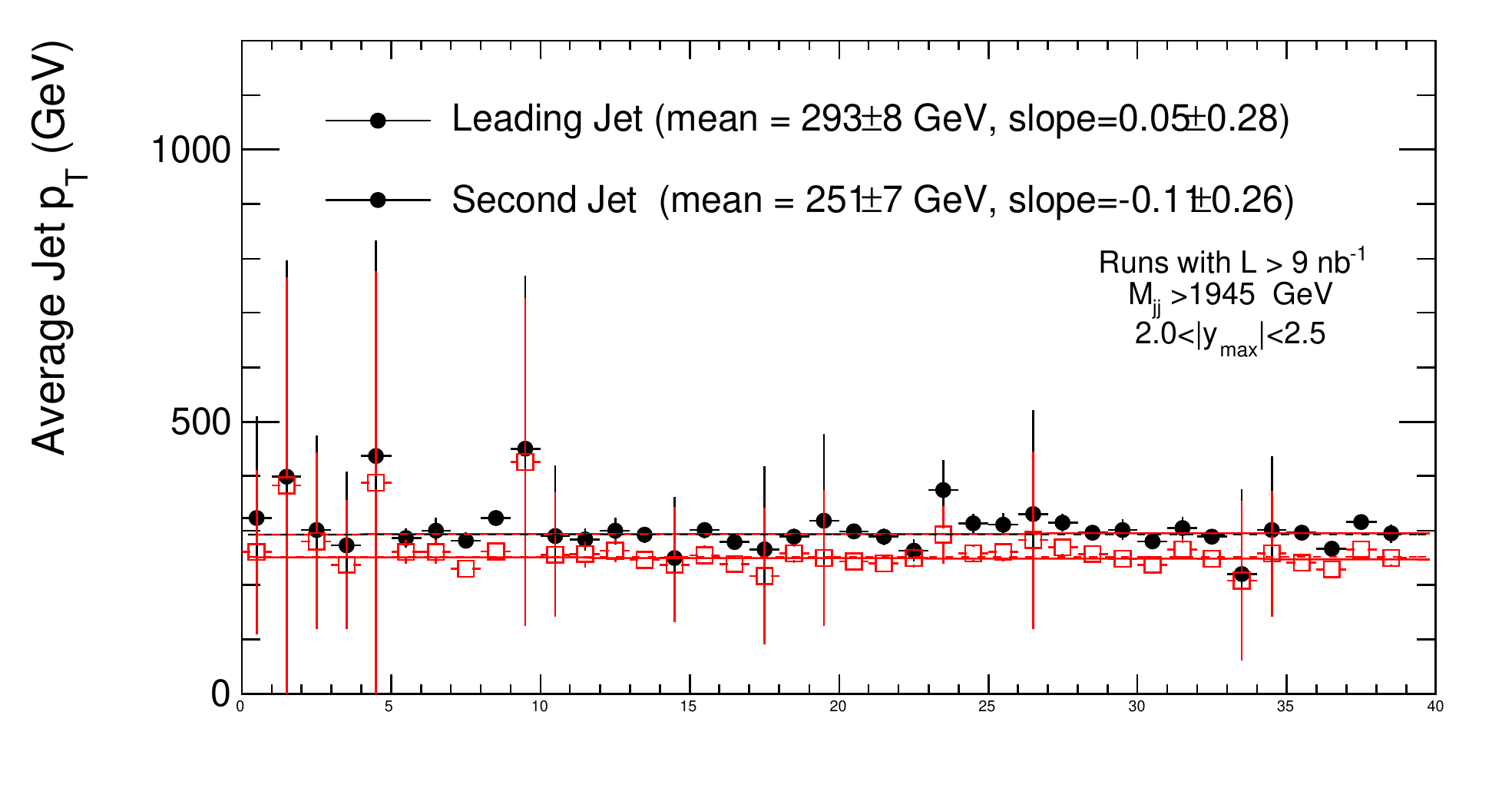} 
  
\capspace
\caption{ The $p_T$ of the leading and second jet  for the five different $y_{max}$ bins and for the
HLT$_{-}$Jet140U trigger as a function of time (run number), fitted with a first degree polynomial. }
\label{fig_appd13}
\end{figure}

\begin{figure}[h]
\centering

\includegraphics[width=0.49\textwidth]{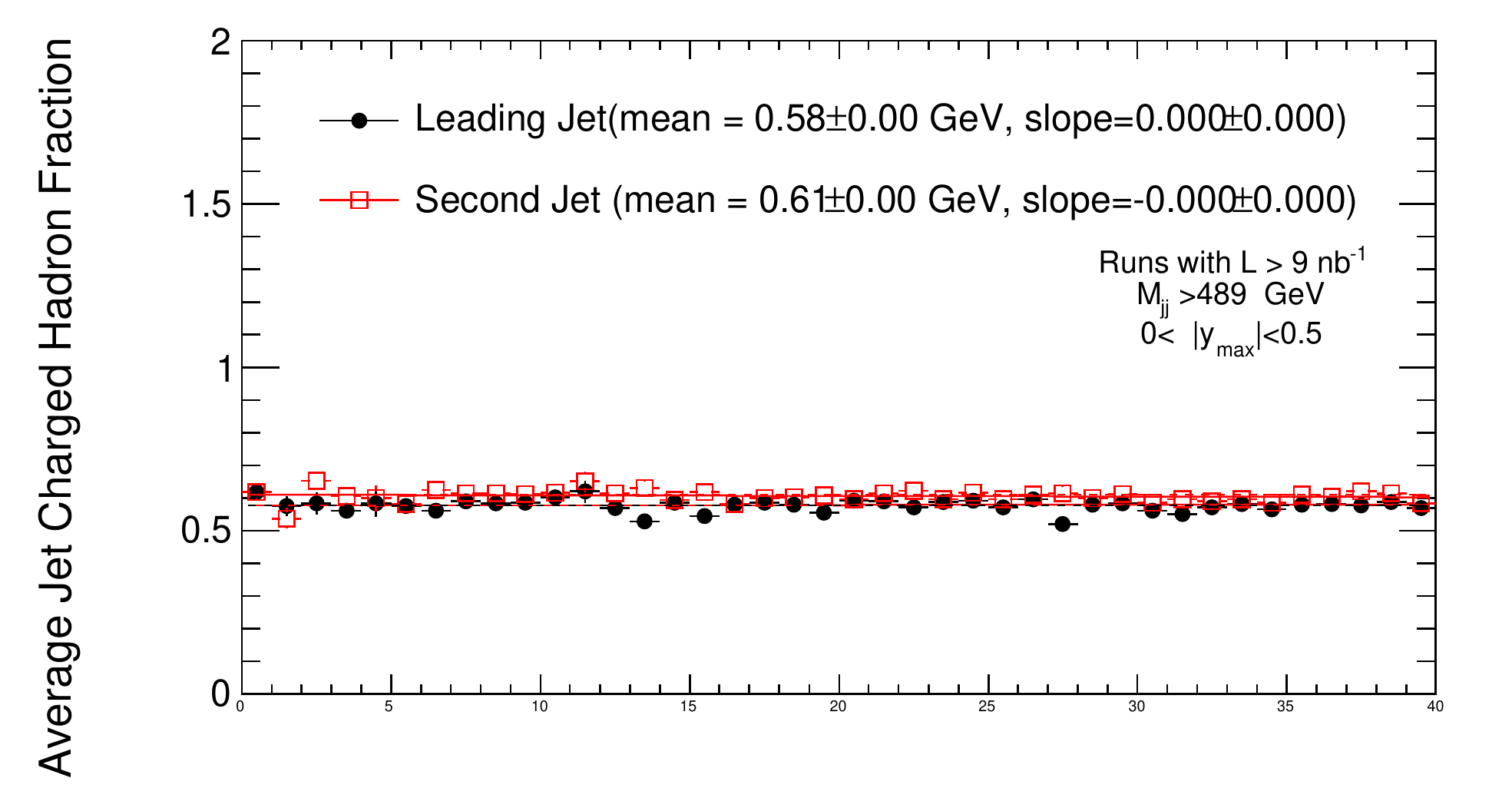} 
\includegraphics[width=0.49\textwidth]{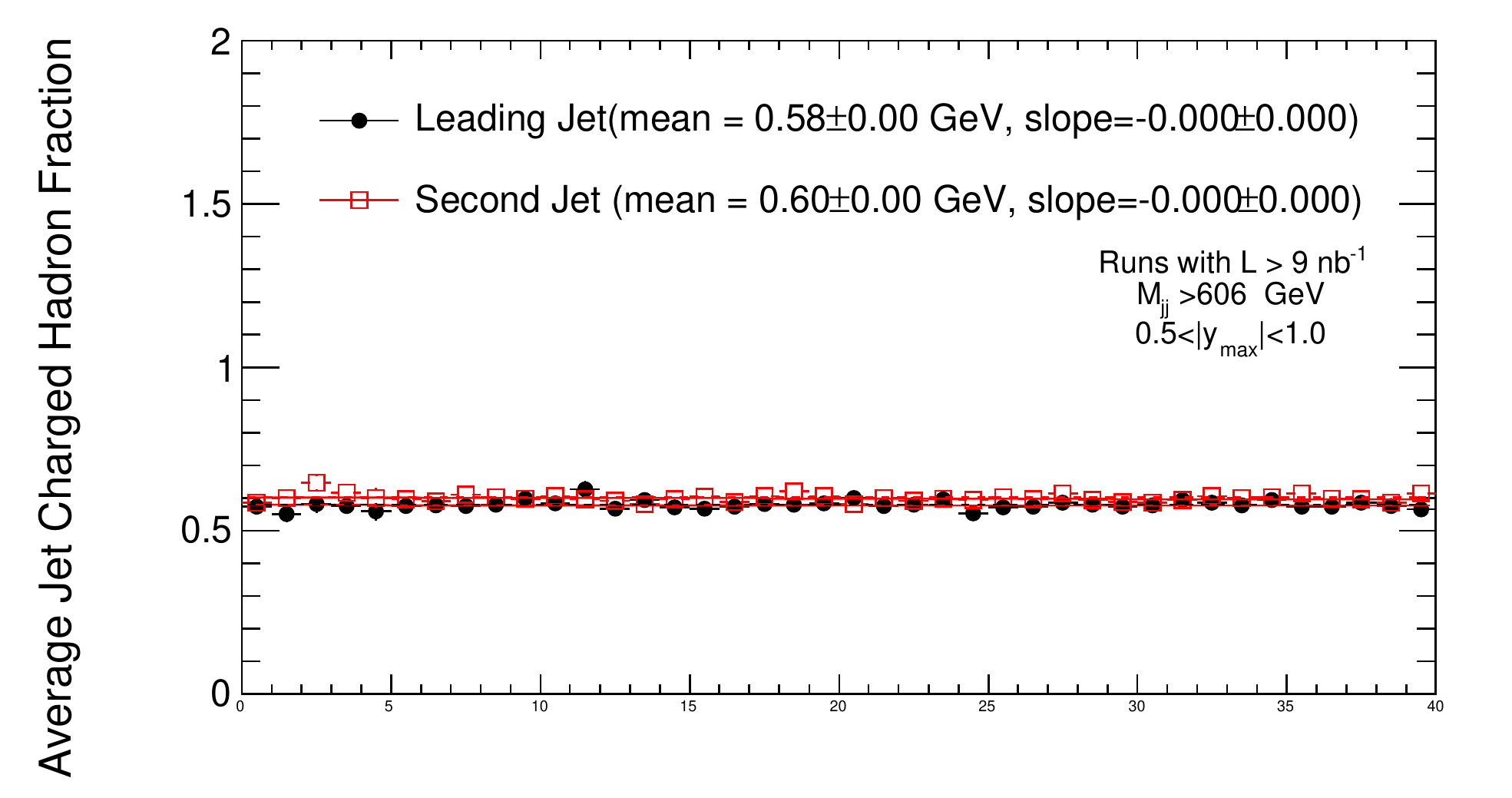} 
\includegraphics[width=0.49\textwidth]{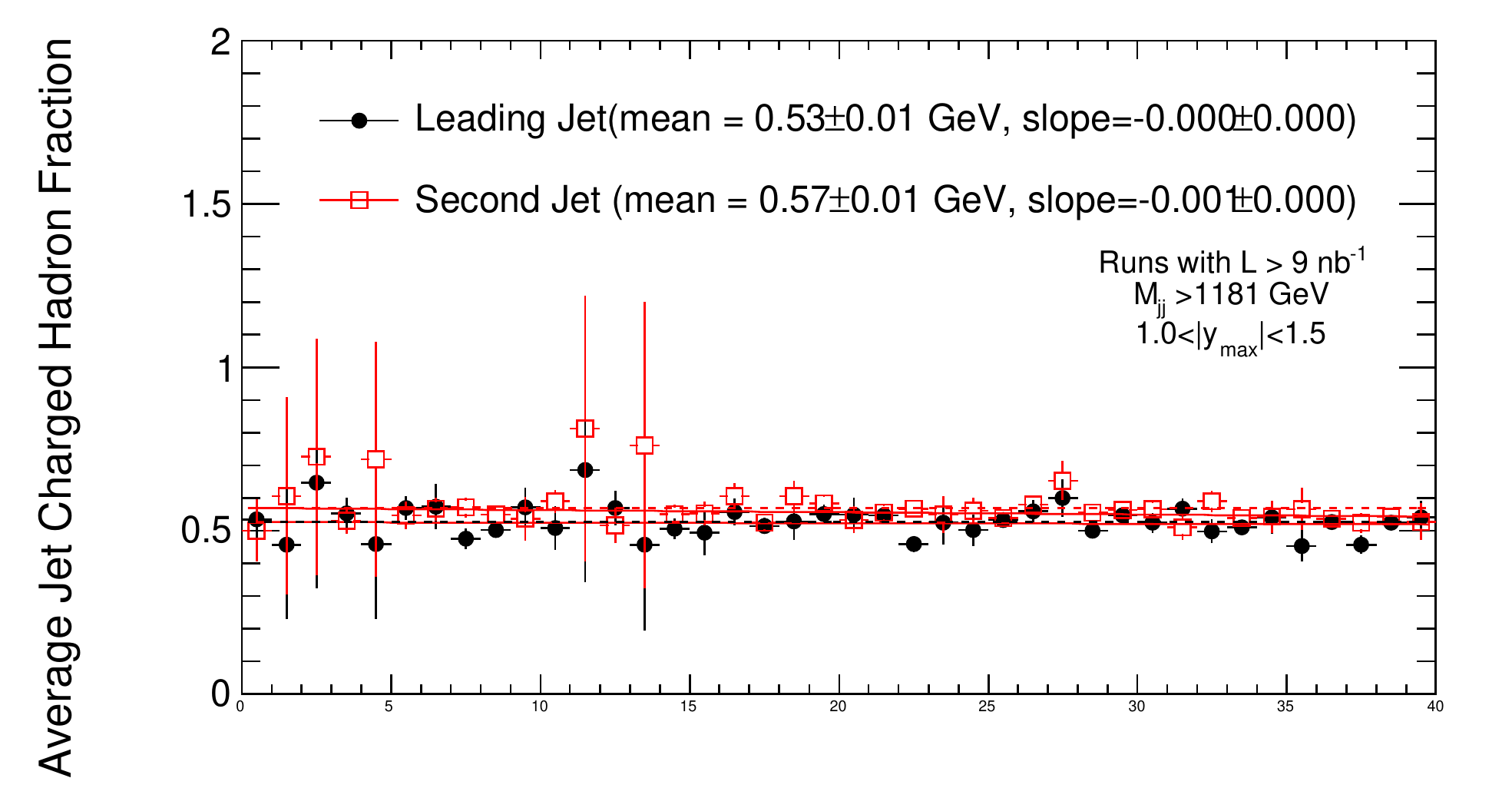} 
\includegraphics[width=0.49\textwidth]{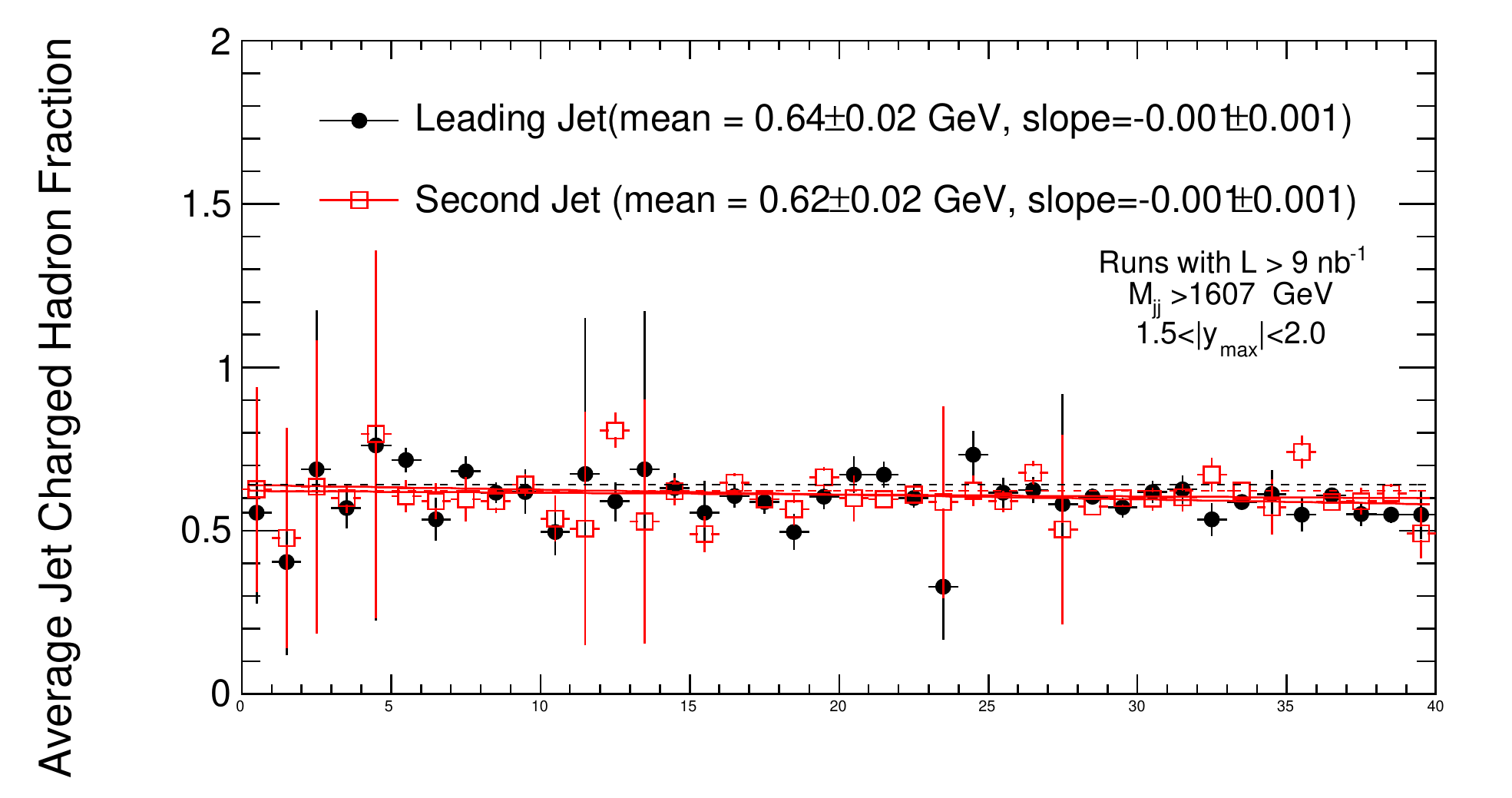} 
\includegraphics[width=0.49\textwidth]{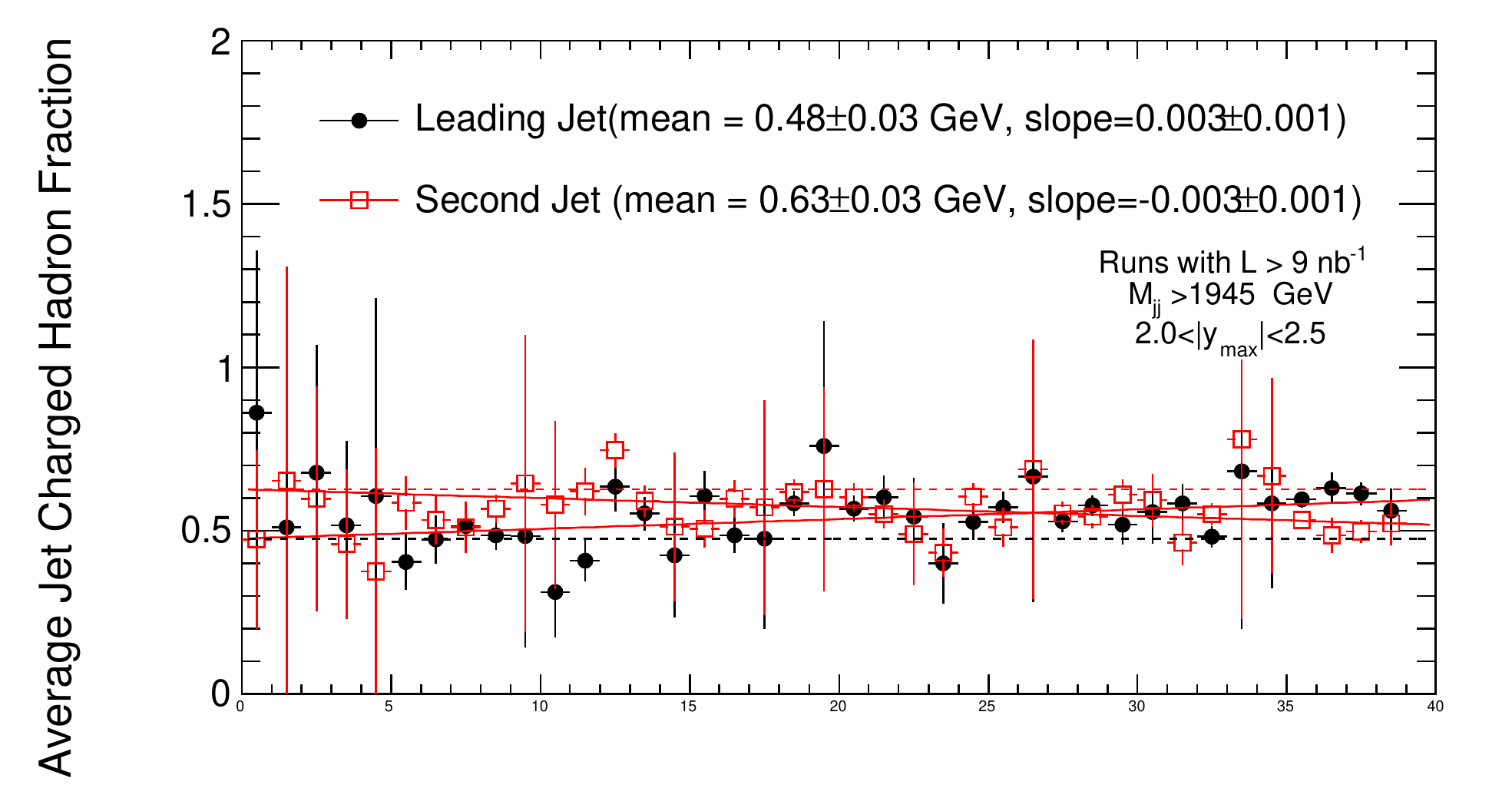} 
 
\capspace
\caption{ The charged hadron fraction  of the leading and second jet  for the five different $y_{max}$ bins and for the
HLT$_{-}$Jet140U trigger as a function of time (run number), fitted with a first degree polynomial. }
\label{fig_appd14}
\end{figure}

\begin{figure}[h]
\centering

\includegraphics[width=0.49\textwidth]{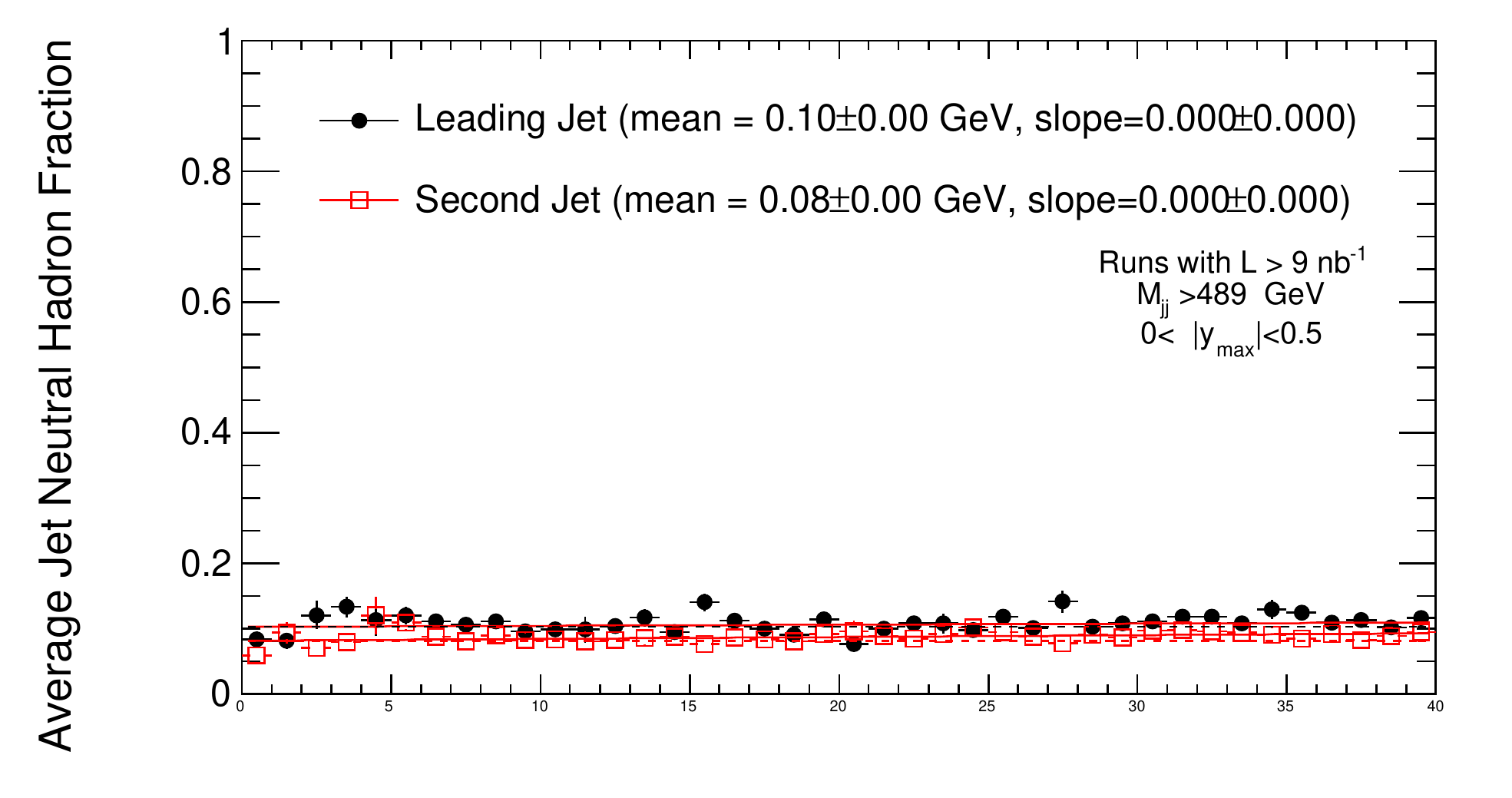} 
\includegraphics[width=0.49\textwidth]{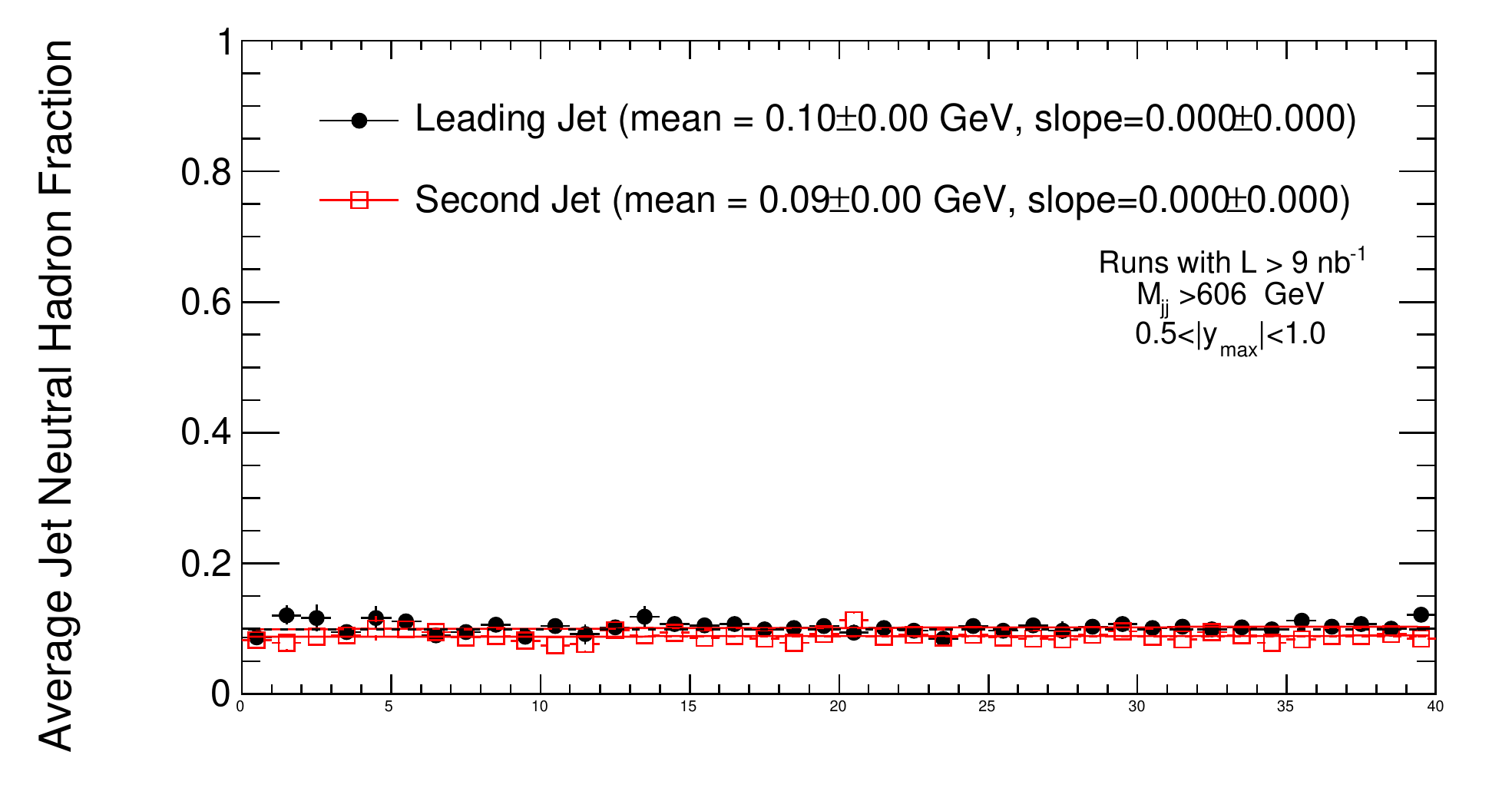} 
\includegraphics[width=0.49\textwidth]{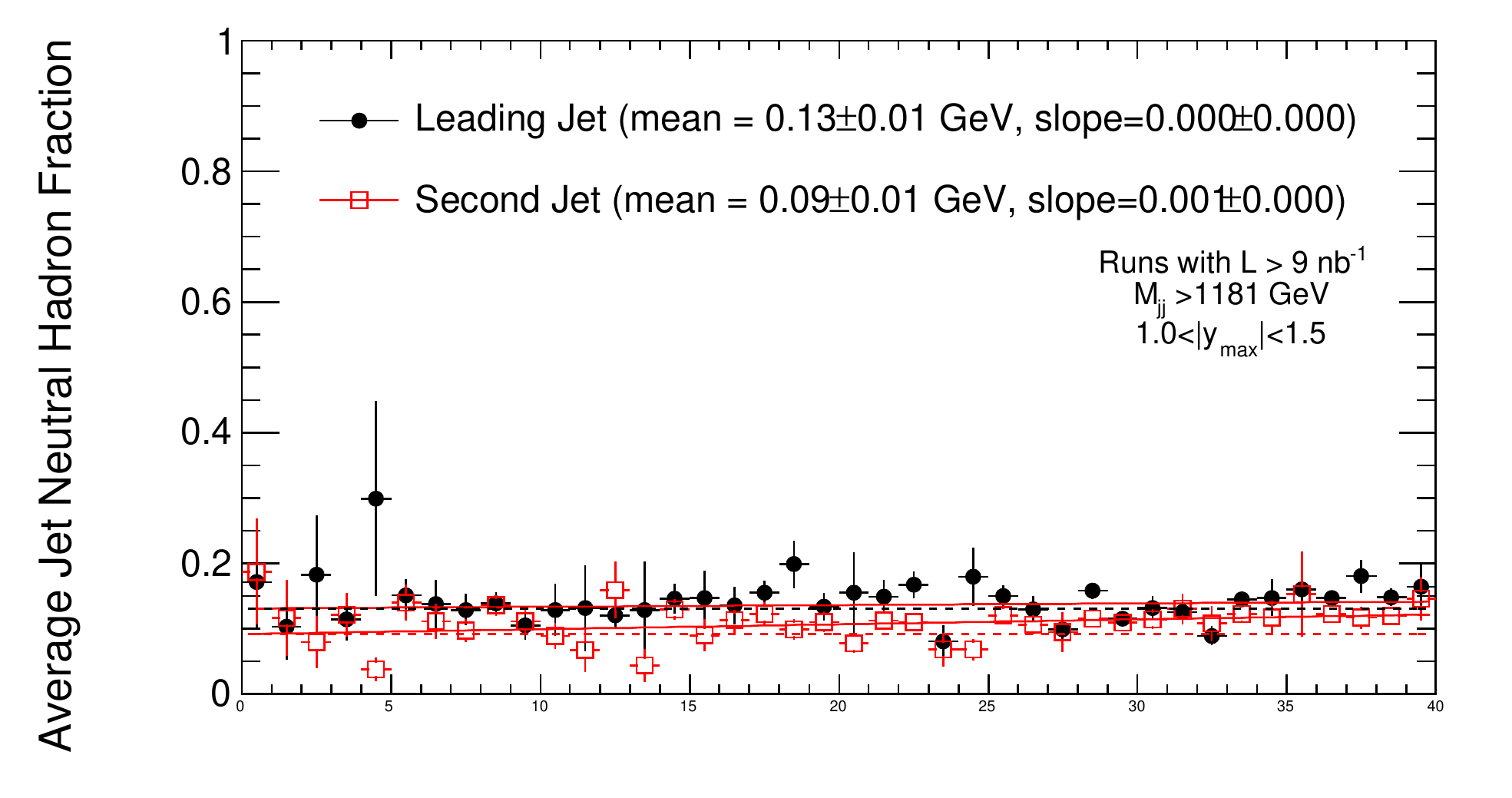} 
\includegraphics[width=0.49\textwidth]{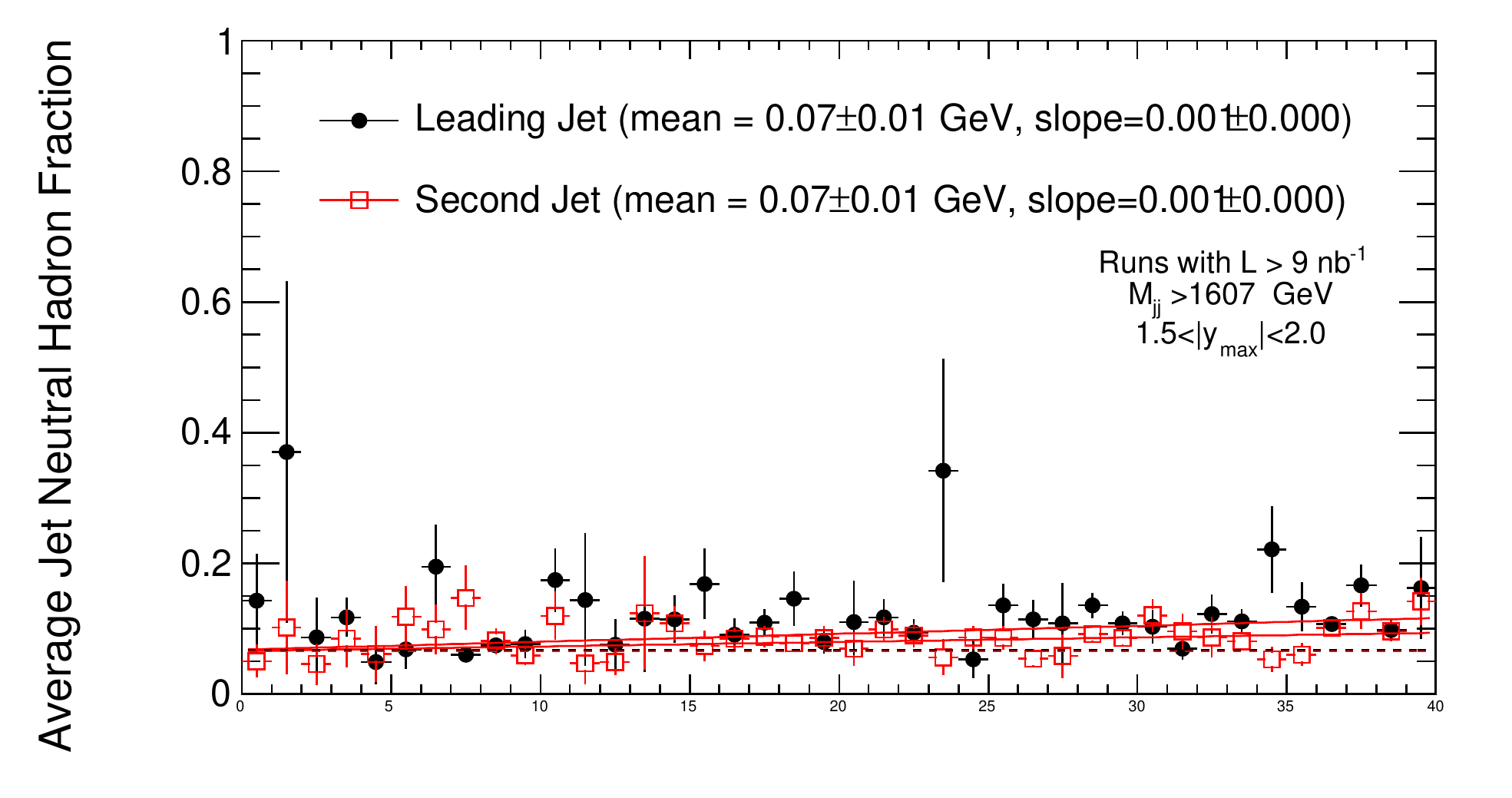} 
\includegraphics[width=0.49\textwidth]{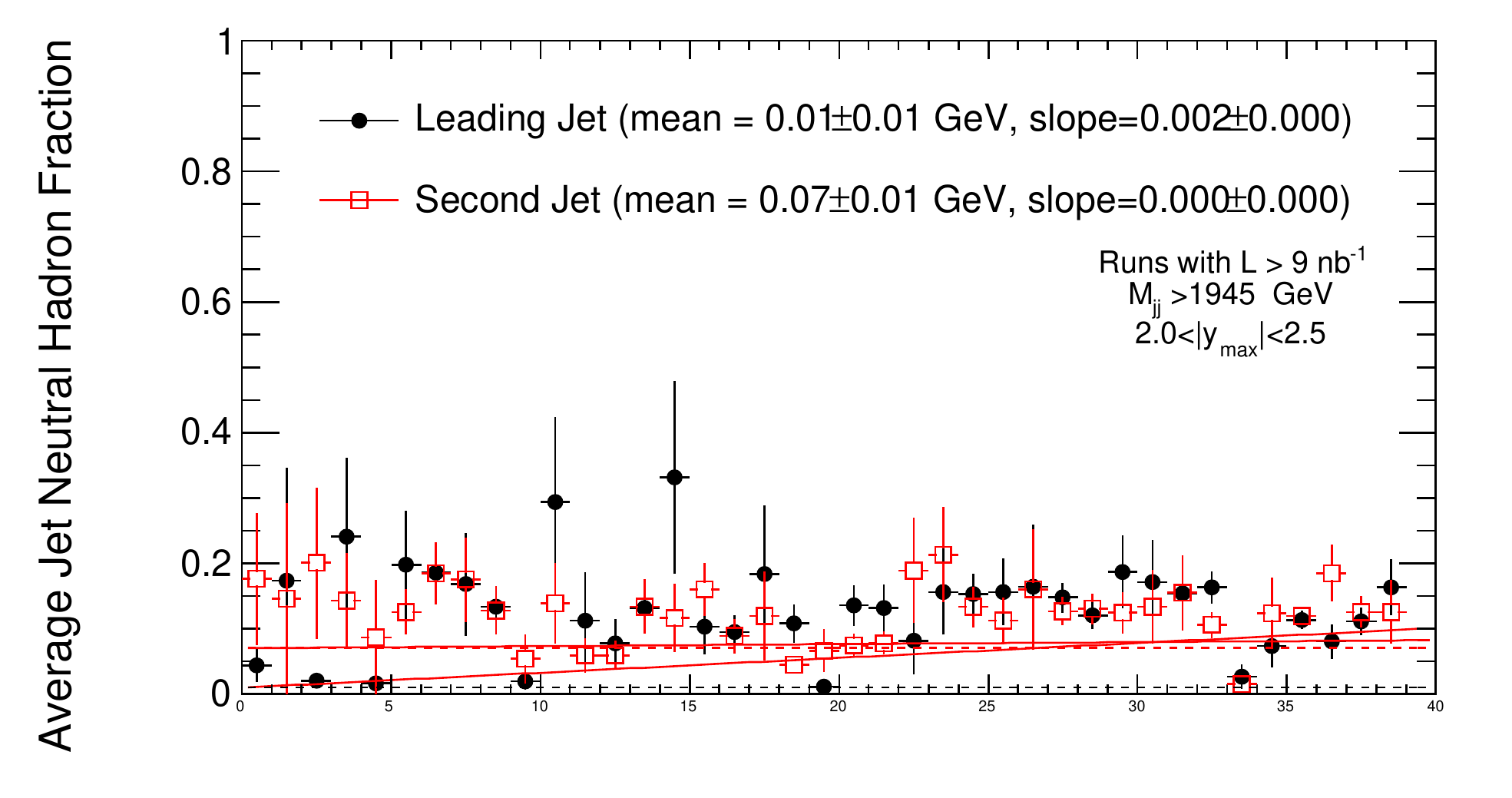} 
  
\capspace
\caption{ The neutral  hadron fraction  of the leading and second jet  for the five different $y_{max}$ bins and for the
HLT$_{-}$Jet140U trigger as a function of time (run number), fitted with a first degree polynomial. }
\label{fig_appd15}
\end{figure}

\clearpage

\begin{figure}[h]
\centering

\includegraphics[width=0.49\textwidth]{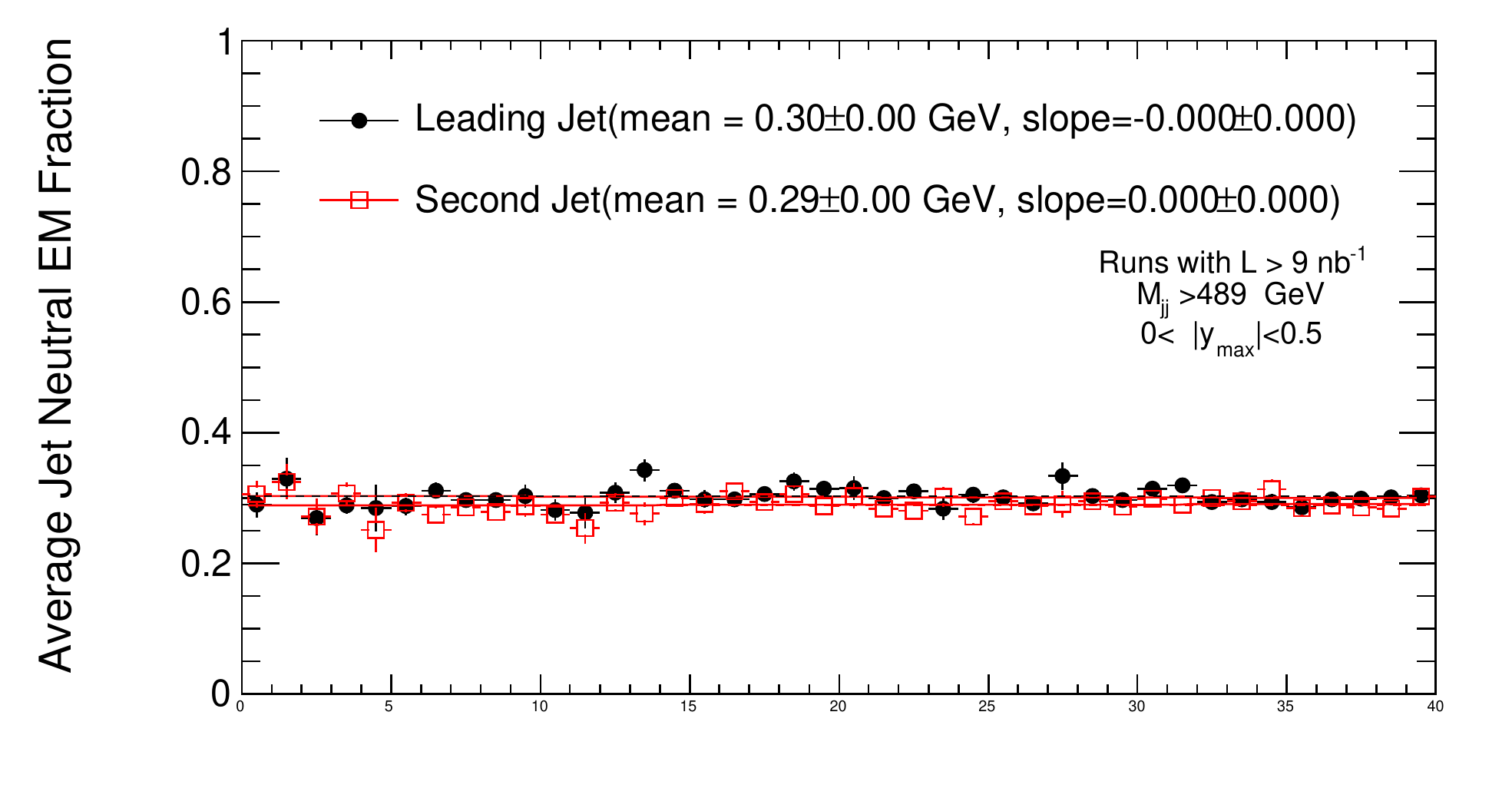} 
\includegraphics[width=0.49\textwidth]{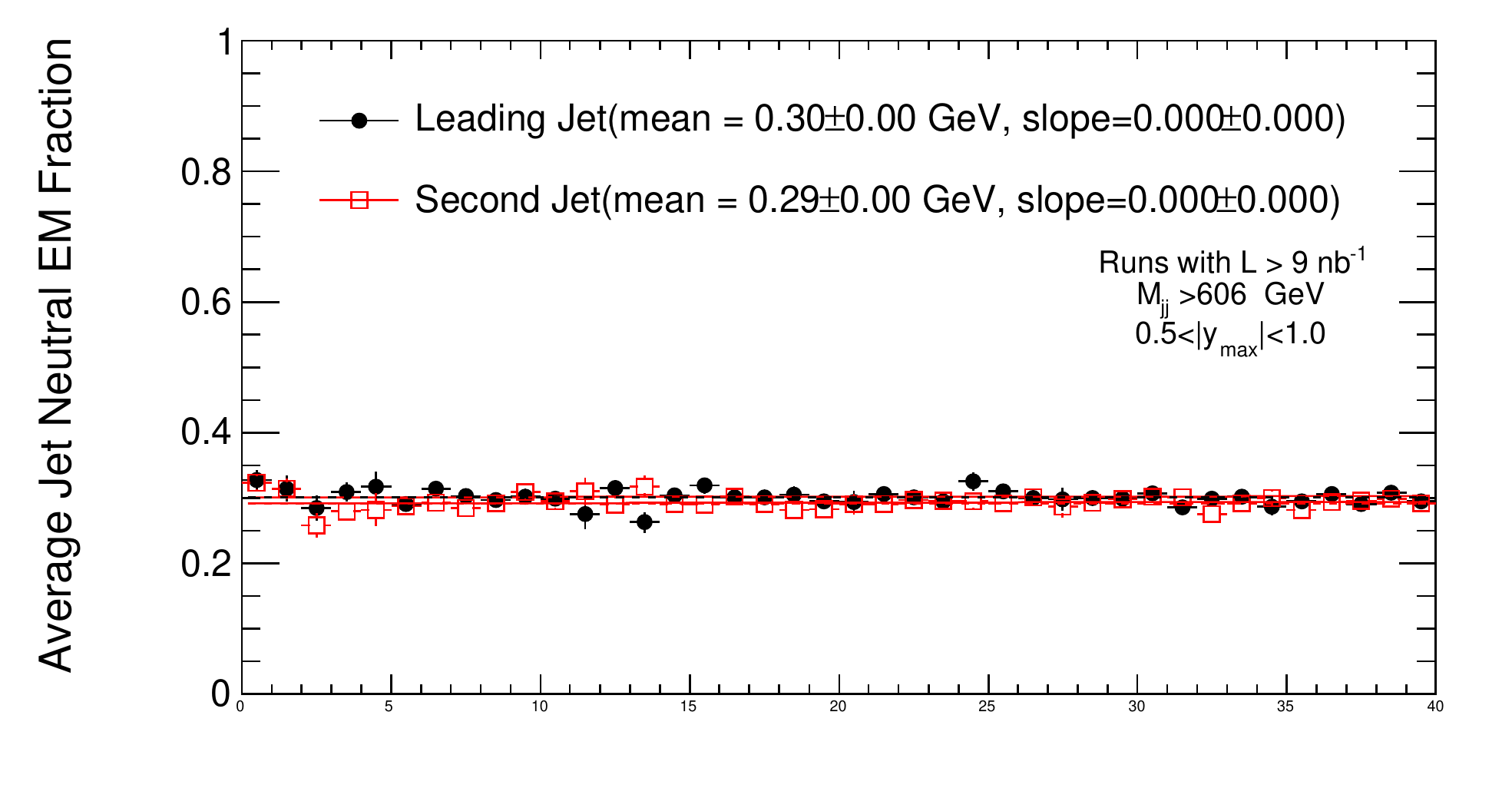} 
\includegraphics[width=0.49\textwidth]{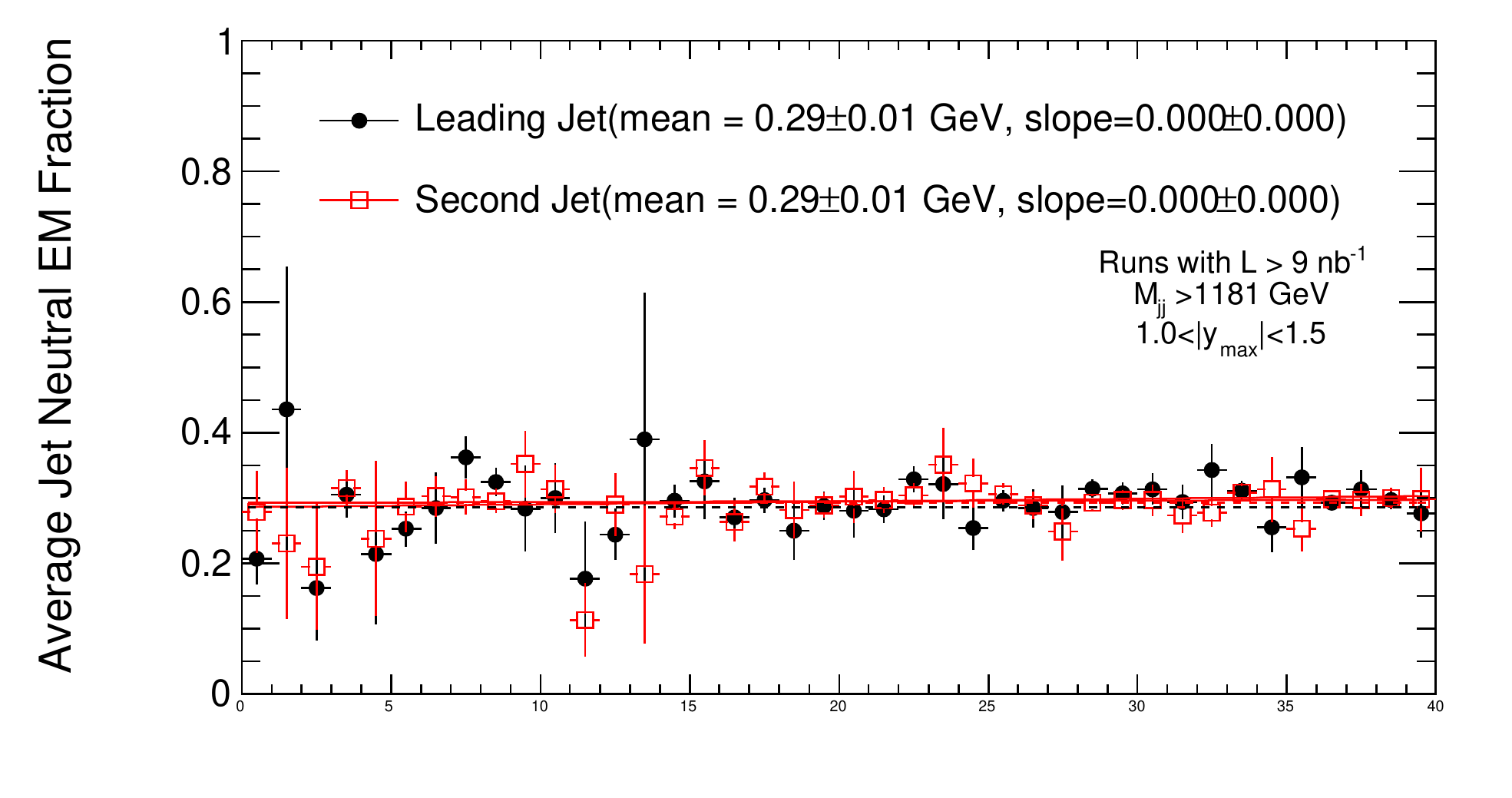} 
\includegraphics[width=0.49\textwidth]{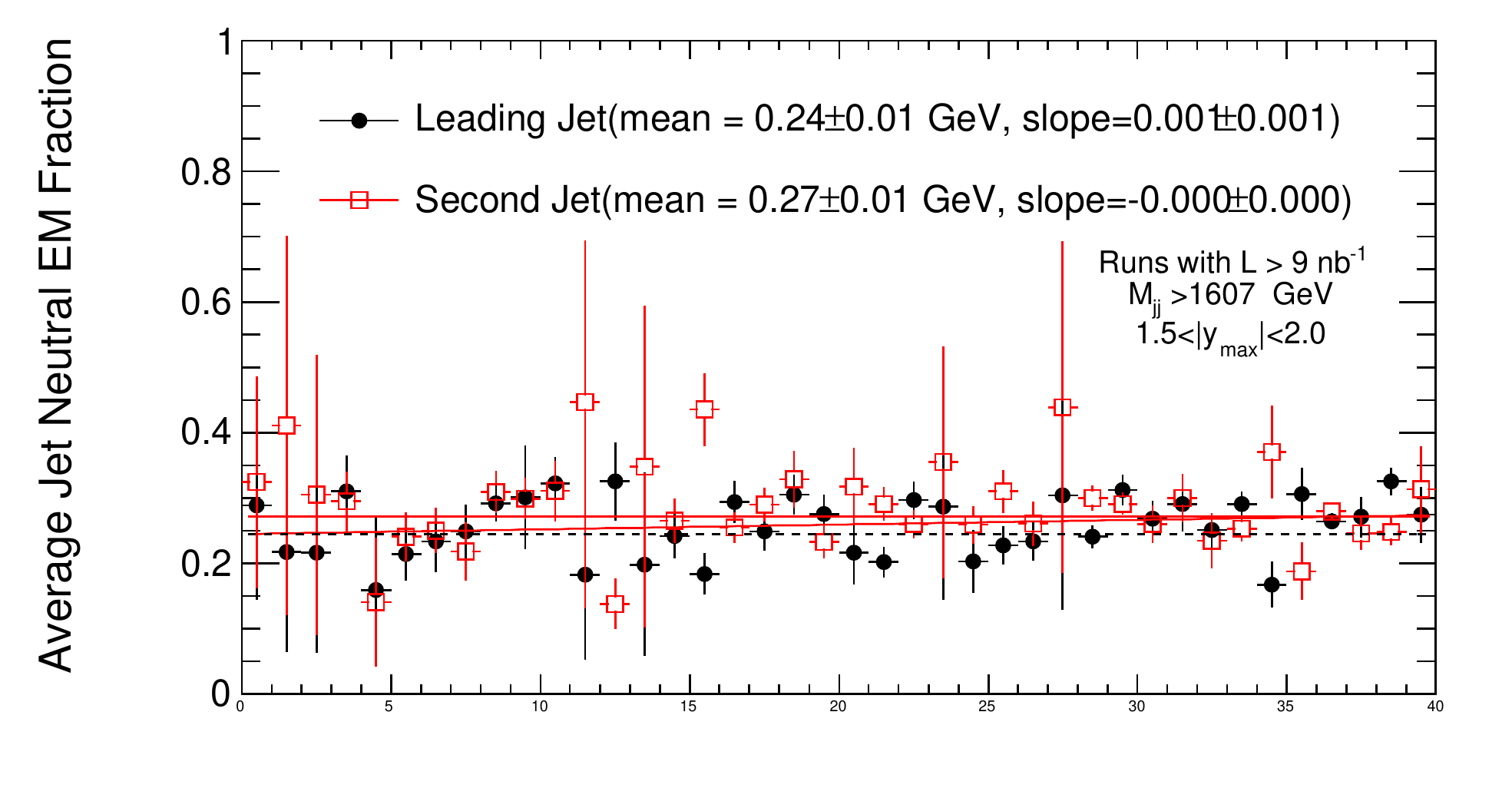} 
\includegraphics[width=0.49\textwidth]{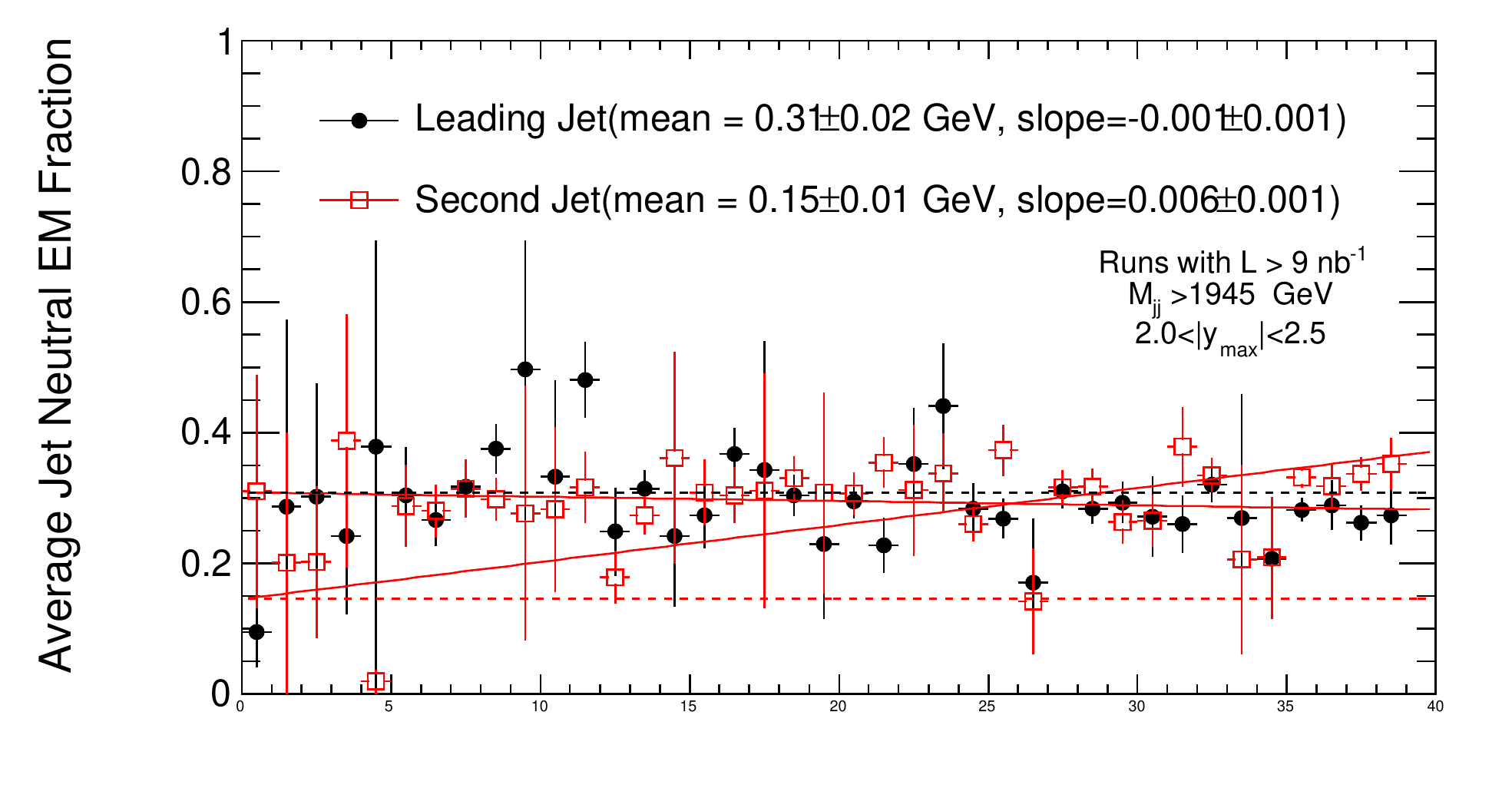} 
  
\capspace
\caption{ The neutral  electromagnetic fraction  of the leading and second jet  for the five different $y_{max}$ bins and for the HLT$_{-}$Jet140U trigger as a function of time (run number), fitted with a first degree polynomial. }
\label{fig_appd16}
\end{figure}

\clearpage
\chapter{Stability Over the Run Period}

\begin{figure}[h]
\centering

\includegraphics[width=0.48\textwidth]{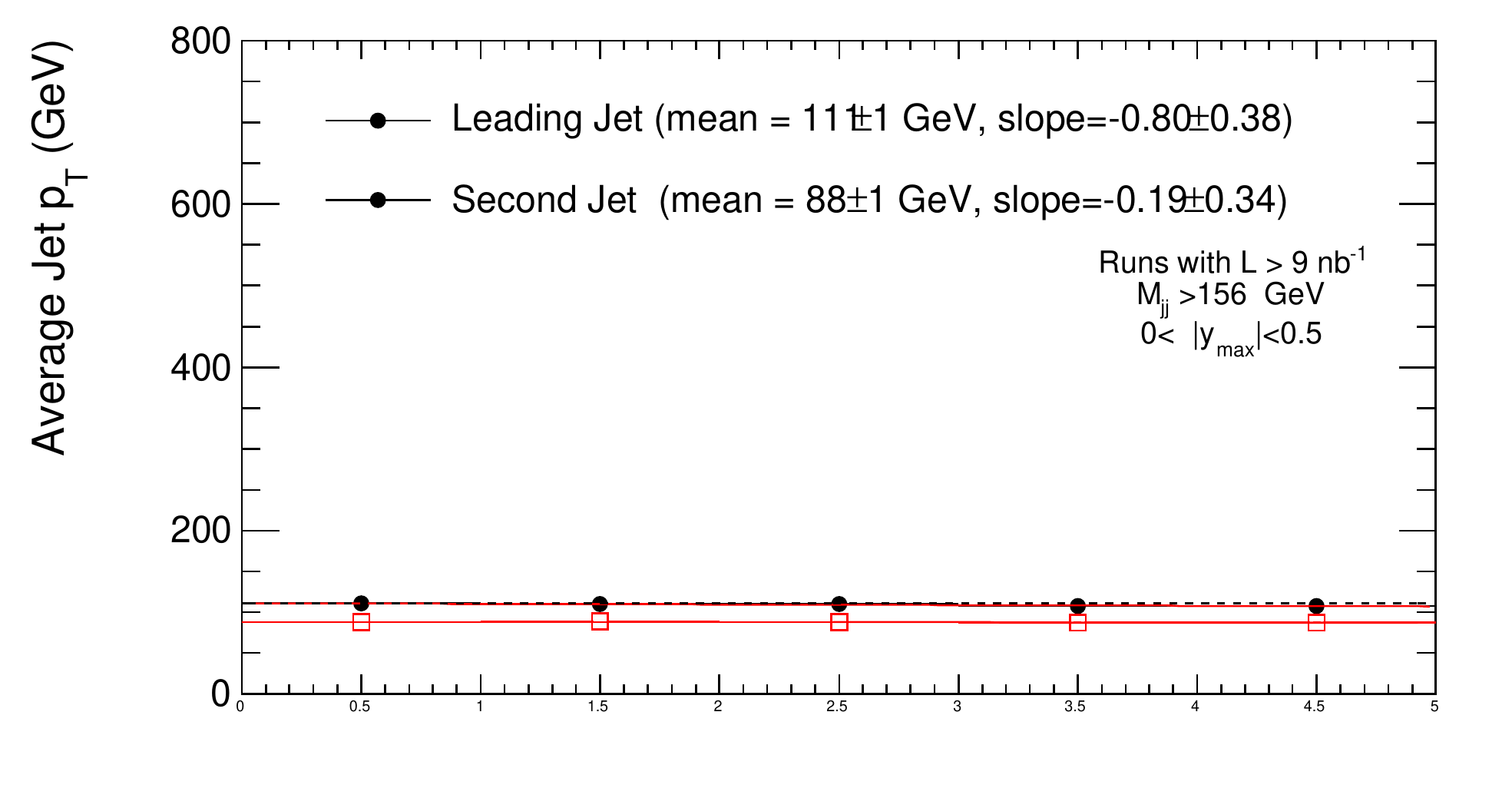} 
\includegraphics[width=0.48\textwidth]{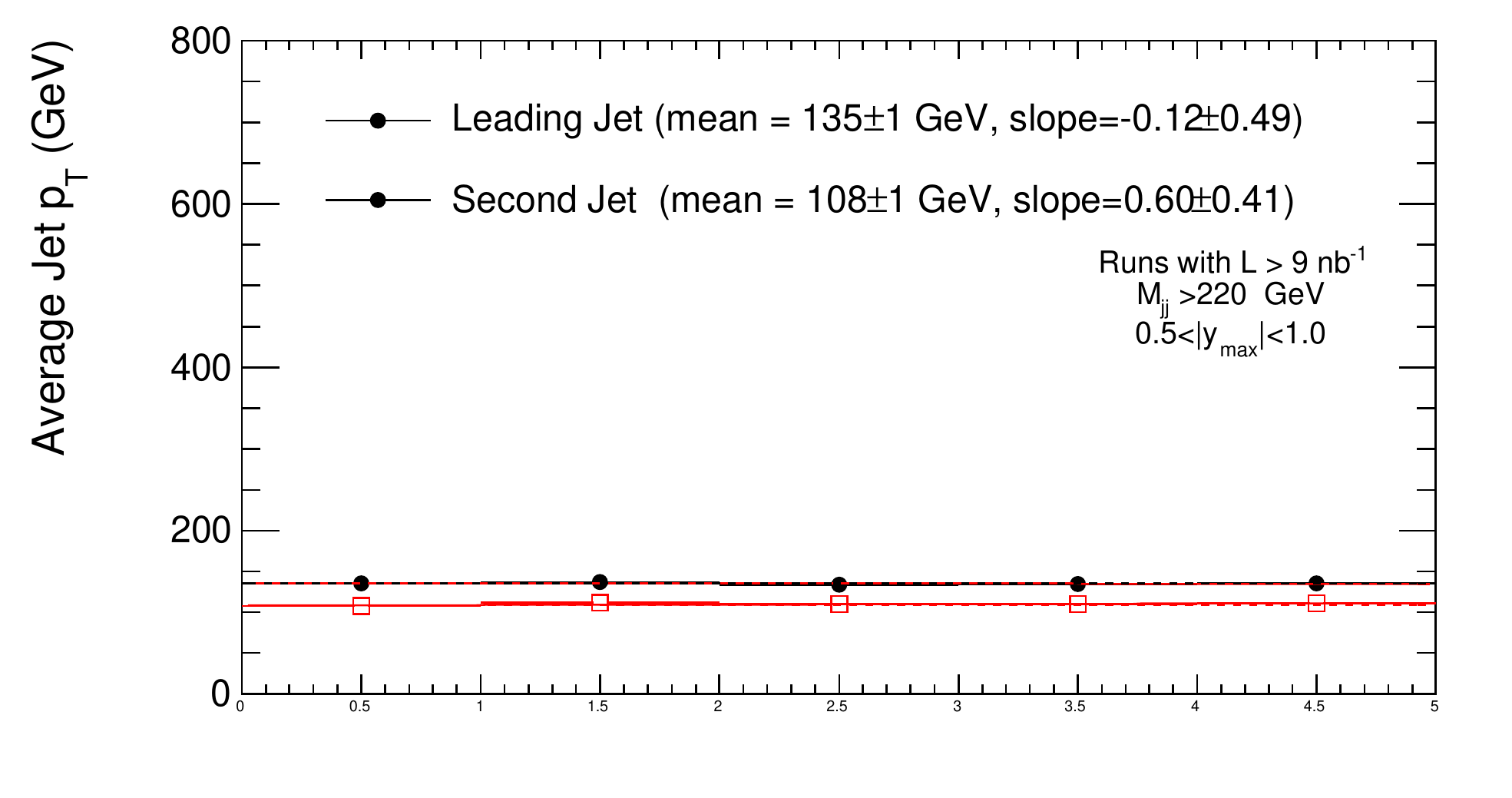}  
\includegraphics[width=0.48\textwidth]{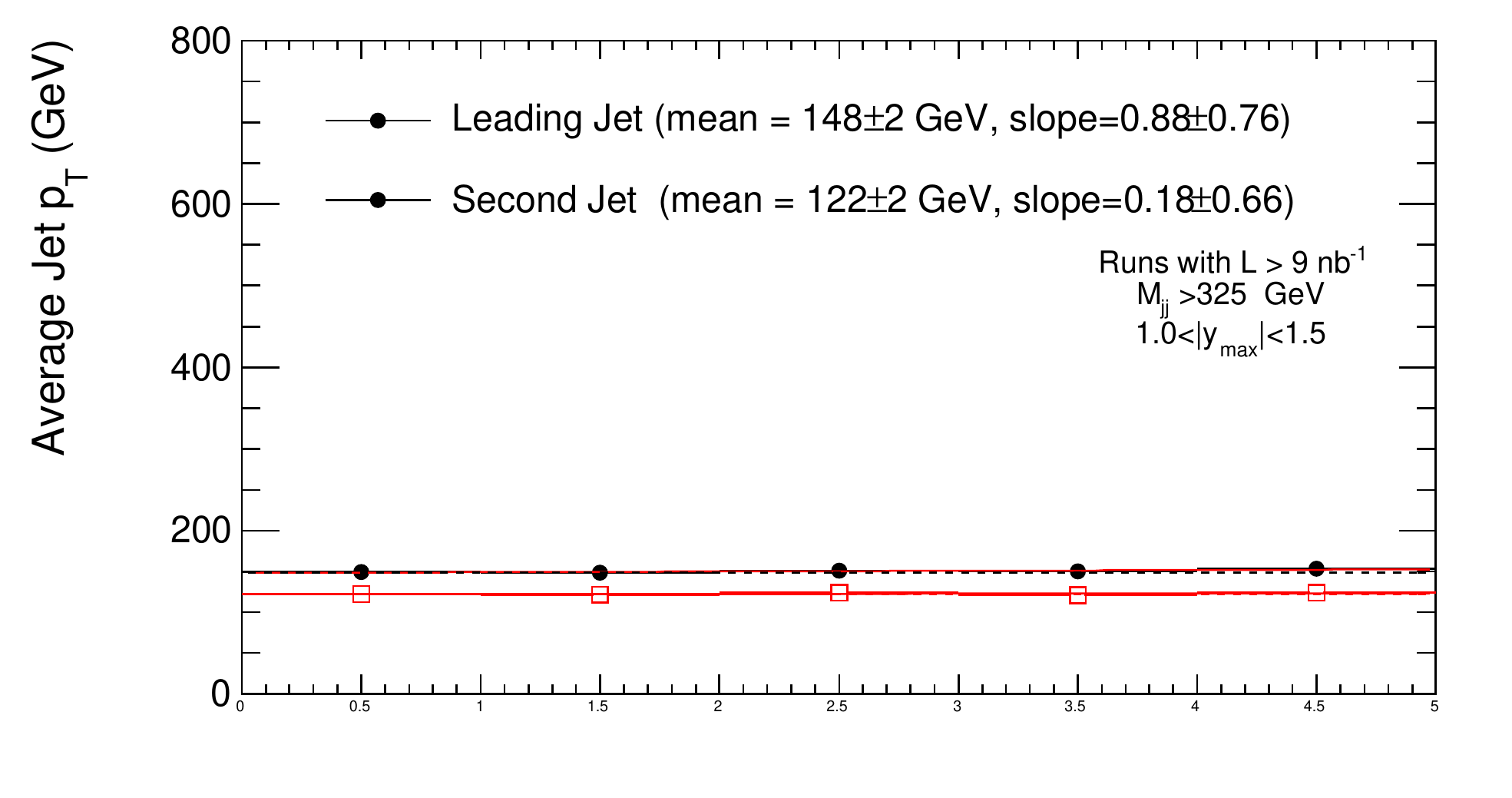}  
\includegraphics[width=0.48\textwidth]{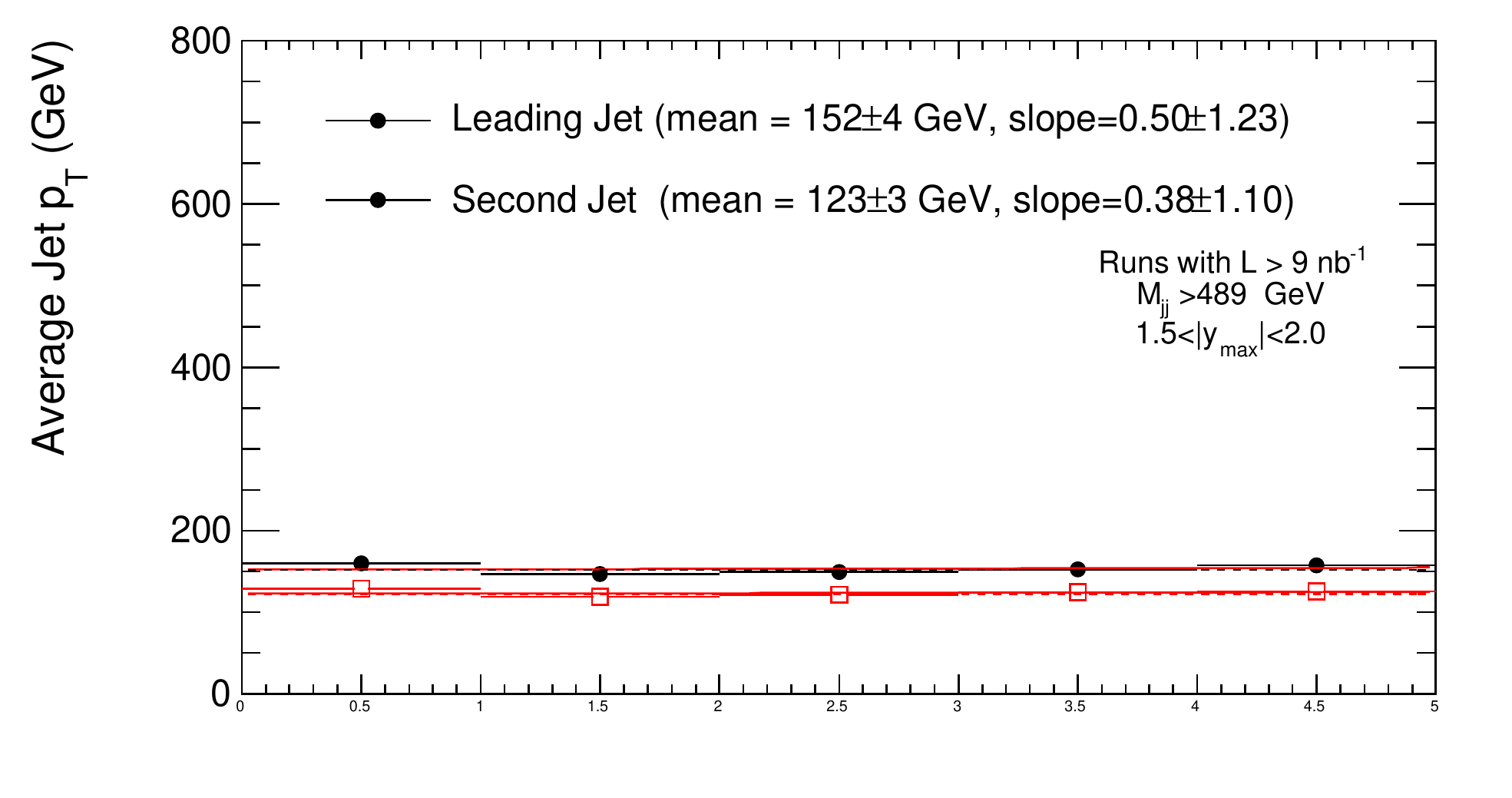} 
\includegraphics[width=0.48\textwidth]{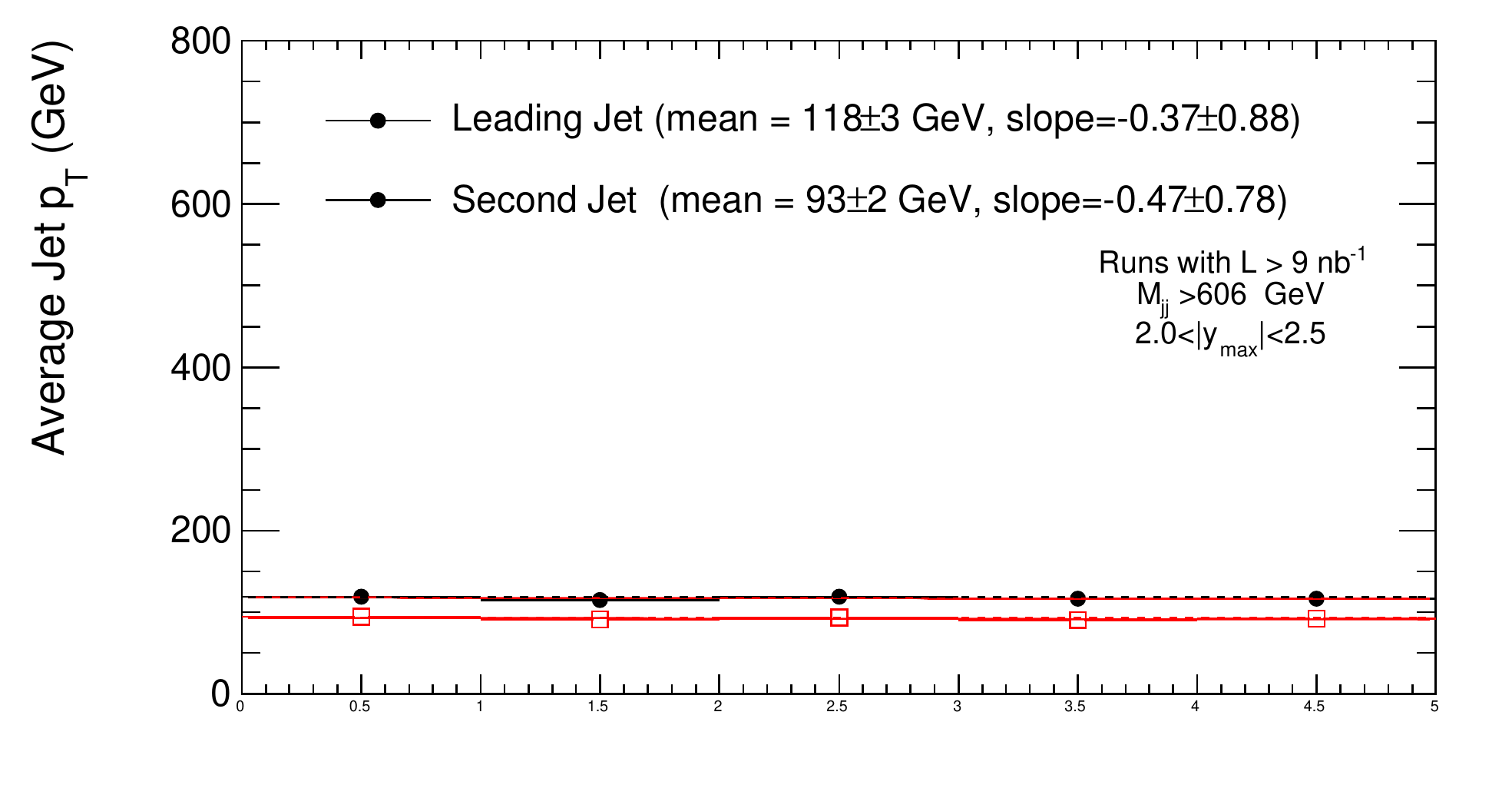} 
   
\capspace
\caption{ The $p_T$ of the leading and second jet  for the five different $y_{max}$ bins and for the 
HLT$_{-}$Jet30U trigger as a function of time (run number), fitted with a first degree polynomial.}
\label{fig_appd1}
\end{figure}

\begin{figure}[h]
\centering

\includegraphics[width=0.52\textwidth]{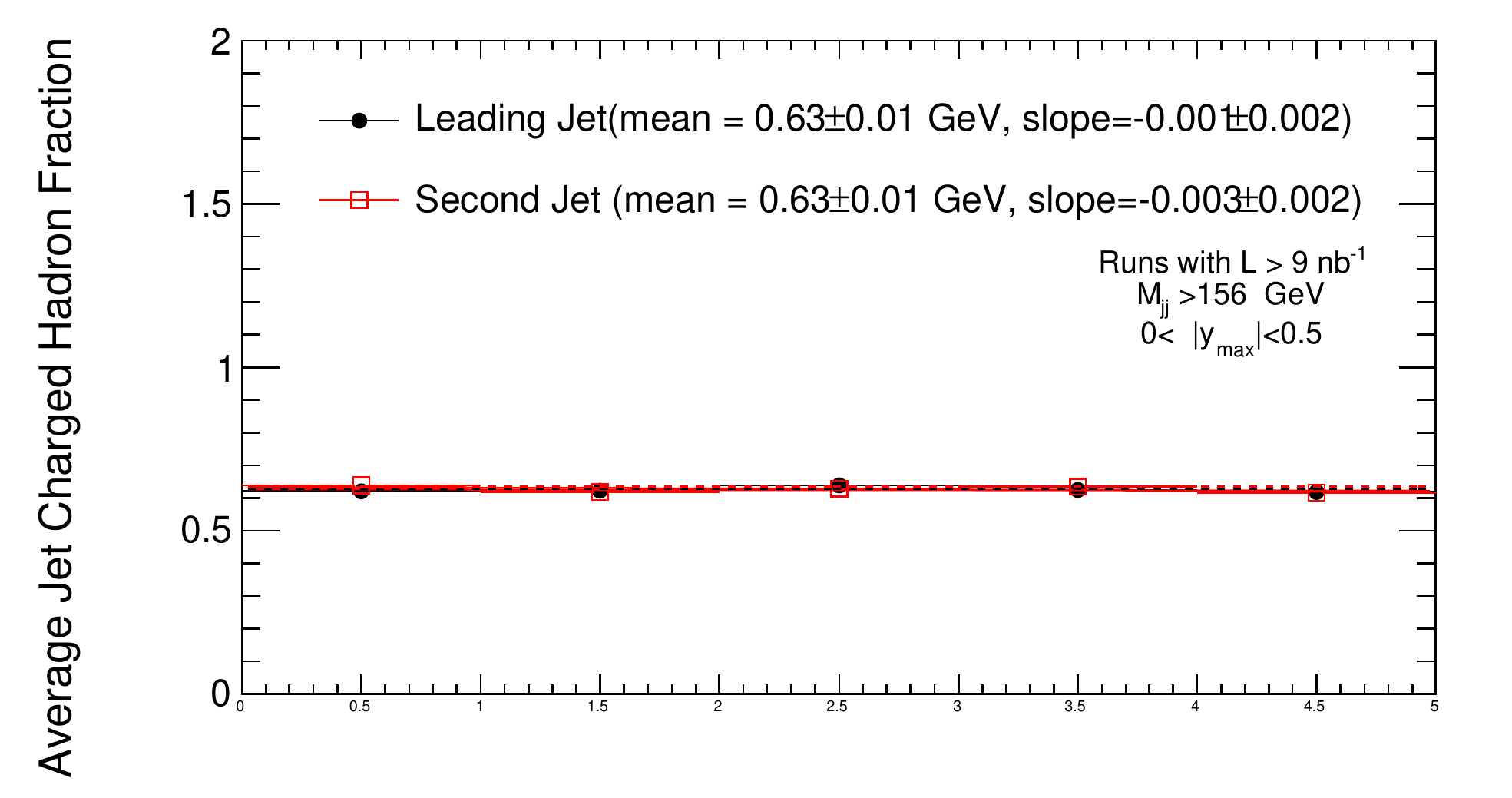} 
\includegraphics[width=0.52\textwidth]{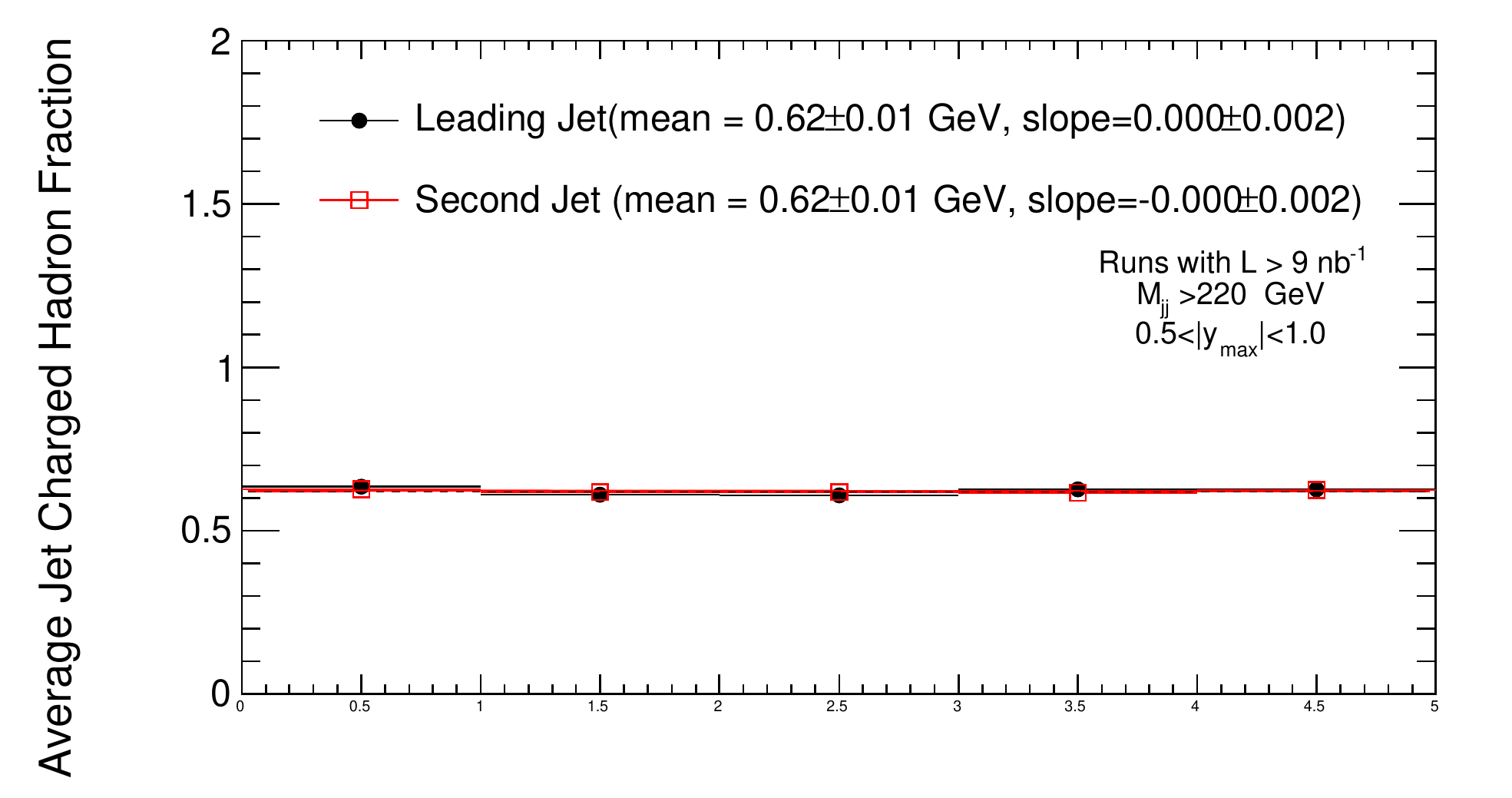}  
\includegraphics[width=0.52\textwidth]{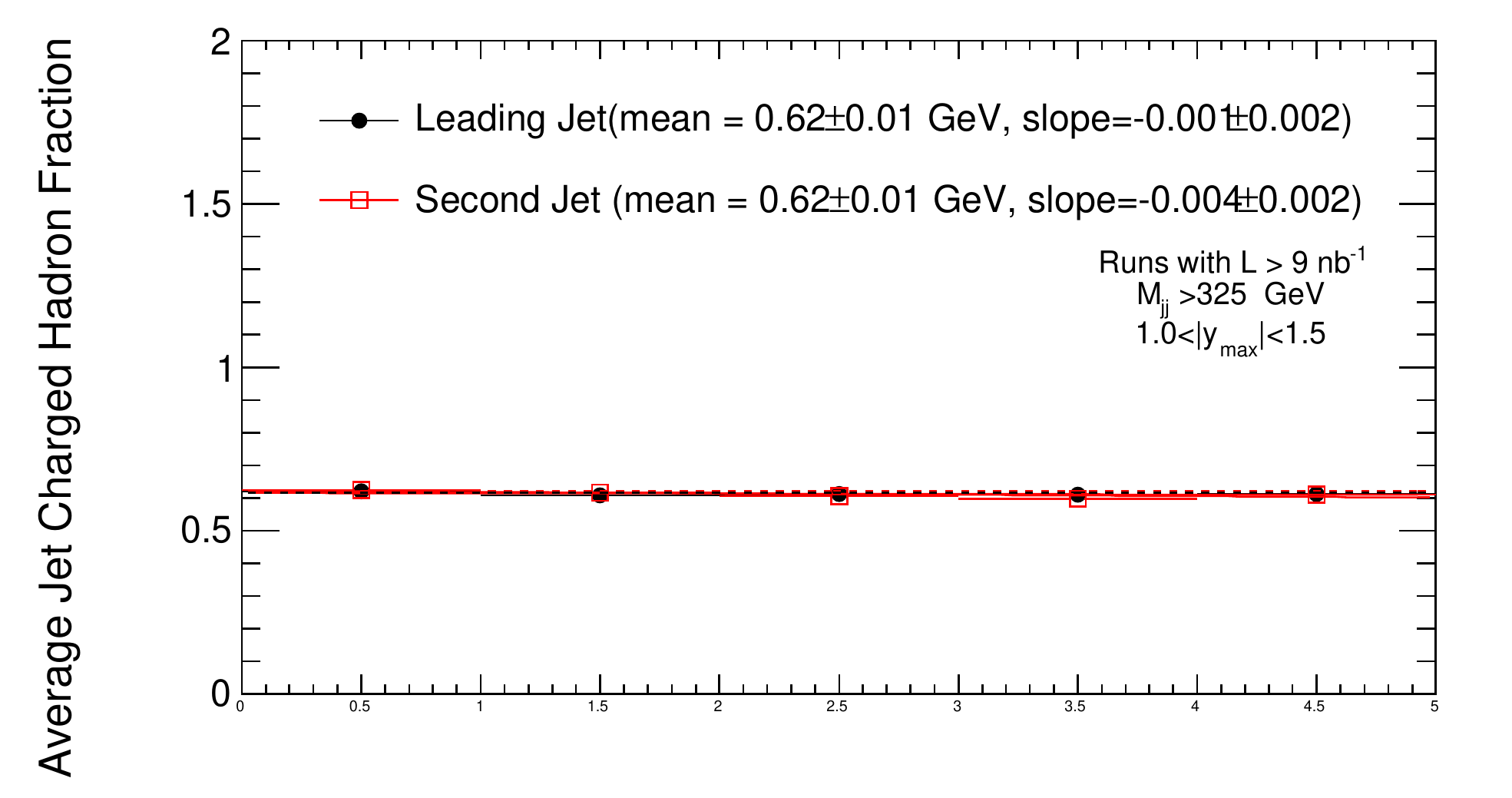}  
\includegraphics[width=0.52\textwidth]{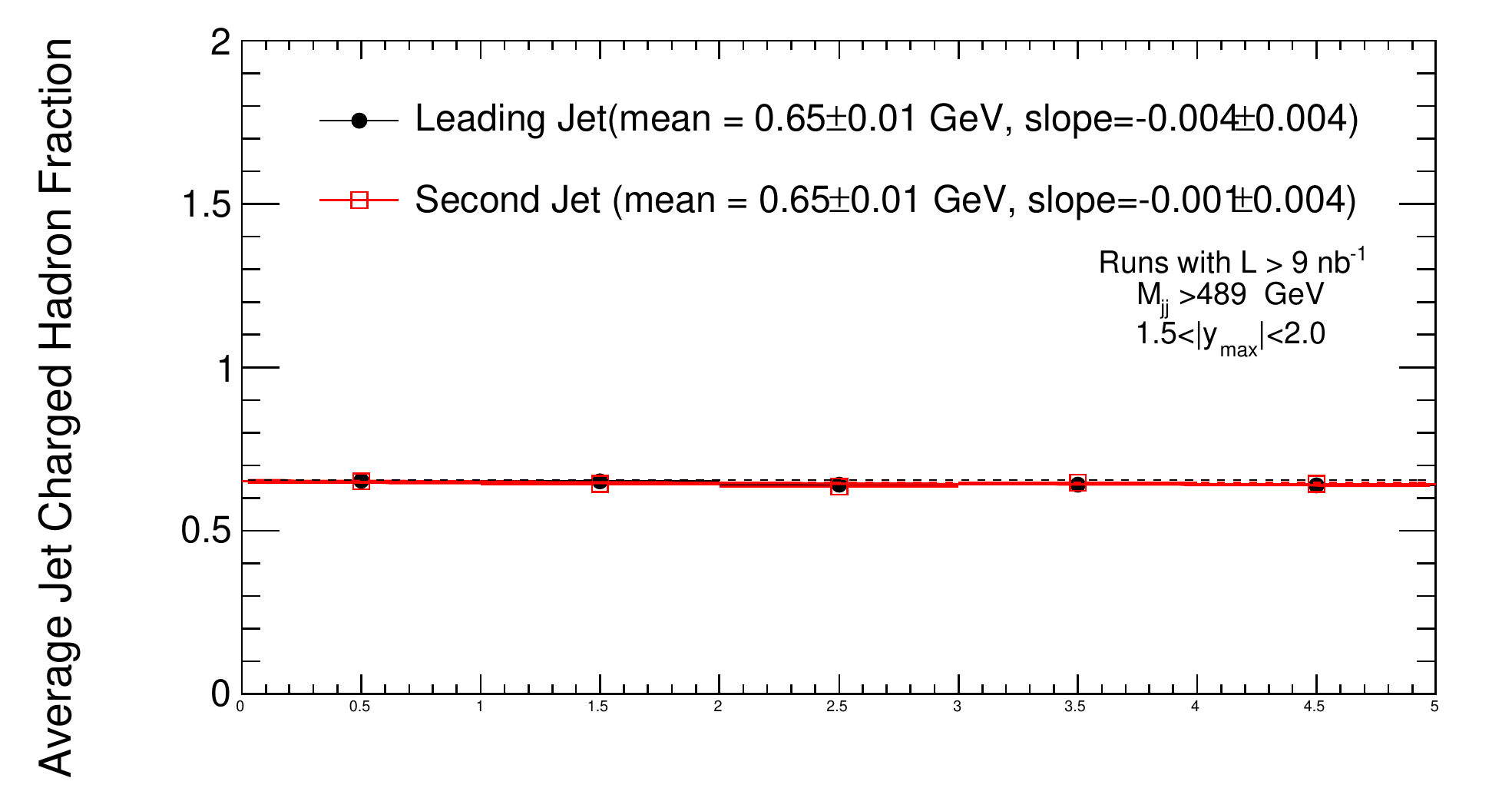} 
\includegraphics[width=0.52\textwidth]{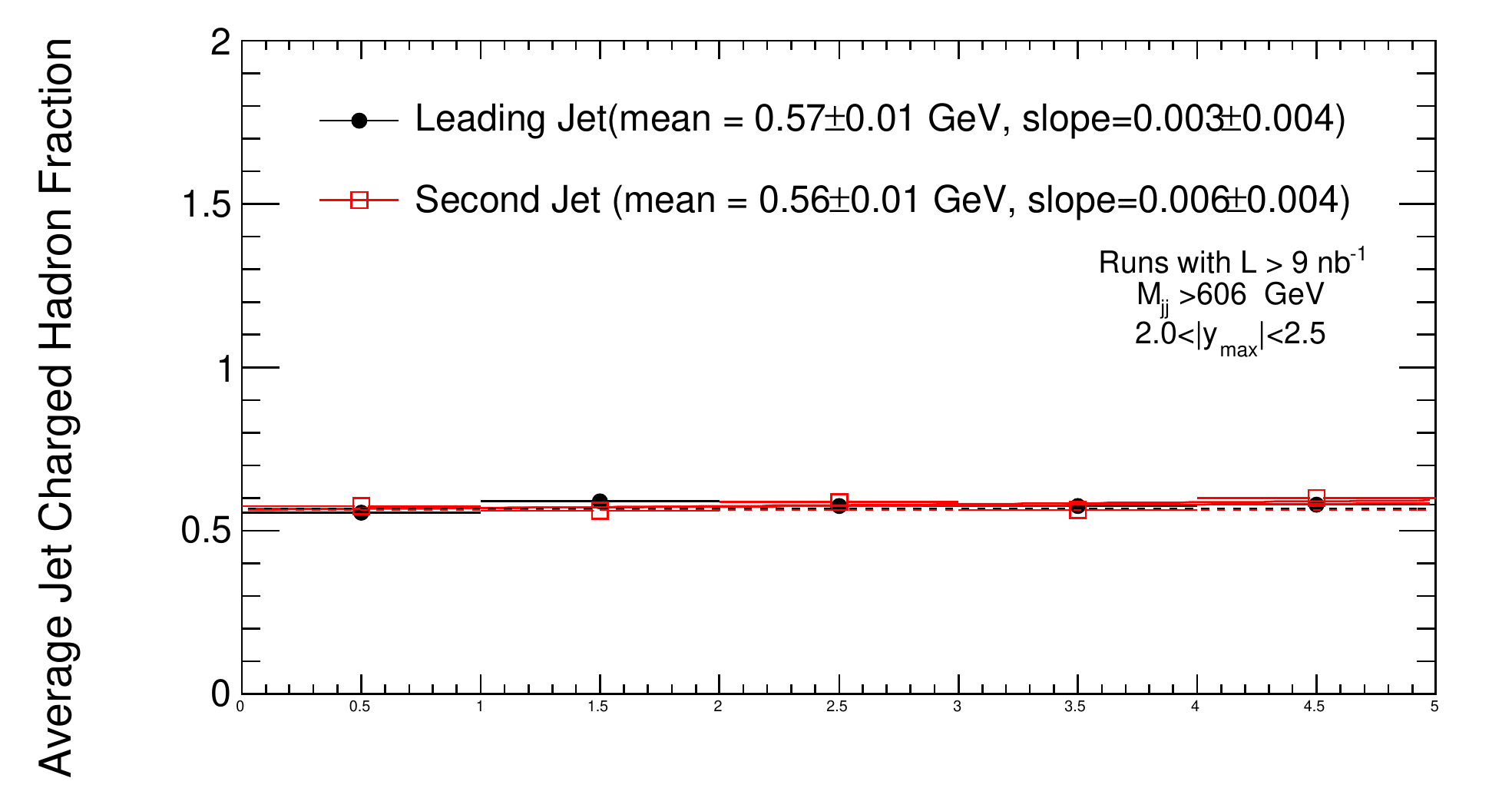} 
   
\capspace
\caption{ The charged hadron fraction  of the leading and second jet  for the five different $y_{max}$ bins and for the HLT$_{-}$Jet30U trigger as a function of time (run number), fitted with a first degree polynomial.}
\label{fig_appd2}
\end{figure}

\begin{figure}[h]
\centering

\includegraphics[width=0.52\textwidth]{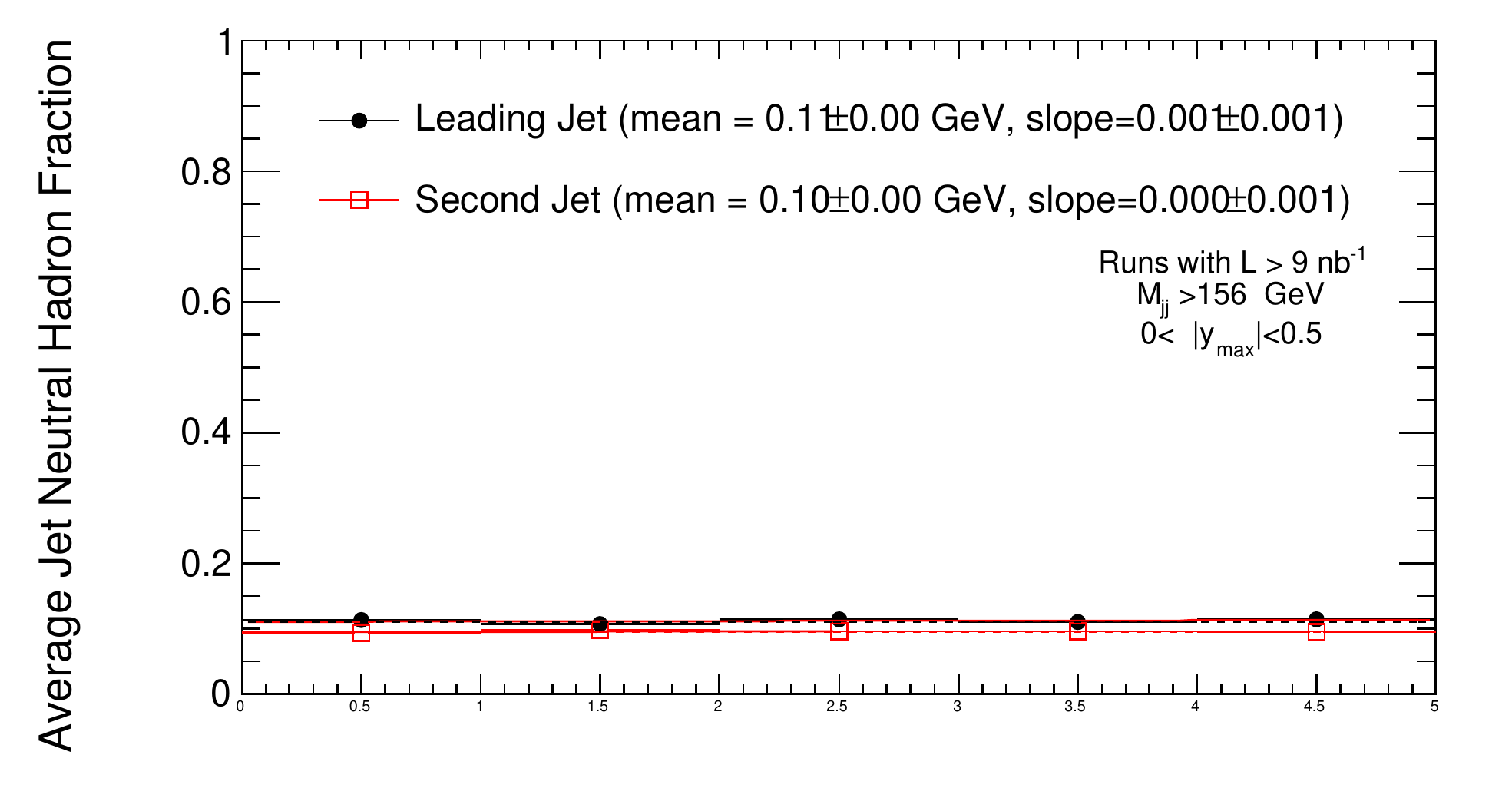} 
\includegraphics[width=0.52\textwidth]{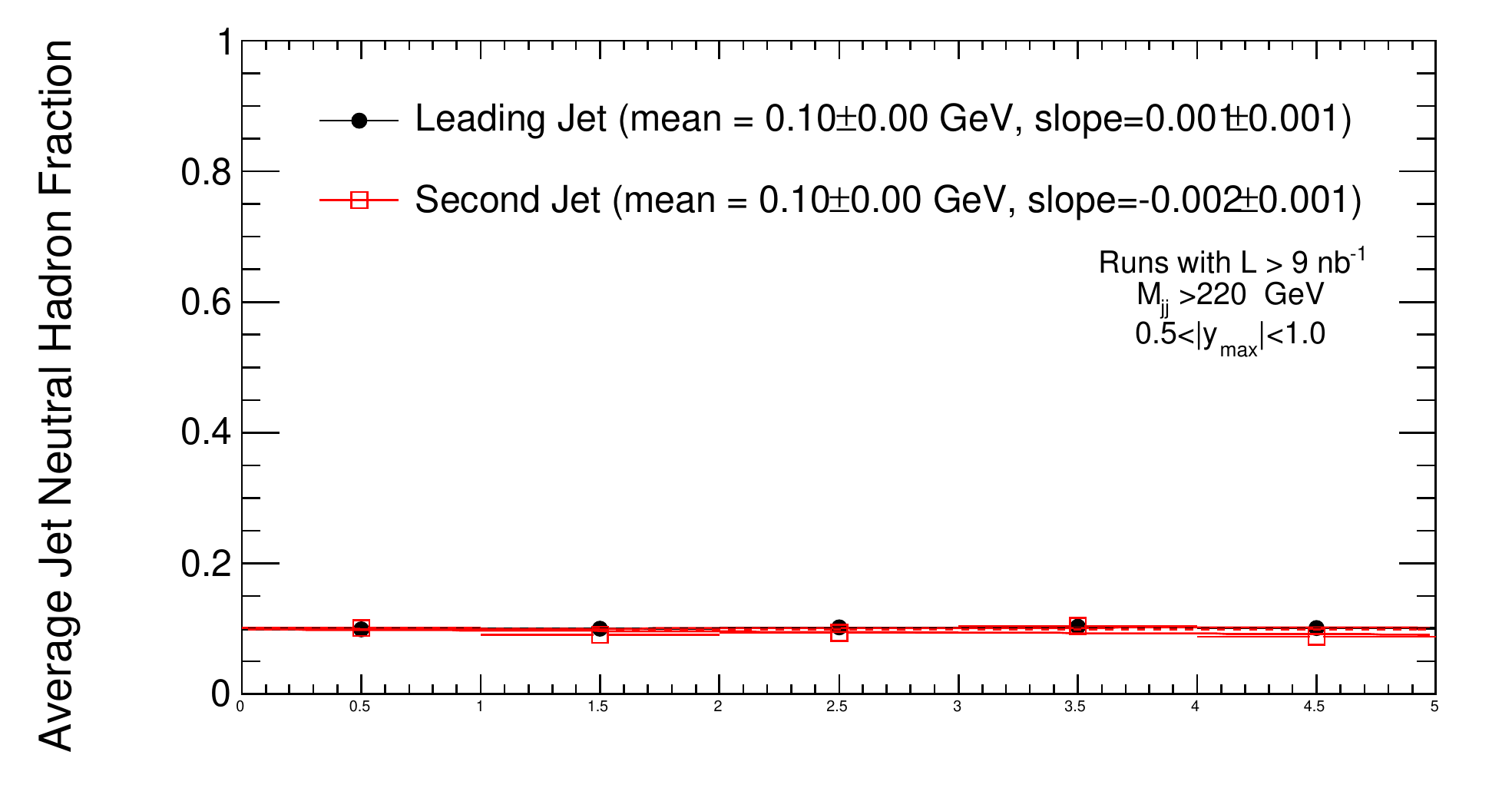}  
\includegraphics[width=0.52\textwidth]{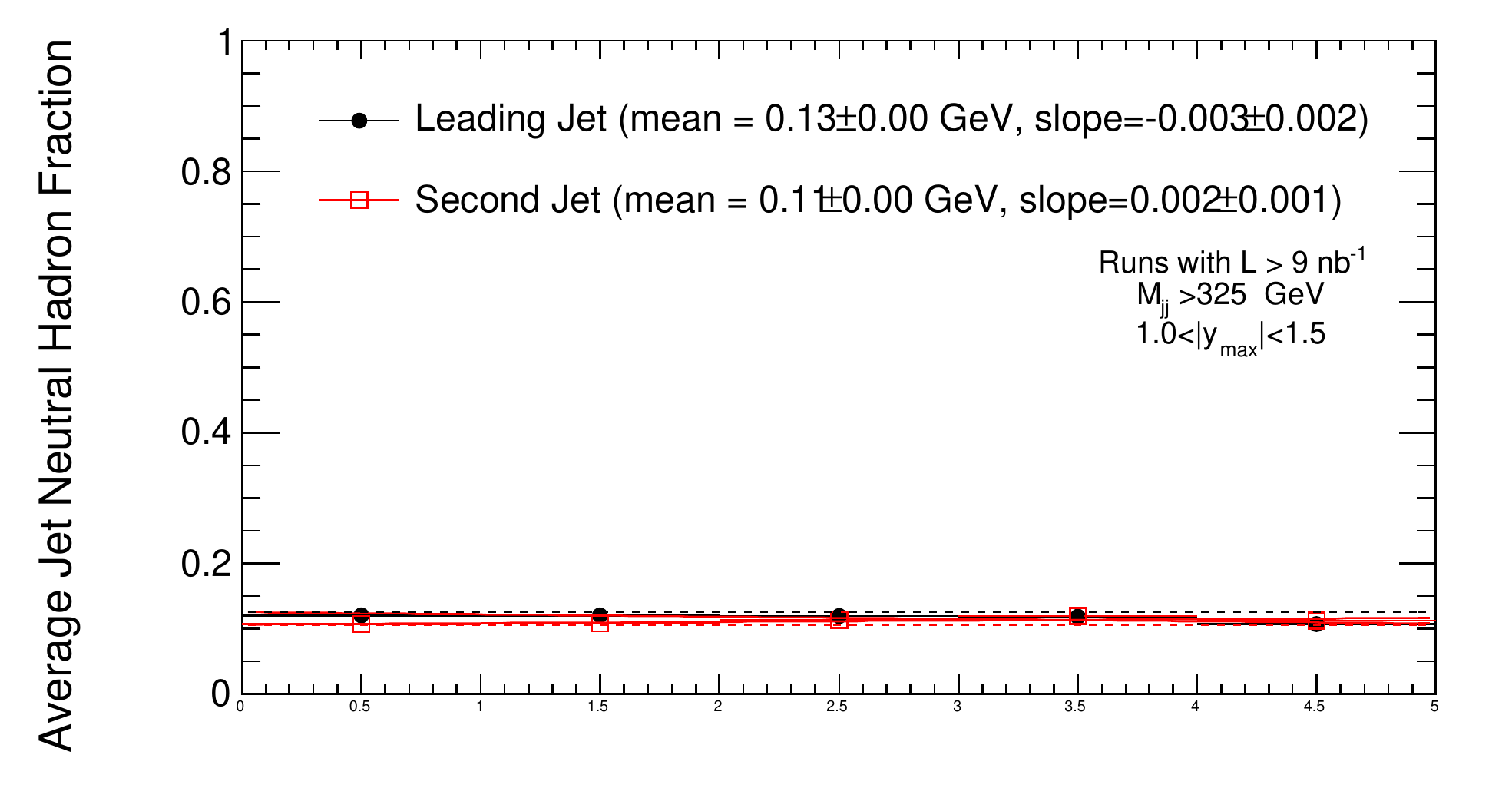}  
\includegraphics[width=0.52\textwidth]{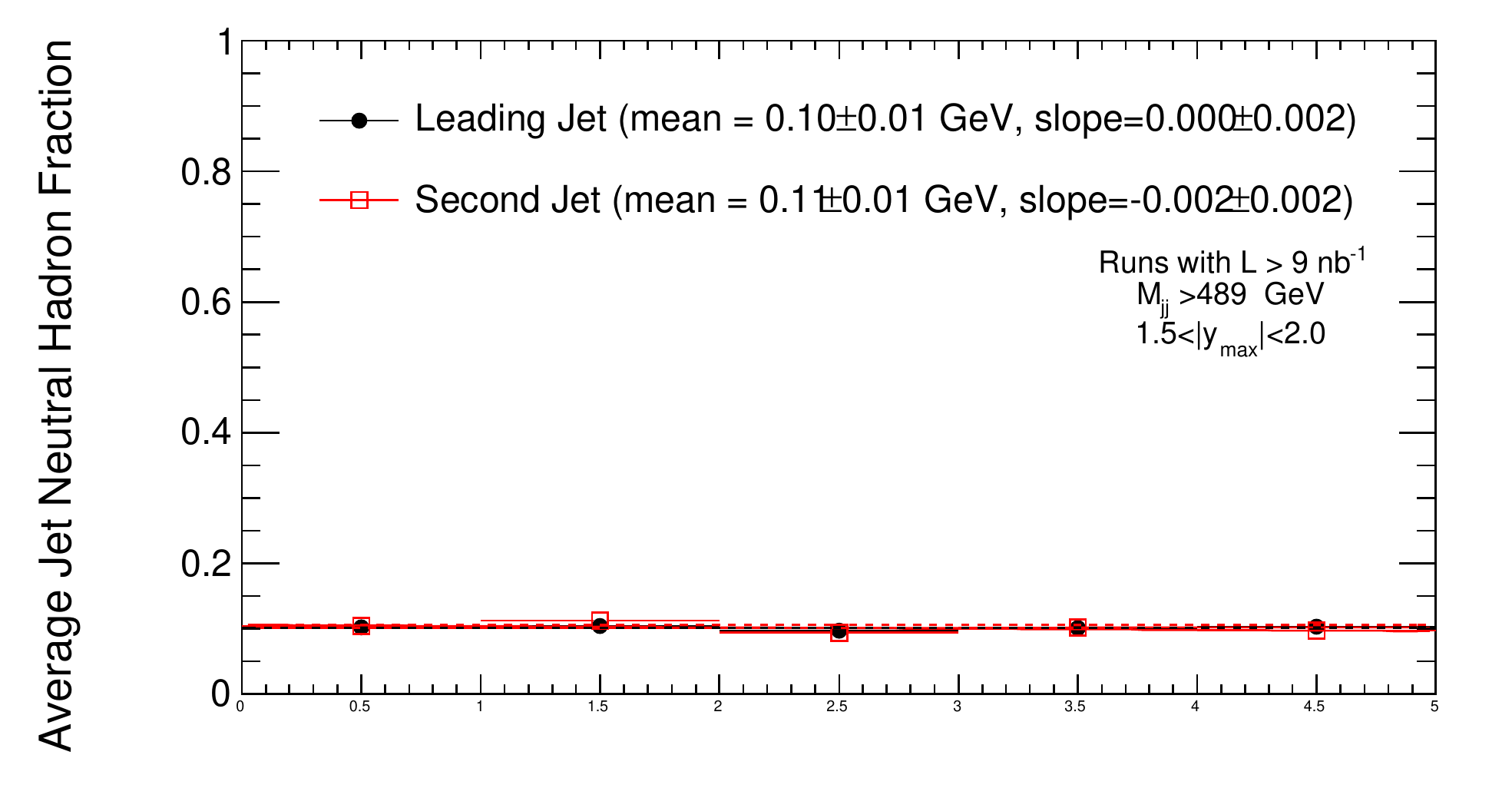} 
\includegraphics[width=0.52\textwidth]{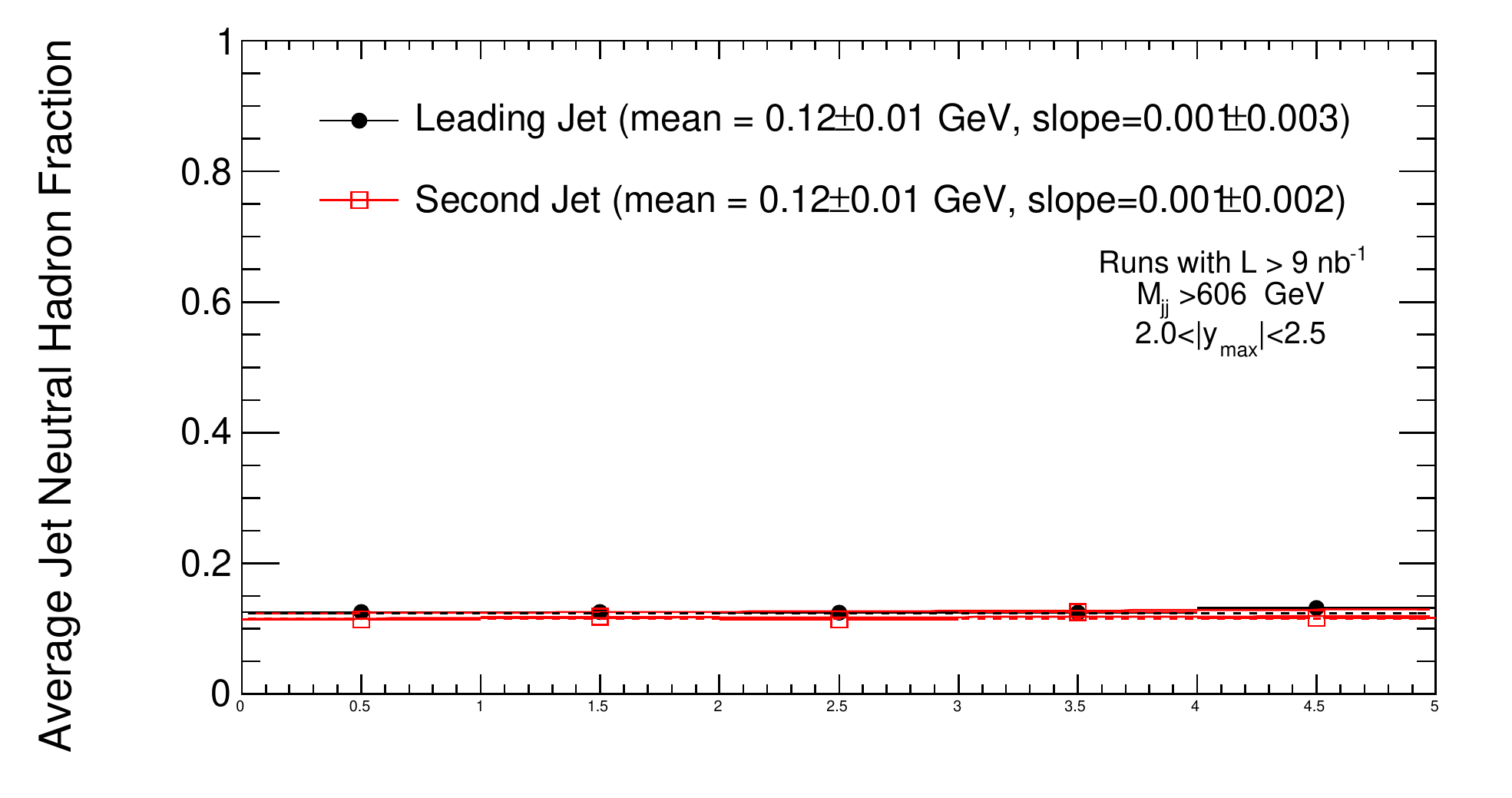} 
   
\capspace
\caption{ The neutral  hadron fraction  of the leading and second jet  for the five different $y_{max}$ bins and for the 
HLT$_{-}$Jet30U trigger as a function of time (run number), fitted with a first degree polynomial.}
\label{fig_appd3}
\end{figure}

 \clearpage

\begin{figure}[h]
\centering

\includegraphics[width=0.52\textwidth]{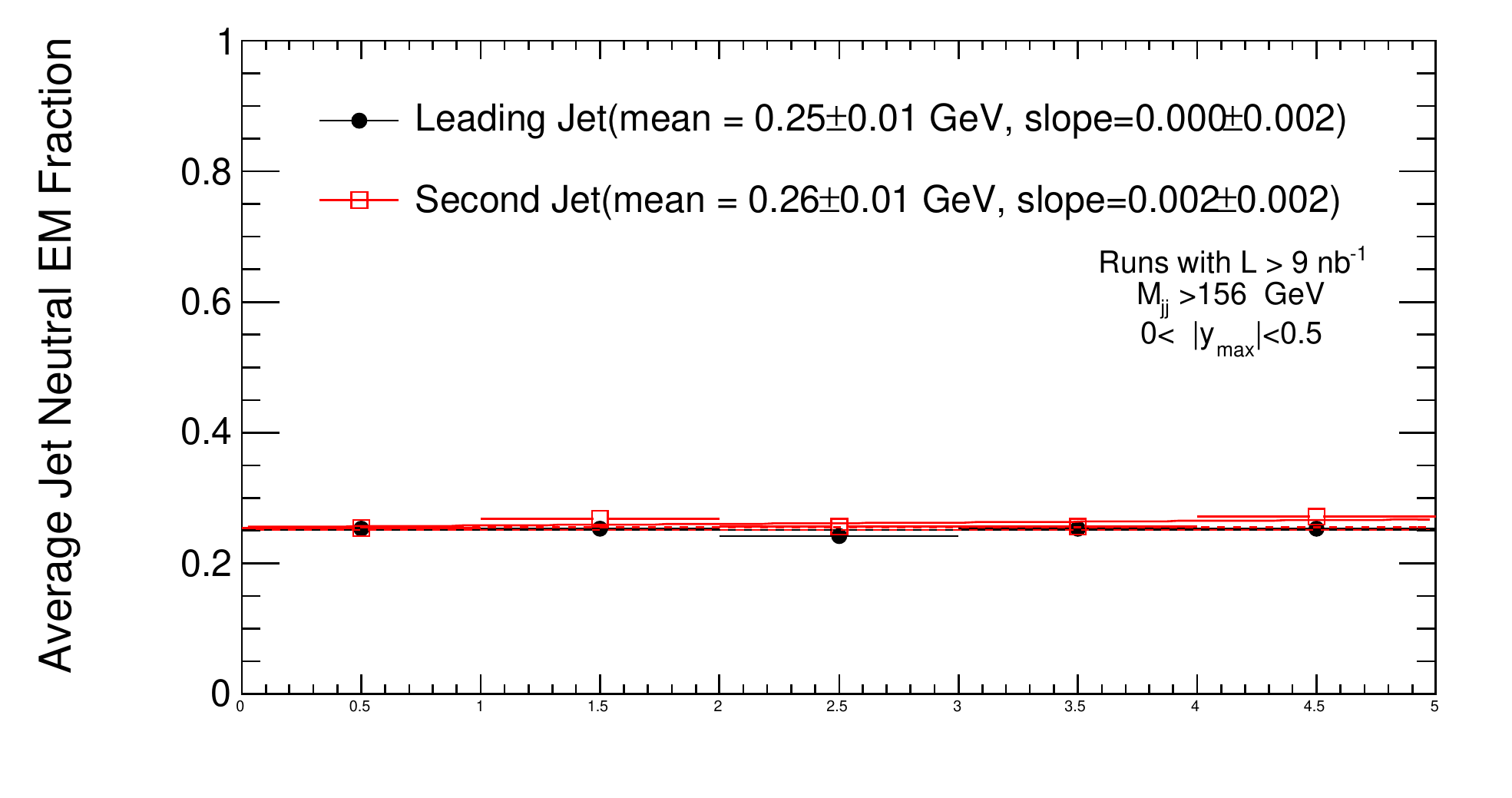} 
\includegraphics[width=0.52\textwidth]{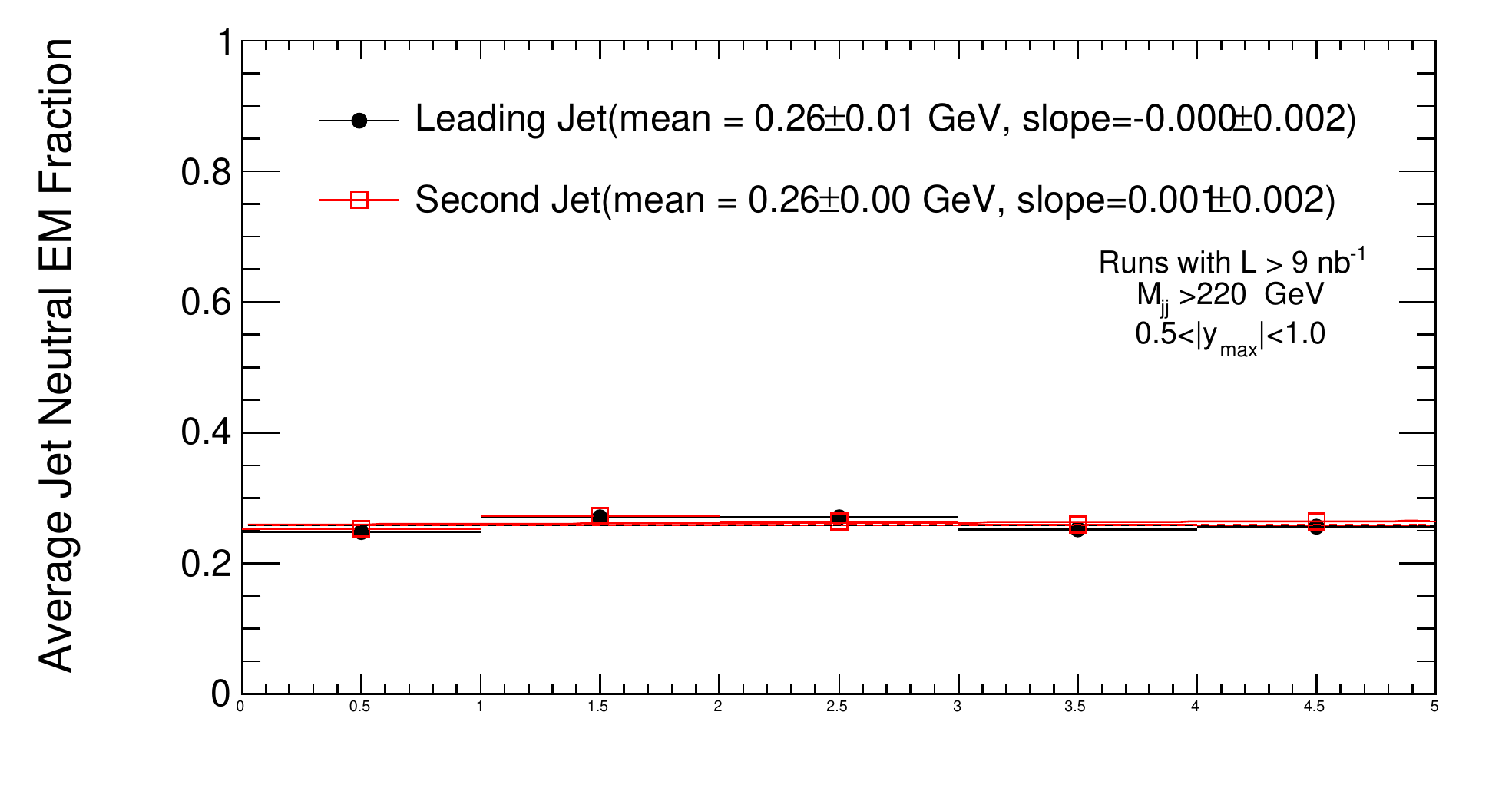}  
\includegraphics[width=0.52\textwidth]{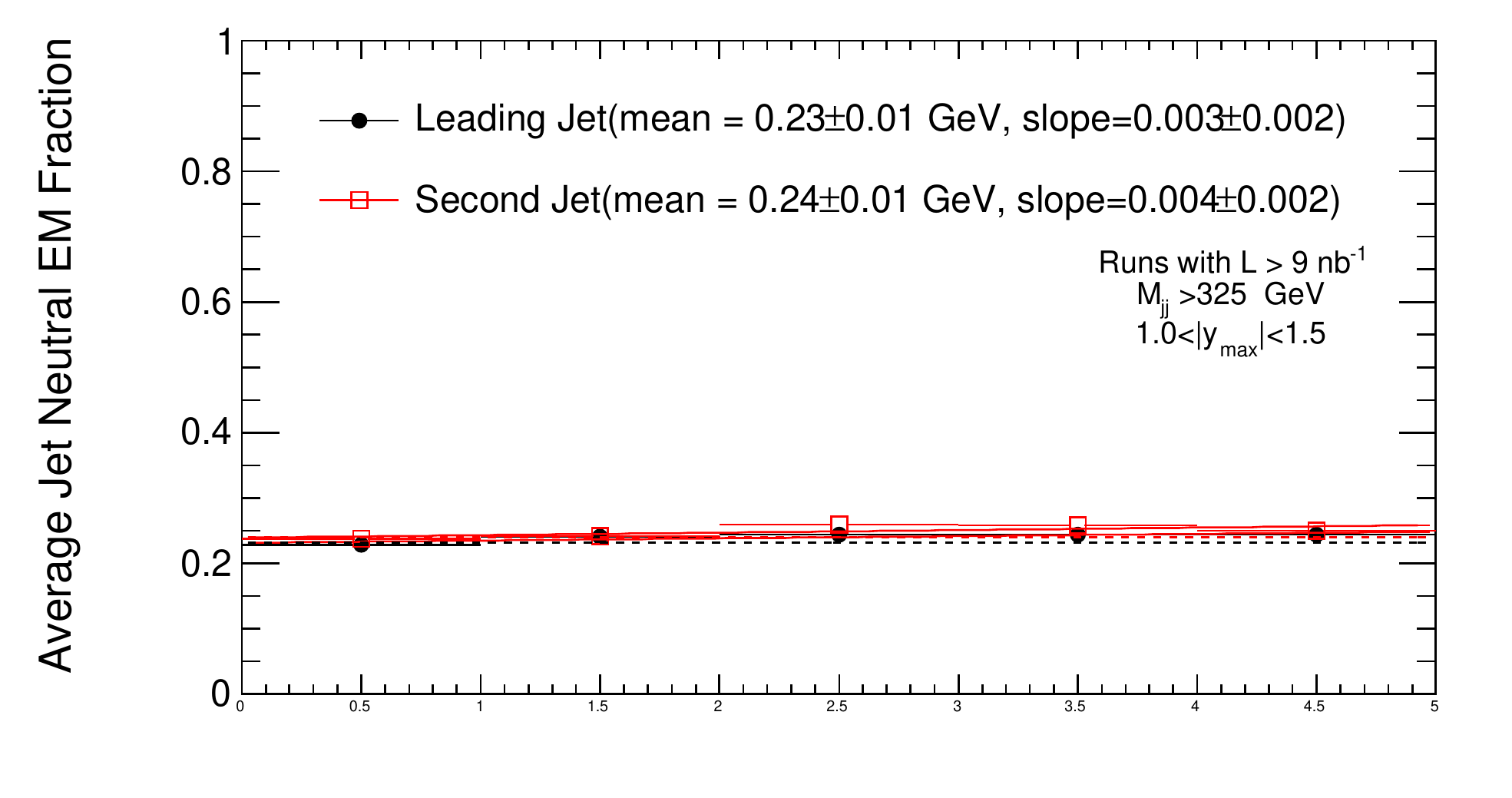}  
\includegraphics[width=0.52\textwidth]{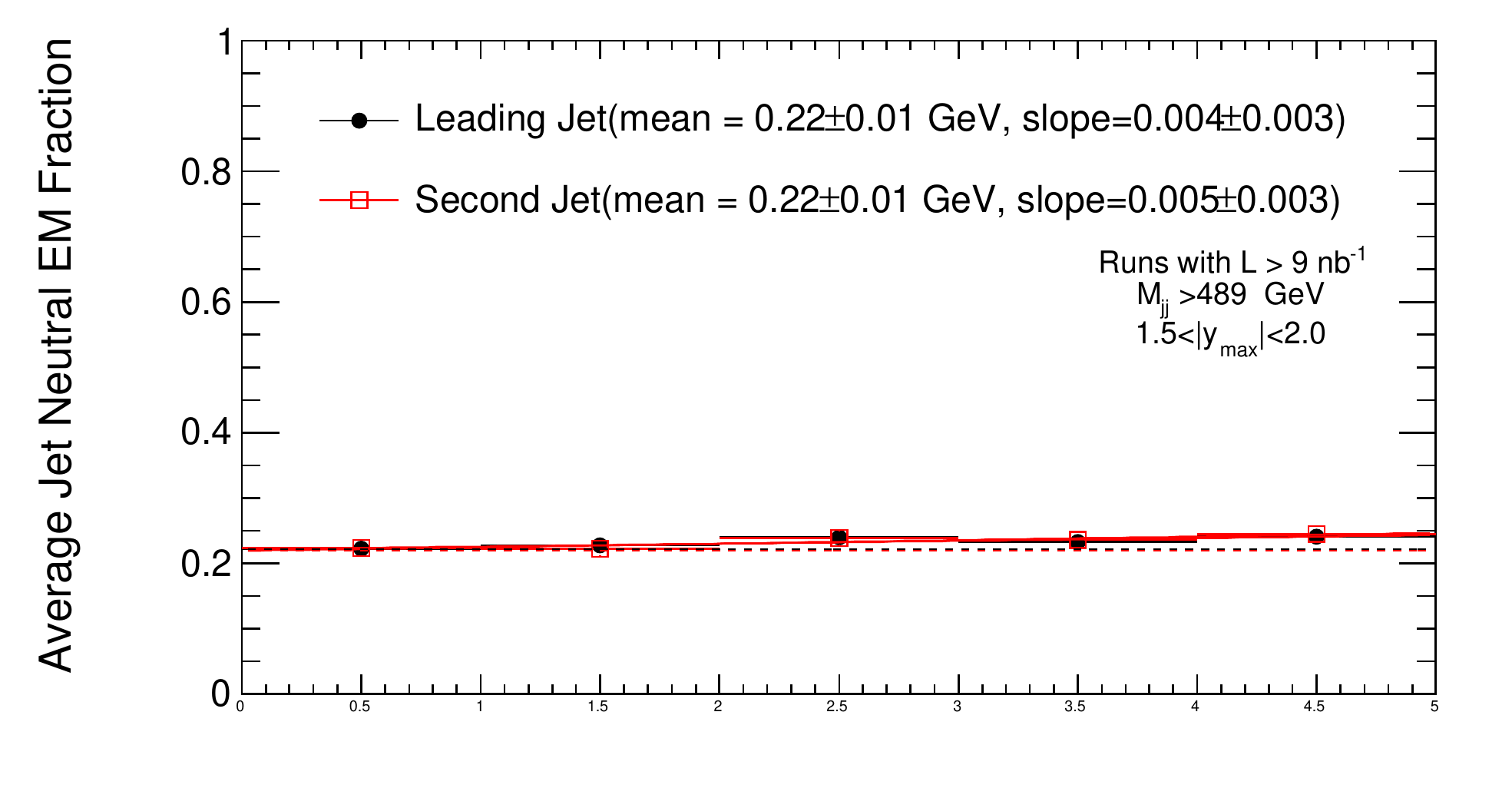} 
\includegraphics[width=0.52\textwidth]{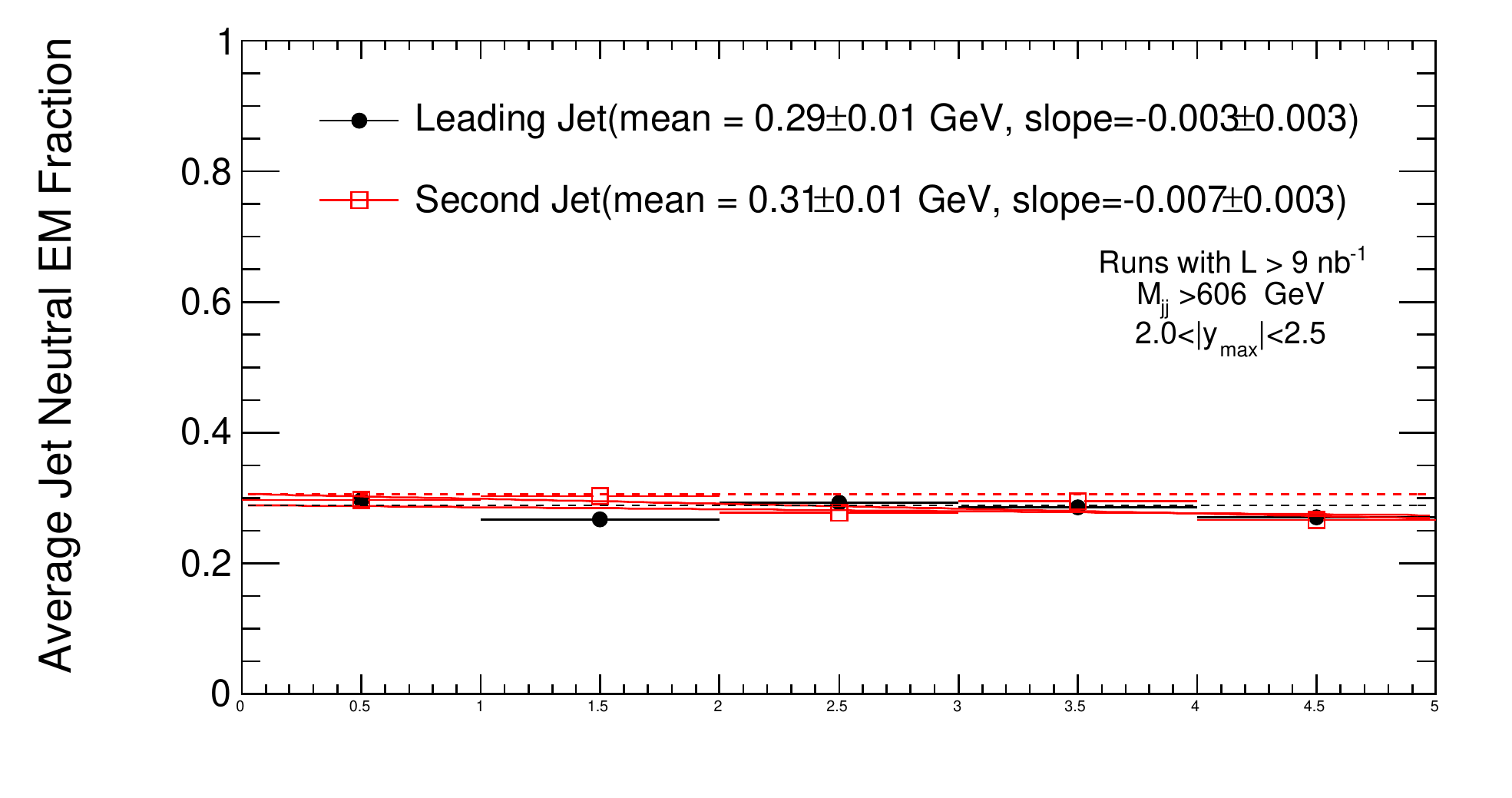} 
   
\capspace
\caption{ The neutral  electromagnetic fraction  of the leading and second jet  for the five different $y_{max}$ bins and for the 
HLT$_{-}$Jet30U trigger as a function of time (run number), fitted with a first degree polynomial.}
\label{fig_appd4}
\end{figure}

%%% jet  50

\begin{figure}[h]
\centering

\includegraphics[width=0.52\textwidth]{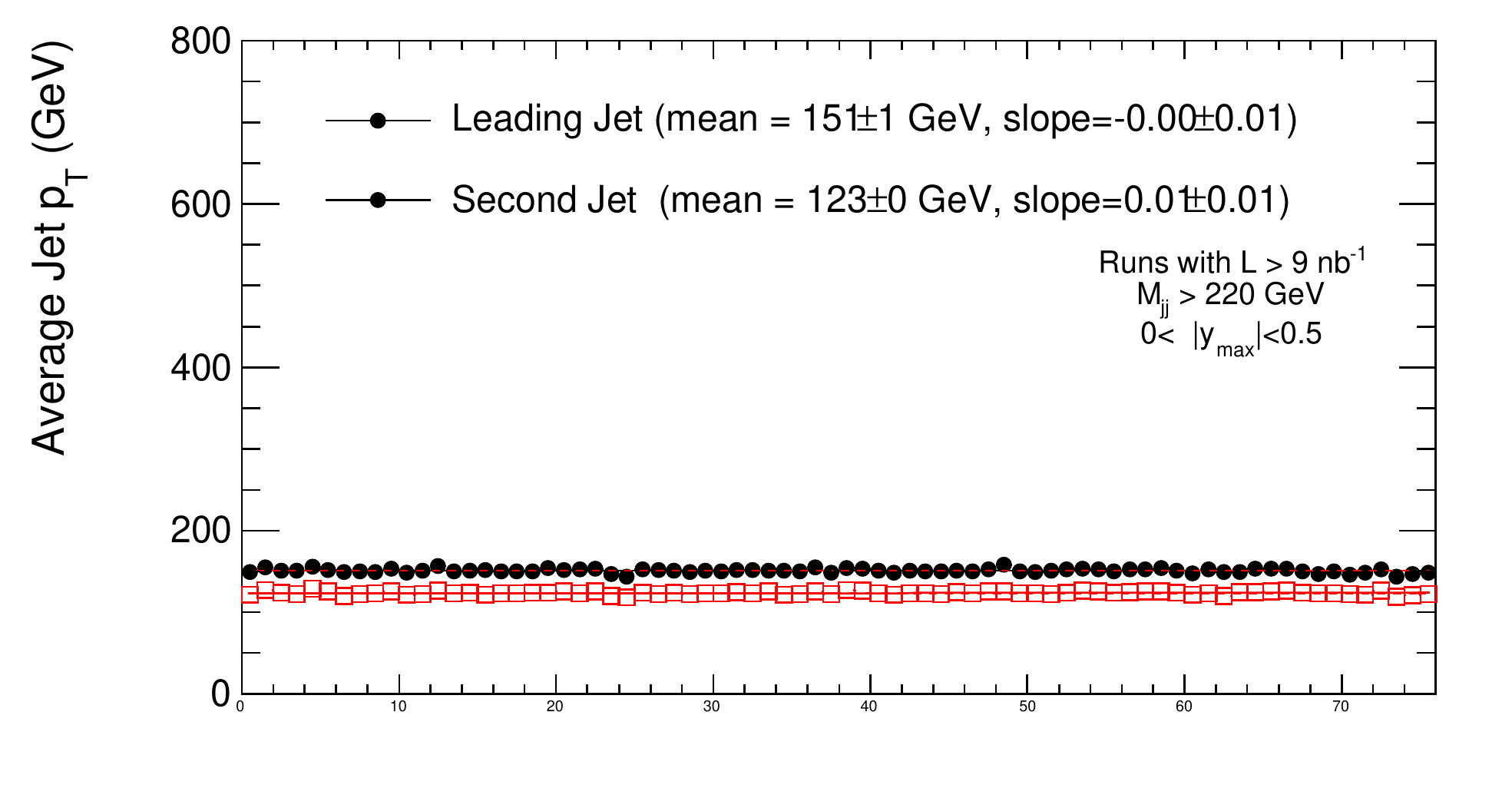} 
\includegraphics[width=0.52\textwidth]{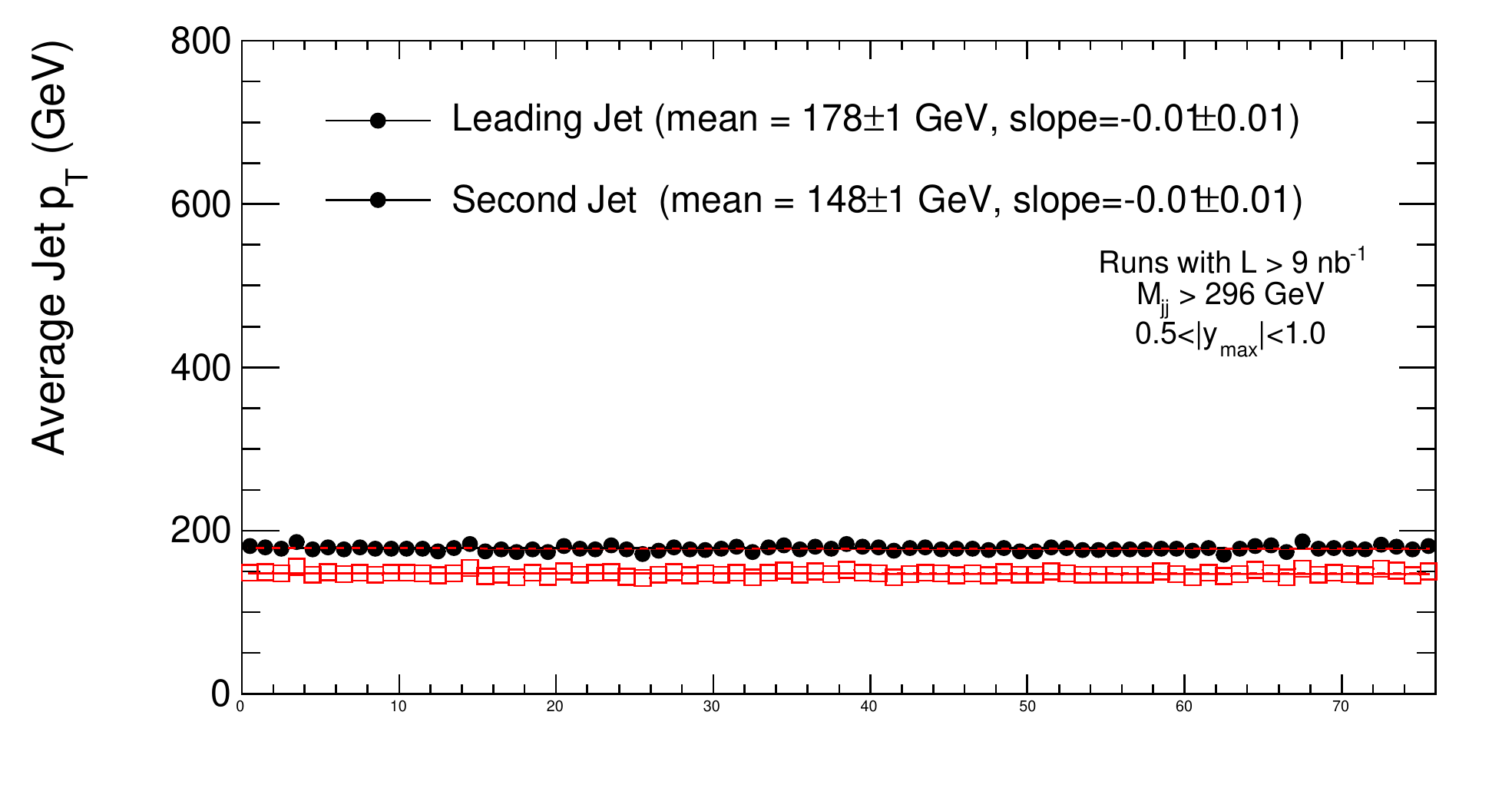} 
\includegraphics[width=0.52\textwidth]{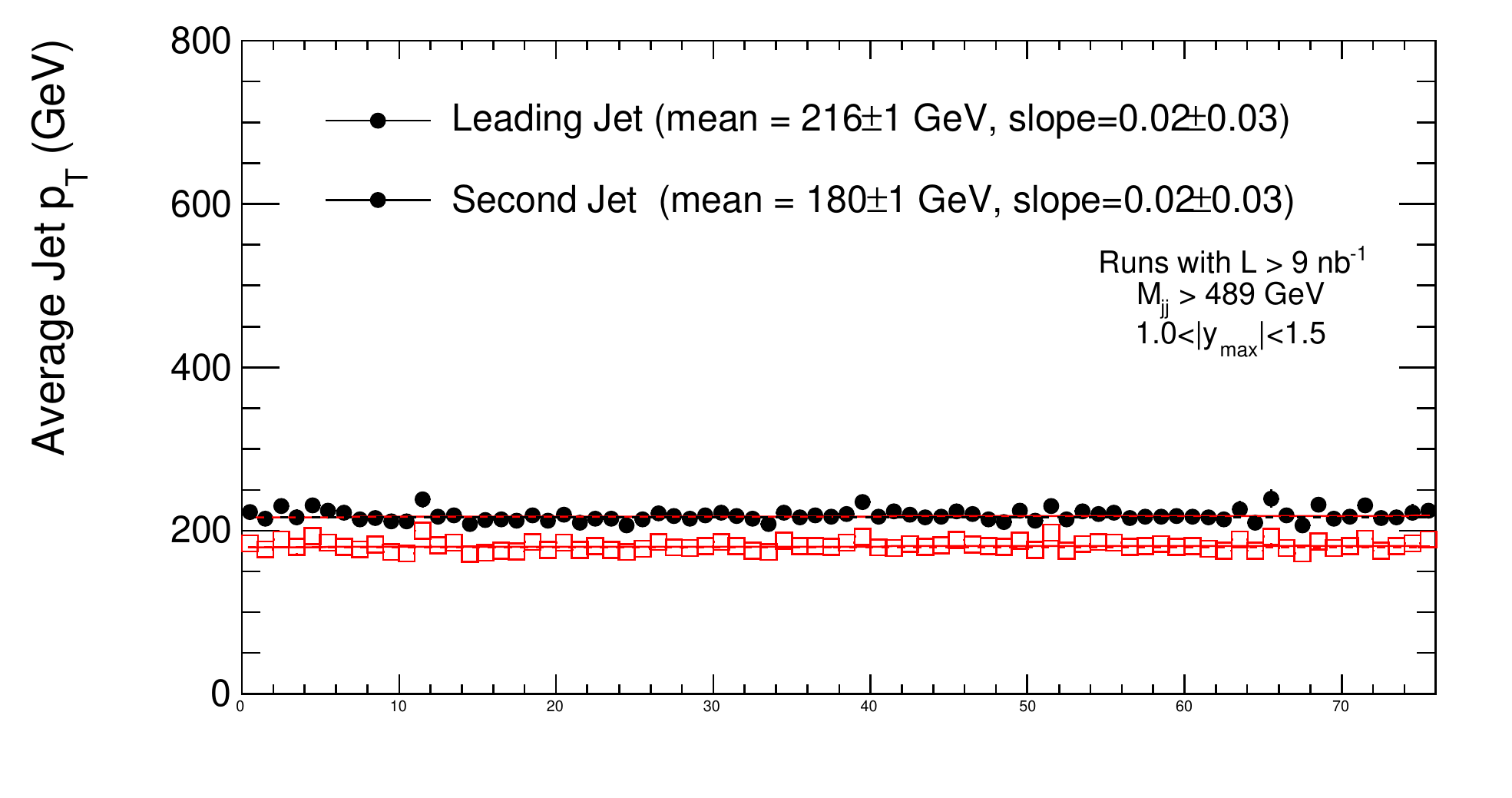} 
\includegraphics[width=0.52\textwidth]{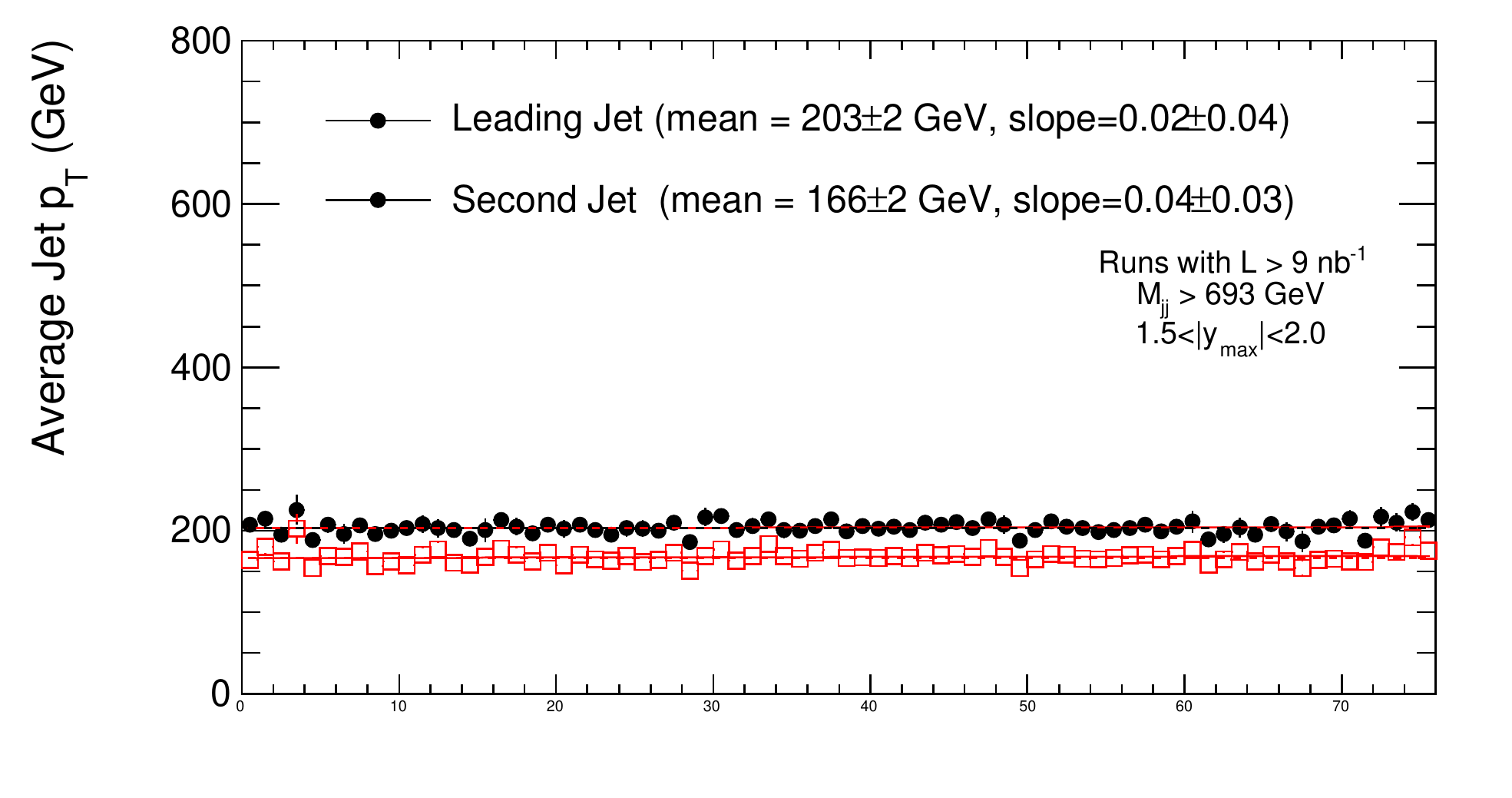} 
\includegraphics[width=0.52\textwidth]{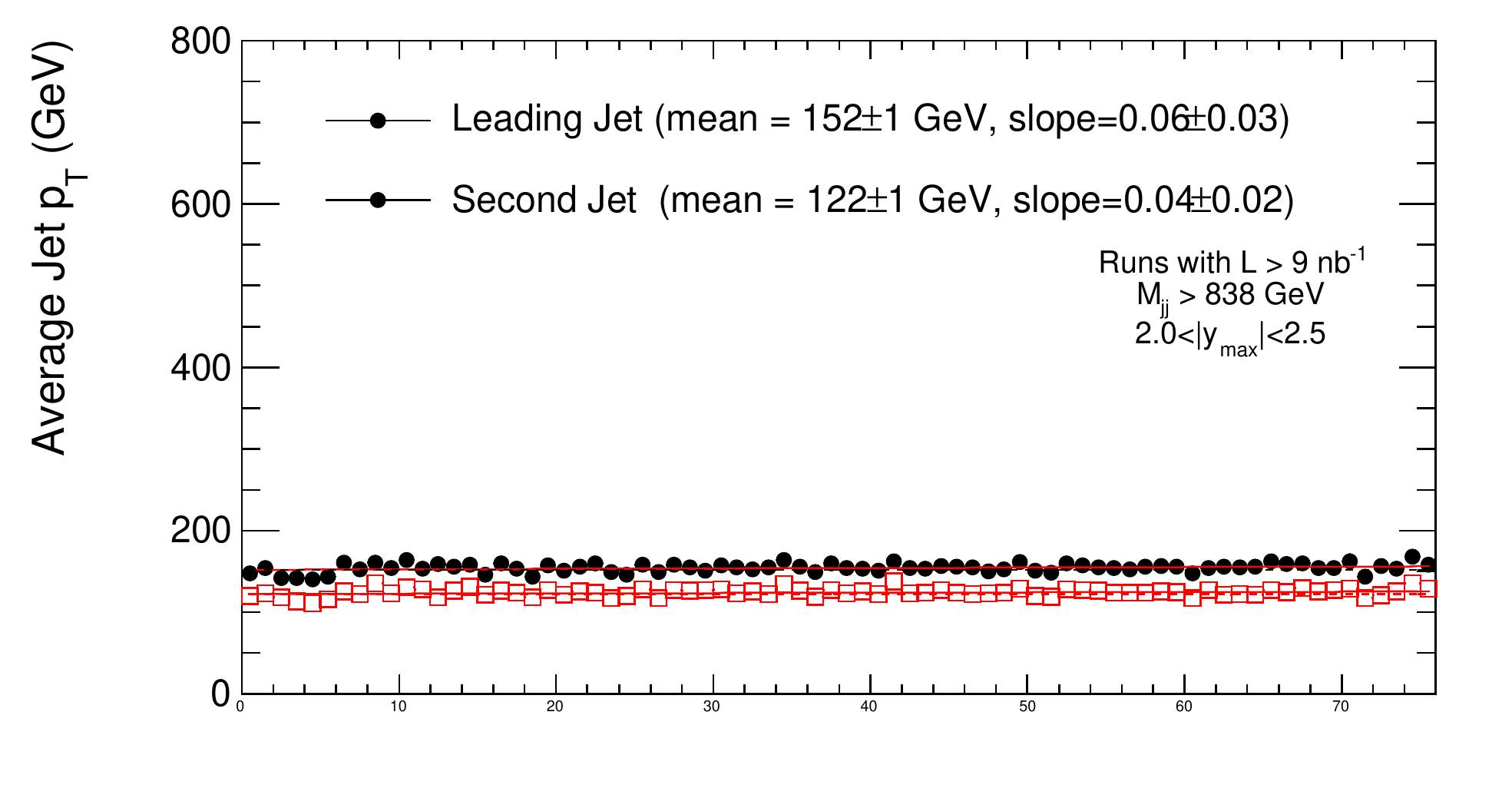} 
  
\capspace
\caption{ The $p_T$ of the leading and second jet  for the five different $y_{max}$ bins and for the
HLT$_{-}$Jet50U trigger as a function of time (run number), fitted with a first degree polynomial.}
\label{fig_appd5}
\end{figure}

\begin{figure}[h]
\centering

\includegraphics[width=0.52\textwidth]{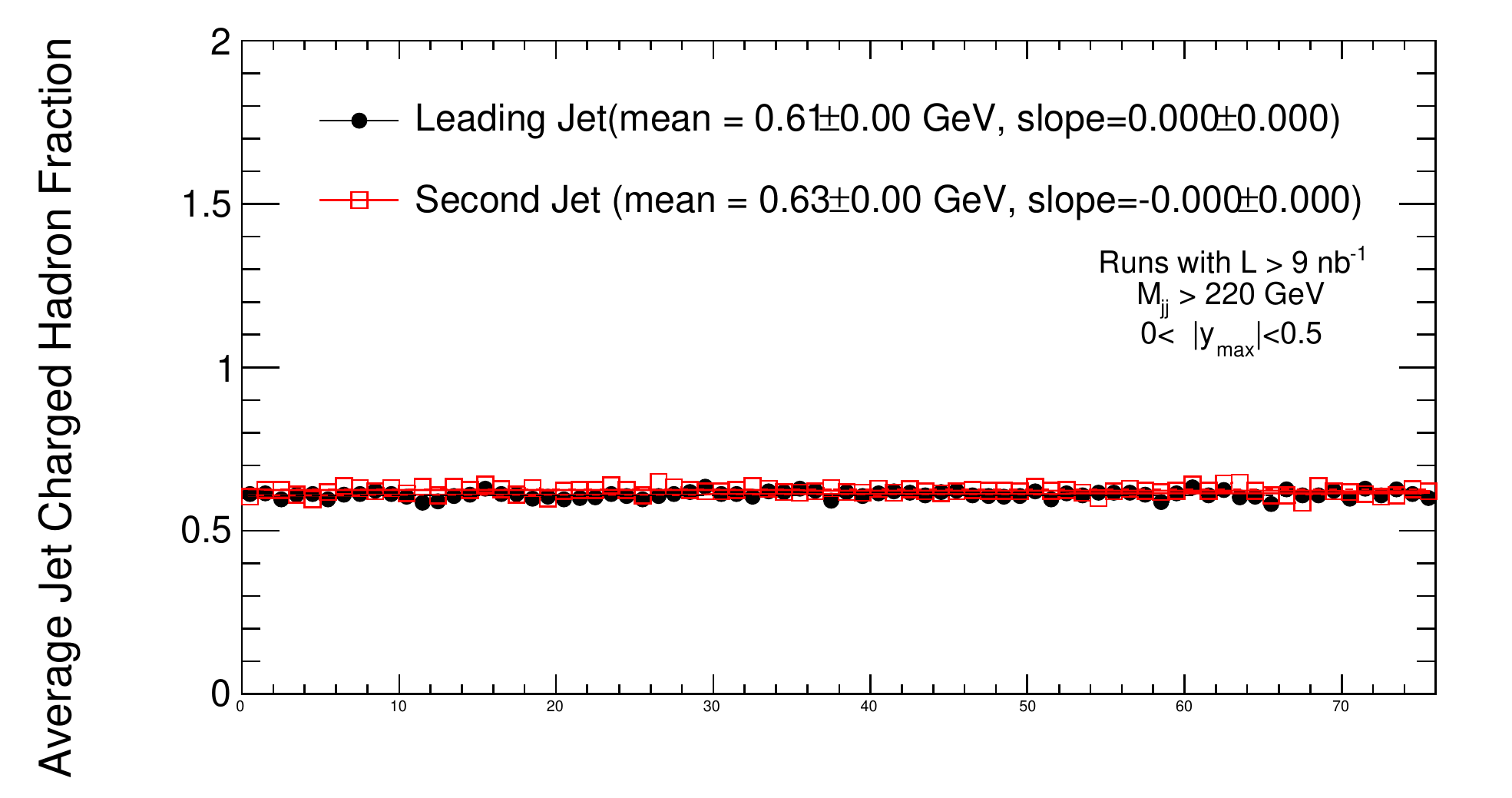} 
\includegraphics[width=0.52\textwidth]{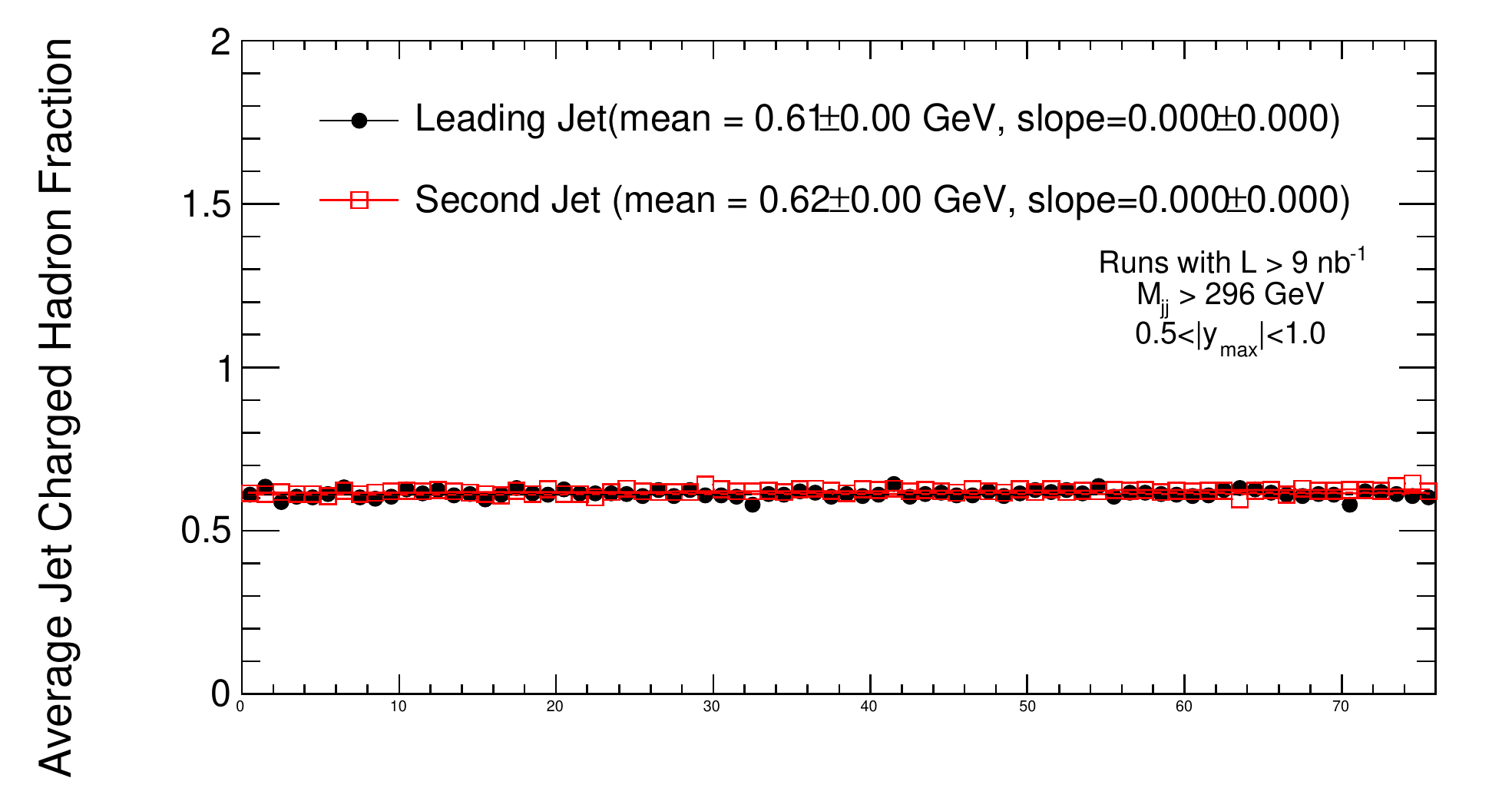} 
\includegraphics[width=0.52\textwidth]{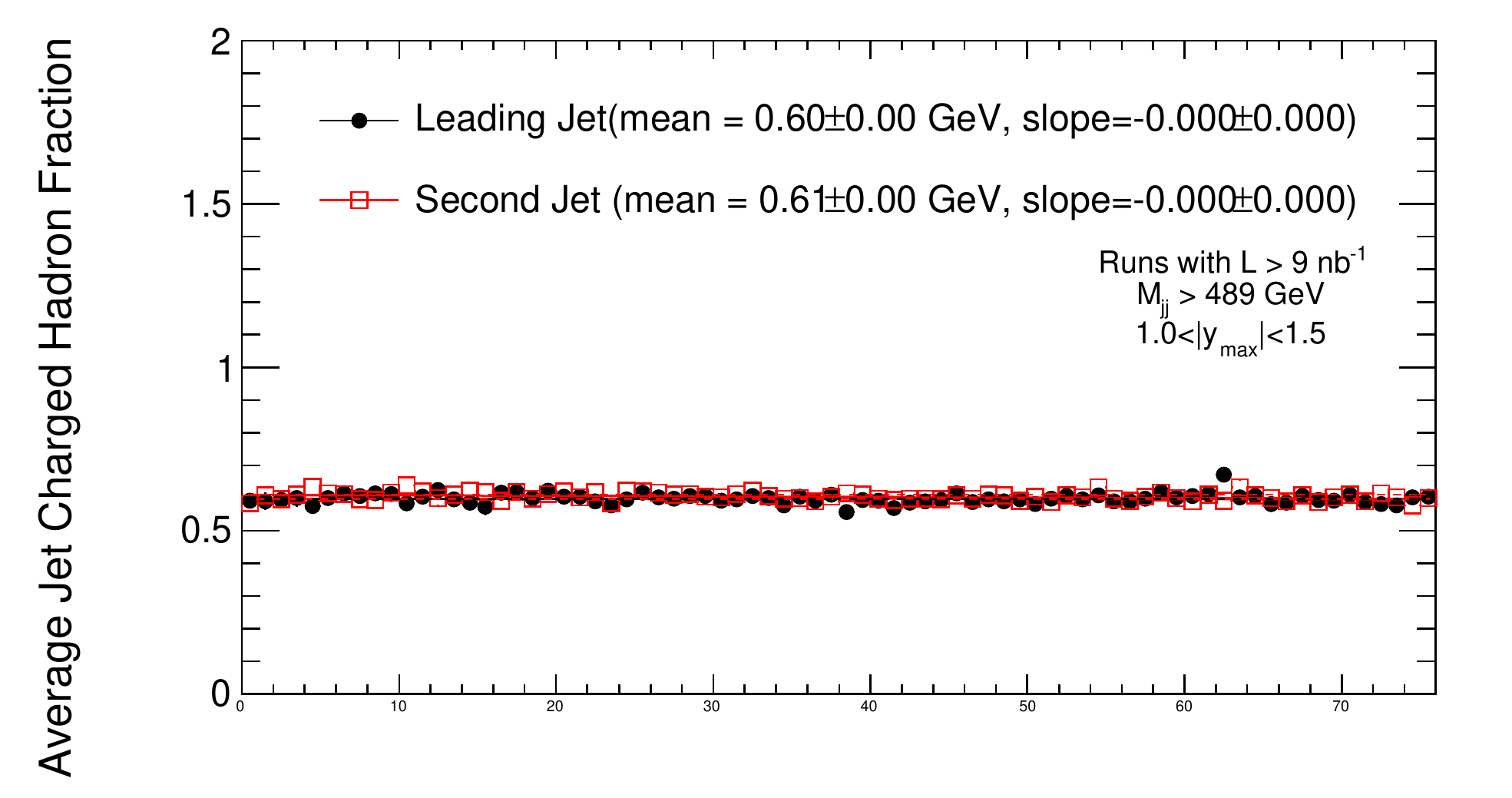} 
\includegraphics[width=0.52\textwidth]{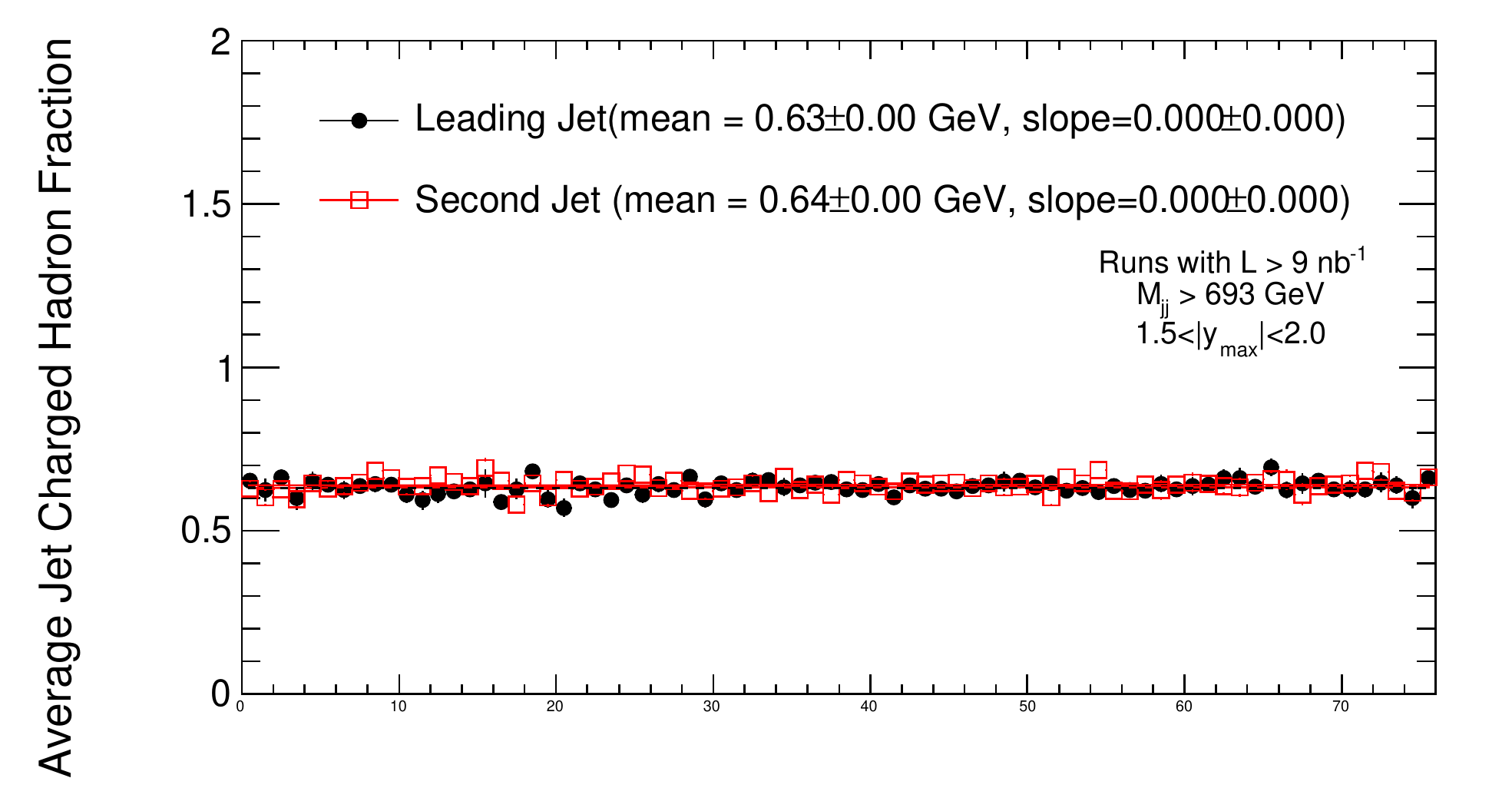} 
\includegraphics[width=0.52\textwidth]{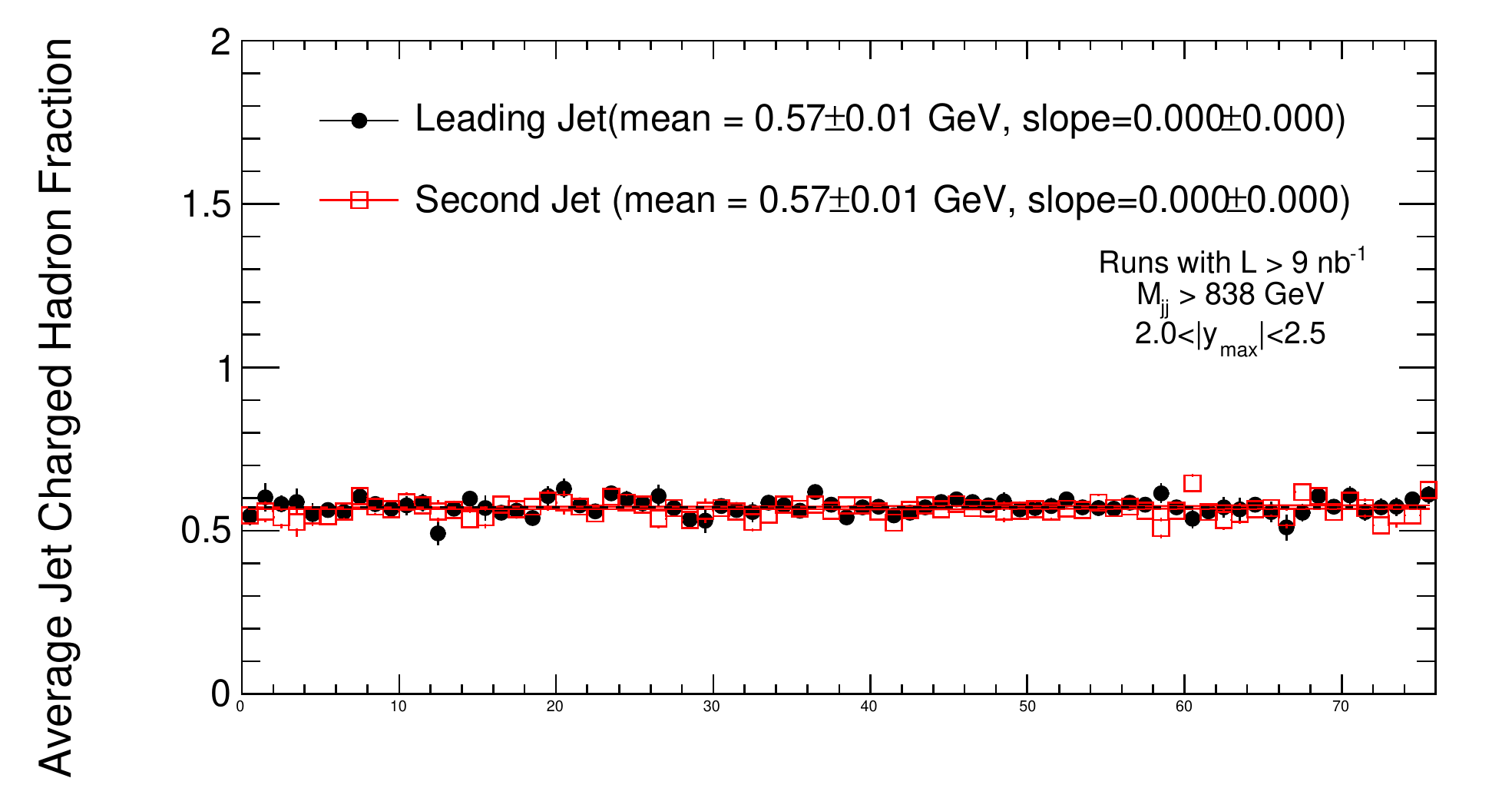} 
  
\capspace
\caption{ The charged hadron fraction  of the leading and second jet  for the five different $y_{max}$ bins and for the
HLT$_{-}$Jet50U trigger as a function of time (run number), fitted with a first degree polynomial.}
\label{fig_appd6}
\end{figure}

\clearpage

\begin{figure}[h]
\centering

\includegraphics[width=0.52\textwidth]{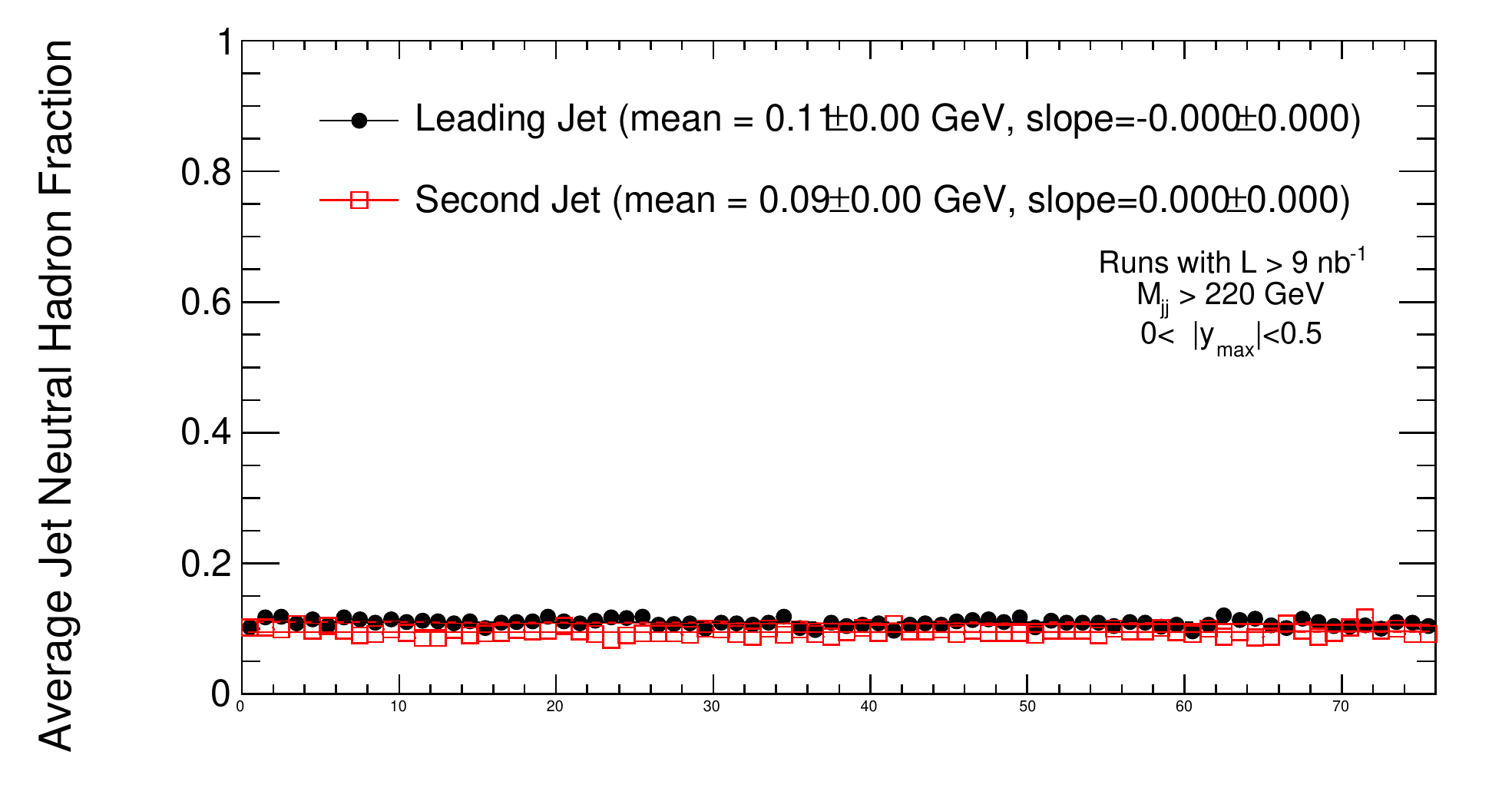} 
\includegraphics[width=0.52\textwidth]{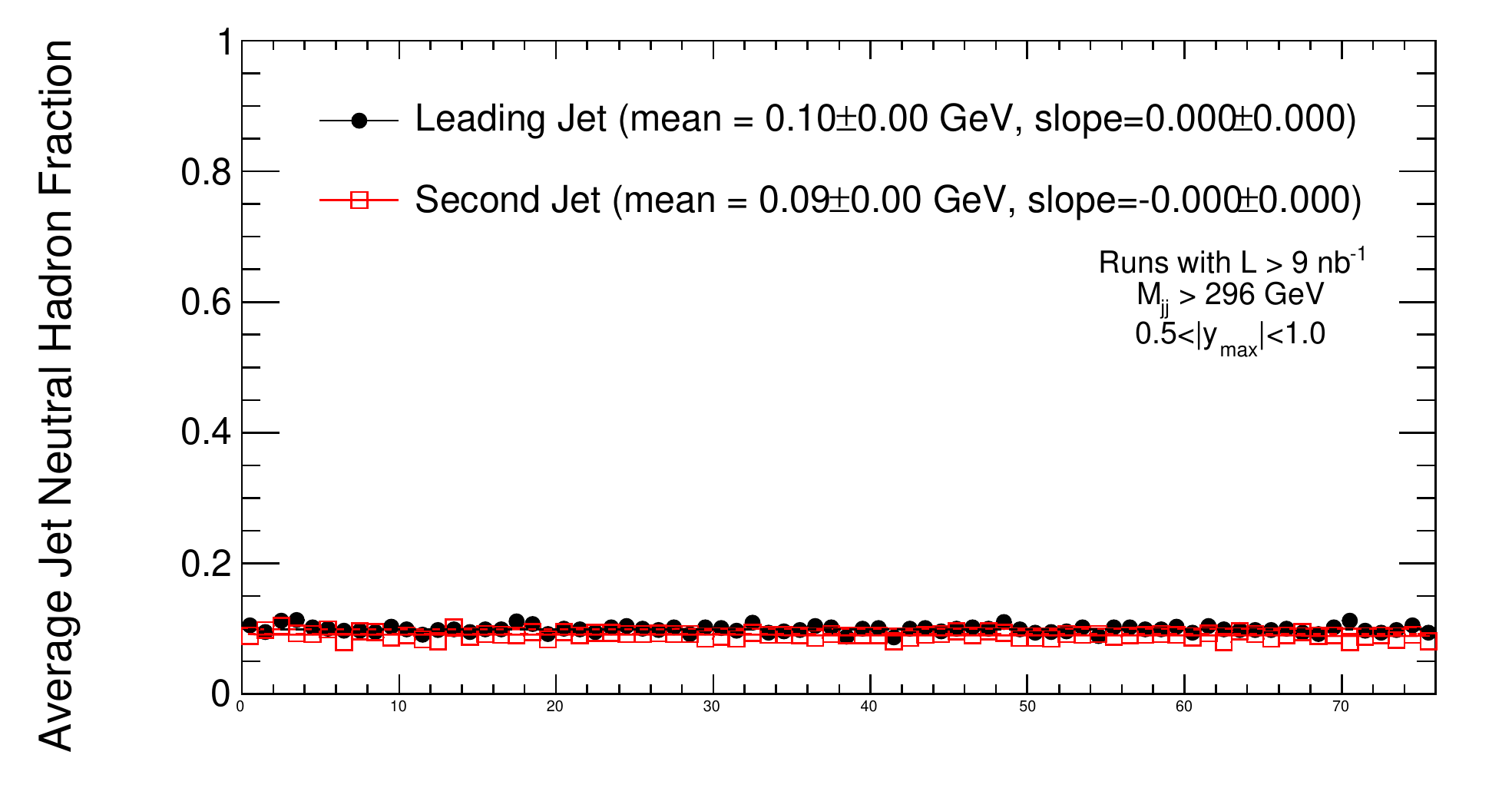} 
\includegraphics[width=0.52\textwidth]{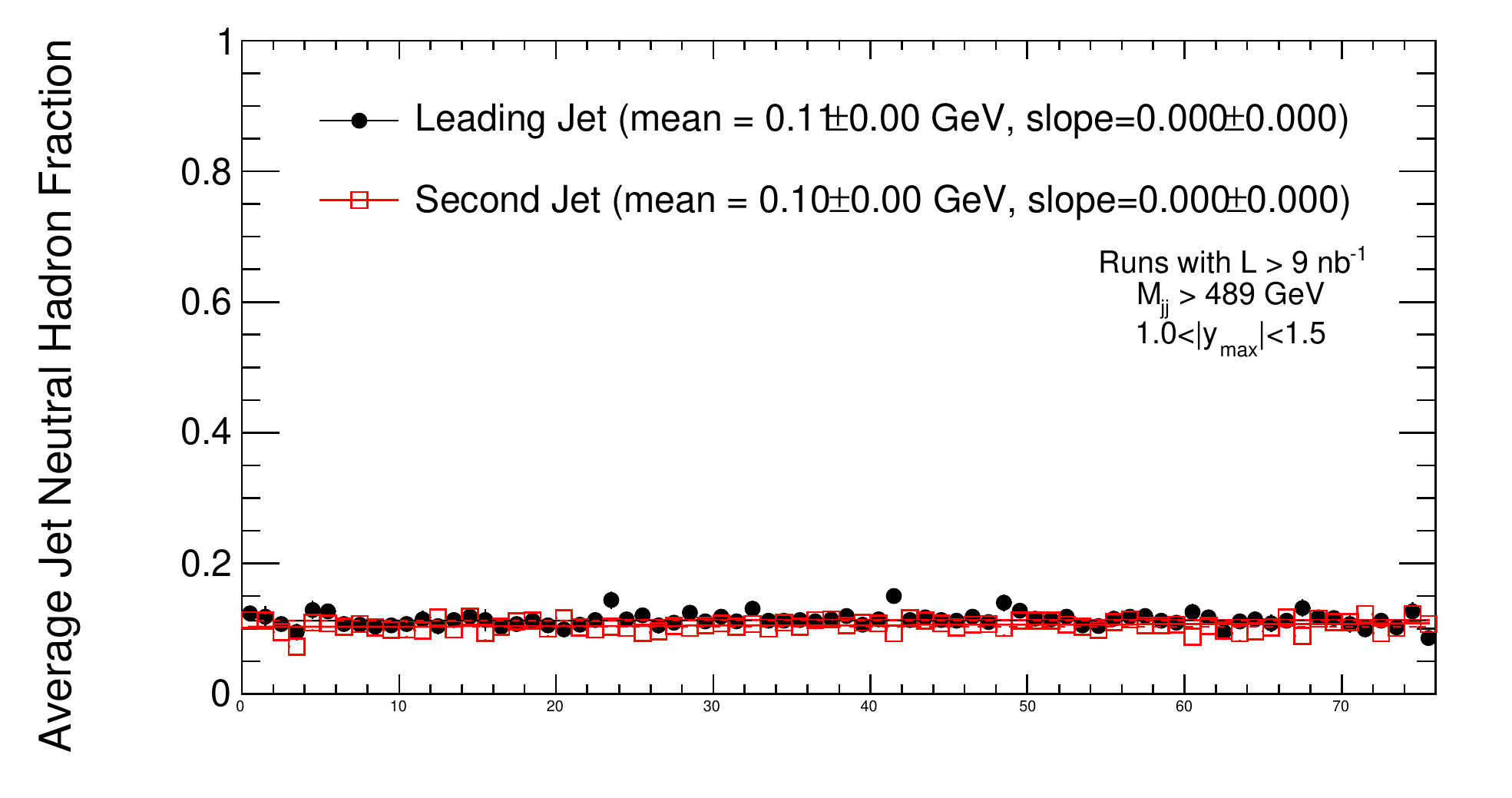} 
\includegraphics[width=0.52\textwidth]{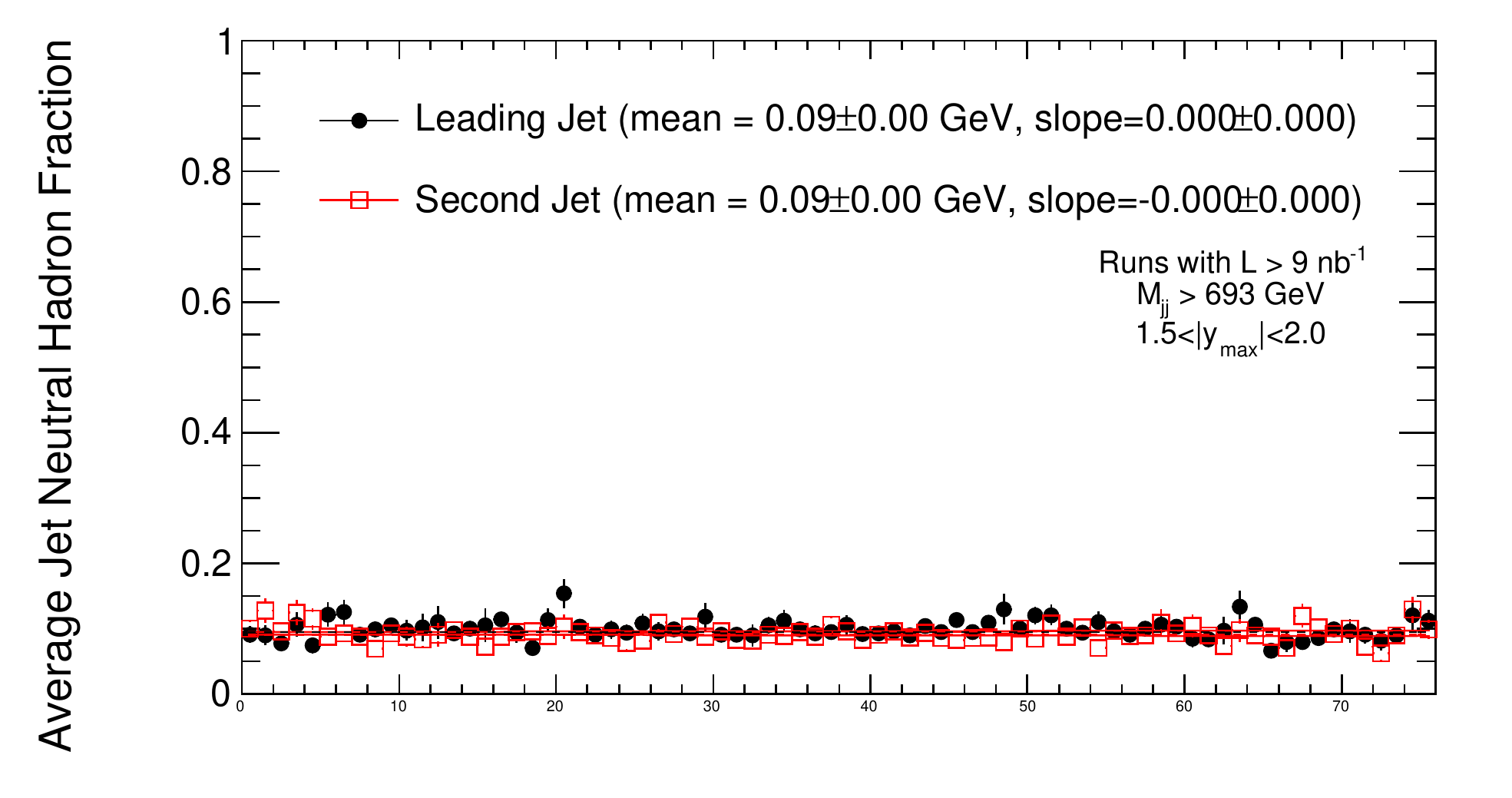} 
\includegraphics[width=0.52\textwidth]{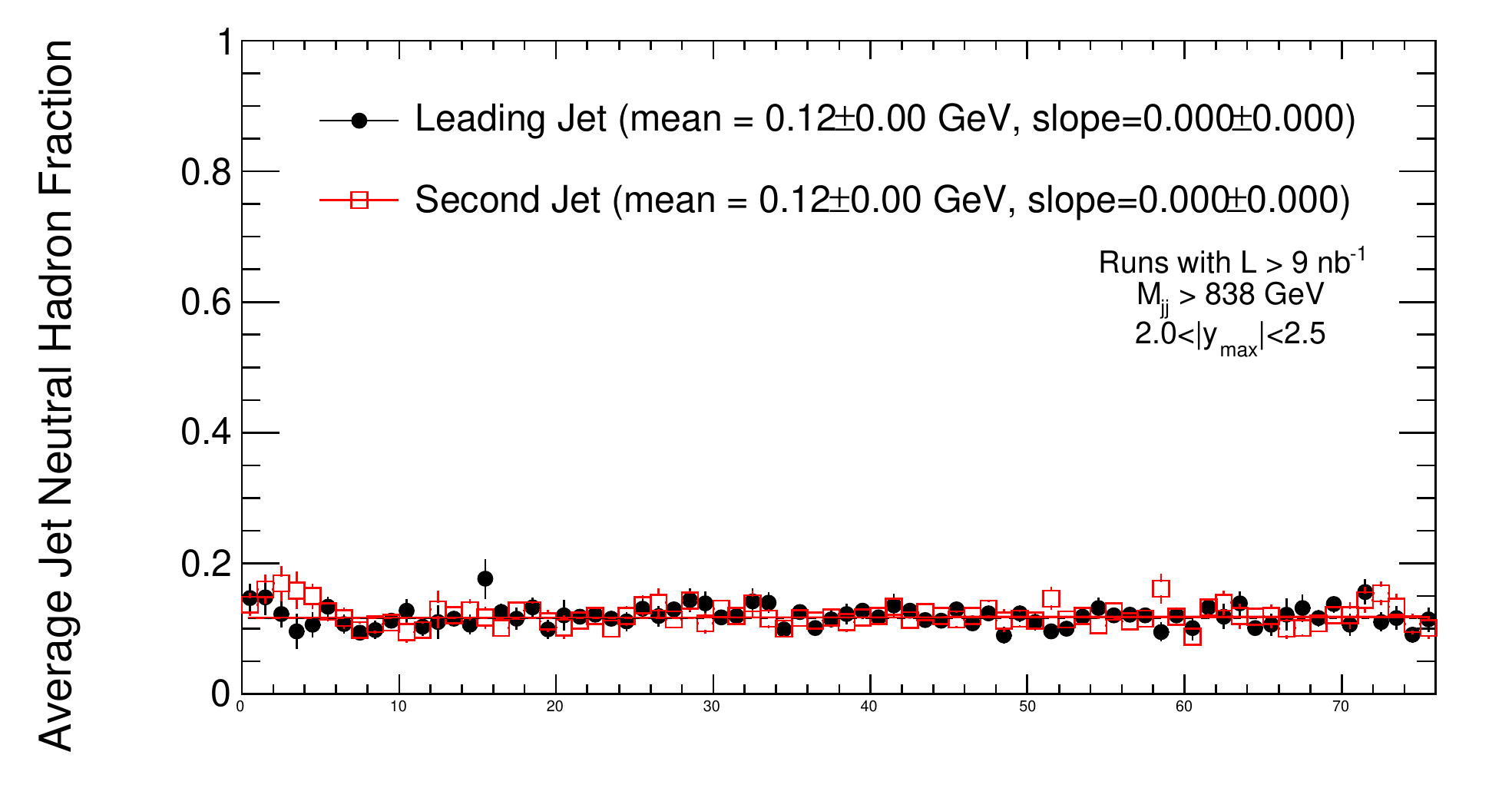} 
  
\capspace
\caption{ The neutral  hadron fraction  of the leading and second jet  for the five different $y_{max}$ bins and for the
HLT$_{-}$Jet50U trigger as a function of time (run number), fitted with a first degree polynomial.}
\label{fig_appd7}
\end{figure}

\begin{figure}[h]
\centering

\includegraphics[width=0.52\textwidth]{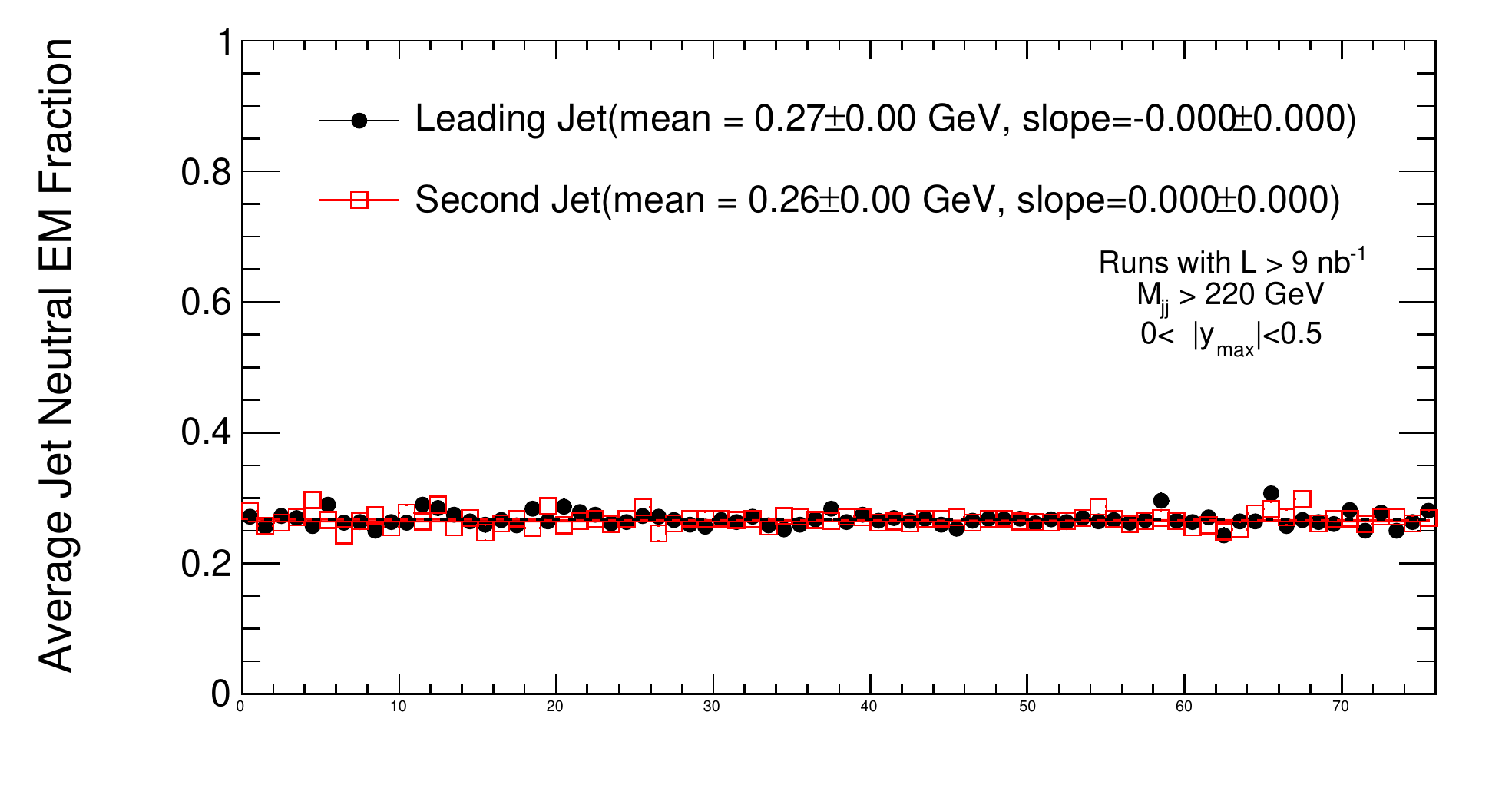} 
\includegraphics[width=0.52\textwidth]{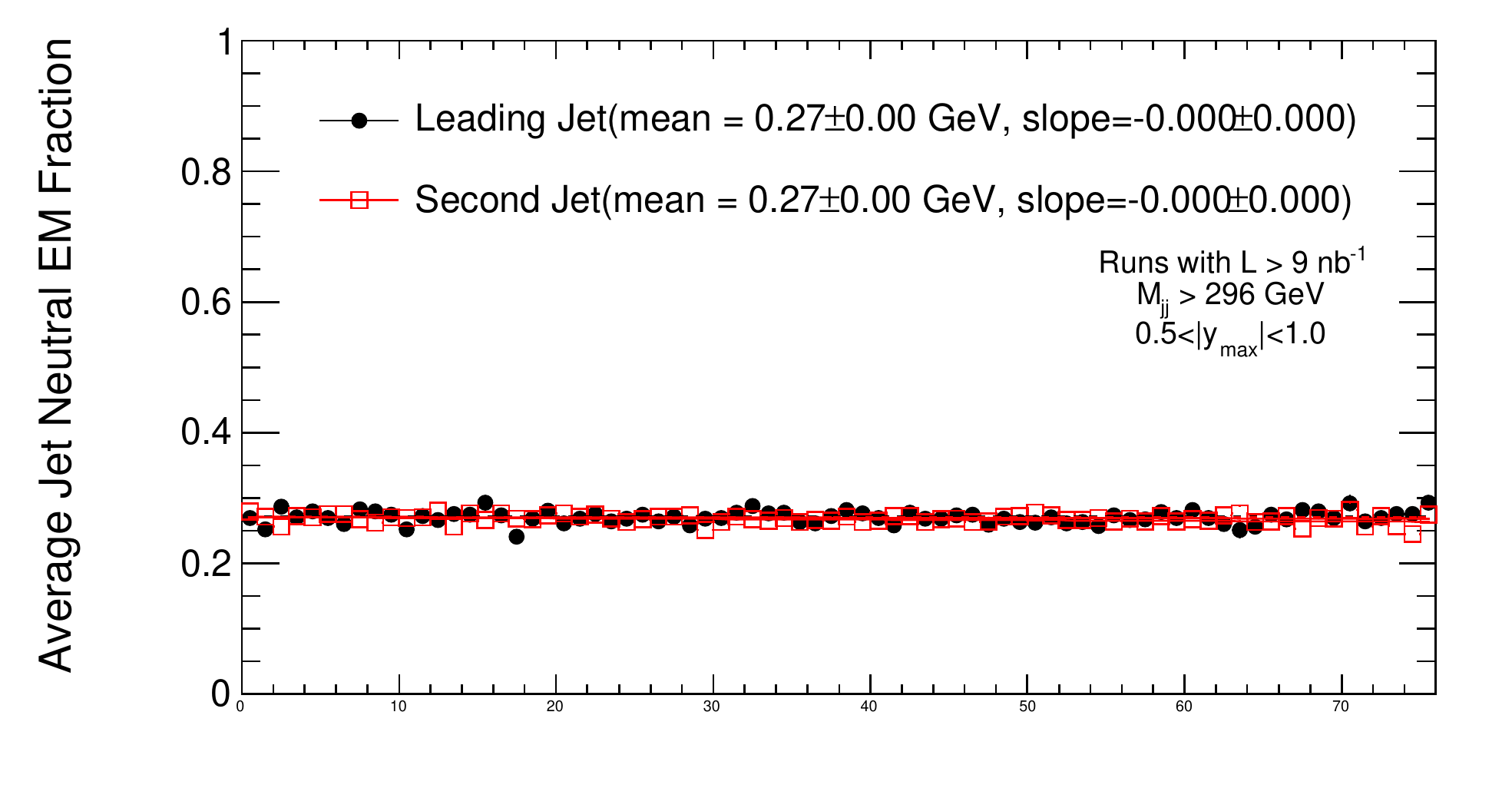} 
\includegraphics[width=0.52\textwidth]{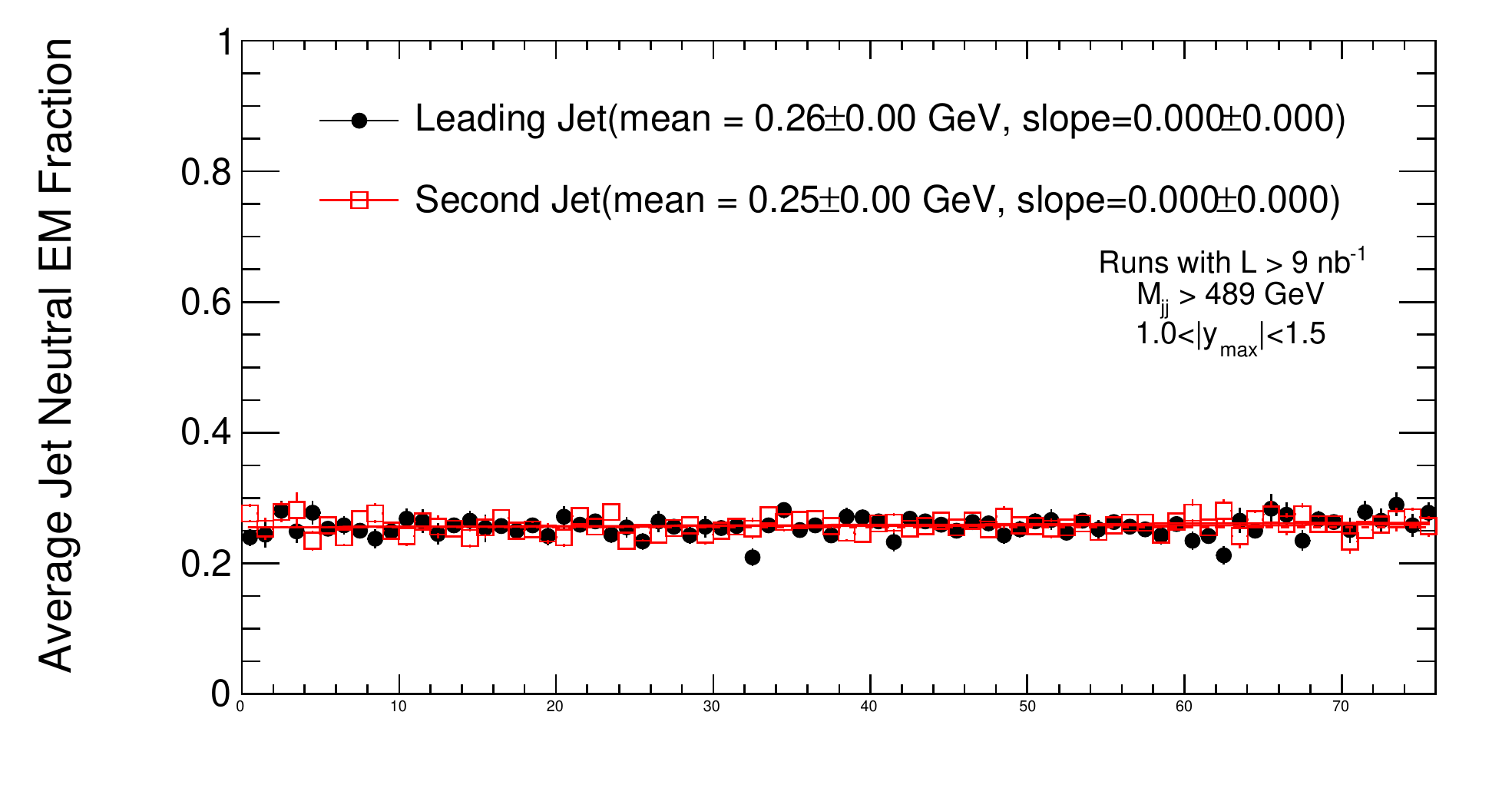} 
\includegraphics[width=0.52\textwidth]{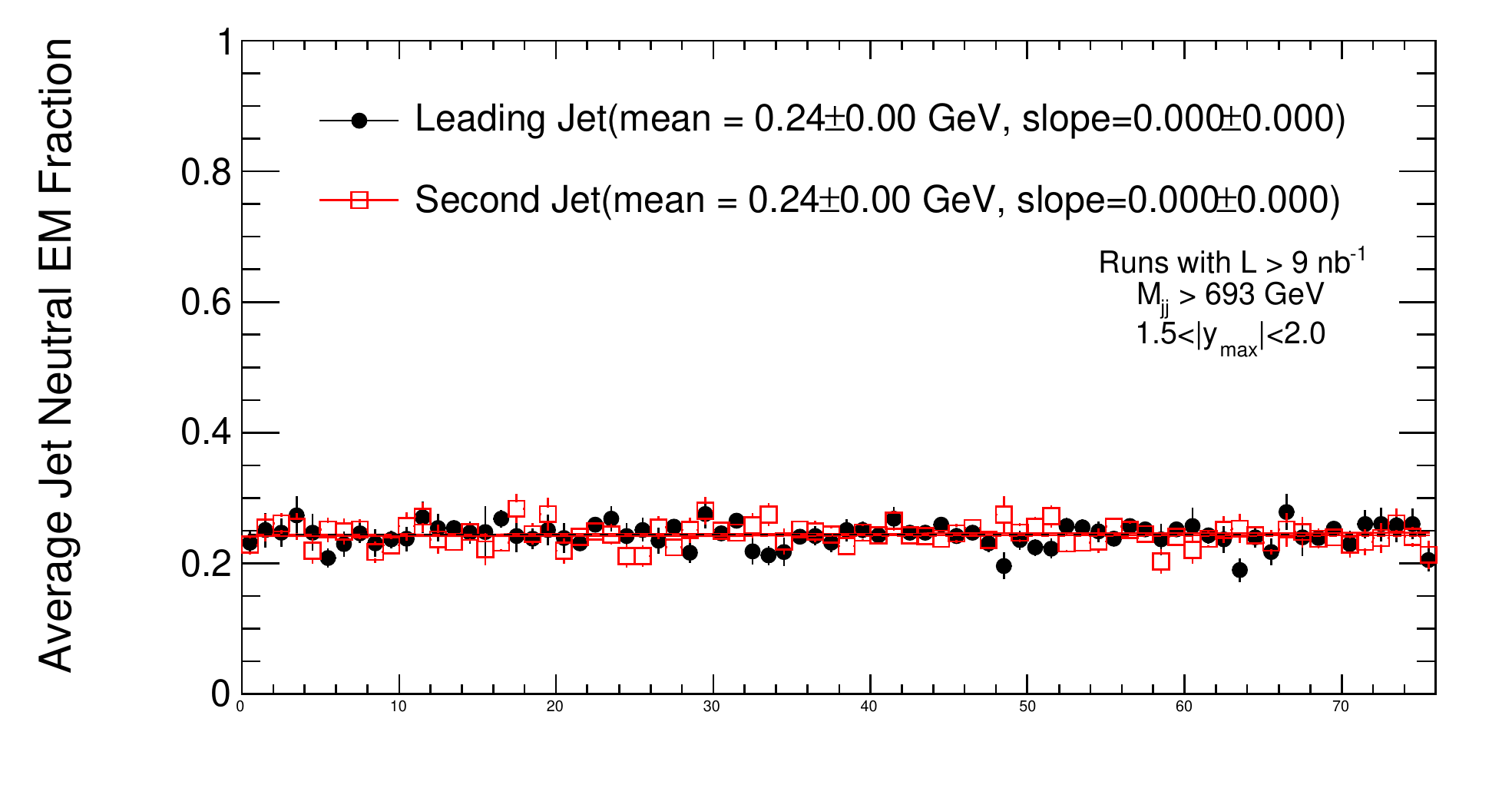} 
\includegraphics[width=0.52\textwidth]{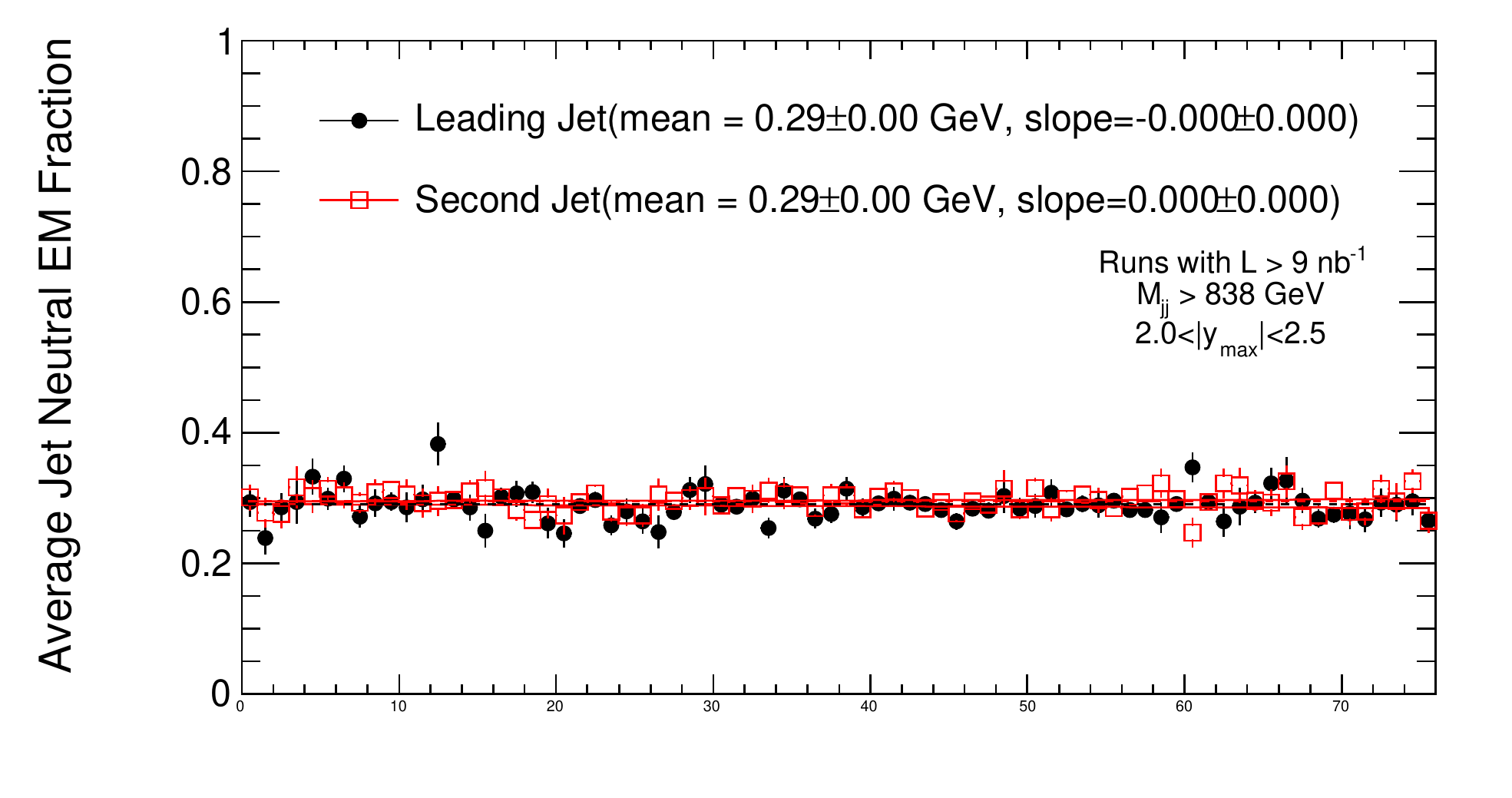} 
  
\capspace
\caption{ The neutral  electromagnetic fraction  of the leading and second jet  for the five different $y_{max}$ bins and for the
HLT$_{-}$Jet50U trigger as a function of time (run number), fitted with a first degree polynomial.}
\label{fig_appd8}
\end{figure}

%%% jet 100

\begin{figure}[h]
\centering

\includegraphics[width=0.52\textwidth]{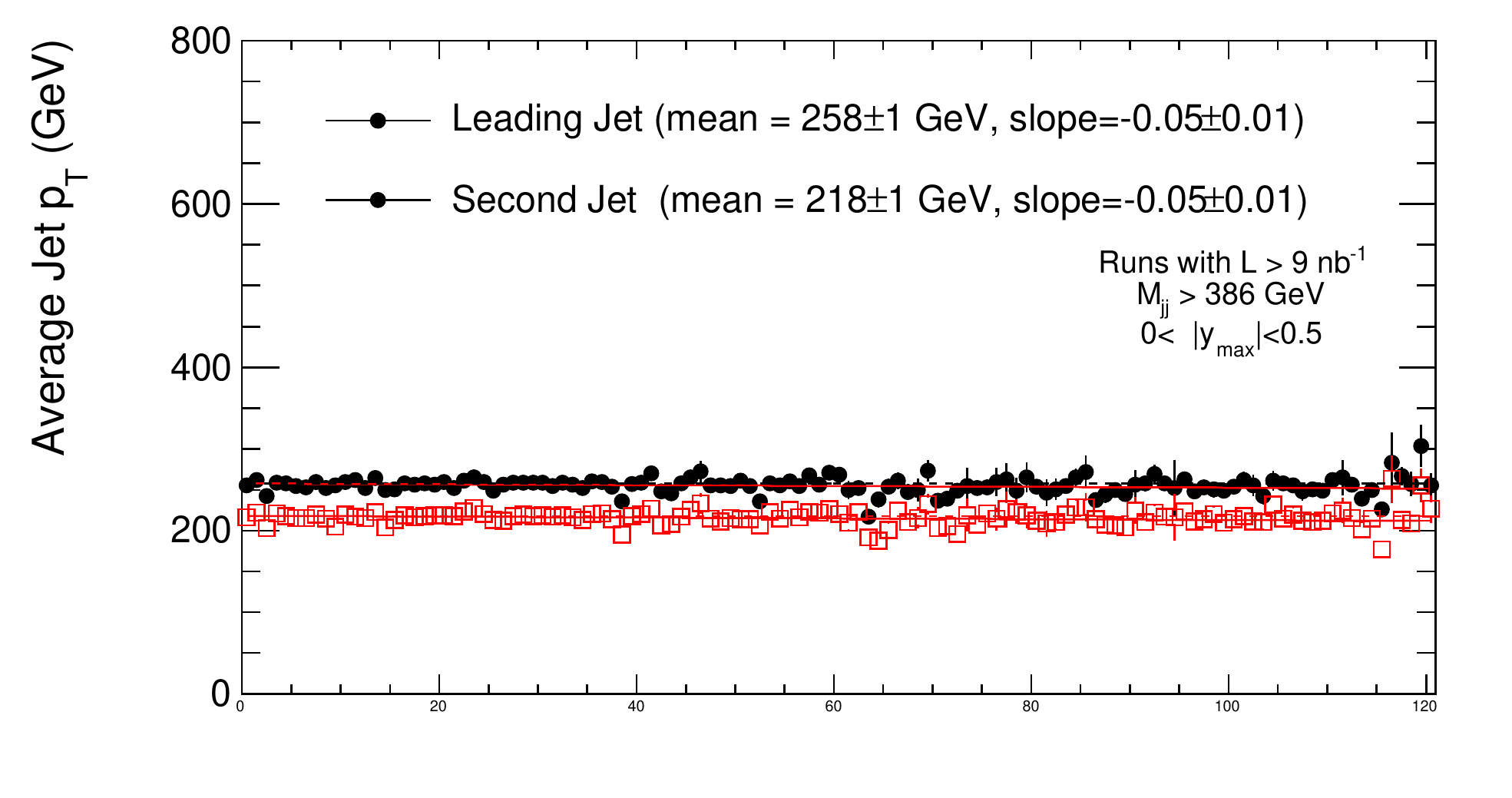} 
\includegraphics[width=0.52\textwidth]{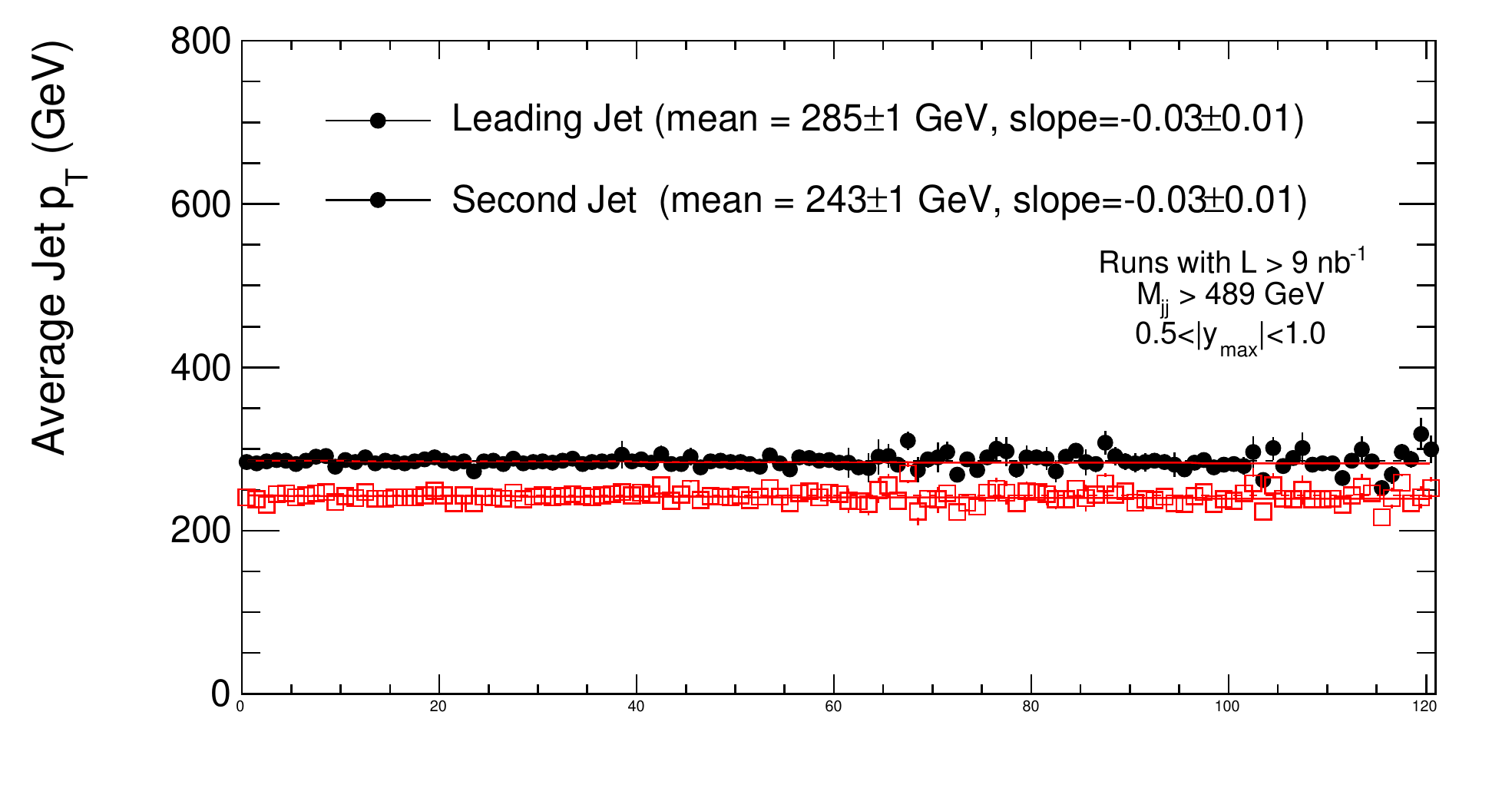} 
\includegraphics[width=0.52\textwidth]{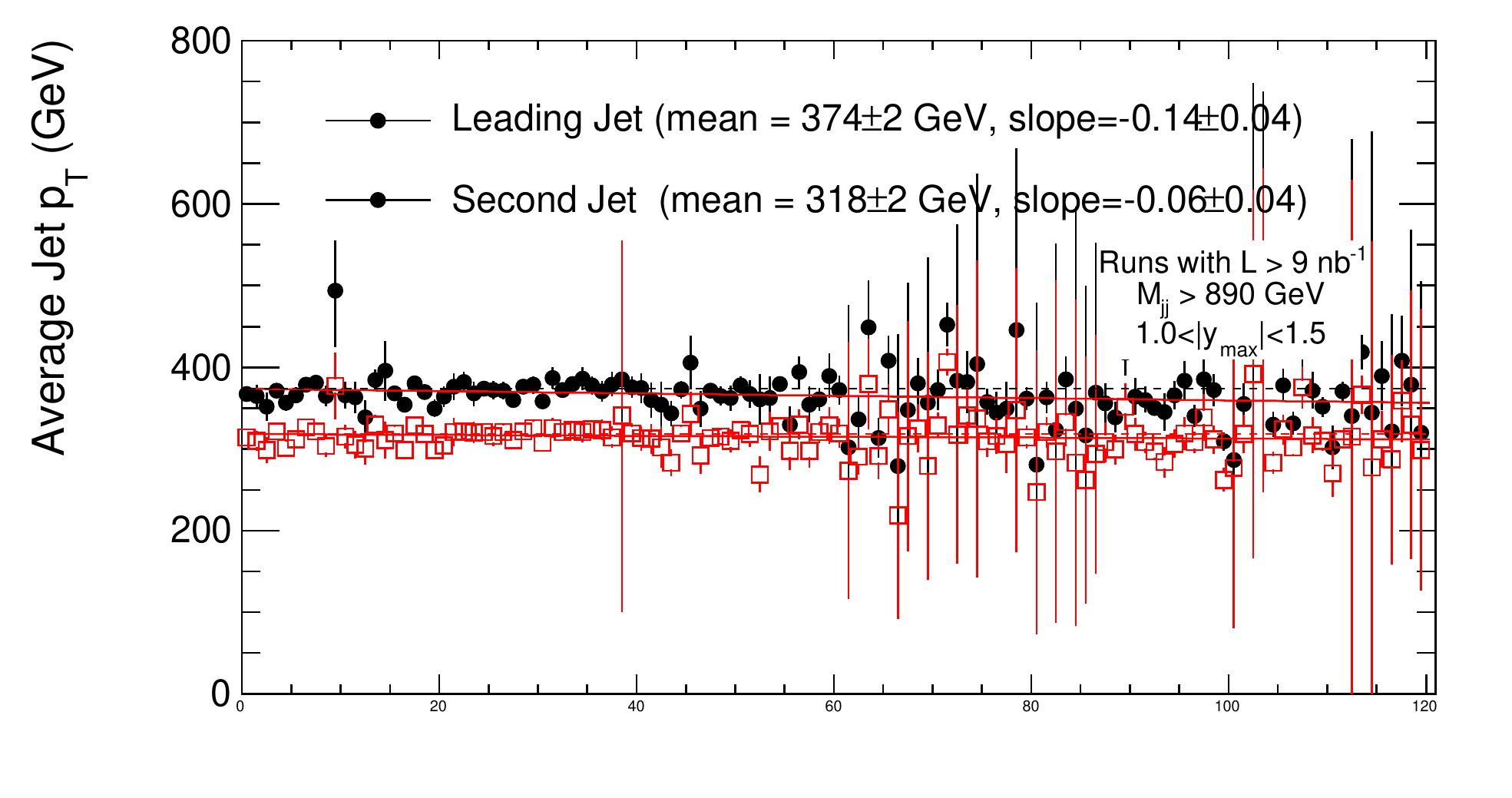} 
\includegraphics[width=0.52\textwidth]{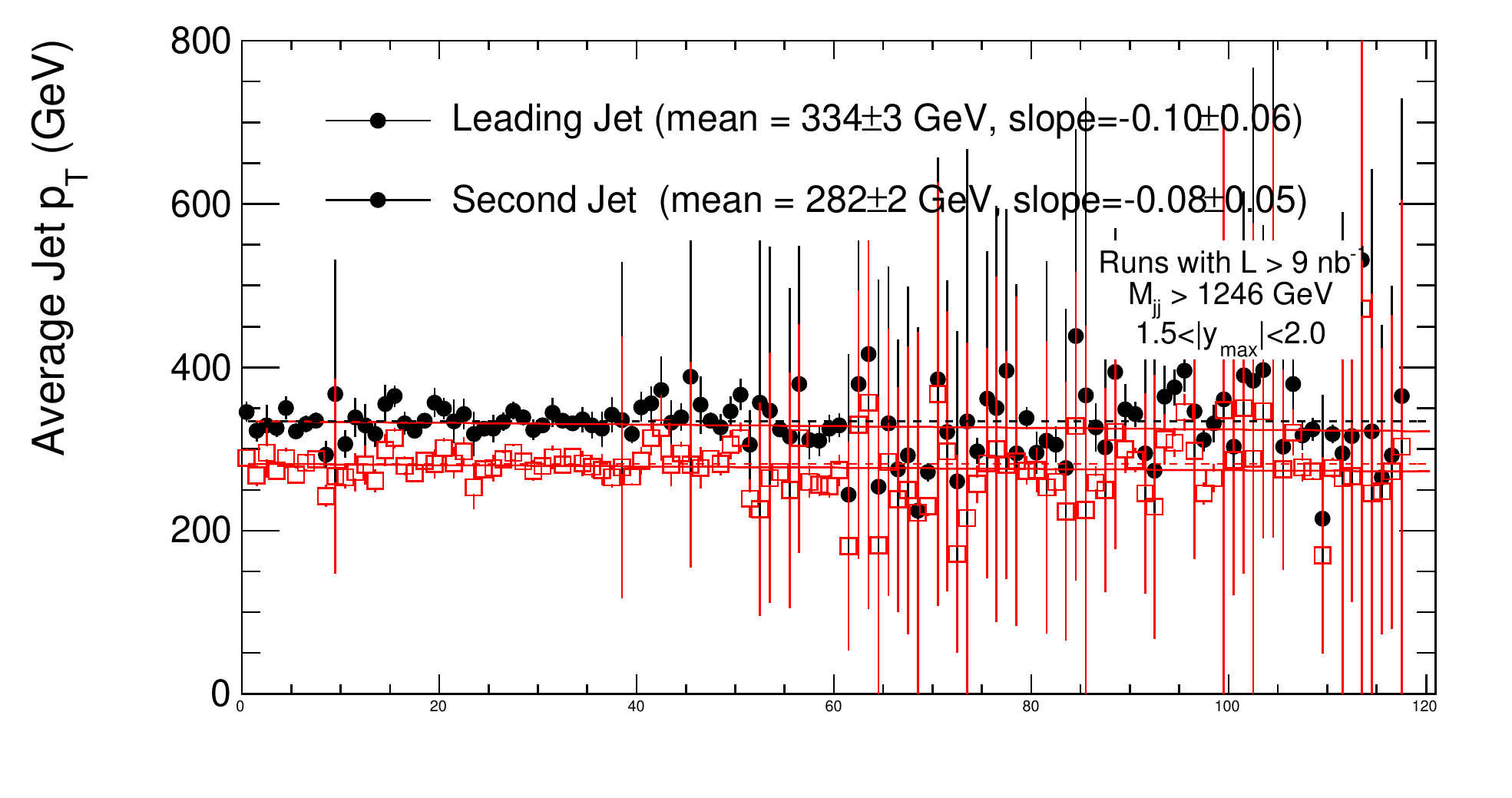} 
\includegraphics[width=0.52\textwidth]{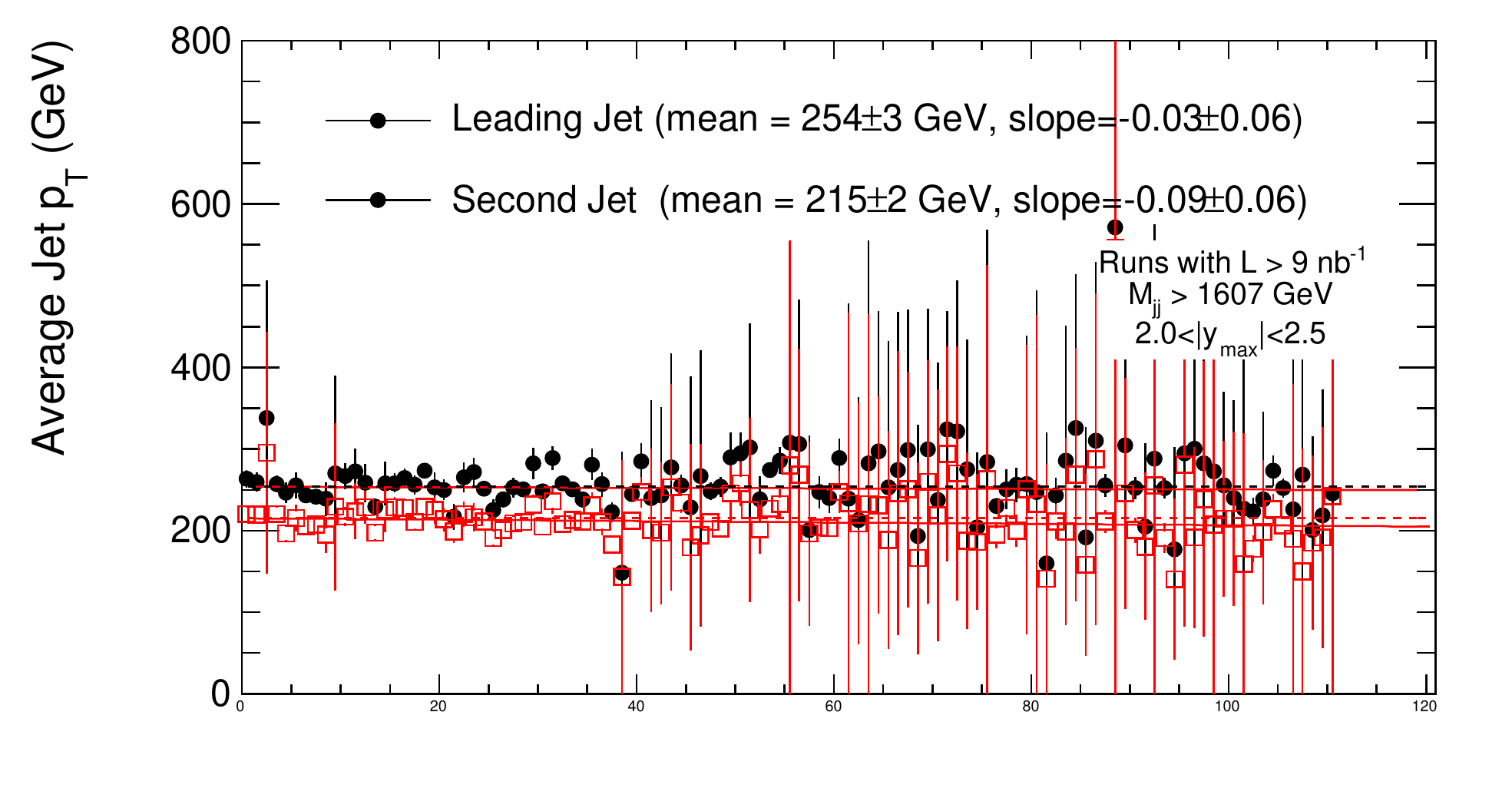} 
  
\capspace
\caption{ The $p_T$ of the leading and second jet  for the five different $y_{max}$ bins and for the
HLT$_{-}$Jet100U trigger as a function of time (run number), fitted with a first degree polynomial.}
\label{fig_appd9}
\end{figure}

\clearpage

\begin{figure}[h]
\centering

\includegraphics[width=0.52\textwidth]{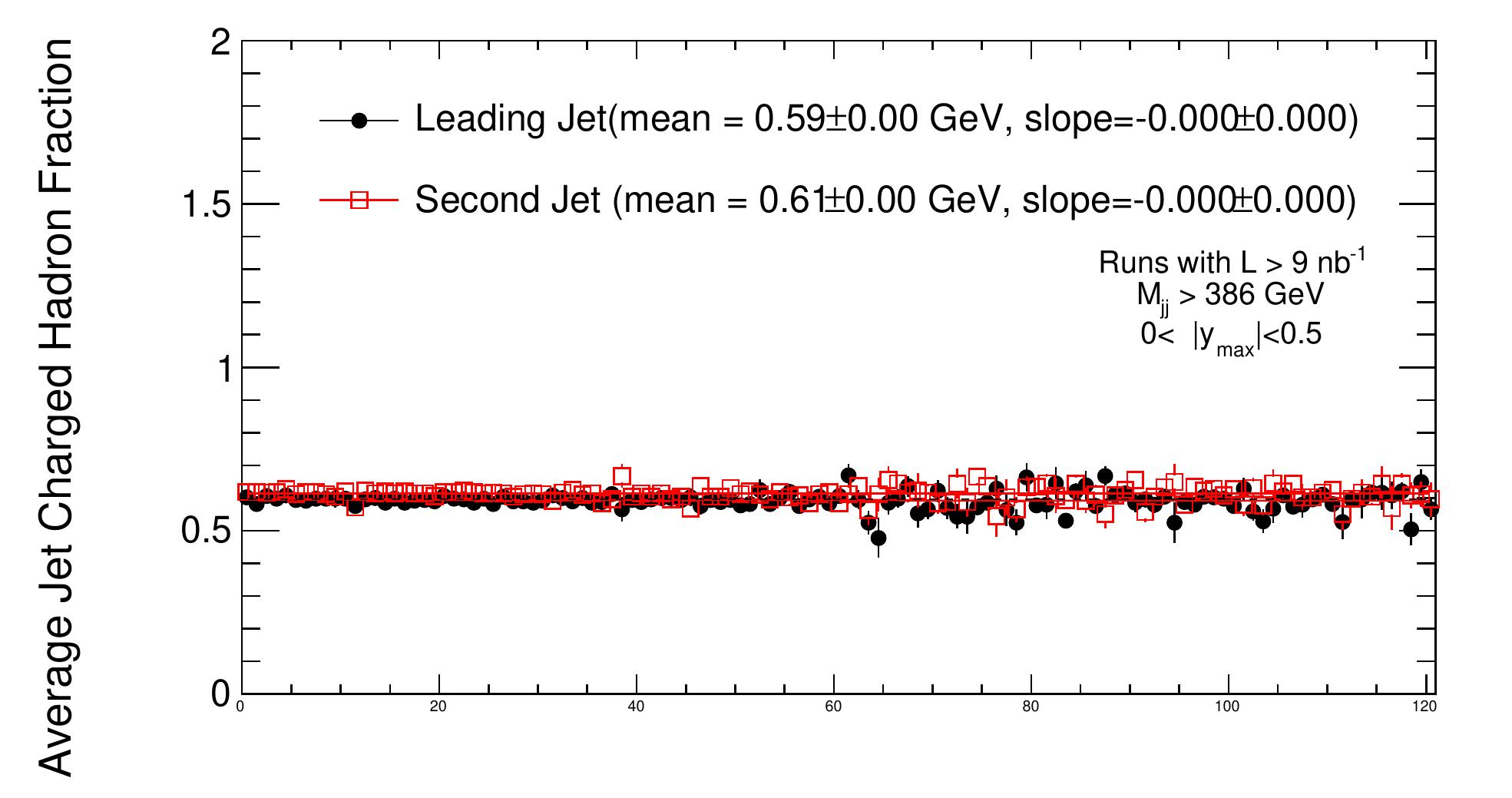} 
\includegraphics[width=0.52\textwidth]{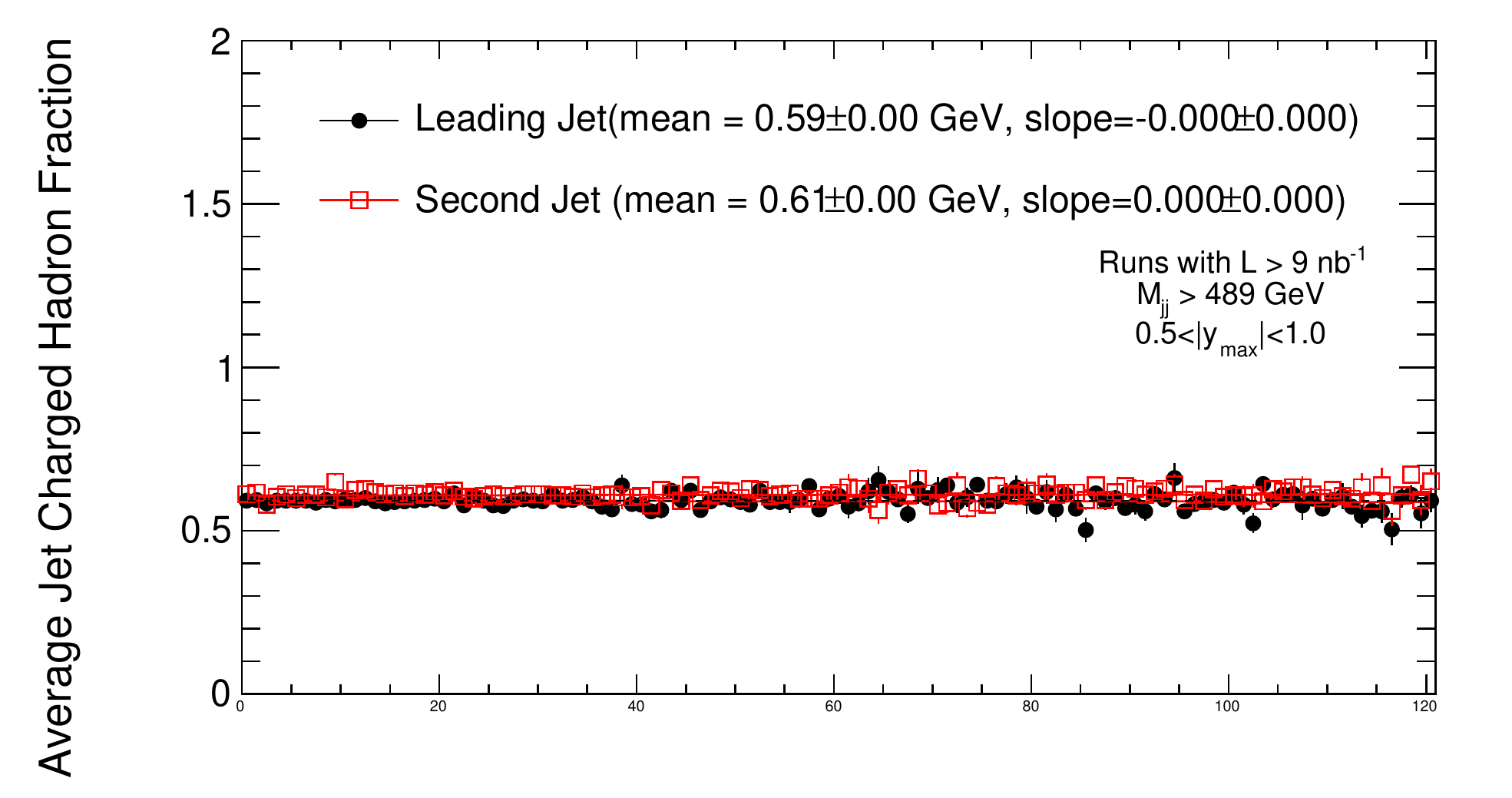} 
\includegraphics[width=0.52\textwidth]{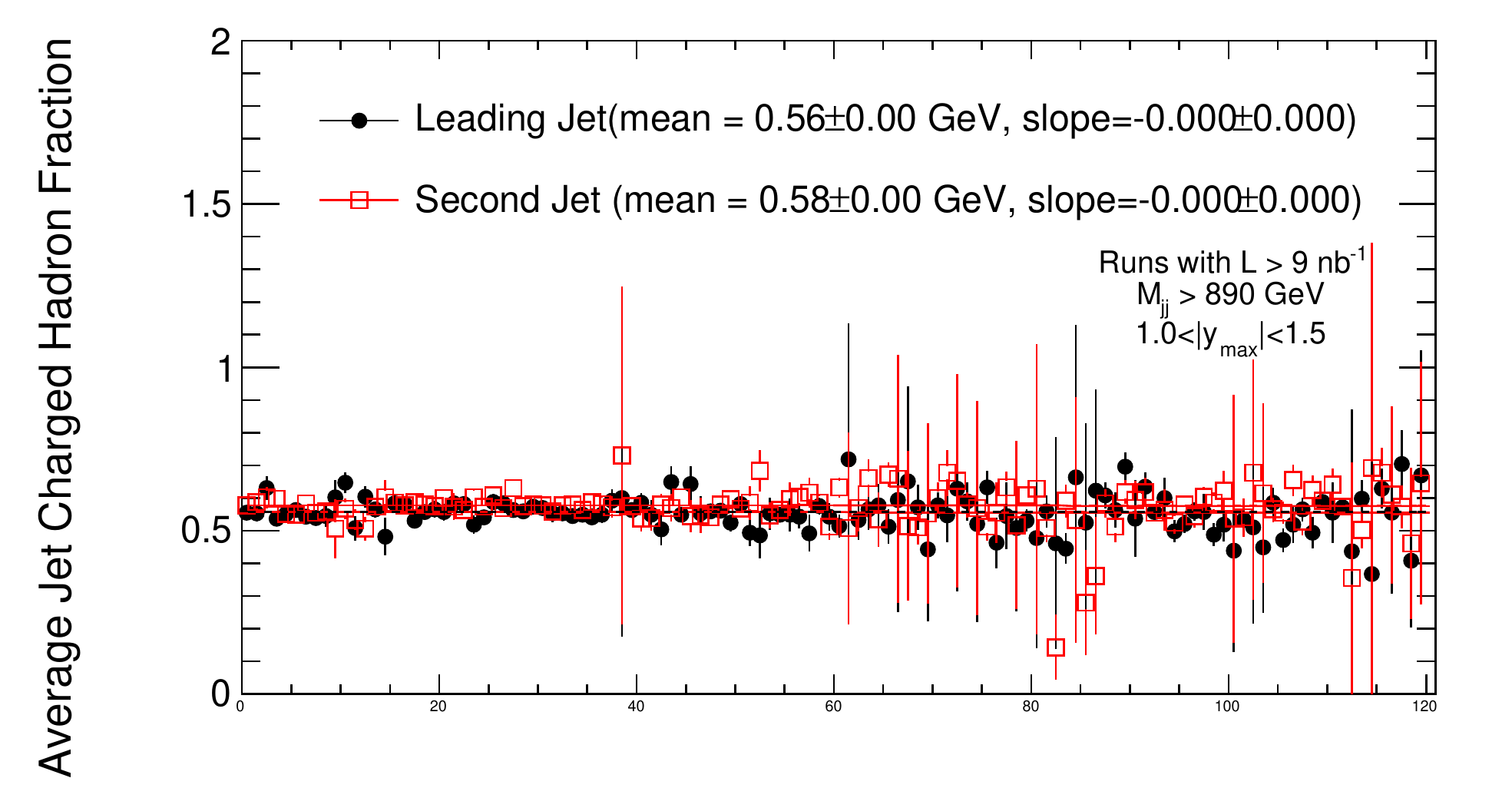} 
\includegraphics[width=0.52\textwidth]{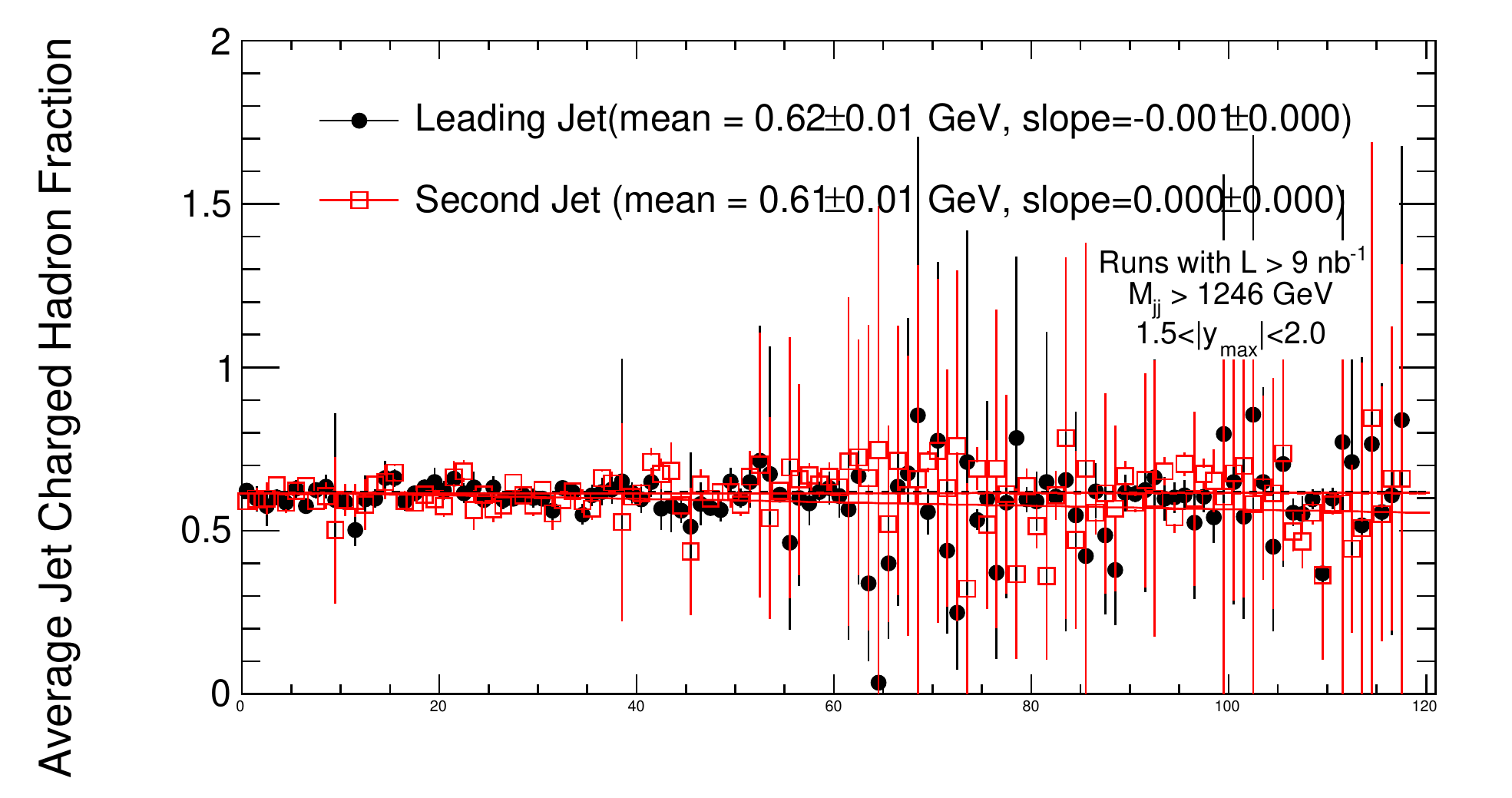} 
\includegraphics[width=0.52\textwidth]{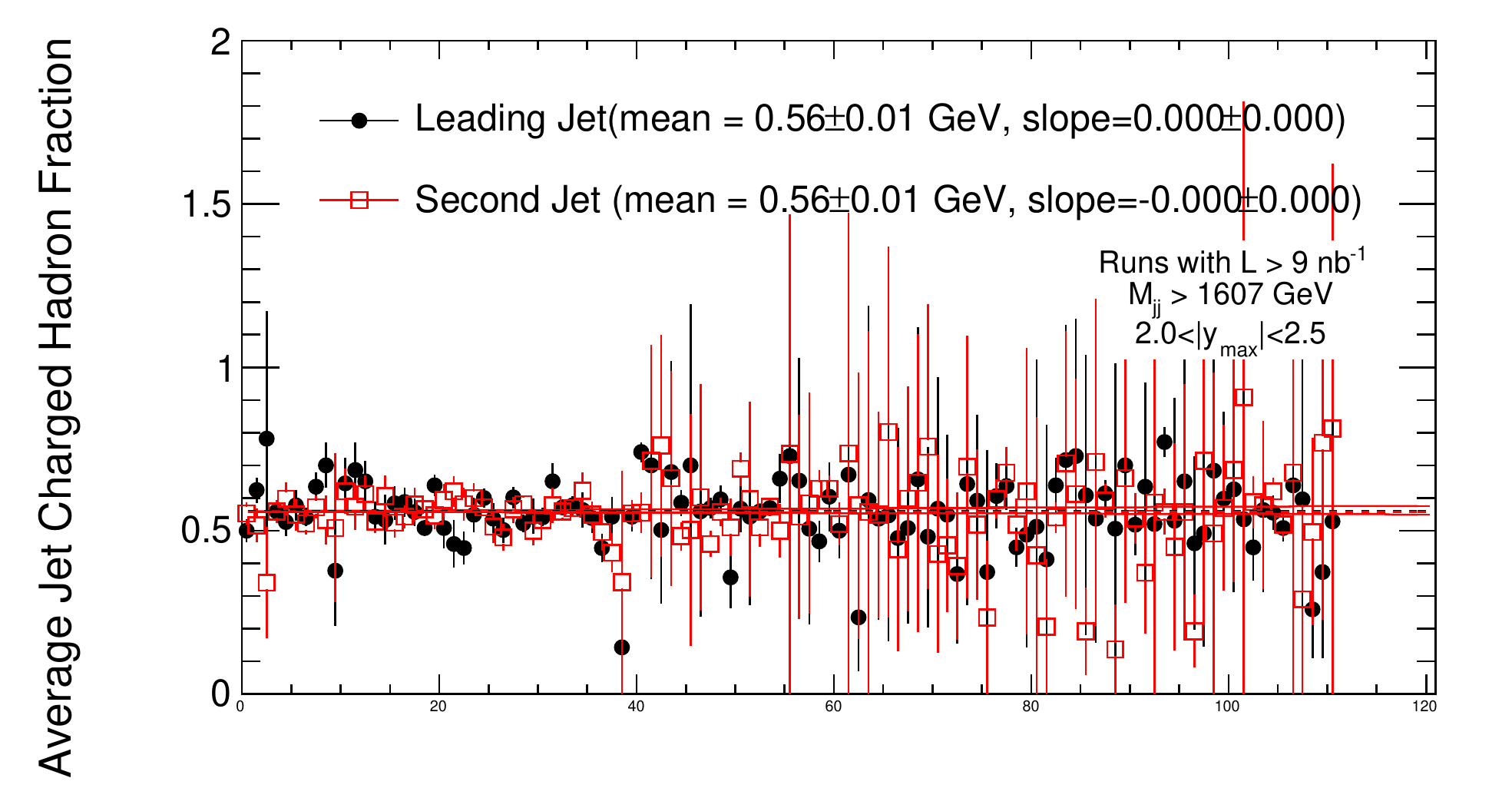} 
  
\capspace
\caption{ The charged hadron fraction  of the leading and second jet  for the five different $y_{max}$ bins and for the
HLT$_{-}$Jet100U trigger as a function of time (run number), fitted with a first degree polynomial.}
\label{fig_appd10}
\end{figure}

\begin{figure}[h]
\centering

\includegraphics[width=0.52\textwidth]{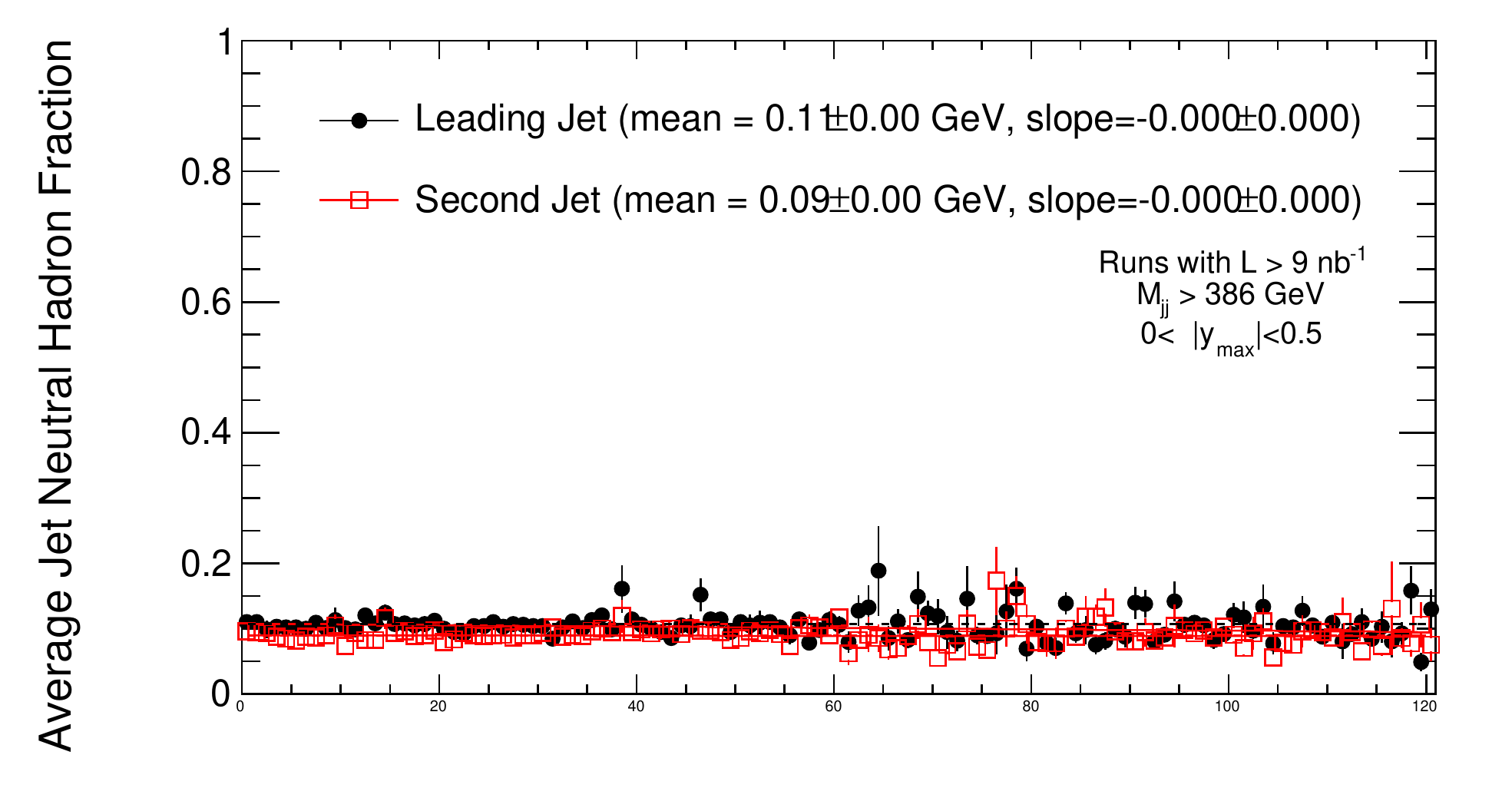} 
\includegraphics[width=0.52\textwidth]{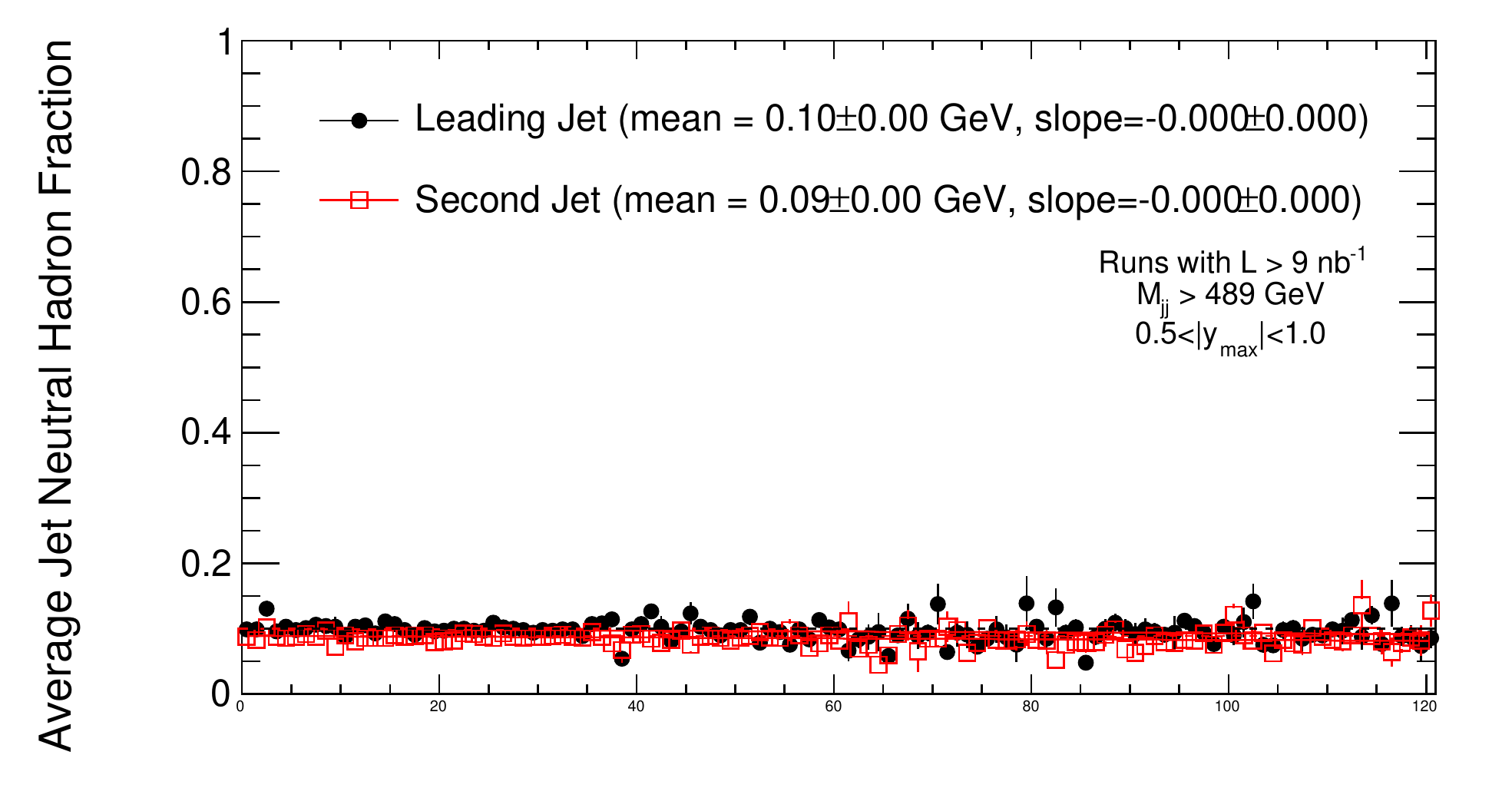} 
\includegraphics[width=0.52\textwidth]{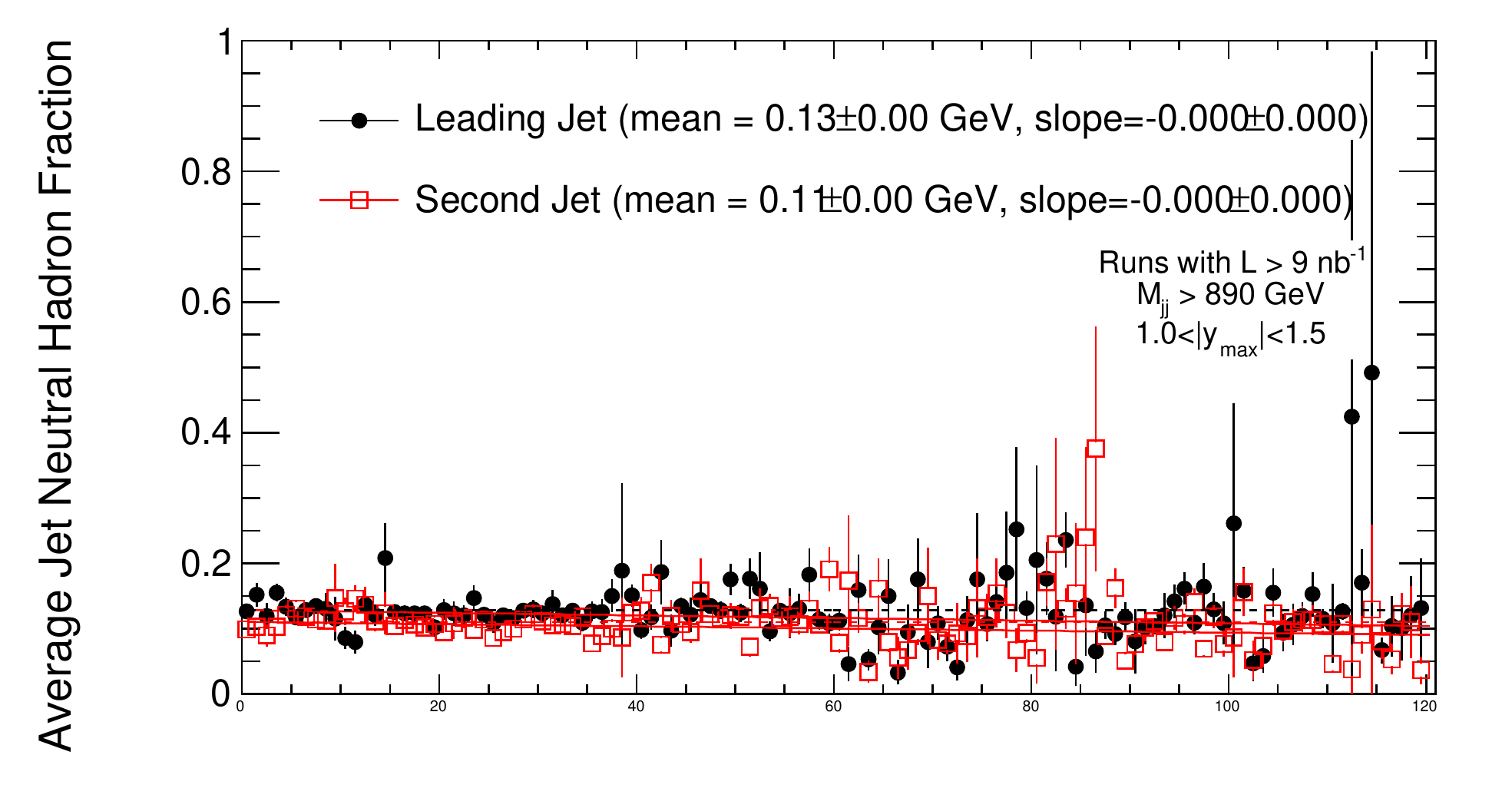} 
\includegraphics[width=0.52\textwidth]{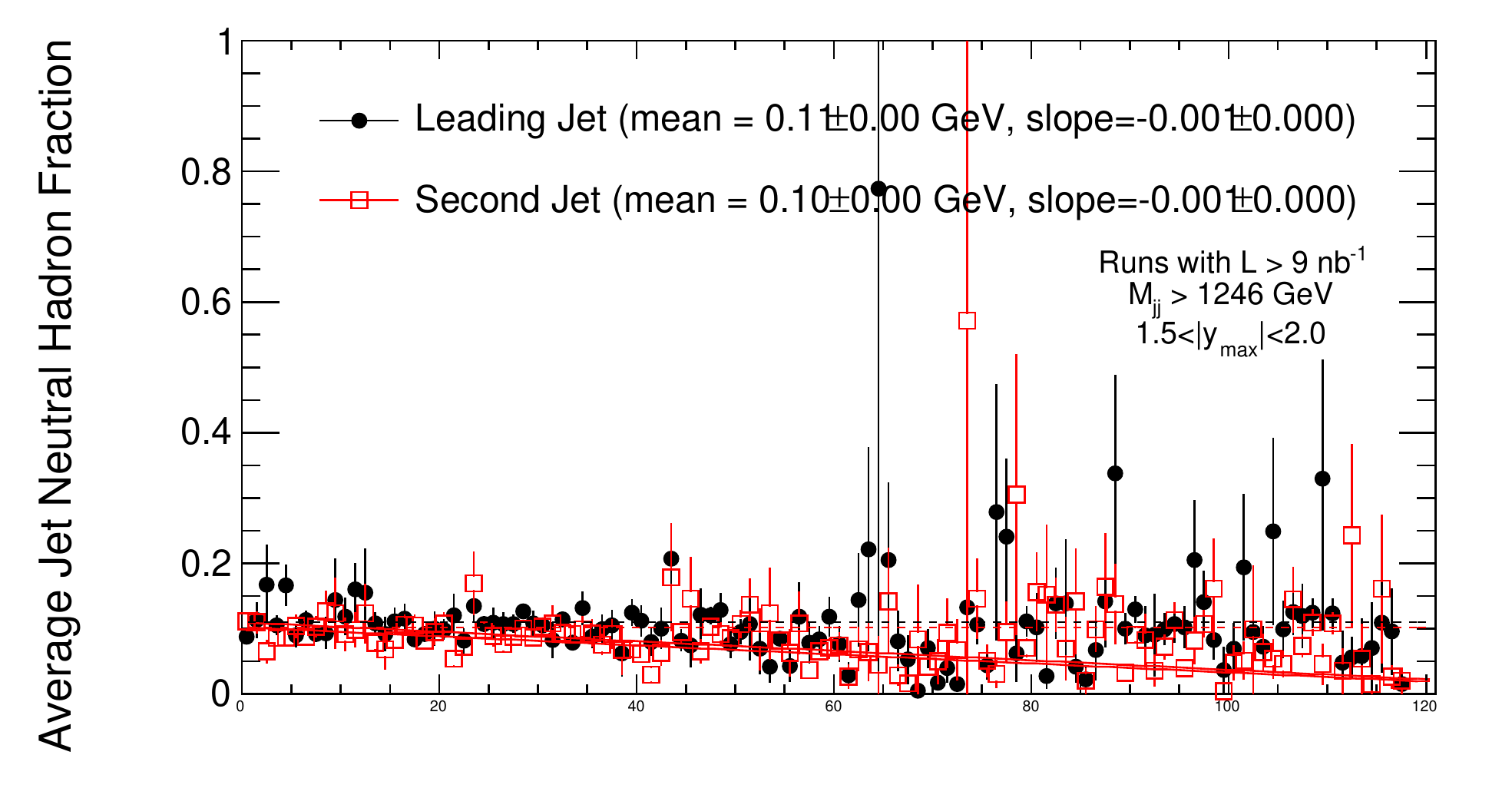} 
\includegraphics[width=0.52\textwidth]{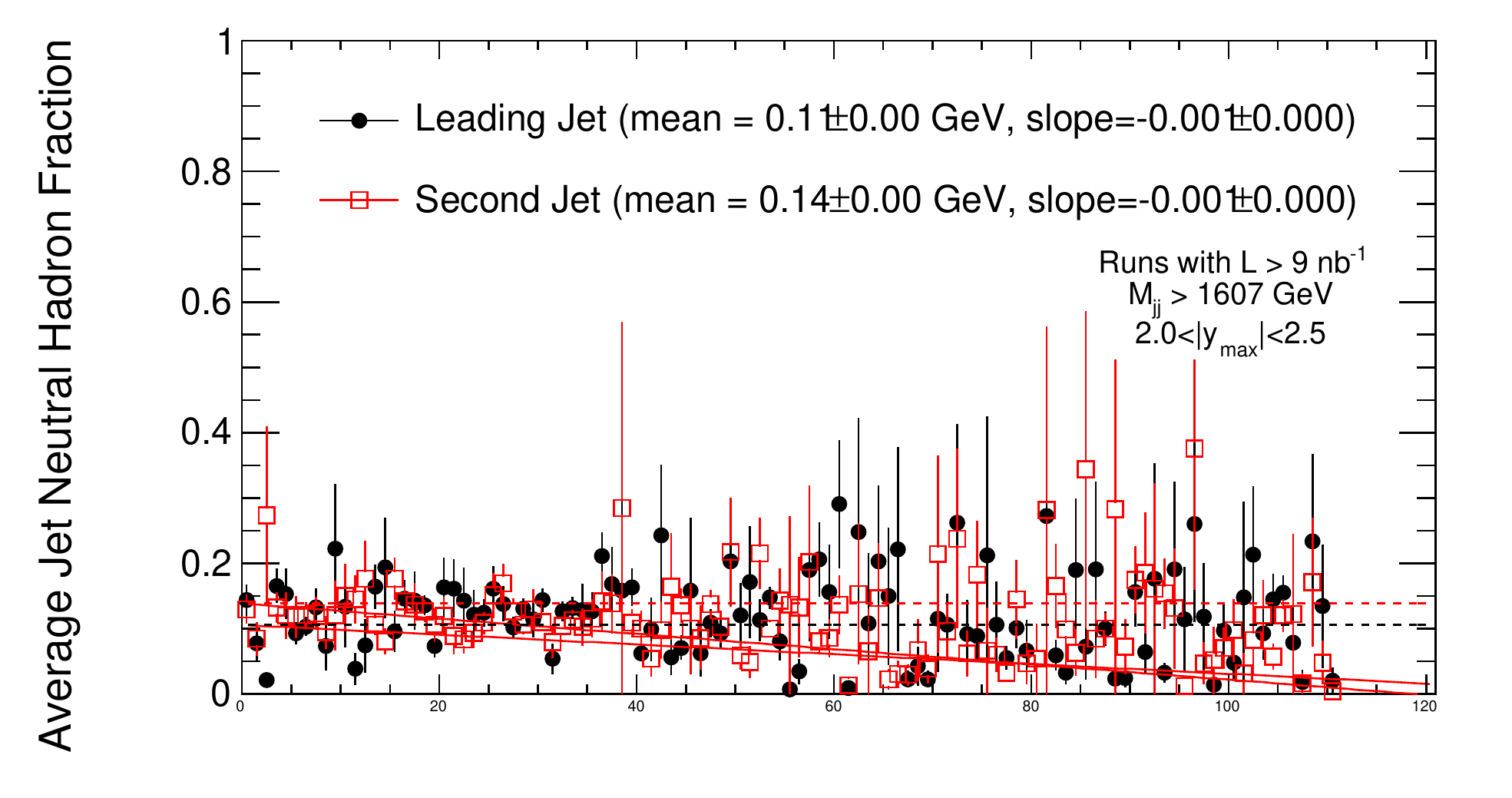} 
  
\capspace
\caption{ The neutral  hadron fraction  of the leading and second jet  for the five different $y_{max}$ bins and for the
HLT$_{-}$Jet100U trigger as a function of time (run number), fitted with a first degree polynomial.}
\label{fig_appd11}
\end{figure}

\begin{figure}[h]
\centering

\includegraphics[width=0.52\textwidth]{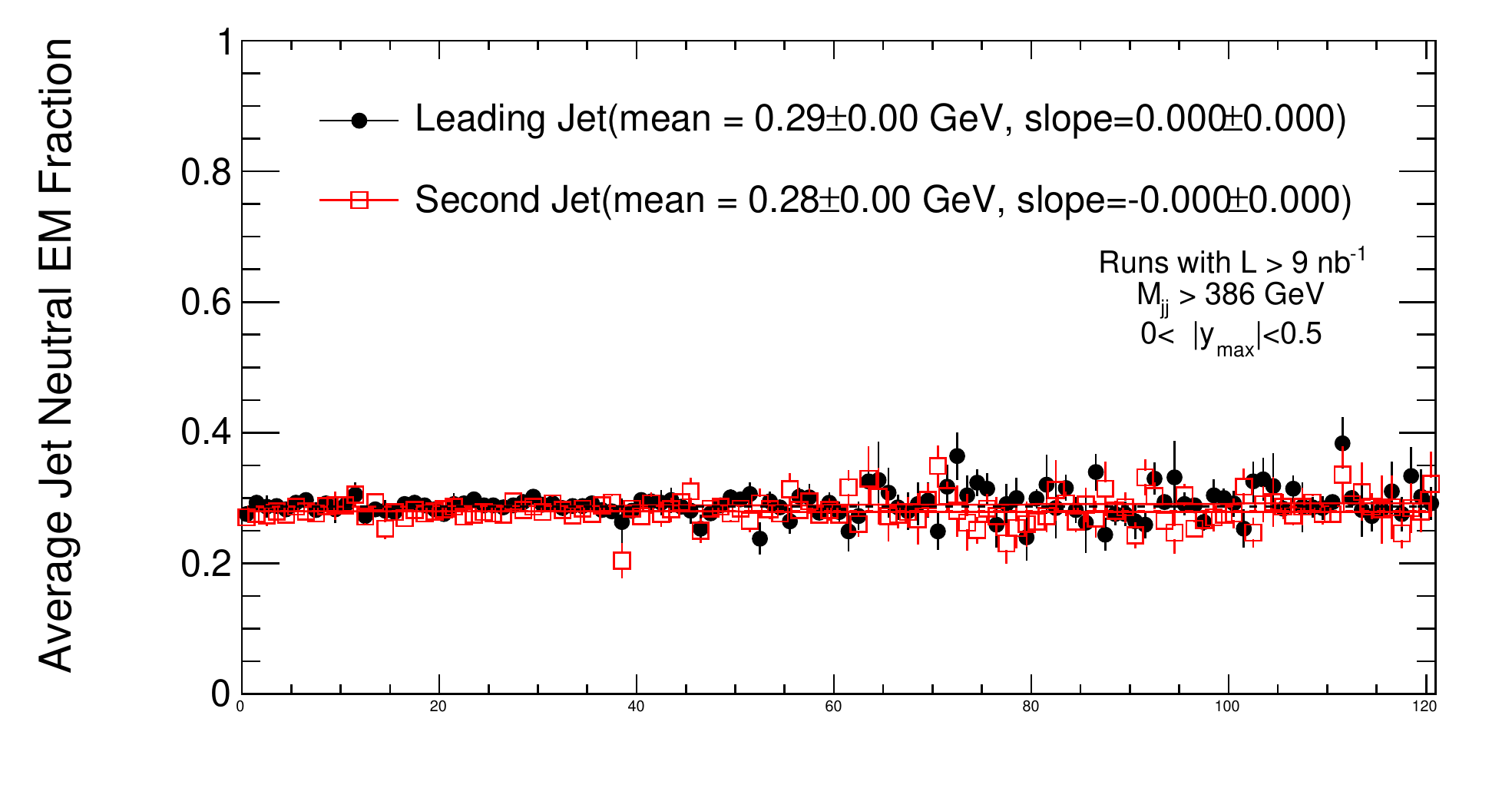} 
\includegraphics[width=0.52\textwidth]{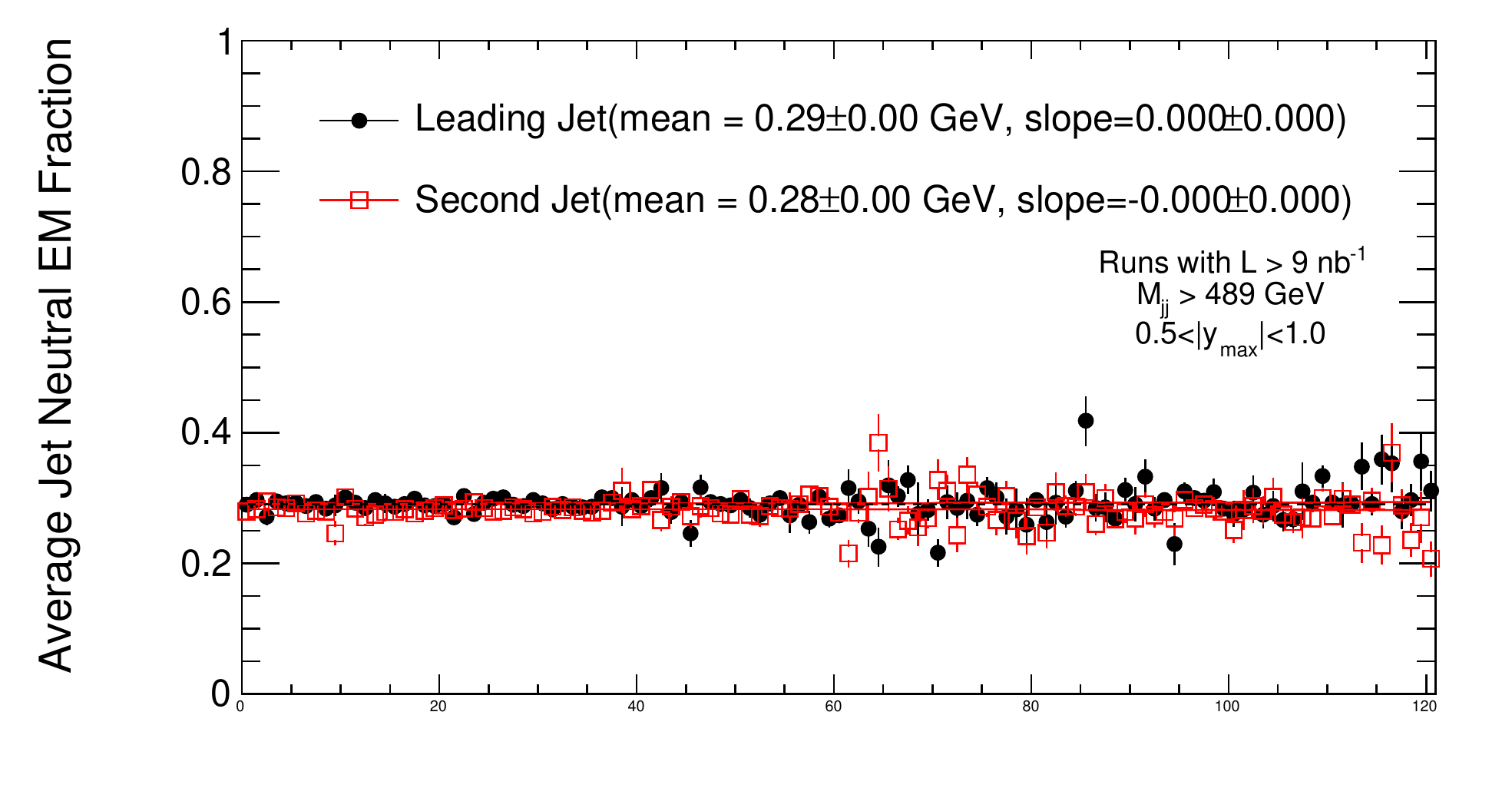} 
\includegraphics[width=0.52\textwidth]{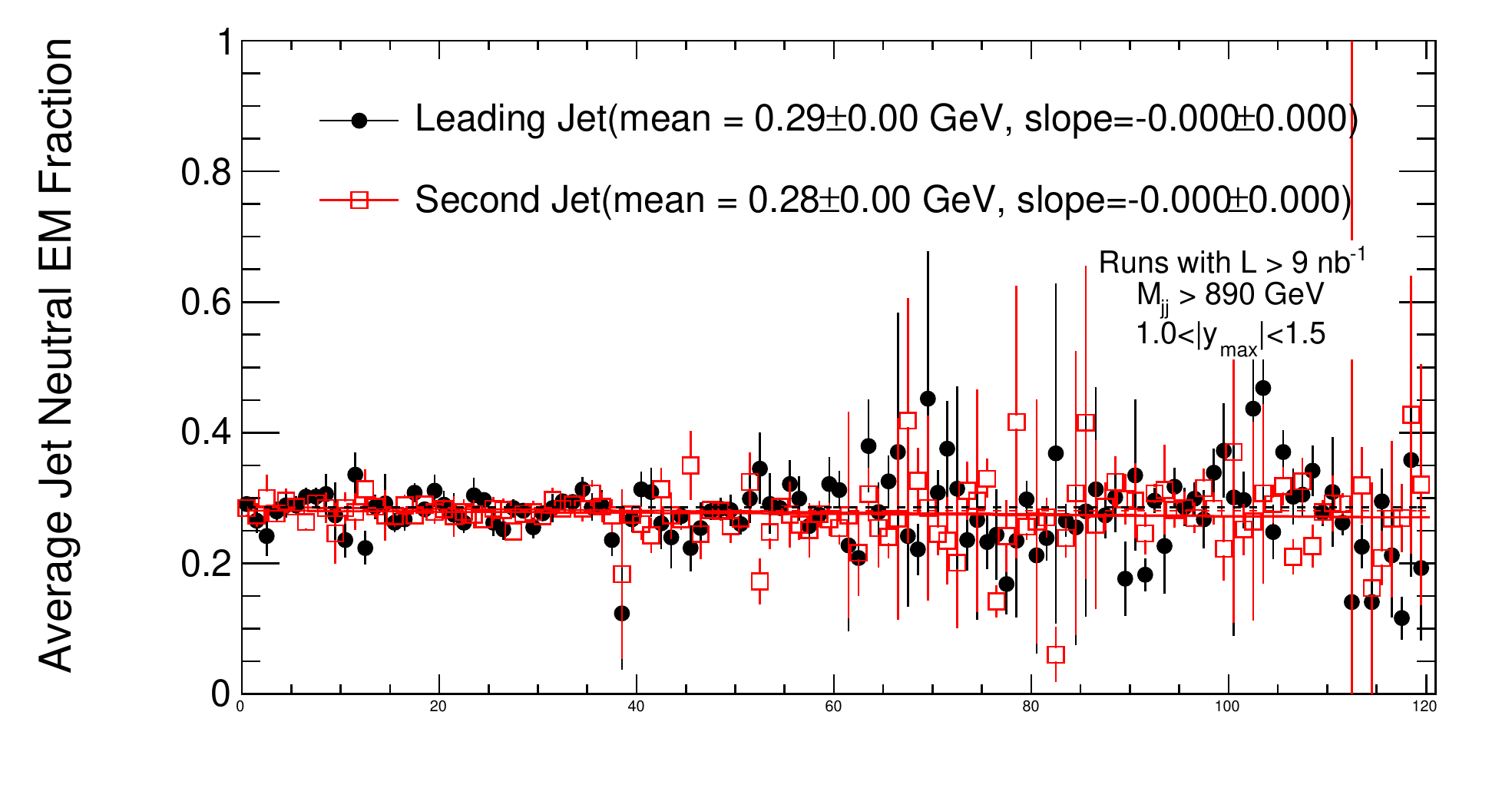} 
\includegraphics[width=0.52\textwidth]{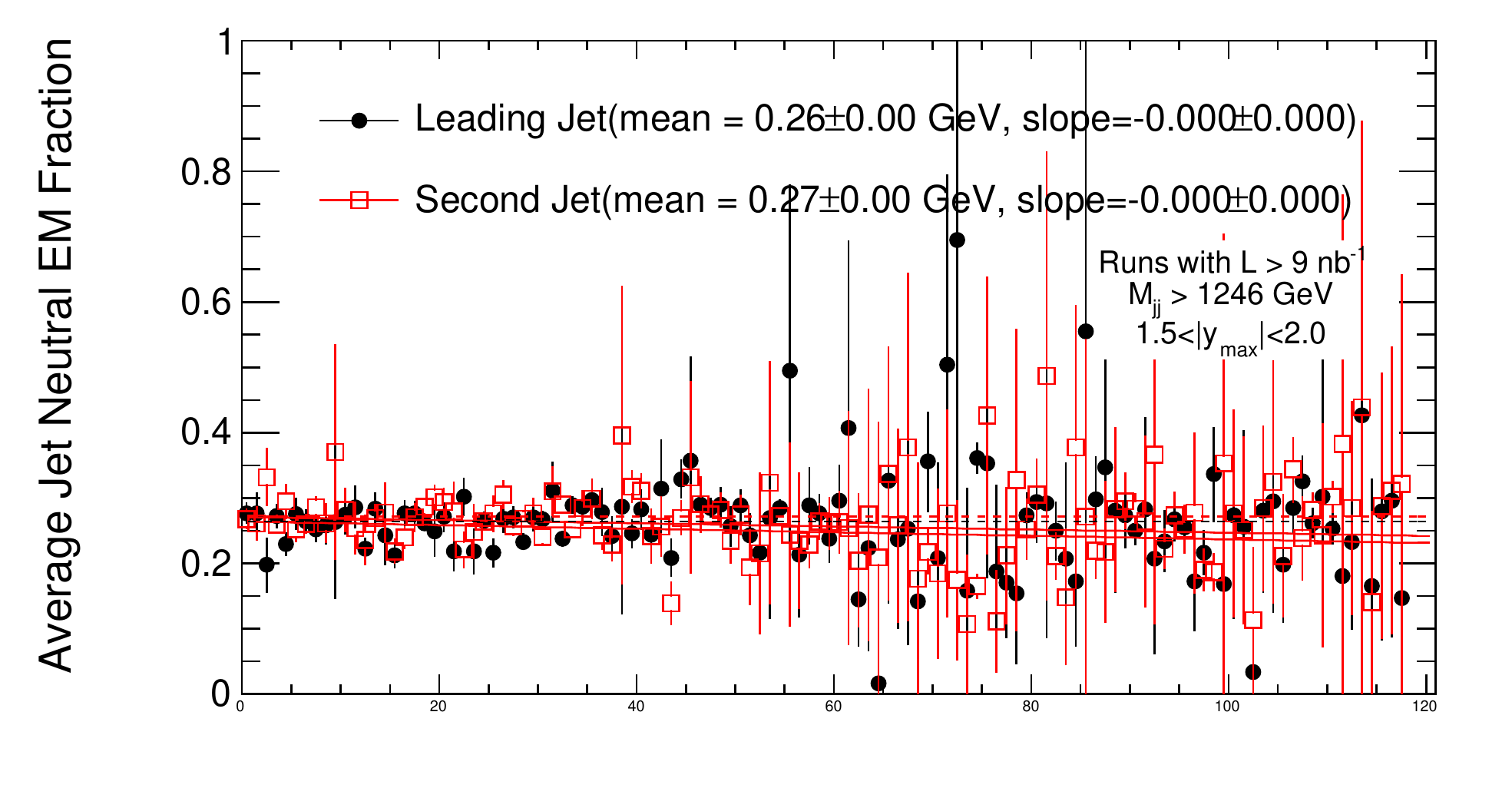} 
\includegraphics[width=0.52\textwidth]{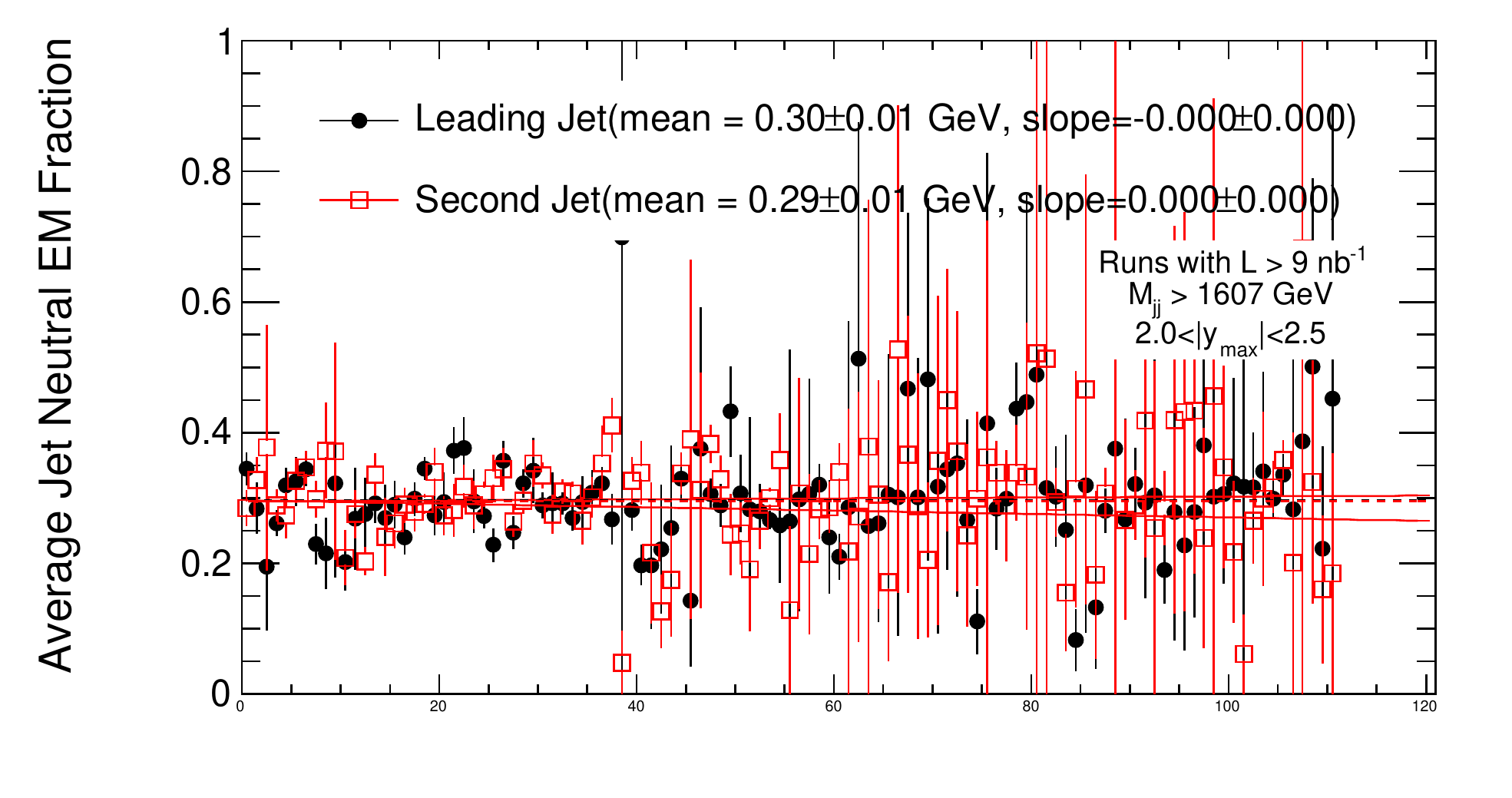} 
  
\capspace
\caption{ The neutral  electromagnetic fraction  of the leading and second jet  for the five different $y_{max}$ bins and for the
HLT$_{-}$Jet100U trigger as a function of time (run number), fitted with a first degree polynomial.}
\label{fig_appd12}
\end{figure}

\clearpage
%%% jet 140

\begin{figure}[h]
\centering

\includegraphics[width=0.52\textwidth]{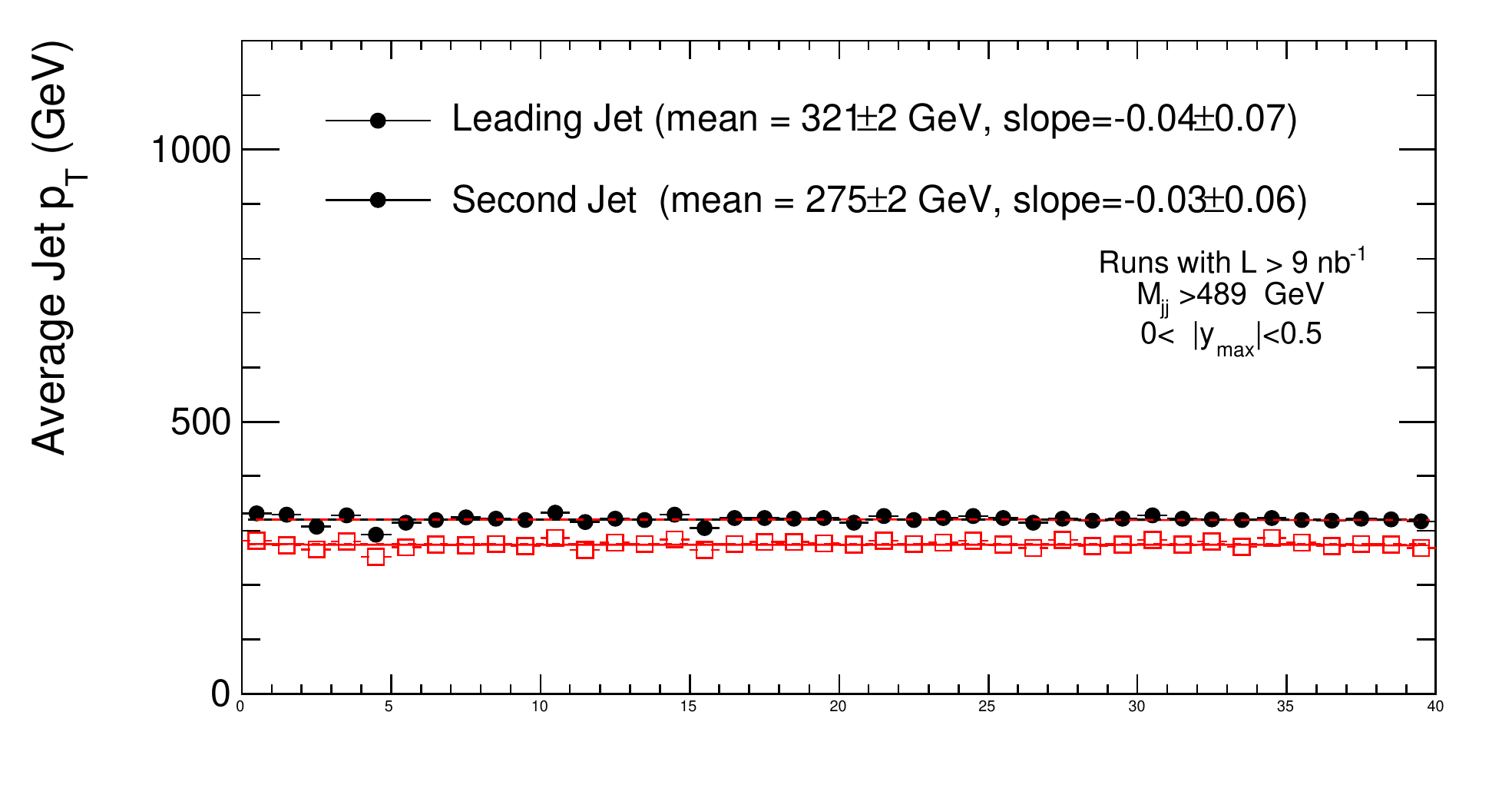} 
\includegraphics[width=0.52\textwidth]{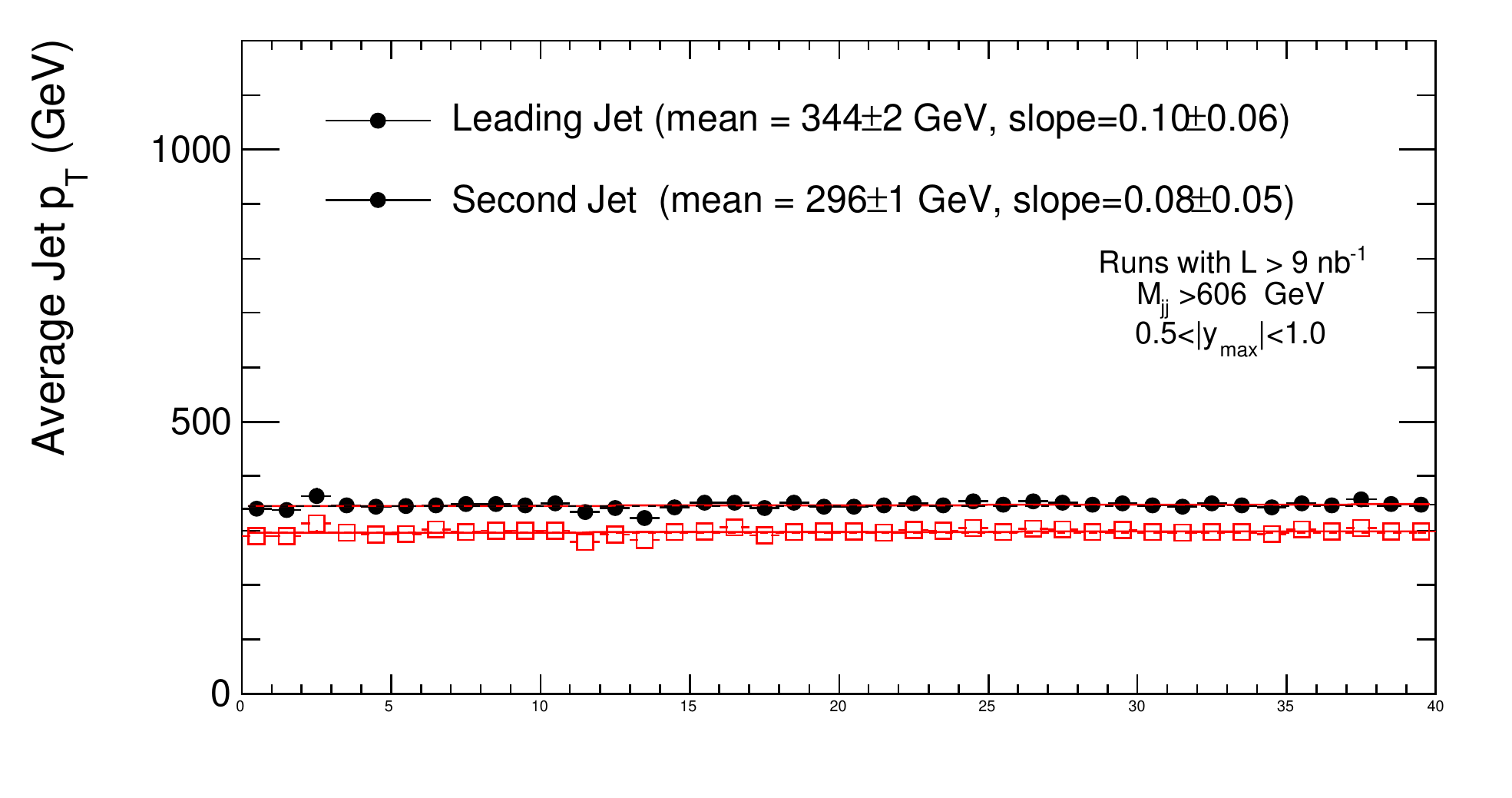} 
\includegraphics[width=0.52\textwidth]{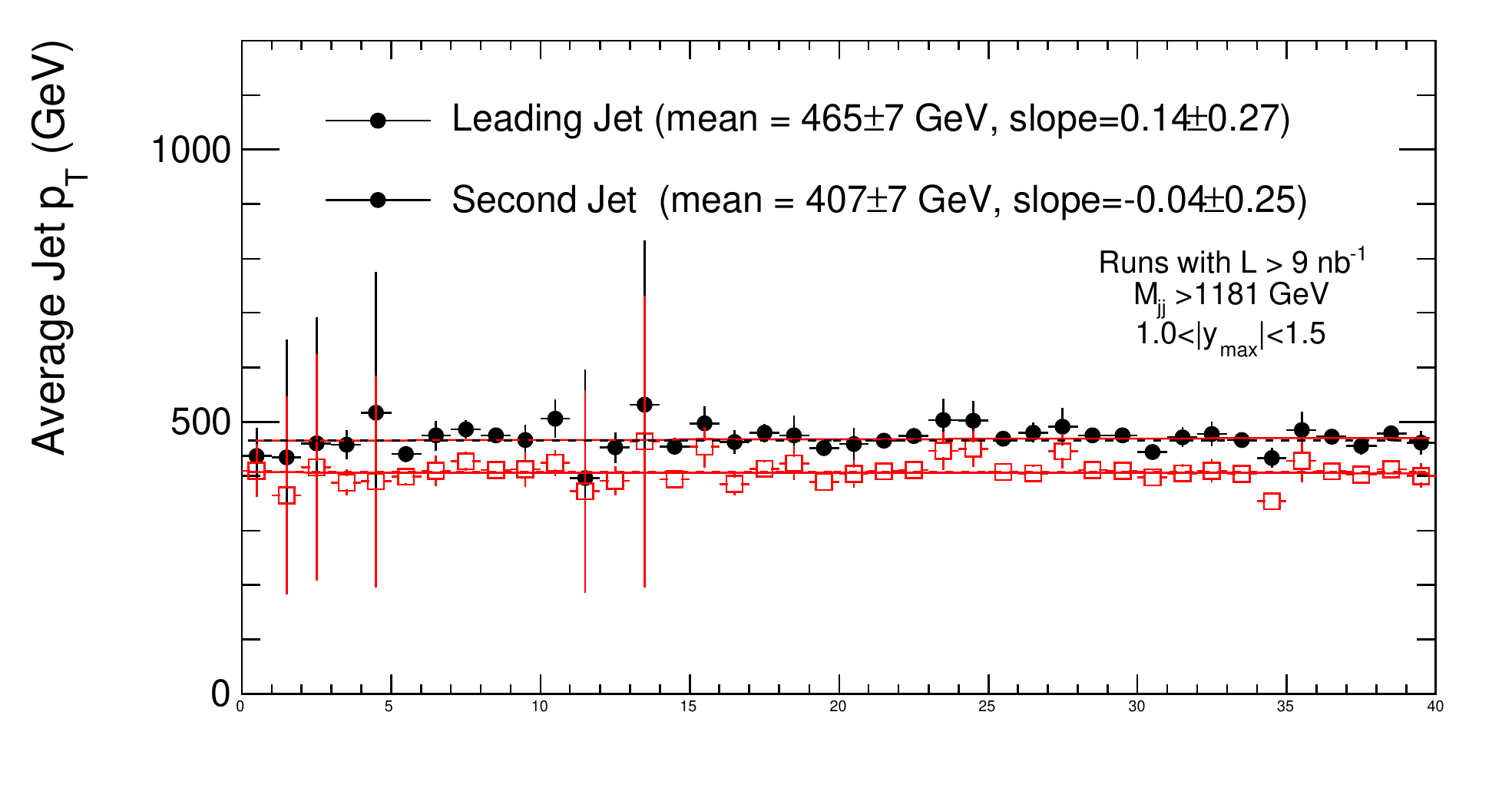} 
\includegraphics[width=0.52\textwidth]{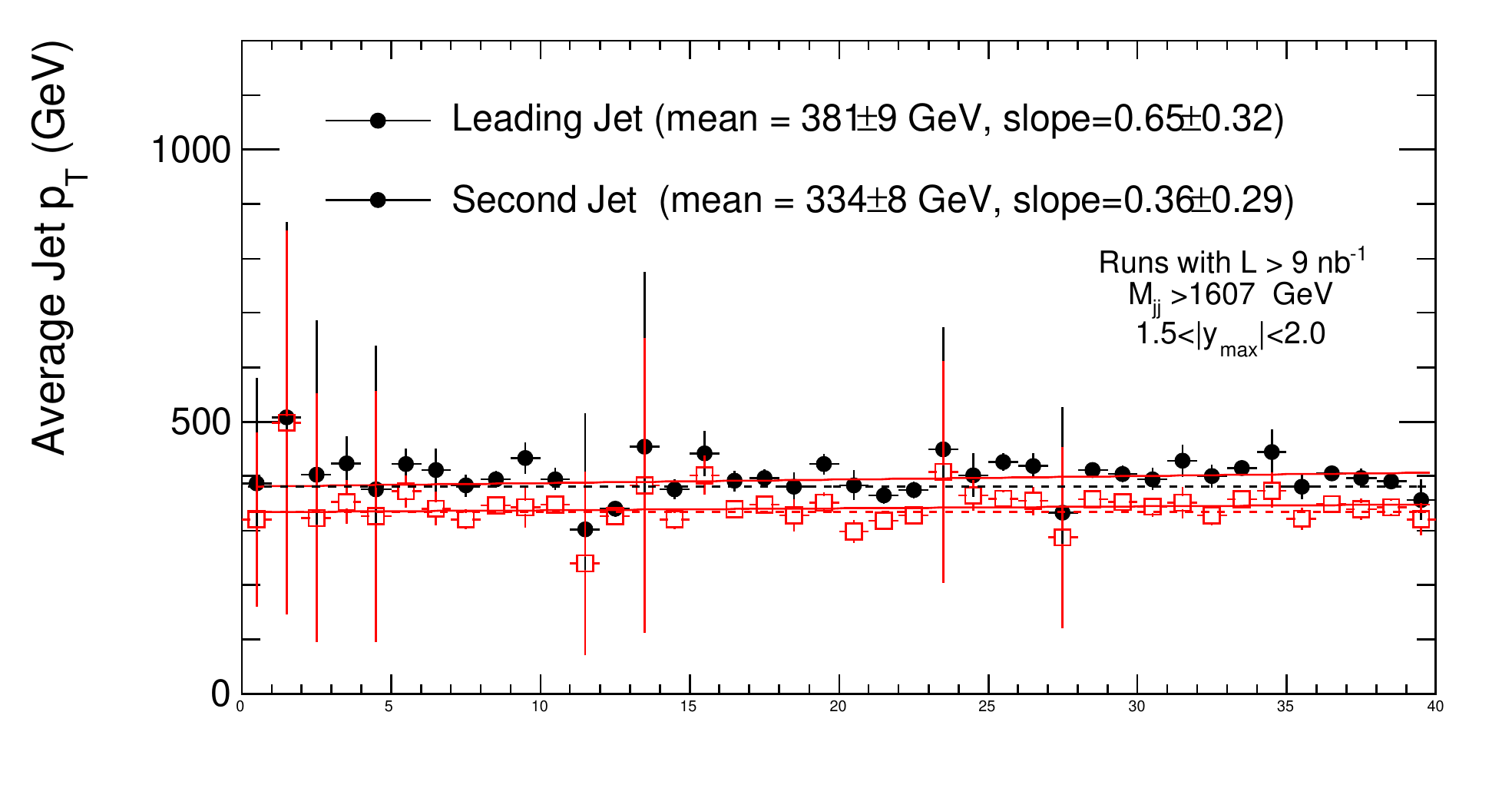} 
\includegraphics[width=0.52\textwidth]{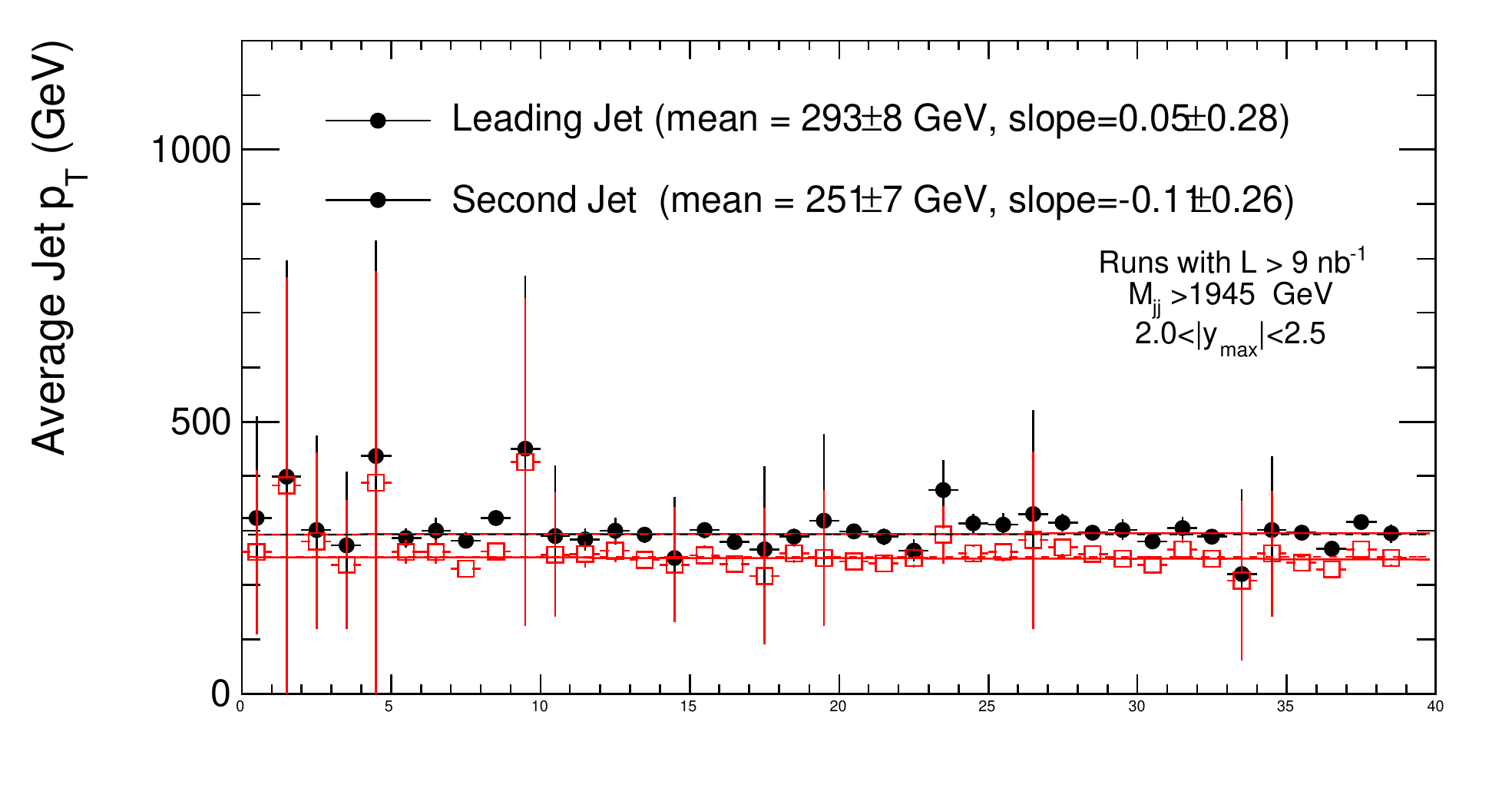} 
  
\capspace
\caption{ The $p_T$ of the leading and second jet  for the five different $y_{max}$ bins and for the
HLT$_{-}$Jet140U trigger as a function of time (run number), fitted with a first degree polynomial.}
\label{fig_appd13}
\end{figure}

\begin{figure}[h]
\centering

\includegraphics[width=0.52\textwidth]{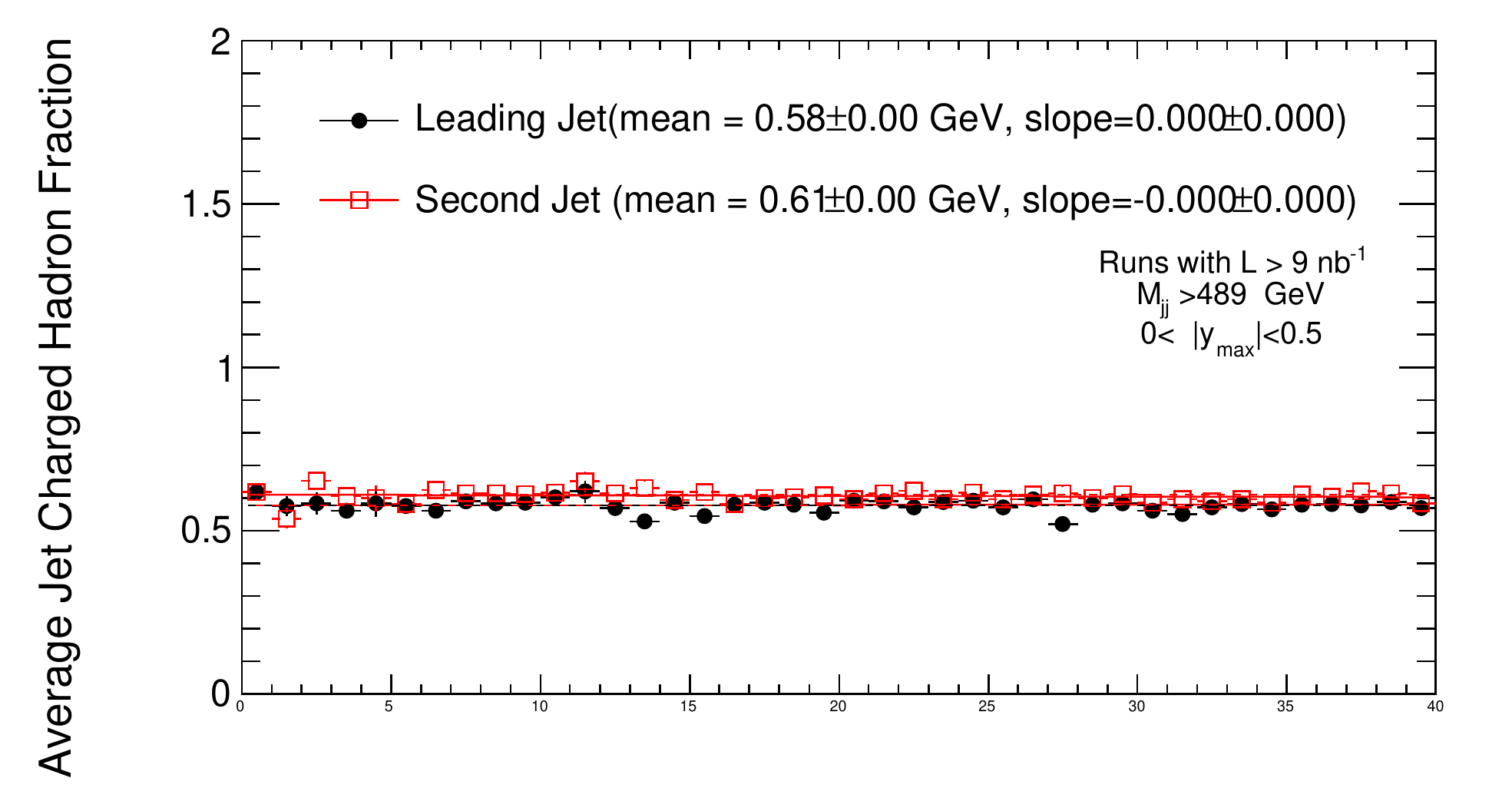} 
\includegraphics[width=0.52\textwidth]{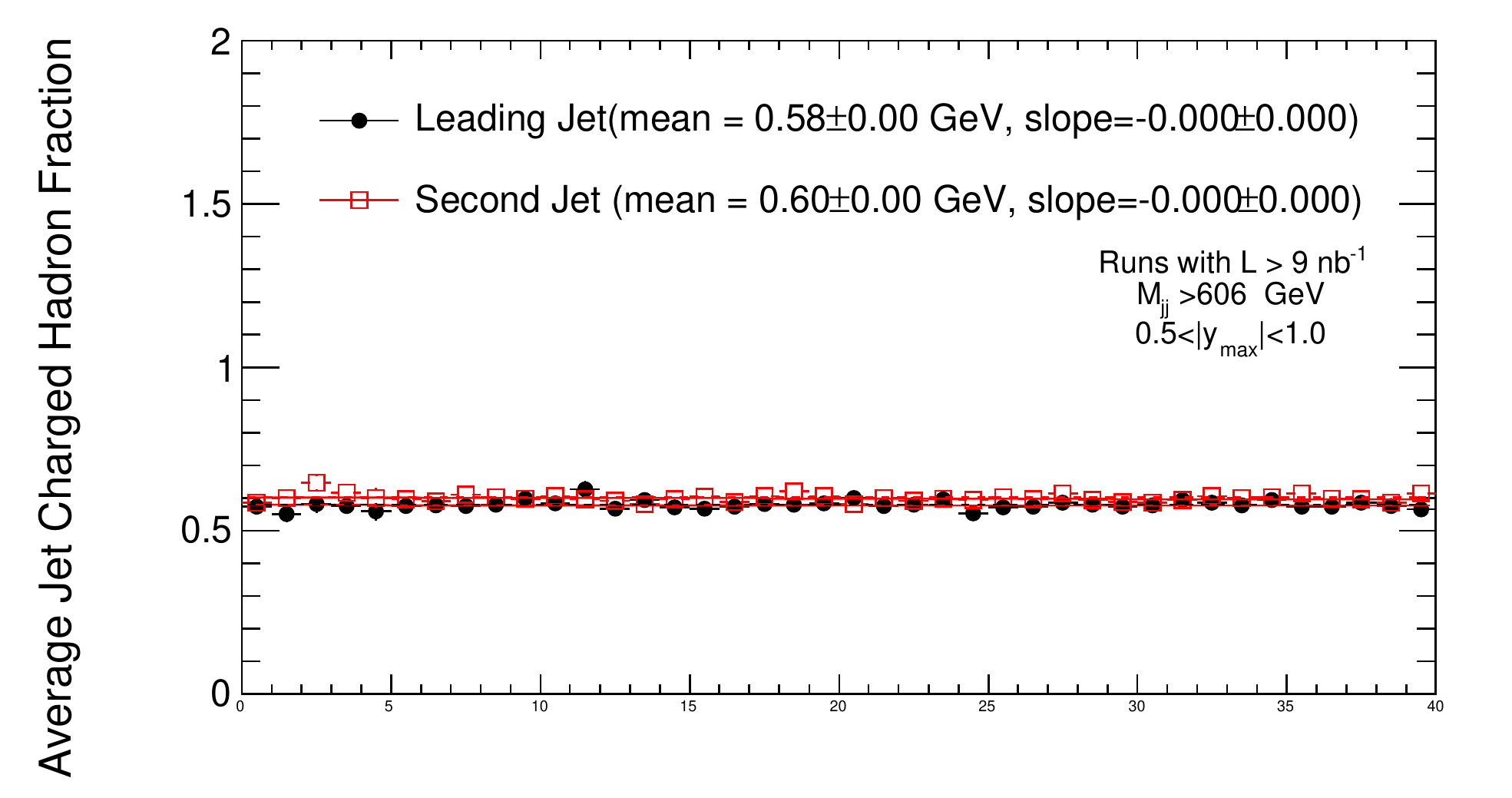} 
\includegraphics[width=0.52\textwidth]{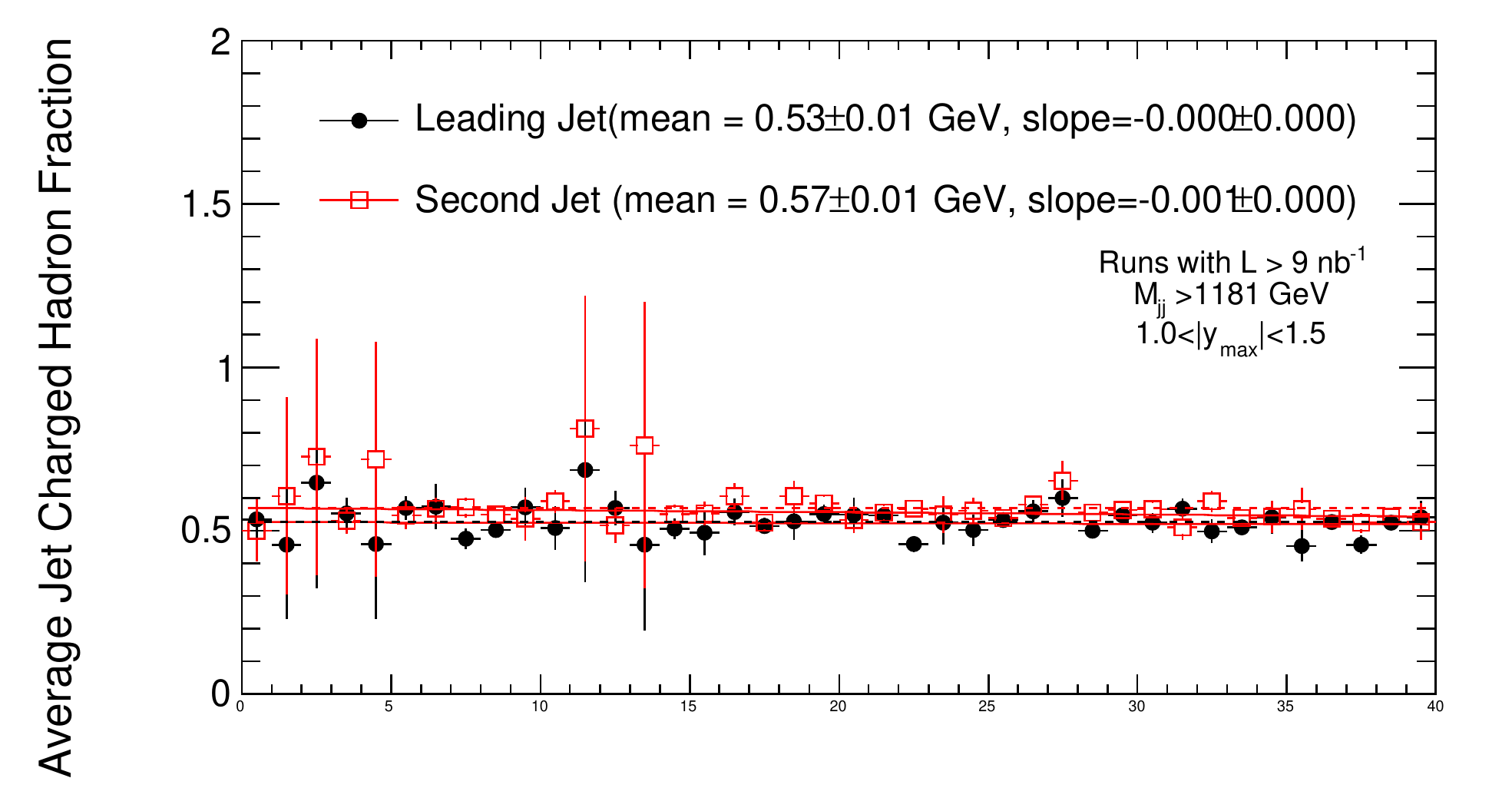} 
\includegraphics[width=0.52\textwidth]{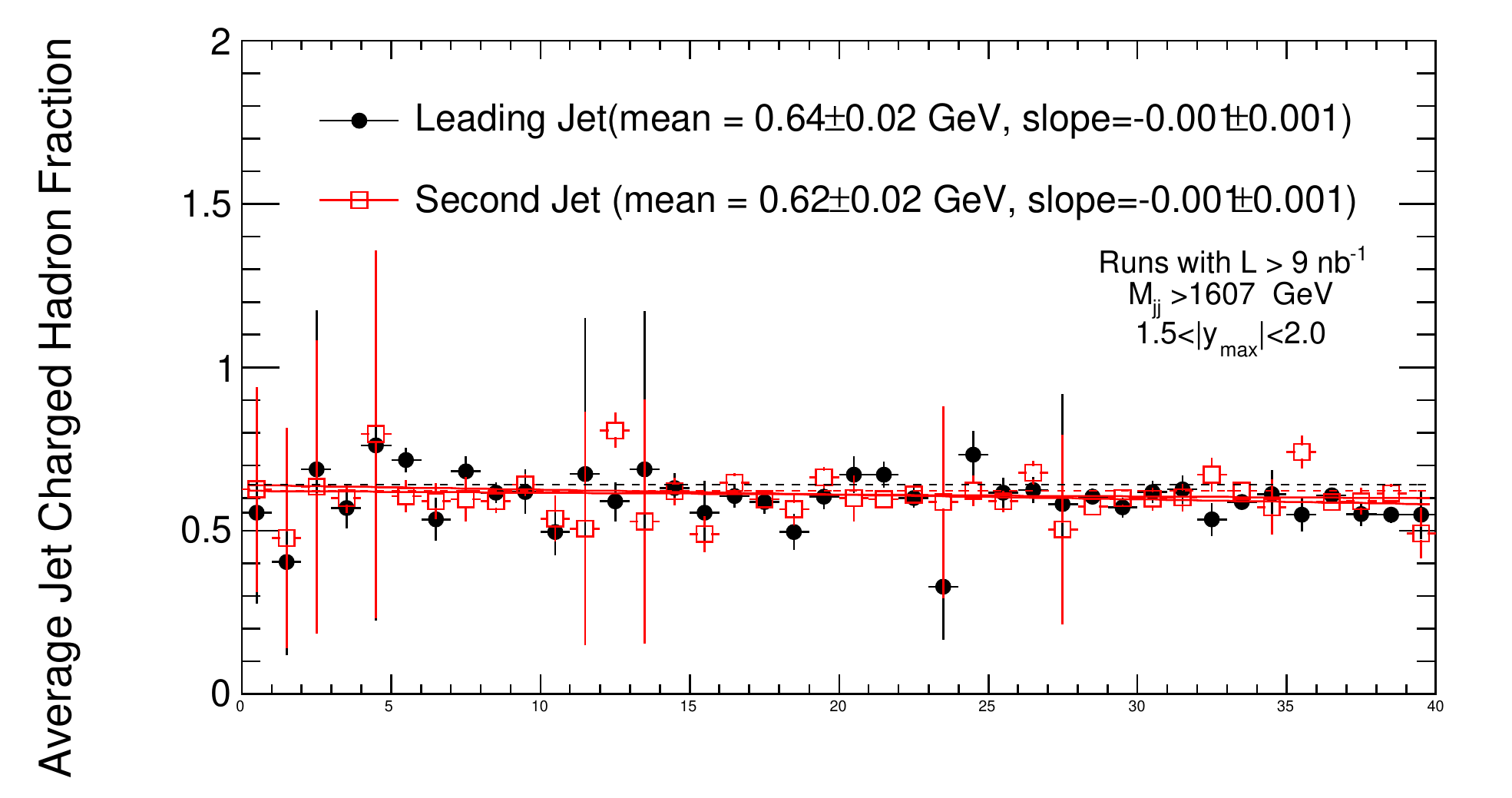} 
\includegraphics[width=0.52\textwidth]{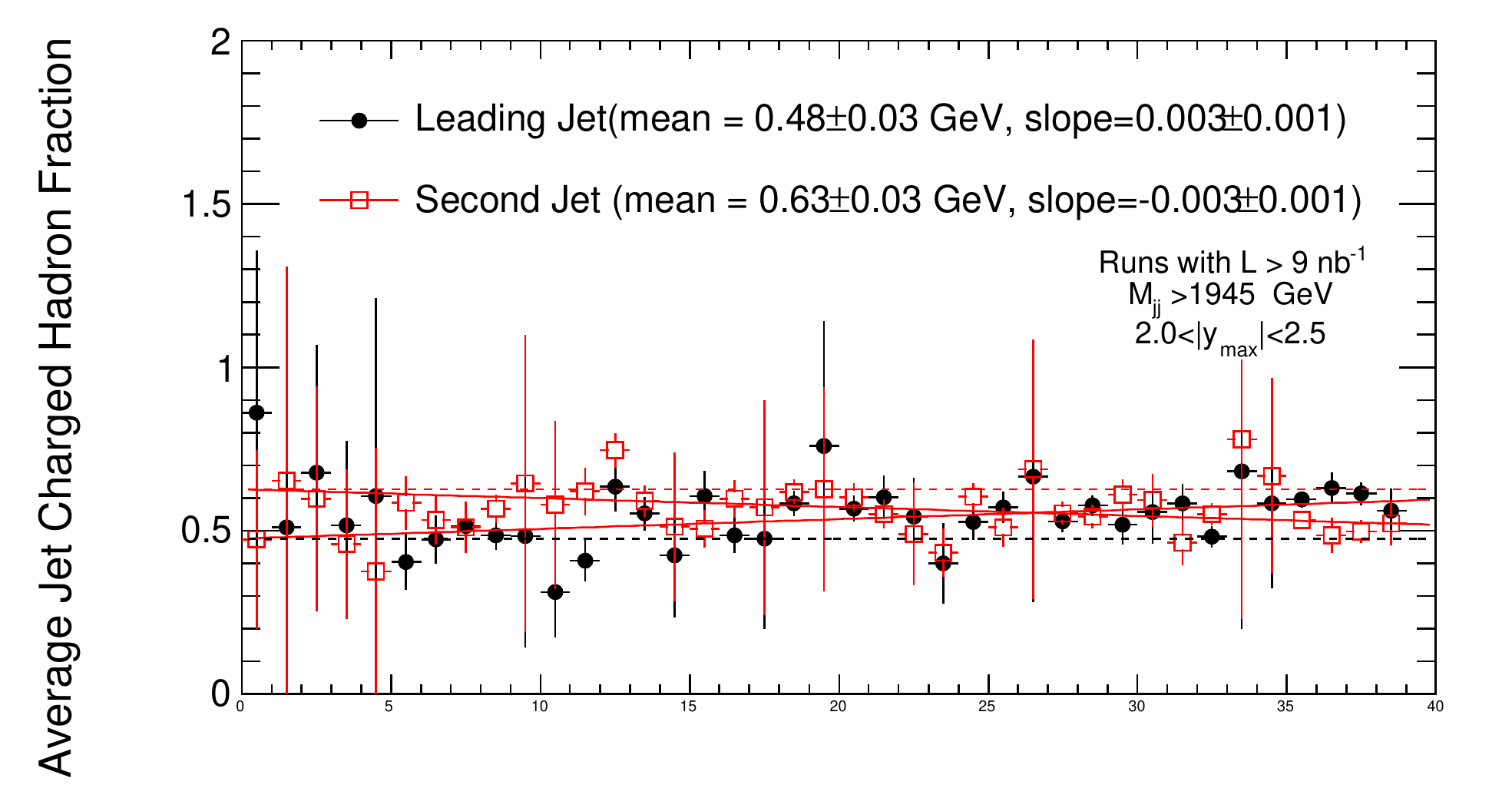} 
 
\capspace
\caption{ The charged hadron fraction  of the leading and second jet  for the five different $y_{max}$ bins and for the
HLT$_{-}$Jet140U trigger as a function of time (run number), fitted with a first degree polynomial.}
\label{fig_appd14}
\end{figure}

\begin{figure}[h]
\centering

\includegraphics[width=0.52\textwidth]{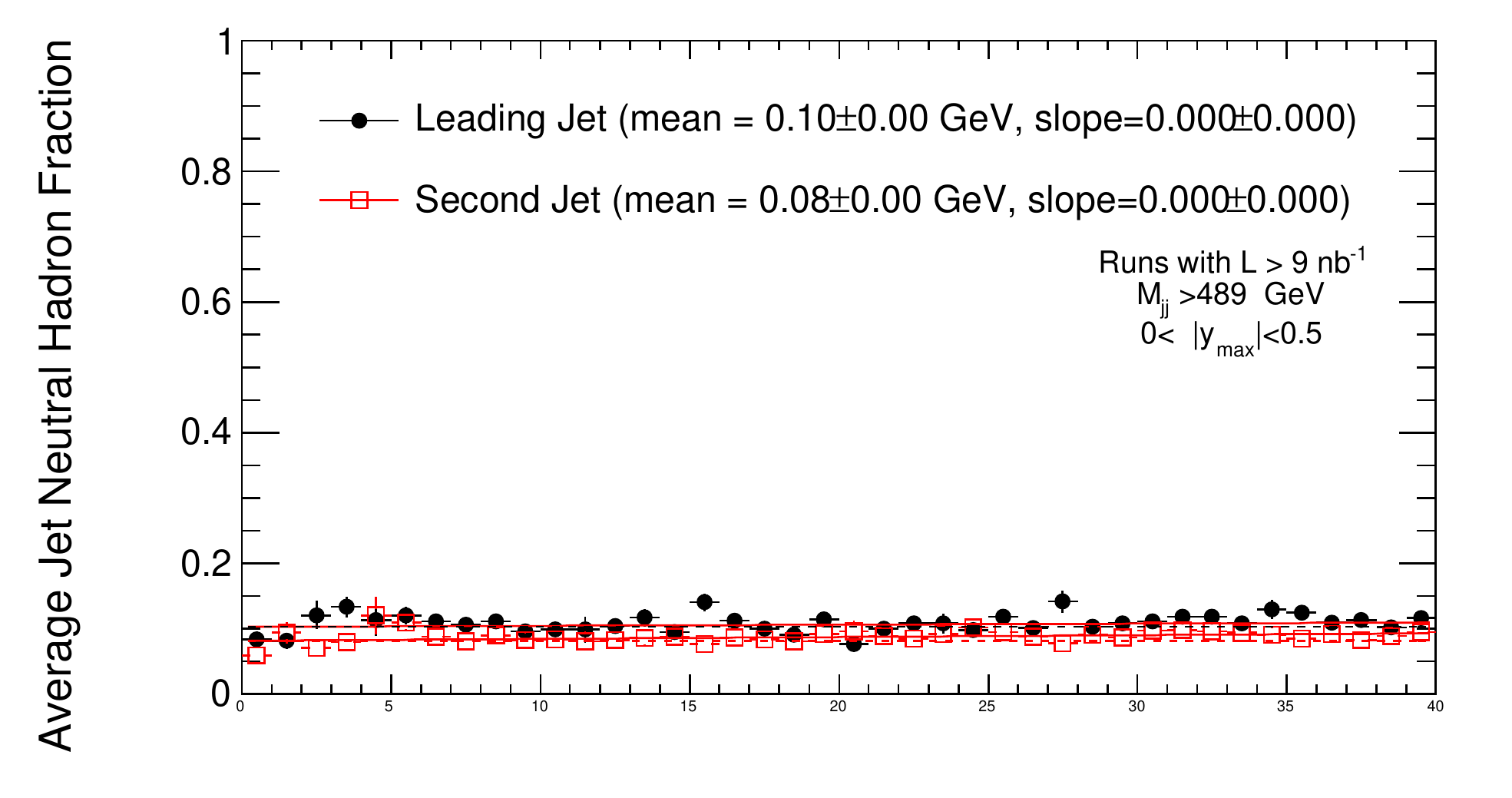} 
\includegraphics[width=0.52\textwidth]{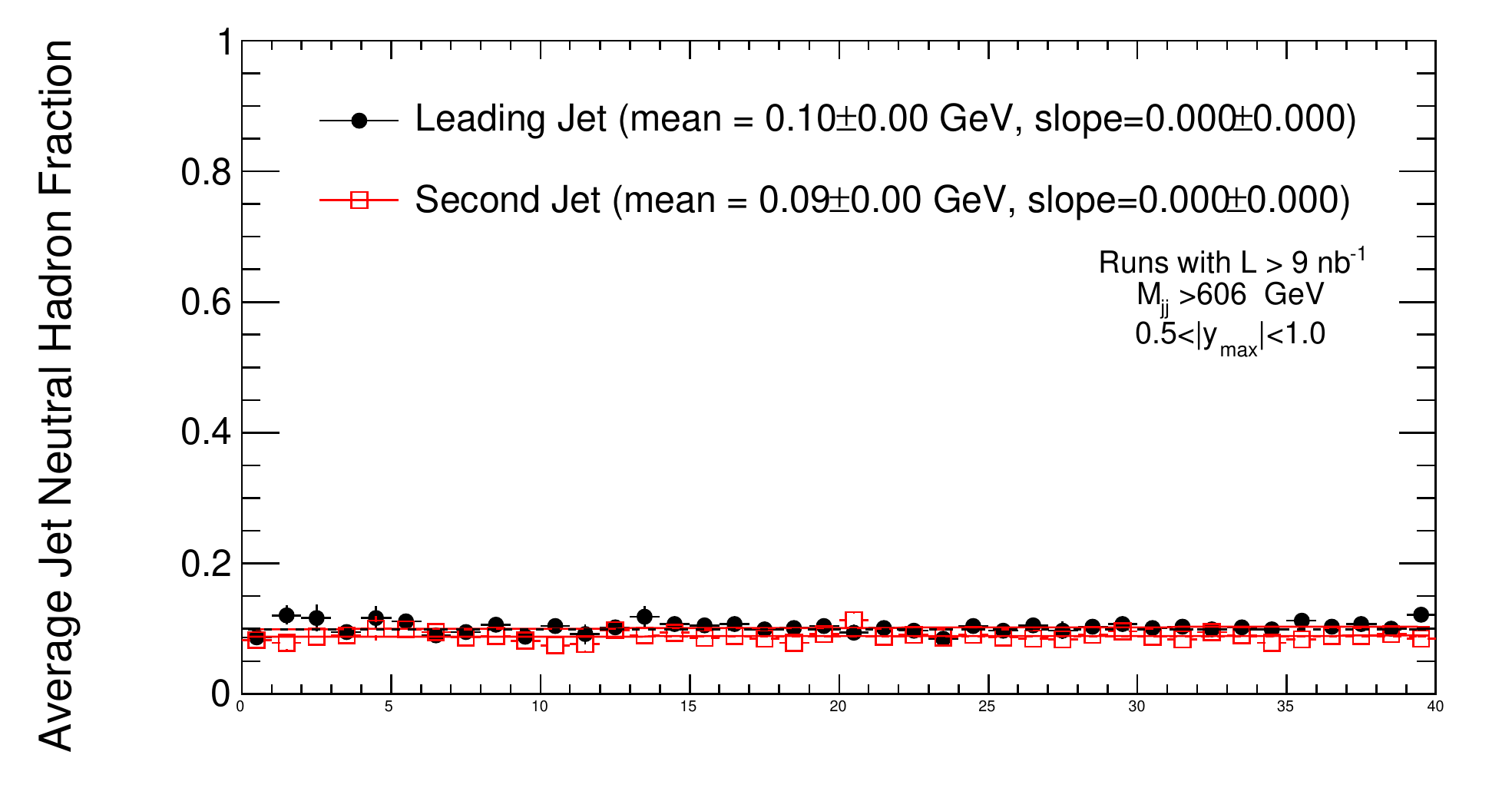} 
\includegraphics[width=0.52\textwidth]{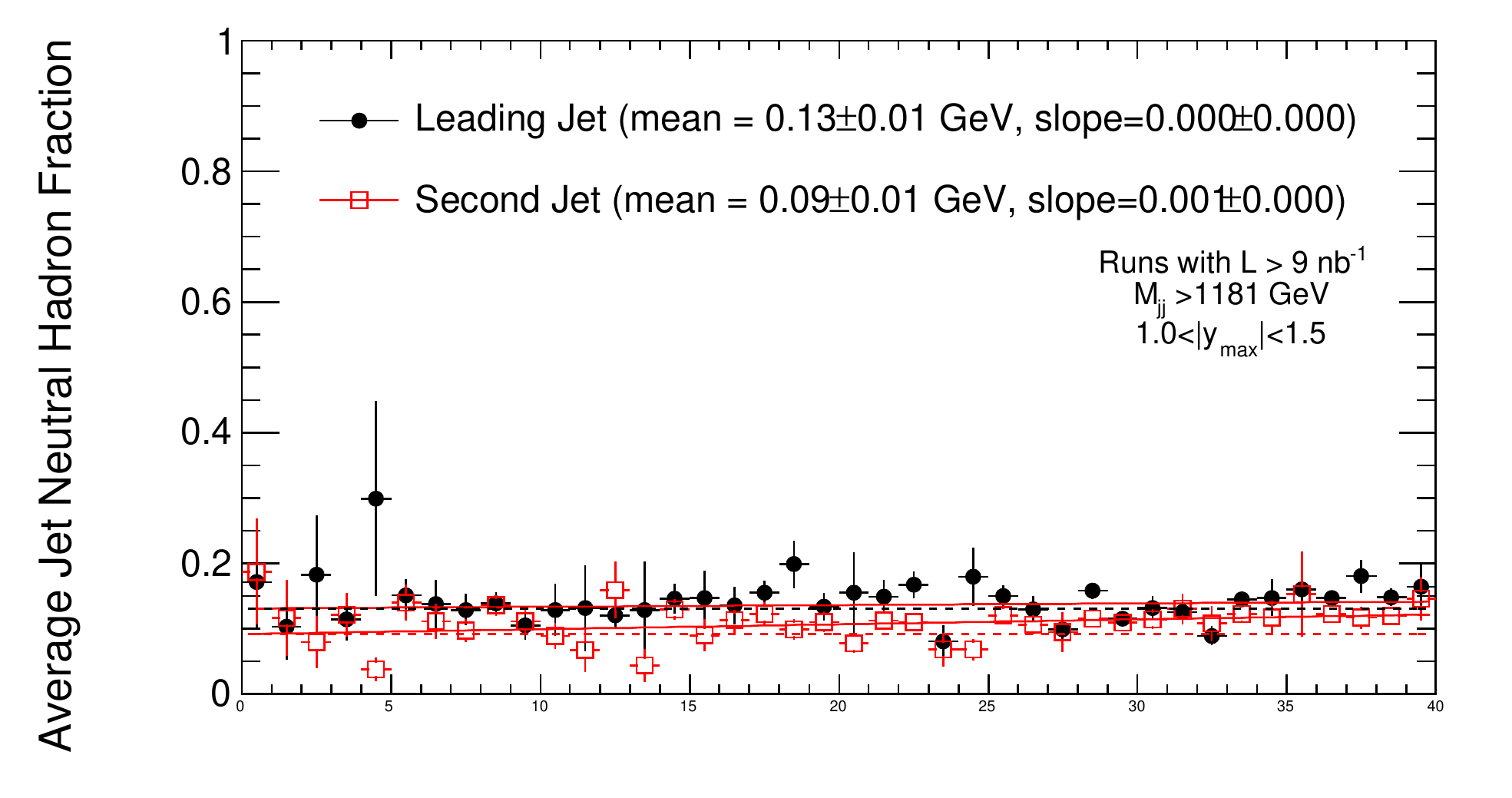} 
\includegraphics[width=0.52\textwidth]{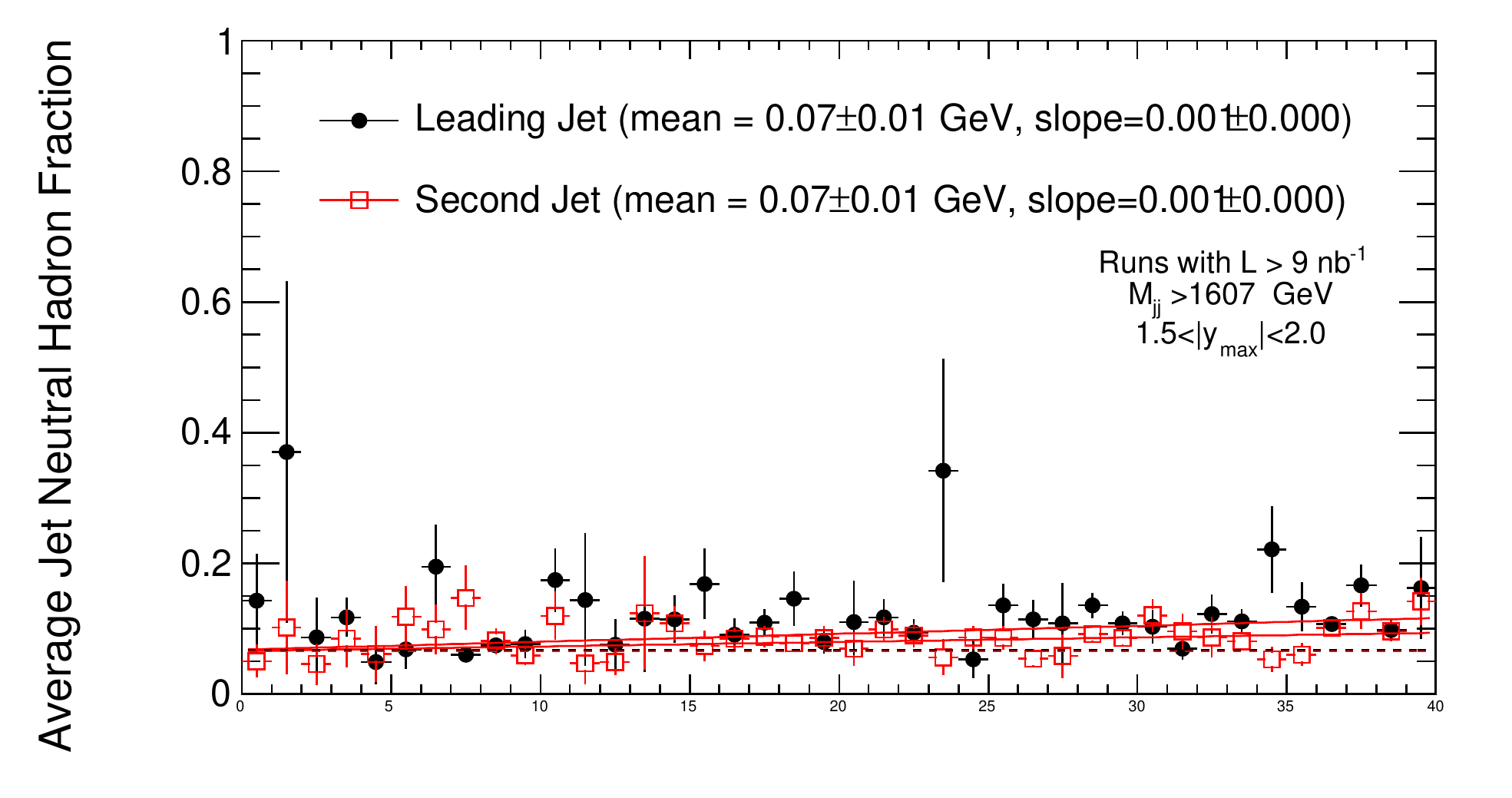} 
\includegraphics[width=0.52\textwidth]{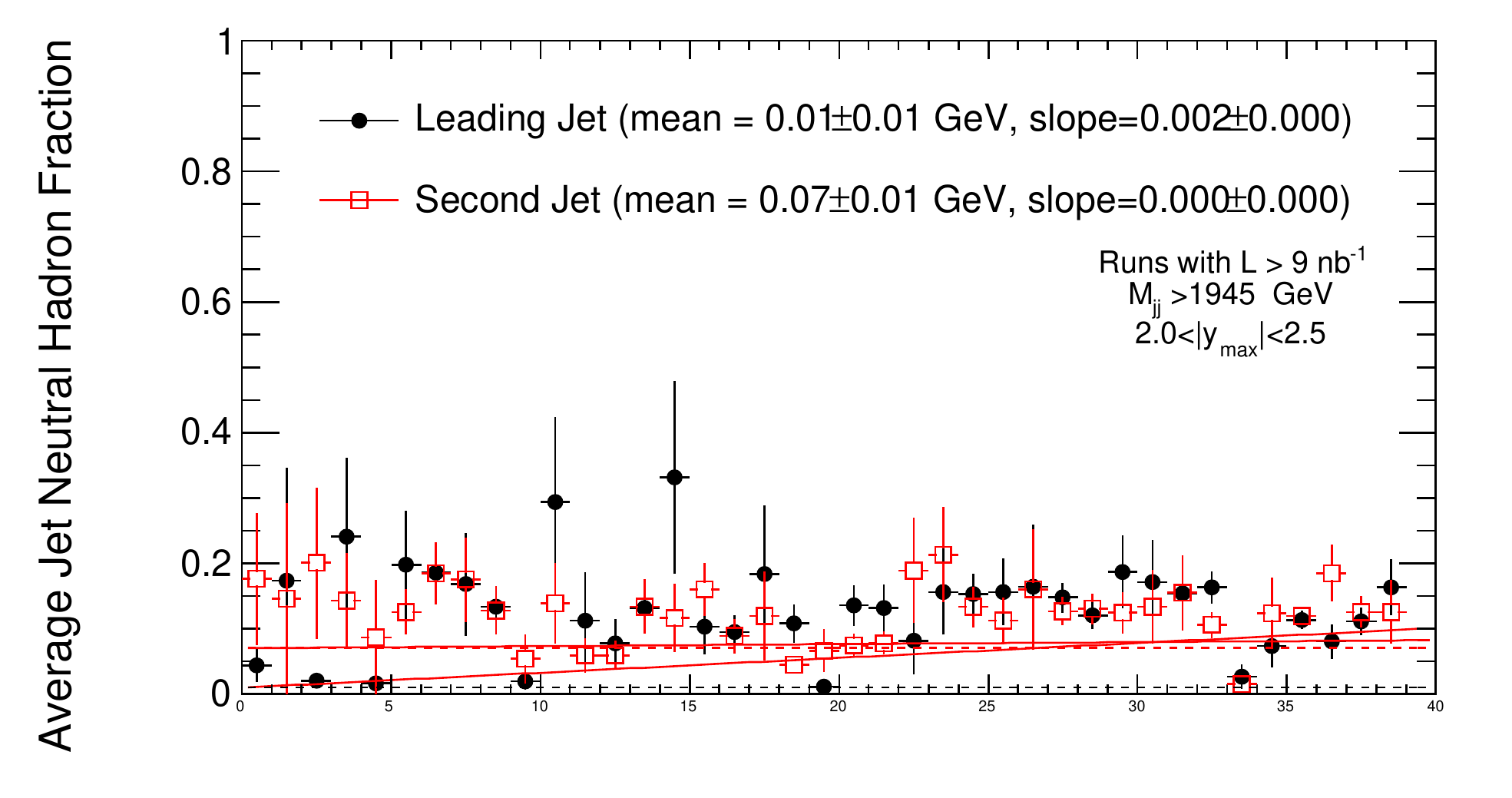} 
  
\capspace
\caption{ The neutral  hadron fraction  of the leading and second jet  for the five different $y_{max}$ bins and for the
HLT$_{-}$Jet140U trigger as a function of time (run number), fitted with a first degree polynomial.}
\label{fig_appd15}
\end{figure}

\clearpage

\begin{figure}[h]
\centering

\includegraphics[width=0.52\textwidth]{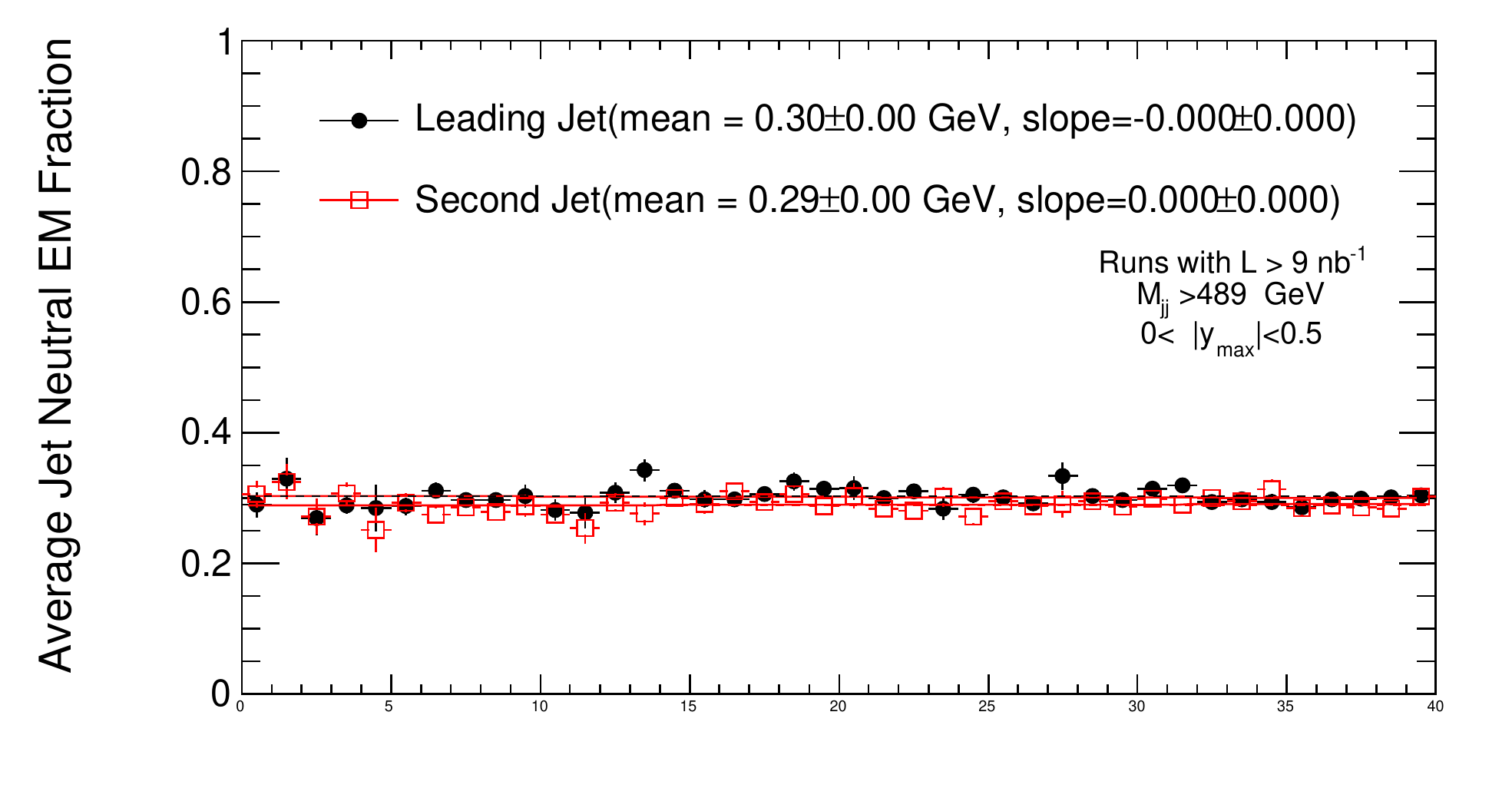} 
\includegraphics[width=0.52\textwidth]{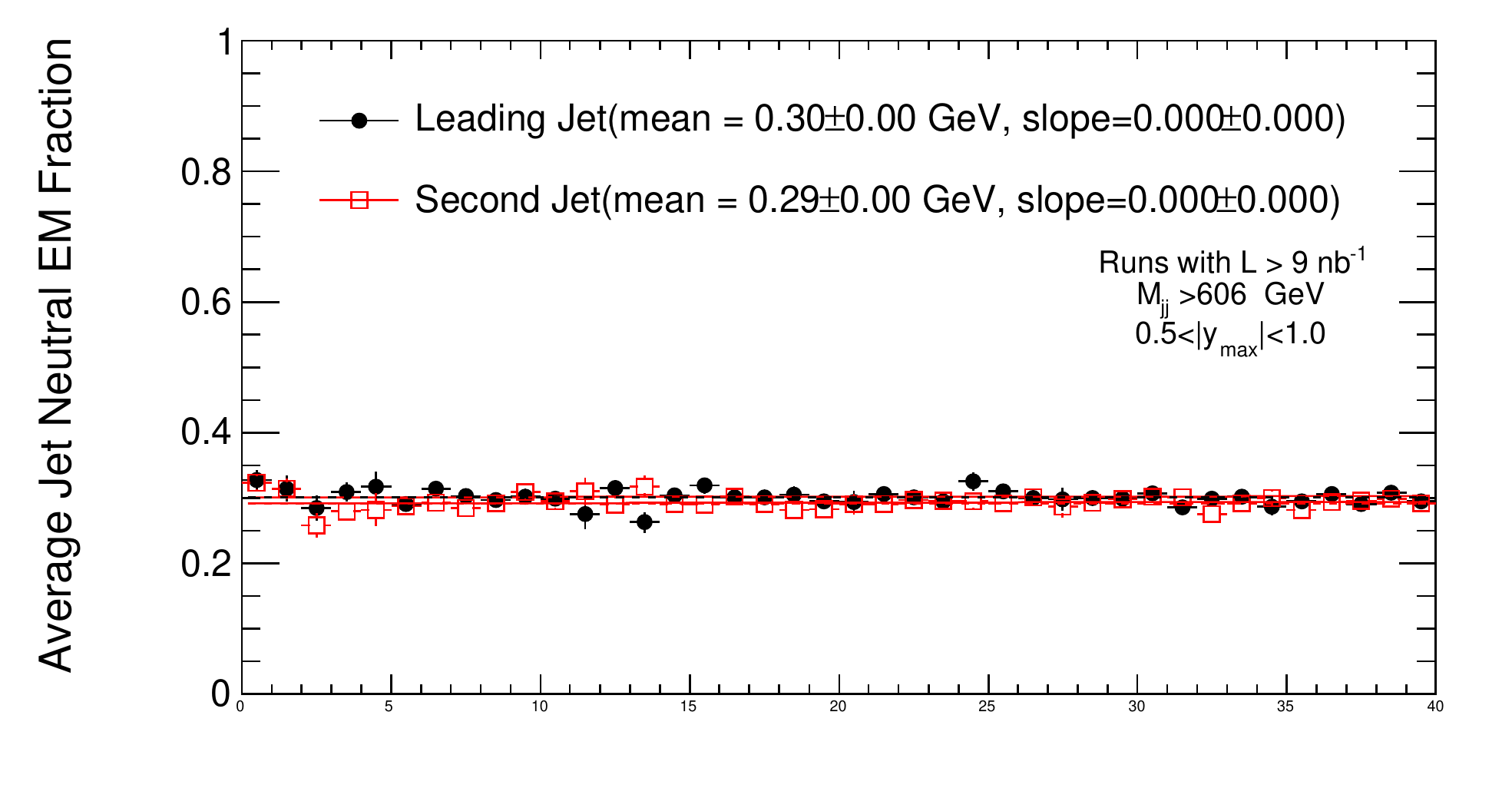} 
\includegraphics[width=0.52\textwidth]{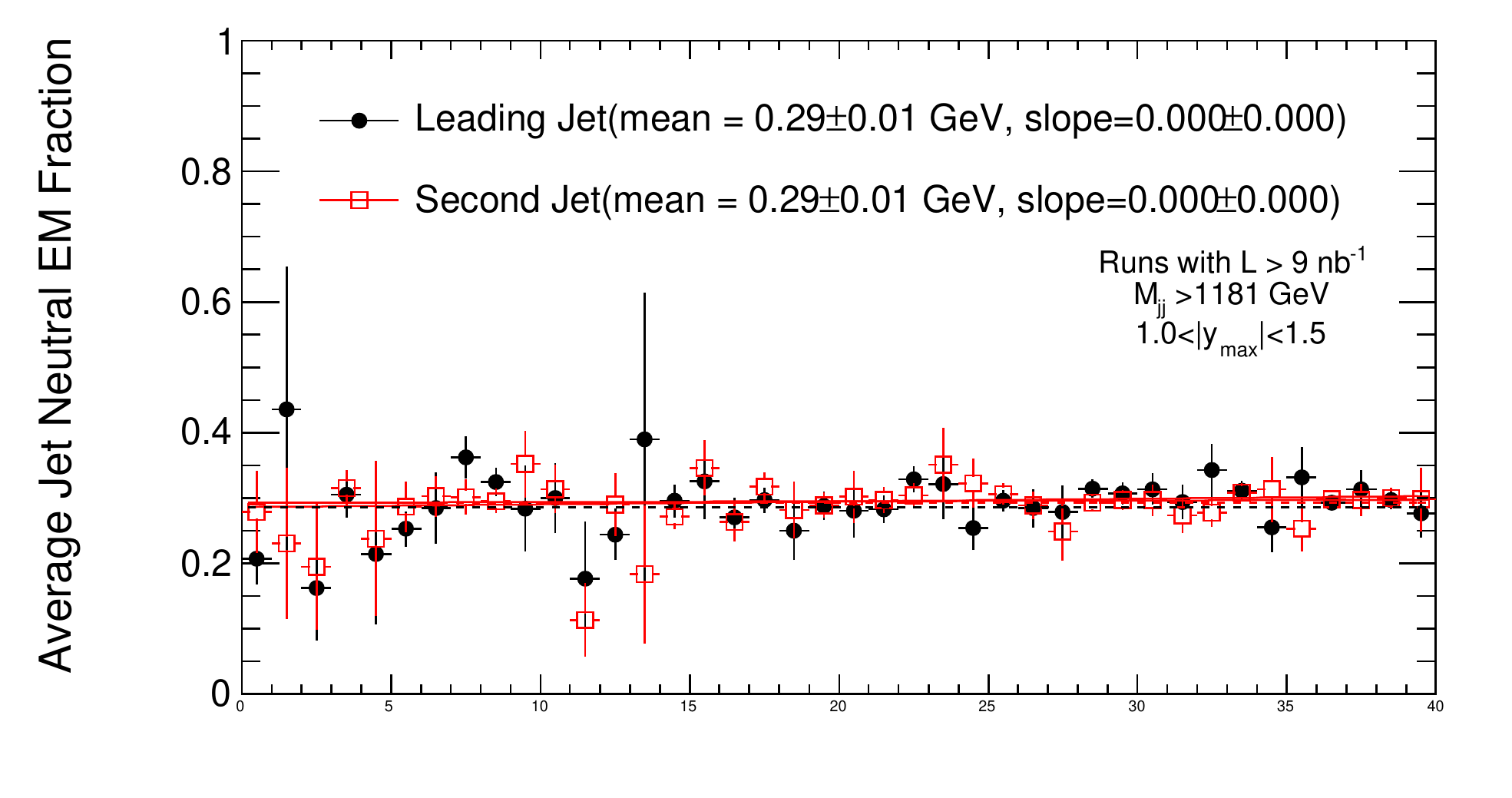} 
\includegraphics[width=0.52\textwidth]{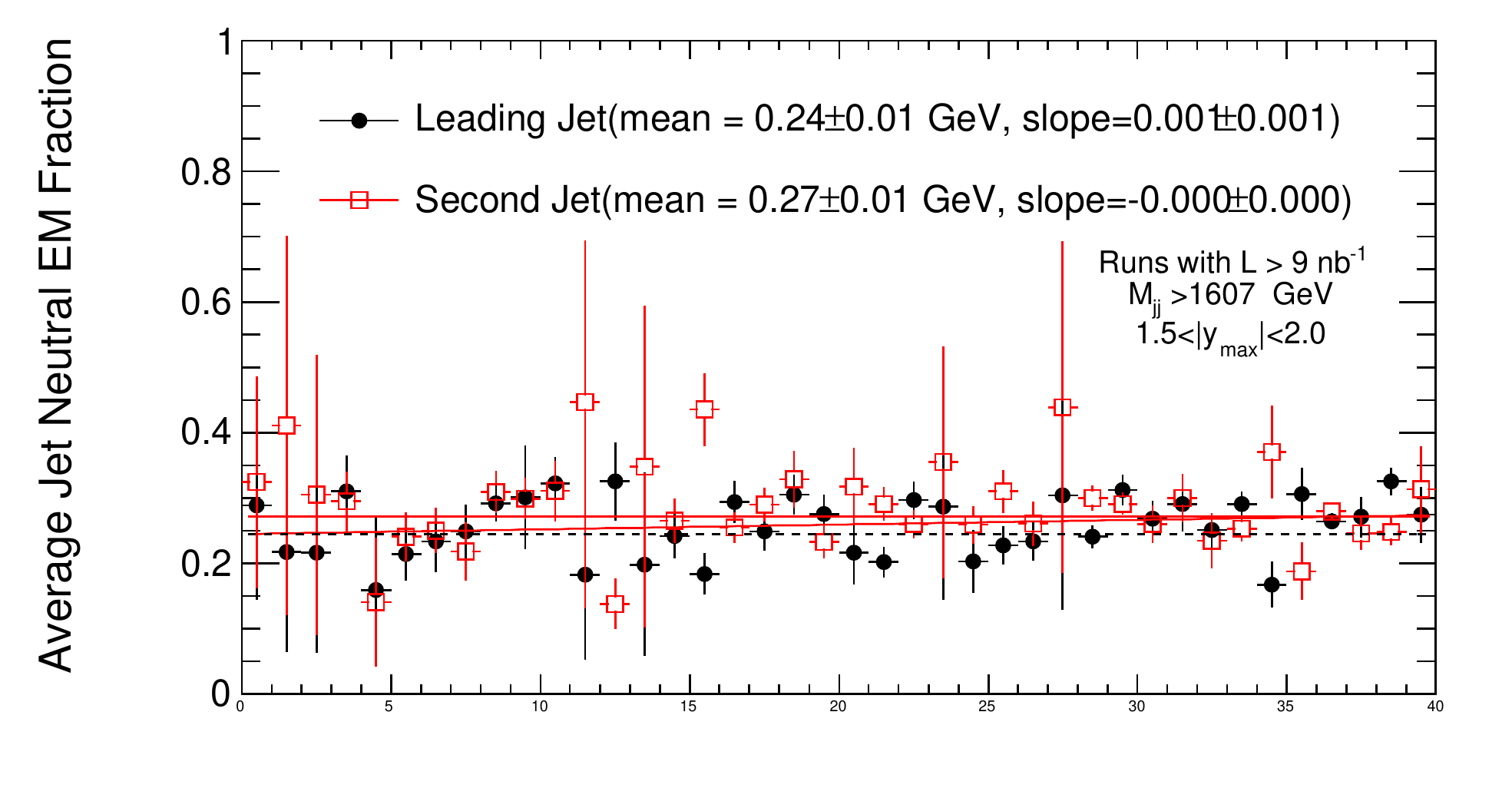} 
\includegraphics[width=0.52\textwidth]{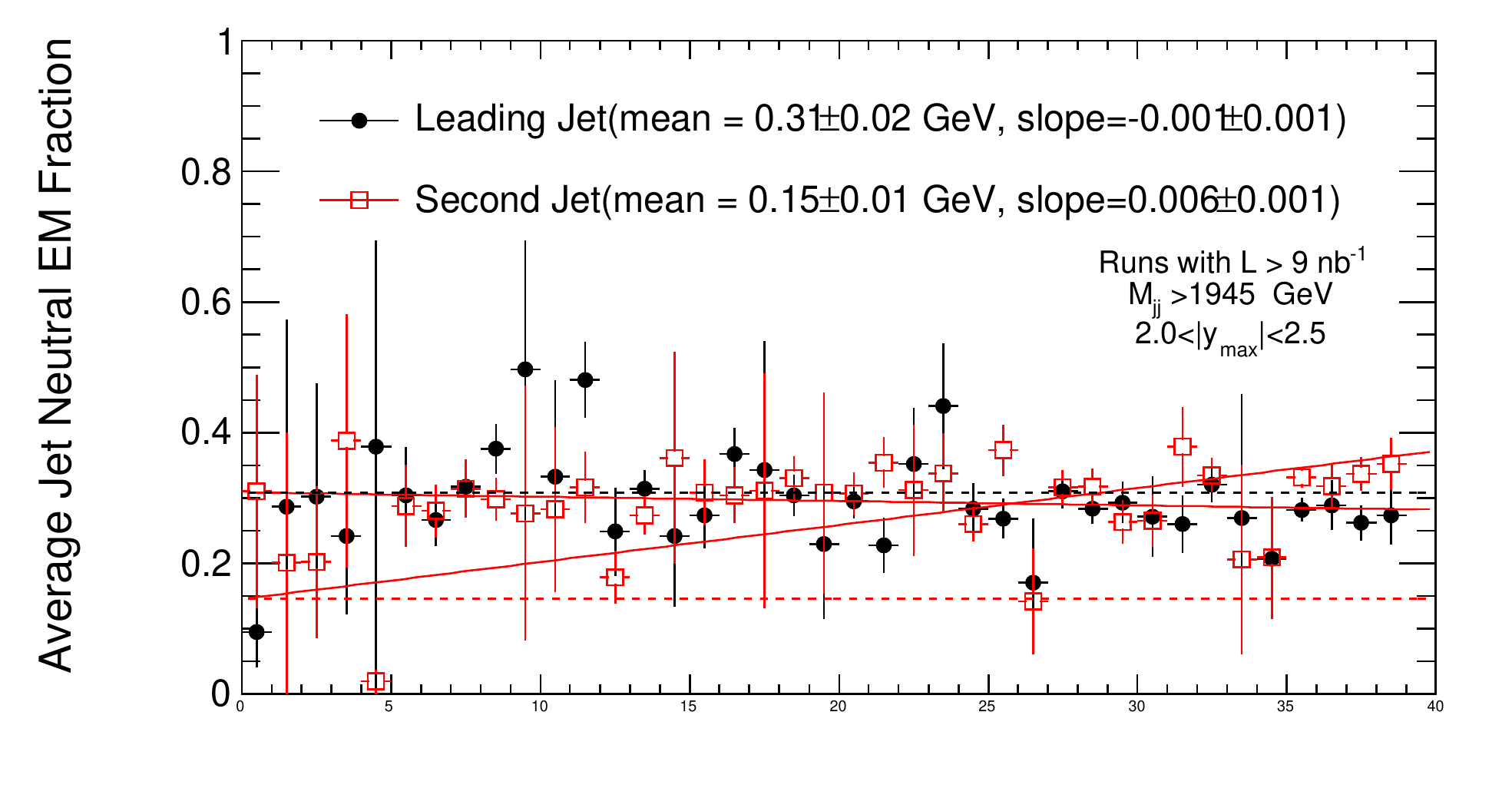} 
  
\capspace
\caption{ The neutral  electromagnetic fraction  of the leading and second jet  for the five different $y_{max}$ bins and for the HLT$_{-}$Jet140U trigger as a function of time (run number), fitted with a first degree polynomial.}
\label{fig_appd16}
\end{figure}

\interlinepenalty=10000

%
% You can also try harvardbibliography, bibnotcited and
% harvardbibnotcited environments for other types of bibliographical
% lists.
%

\end{document}